\documentclass[iop]{emulateapj}
\usepackage{natbib}
\usepackage{subfigure} 
\usepackage[centertags] {amsmath}
\usepackage{rotating}
\usepackage{longtable}

\newcommand{\dex}[1]    {\ensuremath{\times\textrm{10}^{#1}}}
\newcommand{\kms}       {\ensuremath{\textrm{km~s}^{-1}}}
\newcommand{\kpc}       {\textrm{kpc}}
\newcommand{\magas}     {\ensuremath{\textrm{ mag arcsec}^{-2}}}

\newcommand{\muB}       {\ensuremath{\mu_\textrm{{B}}}}
\newcommand{\muI}       {\ensuremath{\mu_\textrm{{I}}}}
\newcommand{\pc}        {\textrm{ pc }}
\newcommand{\BJ}        {\ensuremath{\textrm{B}_\textrm{J}}}
\newcommand{\IN}        {\ensuremath{\textrm{I}_\textrm{N}}}

\newcommand{\Halpha}    {\textrm{H}\ensuremath{_{\alpha}}}
\newcommand{\HI}        {\textrm{H{\sc i}}}
\newcommand{\Lsun}      {\ensuremath{\textrm{L}_{\sun}}}

\newcommand{\Msun}      {\ensuremath{\textrm{M}_{\sun}}}
\newcommand{\Mpc}       {\textrm{Mpc}}

\newcommand{\MB}        {\ensuremath{\textrm{M}_\textrm{B}}}
\newcommand{\MV}        {\ensuremath{\textrm{M}_\textrm{V}}}

\newcommand{\RF}        {\ensuremath{\textrm{R}_\textrm{F}}}

\newcommand{\galex}     {\emph{GALEX}}

\renewcommand{\min}     {\ensuremath{^\prime}}

\renewcommand{\deg}     {\ensuremath{^\circ}}

\begin{document}

\title{Dwarf Galaxies in the Halo of NGC~891}

\author{Earl Schulz}
\affil{60 Mountain Road, North Granby, CT 06060} \email{earlschulz@gmail.com}

\begin{abstract}
We report the results of a survey of the region within 40 arcmin of NGC 891, a nearby nearly perfectly edge-on spiral galaxy.  Candidate ``non-stars" with diameters greater than 15 arcsec were selected from the GSC 2.3.2 catalog and cross-comparison of observations in several bands using archived \galex,  DSS2, WISE, and 2MASS images identified contaminating stars, artifacts and background galaxies, all of which were excluded.  The resulting 71 galaxies, many of which were previously uncataloged, comprise a size limited survey of the region.
A majority of the galaxies are in the background of NGC 891 and are for the most part members of the Abell 347 cluster at a distance  of about 75 \Mpc.  The new finds approximately double the known membership of Abell 347, previously thought to be relatively sparse.
We identify a total of 7 dwarf galaxies, most of which are new discoveries.  The newly discovered dwarf galaxies are dim and gas-poor and may be associated with the previously observed arcs of RGB halo stars in the halo and the prominent HI filament and the lopsided features in the disk of NGC~891. Several of the dwarfs show signs of disruption, consistent with being remnants of an ancient collision.

\end{abstract}
\section{Introduction}
Several continuing surveys are underway to find and catalog galaxies in the Local Volume, LV, defined as the region  D $<10\ \Mpc$. \cite{kai12} describes the latest version of the \cite{kar04} catalog of galaxies of the LV which now contains more than 800 galaxies.  The catalog is complete to nearly 100\% for galaxies within 2 \Mpc\ and is about 70\%-80\%complete  within 8 \Mpc. The inventory of dwarfs within 10 \Mpc\ is much more uncertain and there have been several recent efforts to improve the galaxy census in this region; see especially \cite{gildepaz07,kar07,huc09,kar10,whi07}.  Most recently, \cite{mccon12} presents observational data for all of the approximately 100 dwarf galaxies in the Local Group for which D $<3\ \Mpc$.

More than 80\% of the LV galaxies are dwarfs and nearly all of the  undetected galaxies must be dwarf irregular (dIrr) galaxies and dwarf spheroidal (dSph) galaxies which are the most challenging to detect. dSph galaxies are small $( 300\pc \le D \le 1000 \pc)$ and faint $(M_V > -14)$ and are morphologically distinct from larger galaxies.  dSph's are very gas-poor and do not possess structures such as a spiral arms or bulges and do not have a discernable nucleus.  dSph galaxies are very dim with surface brightness in the range $ 24 \le \muB \le 31$. dIrr galaxies are somewhat larger and brighter than dSph galaxies and are distinguished from dSph galaxies by having a much greater gas content and by their characteristic appearance caused by 'lumpy' star forming regions.

dSph galaxies are usually found within about 100 \kpc\ of a larger galaxy whereas dIrr are usually undisturbed and more distant from the nearest large galaxy.  This is evidence that the two categories may overlap and that some dSph galaxies may be dIrr galaxies which have been stripped of gas by interaction with the halo gas of a larger galaxy.
\subsection{Streams and Dwarfs in the Halo of NGC 891}
In this paper we are concerned with dSph galaxies near NGC~891. NGC~891 is a nearby nearly perfectly edge-on ($\theta >89.8\deg$) spiral galaxy which is classified as Sb/SBb and is a member of the NGC 1023 group.  NGC~891  resembles the MW galaxy and has been been observed in many wavelengths. We take the the distance to NGC~891 to be 9.8 \Mpc\ which is consistent with recent SBF and TRGB measurements as given in \cite{tik05} so that $1\arcsec\approx47.5\pc$.  The heliocentric velocity of NGC 891 is V$_0=528 ~\kms$, suggesting a peculiar motion of about $-180~\kms$. 

NGC~891 is near to the galactic plane at latitude $b_{\rm{gal}}=-17\deg $ and is very near to the supergalactic plane at $b_{\rm{SG}}=-5\deg$. The foreground extinction toward NGC~891 is .28 mag in B which is significant but, even so, detailed observations in the region are possible.

\cite{mou10::1} presents evidence of an ancient accretion event which affected NGC~891. They find features which extend well into the halo including a number of arcing loops extending to about 30 \kpc\ to the east of the disk and 40 \kpc\ to the west. The loops appear to connect and show high metallicity and ages greater than a few gigayears.  Tidal streams of this sort are quite common in the LV.  \cite{mar10}, for instance, observed eight spiral galaxies using very deep $(\mu_V \sim28.5 \magas)$ wide field imaging and found that six of the galaxies had previously undetected stellar structures extending out to $\sim30 \kpc$.

\cite{oos07} also found evidence of an accretion event: a large filament which extends out to about 22 \kpc\ above the plane of the galaxy contains about 1.2 \dex{9} \Msun\ of \HI. \cite{oos07} performs a simple calculation which shows that it is unlikely that this large amount of gas could have been expelled by a galactic fountain type of mechanism and must therefore be due to an accretion event. Finally, \cite{shi10} found asymmetries in the distribution of planetary nebula of NGC~891 which they take as evidence of a collision or fly-by interaction.

If the disturbances to NGC~891 were caused by a collision with another galaxy there should be nearby evidence of the disrupting galaxy. \cite{oos07} proposed that the gas rich dwarf irregular galaxy UGC 1807, which is  29\min\ distant from NGC~891, might have caused the disturbance. However, UGC 1807 appears to be undisturbed and shows no signs of a recent interaction and so there is reason to suppose that even if UGC 1807 was the original disrupting galaxy there might be one or more dwarf galaxies in the near vicinity on NGC~891 consisting of tidal remnants of the original collision which account for the continuing disturbances.

\begin{figure*}[t]
  \fbox{\plotone{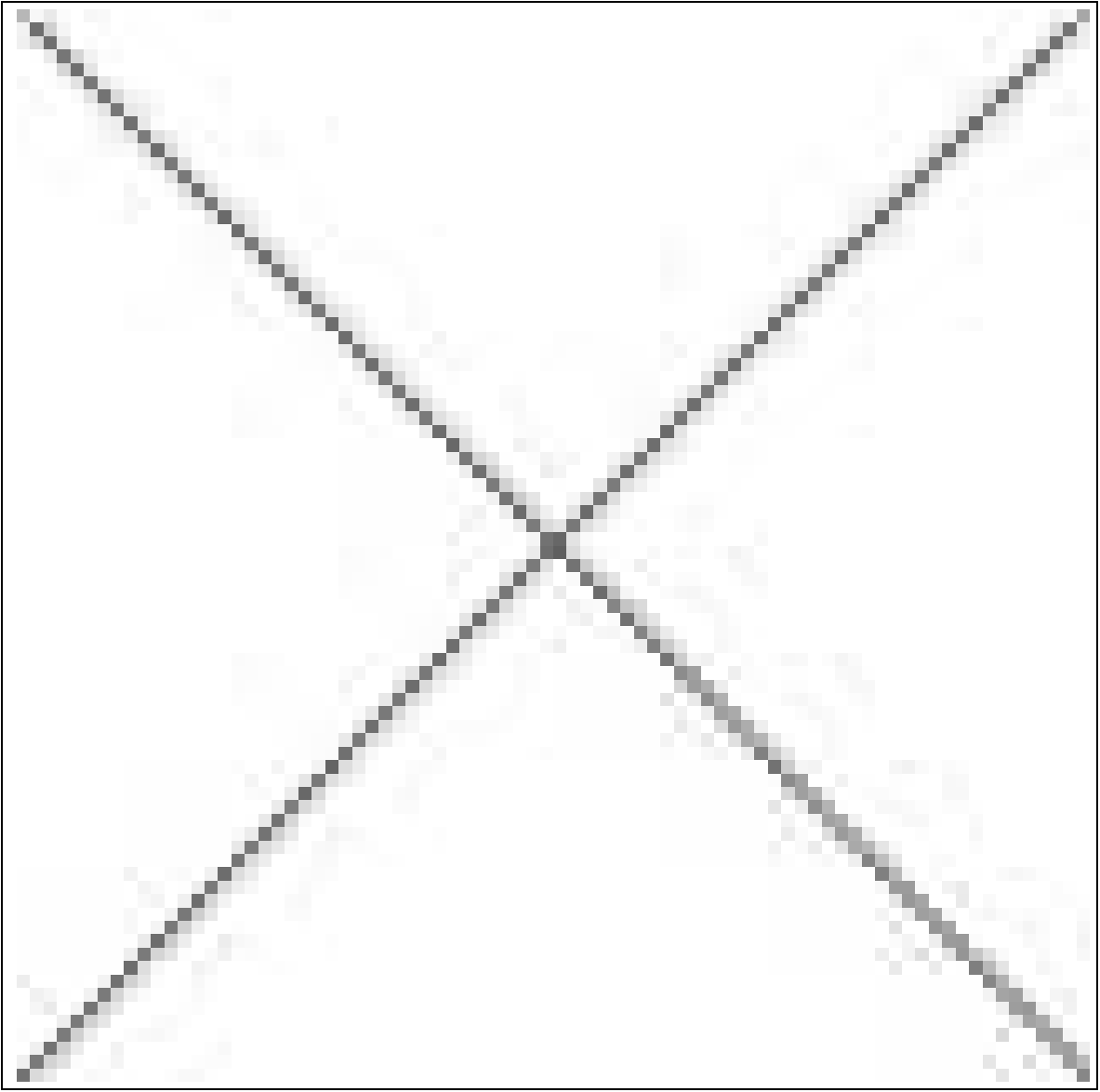}}
  \caption{The over-densities of  RGB stars in the halo of NGC 891 define streams and loops which possibly due to one or more accretion events.  A feature at a height of \~ \kpc\ above the disk is centered on the dSph galaxy HFLLZOA F172. This figure will reproduce figure 1 from \cite{mou10::1} if permission is granted.}
  \label{fig:FromMouhcine}
\end{figure*}

\begin{figure*}[t]
  \fbox{\plotone{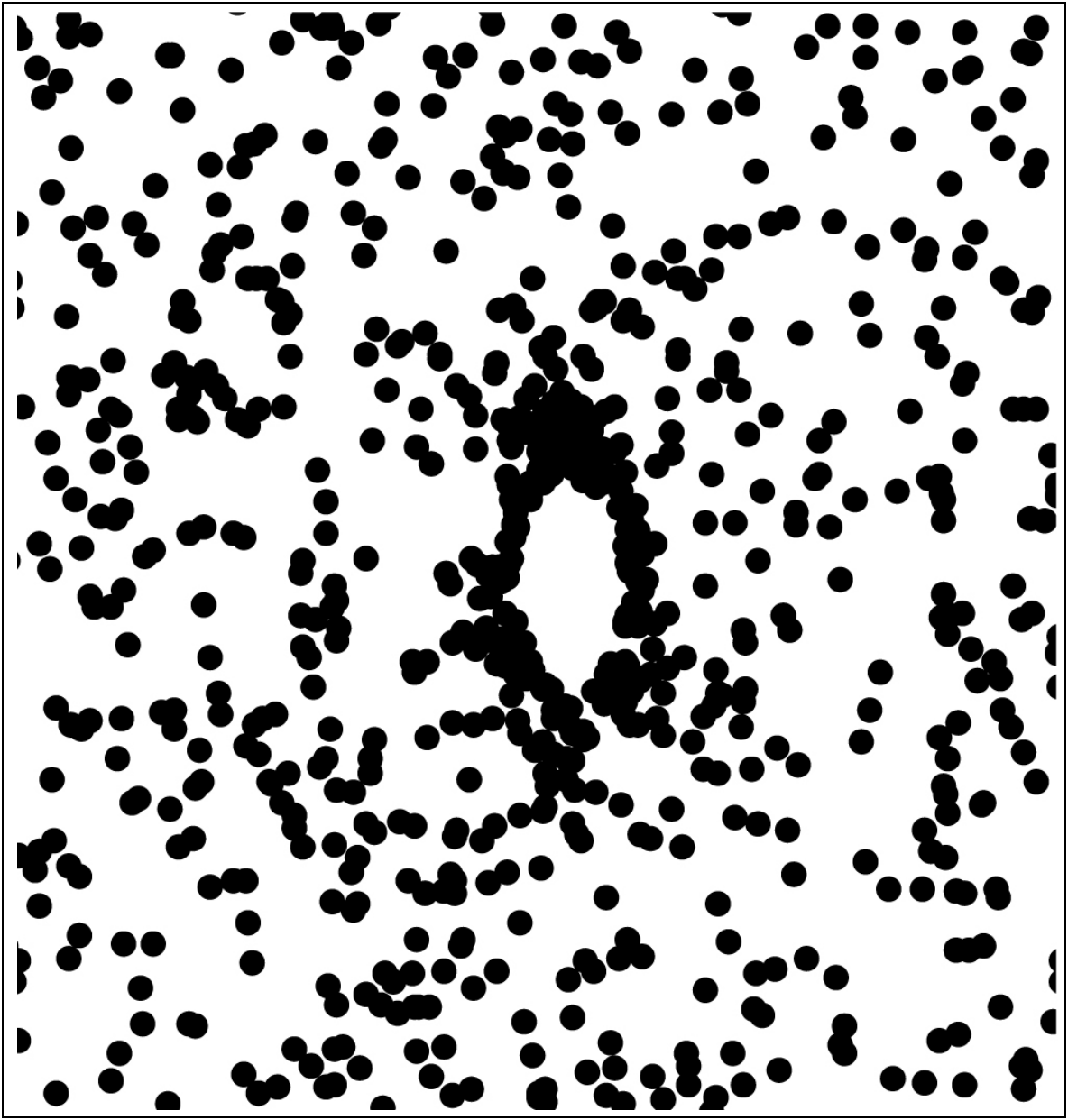}}
  \caption{An expanded view of a small section of \ref{fig:FromMouhcine}.  RGB stars in the halo of NGC 891 outline the tidal radius of HFLLZOA F172.}
  \label{fig:CutFromMouhcine}
\end{figure*}

The panoramic view of the halo region of NGC 891 presented in \cite{mou10::1} was the impetus for the search presented here and their figure 1. \cite{mou10::1} used the Subaru Prime Focus Camera on the 8.2 meter Subaru Telescope to image the brightest 2 mags of Red Giant Branch (RGB) stars in the outer regions of NGC~891. The images were processed to eliminate foreground stars and other artifacts and to isolate RGB halo stars. The resulting panoramic view of NGC~891 covers a 90 \kpc\ x 90 \kpc\ area and consists of points representing RGB stars in the halo.  Unremarked in \cite{mou10::1}, the panoramic view shows an overdensity of RGB stars  in the NW quadrant of the halo of NGC~891 which frame the tidal radius of dwarf galaxy HFLLZOA F172 \citep{hau95,tre09}  at a distance of 15\min~ from NGC 891.  Figure \ref{fig:CutFromMouhcine} is a blowup of a small section of \ref{fig:FromMouhcine}, which is reproduced from \cite{mou10::1}.  Inspection of archived images found that HFLLZOA F172 is certainly a dSph and its association with the halo stars proves that it is in the halo of NGC 891. Archived images of the other galaxies in the \cite{hau95} catalog within 1\deg of NGC~891 were examined and a single additional dSph was found: HFLLZOA F182 at a distance of 20.1\min.  The remaining galaxies in the \cite{hau95} catalog near to NGC~891 were either background galaxies or were ambiguous.
\subsection{The Missing Satellite Problem}
The Cold Dark Matter(CDM) model predicts that galaxy formation arises by a hierarchical combination of small CDM halos which begins at high redshifts($z >\sim 100$) and is essentially complete by about $z \sim 2-3$\citep{whi78,whi91}.  This clustering is purely gravitational and is driven by the dominant CDM component.  Baryonic material which falls into the the CDM gravitational wells builds the visible structures of the galaxies as gas cools and forms stars in a process which continues to the present day.  Simulations predict that the outermost regions of acquired CDM satellites merge with the halo of the accreting galaxy to form a single large CDM halo such as that which is thought to host the MW galaxy.  The cores of most of the CDM satellites are predicted to survive and we observe the baryonic portions of these satellites as dwarf galaxies.

The so-called missing satellite problem \citep{kly99,moo99} is that far fewer dwarf companions of large galaxies are observed than are predicted.  For example, a galaxy the size of the MW way is predicted to have several hundred dwarf companions whereas only 26 have been discovered to date and this discrepancy seems to apply generally. It's possible that the predicted DM subhalos do not, in fact, exist but this would conflict with the concordant model and so much effort is being spent to reconcile the disagreement.

\cite{bul10} summarize the recent work related to the recent discovery of more than 2 dozen new dwarf companions to the Milky Way and M31. Most of the new dwarfs are fainter than any previously known galaxies with the faintest dwarfs having luminosity of only $\textrm{10}^2 - \textrm{10}^4  \Lsun $ which supports the idea that there is a large population of dwarf galaxies that have not yet been detected.  Assuming that the new class of dwarfs can satisfy the numeric deficiency the most pressing problem is to determine the relationship between dwarf luminosity, dwarf baryonic mass, and the mass of the DM subhalo.  Currently there is no useful trend and in some cases the kinetics of the smallest, least luminous, dwarfs imply a DM halo which is as massive as systems 10,000 times more luminous.

A number of studies \citep[e.g.,][]{lar74,dek86} have shown the processes of supernova heating and the reionization of the Universe could have had a strong effect to reduce the baryonic fraction of dwarfs at z $\gtrsim$ 6.  Later, stochastic effects might have resulted in the situation which seem we seem to see today where some very large halos have not acquired significant baryonic component\citep{bar99}.  Alternately, \cite{nic11} proposes that ram-pressure stripping and super-nova heating combined can account for the observed populations of gas-poor dwarfs within about 270 \kpc\ of their primary. \cite{guo10}  supports the idea that the abundance of  DM halos as a function of their mass is well known and supported by basic theory.  They propose that stellar mass as a function of the halo mass is monotonic but not at all linear.  Star formation efficiency is much lower for both the highest and the lowest halo mass systems and so we observe a distribution of galaxies which is biased toward middle values.

\cite{boy11} and \cite{boy12} discuss a further complication in that the kinematics of the brightest dwarf satellites of the Milky Way imply a halo density distribution which is much less dense than demanded by theory.  The apparent lack of high mass subhalos might be a more fundamental problem than the numeric deficiency.

The current work contributes to understanding the ``missing satellite problem" by providing a dwarf galaxy count for a galaxy similar to the MW.  It is part of a growing number of surveys which indicate more and more strongly that there is a basic conflict between theory and observation.
\section{A Size-Limited Galaxy Survey}
A methodical search for visible galaxies in the vicinity of NGC 891 was undertaken by examining candidates taken from The Guide Star Catalog II \citep[GSC 2.3.2,][]{las08}. The GSC 2.3.2 is a deep all-sky catalog derived from the Digitized Sky Surveys, which in turn were created from Palomar and UK Schmidt survey plates, and is based on at least two epochs and three passbands. The catalog classifies objects as one of two types: ``stars", which are classified correctly as such with high confidence, and ``nonstars" which include not only galaxies but also overlapping images and artifacts such as diffraction spikes near bright stars, halos, etc. In general, ``nonstars" are primarily galaxies far from the galactic plane and primarily blends near the plane.
The stellar magnitudes given in GSC 2.3.2 are accurate. However, because the photometric pipeline is tuned for point-like objects, the magnitude of bright galaxies $ \RF < 18$ are systematically overestimated. The \RF~magnitudes are most affected and were found to be overestimated by as much as several magnitudes.

The candidates were 173 ``nonstars" from the GSC 2.3.2 catalog lying within 40\min\ of NGC~891 (113 \kpc\ at the distance of NGC~891) for which the major axis was greater than 15\arcsec.  The search region was limited in order to keep the number of candidates to a reasonable total and because \galex\ observations covered this region.

Each of the candidate ``non-stars" was examined by using the NASA \emph{Skyview} utility to generate $1\min \times 1\min$ false color images from archived data consisting of \galex\ FUV and NUV images\citep{mar05}; DSS2 R, B and NIR images; 2MASS H, J and K images \citep{skr06}; and WISE W1, W2, W3 and W4 images \citep{wri10}.  In addition a standard rgb image constructed from \galex\ NUV, DSS2 R and 2MASS J wavelengths proved useful because of the large UV excess exhibited by the dSph galaxies in the sample.  These observations in several wavelengths eliminated overlapping images and artifacts with high confidence leaving 71 confirmed galaxies which comprise a complete size-limited survey of the region.
\subsection{Distinguishing the Dwarf Galaxies}
The goal of the present study is to differentiate foreground dSph galaxies from normally sized galaxies at a much greater distance in the background.  This task is similar to that described in \cite{con02} which defined photometric and structural properties of galaxies in clusters in order to discriminate members of the Abell 426 group (at at distance of 77 \Mpc) from background galaxies at distances up to z=0.5.  However, it is easier to distinguish dSph's because they are very different in morphology from the larger galaxies whereas the differences which separate cluster galaxies from isolated galaxies are relatively subtle.  In addition, the availability of images at many wavelengths is invaluable.

dSph galaxies have diameters in the range of about 200 pc $< D <$ 1,800 pc.  By design, the minimum diameter of the survey galaxies is 15\arcsec\ and the diameter of largest galaxies for which there is no velocity data is about 40\arcsec.  At the distance NGC 891 this range of angular size corresponds to a range of actual size of about 700 pc $< D <$  1,900 pc compared  to a range of actual size of about 5,000 pc $< D <$ 15,000 pc at the distance of the Abell 347 cluster.   Thus, based on the size, the foreground galaxies consist solely of dSph galaxies whereas the background galaxies are normally-sized; two distinctly different populations.

Since many of the survey galaxies lack measured distances, it is not possible to use absolute magnitude to identify likely dSph galaxies.  Instead we ordered the candidates using a \IN\ surface brightness metric  calculated from the apparent NIR magnitude and angular size as given in the GSC 2.3.2 catalog.  Testing showed that this parameter is the best discriminator to sort out foreground from background galaxies, presumably because NIR magnitude is a better measure of stellar surface mass density than the R or B magnitudes, which show much more scatter.  Note that the magnitudes given in GSC 2.3.2 for extended sources are known to be systematically too bright by 2 mag or more and the cataloged radial size (which is the projected radius at which the R magnitude surface brightness reaches the background level) differs significantly from the tidal radius which would be more appropriate for this use. Despite these known errors, the metric is useful because it is a monotonic measure sufficient for the purpose of sorting and comparing the candidates.

The survey galaxies were ordered by \IN\ surface brightness metric described above and it was found that \emph{none} of the lowest surface brightness candidates $ \muI > \sim21.8 \magas$ showed signs of structure such as spiral arms or a discernable core. In contrast, \emph{all} of the higher surface brightness candidate galaxies for which $ \muI < \sim21\magas$ showed signs of structure such spiral arms and most commonly a distinct core and so are not dSph galaxies. This is consistent with previous work; see for example \cite{gal94}.

\cite{gre03} found that nondetections of HI occur in low-mass dwarfs that are within about 300 \kpc\ of their host. Similarly, \cite{grc09} reports that galaxies within about 270 \kpc\ of the Milky Way or Andromeda are undetected in \HI\ (i.e. \HI\ mass is less than about $10^4 \Msun$ for Milky Way dwarfs), while those further than ~270 \kpc\ are predominantly detected with \HI\ masses in the range $10^5\Msun$ to $10^8\Msun$.  The NGC~891 companion dSph galaxies  are consistent with this finding in that none of the dSph galaxies near NGC 891 were found by the 2MASX survey and the candidate galaxies which were dimmest in NIR were not detected in 2MASS J, H, or K bands or in WISE W3 or W4 bands which are sensitive to secondary emissions from the dust and gas in the ISM. The 2MASX survey \citep{jar00}  extended source sensitivity is about 14.7, 13.9, and 13.1 mag at $J$, $H$,  $K_s$.  The detection thresholds were chosen to assure complete detection of galaxies brighter than $K_s\sim13.5$ and $J\sim15$ mag away from the Galactic plane and the limit is somewhat brighter close to the plane.  Because the lowest surface brightness galaxies are bluer within the NIR bands, the faintest galaxies are observed only in J. Some previous surveys of the LV galaxies took candidates from the 2MASX Catalog \citep[e.g.,][]{gildepaz07}  and these surveys missed the gas-free dSph galaxies reported here even though they are readily apparent UV and visible wavelengths.

Here we make the assumption that background galaxies are members of the Abell 347 cluster in the background of NGC 891.  Abell 347 is at a nominal recession velocity of V=5516 \kms\ and is a member of the Perseus-Pisces supercluster. As argued in \cite{tre09}, the void between the LV and the Abell 347 cluster makes it possible to distinguish between the foreground and background galaxies based on morphology with some degree of confidence. Accordingly, a goal of the present study is  to differentiate between dSph galaxies in the foreground at a distance of about 9.8 \Mpc\ and background galaxies at a distance of about 72 \Mpc\ or more.  In particular, the velocity  histogram presented as Figure 4 of \cite{tre09} supports the assumption that contamination along the line of sight  to the Abell 347 cluster is insignificant. The \cite{sak12::1} catalog identifies a total 37 members of Abell 347, 12 of which are in the region $R \leq 40\arcmin$ around NGC 891, and the recent 2MRS survey \citep{huc12} approximately doubles these counts. All but four of the background galaxies for which there are measured velocities lie between 5,500 \kms\ and 6,500 \kms\ and there no galaxies in the background of NGC 891 with measured velocities between 630\kms\ and 4450\kms.
\section{Discussion}
Table \ref{tbl:allgal} gives data for the 71 galaxies from the GSC 2.3.2 catalog.  Table \ref{tbl:allgal} provides the following information:

\begin{raggedright}
\newcounter{Lcount}
\begin{list}{Col: {\arabic{Lcount}}:}{}
\usecounter{Lcount}
\setlength{\itemsep}{-0.8ex}
  \item  ID Number
  \item  GSC 2.3.2 ID
  \item  Position (J2000 RA/DEC)
  \item  Distance from NGC 891 (arcmin)
  \item  Most commonly used alternate ID.
  \item  ID number if listed in the \cite{sak12::1} Catalog.
  \item  Indicates if the galaxy is listed in the 2MASX Catalog \citep{jar00}.
  \item  Radial velocity (km s$^{-1}$) if available.
  \item  Apparent \RF\ magnitude from GSC 2.3.2 (mag).
  \item  Apparent \BJ\ magnitude from GSC 2.3.2 (mag).
  \item  Apparent \IN\ magnitude from GSC 2.3.2 (mag).
  \item  Semi-major axis from GSC 2.3.2 (arcsec).
  \item  Calculated NIR surface brightness (mag/arcsec$^2$).
  \item  "F" indicates that this is a foreground dwarf.
\end{list}
\end{raggedright}

Table \ref{tbl:allgal} identifies 7 dwarf companions of NGC 891 all but two of which are new discoveries. Note that the boundary which defines the dwarf galaxies is uncertain by at least 0.5 mag  which might have caused  a few of the galaxies in Table \ref{tbl:allgal} to be mis-classified.

Figure \ref{fig:NGC891_with_dSphs} shows the distribution of the dwarf galaxies relative to NGC 891.  The dwarfs appear to  be distributed isotropically around NGC 891, the host galaxy.  The dSph galaxies which were determined to be companions of NGC 891 showed weak indication in WISE W1 and W2 bands which are sensitive to emission from the stellar disk.  In contrast, none of the dSph galaxies are visible in the Wise W3 and W4 bands which primarily measure emission from dust the ISM.  We take this to mean that these galaxies have been stripped of essentially all gas and dust.

Figure \ref{fig:allgal} provides a set of 13 false color images generated using the Skyview utility for of each of the Table \ref{tbl:allgal} entries.  The  images are sorted by DSS2 NIR surface brightness in the same way as the Table \ref{tbl:allgal} entries. Each of the images cover a $ 1\min \times 1\min$ field of view and is centered at the coordinates given in Table \ref{tbl:allgal}. The first image of each set is a rgb overlay and the remaining images use linear brightness scaling and a preset false color scheme (the "Stern Special")  which was found to best bring out the faintest images.  The images are as follows:

\begin{tabular}{l l r r}
1     & \multicolumn{3}{l}{rgb overlay: R, NUV, and J}      \\
2     &  Galex FUV   &0.155 \\
3     &  Galex NUV   &0.223 \\
4     &  DSS2  B     &0.442 \\
5     &  DSS2  R     &0.647 \\
6     &  DSS2  NIR   &0.786 \\
7     &  2MASS J     &1.235 \\
8     &  2MASS H     &1.662 \\
9     &  2MASS K     &2.159 \\
10    &  WISE W1     &3.4   \\
11    &  WISE W2     &4.6   \\
12    &  WISE W3     &12    \\
13    &  WISE W4     &22    \\
\end{tabular}\\
where the third column gives the central wavelength of each of the images.

Table \ref{tbl:allgal} and Figure \ref{fig:allgal} show the trend that the brighter galaxies possess internal structure such as spiral arms and distinct cores whereas the dimmest galaxies do not.  A few of the dimmer galaxies, Galaxies No. 5 (NCIA019215) and No. 13 (NBZ5004290),  seem to consist of two or three separate bodies spread out over an area of about 30-40\arcsec, suggesting that these dwarfs may consist of tidal remnants which are still coalescing.

Velocity data taken from NED is given for 24 of the candidate galaxies.  Most of the background galaxies are seen to be members of the Abell 347 group except that four galaxies( 31, 40, 50, and 63) are apparently located at greater recession velocity, ranging from 10,000 \kms\ to 20,000 \kms.

As indicated in Table \ref{tbl:allgal}, 11 of the galaxies were identified by \cite{sak12::1} as being members of Abell 347. \cite{sak12::1} used the MOSAIC-1 CCD Imager on the Kitt Peak National Observatory 0.9 m telescope to survey several clusters.  The FOV was $59\min \times 59\min$ and two pointings were made in the Abell 347 region in the background of NGC 891.  The survey determined group membership using narrowband filters corresponding to \Halpha\ redshifted by various amounts.  The pointings in the Abell 347 field used the 120\AA-shifted filter which corresponds to $\Delta V \sim 5516 ~\kms $.

32 galaxies were not identified as being in the foreground, do not have measured velocities, and are not included in the \cite{sak12::1} catalog.  Based on morphology and size most or all of these galaxies are members of the Abell 147 cluster which approximately doubles the known membership of the cluster.

The association of the dSph galaxies with the tidal streams and the signs of disruption near NGC891 suggest that some or all of the dwarfs might be of tidal origin. This is born out by the appearance of several of the dSph galaxies which appear to be disturbed and "raggedy" with a few seeming to consist of 2 or 3 distinct regions.  A possible explanation of situation is that the largest dwarf, UGC 1807, had a fly-by interaction with NGC 891 a few Gyr ago and has since recovered so that it now appears to be an undisturbed dI.  The other dwarf galaxies are tidal remnants of the original interaction which are bound to NGC 891 and some of this family periodically passes through the plane of NGC 891, disrupting the disk and creating new tidal streams.  This scenario is consistent with, for example, \cite{pur11} who suggests that the evolution of galaxy morphology is not entirely secular and that low-mass minor mergers  probably have an important role in shaping galactic structure. On the other hand, it is not certain that the dwarf galaxies originated from a single disruption event.  For instance, a common origin of many small galaxies in an infalling group or within a large-scale filament the feeds a large galaxy is often given as an explanation of the satellites of the Milky Way and M31.
\subsection{UV and NIR Magnitudes}
APT, the Aperture Photometry Tool, v2.4.2 \citep{lah12} was used to calculate the apparent magnitude of the galaxies in WISE W1, W2, W3 and W4 bands and in  \galex\ FUV, NUV bands.  The resulted are given in in Table \ref{tbl:allmag}.  APT is well suited for use in crowded fields in that it is possible to fit an elliptical aperture to a source to avoid nearby stars and because the calculation of the local sky background is not restricted to a simple annulus.

Adjacent WISE images id 0364p424\_ac51 and id 343p424\_ab41, each consisting of 4 images for W1, W2, W3 and W4, were used to calculate WISE magnitudes.  The resolution of these images is  1.375 /arcsec / pixel.  The rectangular images are 93 \arcmin on a side and entirely cover the search region with minimum overlap.

\galex\ NUV magnitudes were calculated using one of three overlapping circular images which combine to cover the search region.  These are NGA\_NGC0891-nd-int(exposure time=1704), GI2\_019004\_3C66B-nd-int (exposure time =6823 sec), and GI5\_063003\_A347\_FIELD1-nd-int (exposure time = 3144 sec).  Each of the images is 1.25 \deg in diameter.  Even though the first of the three images covers almost all of the search region, the other two images which are were used when possible because the exposure times were much longer.  Cross checking showed that using the longer exposure time resulted in magnitudes which were brighter by about 0.5 mag.

\galex\ FUV magnitudes were calculated using two images: NGA\_NGC0891-fd-int(exposure time=1704) and GI2\_019004\_3C66B-nd-int (exposure time = 6047 sec).  Once again the second image was used preferentially because of the longer exposure time.

Table \ref{tbl:allmag} provides the calculated magnitudes for the 71 galaxies described in  Table \ref{tbl:allgal}. Table \ref{tbl:allmag}  provides the following information:
\begin{raggedright}
\newcounter{Tcount}
\begin{list}{Col: {\arabic{Tcount}}:}{}
\usecounter{Tcount}
\setlength{\itemsep}{-0.8ex}
  \item  ID Number
  \item  GSC 2.3.2 ID
  \item  \galex FUV apparent magnitude
  \item  \galex FUV apparent magnitude
  \item  WISE W1 apparent magnitude
  \item  WISE W2 apparent magnitude
  \item  WISE W3 apparent magnitude
  \item  WISE W4 apparent magnitude
  \item  WISE W1 absolute magnitude assuming a distance of 9.8 Mpc for the "Foreground" galaxies identified in Table \ref{tbl:allgal} and a distance of 75 Mpc for the remaining galaxies.
  \item  For comparison, WISE W1 absolute magnitude is provided for the "Foreground" galaxies assuming a distance  distance of 75 Mpc.
  \item  WISE W1 luminosity assuming a distance of 9.8 Mpc for the "Foreground" galaxies identified in Table \ref{tbl:allgal} and a distance of 75 Mpc for the remaining galaxies.
  \item  For comparison, WISE W1 luminosity is provided for the "Foreground" galaxies assuming a distance  distance of 75 Mpc.
 \end{list}
\end{raggedright}

UGC 1807 (ID=7 in Table \ref{tbl:allmag}) is  a well-known dIrr galaxy which was found to have a W1 luminosity of 4.1+08 \Lsun.  The mass-to-light ratio for the W1 band is near unity and so this implies a mass which is typical of dIrr galaxies.  The W1 luminosity of the remainder of the ``Foreground" galaxies ranges from 7.8E+06 to 7.6E+07.  This range is typical for dSph galaxies, as discussed in \cite{gre03}.
\subsection{NIR Surface Brightness vs W1 magnitude}
Figure \ref{fig:NIRmu} plots the DSS2 NIR brightness metric which was used to order the candidates of Table \ref{tbl:allgal} vs WISE W1 magnitude.  The trend is clearly monotonic with a scatter of about 1 mag.   This shows that the information in the GSC 2.3.2, although imprecise, could be used to search for low-mass galaxies in an automated process.

A detailed interpretation of Figure \ref{fig:NIRmu} is somewhat involved. The trend of the background galaxies is consistent with that shown in Figure 9a of \cite{gra03} which plots \MB vs $<\mu>_e$ for a large collection of dE and E galaxies which covers the range of values $-23 <\MB <  -13 $  and $20< \mu_e < 26$.  The single obvious outlying point shown in Figure \ref{fig:NIRmu} is the dIrr galaxy UGC 1807 (ID \#7 in Table \ref{tbl:allgal}.  This point is offset from the trend line by the difference in distance mag $(\Delta m \simeq 34 - 30  )$ and would lie near the trend line of the background galaxies if all were plotted versus absolute magnitude.

The question arises of why the dwarf galaxies identified in Table \ref{tbl:allgal} are not displaced from the trend line in the same way as UGC 1807.  The reason is found in the Figure 2(c) of \cite{derij09} which  plots M vs $\mu$ for many dE's and dSph galaxies.  \cite{derij09} covers the span of $-24 < M_V < -8$ and the dSph galaxies lie in the range $-14 <M_V < -8$. \cite{derij09} find that dSph galaxies do not continue the trendline of dE and E galaxies. The relationship  changes slope at about $\MV > 15$ (i.e. at $W1\sim15$ at the distance of NGC 891) at which point the slope changes by a factor of slightly more than 2.  The offset is approximately $\Delta \MV \sim +4$ for galaxies with surface brightness $\mu_{\textrm{0,V}} \sim 25$ which offsets the change due to the different distance mag.
\subsection{UV Excess}
\cite{LeeJ11} reports that the ratio of \galex\ UV to 2MASS K magnitude is much higher for galaxies which are dim in \IN\ and also that detection in UV is more reliable for dSph and dIrr galaxies than detection in the B or R bands. Similarly, \cite{LeeJ09} reports that the FUV-to-\Halpha ratio is larger than expected, especially for lower luminosity dwarf galaxies.  The FUV/\Halpha\ and NUV/\Halpha\ ratios of the lowest luminosity galaxies are larger than expected by an order of magnitude or more compared to larger galaxies.

\cite{bua05} compared a NUV-selected sample of galaxies to a FIR-selected sample and found that the the average dust attenuation of the NUV-selected sample was 0.8 whereas the average attenuation of the FIR sample was 2.1 and was larger than about 5.0 for some galaxies not detected in FUV.  Strikingly, they found a cutoff so that there is \emph{no} dust attenuation in the NUV band for galaxies with  luminosity less than about 5\dex{8} \Lsun, i.e., the dSph's.

\cite{cor08} modeled the dependance of the attenuation of galaxy FUV emissions, A(FUV), to the ratio of total IR emission(TIR) to FUV and to the age of the underlying stellar distribution.  They find that A(FUV) increases dramatically as the TIR/FUV ratio increases and also increases as the age of the underlying stellar population increases, especially from 3 Gyr to 4 Gyr.  The FUV attenuation can be as large as several mags but is negligible, and age is not significant, for galaxies with a small TIR/FUV ratio.

This effect was seen in the present study and is strong enough to help distinguish foreground dSph's from background galaxies.  All of the identified dSph galaxies are confirmed in NUV and in FUV and none are visible in 2MASS J, H, or K.  The UV excess  of the dSph galaxies shown in Figure \ref{fig:allgal} is estimated to be more than two magnitudes compared to background galaxies.

Figure \ref{fig:NUV-W1} plots \galex NUV - WISE W1 vs W1 magnitude.  This figure clearly shows that there is a strong trend of NUV excess with decreasing mass.

The present results are consistent with \cite{bua05} and suggest that the observed UV excess is due to the absence of gas and dust in the dSph galaxies.  In normally sized galaxies, most of the UV emitted by stars in the observed galaxy is absorbed locally by dust and re-emitted in NIR. In contrast, the UV emitted by the stars of the dust-free dSph galaxies is observed with very little attenuation and the NIR emission is negligible.  Note however that \cite{LeeJ09} argues the contrary and asserts that dust attenuation effects cannot explain the UV excess.

Again \ref{fig:NUV-W1} includes galaxies which are at very different distances and so the apparent W1 Magnitude is not a proxy for mass.
\subsection{Completeness}
\cite{las08} claims that the GSC 2.3 includes ``almost all" of the objects down to the plate limits which are $ \BJ<22.5$, $\RF<20.8$, and $\IN<19.5$.  Table \ref{tbl:allgal} is complete to these plate limits with the added constaint on size of $D > 15\arcsec$.

With one exception, the survey found all of the previously cataloged galaxies in the search region which met the size limit. This good result implies that the current survey is complete to the plate accuracy of the DSS2.  The single exception was that we were unable to verify the dSph galaxy [TT2009]30 reported by \cite{tre09}.  This good result demonstrates that the current survey is complete in the target region to within the defined size limit.
\section{Summary}
We report the results of a size-limited survey of the region near NGC 891.  71 galaxies with apparent diameters less than 15\arcsec\ were found in the region $R \leq 40\arcmin$ around NGC 891.  7 of these galaxies were identified as likely dwarf galaxies in the halo of NGC 891 and several others are likely candidates.  Most of the remaining galaxies are members of the Abell 347 cluster, which approximately doubles the known membership of the cluster.

Although the present survey reports an increase in the number of dwarf galaxies near NGC 891 the number of new discoveries falls far short of that needed to account for the "missing satellites" discussed above.  This evidence supports the current status based on counts of the dwarf companions of the Milky Way that there is a basic conflict between observations and the accepted theory of galaxy formation.

The methodology described here of examining candidates taken from the Guide Star Catalog in several different bands is much more efficient and thorough than previous surveys of the region.  Searches which depend on a single band, especially near the galactic plane, are plagued by contamination, diffraction effects, and various sorts of artifacts which are easily eliminated when the same region is viewed at many wavelengths.
\section*{Acknowledgements}
The Guide Star Catalogue-II is a joint project of the Space Telescope Science Institute and the Osservatorio Astronomico di Torino.  The Digitized Sky Surveys were produced at the Space Telescope Science Institute under U.S. Government grant NAG W-2166. The images of these surveys are based on photographic data obtained using the Oschin Schmidt Telescope on Palomar Mountain and the UK Schmidt Telescope.

This publication makes use of data products from the Two Micron All Sky Survey, which is a joint project of the University of Massachusetts and the Infrared Processing and Analysis Center/California Institute of Technology. 

This research has made use of the Aladin interactive sky atlas, developed at CDS, Strasbourg, France.

We acknowledge the use of NASA's SkyView virtual observatory \citep{mcgly98}    

This research has made use of the VizieR catalogue access tool, CDS, Strasbourg, France.  

This research has made use of the NASA/IPAC Extragalactic Database (NED) which is operated by the Jet Propulsion Laboratory, California Institute of Technology, under contract with the National Aeronautics and Space Administration.  


\clearpage

\renewcommand{\thefootnote}{\alph{footnote}}
\setcounter{footnote}{0}

\newcommand {\fta} {\footnote{ This data from he GSC 2.3.2 catalog is affected by relatively large systematic errors. }}
\newcommand {\fma} {\footnotemark[1]}

\newcommand {\ftb} {\footnote{Estimated}}
\newcommand {\fmb} {\footnotemark[2]}
\newcommand {\ftc} {\footnote{ \#63 was incorrectly cataloged as a star (type=0) in GSC2.3}}

\renewcommand{\arraystretch}{0.92}
{\begin{center}
\begin{longtable}{|r l c c r r c r r r r r r c|}
\caption{Galaxies Near NGC~891\label{tbl:allgal}}\\
\hline
~&&&&&&&&&&&&& \\
 ID &  GSC2.3 ID     &   RA/DE J2000       &  Dist &  alt ID~~~~~~~~& SID & 2MX &   Vel &R$_F$\fta& BJ\fma & IN \fma&Rad\fma   & NIR $\mu$  & Fore \\
    &                &    h:m:s~~~~d:m:s   &arcmin &                & num &     &   kms &   mag   & mag    & mag    & asec     &  m/asc2    &      \\
--- &-----------     &  ------------------ &-------&----------------& --- & --- &------ &------   &------  &------  &  -----   &    -----   & ---- \\
  1 & NCIA029460     &   02:24:42+42:49:54 & 37.44 &                &     &     &       & 16.71   & 17.87  & 19\ftb &   10.6   &    25\fmb  &    F \\
  2 & NCIA018969     &   02:21:12+42:21:50 & 14.93 &HFLLZOA F172    &     &     &       & 15.76   & 17.97  & 18.58  &   12.1   &      24.9  &    F \\
  3 & NCIA030805     &   02:23:35+42:55:27 & 36.37 &                &     &     &       & 16.35   & 18.20  & 18.78  &   11.2   &      24.8  &    F \\
  4 & NCIA019215     &   02:24:24+42:21:10 & 20.49 &                &     &     &       & 17.56   & 18.35  & 17.98  &    9.3   &      23.5  &    F \\
  5 & NBZ5012013     &   02:20:07+42:45:44 & 36.56 &                &     &     &       & 16.33   & 16.73  & 17.95  &    7.7   &      23.3  &    F \\
  6 & NCIB031541     &   02:22:55+41:43:41 & 37.43 &                &     &     &       & 16.08   & 17.36  & 17.62  &    8.4   &      23.3  &    F \\
  7 & NBZ5012371     &   02:21:13+42:45:46 & 28.87 & UGC 1807       &     &     &   629 & 10.22   & 11.26  & 14.36  &   27.6   &      22.7  &    F \\
  8 & NCIA035836     &   02:24:05+42:08:15 & 21.26 &                &     &     &       & 15.34   & 17.44  & 16.39  &   10.2   &      22.5  &    - \\
  9 & NCIA025916     &   02:24:29+42:38:37 & 27.83 &                &     &     &       & 16.47   & 17.34  & 16.96  &    7.7   &      22.3  &    - \\
 10 & NCIA030487     &   02:24:15+42:53:59 & 38.09 &                &     &     &       & 15.50   & 16.29  & 16.50  &    8.8   &      22.1  &    - \\
 11 & NCIA005660     &   02:25:14+41:54:47 & 39.62 &                &     &     &       & 15.51   & 17.33  & 16.24  &    8.7   &      21.9  &    - \\
 12 & NBZ5004290     &   02:19:50+42:15:52 & 30.61 &                &     &     &       & 14.40   & 15.64  & 15.75  &   10.1   &      21.8  &    - \\
 13 & NCIA023873     &   02:24:01+42:32:41 & 20.03 & HFLLZOA F182   &     &     &       & 12.67   & 13.69  & 15.02  &   13.6   &      21.7  &    - \\
 14 & NCIA004638     &   02:23:49+41:53:03 & 31.24 &                &  33 &     &       & 13.26   & 16.17  & 14.57  &   17.6   &      21.7  &    - \\
 15 & NBZ5003674     &   02:19:55+42:11:42 & 30.57 &                &     &     &       & 14.53   & 16.16  & 15.62  &    9.5   &      21.6  &    - \\
 16 & NBZ5015034     &   02:21:17+42:52:44 & 34.76 & PGC 2187376    &   2 &     & 6,805 & 13.81   & 14.97  & 15.28  &    9.4   &      21.3  &    - \\
 17 & NCIA012578     &   02:25:11+42:07:35 & 32.10 & PGC 2192879    &  65 &     &       & 15.44   & 17.45  & 15.79  &    7.7   &      21.2  &    - \\
 18 & NCIA009669     &   02:24:20+42:02:35 & 26.95 &                &  46 &     & 5,680 & 14.62   & 16.39  & 15.43  &    9.3   &      21.2  &    - \\
 19 & NCIA010240     &   02:24:39+42:03:24 & 29.23 &                &  51 &     & 5,910 & 15.13   & 17.01  & 15.55  &    8.3   &      21.1  &    - \\
 20 & NBZ5004664     &   02:20:25+42:17:43 & 23.82 & PGC 2196068    &     &   Y &       & 12.69   & 14.44  & 14.24  &   15.7   &      21.1  &    - \\
 21 & NCIA007657     &   02:21:16+42:00:06 & 25.23 &                &     &   Y &       & 16.33   & 19.21  & 15.68  &    7.8   &      21.0  &    - \\
 22 & NCIA010627     &   02:23:49+42:04:40 & 21.47 &                &     &     &       & 15.32   & 17.66  & 15.36  &    8.2   &      20.9  &    - \\
 23 & NCIA006643     &   02:25:13+41:56:42 & 38.25 &                &  66 &   Y &       & 15.03   & 17.23  & 15.34  &    8.0   &      20.9  &    - \\
 24 & NCIA010131     &   02:21:32+42:04:28 & 19.94 &                &     &     &       & 15.28   & 17.41  & 15.26  &    7.7   &      20.6  &    - \\
 25 & NCIA012182     &   02:22:21+42:08:00 & 13.09 &                &     &     &       & 15.18   & 17.04  & 15.08  &    7.7   &      20.5  &    - \\
 26 & NCIA012187     &   02:25:28+42:06:45 & 35.29 & PGC 2192614    &     &   Y &       & 14.59   & 16.92  & 14.42  &    9.1   &      20.1  &    - \\
 27 & NCIA019867     &   02:26:03+42:22:08 & 38.88 & PGC 2197501    &     &   Y & 4,451 & 14.72   & 16.22  & 14.60  &    7.8   &      20.0  &    - \\
 28 & NCIA026109     &   02:25:13+42:38:55 & 34.57 &                &     &   Y &       & 14.09   & 15.27  & 14.52  &    7.6   &      19.9  &    - \\
 29 & NCIA016387     &   02:20:42+42:16:35 & 20.90 &                &     &   Y &       & 13.38   & 15.69  & 13.81  &   10.1   &      19.9  &    - \\
 30 & NCIA031935     &   02:22:28+43:00:49 & 39.93 & PGC 2211100    &     &   Y &20,006 & 13.15   & 14.50  & 13.81  &   10.1   &      19.9  &    - \\
 31 & NCIA016766     &   02:24:18+42:16:13 & 19.94 &                &     &   Y &       & 13.29   & 15.11  & 13.53  &   10.6   &      19.7  &    - \\
 32 & NCIA029870     &   02:23:39+42:51:56 & 33.32 & HFLLZOA F179   &     &   Y &       & 12.79   & 14.22  & 13.41  &   10.9   &      19.6  &    - \\
 33 & NCIA010841     &   02:20:51+42:05:57 & 24.05 & PGC 2192372    &     &   Y &       & 13.68   & 15.45  & 13.71  &    9.7   &      19.6  &    - \\
 34 & NCIA008246     &   02:23:44+42:00:23 & 24.43 &                &     &   Y &       & 13.74   & 15.65  & 13.95  &    7.7   &      19.4  &    - \\
 35 & NCIA031681     &   02:22:51+42:59:16 & 38.52 & PGC 2210478    &     &   Y &       & 12.85   & 14.45  & 13.31  &   10.6   &      19.4  &    - \\
 36 & NCIA010315     &   02:20:49+42:05:02 & 24.97 & PGC 2192089    &     &     &       & 13.19   & 14.38  & 13.57  &    8.3   &      19.3  &    - \\
 37 & NCIA004881     &   02:23:46+41:53:38 & 30.46 & PGC 2188688    &     &   Y &       & 11.93   & 15.01  & 12.36  &   14.2   &      19.2  &    - \\
 38 & NCIA027059     &   02:23:59+42:42:28 & 26.82 & HFLLZOA F181   &     &   Y &       & 12.87   & 14.05  & 12.91  &   12.0   &      19.1  &    - \\
 39 & NCIA031605     &   02:22:58+42:58:55 & 38.29 & PGC 2210361    &     &   Y &20,296 & 11.69   & 13.54  & 12.20  &   14.7   &      19.0  &    - \\
 40 & NBZ5010876     &   02:21:32+42:41:36 & 23.51 & HFLLZOA F176   &     &   Y &       & 12.71   & 12.86  & 13.04  &    9.3   &      19.0  &    - \\
 41 & NCIA027303     &   02:22:50+42:43:28 & 22.79 & HFLLZOA F210   &     &   Y &       & 12.22   & 13.43  & 12.58  &   12.2   &      18.9  &    - \\
 42 & NCIA029804     &   02:23:24+42:51:48 & 32.28 & HFLLZOA F178   &     &   Y &       & 12.20   & 13.49  & 12.58  &   11.8   &      18.9  &    - \\
 43 & NCIA004956     &   02:23:33+41:53:57 & 29.19 & PGC 212966     &     &   Y &       & 13.23   & 15.92  & 13.05  &    8.8   &      18.9  &    - \\
 44 & NCIA006145     &   02:22:36+41:56:45 & 24.16 & HFLLZOA F187   &     &   Y &       & 13.16   & 15.66  & 13.01  &    9.2   &      18.9  &    - \\
 45 & NCIA001071     &   02:24:01+42:04:02 & 23.44 & HFLLZOA F191   &     &   Y &       &    na   & 13.80  & 11.97  &   15.2   &      18.6  &    - \\
 46 & NCIA001356     &   02:21:06+41:49:13 & 35.59 & PGC 8939       &     &   Y &       & 11.29   & 13.83  & 11.70  &   13.8   &      18.5  &    - \\
 47 & NCIA015505     &   02:24:08+42:13:42 & 18.95 & PGC 2194768    &     &   Y &       & 12.15   & 14.23  & 11.81  &   13.7   &      18.4  &    - \\
 48 & NBZ5000254     &   02:19:46+42:43:15 & 38.08 & HFLLZOA F170   &     &   Y &       & 11.93   & 13.27  & 12.11  &   11.1   &      18.4  &    - \\
 49 & NCIA000887     &   02:23:54+42:12:22 & 17.19 & PGC 9101       &     &   Y &13,041 & 11.00   & 12.71  & 11.18  &   16.0   &      18.3  &    - \\
 50 & NCIA001268     &   02:24:49+41:54:22 & 36.56 & HFLLZOA F206   &     &   Y &       & 11.94   & 13.91  & 11.63  &   14.7   &      18.3  &    - \\
 51 & NCIA001160     &   02:23:30+41:59:54 & 23.51 & NGC 2190563    &  25 &   Y & 6,895 & 12.54   & 14.47  & 12.28  &    9.7   &      18.3  &    - \\
 52 & NCIA000452     &   02:24:32+42:34:45 & 26.03 & PGC 212970     &     &   Y & 6,661 & 11.18   & 12.47  & 11.35  &   15.8   &      18.2  &    - \\
 53 & NCIA001126     &   02:24:47+42:01:27 & 31.56 & PGC009151      &  55 &   Y & 6,086 & 11.02   & 12.57  & 10.86  &   21.5   &      18.2  &    - \\
 54 & NCIA000277     &   02:25:10+42:46:33 & 38.61 &                &     &   Y &       & 10.50   & 11.80  & 10.92  &   17.4   &      18.1  &    - \\
 55 & NCIA001249     &   02:23:16+41:55:21 & 26.74 & PGC 212965     &     &   Y &       & 12.73   & 14.63  & 12.20  &    9.8   &      18.1  &    - \\
 56 & NCIA000313     &   02:23:15+42:43:53 & 24.27 & PGC 2204990    &     &   Y &       & 11.60   & 12.99  & 11.84  &   10.3   &      18.0  &    - \\
 57 & NCIA000956     &   02:22:50+42:09:29 & 11.87 & PGC 9042       &  10 &   Y & 6,390 & 11.50   & 13.06  & 11.51  &   12.1   &      18.0  &    - \\
 58 & NCIA030617     &   02:23:03+42:54:45 & 34.31 & PGC 3097117    &  19 &   Y & 6,895 & 11.02   & 12.84  & 11.10  &   14.2   &      17.9  &    - \\
 59 & NCIA001042     &   02:22:59+42:05:38 & 16.02 & PGC 2192261    &     &   Y &       & 10.98   & 12.83  & 10.71  &   18.8   &      17.9  &    - \\
 60 & NCIA001337     &   02:23:03+41:50:53 & 30.53 & HFLLZOA F261   &     &   Y &       & 11.77   & 13.69  & 11.39  &   12.9   &      17.9  &    - \\
 61 & NBZ5000678     &   02:21:24+42:52:34 & 34.14 & PGC 8955       &   3 &   Y & 6,639 & 10.03   & 12.05  & 10.13  &   23.9   &      17.8  &    - \\
 62 & NBZ5000685     &   02:21:12+42:51:48 & 34.29 & PGC 8948       &     &   Y &10,073 & 10.10   & 11.82  & 10.32  &   19.4   &      17.8  &    - \\
 63 & NCIA000991\ftc &   02:25:33+42:08:03 & 35.87 & PGC 2193030    &     &   Y & 4,451 & 10.90   & 12.84  & 10.67  &   16.5   &      17.7  &    - \\
 64 & NCIA000260     &   02:23:52+42:47:34 & 30.39 & PGC 220681     &     &   Y & 5,984 & 10.38   & 11.74  & 10.57  &   15.4   &      17.6  &    - \\
 65 & NCIA001051     &   02:25:16+42:05:22 & 34.00 & NGC 906        &     &   Y & 4,680 &  8.83   &  9.99  & 8.75   &   34.2   &      17.5  &    - \\
 66 & NCIA000248     &   02:22:18+42:48:19 & 27.55 & PGC 9017       &     &   Y & 6,354 & 10.12   & 11.51  & 10.11  &   19.0   &      17.4  &    - \\
 67 & NCIA001220     &   02:23:20+41:57:05 & 25.39 & NGC 898        &     &   Y & 5,495 &  8.54   &  9.76  & na     &   41.8   &    17\fmb  &    - \\
 68 & NCIA000407     &   02:24:44+42:37:23 & 29.32 & UGC 1859       &     &   Y & 6,087 &  8.95   & 10.05  & 8.41   &   30.6   &      16.9  &    - \\
 69 & NCIA001111     &   02:25:23+42:02:08 & 36.64 & NGC 909        &     &   Y & 4,978 &  9.05   & 10.48  & 8.54   &   24.0   &      16.6  &    - \\
 70 & NCIA001162     &   02:24:02+41:59:44 & 26.83 & PGC 9108       &     &   Y & 5,659 &  9.80   & 11.13  & 9.39   &   16.1   &      16.5  &    - \\
 71 & NCIA000110     &   02:23:13+42:59:16 & 39.07 & PGC 212964     &     &   Y & 6,595 &  na     & na     & 9.98   &    8.0   &      15.6  &    - \\
\hline
\end{longtable}
\end{center}}

\clearpage


\renewcommand{\thefootnote}{\alph{footnote}}
\setcounter{footnote}{0}

\renewcommand {\fta} {\footnote{Assuming a distance of 9.8 Mpc for Foreground galaxies indicated on Table \ref{tbl:allgal} and 75 Mpc for the remainder.}}
\renewcommand {\fma} {\footnotemark[1]}
\renewcommand {\ftb} {\footnote{Assuming a distance of 75 Mpc for Foreground galaxies for comparison}}
\renewcommand {\fmb} {\footnotemark[2]}

\renewcommand{\arraystretch}{0.92}
{\begin{center}
\begin{longtable}{| r r r r r r r r r r r r|}
\caption{\galex and WISE magnitudes \label{tbl:allmag}}\\
\hline
~&&&&&&&&&&& \\
 ID &  GSC2.3 ID & FUV  & NUV  & W1   & W2   & W3   & W4   & W1\fta & W1\ftb & W1 L\fma & W1 L\fmb\\
    &            & mag  & mag  & mag  & mag  & mag  & mag  & Mag    & Mag    & \Lsun~~  & \Lsun~~ \\
--- &----------- & ---- & ---- & ---- & ---- & ---- & ---- &  ----  &  ----  & ----     &  ----   \\
  1 & NCIA029460 & 20.4 & 20.2 & 16.0 & 15.6 & 14.6 & 11.8 & -14.0  & -18.4  & 7.8E+06  & 4.5E+08 \\
  2 & NCIA018969 & 21.4 & 21.0 & 15.2 & 16.4 & ---  & ---  & -14.8  & -19.2  & 1.7E+07  & 9.7E+08 \\
  3 & NCIA030805 & 20.7 & 20.5 & 15.4 & 15.4 & ---  & 13.1 & -14.6  & -19.0  & 1.4E+07  & 8.2E+08 \\
  4 & NCIA019215 & 20.9 & 21.0 & 15.1 & 15.7 & 16.5 & ---  & -14.9  & -19.3  & 1.7E+07  & 1.0E+09 \\
  5 & NBZ5012013 & 19.7 & 20.0 & 15.3 & 14.9 & 13.8 & ---  & -14.7  & -19.1  & 1.5E+07  & 8.6E+08 \\
  6 & NCIB031541 & 19.2 & 19.4 & 15.2 & 15.3 & 14.3 & ---  & -14.8  & -19.2  & 1.6E+07  & 9.2E+08 \\
  7 & NBZ5012371 & 17.0 & 16.7 & 11.7 & 11.7 & 10.8 & ---  & -18.3  & -22.7  & 4.1E+08  & 2.3E+10 \\
  8 & NCIA035836 & 20.2 & 20.5 & 14.6 & 14.6 & 11.9 & 11.8 & -19.8  &        & 1.7E+09  &         \\
  9 & NCIA025916 & 20.7 & 20.6 & 15.0 & 14.6 & 11.9 & 12.2 & -19.4  &        & 1.1E+09  &         \\
 10 & NCIA030487 & 21.8 & 21.1 & 15.0 & 15.0 & 12.9 & 10.6 & -19.4  &        & 1.1E+09  &         \\
 11 & NCIA005660 & 20.2 & 20.2 & 15.0 & 14.8 & 12.6 & 12.6 & -19.4  &        & 1.2E+09  &         \\
 12 & NBZ5004290 & 18.5 & 18.4 & 14.2 & 14.3 & 11.9 & 11.1 & -20.2  &        & 2.3E+09  &         \\
 13 & NCIA023873 & 18.4 & 18.0 & 13.5 & 13.5 & 10.7 & 8.6  & -20.9  &        & 4.4E+09  &         \\
 14 & NCIA004638 & 19.6 & 19.4 & 13.9 & 13.9 & 11.7 & 10.0 & -20.5  &        & 3.0E+09  &         \\
 15 & NBZ5003674 & 19.2 & 19.5 & 14.5 & 14.5 & 11.9 & 12.0 & -19.9  &        & 1.8E+09  &         \\
 16 & NBZ5015034 & 20.3 & 19.9 & 14.1 & 14.1 & 11.6 & 10.7 & -20.3  &        & 2.6E+09  &         \\
 17 & NCIA012578 & 20.8 & 20.7 & 14.5 & 14.7 & 13.0 & 10.3 & -19.9  &        & 1.8E+09  &         \\
 18 & NCIA009669 & 19.8 & 19.7 & 14.4 & 14.4 & 11.8 & 10.0 & -20.0  &        & 1.9E+09  &         \\
 19 & NCIA010240 & 20.3 & 20.2 & 14.3 & 14.2 & 11.3 & 9.8  & -20.1  &        & 2.2E+09  &         \\
 20 & NBZ5004664 & 18.9 & 18.9 & 13.1 & 13.1 & 10.3 & 14.5 & -21.3  &        & 6.4E+09  &         \\
 21 & NCIA007657 & 20.3 & 20.6 & 13.3 & 13.3 & 11.3 & 10.1 & -21.1  &        & 5.2E+09  &         \\
 22 & NCIA010627 & 20.6 & 20.8 & 14.0 & 14.0 & 11.0 & 11.0 & -20.4  &        & 2.9E+09  &         \\
 23 & NCIA006643 & 21.1 & 22.0 & 14.7 & 14.7 & 15.5 & 10.9 & -19.7  &        & 1.6E+09  &         \\
 24 & NCIA010131 & 20.9 & 21.0 & 14.5 & 14.4 & 12.8 & ---  & -19.9  &        & 1.8E+09  &         \\
 25 & NCIA012182 & 19.7 & 19.5 & 14.1 & 13.9 & 11.0 & 9.9  & -20.3  &        & 2.6E+09  &         \\
 26 & NCIA012187 & 20.4 & 21.3 & 13.0 & 12.9 & 10.1 & 8.7  & -21.4  &        & 7.4E+09  &         \\
 27 & NCIA019867 & 20.6 & 21.6 & 13.0 & 12.8 & 11.0 & 11.2 & -21.4  &        & 7.1E+09  &         \\
 28 & NCIA026109 & 22.5 & 22.0 & 13.8 & 14.0 & 13.3 & ---  & -20.6  &        & 3.6E+09  &         \\
 29 & NCIA016387 & ---  & 19.0 & 13.1 & 13.0 & 10.4 & 9.2  & -21.3  &        & 6.7E+09  &         \\
 30 & NCIA031935 & 20.4 & 19.9 & 12.5 & 12.3 & 8.7  & 6.5  & -21.9  &        & 1.1E+10  &         \\
 31 & NCIA016766 & 19.3 & 19.2 & 13.3 & 13.2 & 10.4 & 9.3  & -21.1  &        & 5.4E+09  &         \\
 32 & NCIA029870 & 22.6 & 21.4 & 13.0 & 13.0 & 13.5 & 10.3 & -21.4  &        & 7.4E+09  &         \\
 33 & NCIA010841 & 19.0 & 19.0 & 12.9 & 12.8 & 9.5  & 8.4  & -21.5  &        & 7.8E+09  &         \\
 34 & NCIA008246 & 19.7 & 19.7 & 13.6 & 13.6 & 10.2 & 8.5  & -20.8  &        & 4.3E+09  &         \\
 35 & NCIA031681 & 20.9 & 20.7 & 12.5 & 12.5 & 10.4 & 9.6  & -21.9  &        & 1.1E+10  &         \\
 36 & NCIA010315 & 18.9 & 18.6 & 13.7 & 13.9 & 11.2 & 8.9  & -20.7  &        & 3.8E+09  &         \\
 37 & NCIA004881 & 20.5 & 21.6 & 12.3 & 12.3 & 11.8 & 10.2 & -22.1  &        & 1.4E+10  &         \\
 38 & NCIA027059 & 21.8 & 21.5 & 12.7 & 12.7 & 12.4 & 11.3 & -21.7  &        & 9.9E+09  &         \\
 39 & NCIA031605 & 21.9 & 21.6 & 11.8 & 11.8 & 10.4 & 9.9  & -22.6  &        & 2.1E+10  &         \\
 40 & NBZ5010876 & 19.3 & 18.9 & 12.8 & 12.6 & 9.3  & 8.1  & -21.6  &        & 8.8E+09  &         \\
 41 & NCIA027303 & 21.7 & 20.7 & 12.4 & 12.5 & 10.8 & 9.9  & -22.0  &        & 1.2E+10  &         \\
 42 & NCIA029804 & 23.0 & 21.2 & 12.5 & 12.5 & ---  & ---  & -21.9  &        & 1.1E+10  &         \\
 43 & NCIA004956 & 20.5 & 22.2 & 12.7 & 12.7 & 12.0 & ---  & -21.7  &        & 9.8E+09  &         \\
 44 & NCIA006145 & 20.4 & 20.6 & 12.7 & 12.7 & 10.9 & 8.6  & -21.7  &        & 9.6E+09  &         \\
 45 & NCIA001071 & 19.0 & 18.8 & 12.0 & 11.9 & 9.2  & 7.3  & -22.4  &        & 1.8E+10  &         \\
 46 & NCIA001356 & 17.6 & 17.4 & 12.1 & 12.0 & 8.5  & 5.8  & -22.3  &        & 1.6E+10  &         \\
 47 & NCIA015505 & 20.6 & 20.2 & 11.9 & 11.9 & 10.3 & 9.7  & -22.5  &        & 1.9E+10  &         \\
 48 & NBZ5000254 & 26.1 & 19.6 & 11.8 & 11.7 & 9.1  & 6.9  & -22.6  &        & 2.1E+10  &         \\
 49 & NCIA000887 & 18.8 & 18.5 & 11.7 & 11.6 & 8.8  & 7.1  & -22.7  &        & 2.3E+10  &         \\
 50 & NCIA001268 & 20.4 & 20.8 & 12.2 & 12.2 & 11.1 & 11.3 & -22.2  &        & 1.5E+10  &         \\
 51 & NCIA001160 & 21.2 & 20.7 & 12.4 & 12.5 & 12.7 & 13.9 & -22.0  &        & 1.2E+10  &         \\
 52 & NCIA000452 & 20.2 & 19.5 & 11.8 & 11.8 & 8.8  & 7.7  & -22.6  &        & 2.2E+10  &         \\
 53 & NCIA001126 & 18.6 & 18.5 & 11.4 & 11.3 & 7.8  & 6.3  & -23.0  &        & 3.1E+10  &         \\
 54 & NCIA000277 & 21.0 & 20.2 & 11.7 & 11.7 & 10.1 & 8.7  & -22.7  &        & 2.4E+10  &         \\
 55 & NCIA001249 & 20.6 & 21.6 & 12.4 & 12.5 & 12.4 & 10.1 & -22.0  &        & 1.3E+10  &         \\
 56 & NCIA000313 & 22.1 & 20.9 & 11.6 & 11.6 & 11.8 & 10.4 & -22.8  &        & 2.7E+10  &         \\
 57 & NCIA000956 & 18.6 & 18.3 & 12.1 & 12.0 & 8.8  & 7.5  & -22.3  &        & 1.6E+10  &         \\
 58 & NCIA030617 & 22.2 & 20.8 & 11.5 & 11.4 & 9.9  & 8.0  & -22.9  &        & 2.8E+10  &         \\
 59 & NCIA001042 & 20.2 & 19.5 & 11.5 & 11.4 & 8.2  & 5.5  & -22.9  &        & 2.8E+10  &         \\
 60 & NCIA001337 & 20.1 & 20.6 & 12.1 & 12.1 & 12.1 & ---  & -22.3  &        & 1.7E+10  &         \\
 61 & NBZ5000678 & 18.4 & 18.0 & 10.5 & 10.2 & 6.3  & 4.4  & -23.9  &        & 7.2E+10  &         \\
 62 & NBZ5000685 & 19.0 & 18.7 & 11.2 & 11.2 & 9.4  & 8.2  & -23.2  &        & 3.9E+10  &         \\
 63 & NCIA000991 & 20.6 & 20.7 & 11.6 & 11.6 & 11.4 & 9.6  & -22.8  &        & 2.6E+10  &         \\
 64 & NCIA000260 & 21.5 & 20.5 & 11.3 & 11.3 & 11.2 & ---  & -23.1  &        & 3.5E+10  &         \\
 65 & NCIA001051 & 17.4 & 16.9 & 9.8  & 9.8  & 7.0  & 5.5  & -24.6  &        & 1.4E+11  &         \\
 66 & NCIA000248 & 21.3 & 20.4 & 11.0 & 11.1 & 10.9 & 12.0 & -23.4  &        & 4.4E+10  &         \\
 67 & NCIA001220 & 18.7 & 18.5 & 9.2  & 9.2  & 7.2  & 5.9  & -25.2  &        & 2.4E+11  &         \\
 68 & NCIA000407 & 19.6 & 19.2 & 9.6  & 9.6  & 9.3  & 7.9  & -24.8  &        & 1.6E+11  &         \\
 69 & NCIA001111 & 19.7 & 19.4 & 10.3 & 10.3 & 9.9  & 9.6  & -24.1  &        & 9.0E+10  &         \\
 70 & NCIA001162 & 19.1 & 18.6 & 11.3 & 11.3 & 8.9  & 7.6  & -23.1  &        & 3.5E+10  &         \\
 71 & NCIA000110 & 20.3 & 19.3 & 10.8 & 10.8 & 10.4 & 8.9  & -23.6  &        & 5.6E+10  &         \\
\hline
\end{longtable}
\end{center}}


\clearpage

  \newcommand{\HDb}[1] {\subfigure{\makebox[0.07\textwidth]{#1}}}
  \newcommand{\FGb}[1] {\includegraphics[width=0.07\textwidth]{#1}}


\clearpage
\begin{figure}[ht]
  \HDb{rgb}    \HDb{FUV}      \HDb{NUV}     \HDb{B}       \HDb{R}       \HDb{IR}      \HDb{J}       \HDb{H}       \HDb{K}       \HDb{W1}      \HDb{W2}      \HDb{W3}      \HDb{W4}    \\
  \FGb{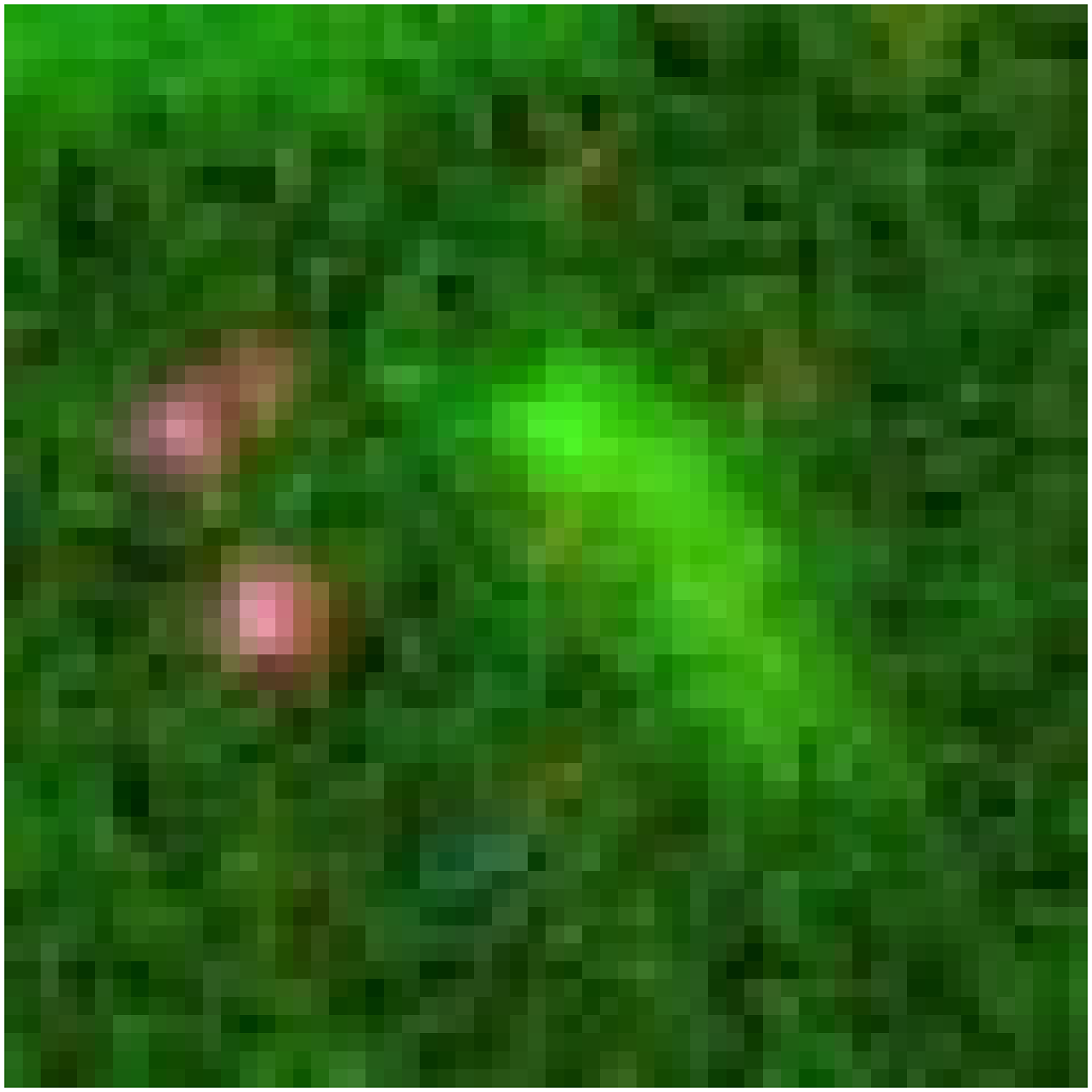}   \FGb{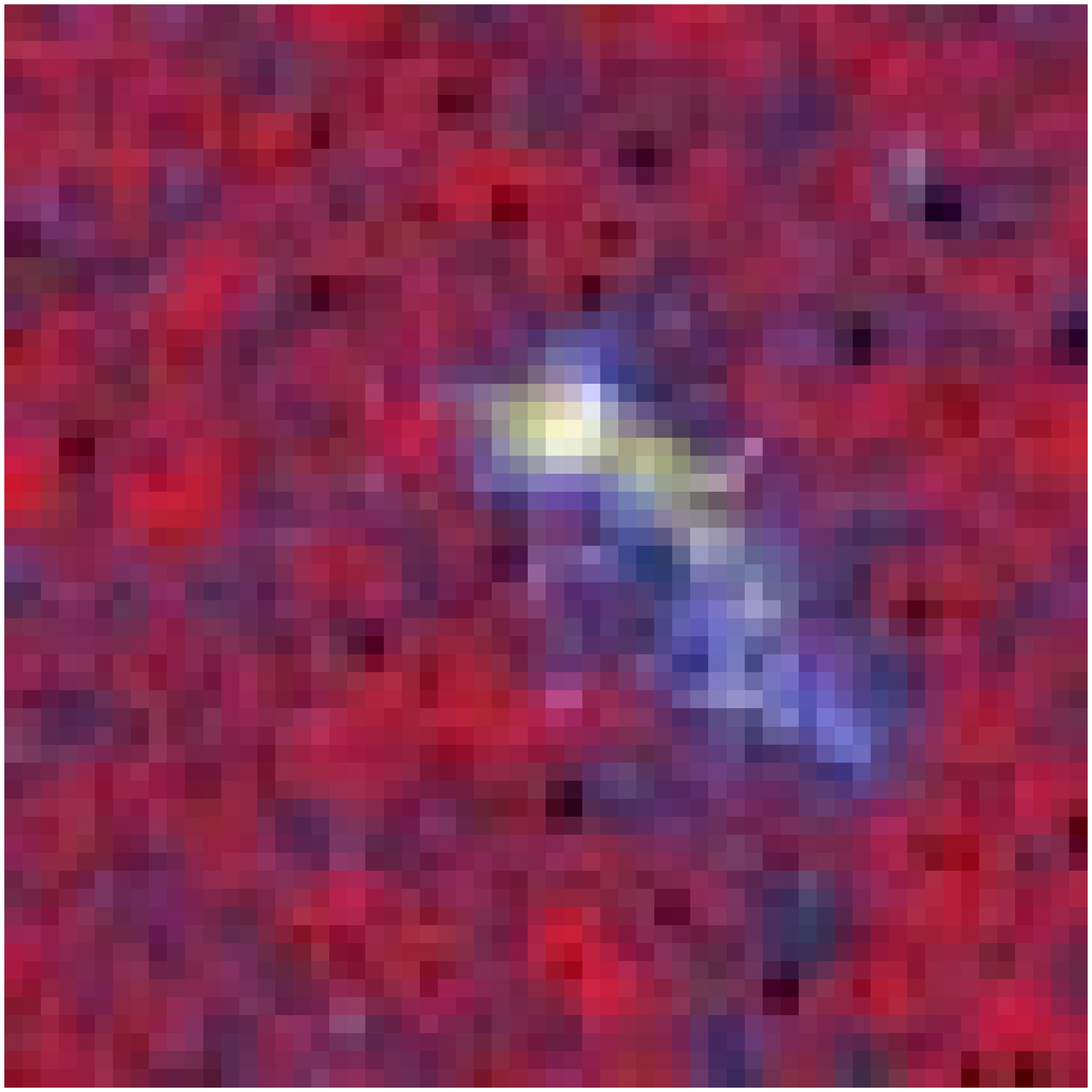}     \FGb{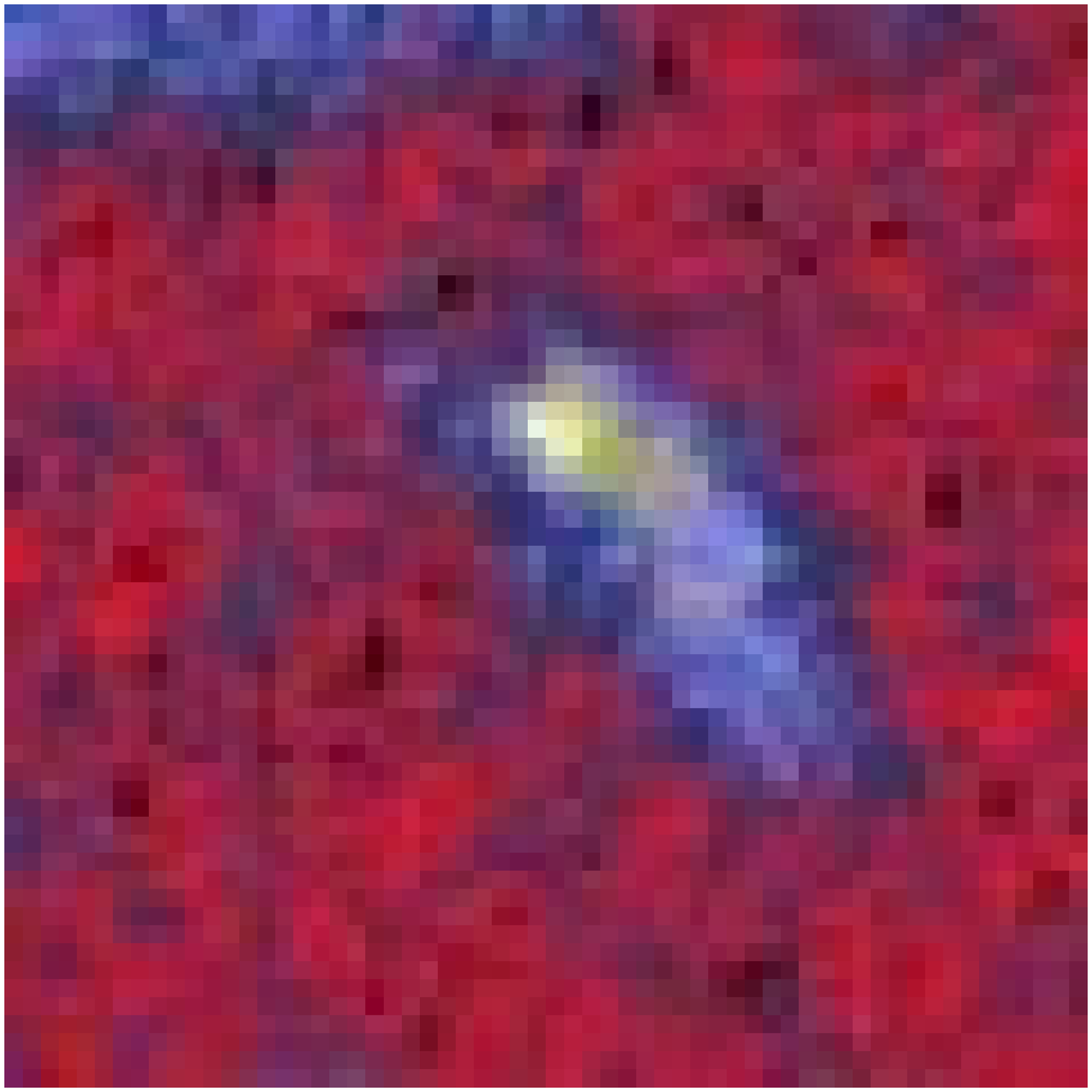}    \FGb{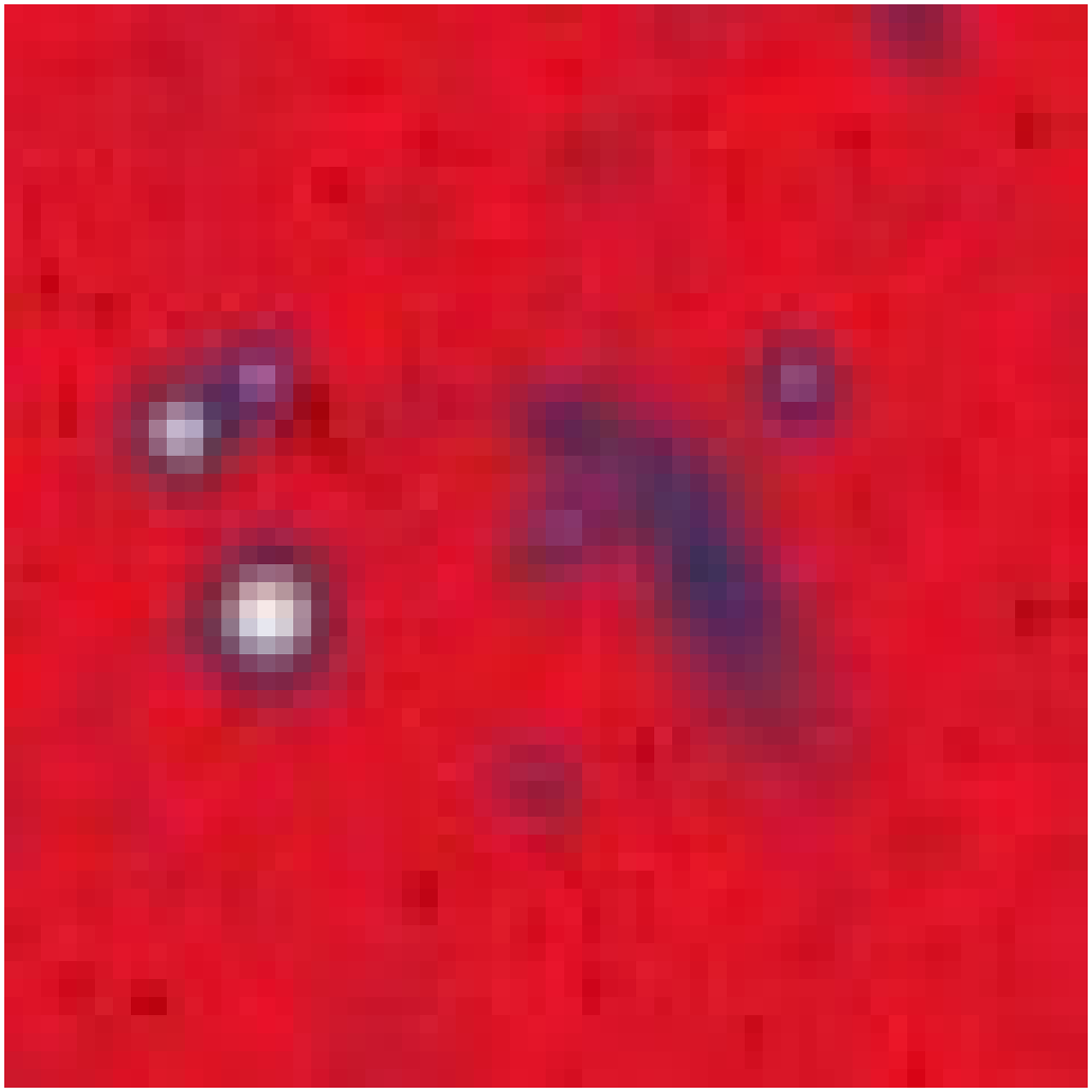}    \FGb{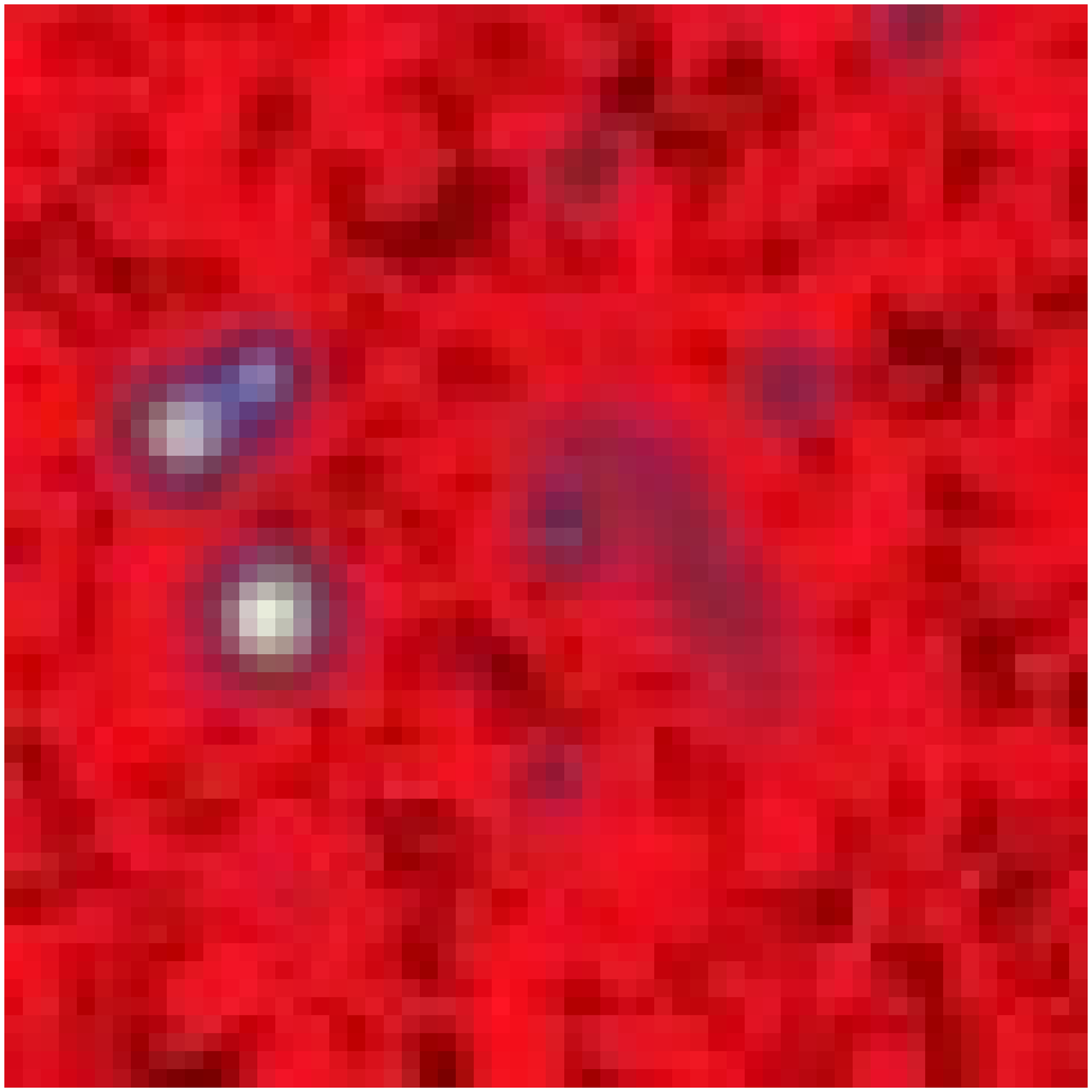}    \FGb{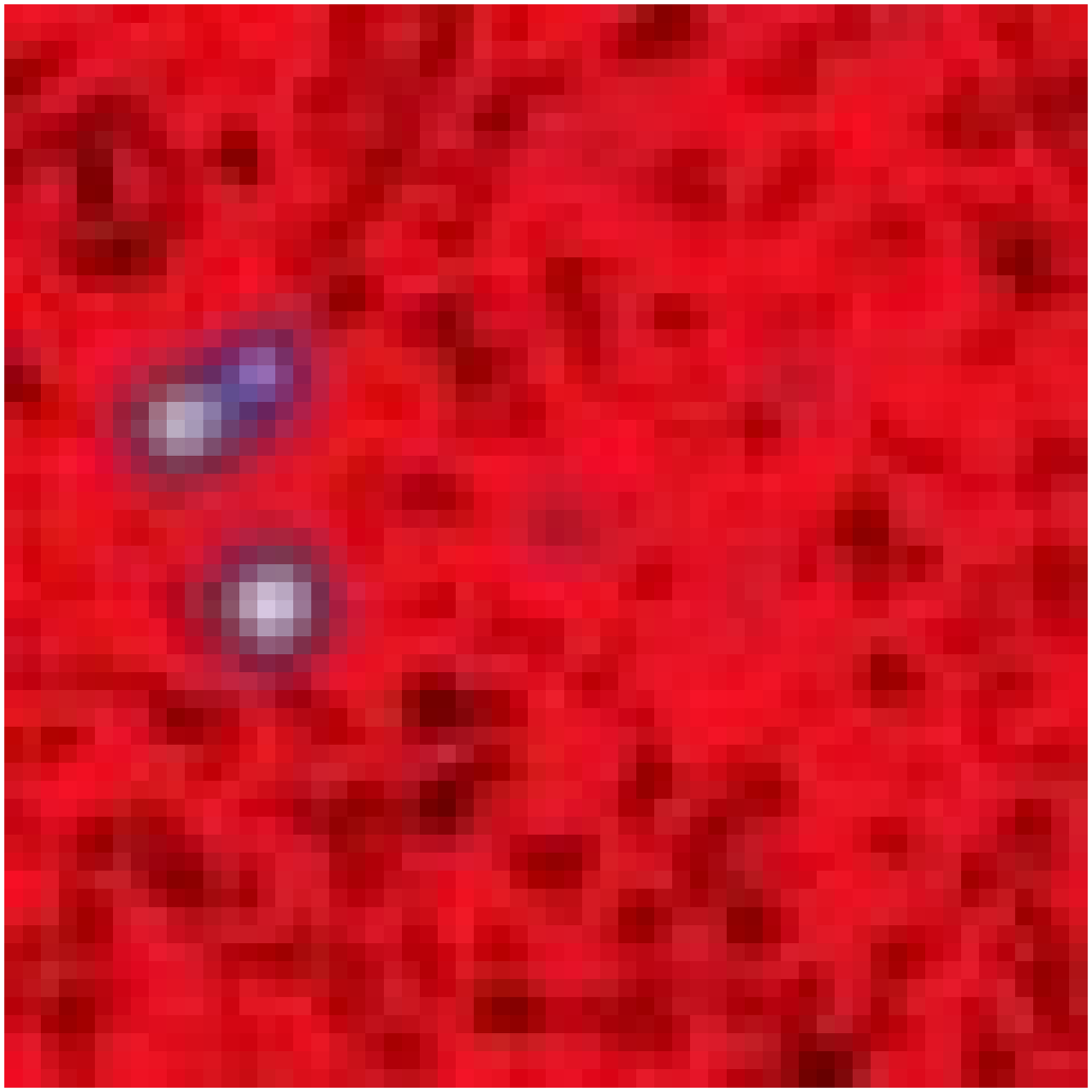}    \FGb{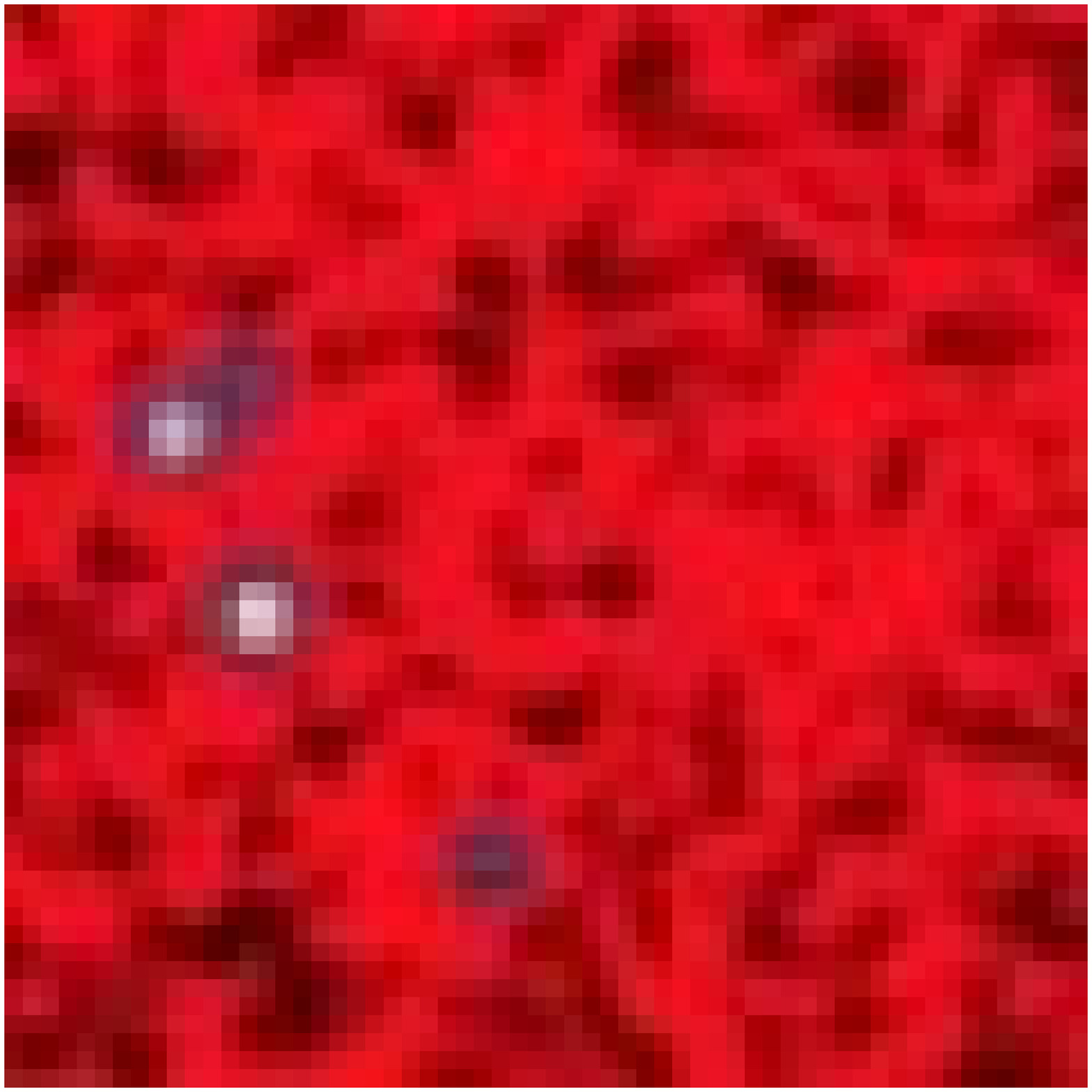}    \FGb{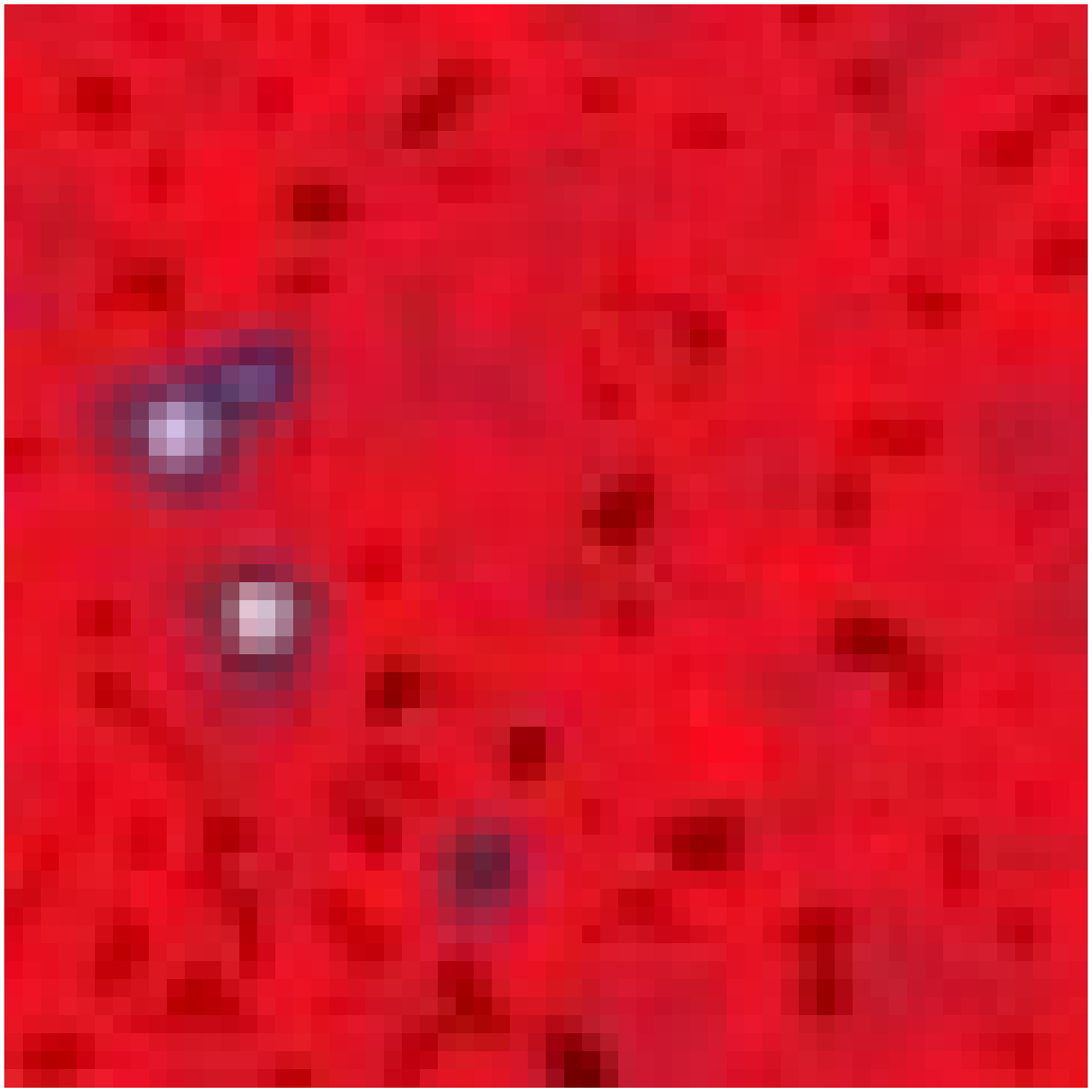}    \FGb{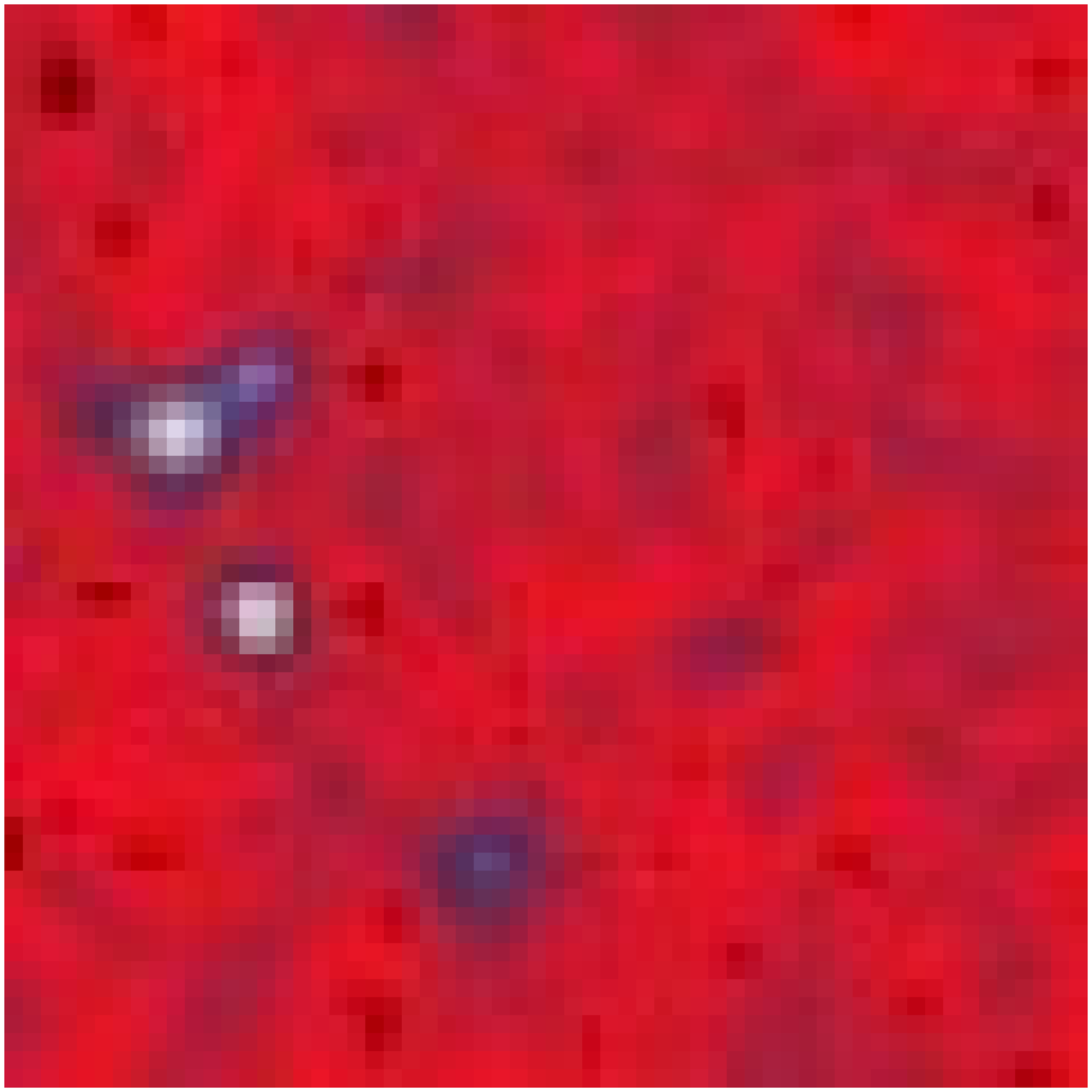}   \FGb{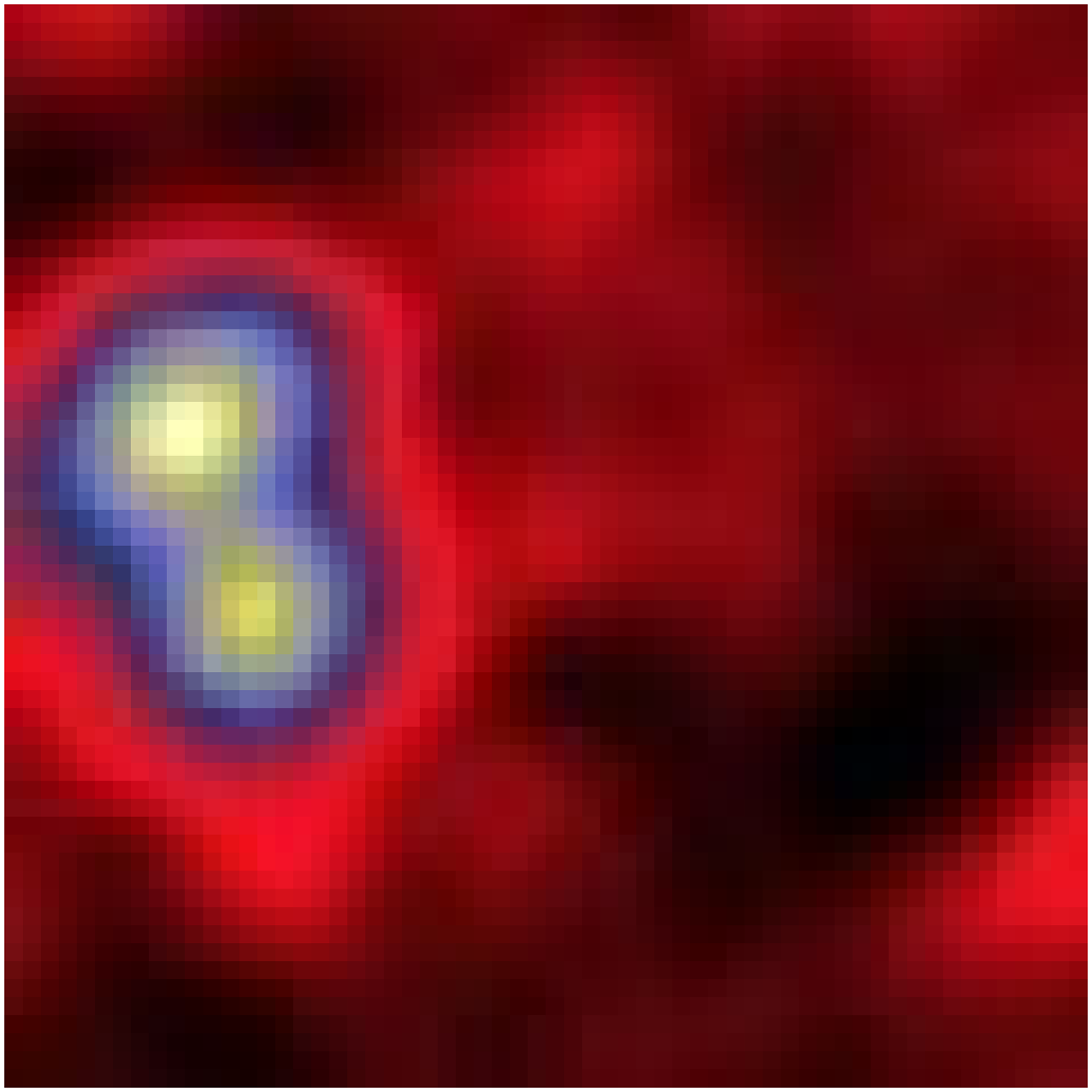}   \FGb{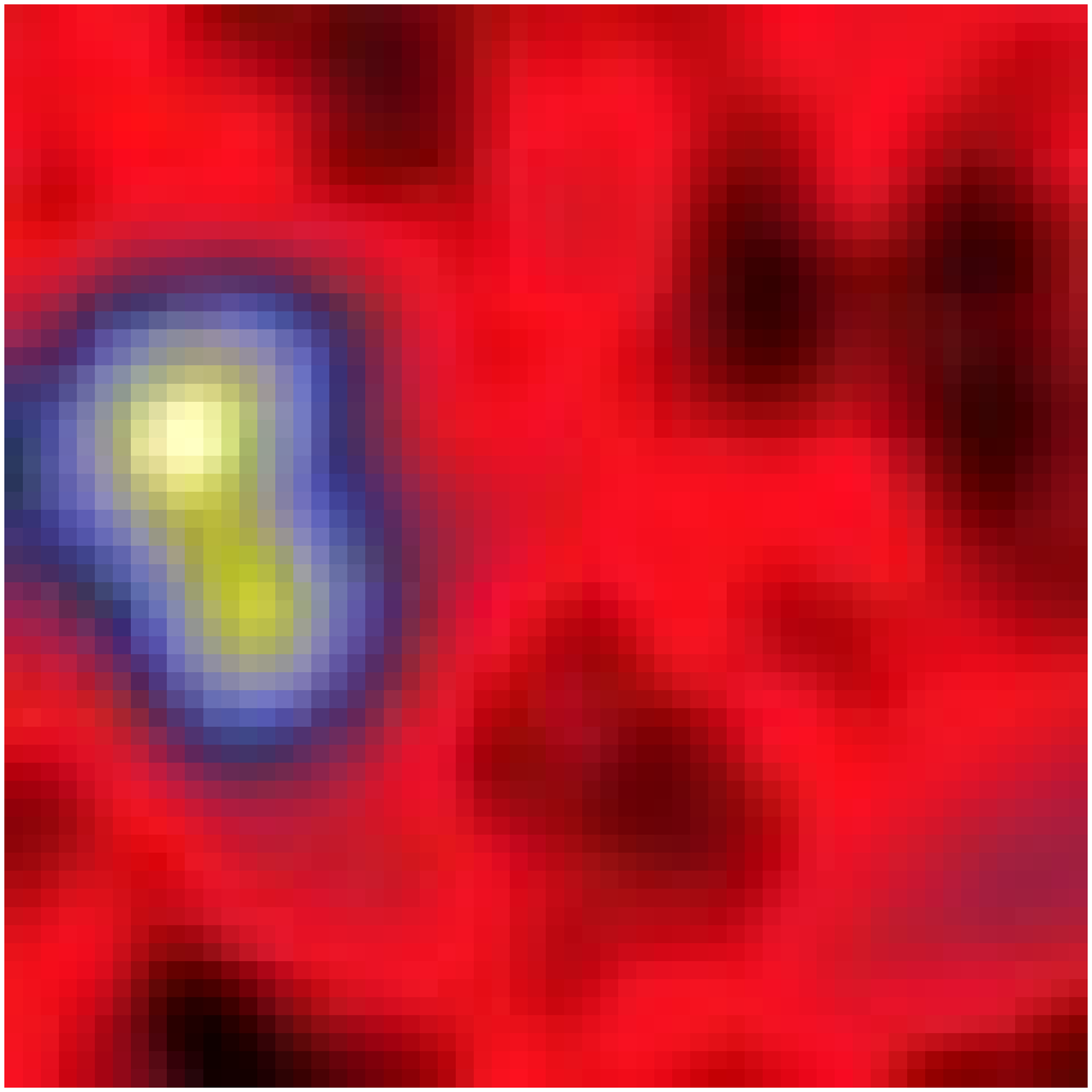}   \FGb{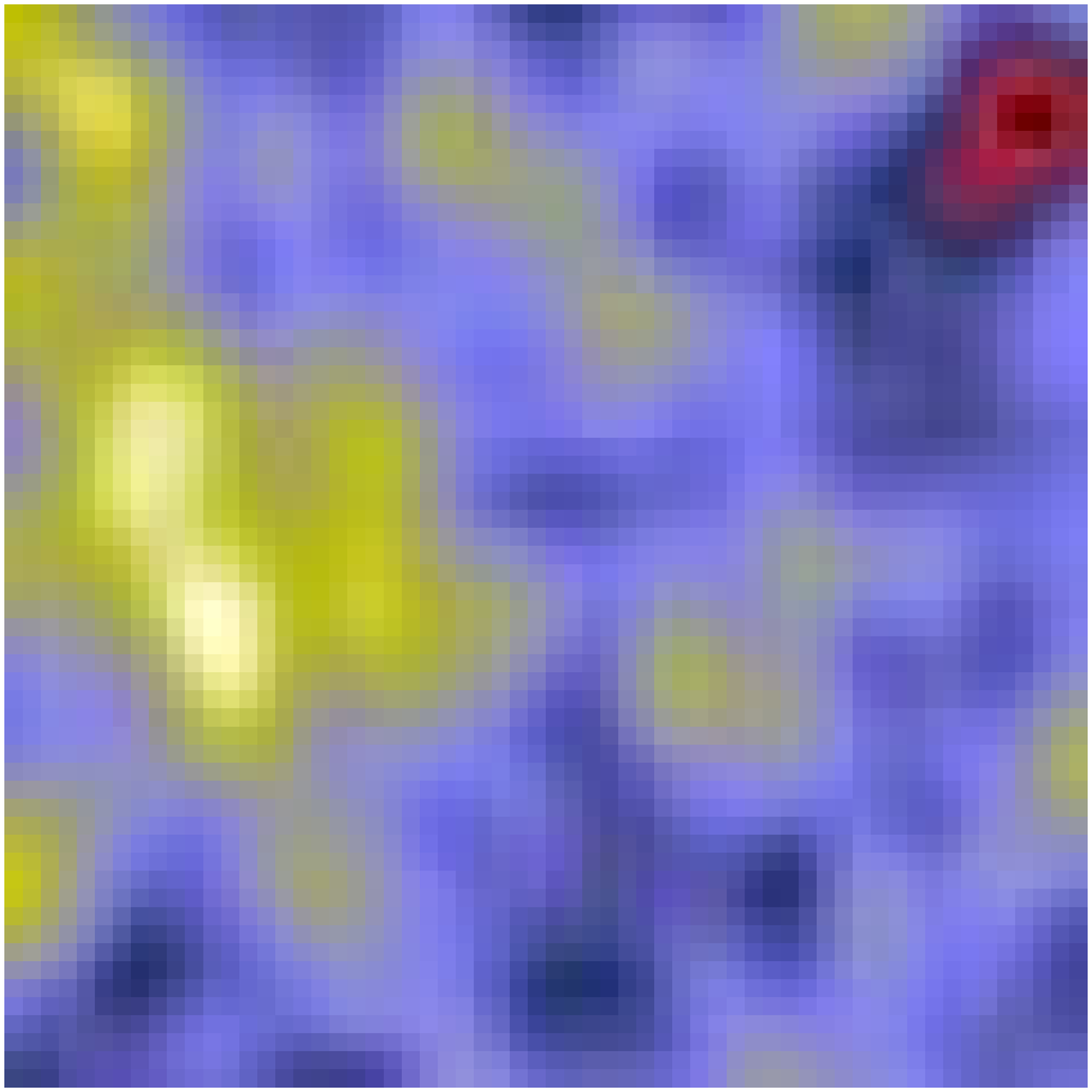}   \FGb{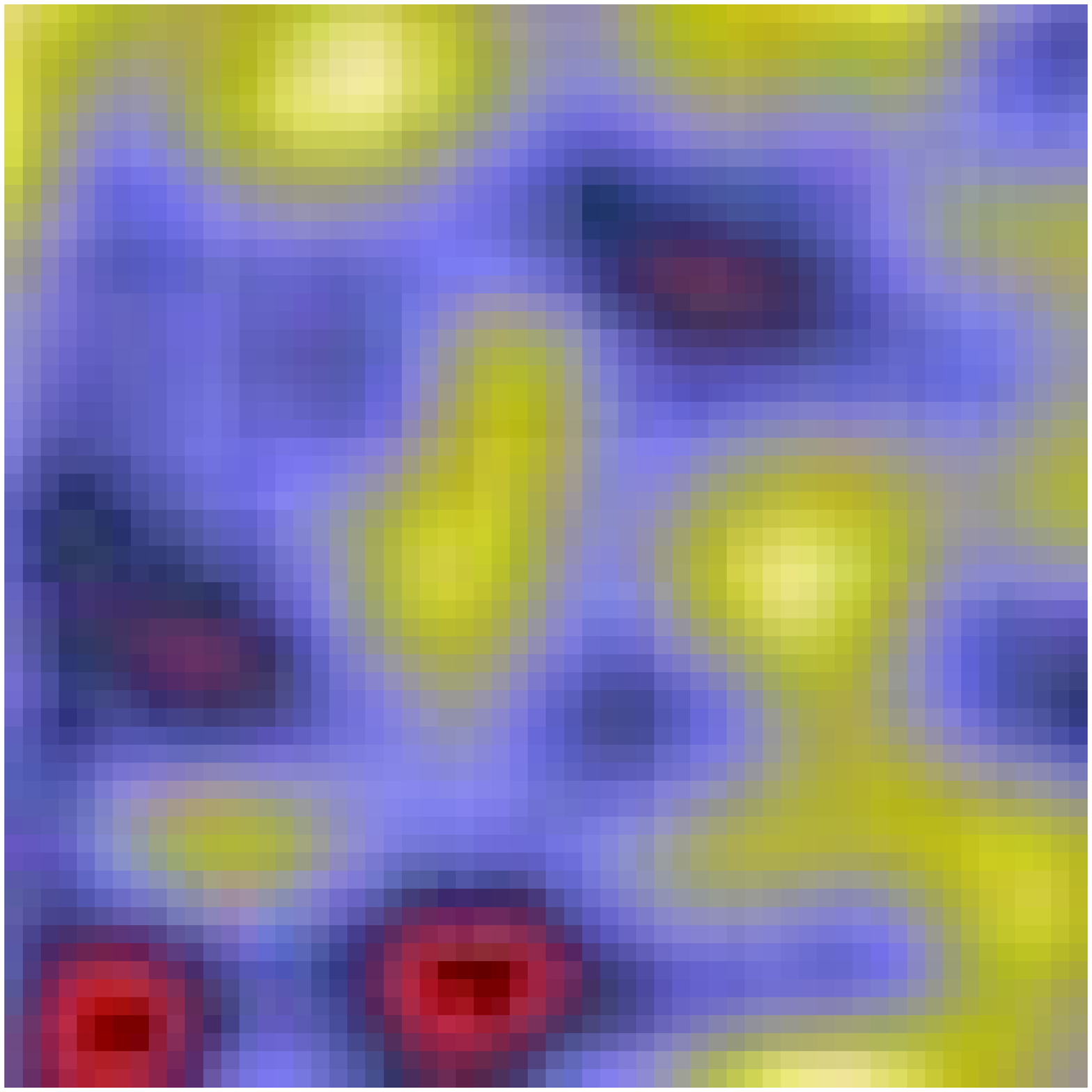} \\  {\#1   NCIA029460}  \\
  \FGb{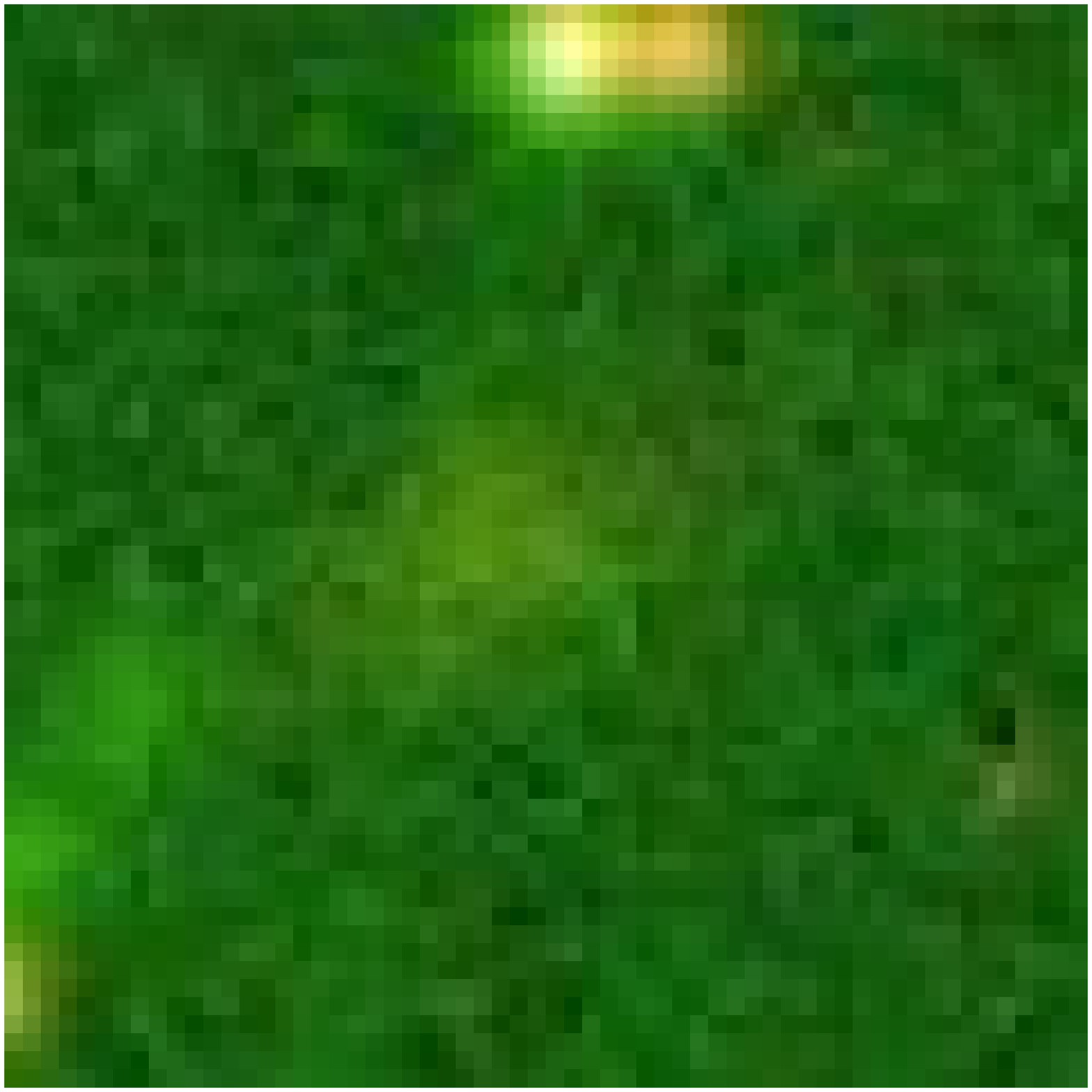}  \FGb{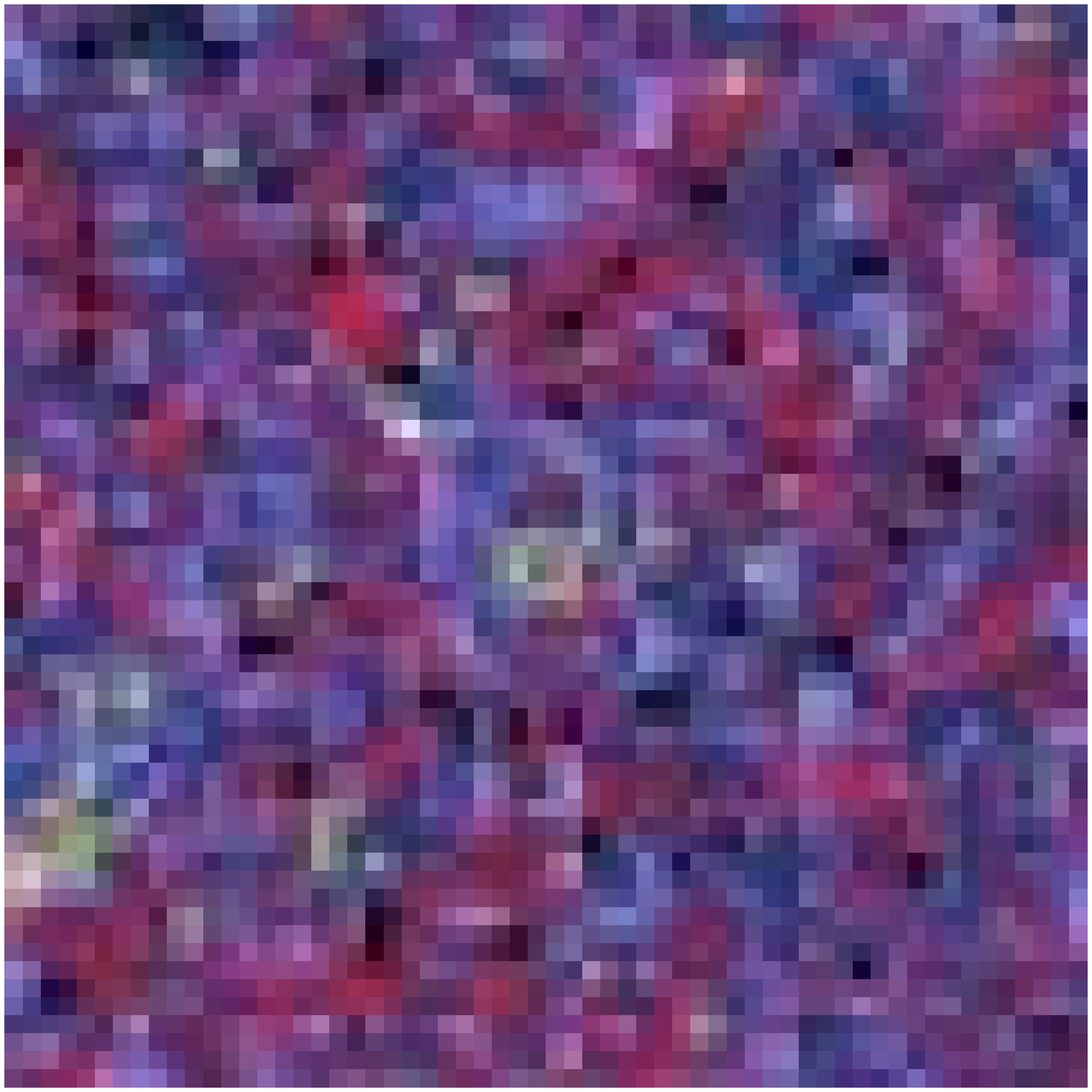}    \FGb{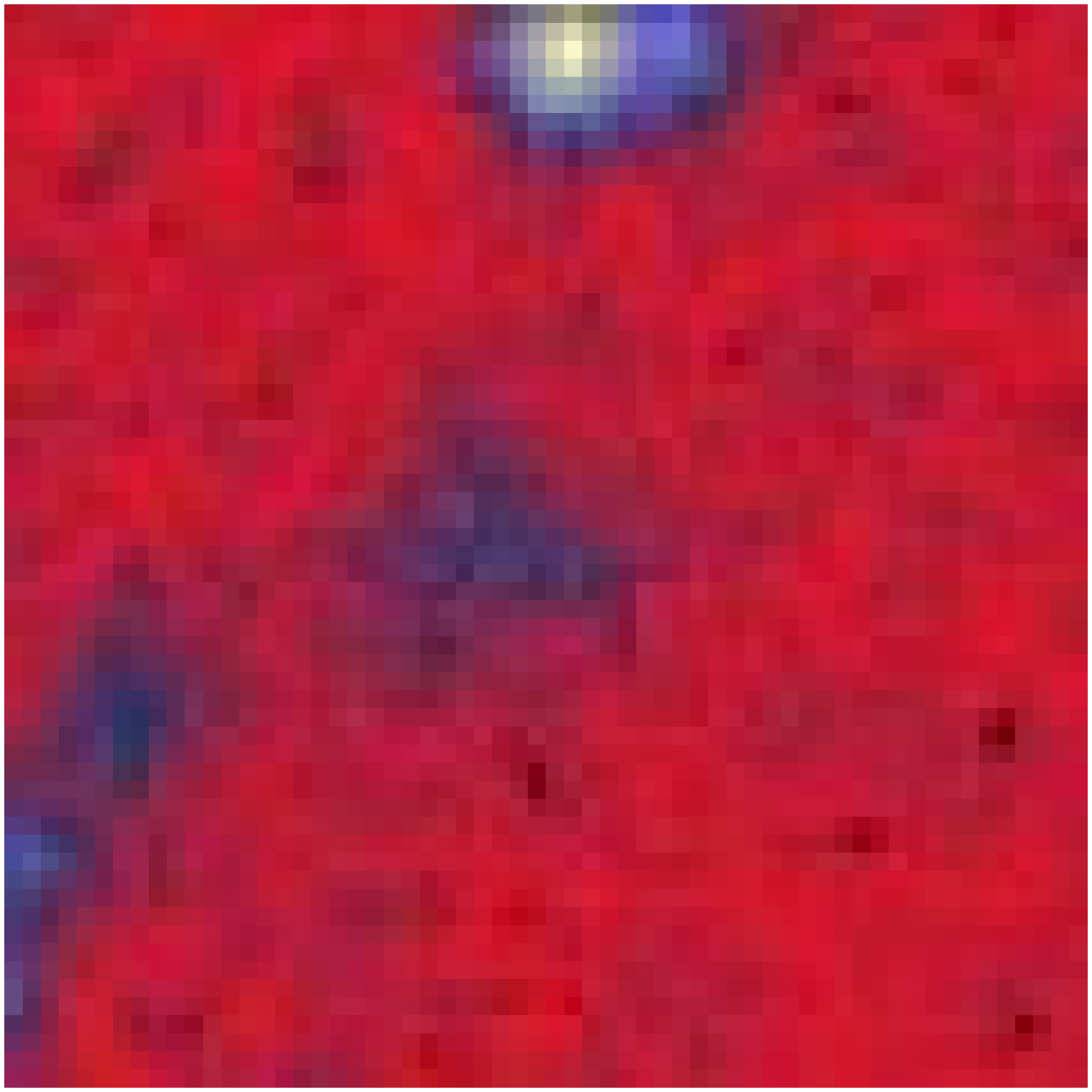}   \FGb{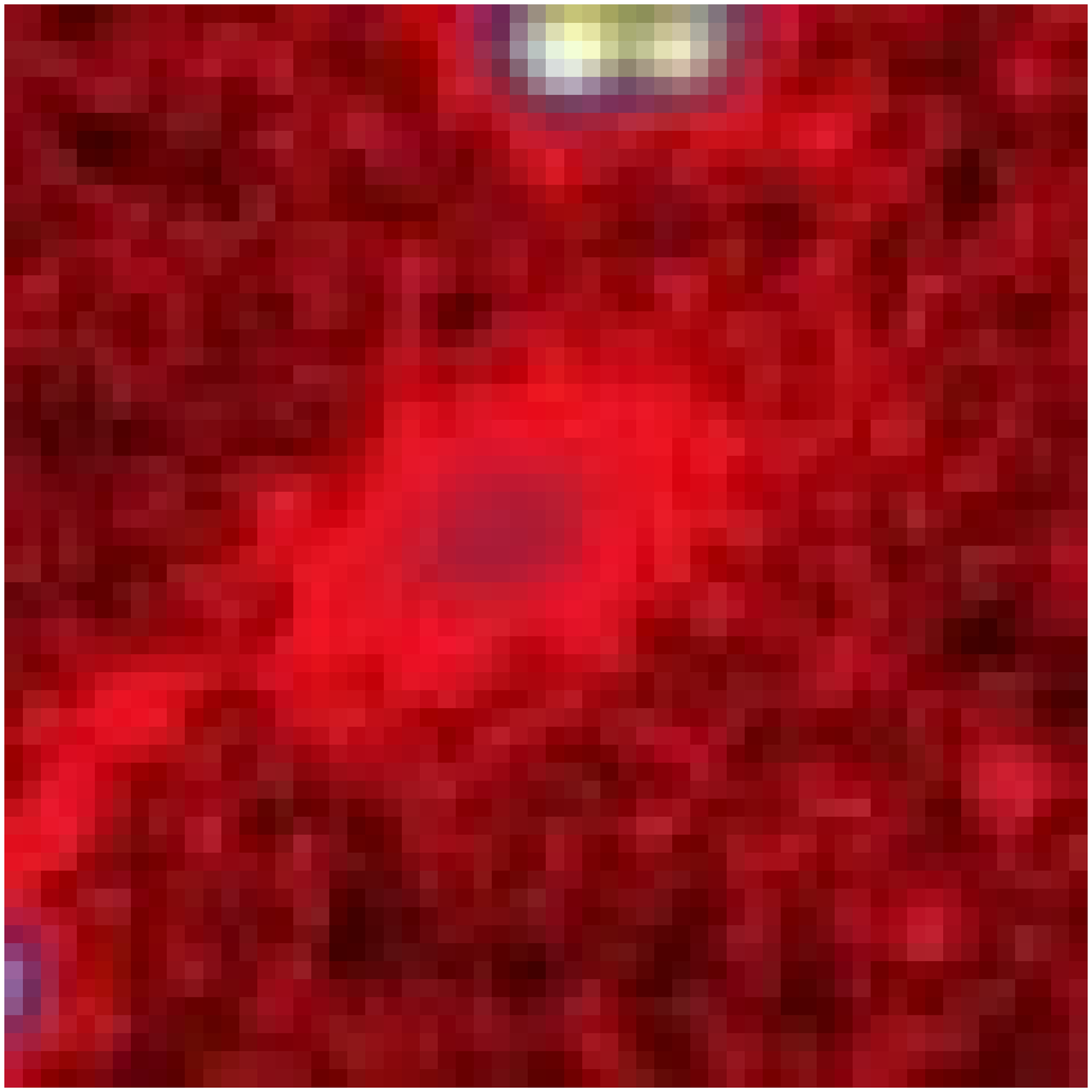}   \FGb{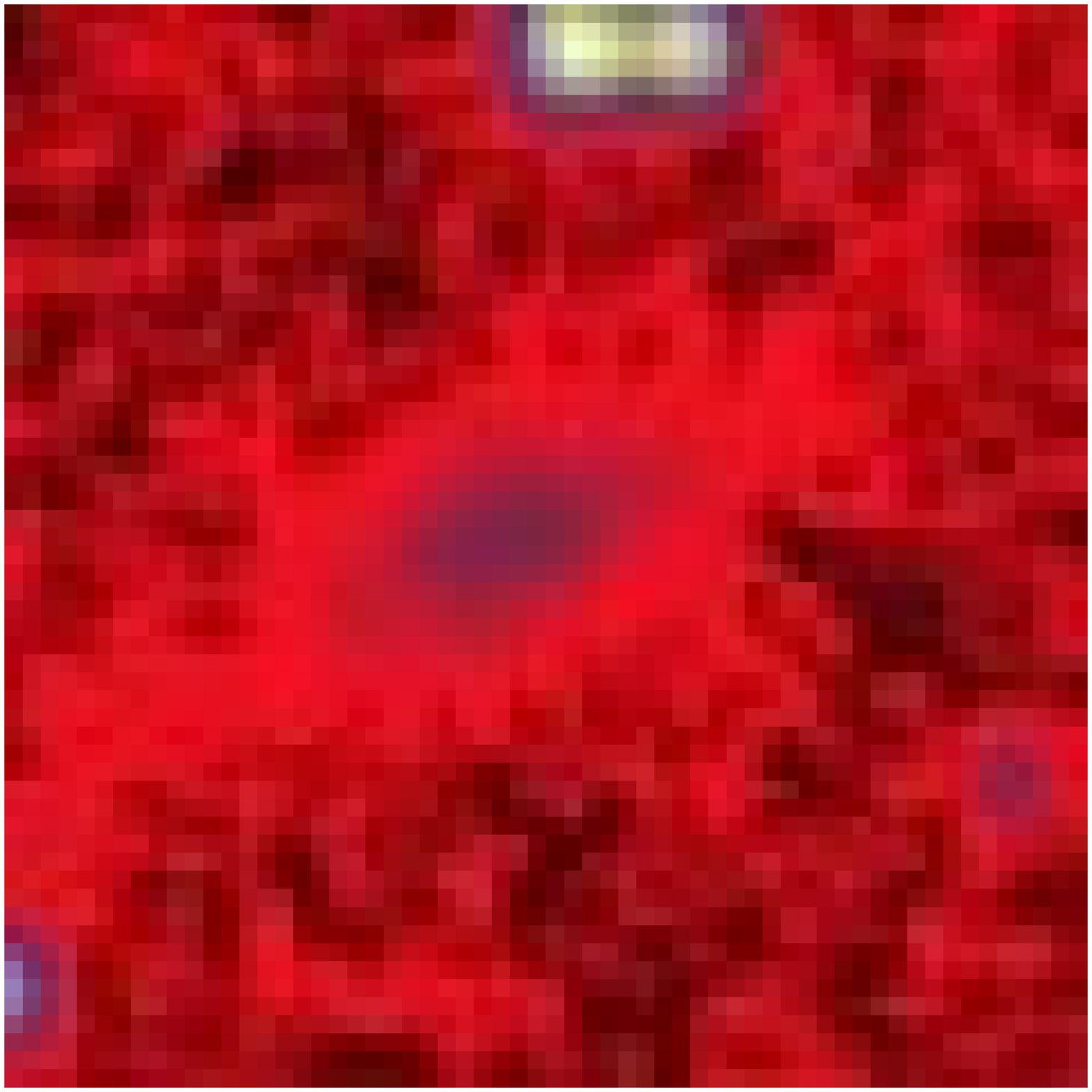}   \FGb{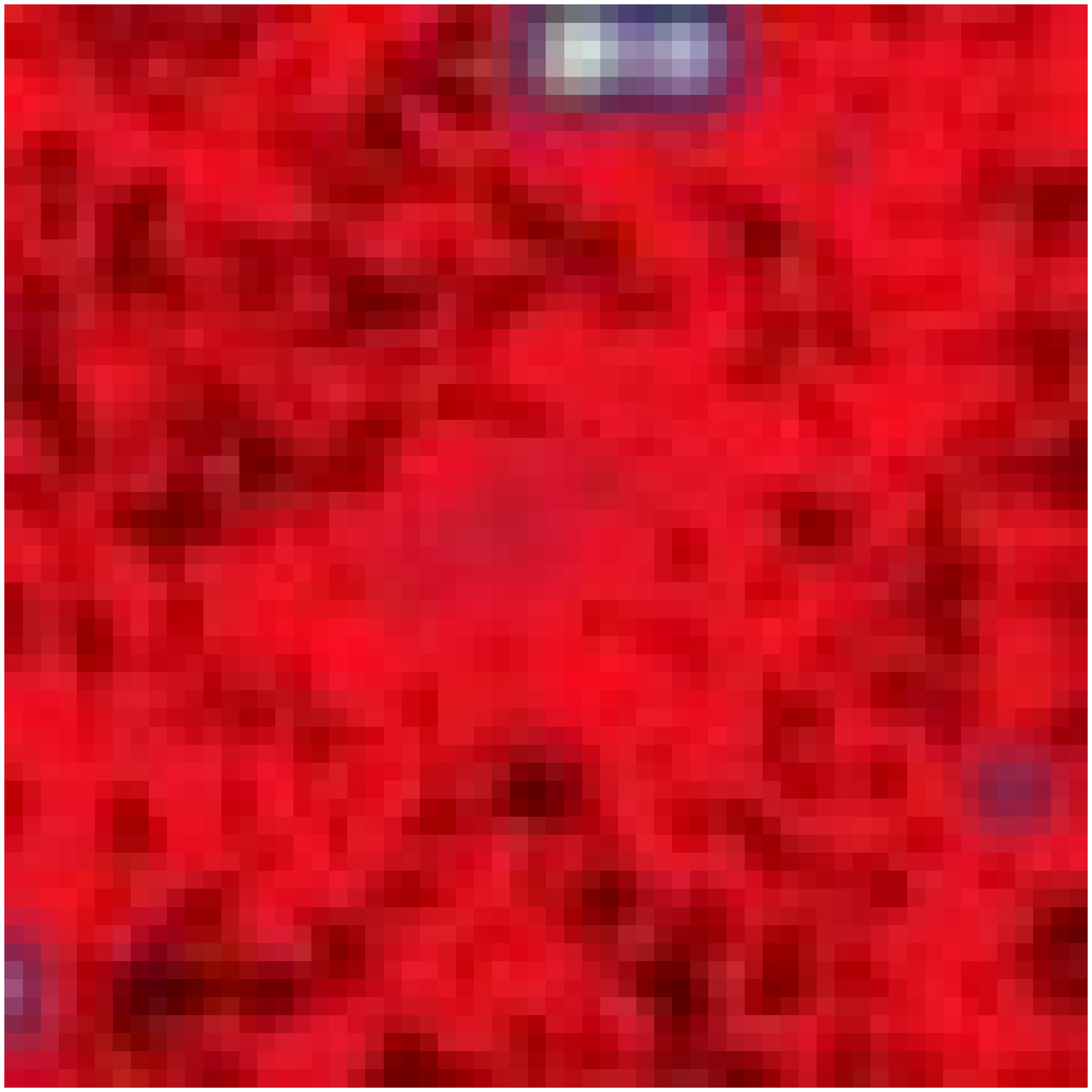}   \FGb{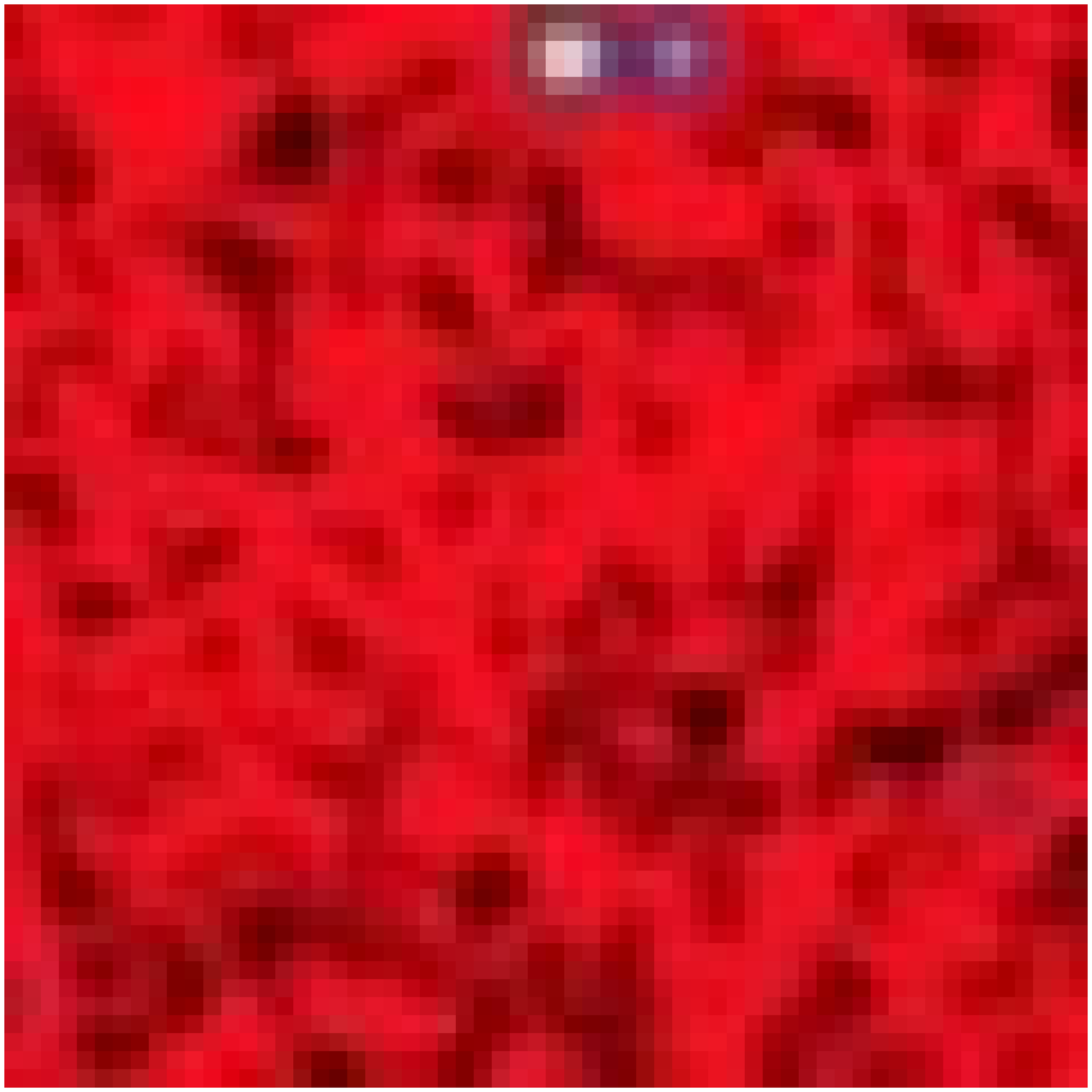}   \FGb{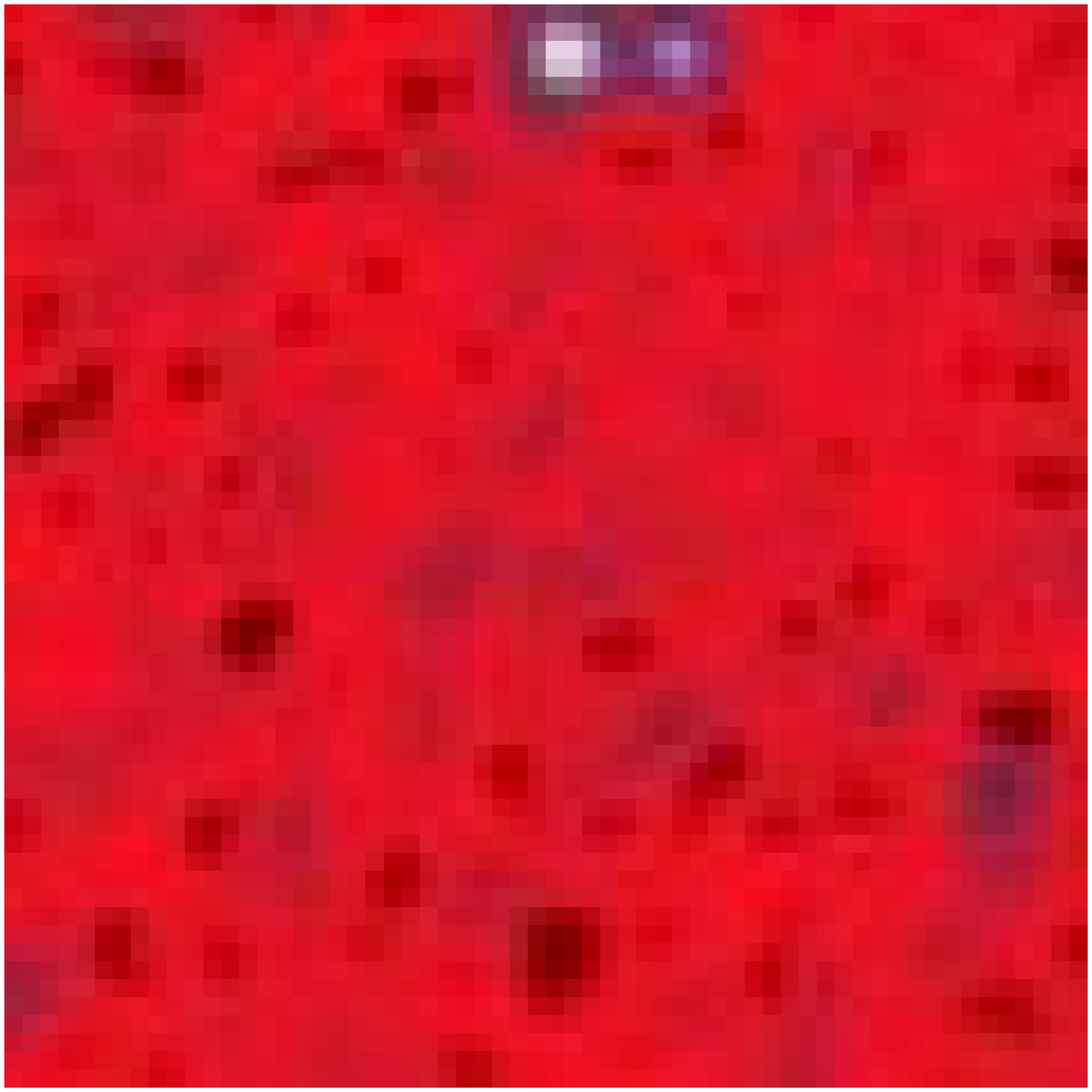}   \FGb{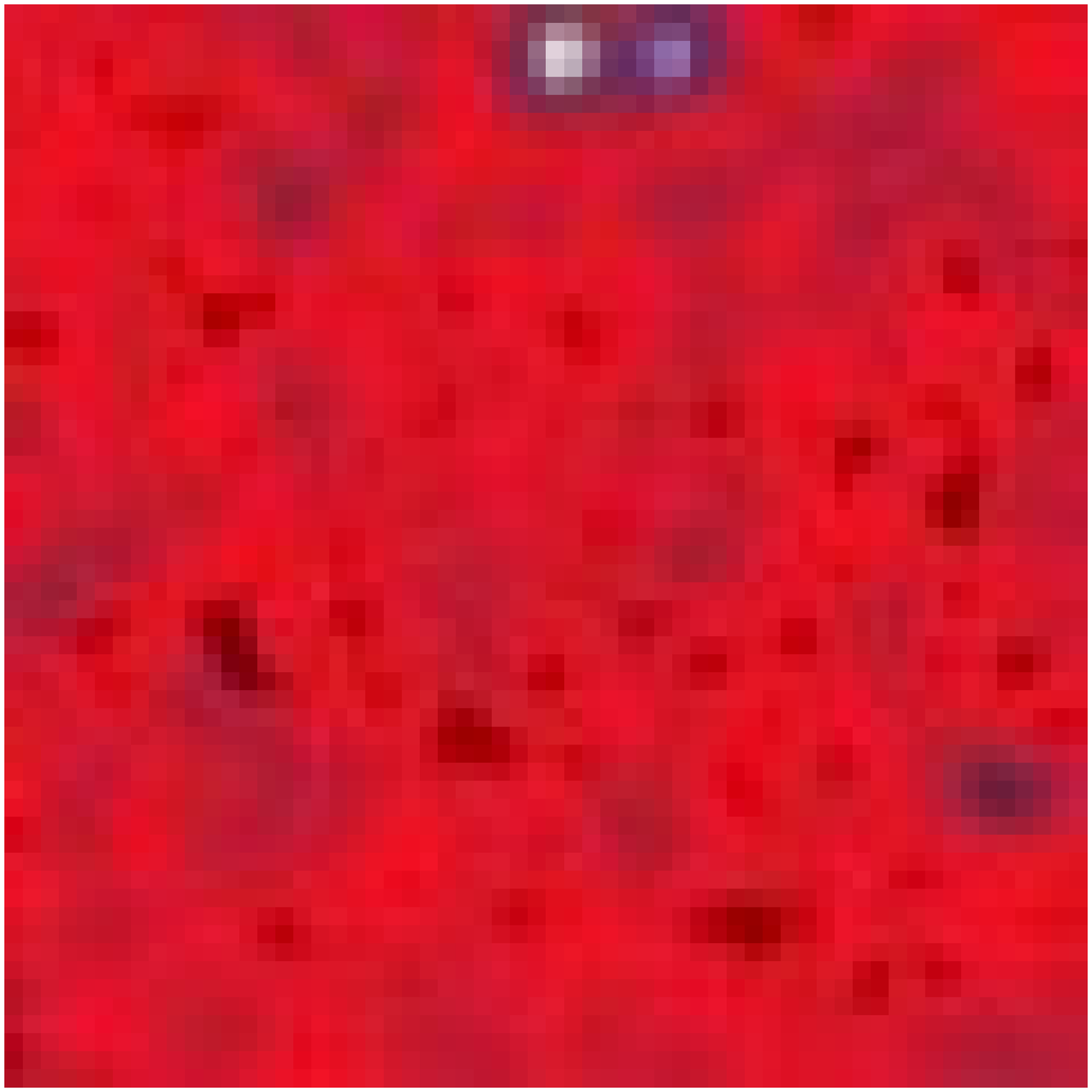}   \FGb{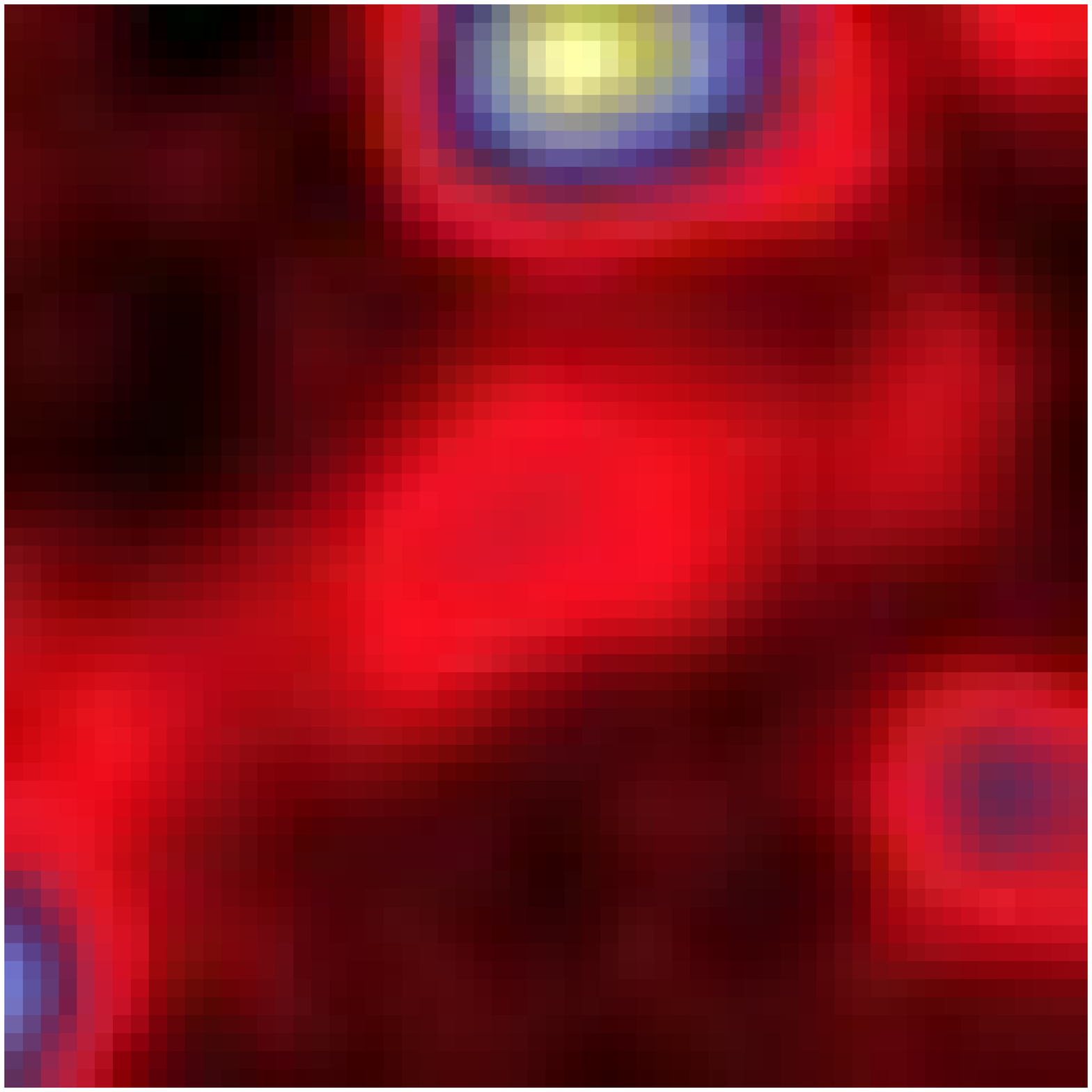}   \FGb{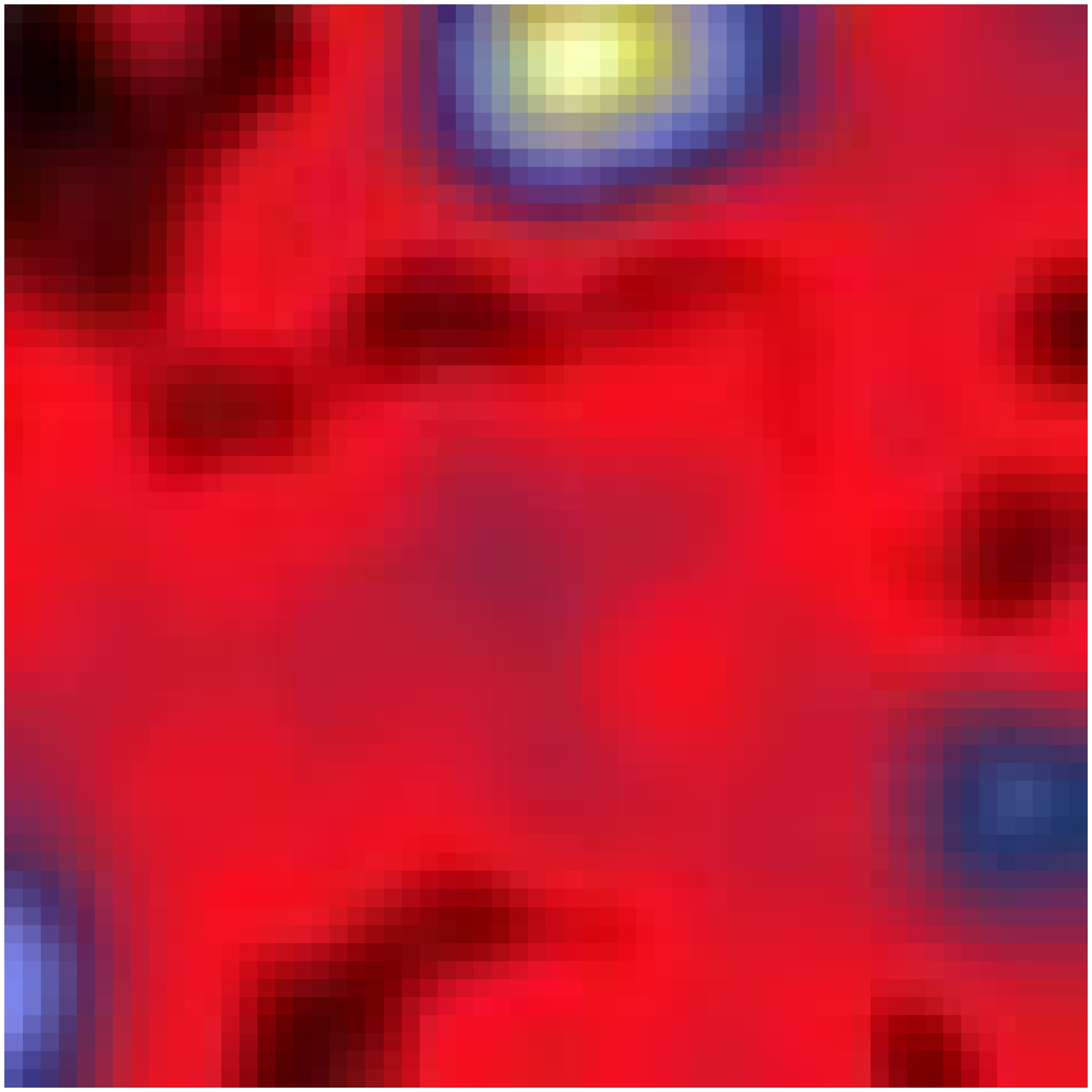}   \FGb{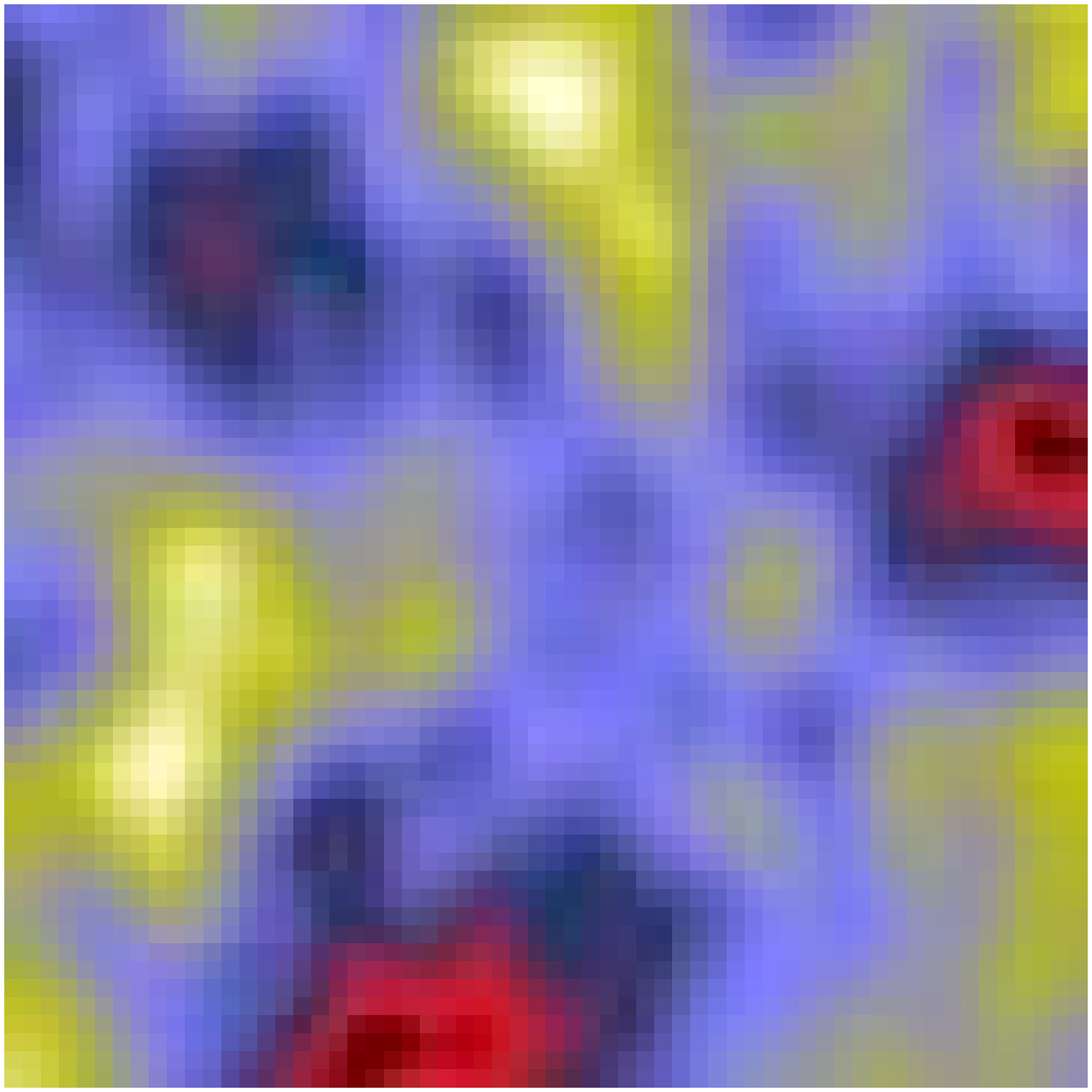}   \FGb{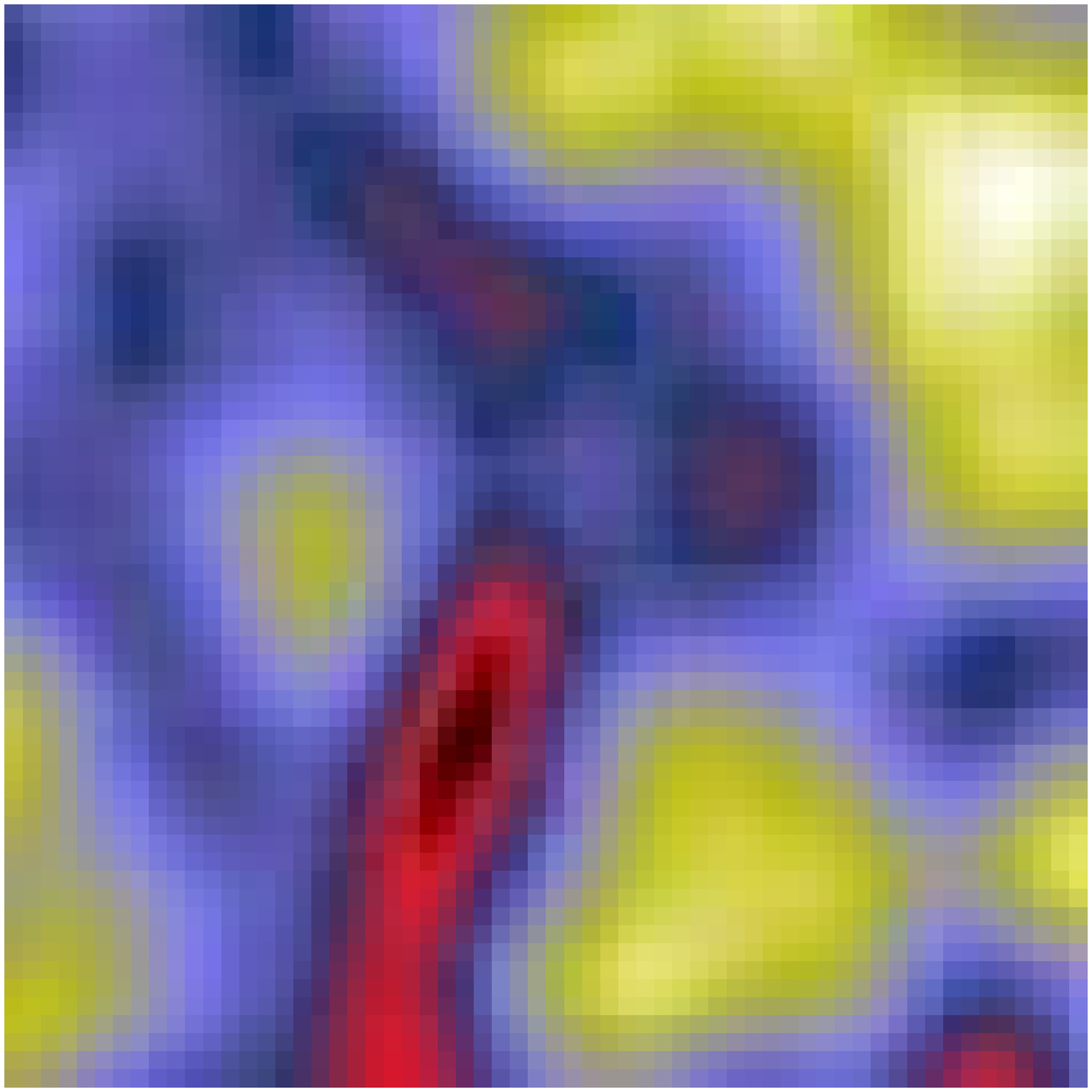} \\  {\#2   NCIA018969}  \\
  \FGb{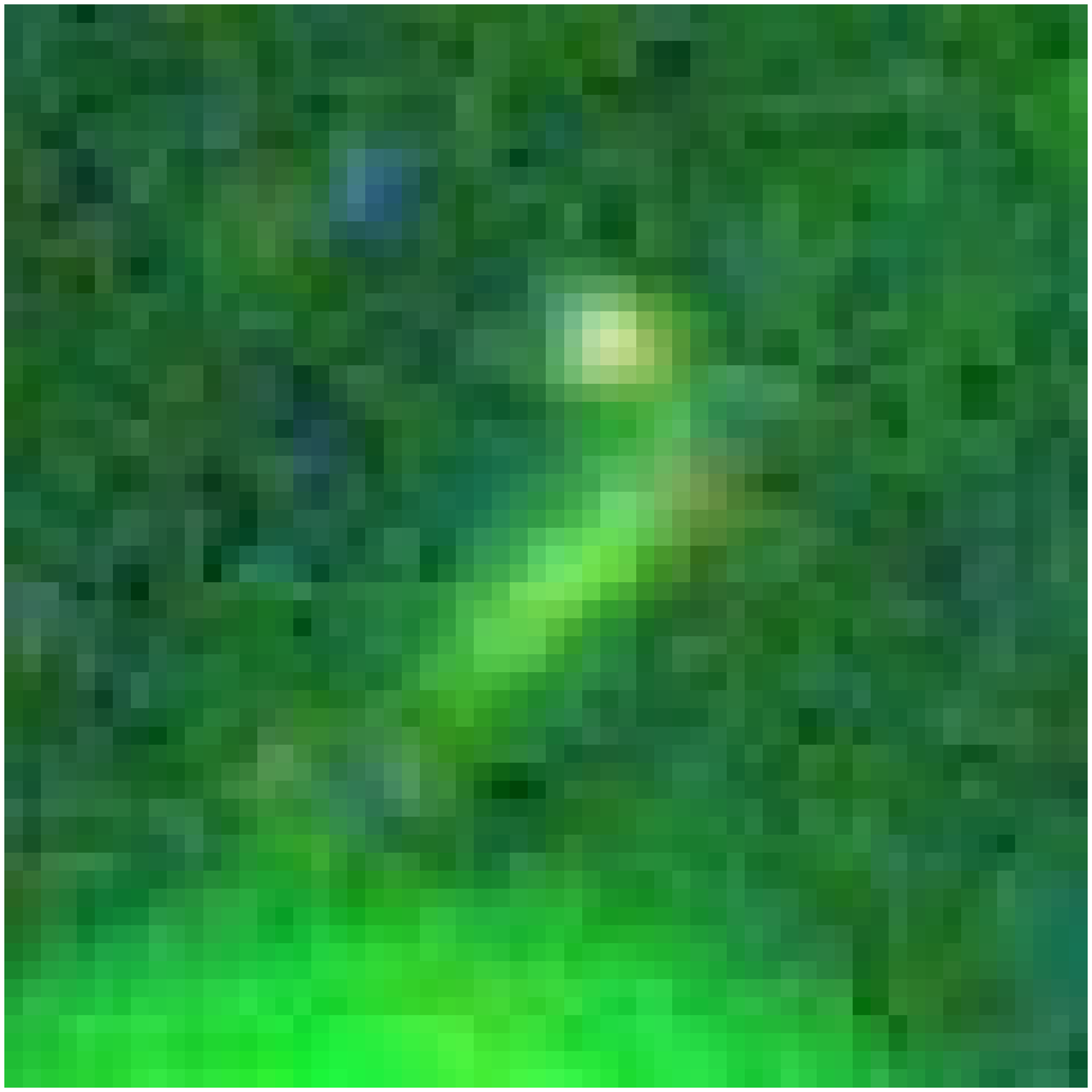}  \FGb{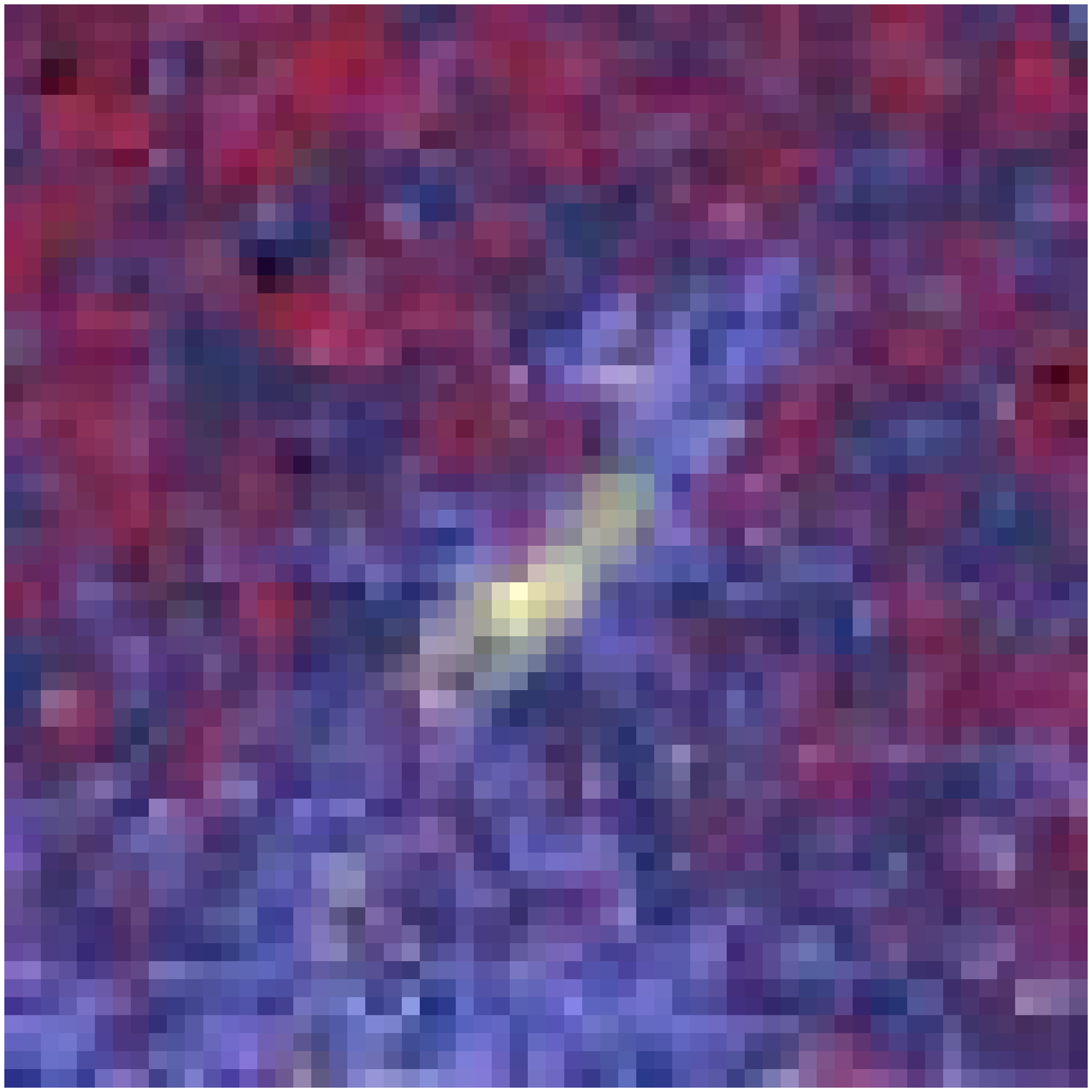}    \FGb{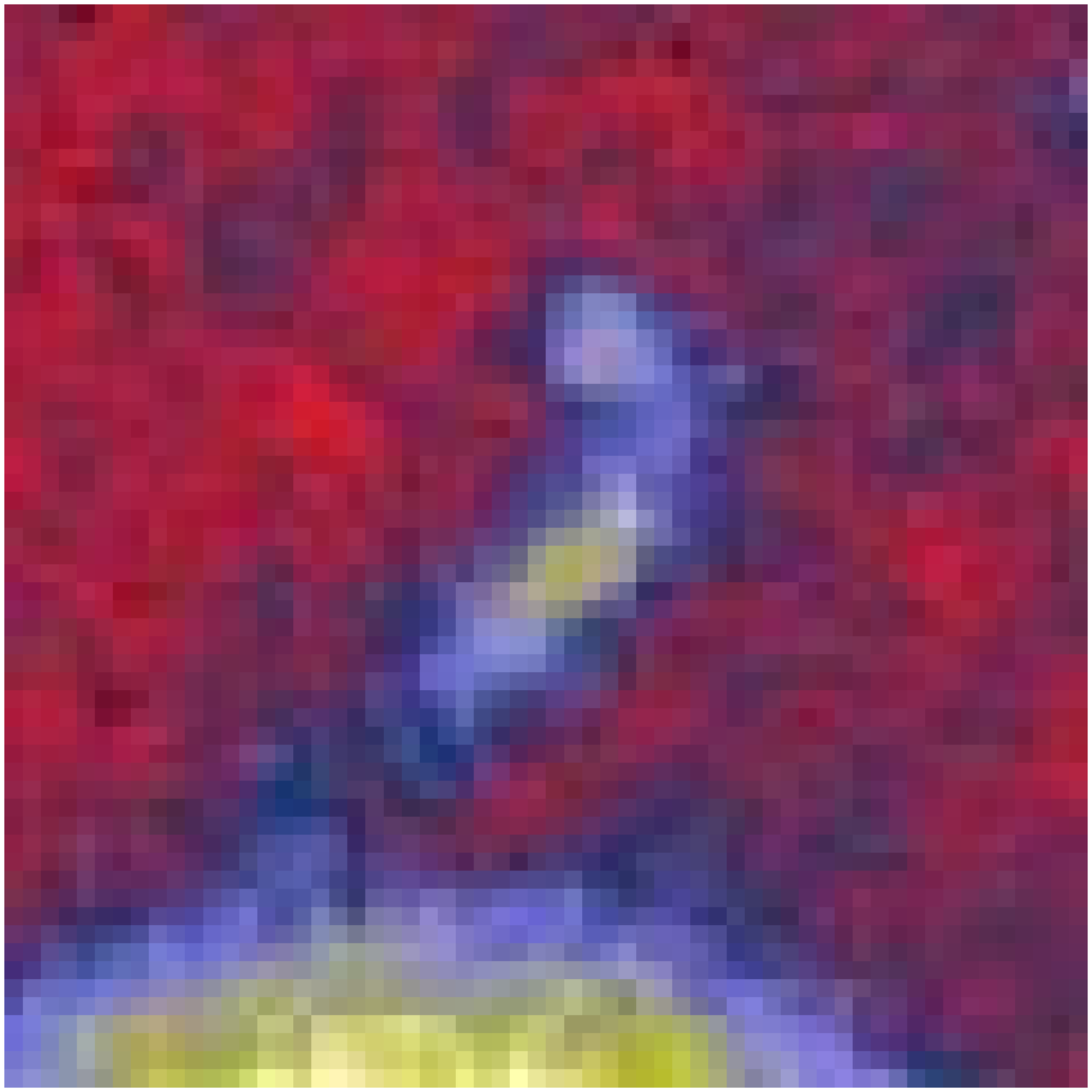}   \FGb{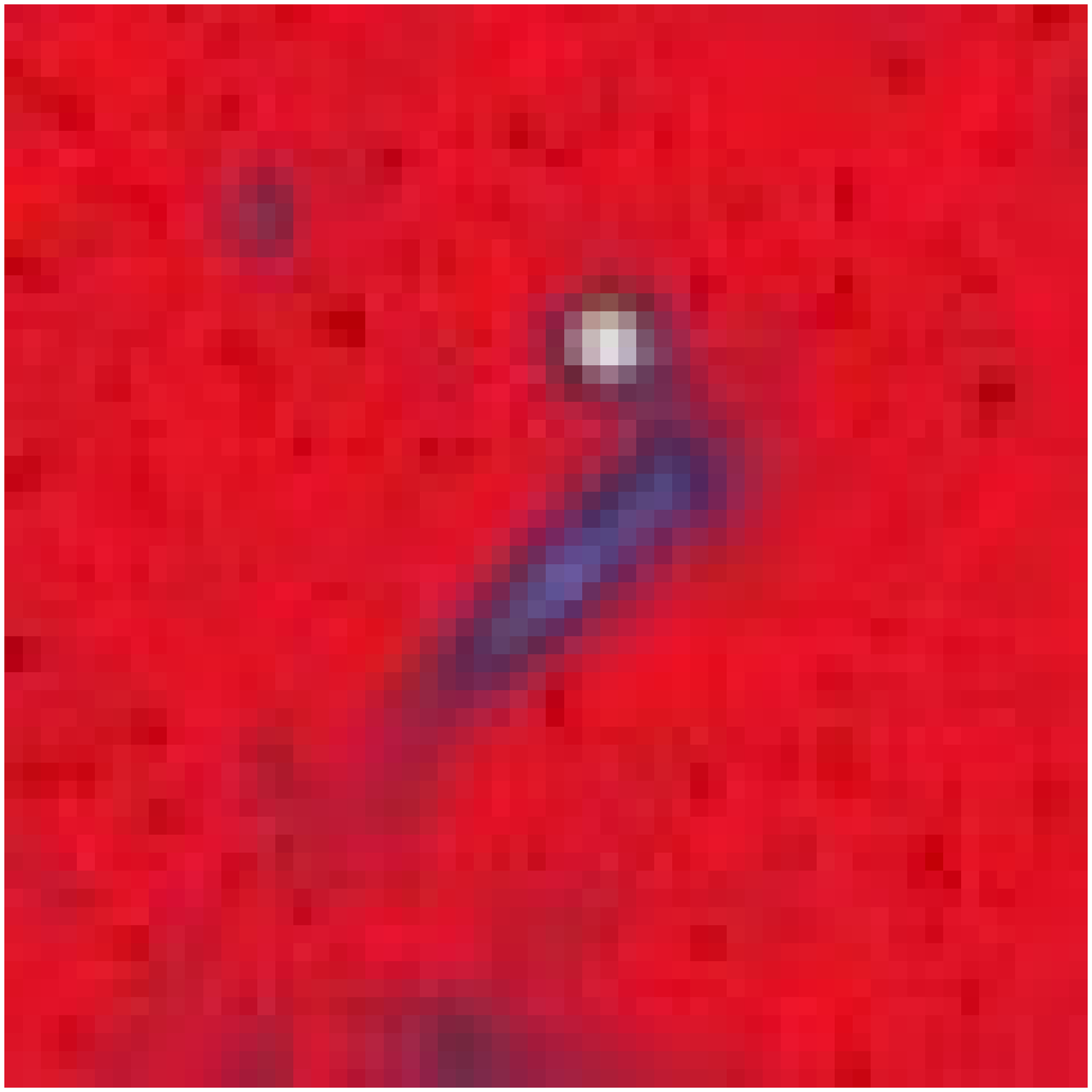}   \FGb{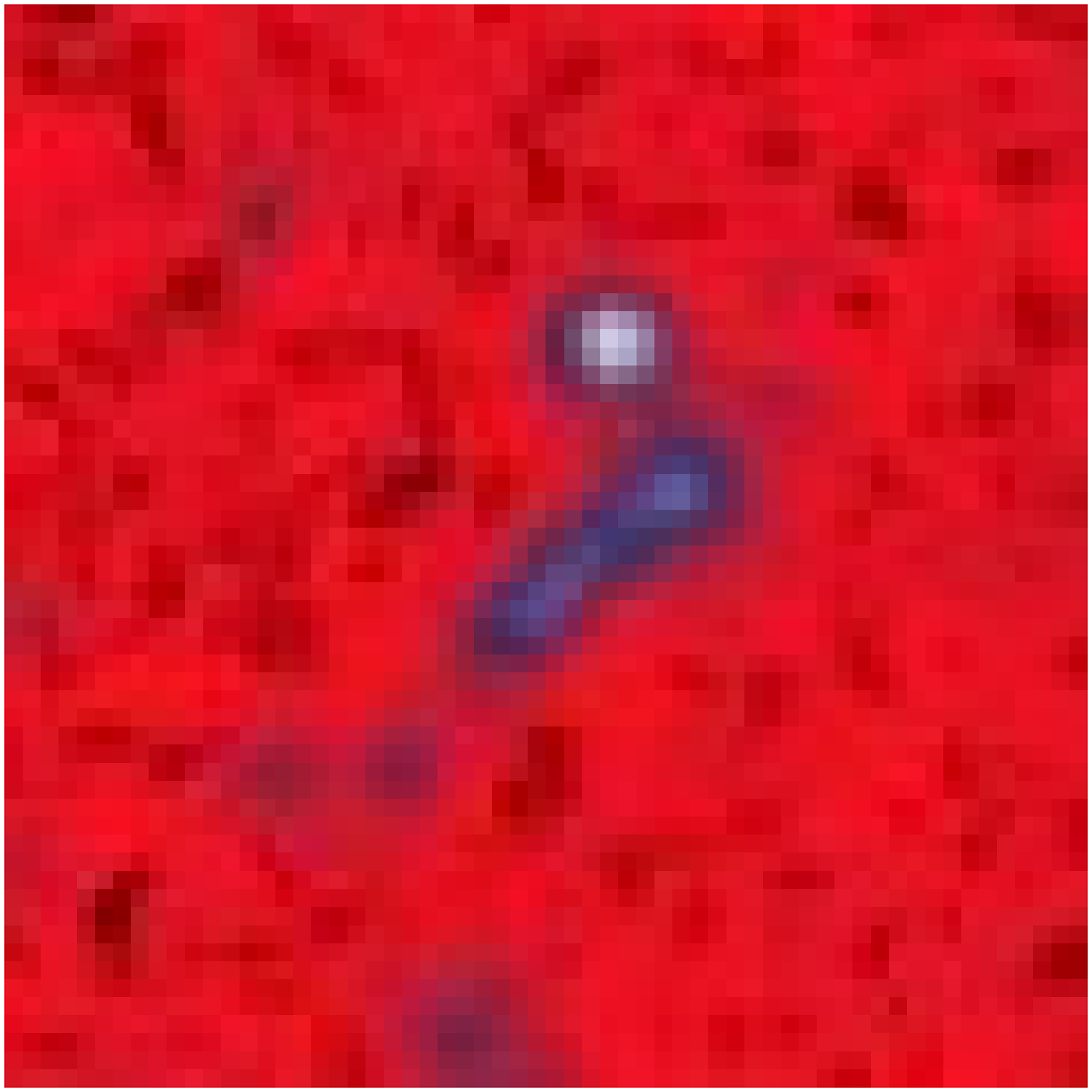}   \FGb{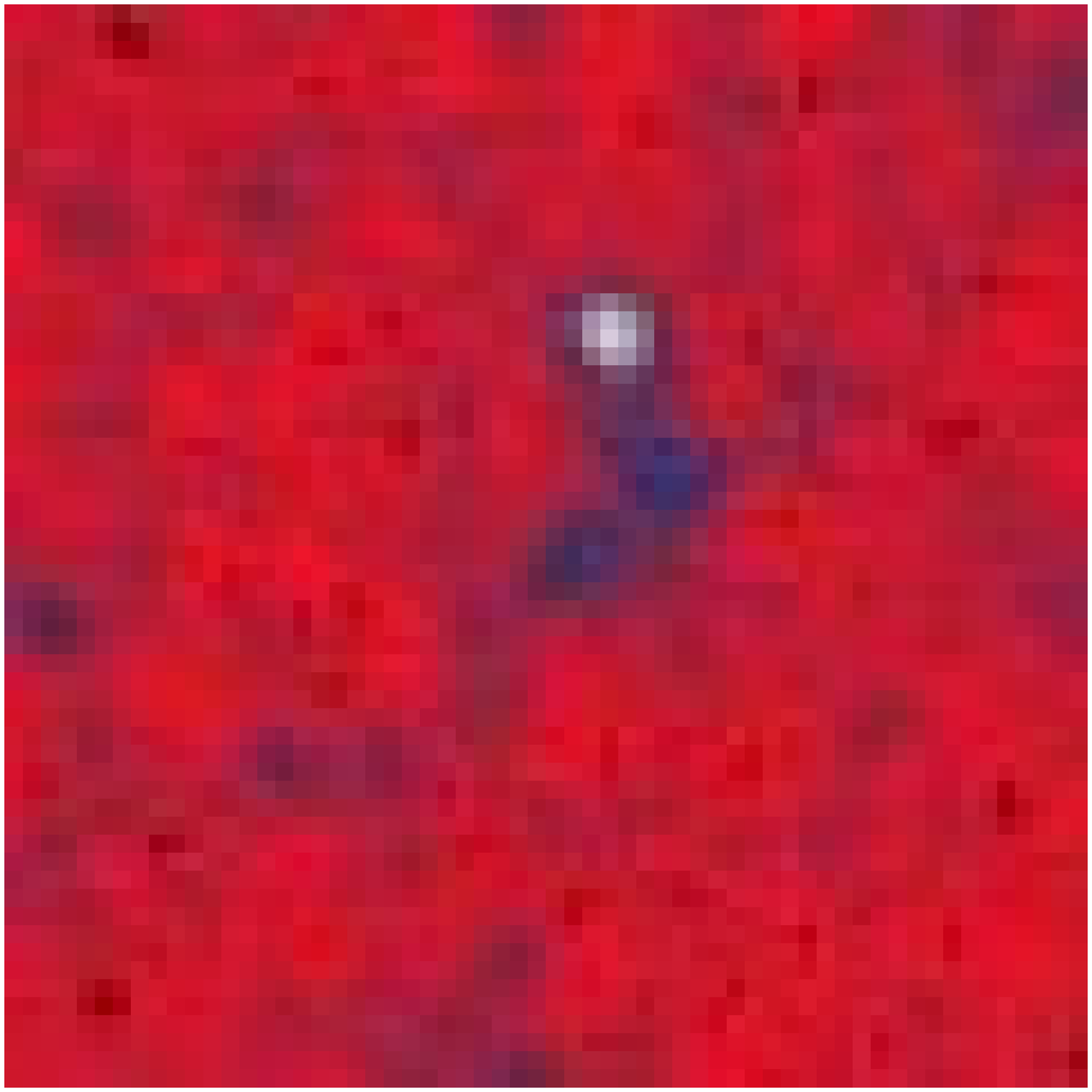}   \FGb{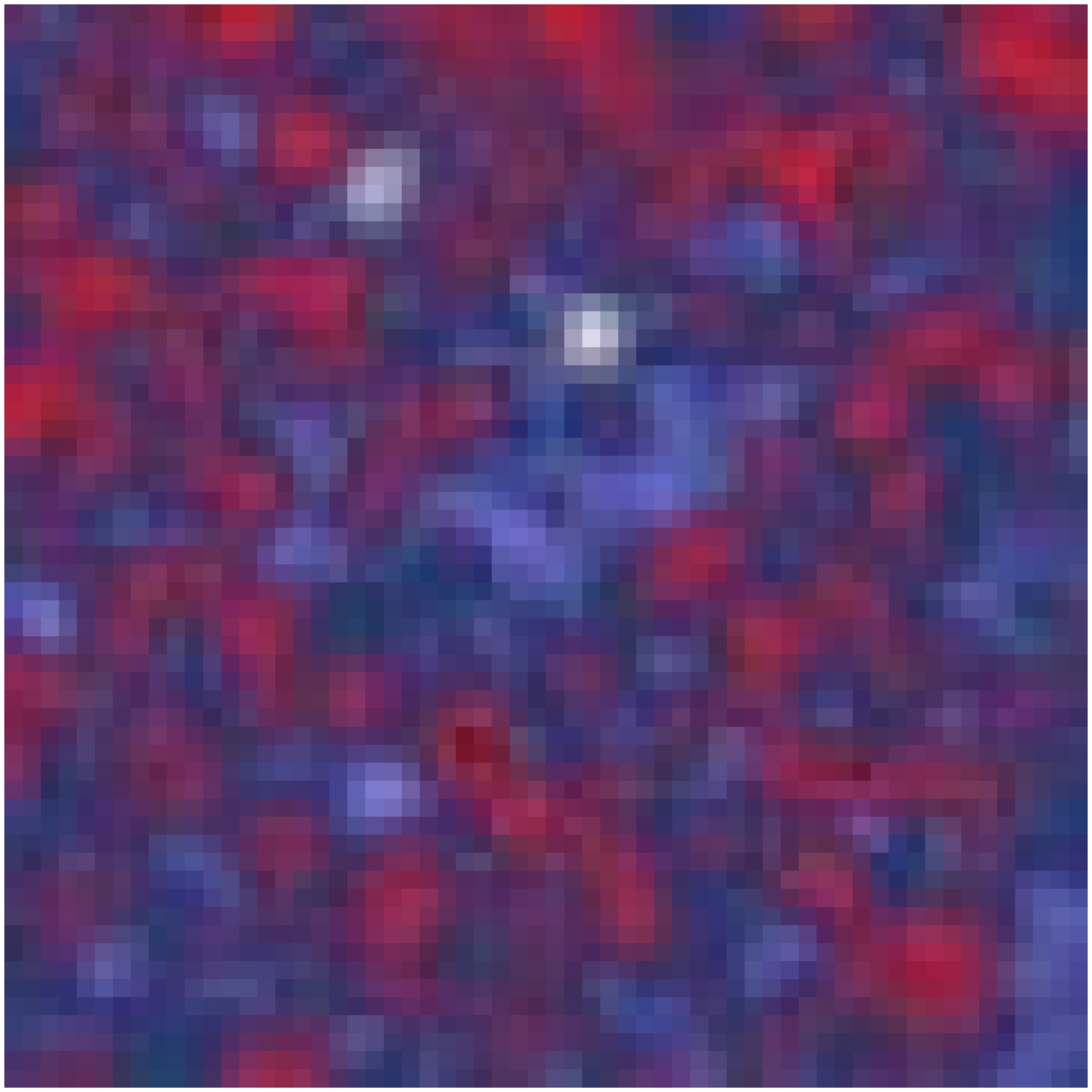}   \FGb{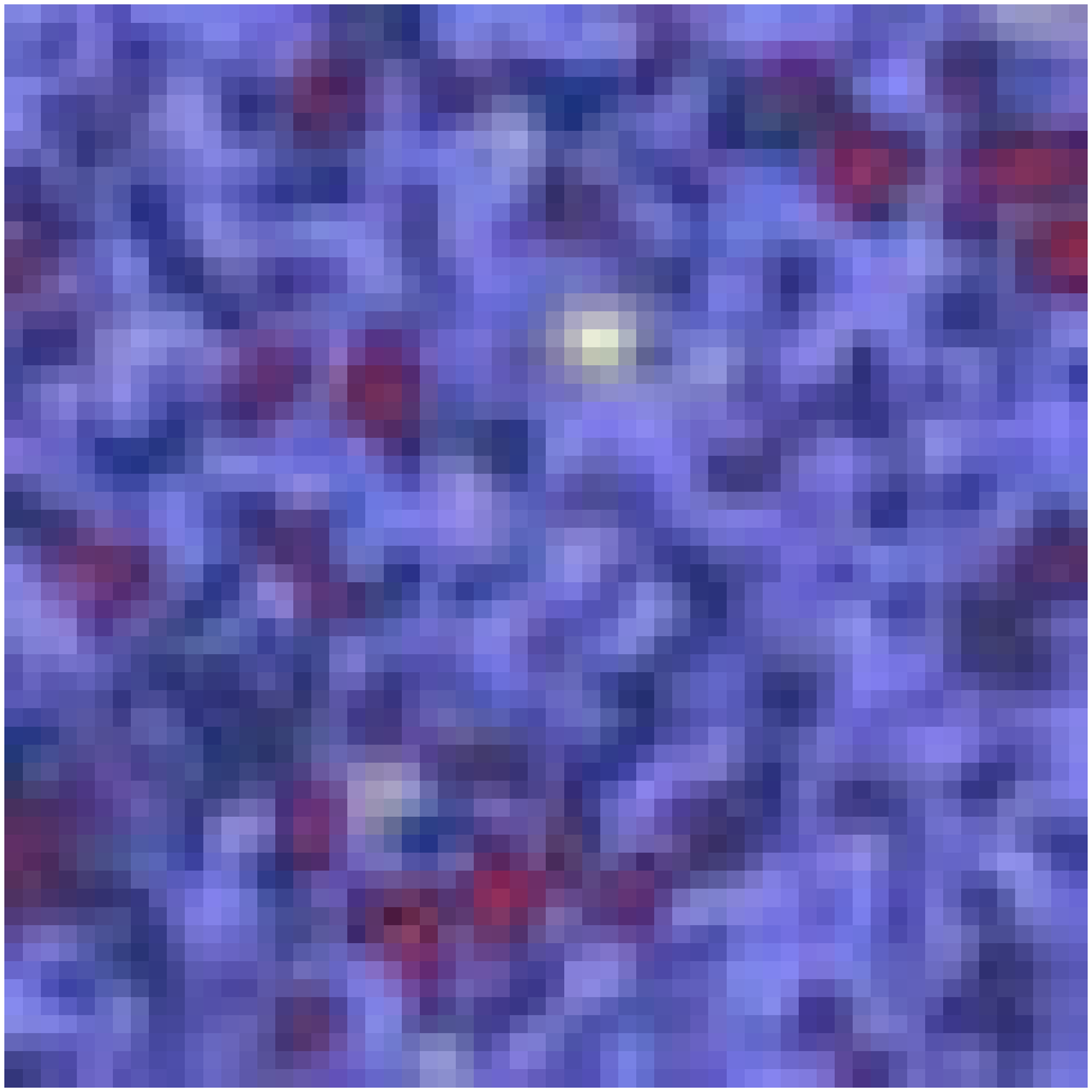}   \FGb{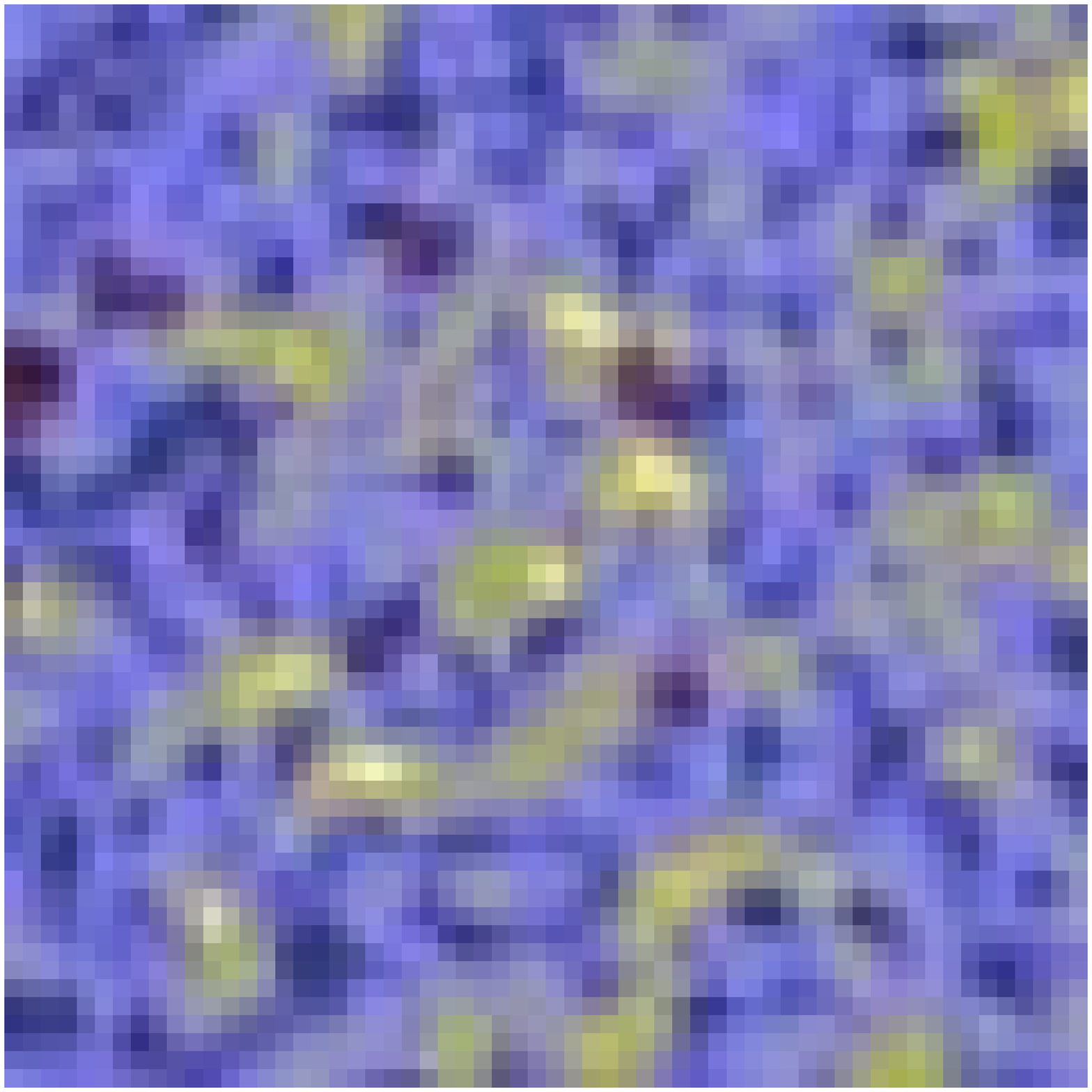}   \FGb{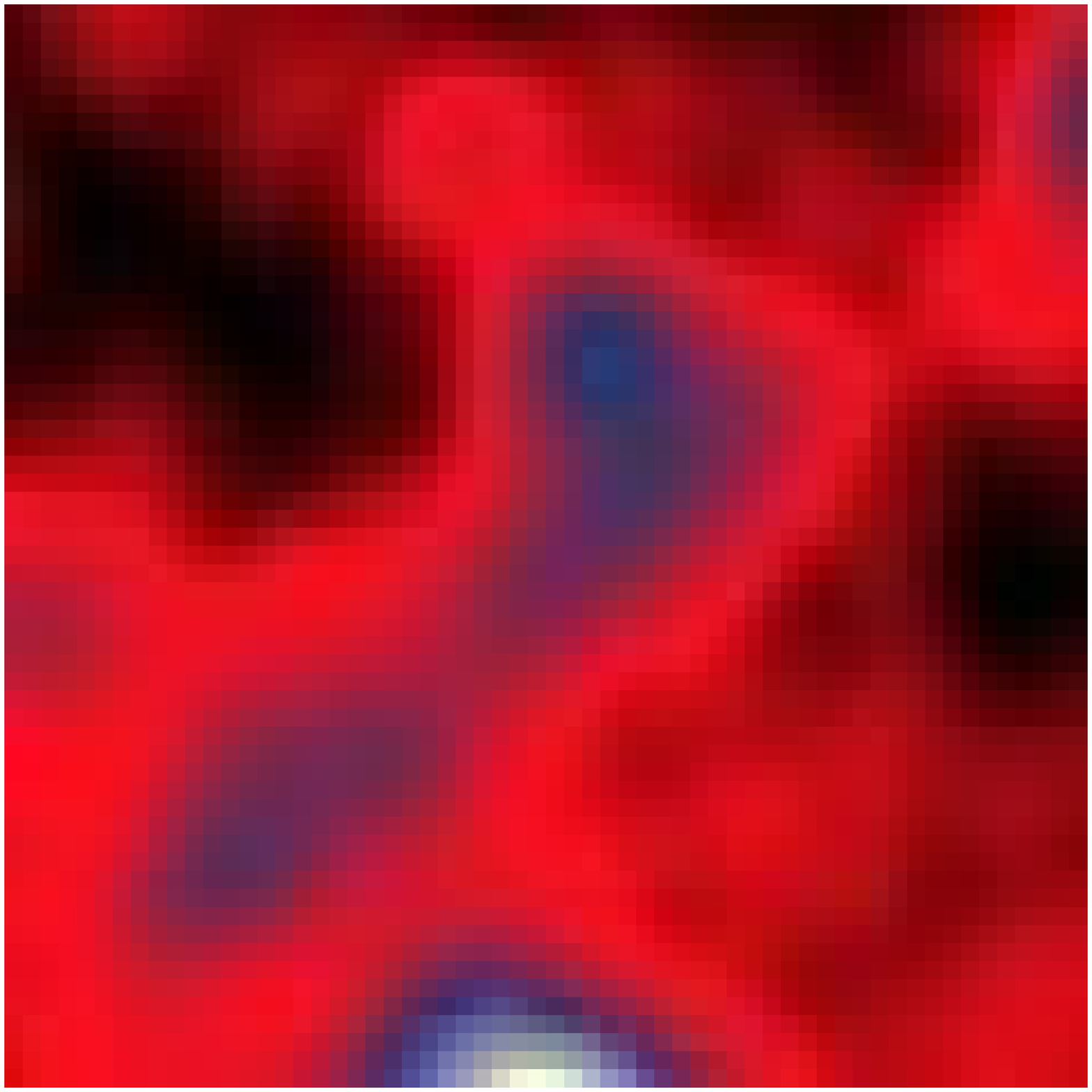}   \FGb{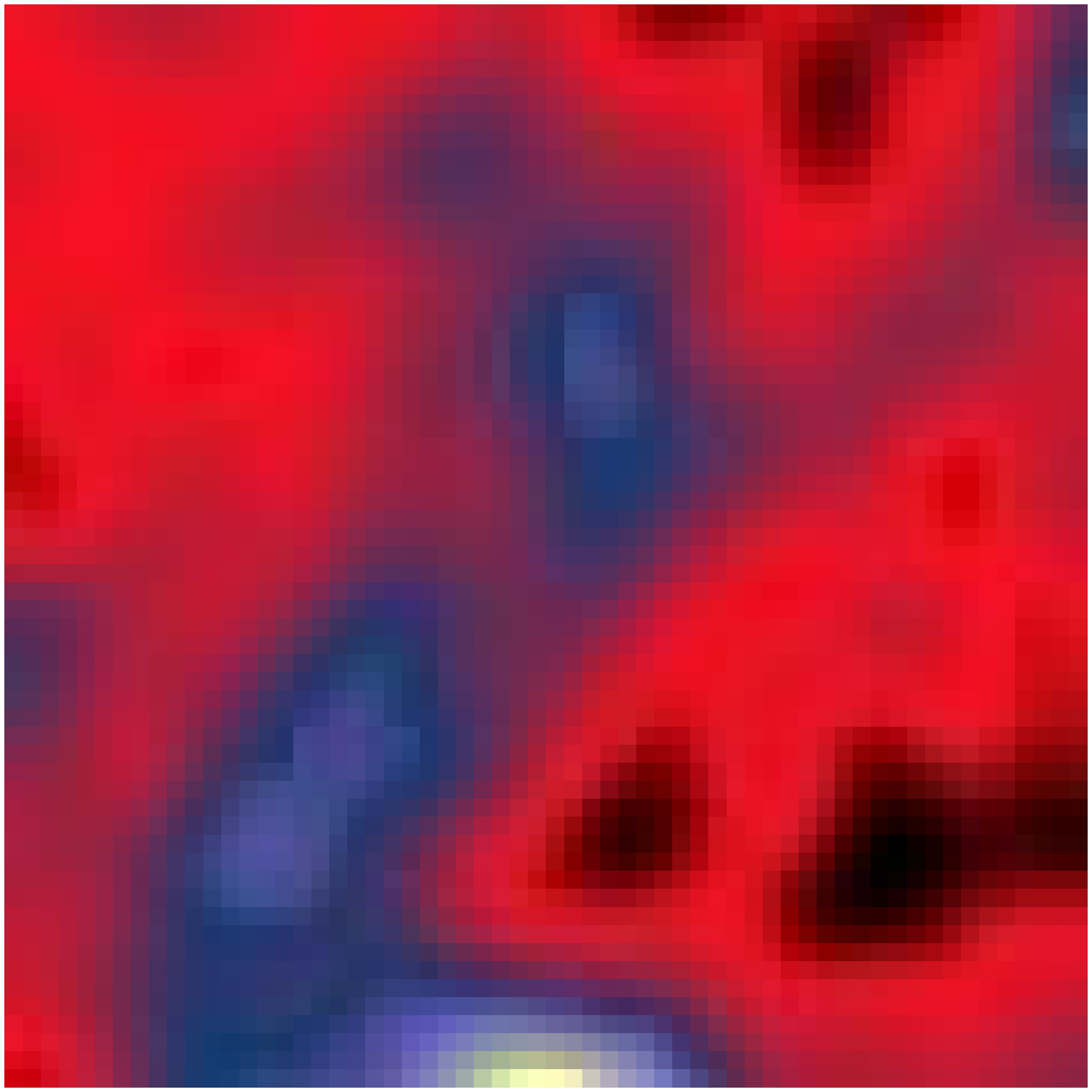}   \FGb{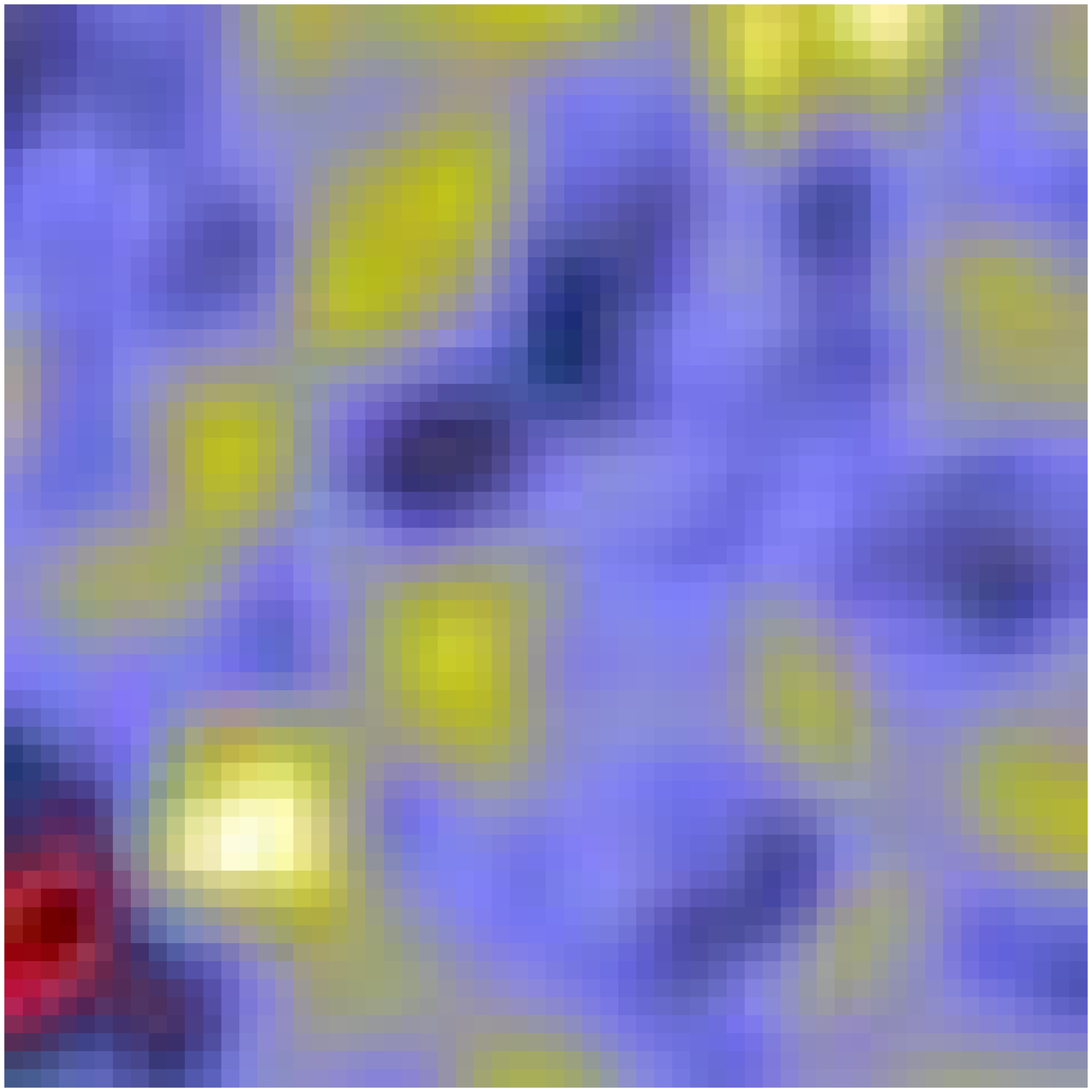}   \FGb{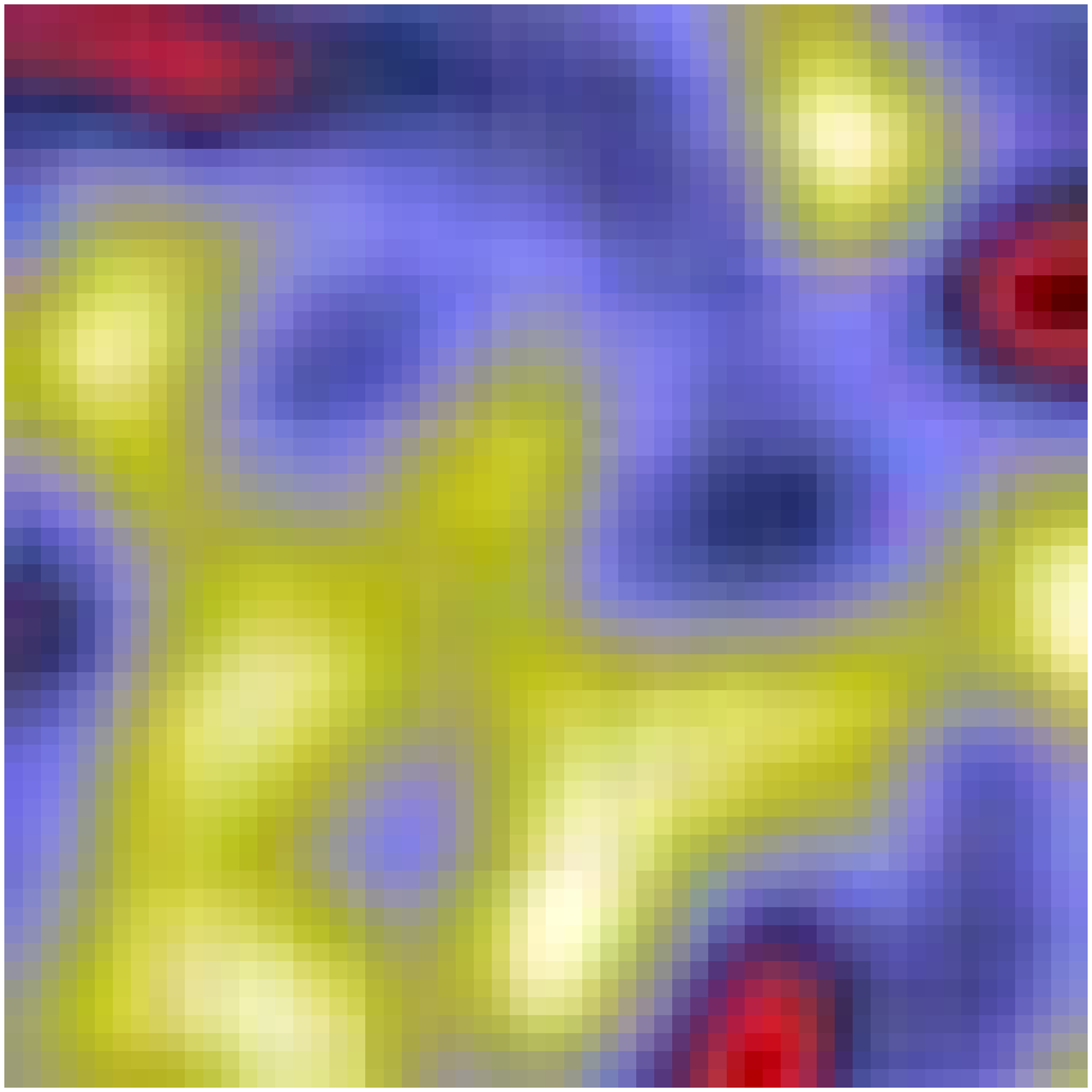} \\  {\#3   NCIA030805}  \\
  \FGb{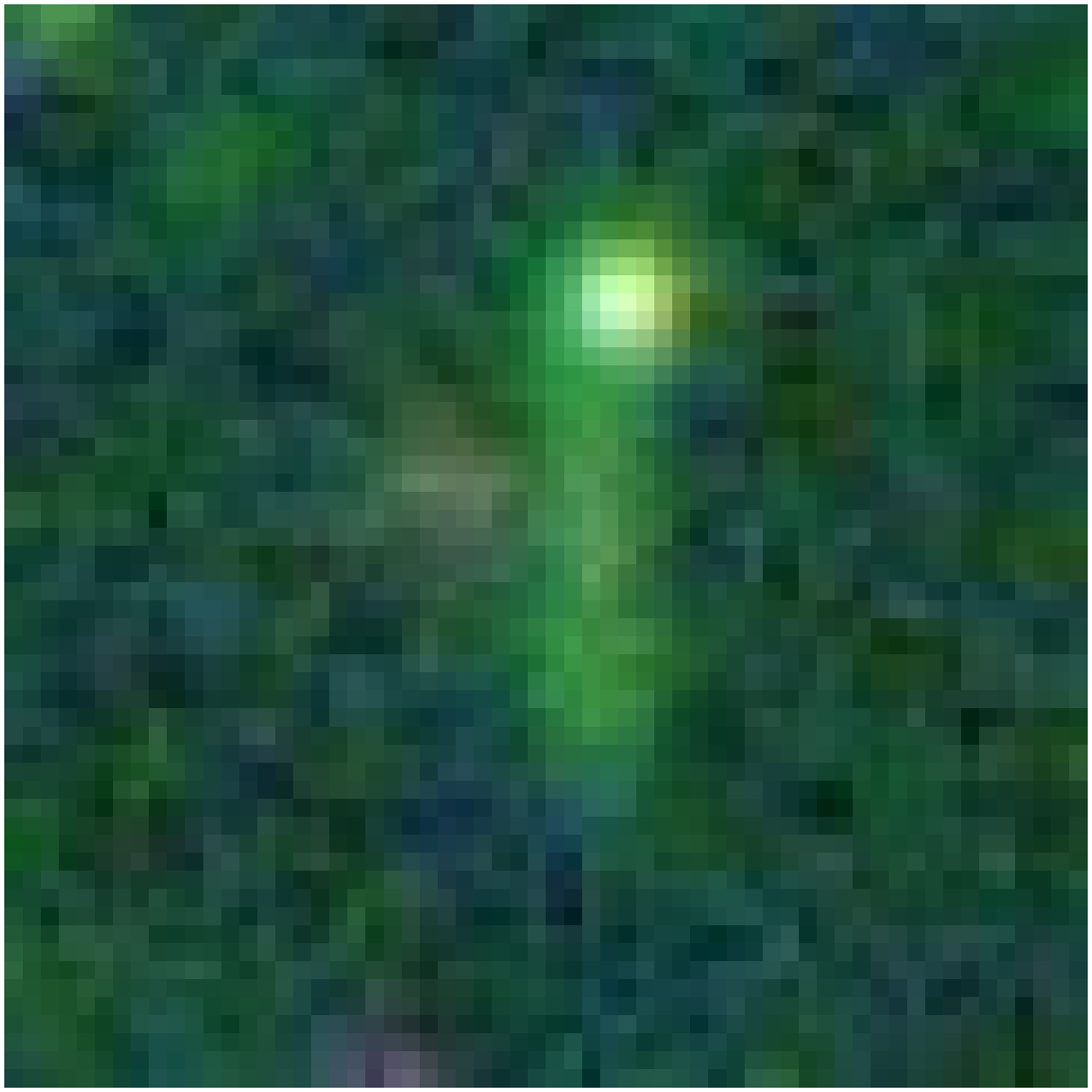}  \FGb{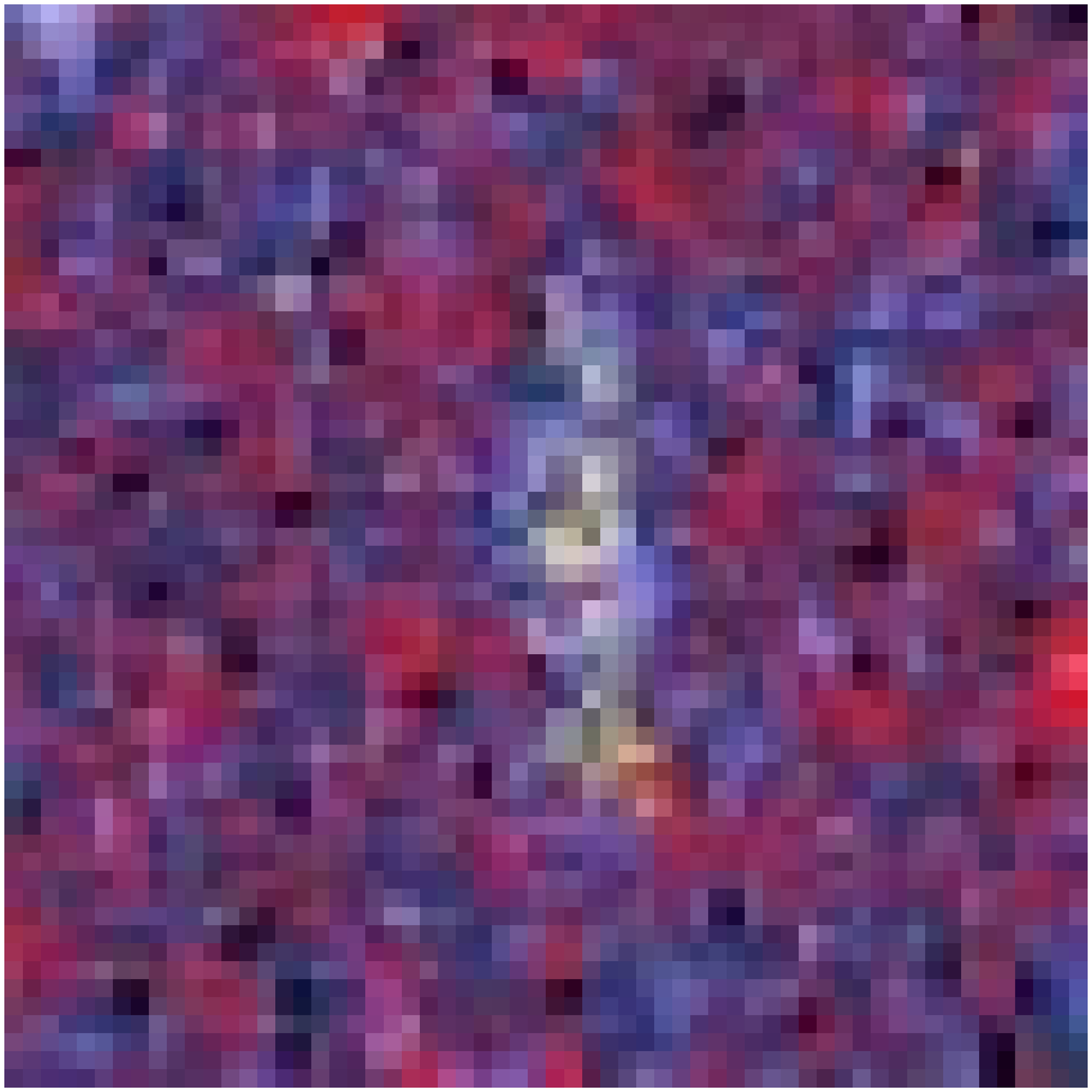}    \FGb{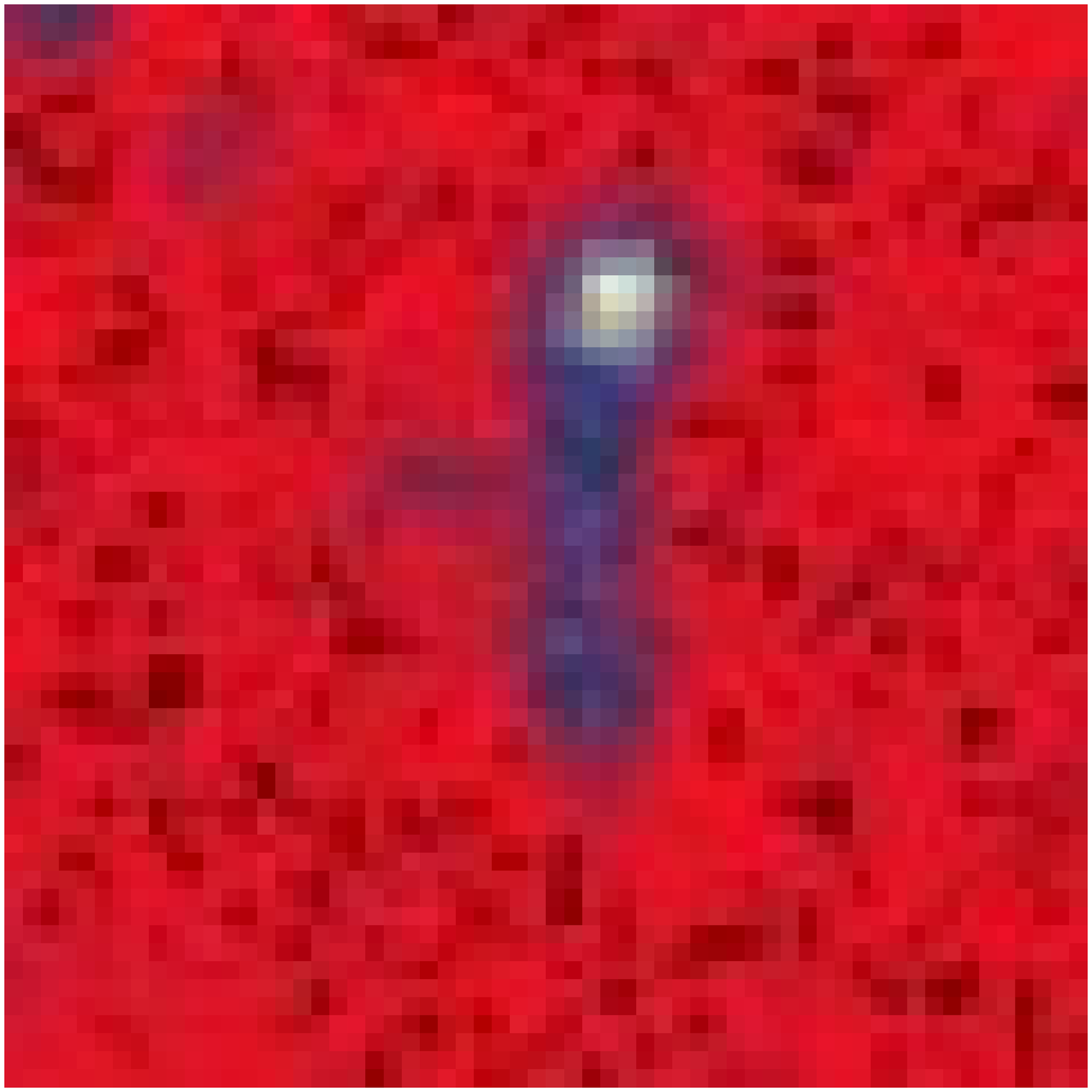}   \FGb{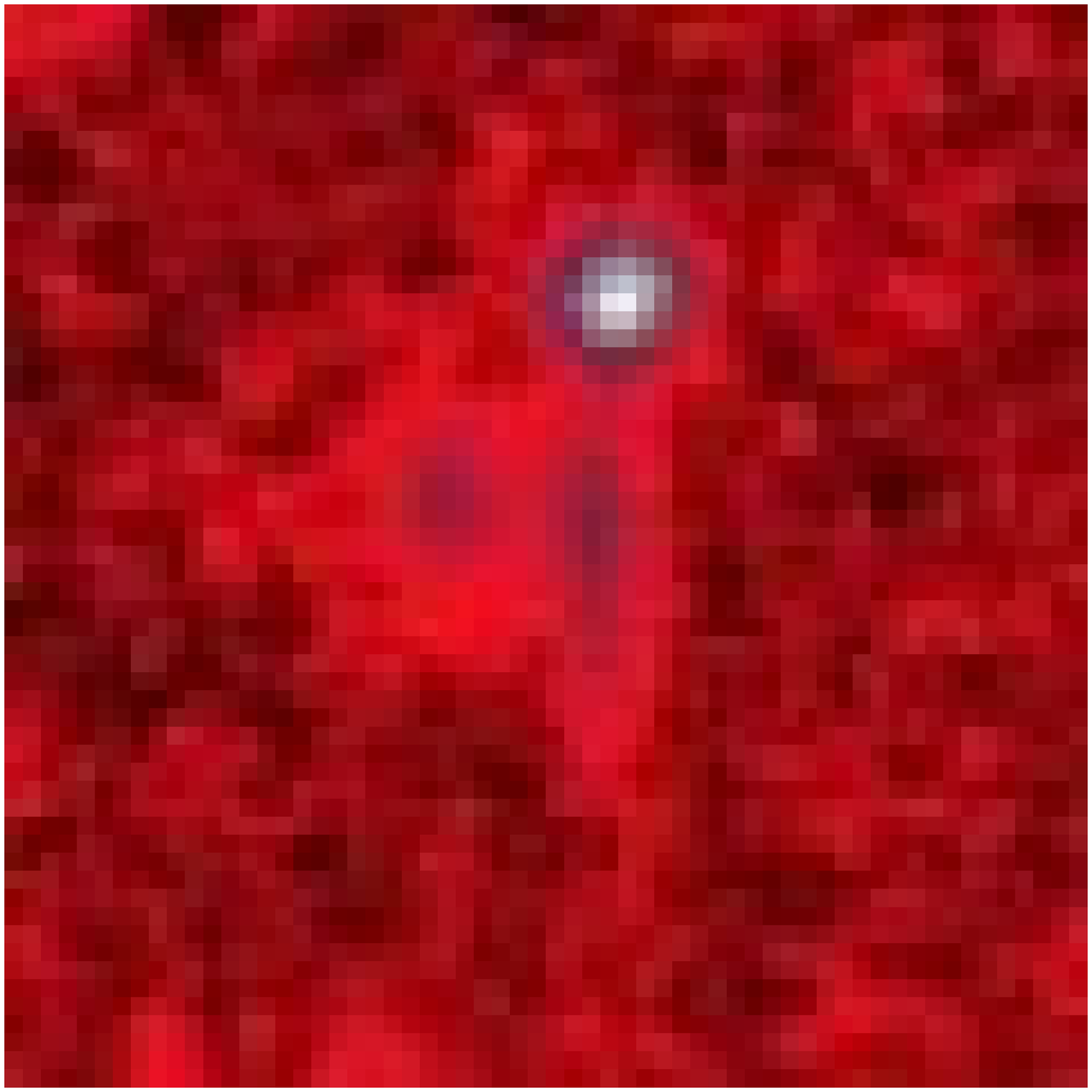}   \FGb{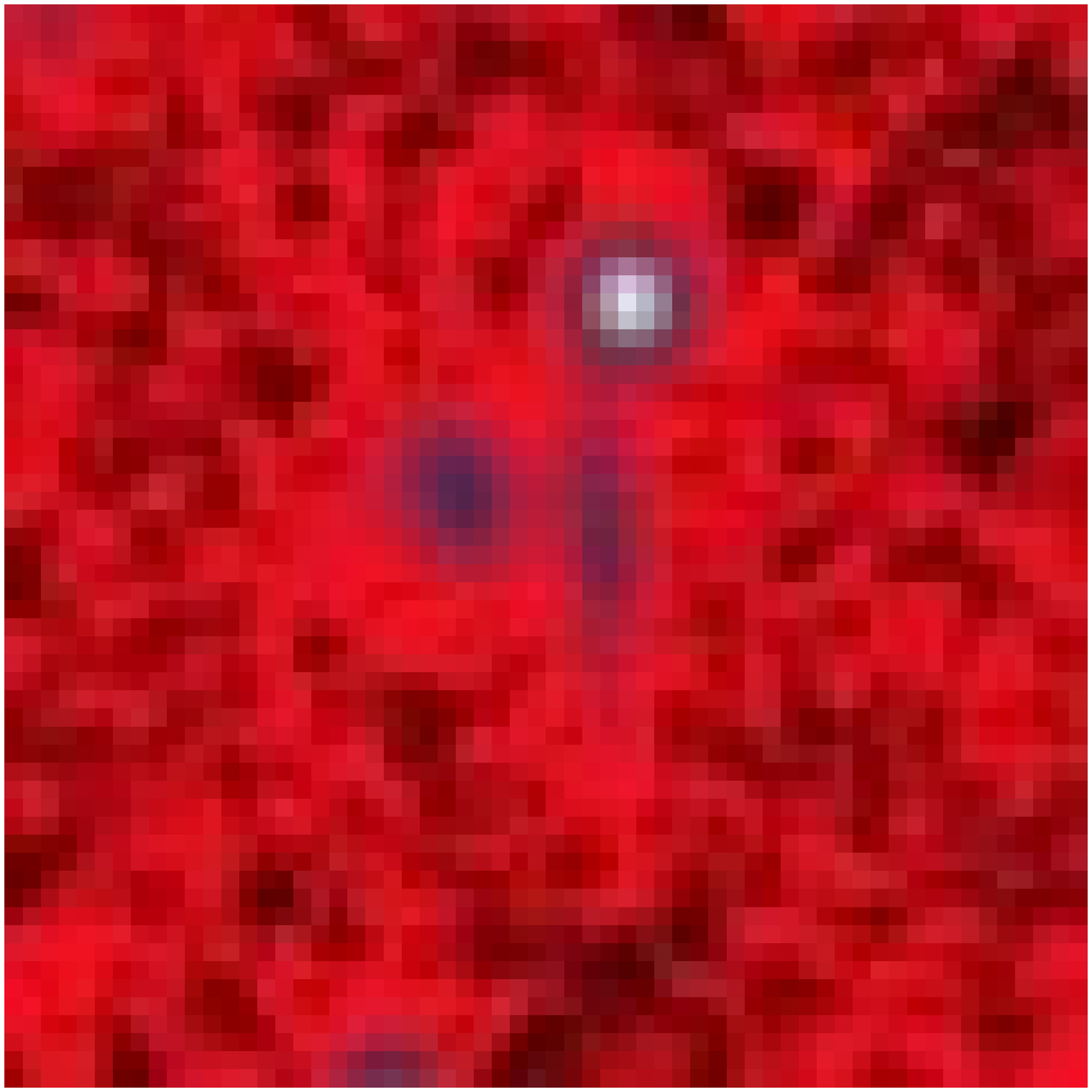}   \FGb{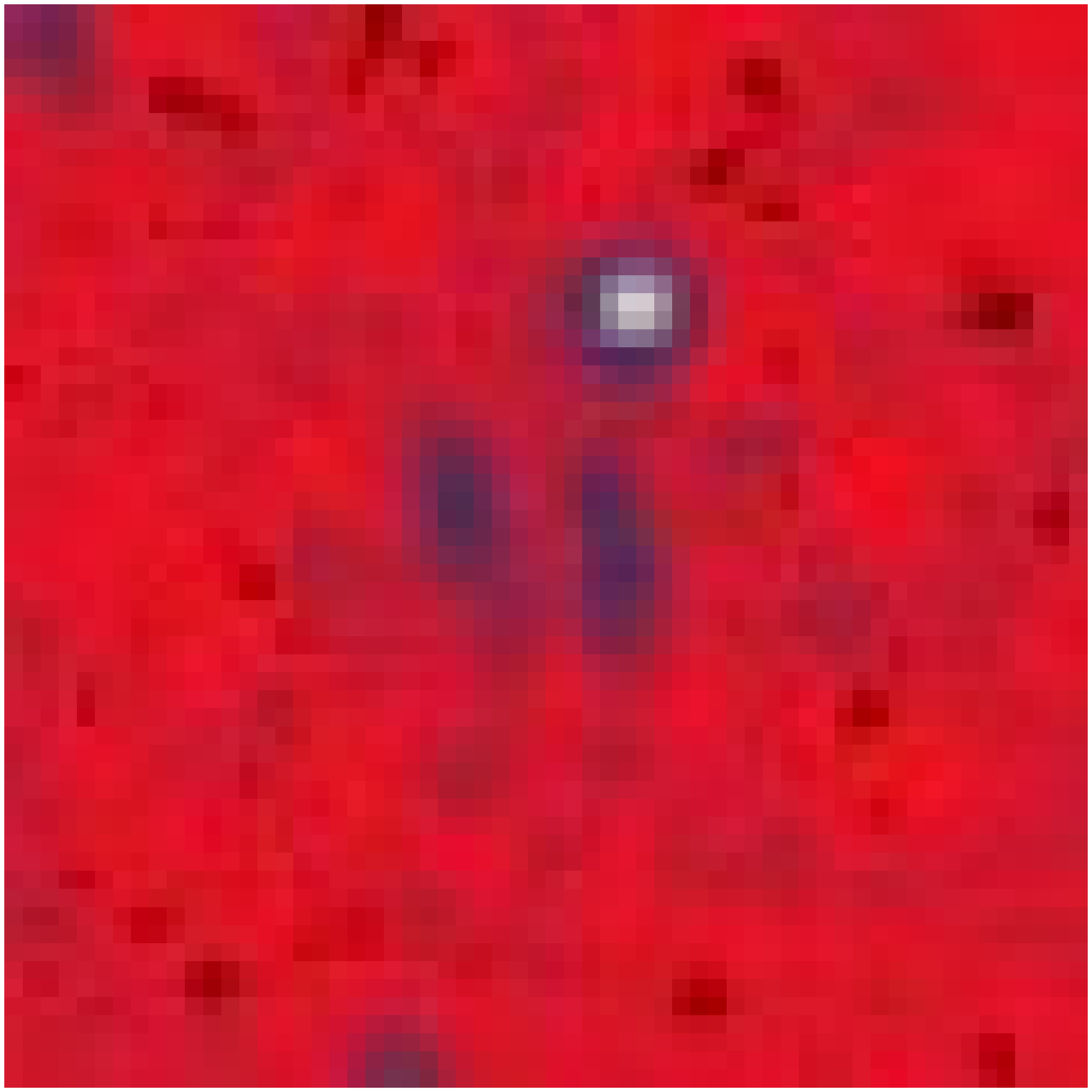}   \FGb{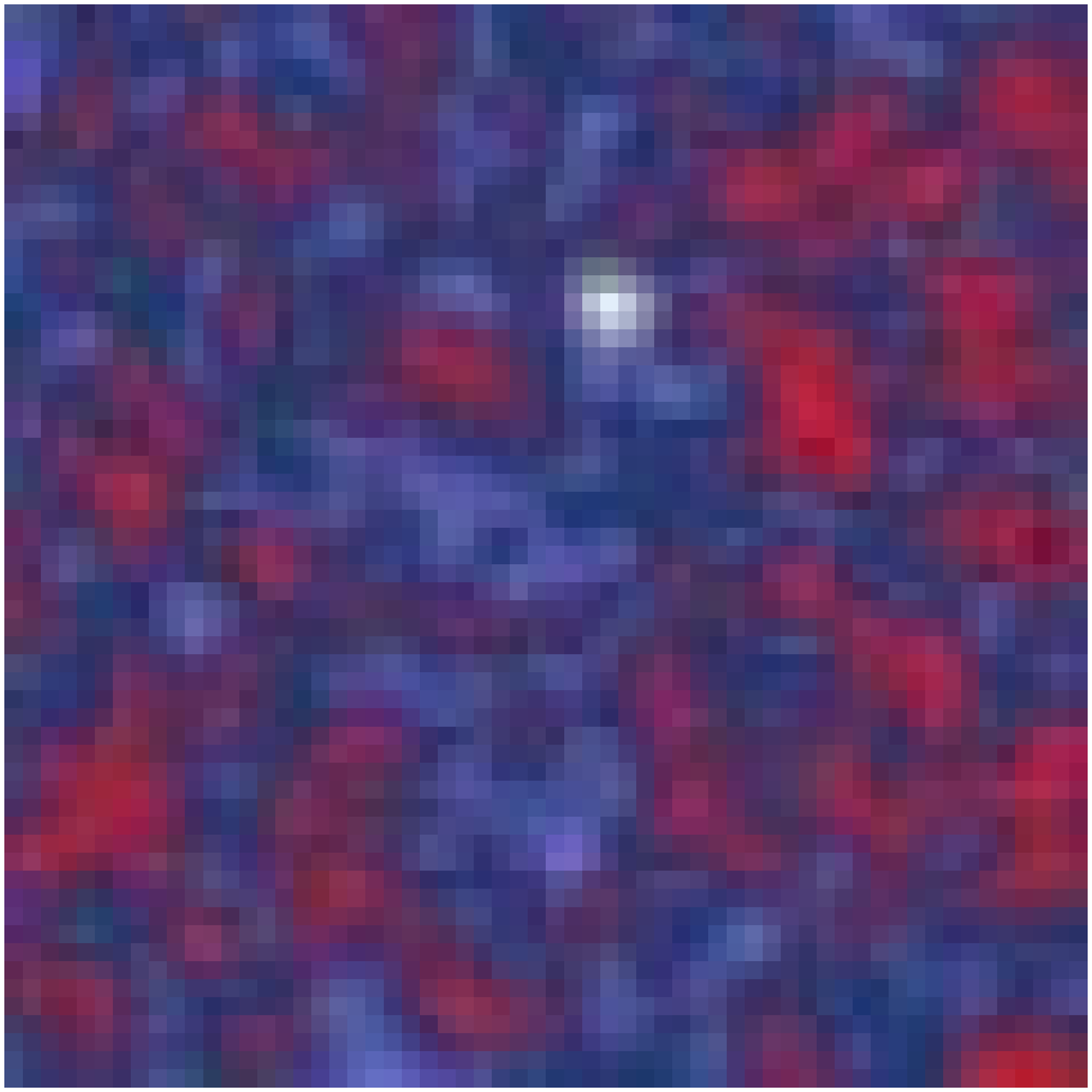}   \FGb{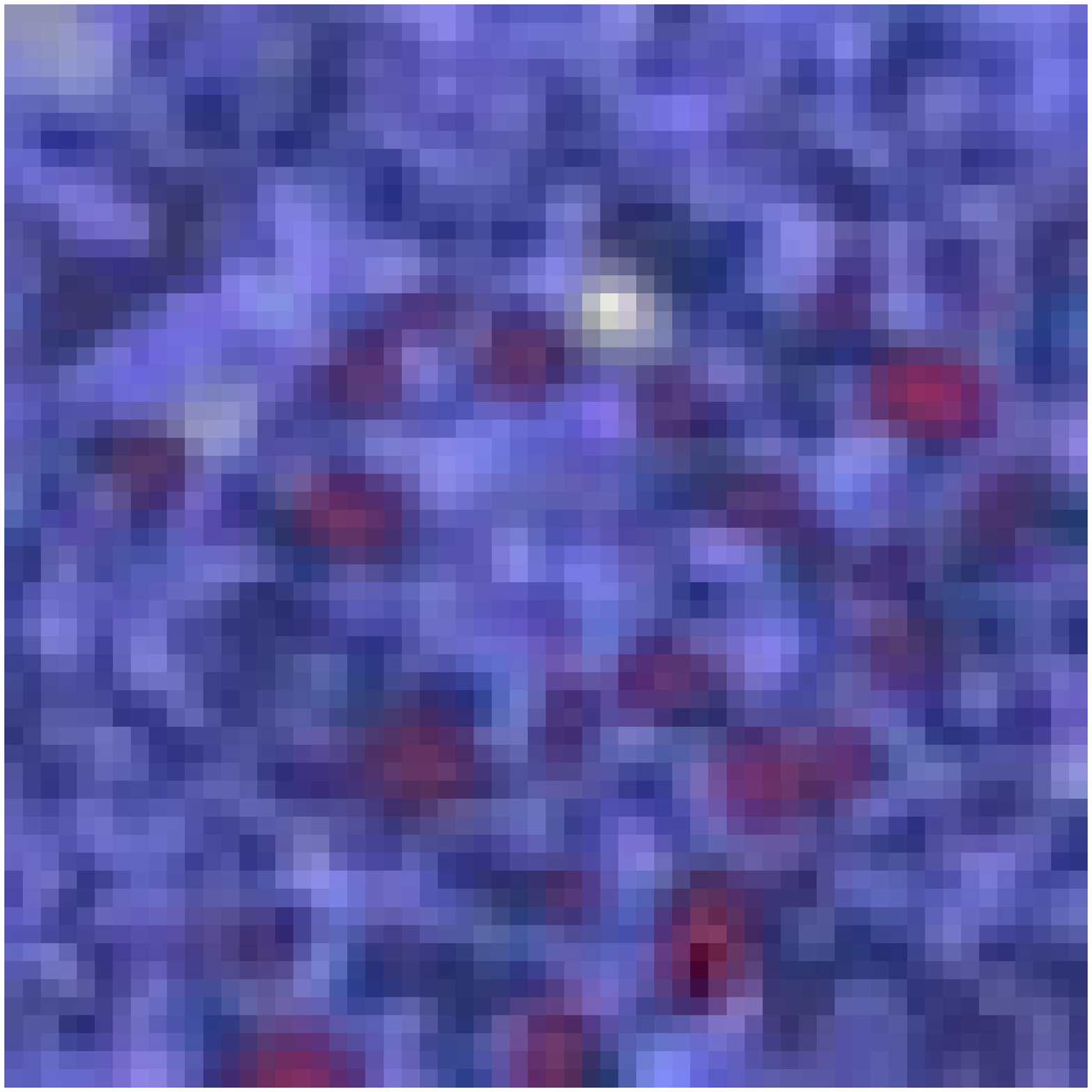}   \FGb{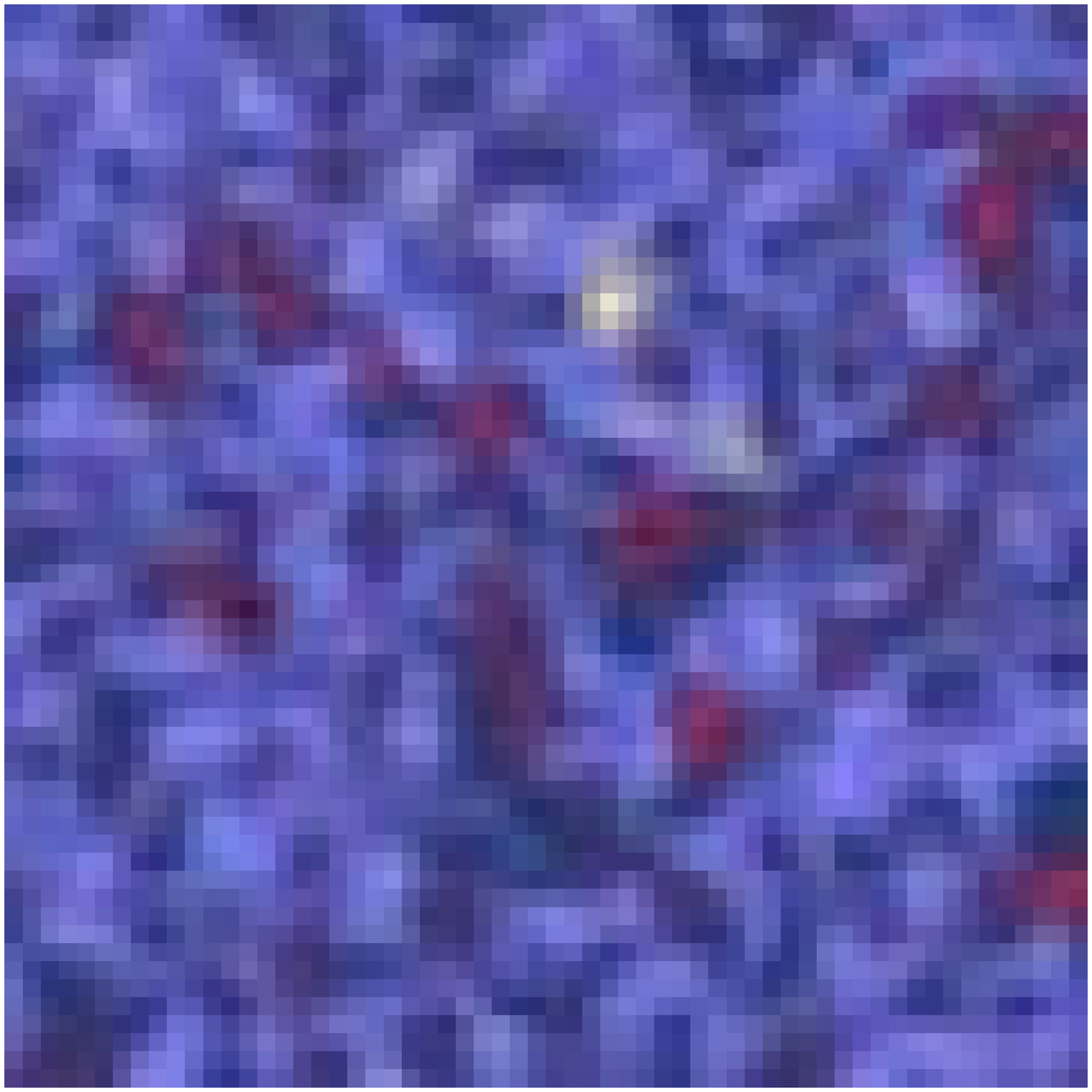}   \FGb{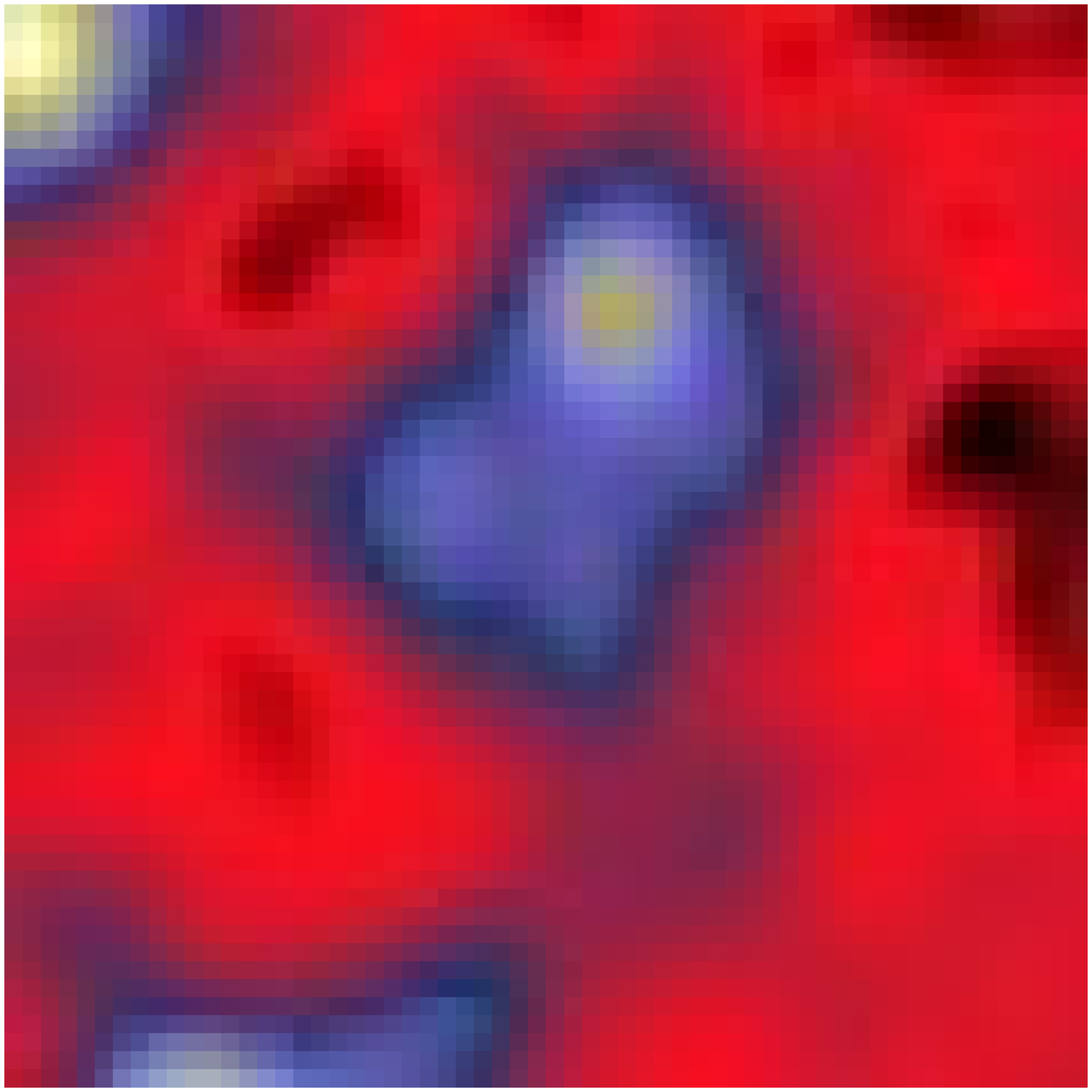}   \FGb{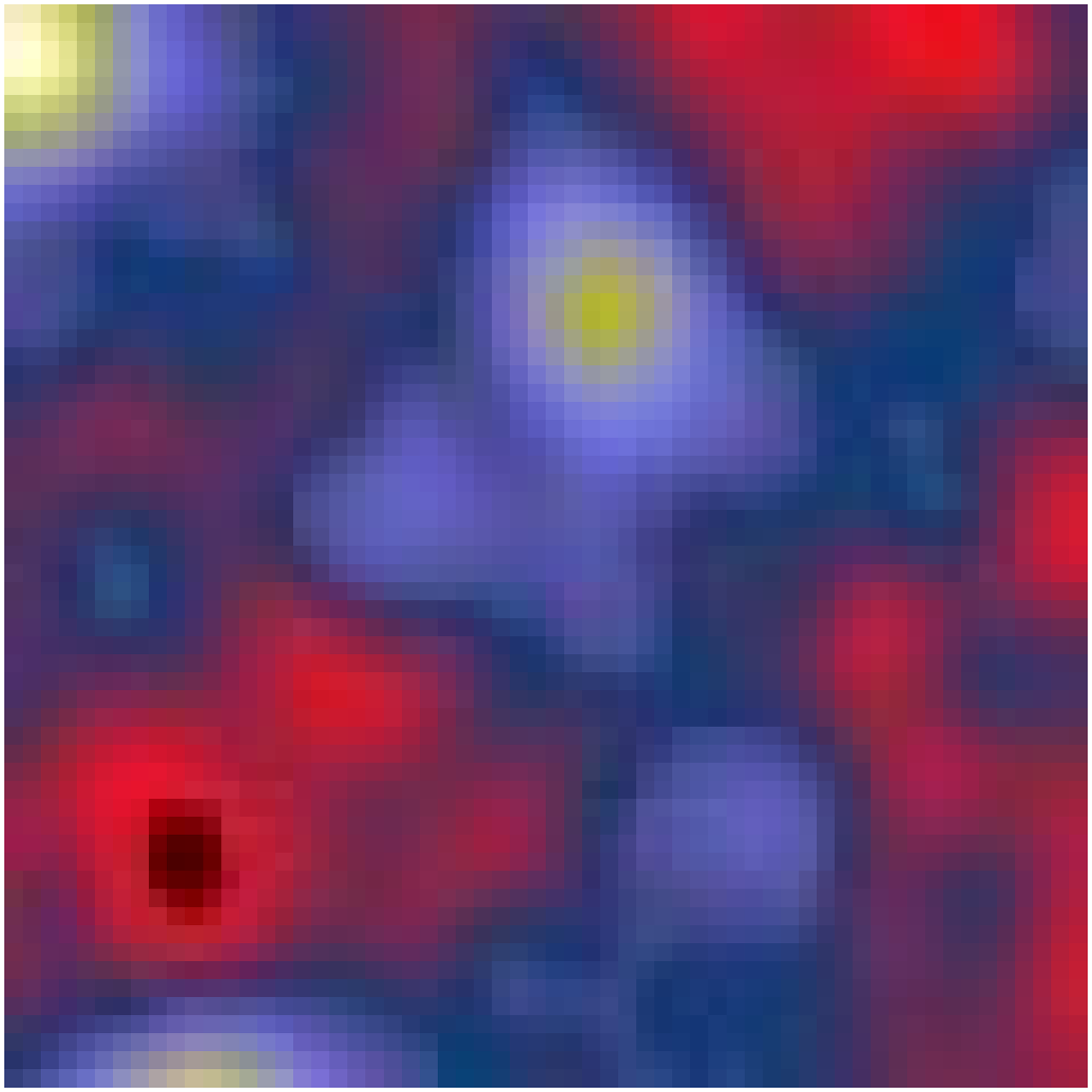}   \FGb{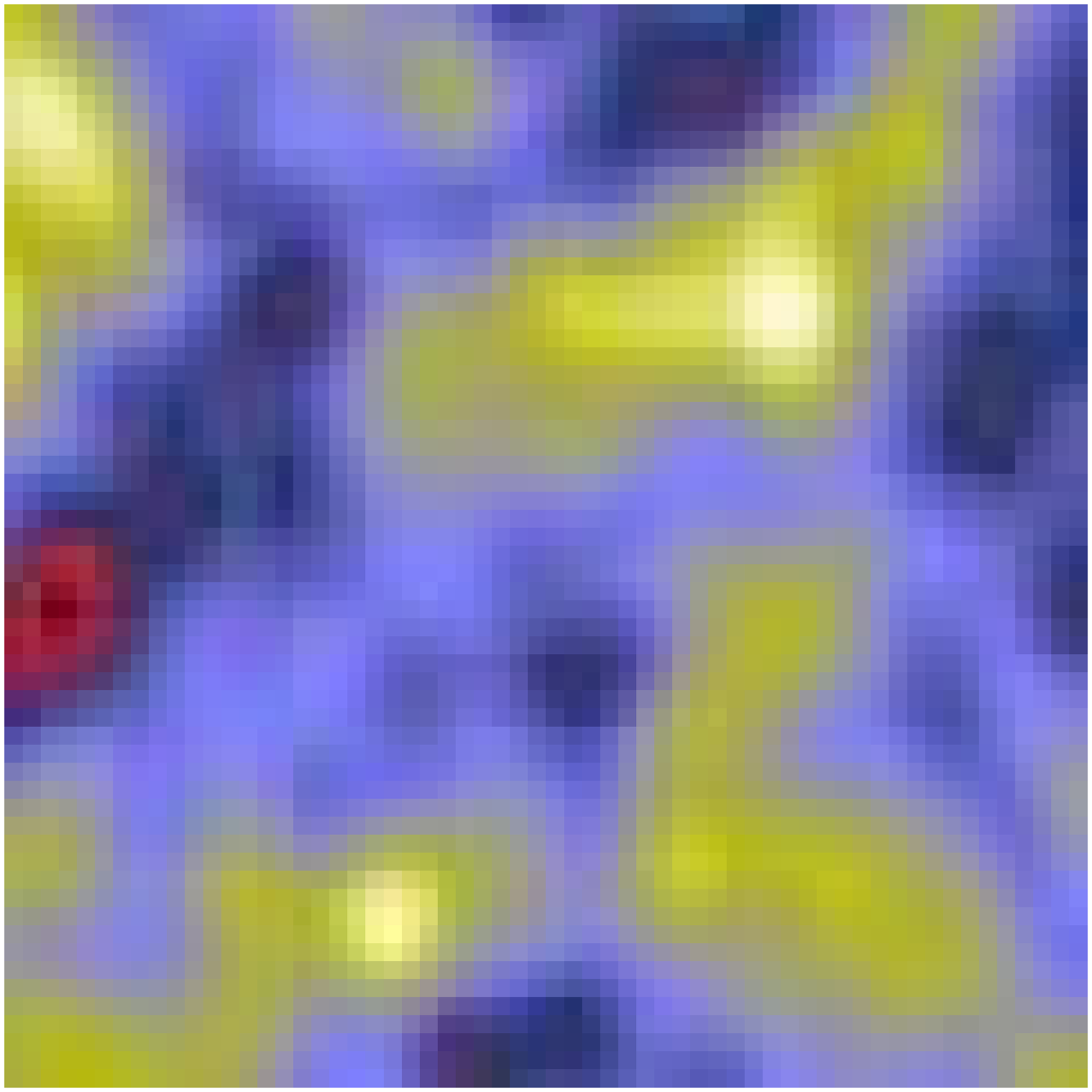}   \FGb{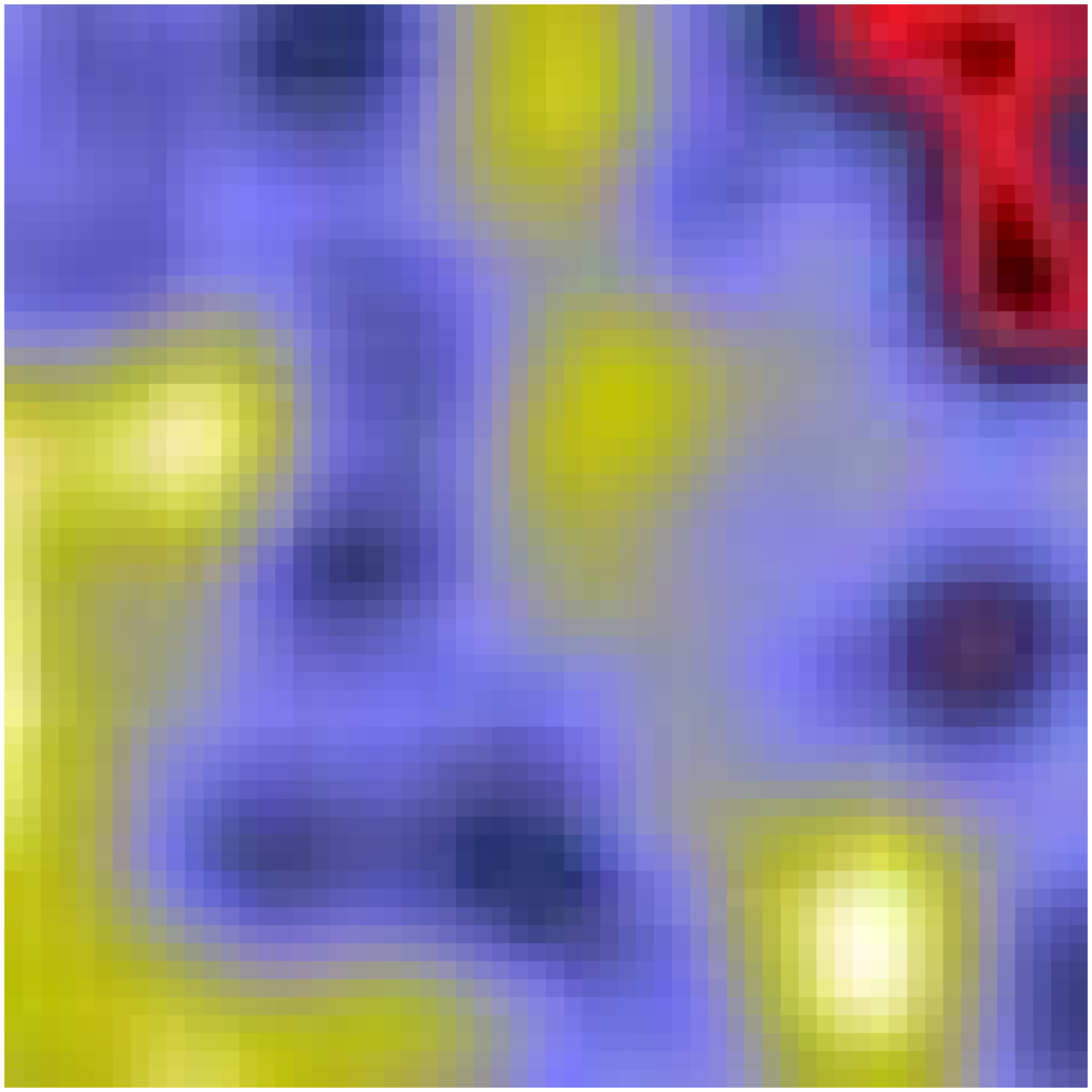} \\  {\#4   NCIA019215}  \\
  \FGb{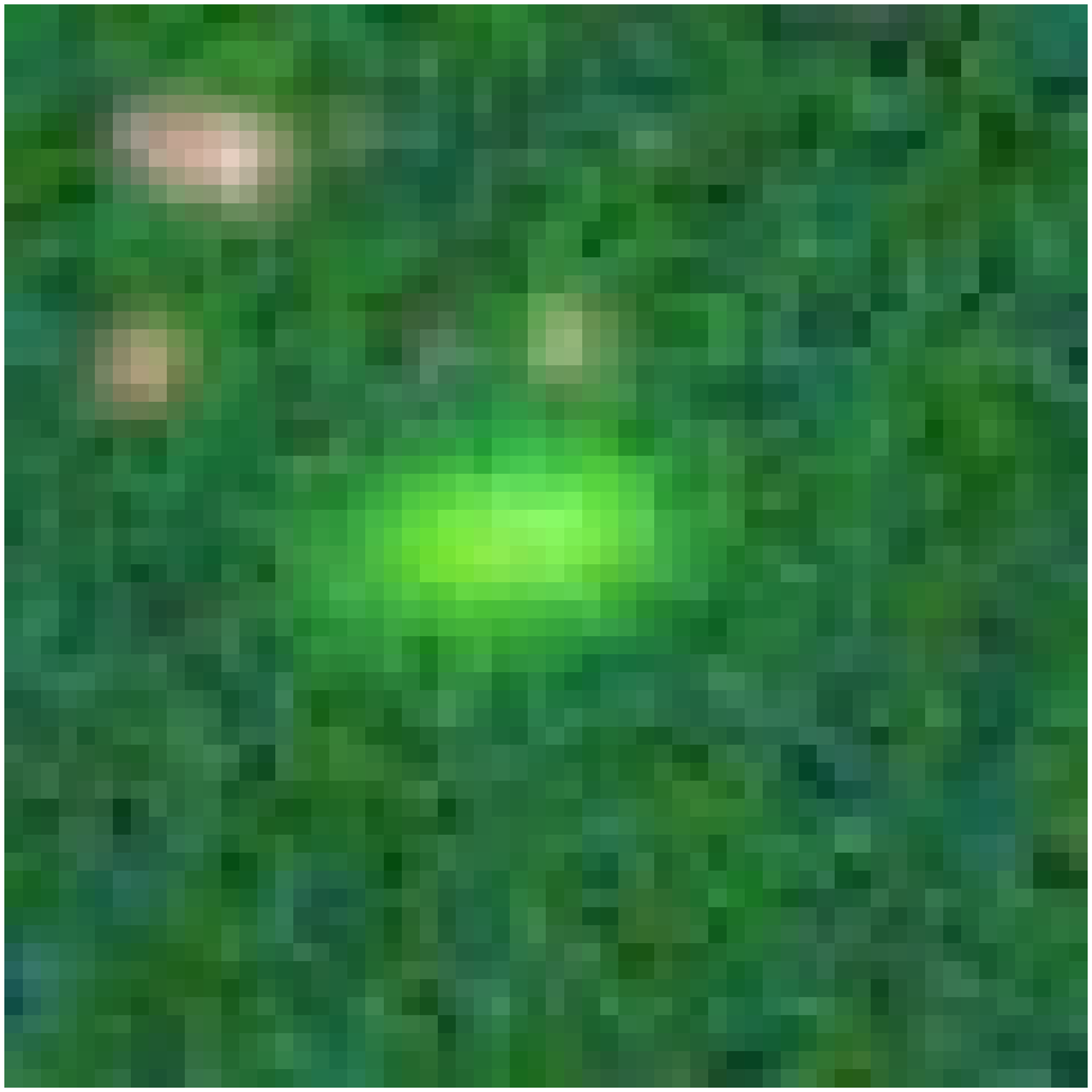}  \FGb{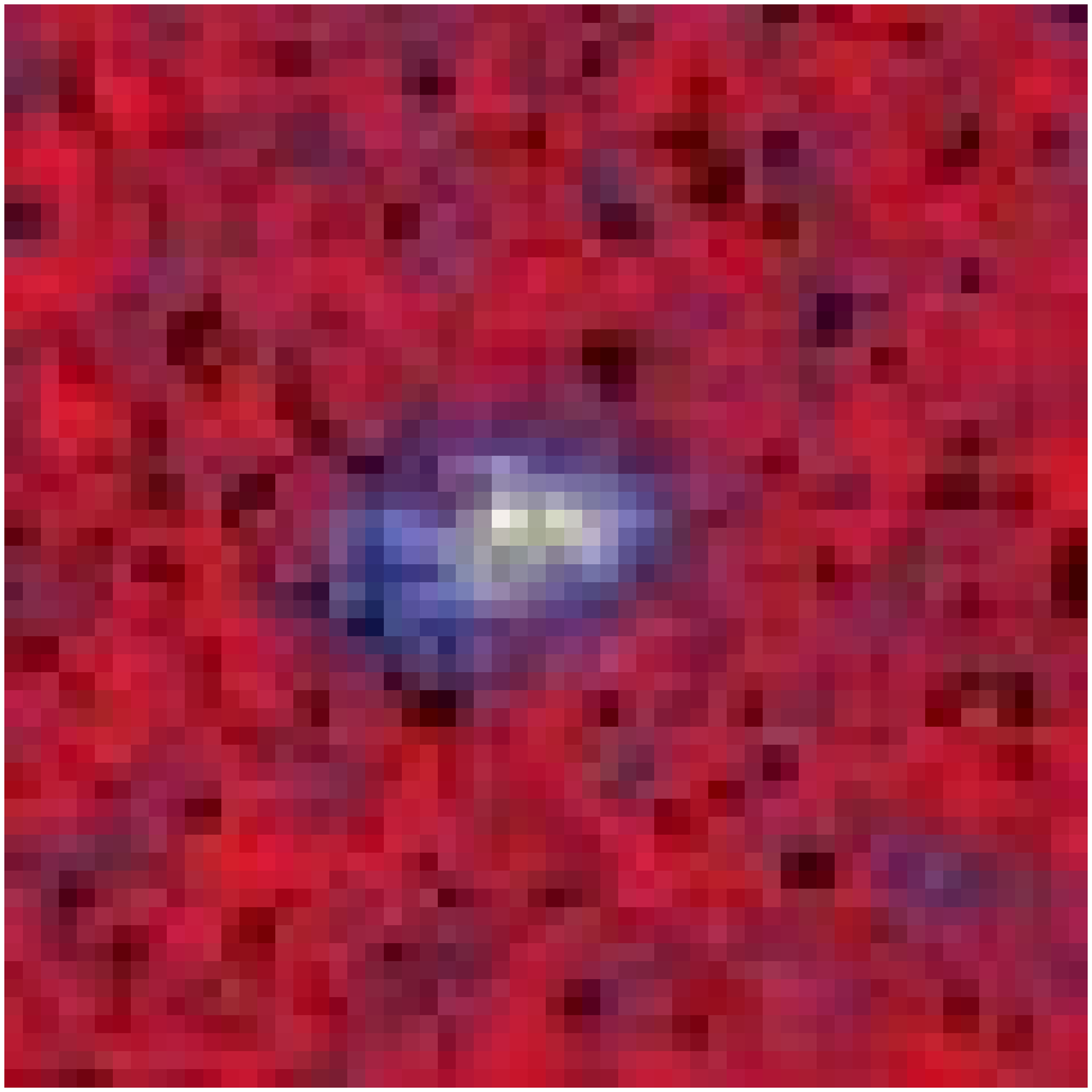}    \FGb{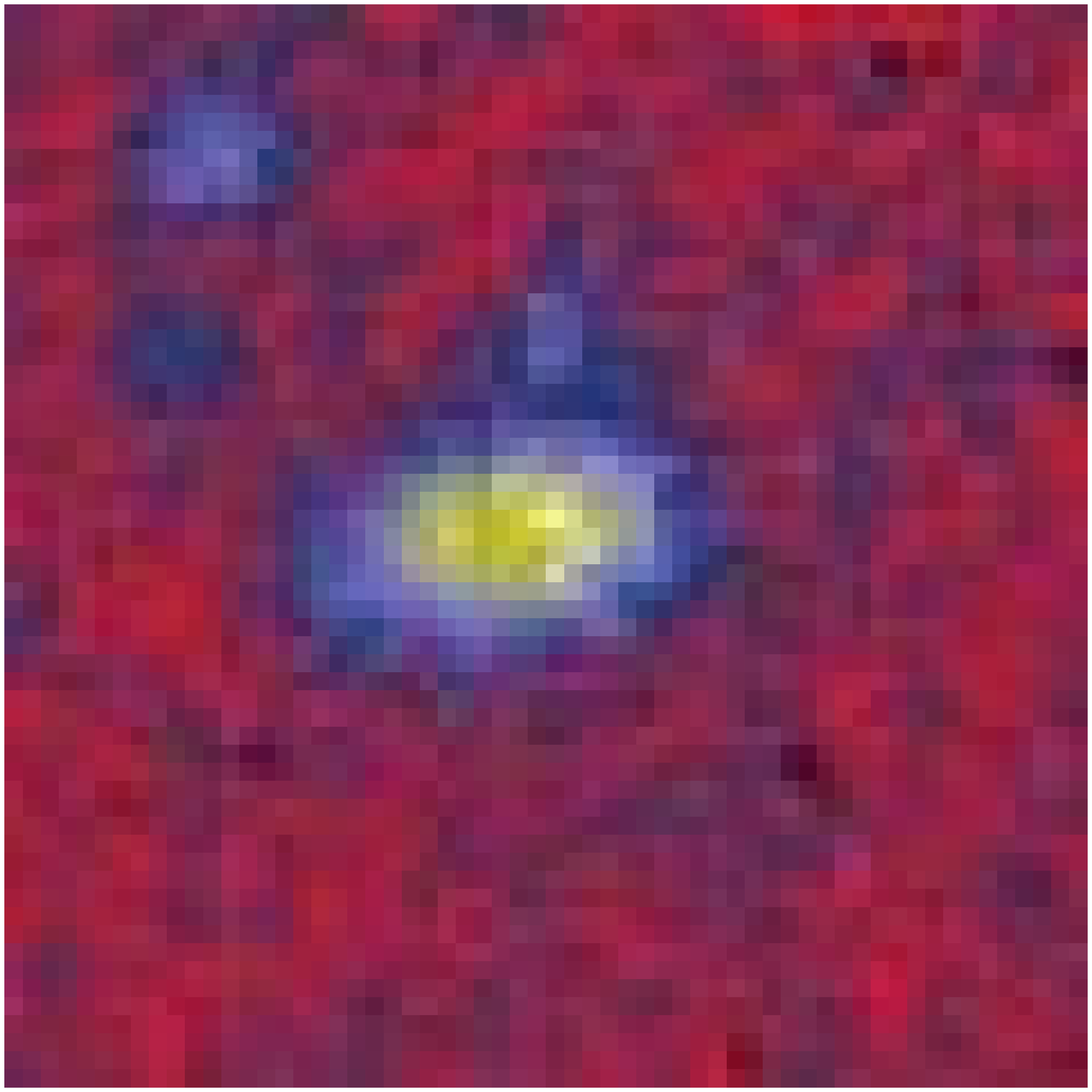}   \FGb{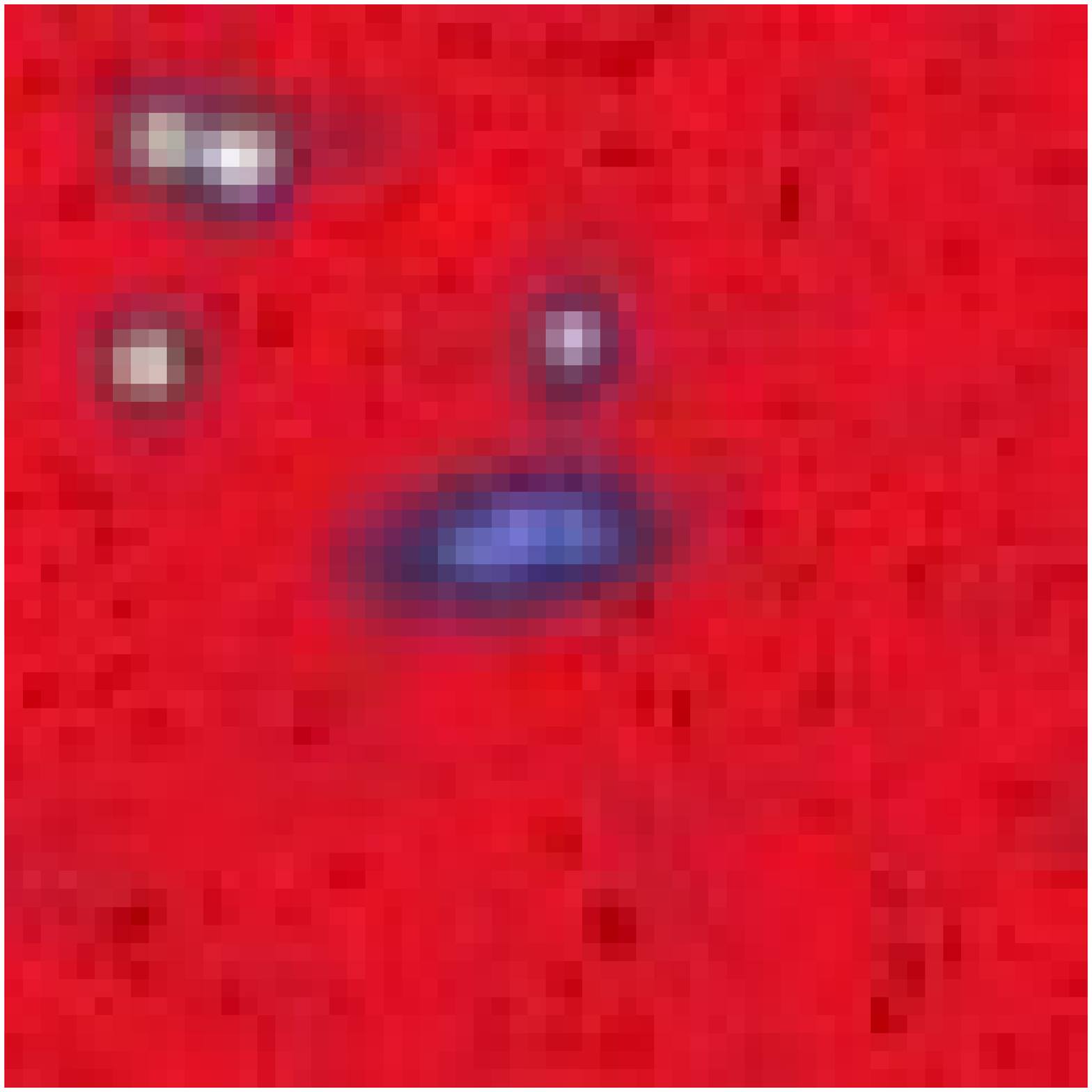}   \FGb{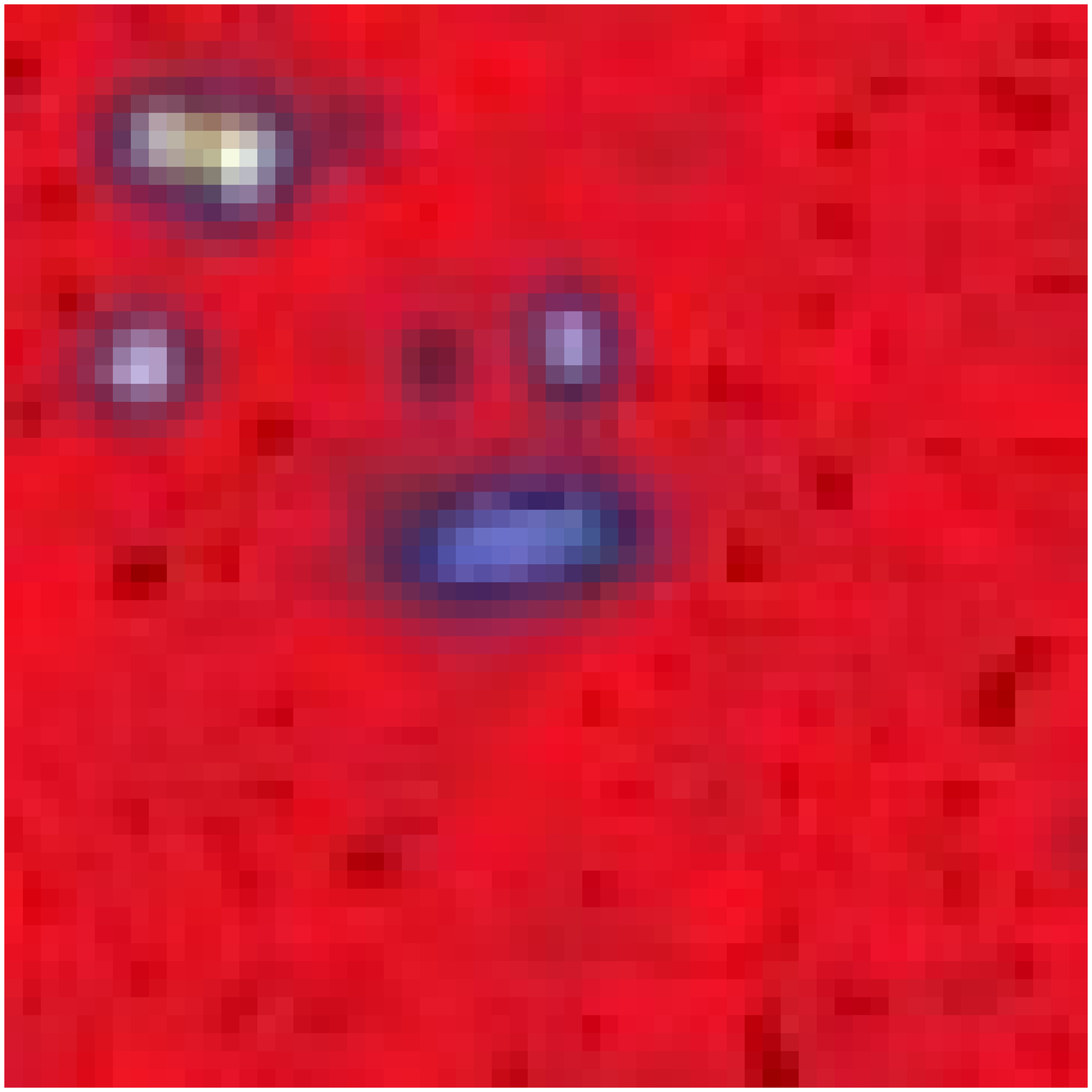}   \FGb{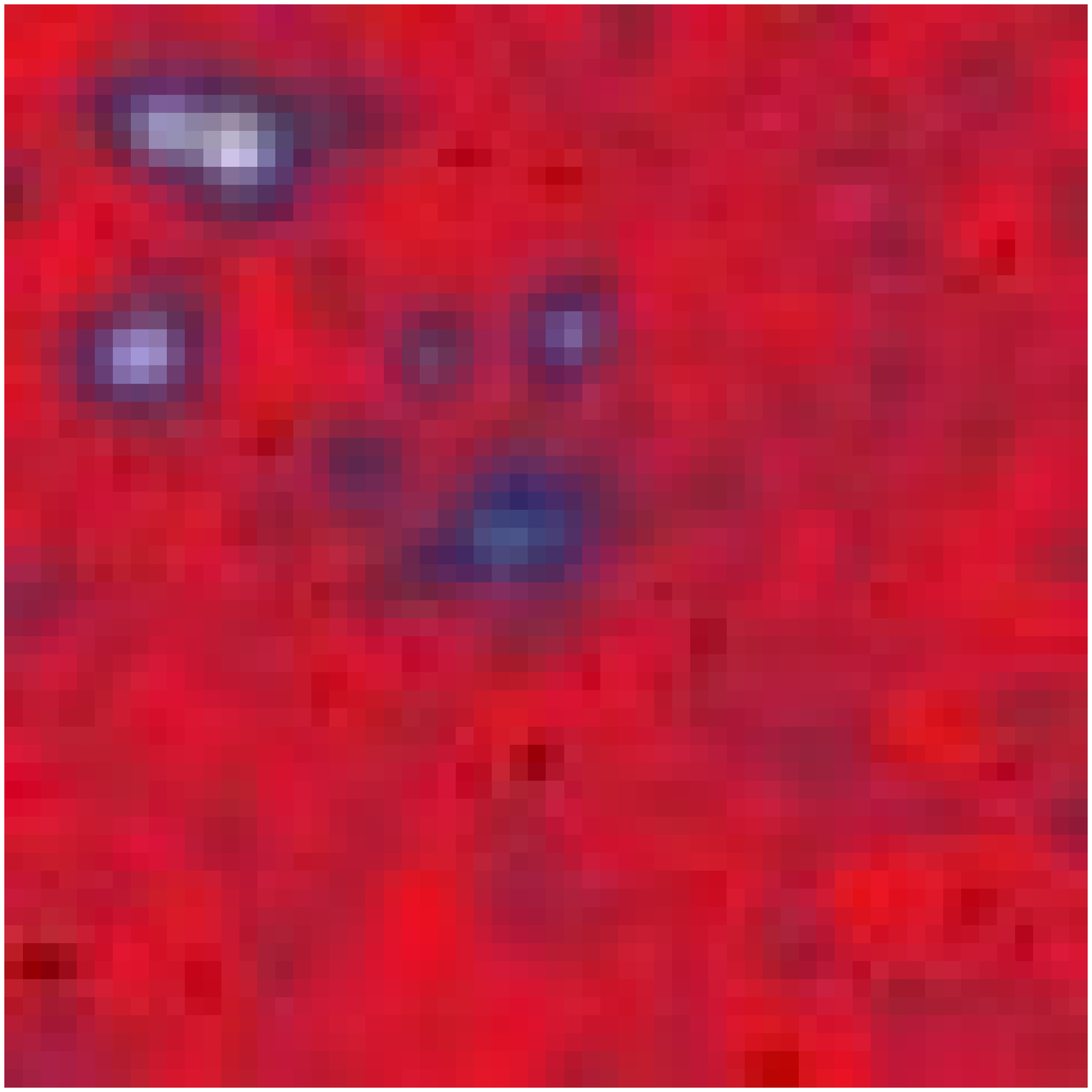}   \FGb{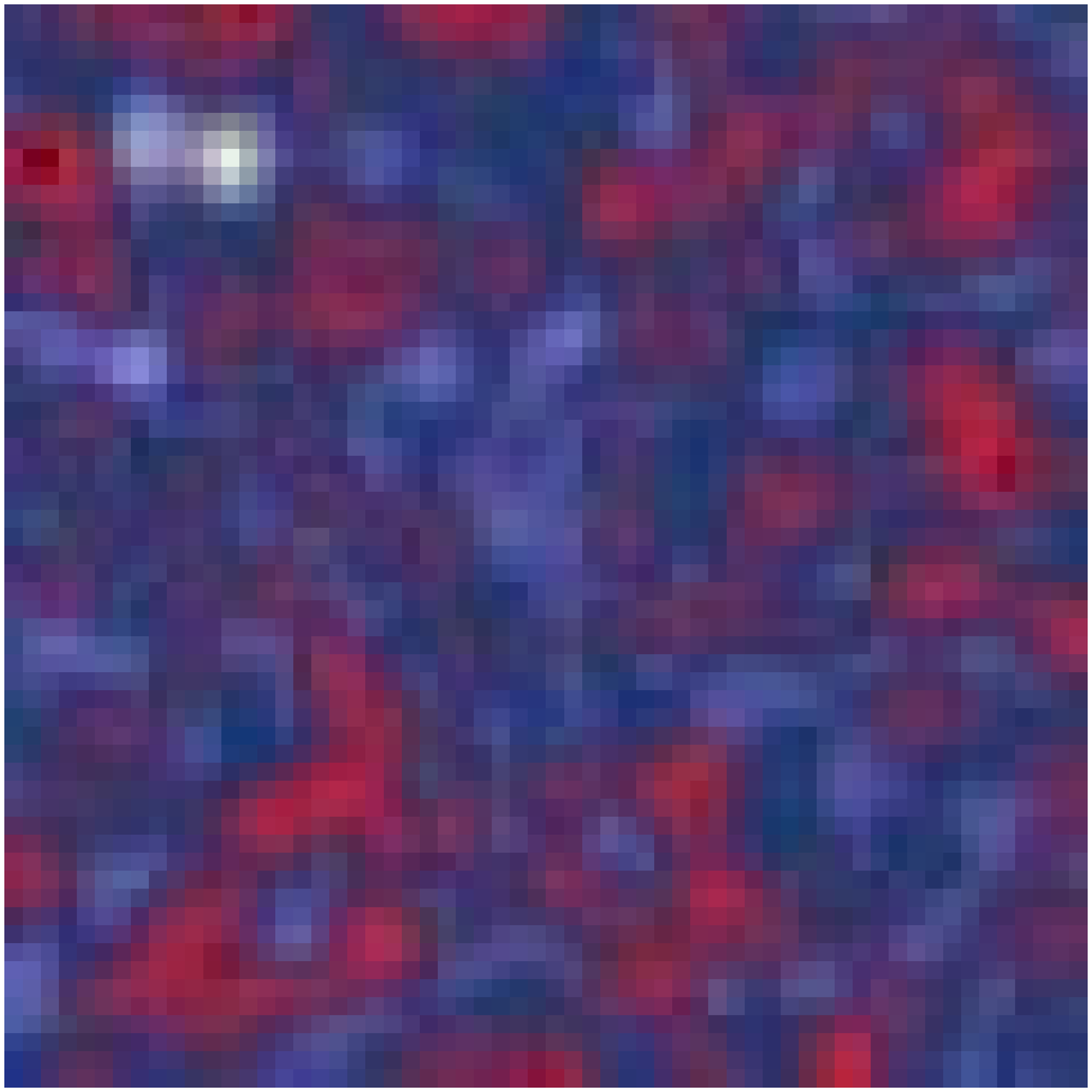}   \FGb{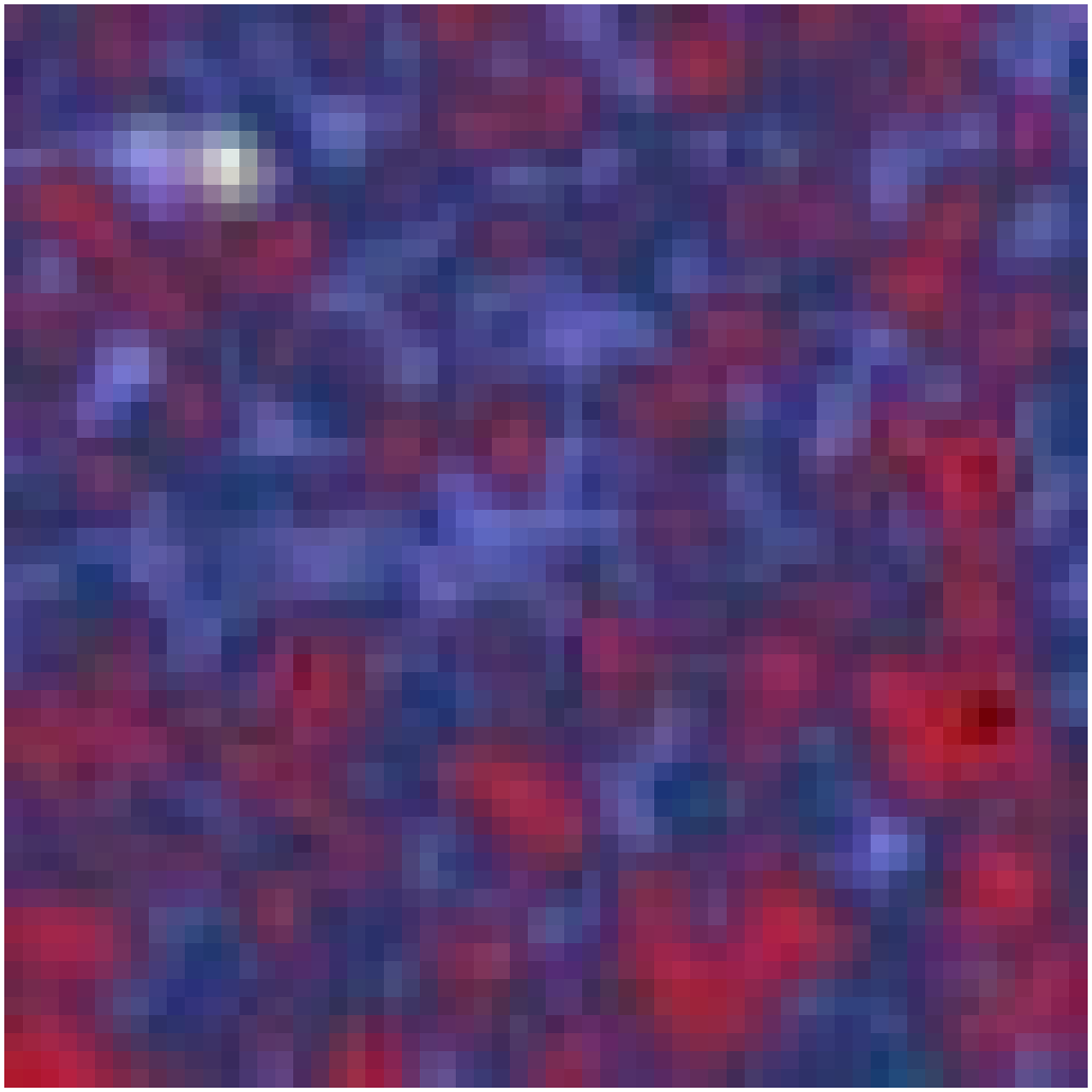}   \FGb{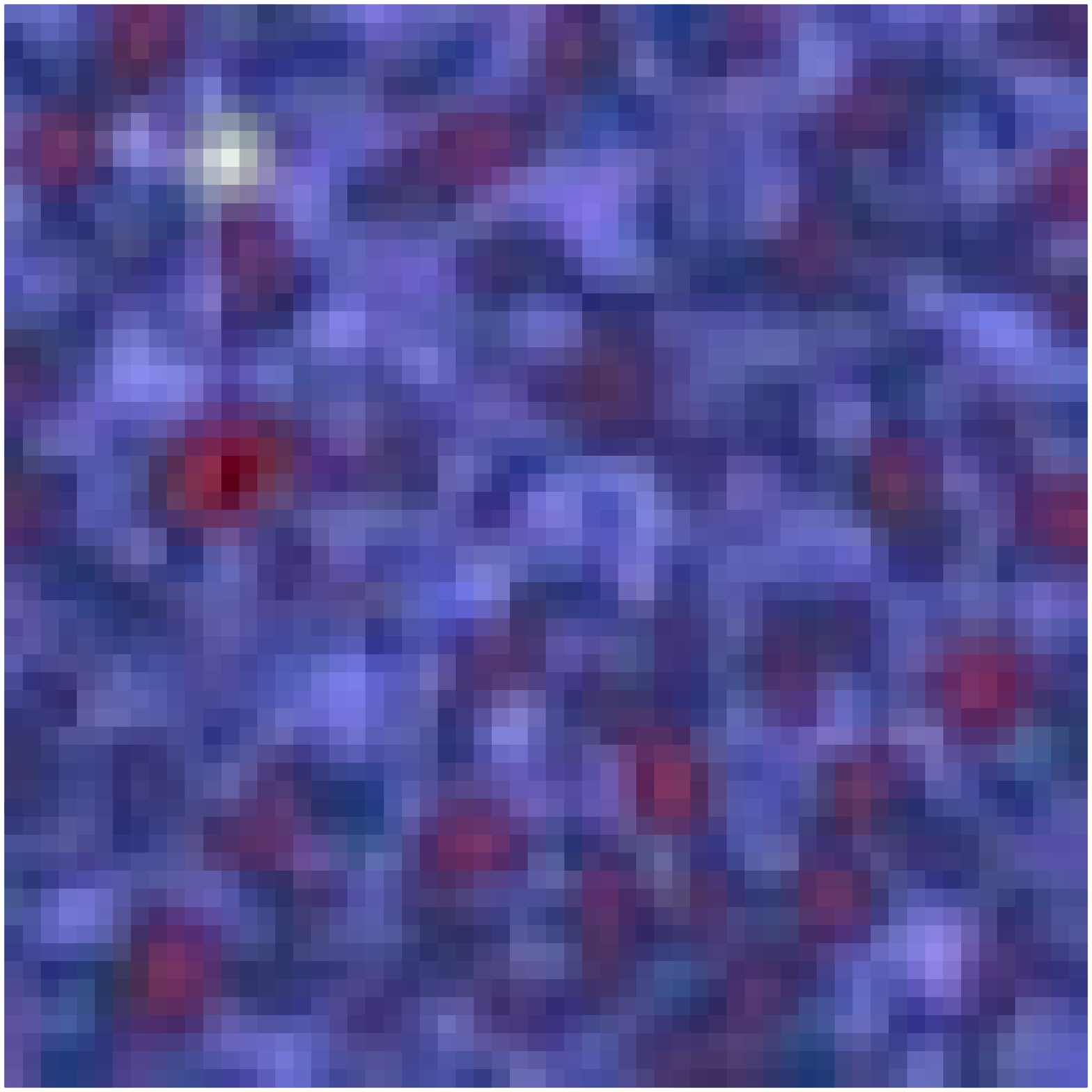}   \FGb{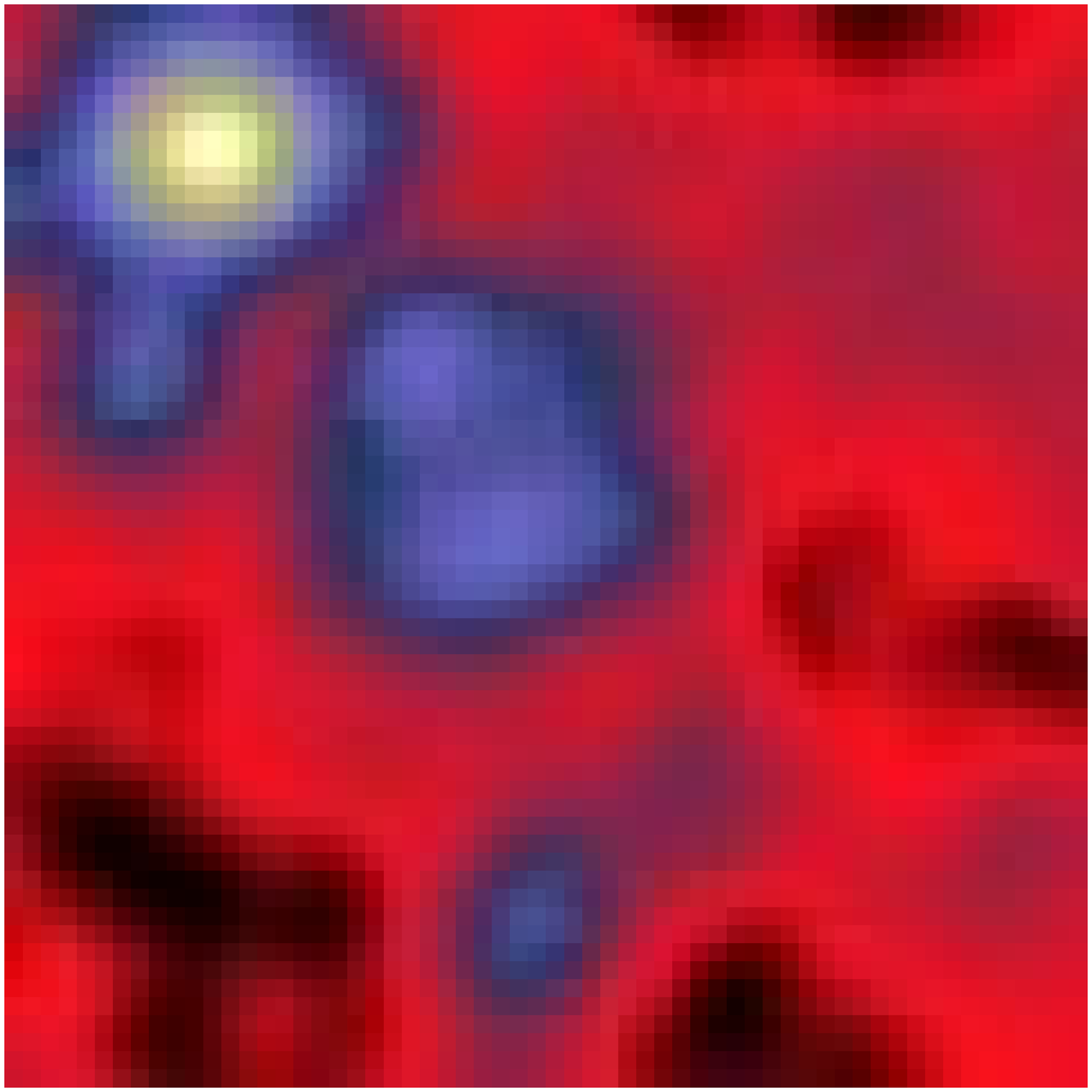}   \FGb{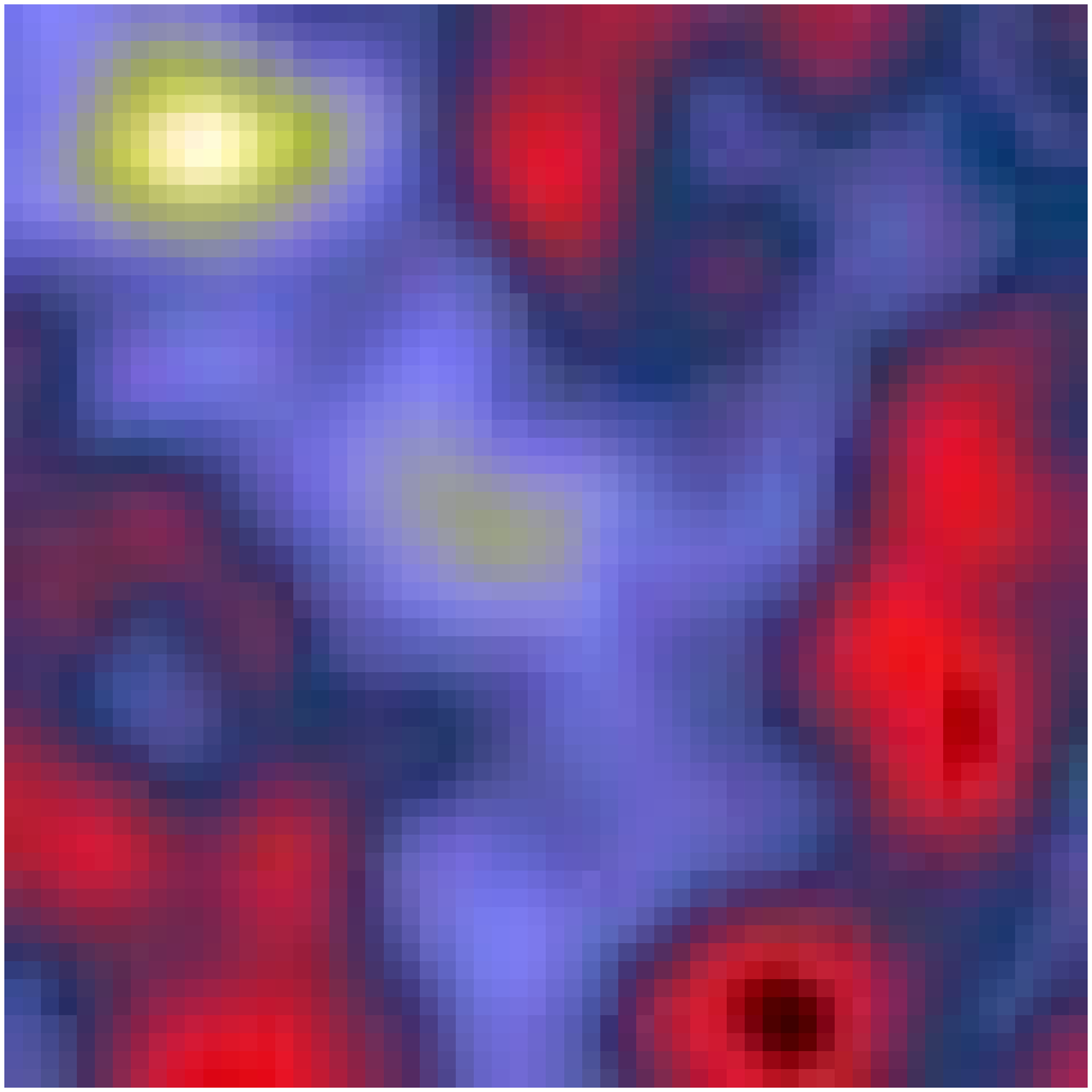}   \FGb{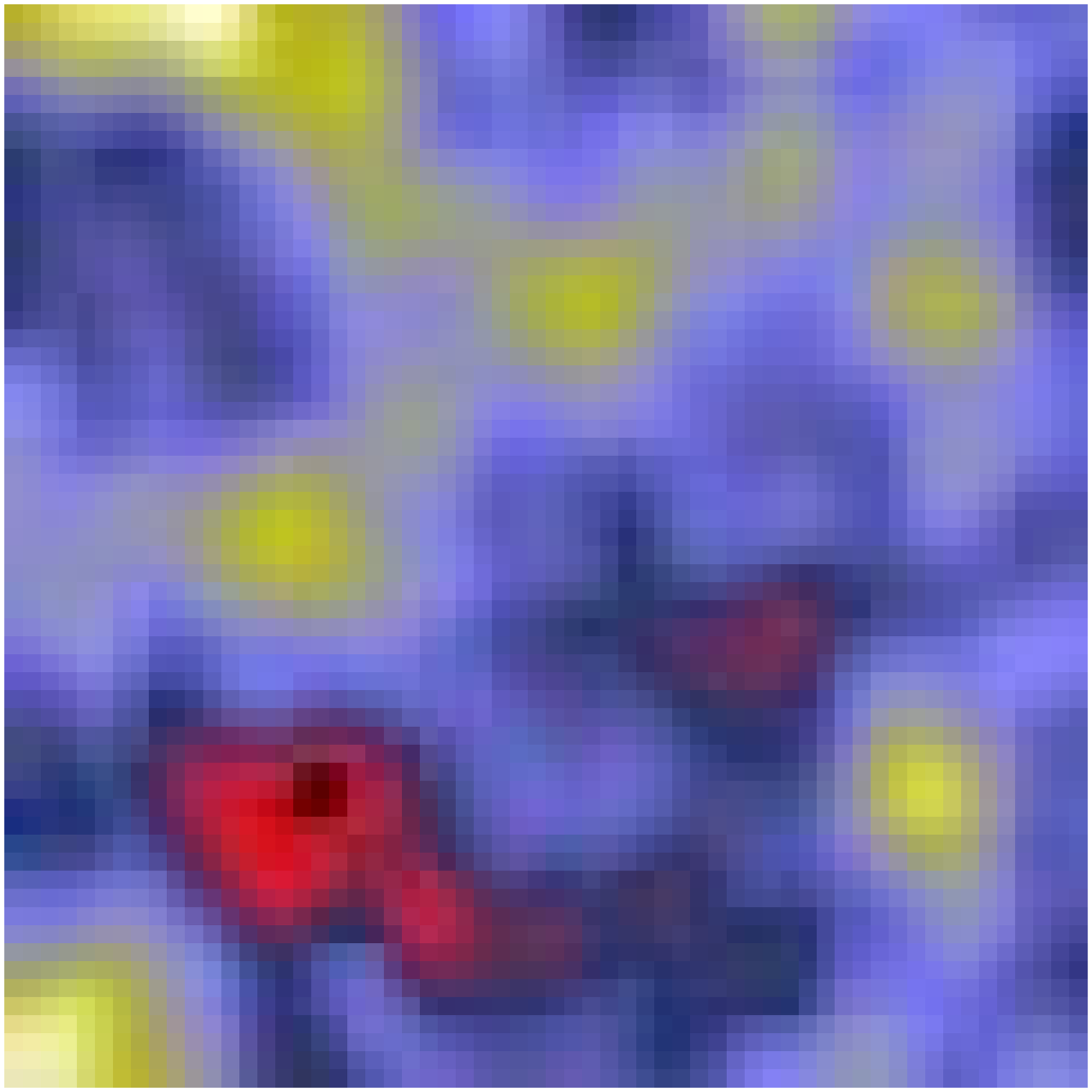}   \FGb{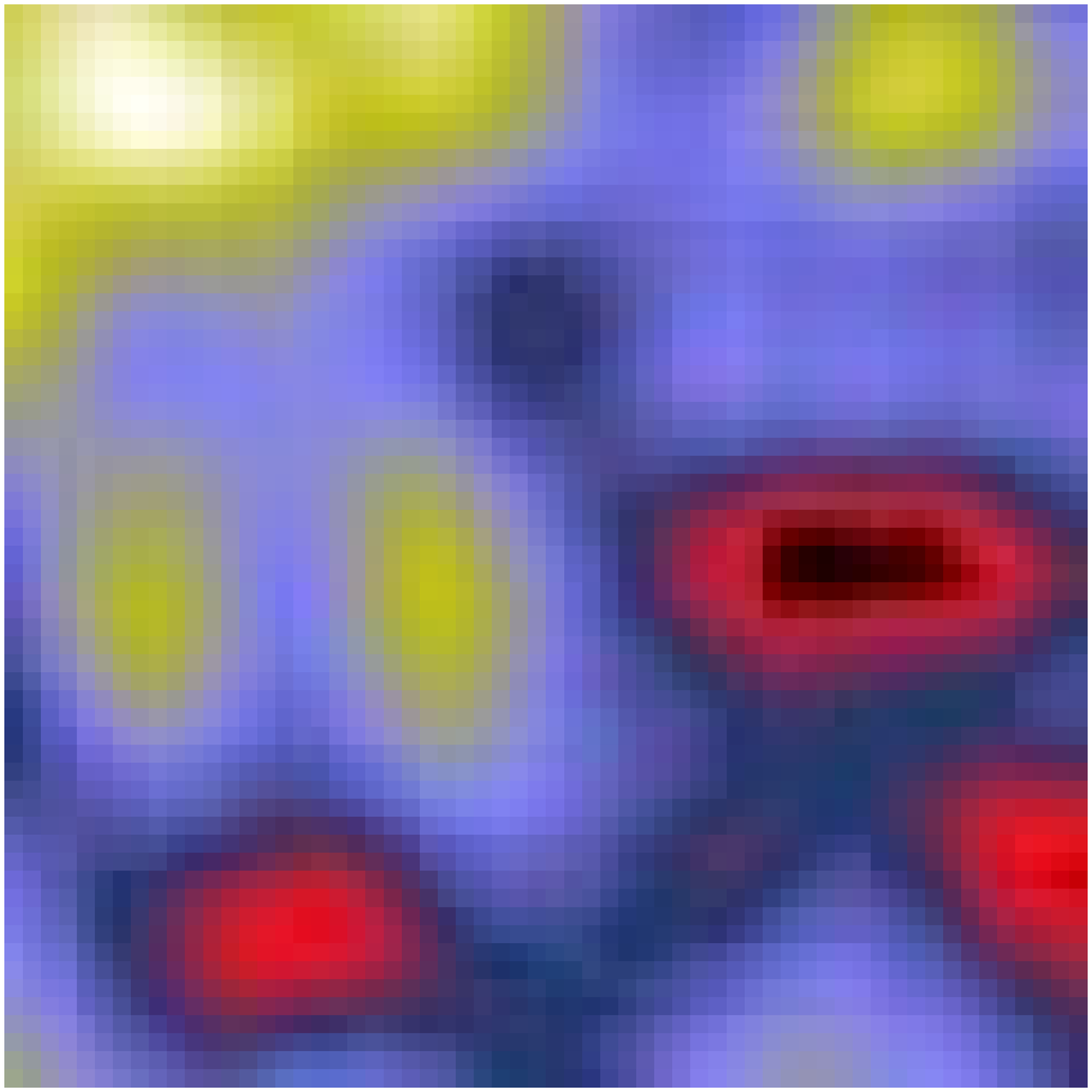} \\  {\#5   NBZ5012013}  \\
  \FGb{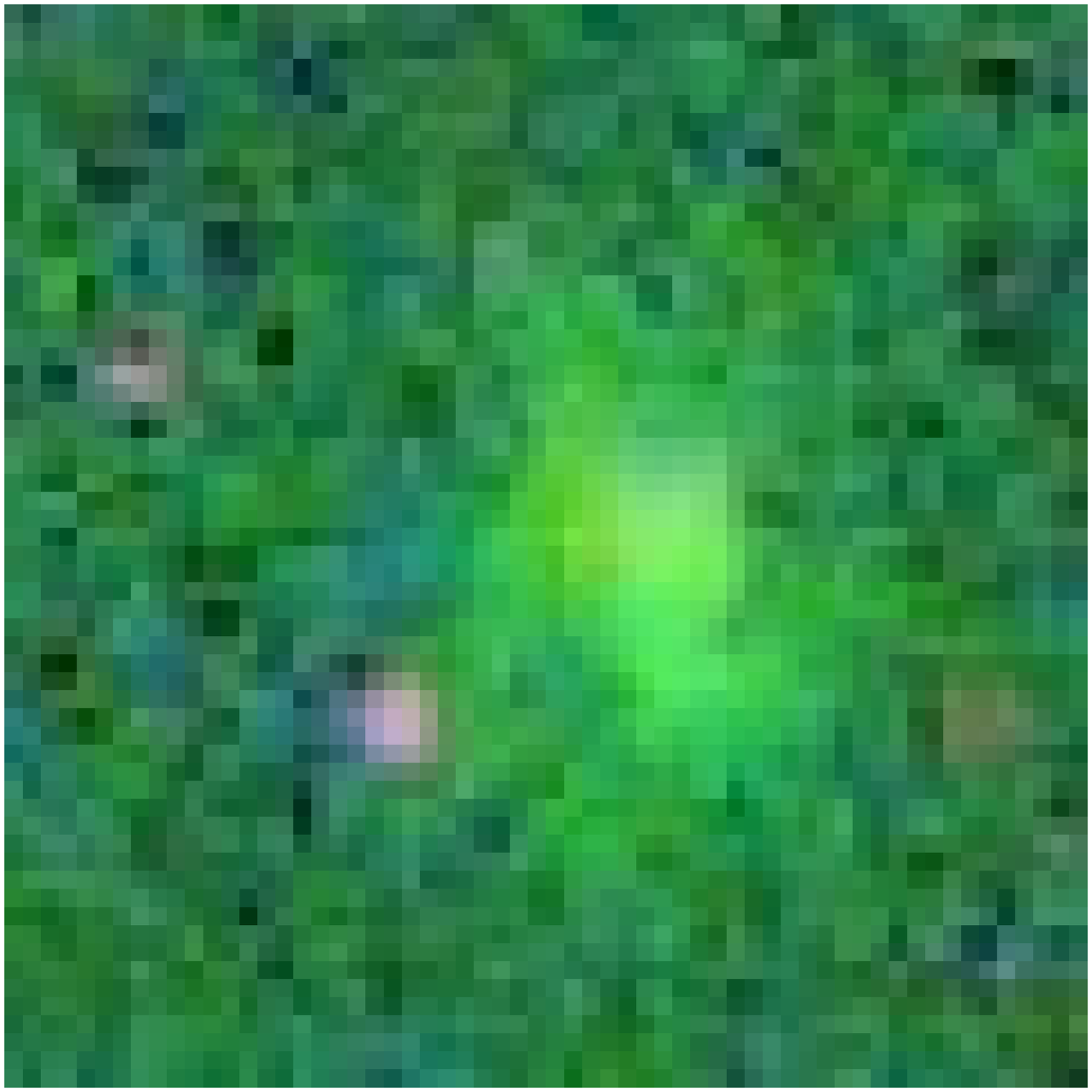}  \FGb{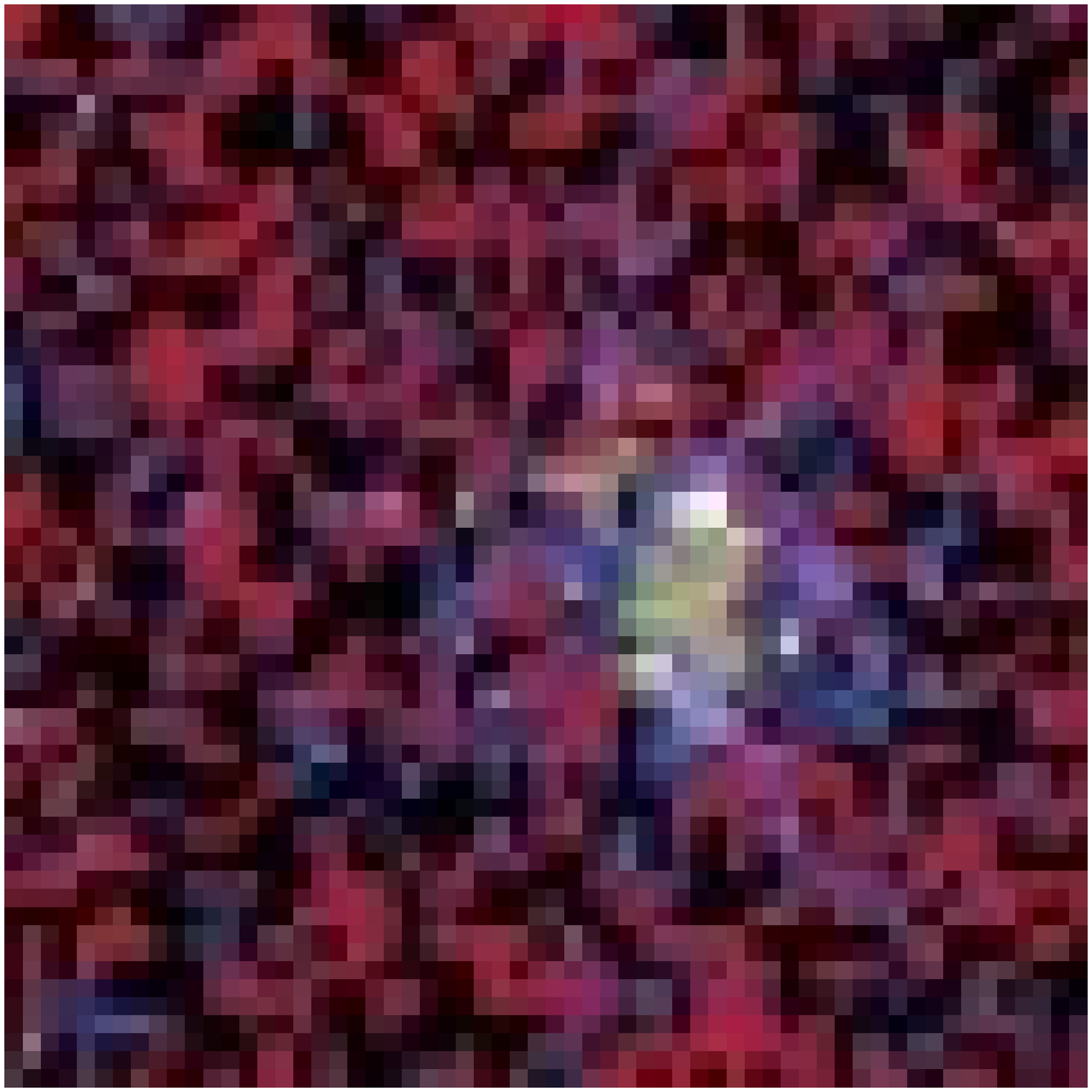}    \FGb{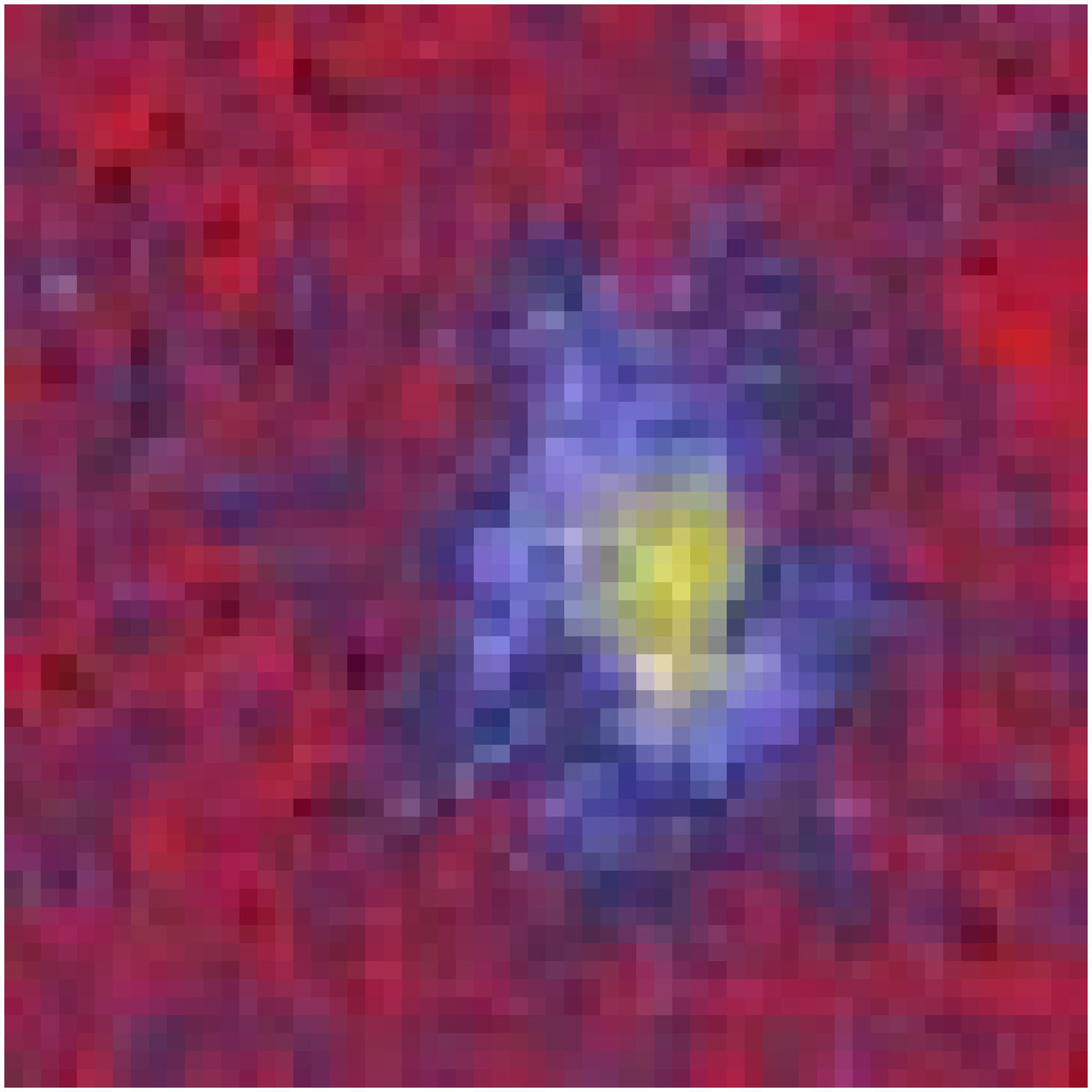}   \FGb{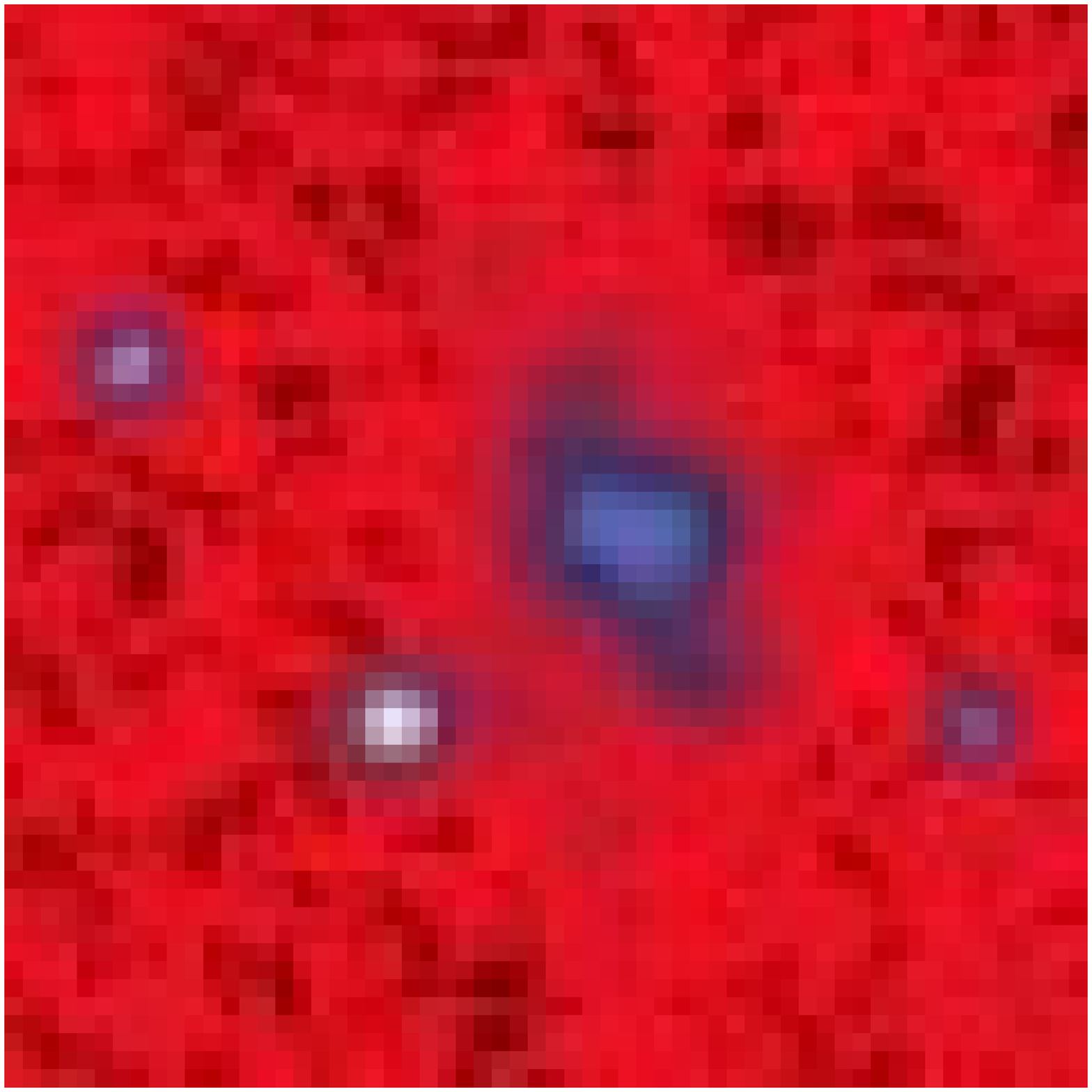}   \FGb{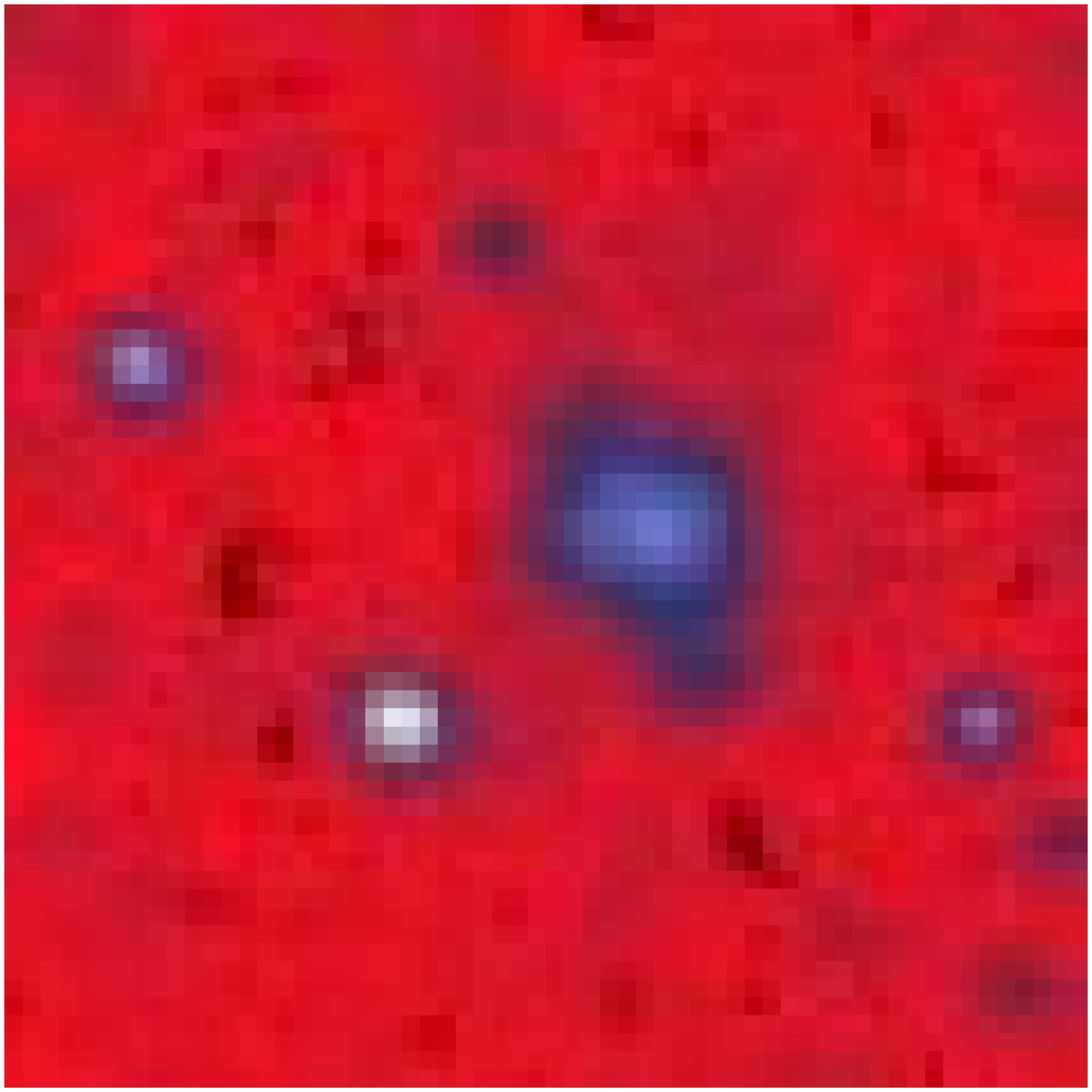}   \FGb{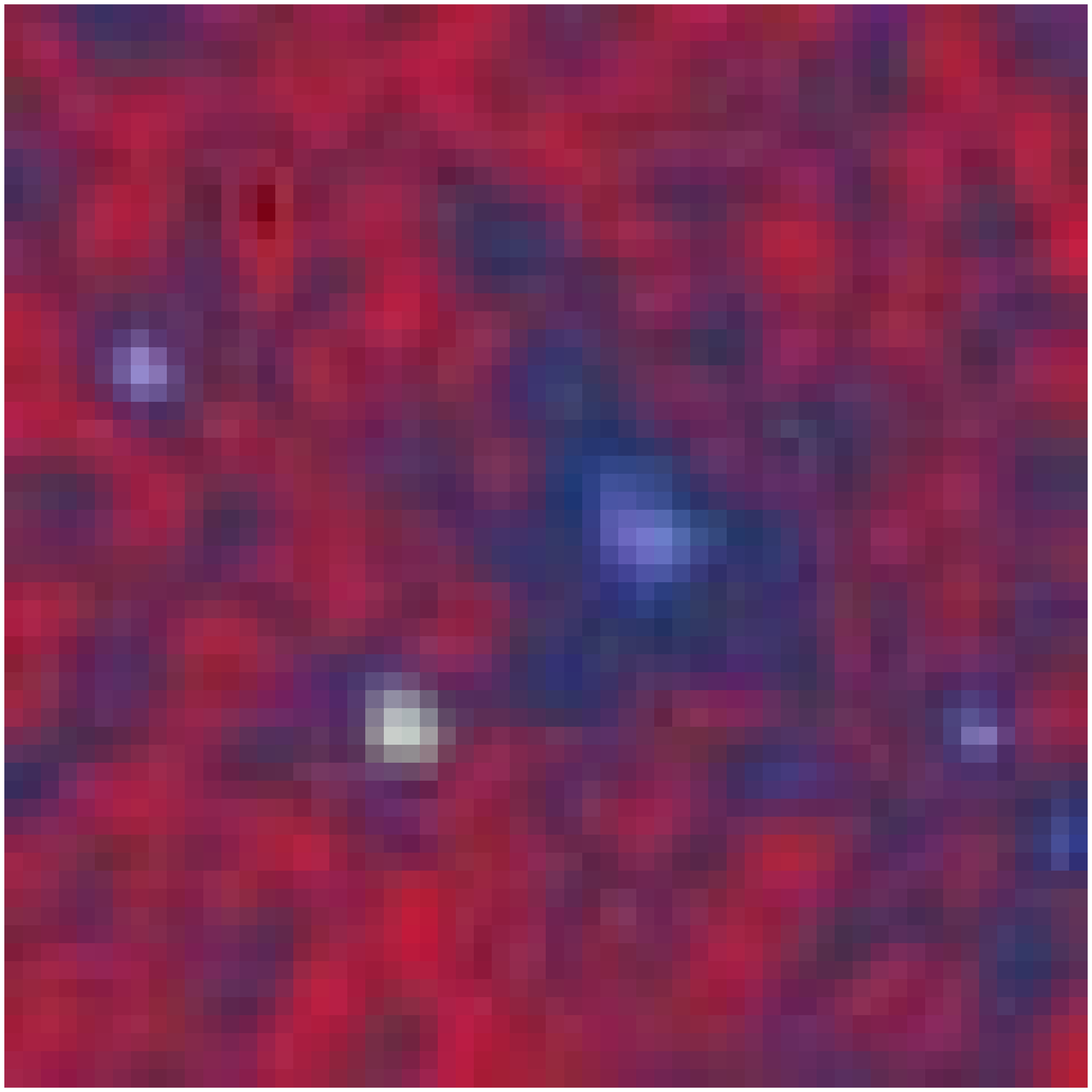}   \FGb{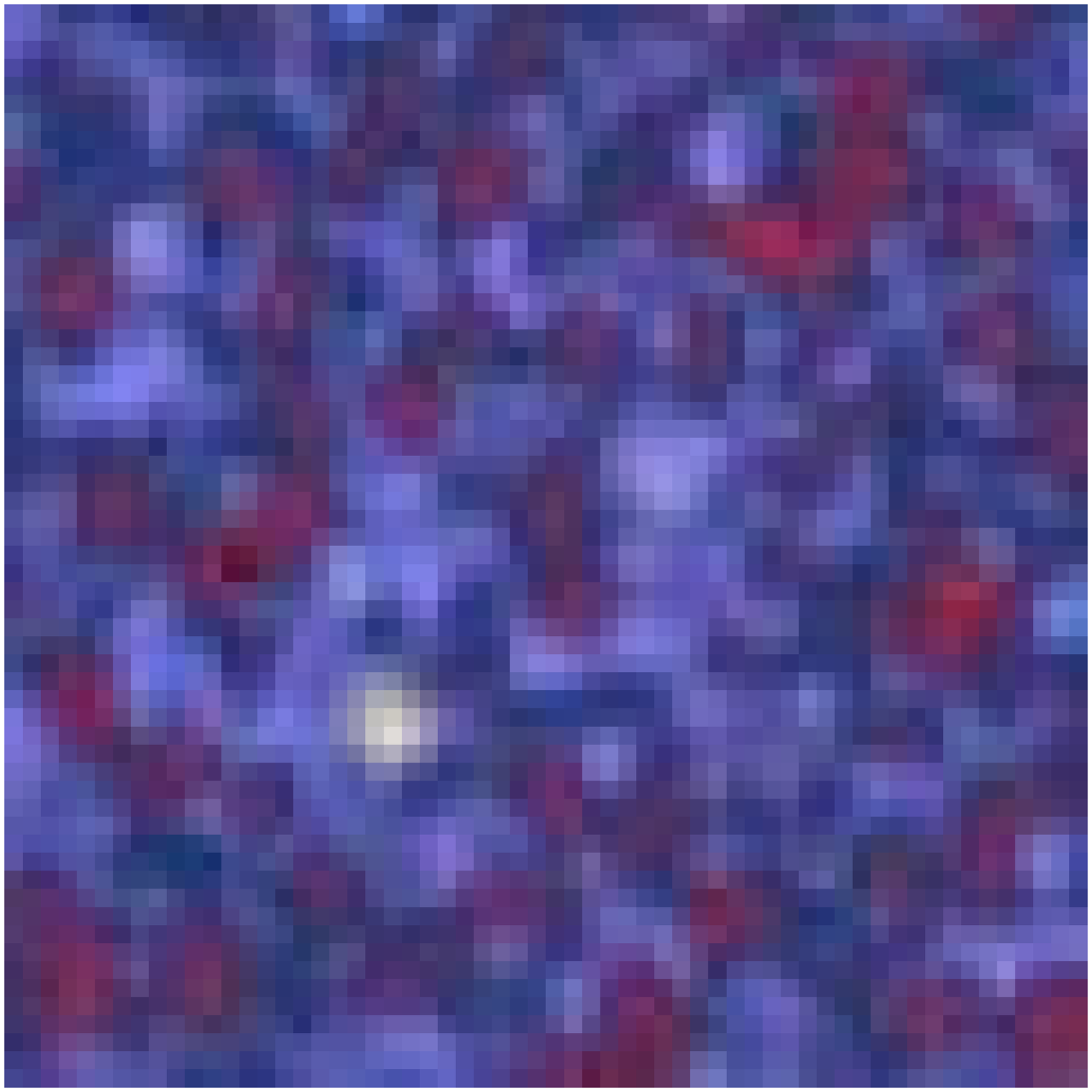}   \FGb{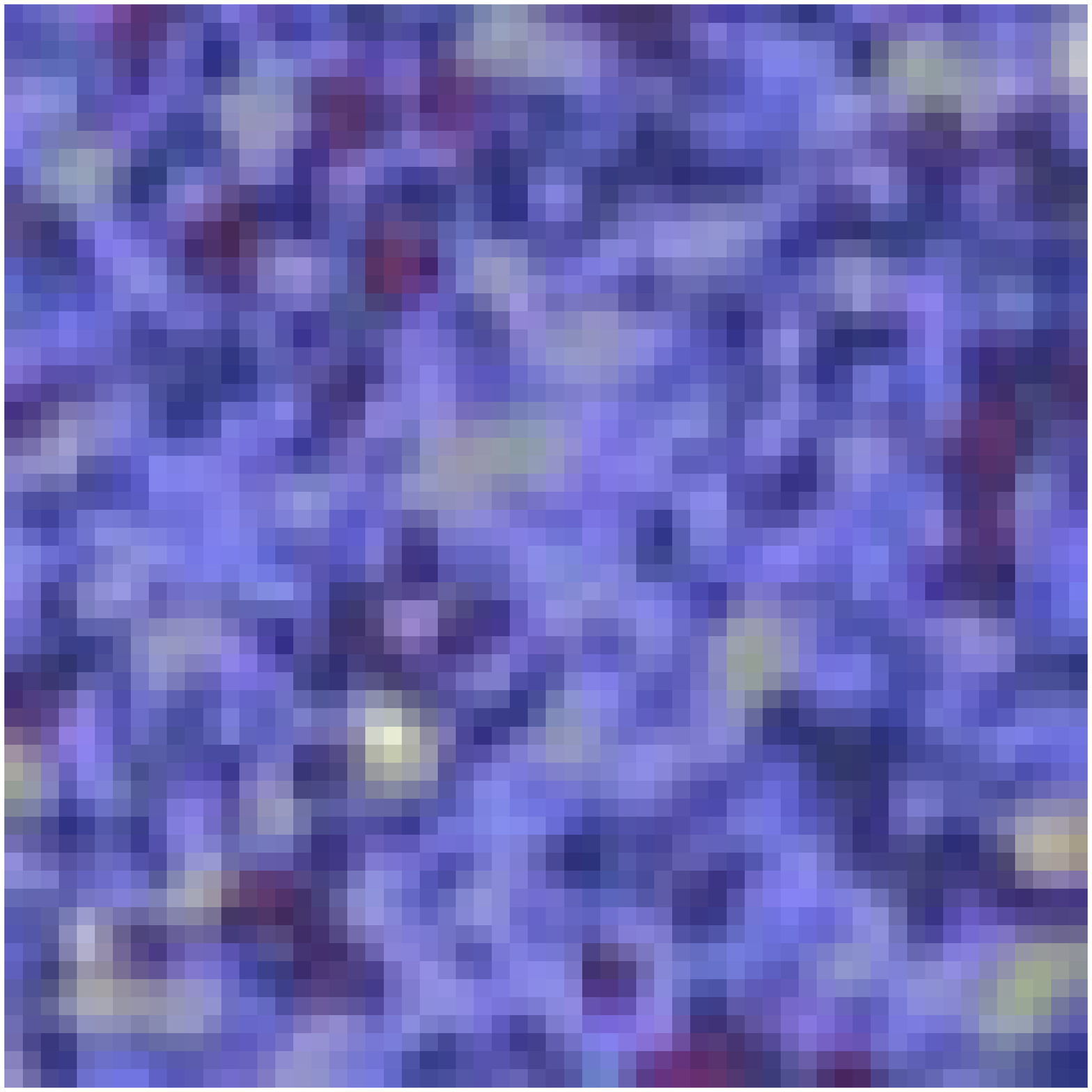}   \FGb{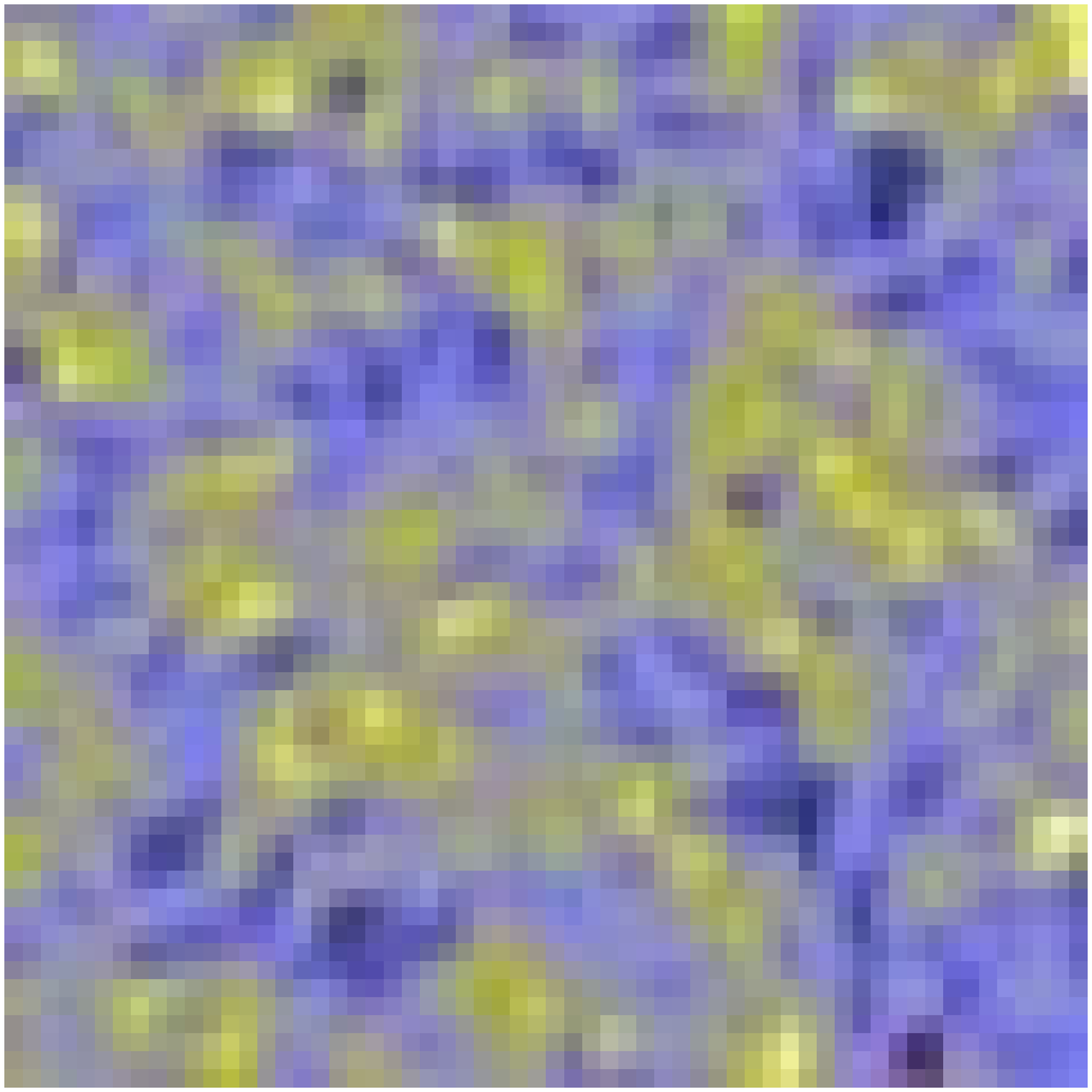}   \FGb{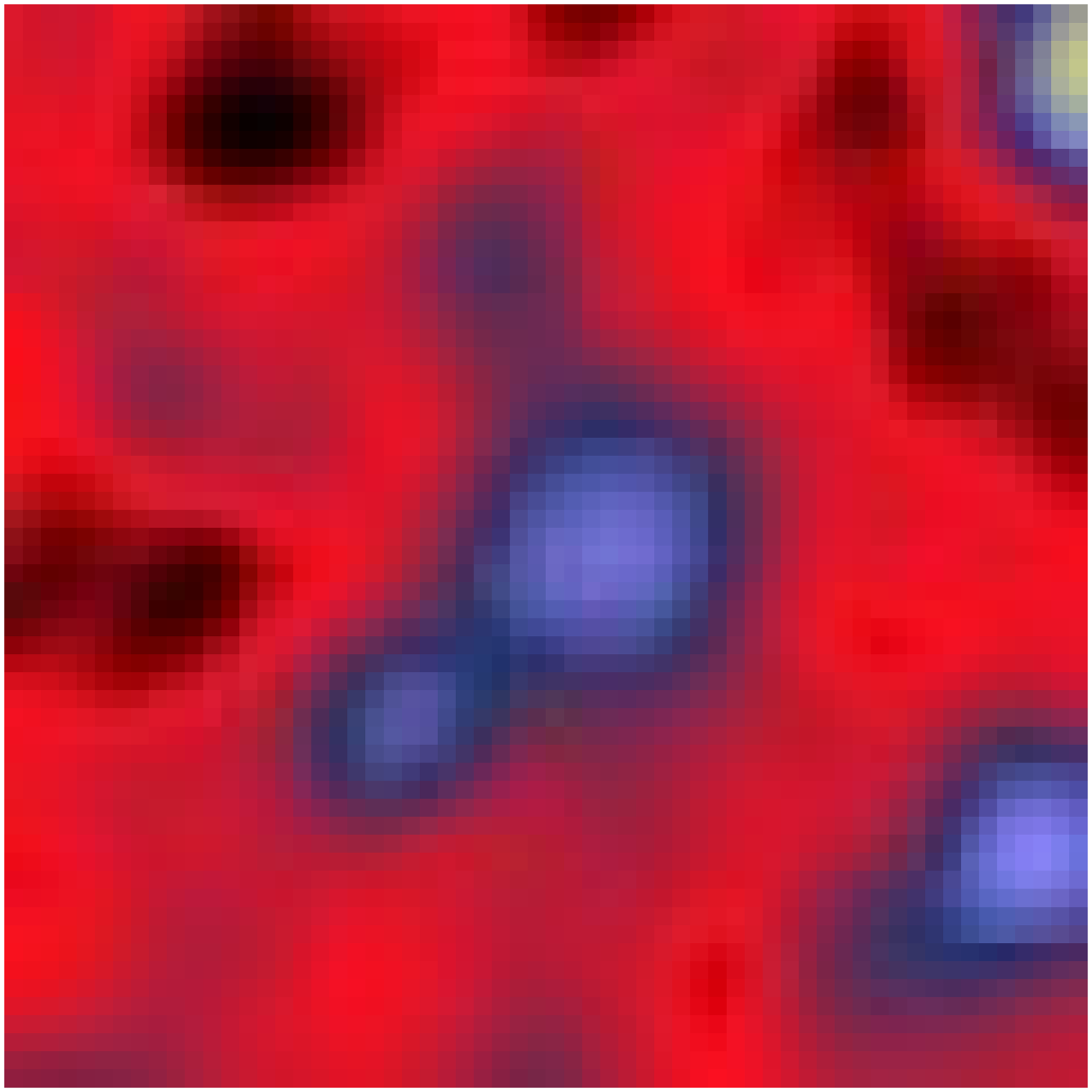}   \FGb{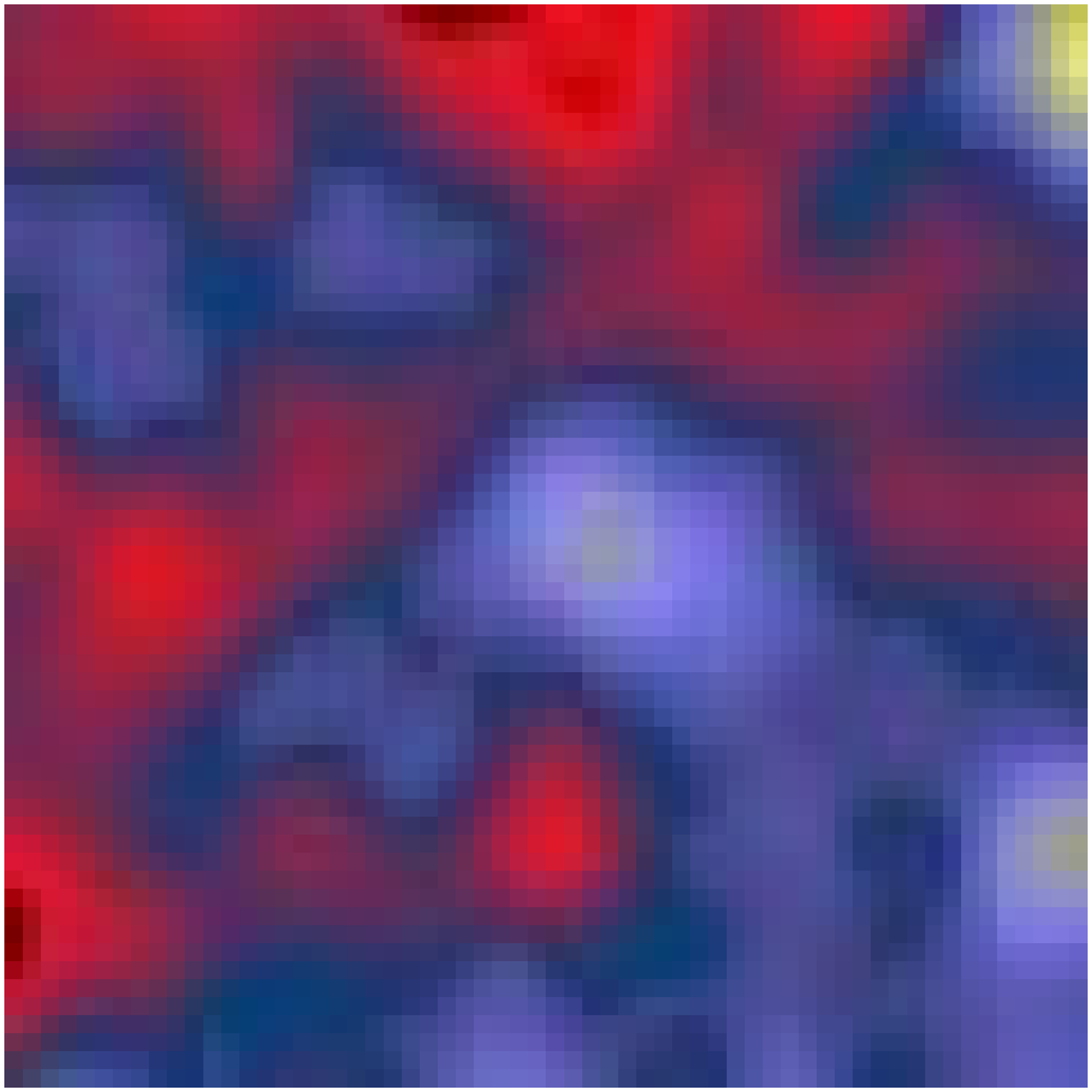}   \FGb{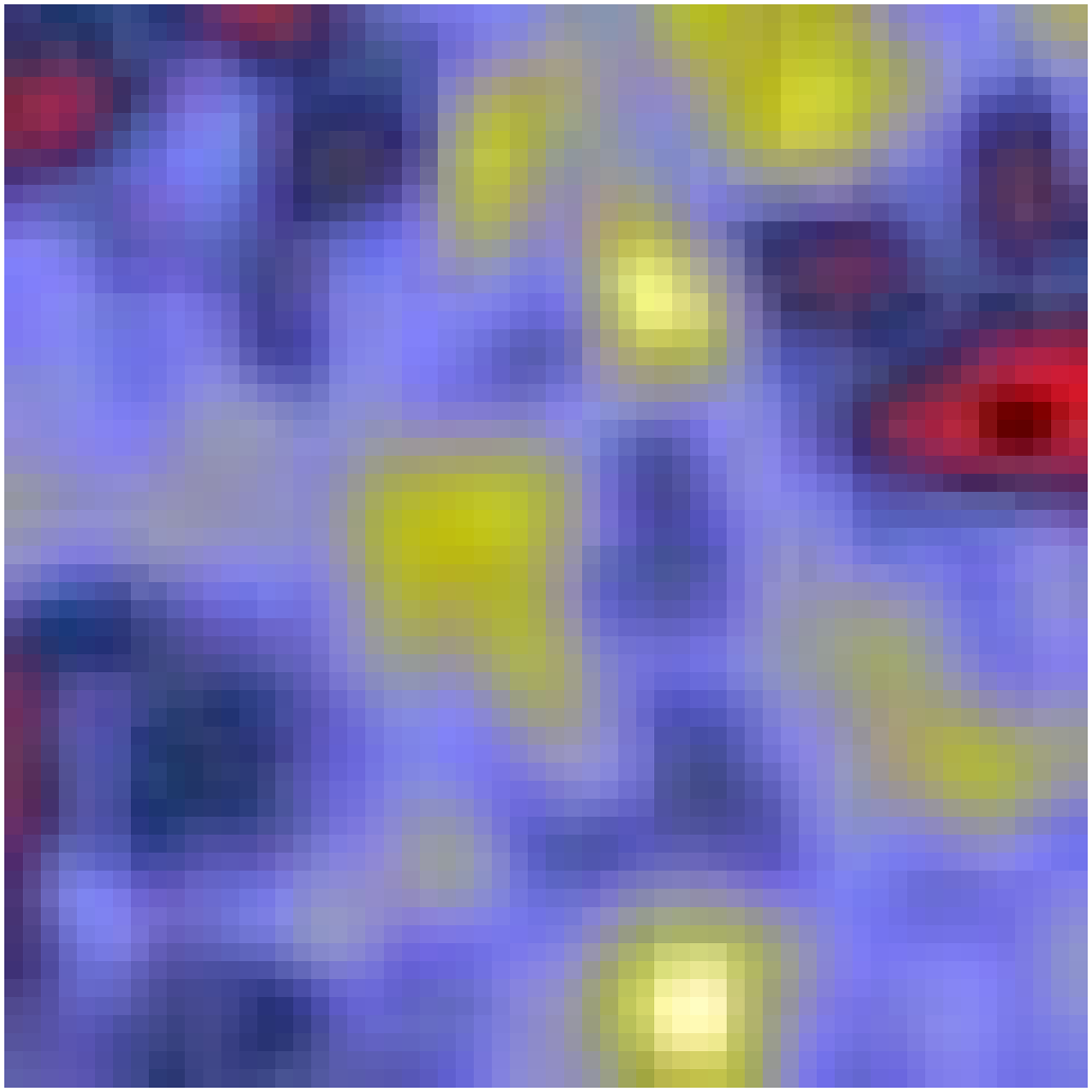}   \FGb{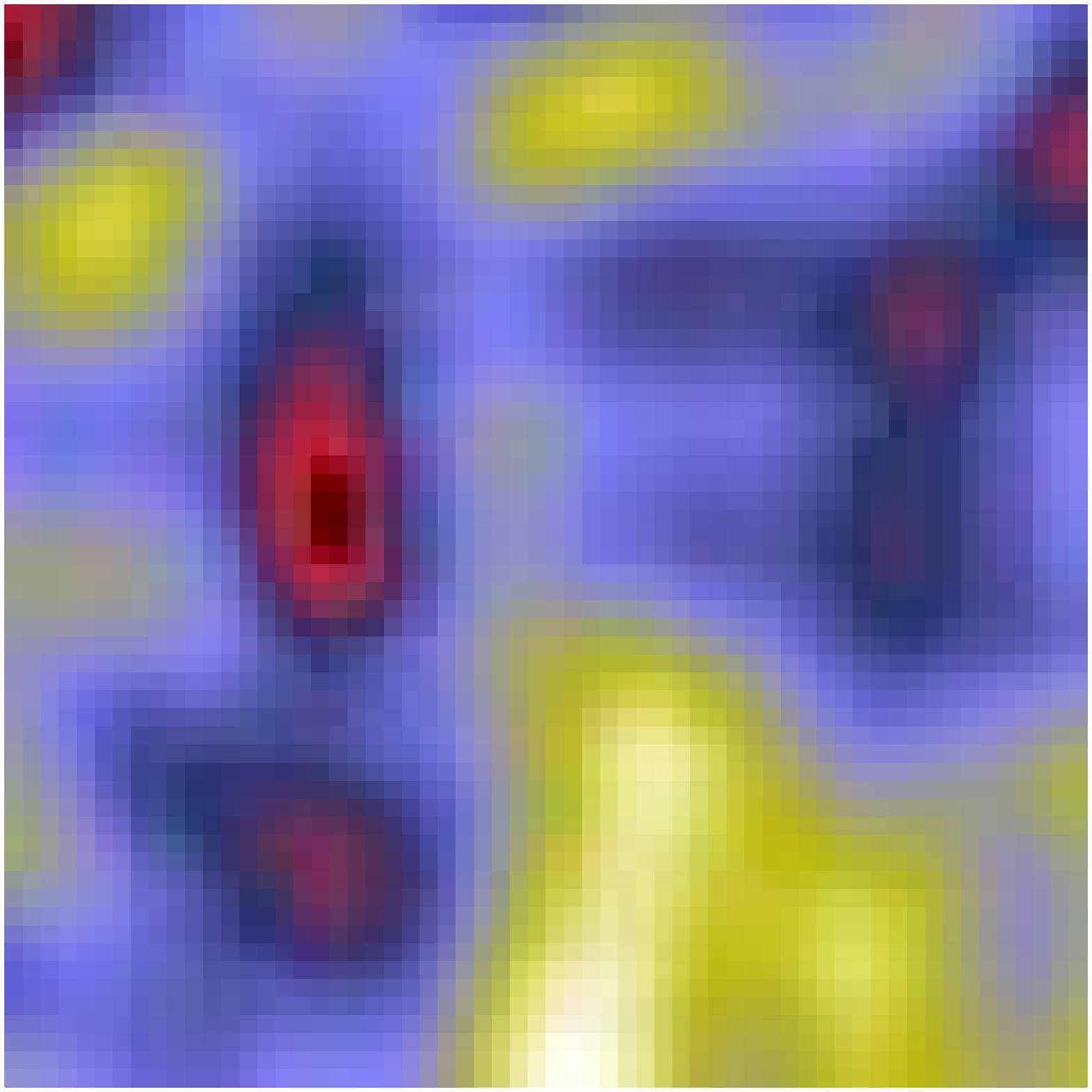} \\  {\#6   NCIB031541}  \\
  \FGb{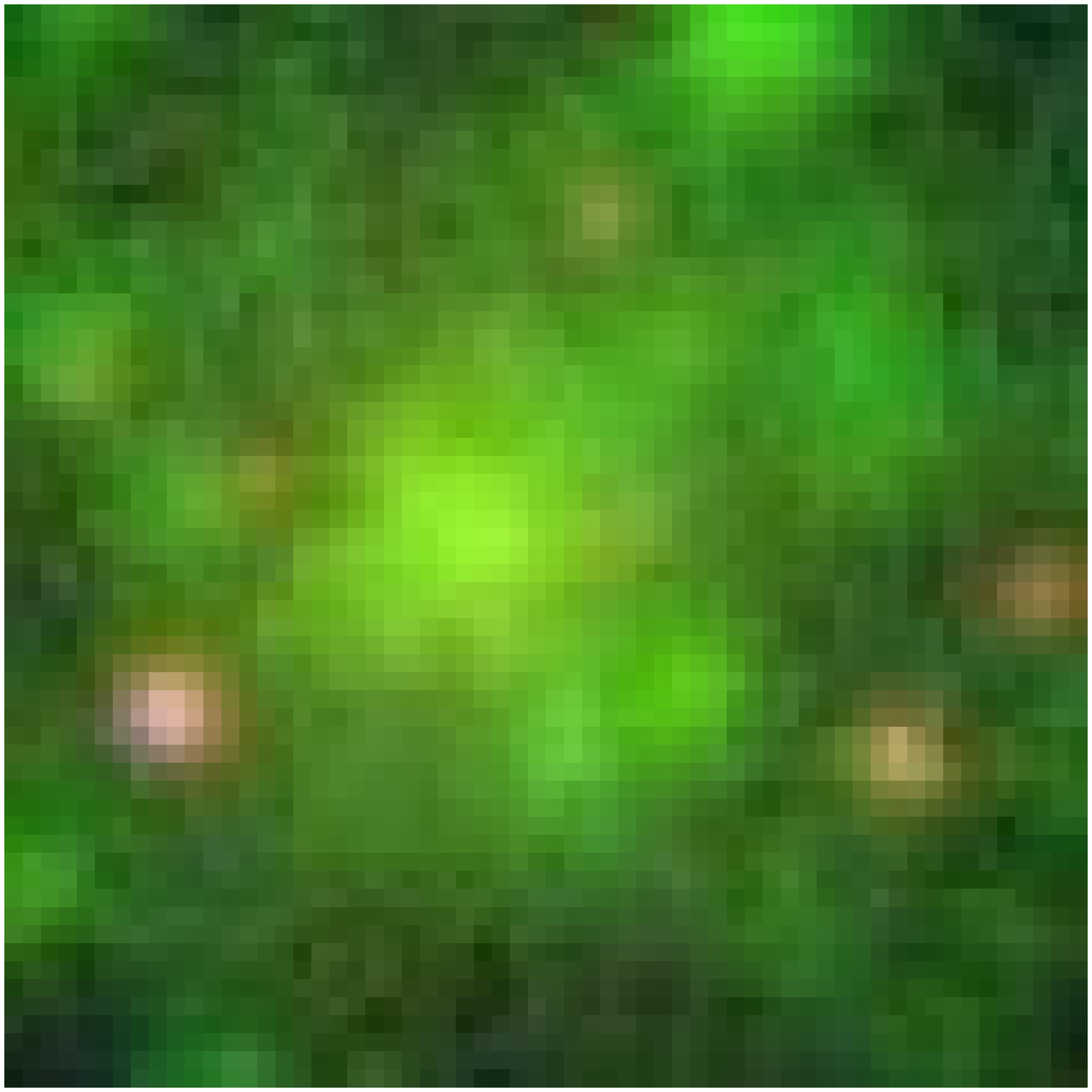}  \FGb{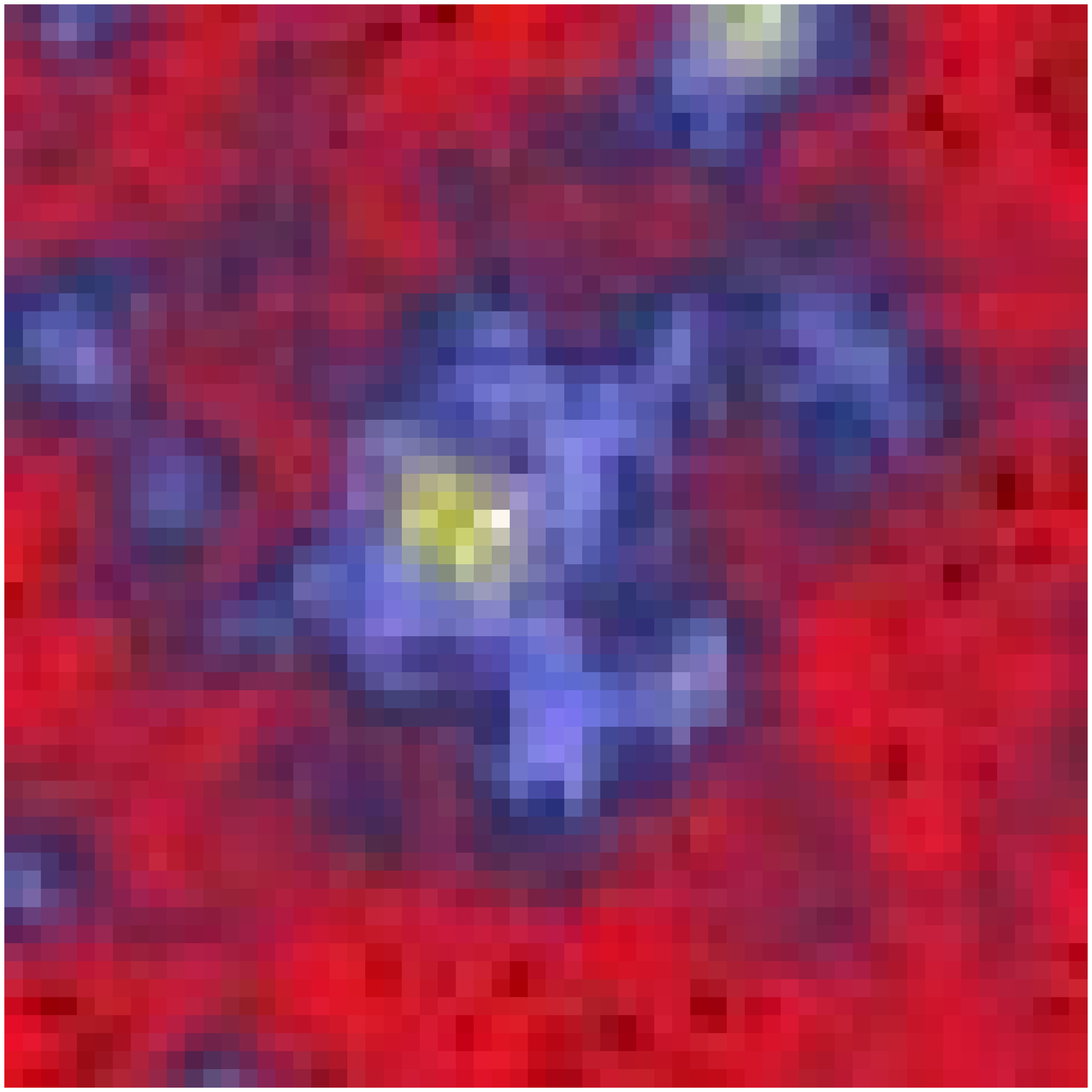}    \FGb{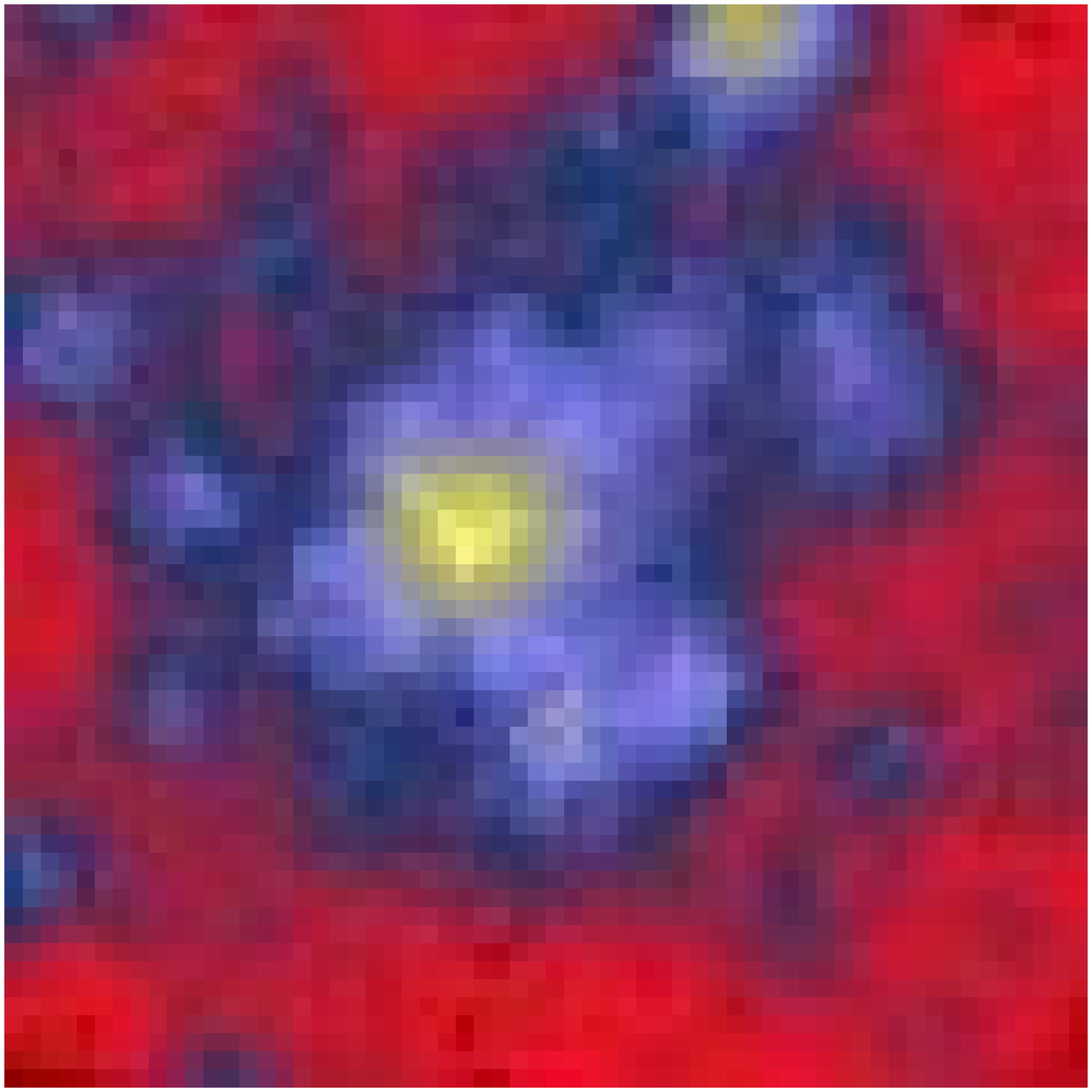}   \FGb{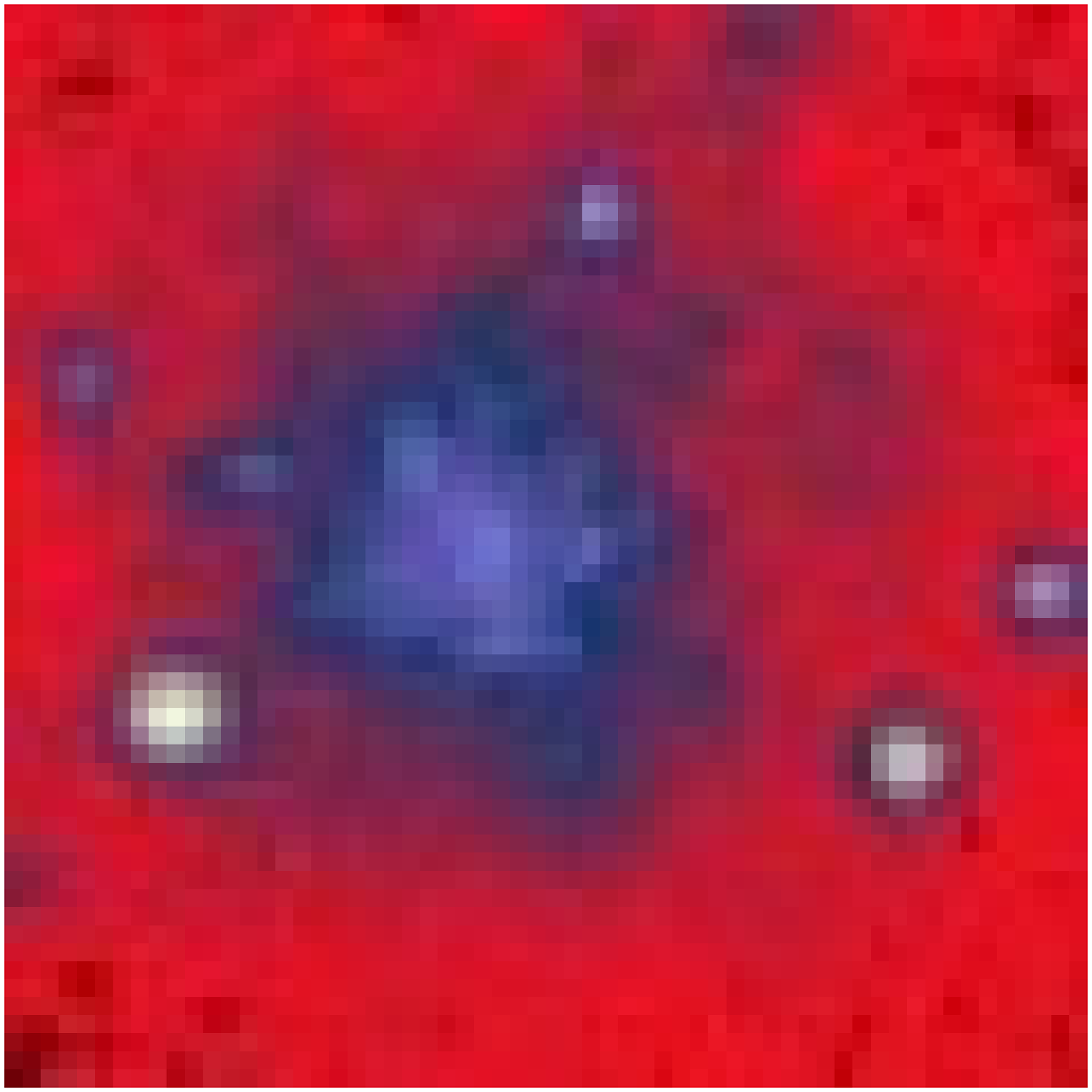}   \FGb{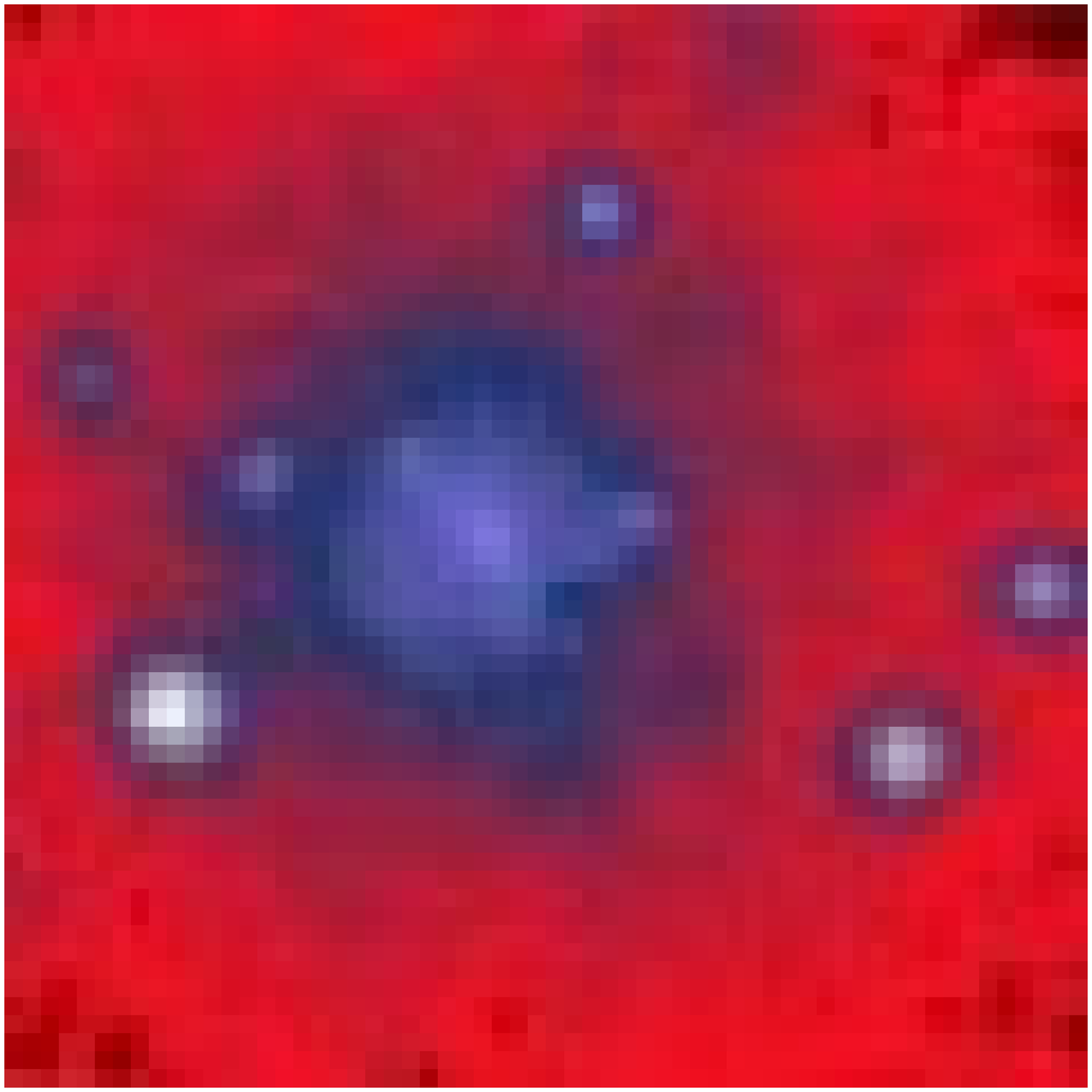}   \FGb{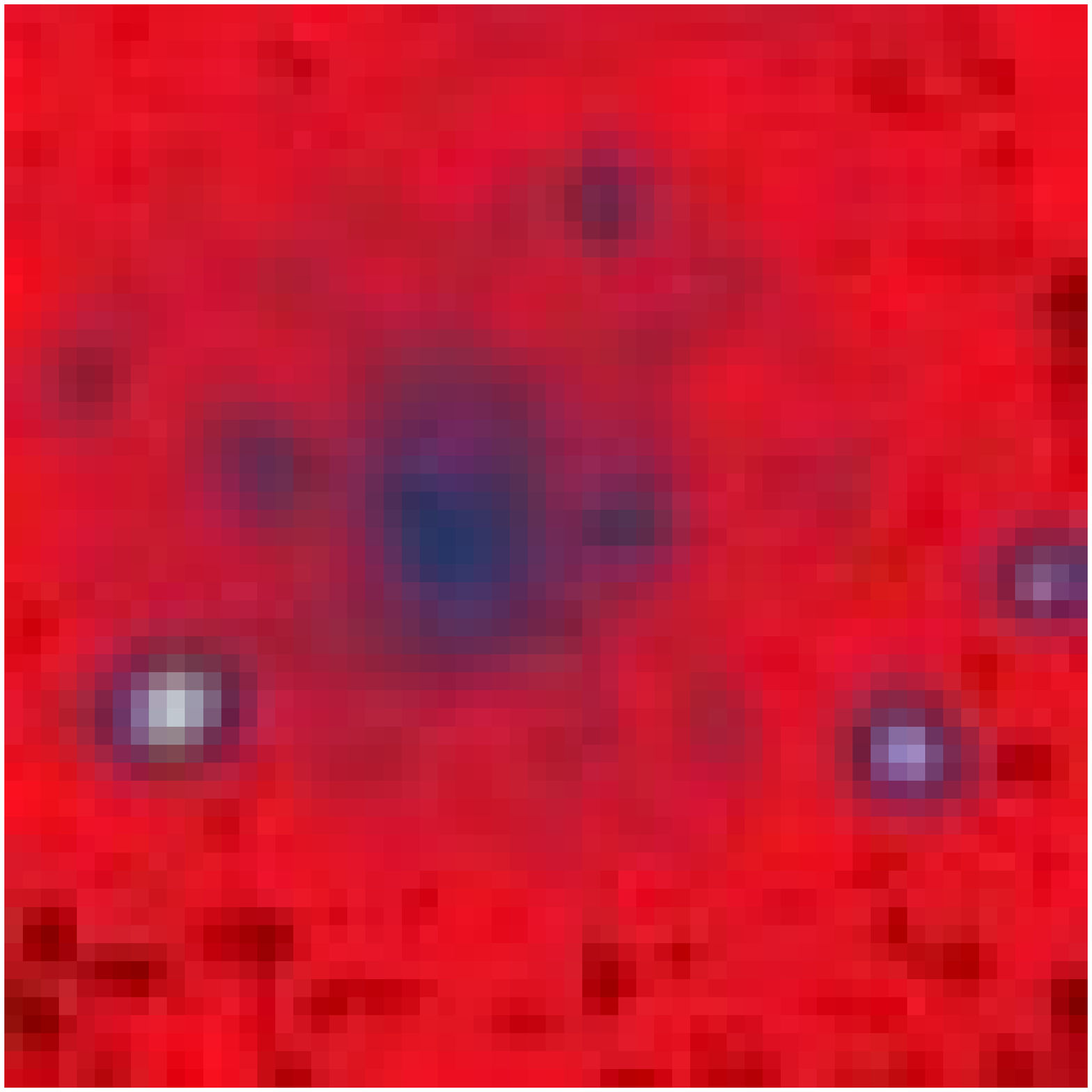}   \FGb{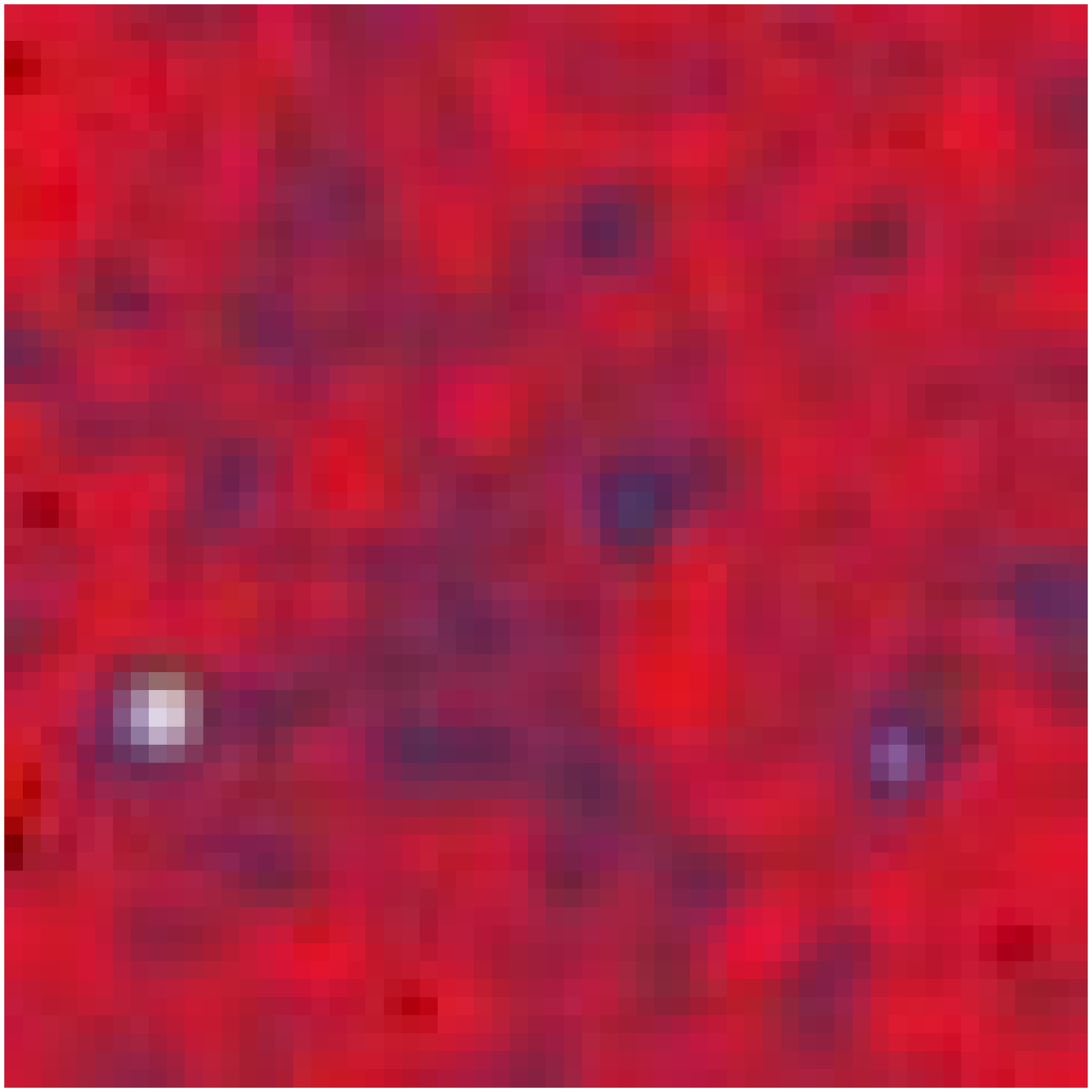}   \FGb{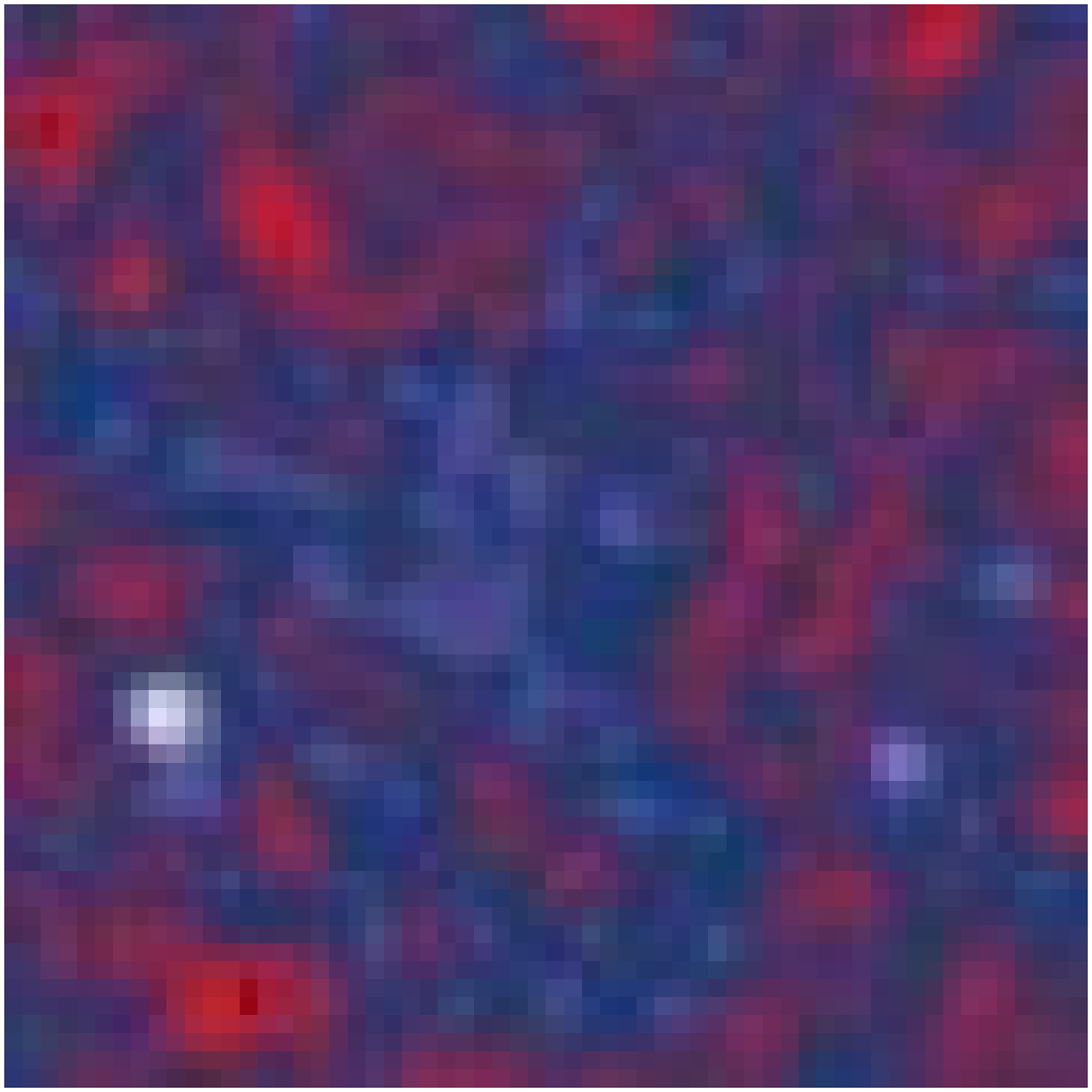}   \FGb{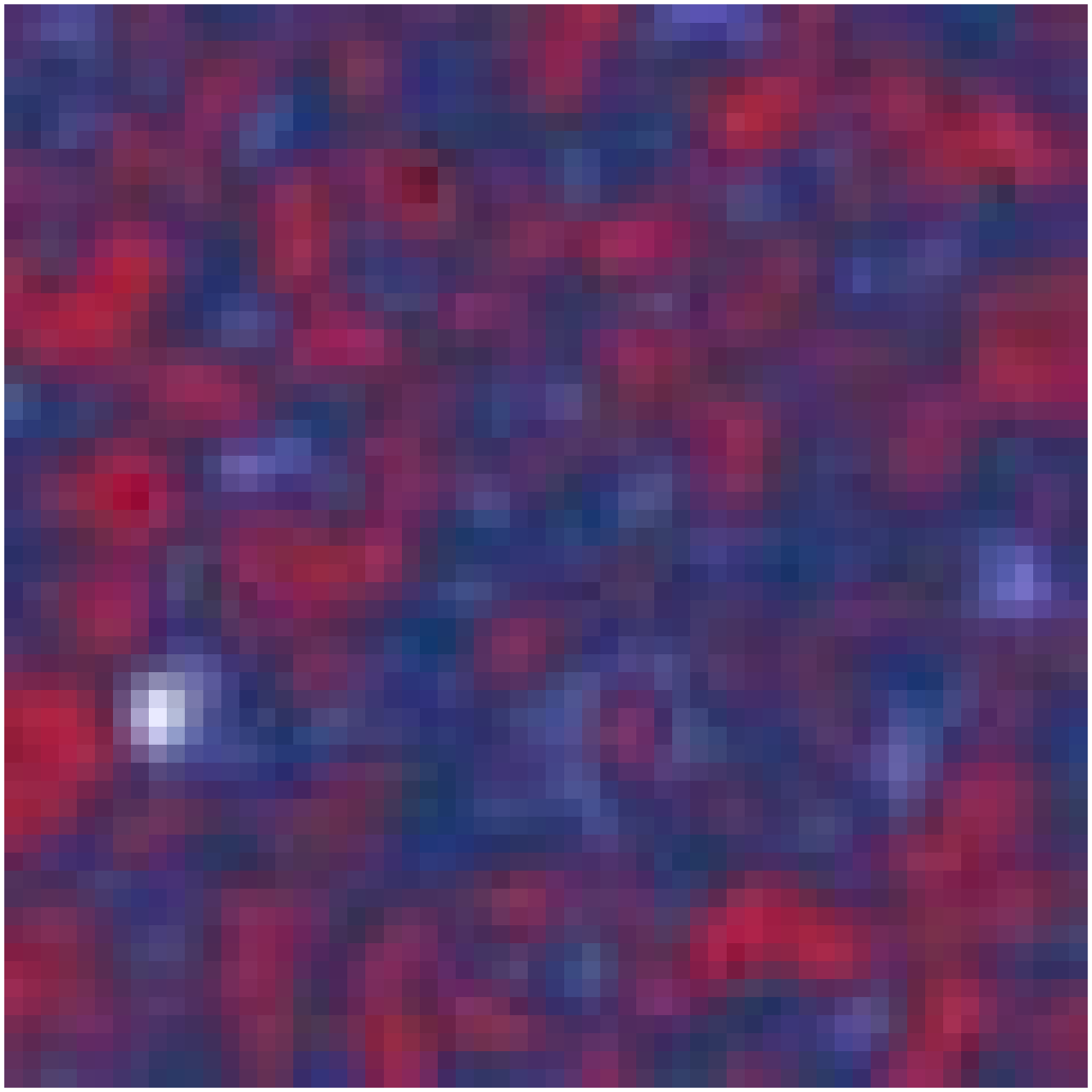}   \FGb{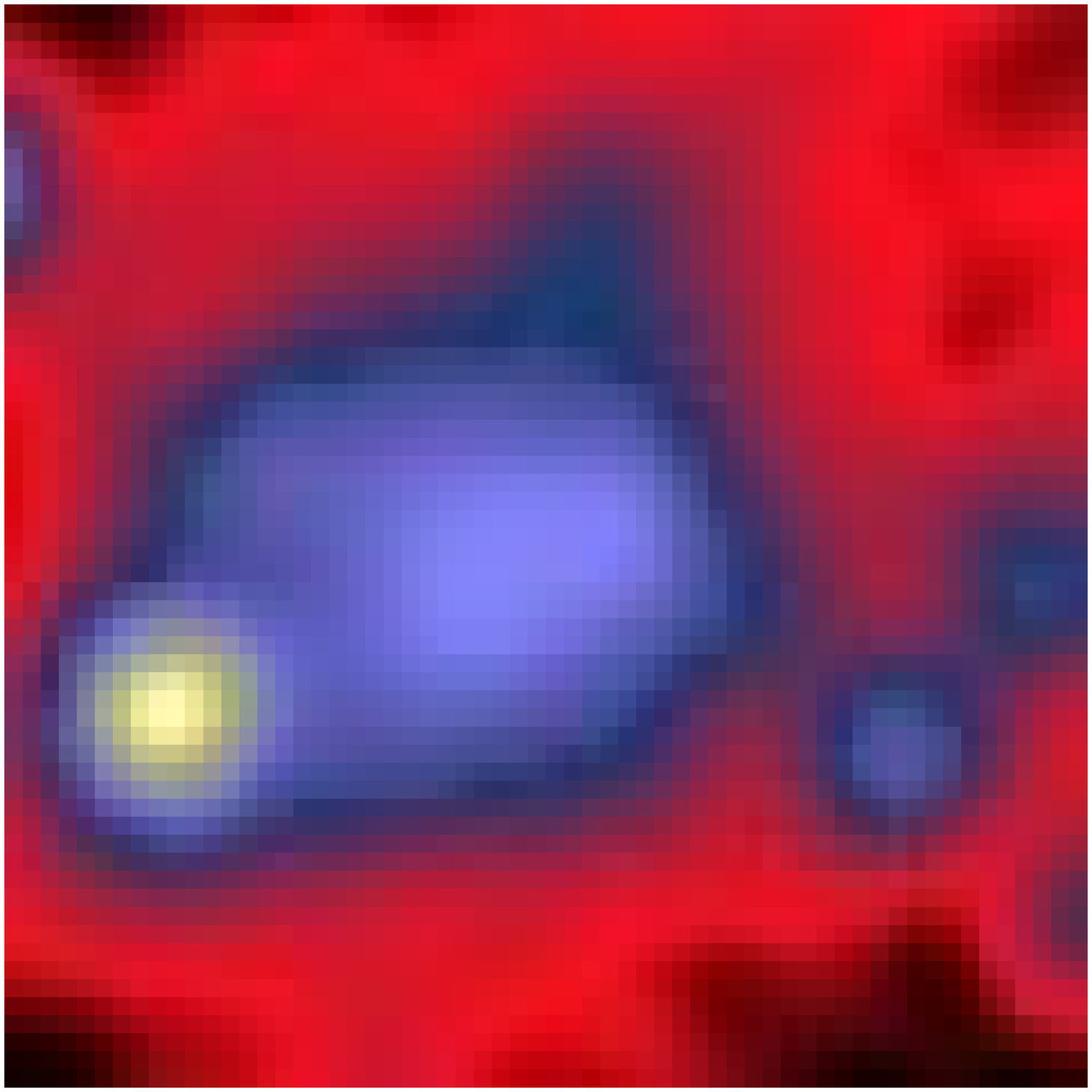}   \FGb{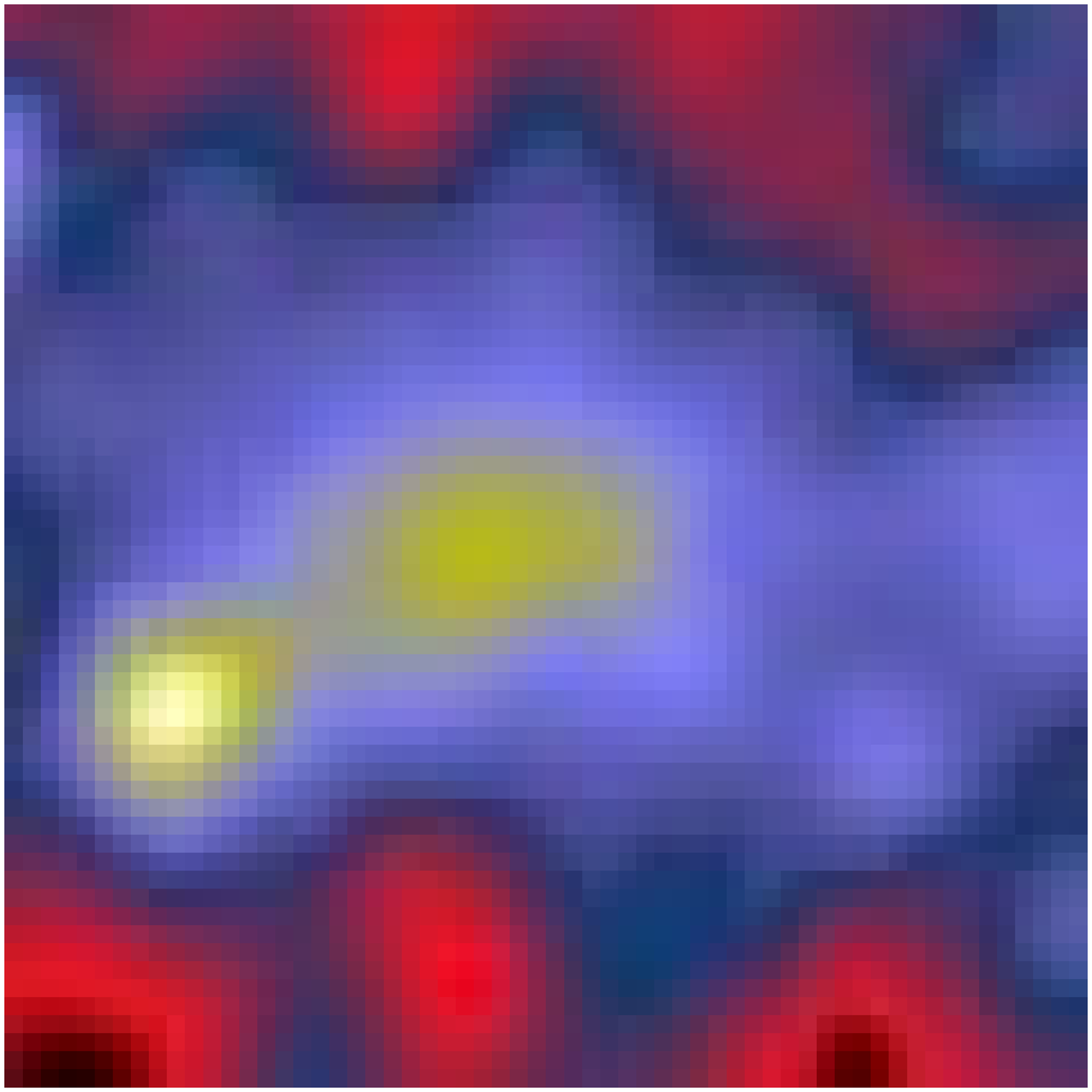}   \FGb{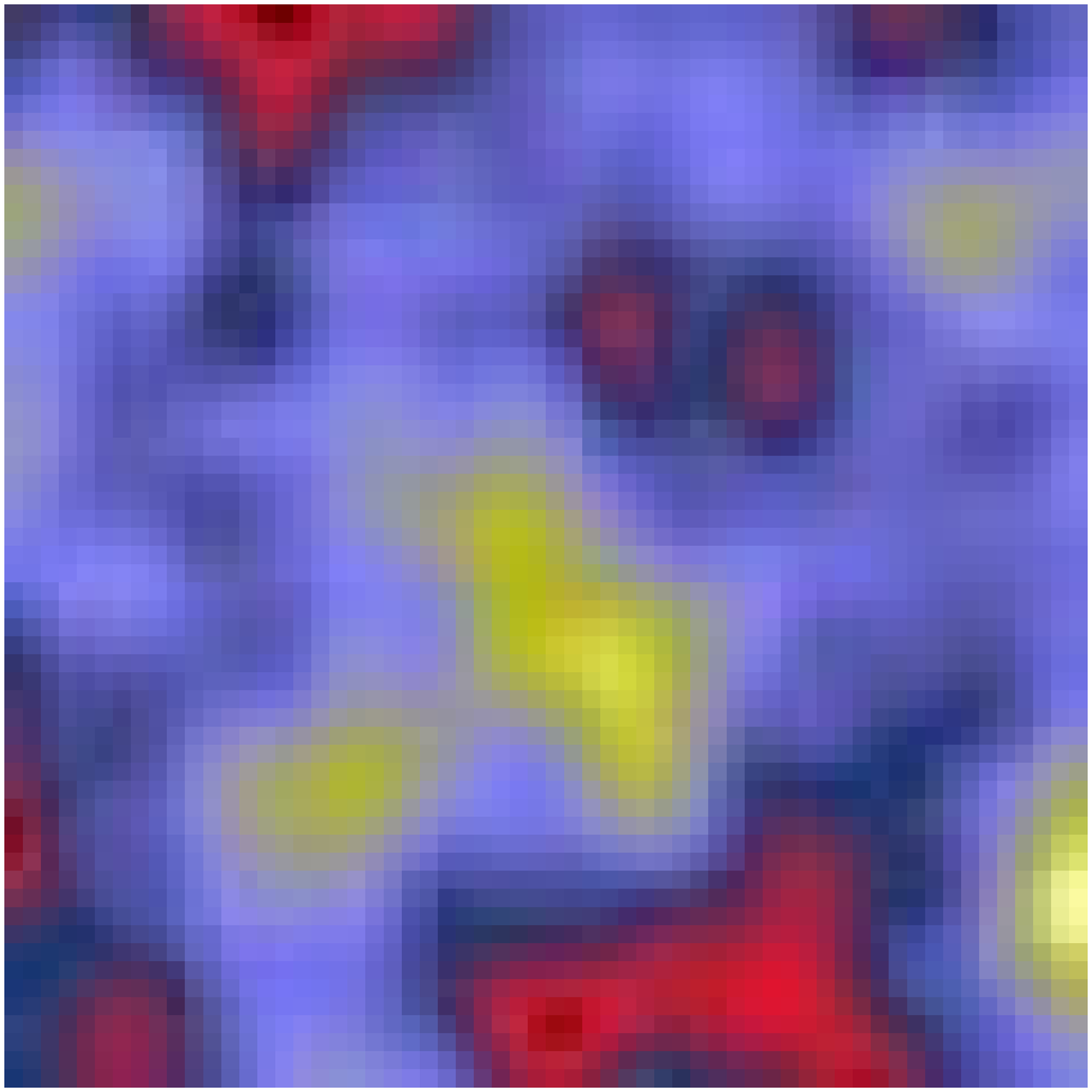}   \FGb{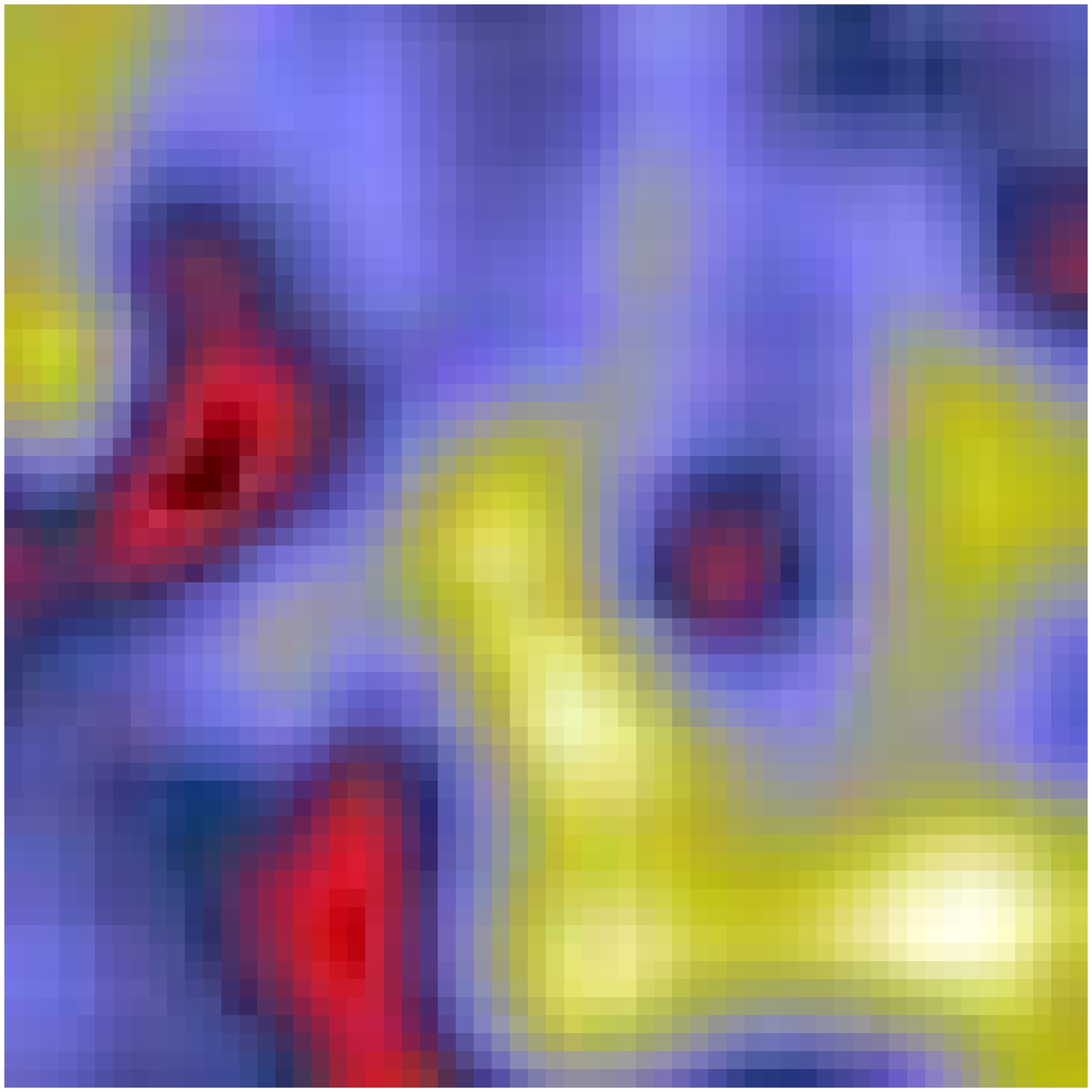} \\  {\#7   NBZ5012371}  \\
  \FGb{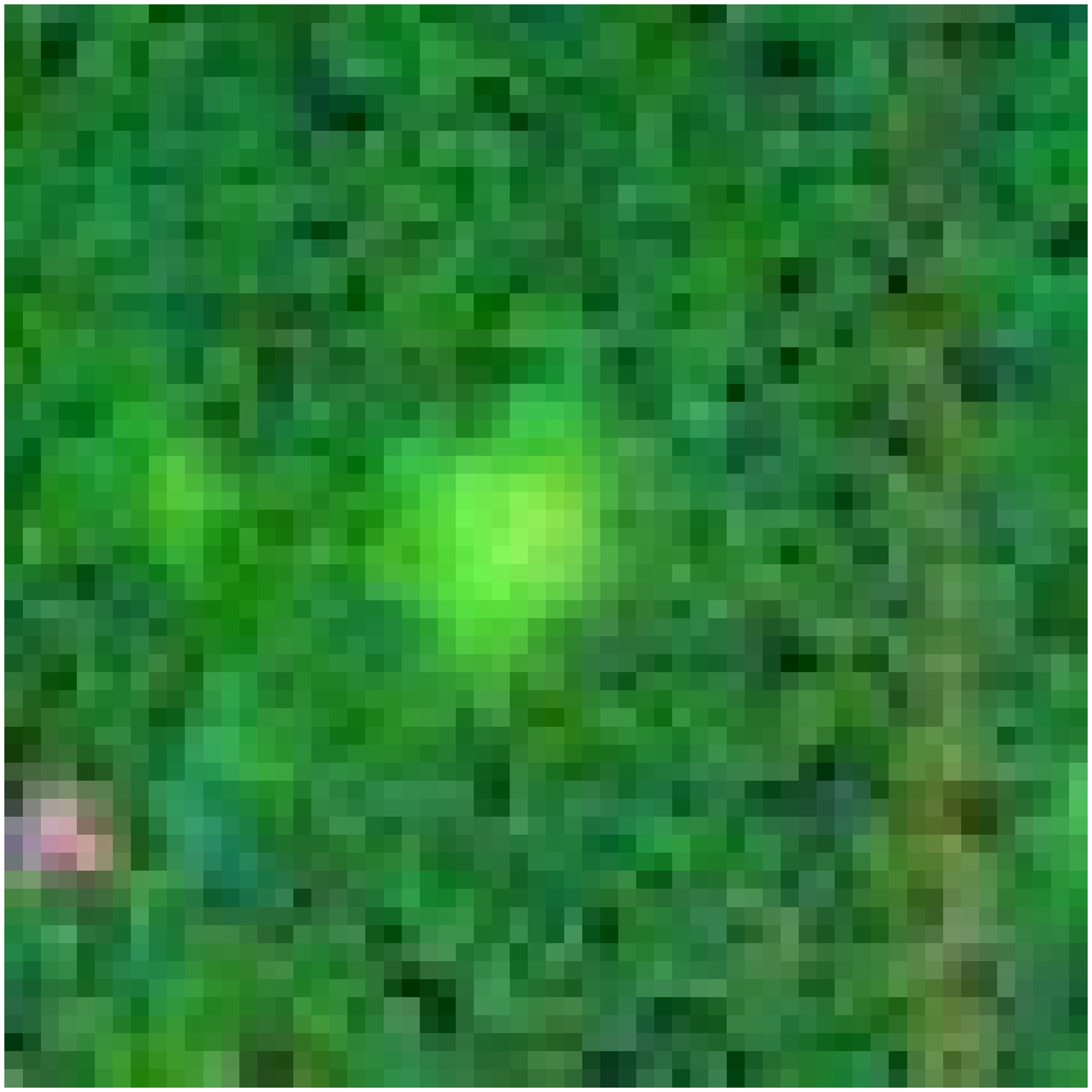}  \FGb{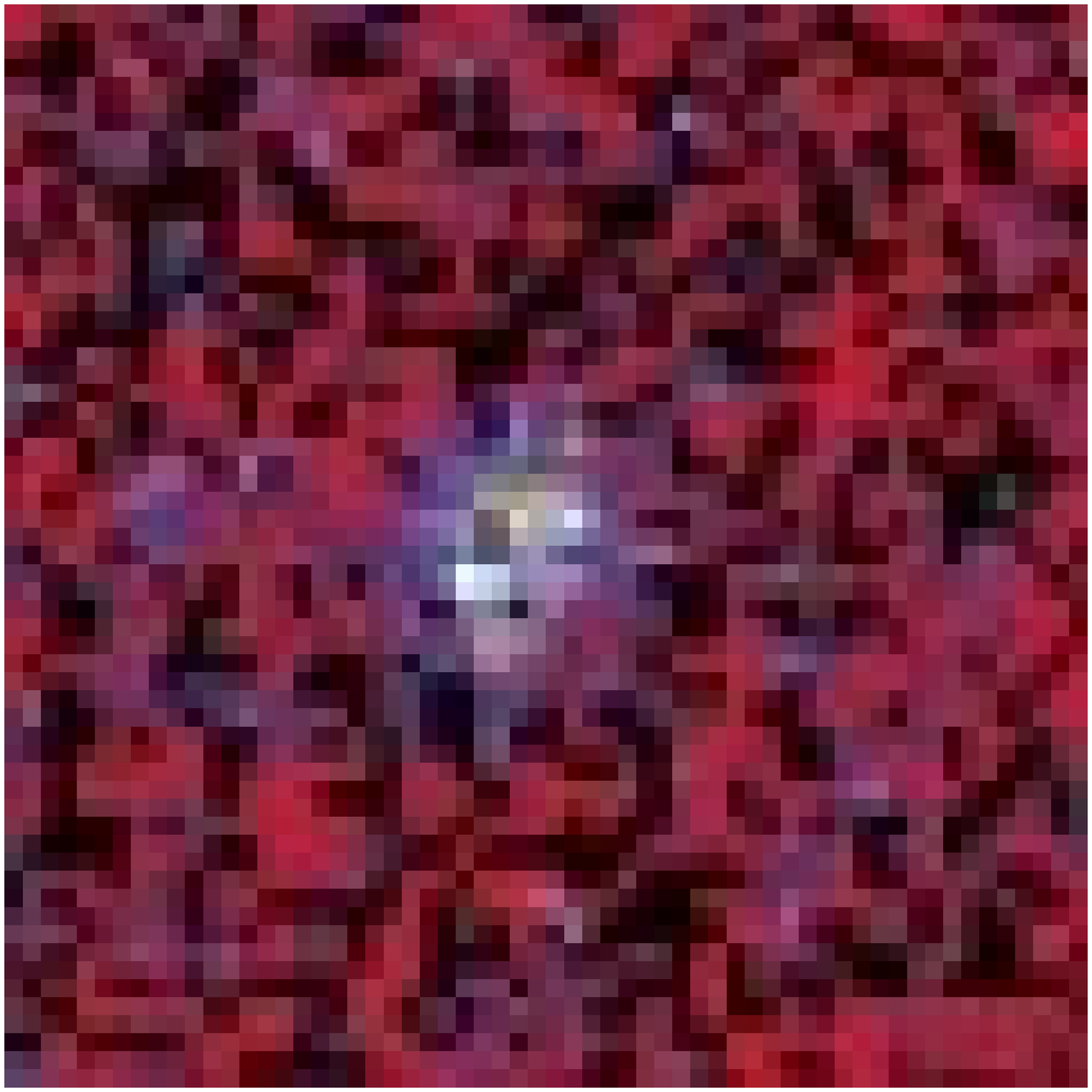}    \FGb{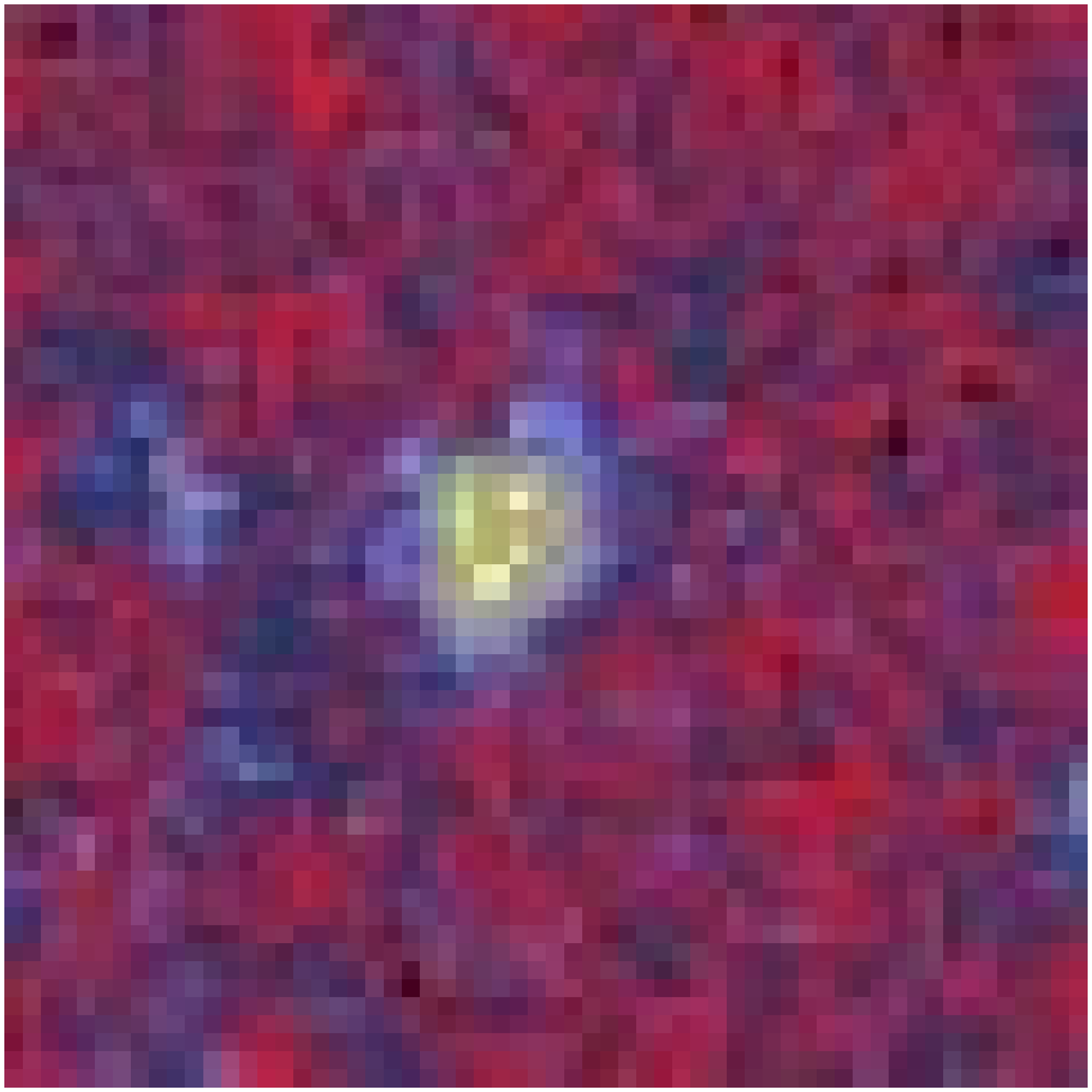}   \FGb{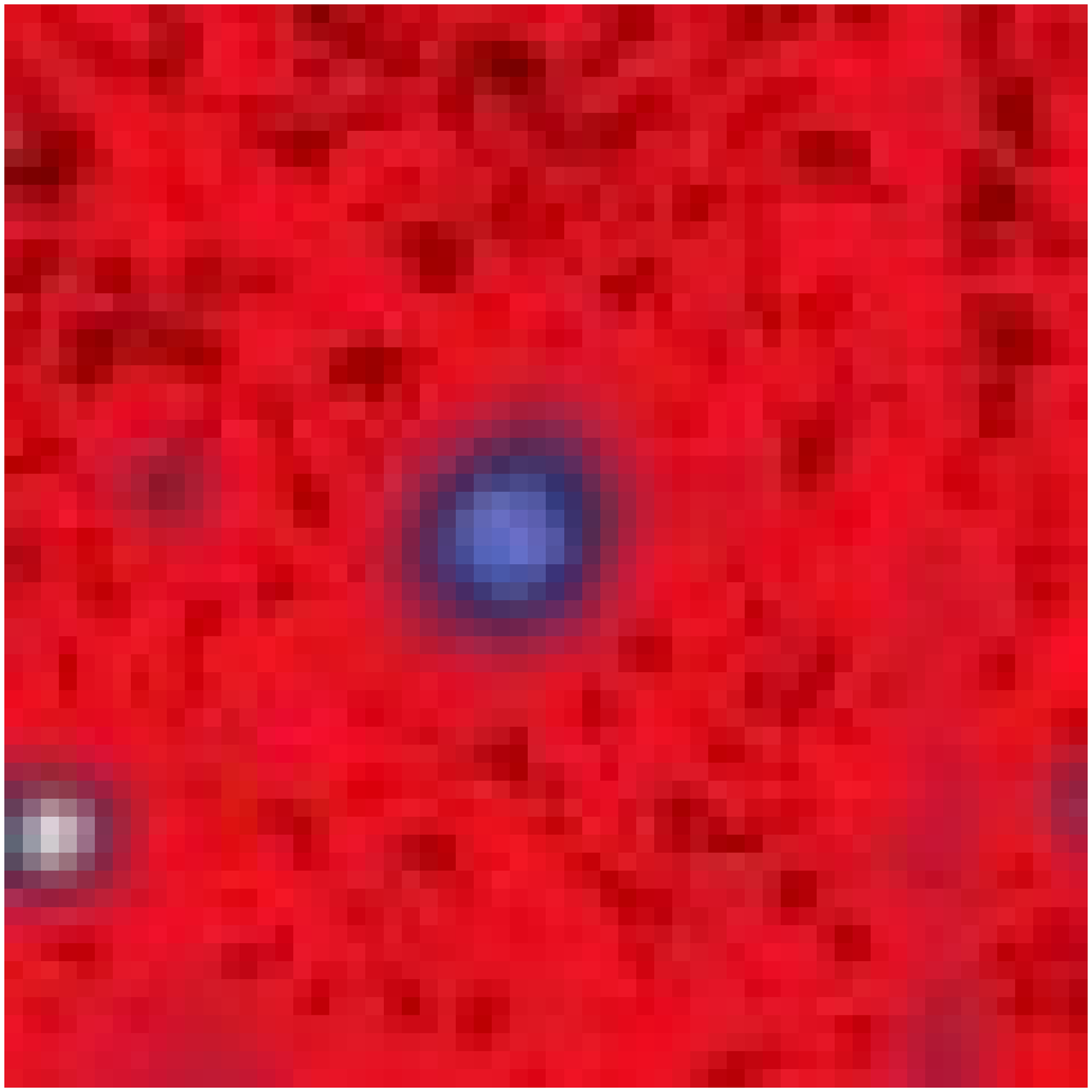}   \FGb{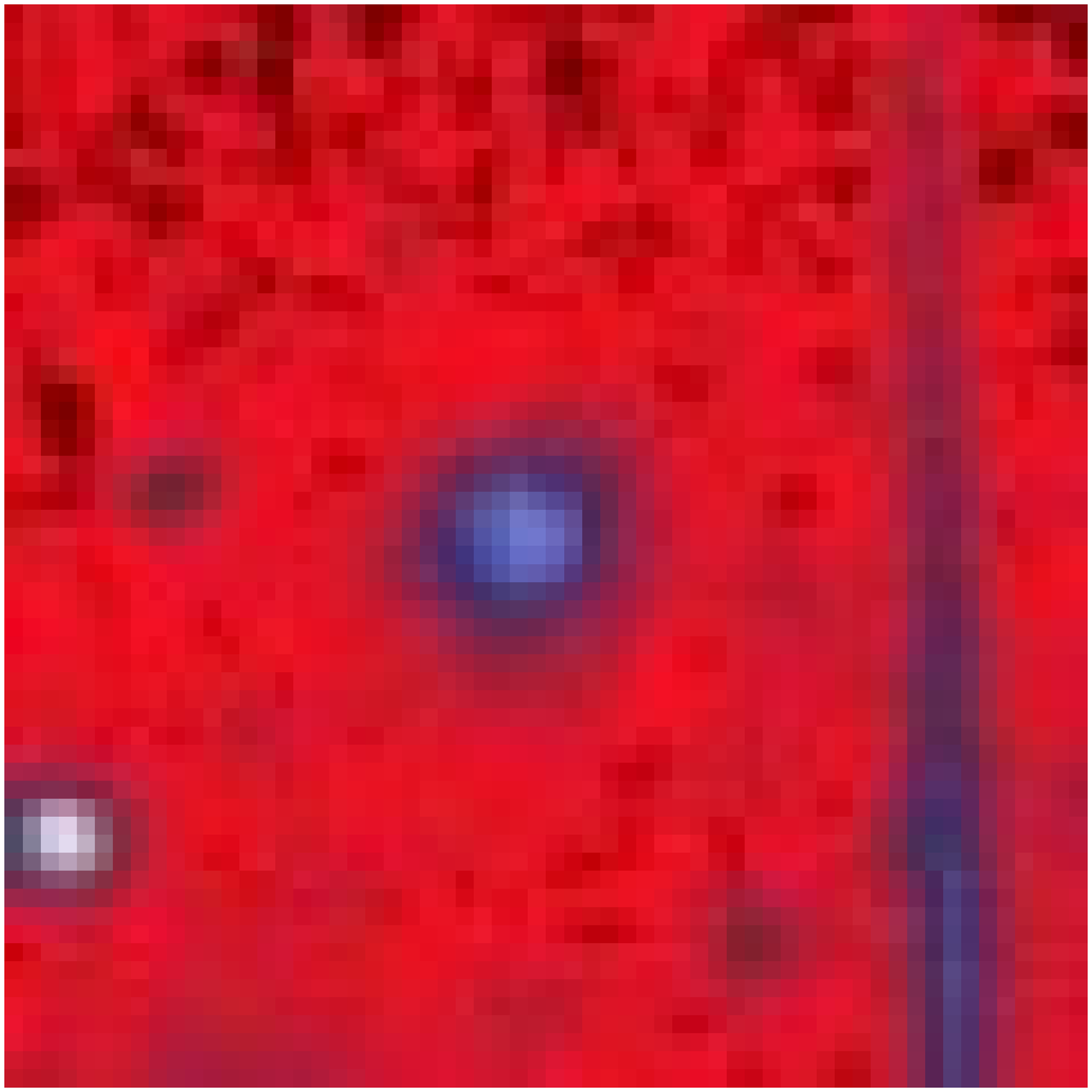}   \FGb{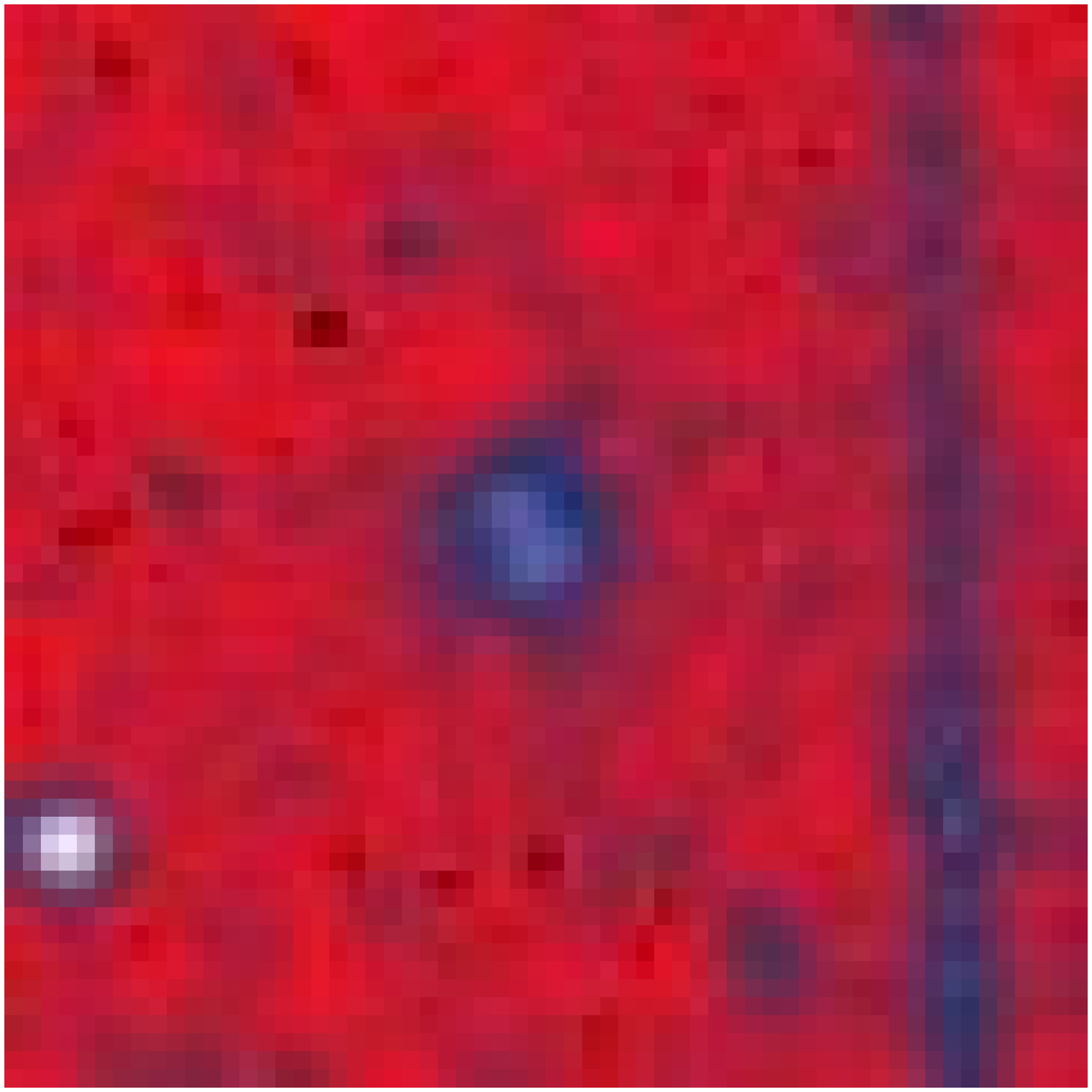}   \FGb{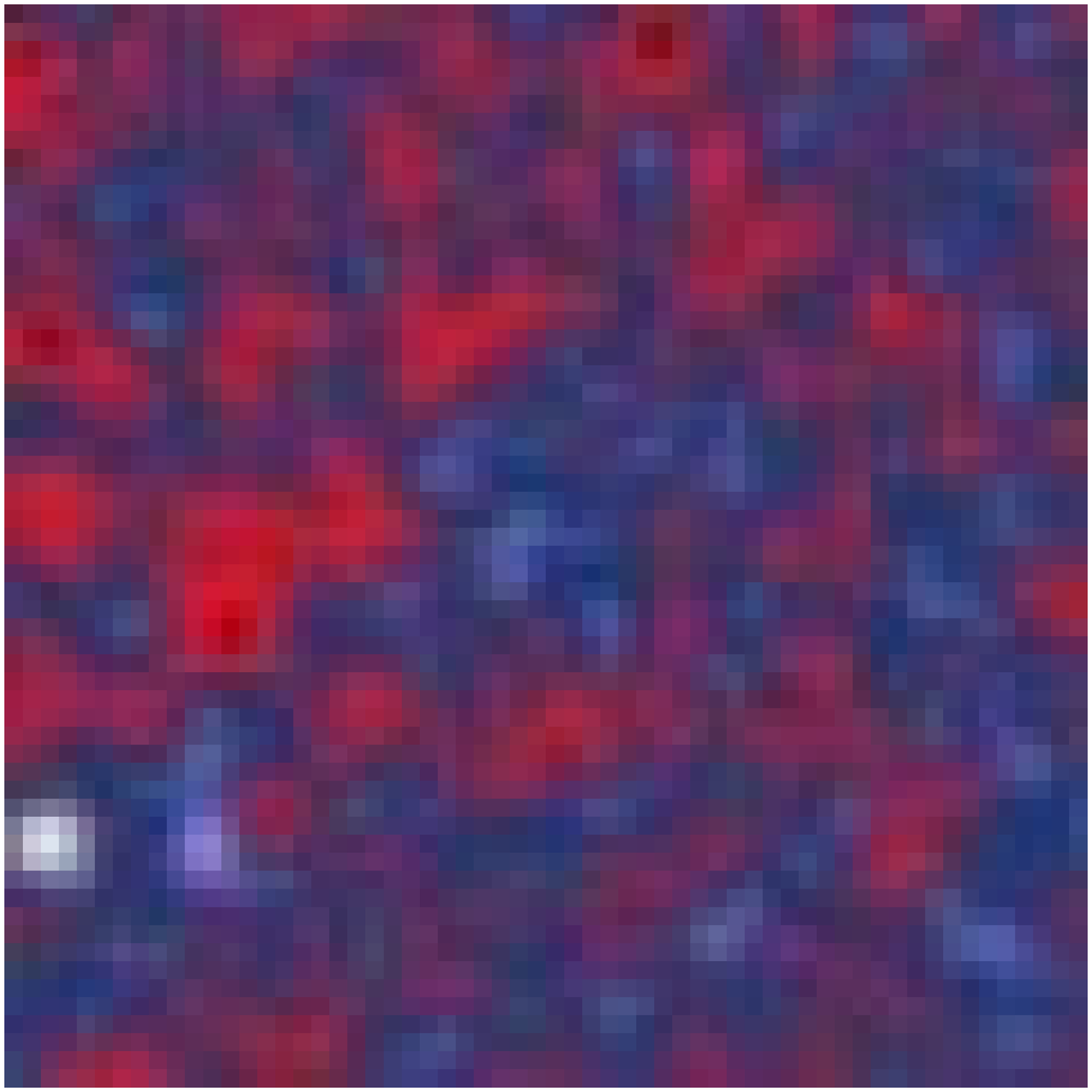}   \FGb{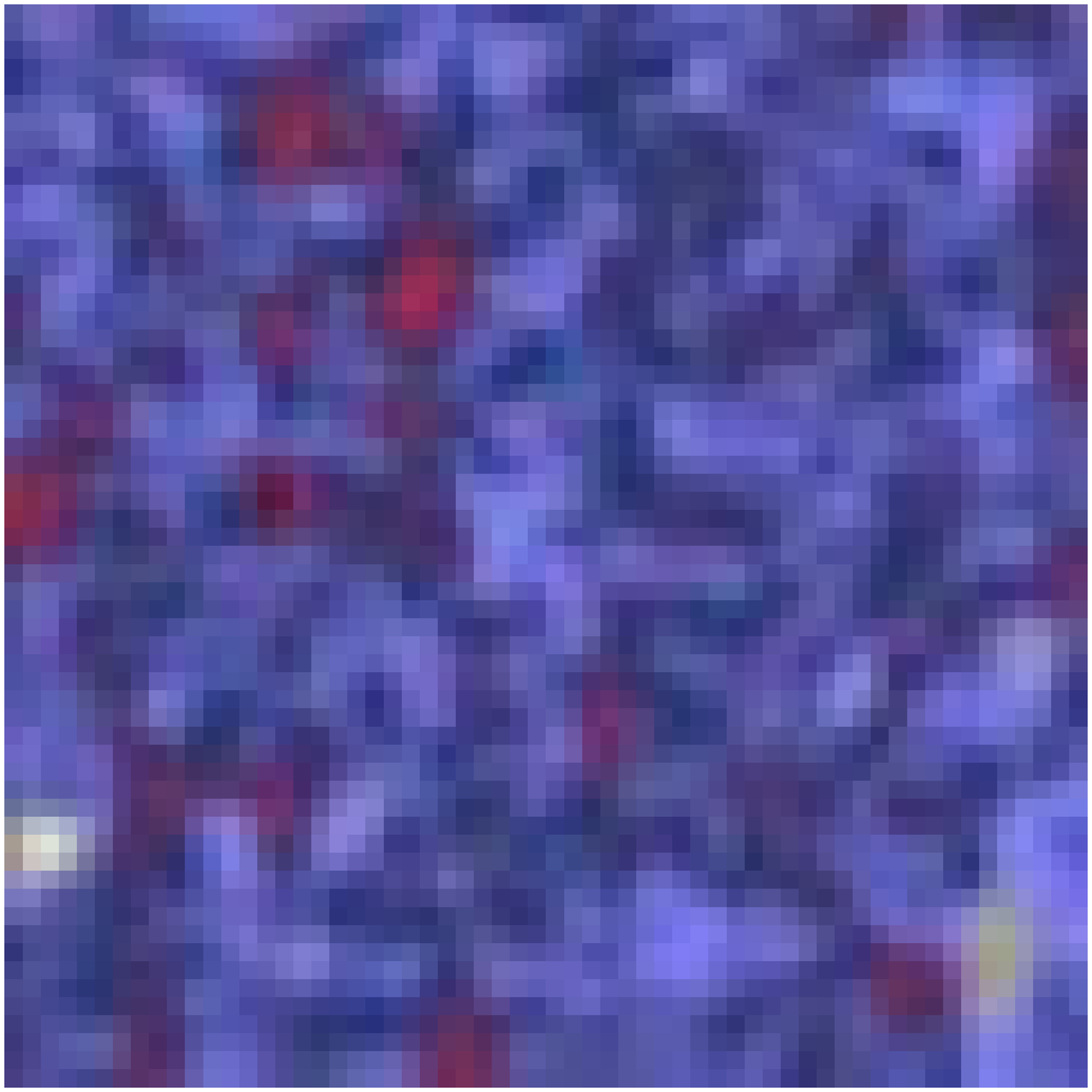}  \FGb{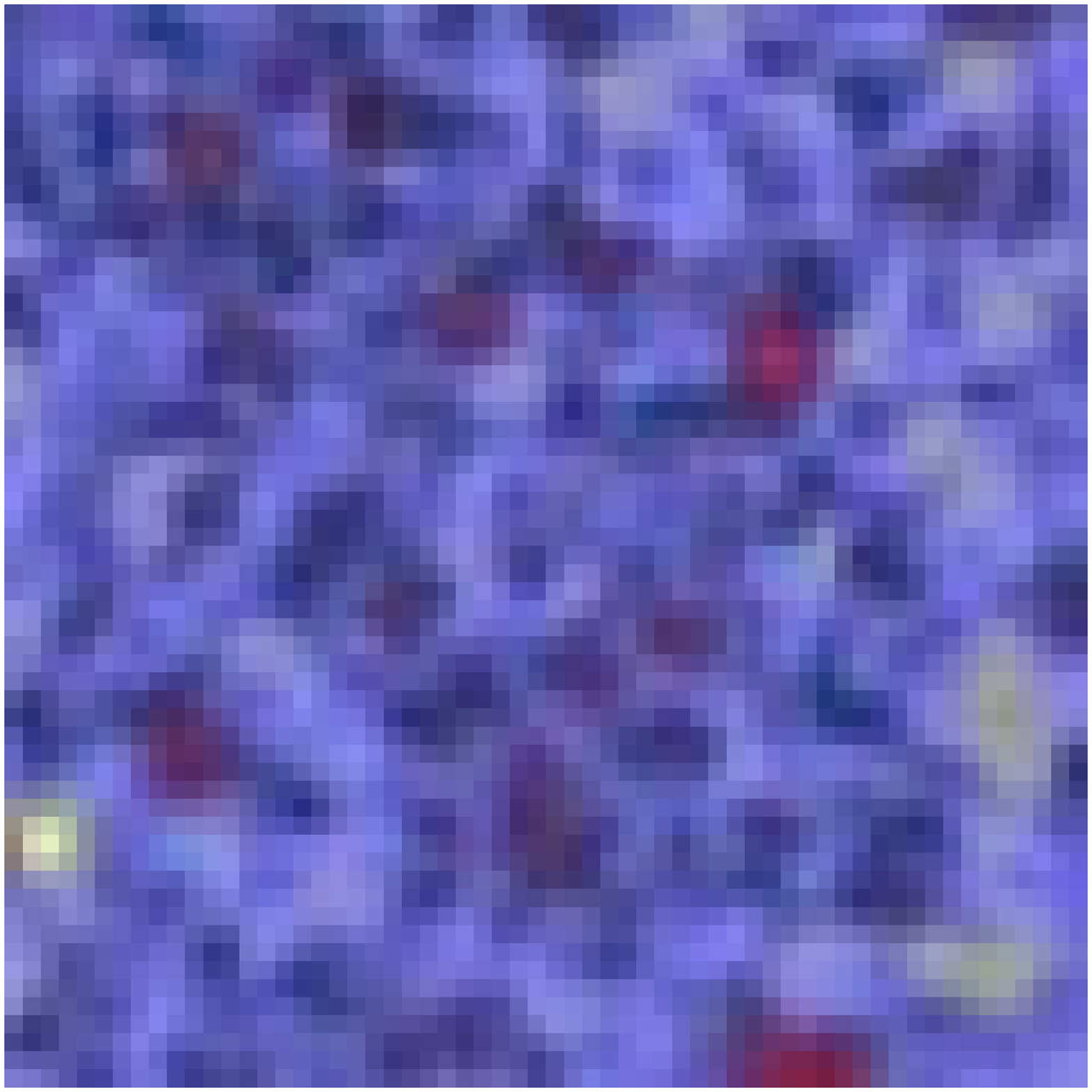}  \FGb{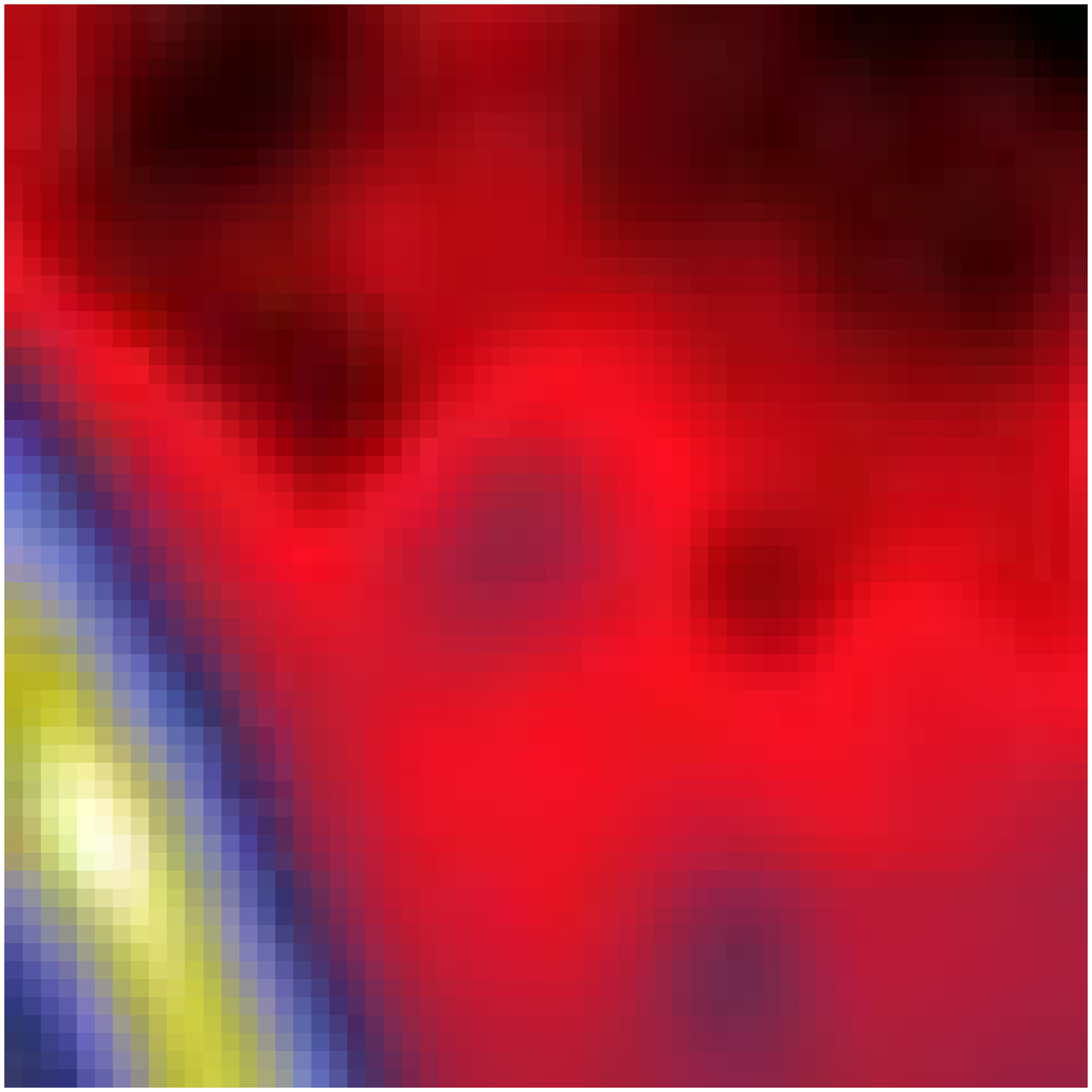}  \FGb{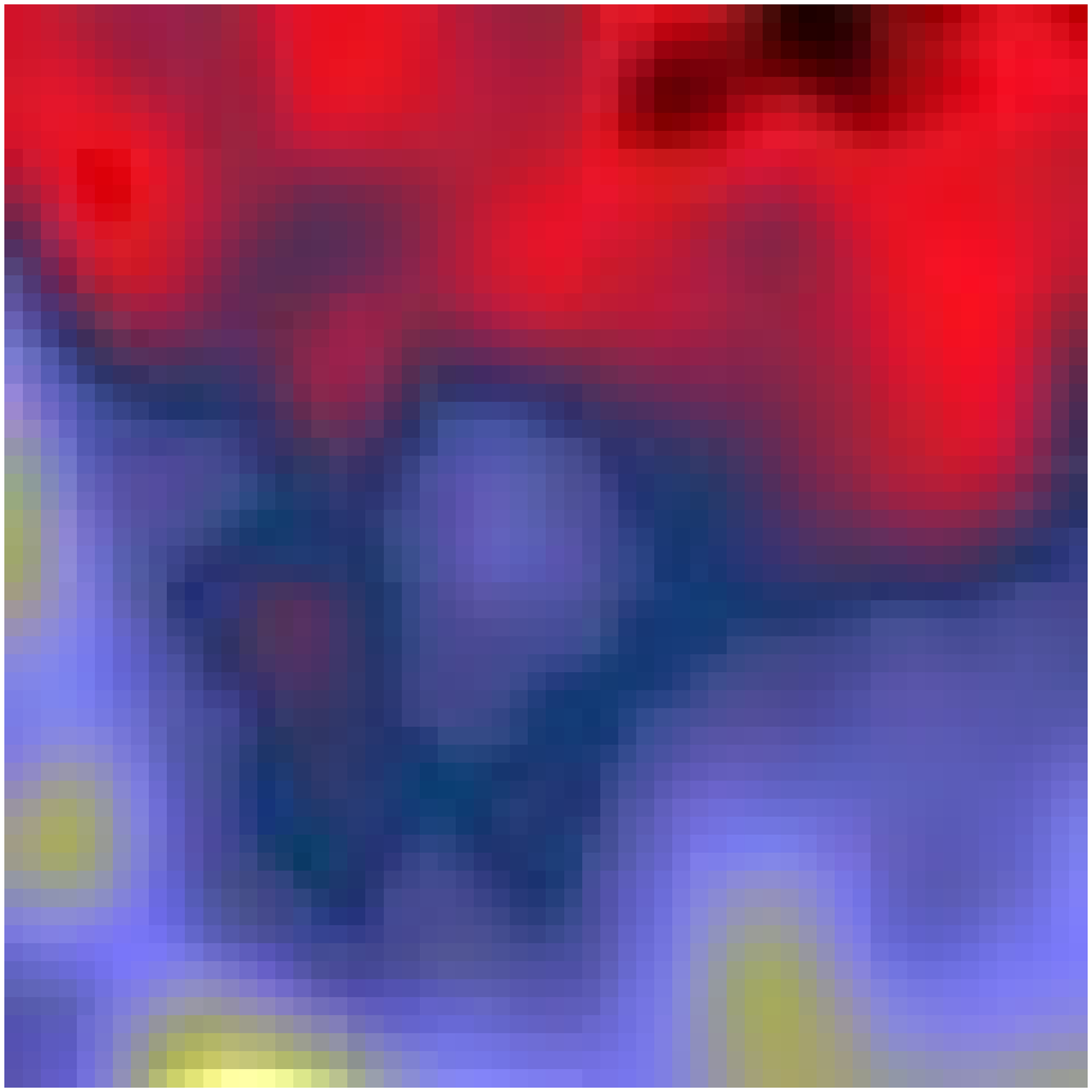}  \FGb{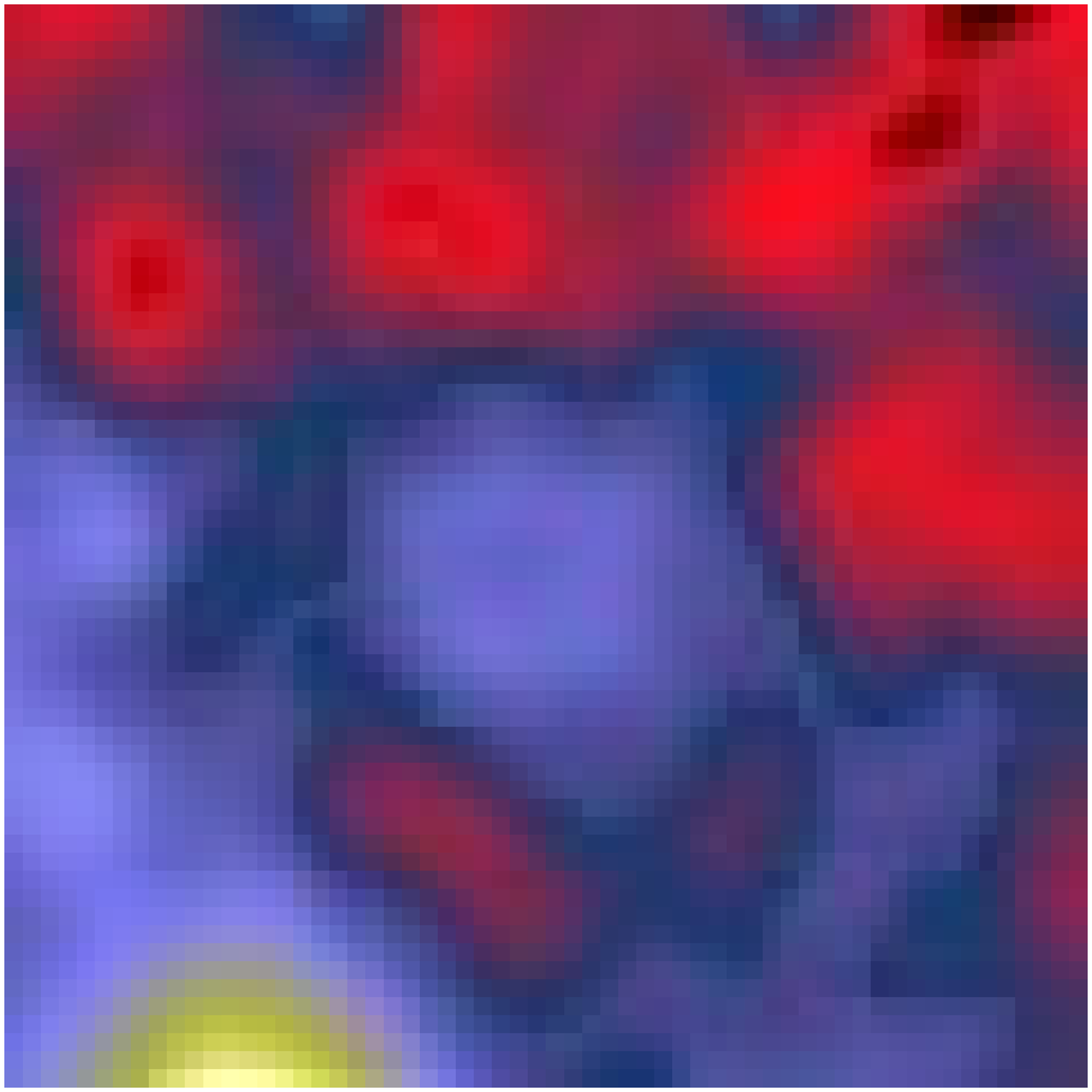}  \FGb{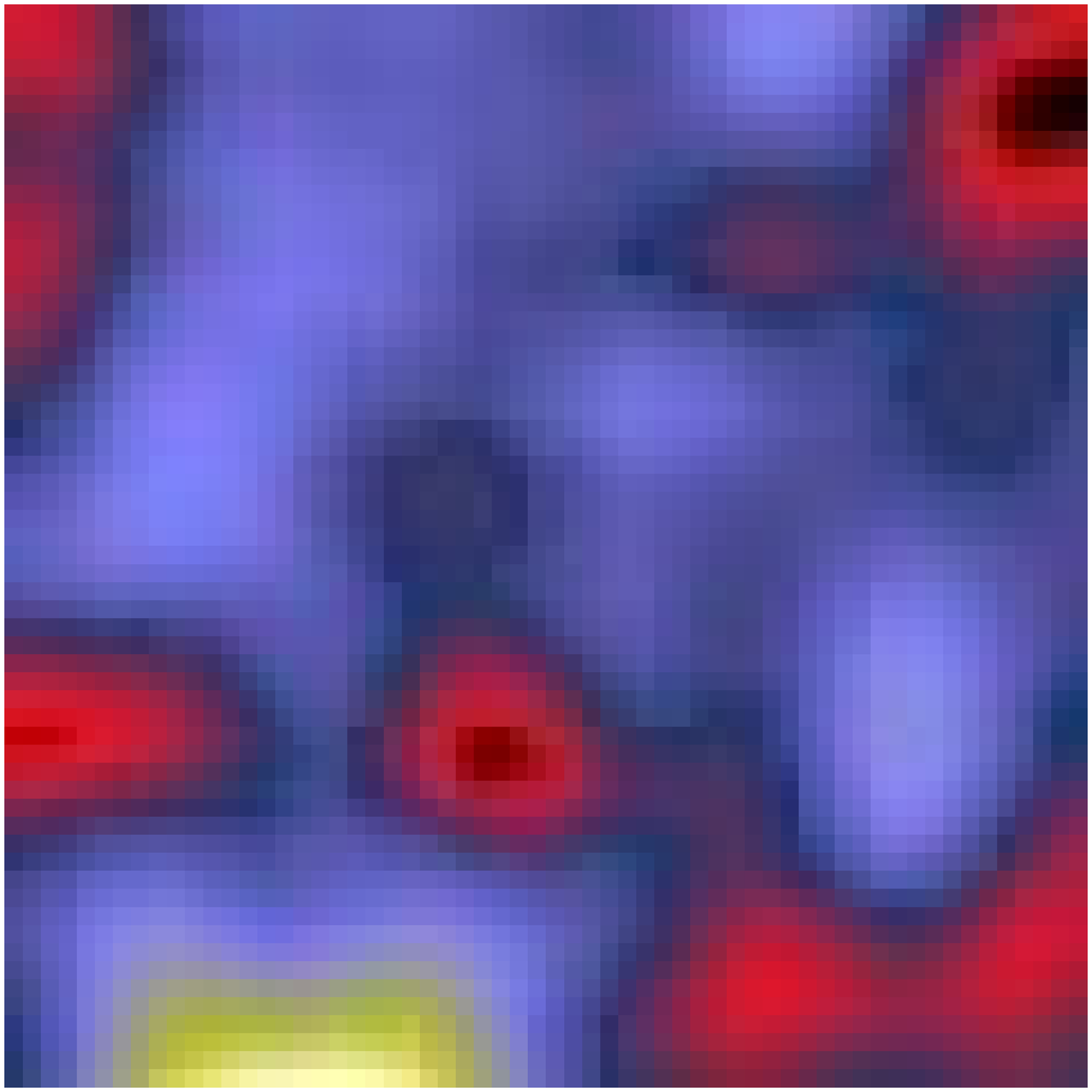}\\  {\#8   NCIA035836}  \\
  \FGb{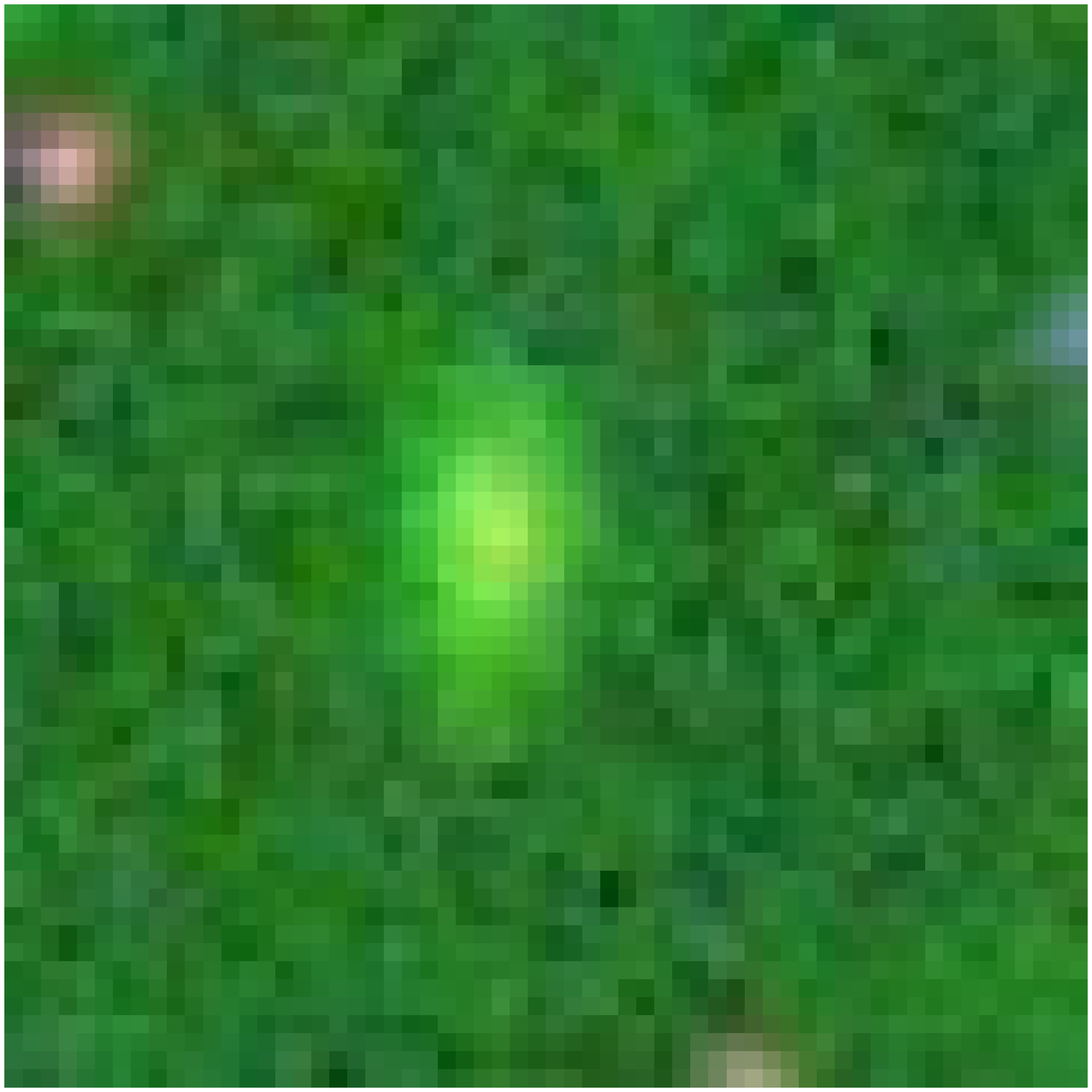} \FGb{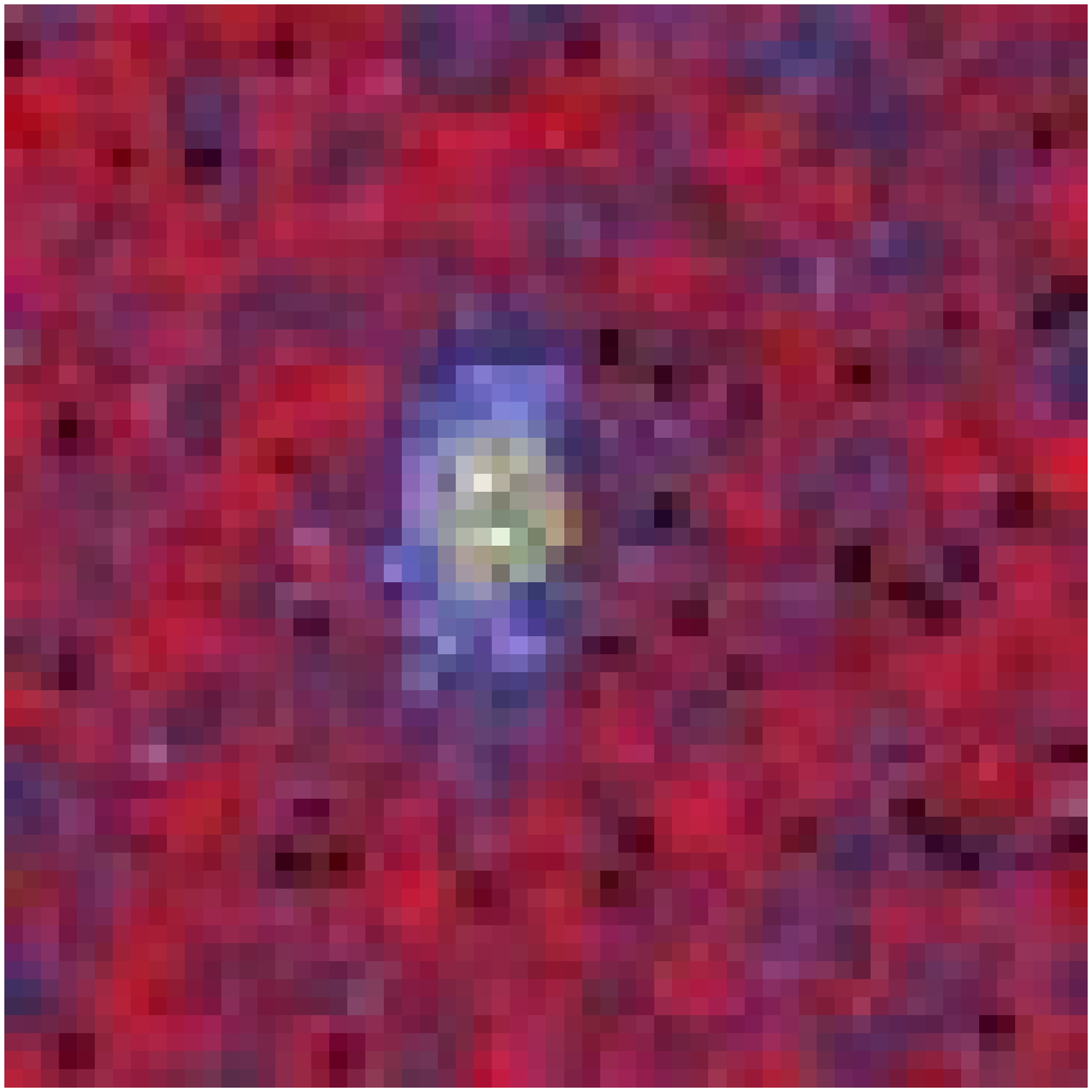}   \FGb{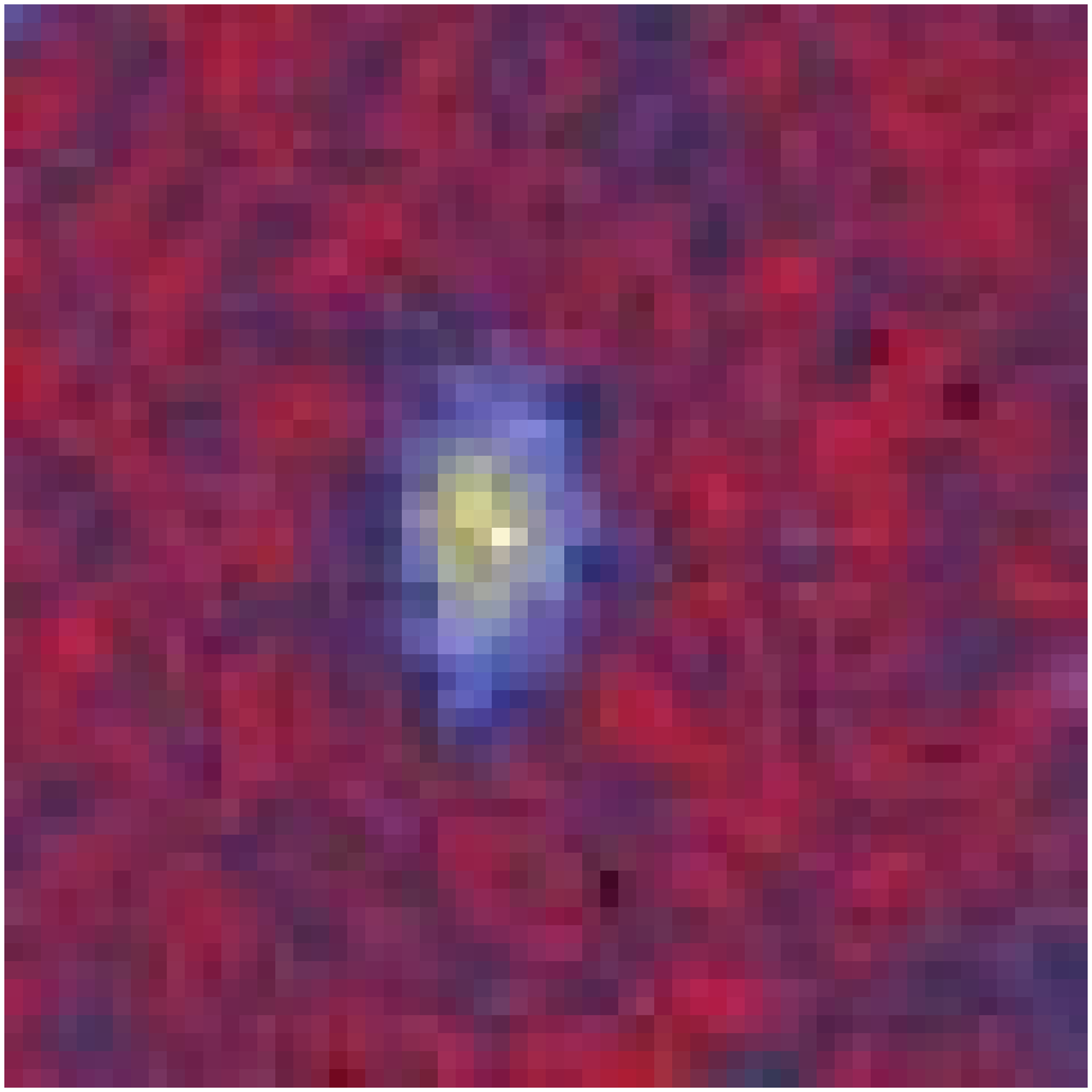}  \FGb{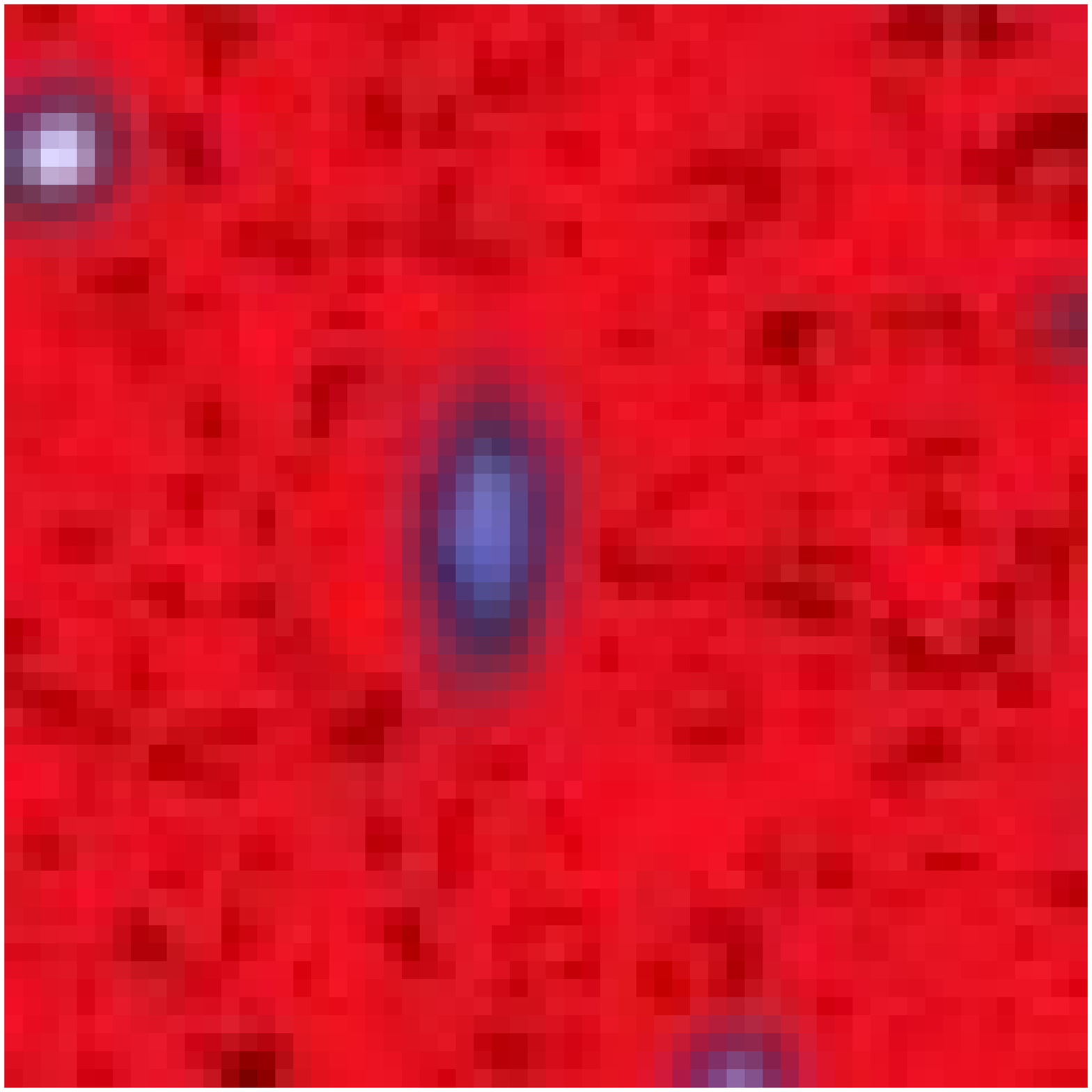}  \FGb{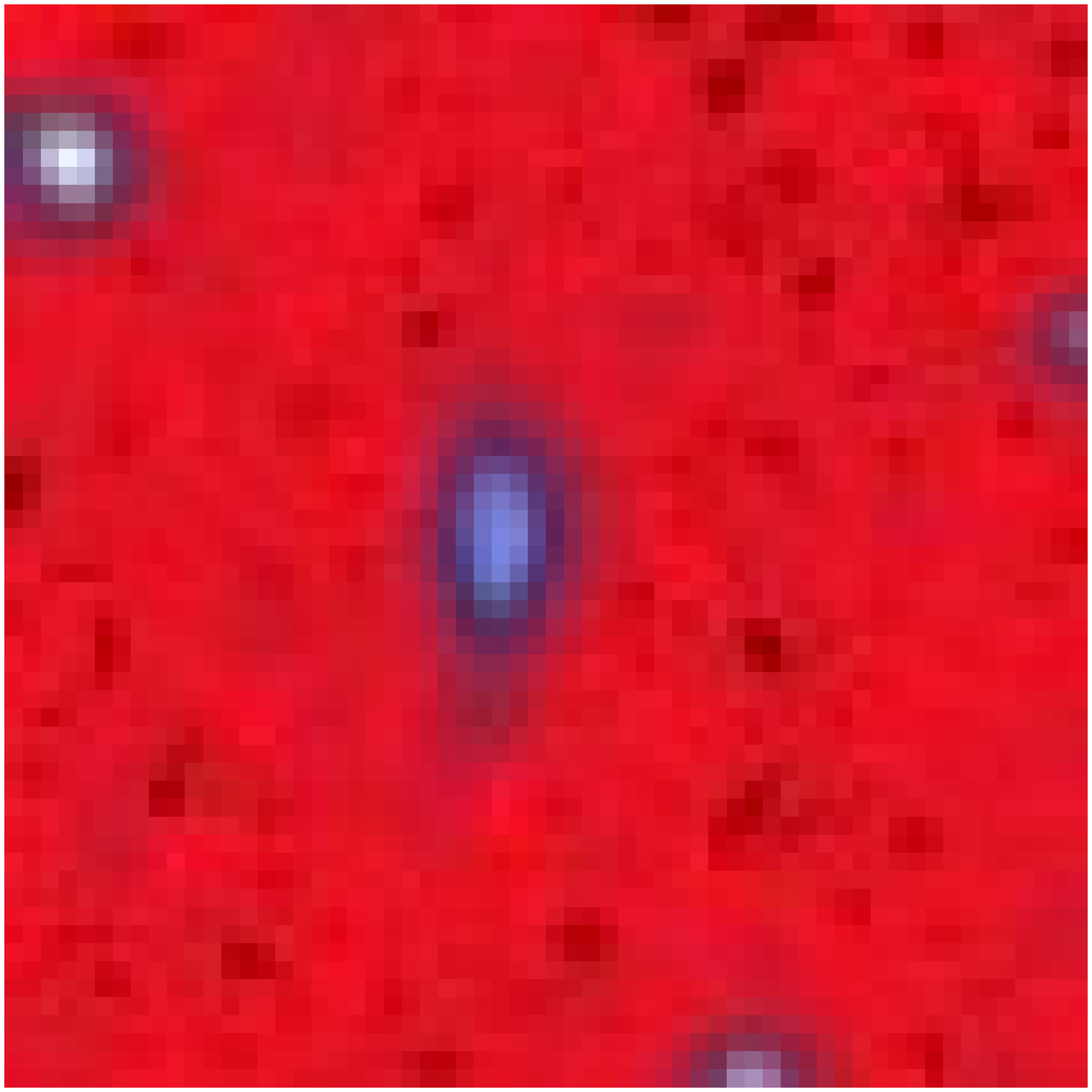}  \FGb{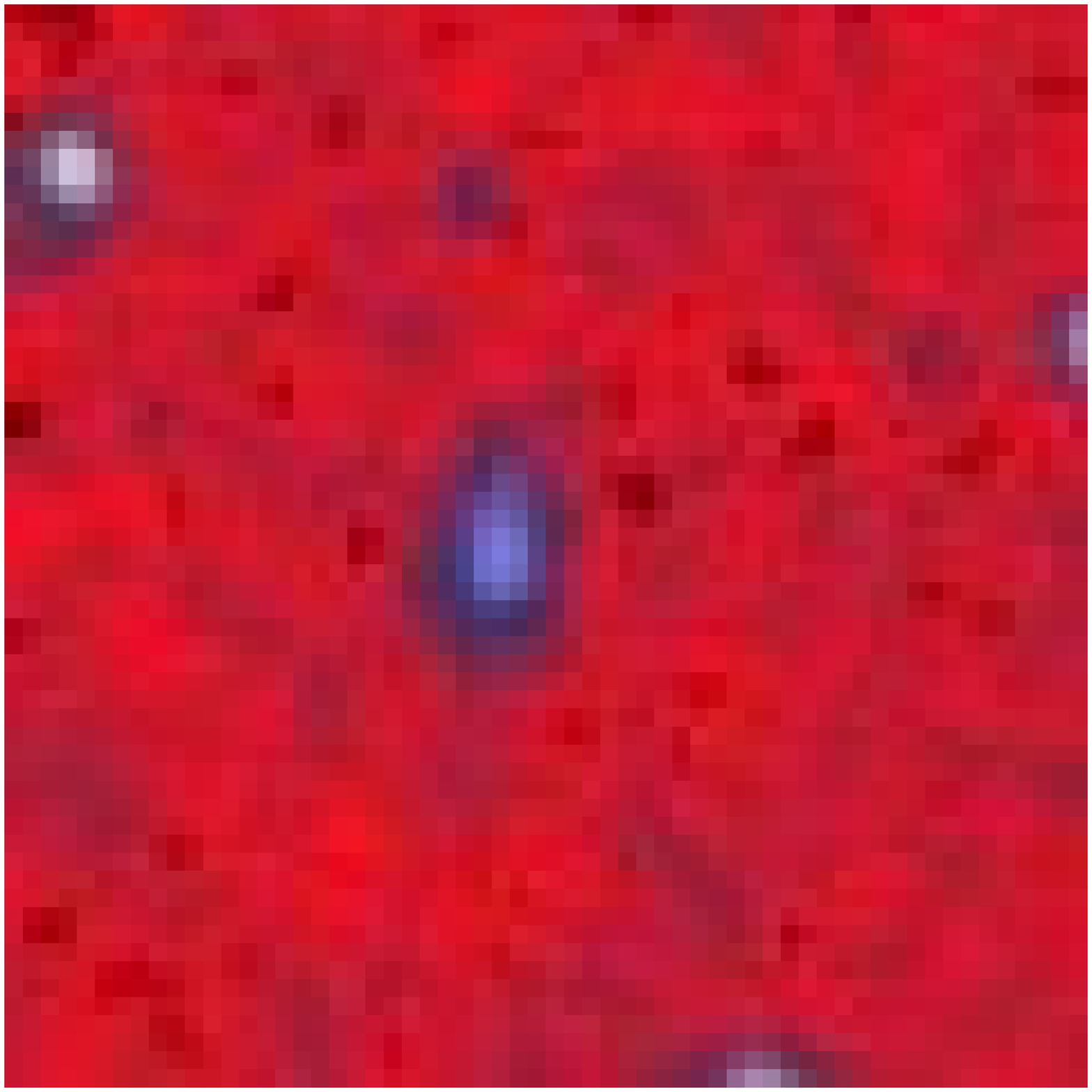}  \FGb{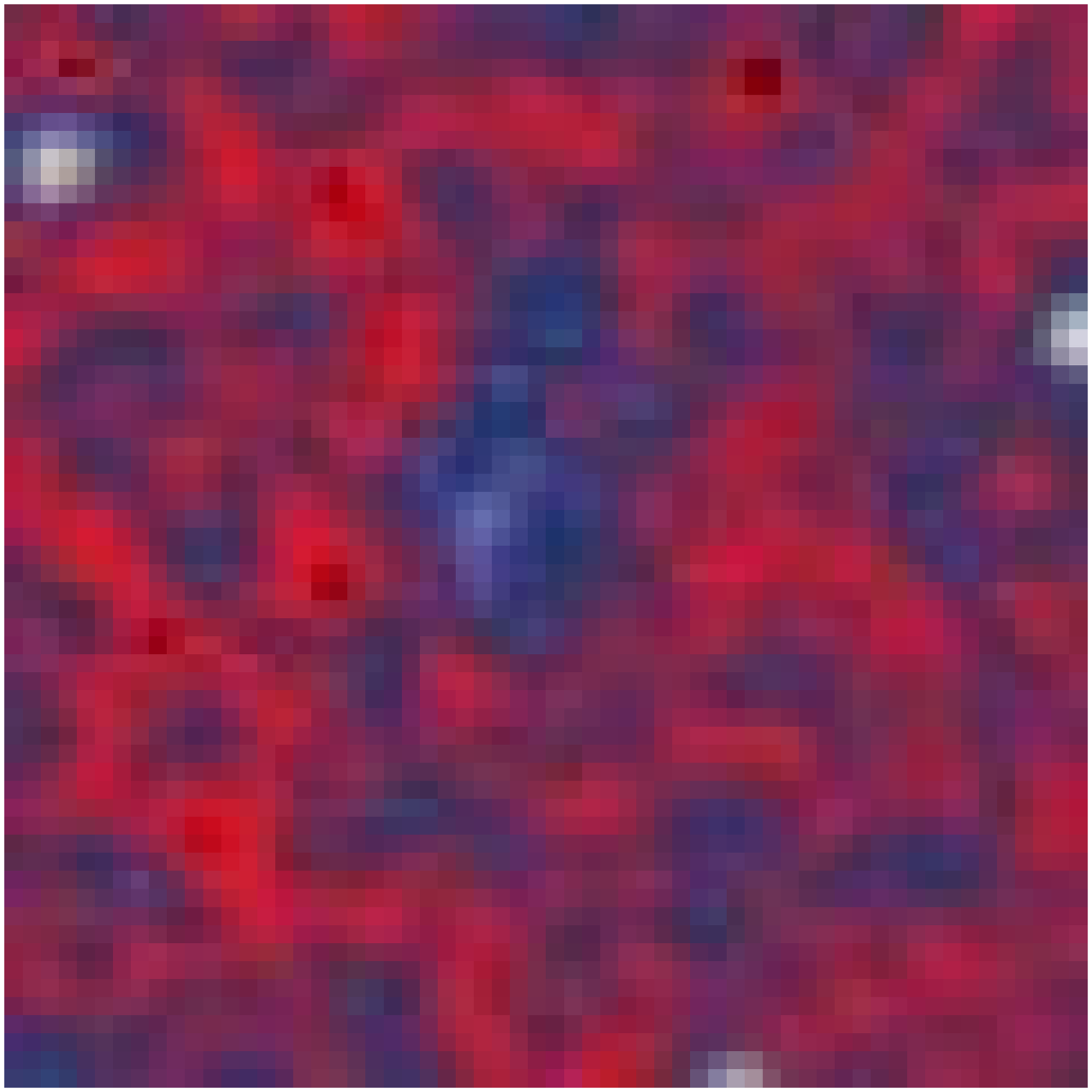}  \FGb{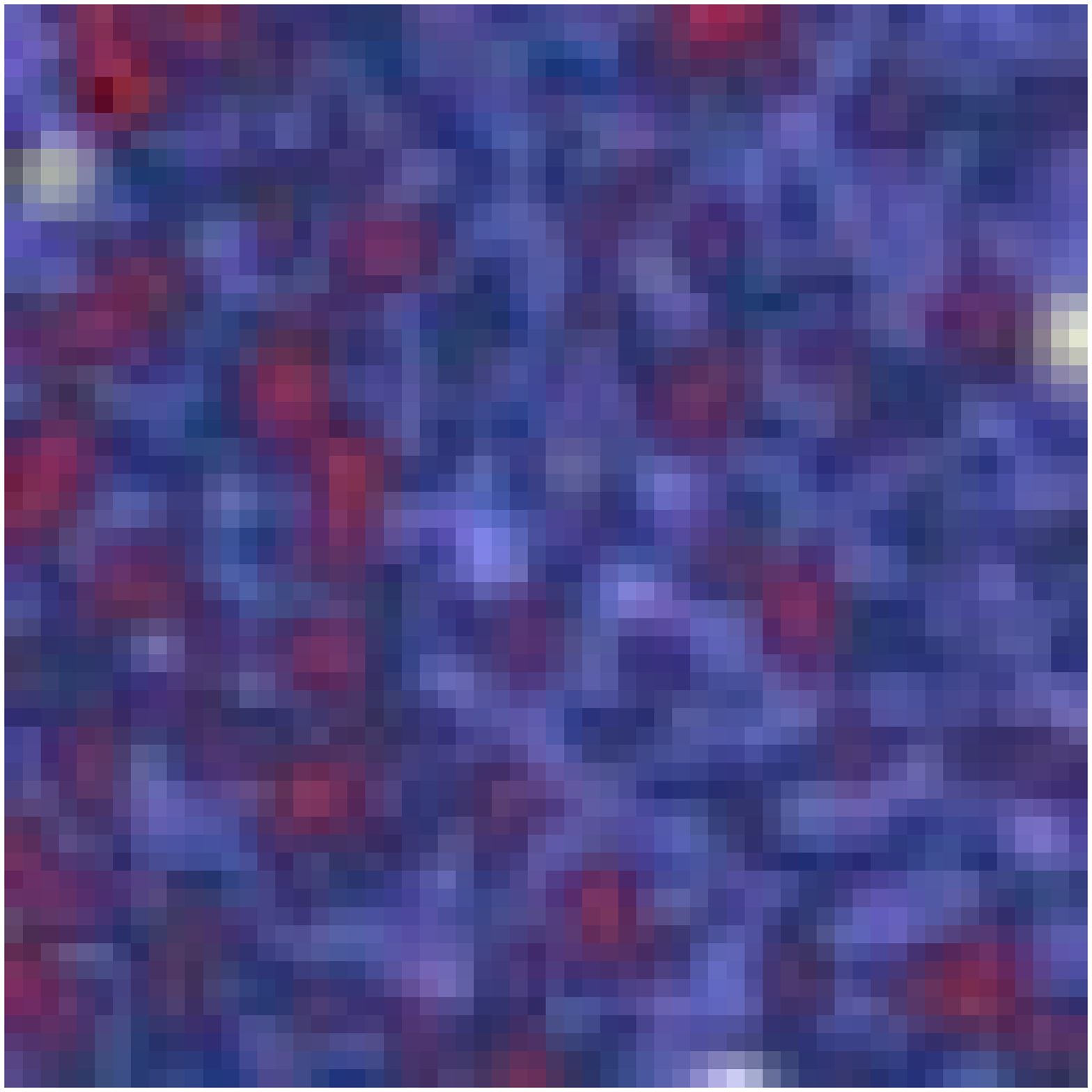}  \FGb{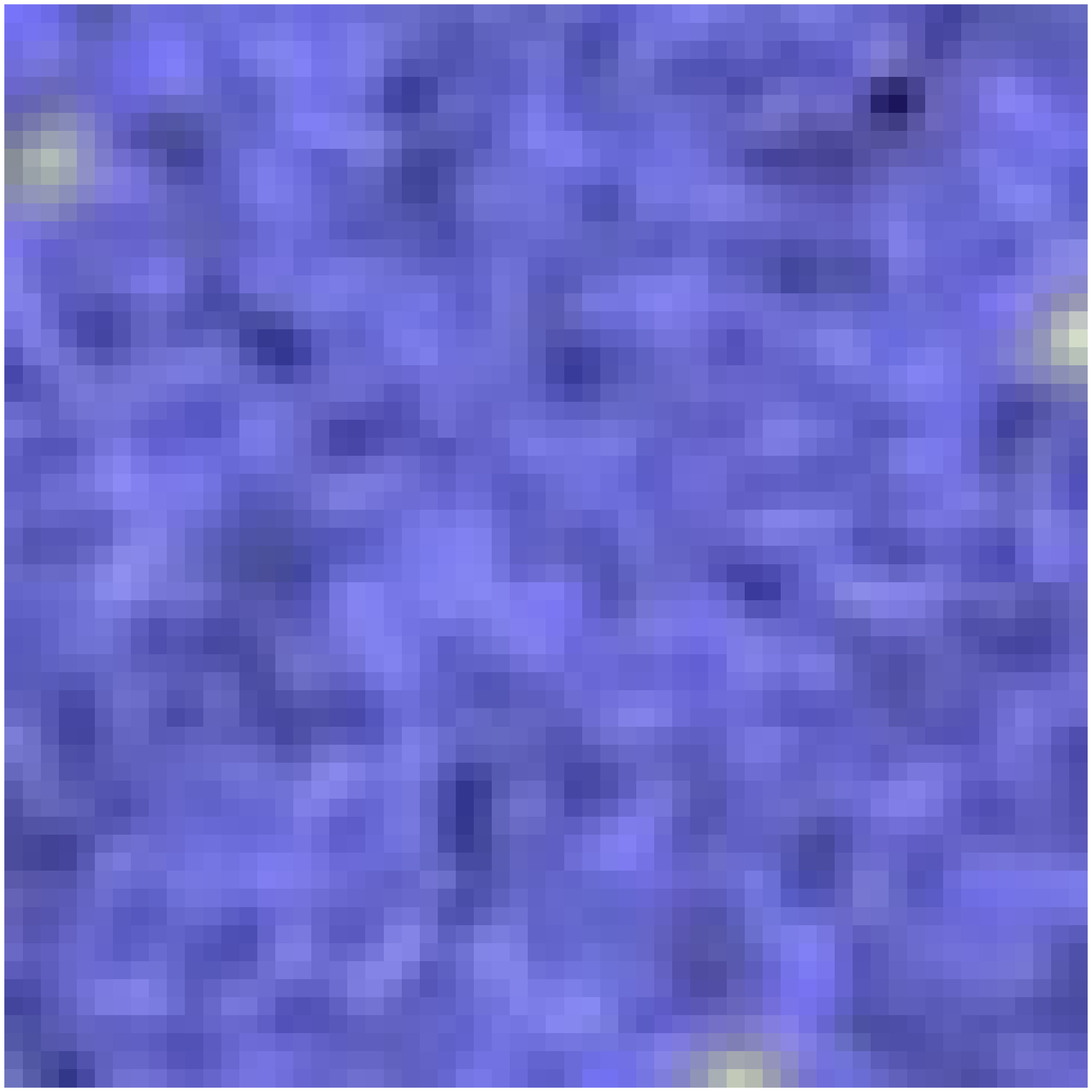}  \FGb{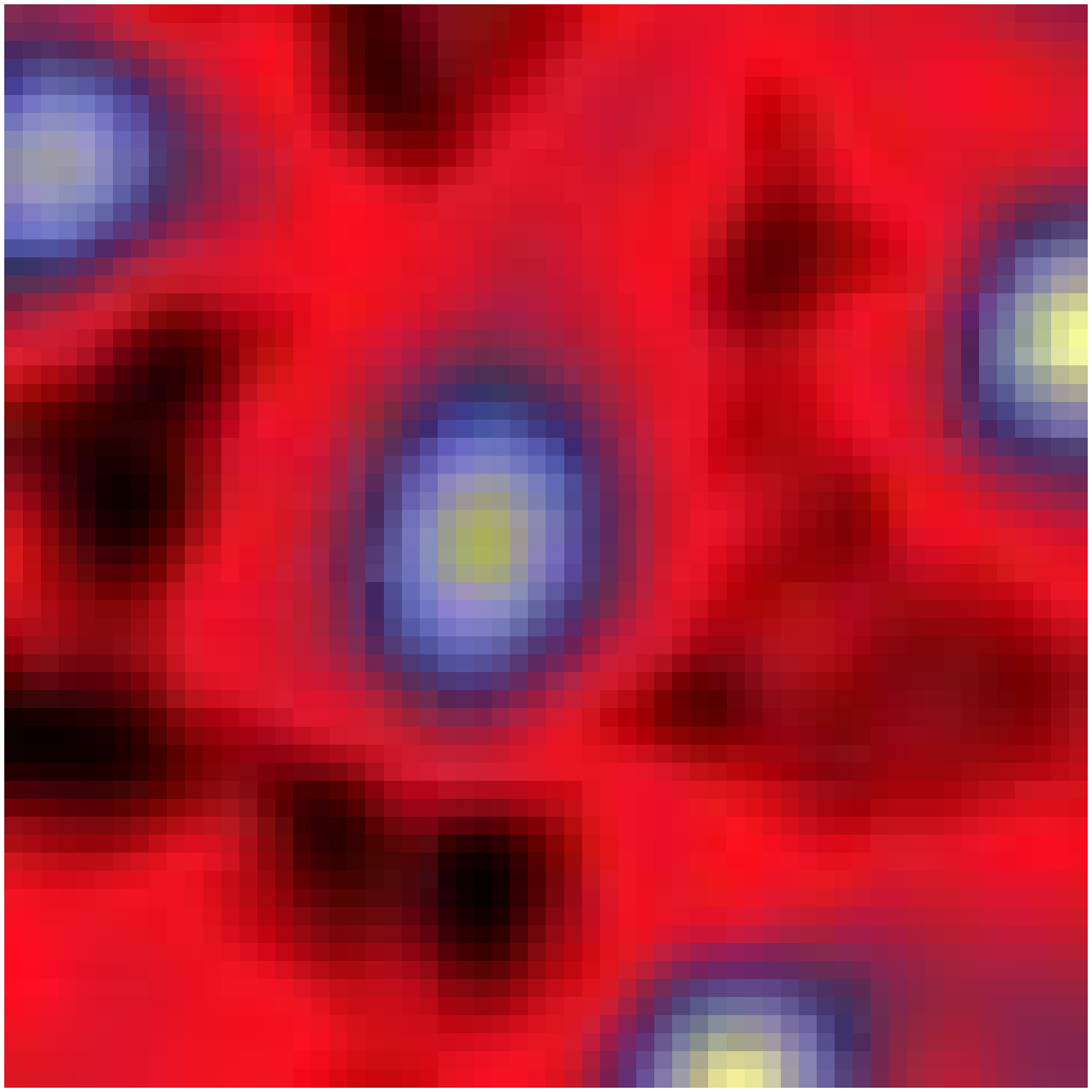}  \FGb{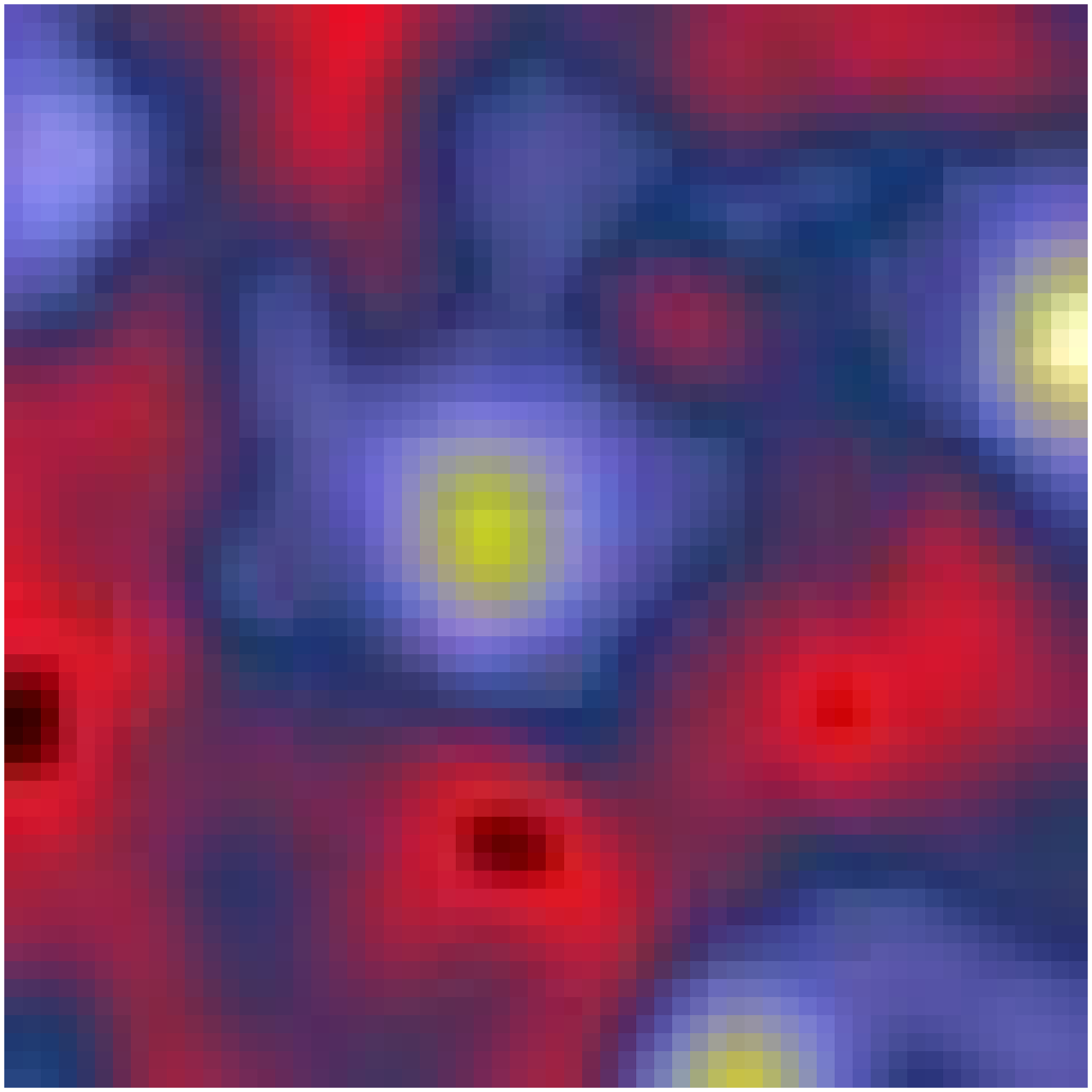}  \FGb{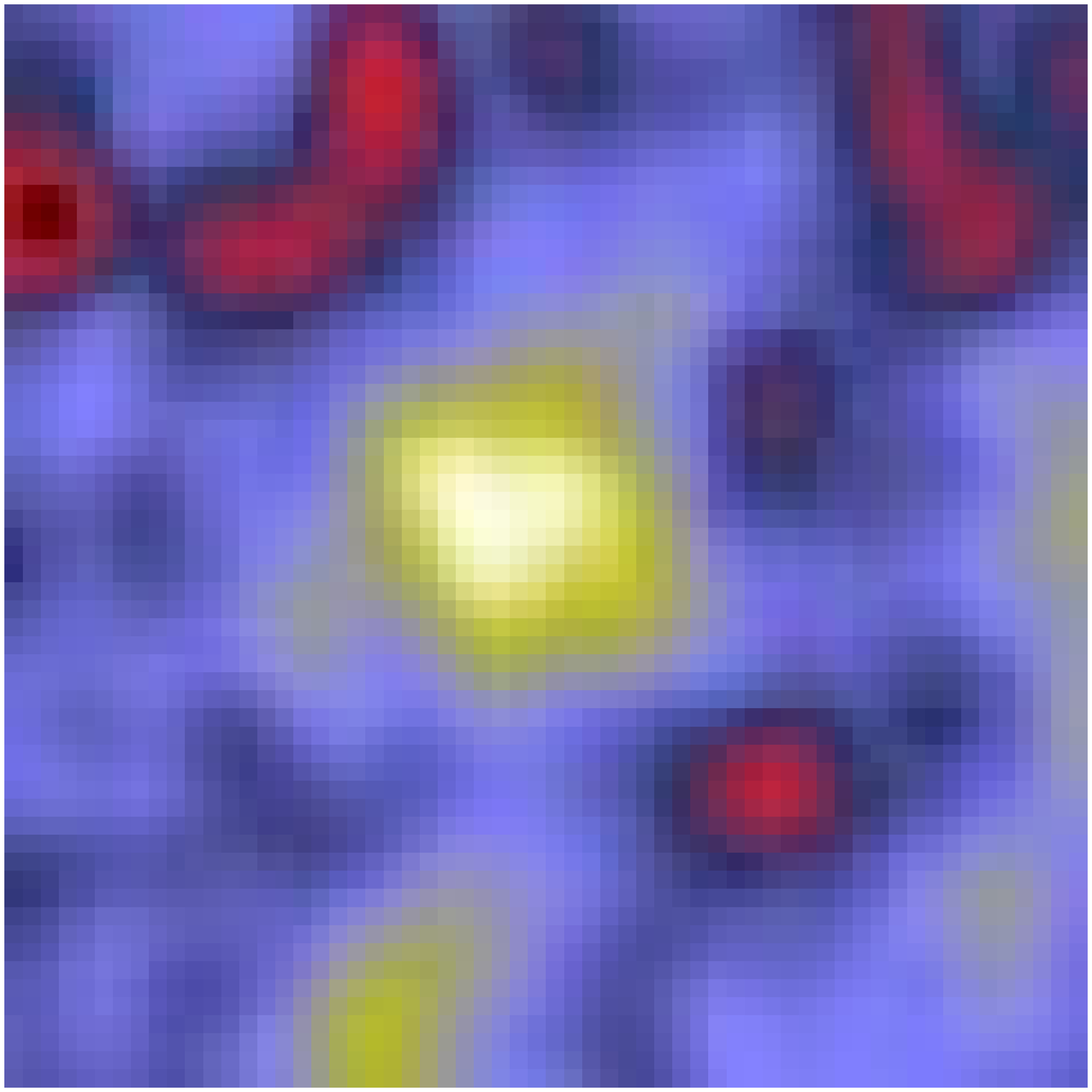}  \FGb{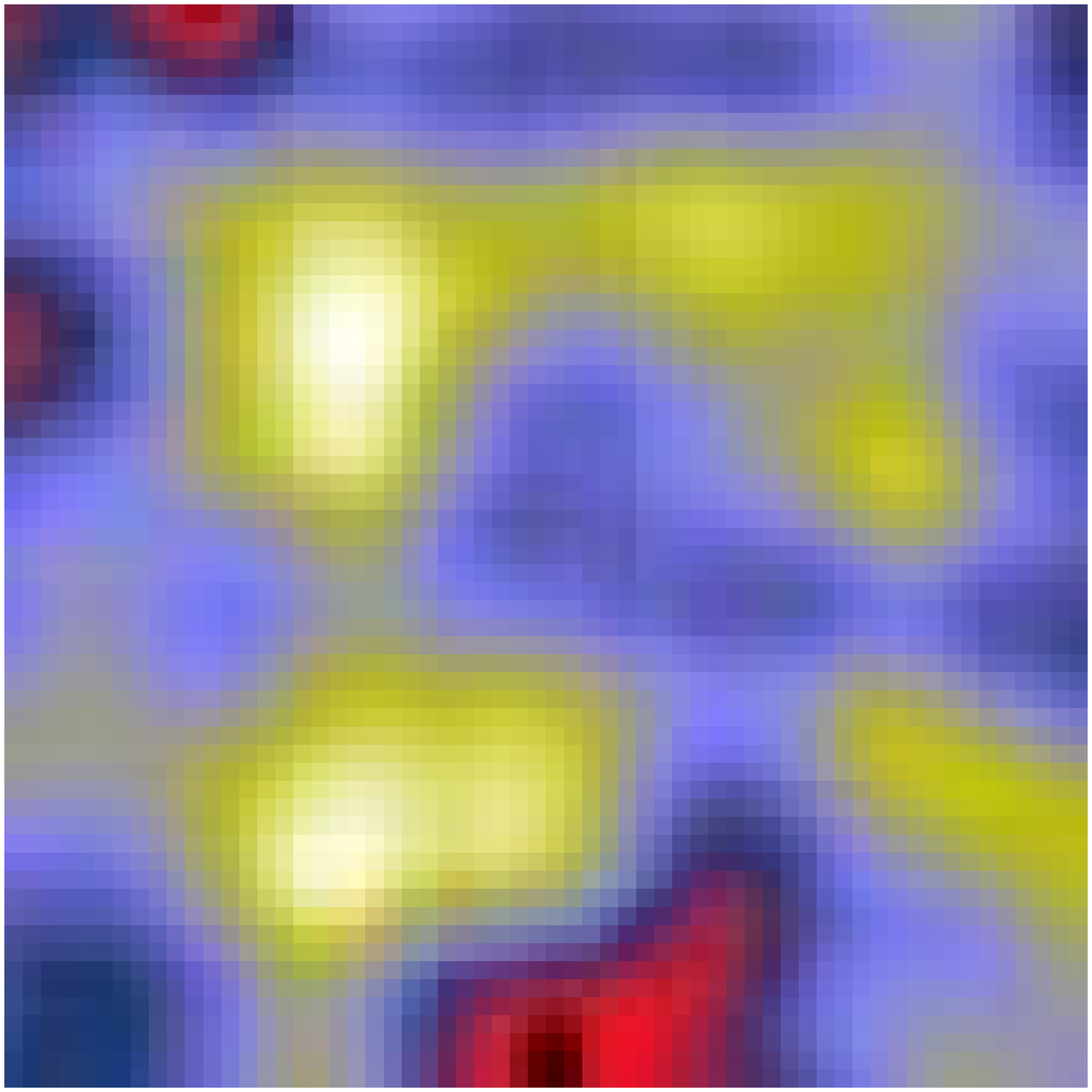}\\  {\#9   NCIA025916}  \\
  \FGb{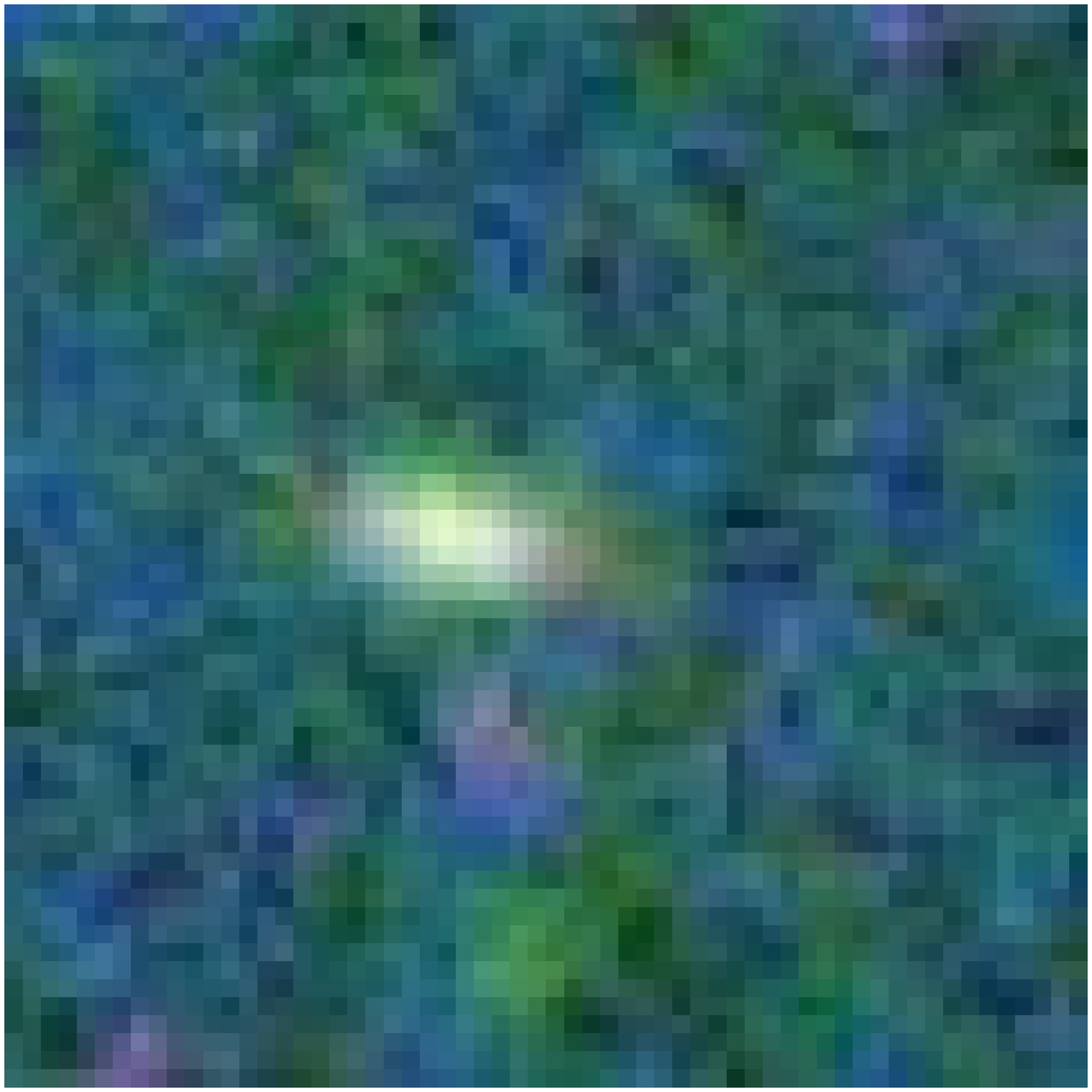} \FGb{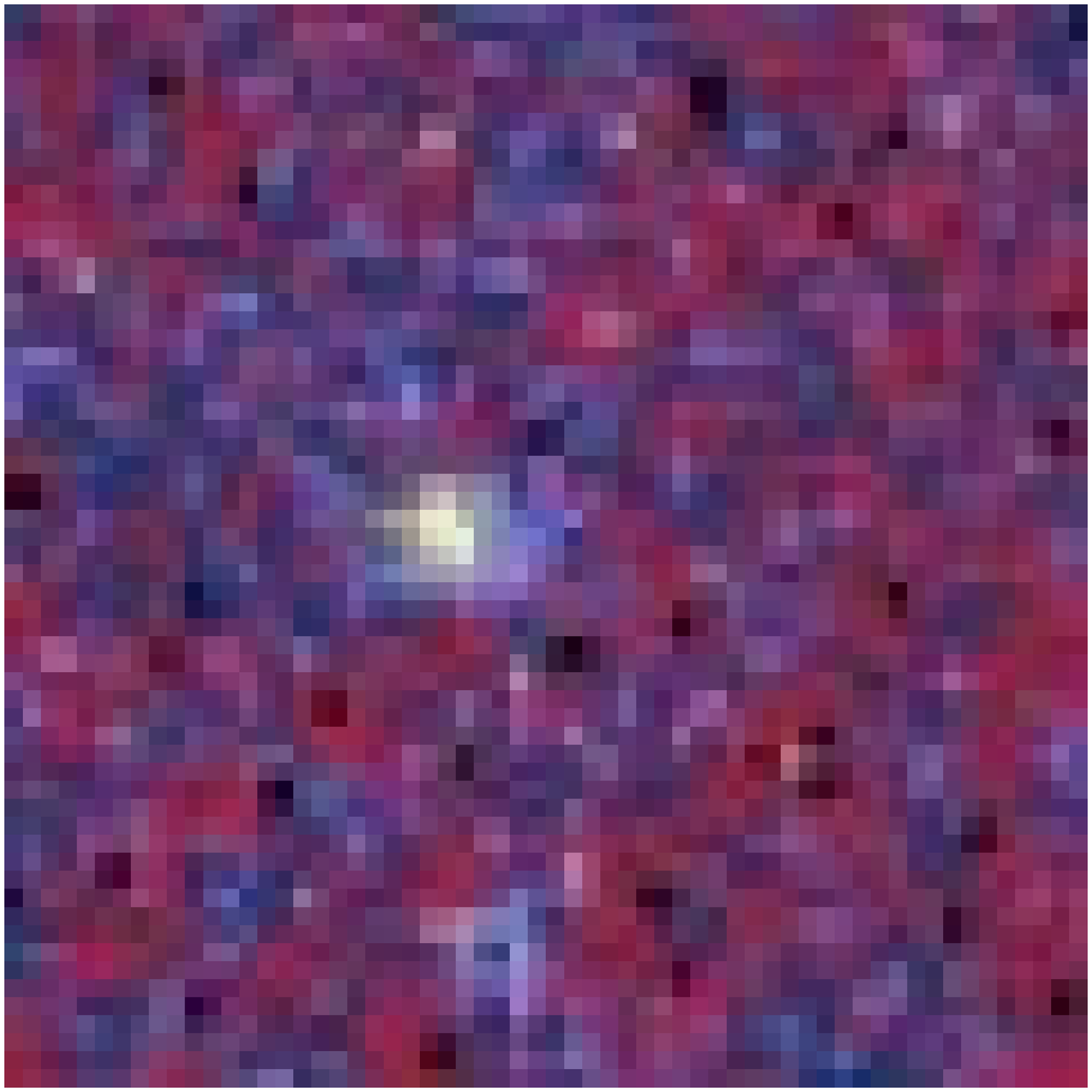}   \FGb{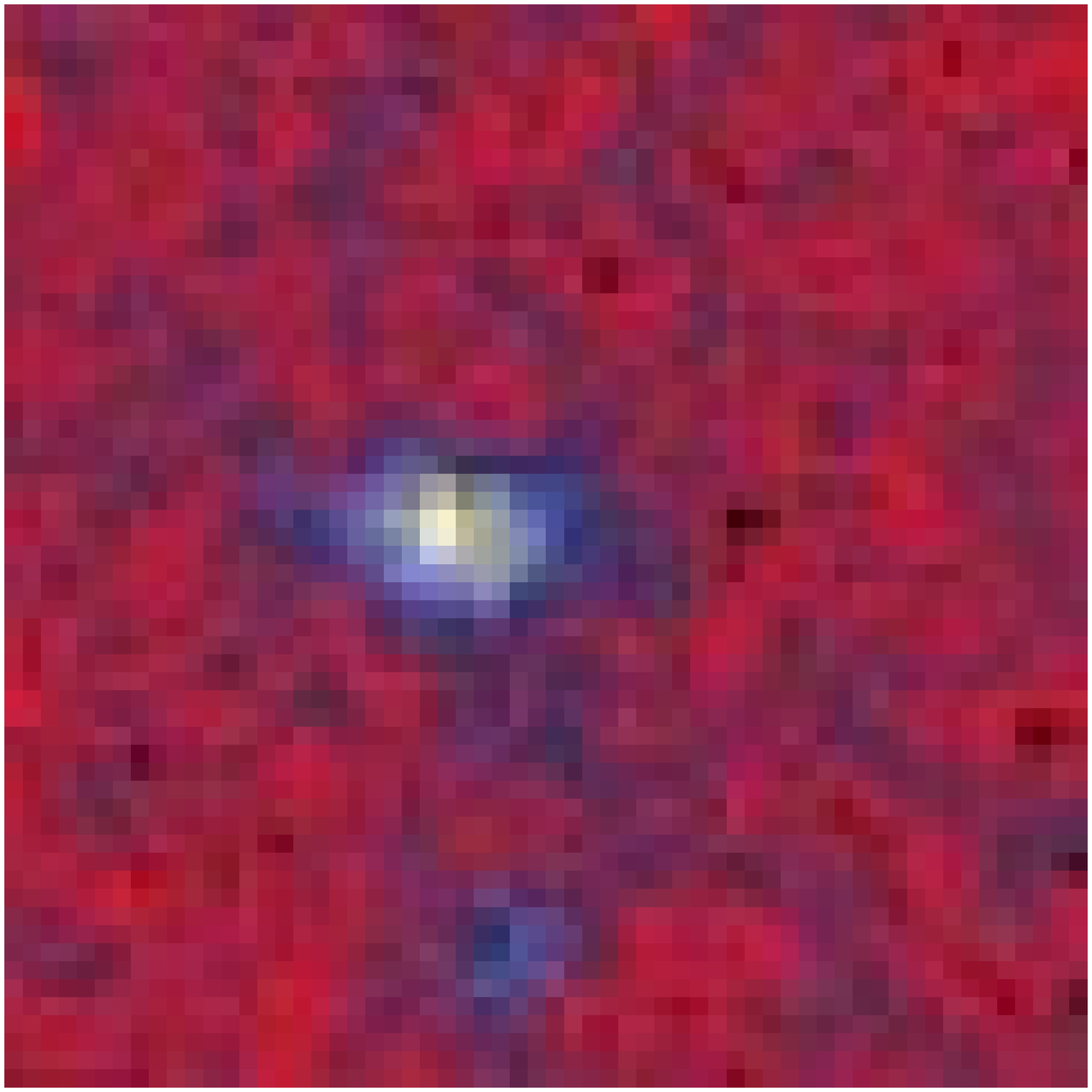}  \FGb{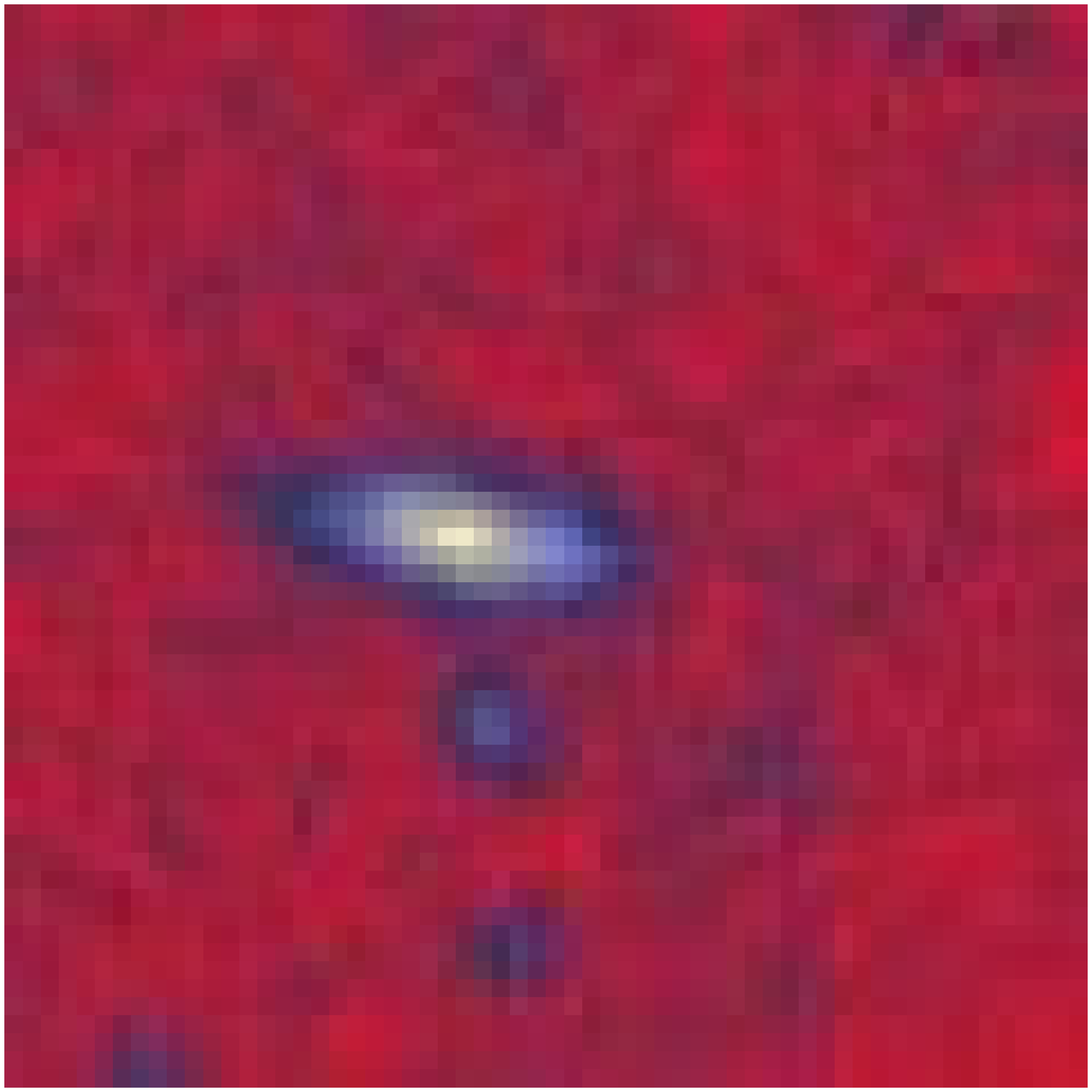}  \FGb{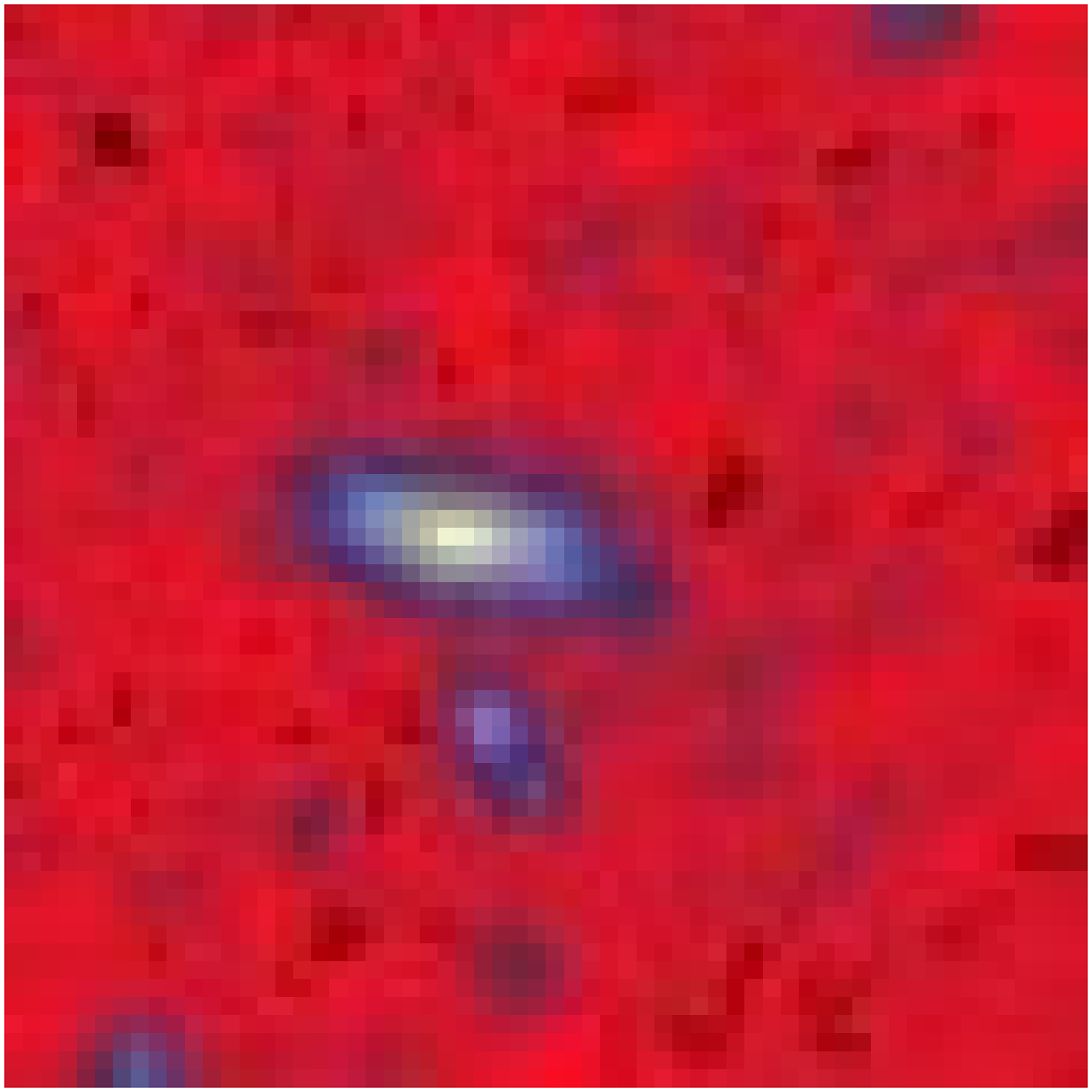}  \FGb{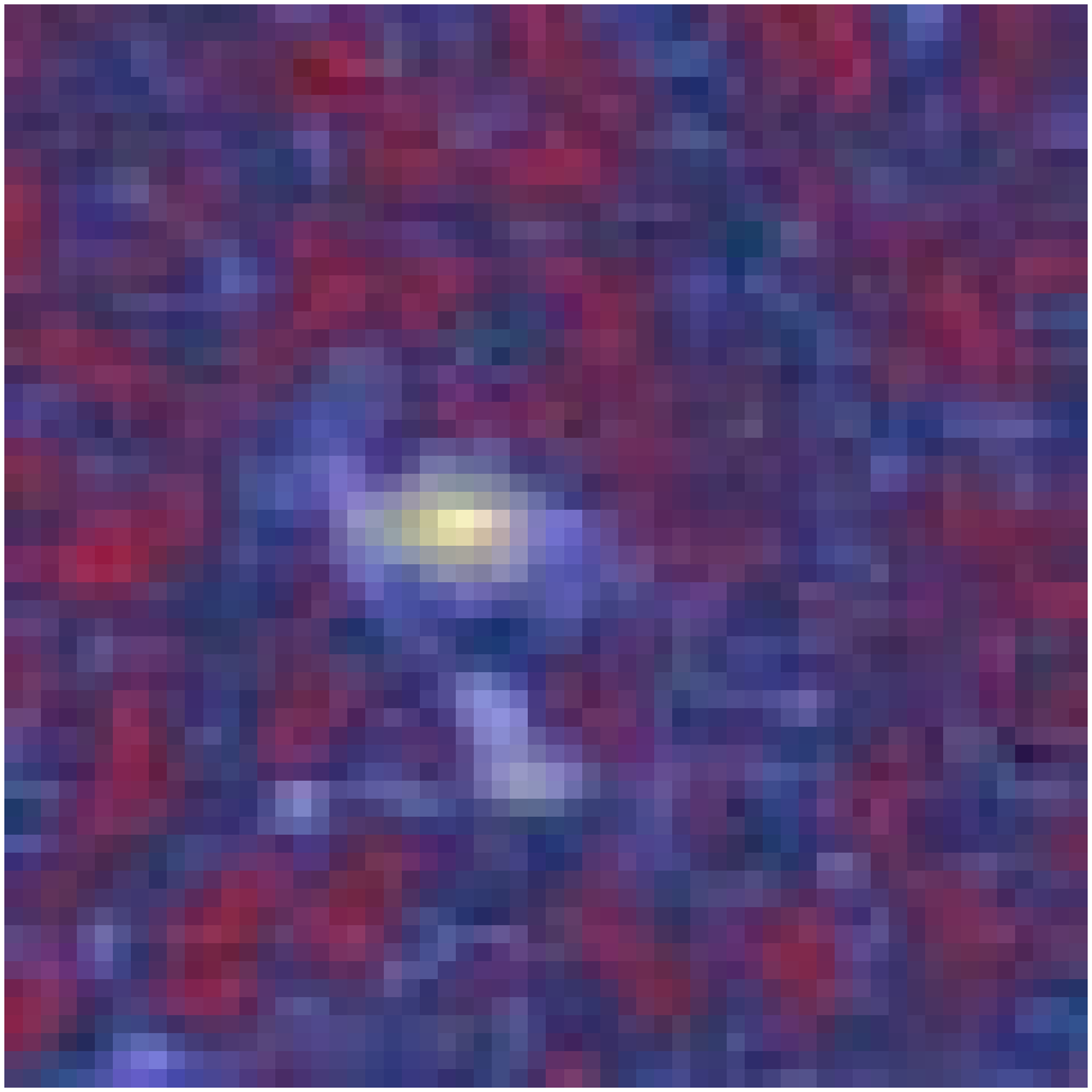}  \FGb{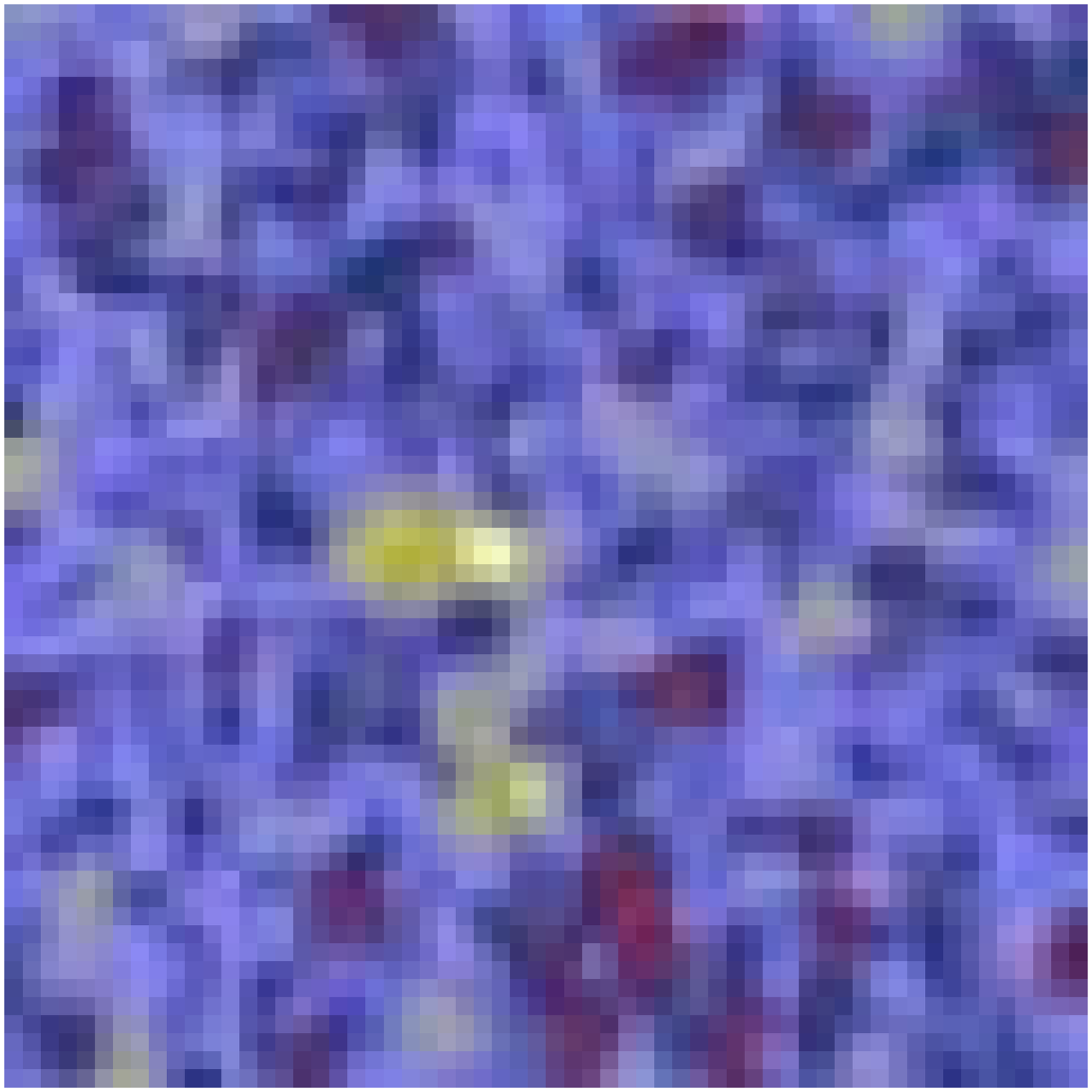}  \FGb{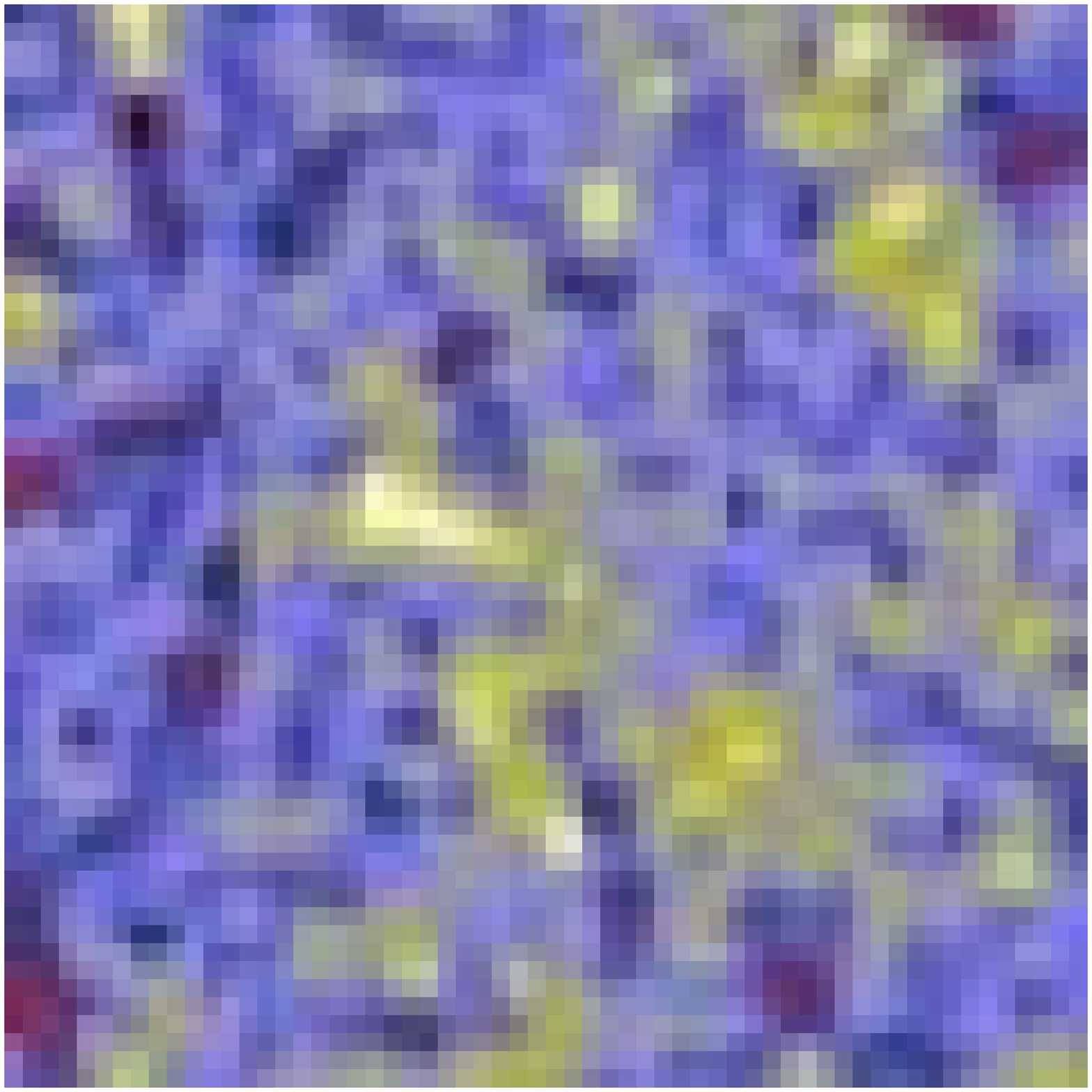}  \FGb{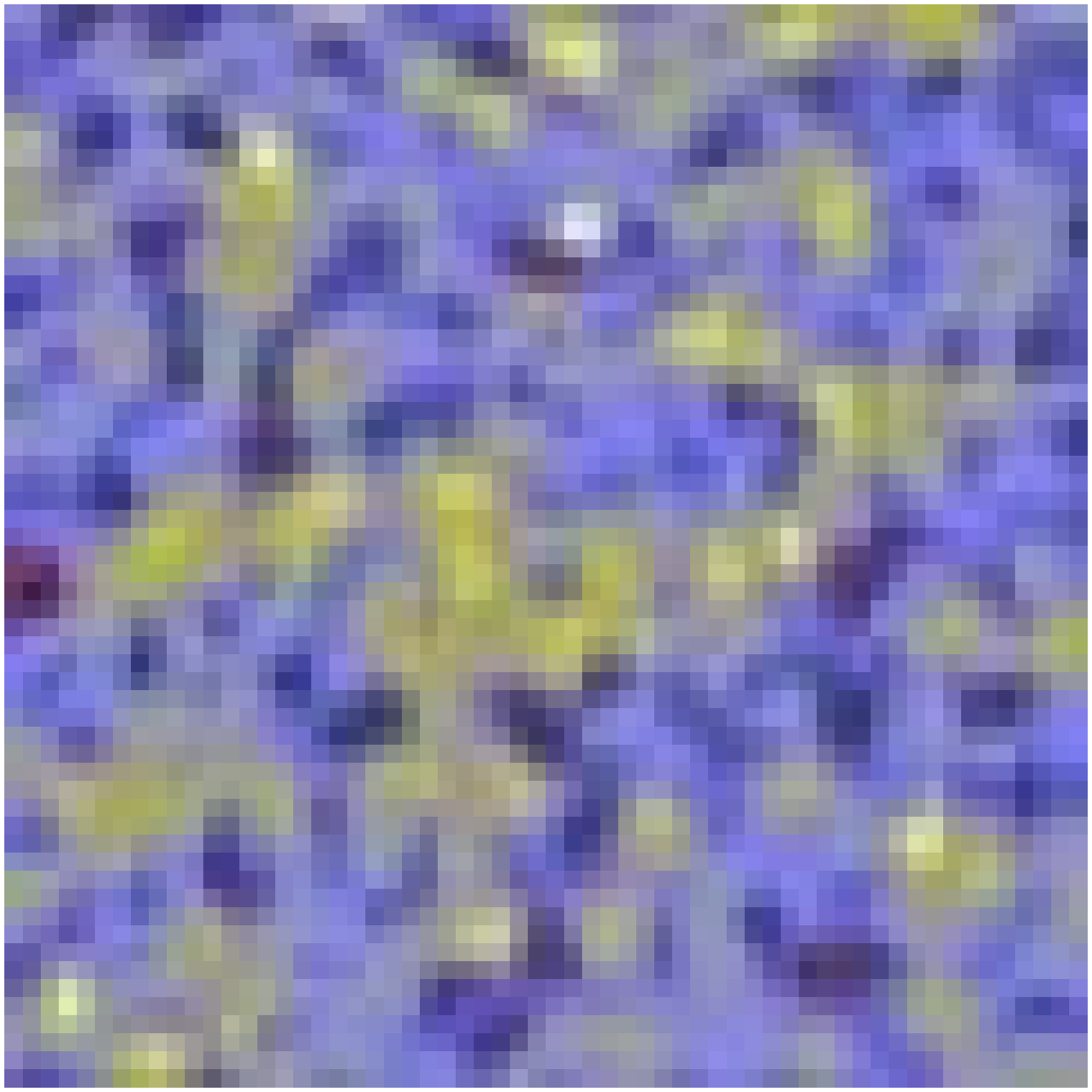}  \FGb{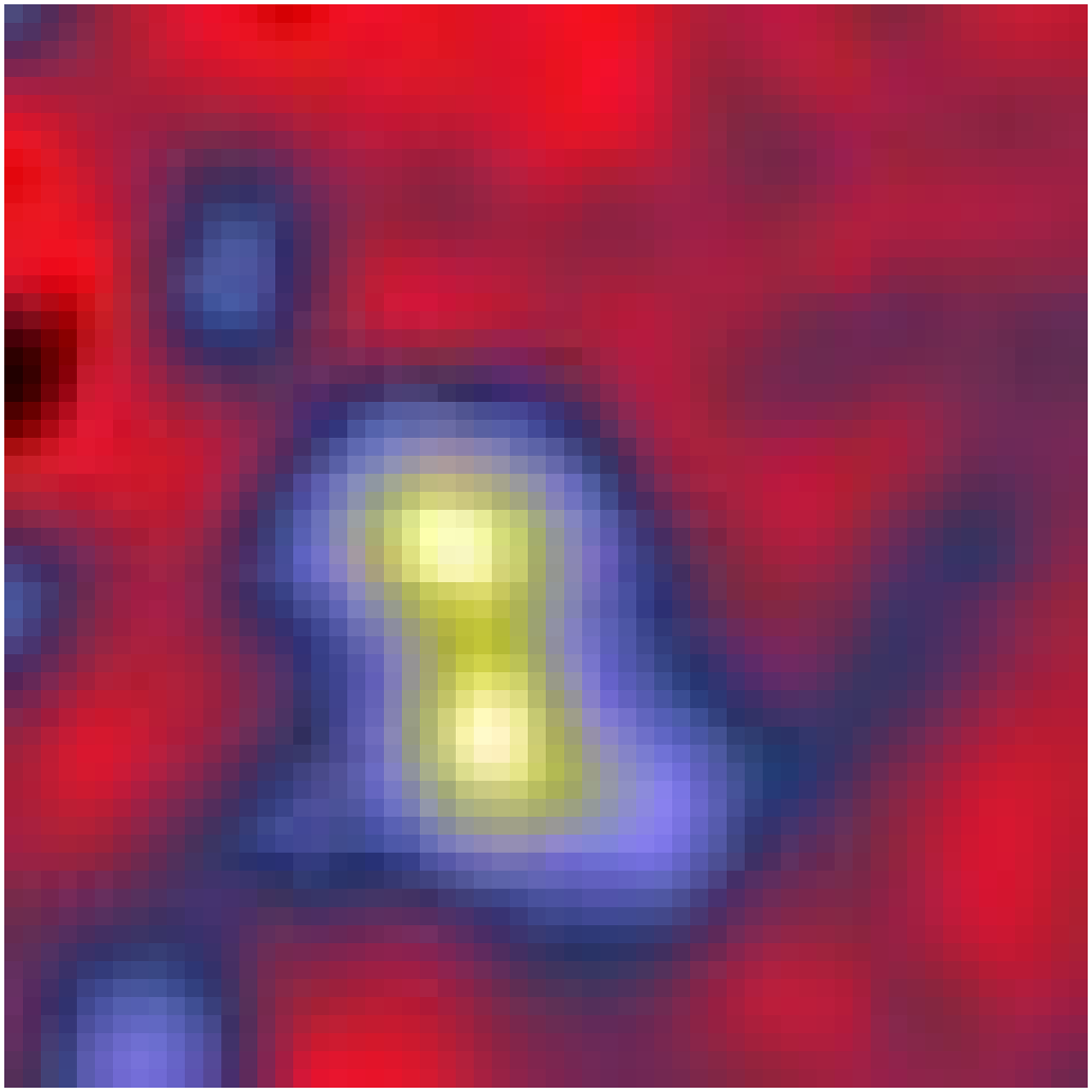}  \FGb{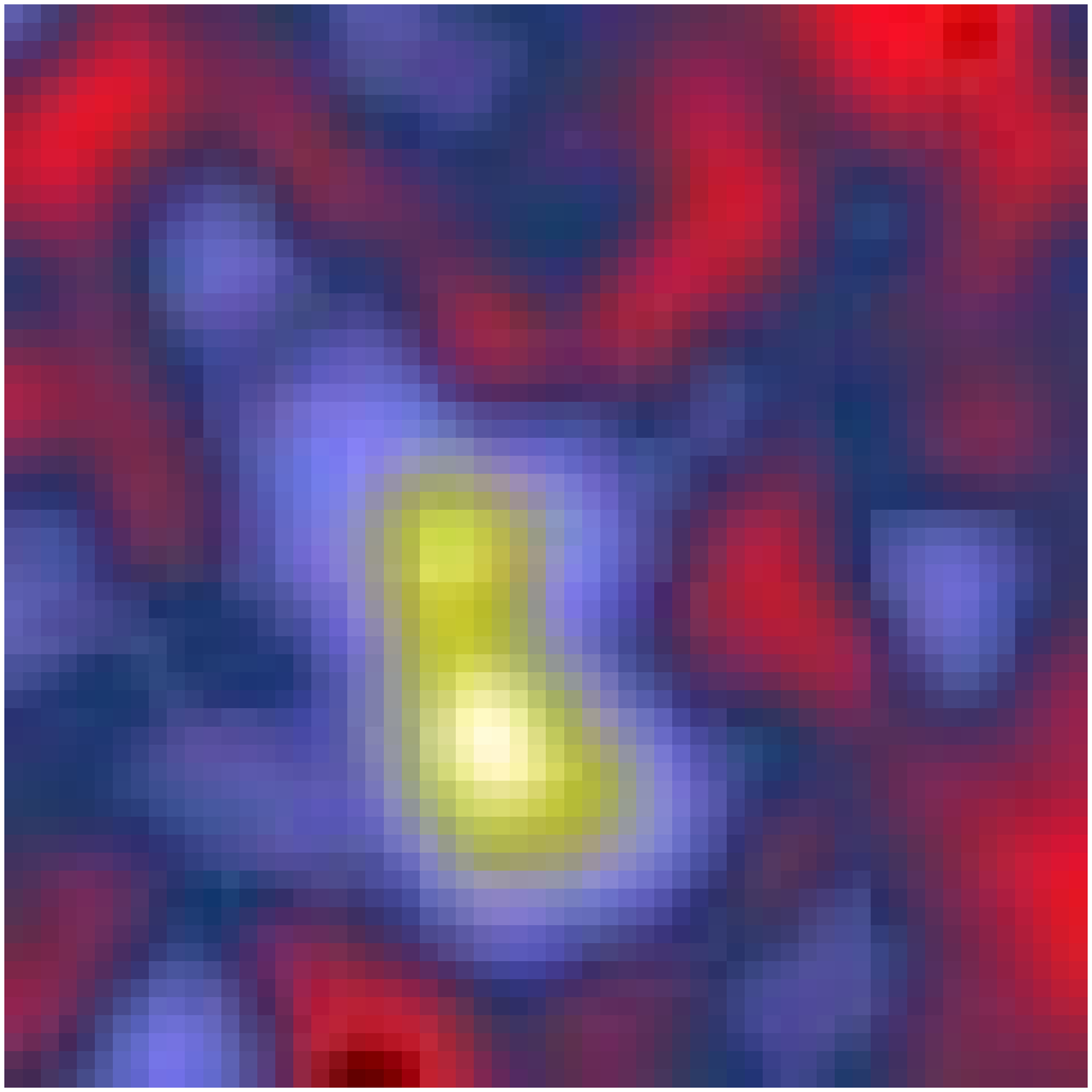}  \FGb{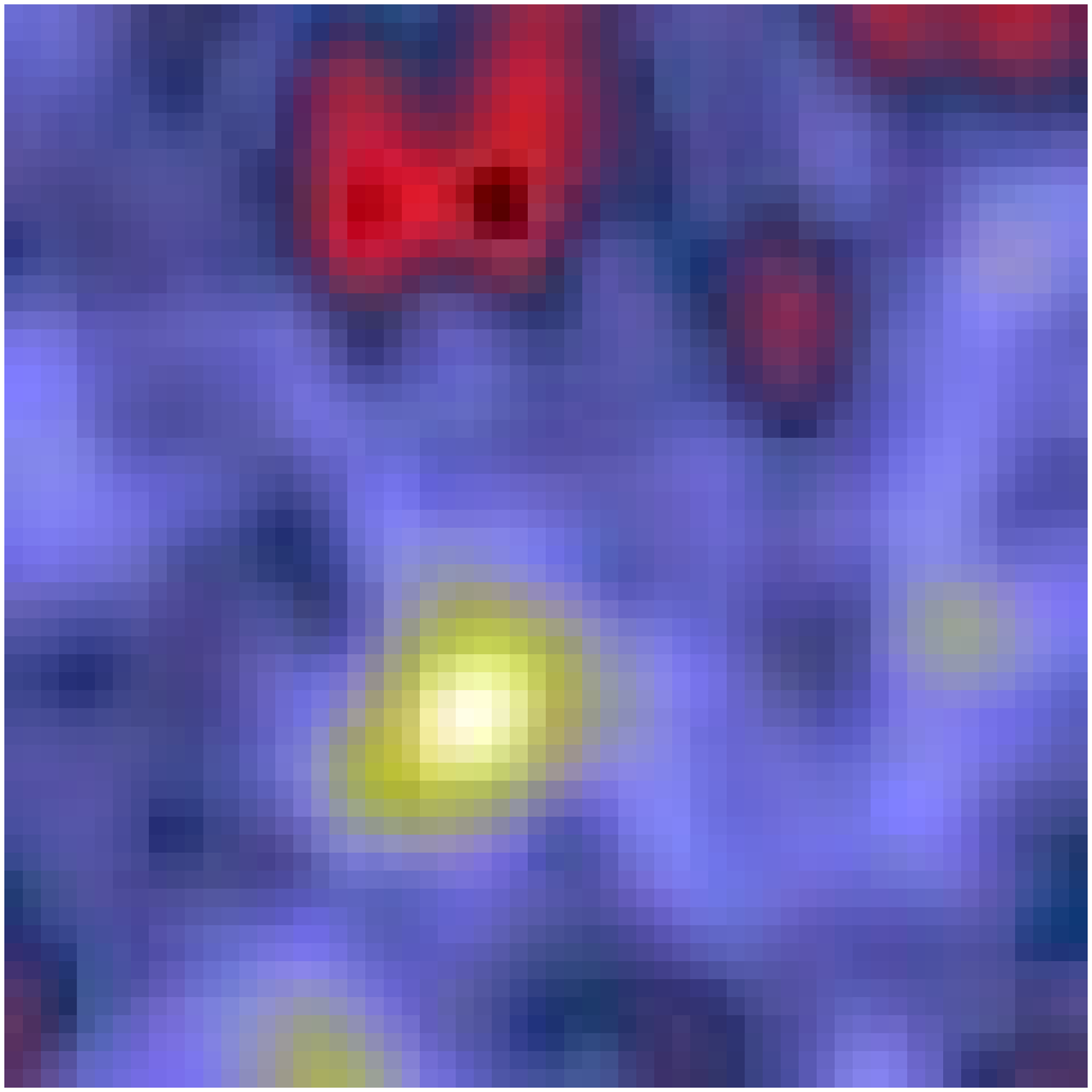}  \FGb{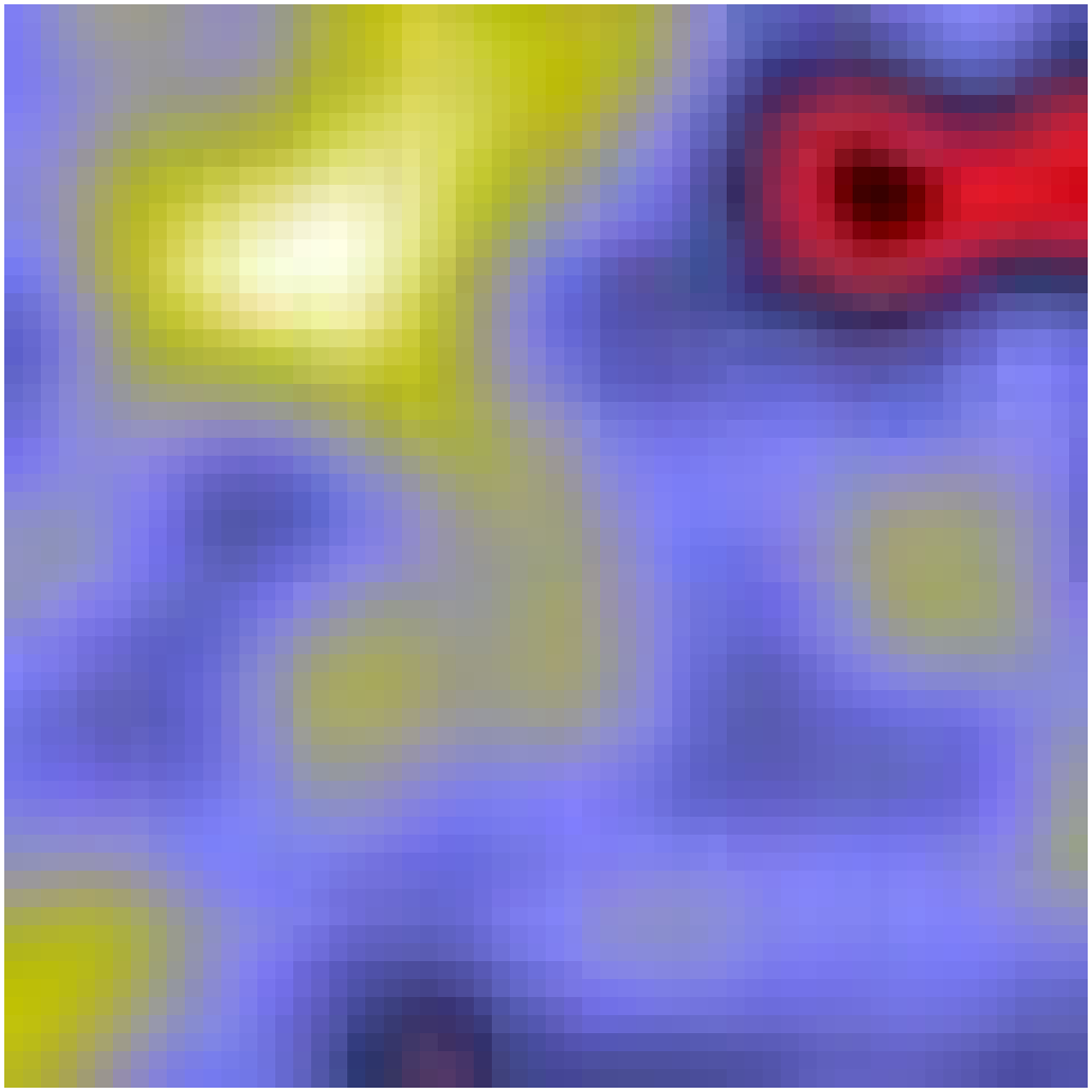}\\  {\#10  NCIA030487}  \\
  \FGb{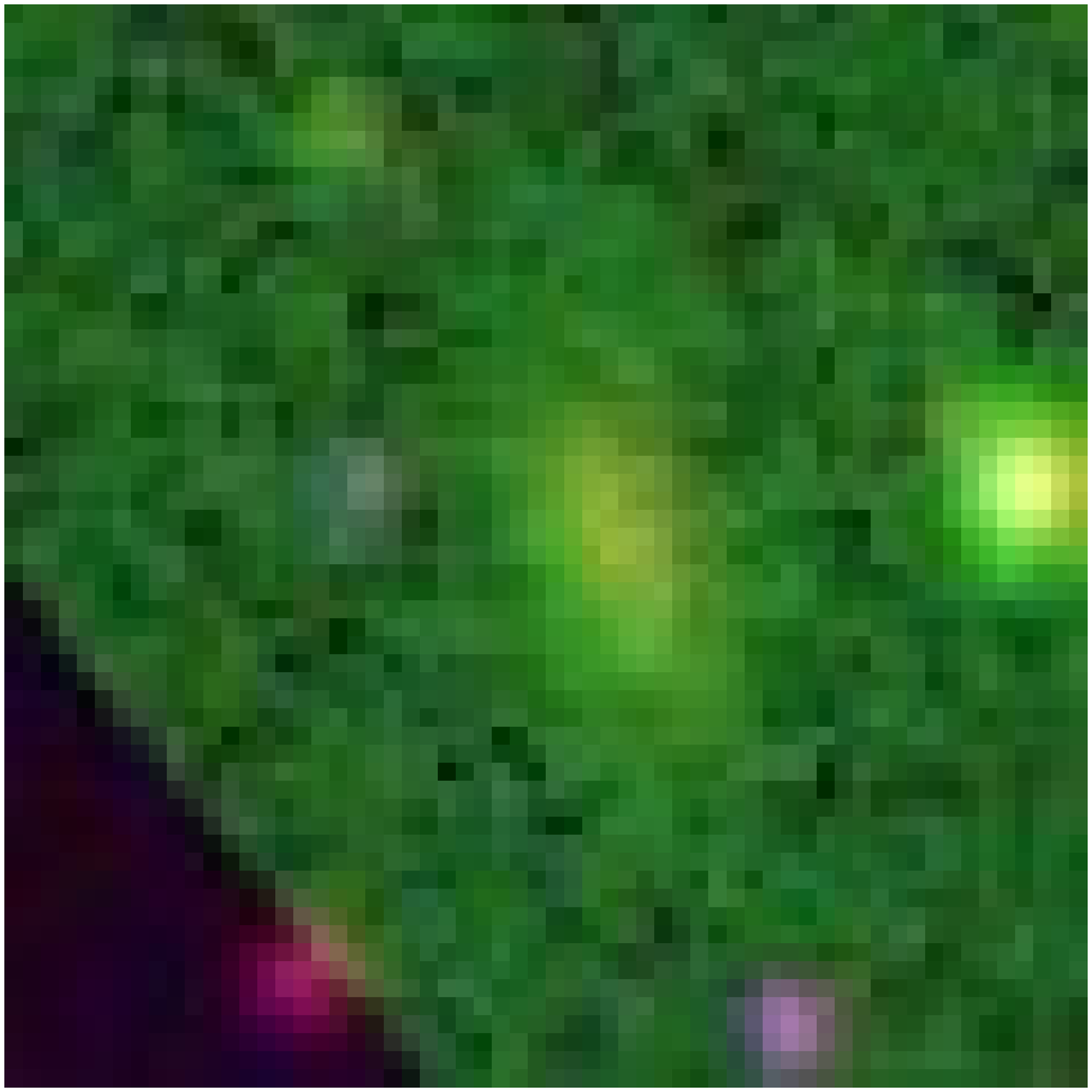} \FGb{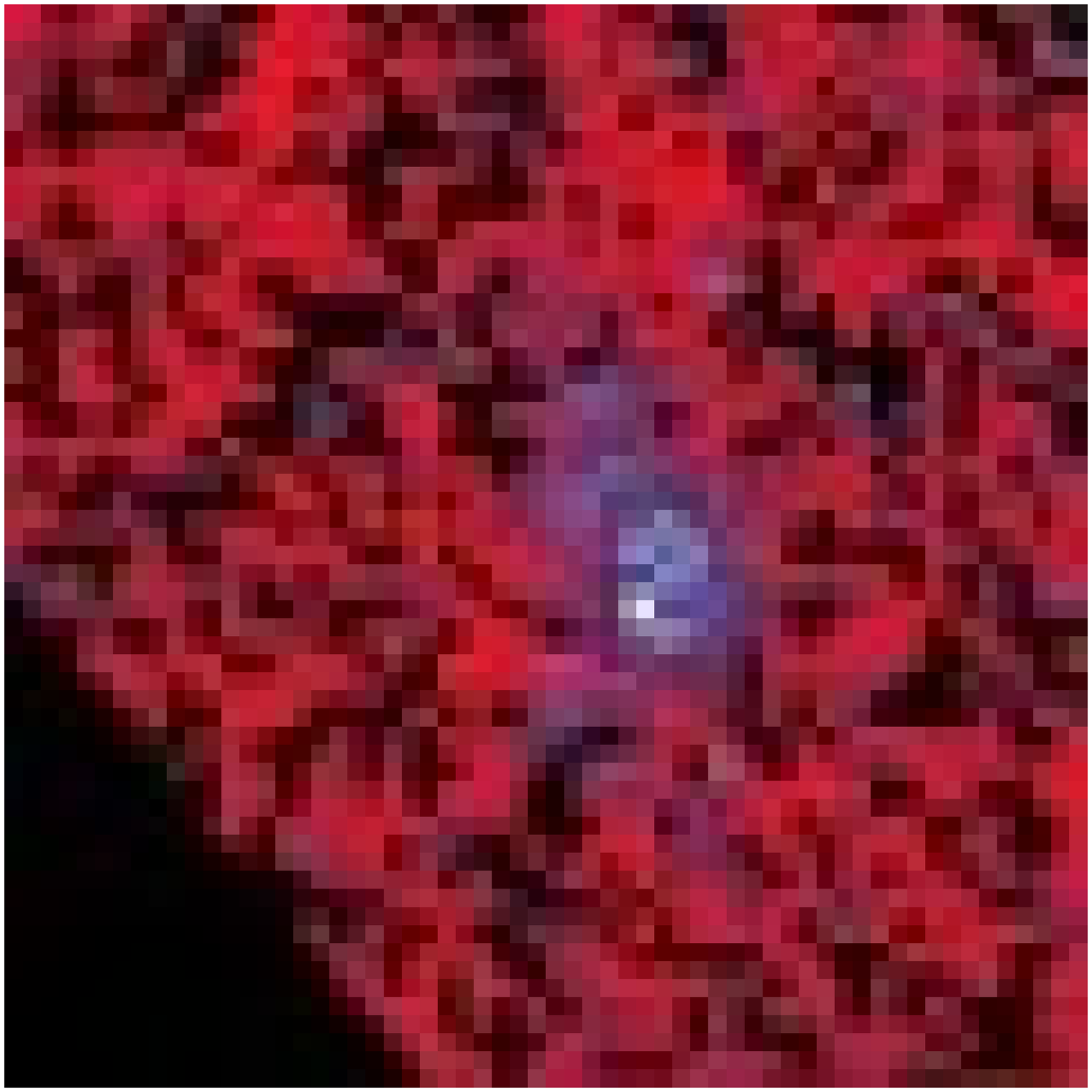}   \FGb{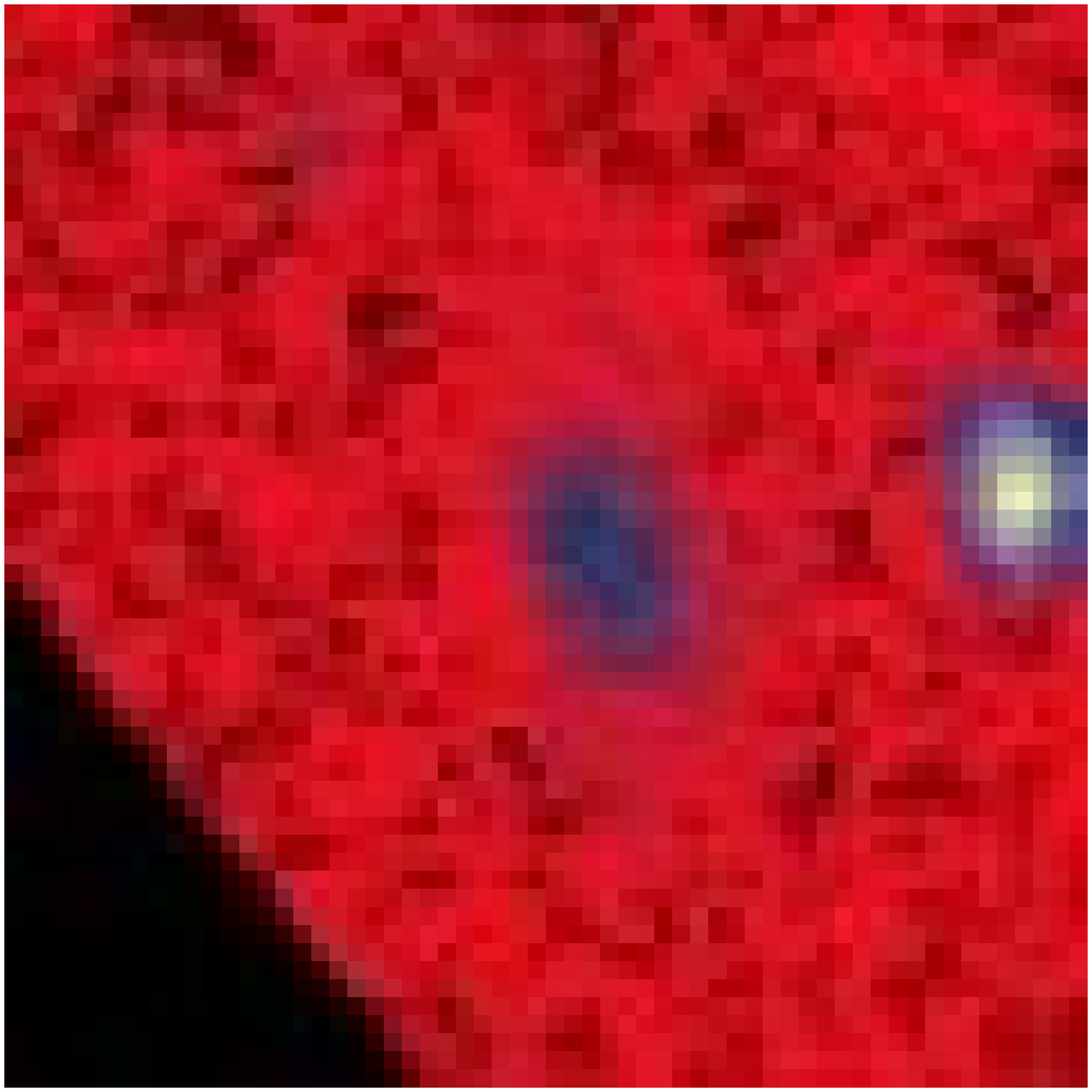}  \FGb{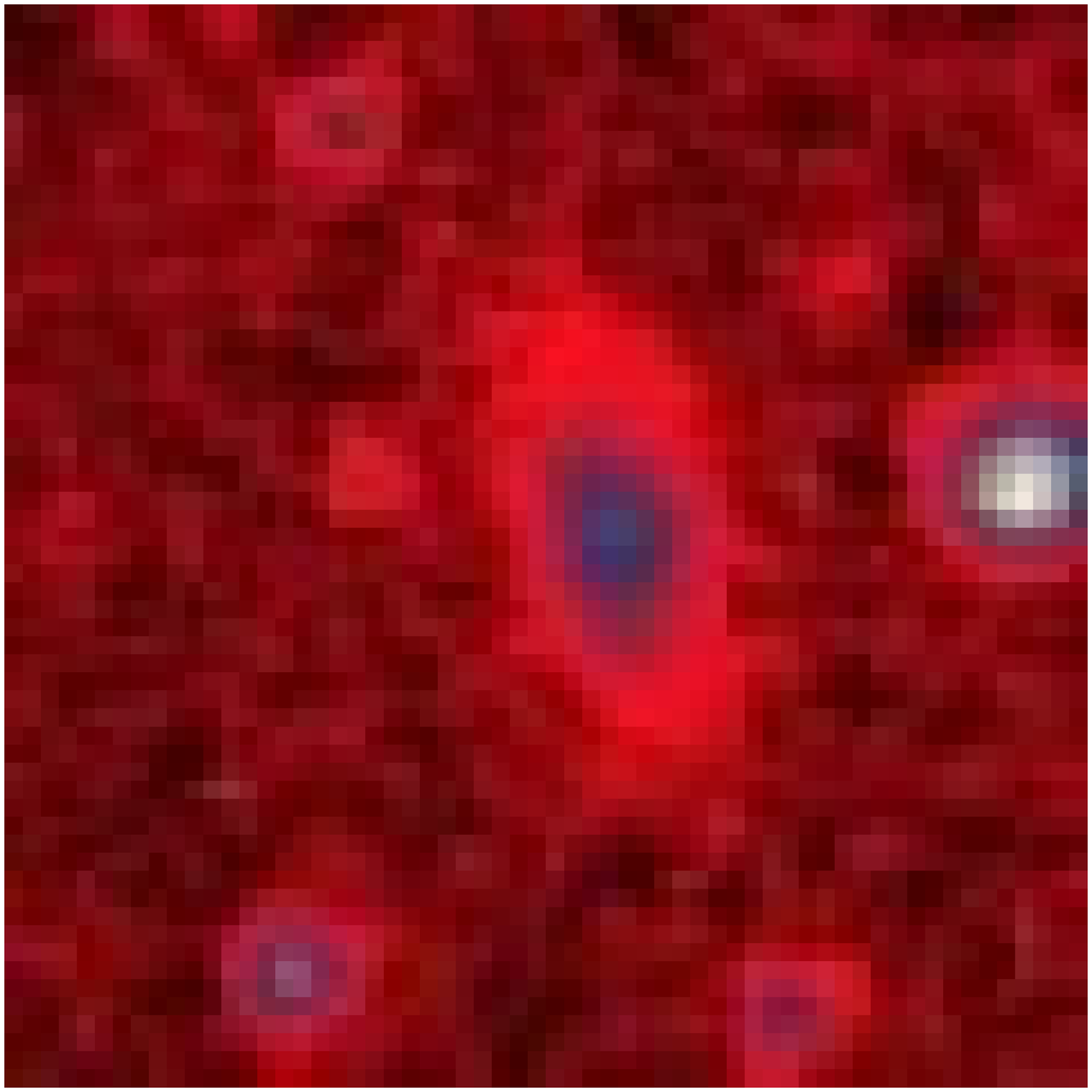}  \FGb{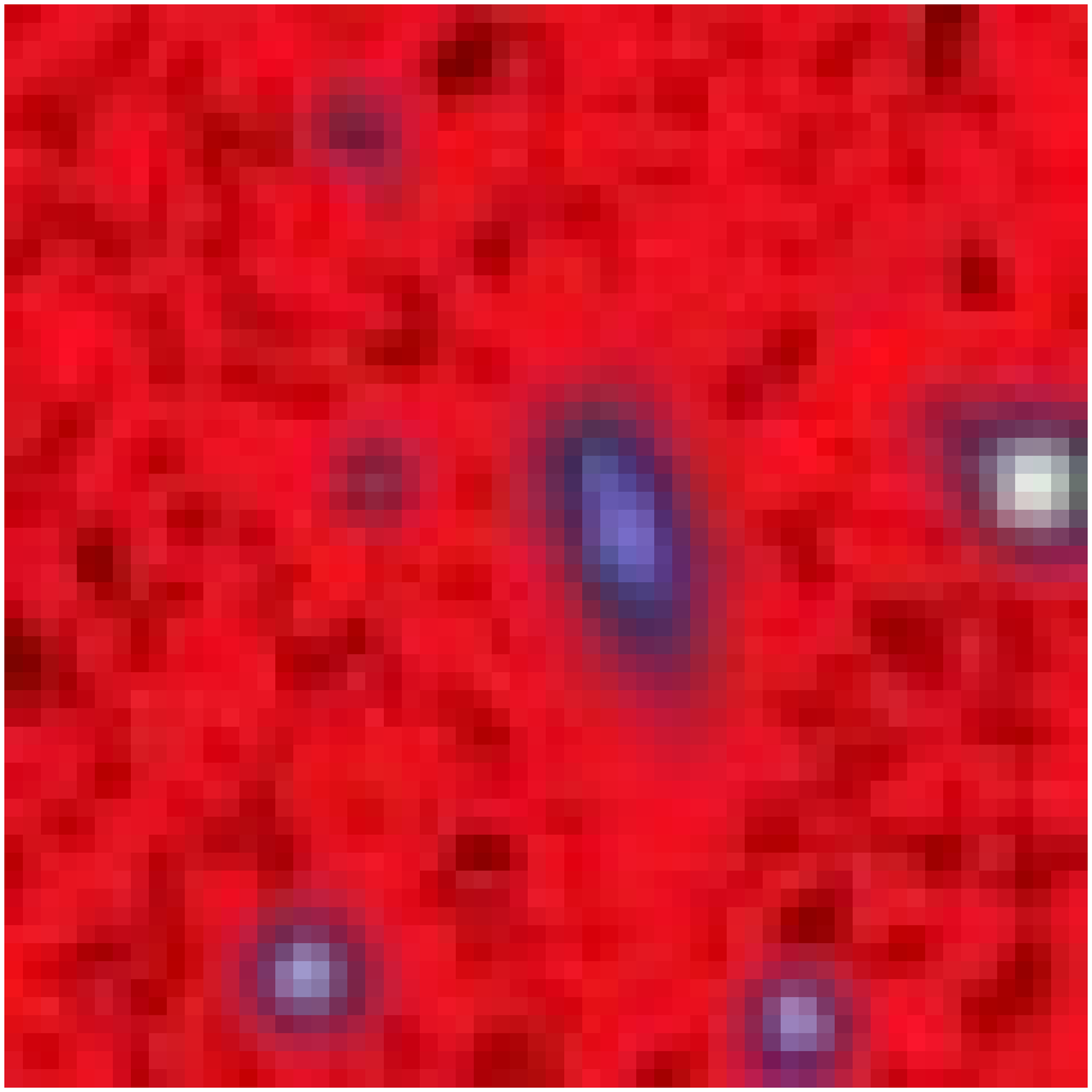}  \FGb{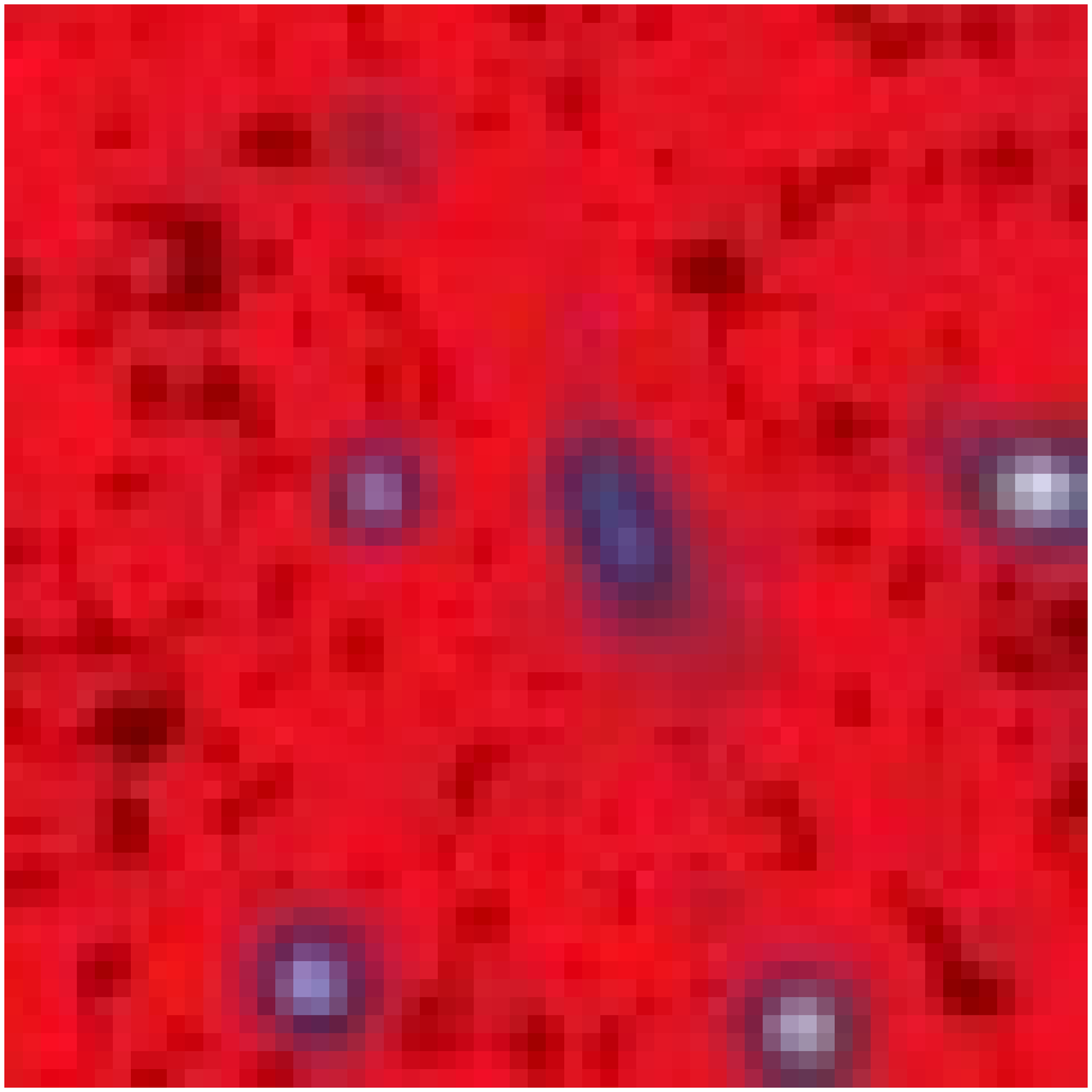}  \FGb{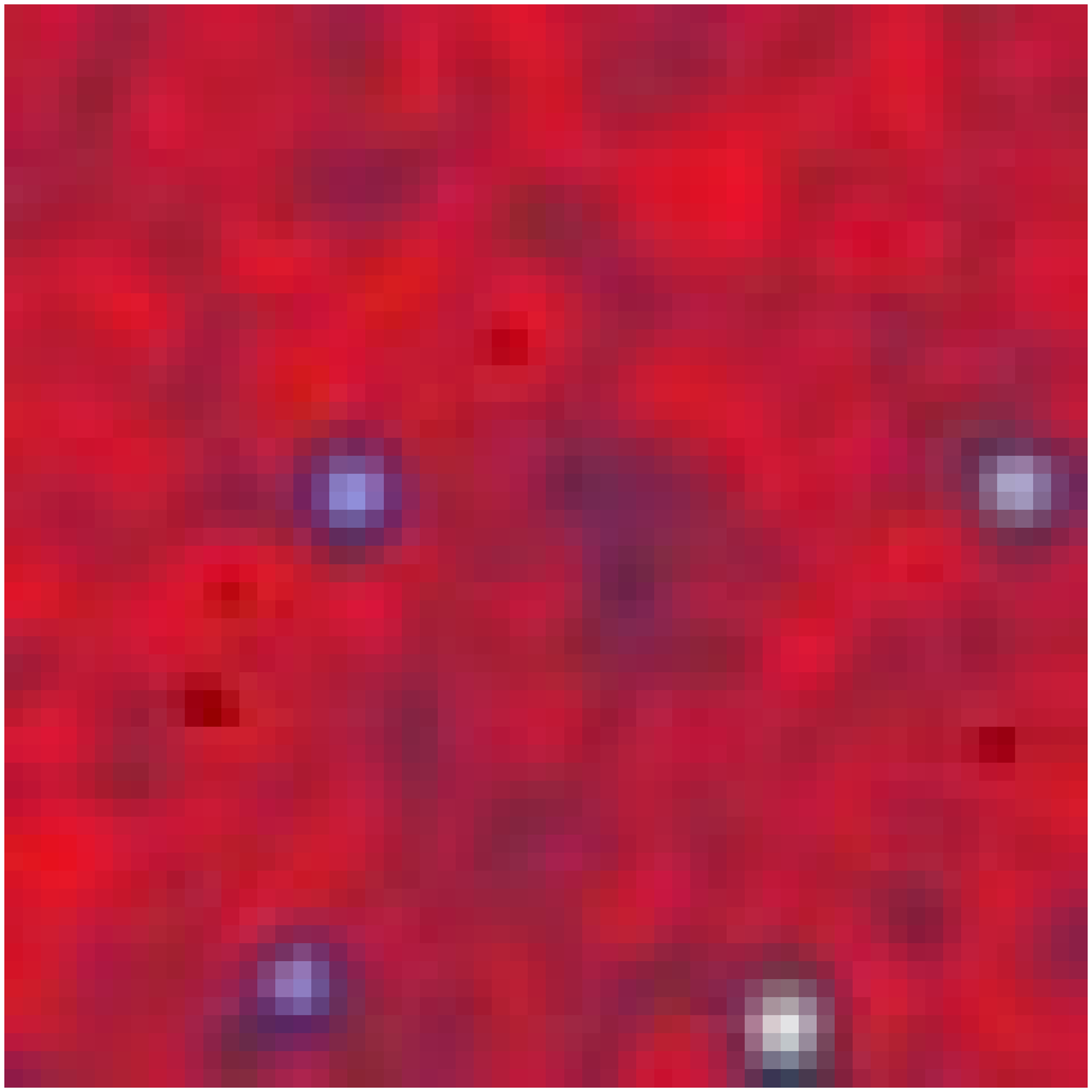}  \FGb{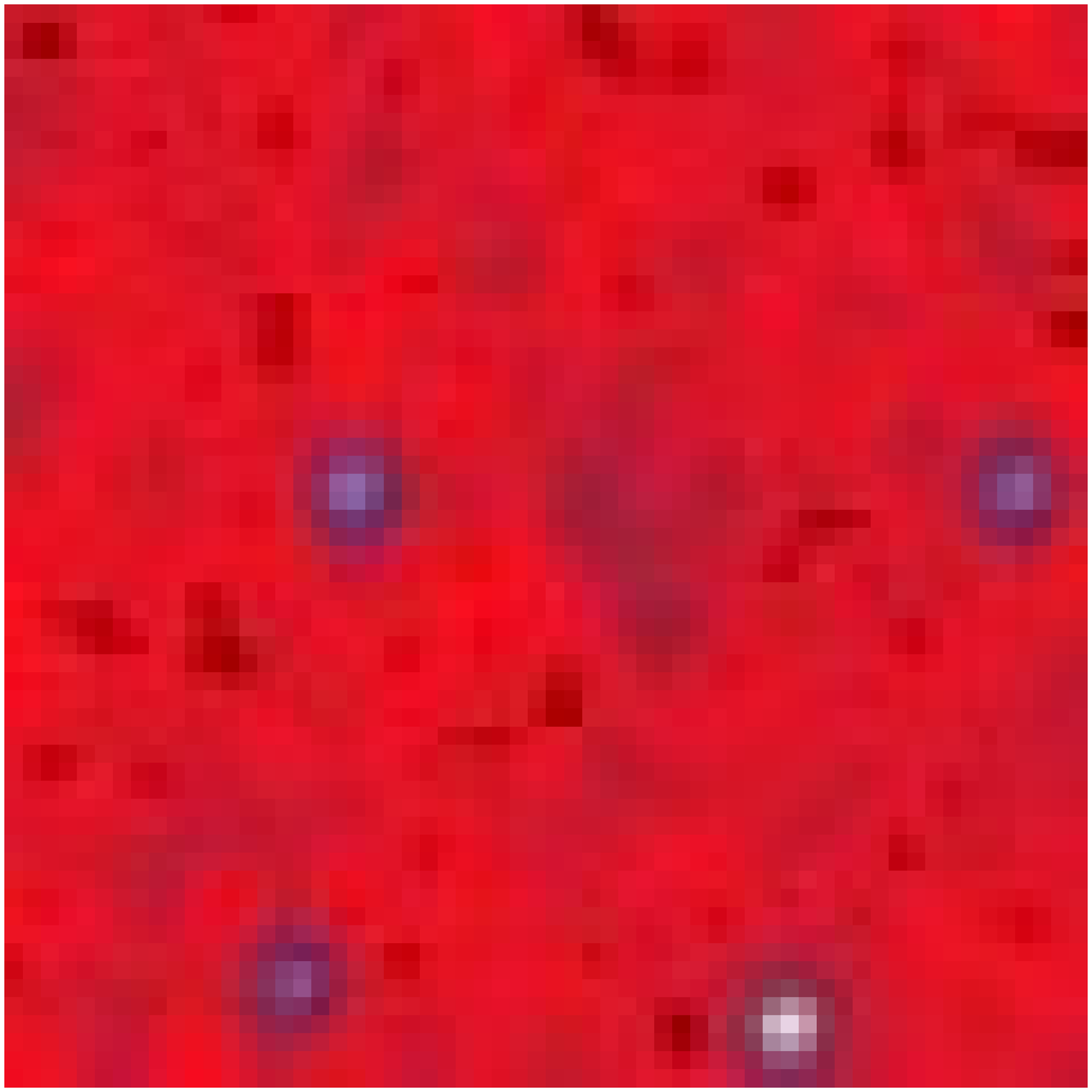}  \FGb{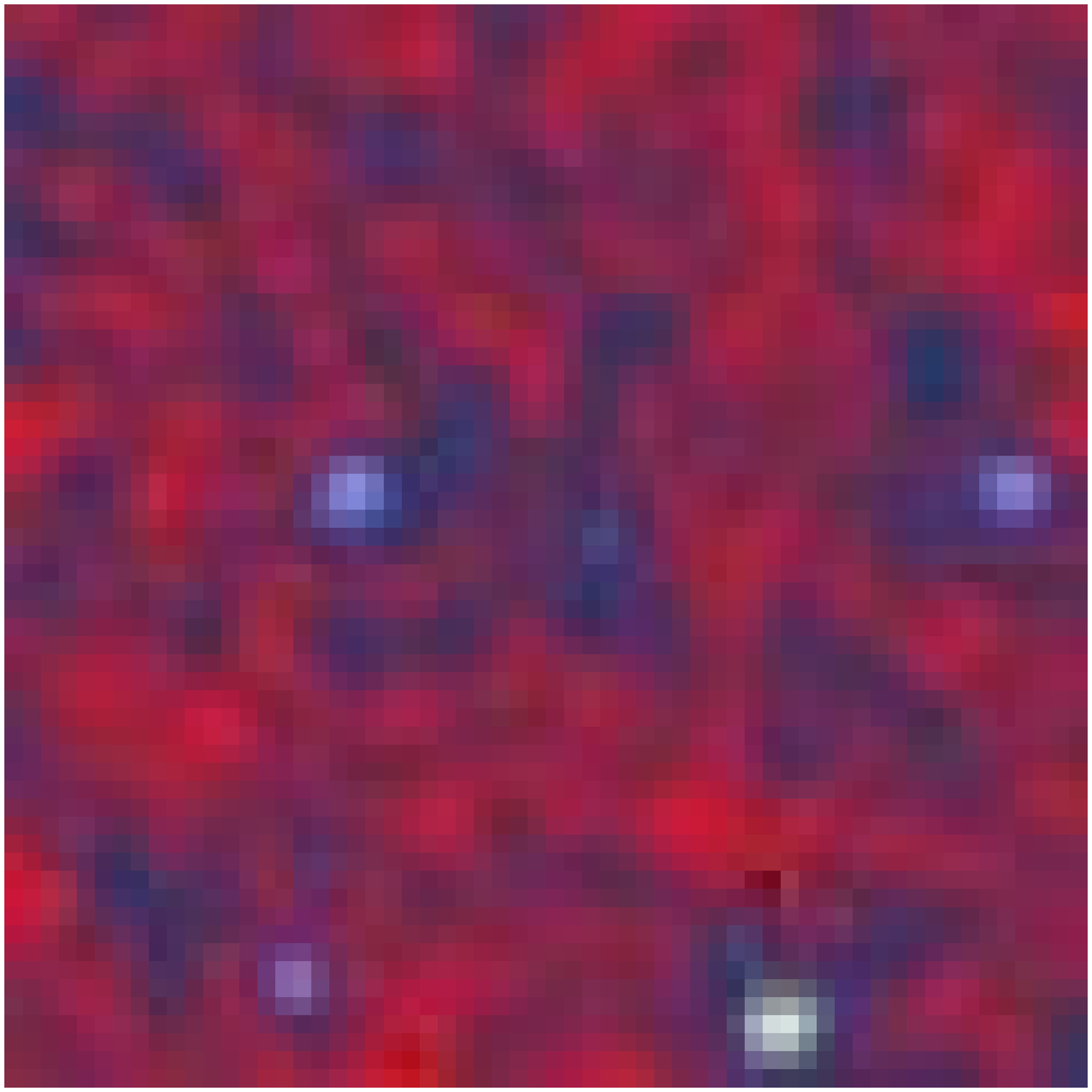}  \FGb{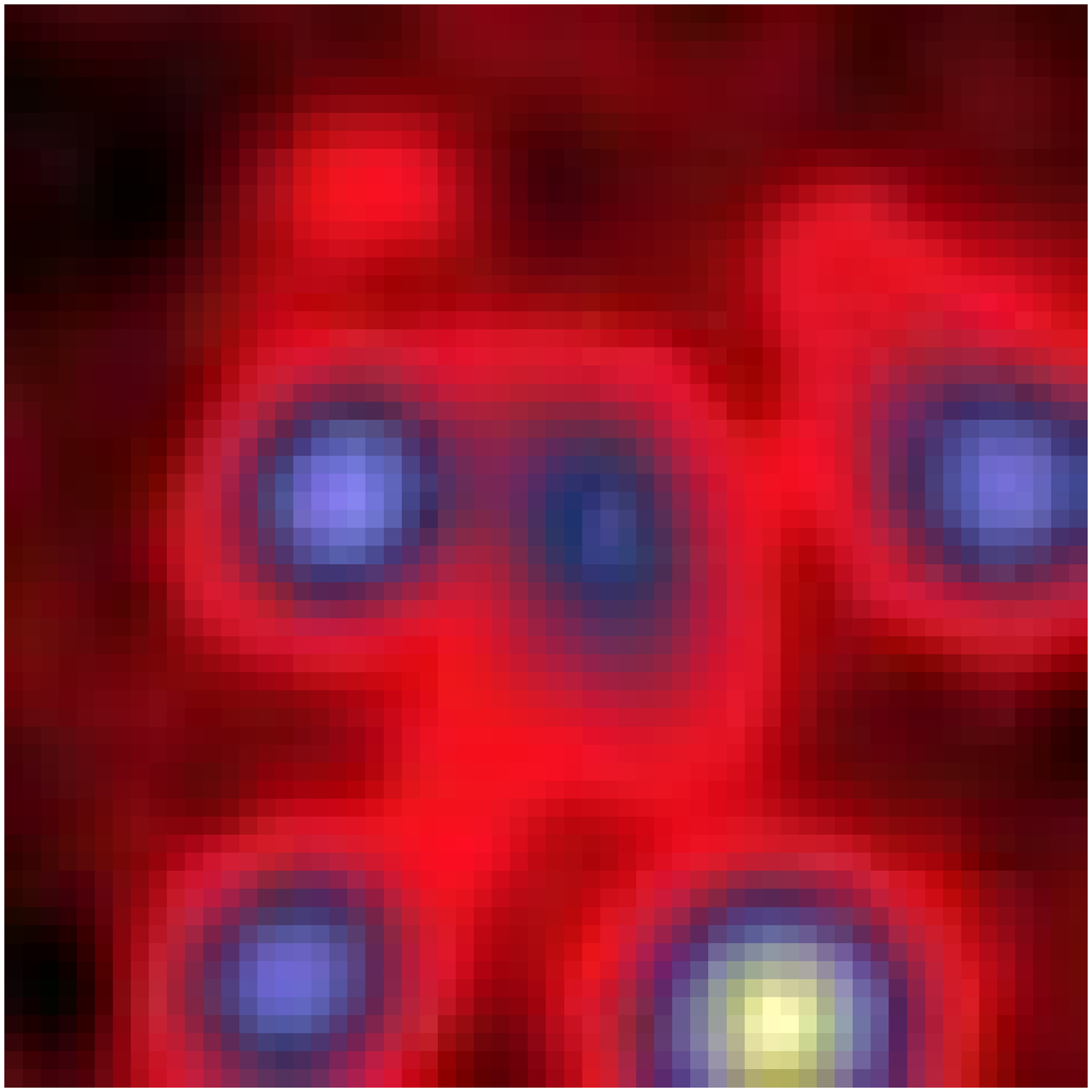}  \FGb{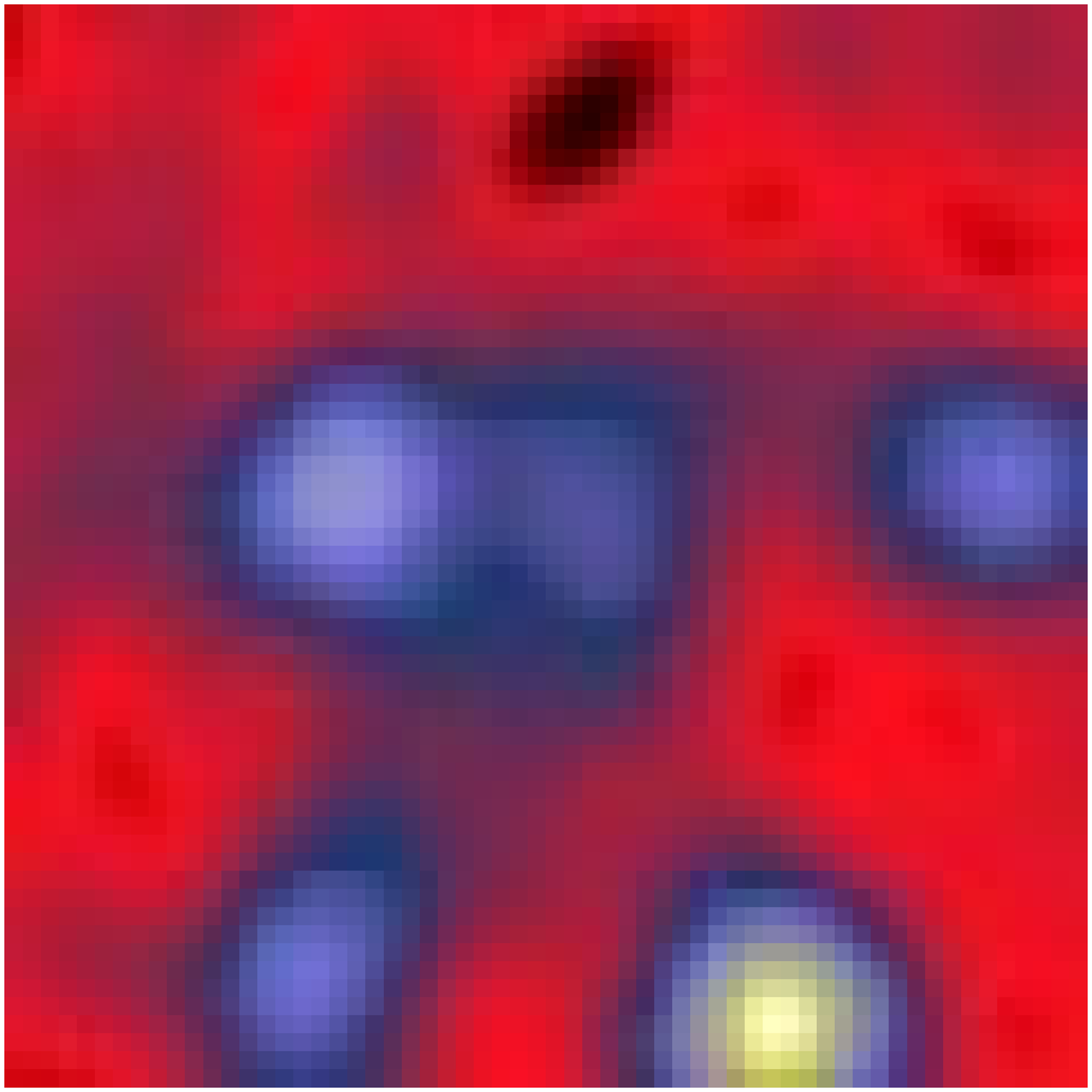}  \FGb{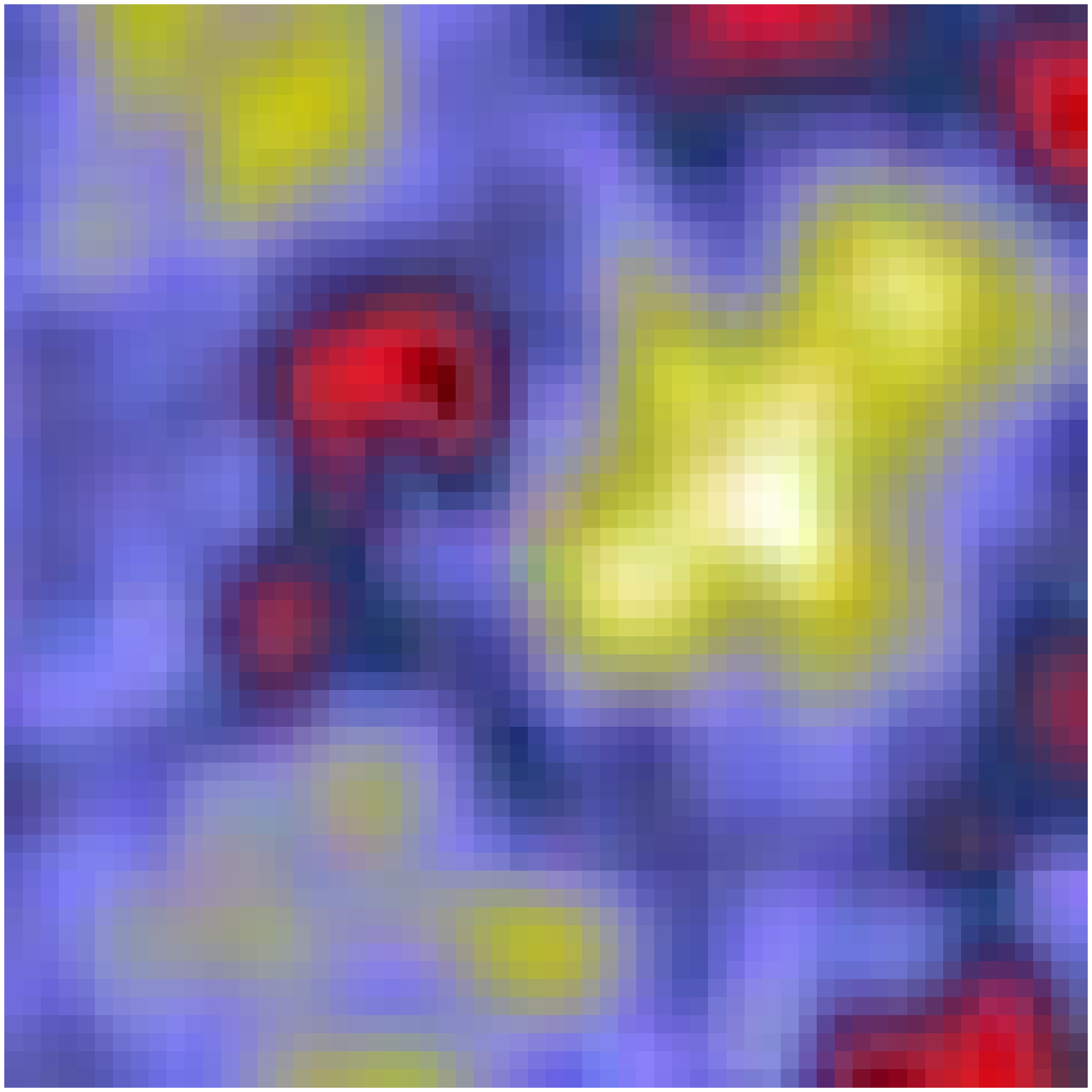}  \FGb{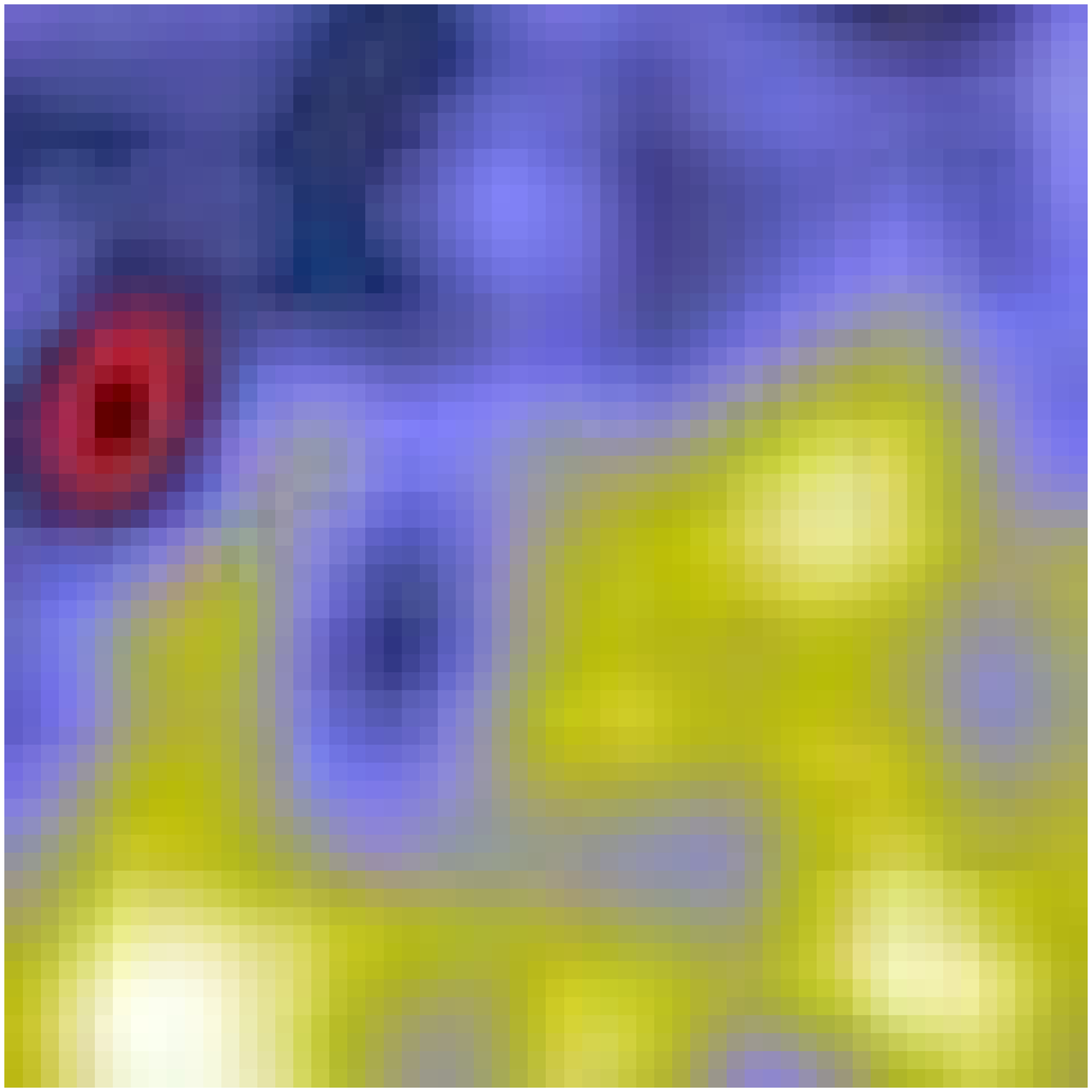}\\  {\#11  NCIA005660}  \\
  \FGb{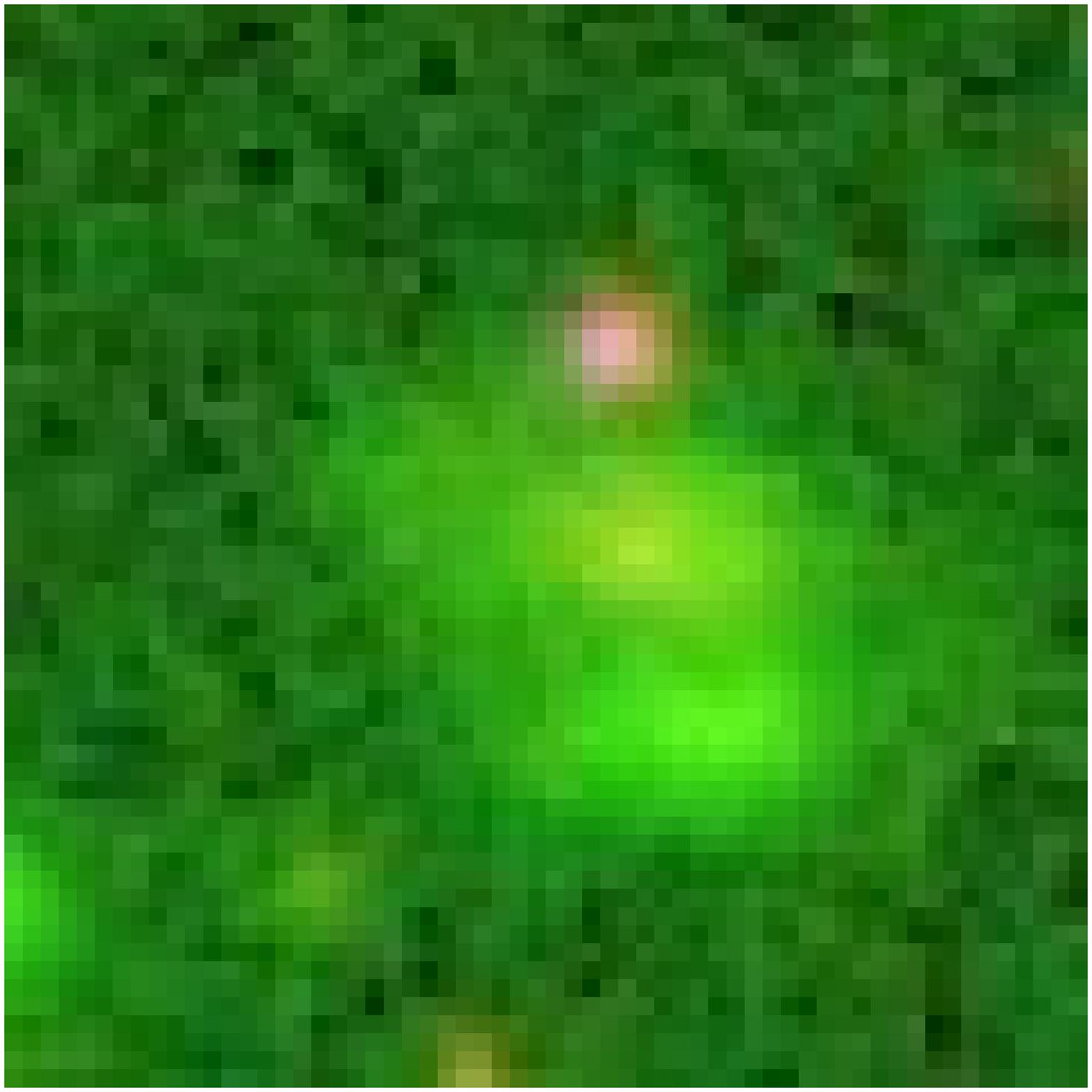} \FGb{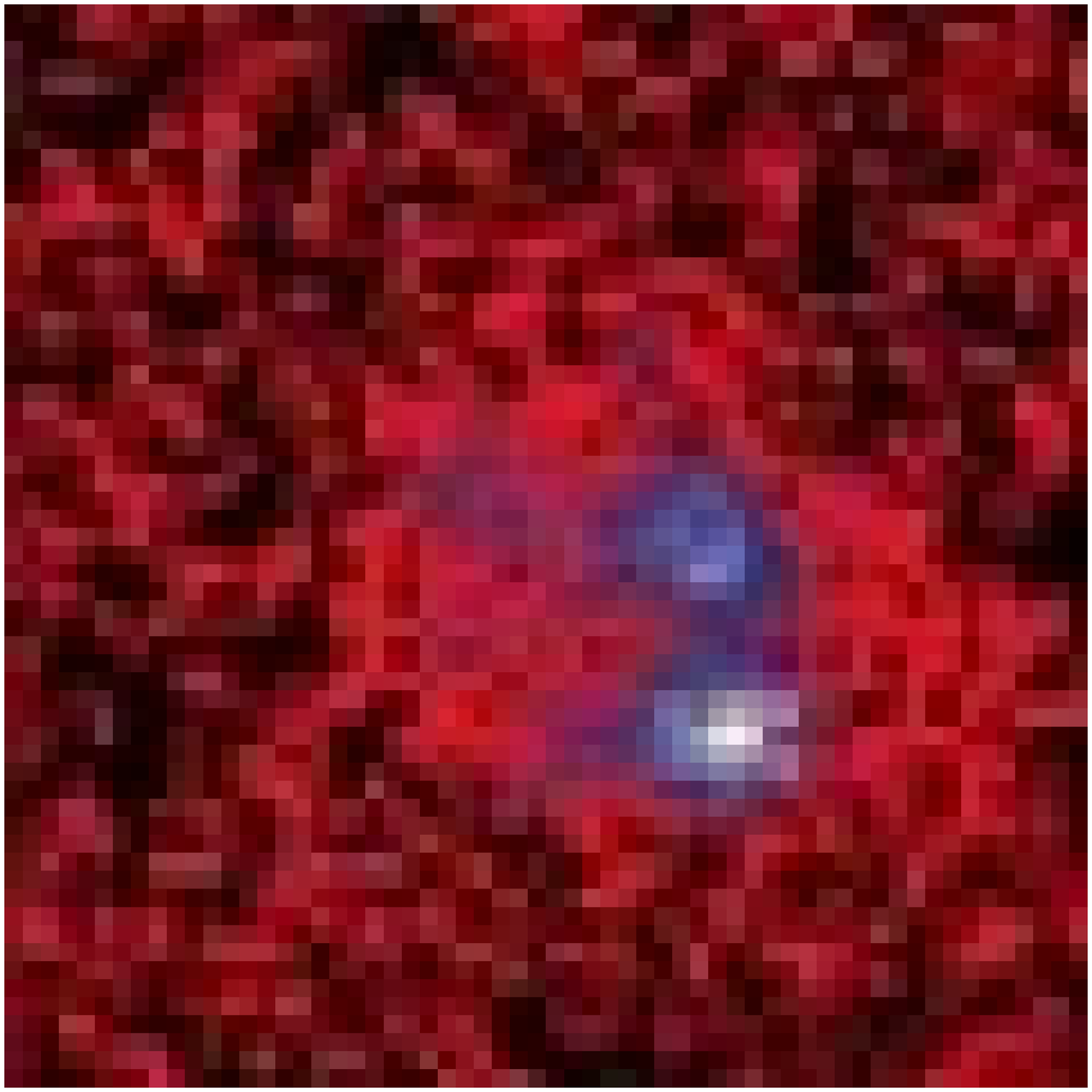}   \FGb{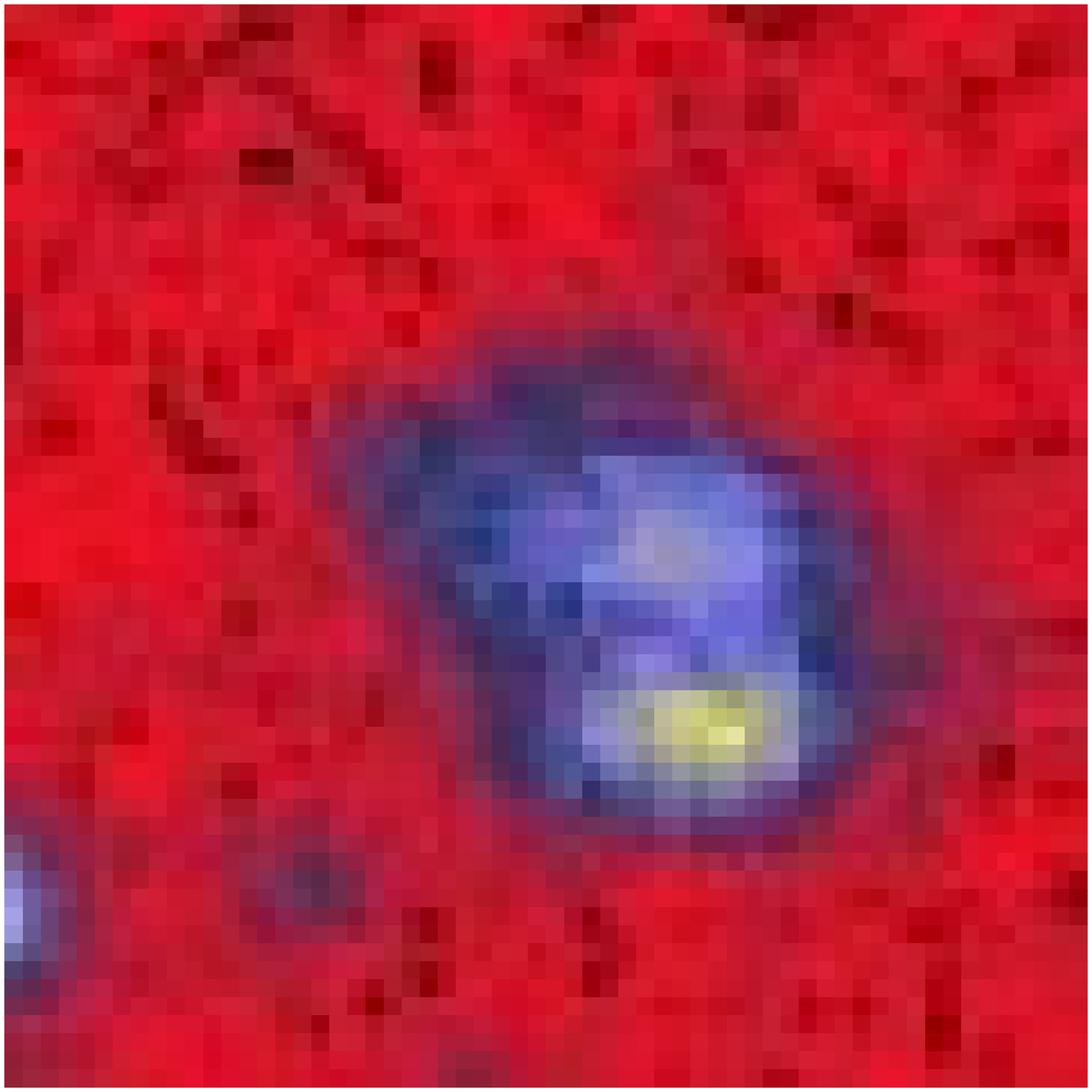}  \FGb{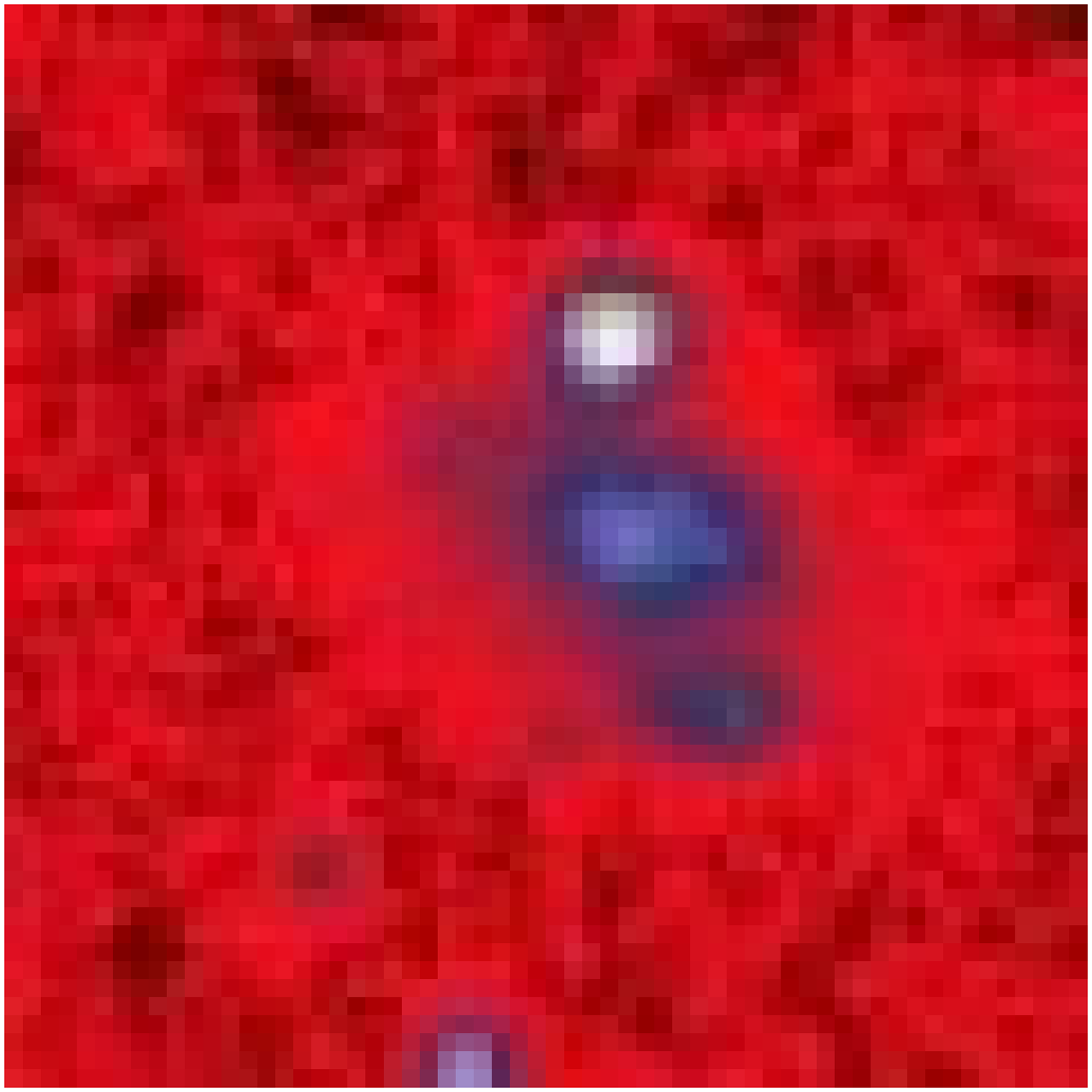}  \FGb{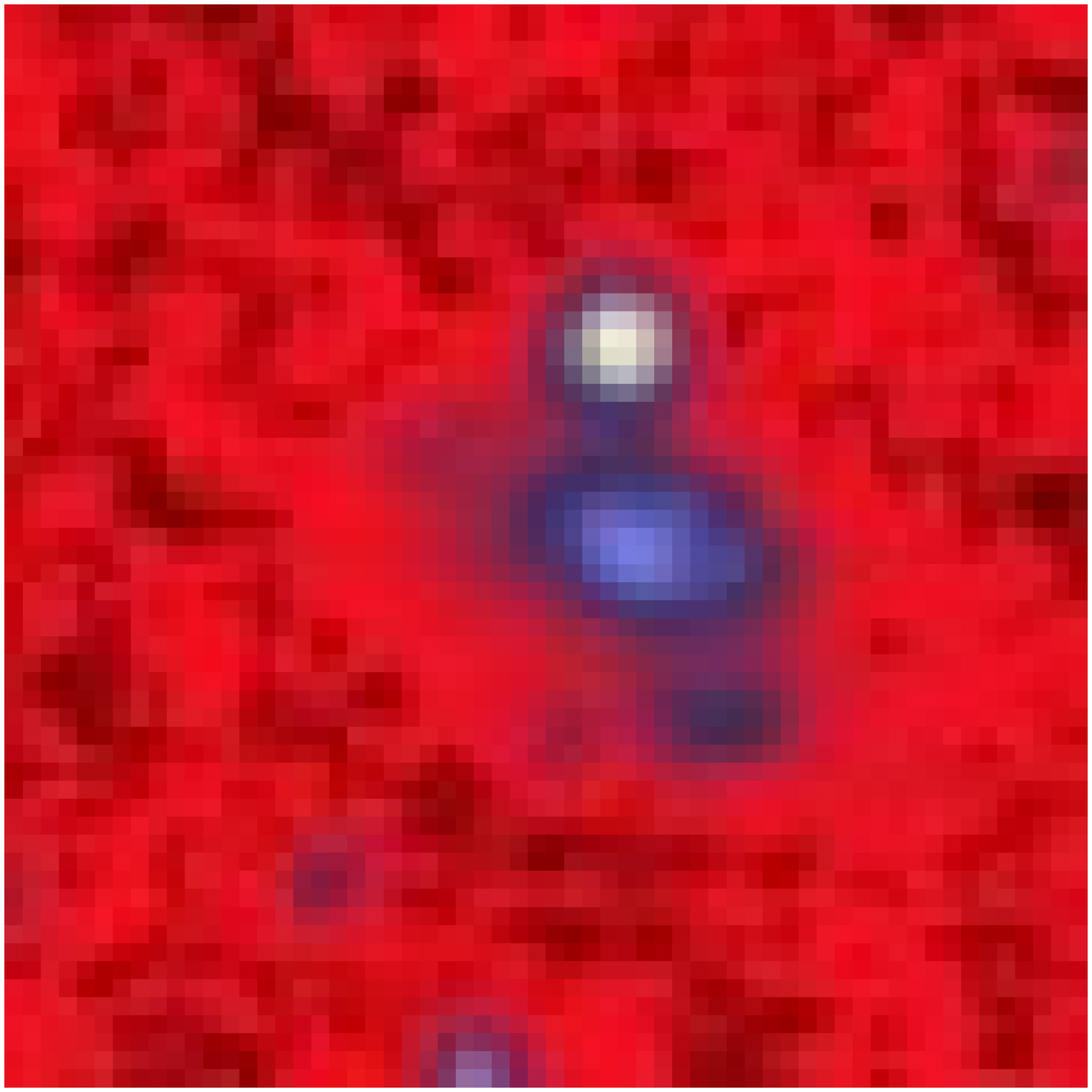}  \FGb{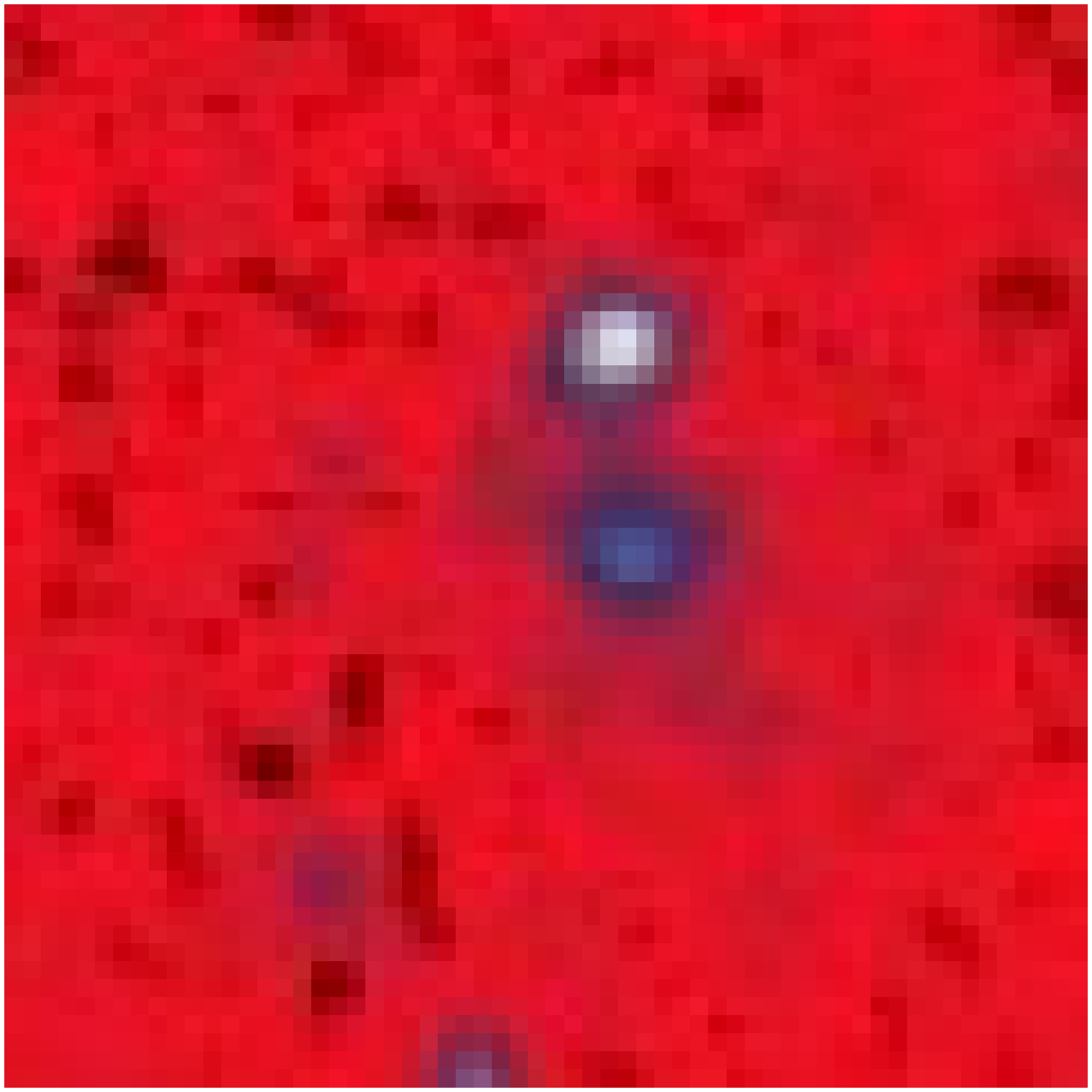}  \FGb{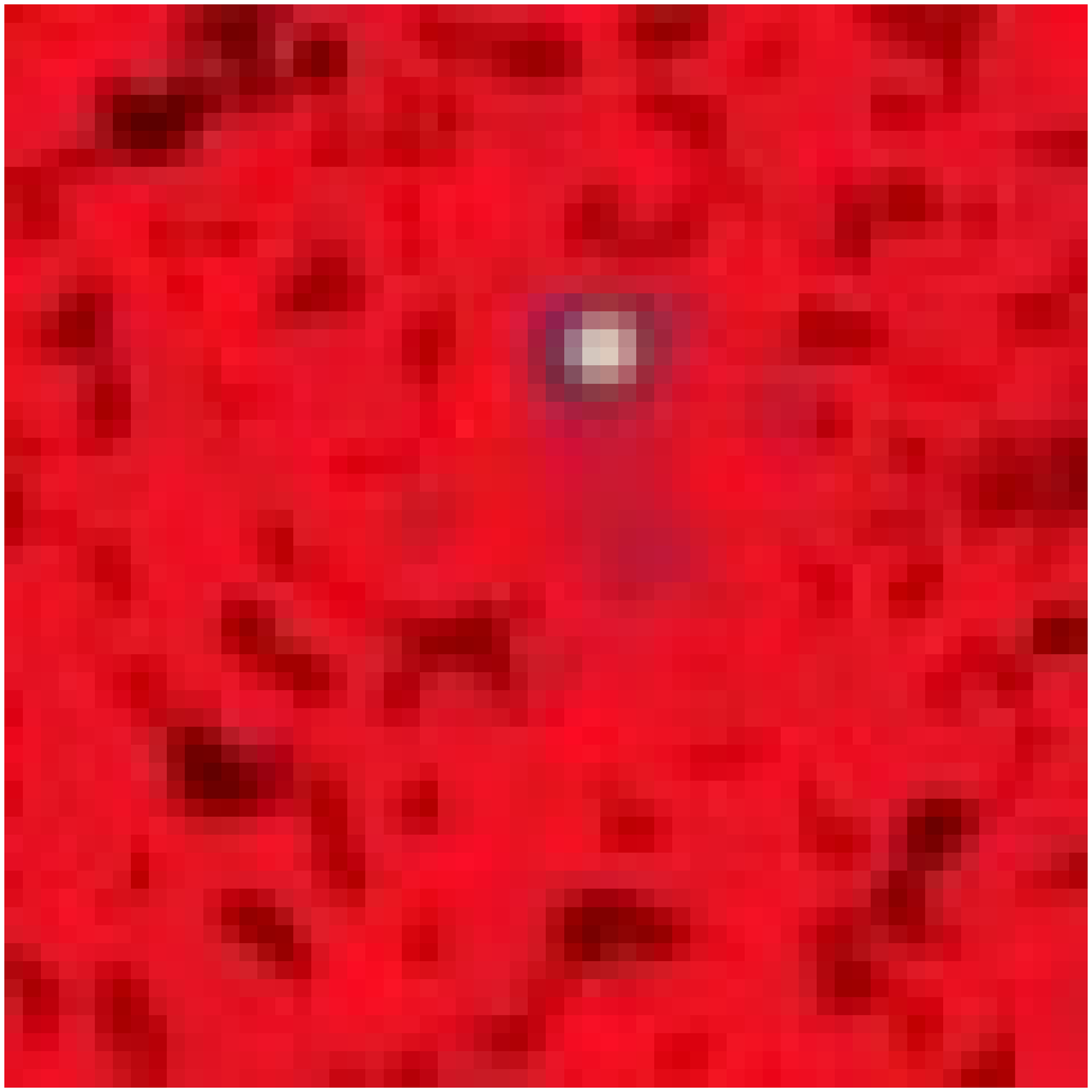}  \FGb{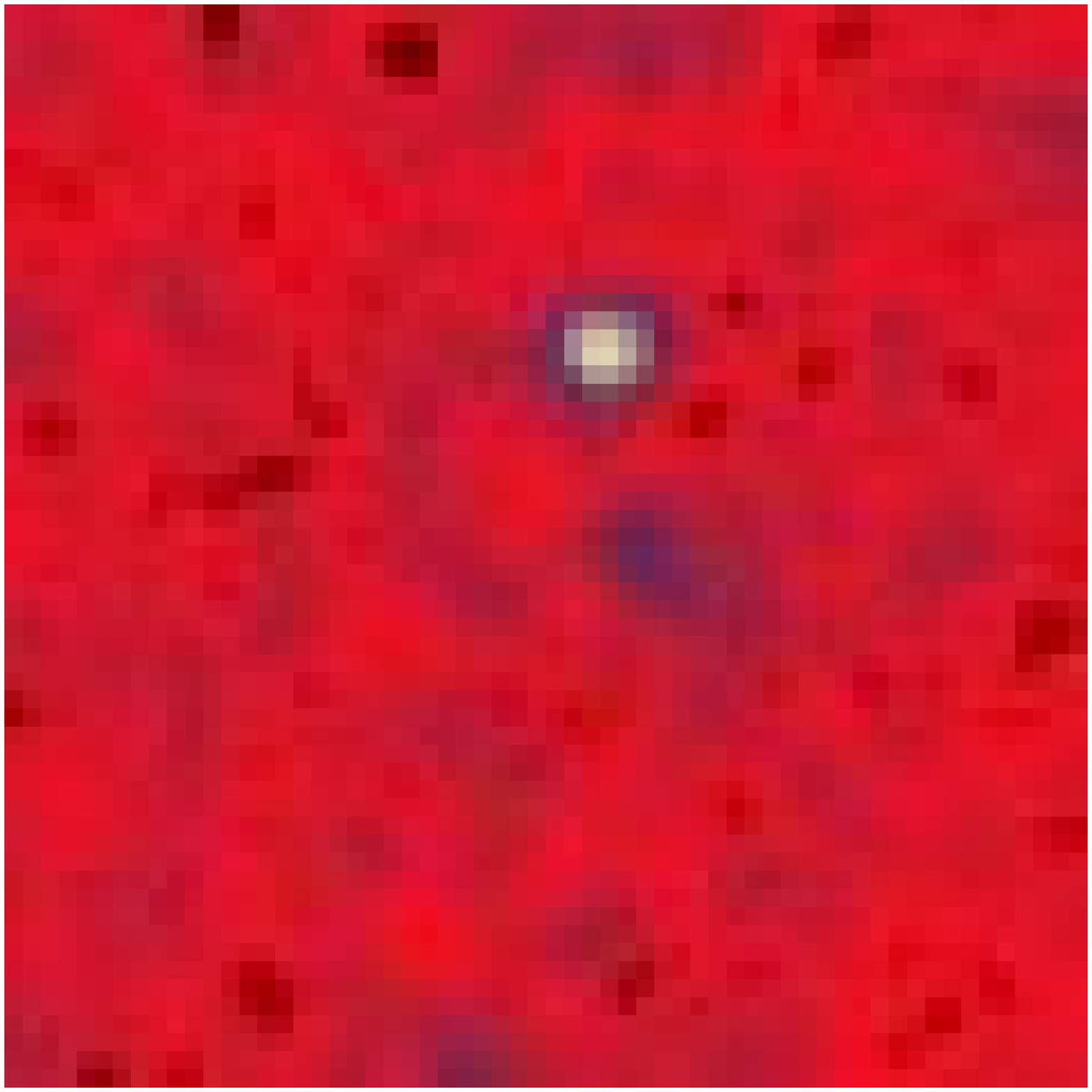}  \FGb{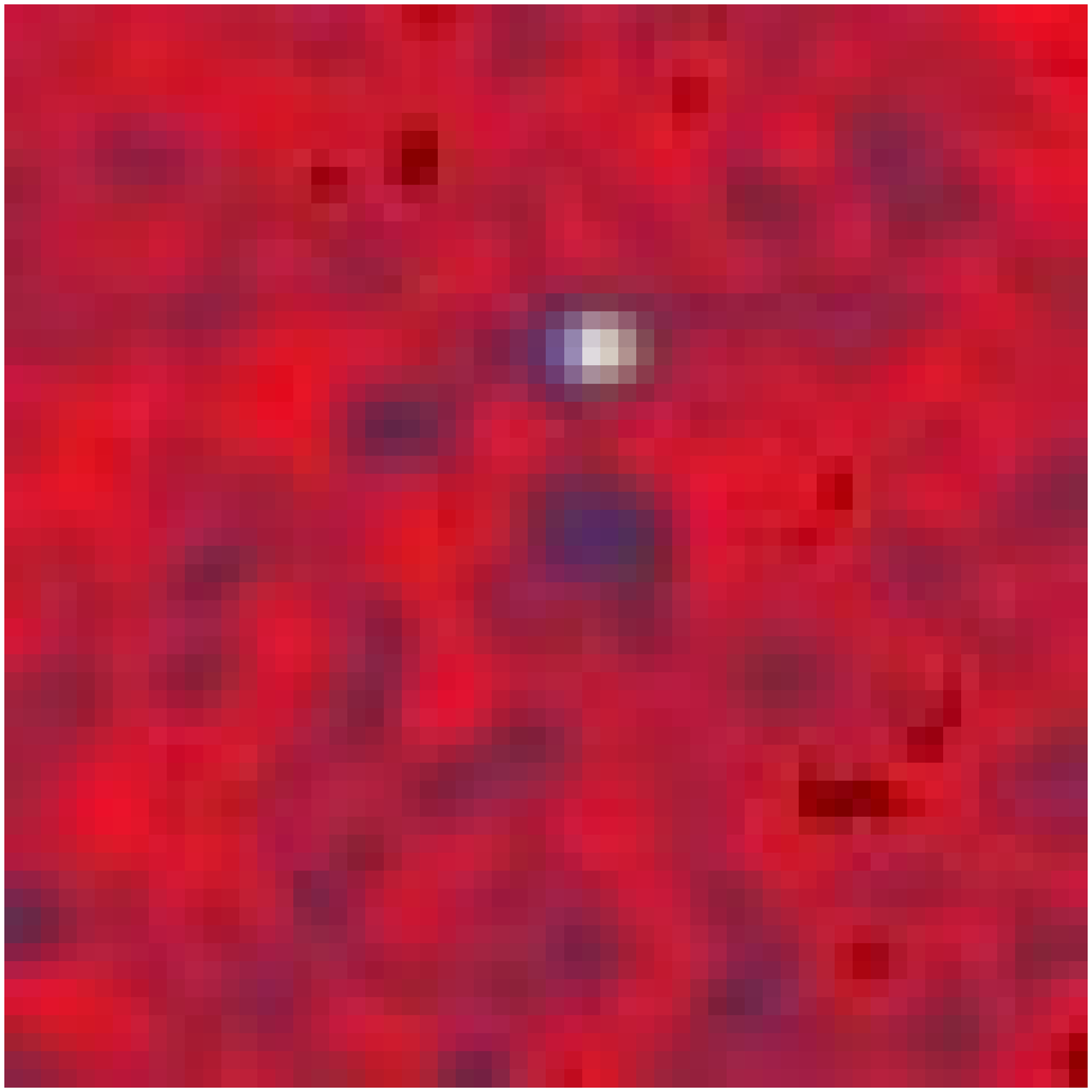}  \FGb{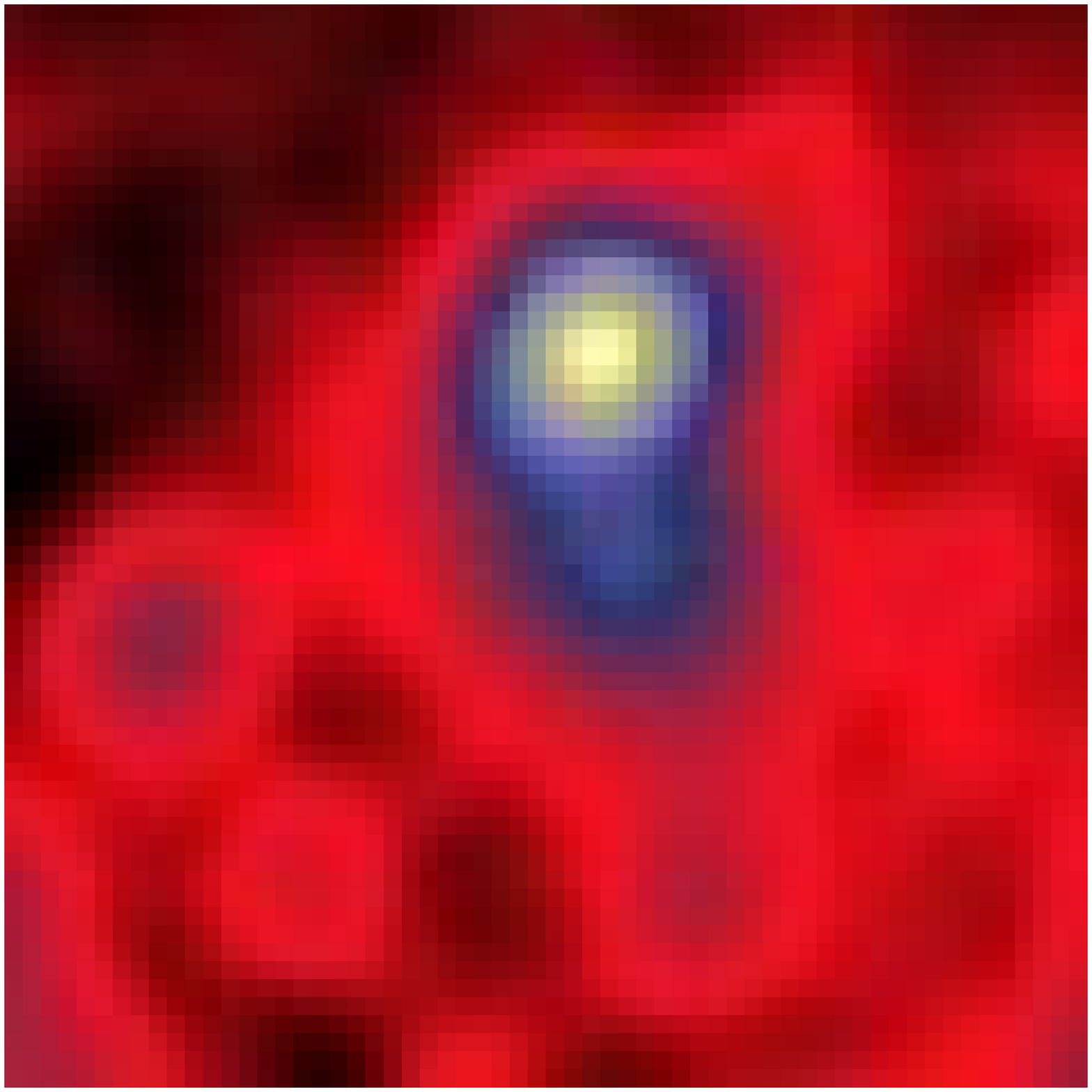}  \FGb{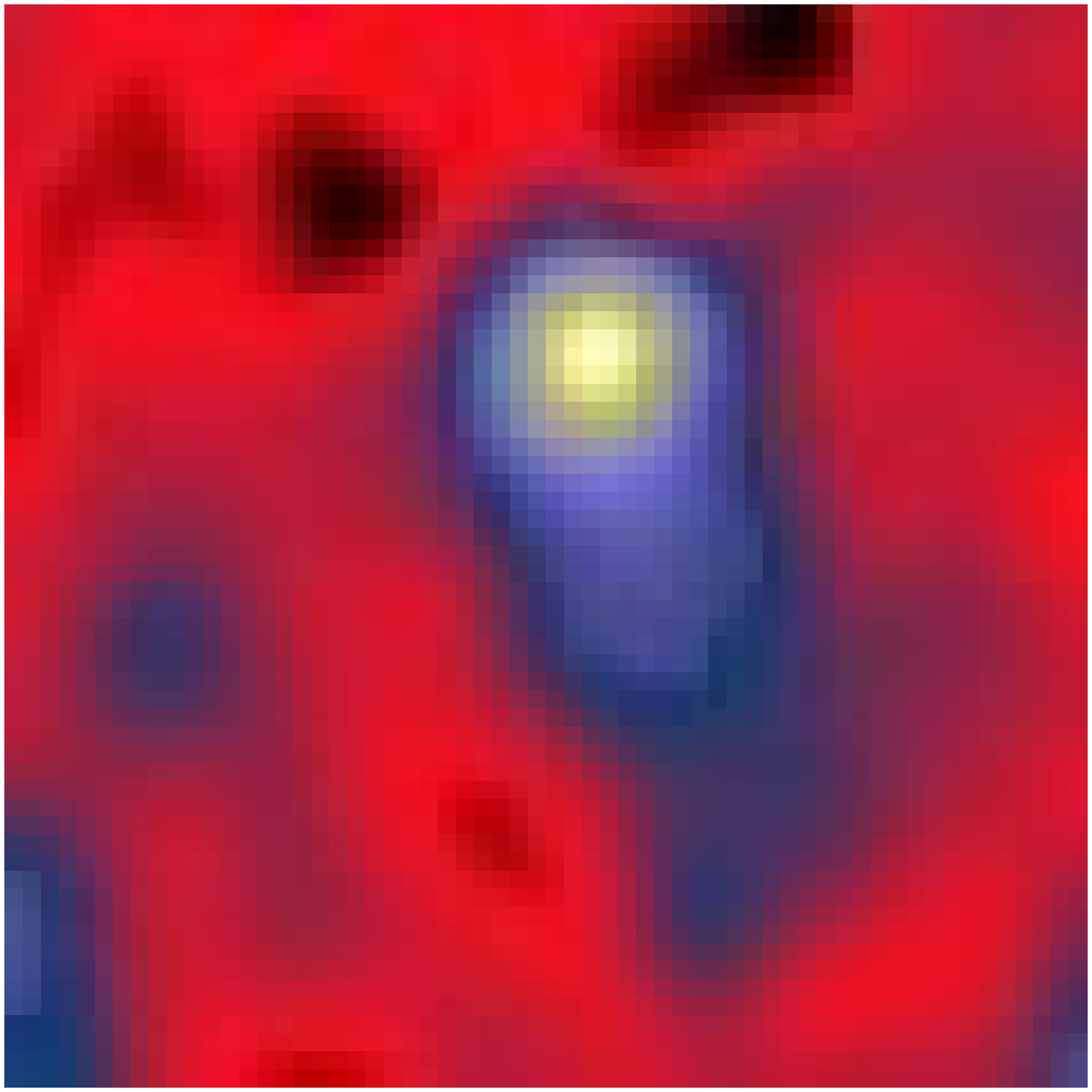}  \FGb{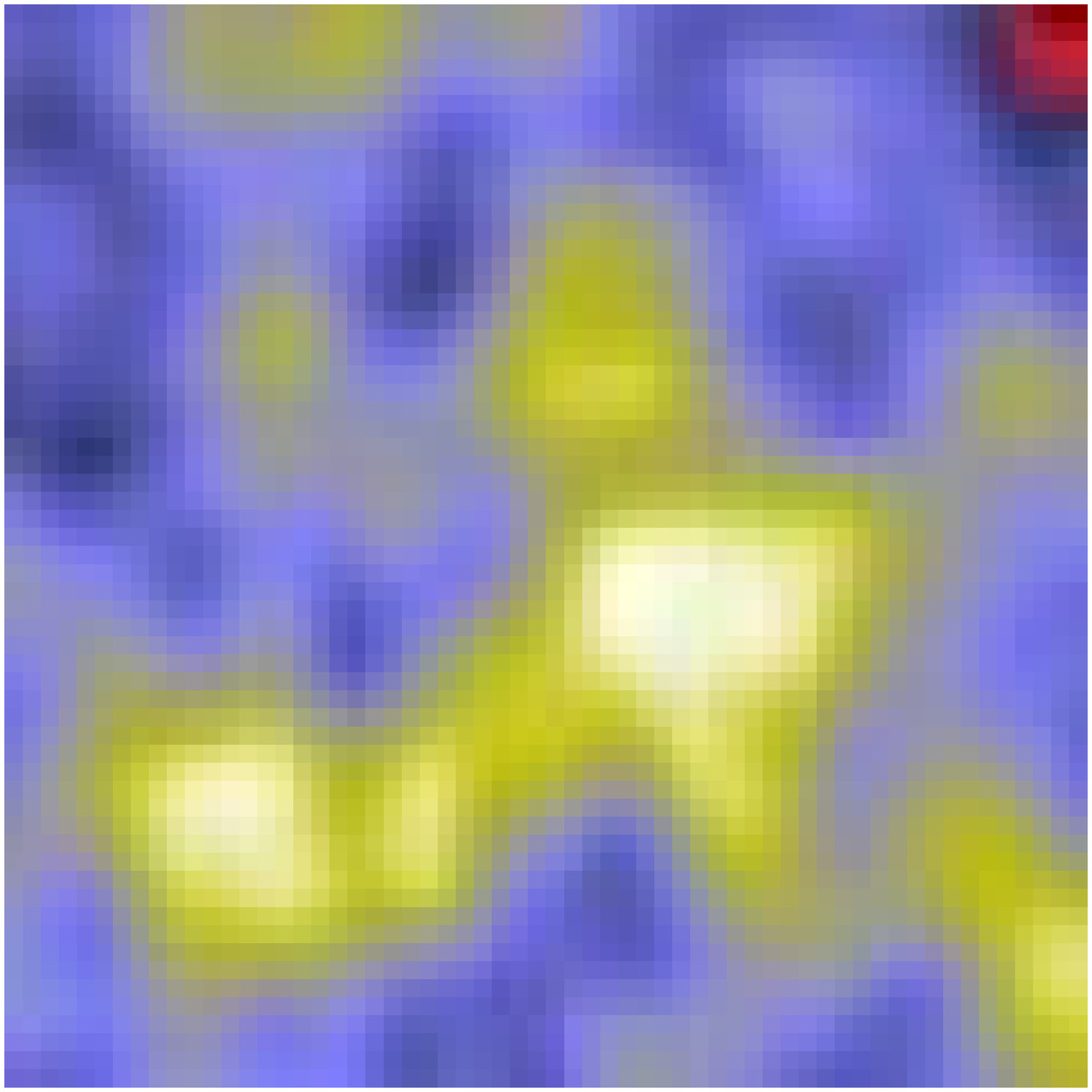}  \FGb{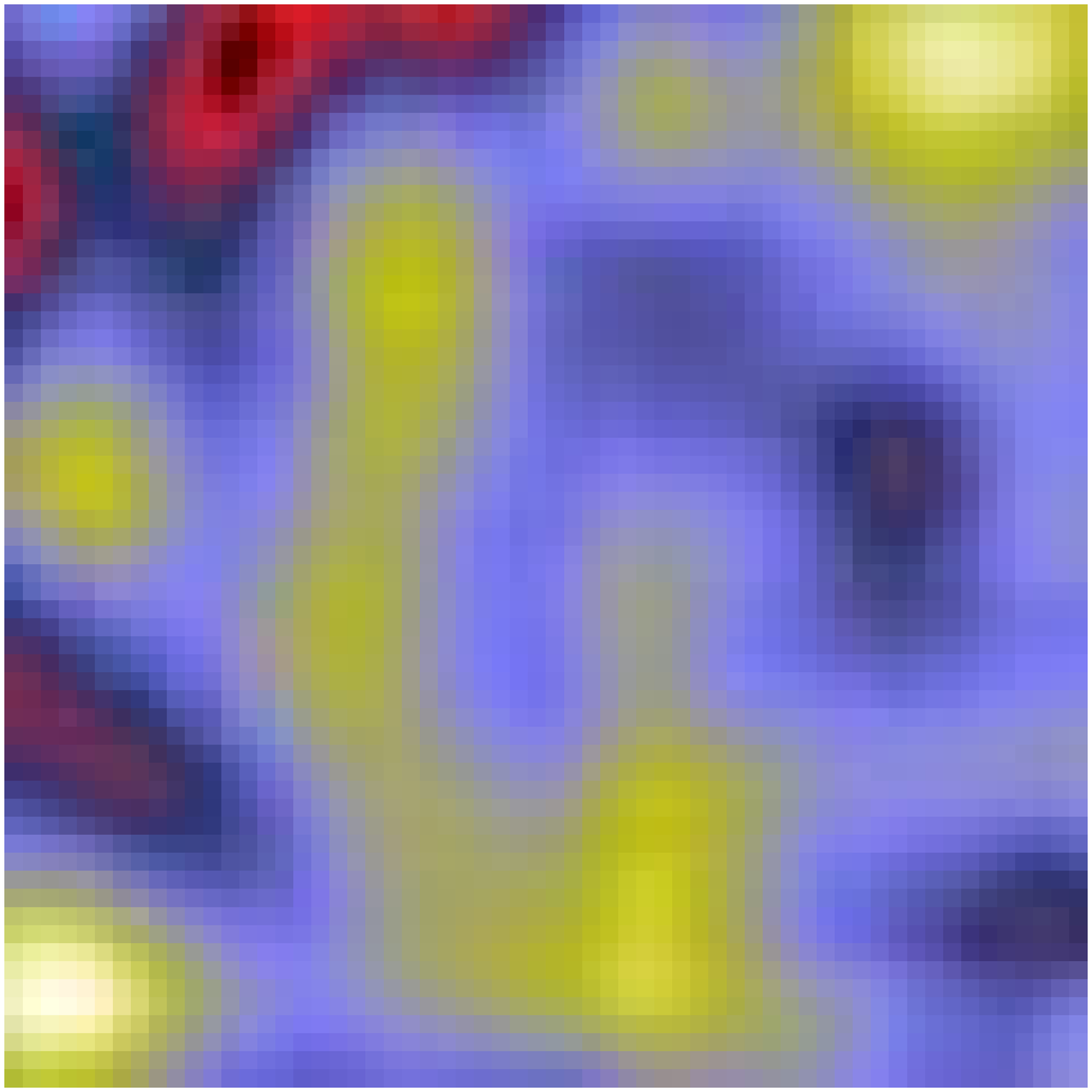}\\  {\#12  NBZ5004290}  \\
  \FGb{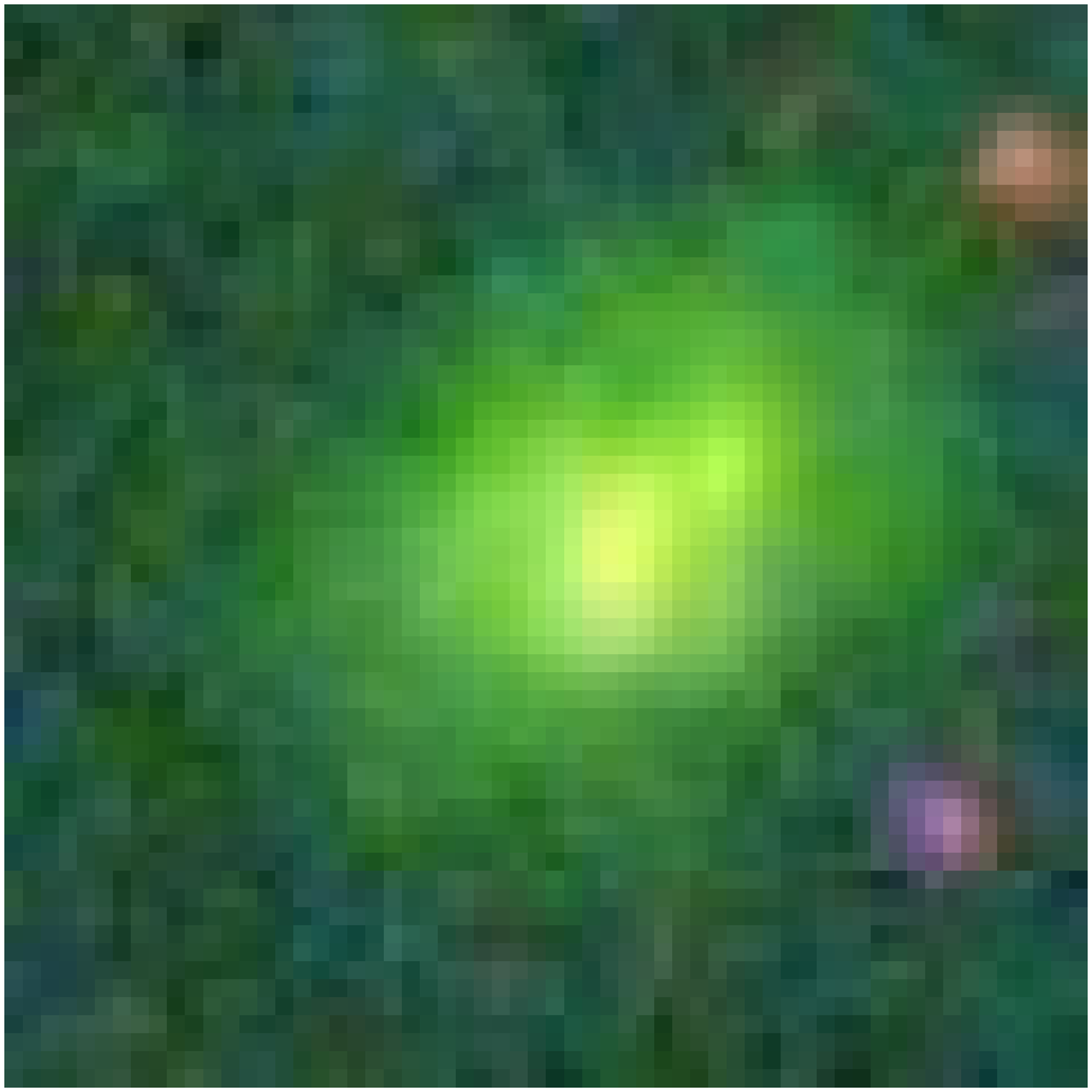} \FGb{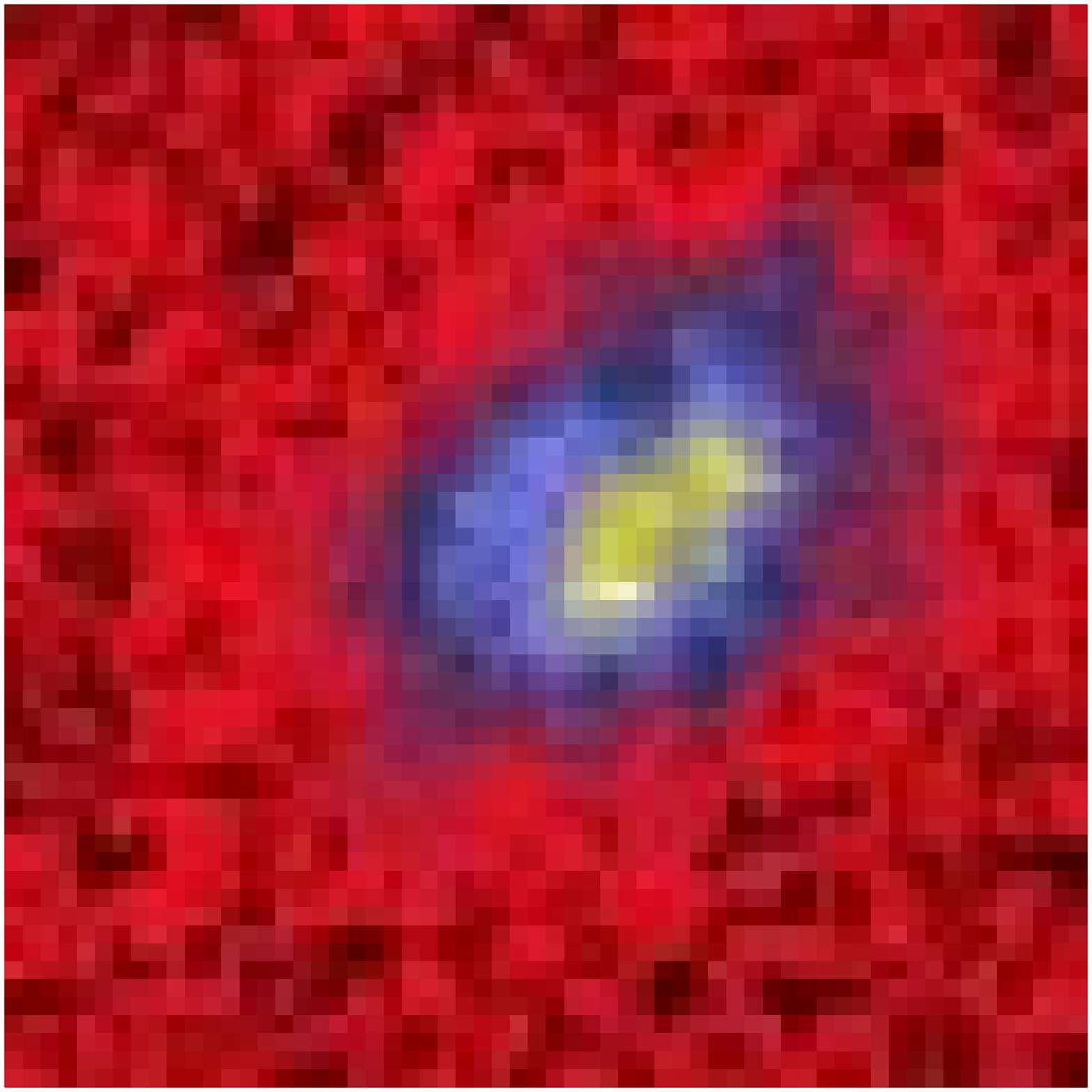}   \FGb{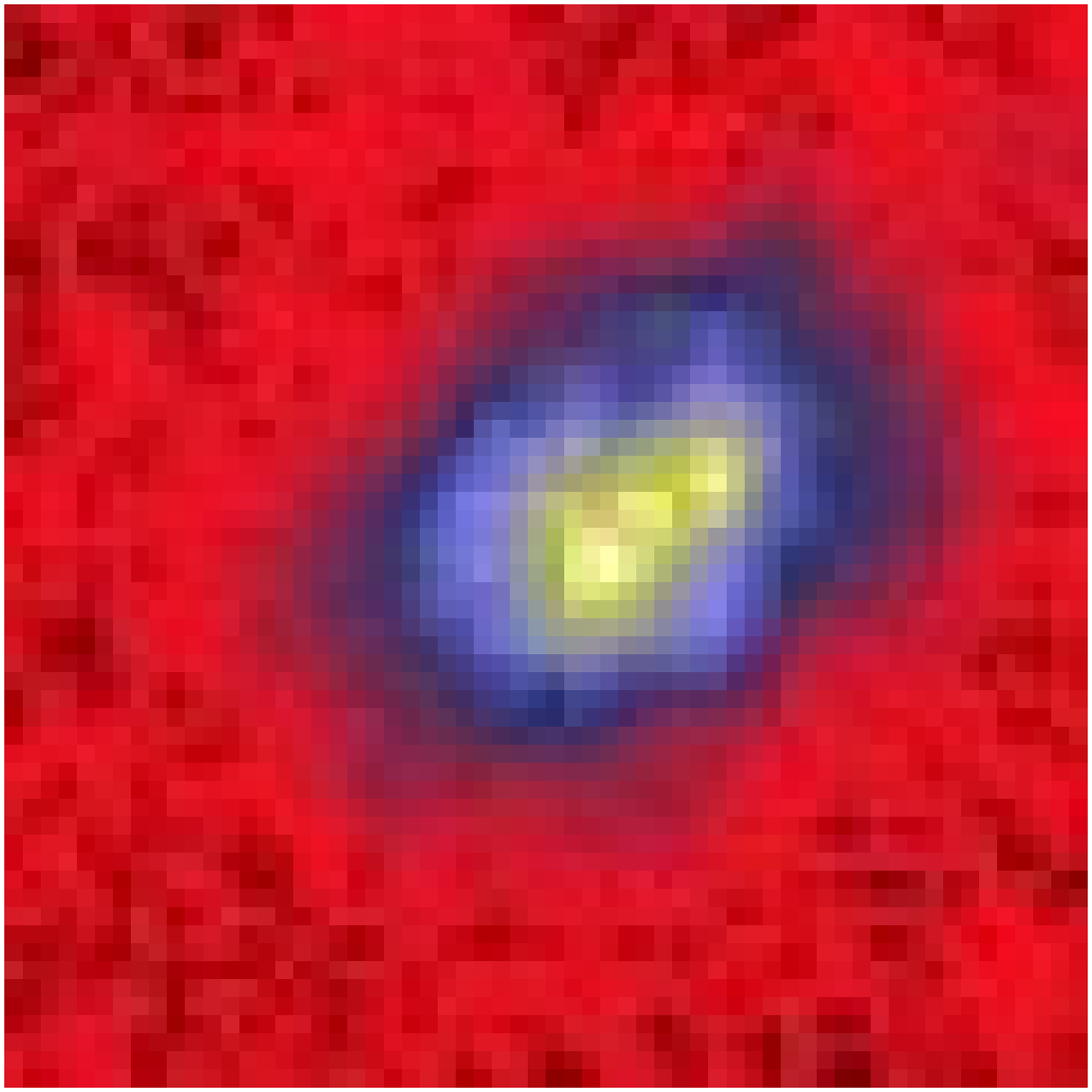}  \FGb{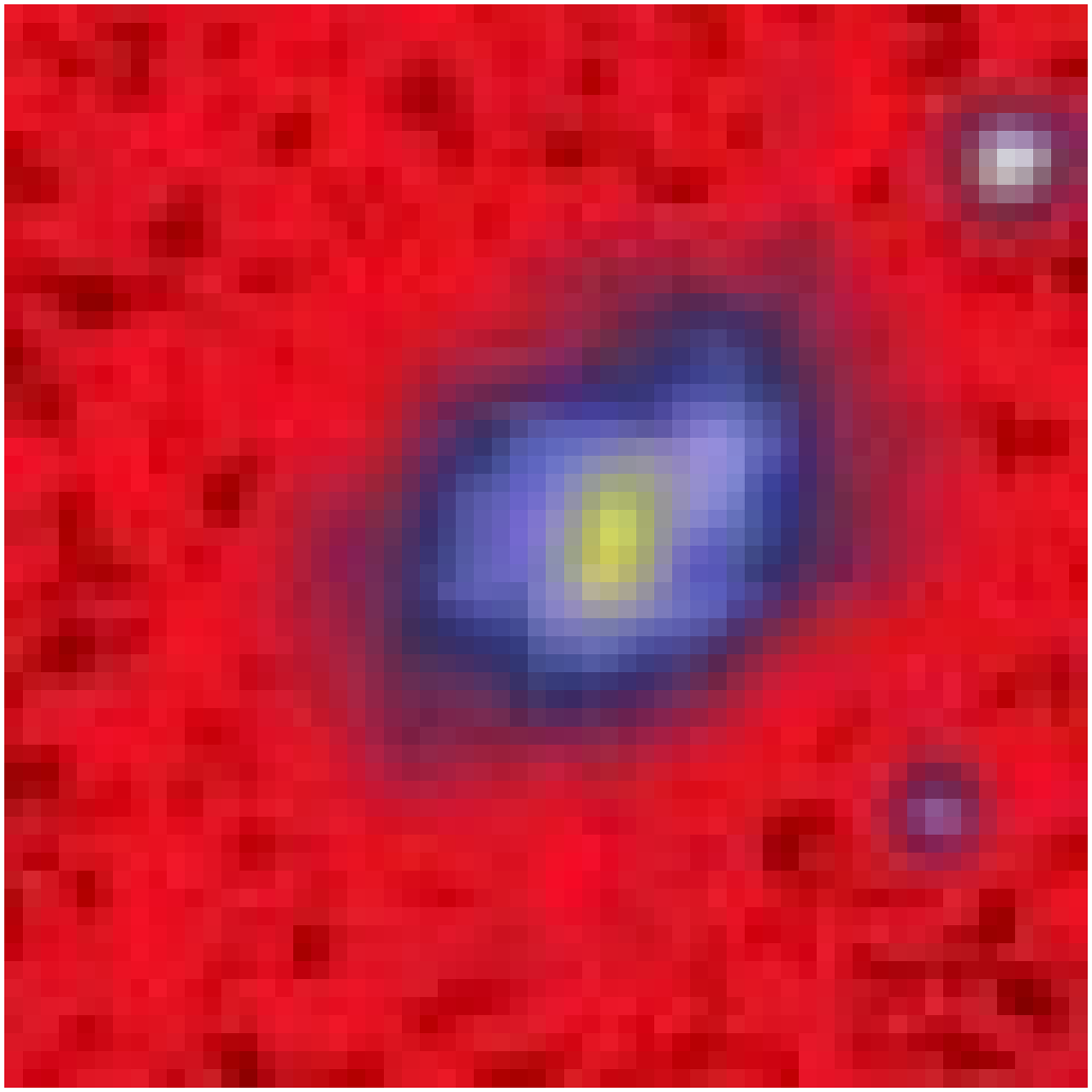}  \FGb{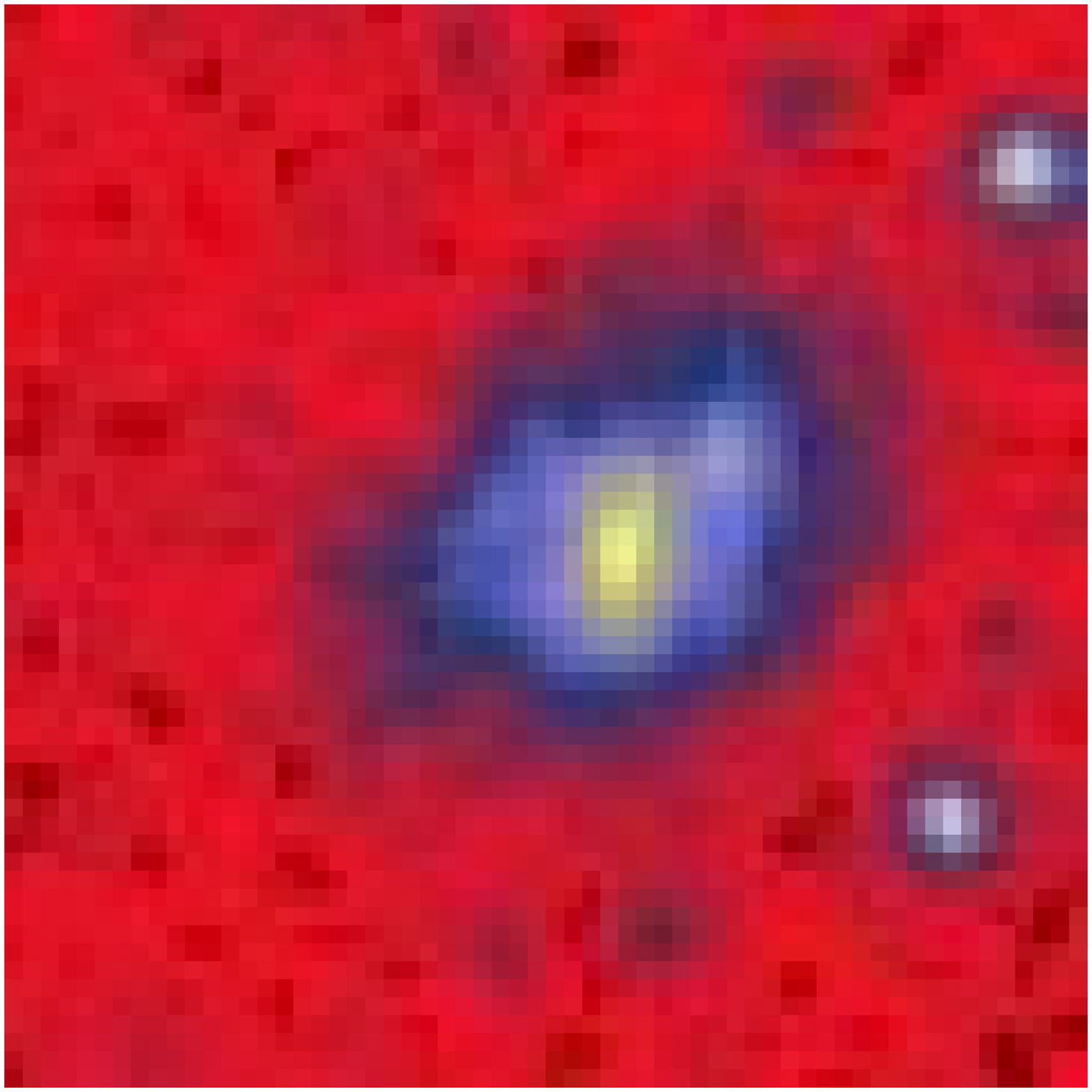}  \FGb{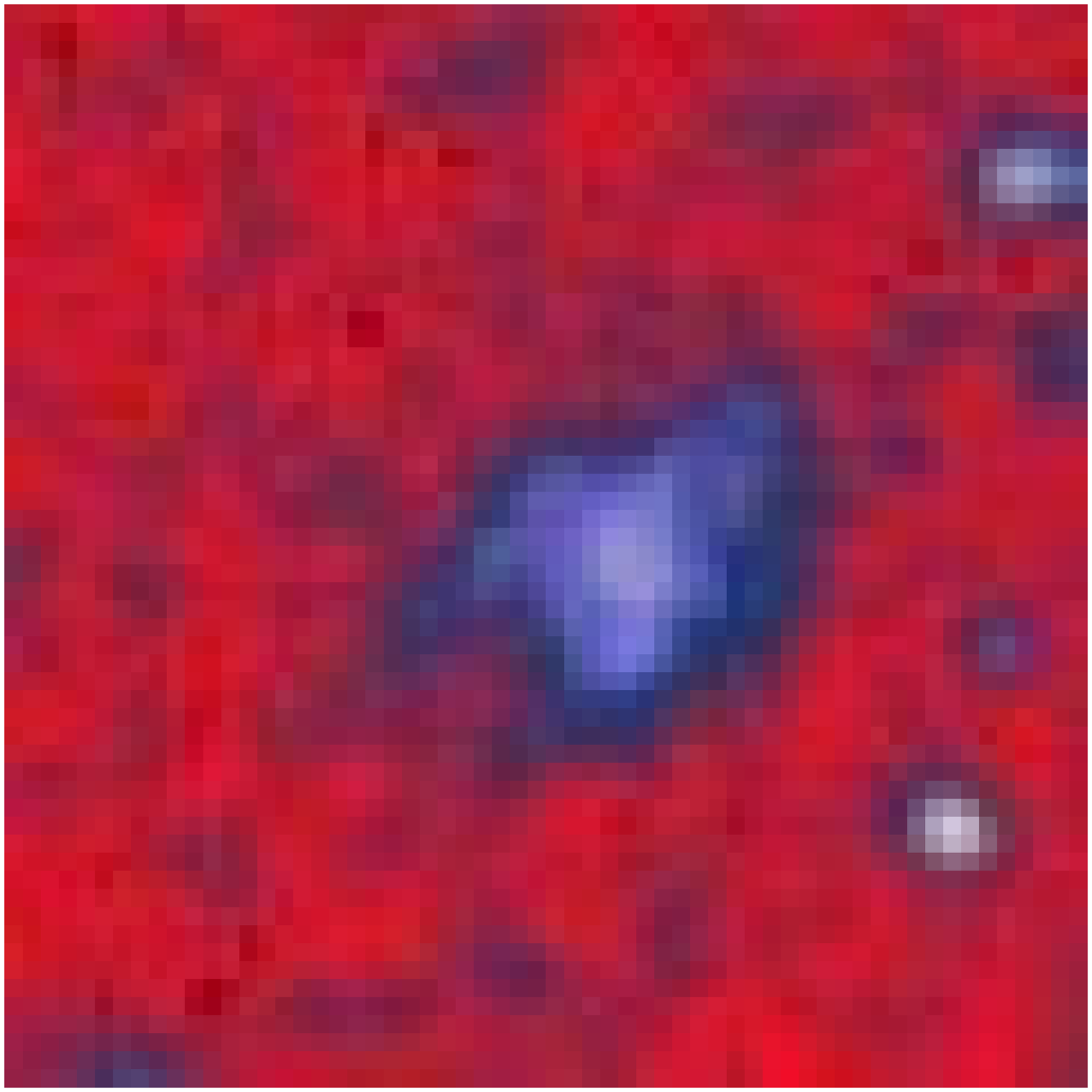}  \FGb{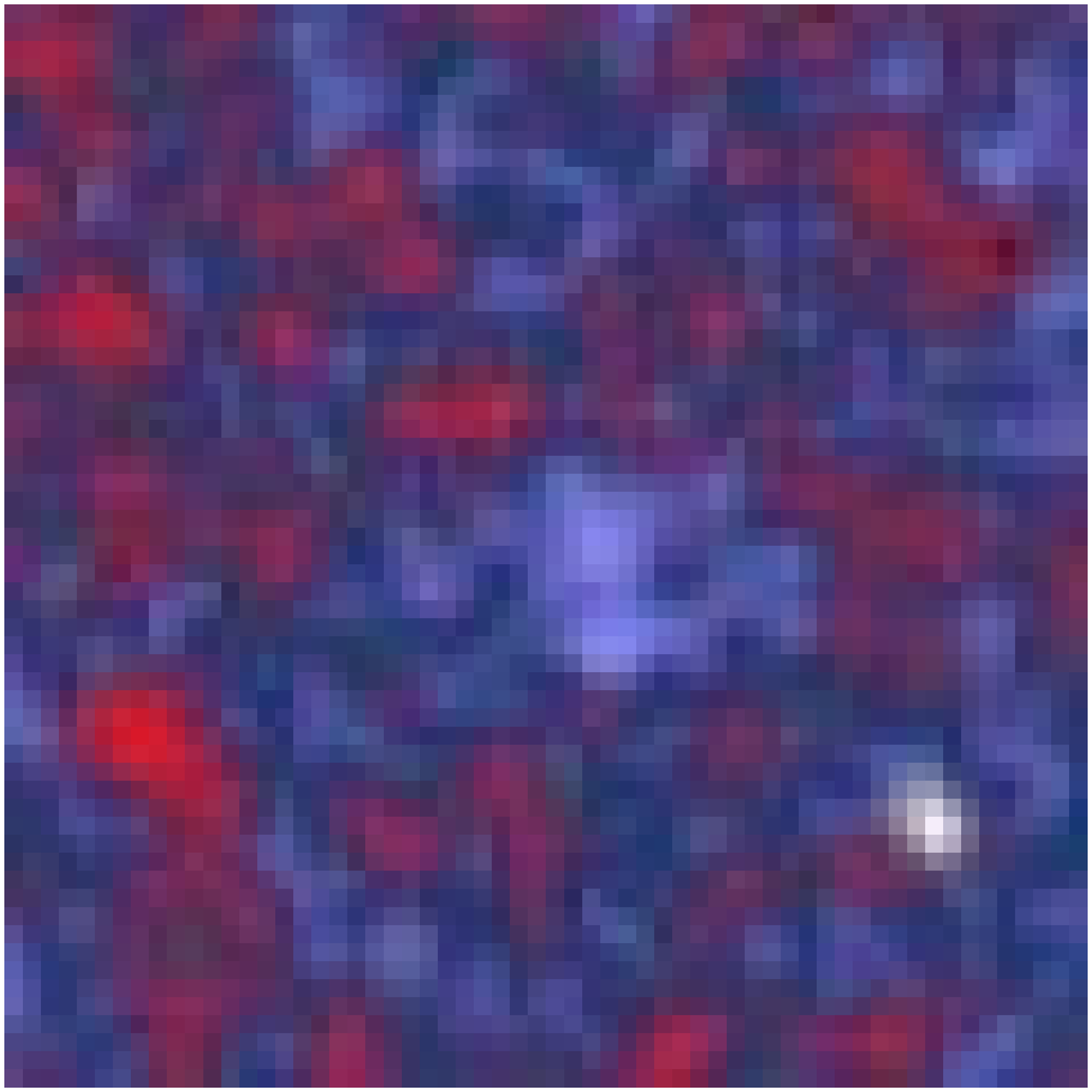}  \FGb{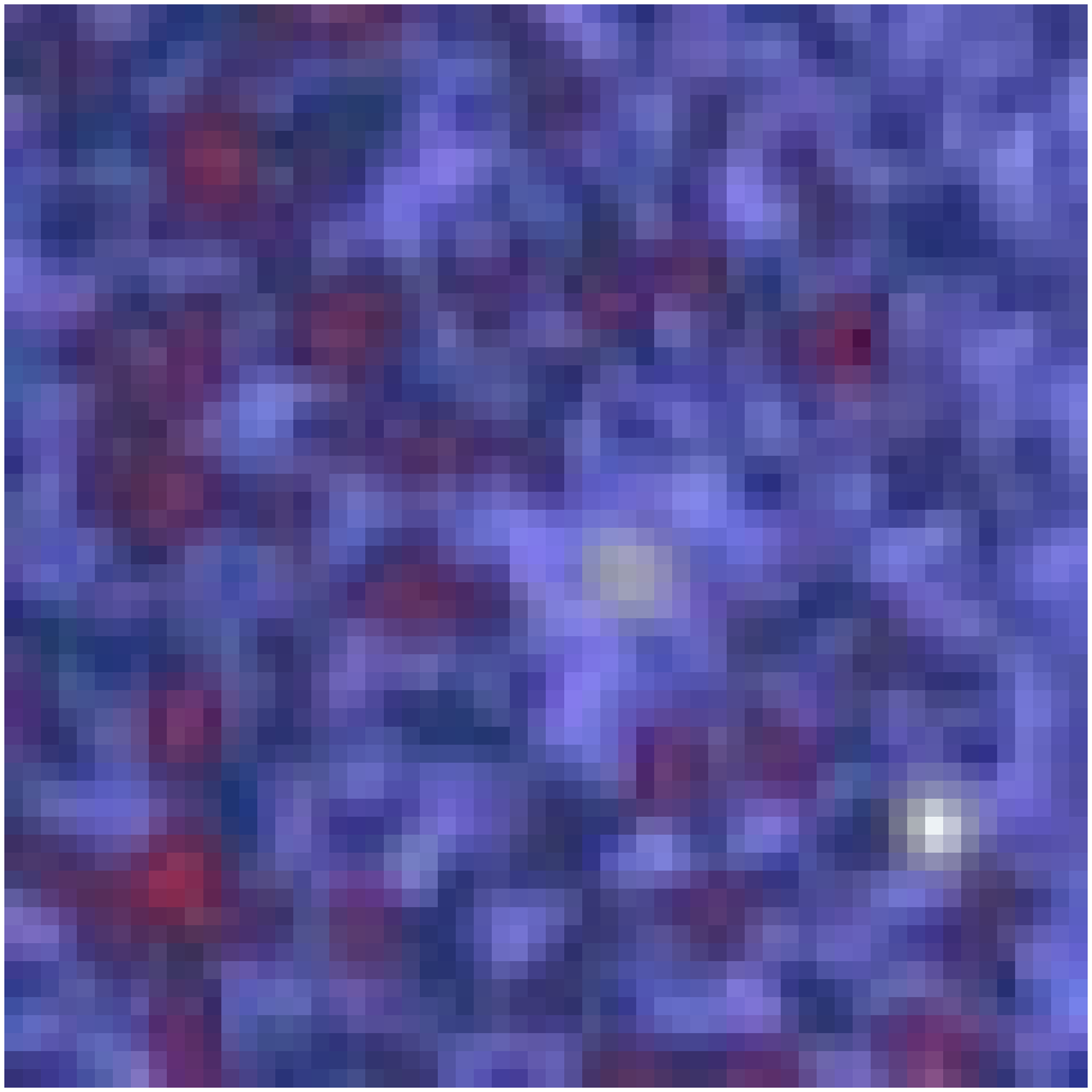}  \FGb{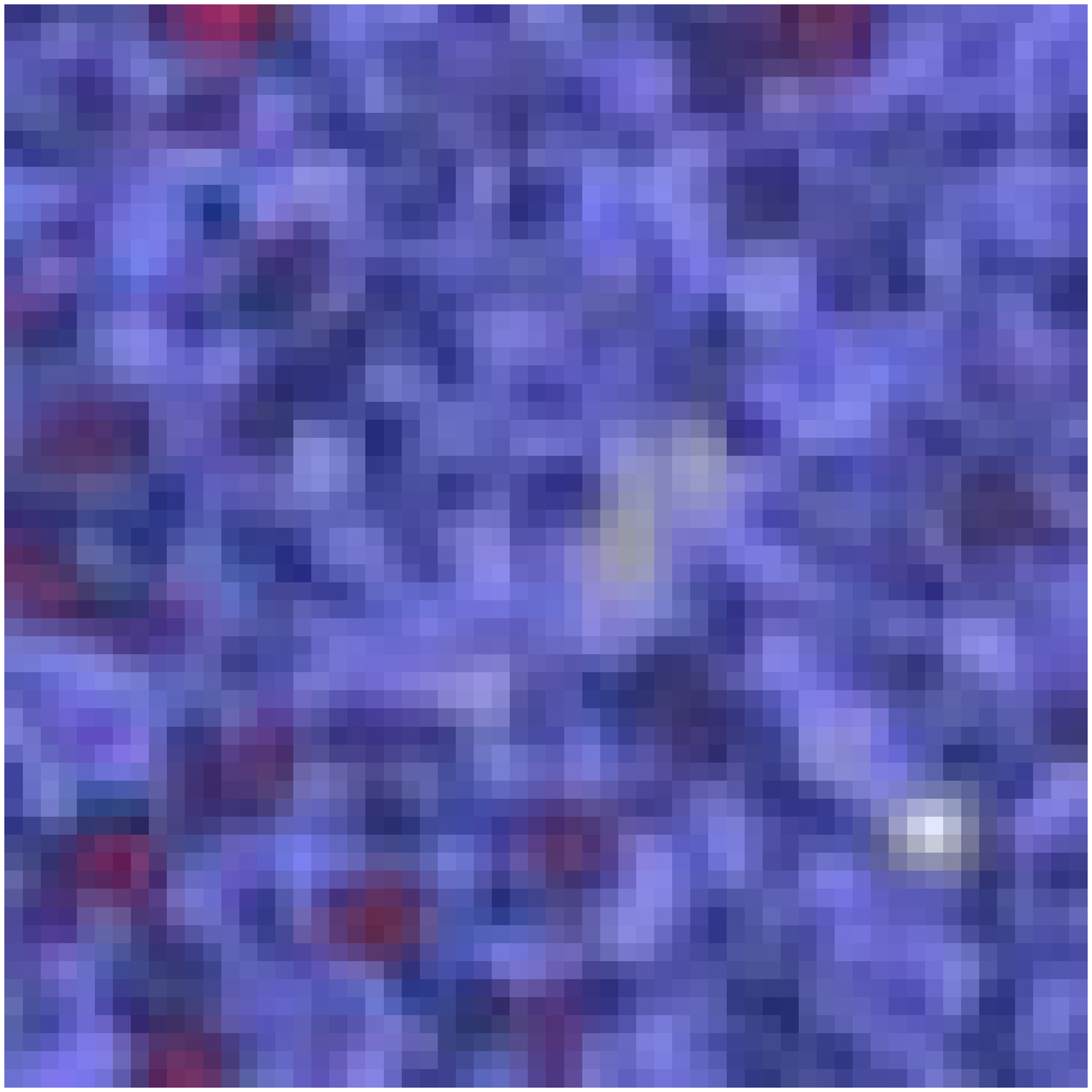}  \FGb{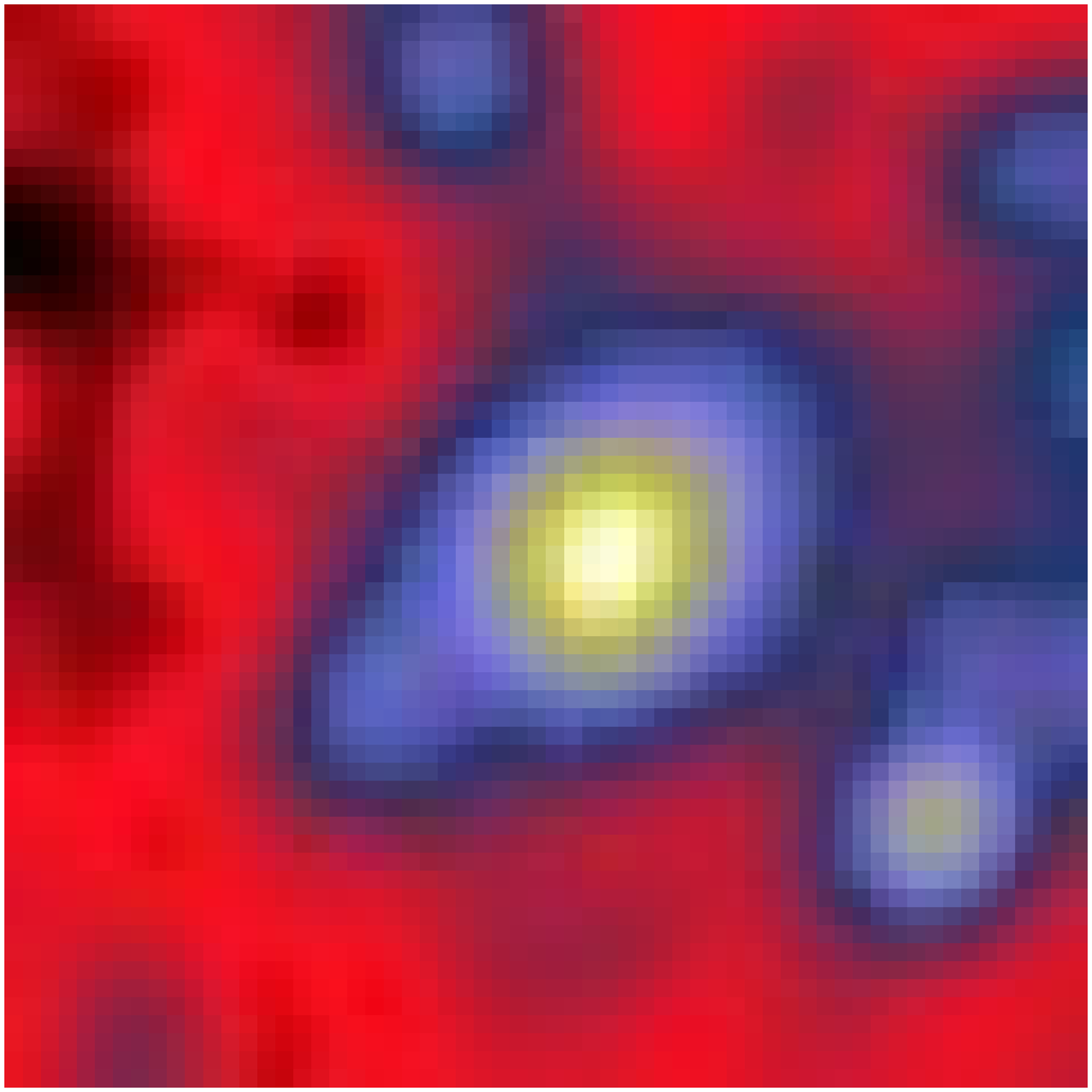}  \FGb{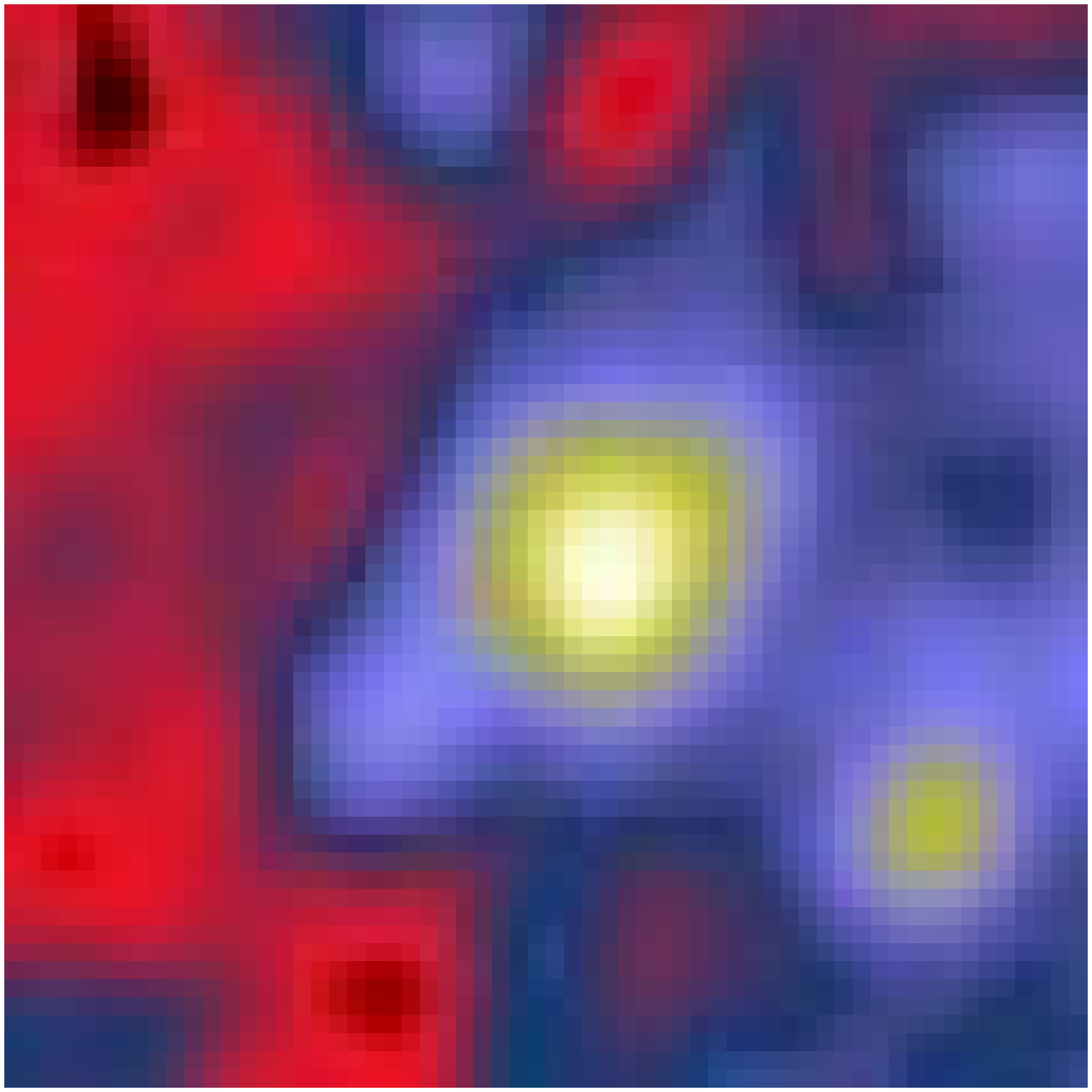}  \FGb{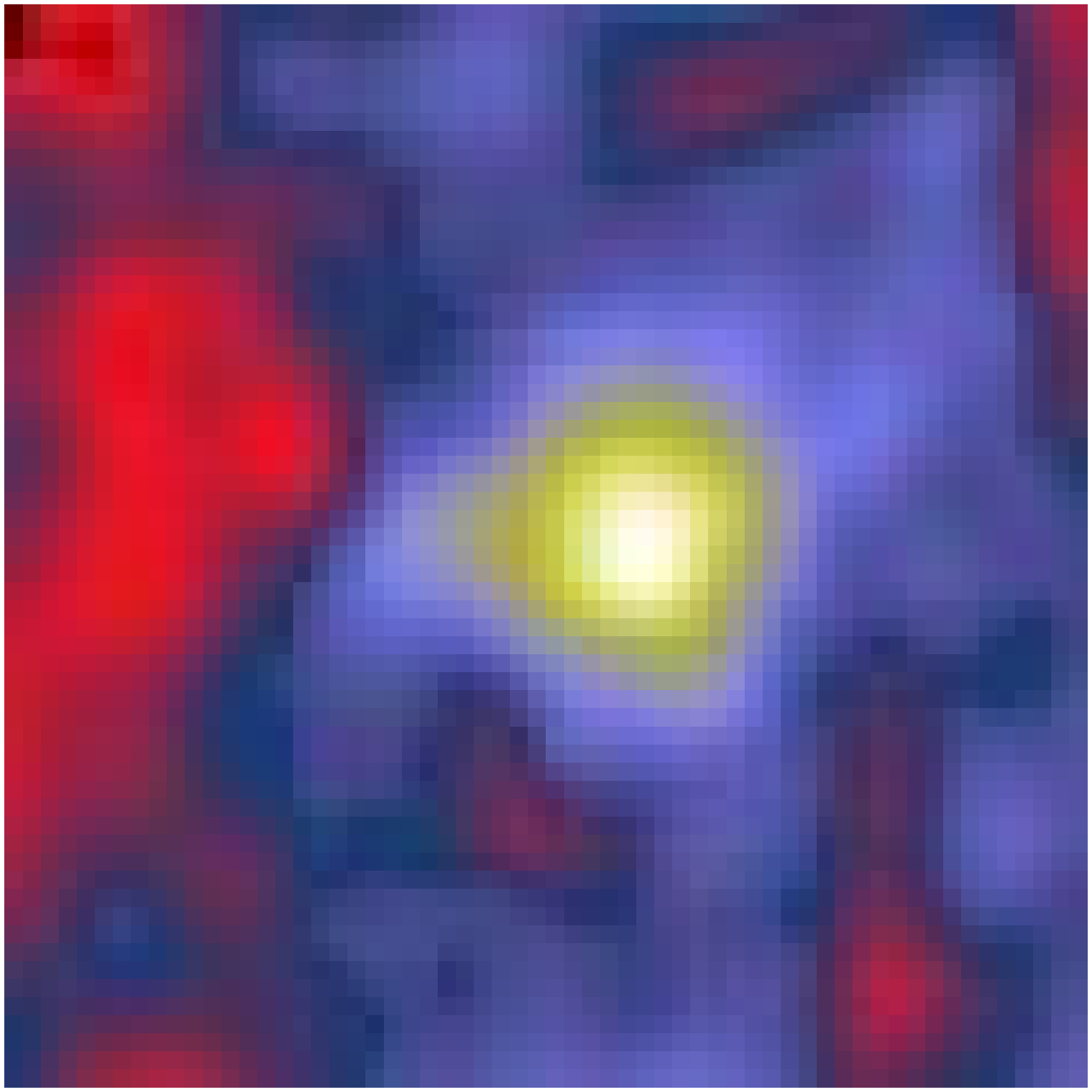}  \FGb{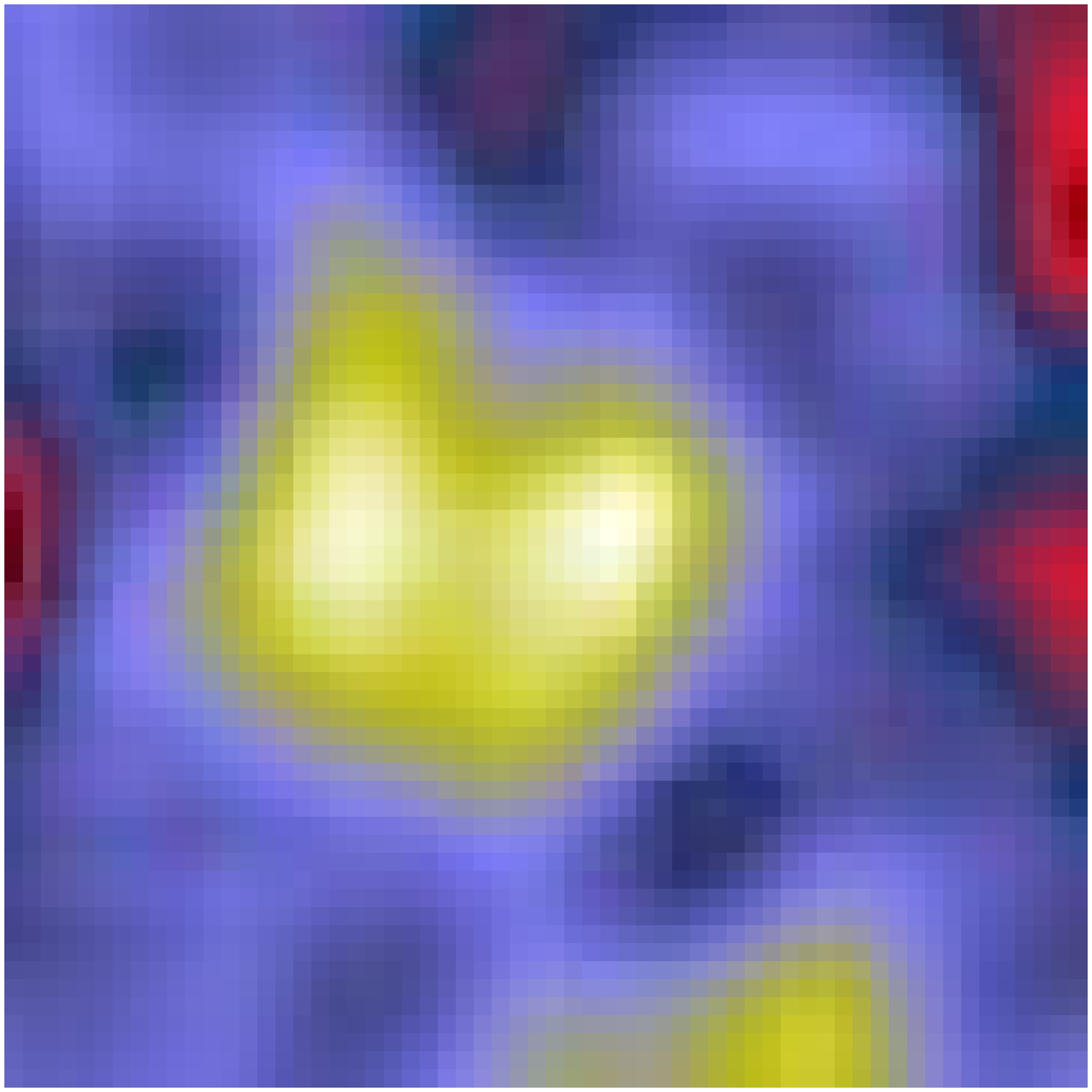}\\  {\#13  NCIA023873}  \\
\caption{Visible Galaxies within 40\min\ of NGC~891 }\label{fig:allgal}
\end{figure}

\clearpage
\begin{figure}[ht]
  \HDb{rgb}     \HDb{FUV}     \HDb{NUV}     \HDb{B}       \HDb{R}       \HDb{IR}      \HDb{J}       \HDb{H}       \HDb{K}       \HDb{W1}      \HDb{W2}      \HDb{W3}      \HDb{W4}     \\
  \FGb{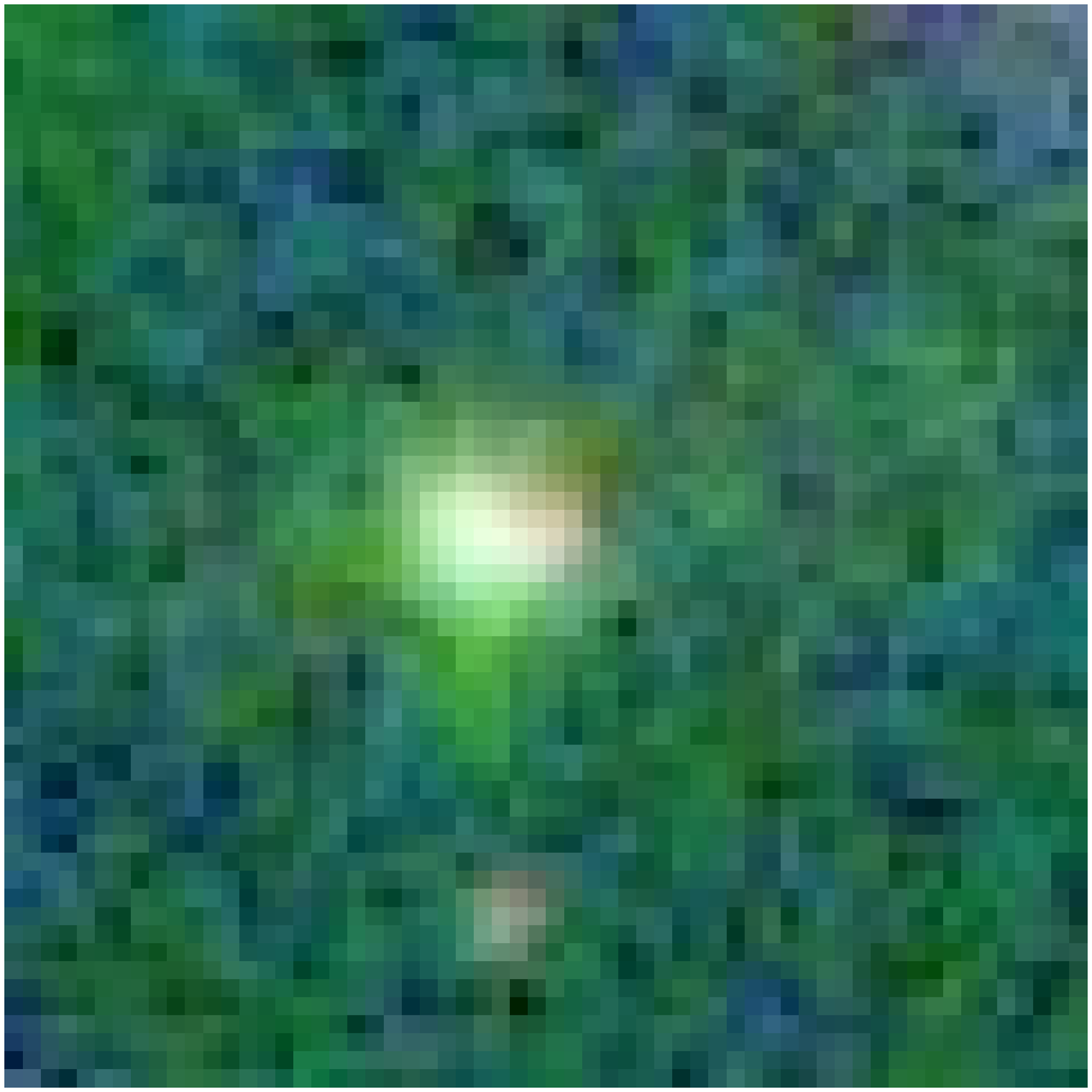}  \FGb{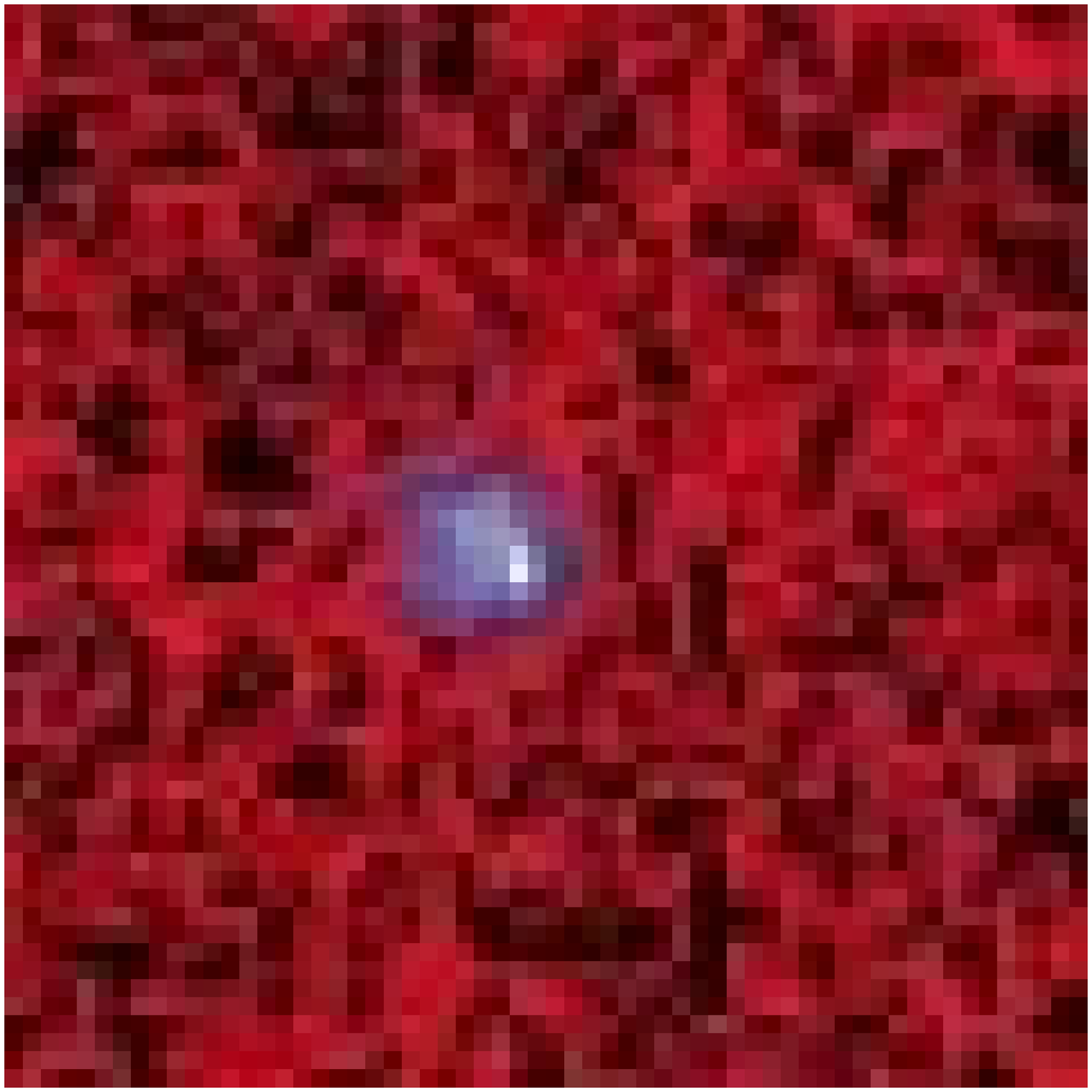}  \FGb{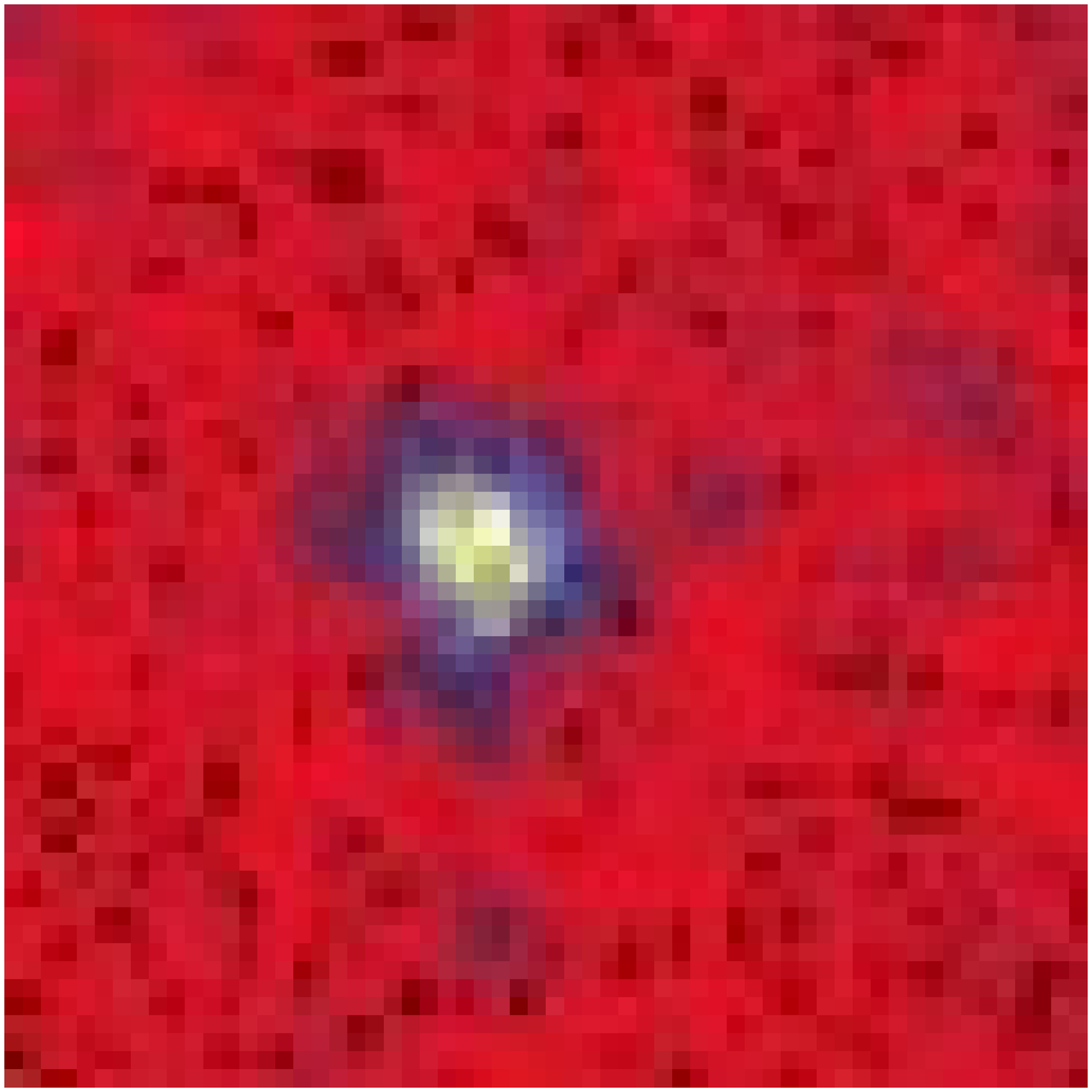}  \FGb{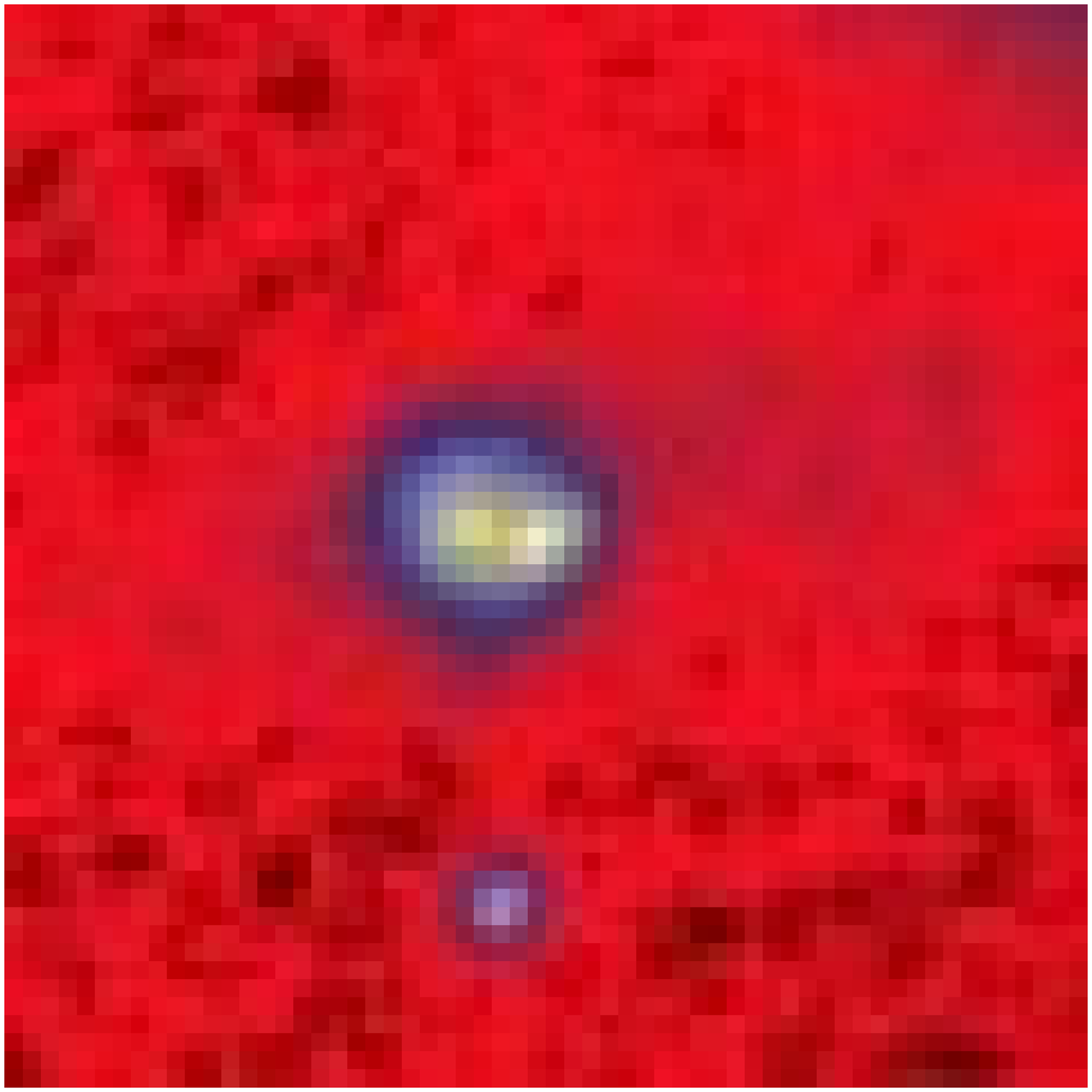}  \FGb{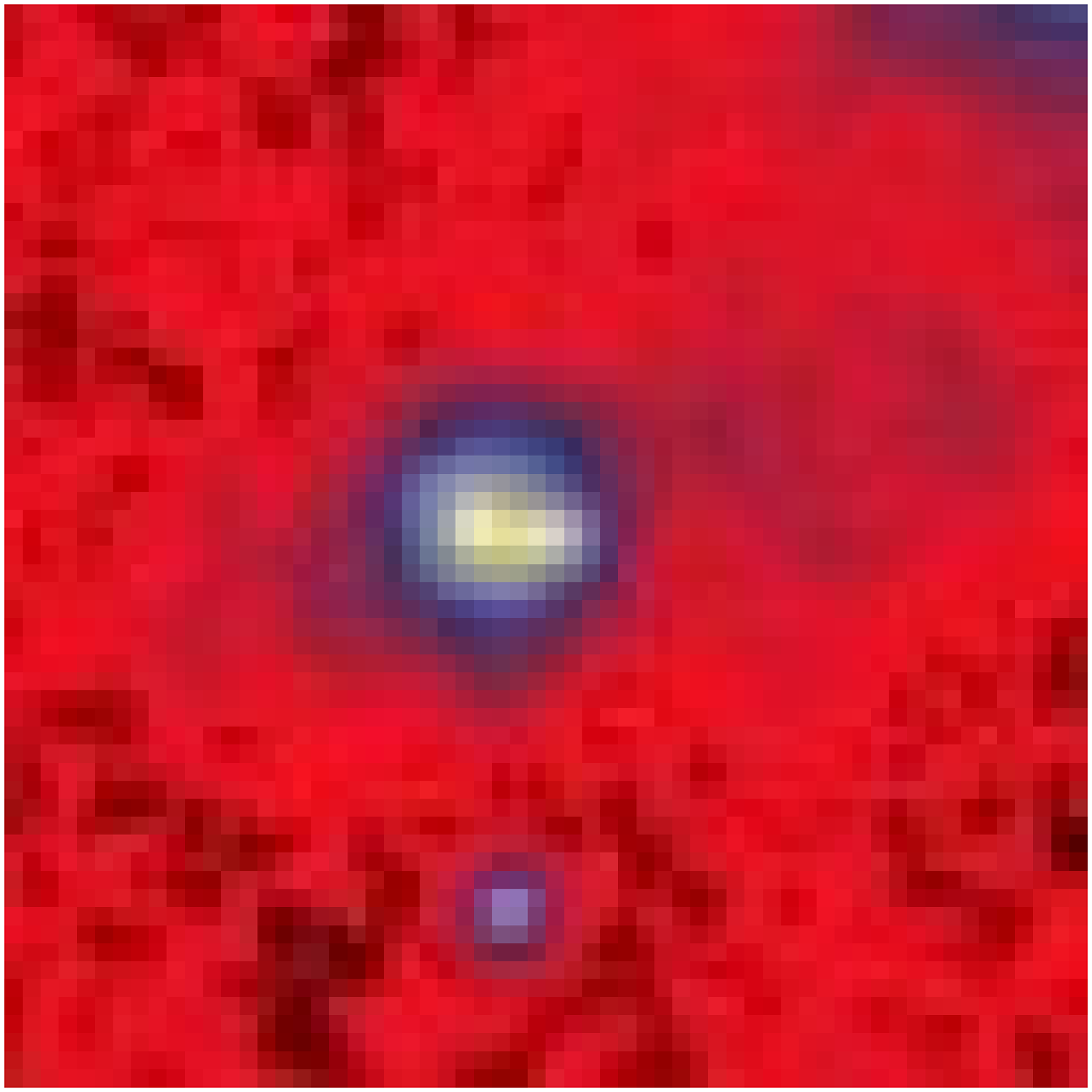}  \FGb{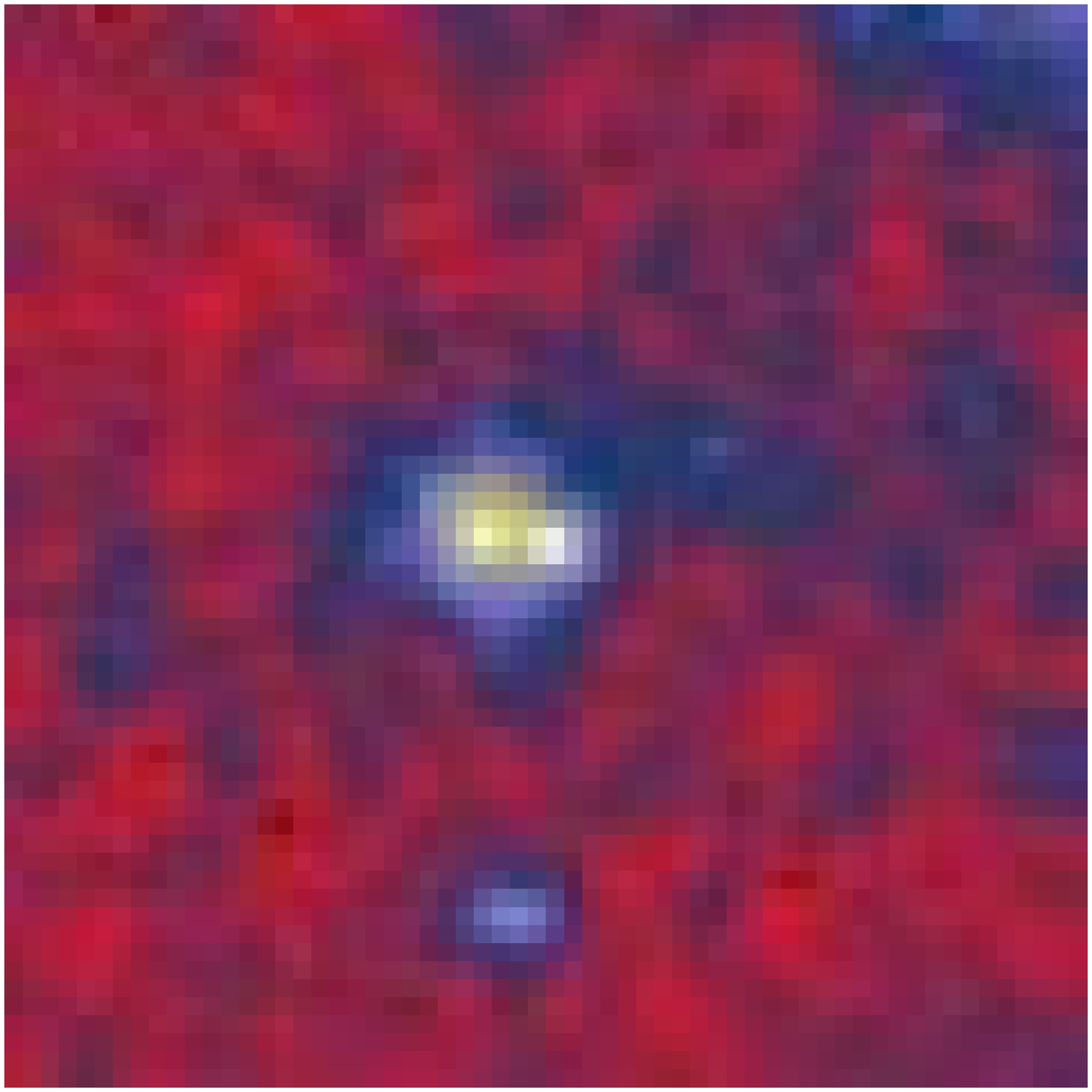}  \FGb{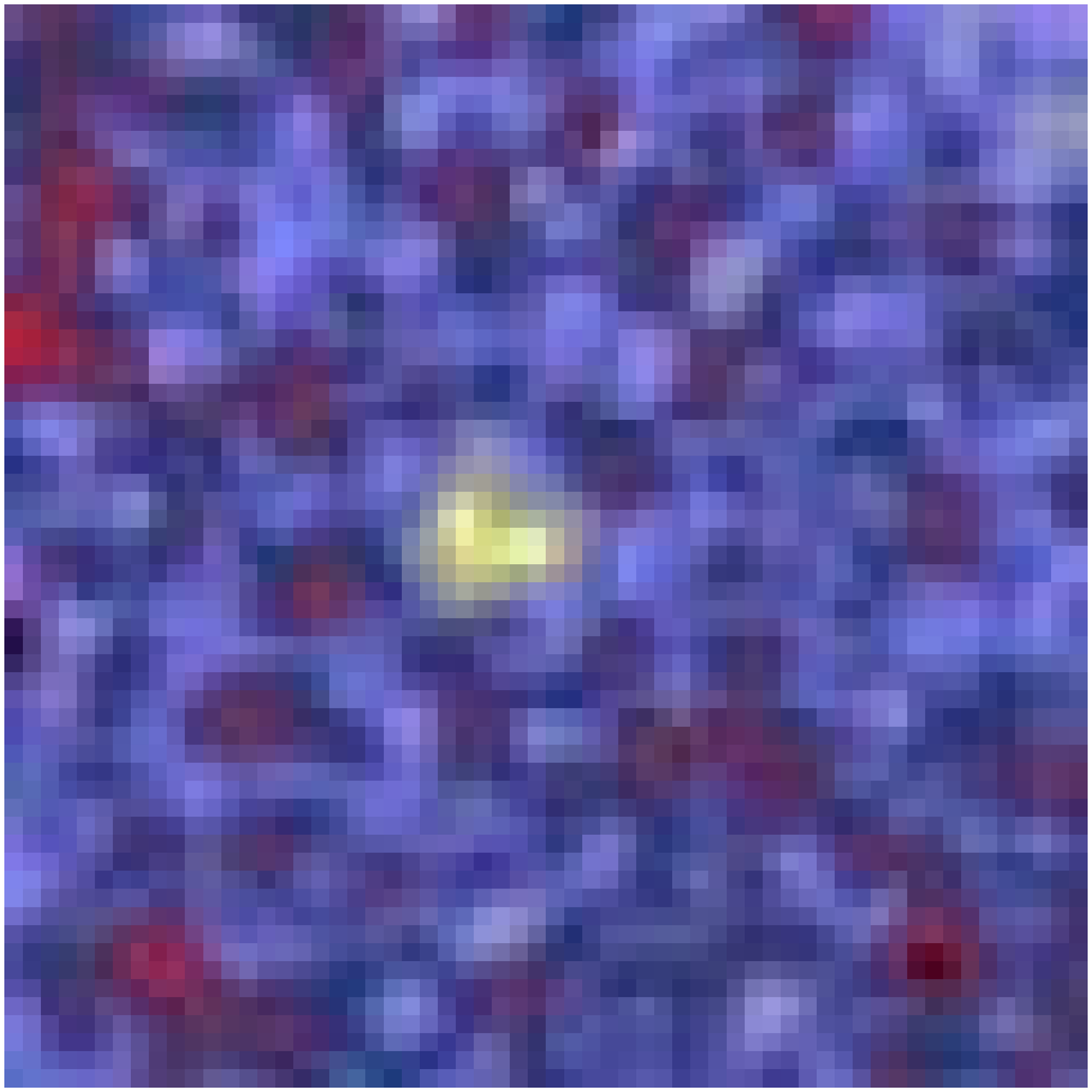}  \FGb{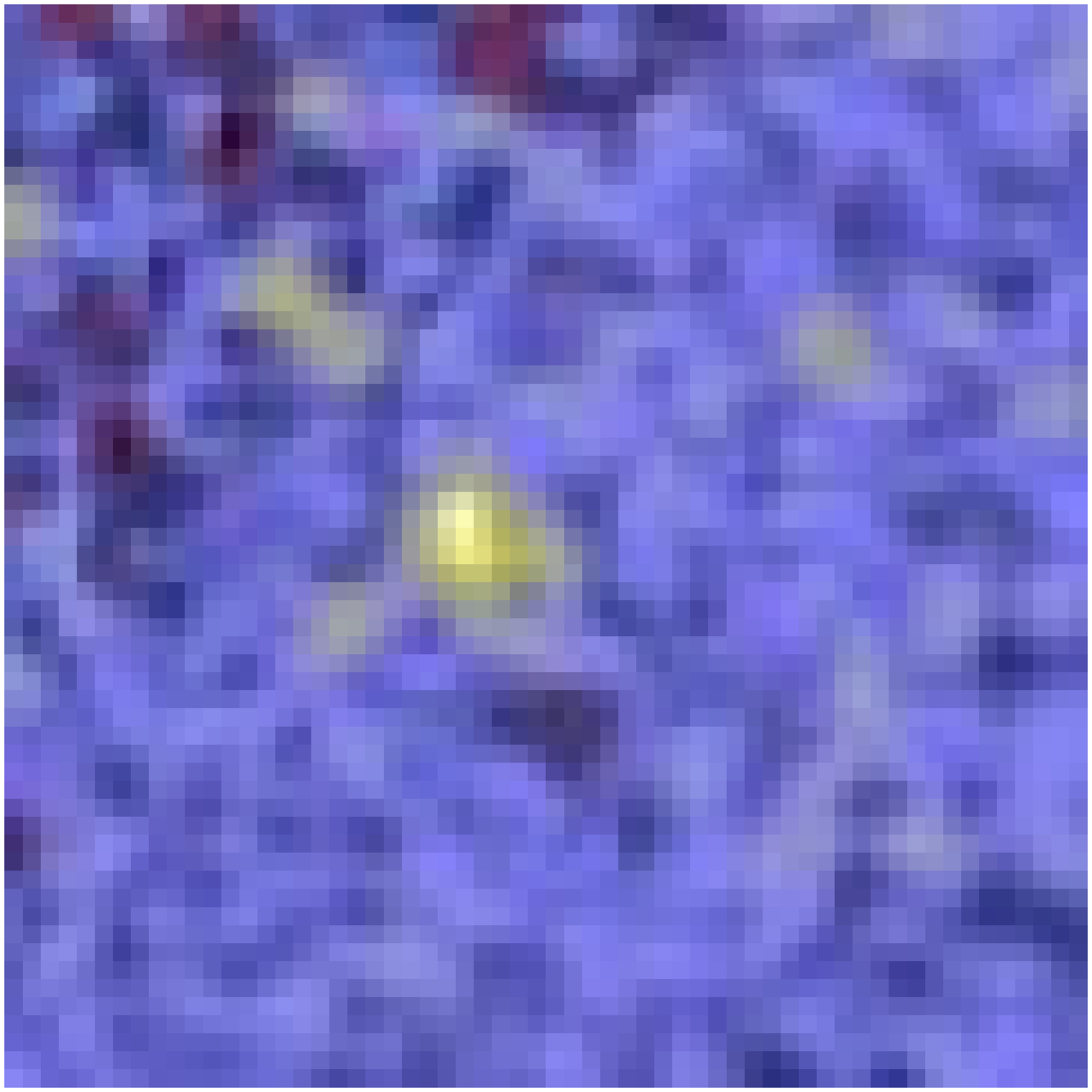}  \FGb{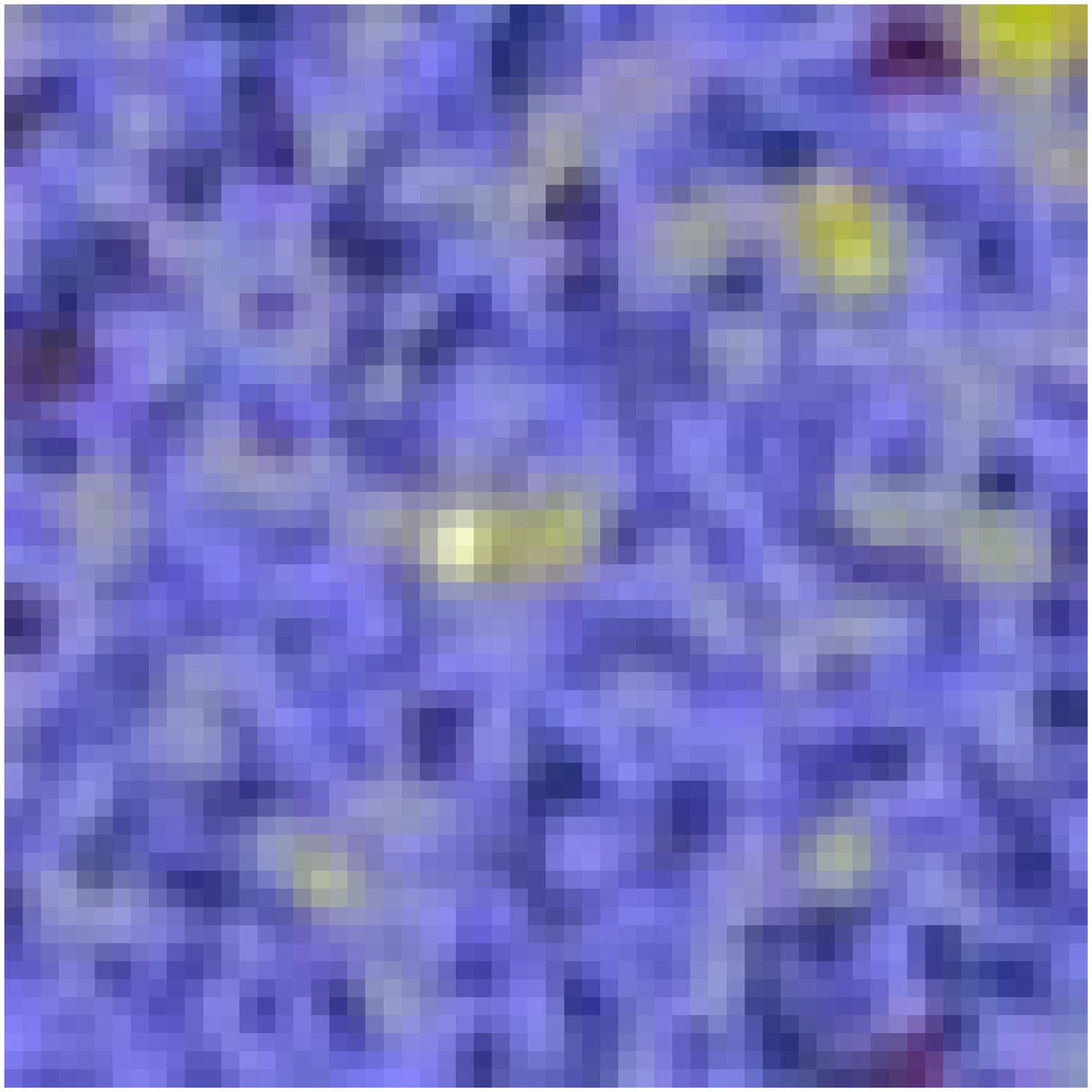}  \FGb{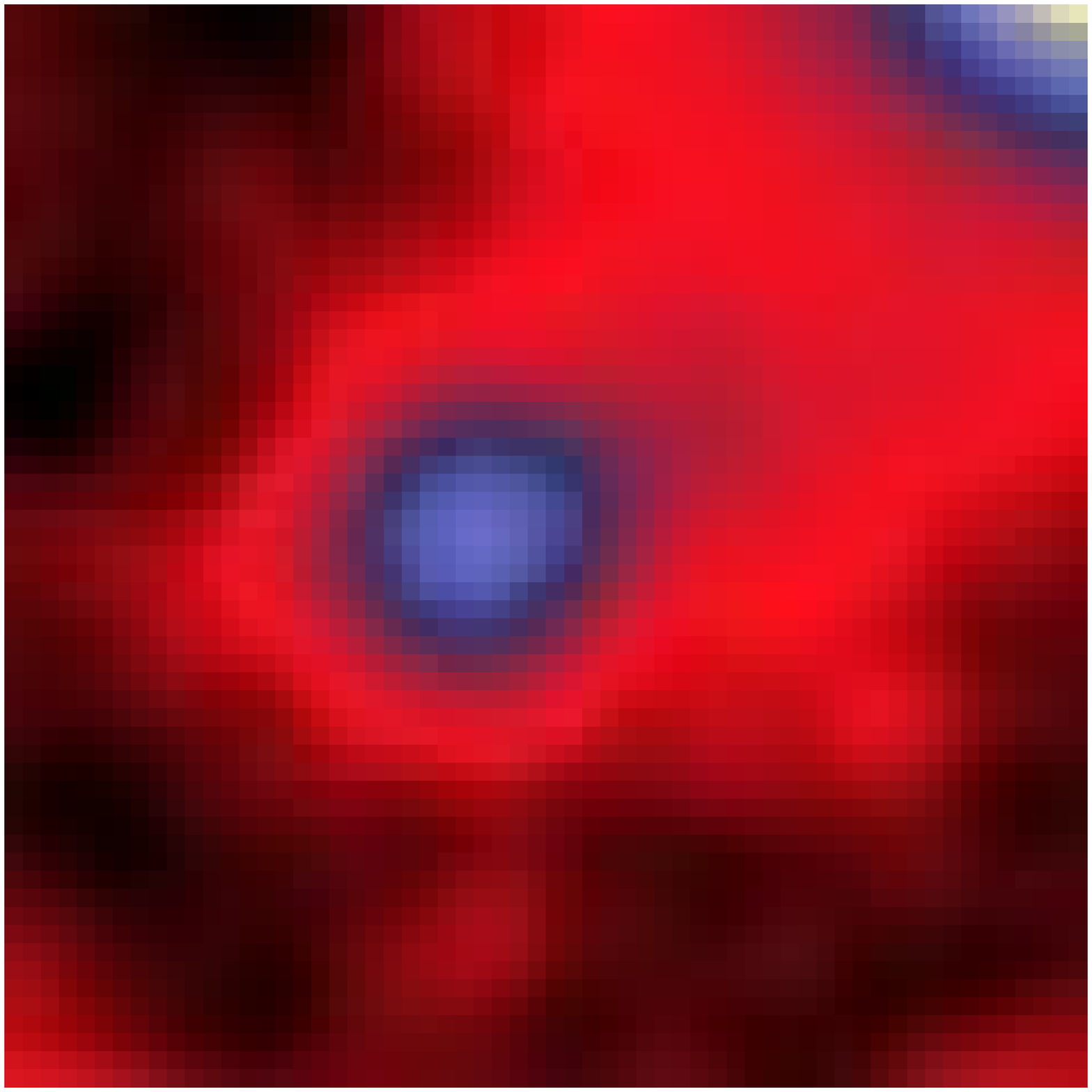}  \FGb{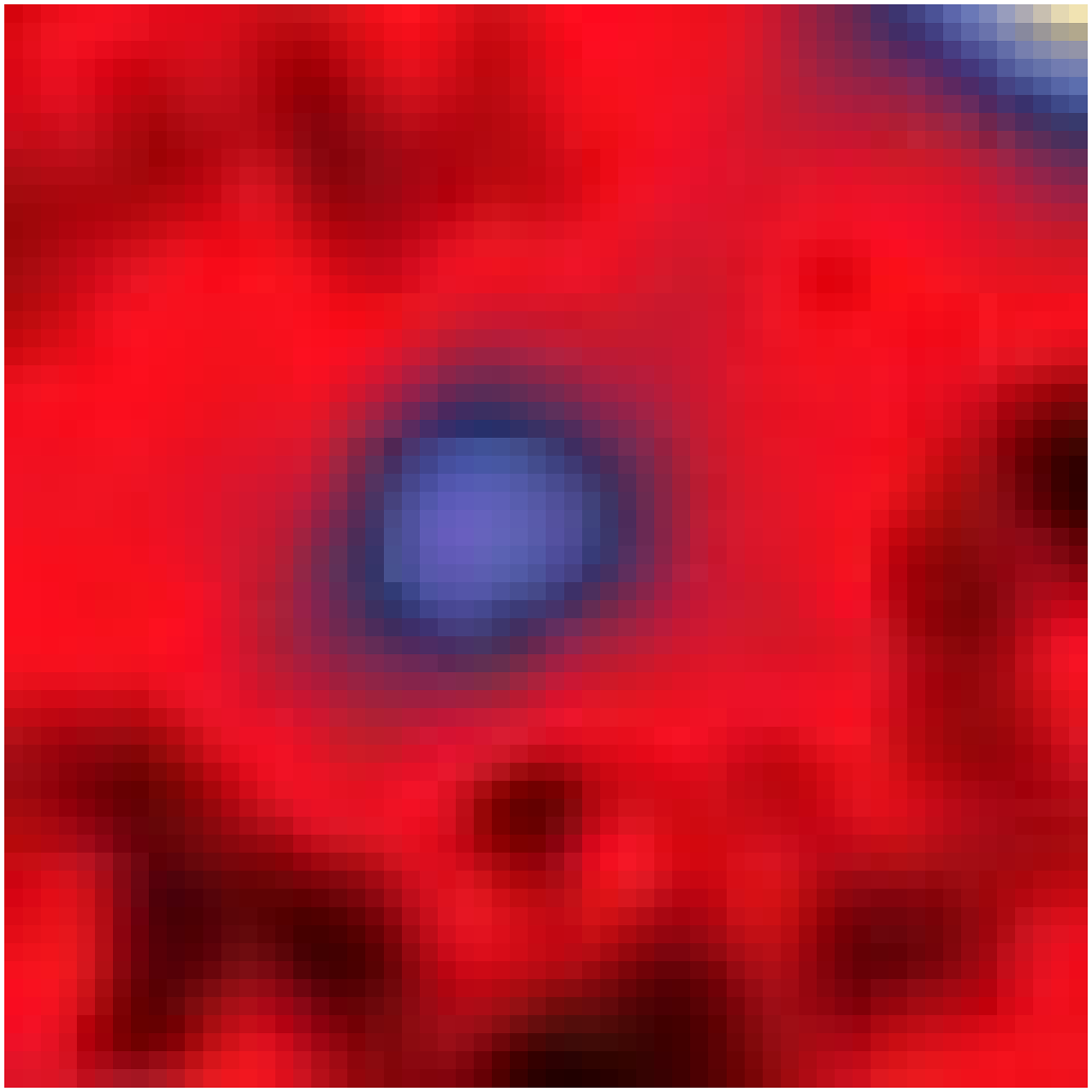}  \FGb{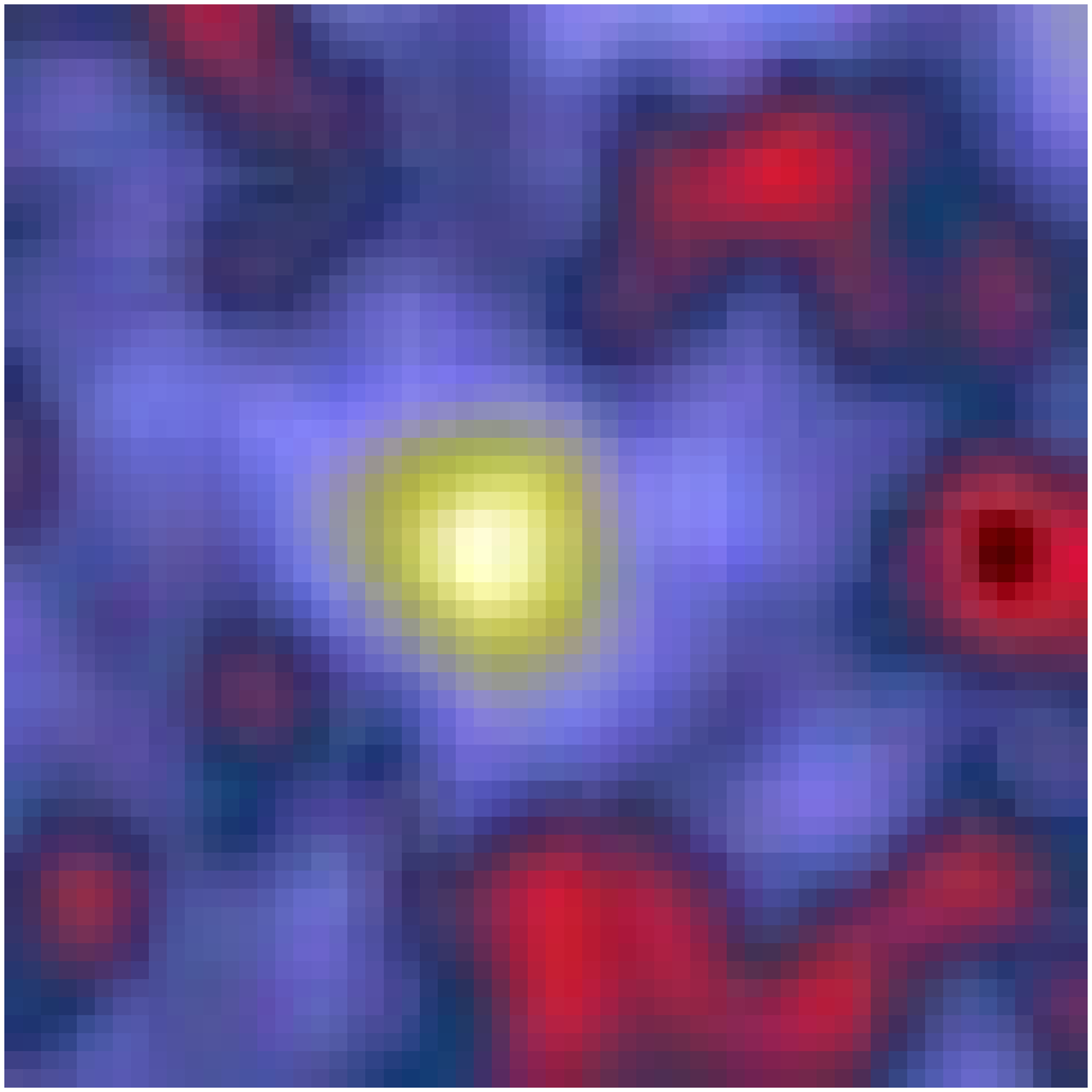}  \FGb{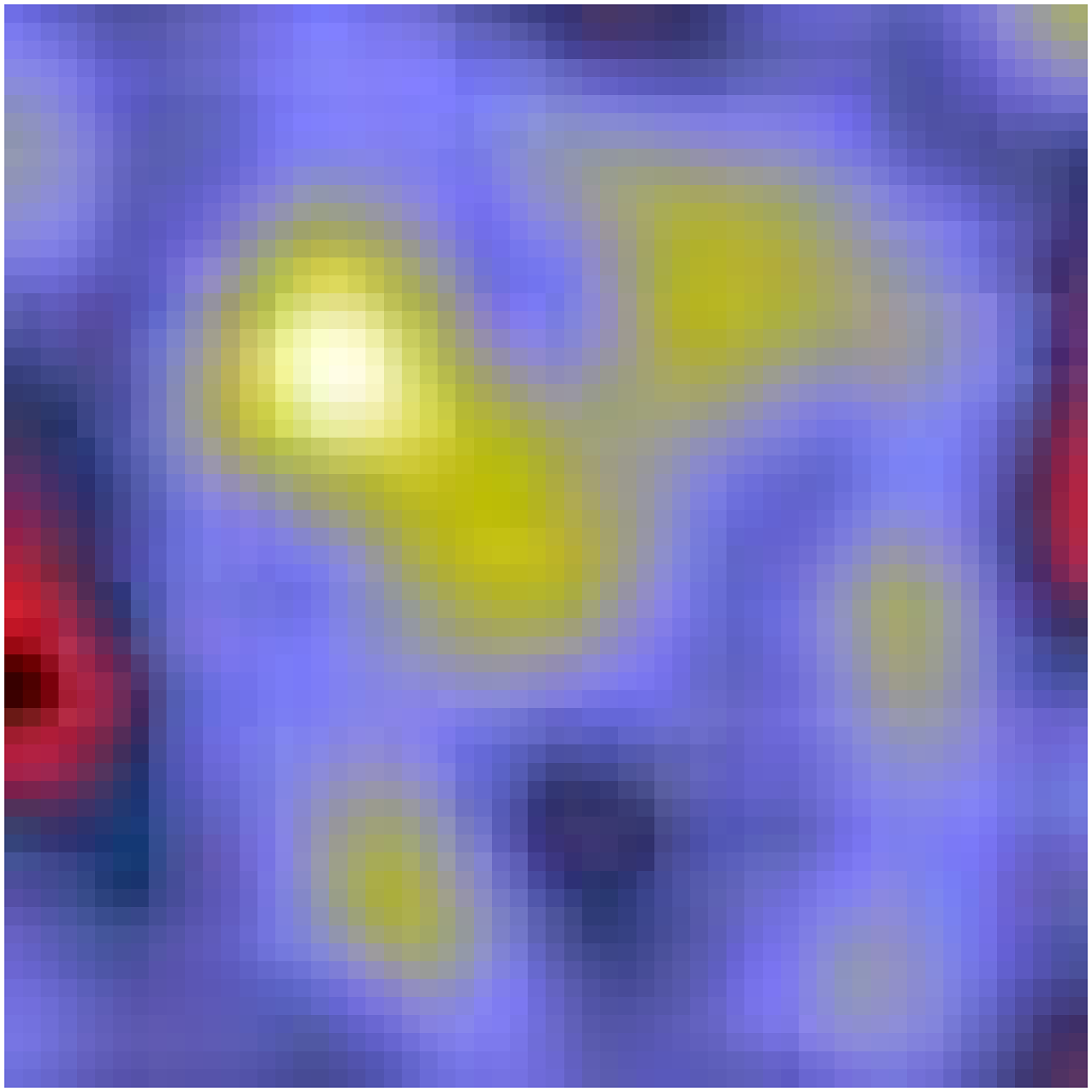} \\  {\#14   NCIA004638}  \\
  \FGb{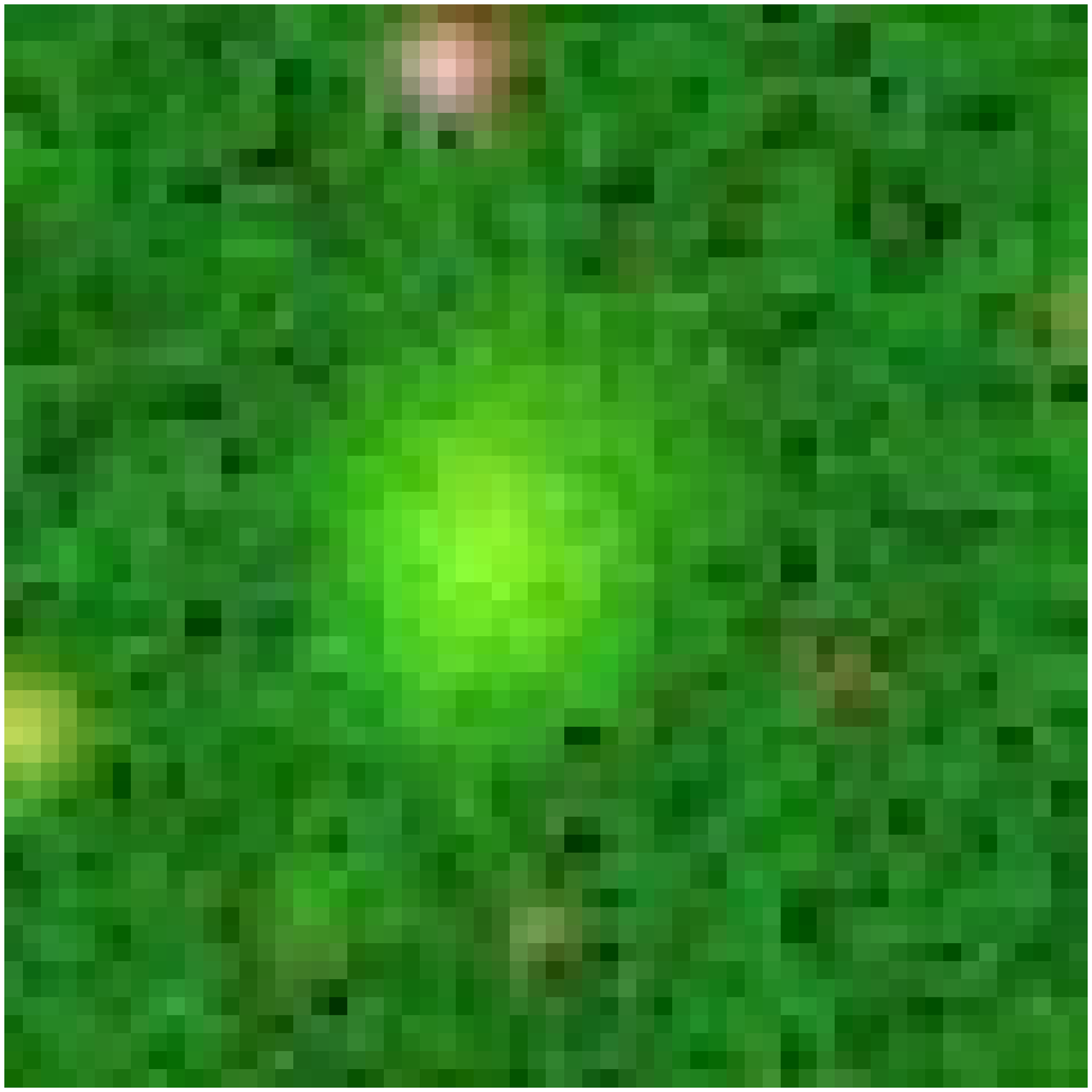}  \FGb{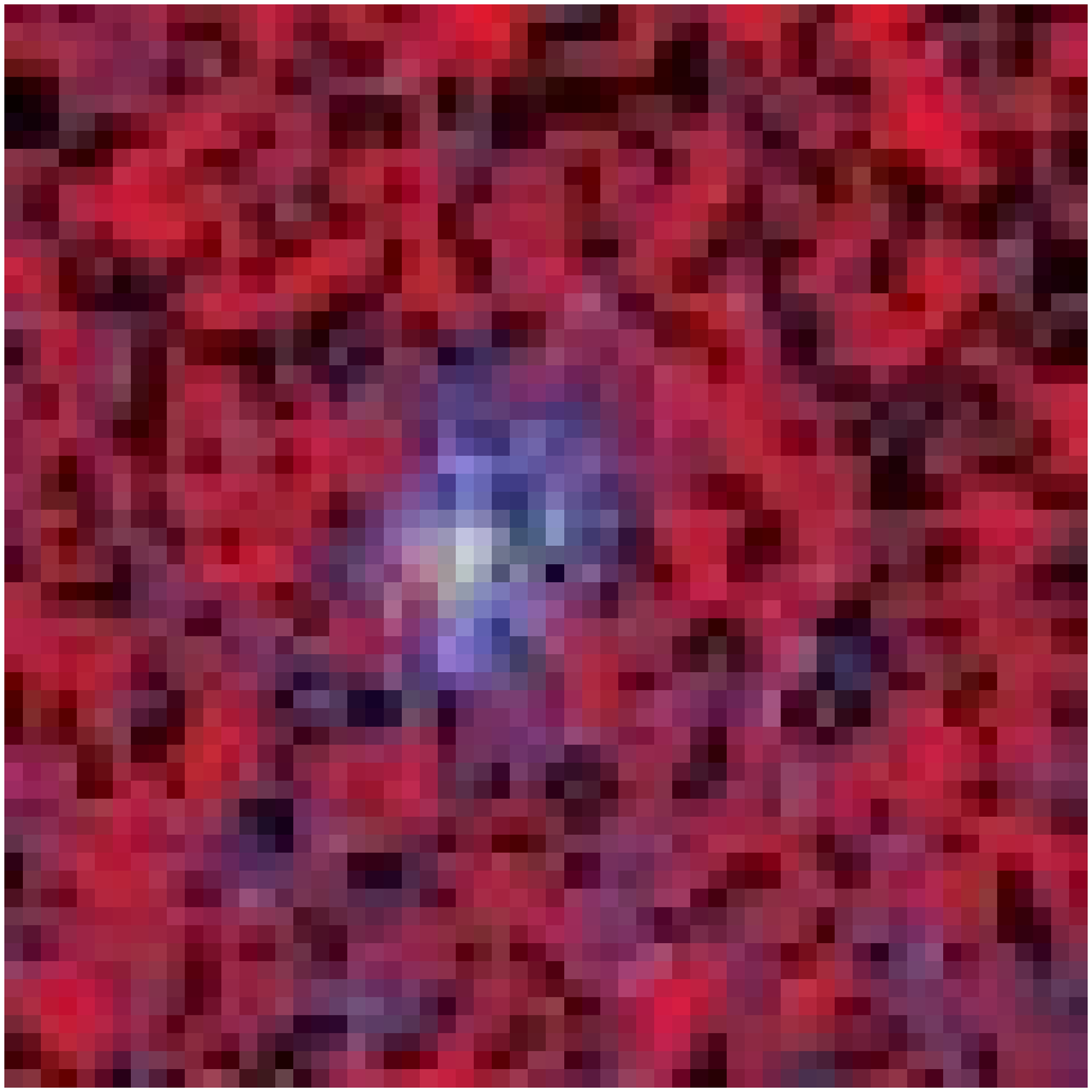}  \FGb{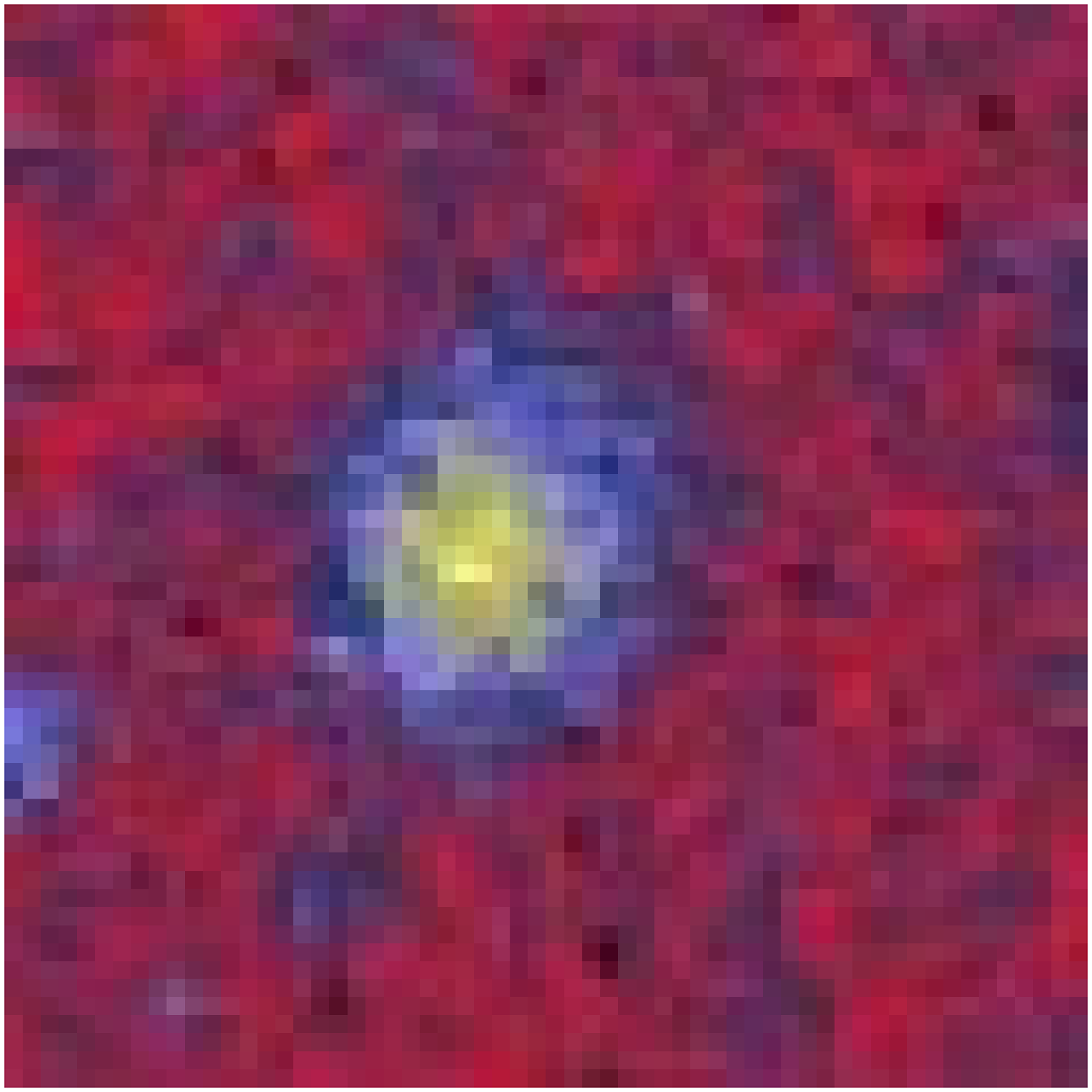}  \FGb{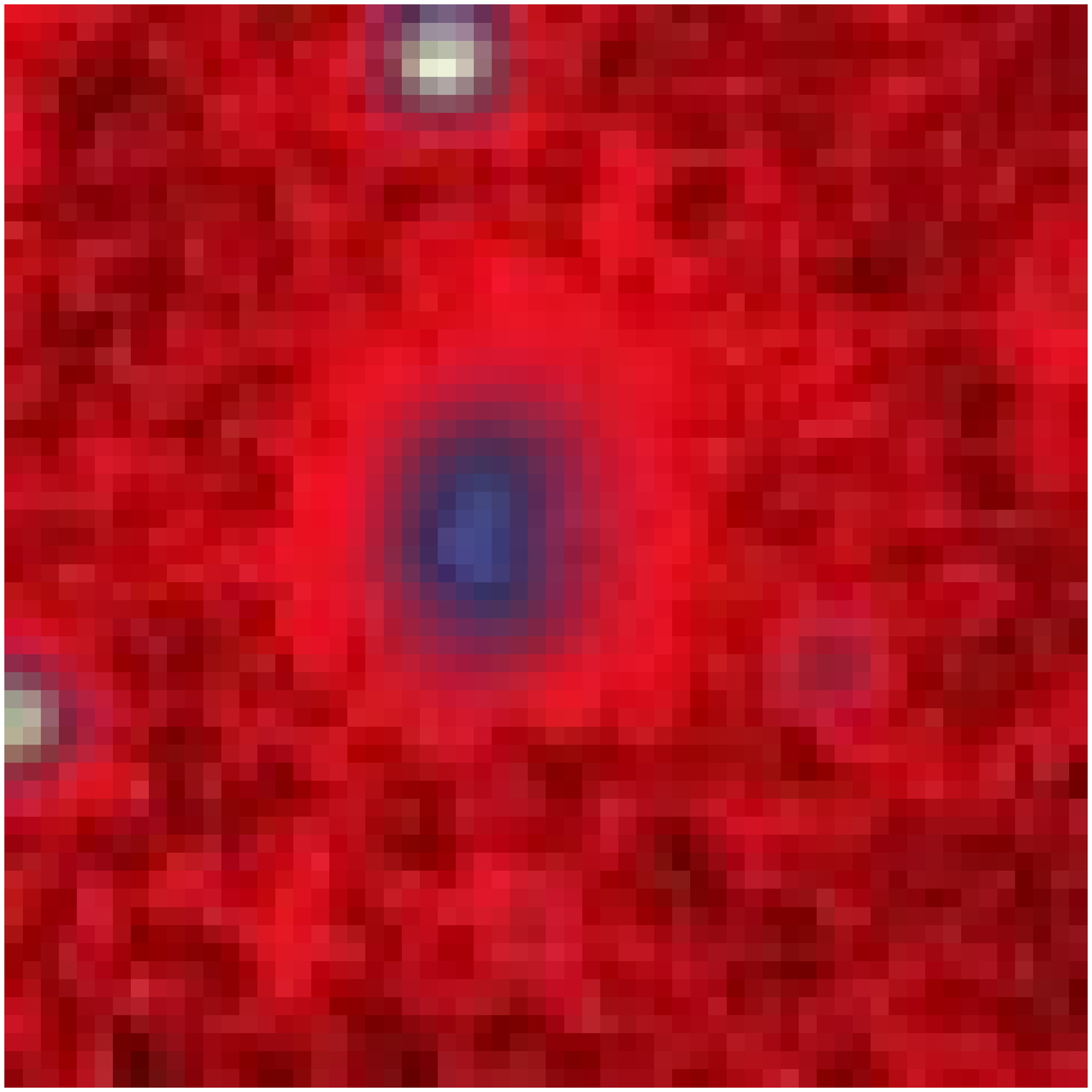}  \FGb{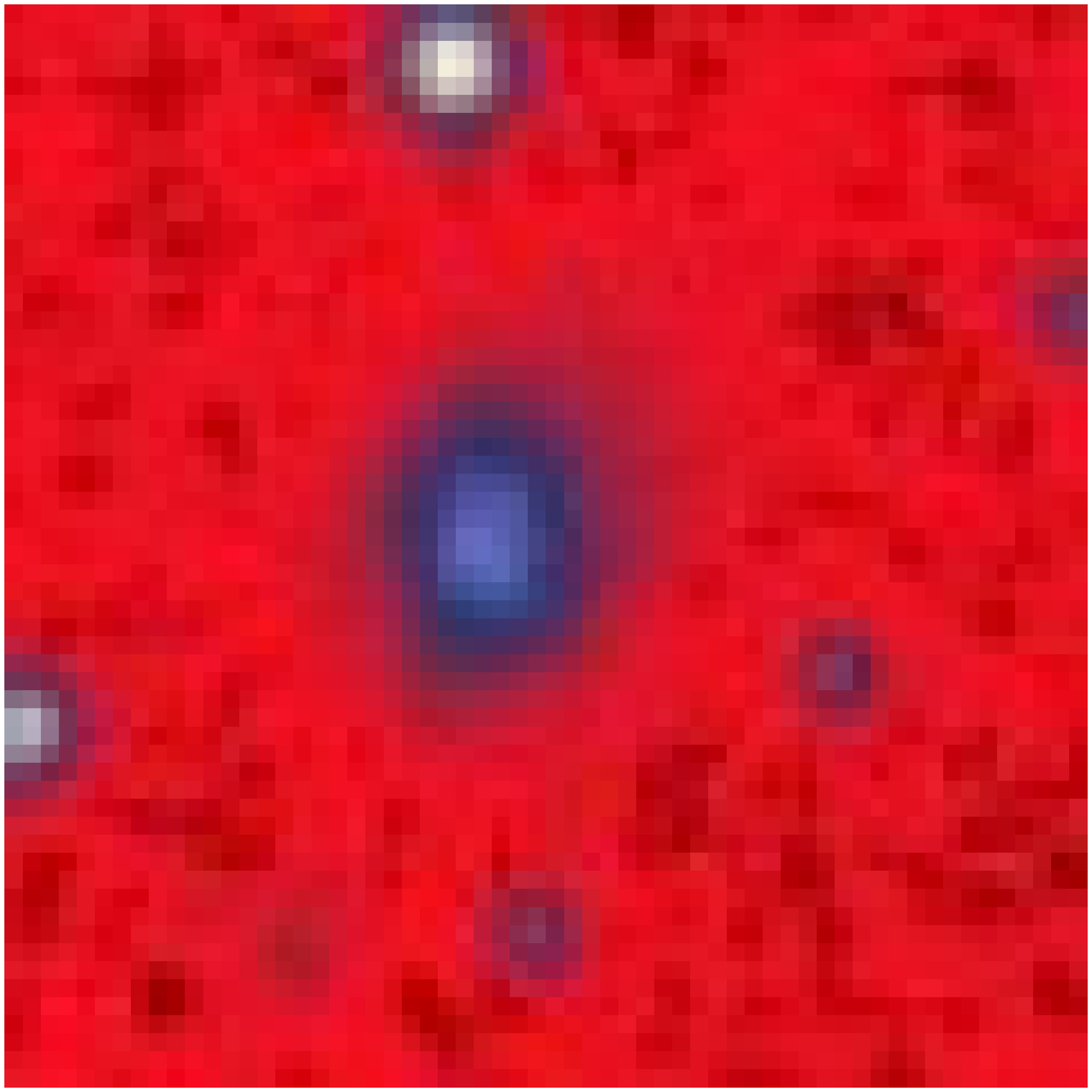}  \FGb{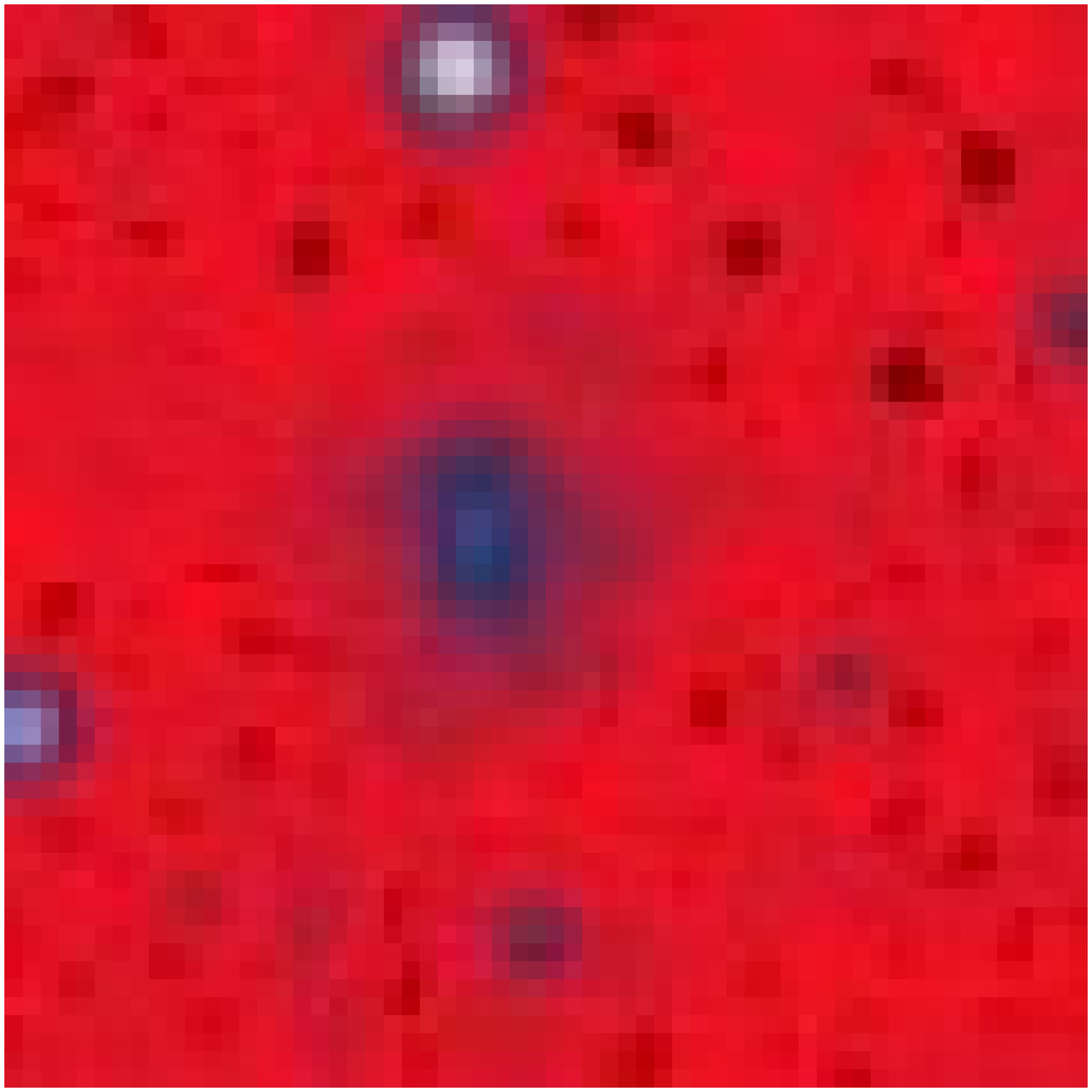}  \FGb{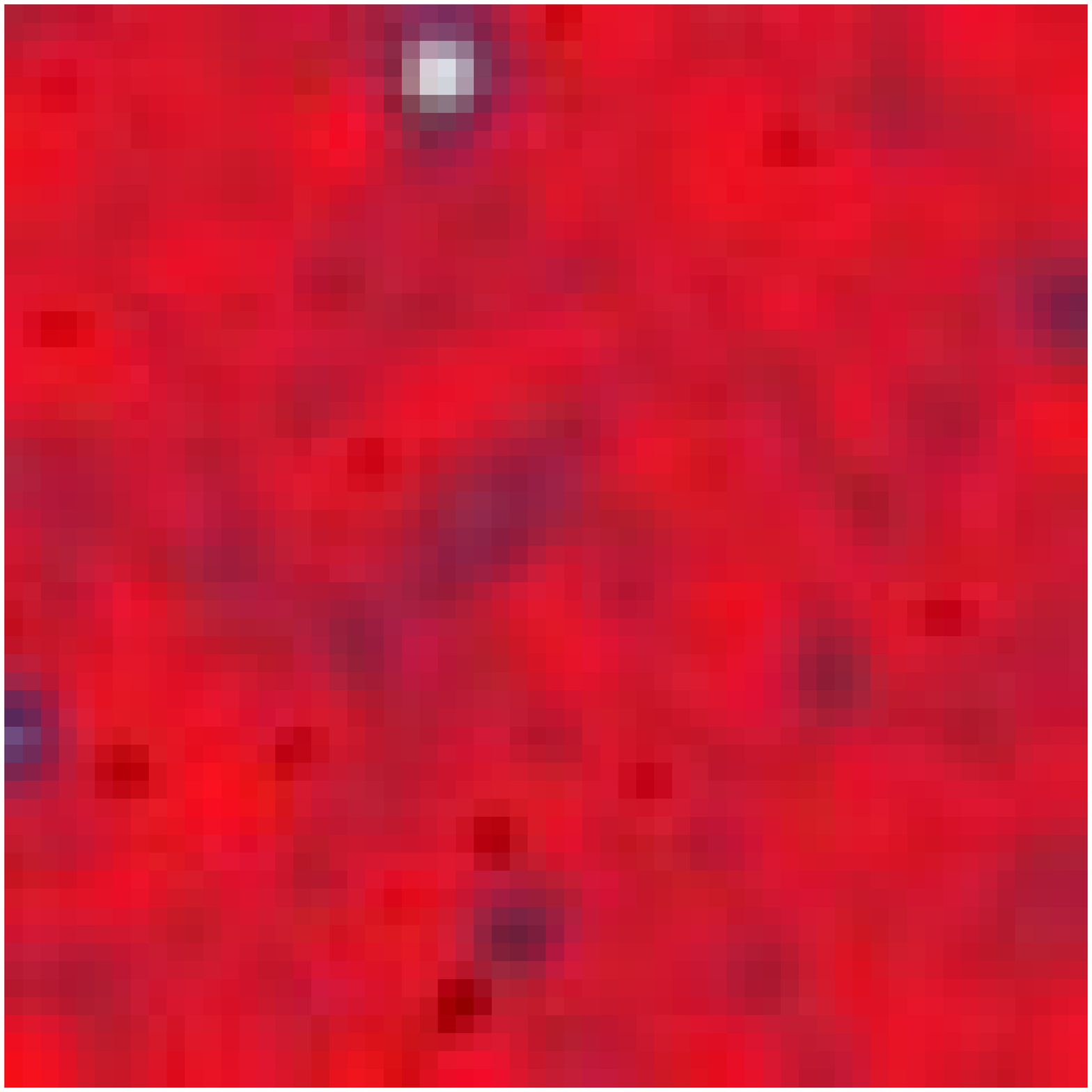}  \FGb{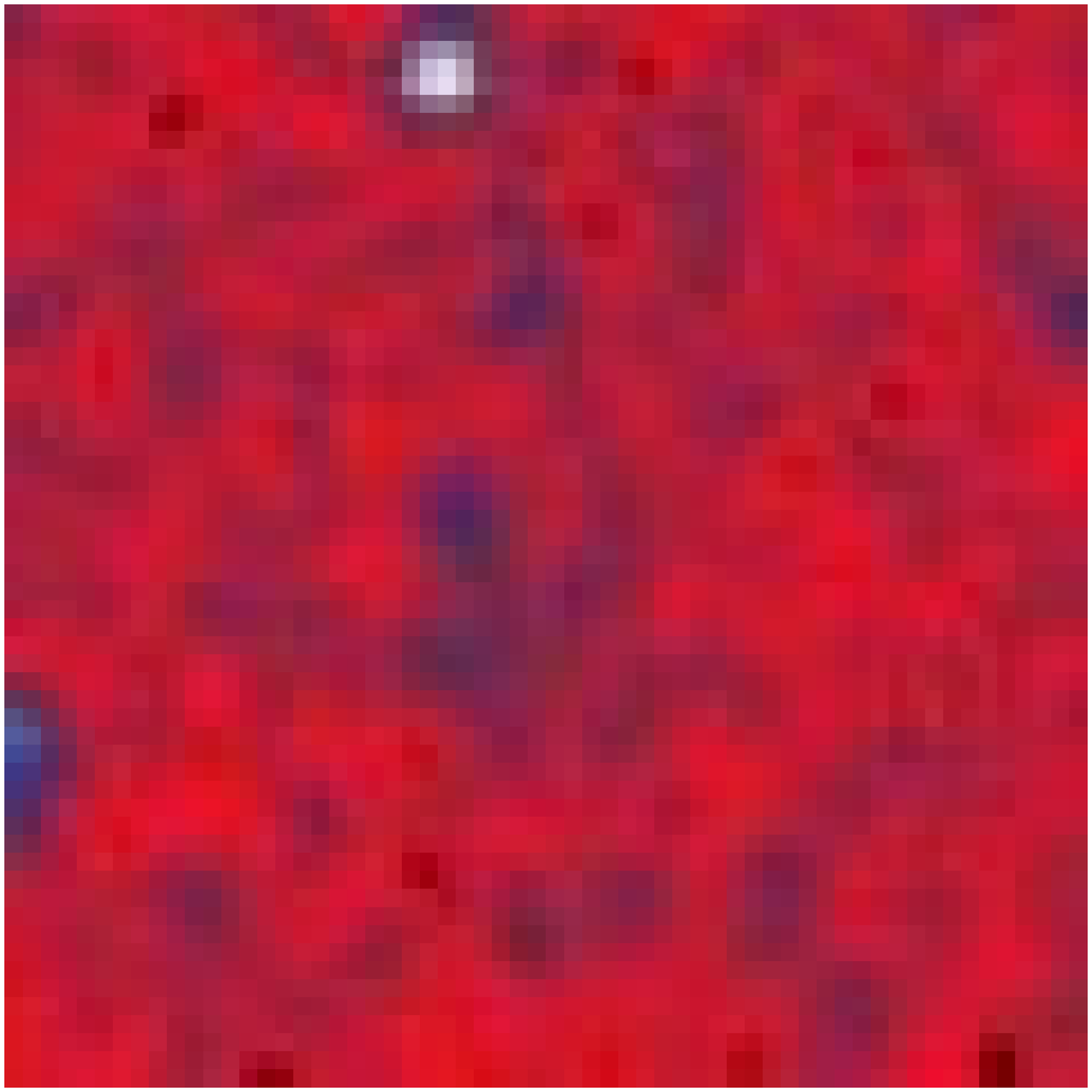}  \FGb{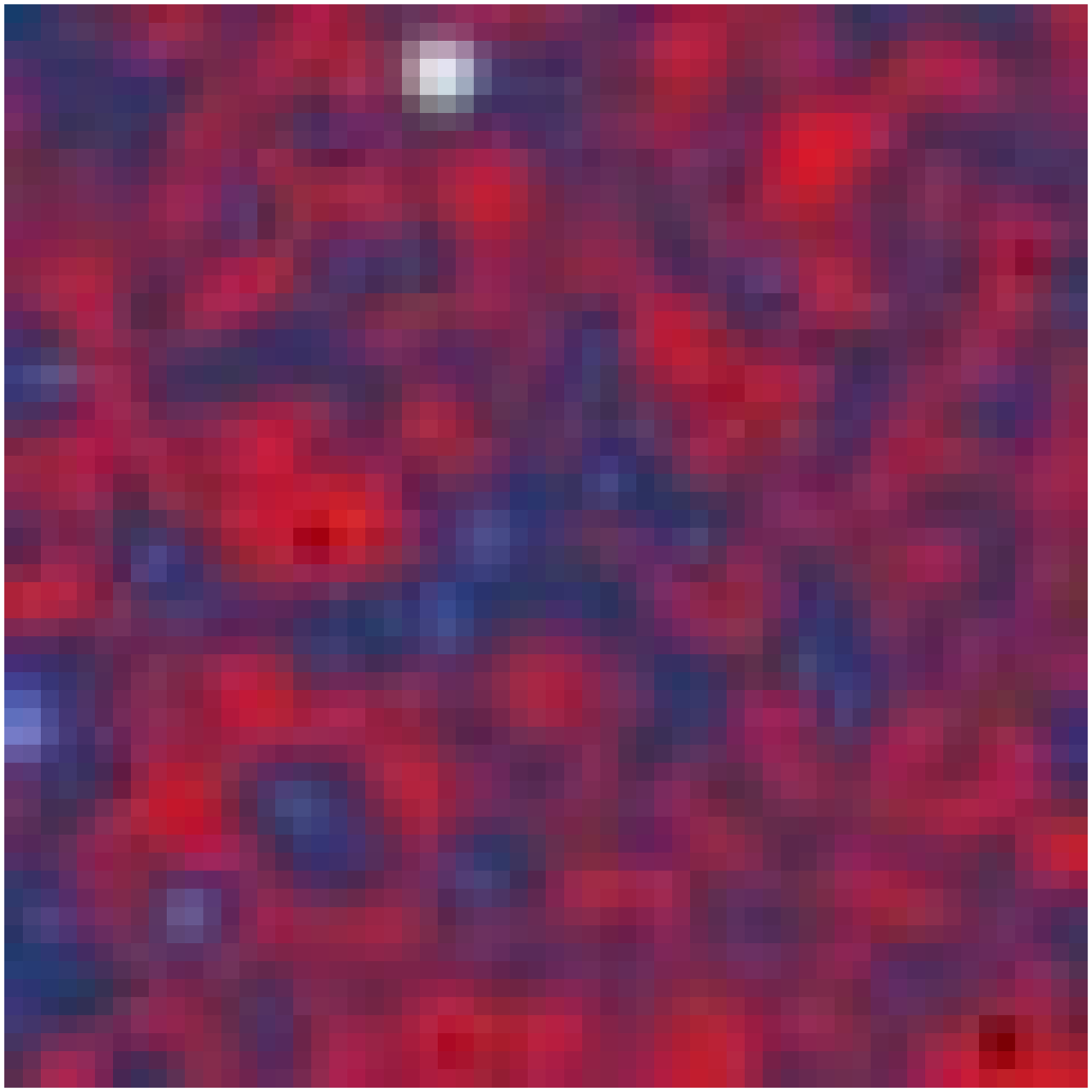}  \FGb{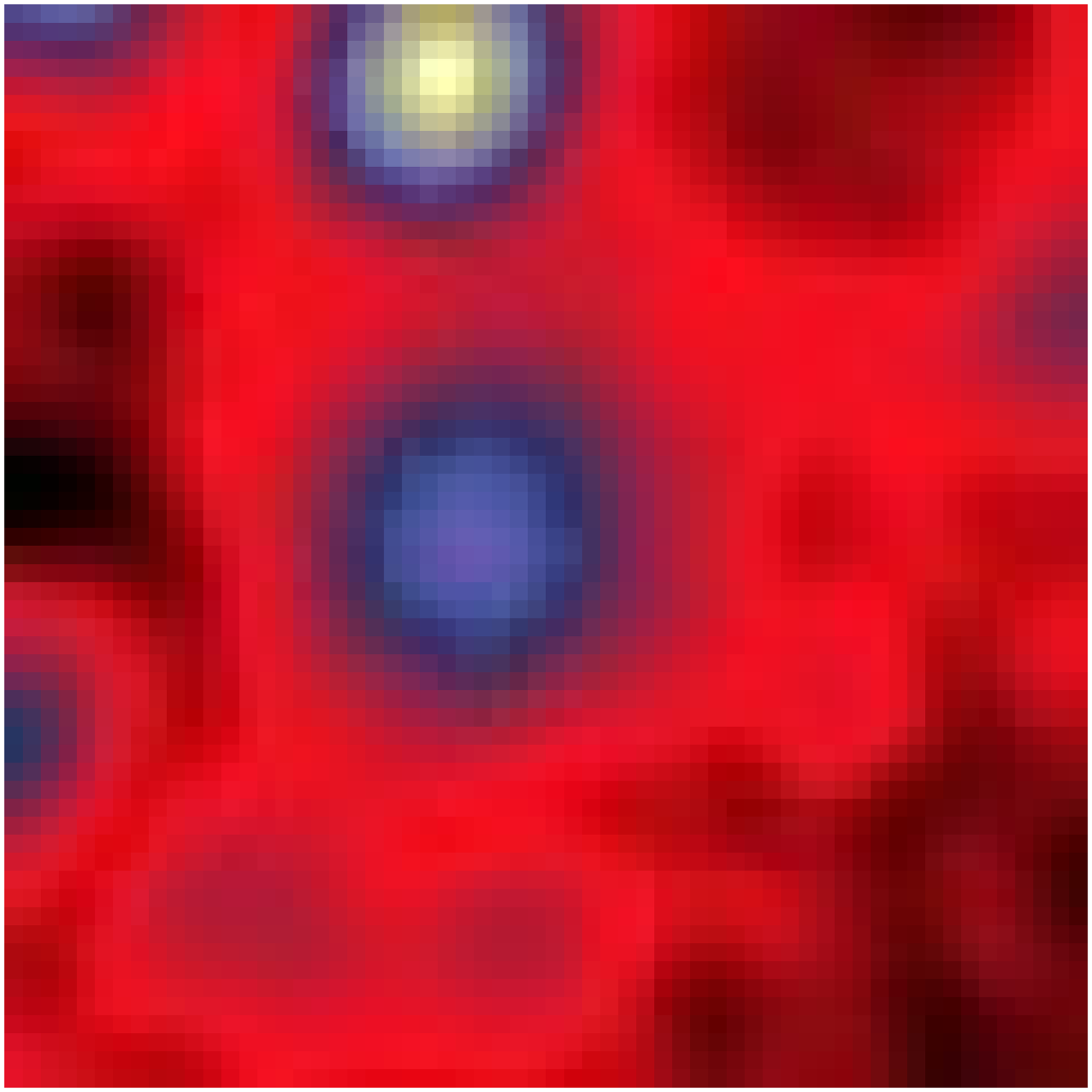}  \FGb{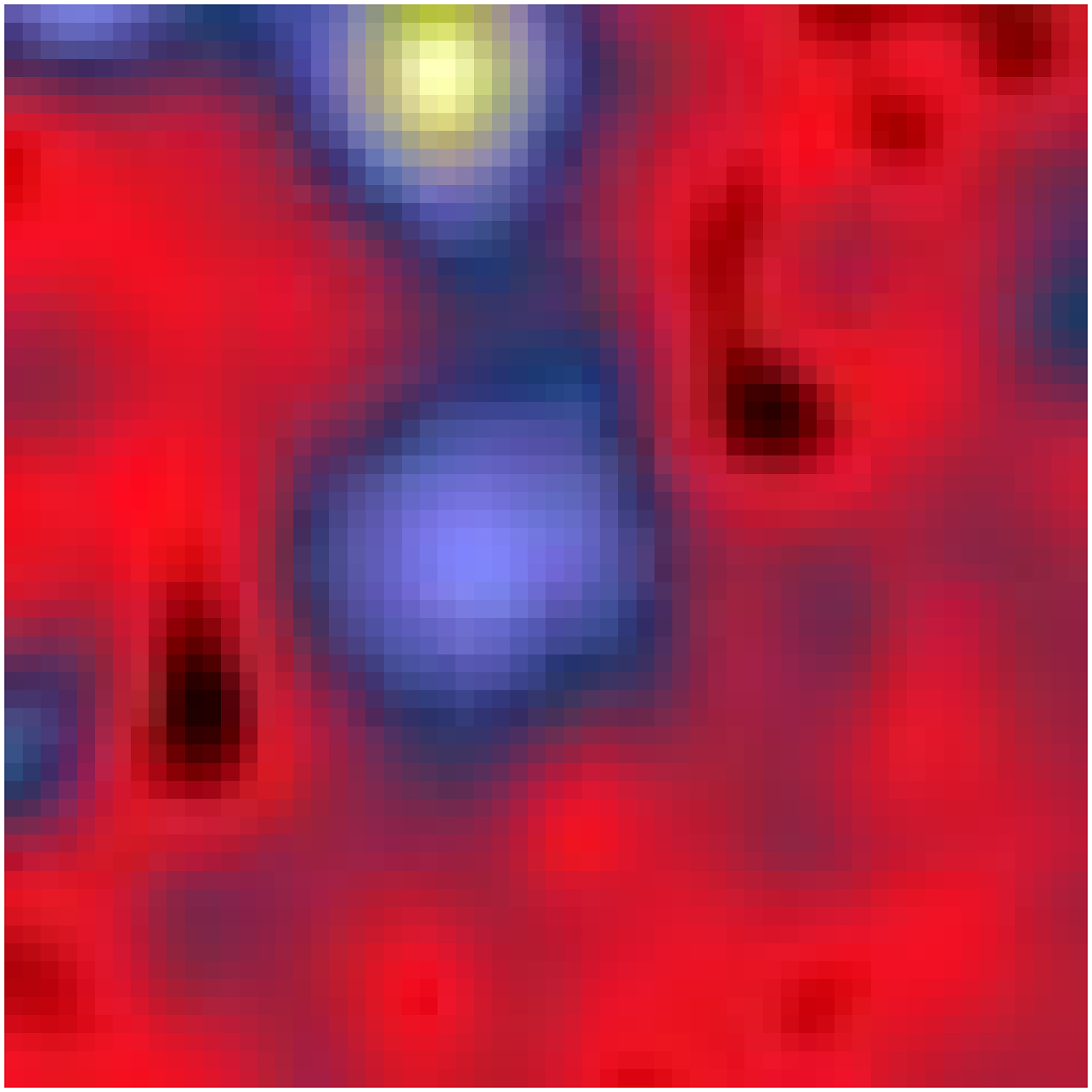}  \FGb{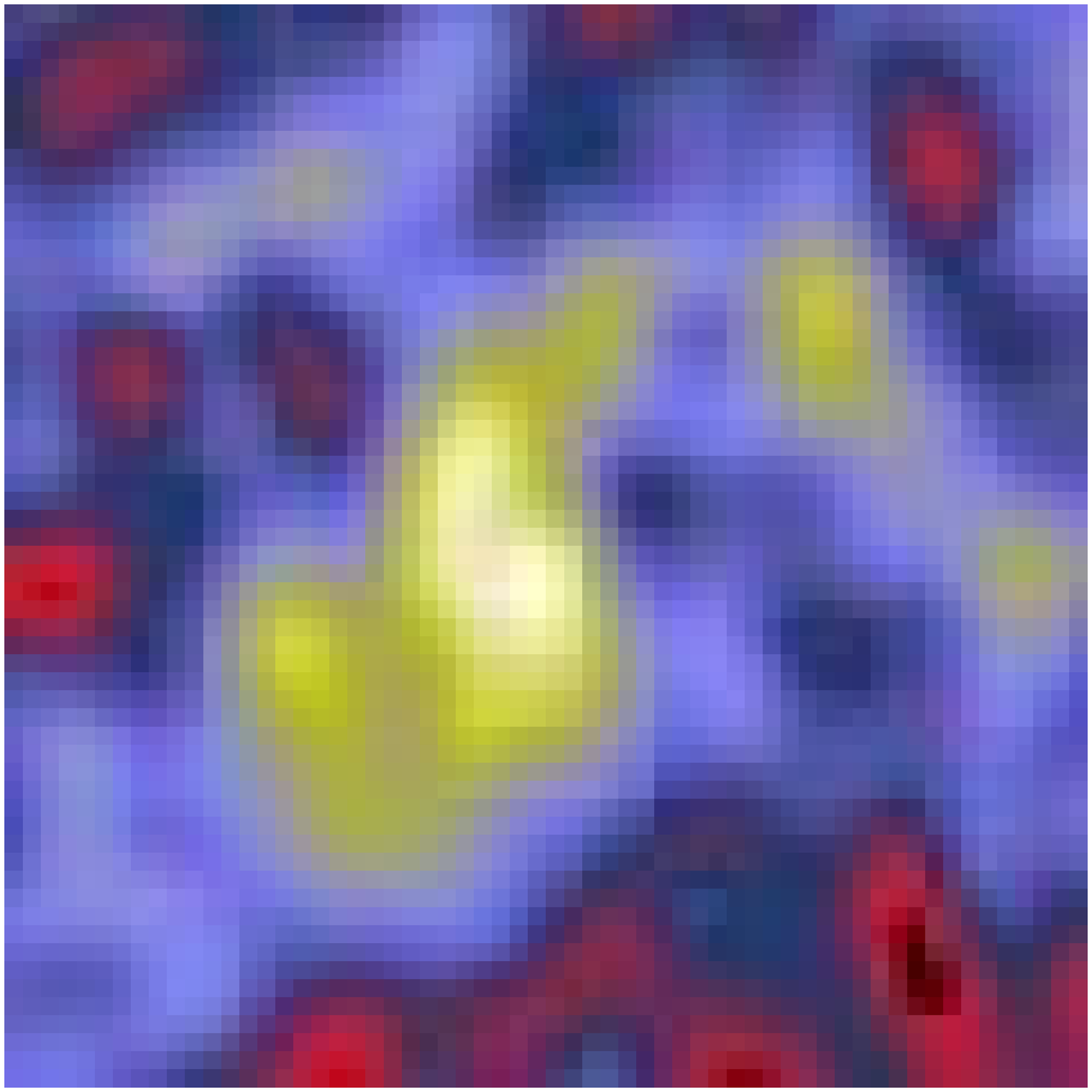}  \FGb{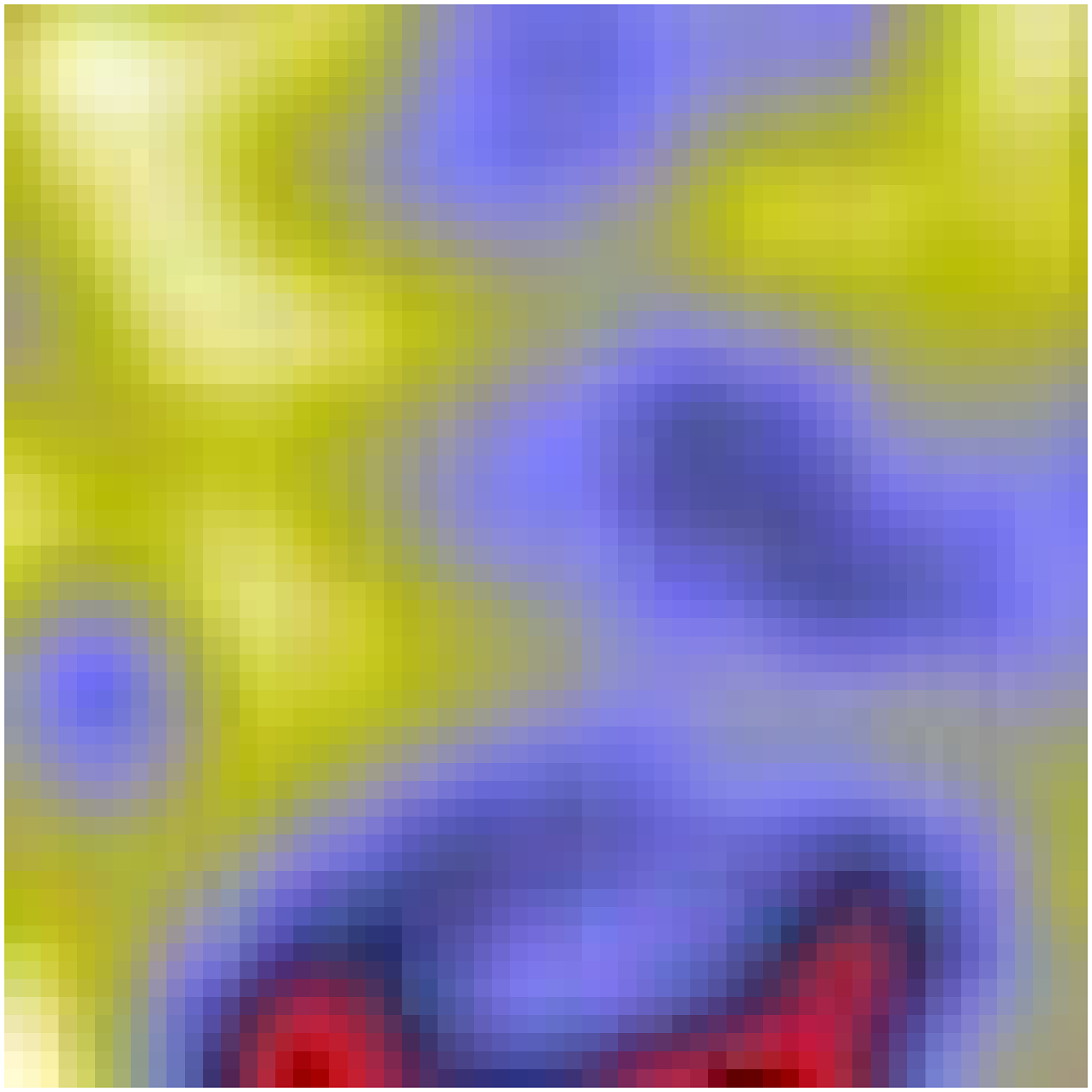} \\  {\#15   NBZ5003674}  \\
  \FGb{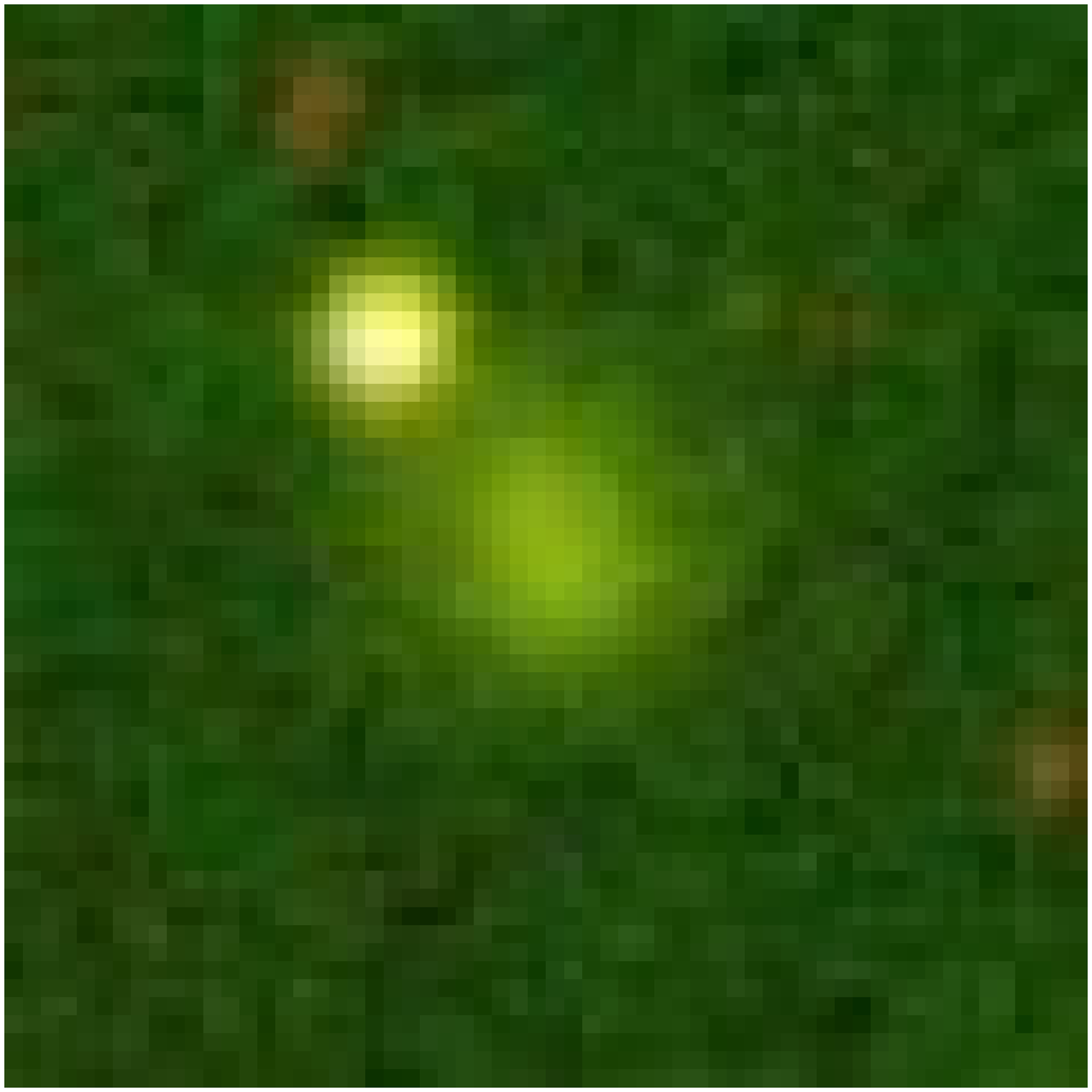}  \FGb{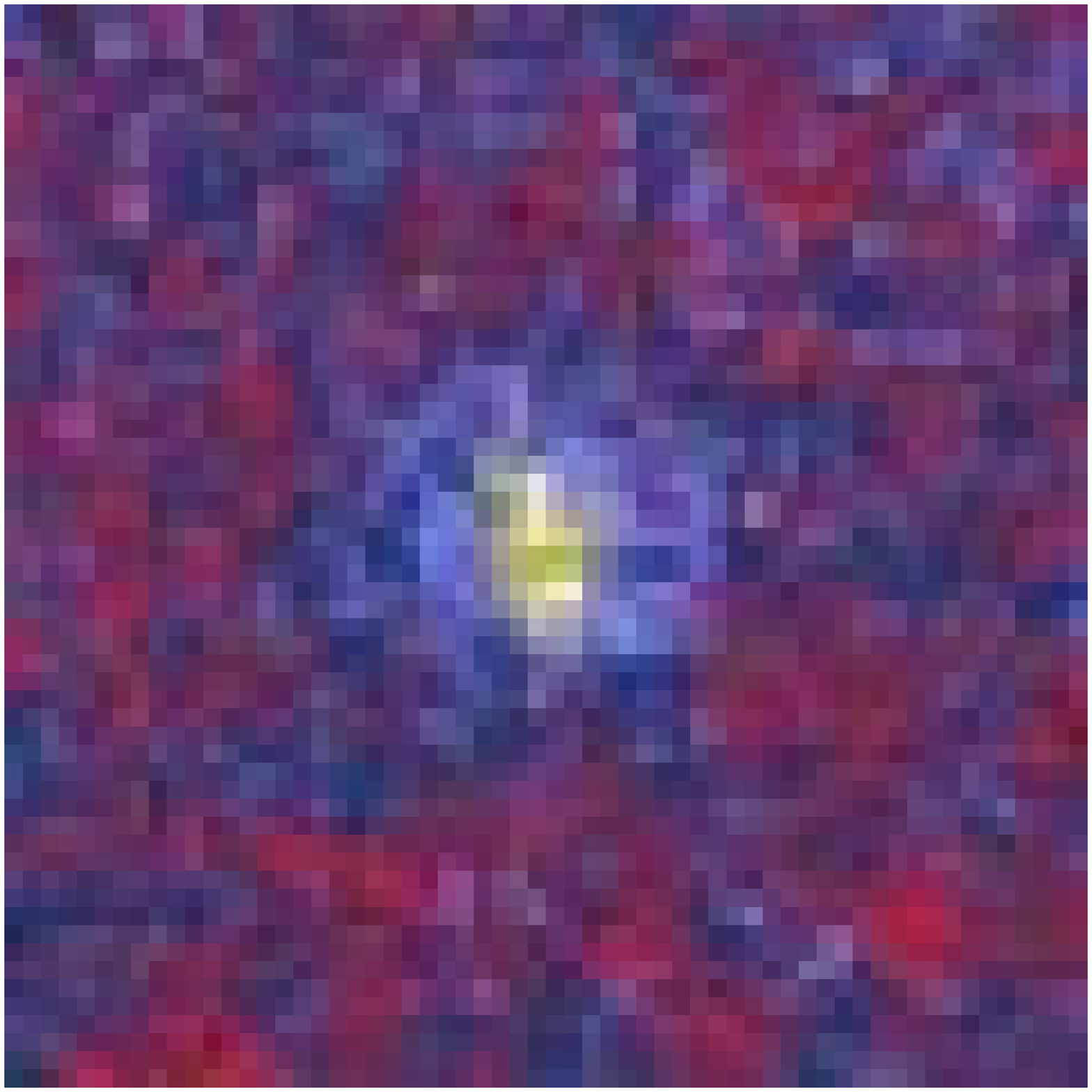}  \FGb{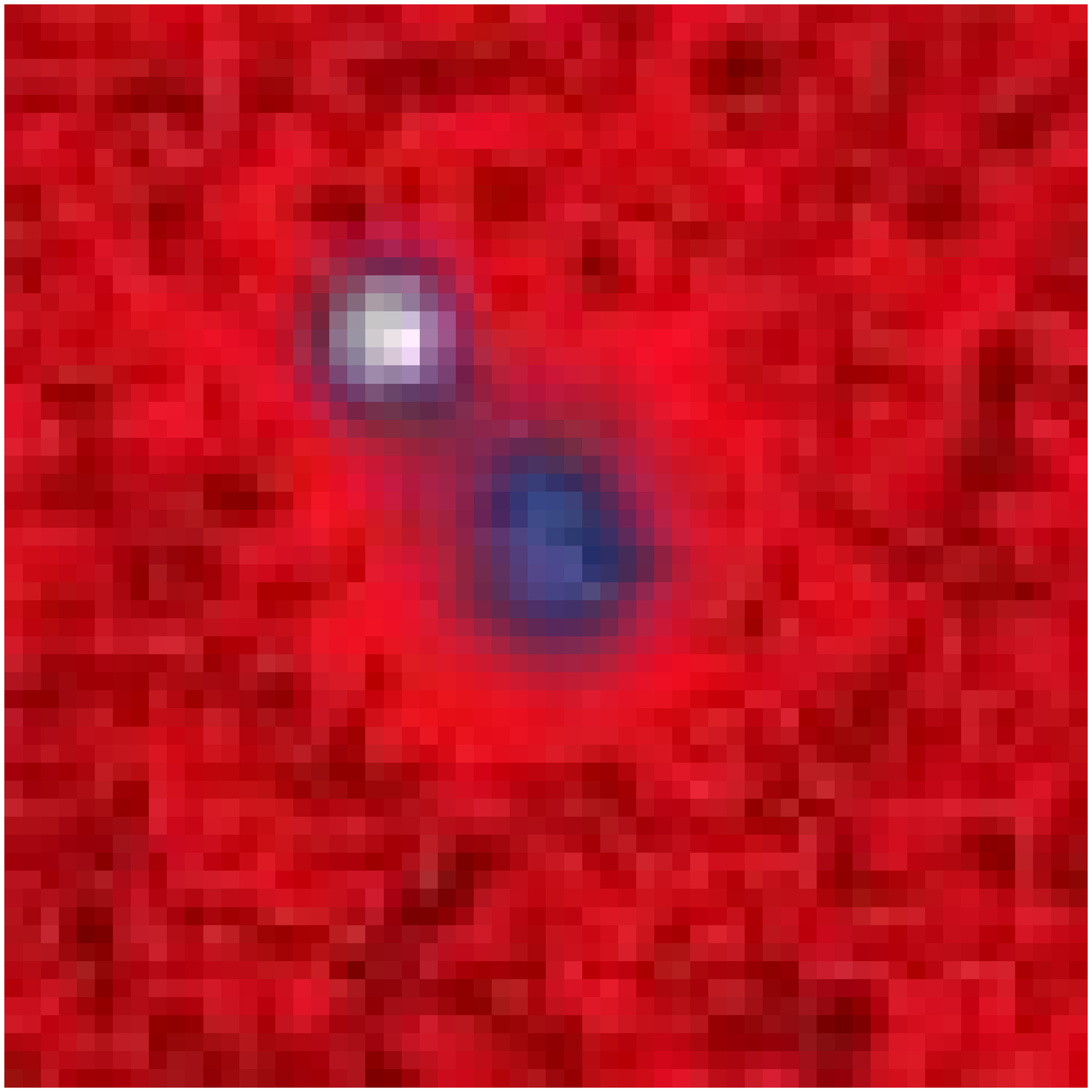}  \FGb{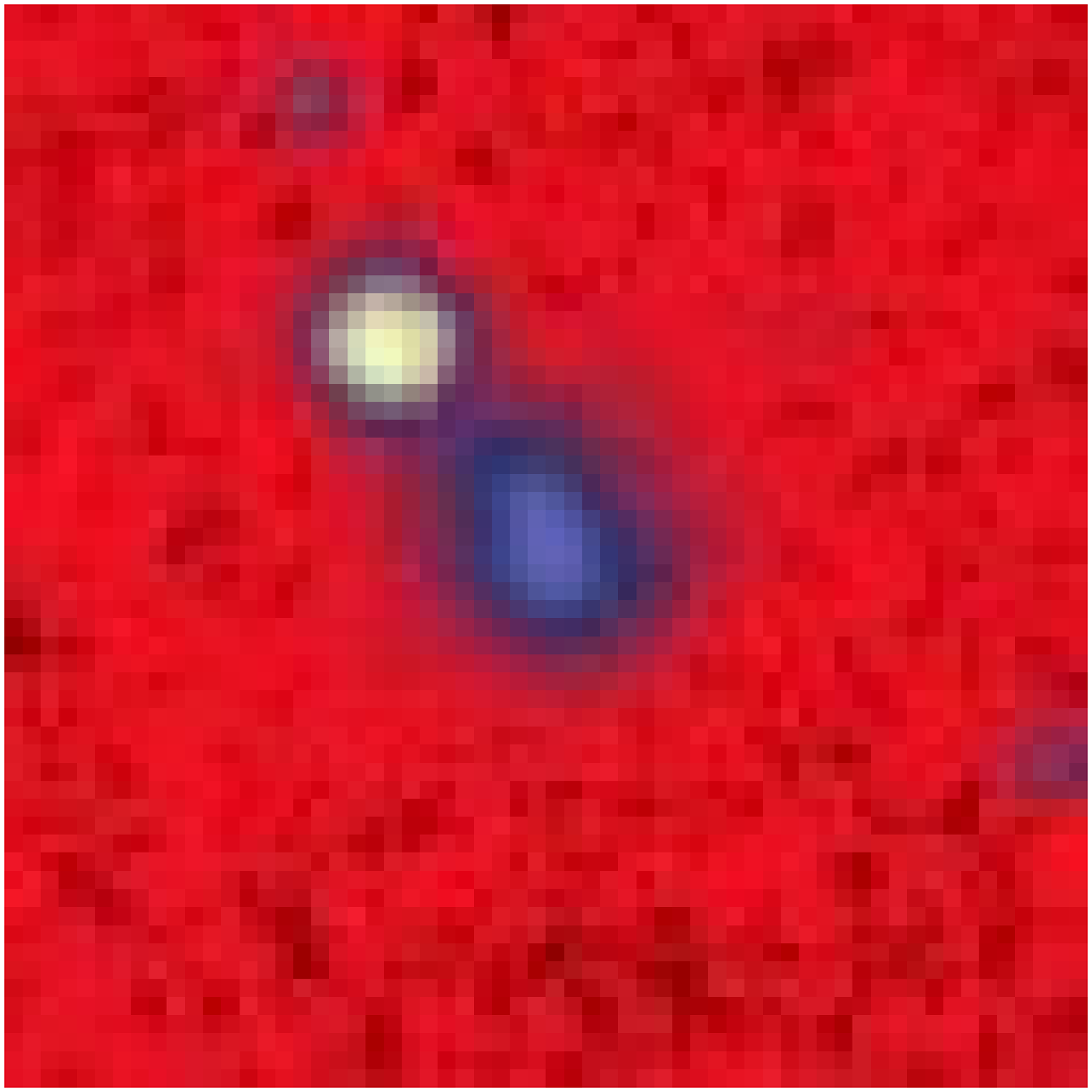}  \FGb{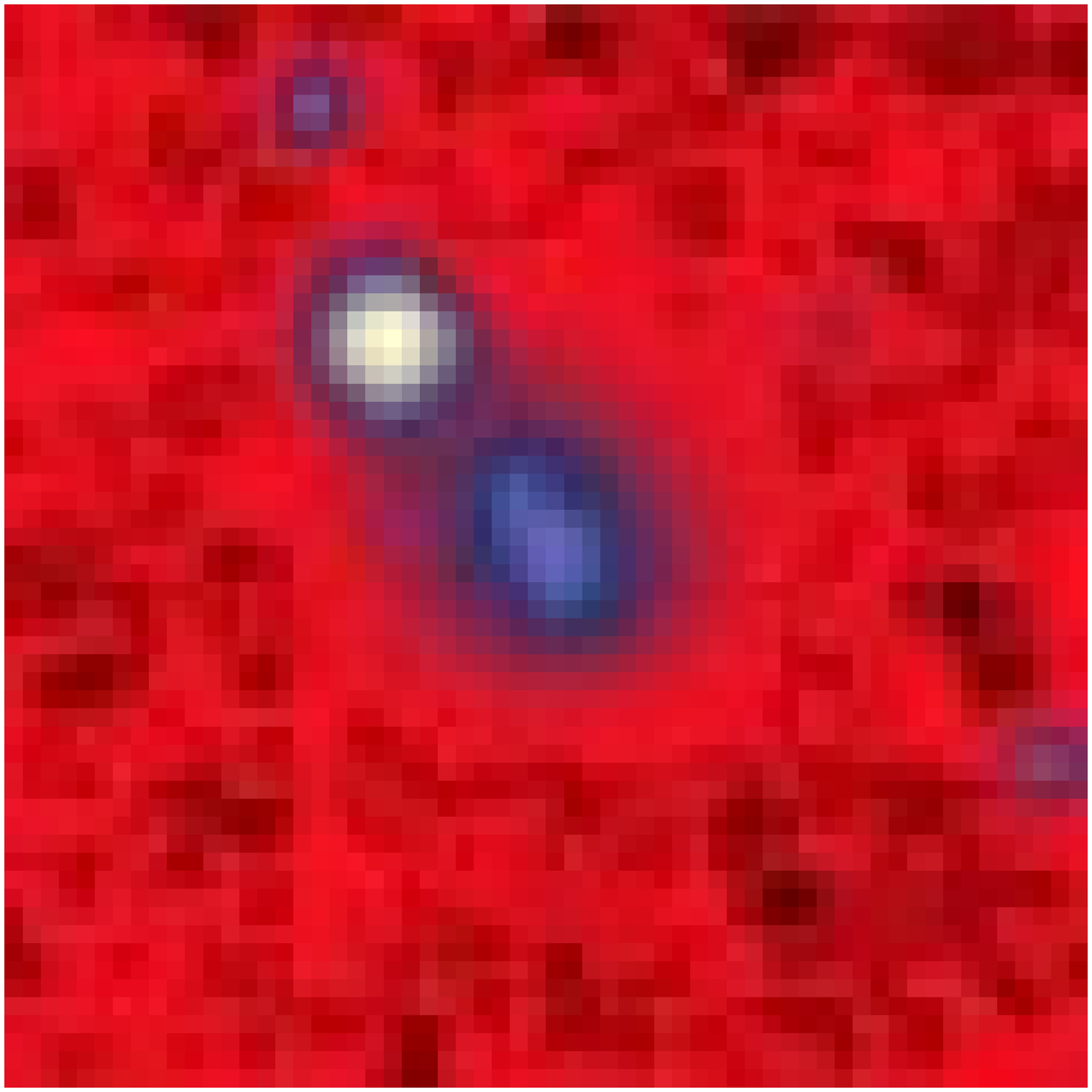}  \FGb{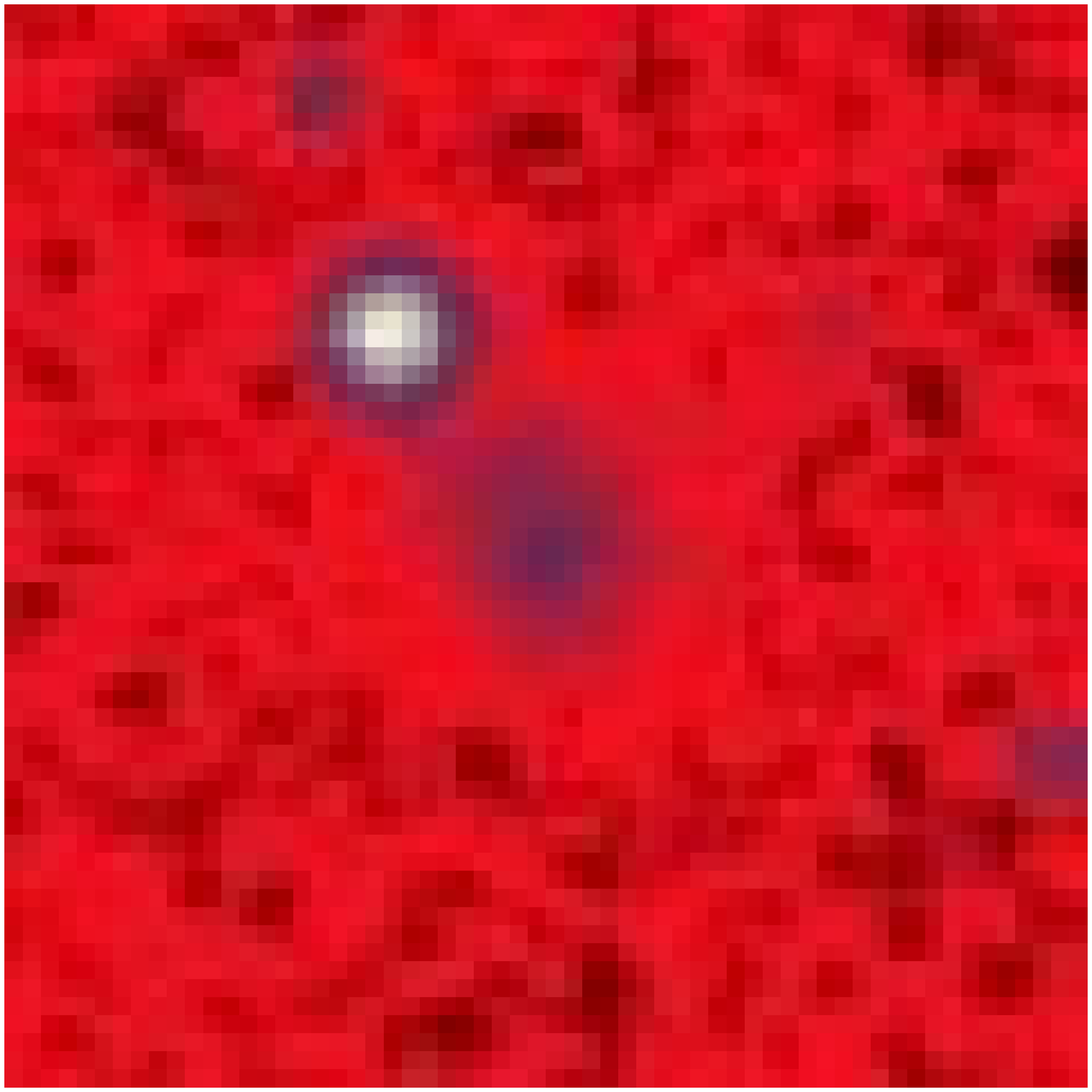}  \FGb{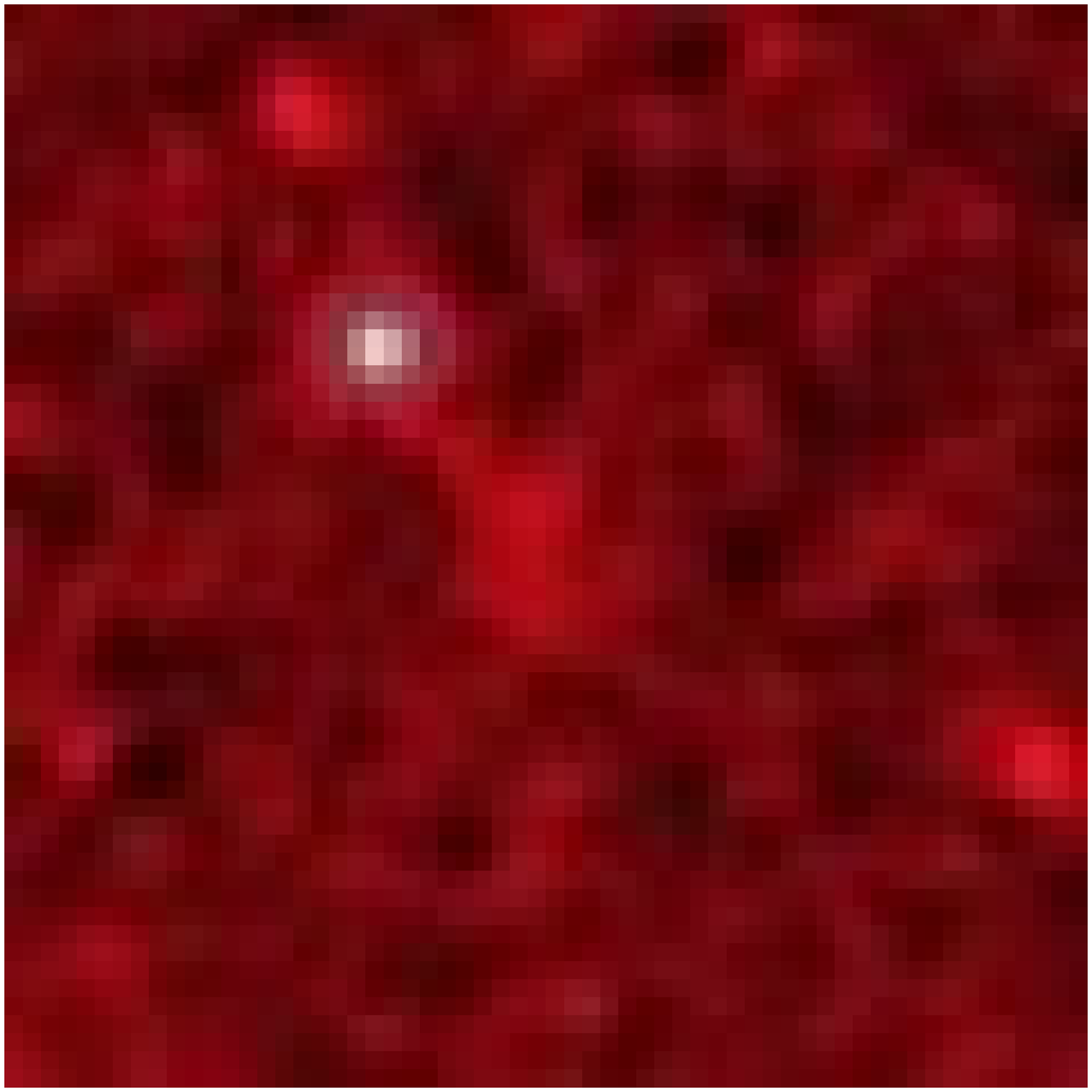}  \FGb{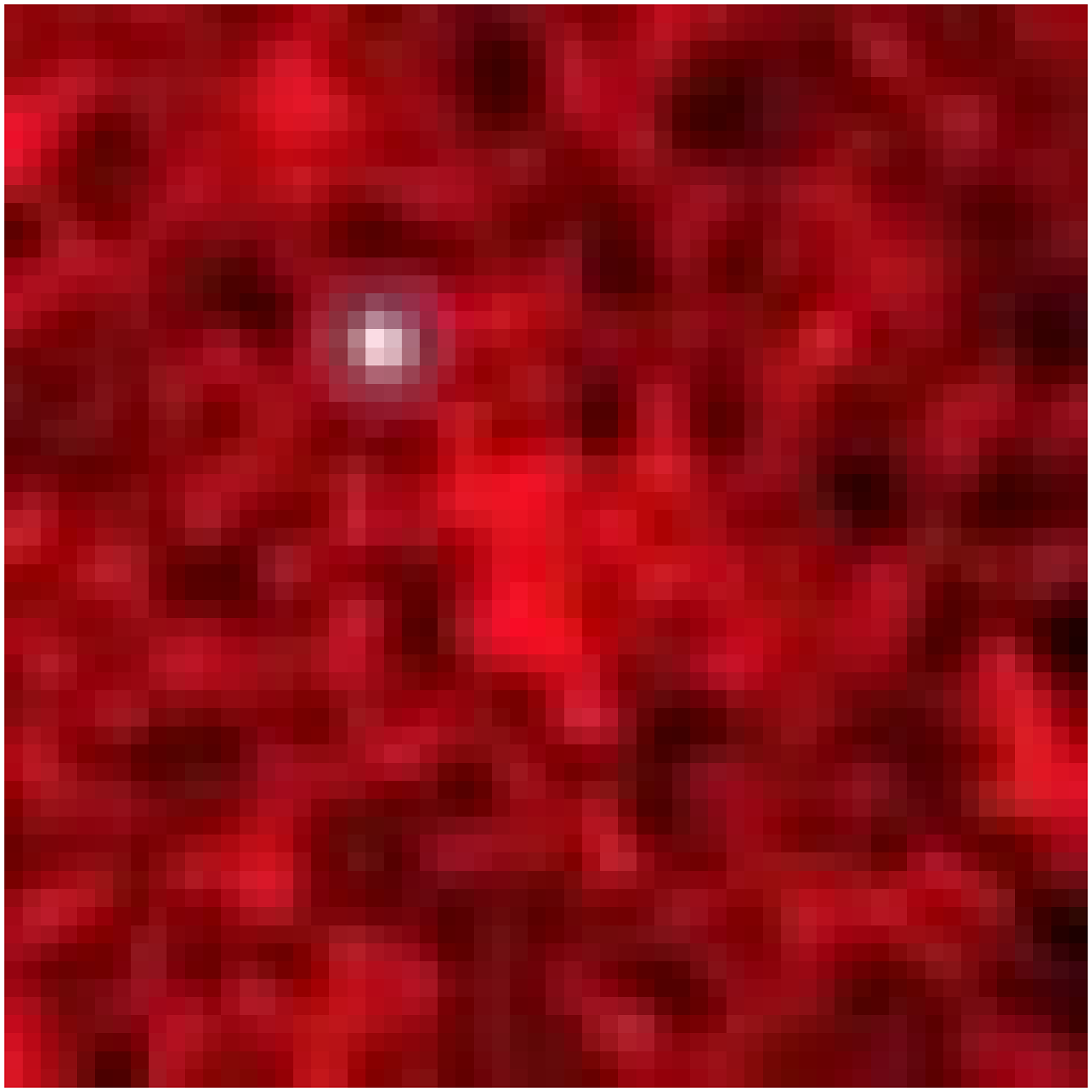}  \FGb{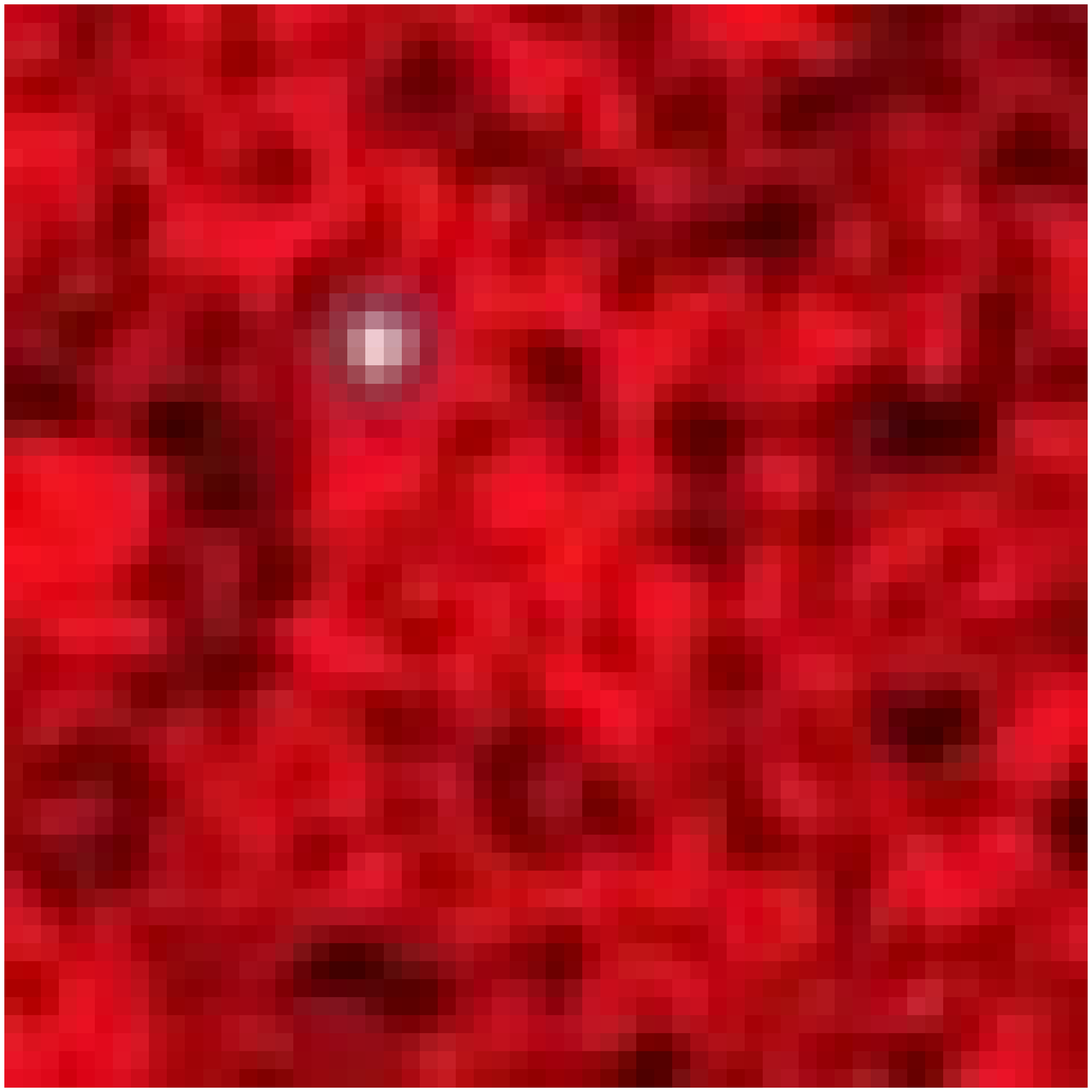}  \FGb{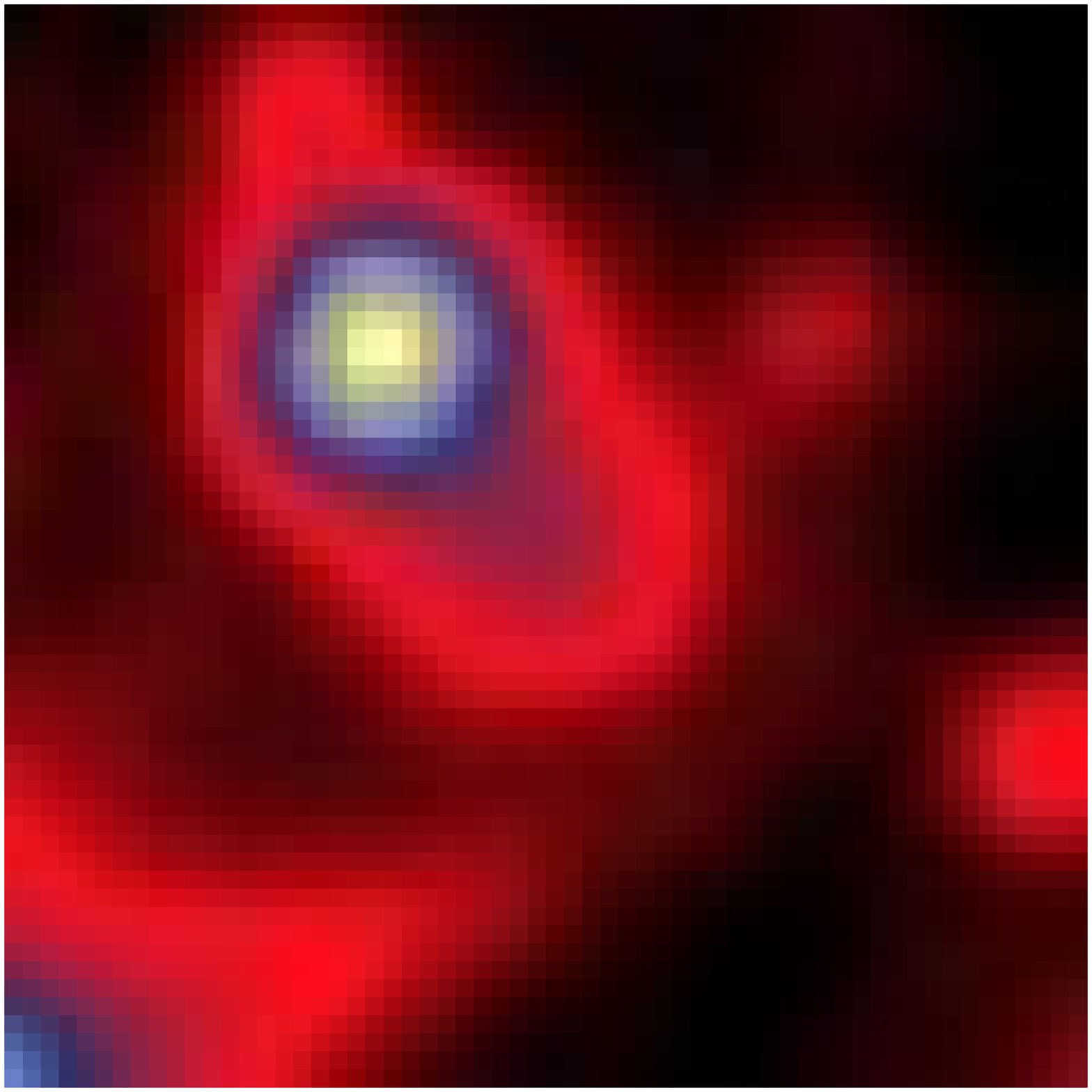}  \FGb{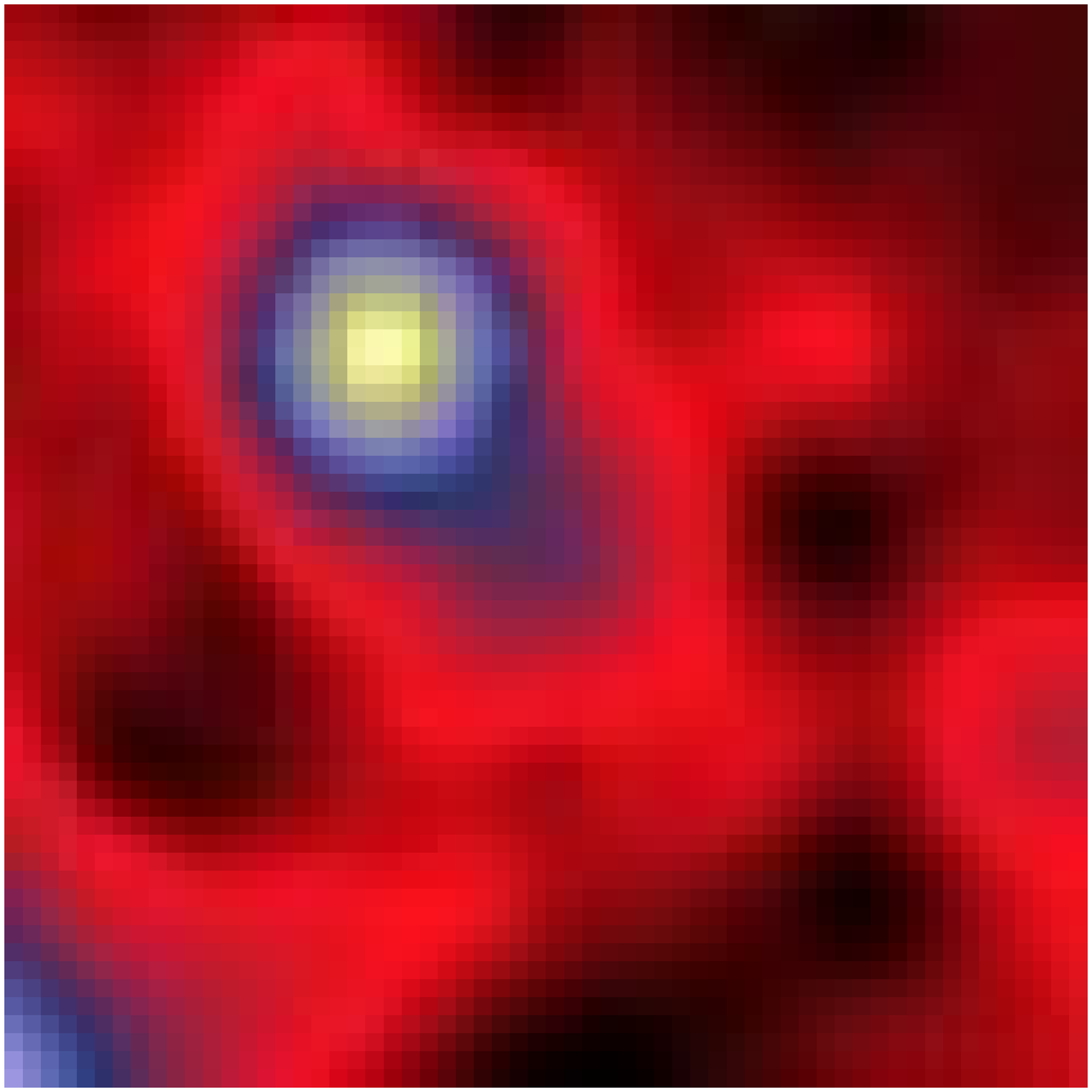}  \FGb{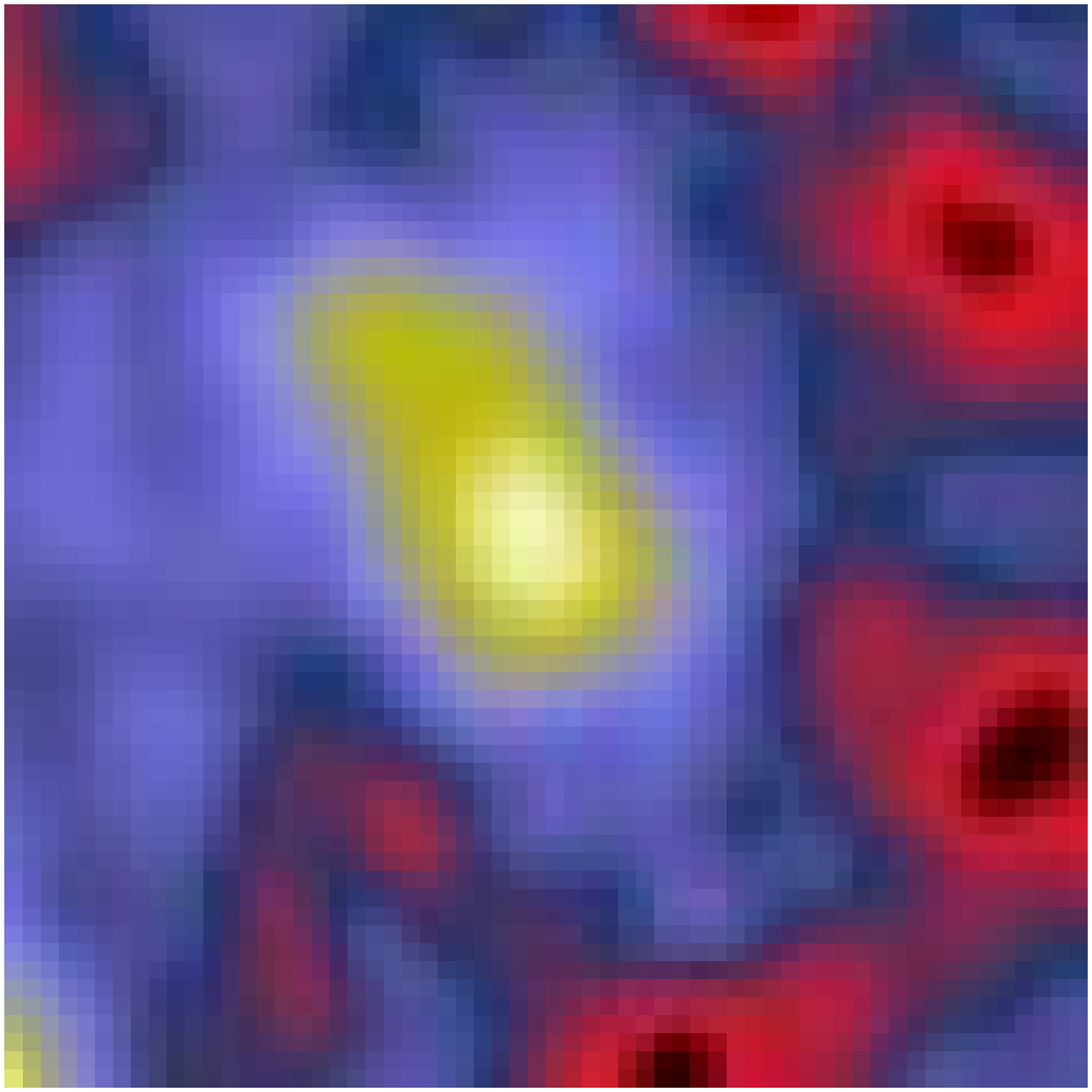}  \FGb{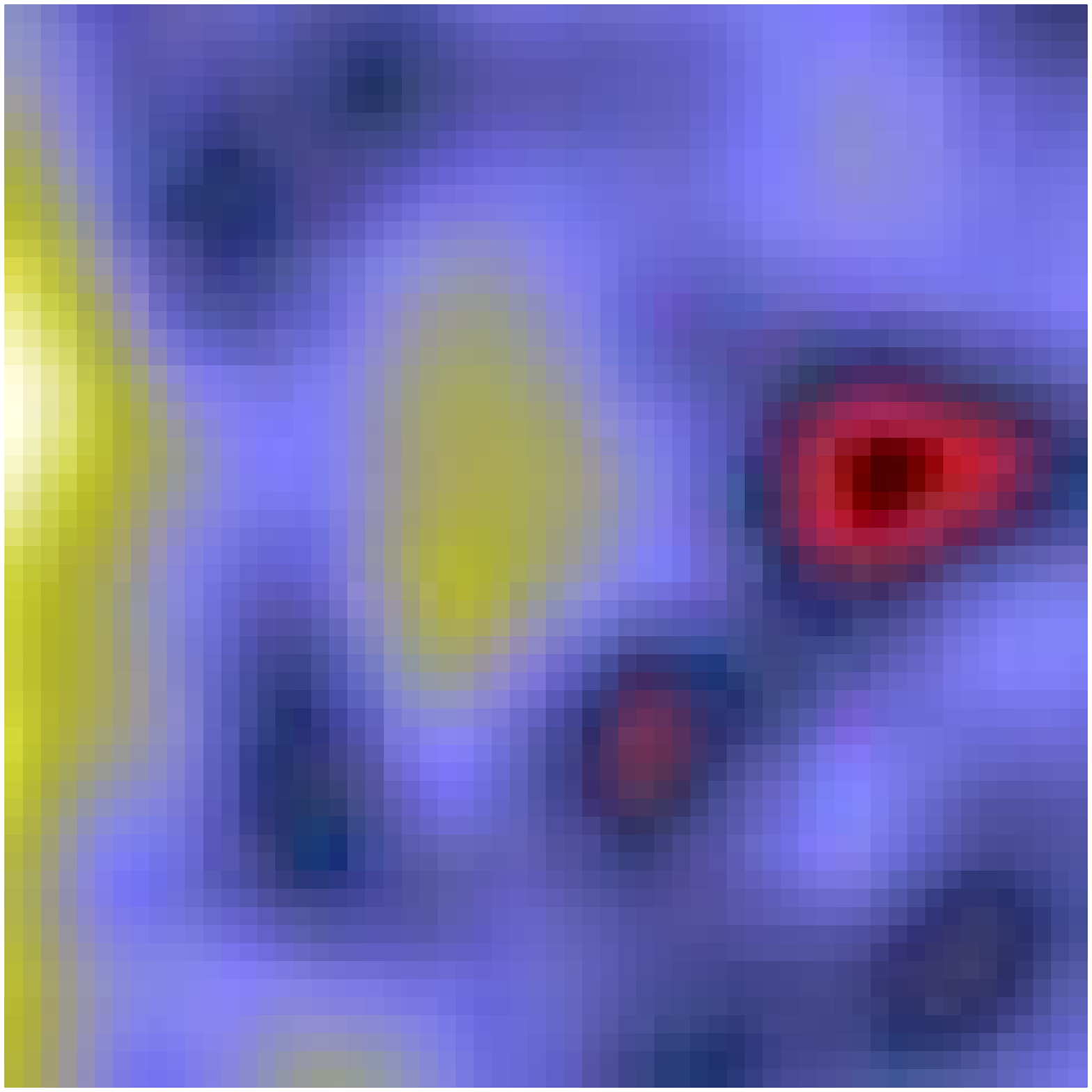} \\  {\#16   NBZ5015034}  \\
  \FGb{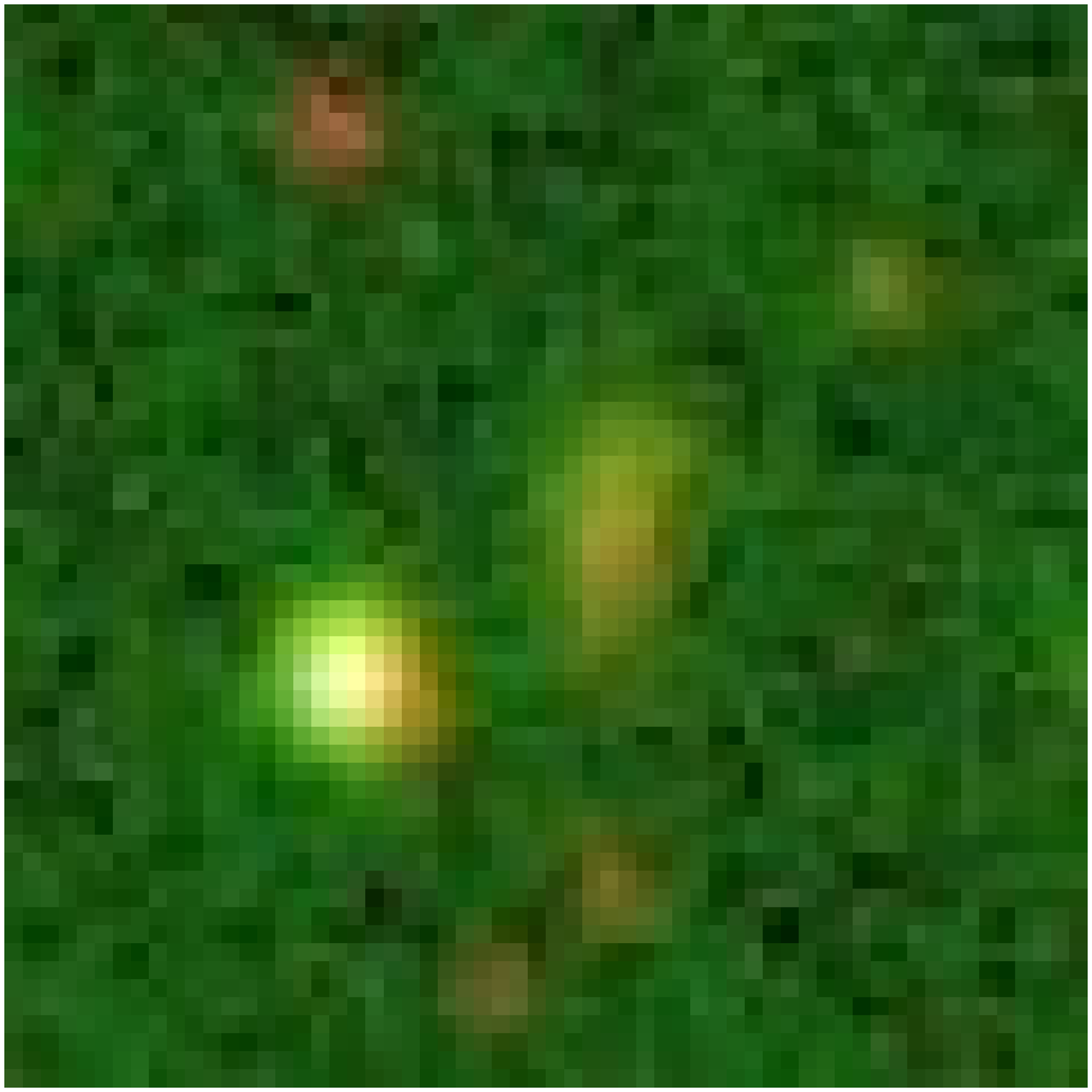}  \FGb{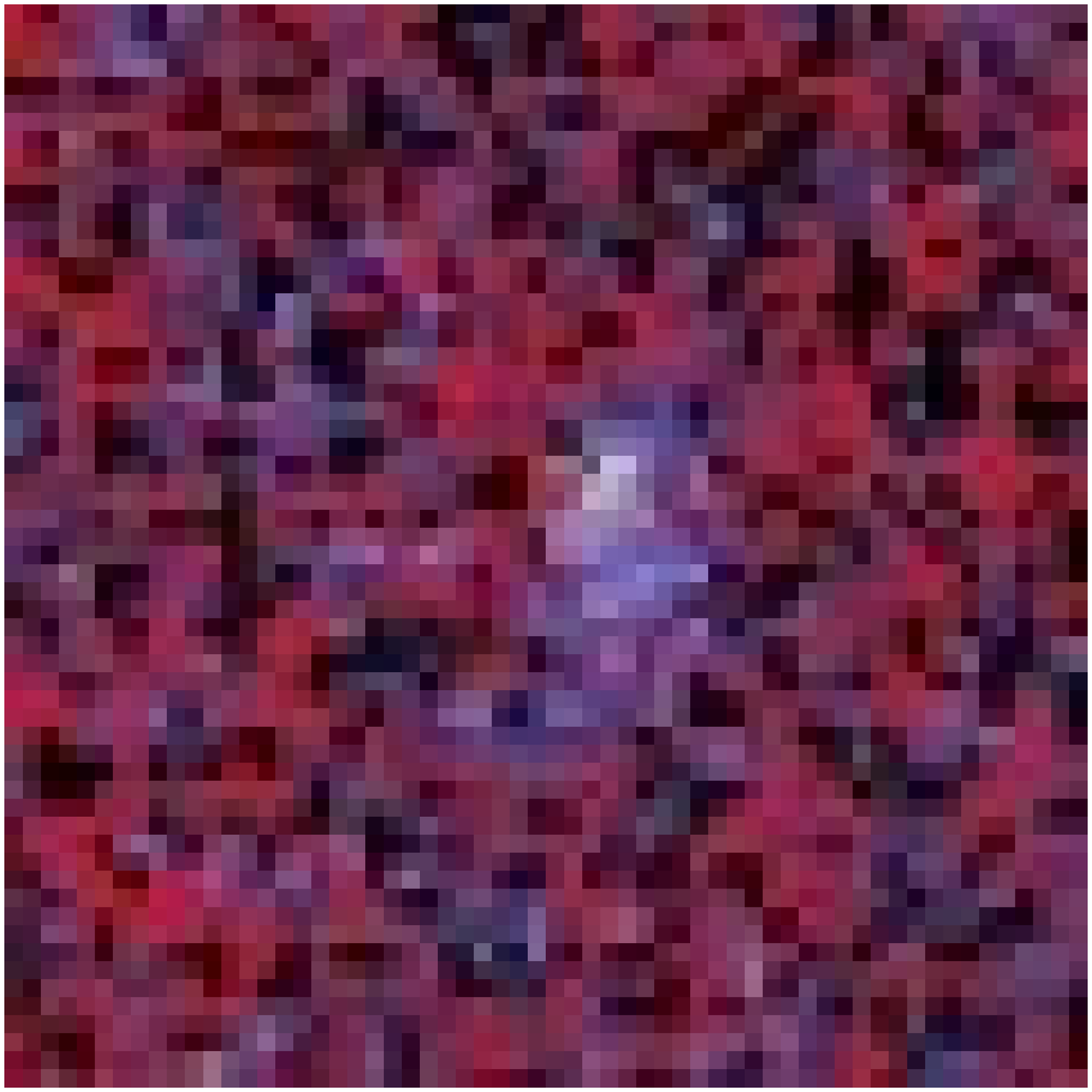}  \FGb{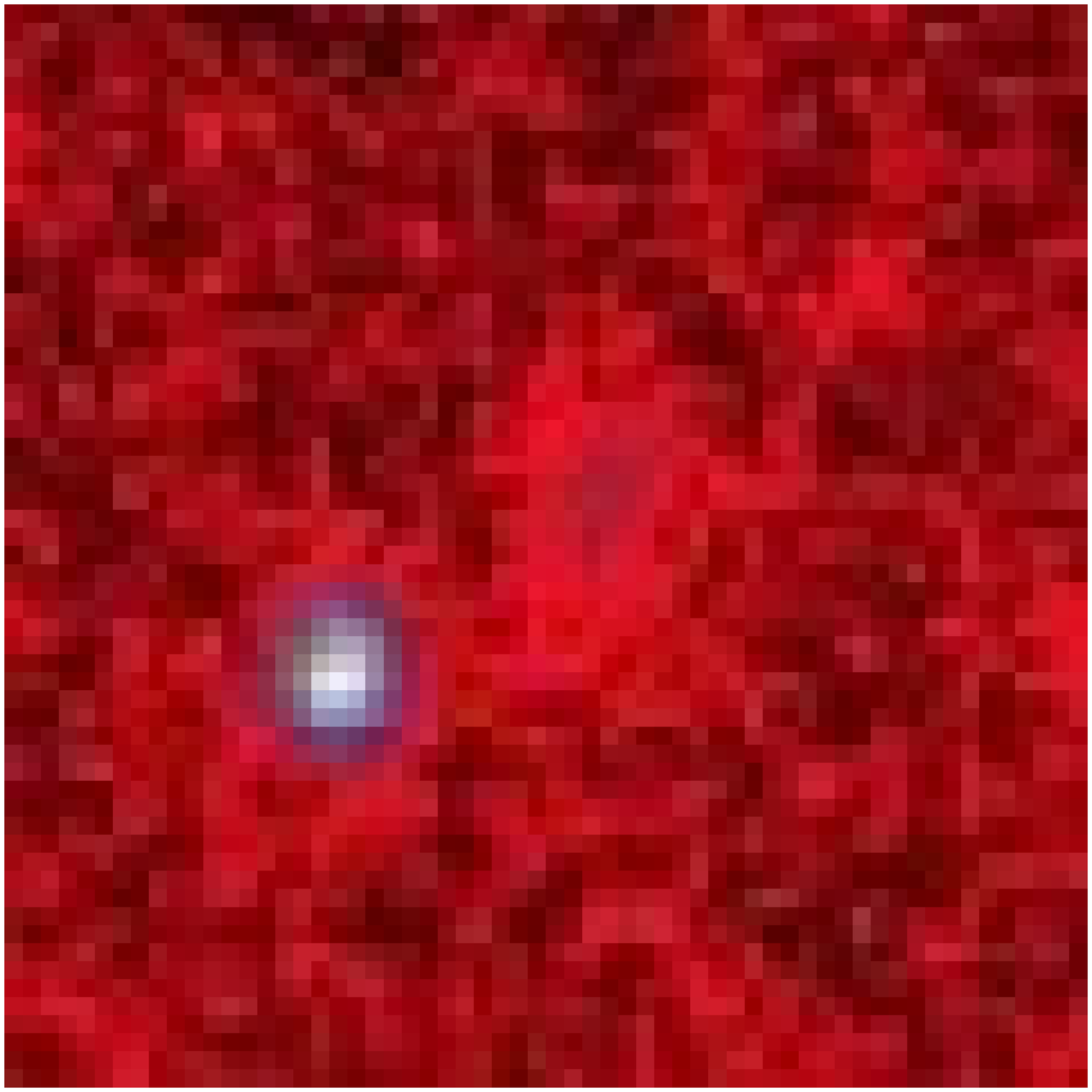}  \FGb{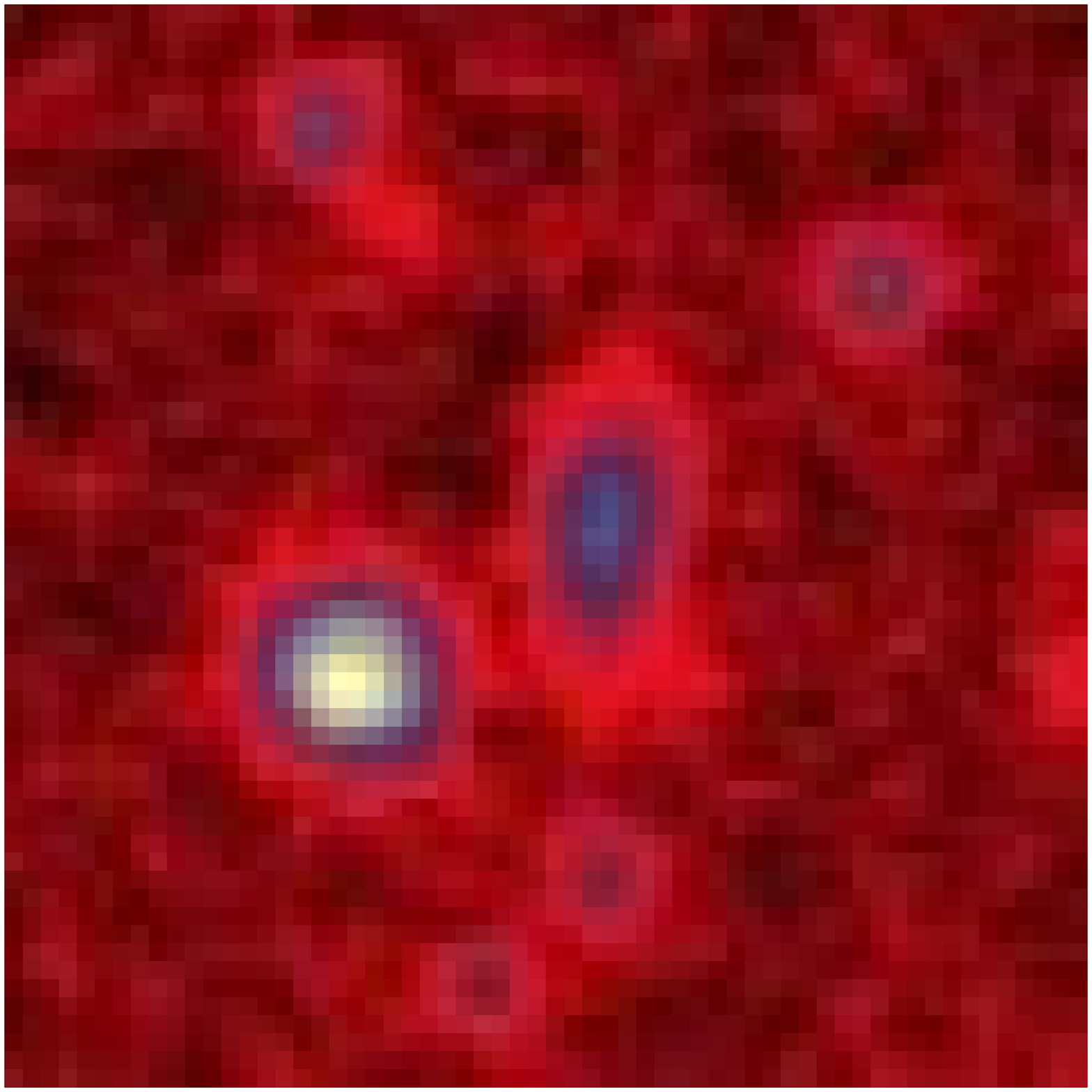}  \FGb{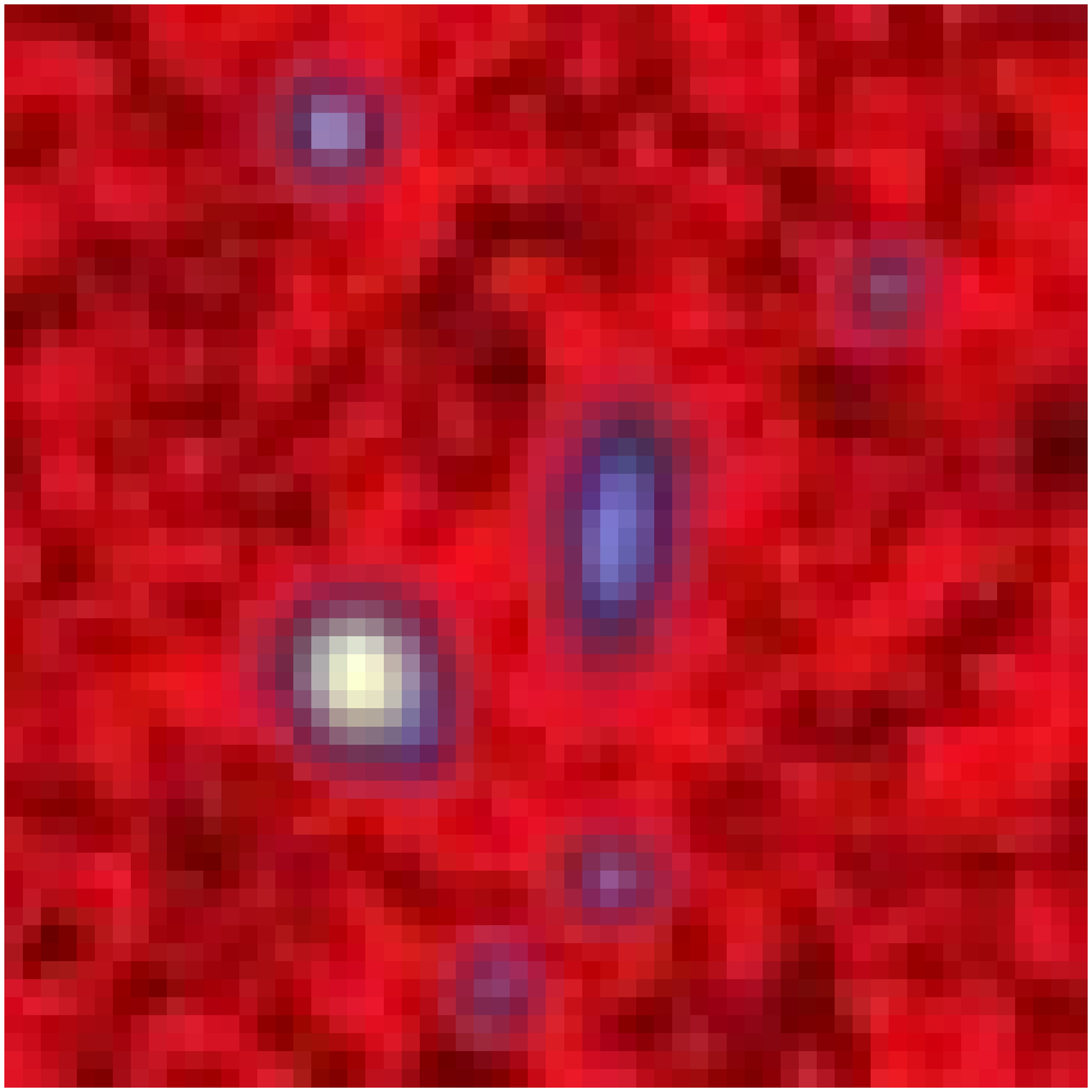}  \FGb{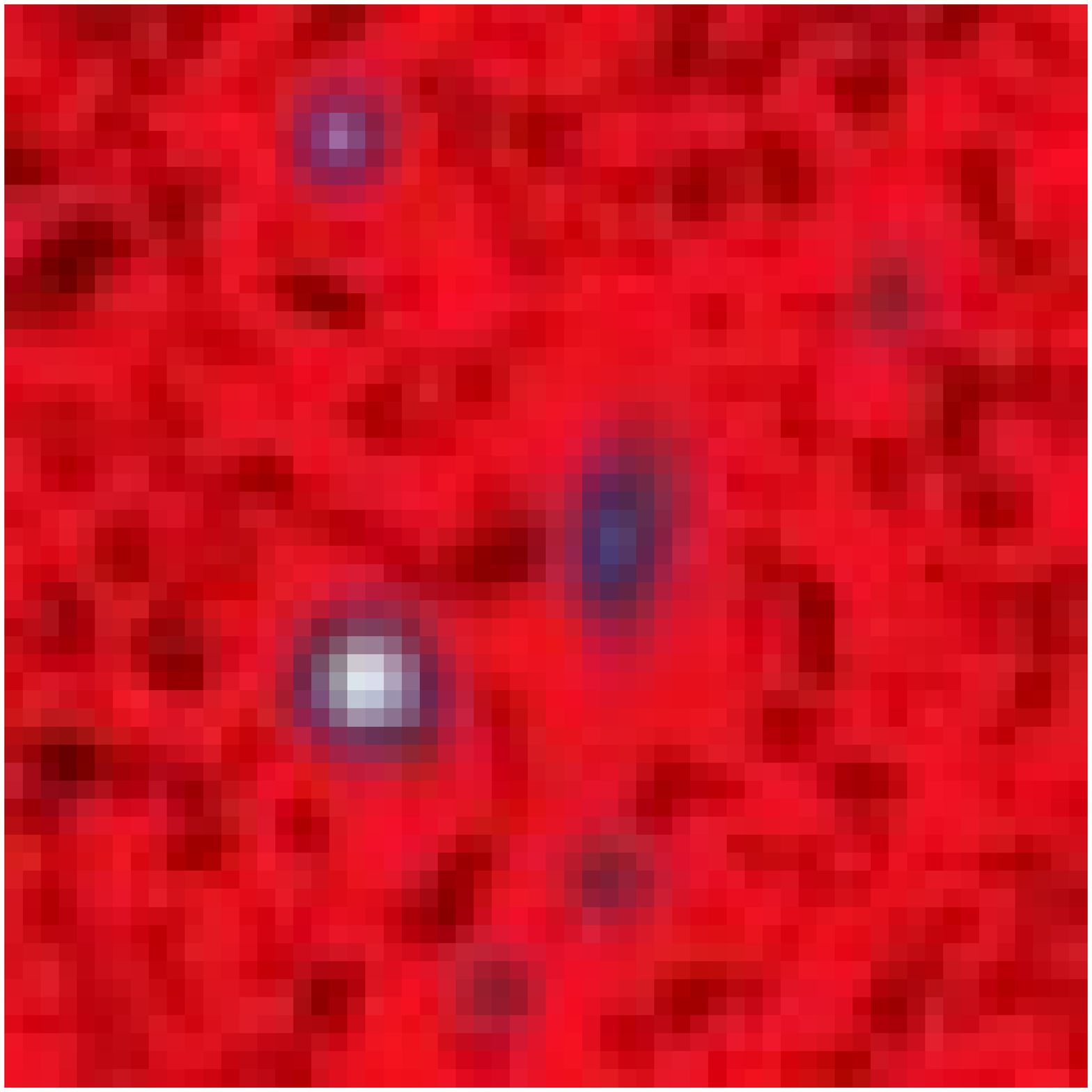}  \FGb{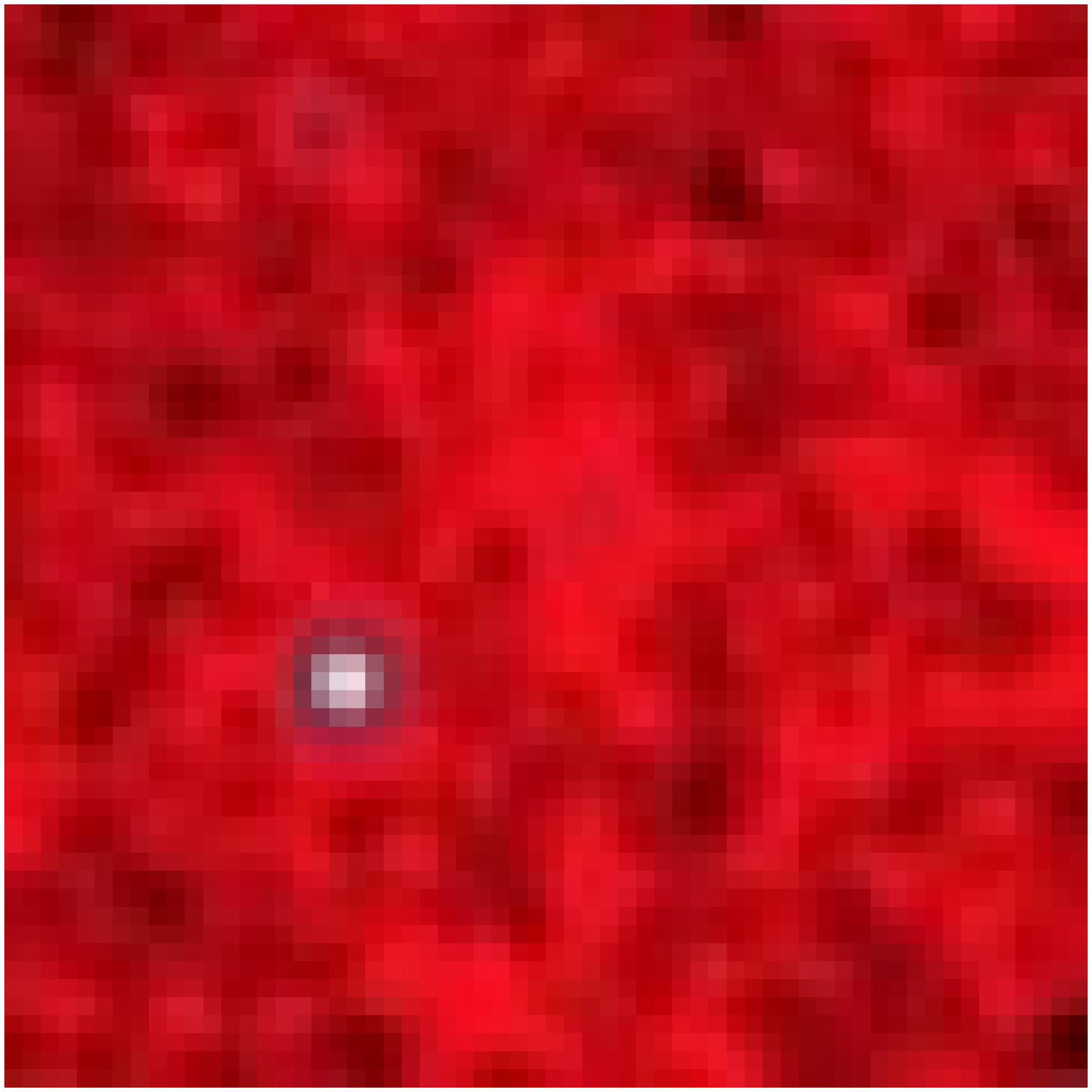}  \FGb{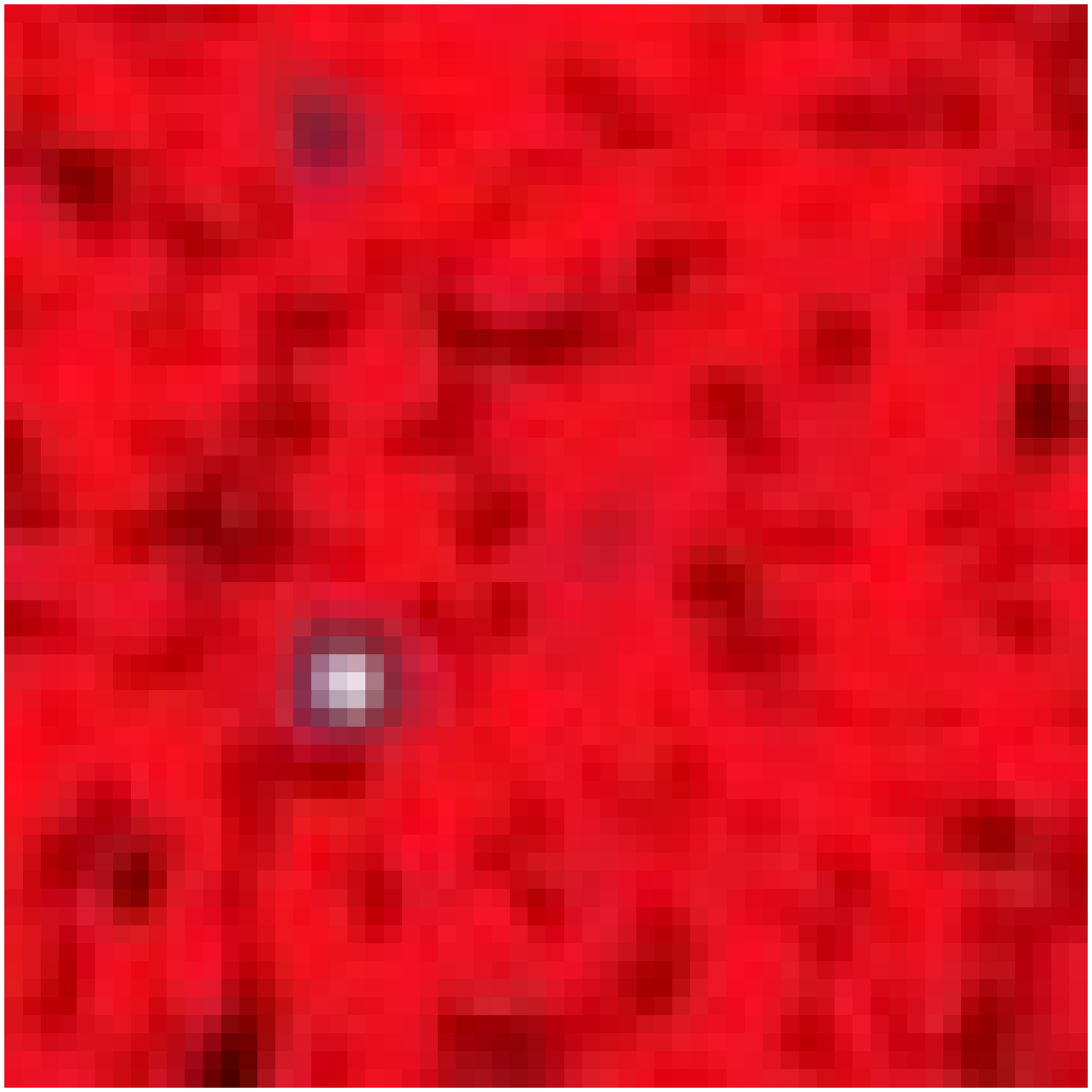}  \FGb{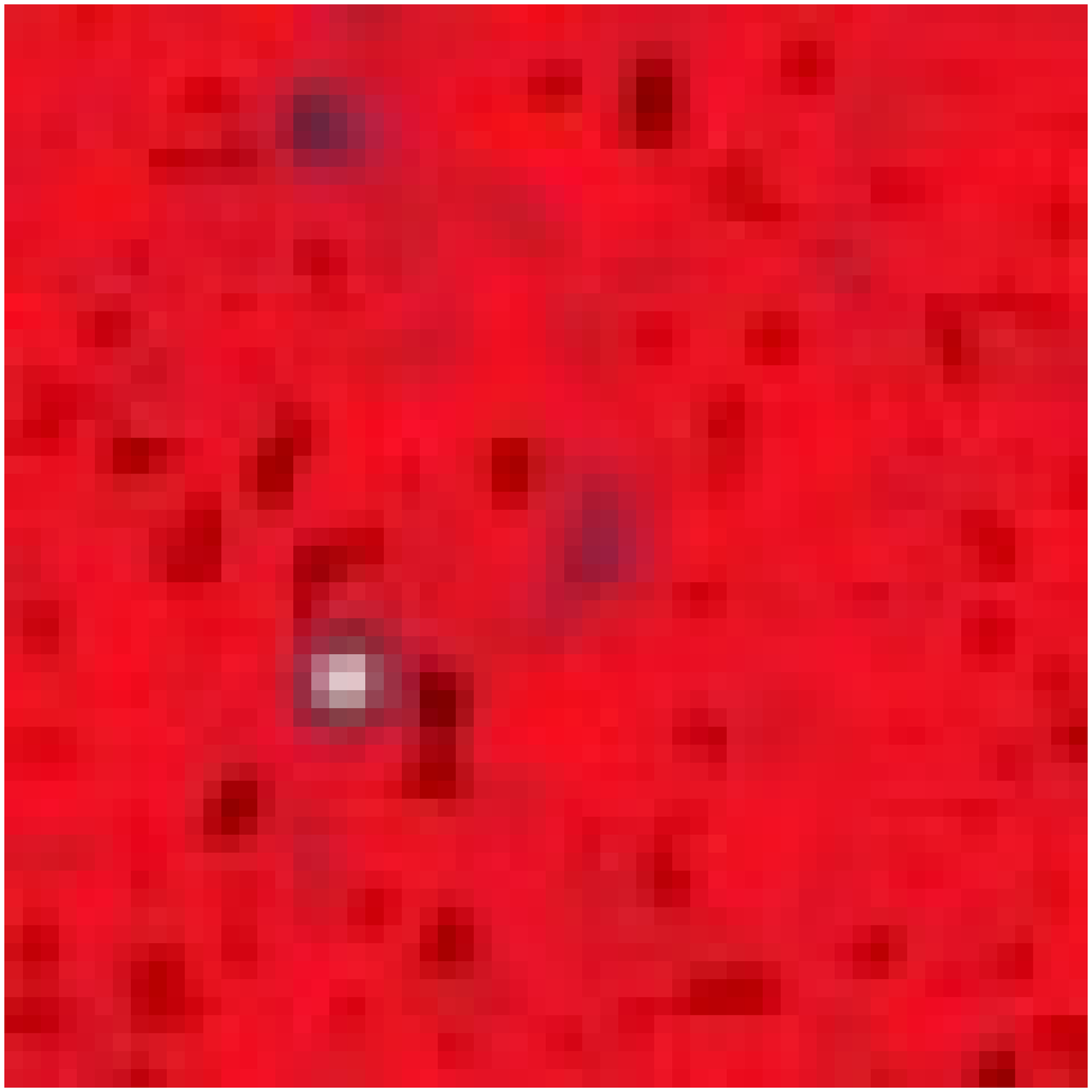}  \FGb{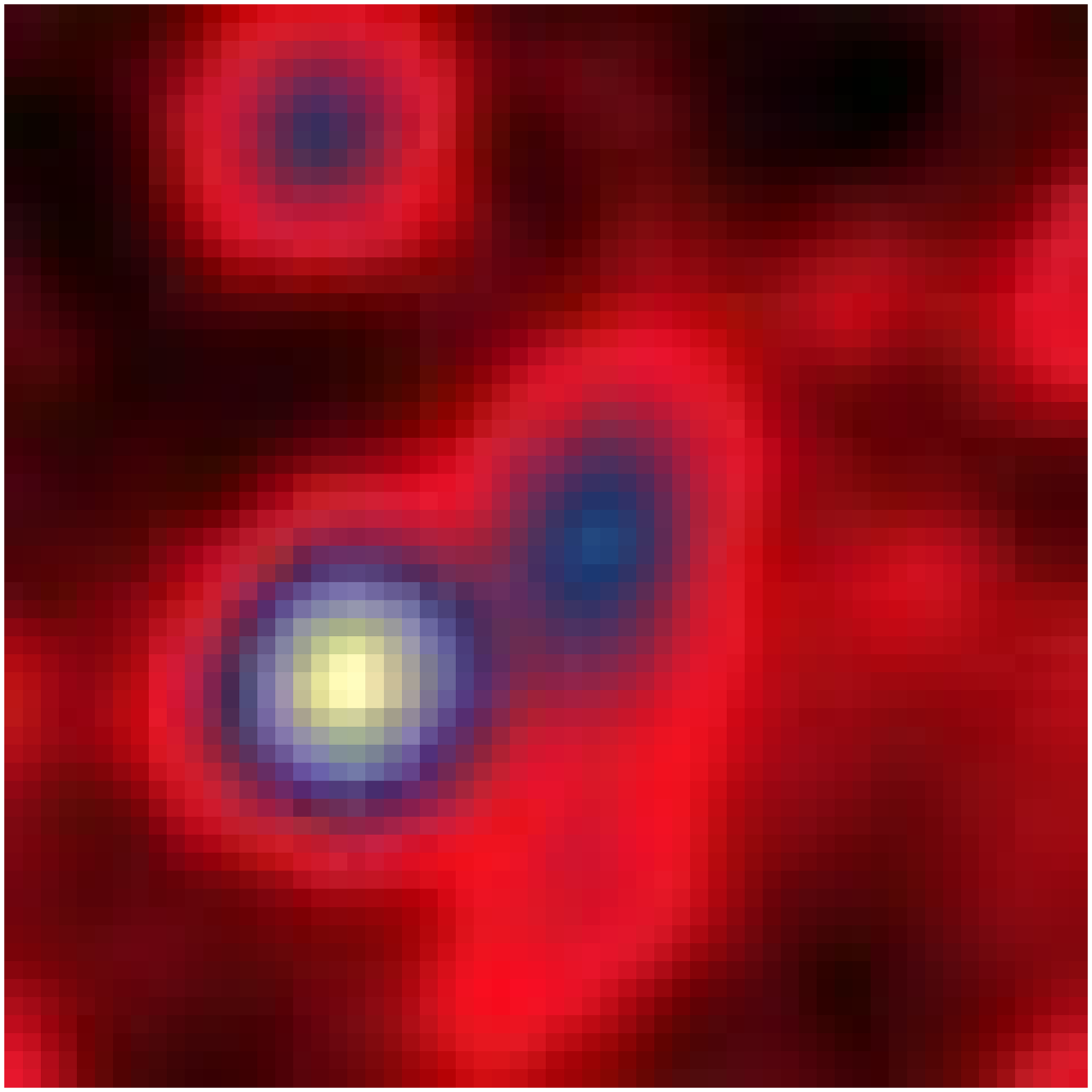}  \FGb{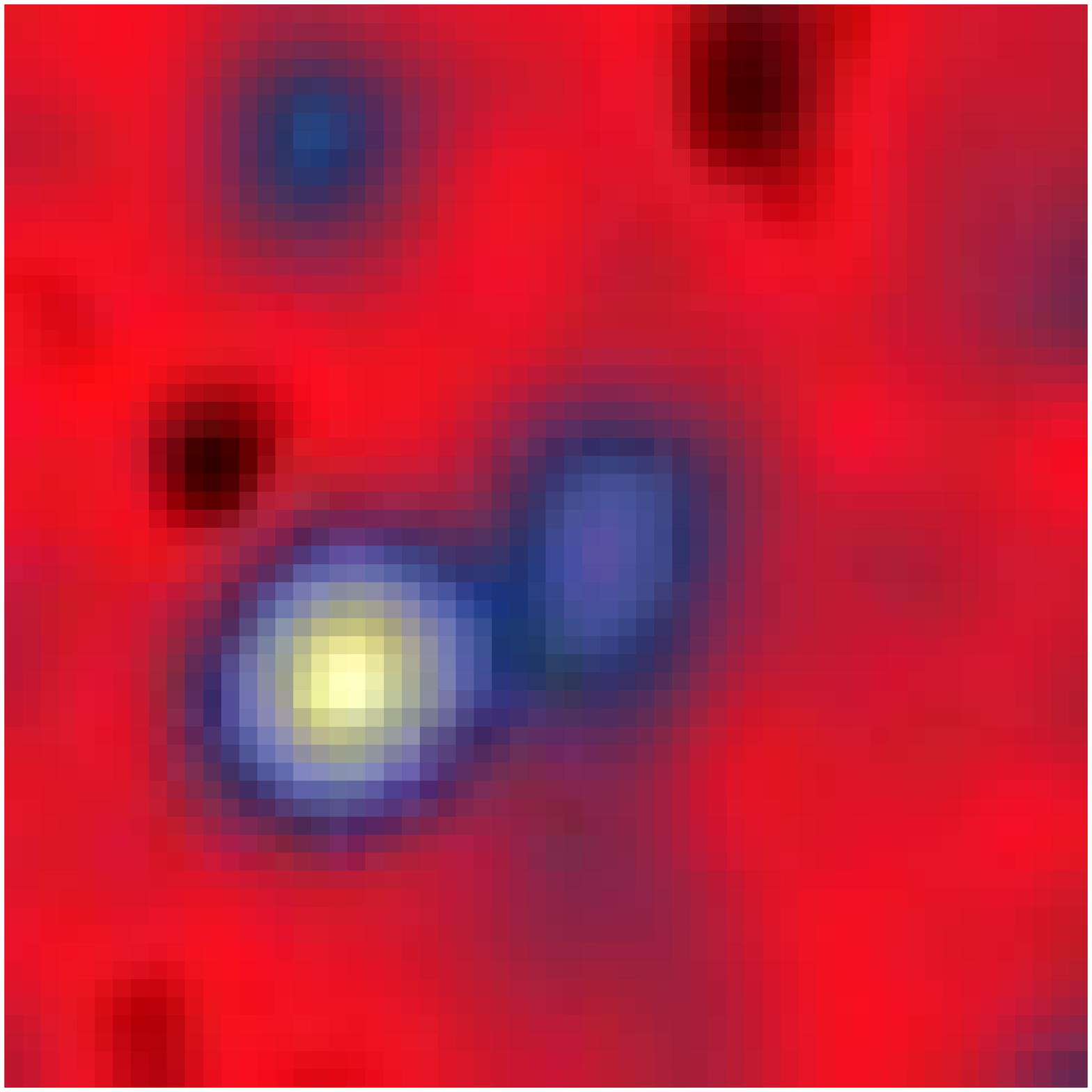}  \FGb{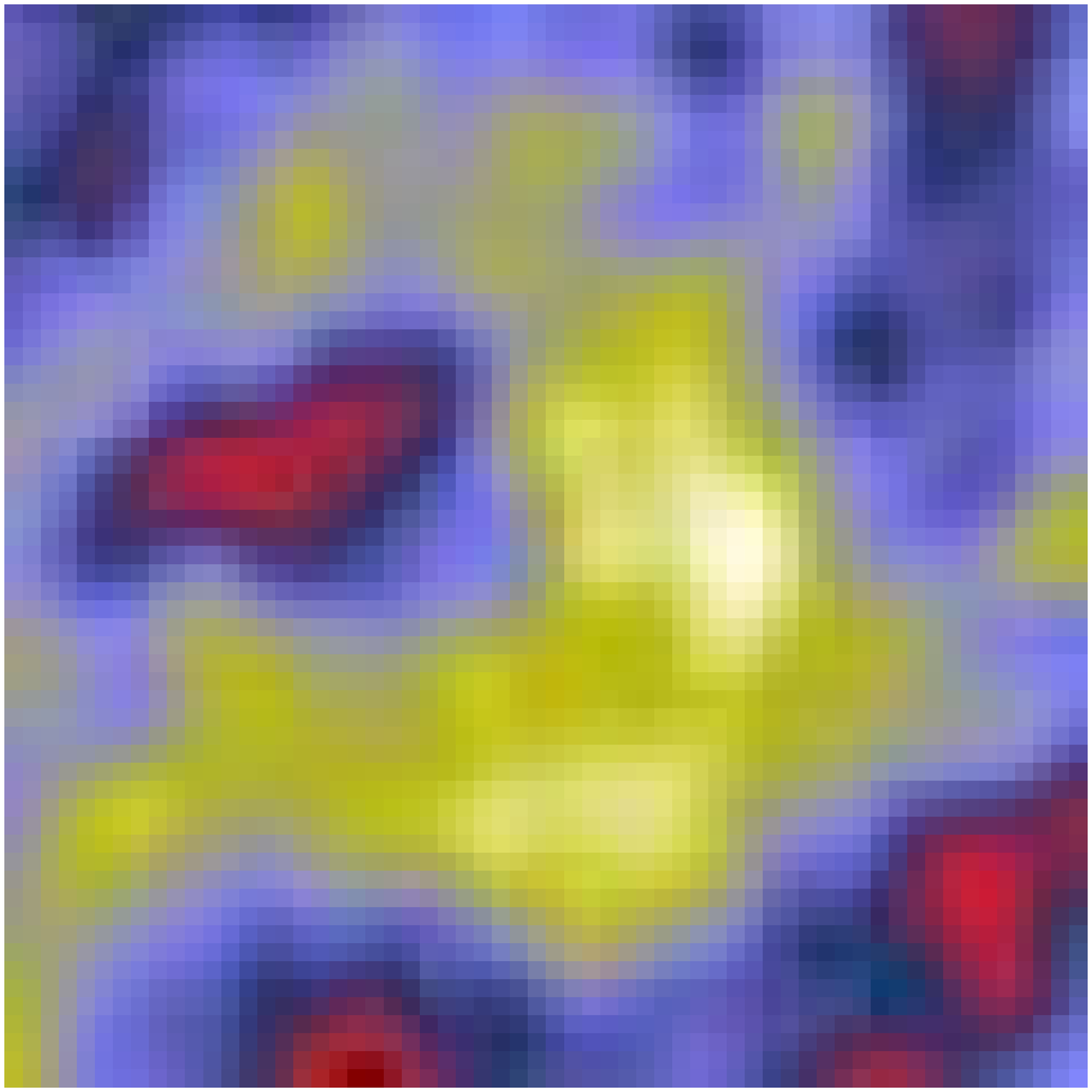}  \FGb{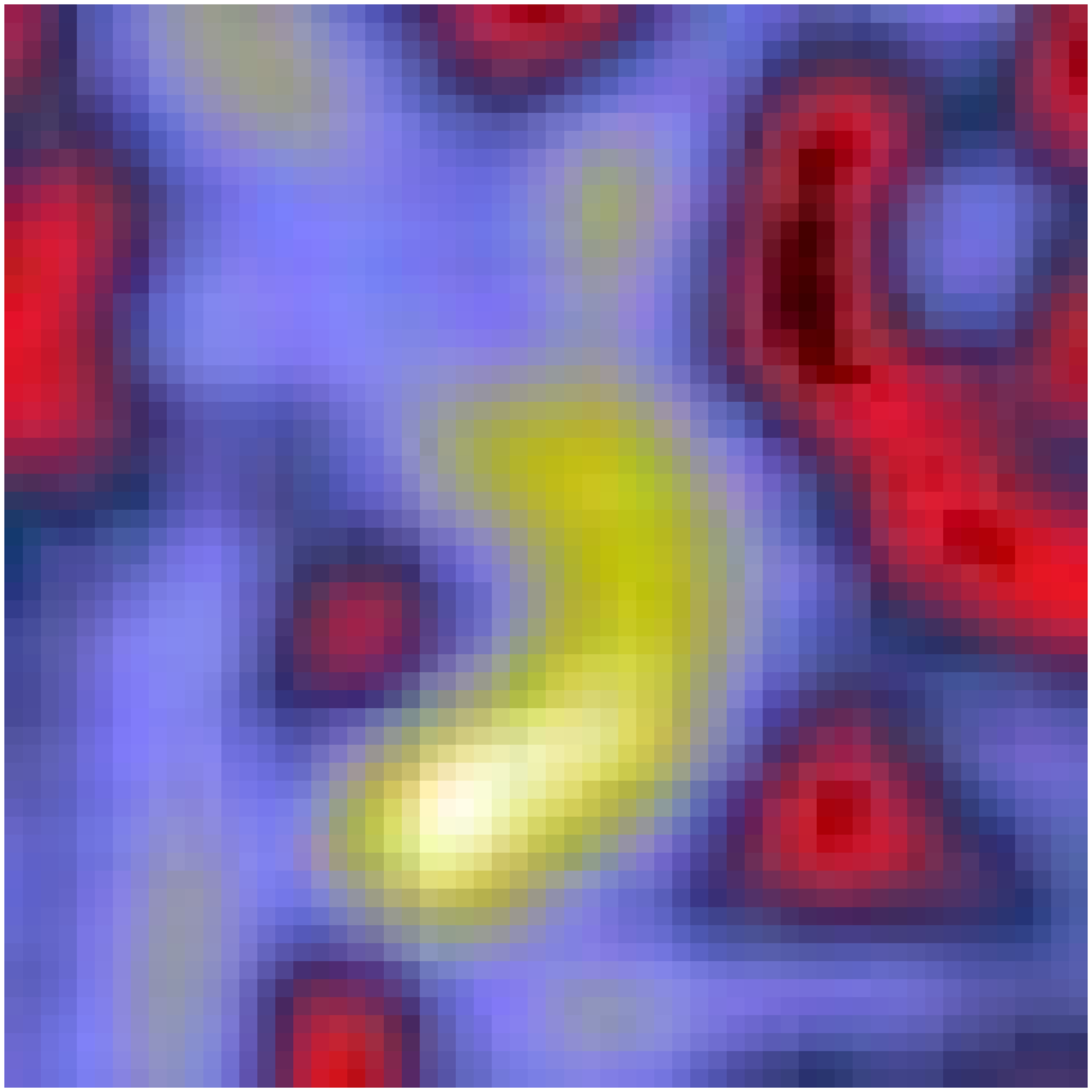} \\  {\#17   NCIA012578}  \\
  \FGb{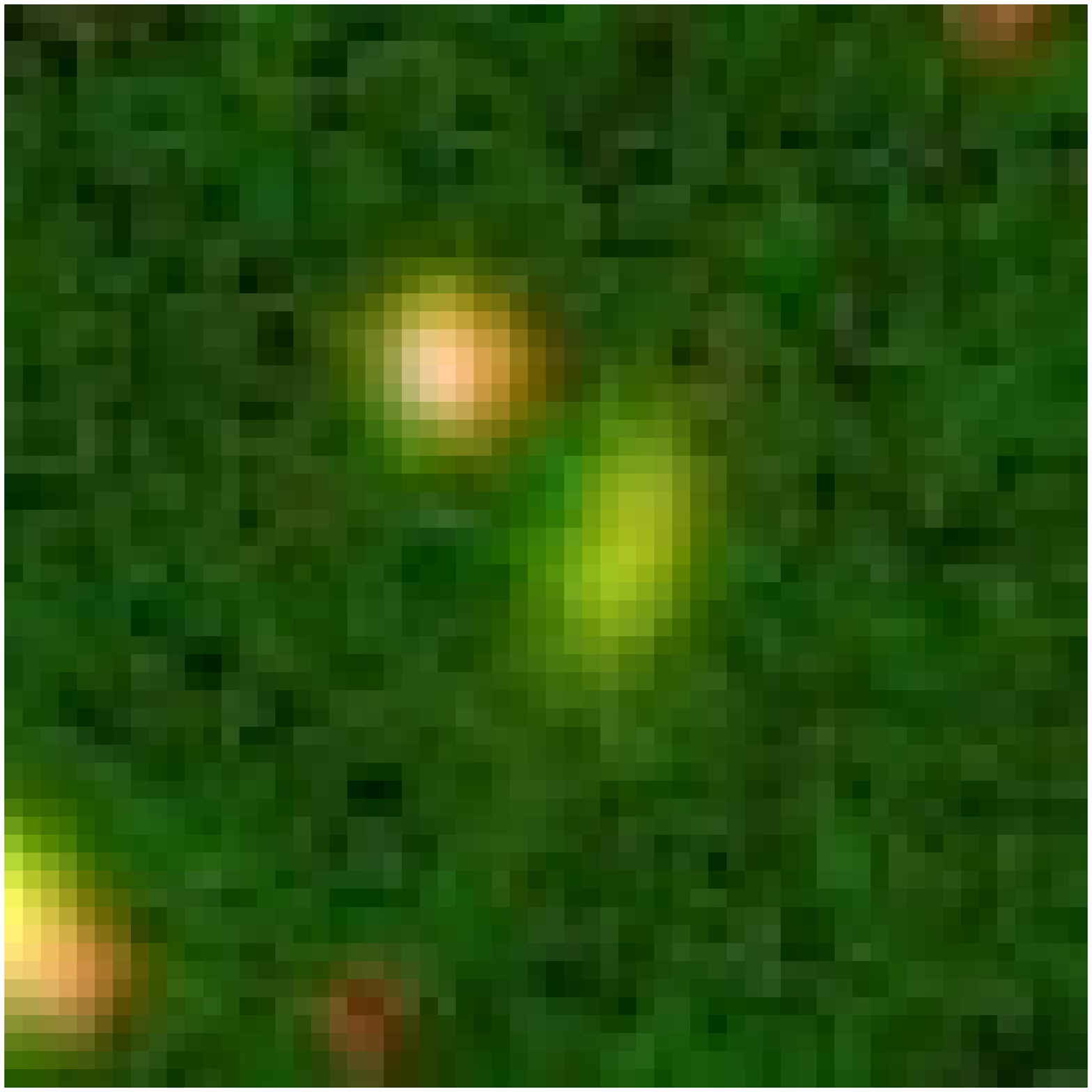}  \FGb{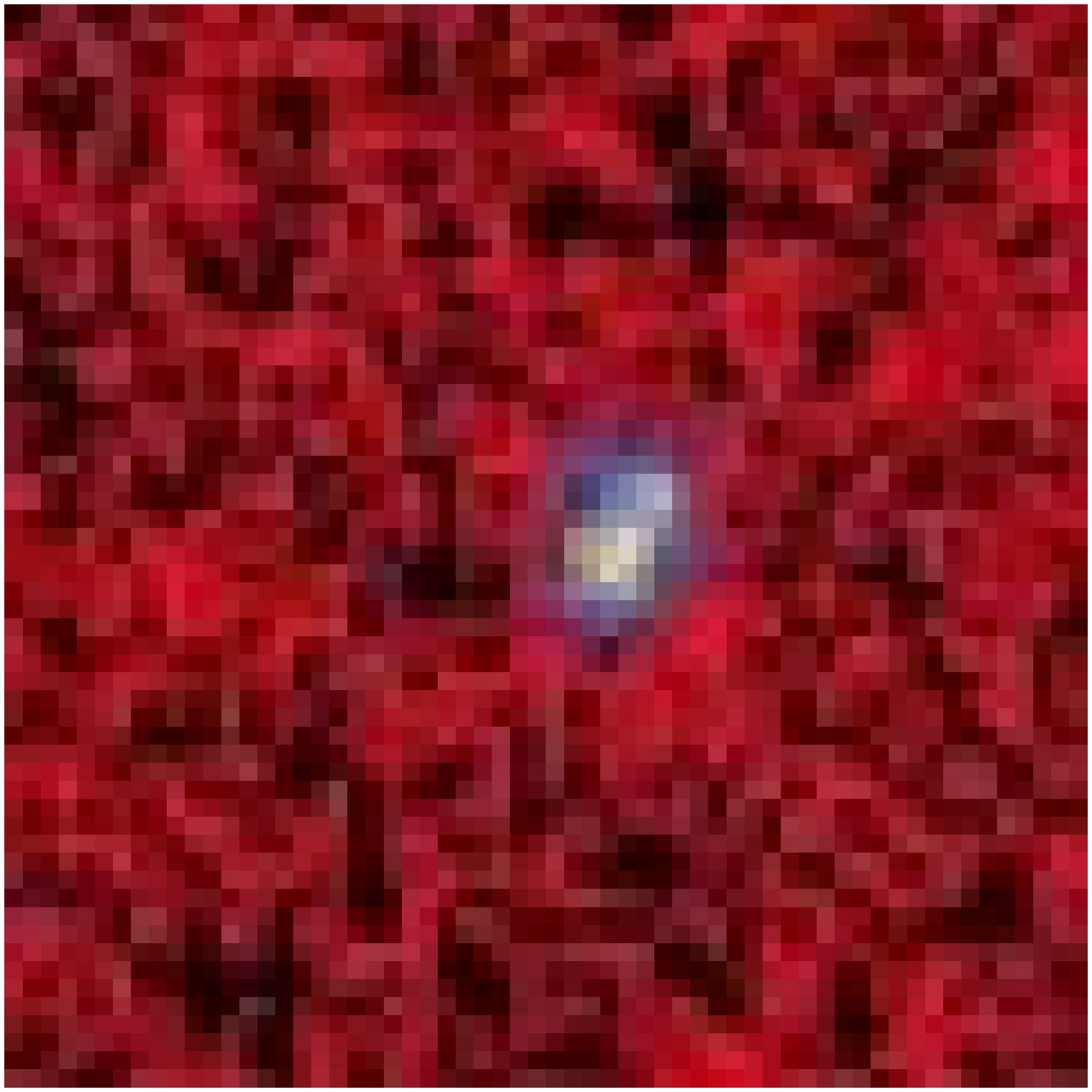}  \FGb{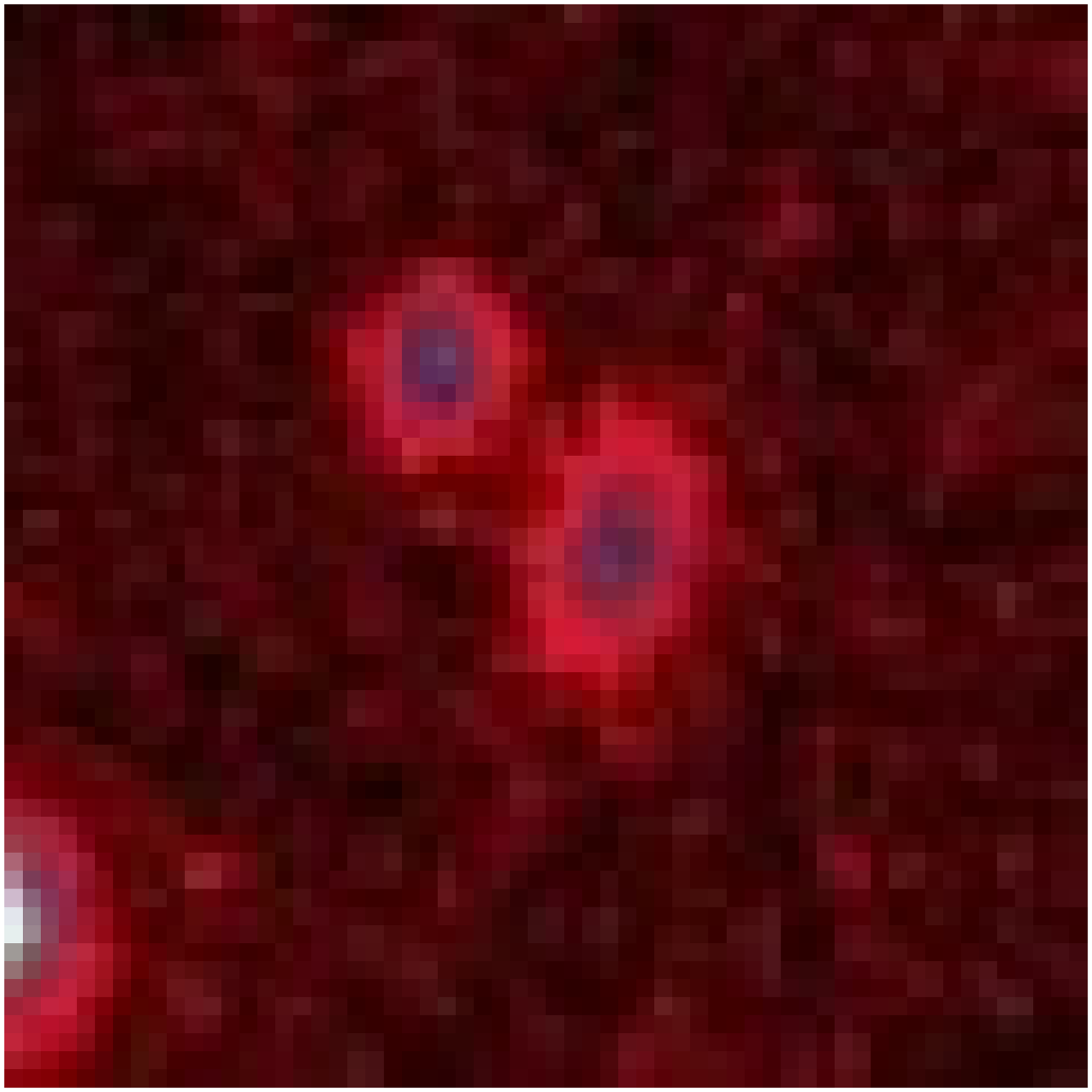}  \FGb{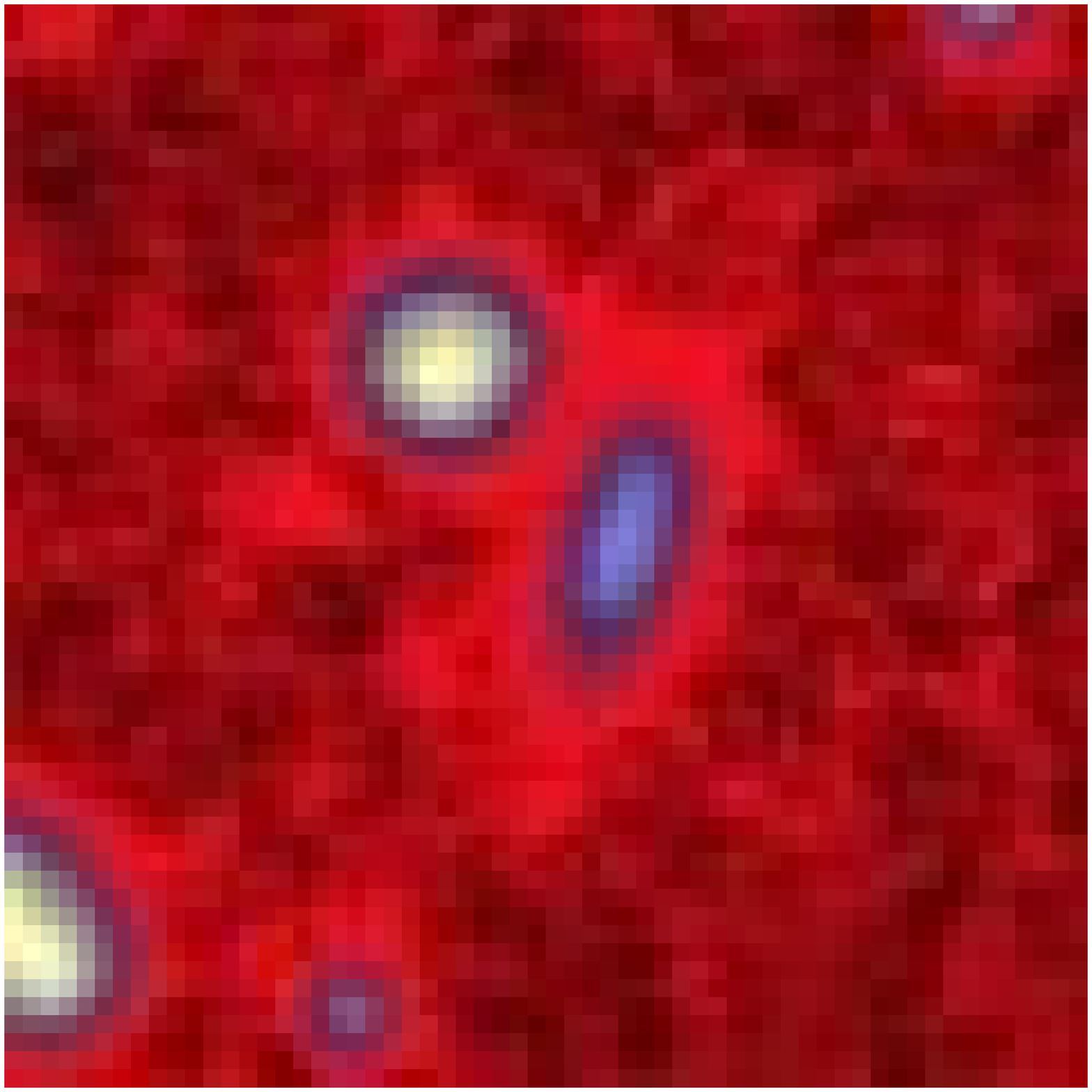}  \FGb{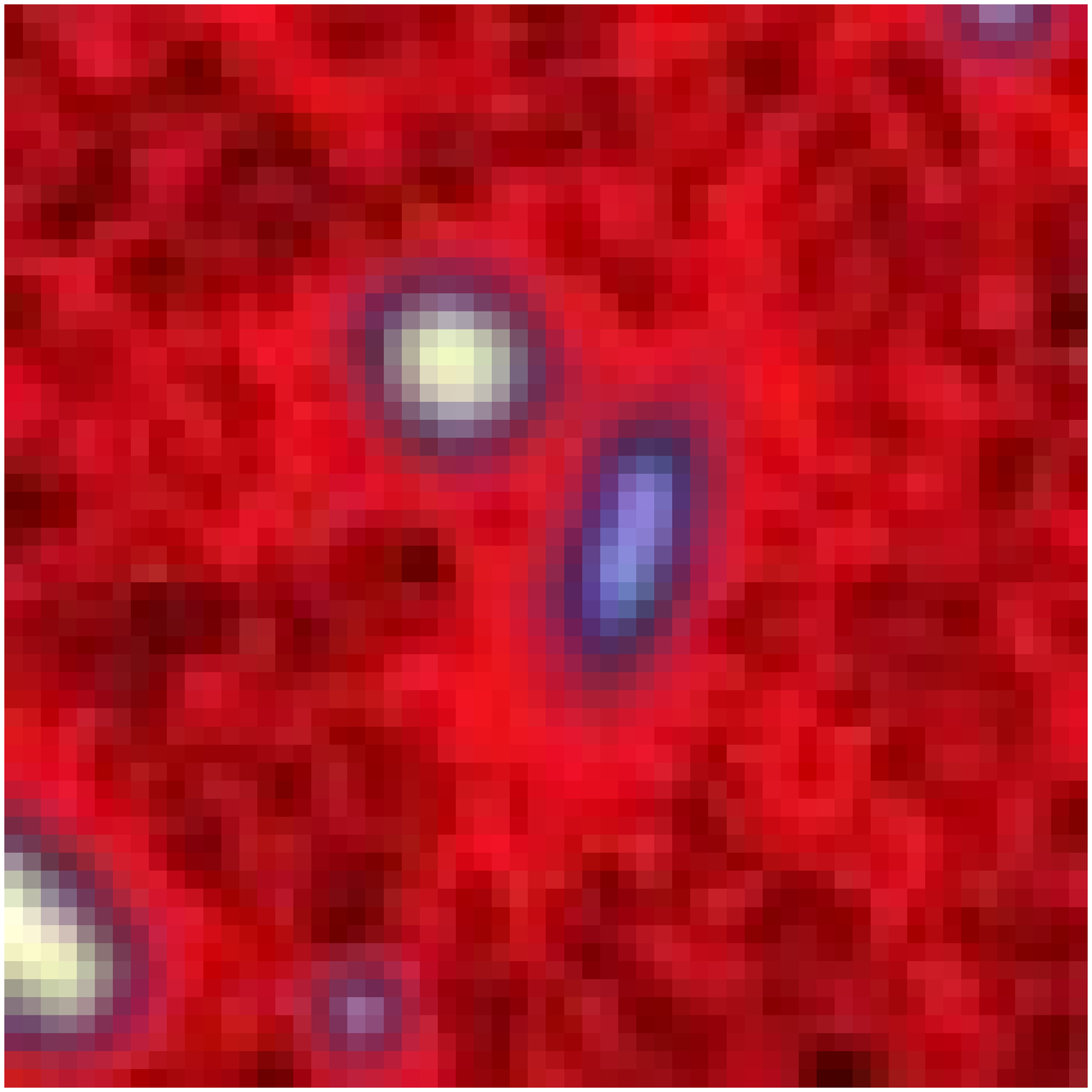}  \FGb{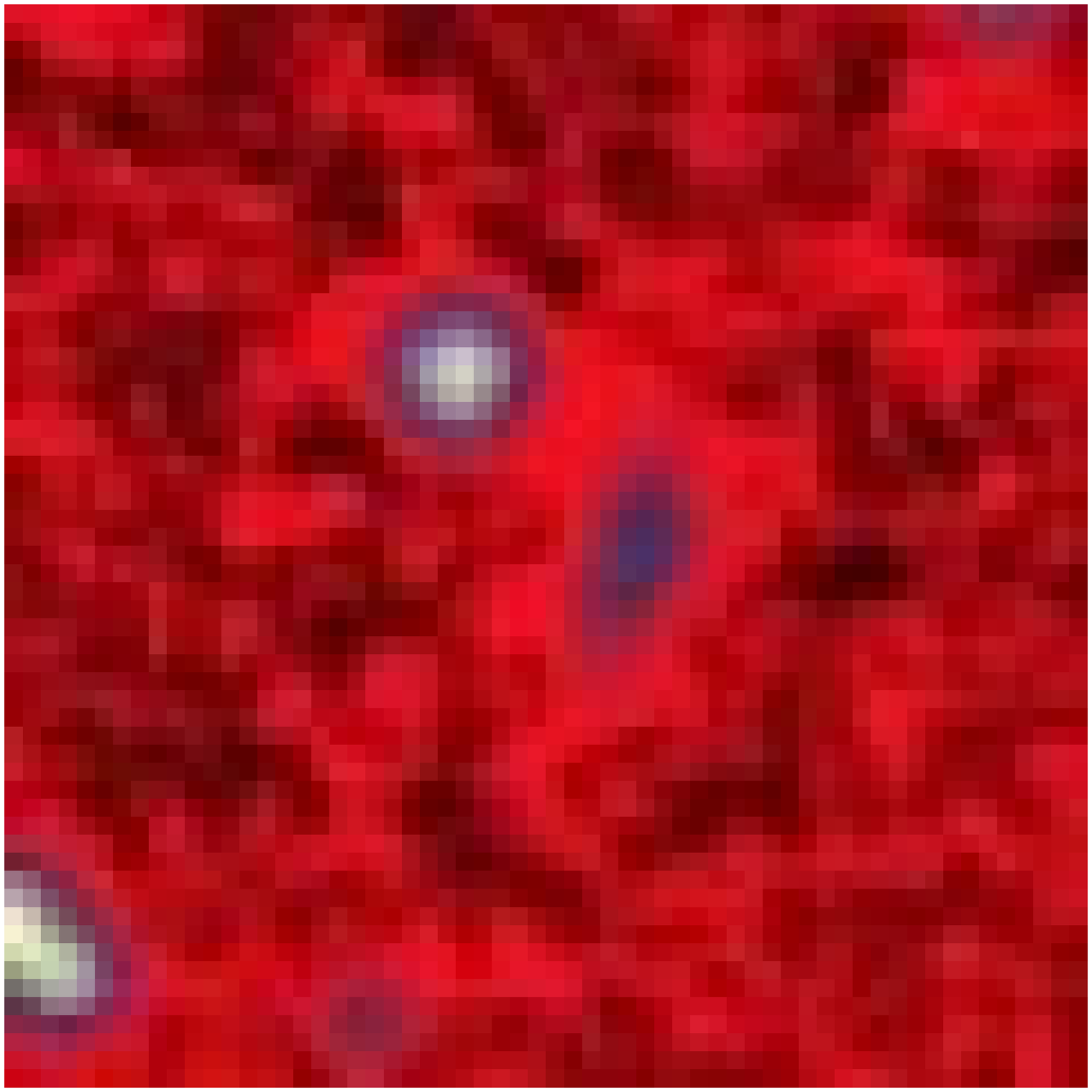}  \FGb{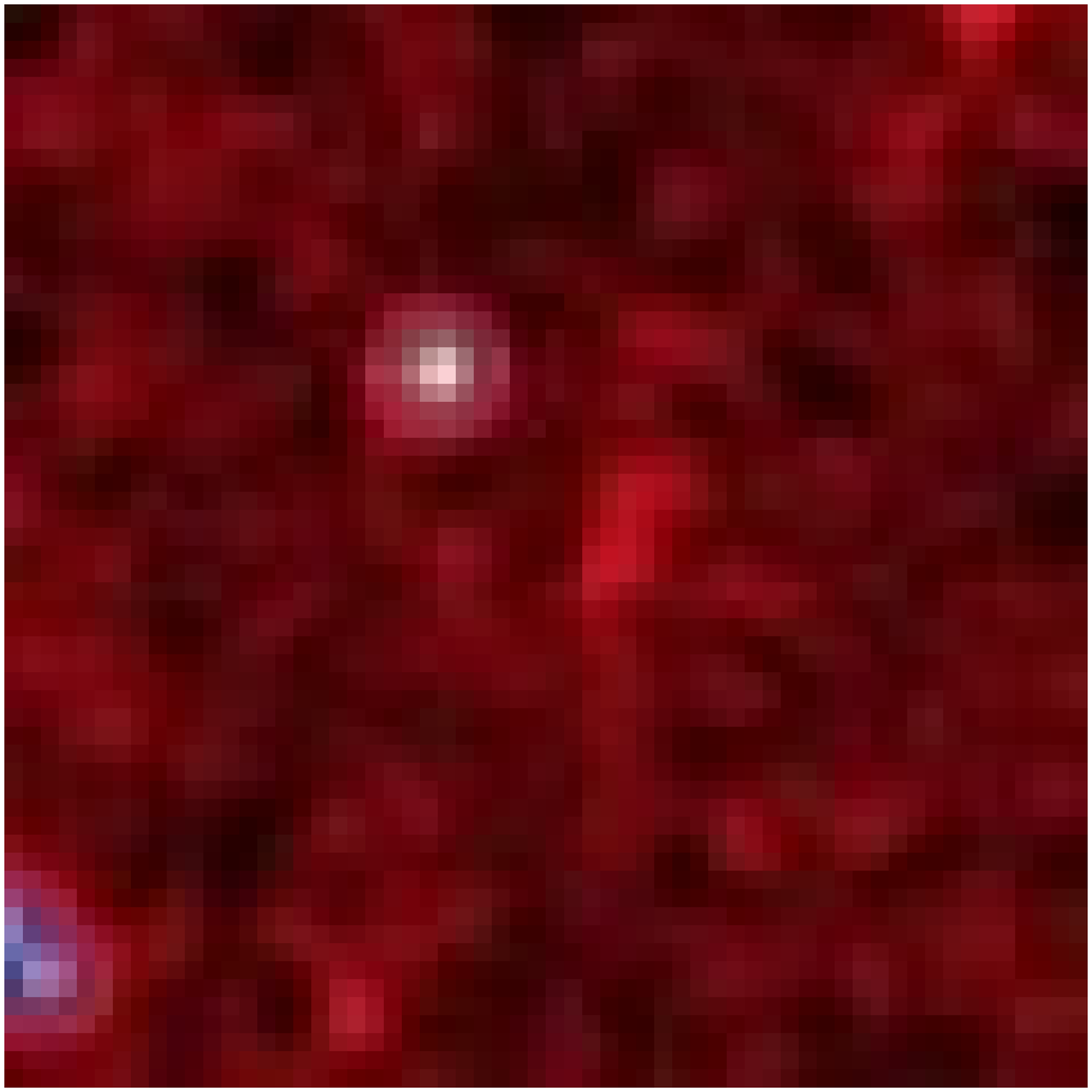}  \FGb{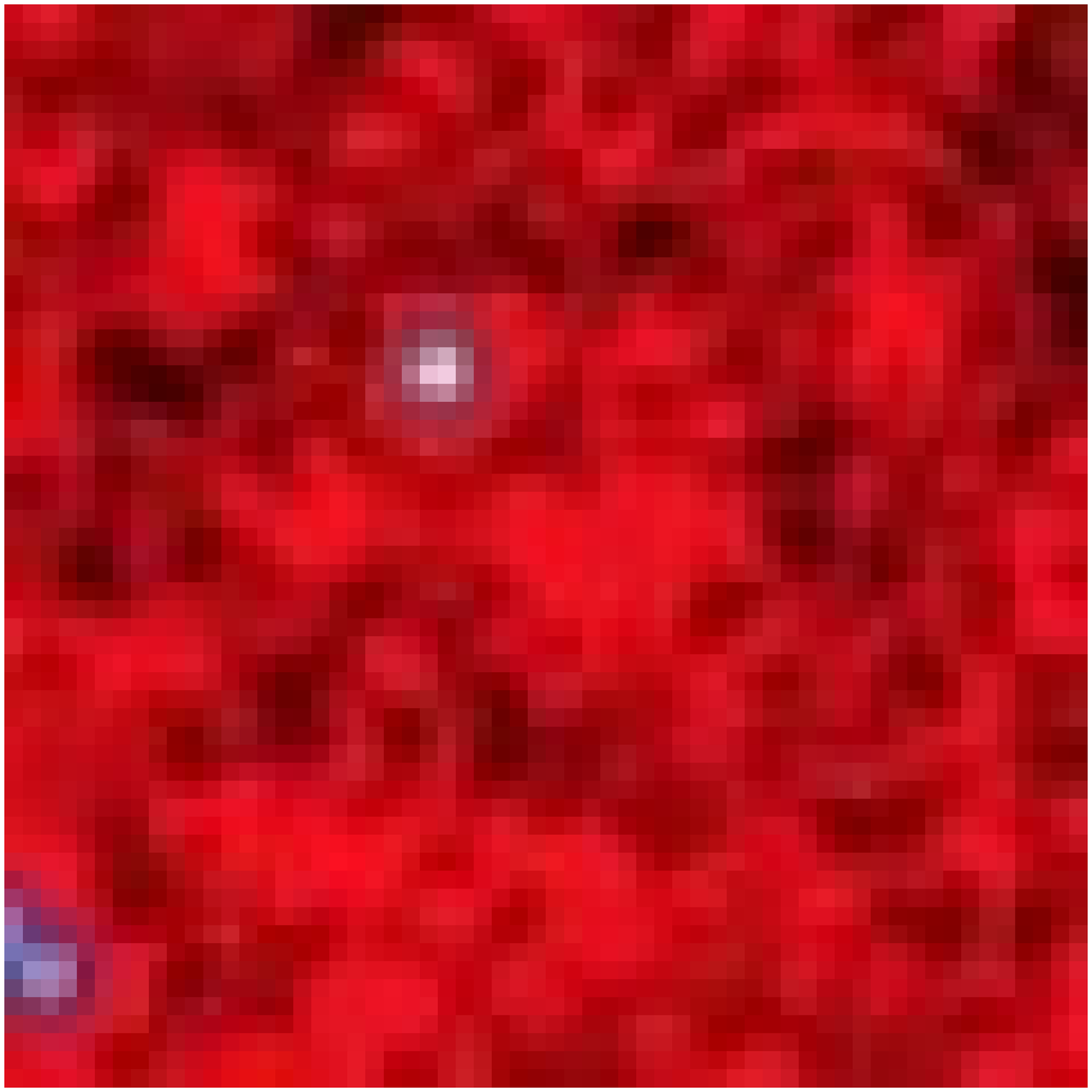}  \FGb{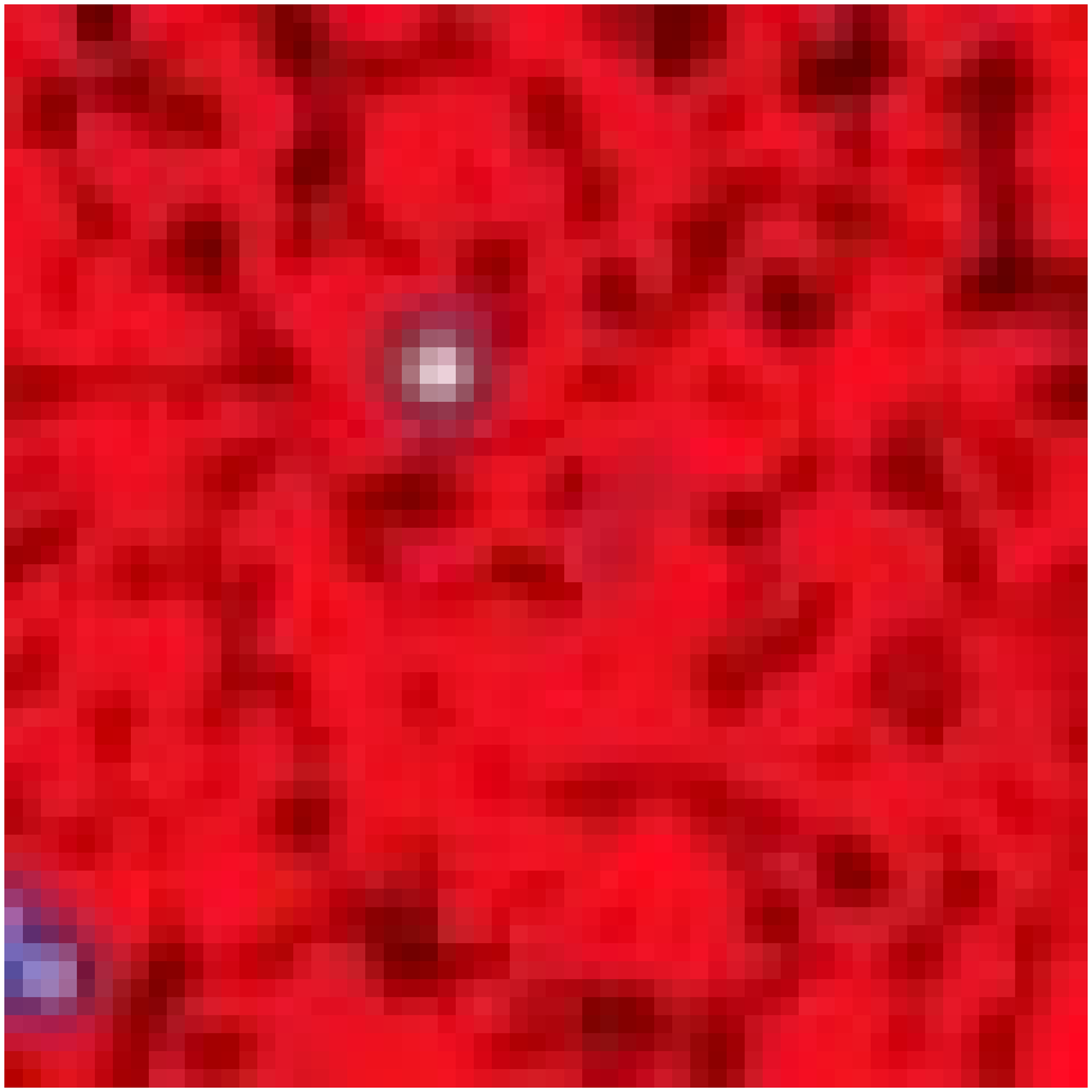}  \FGb{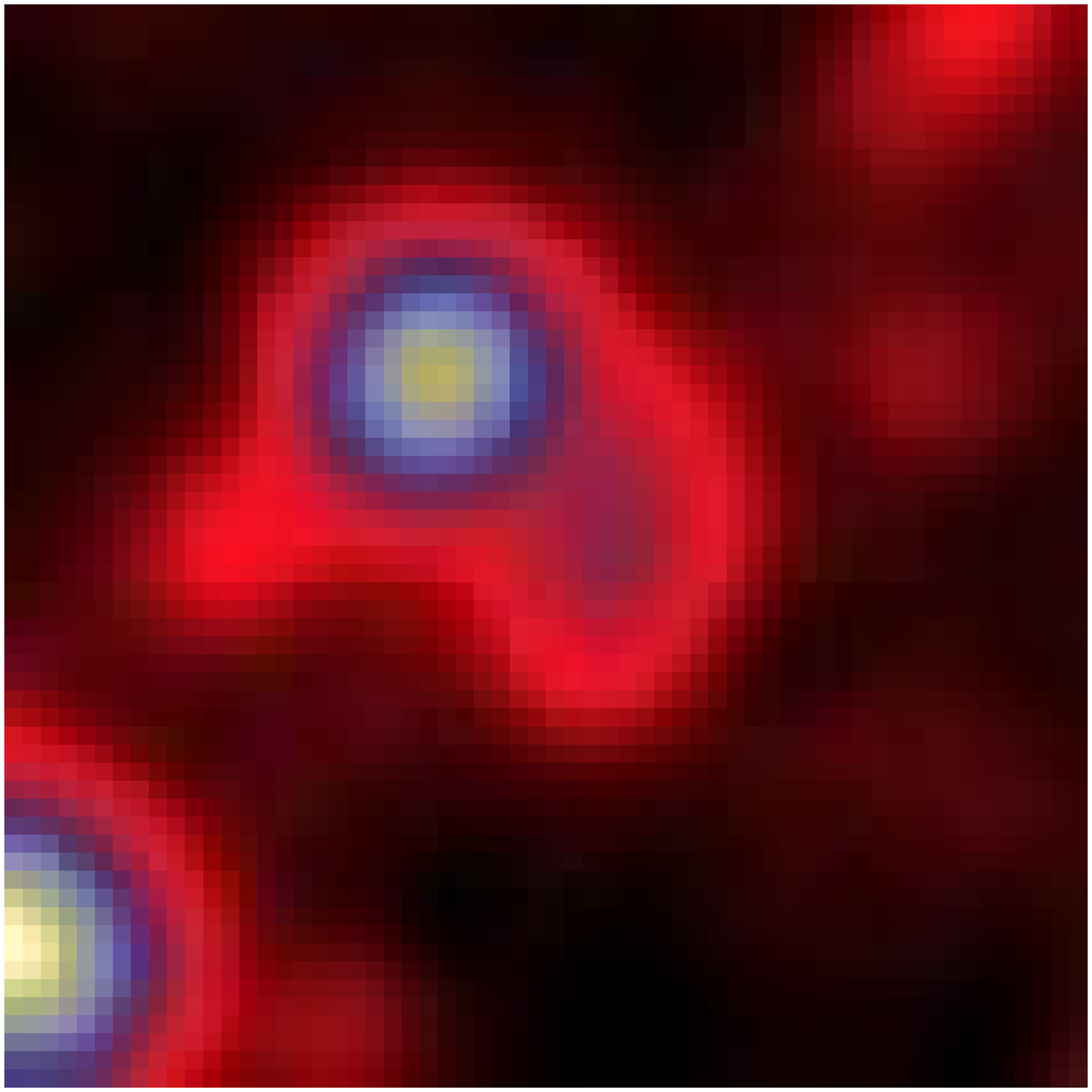}  \FGb{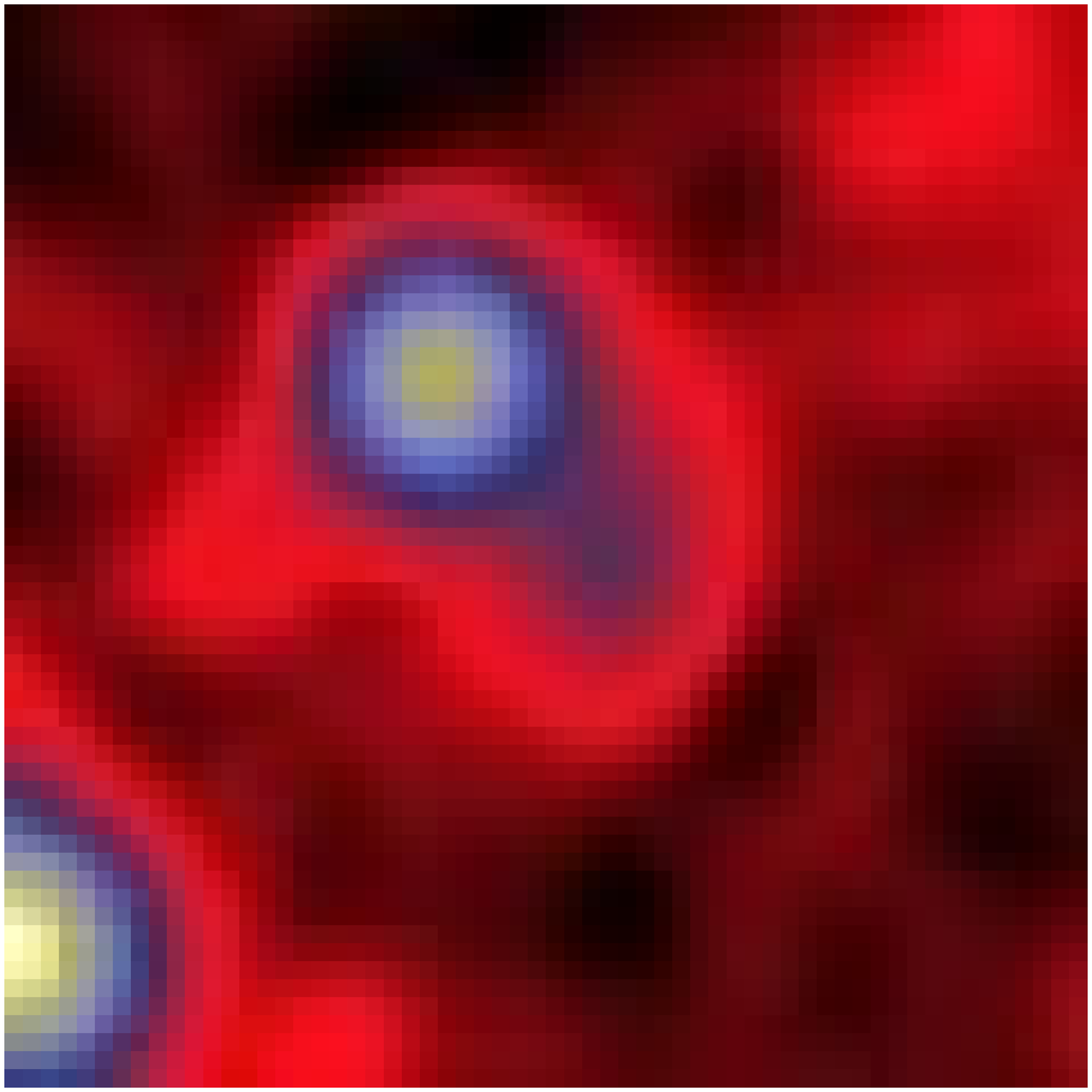}  \FGb{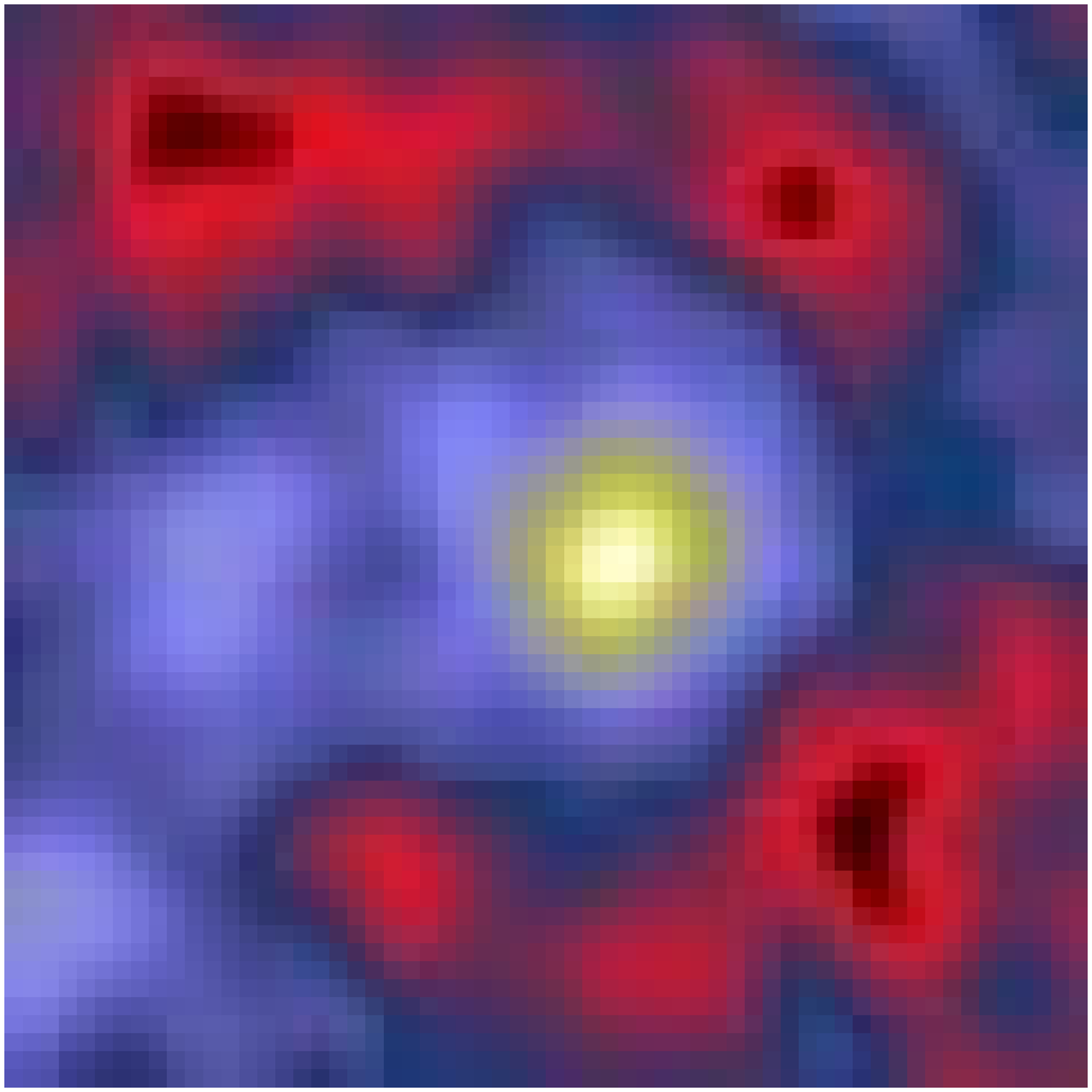}  \FGb{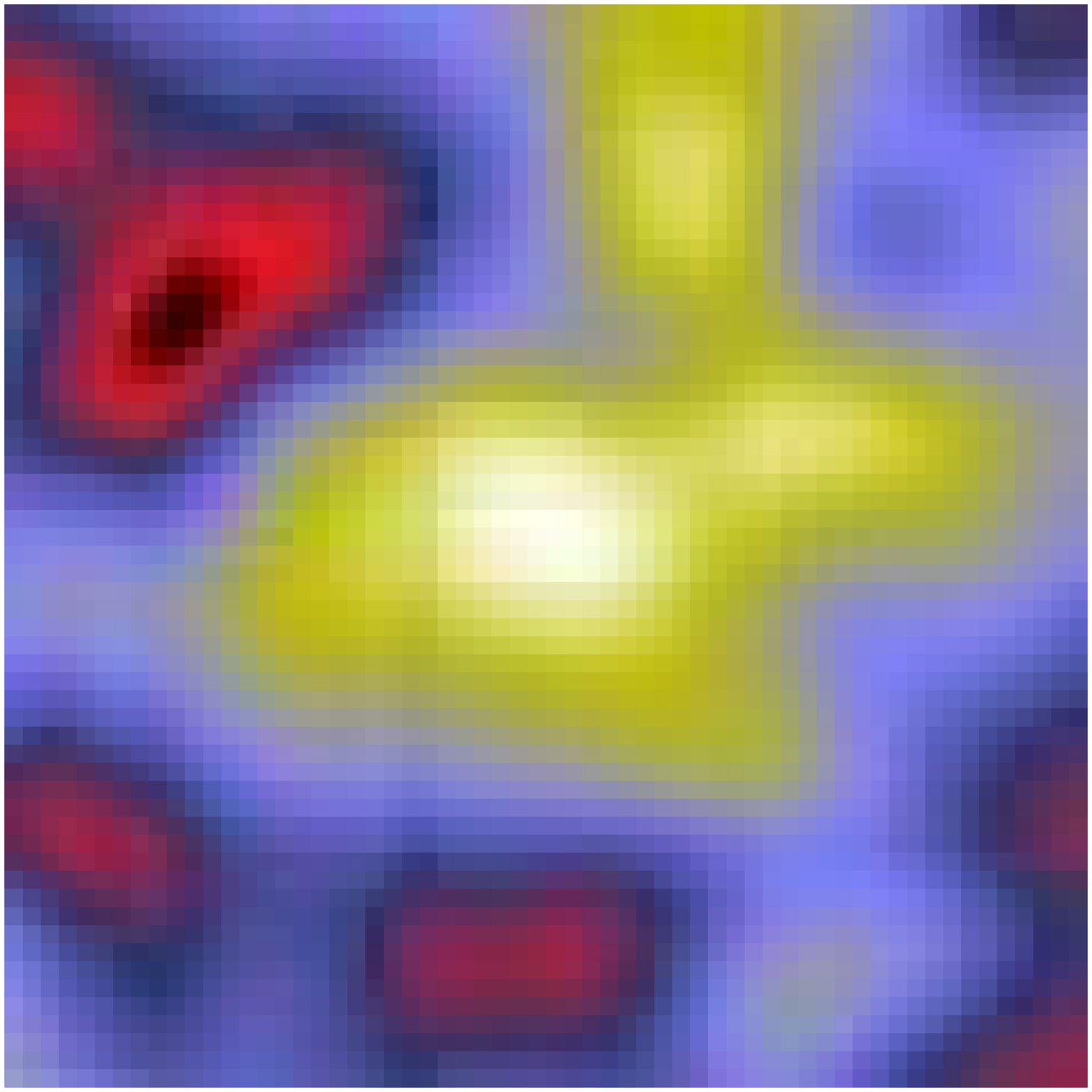} \\  {\#18   NCIA009669}  \\
  \FGb{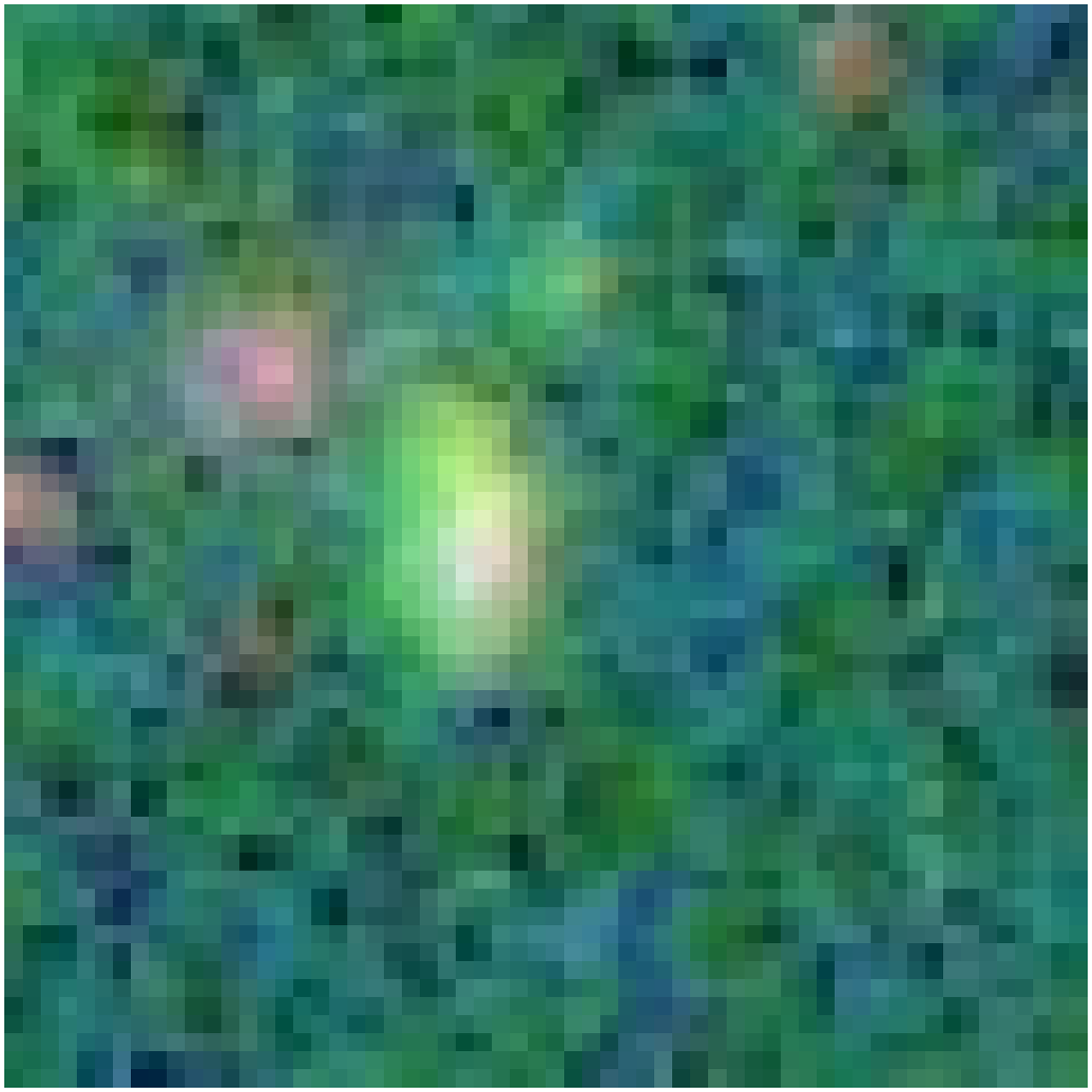}  \FGb{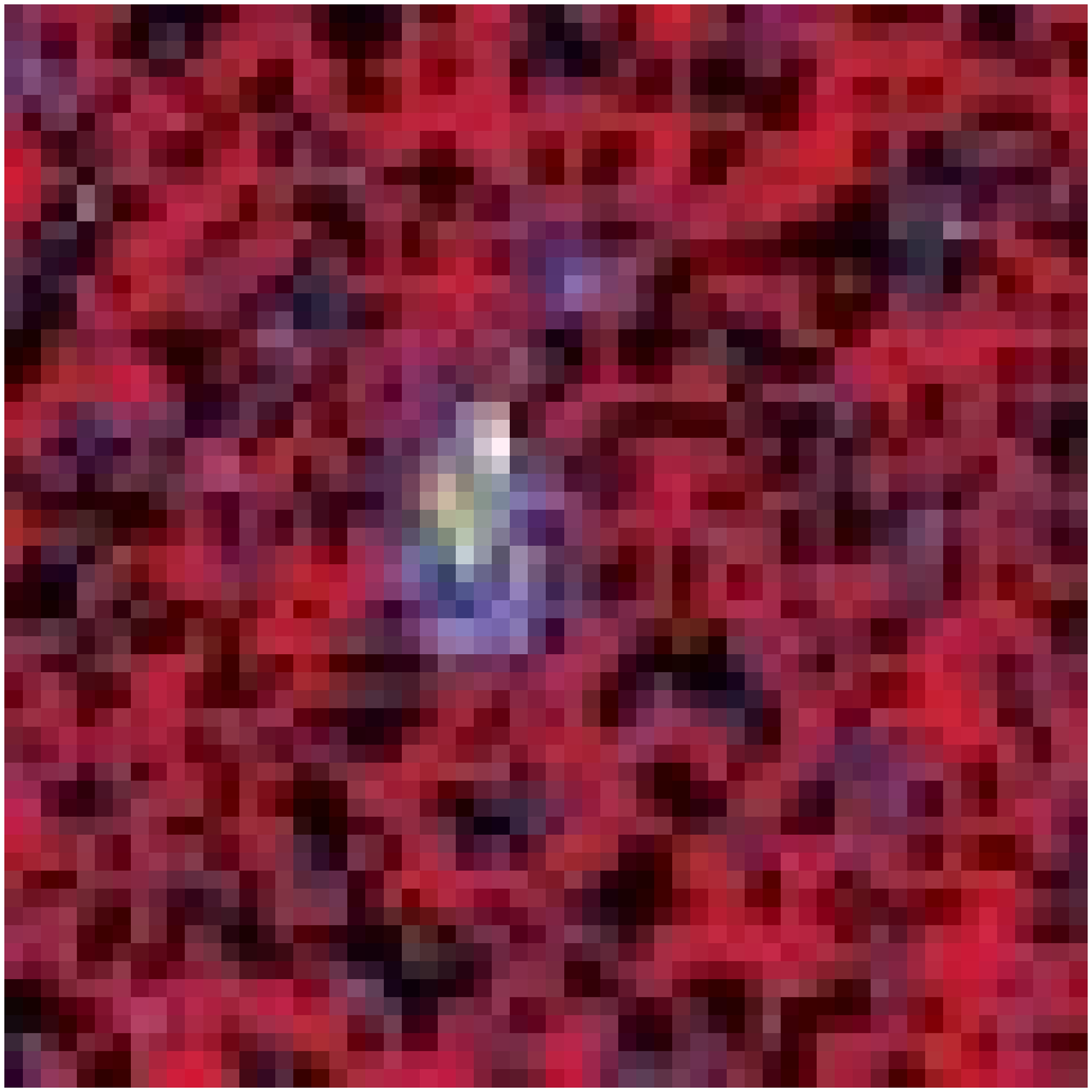}  \FGb{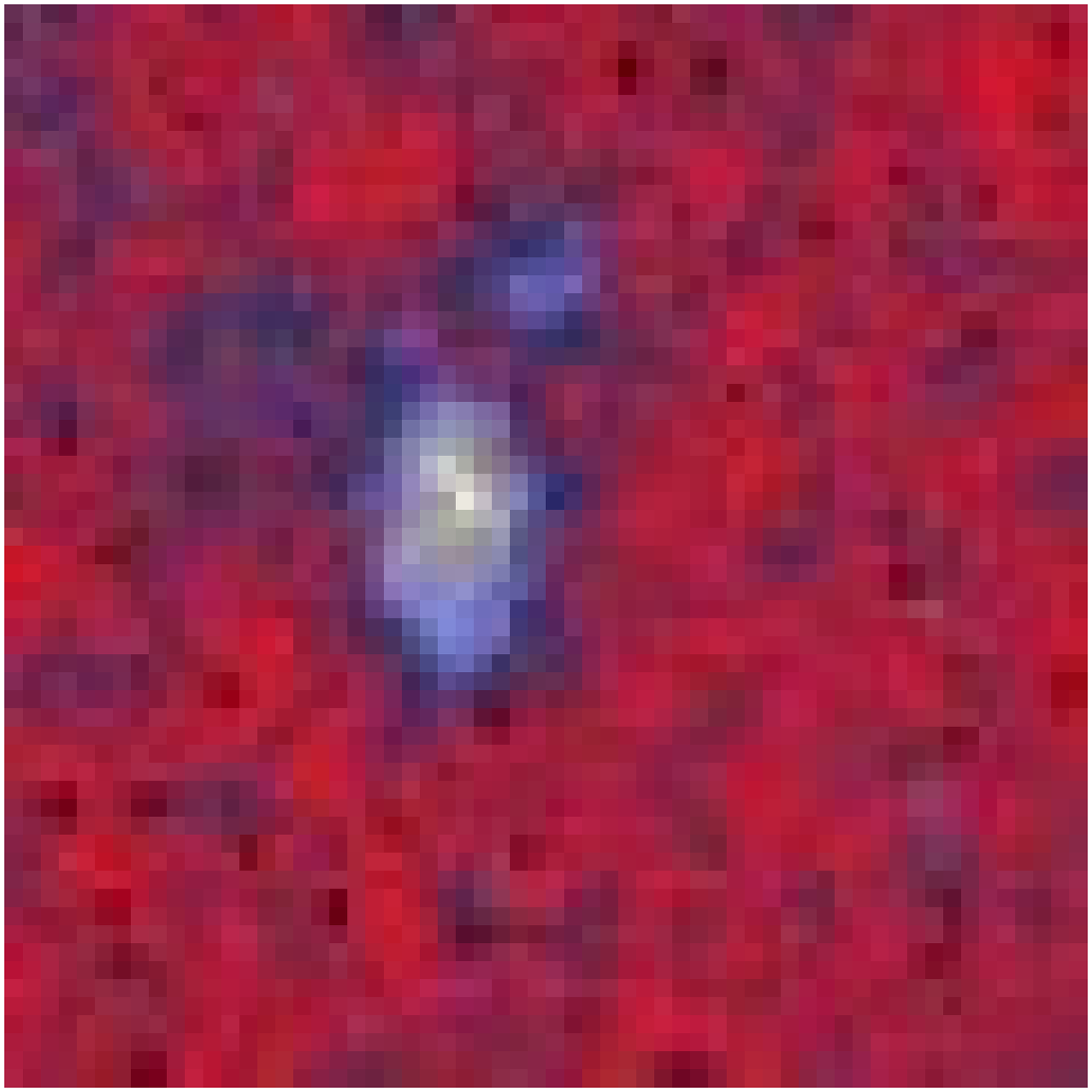}  \FGb{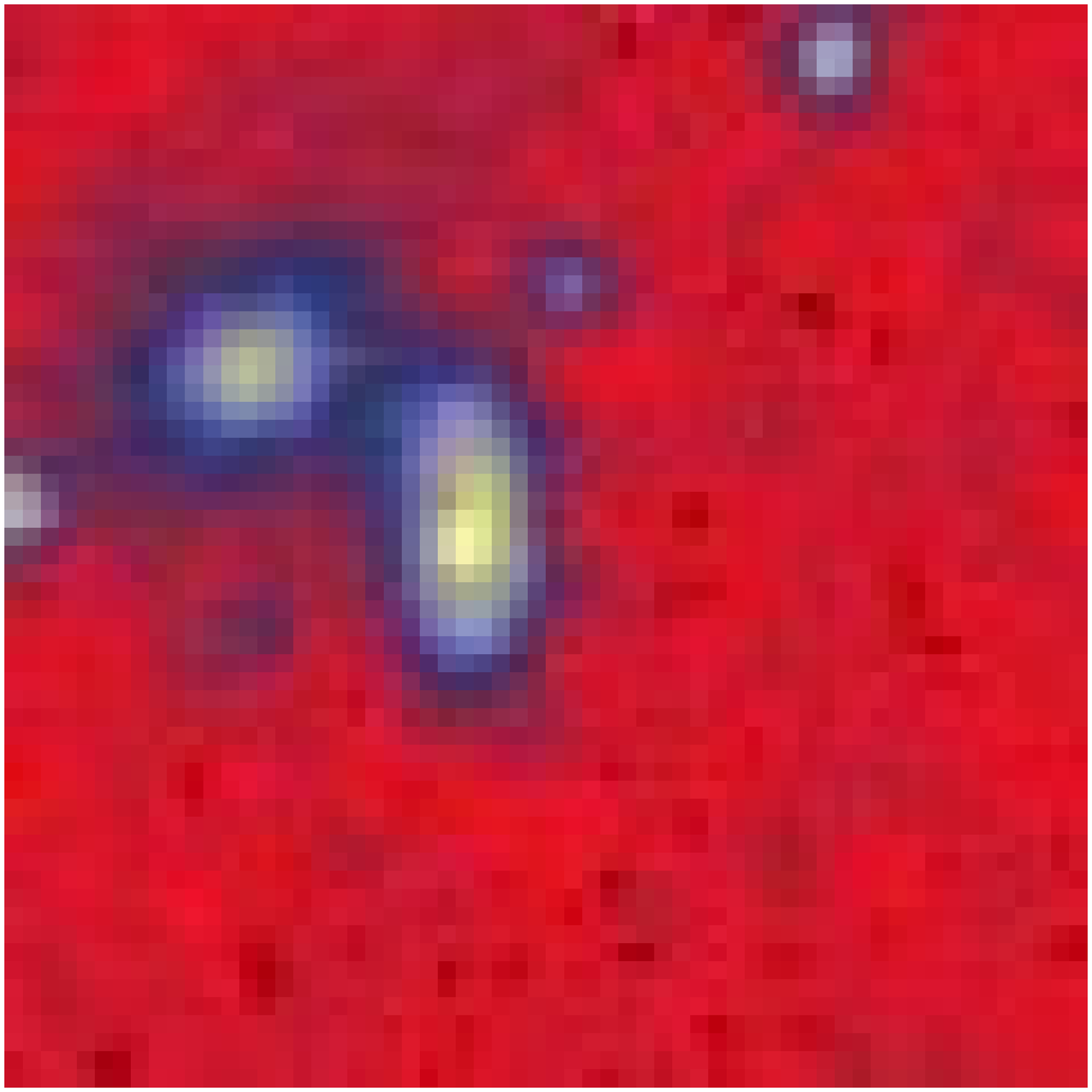}  \FGb{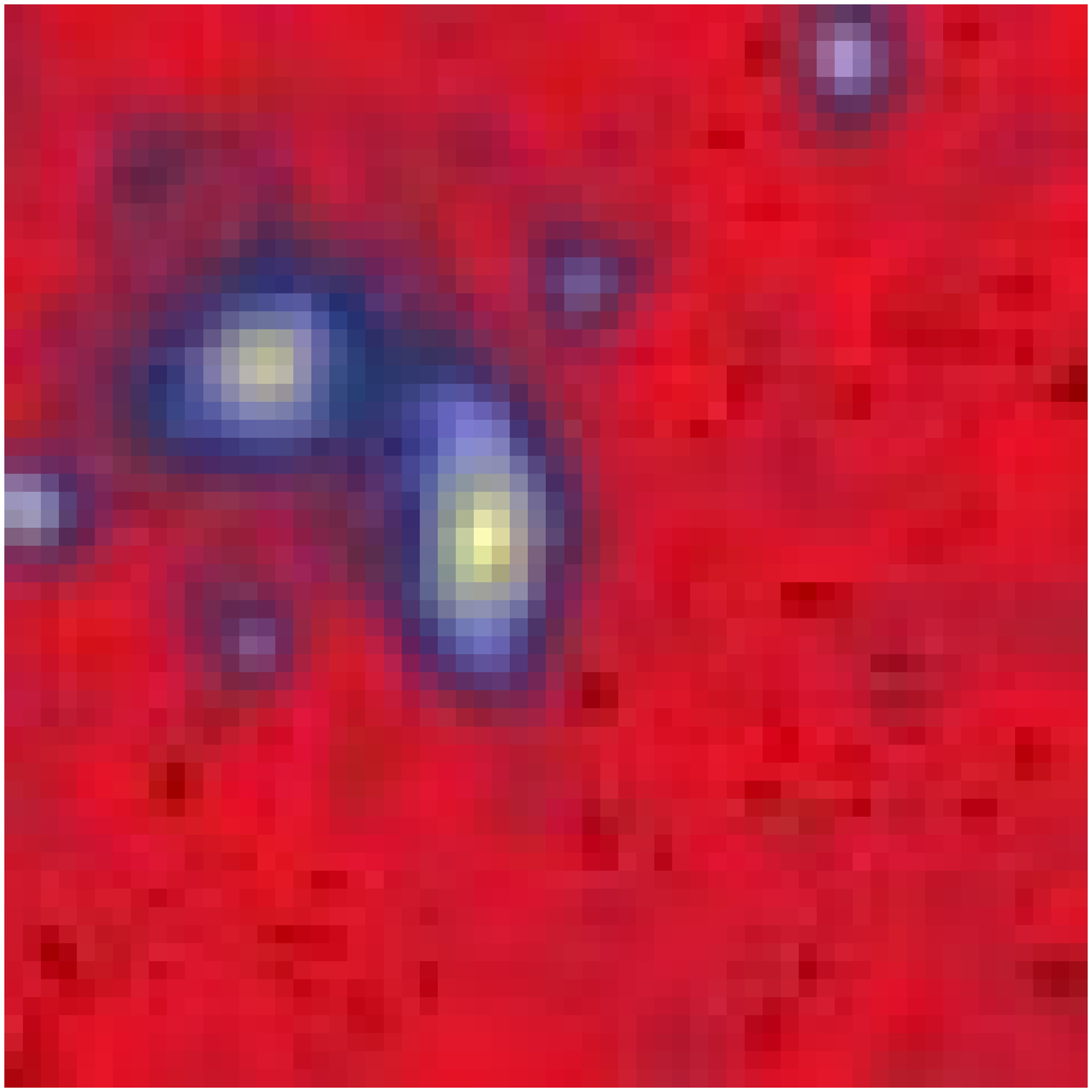}  \FGb{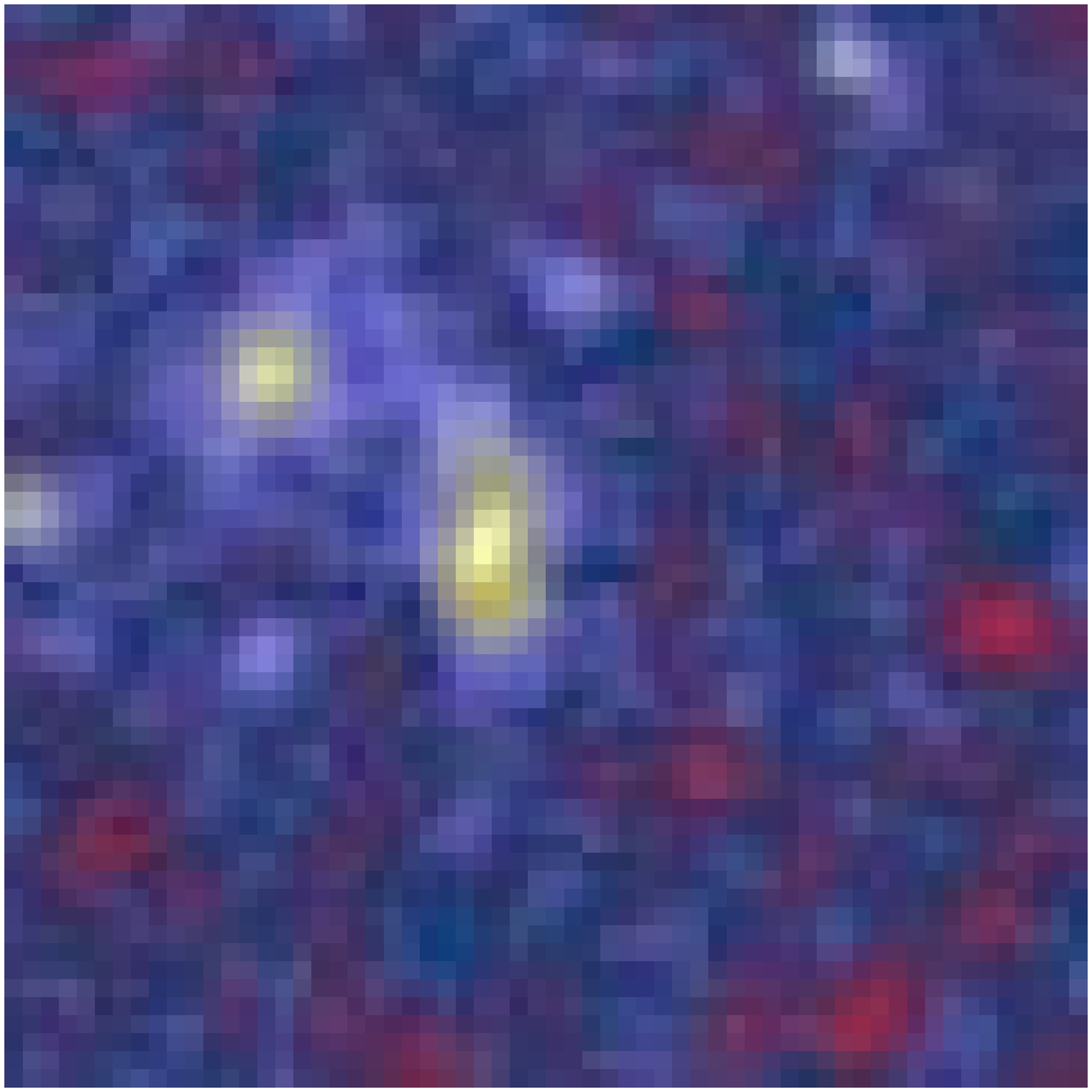}  \FGb{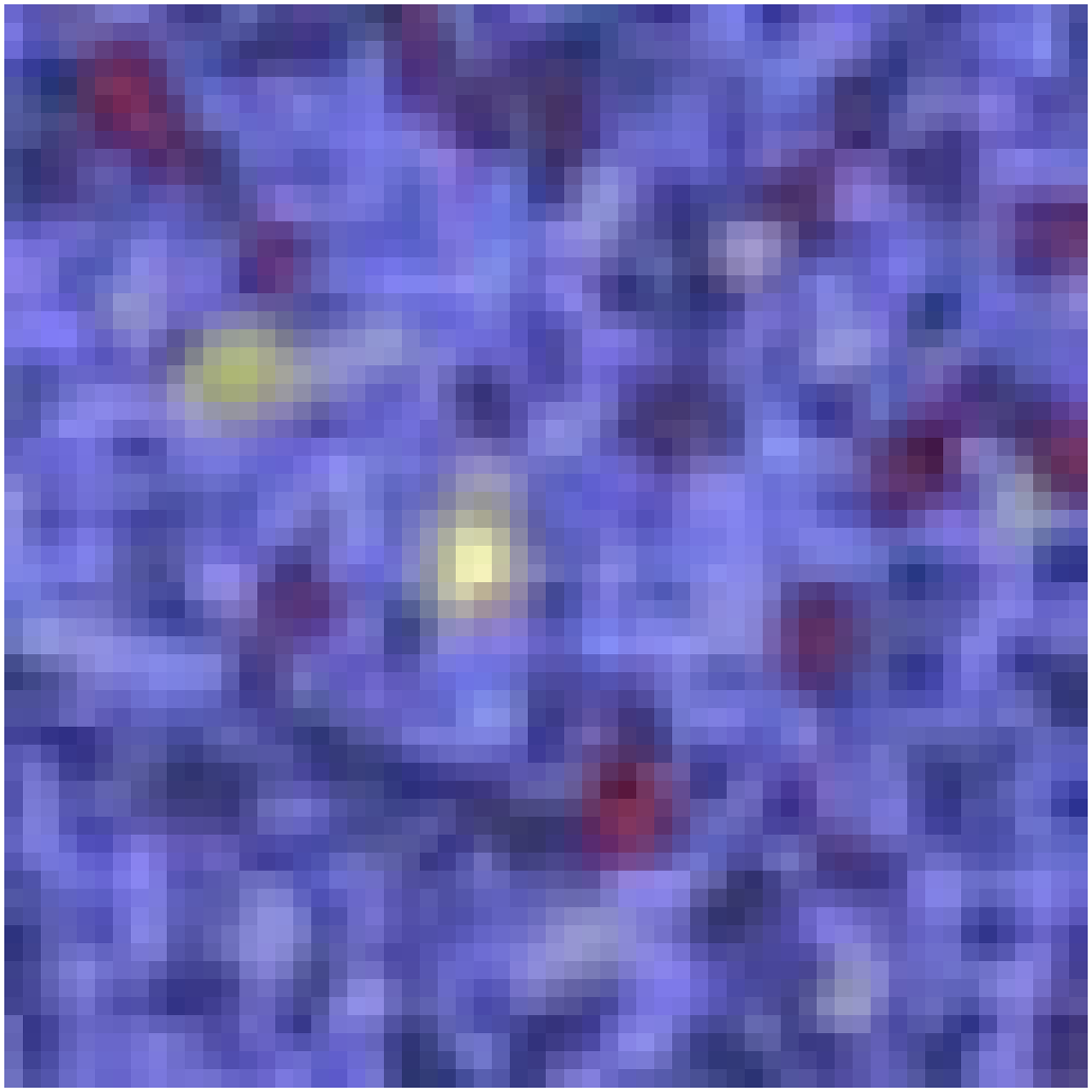}  \FGb{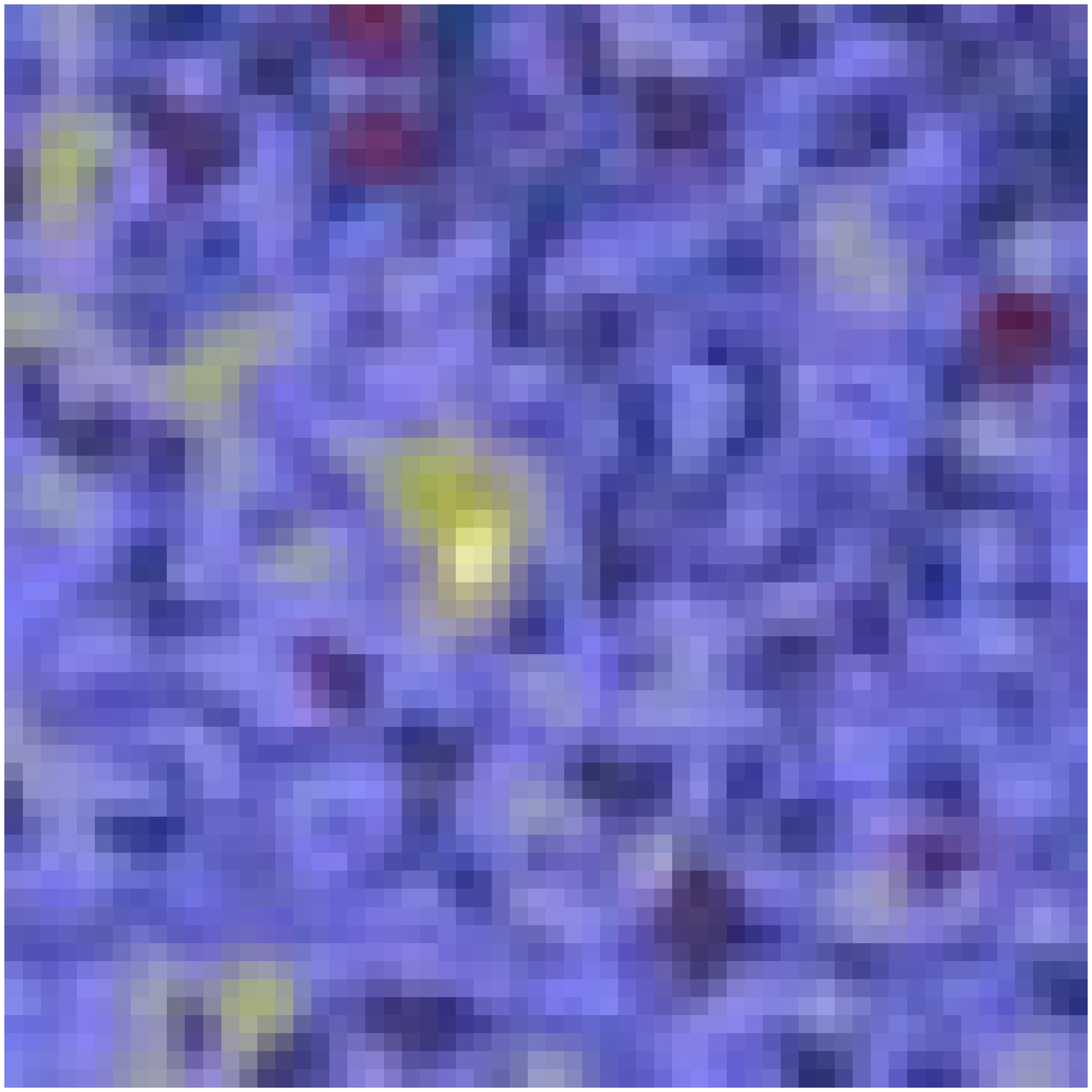}  \FGb{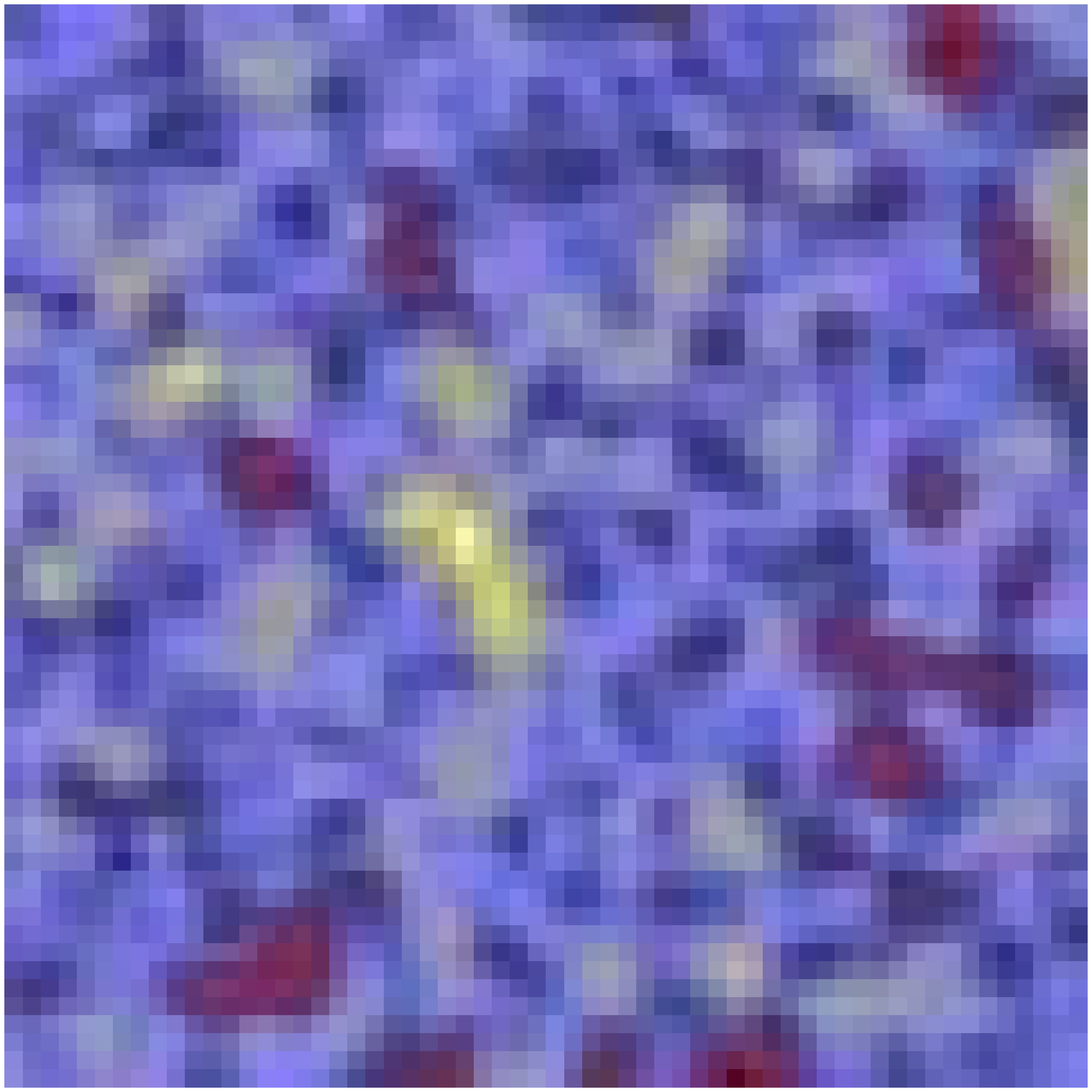}  \FGb{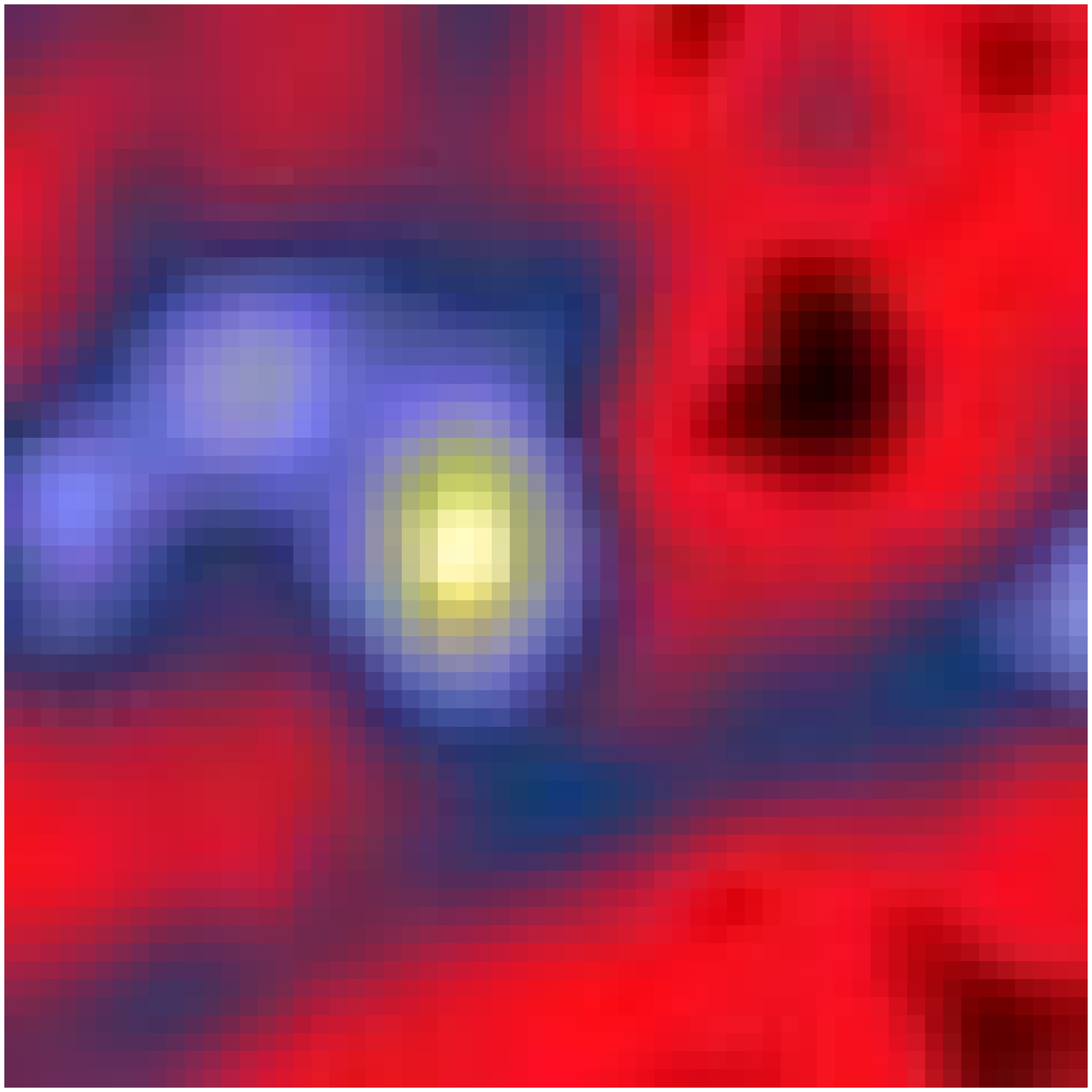}  \FGb{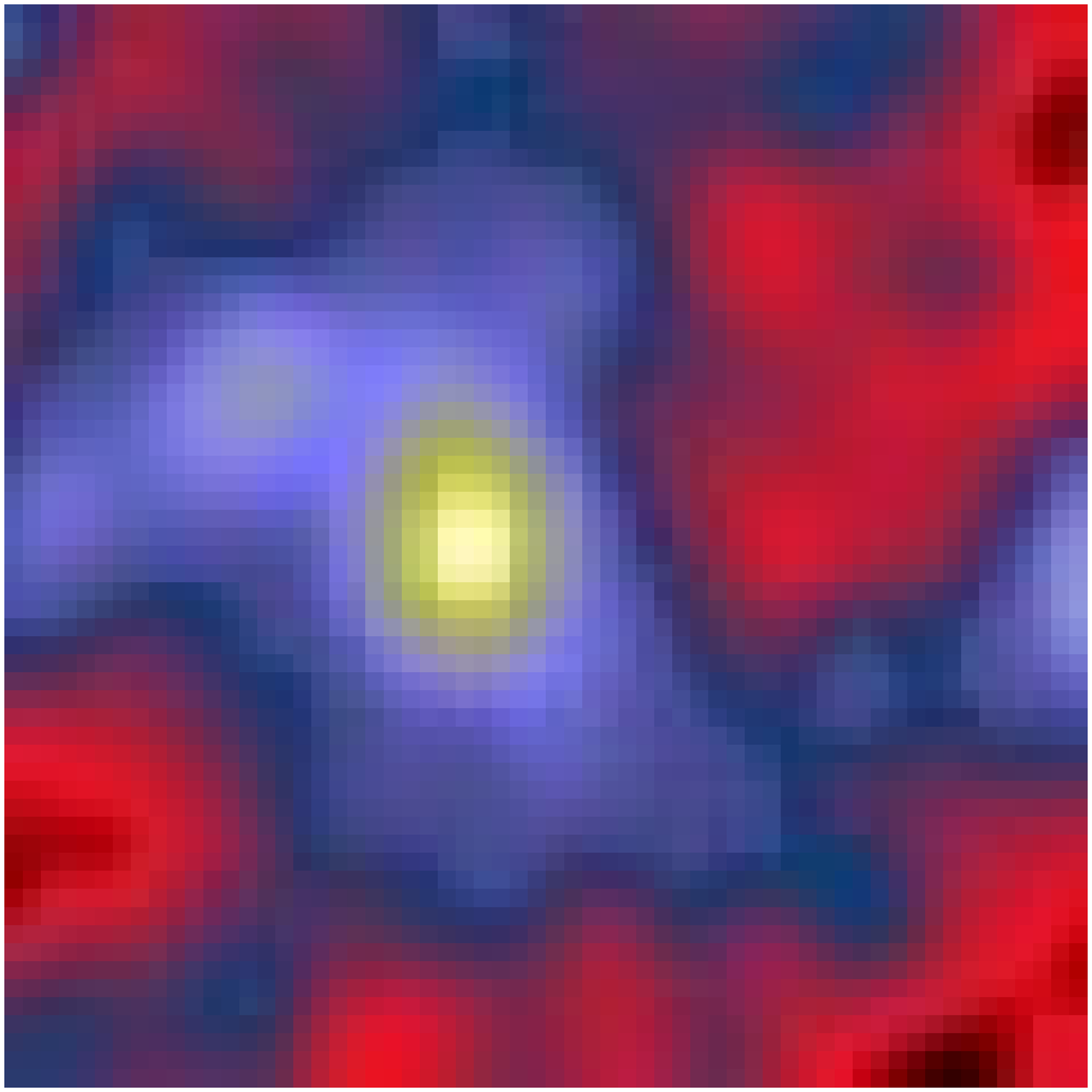}  \FGb{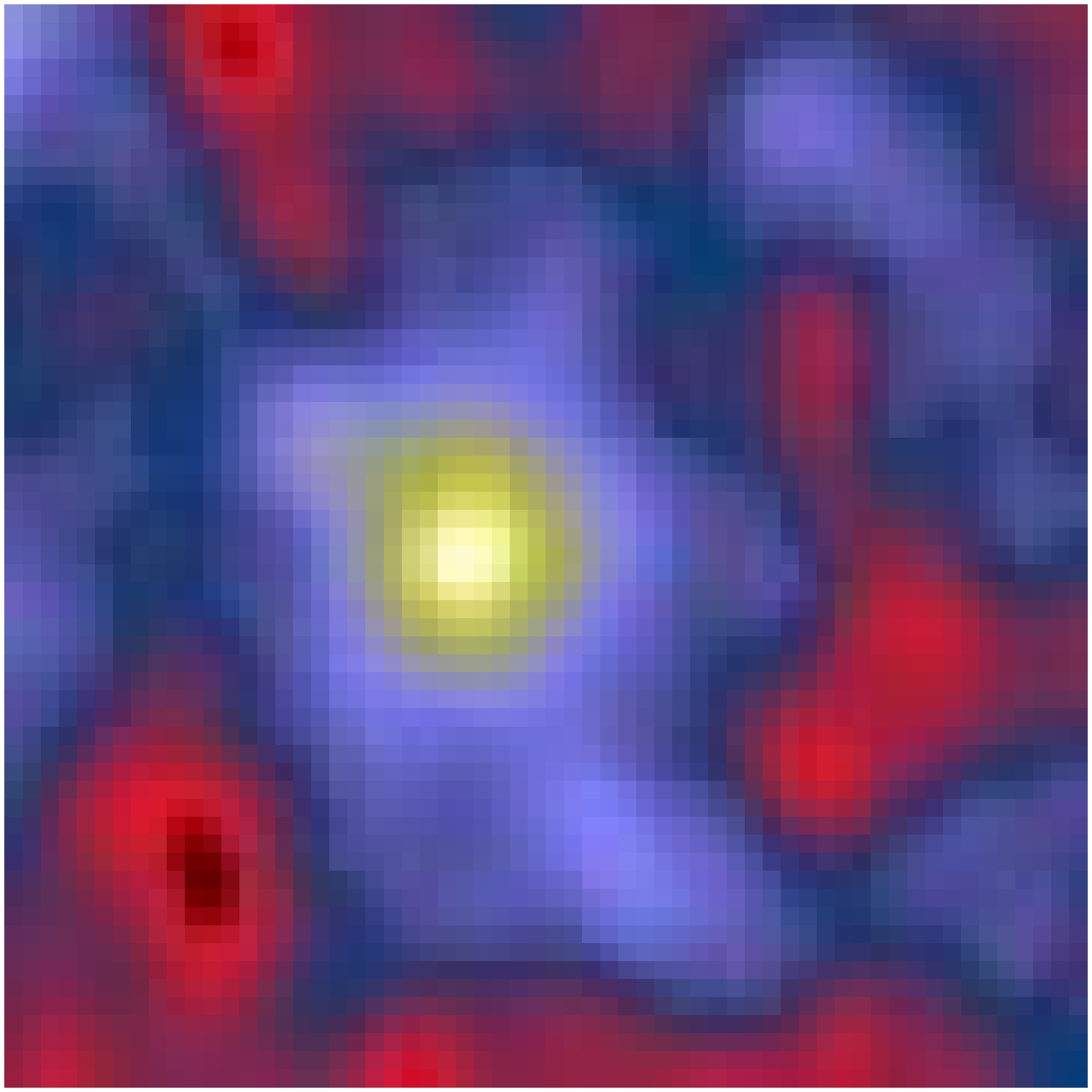}  \FGb{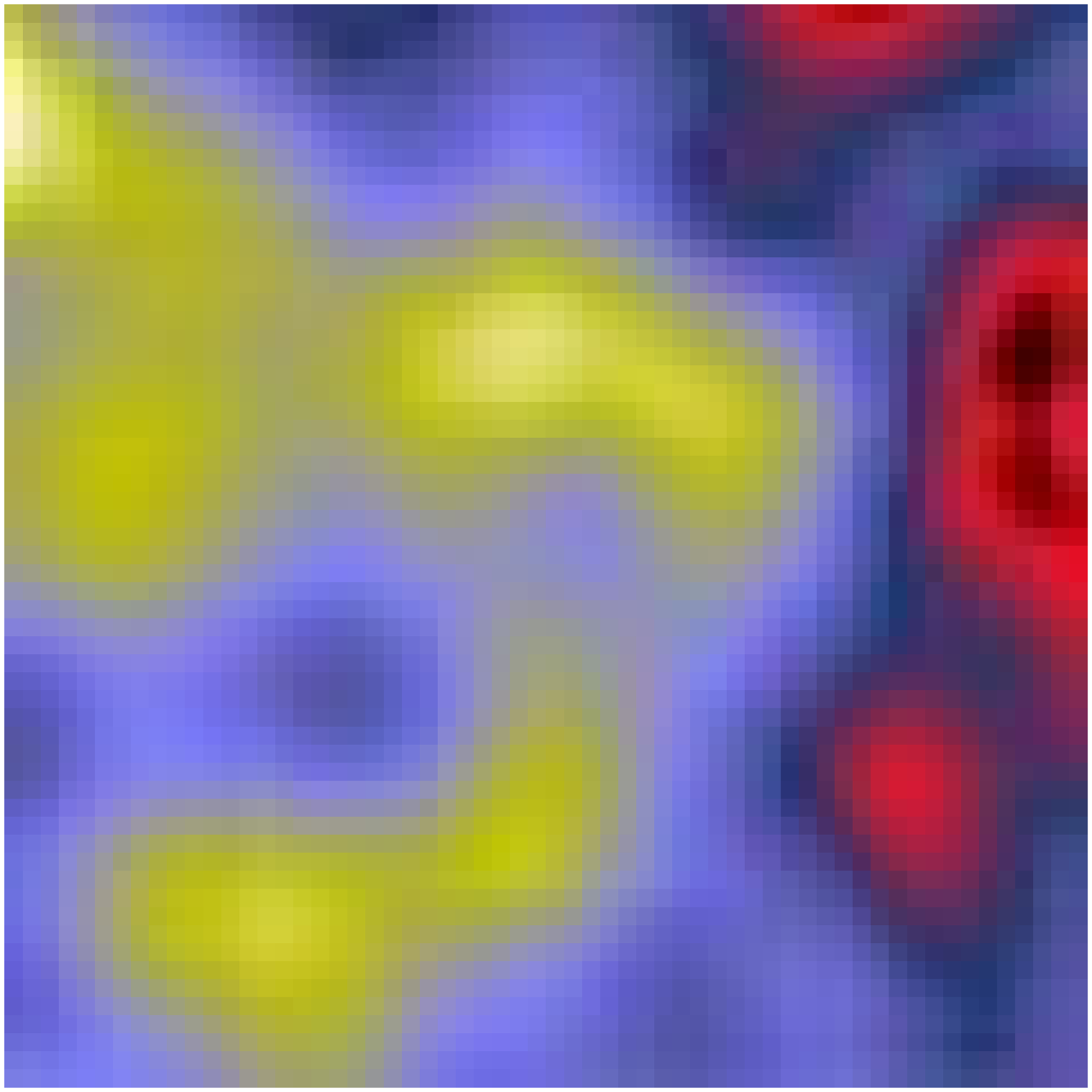} \\  {\#19   NCIA010240}  \\
  \FGb{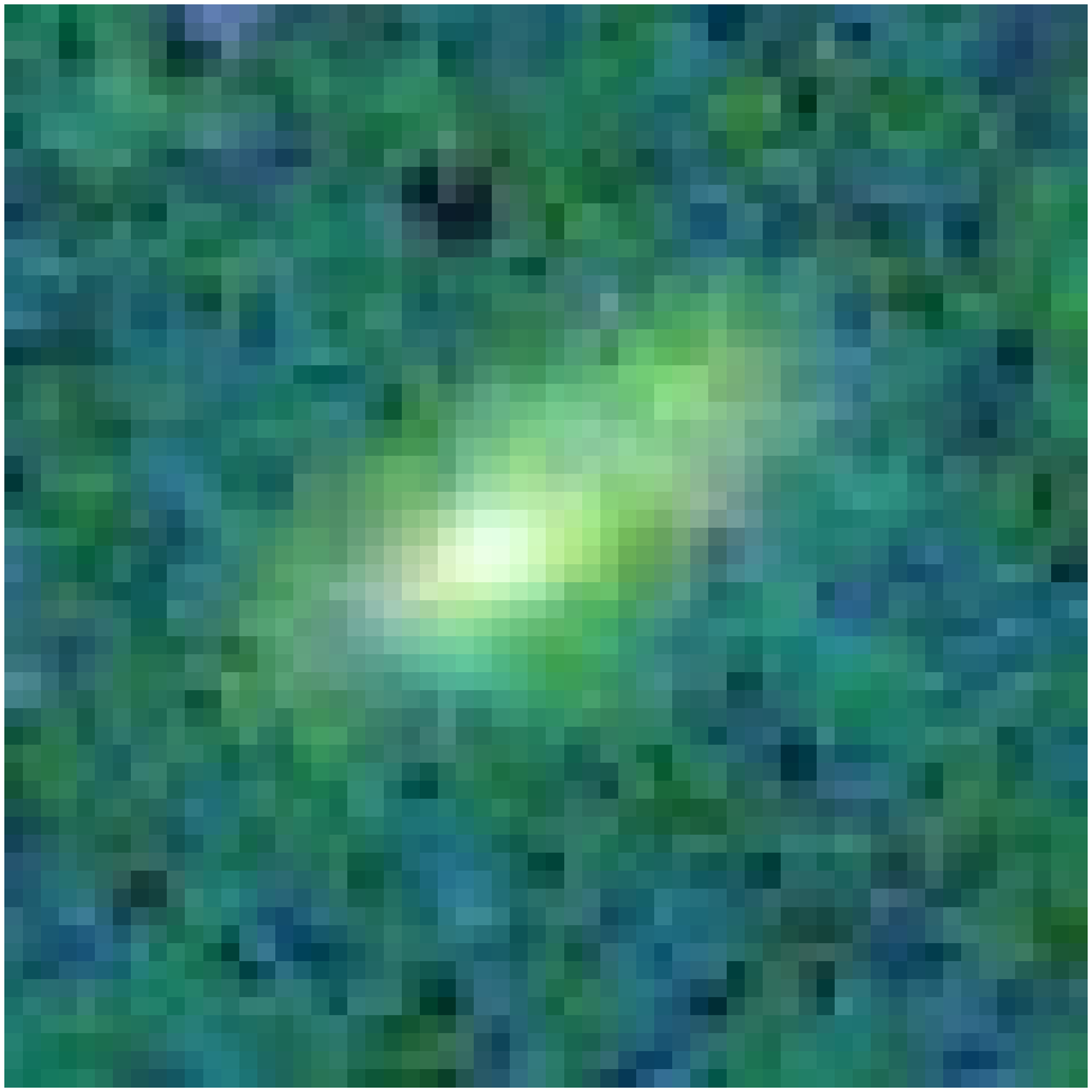}  \FGb{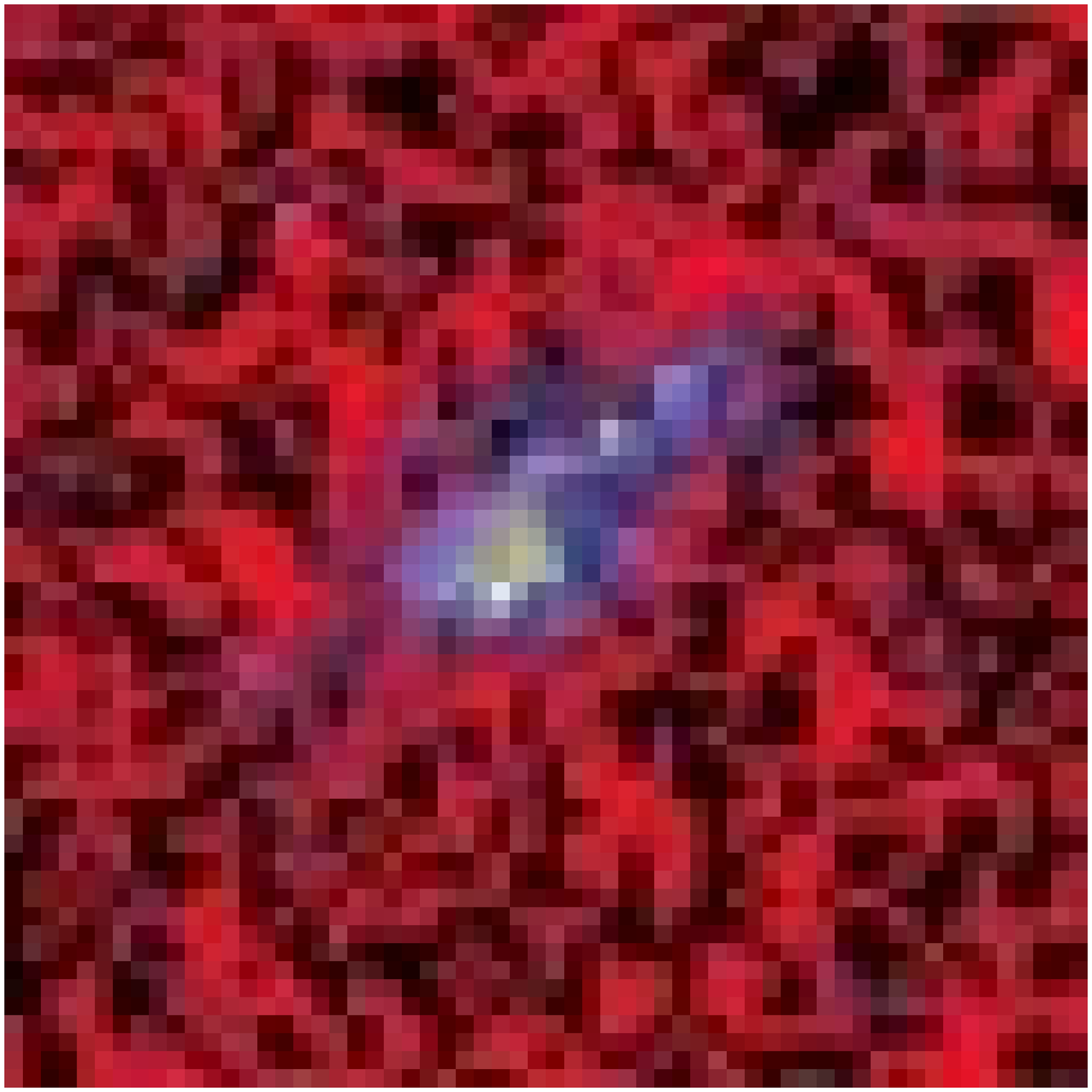}  \FGb{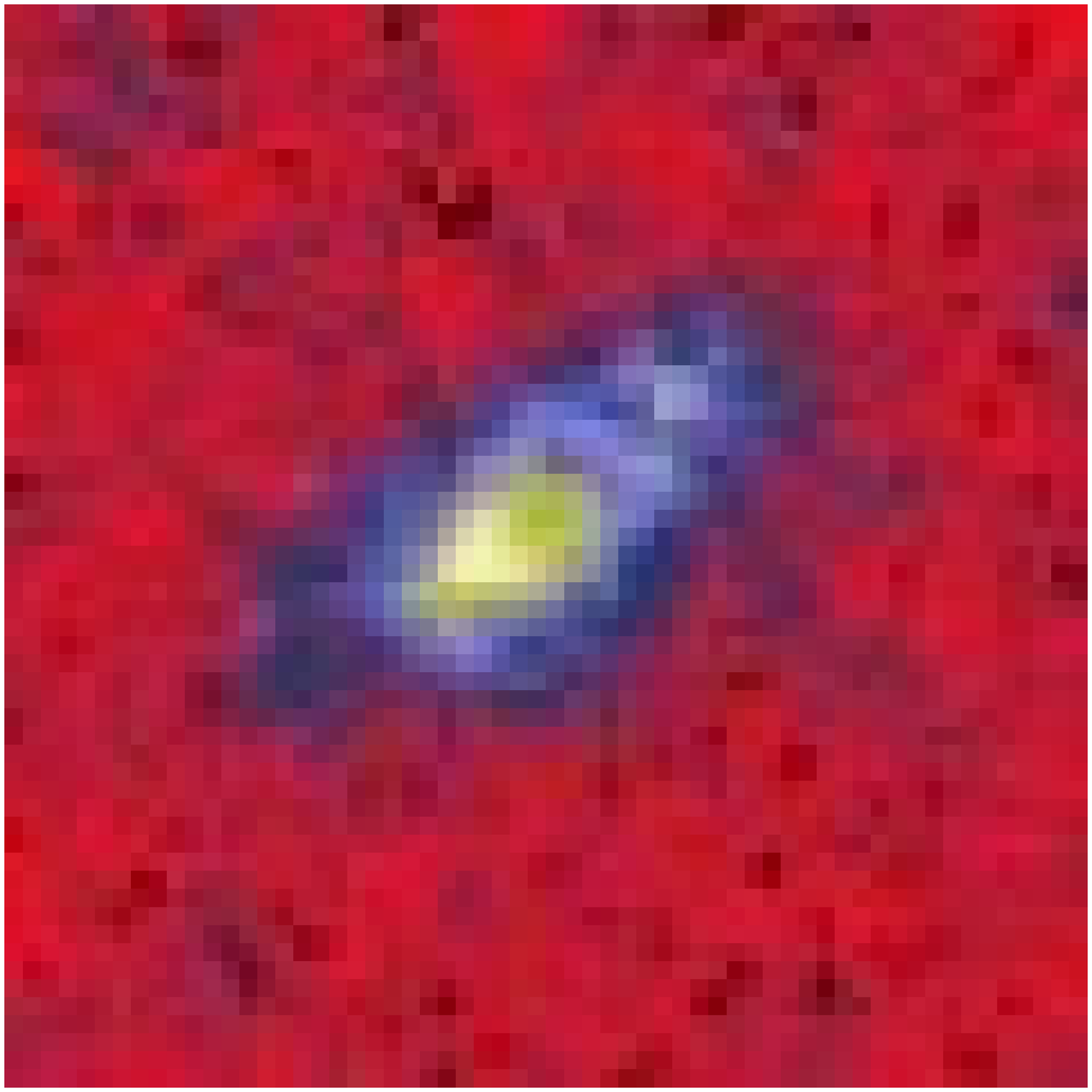}  \FGb{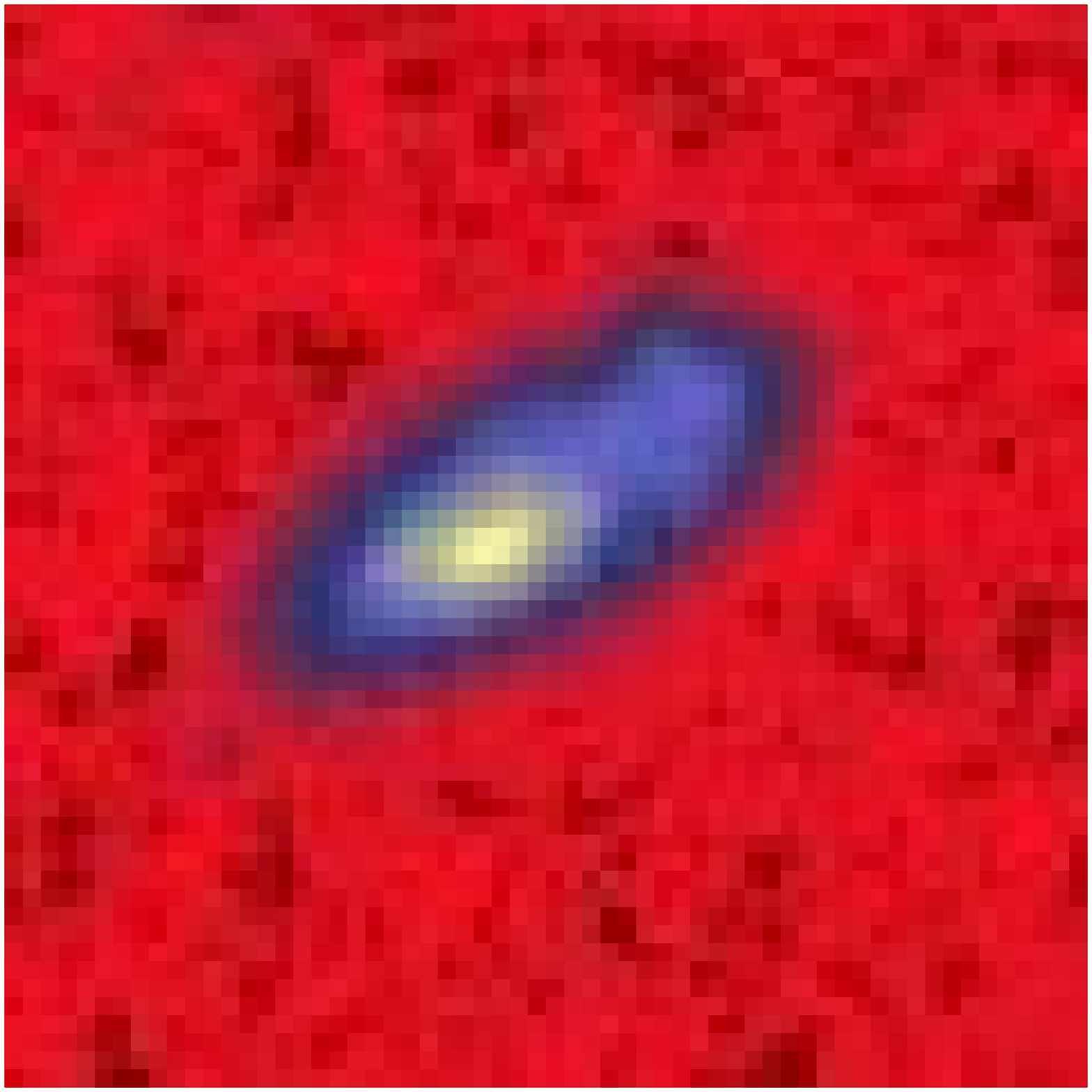}  \FGb{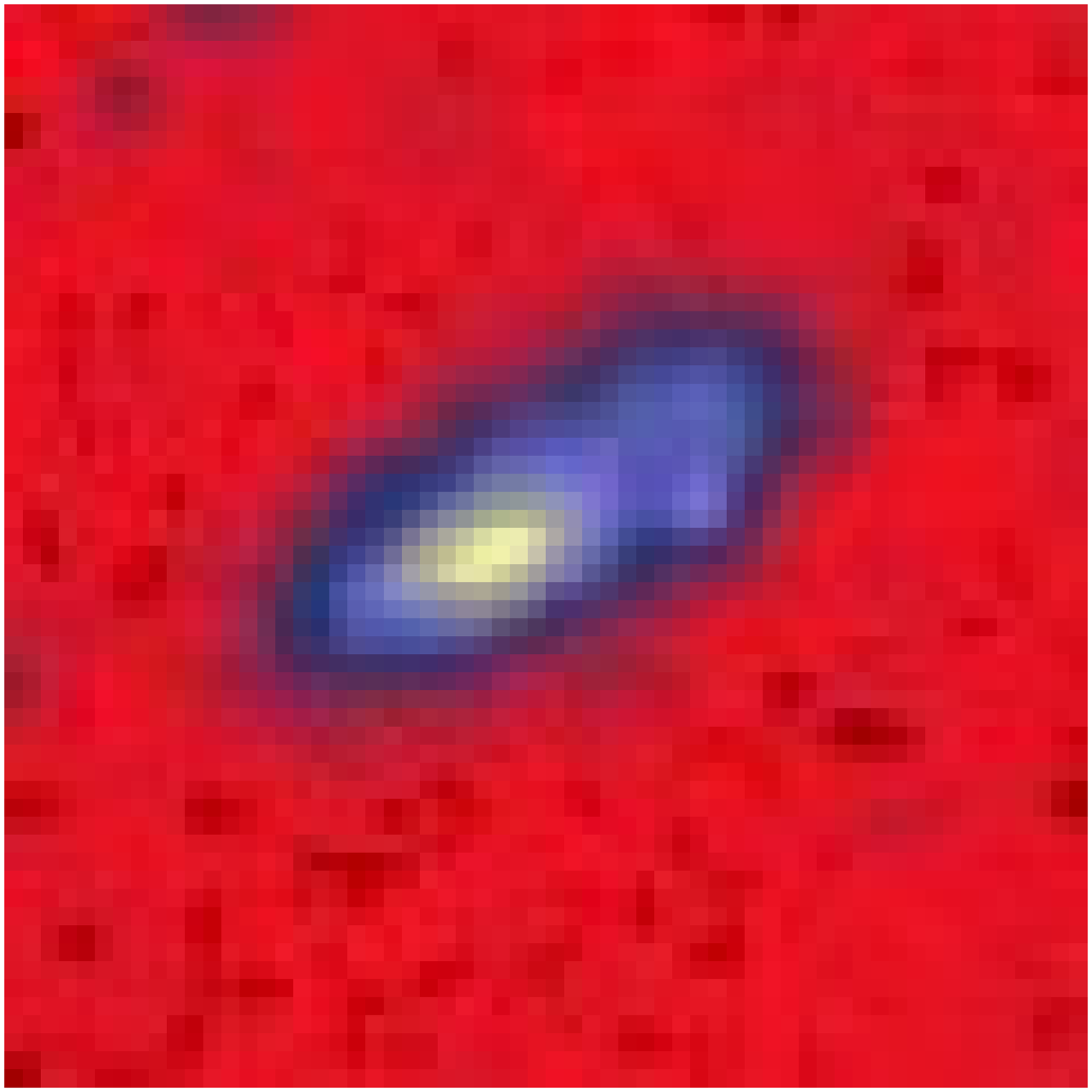}  \FGb{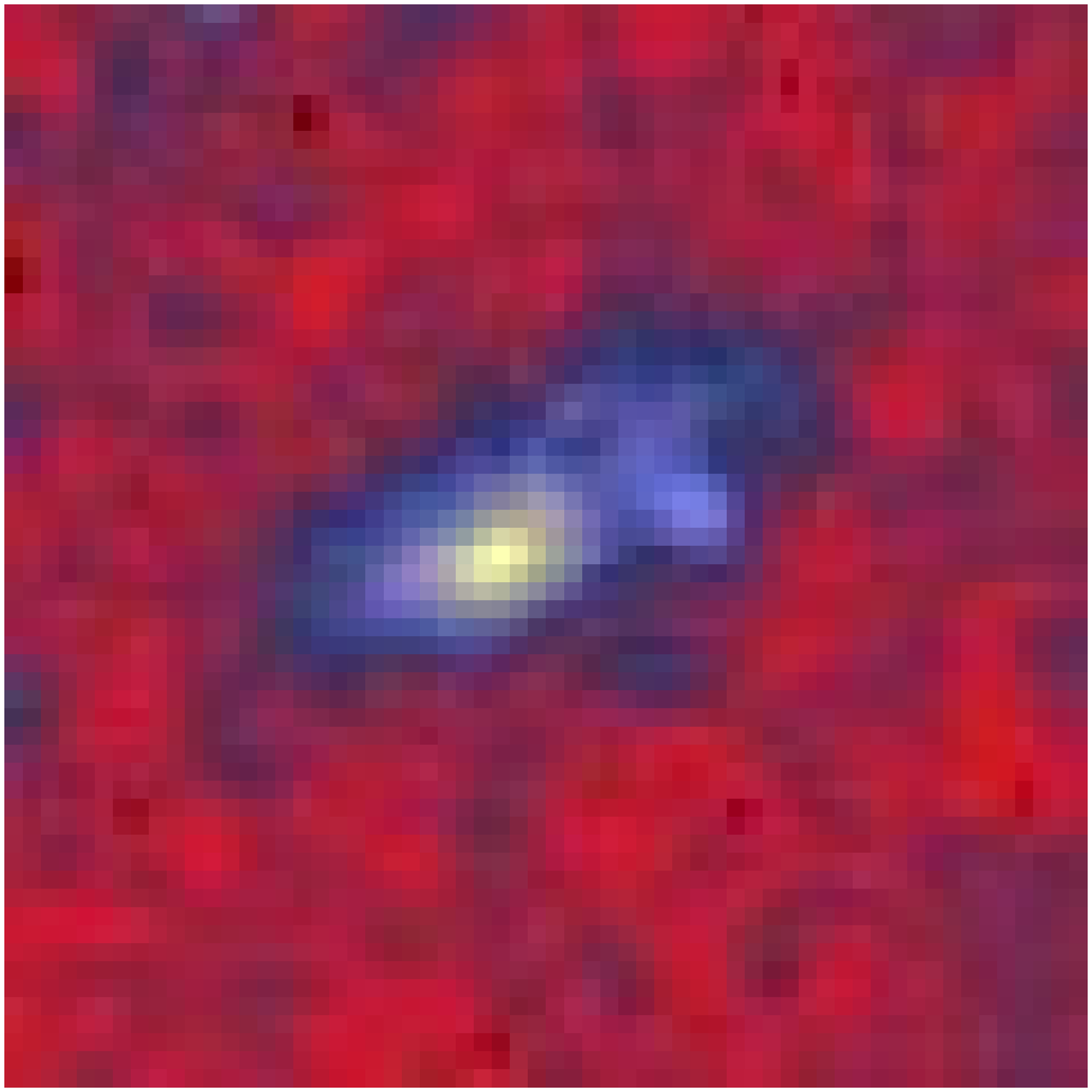}  \FGb{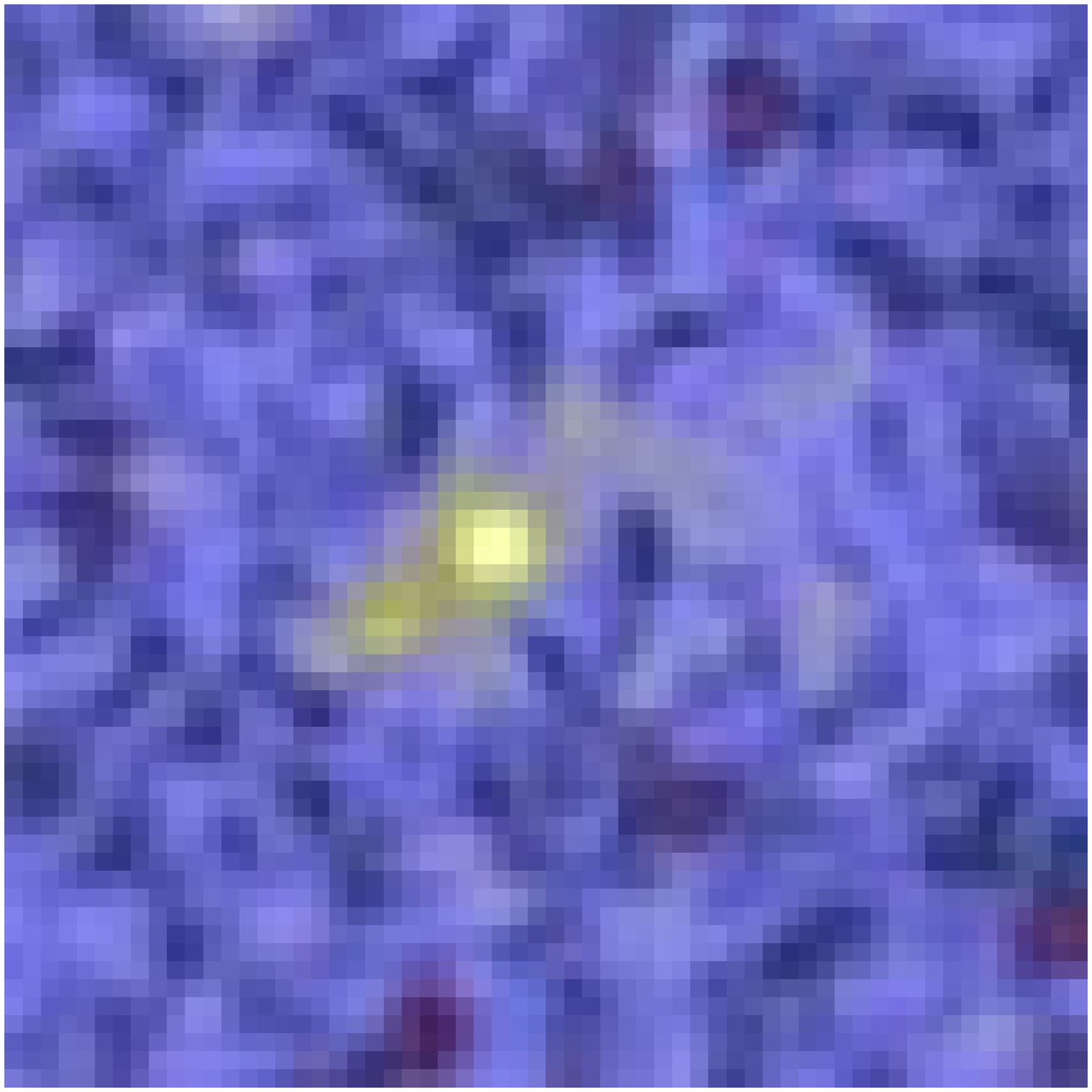}  \FGb{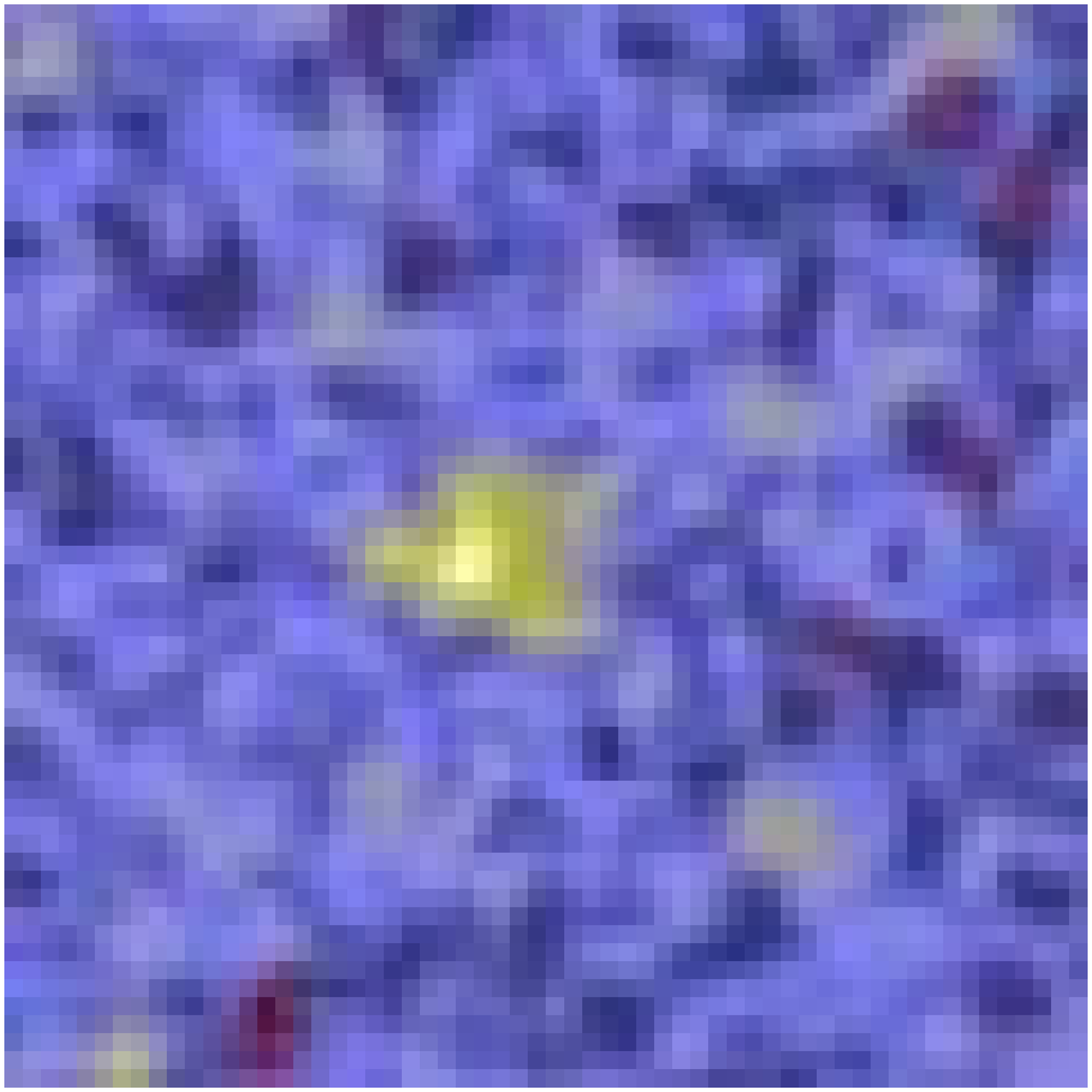}  \FGb{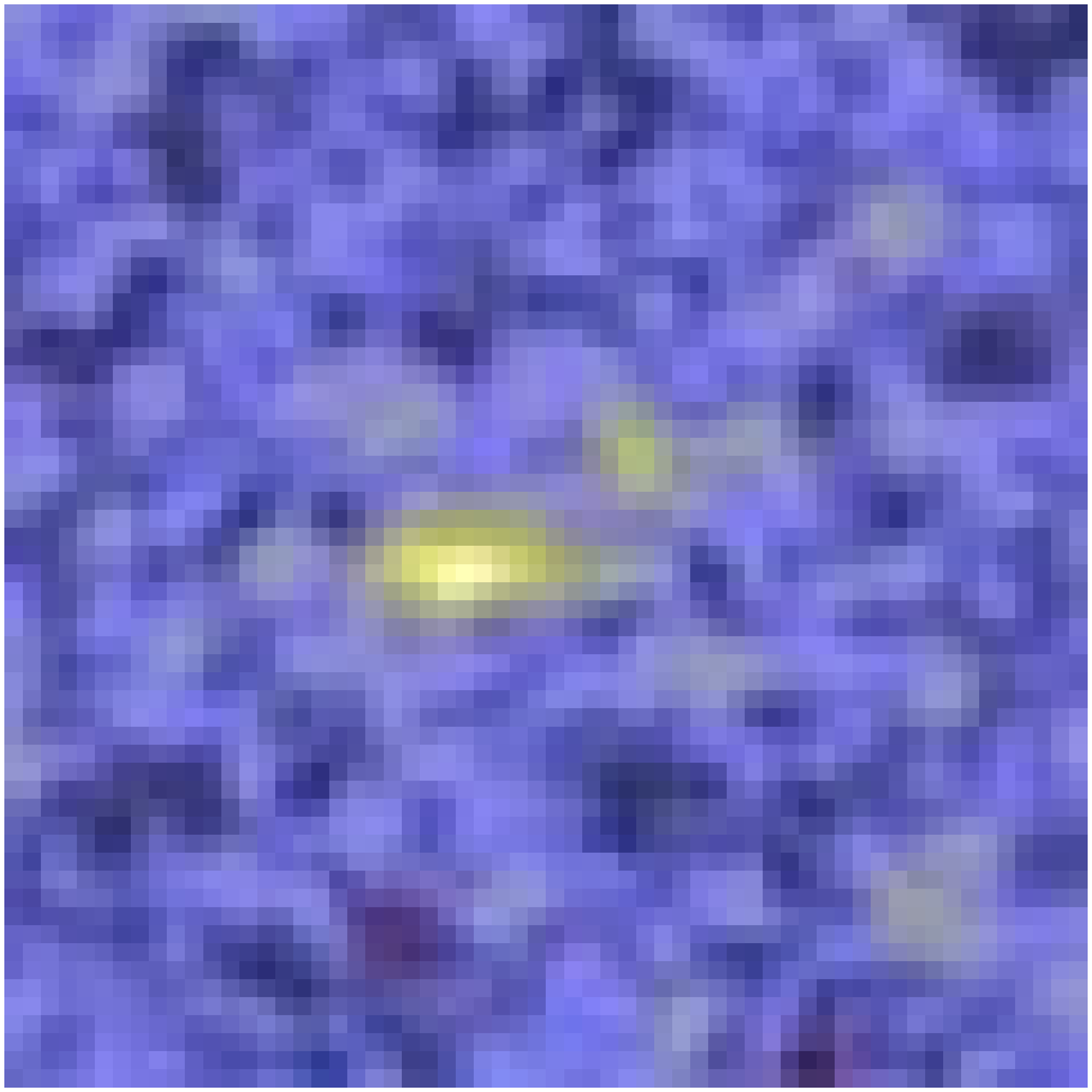}  \FGb{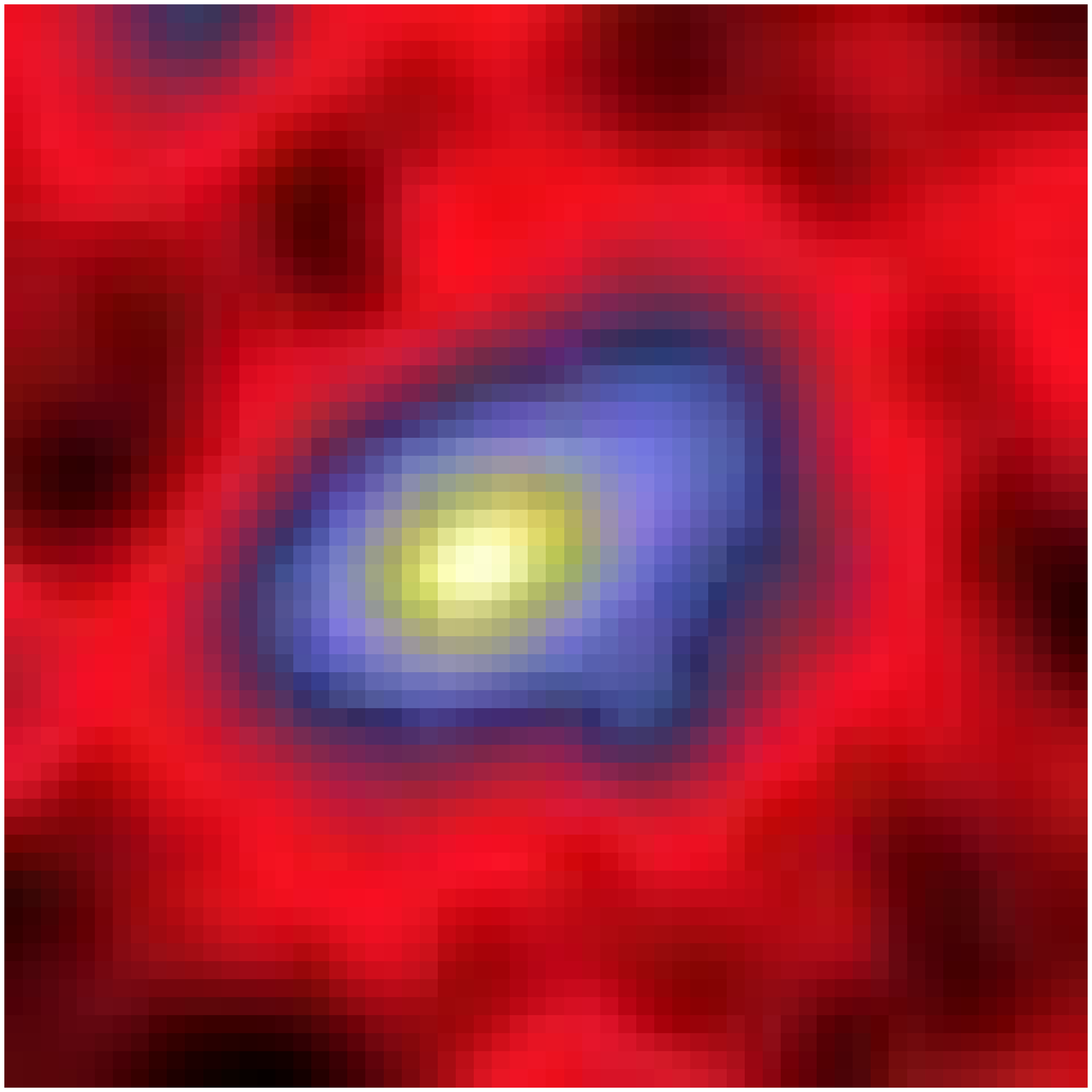}  \FGb{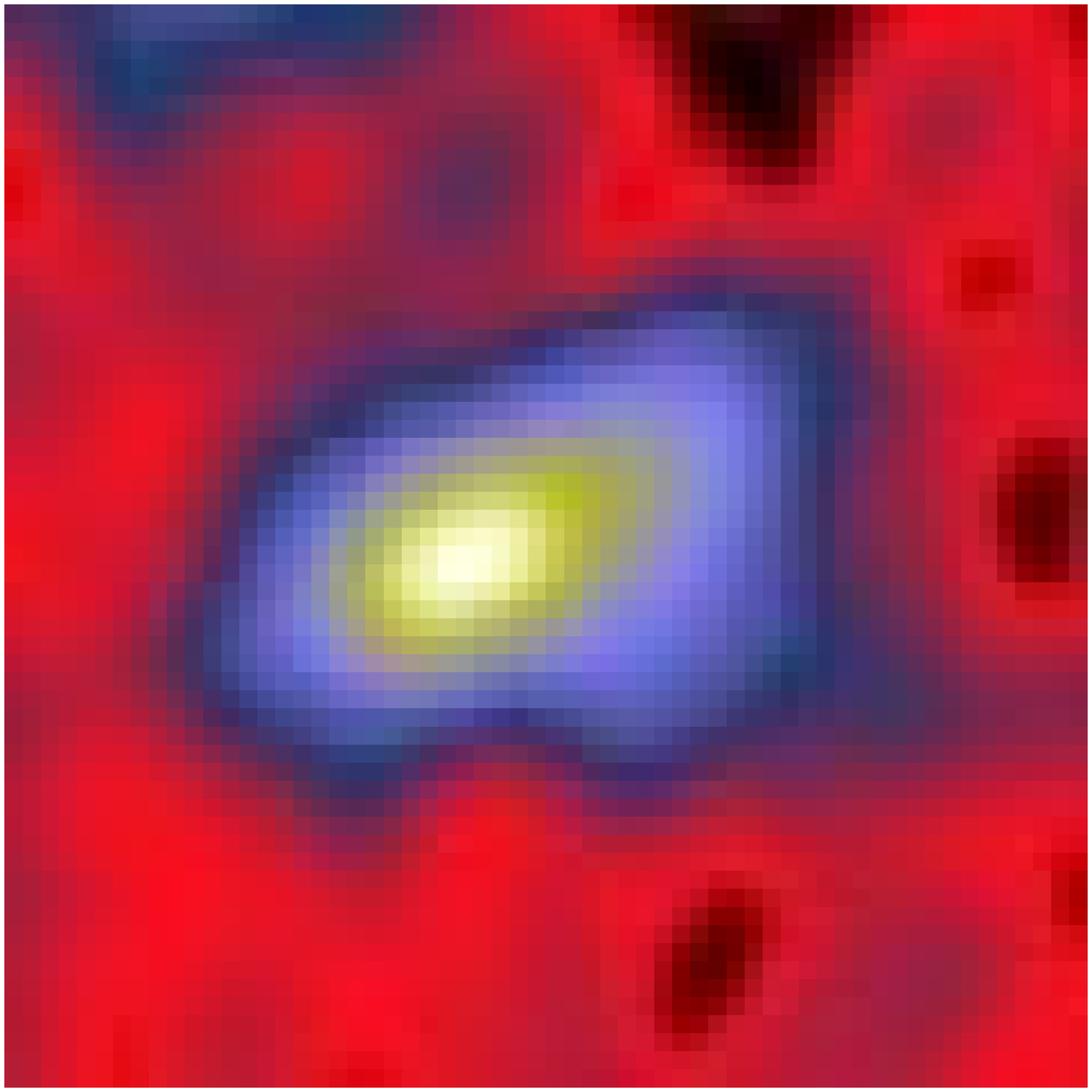}  \FGb{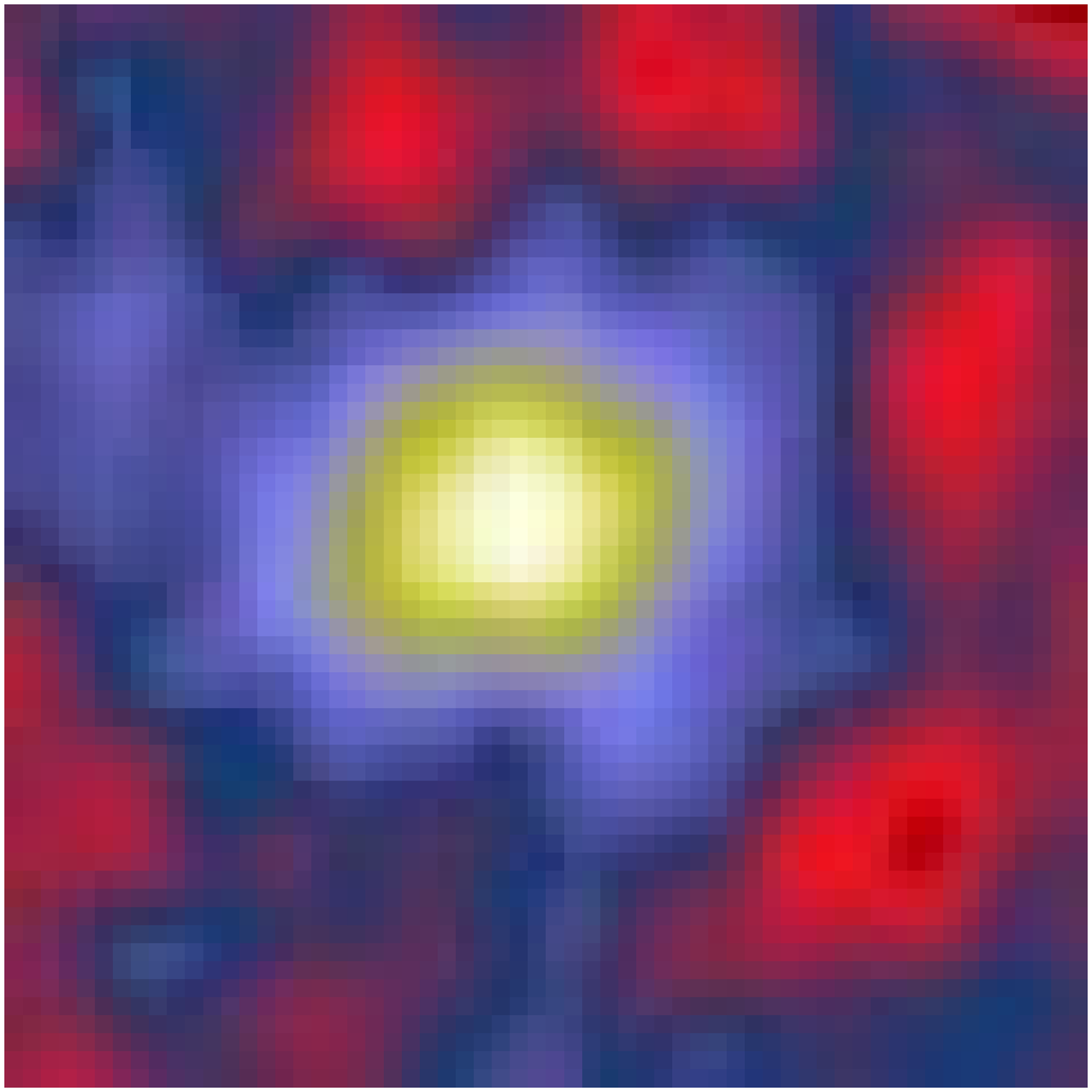}  \FGb{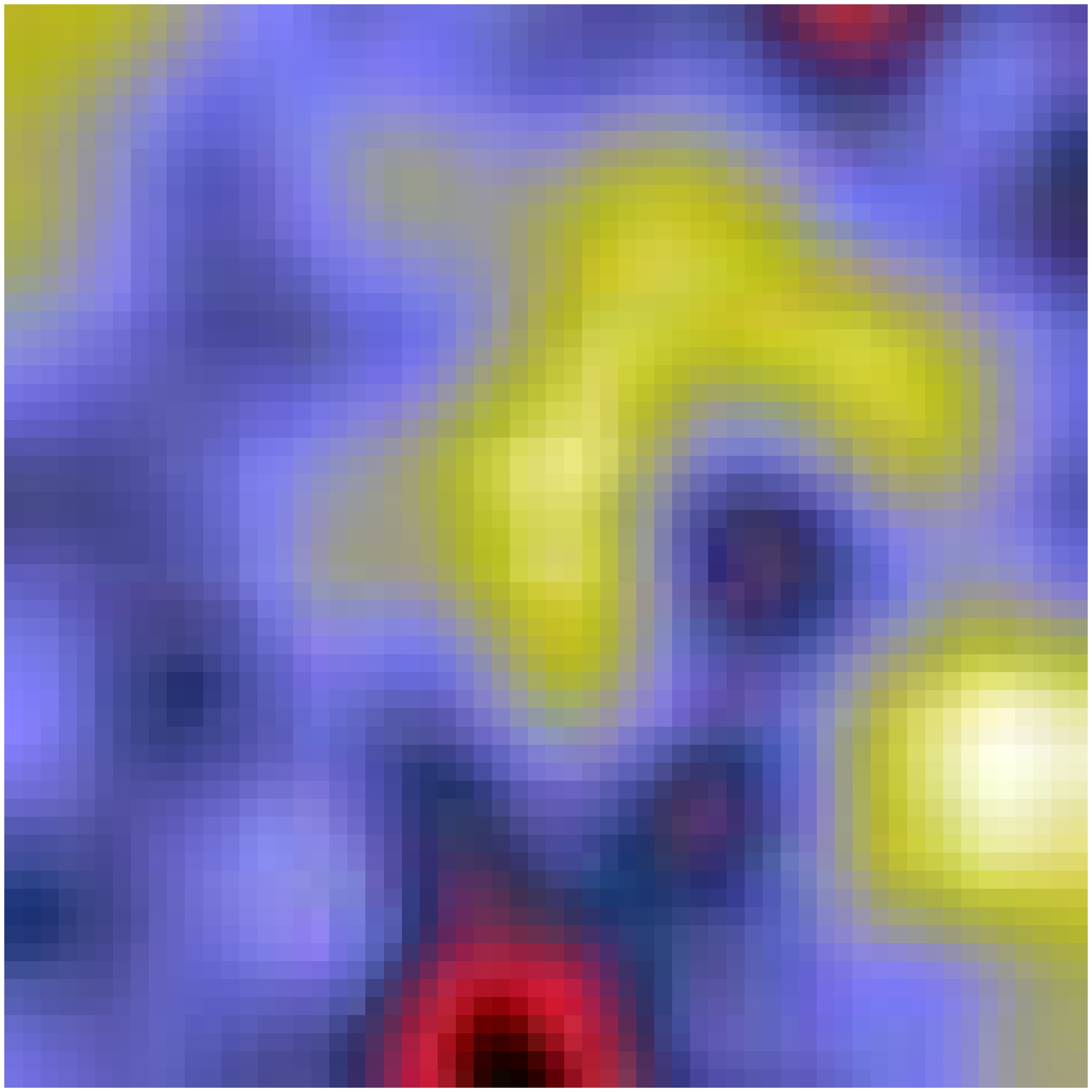} \\  {\#20   NBZ5004664}  \\
  \FGb{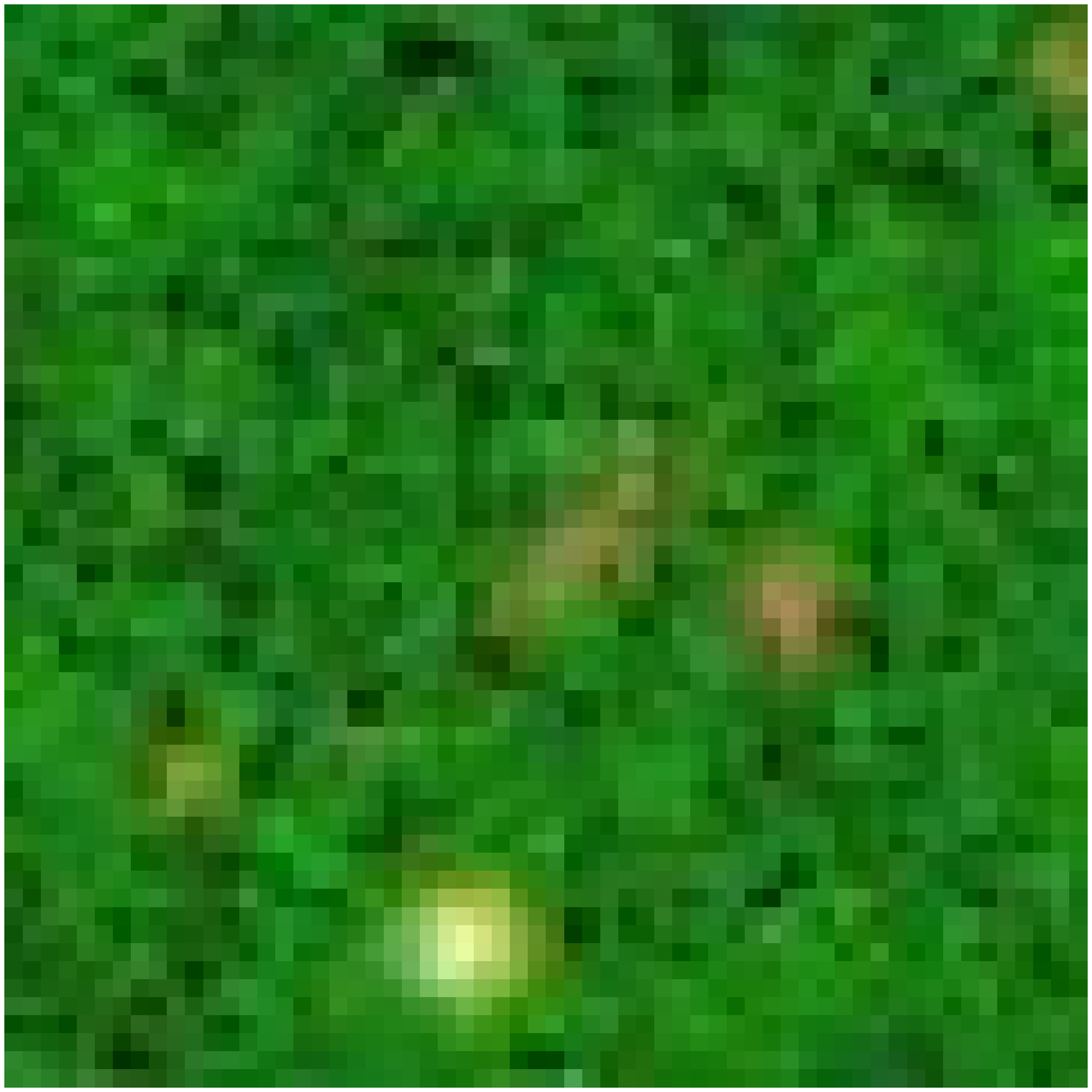}  \FGb{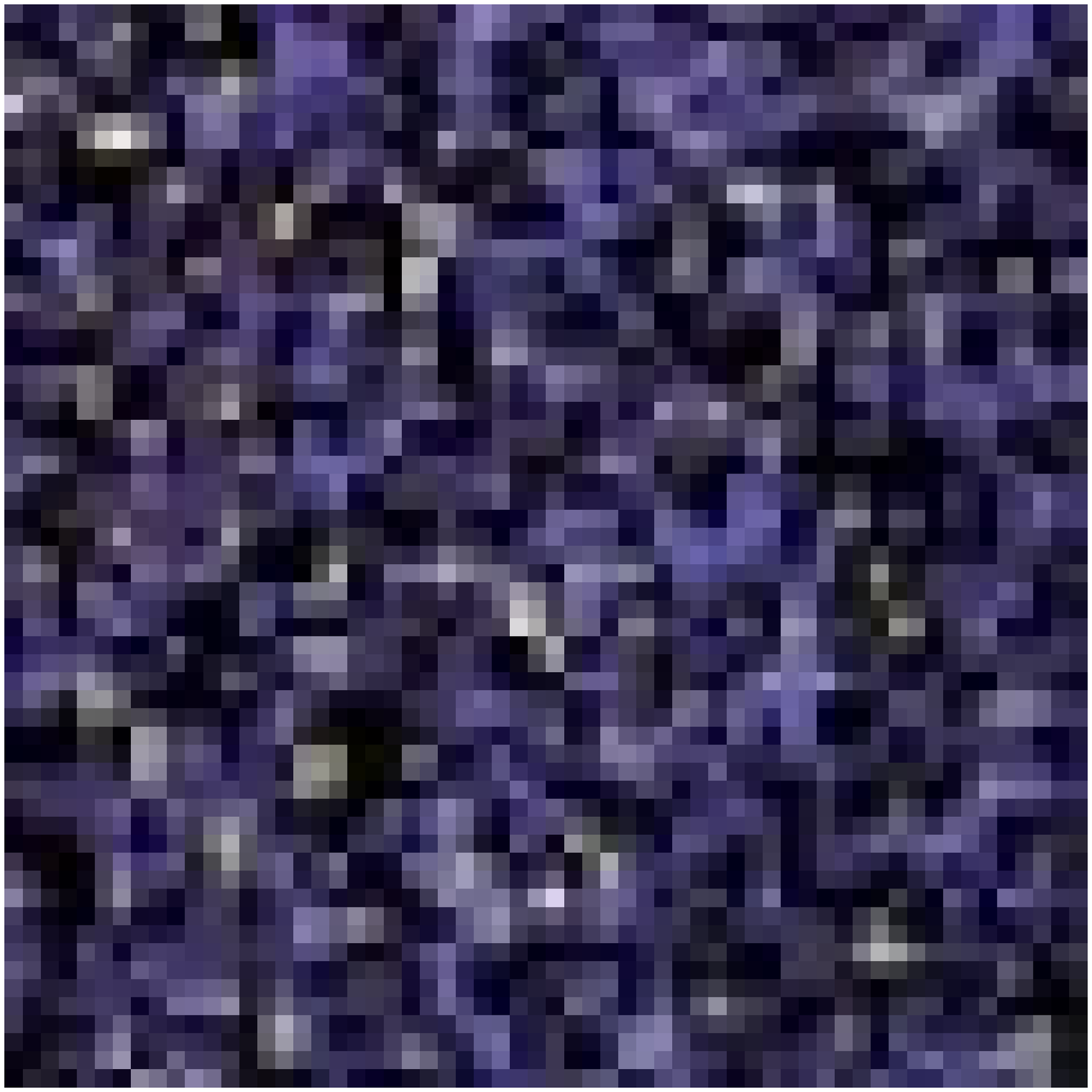}  \FGb{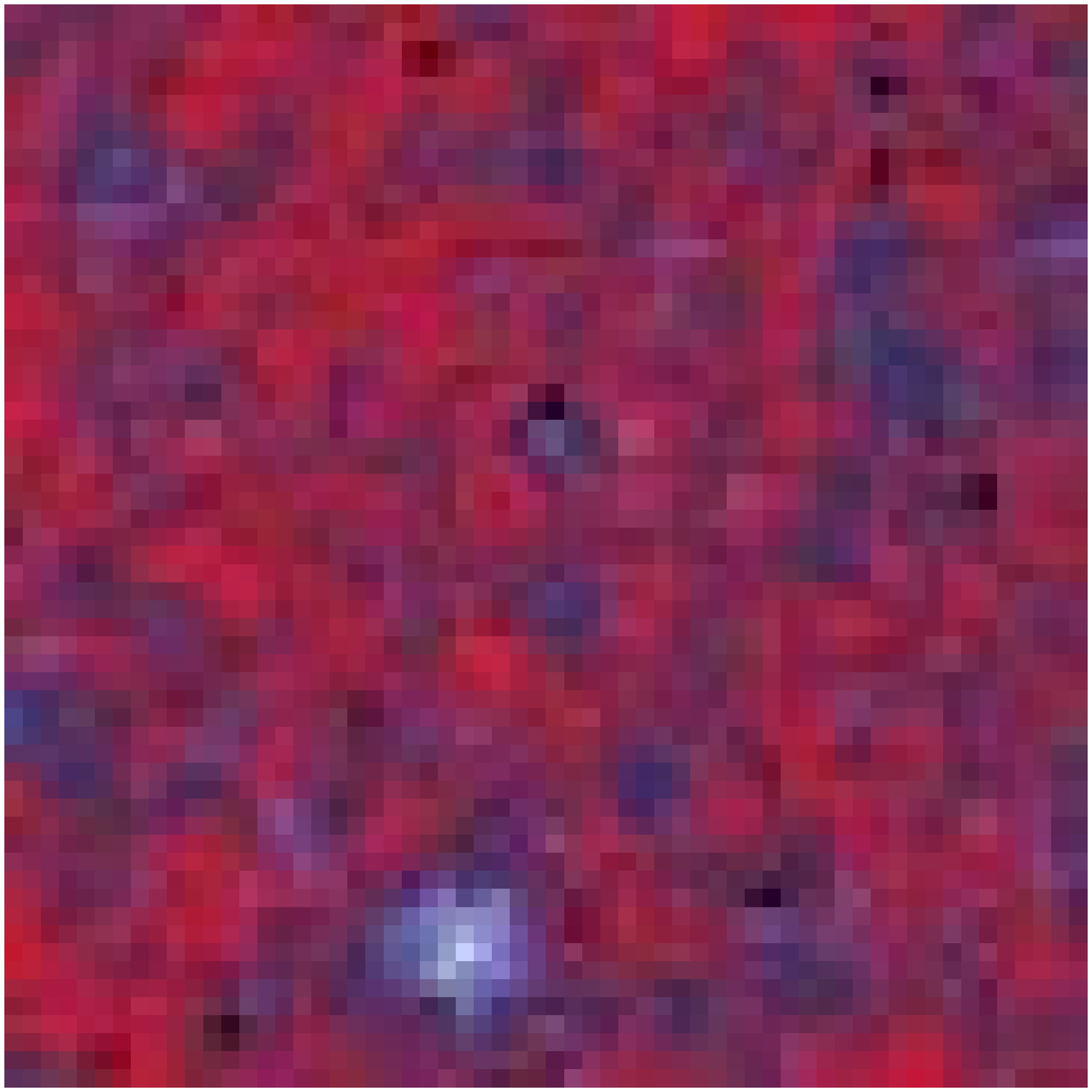}  \FGb{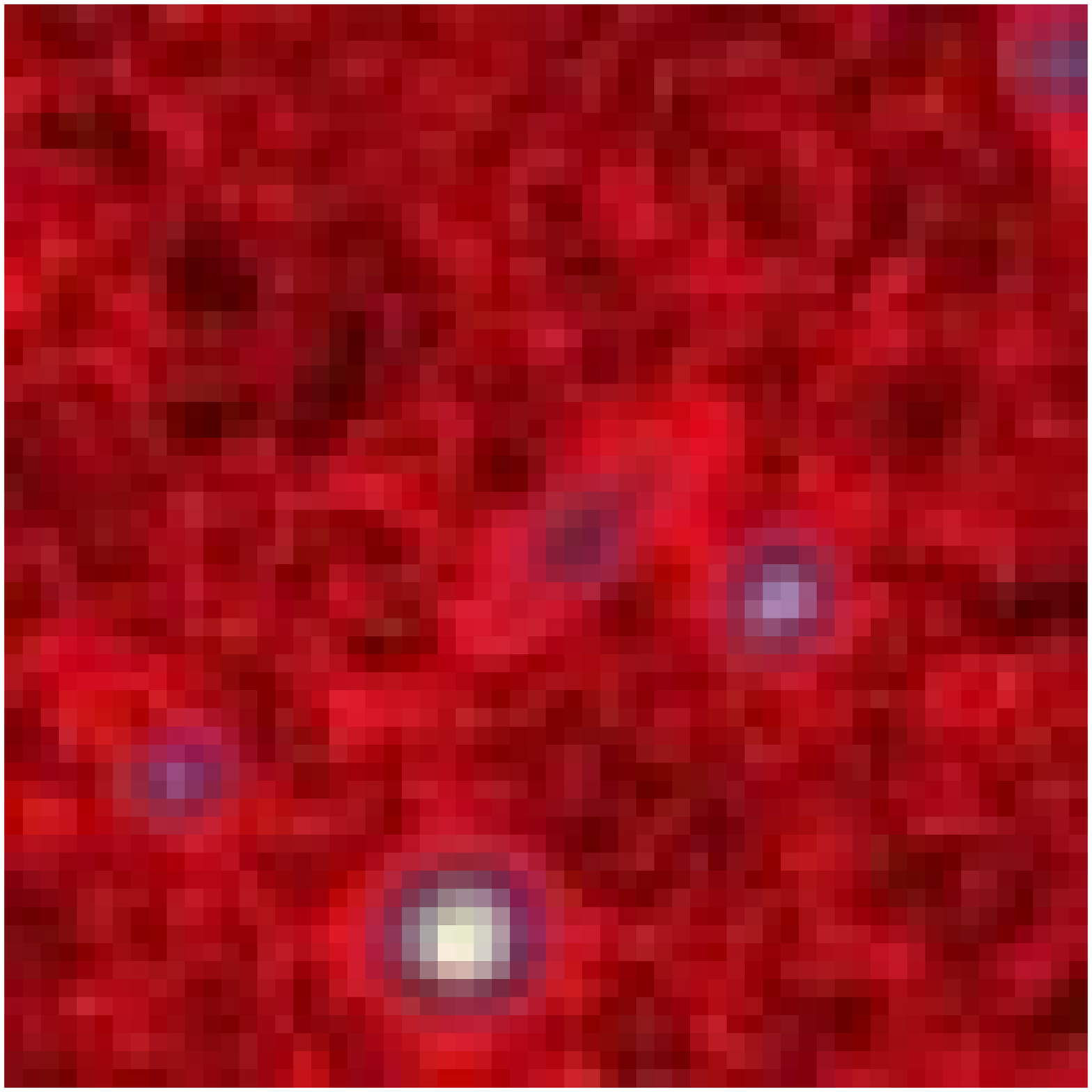}  \FGb{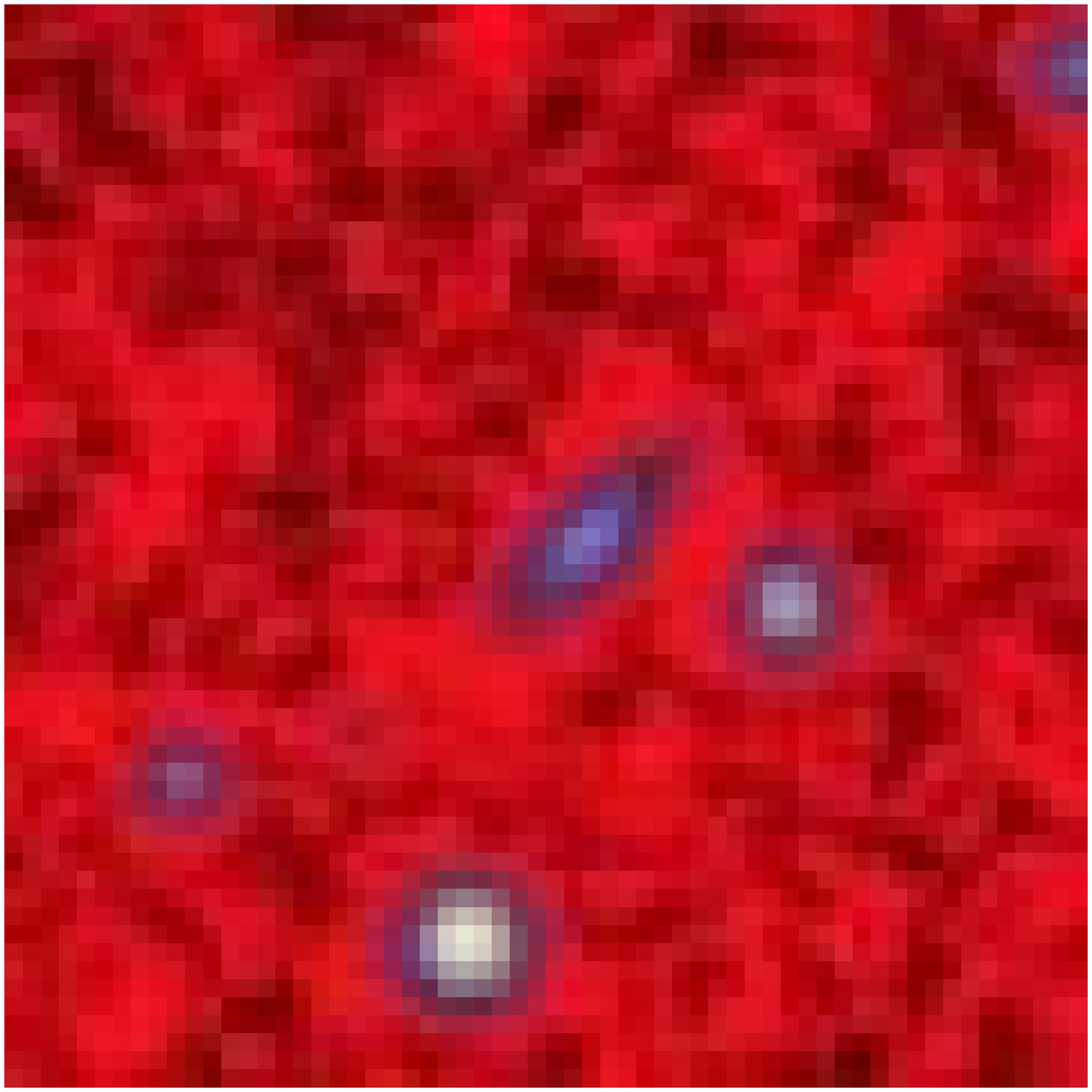}  \FGb{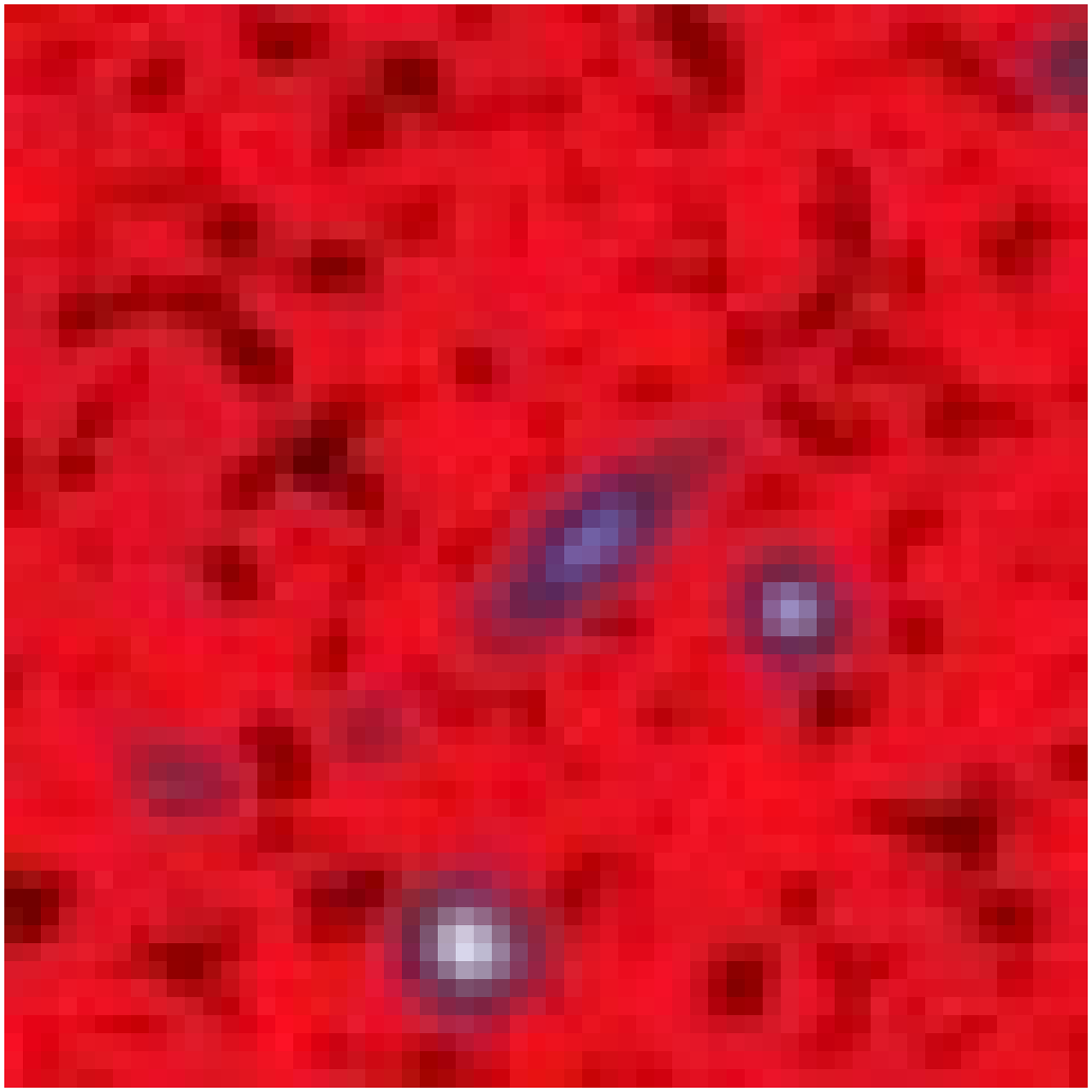}  \FGb{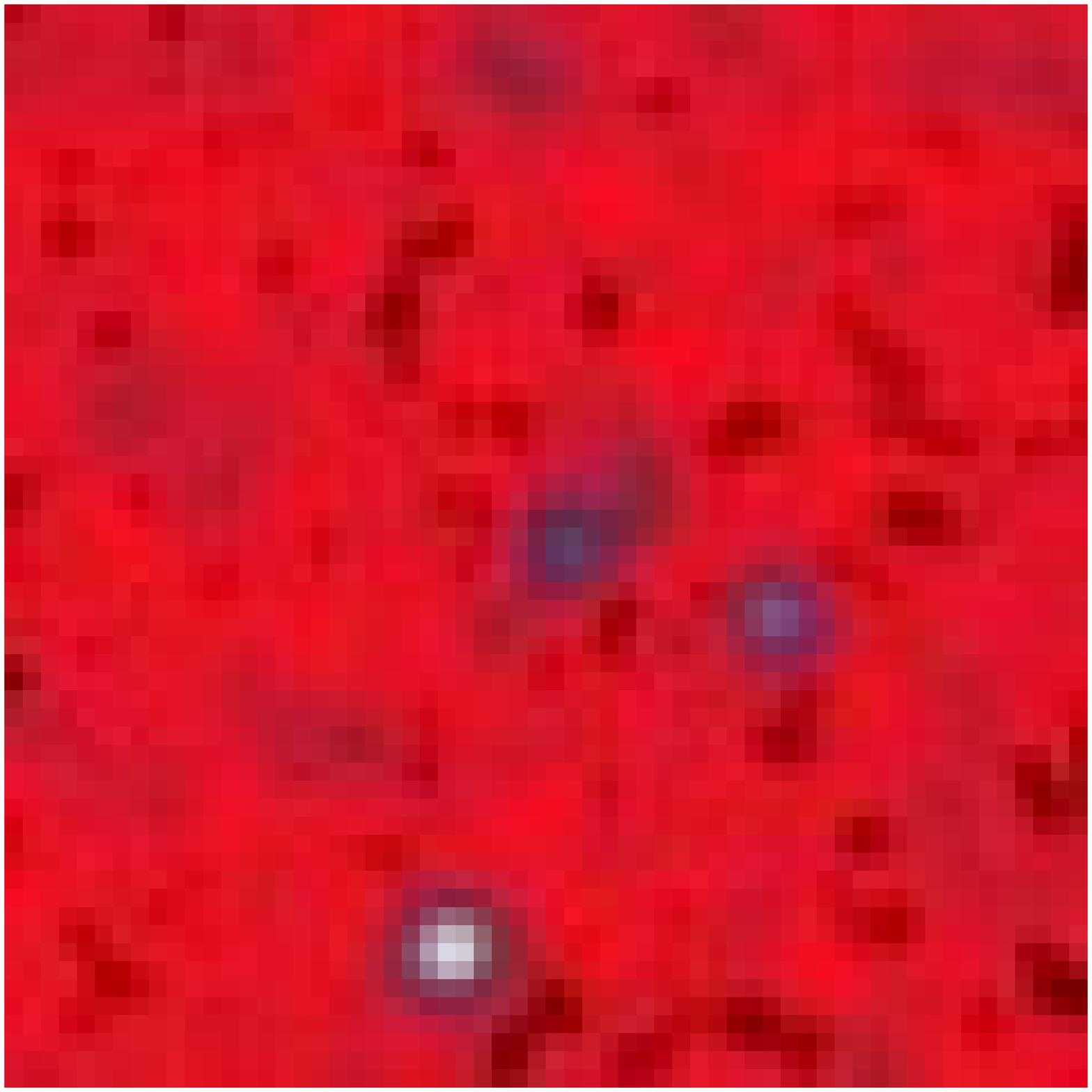}  \FGb{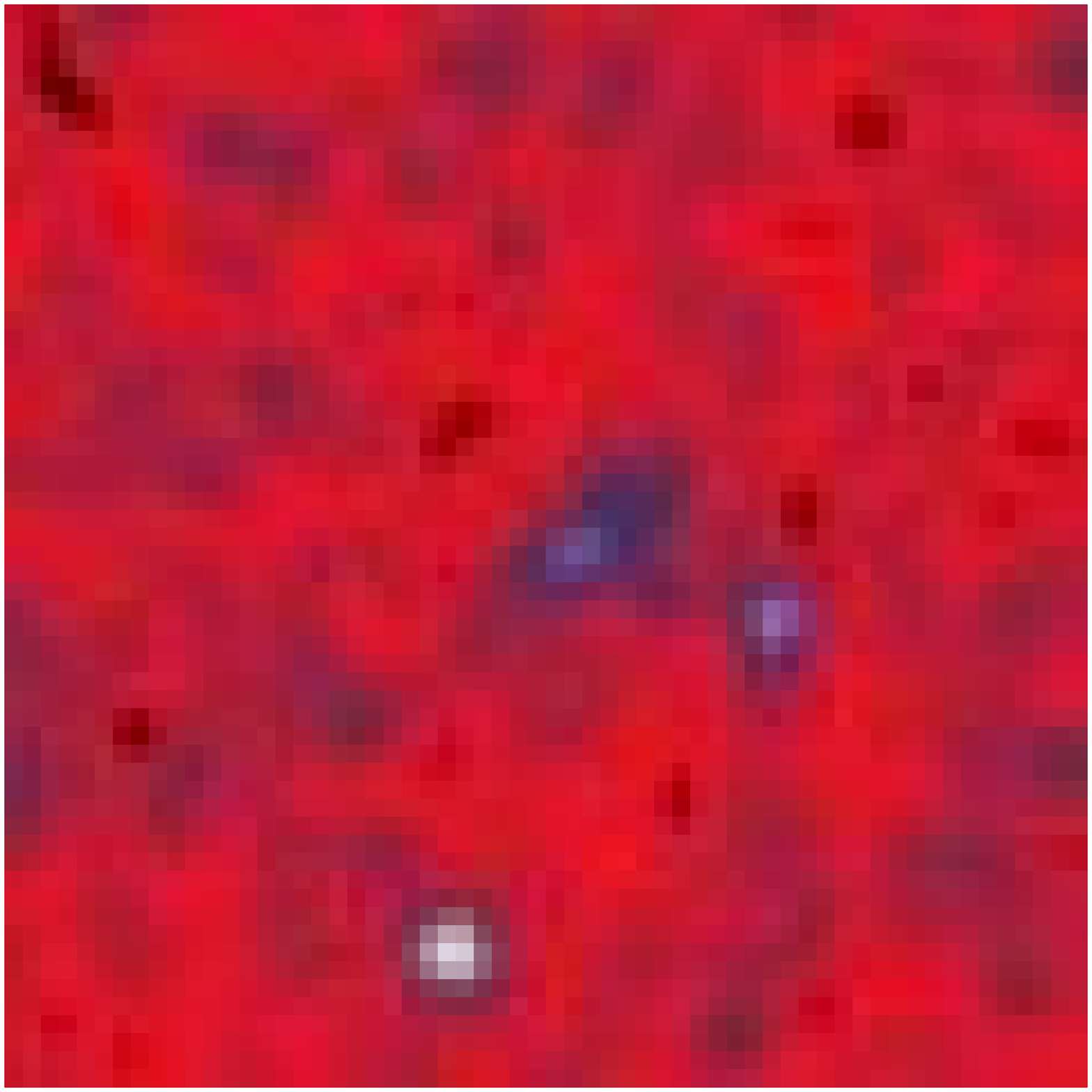}  \FGb{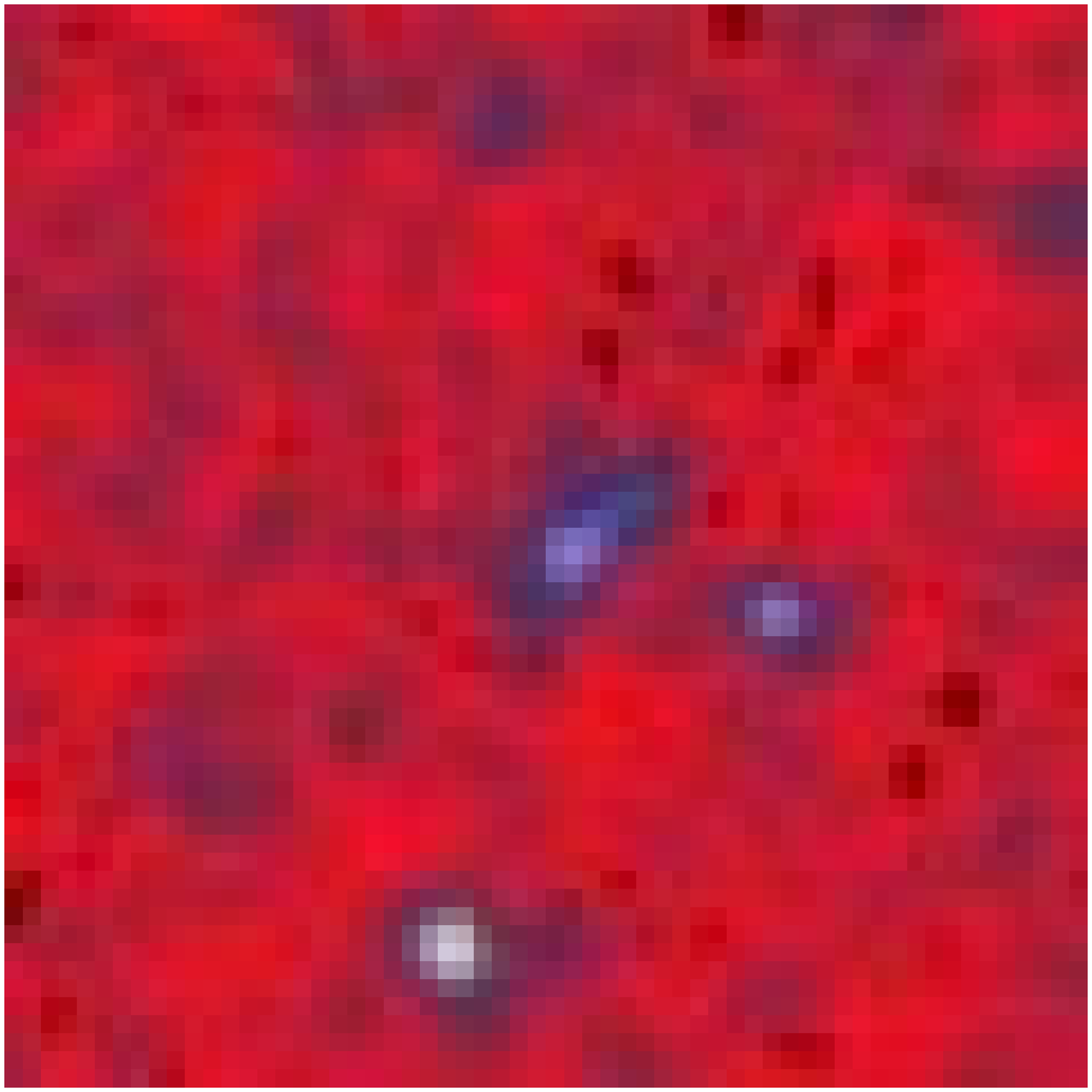}  \FGb{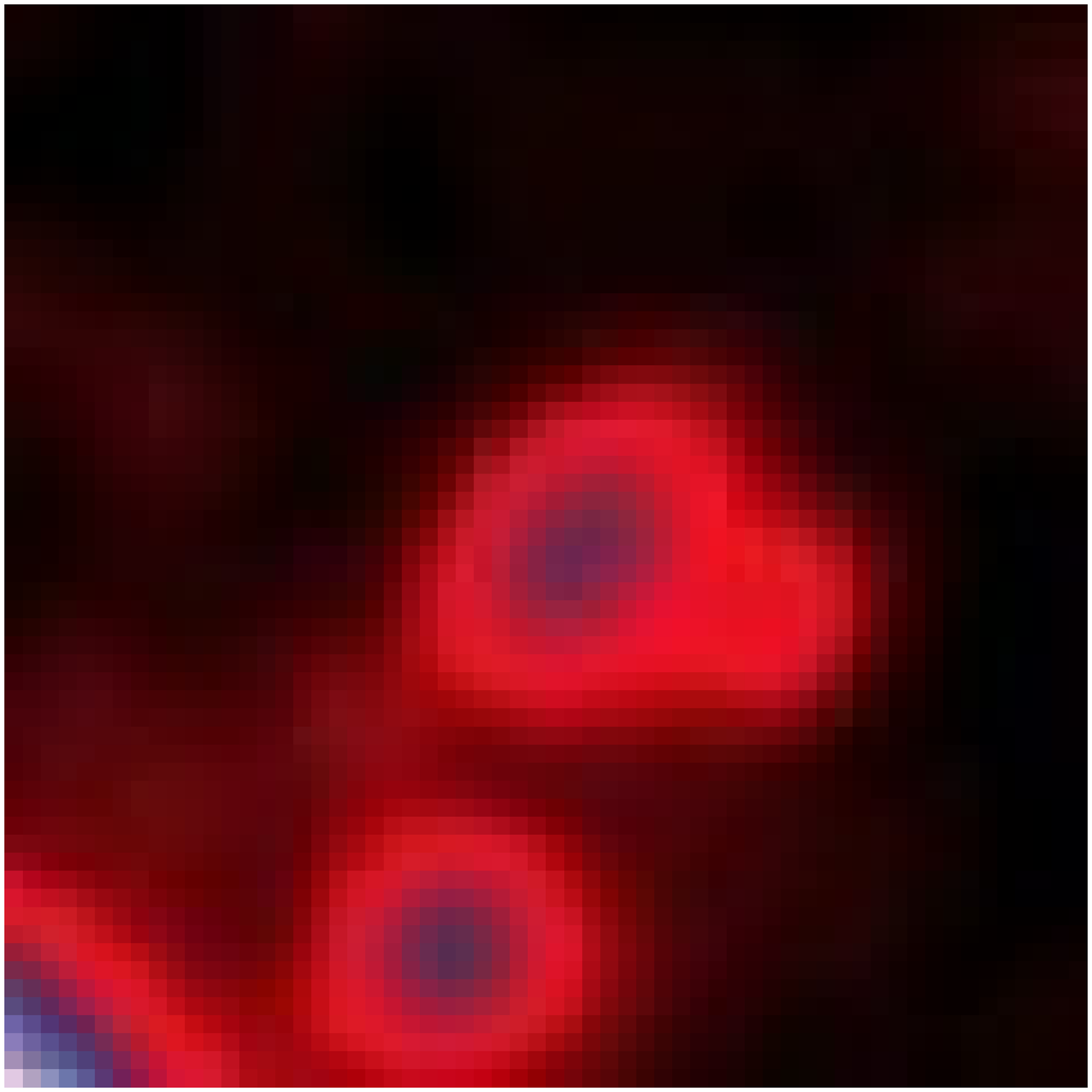}  \FGb{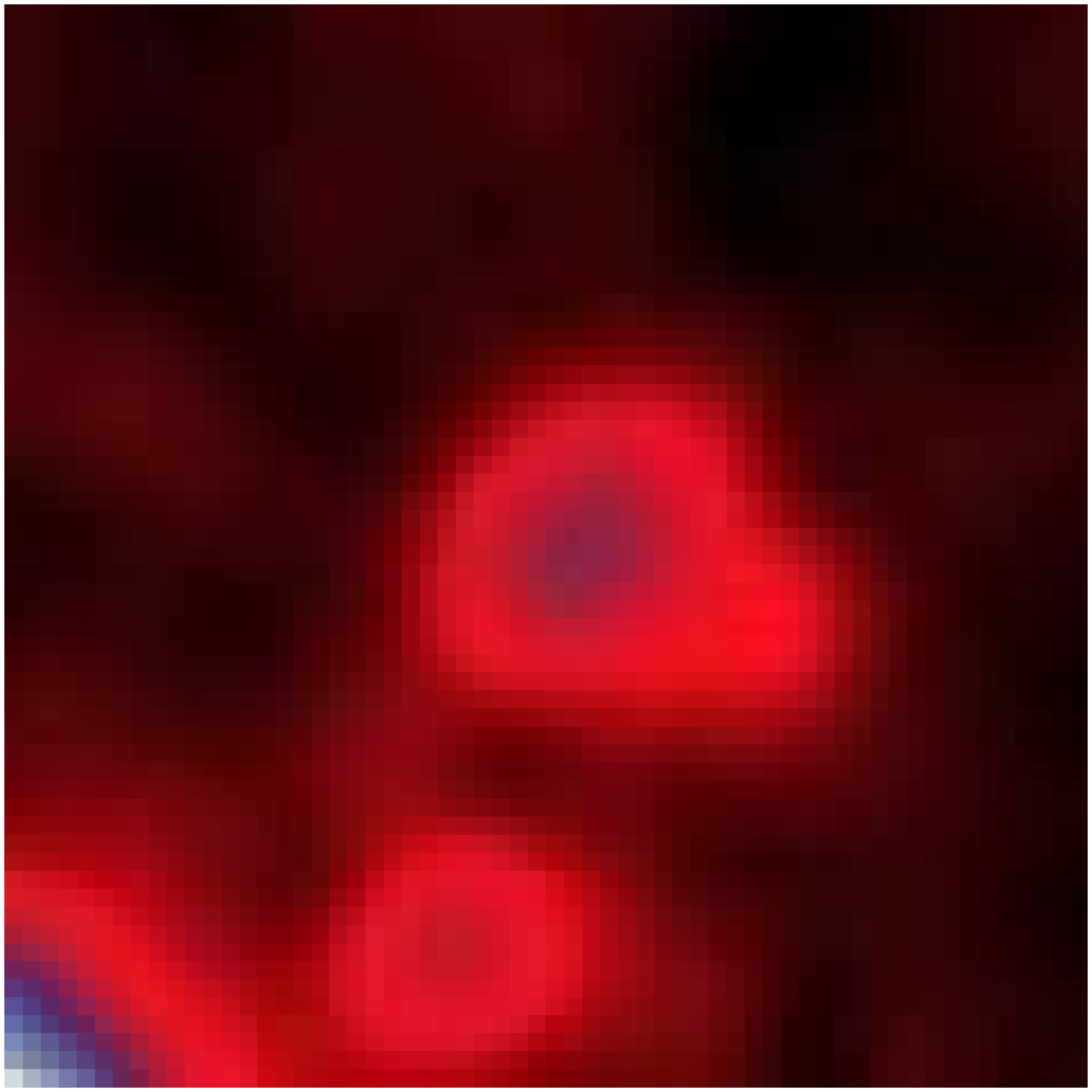}  \FGb{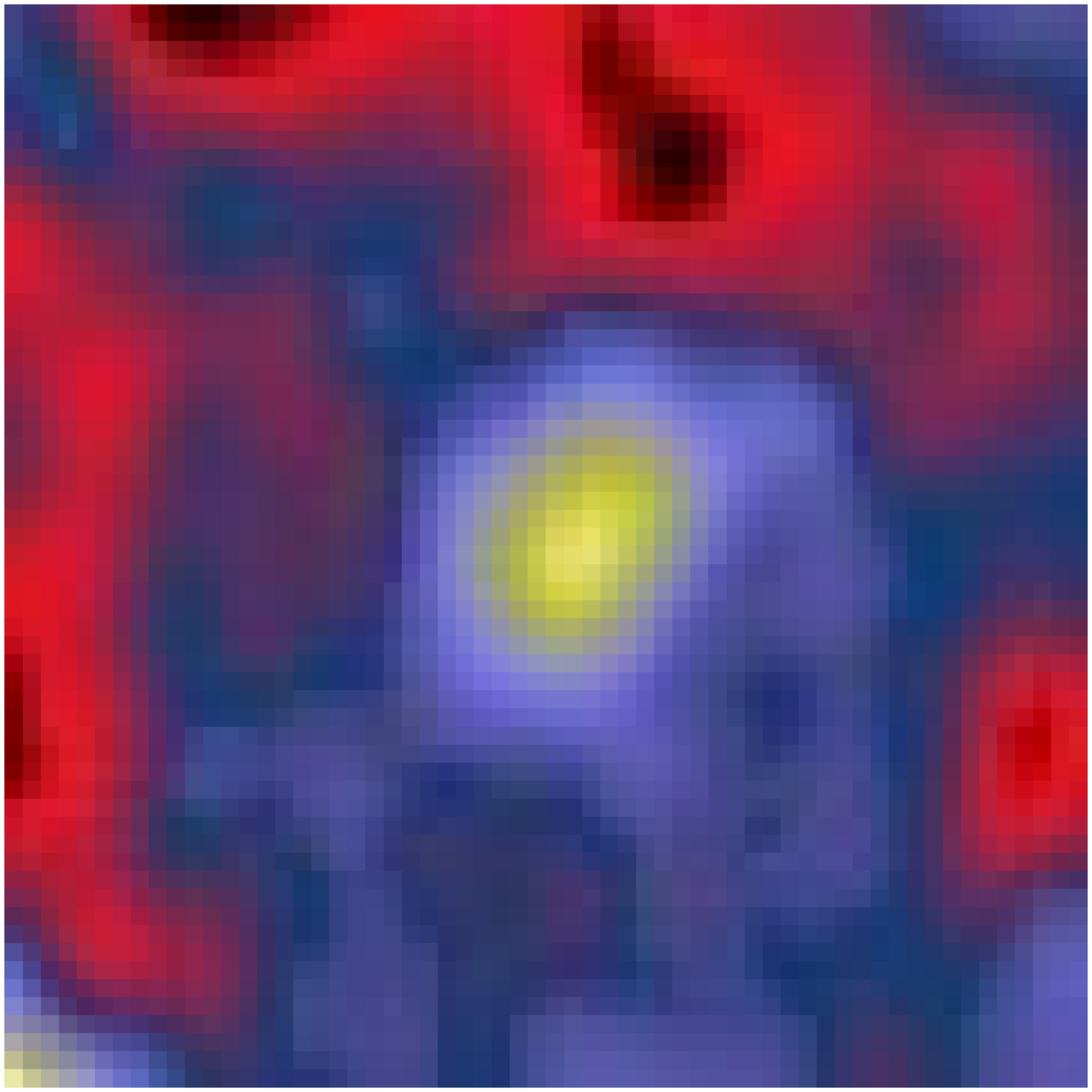}  \FGb{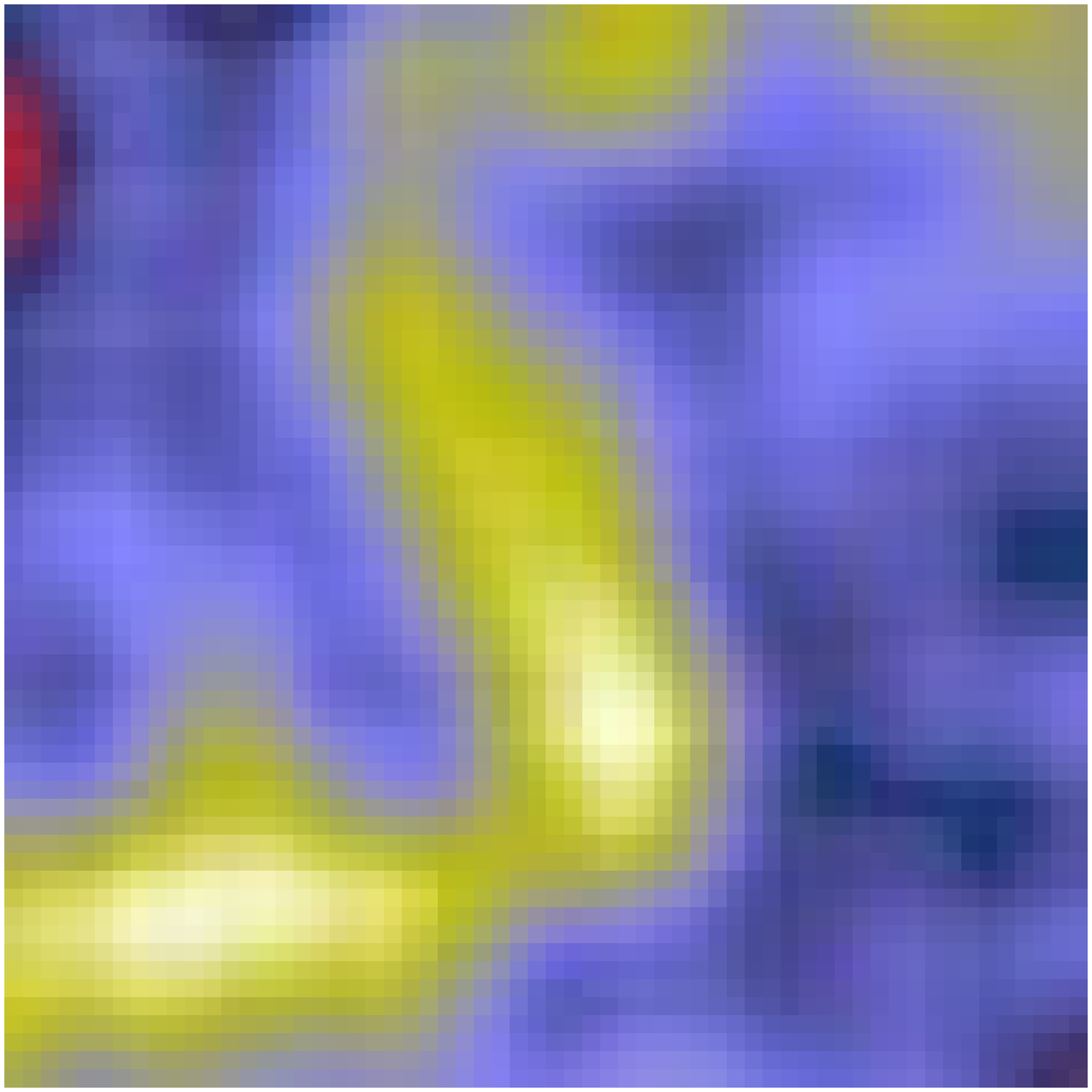} \\  {\#21   NCIA007657}  \\
  \FGb{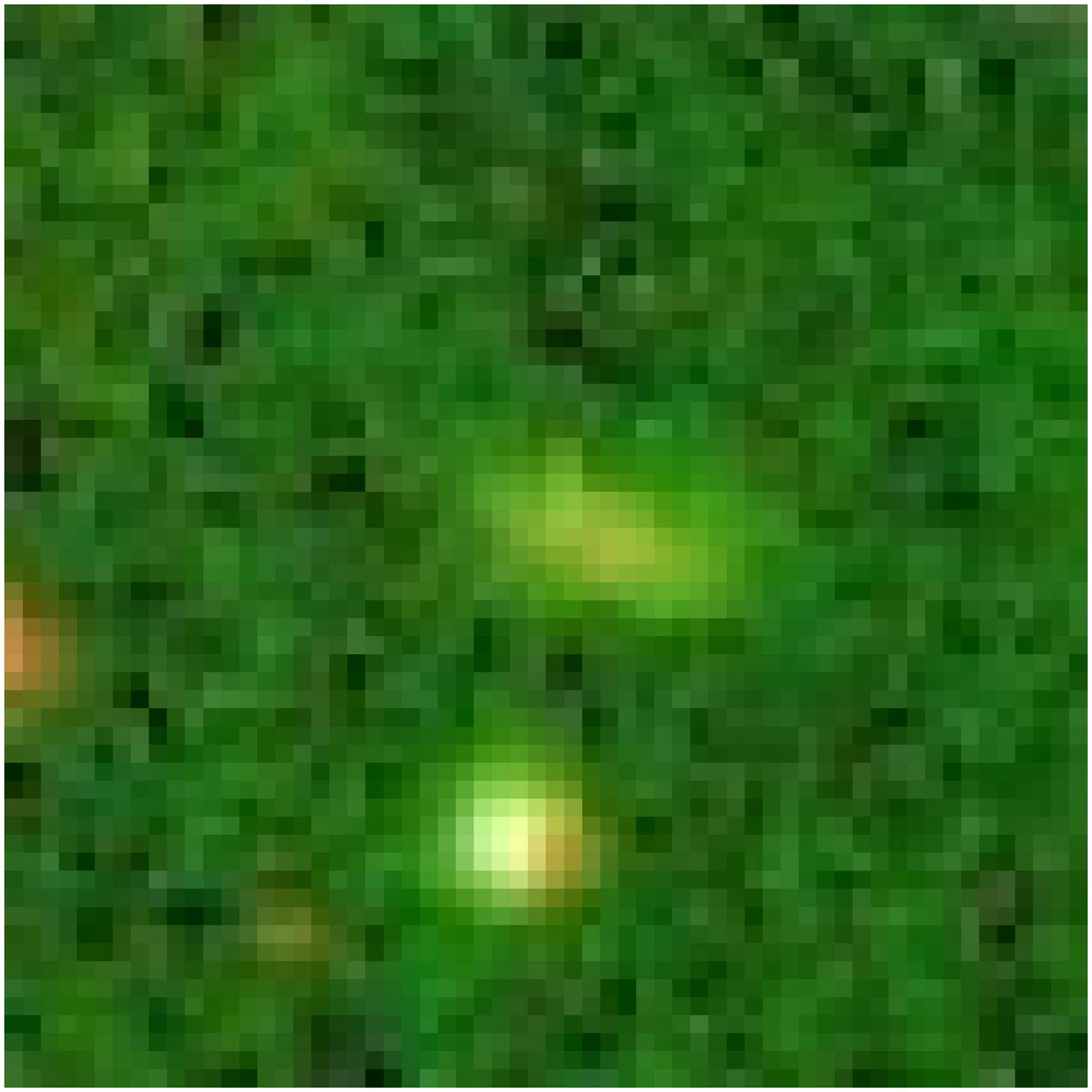}  \FGb{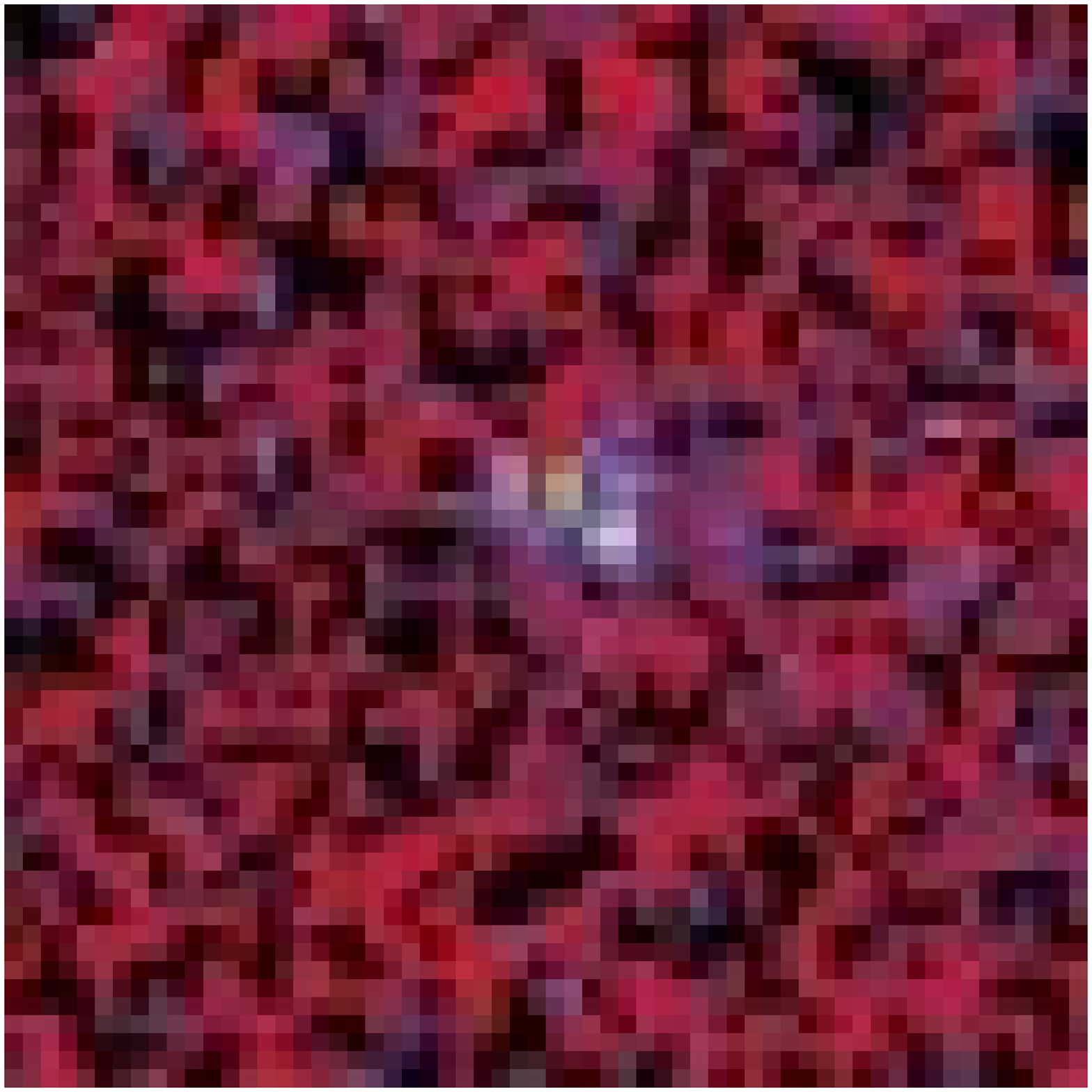}  \FGb{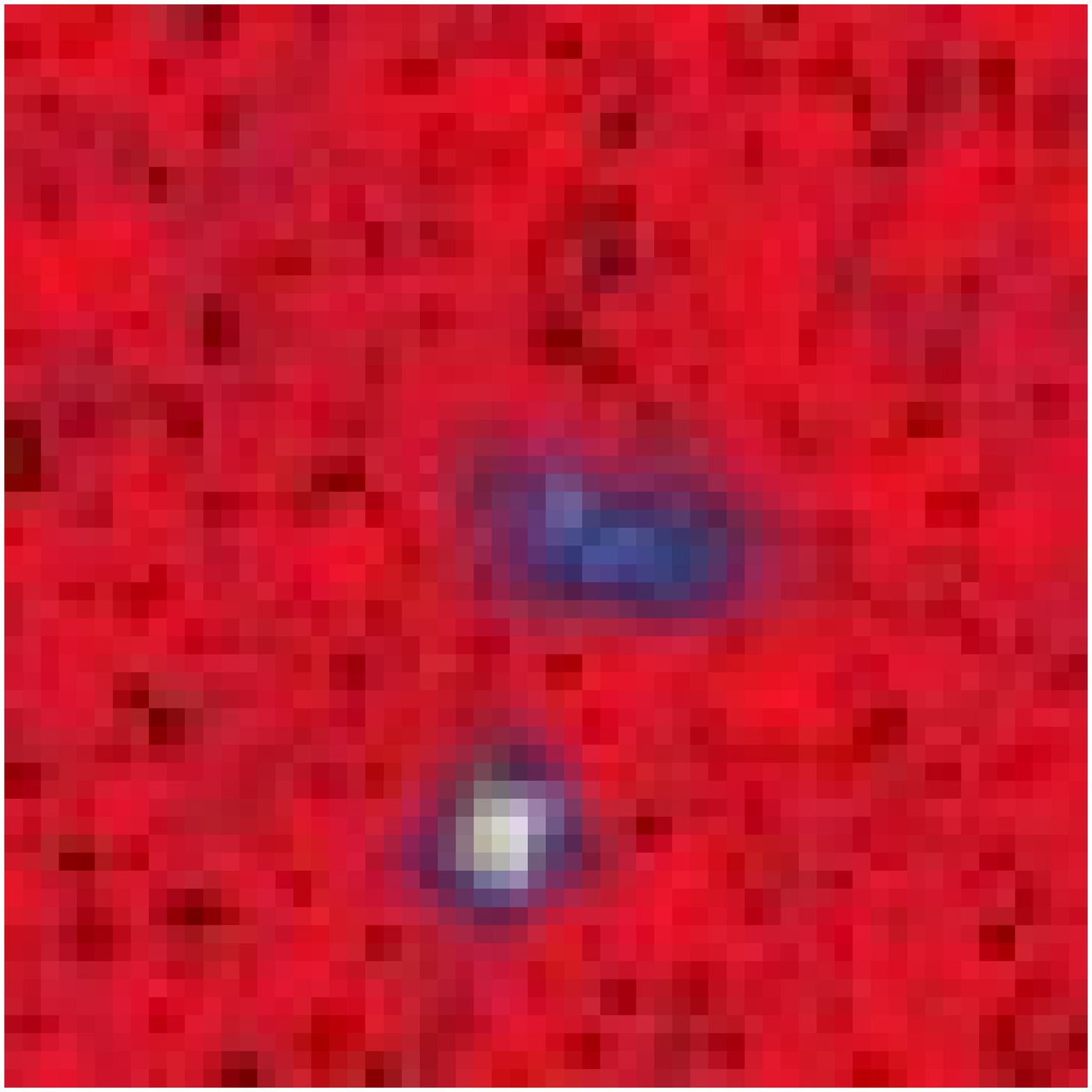}  \FGb{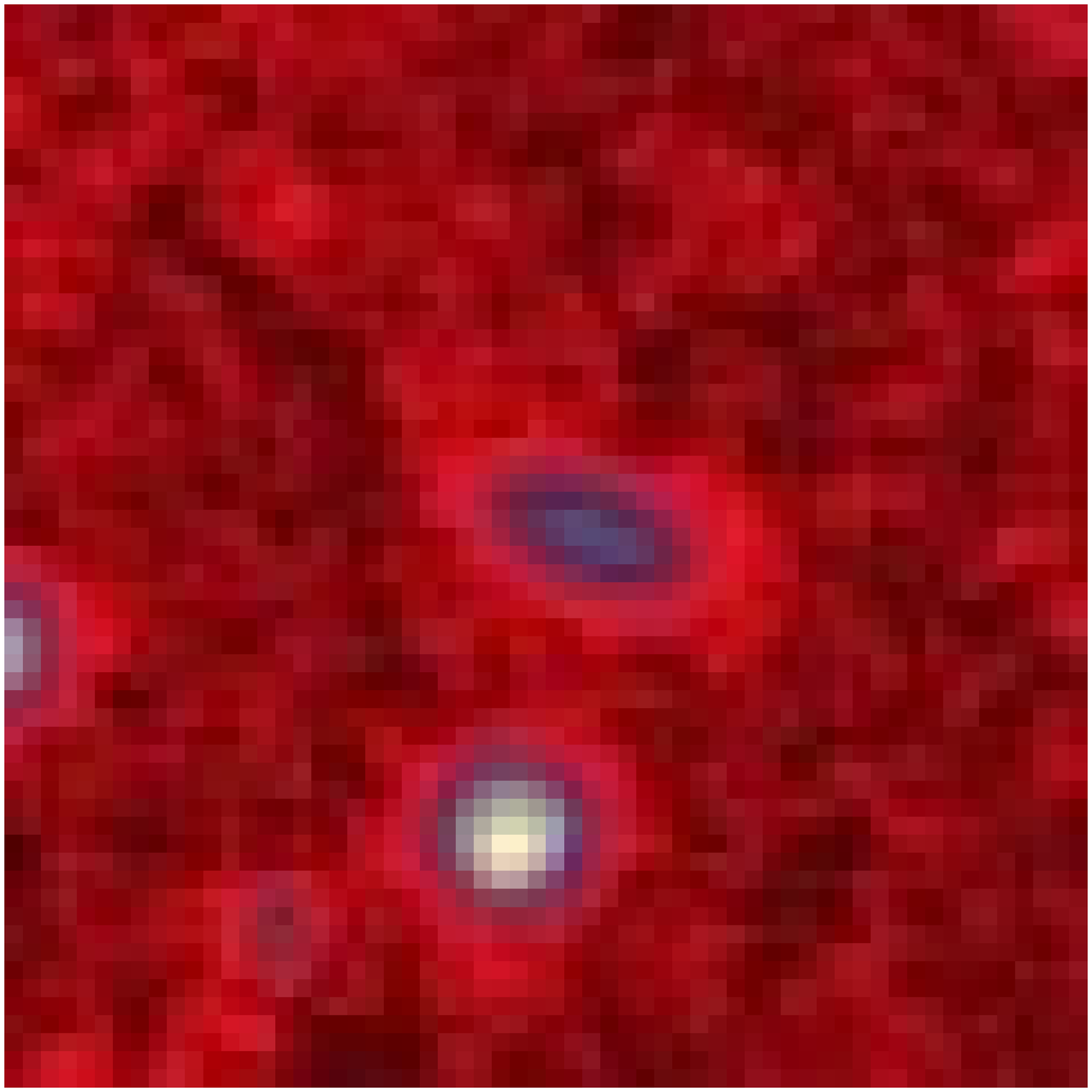}  \FGb{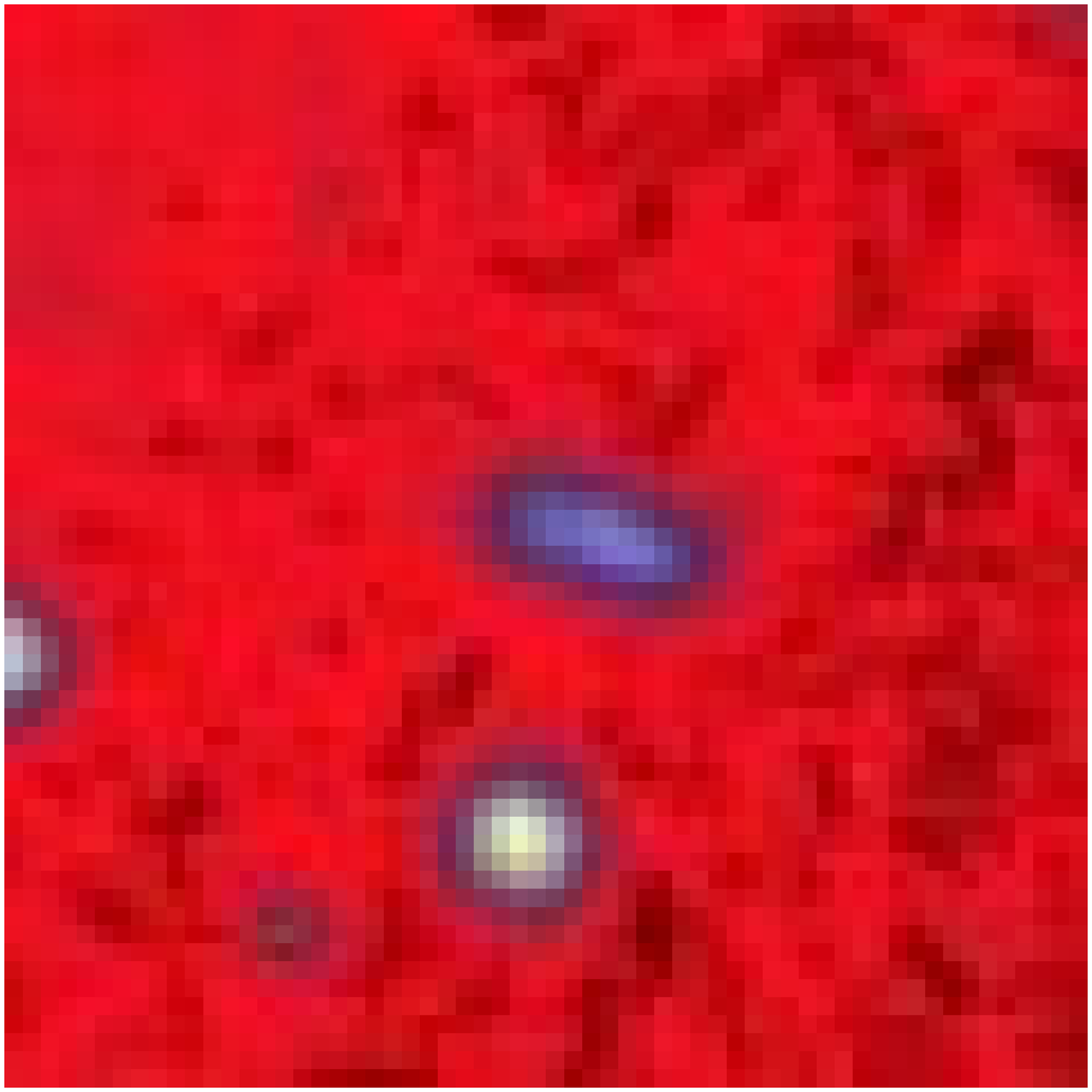}  \FGb{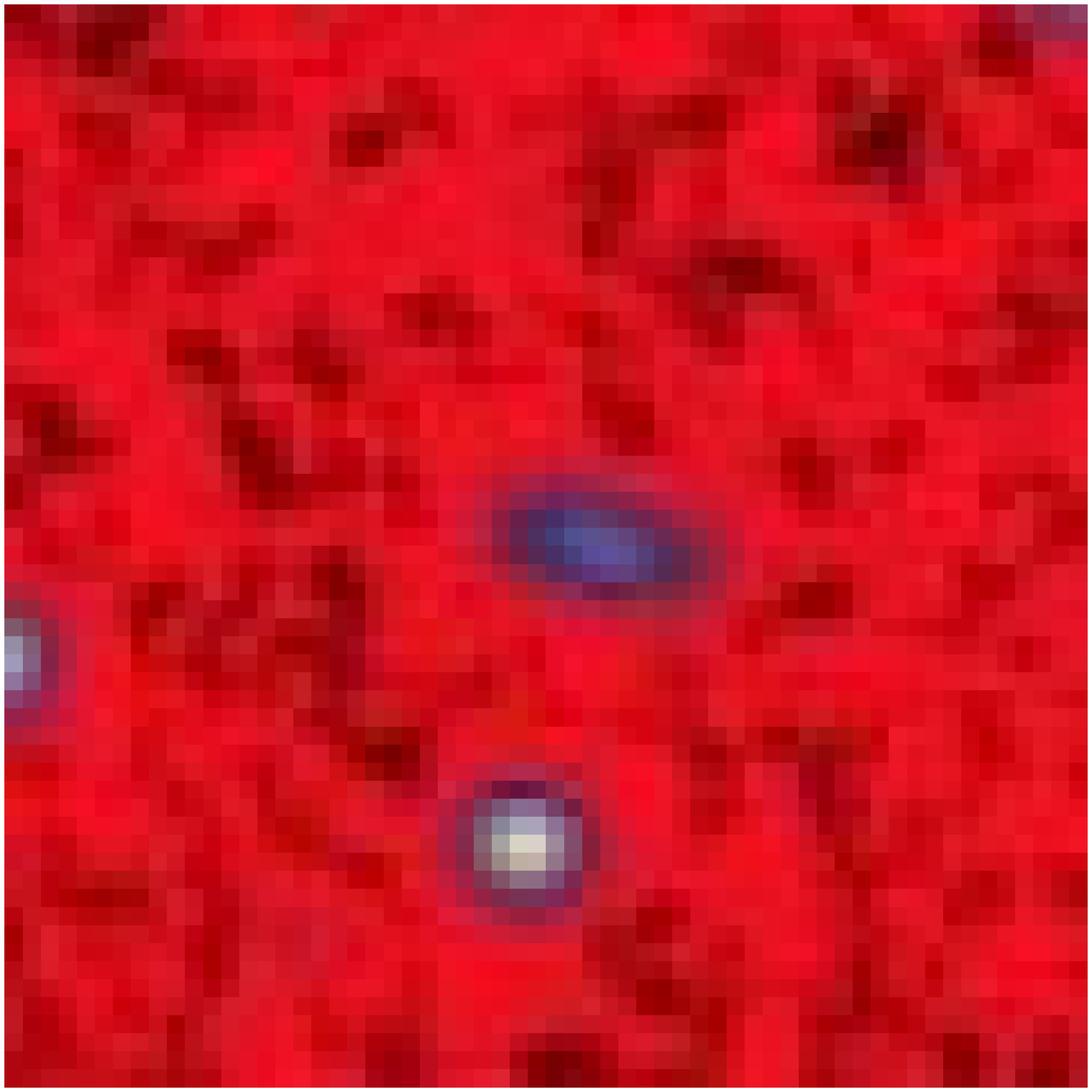}  \FGb{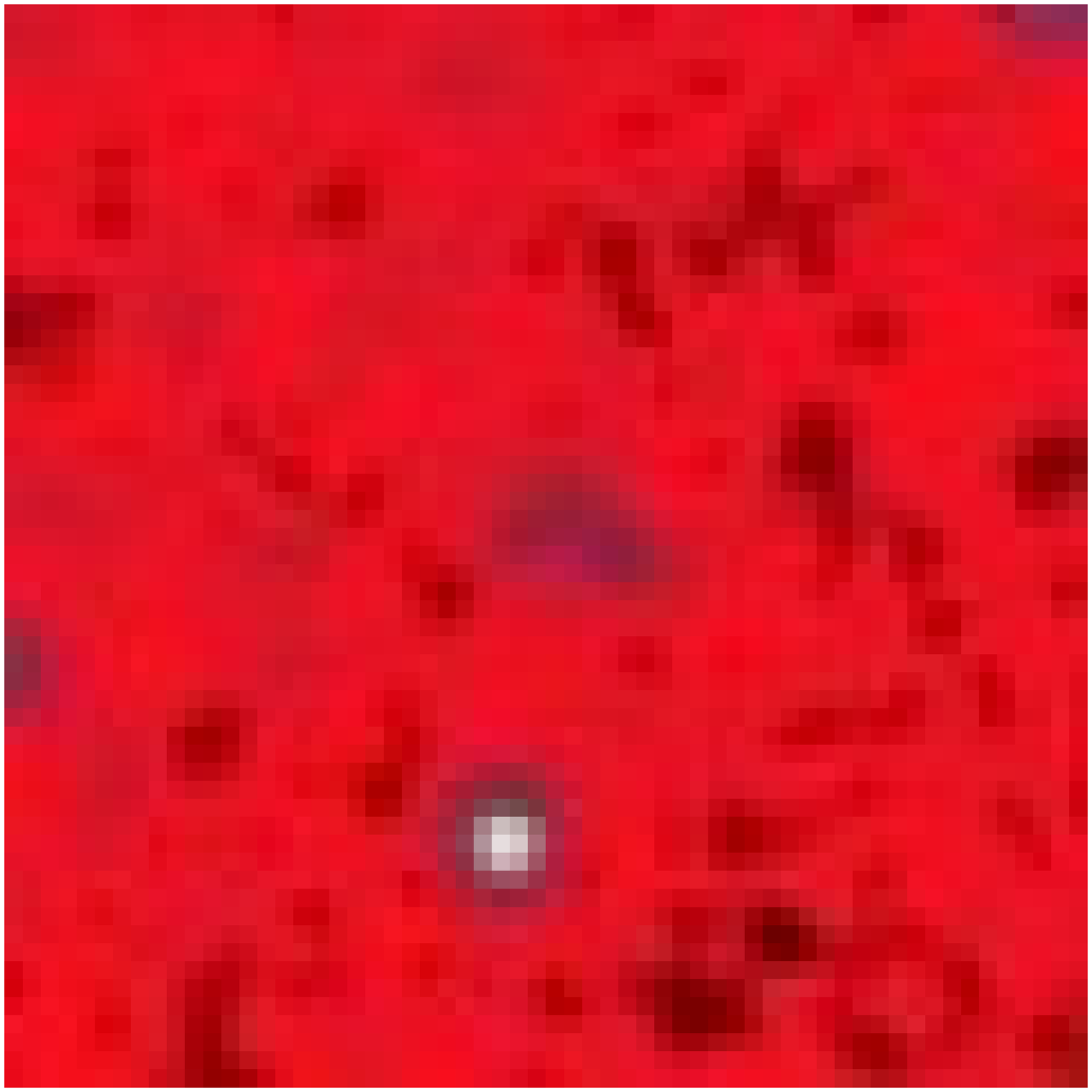}  \FGb{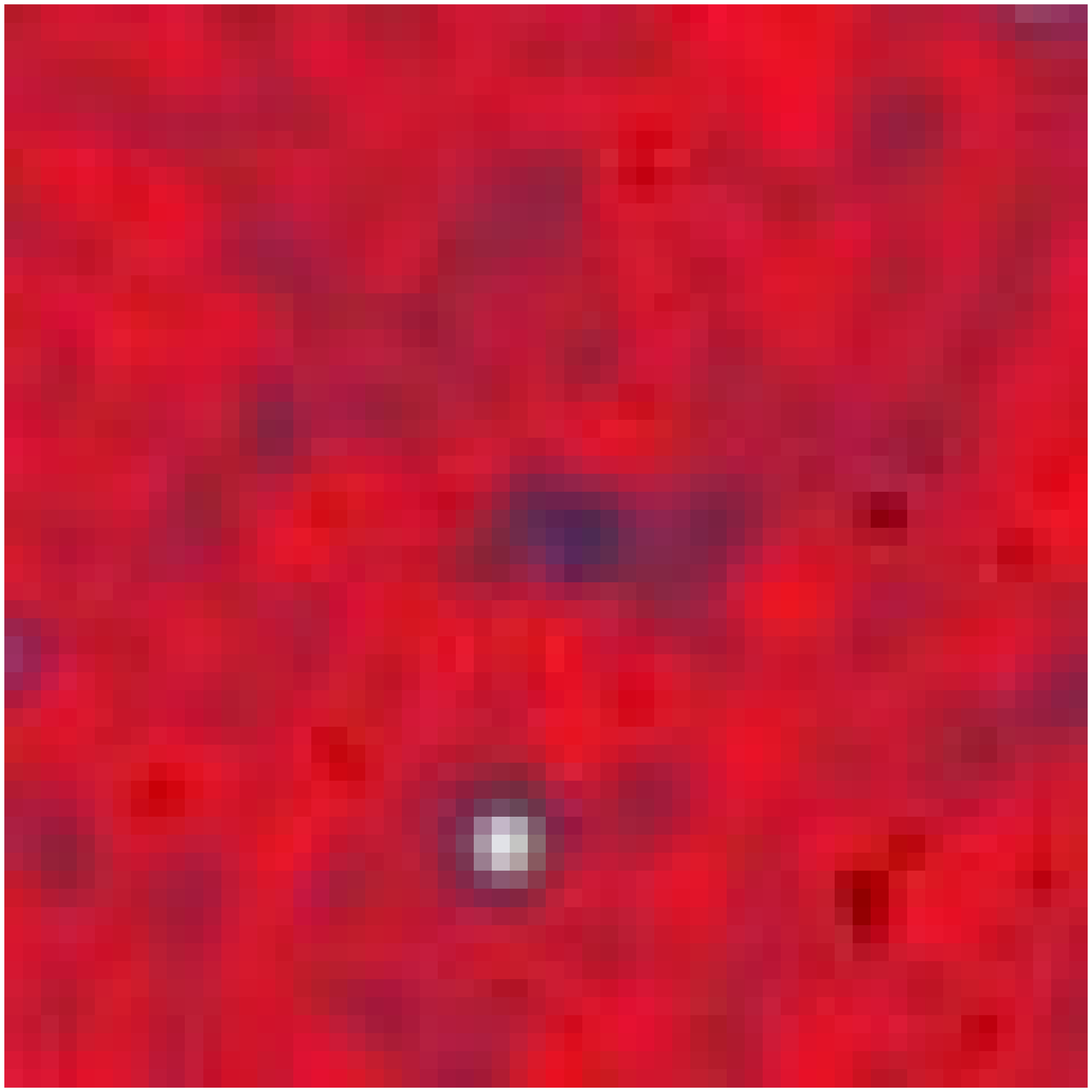}  \FGb{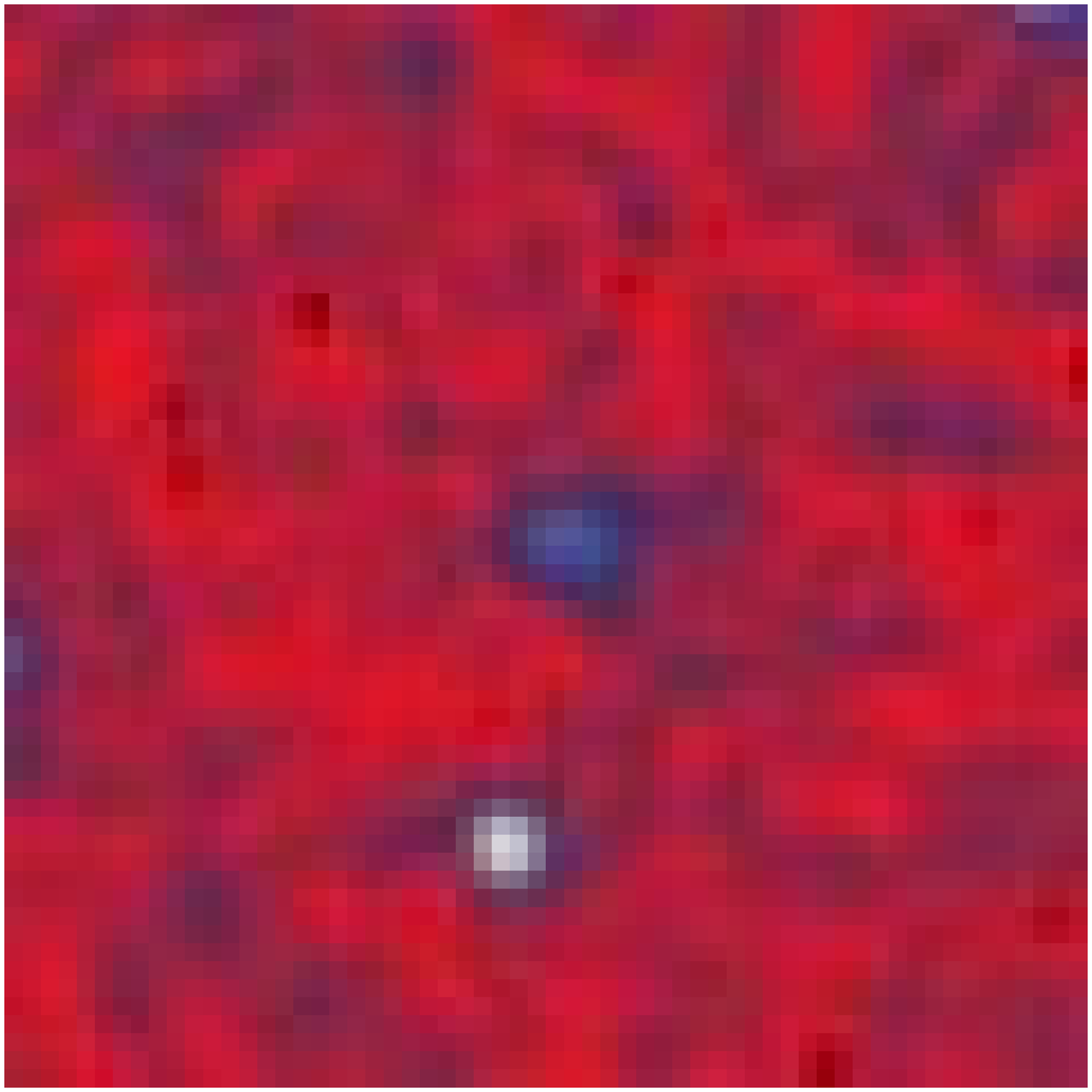}  \FGb{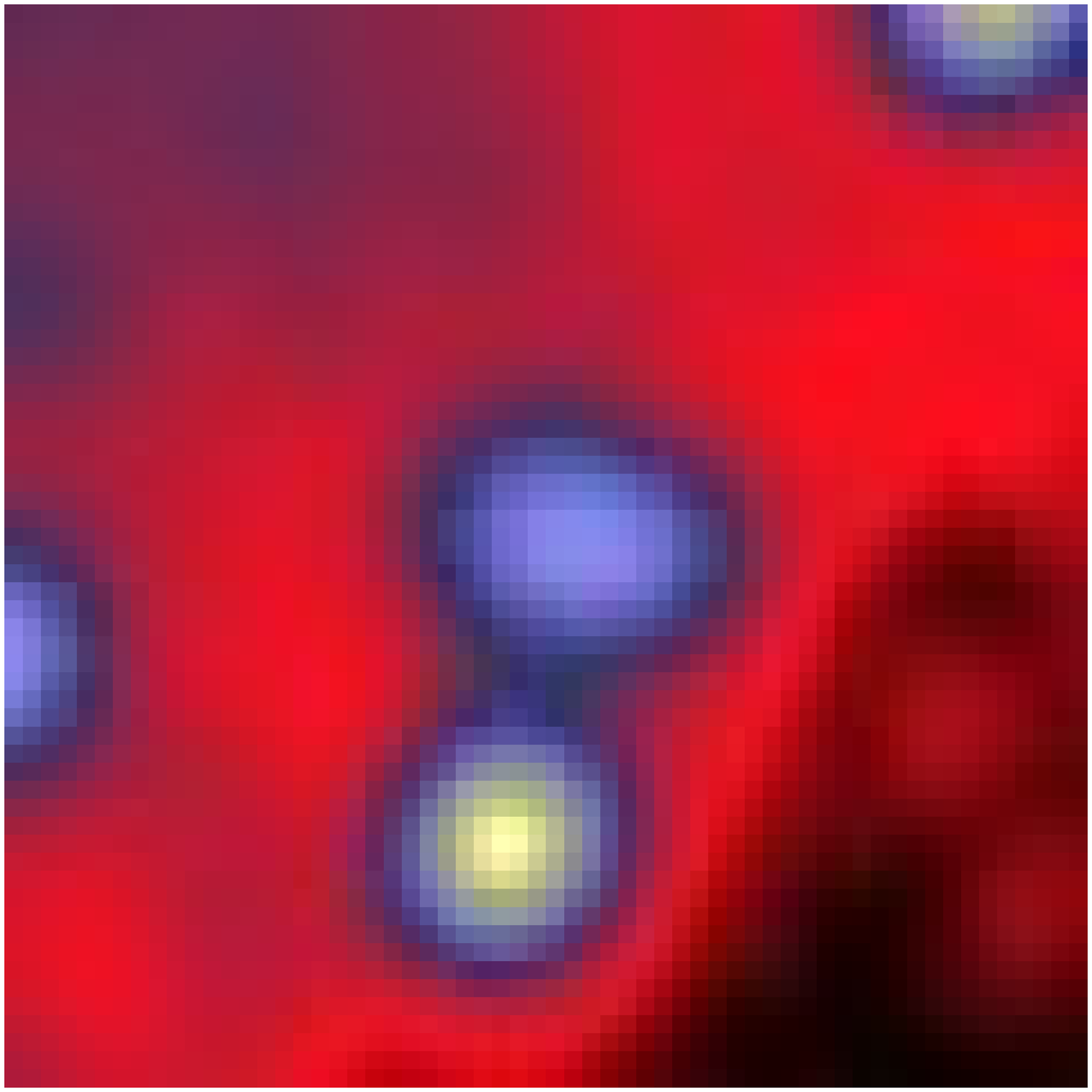}  \FGb{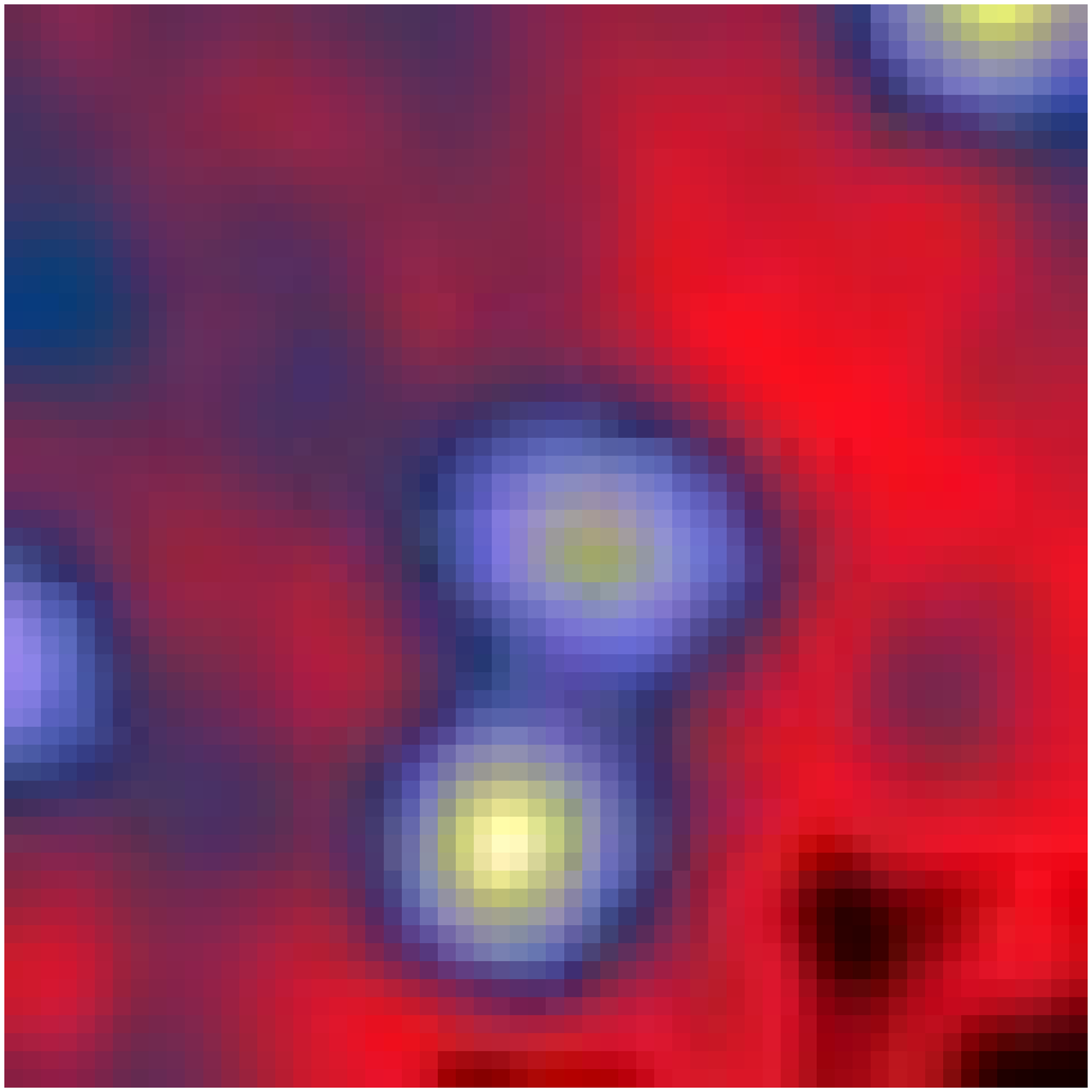}  \FGb{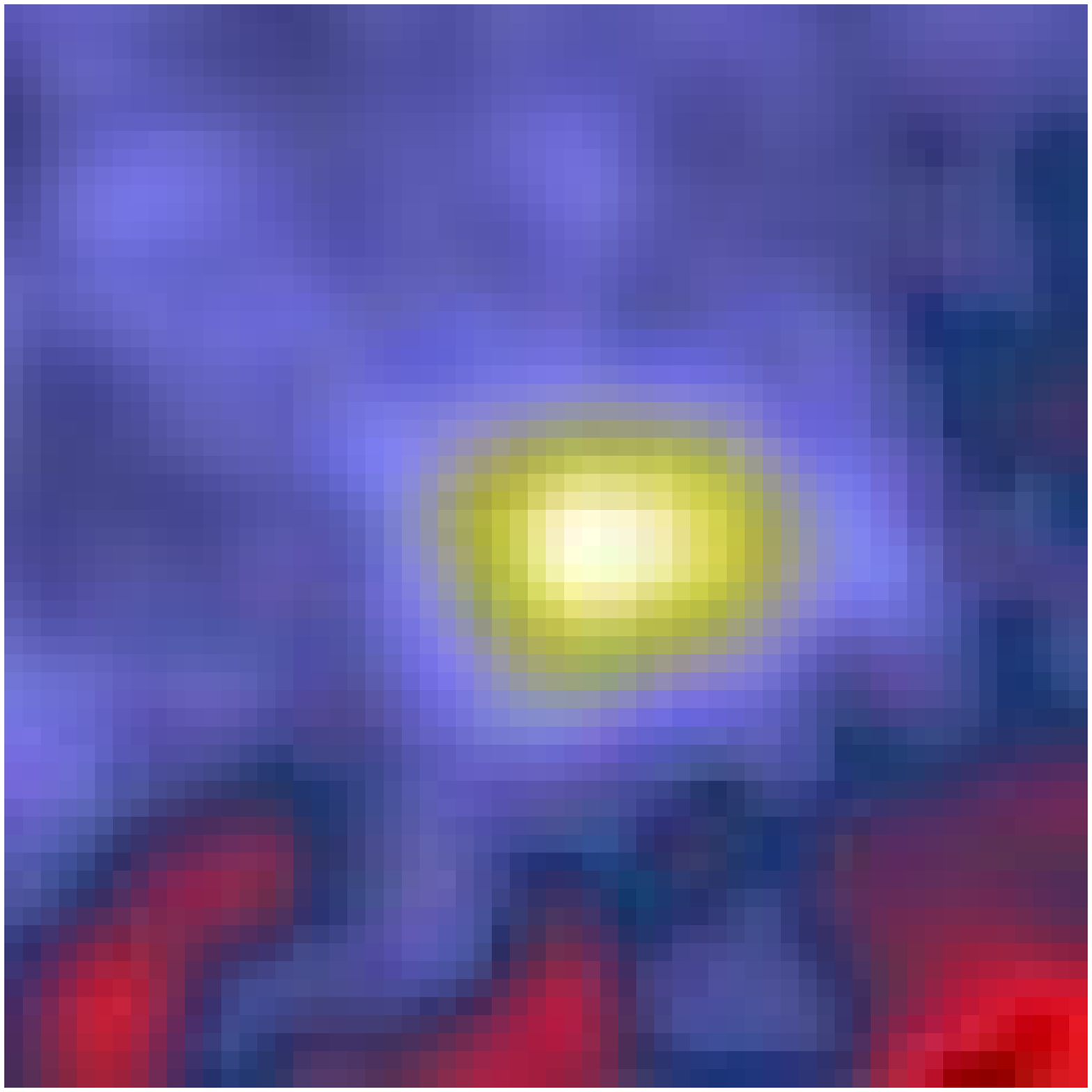}  \FGb{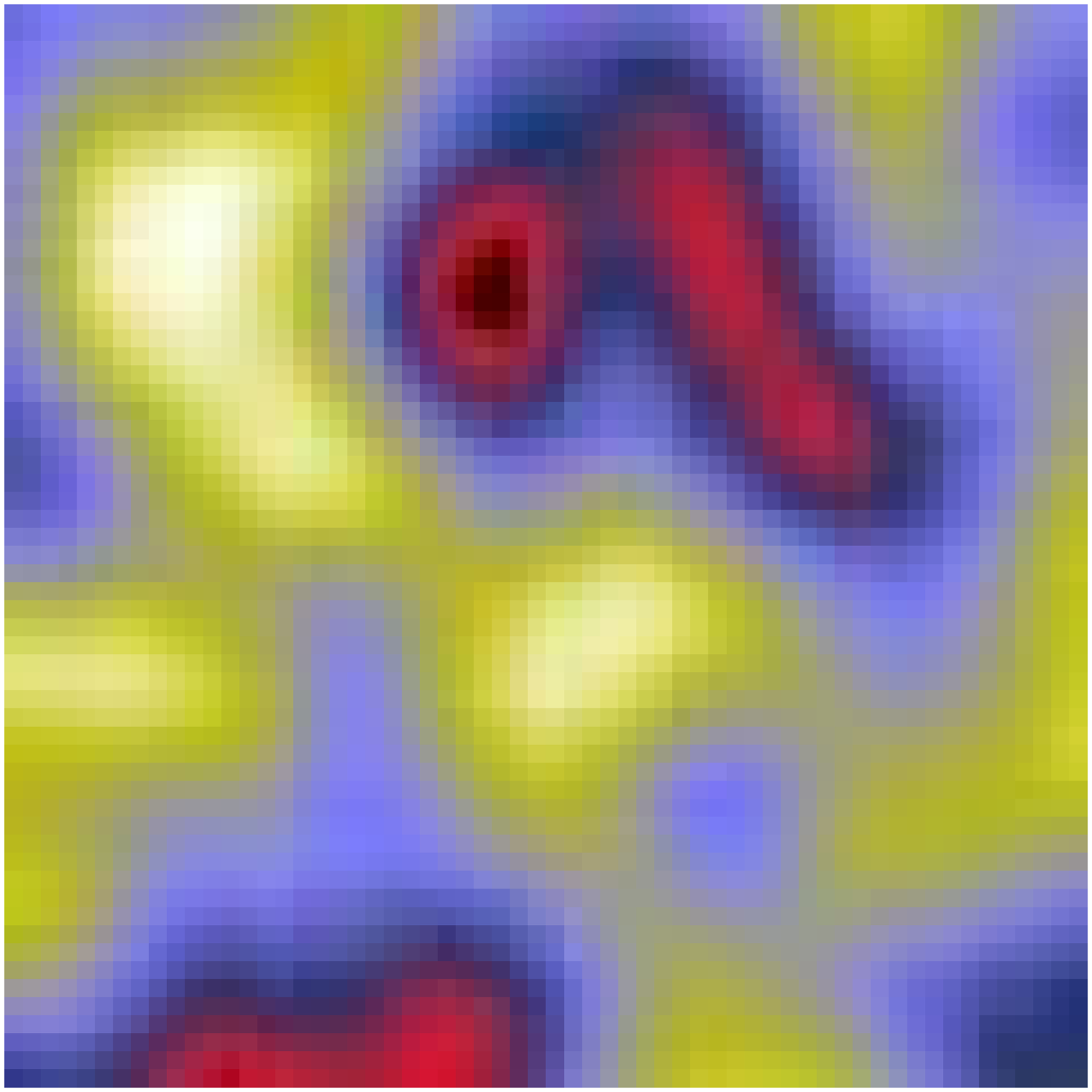} \\  {\#22   NCIA010627}  \\
  \FGb{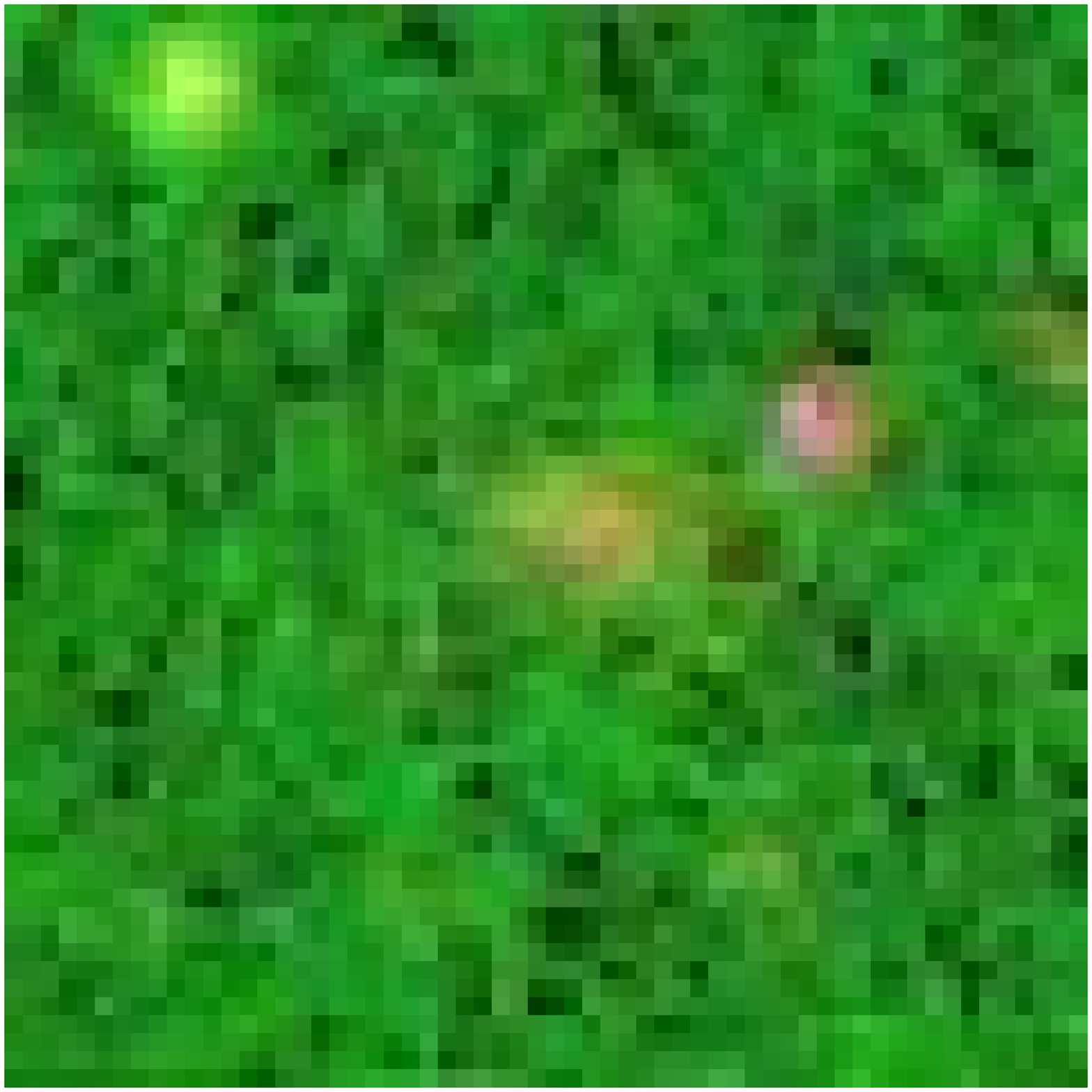}  \FGb{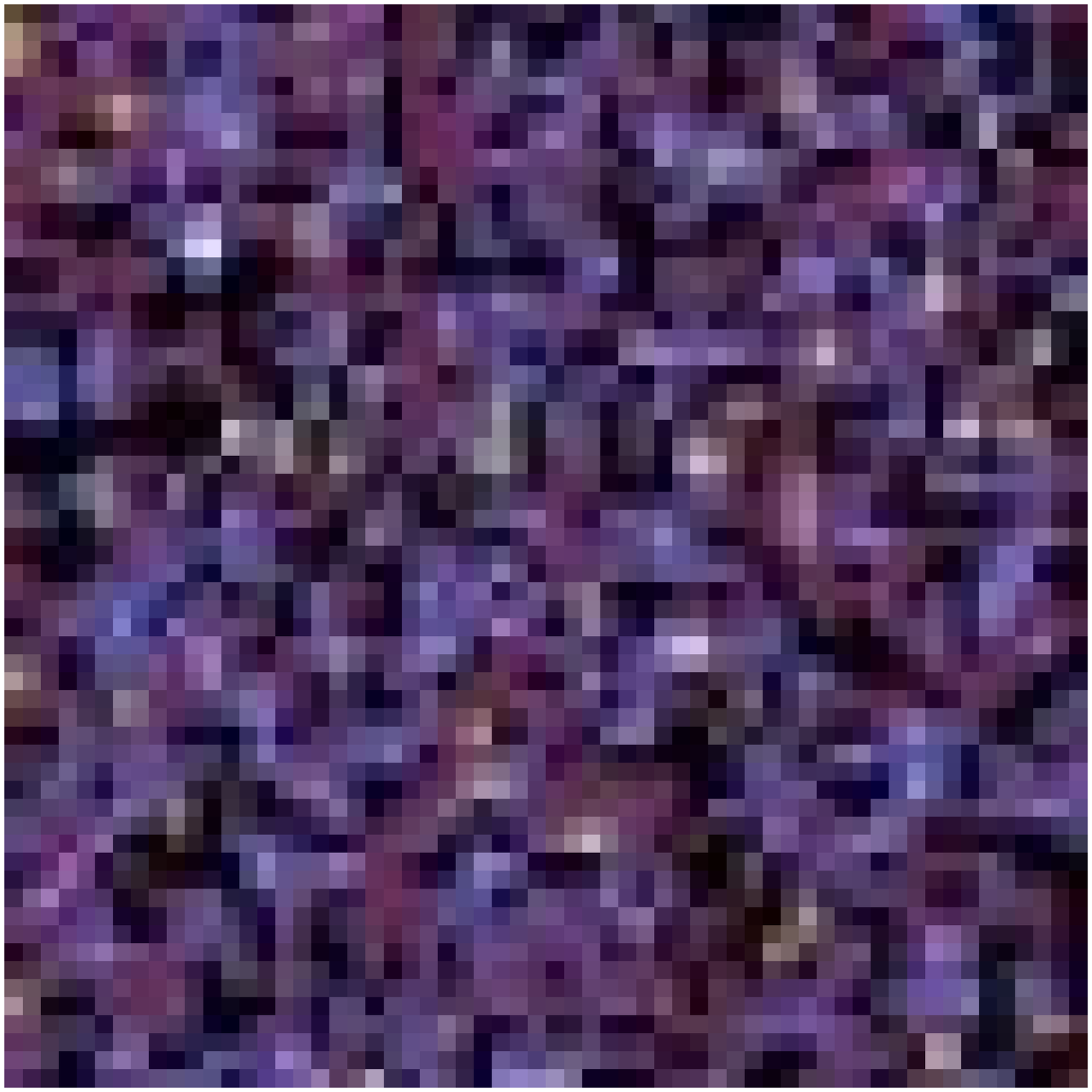}  \FGb{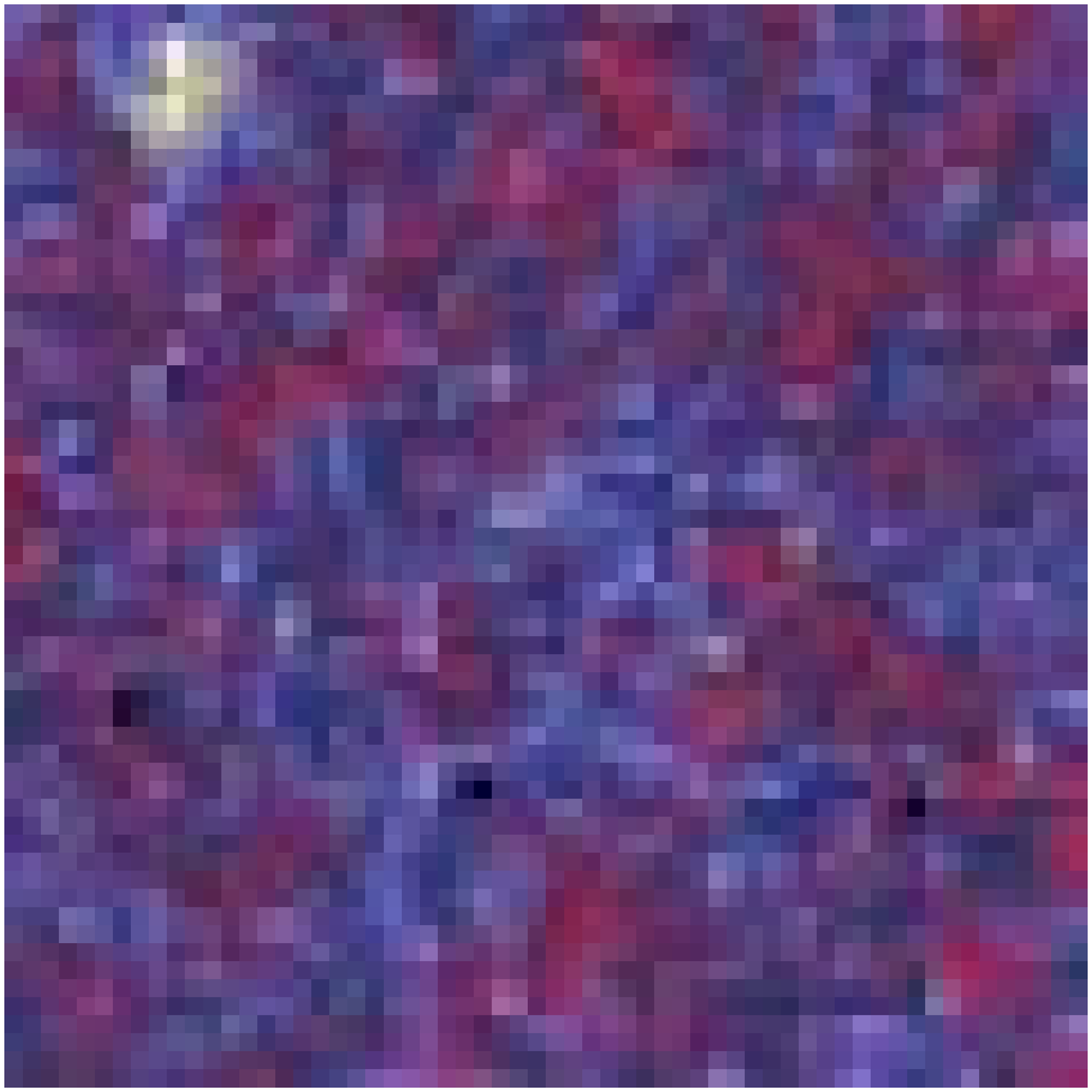}  \FGb{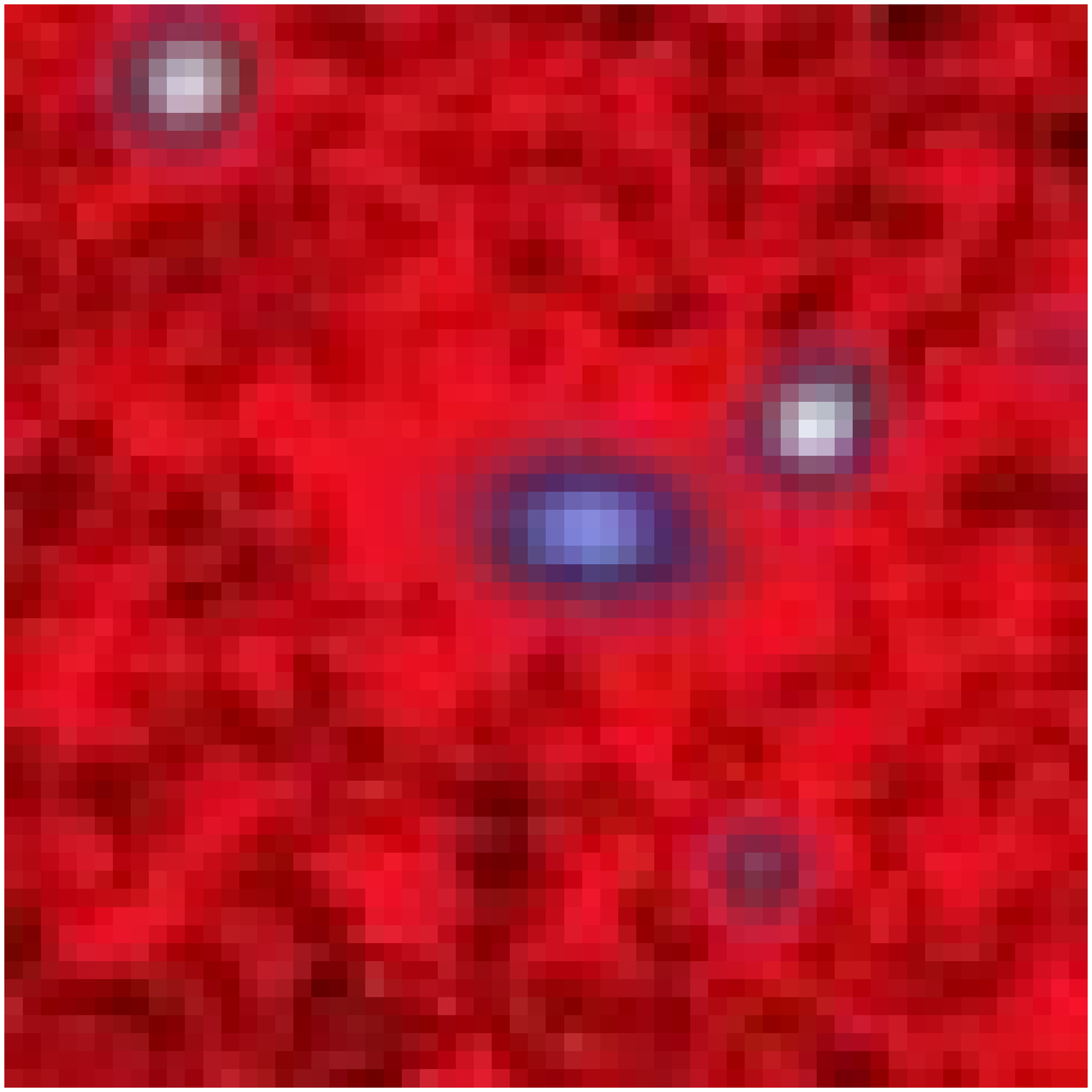}  \FGb{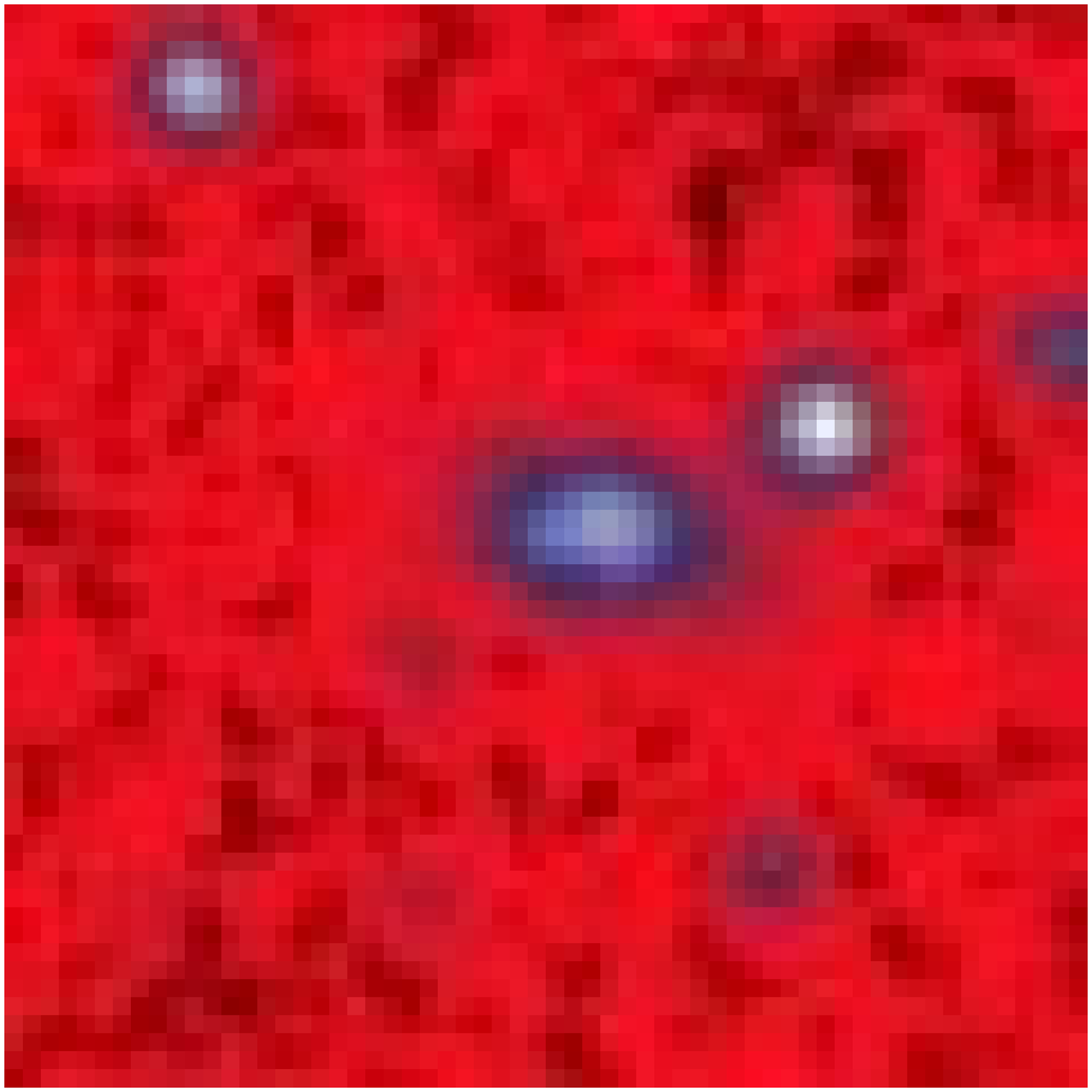}  \FGb{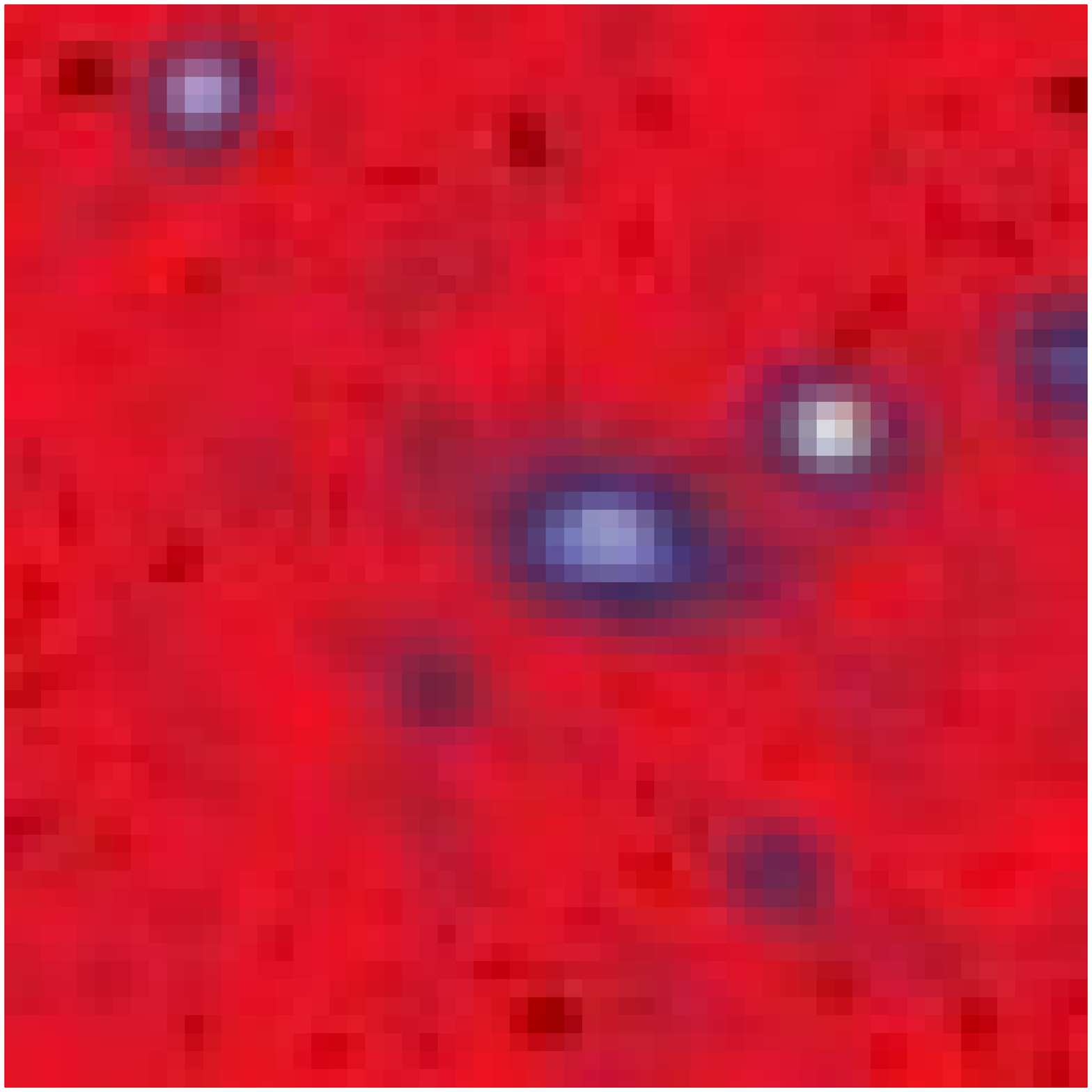}  \FGb{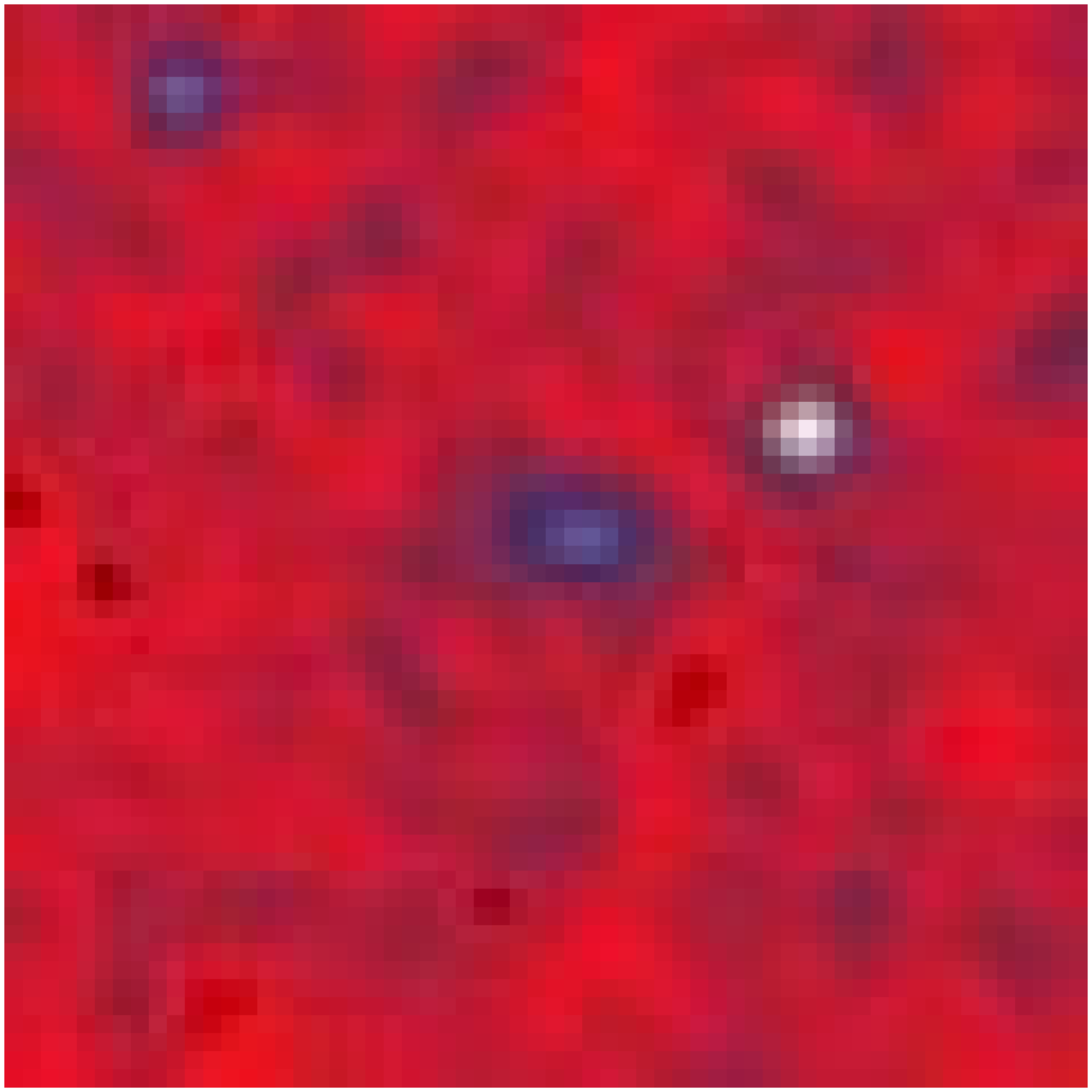}  \FGb{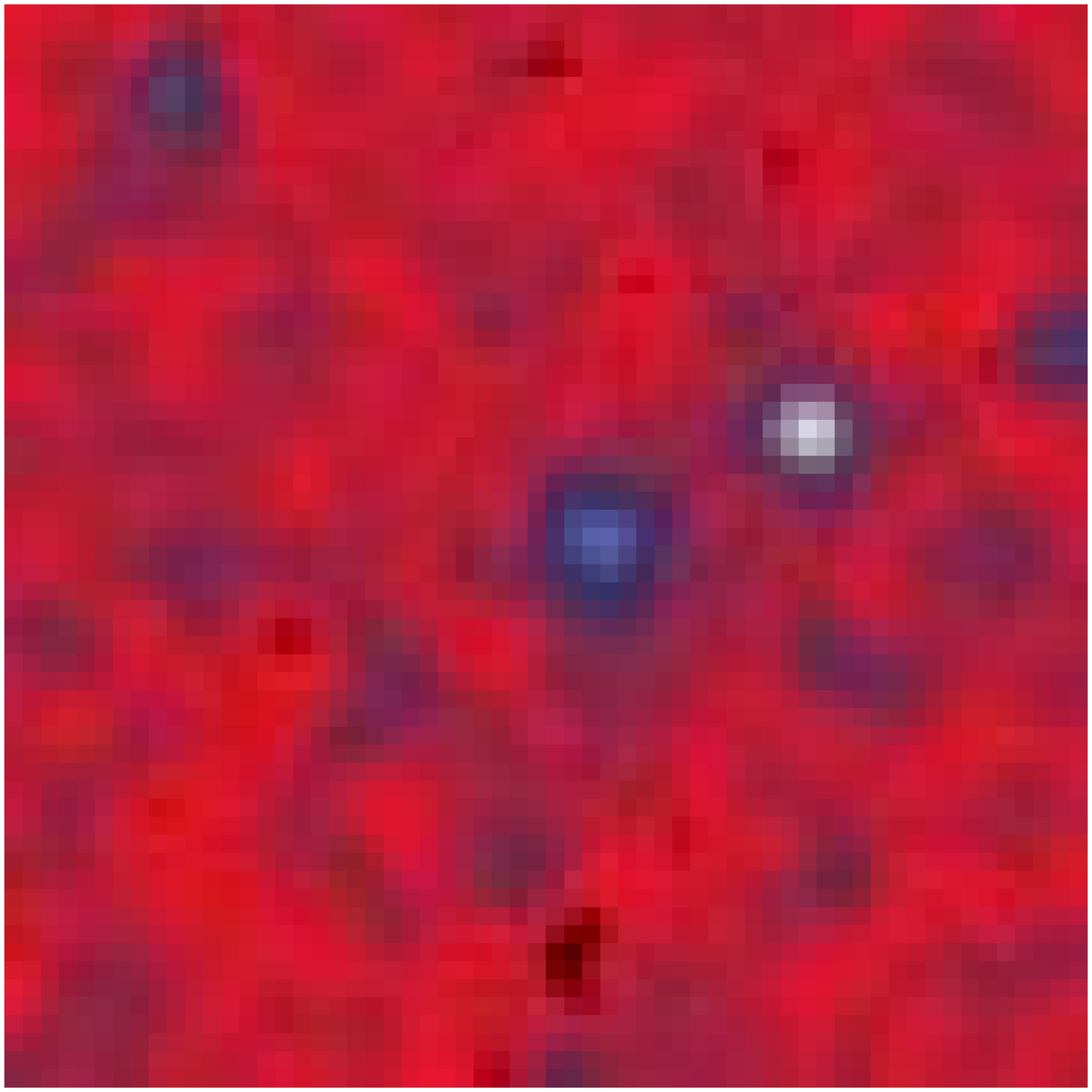}  \FGb{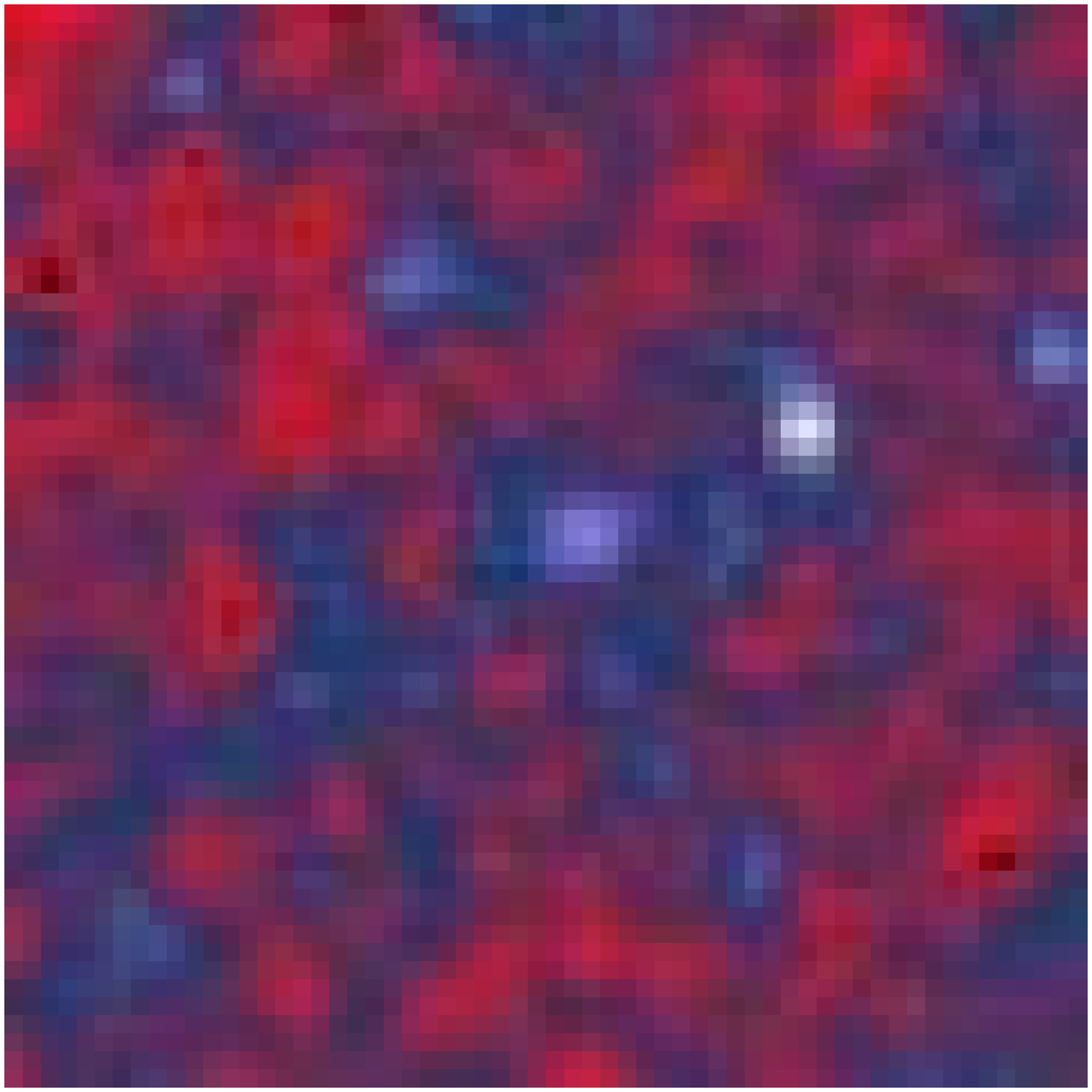}  \FGb{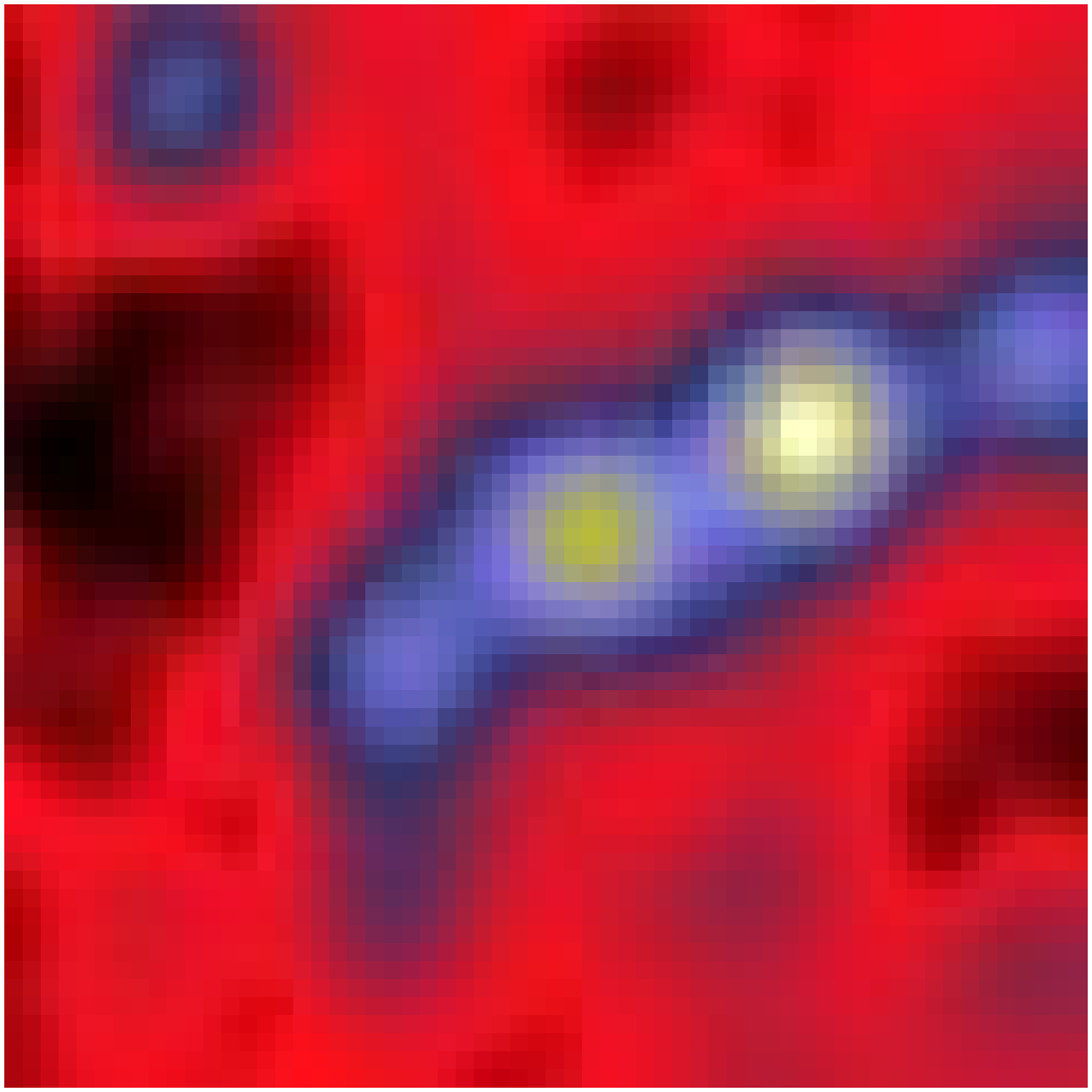}  \FGb{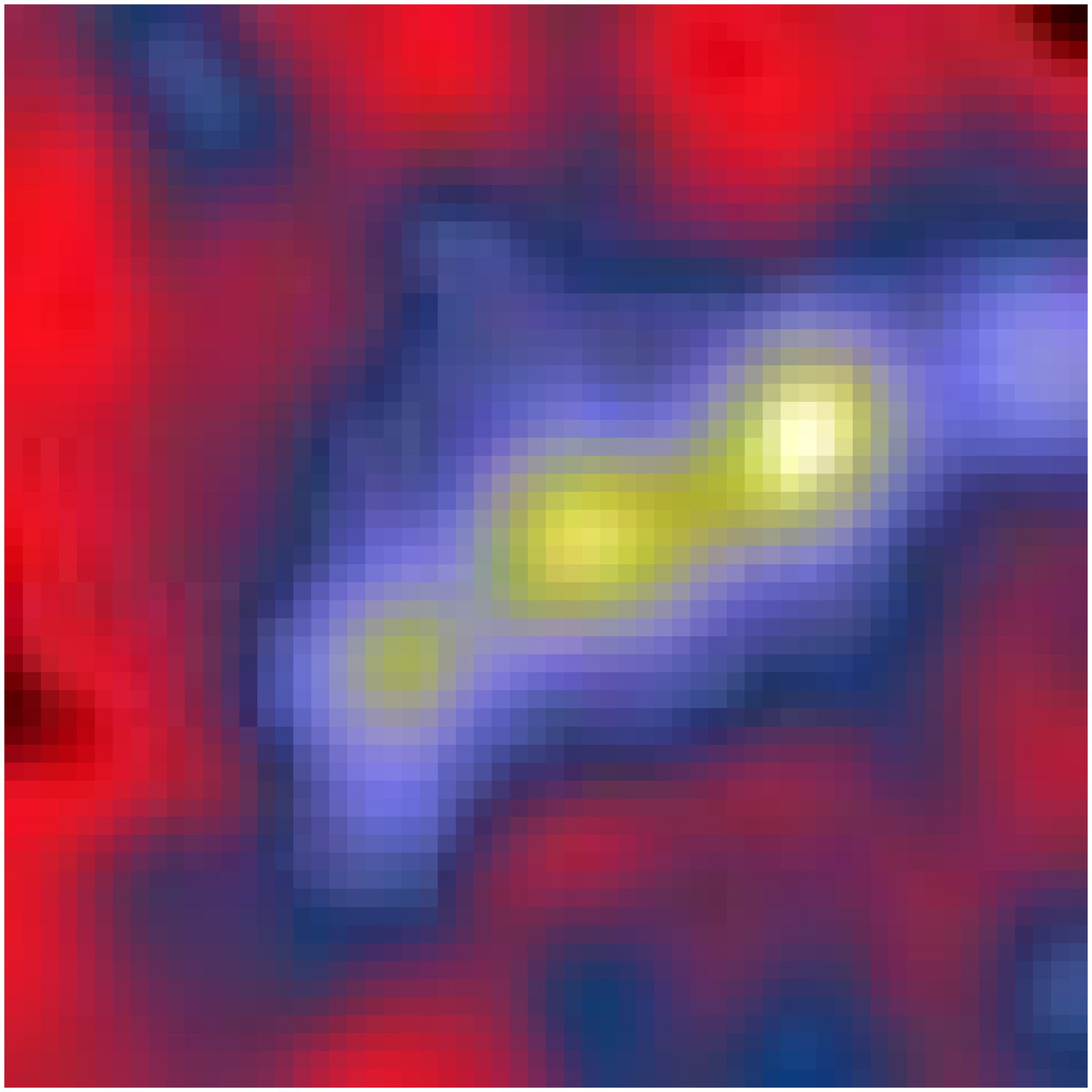}  \FGb{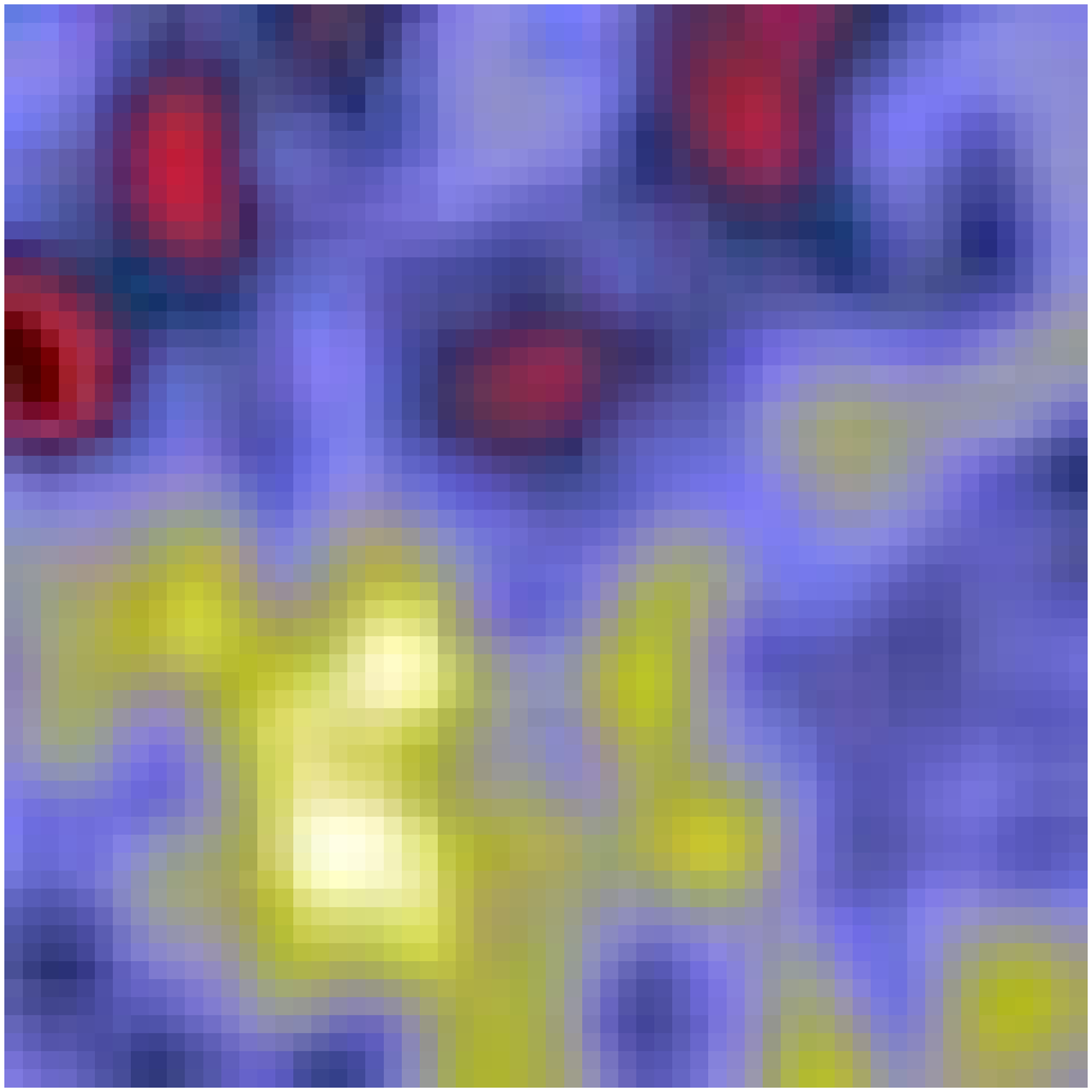}  \FGb{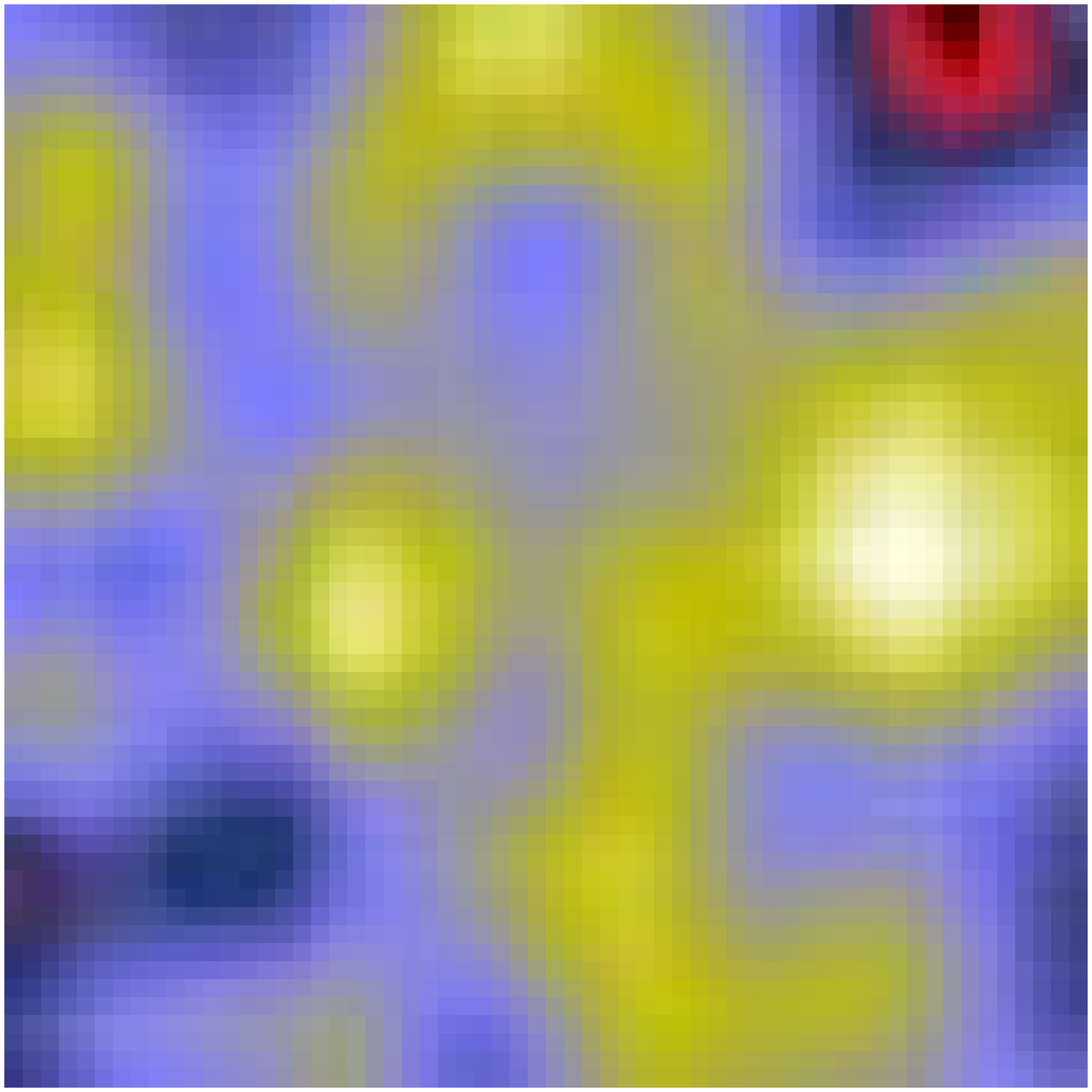} \\  {\#23   NCIA006643}  \\
  \FGb{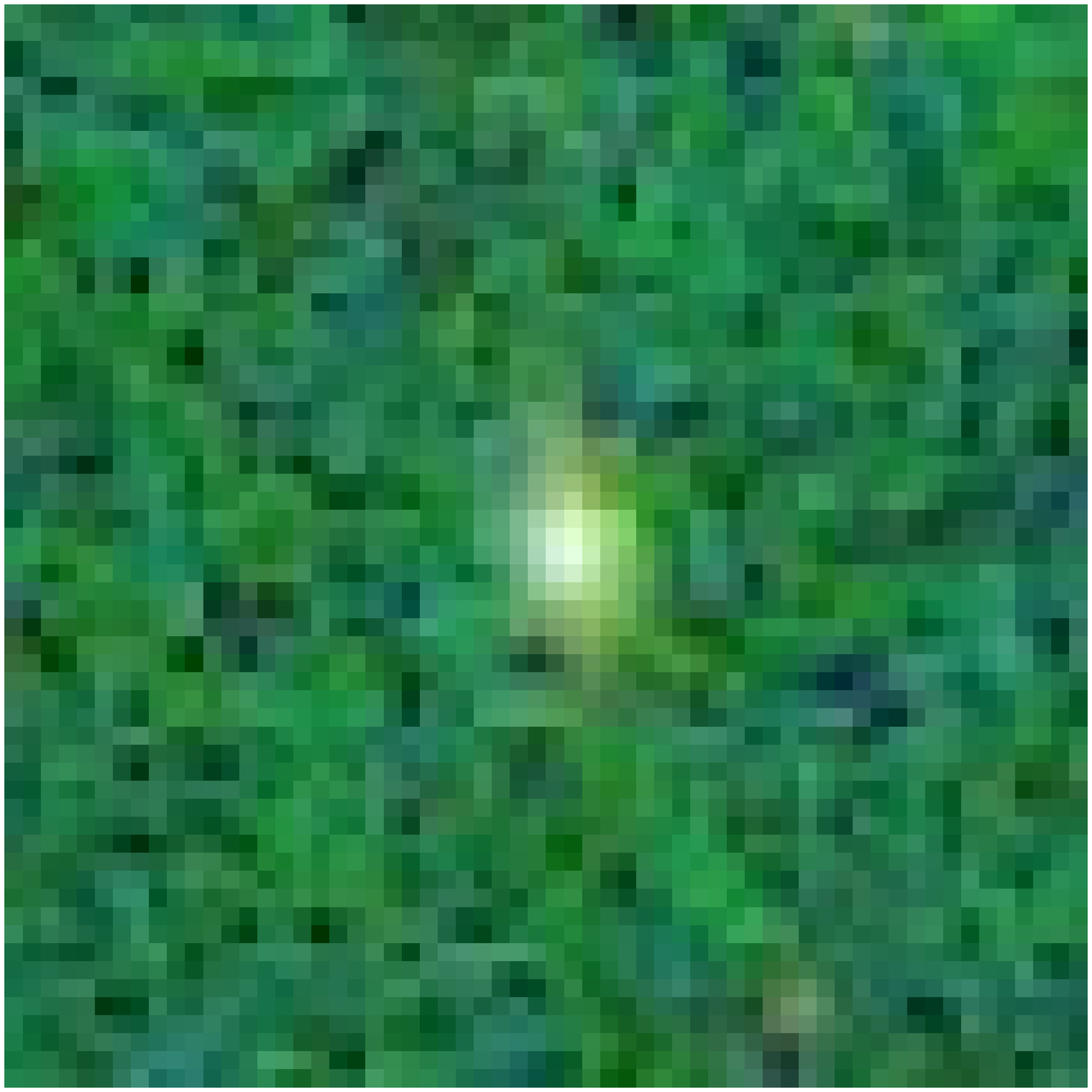}  \FGb{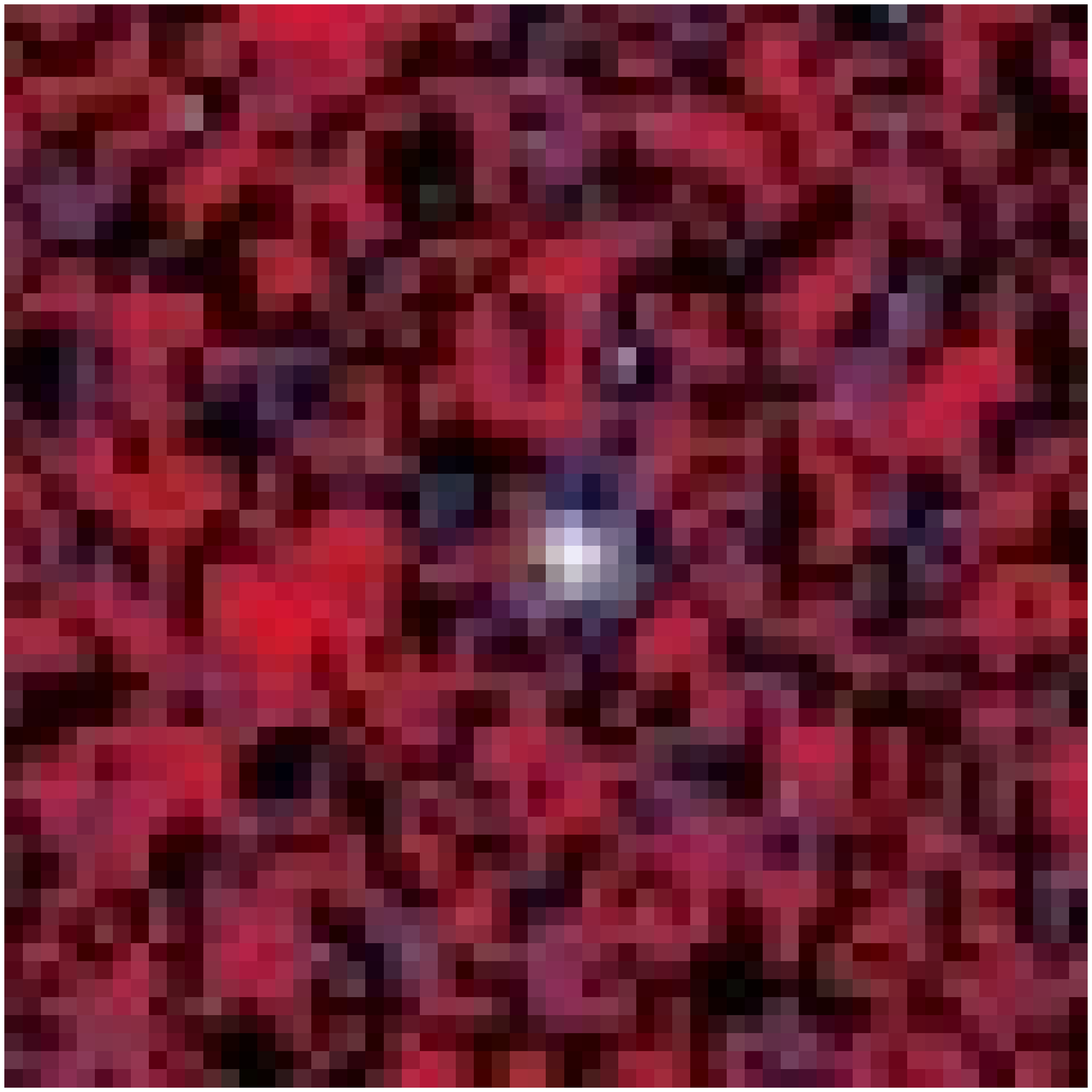}  \FGb{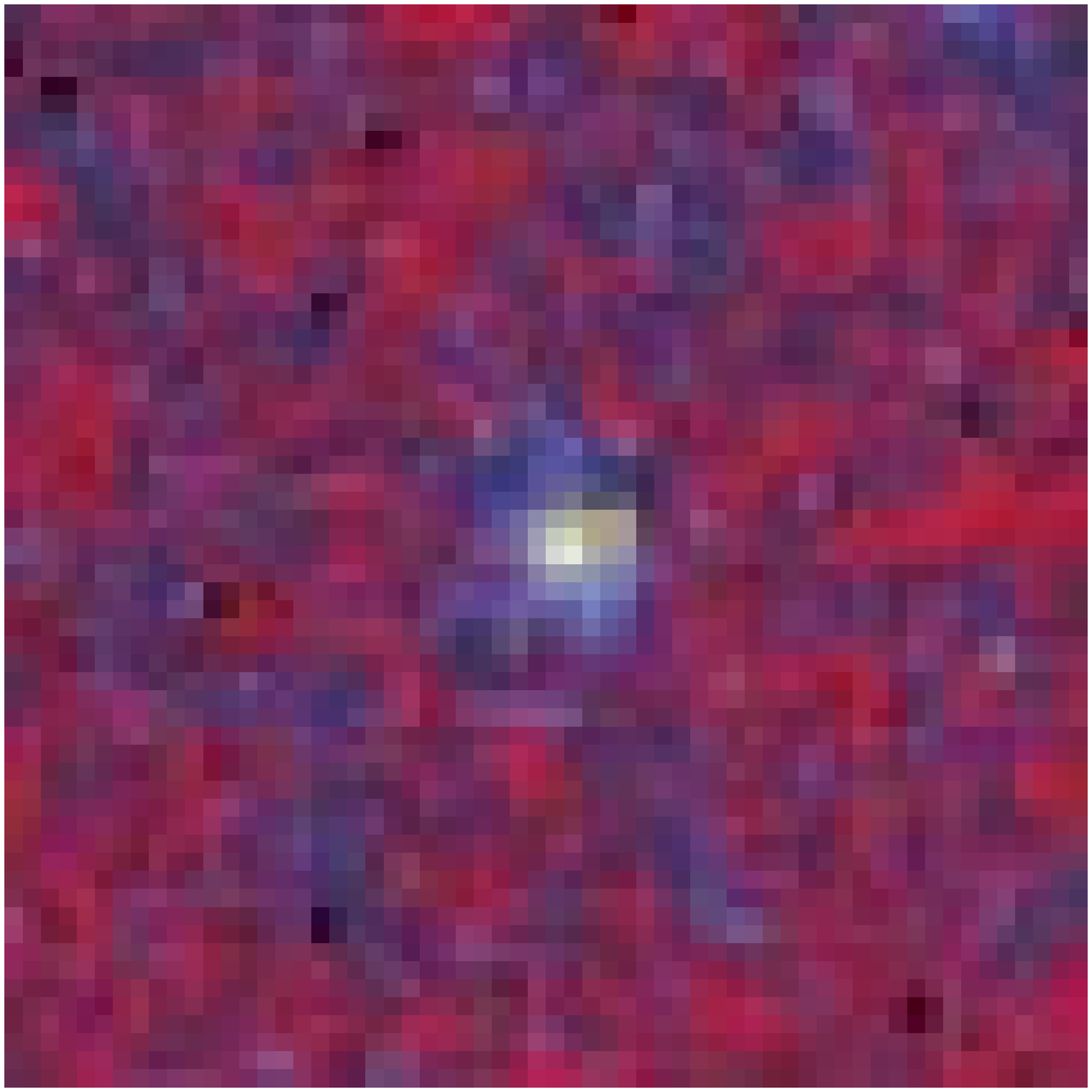}  \FGb{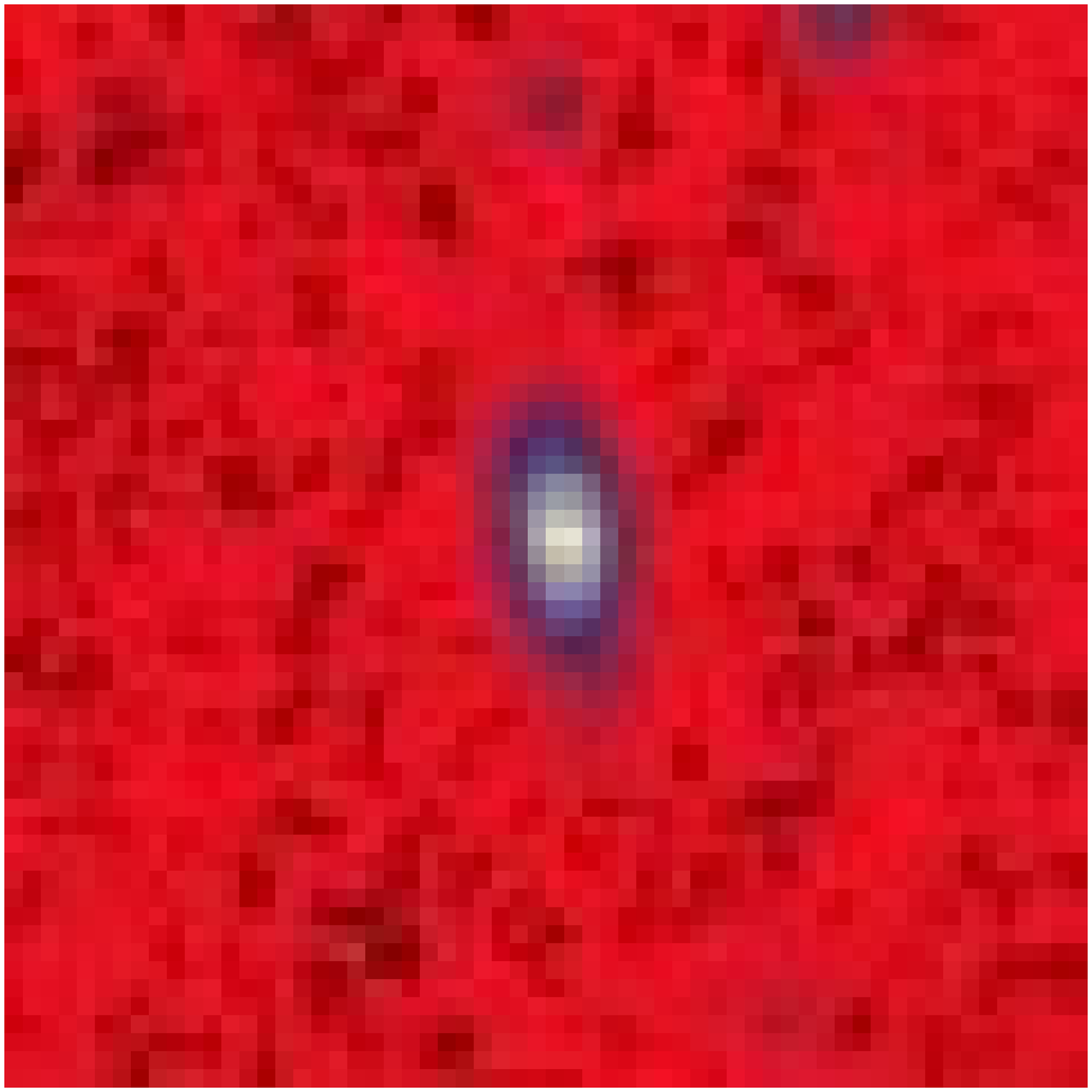}  \FGb{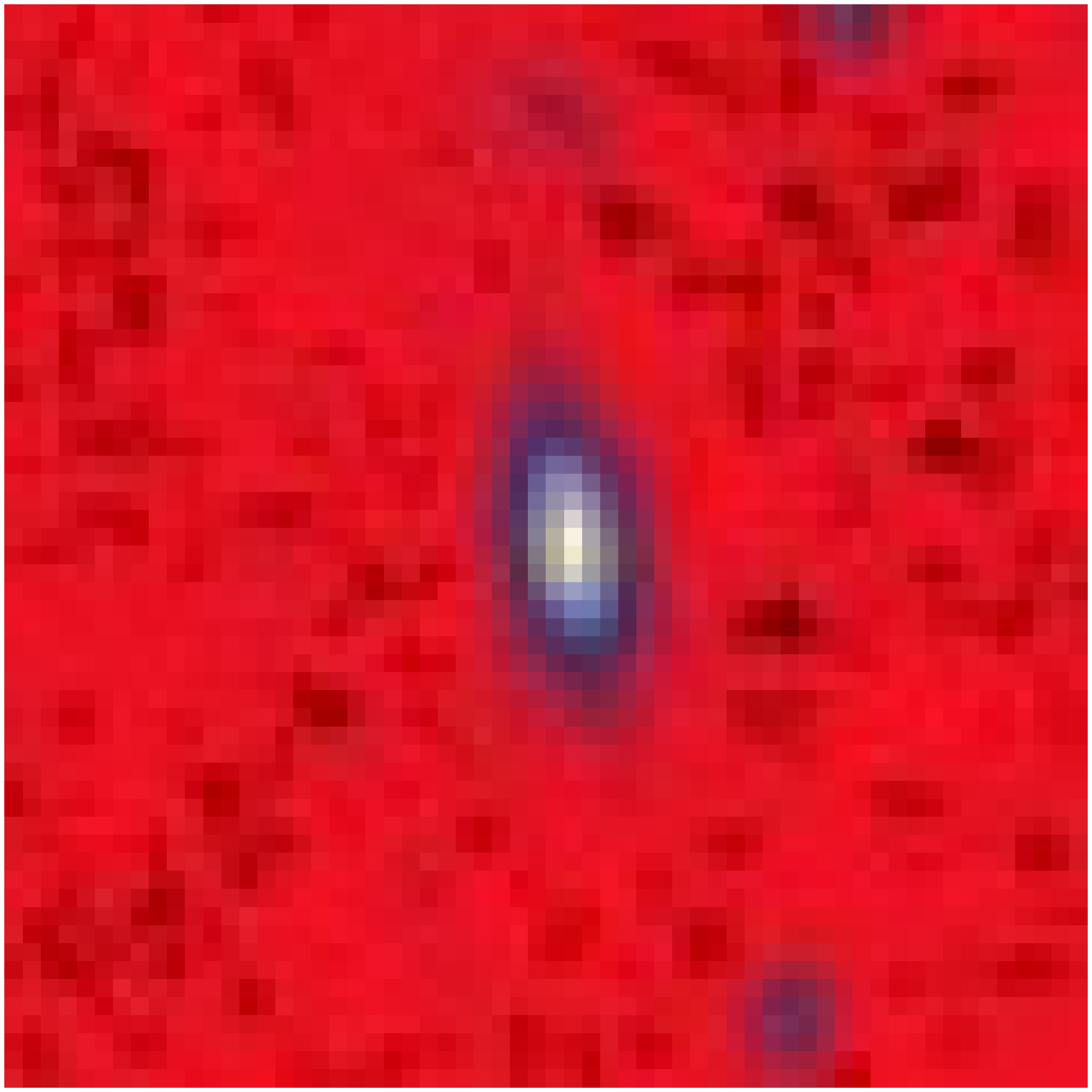}  \FGb{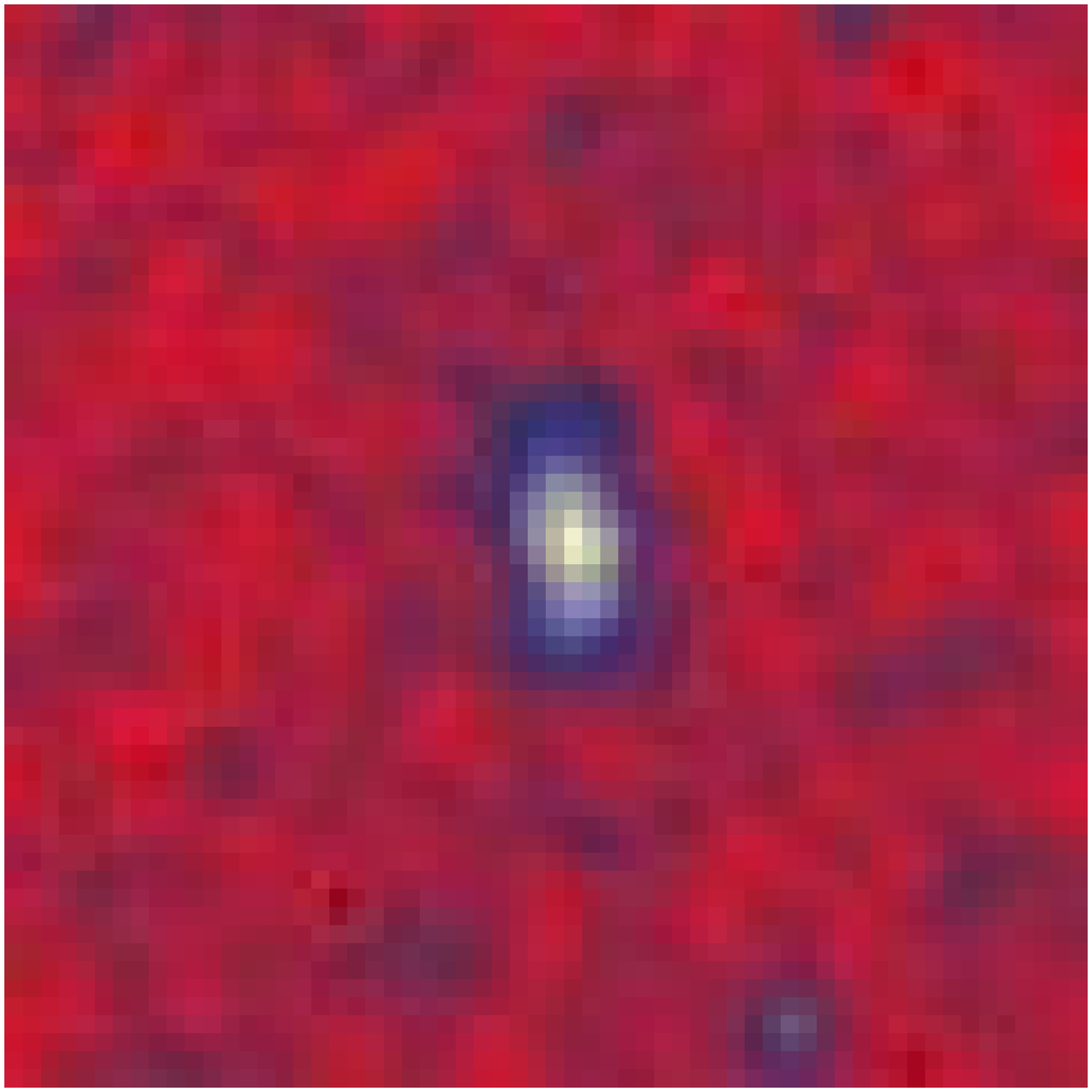}  \FGb{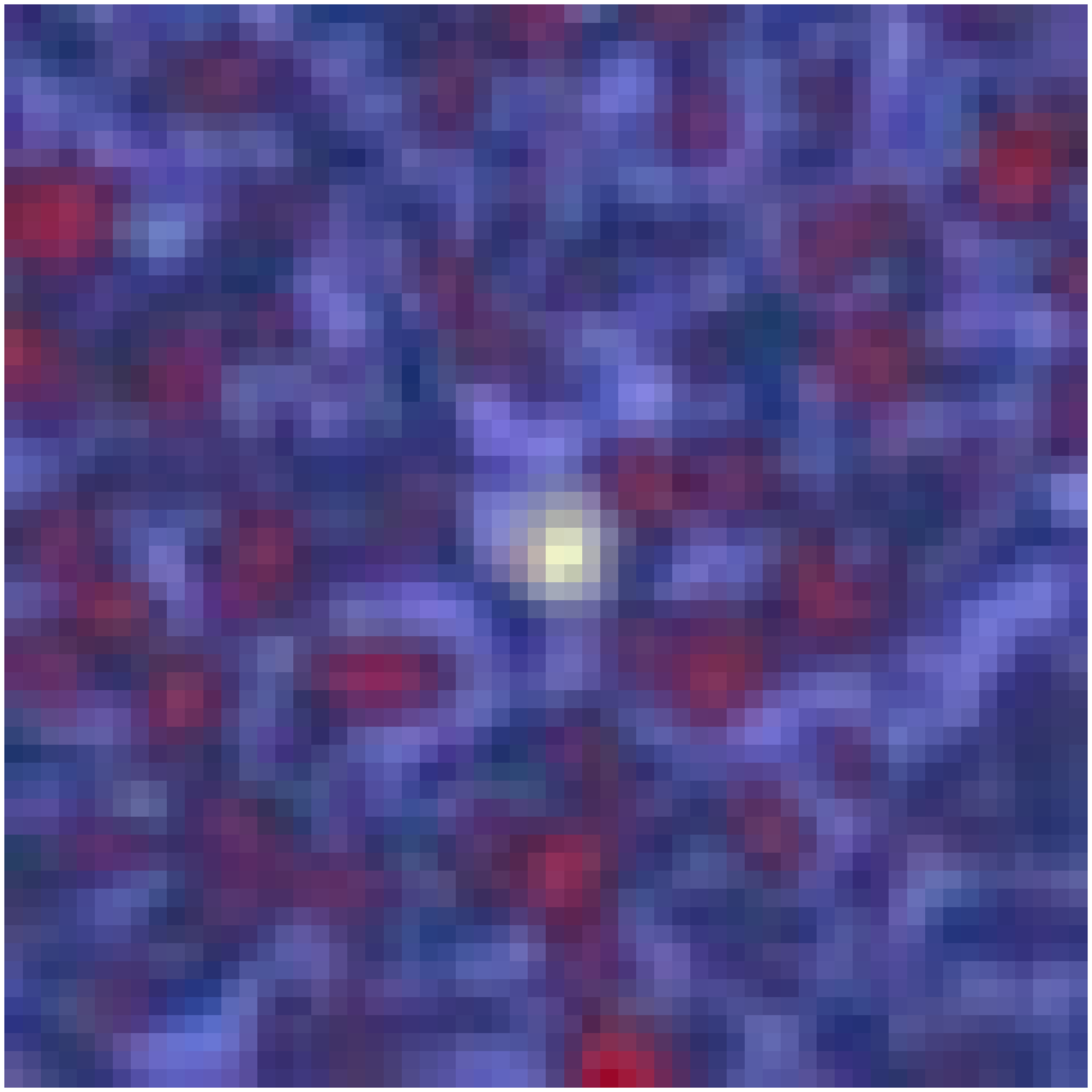}  \FGb{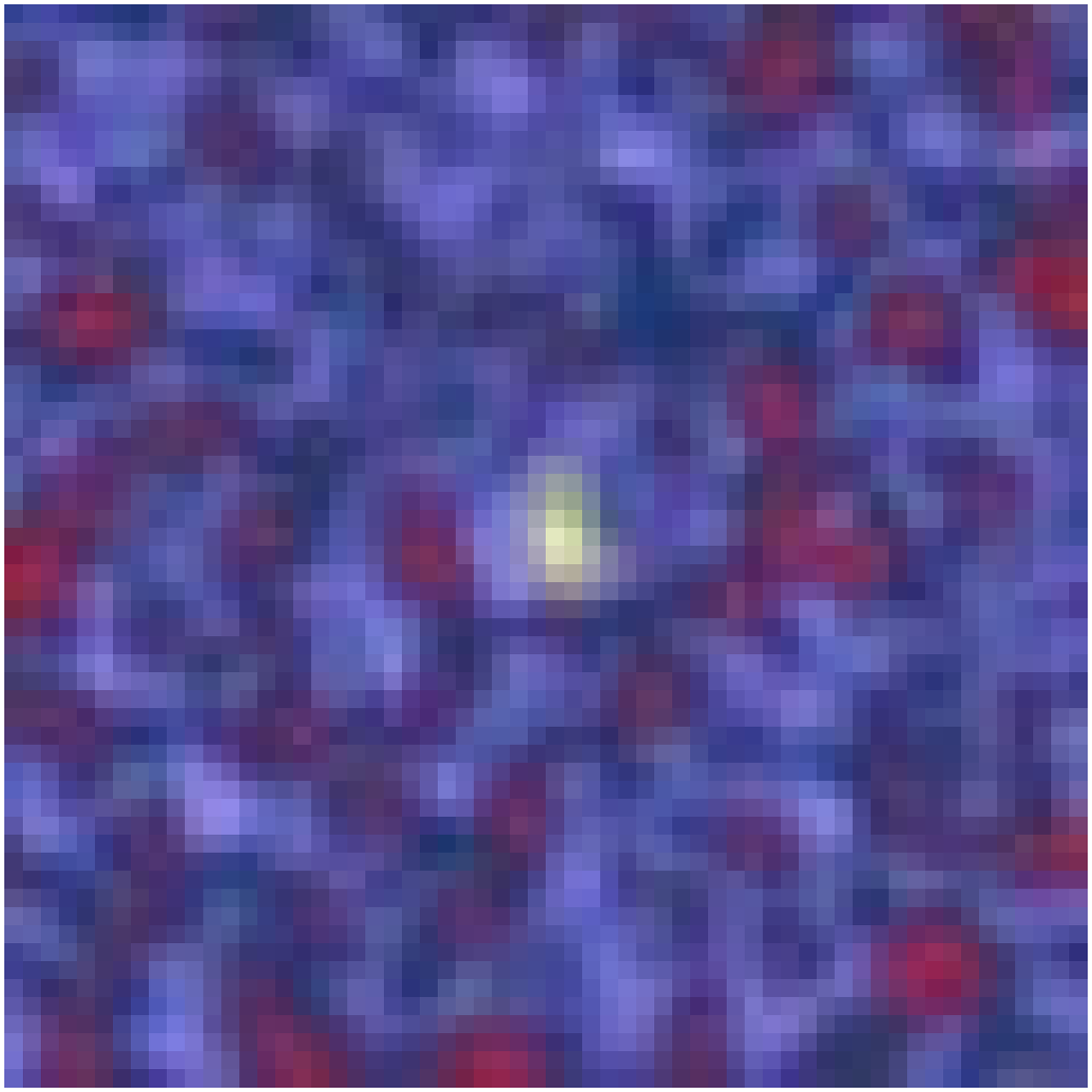}  \FGb{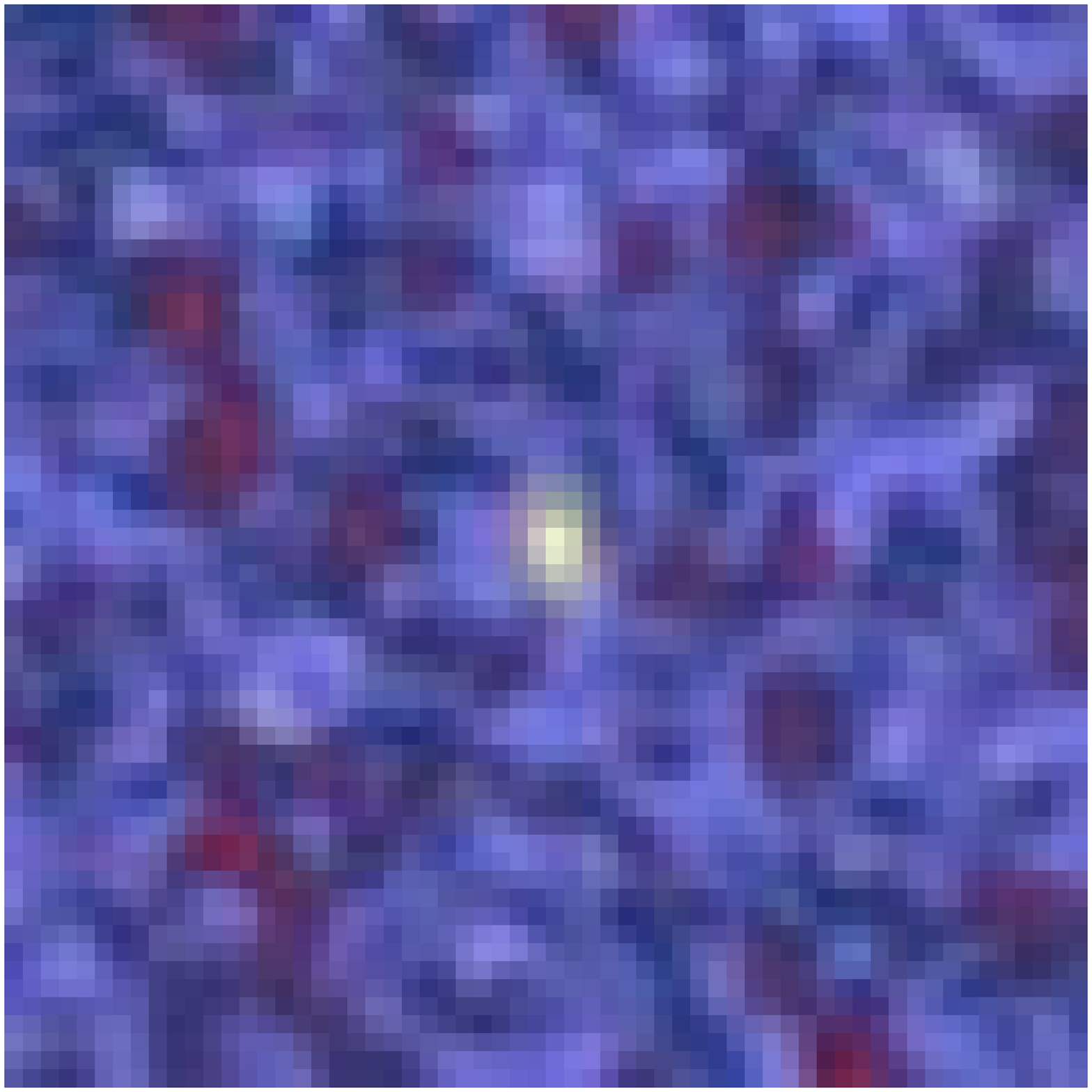}  \FGb{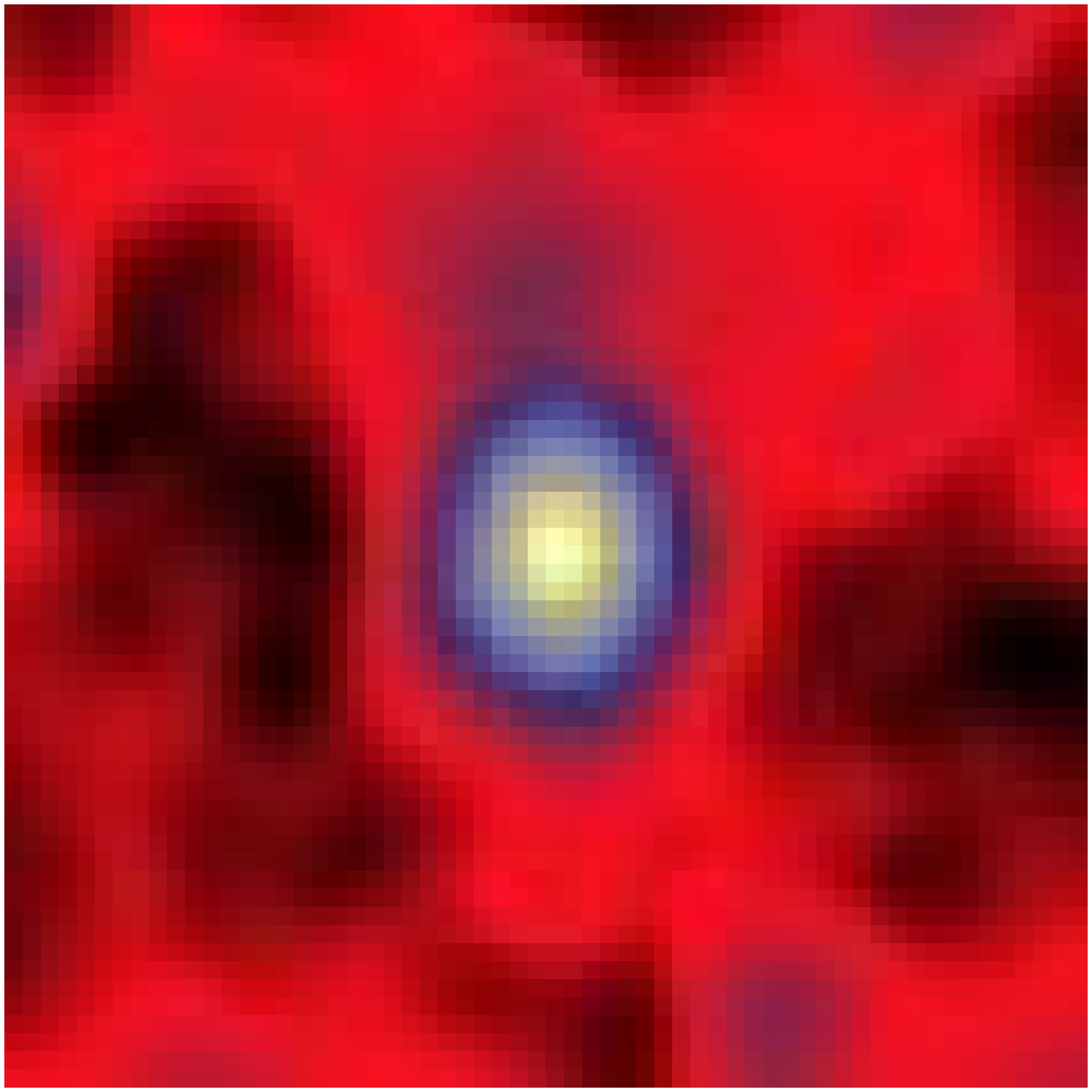}  \FGb{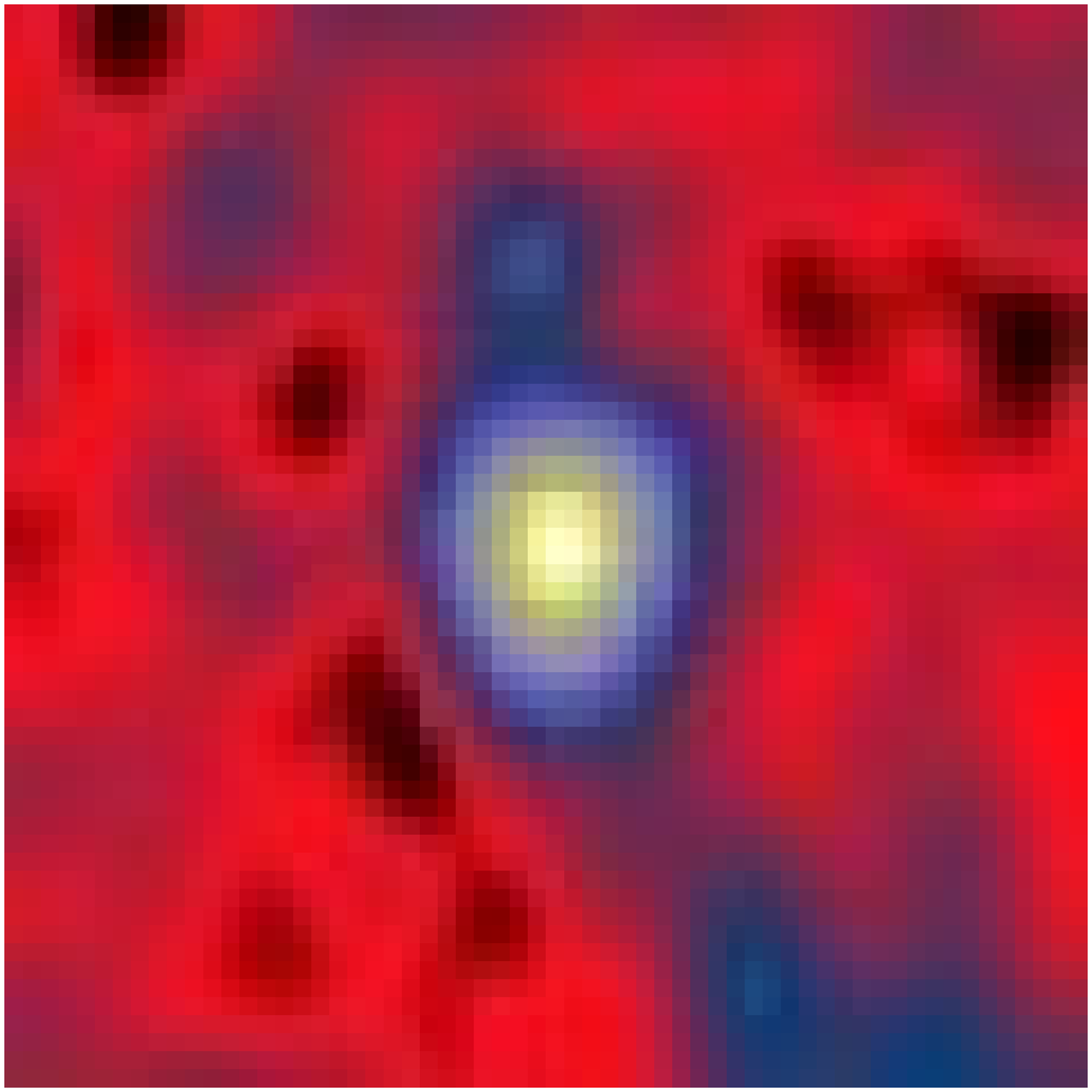}  \FGb{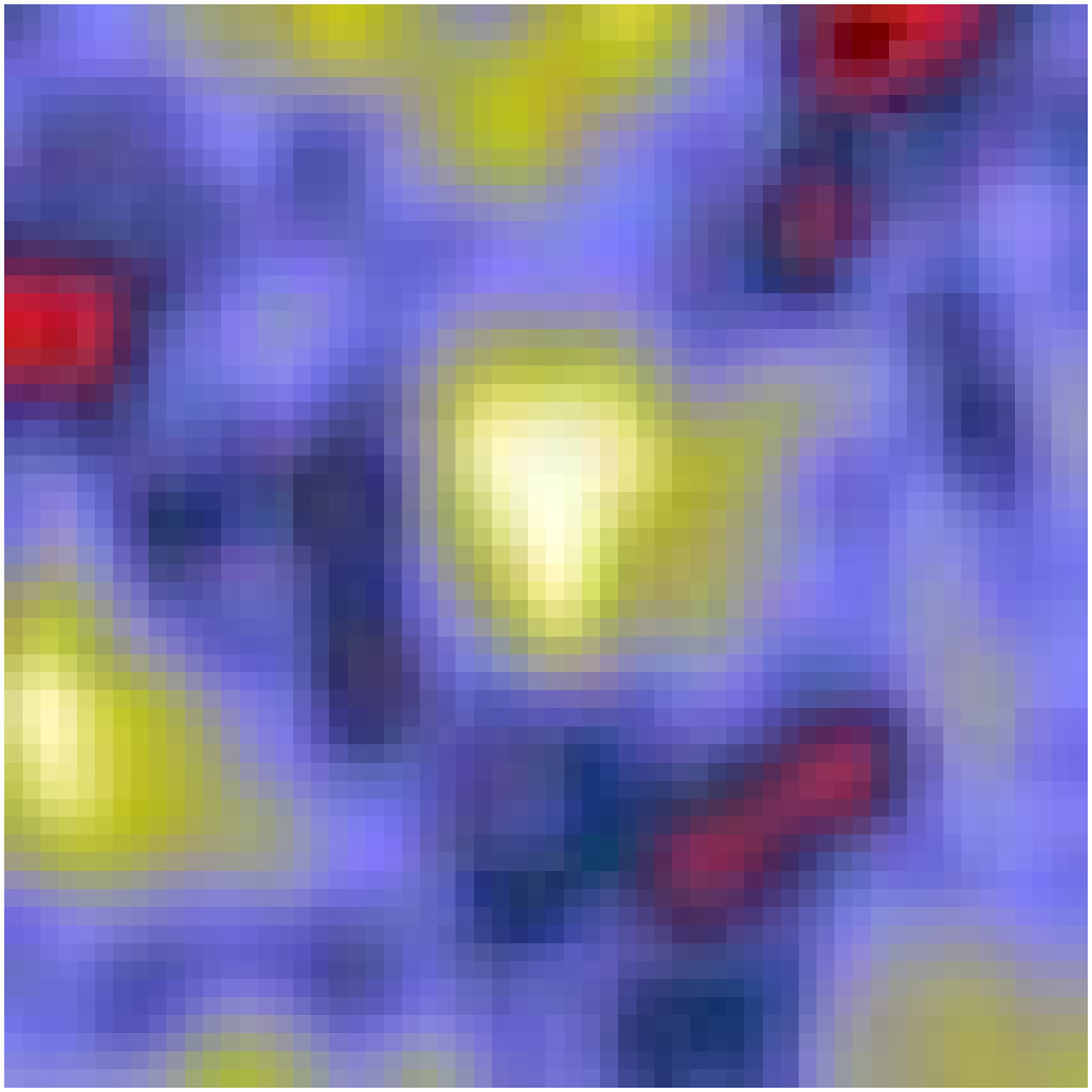}  \FGb{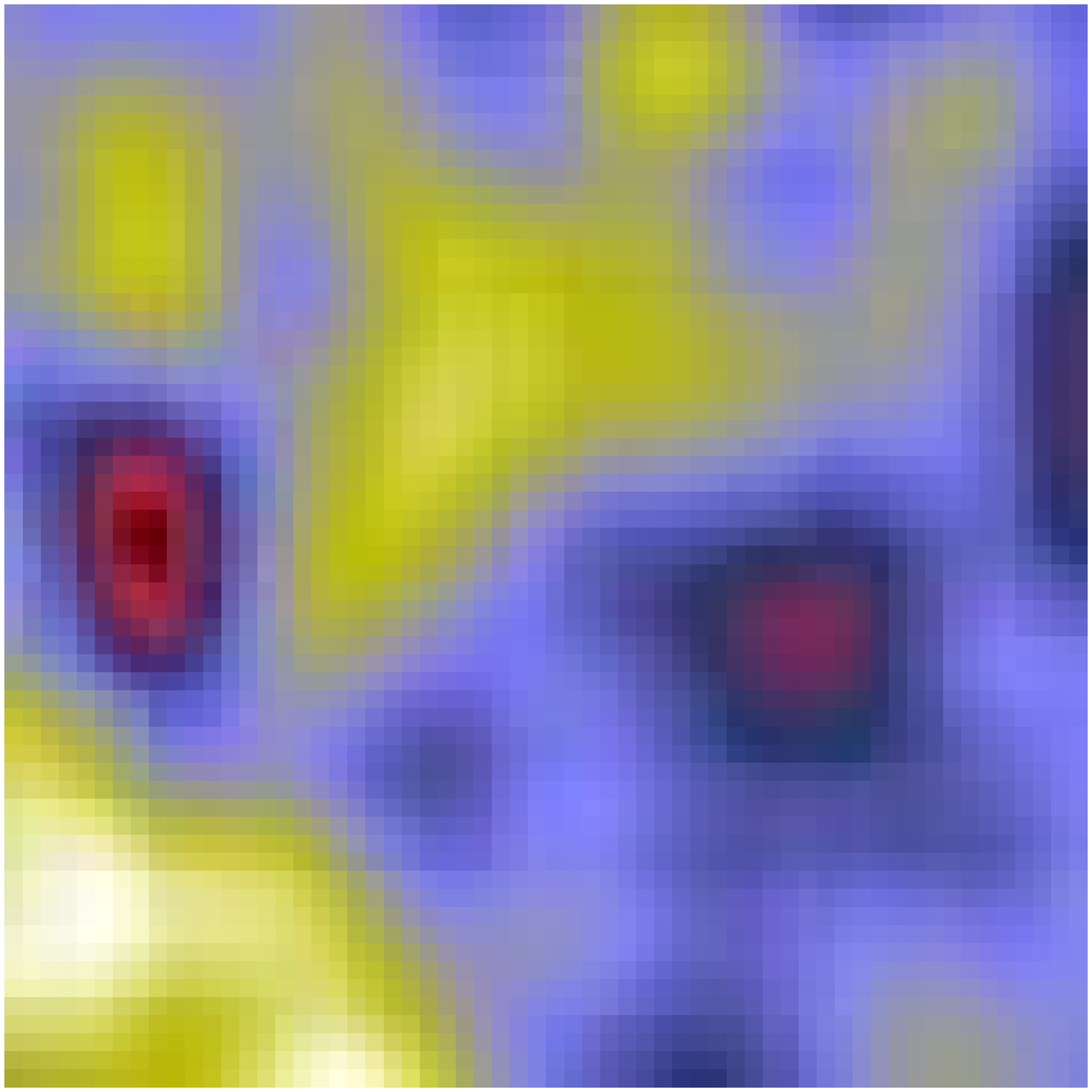} \\  {\#24   NCIA010131}  \\
  \FGb{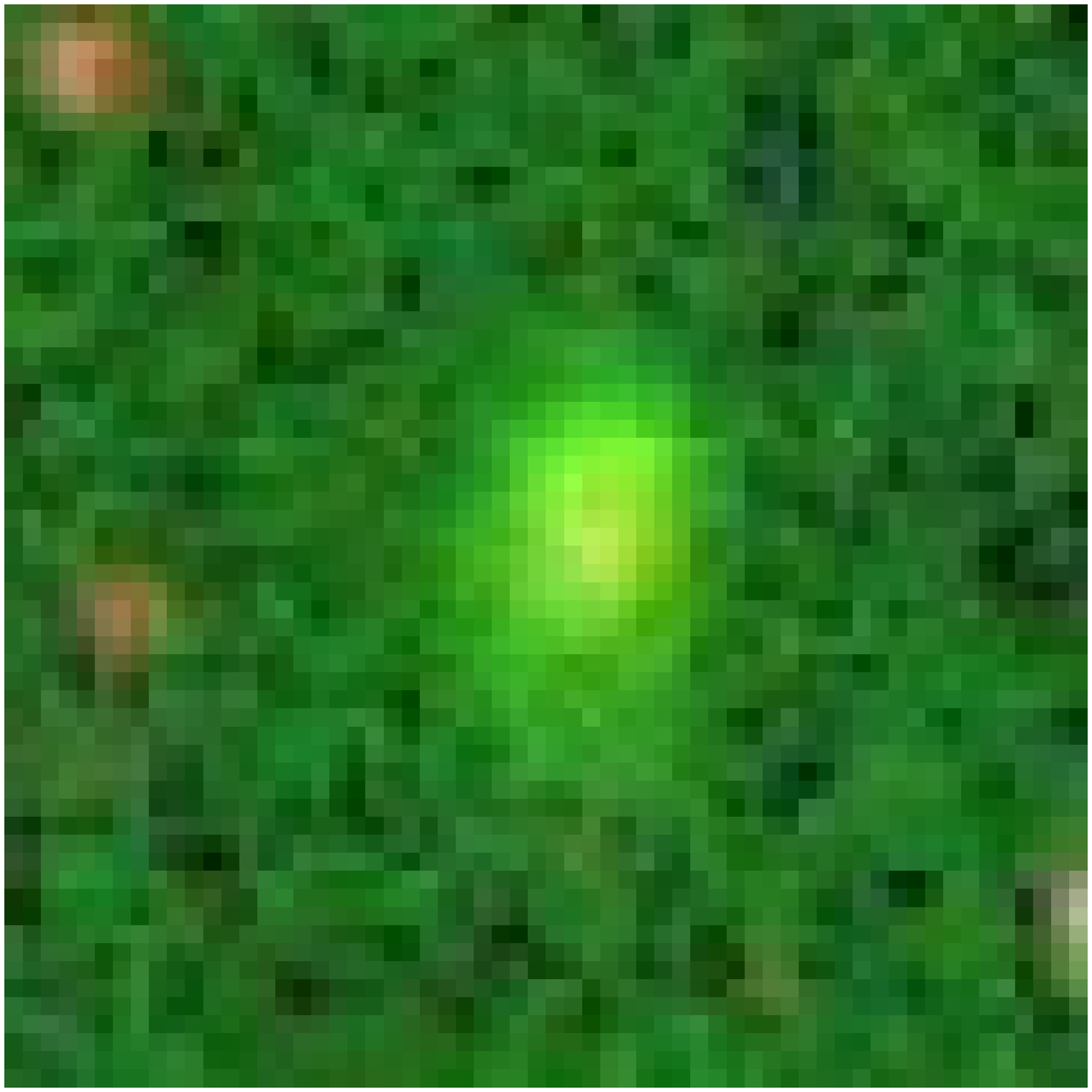}  \FGb{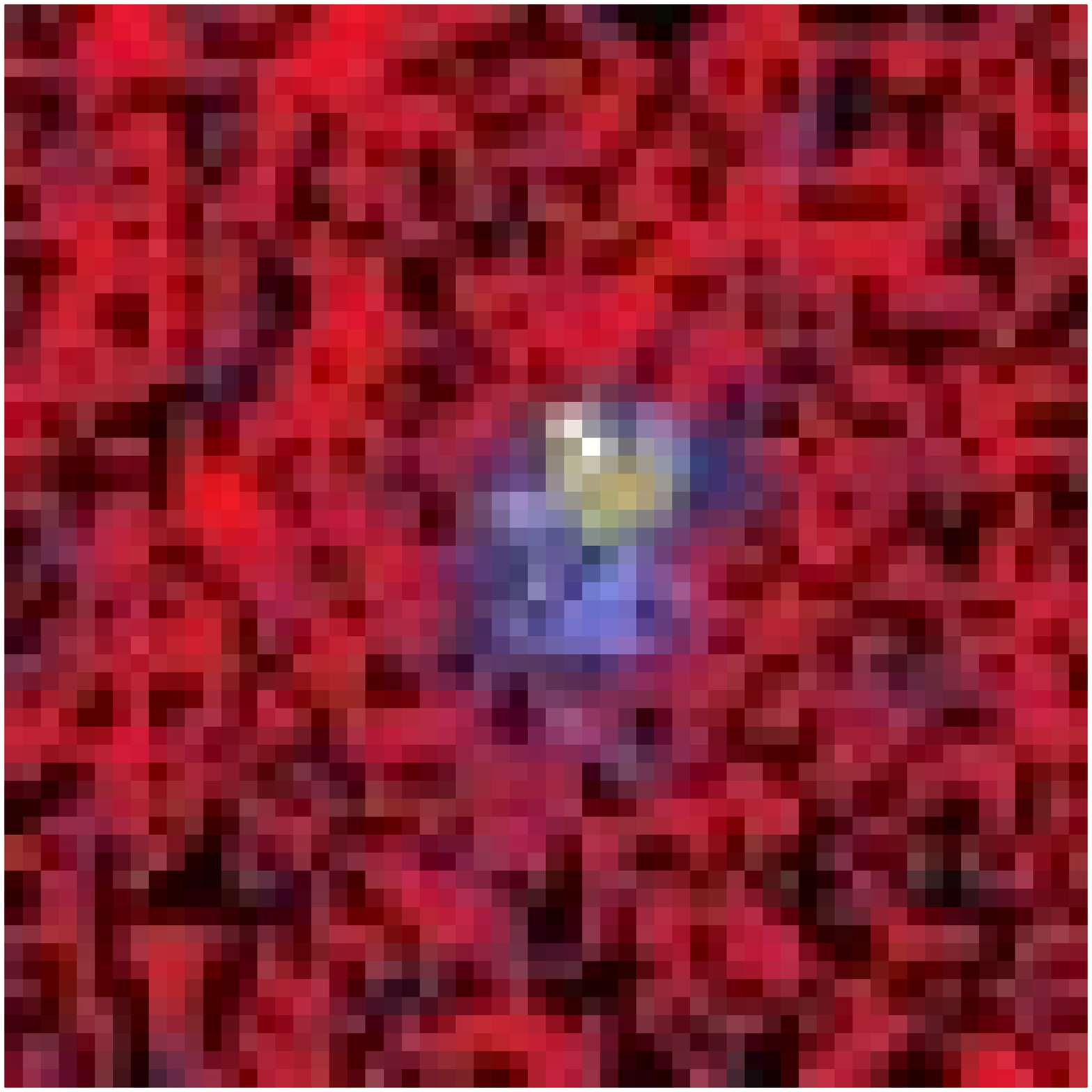}  \FGb{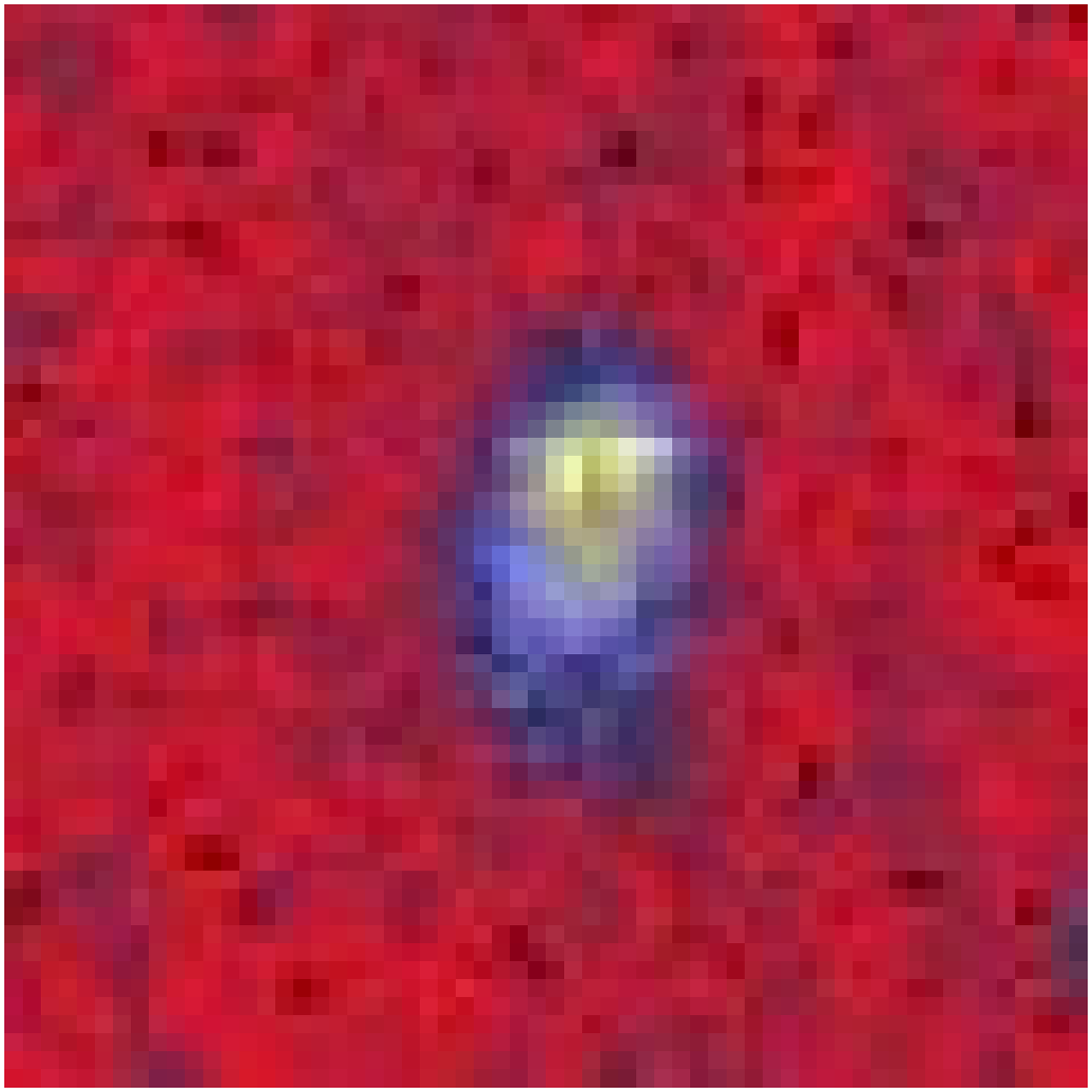}  \FGb{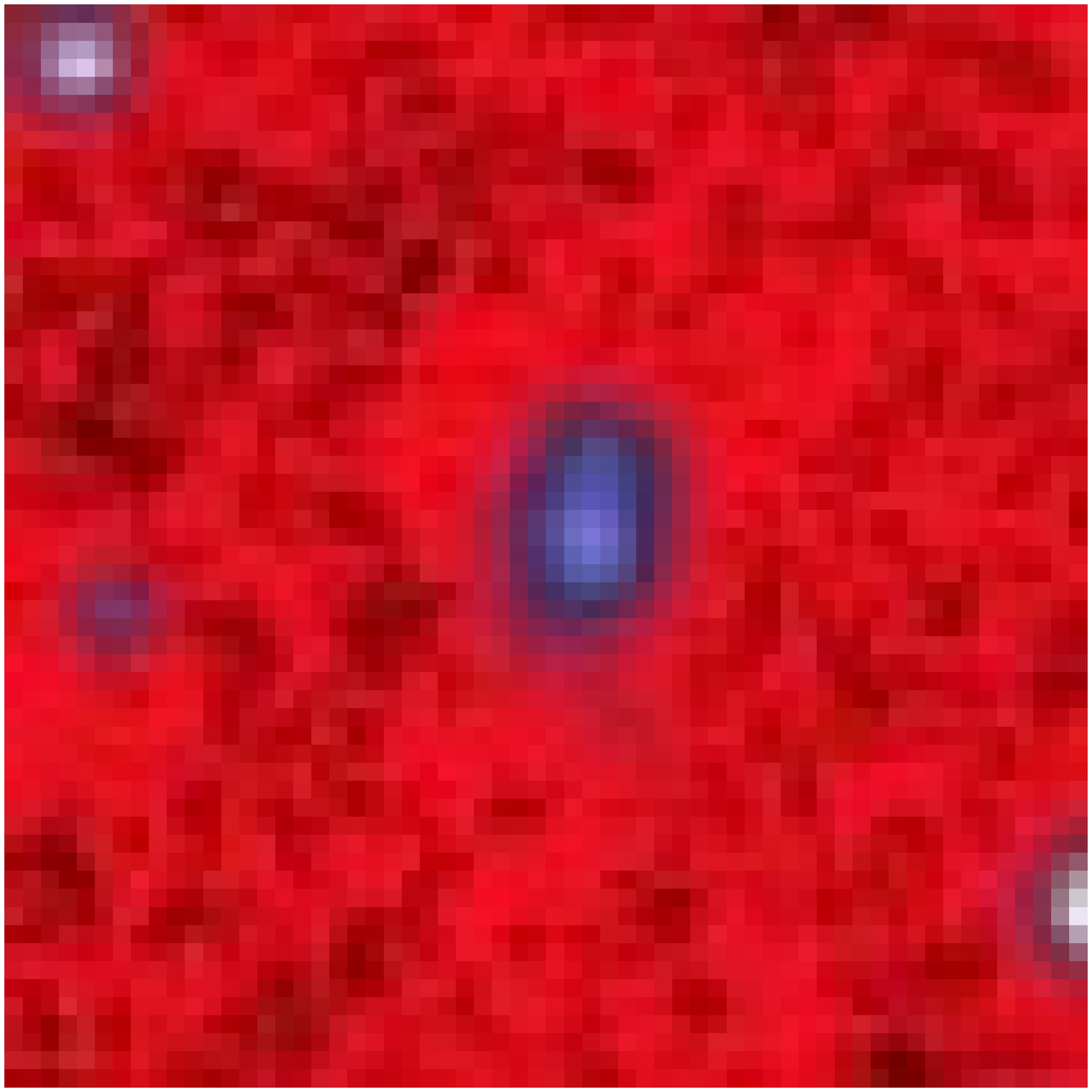}  \FGb{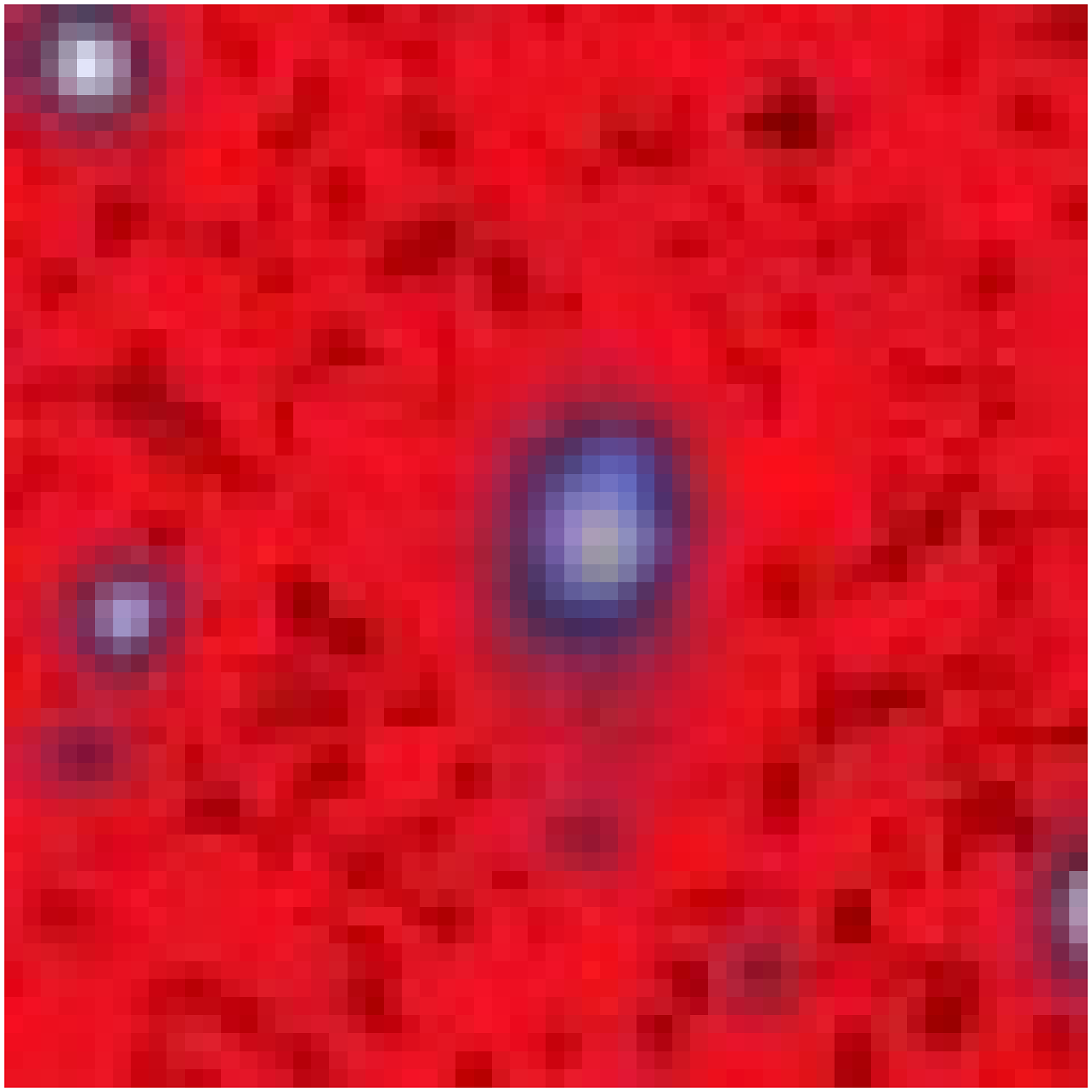}  \FGb{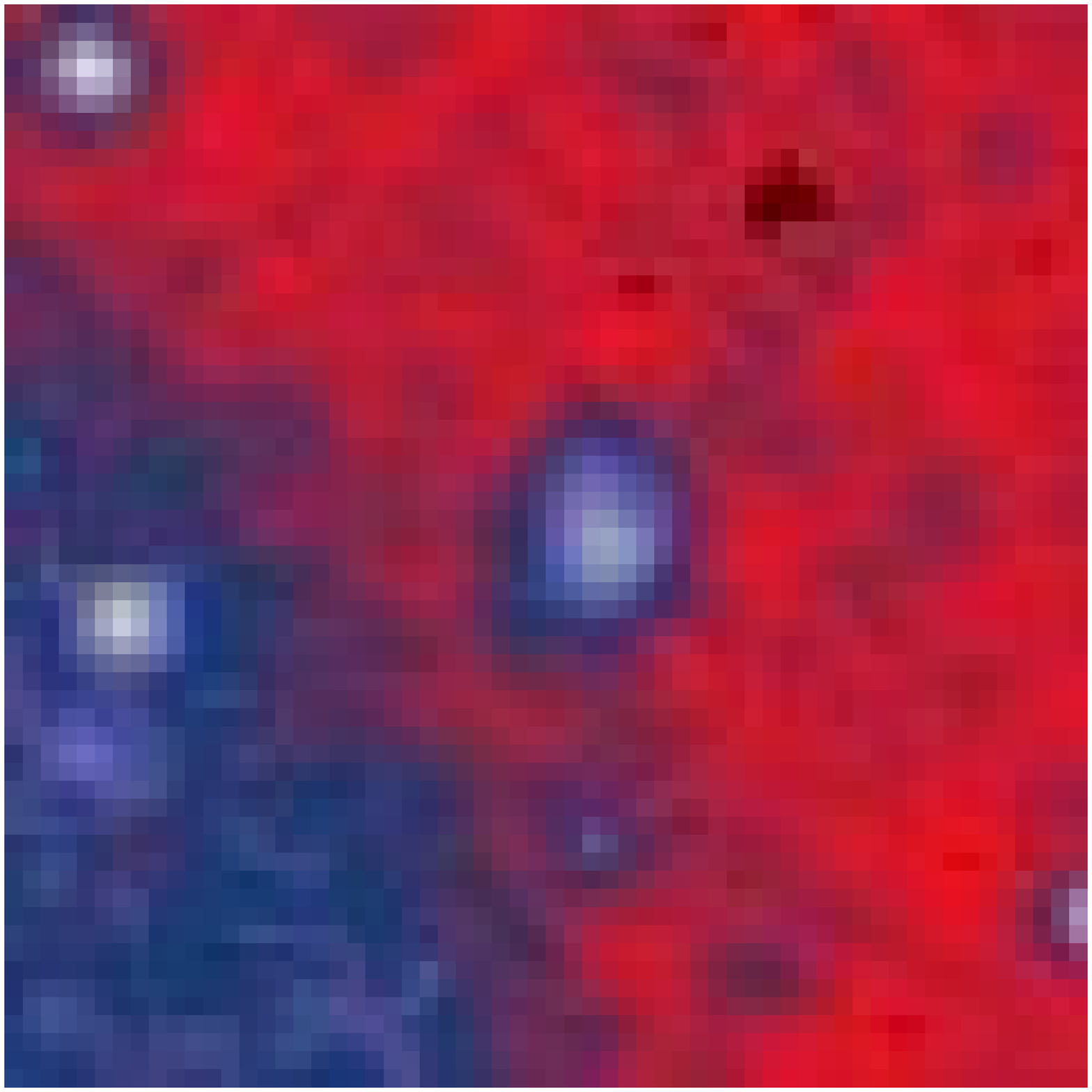}  \FGb{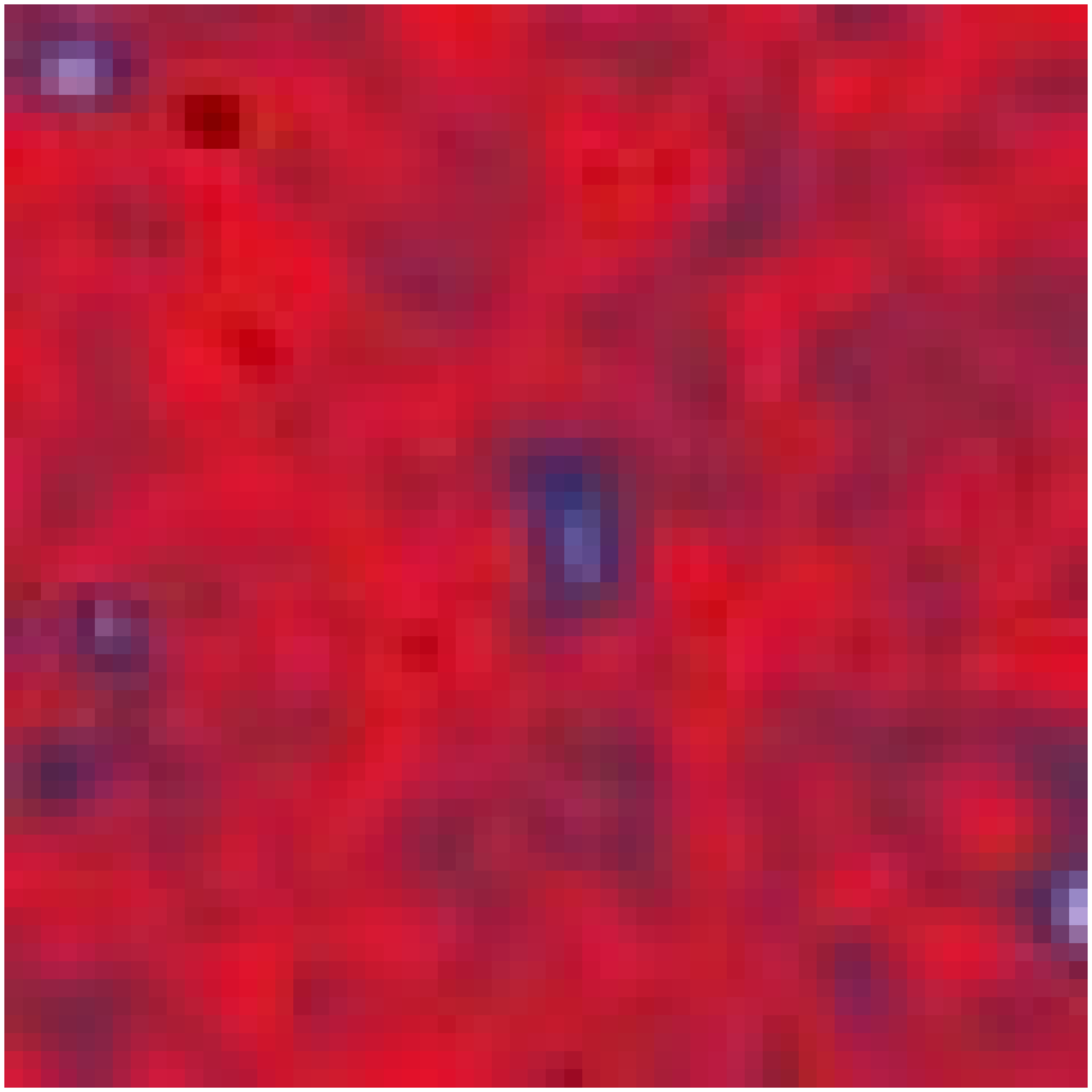}  \FGb{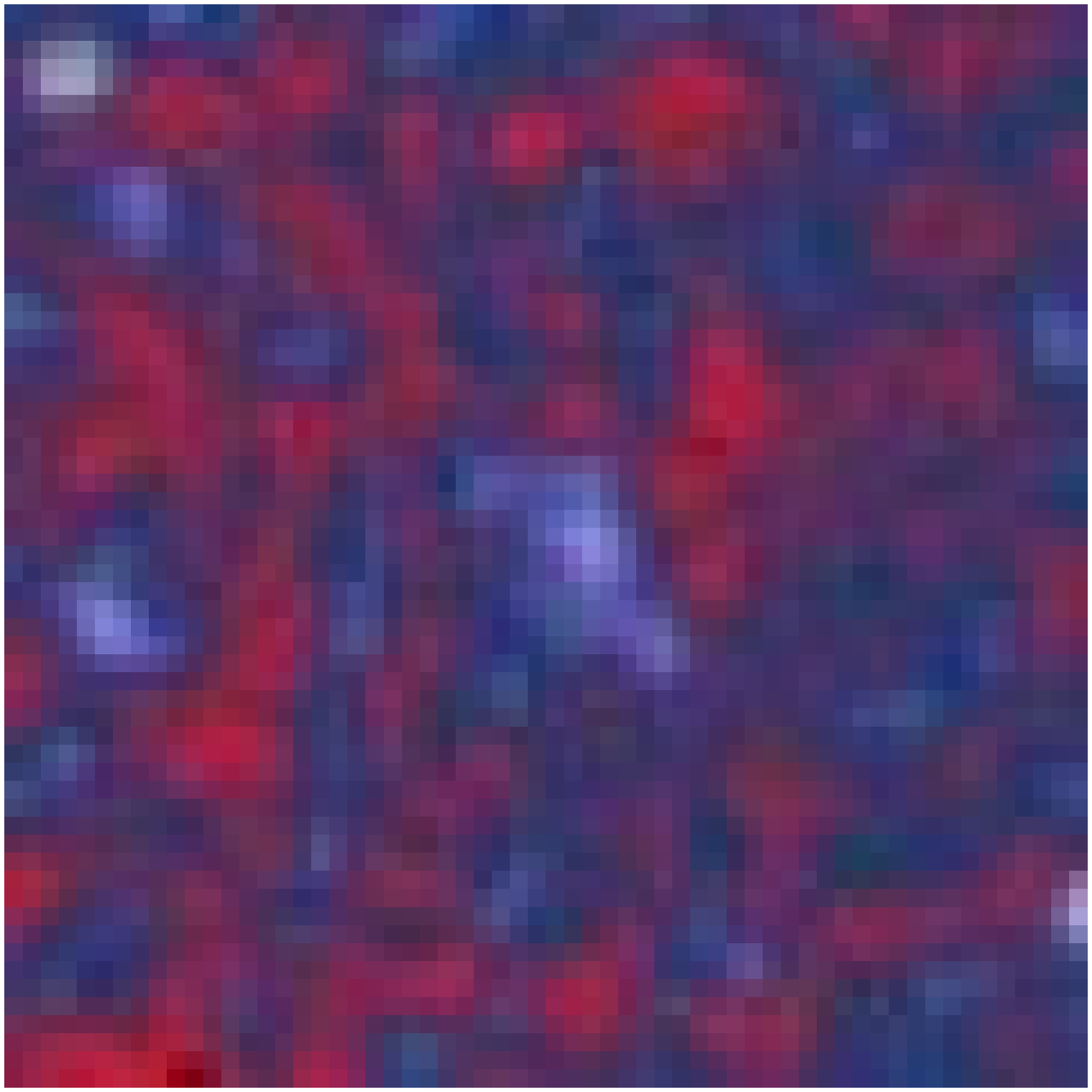}  \FGb{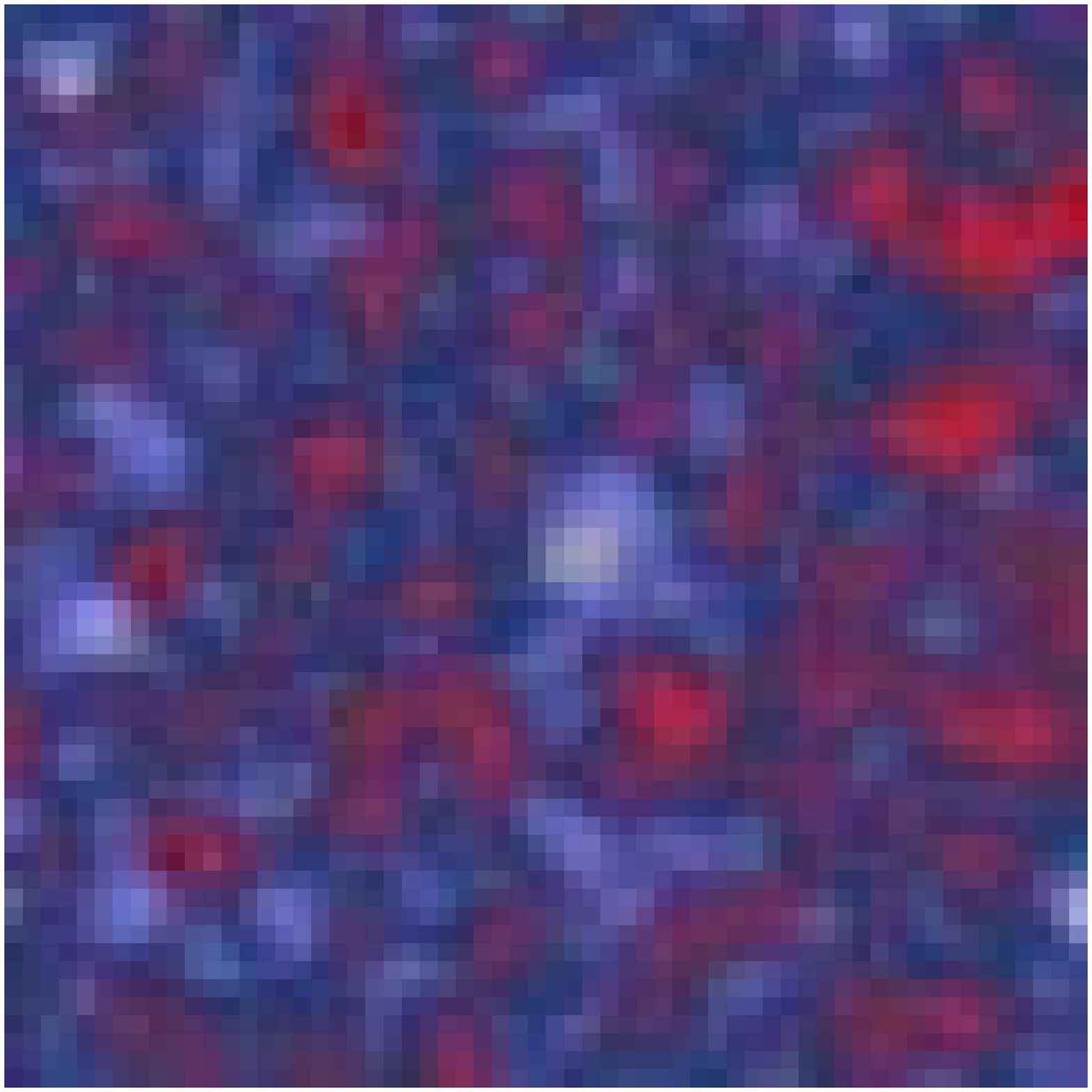}  \FGb{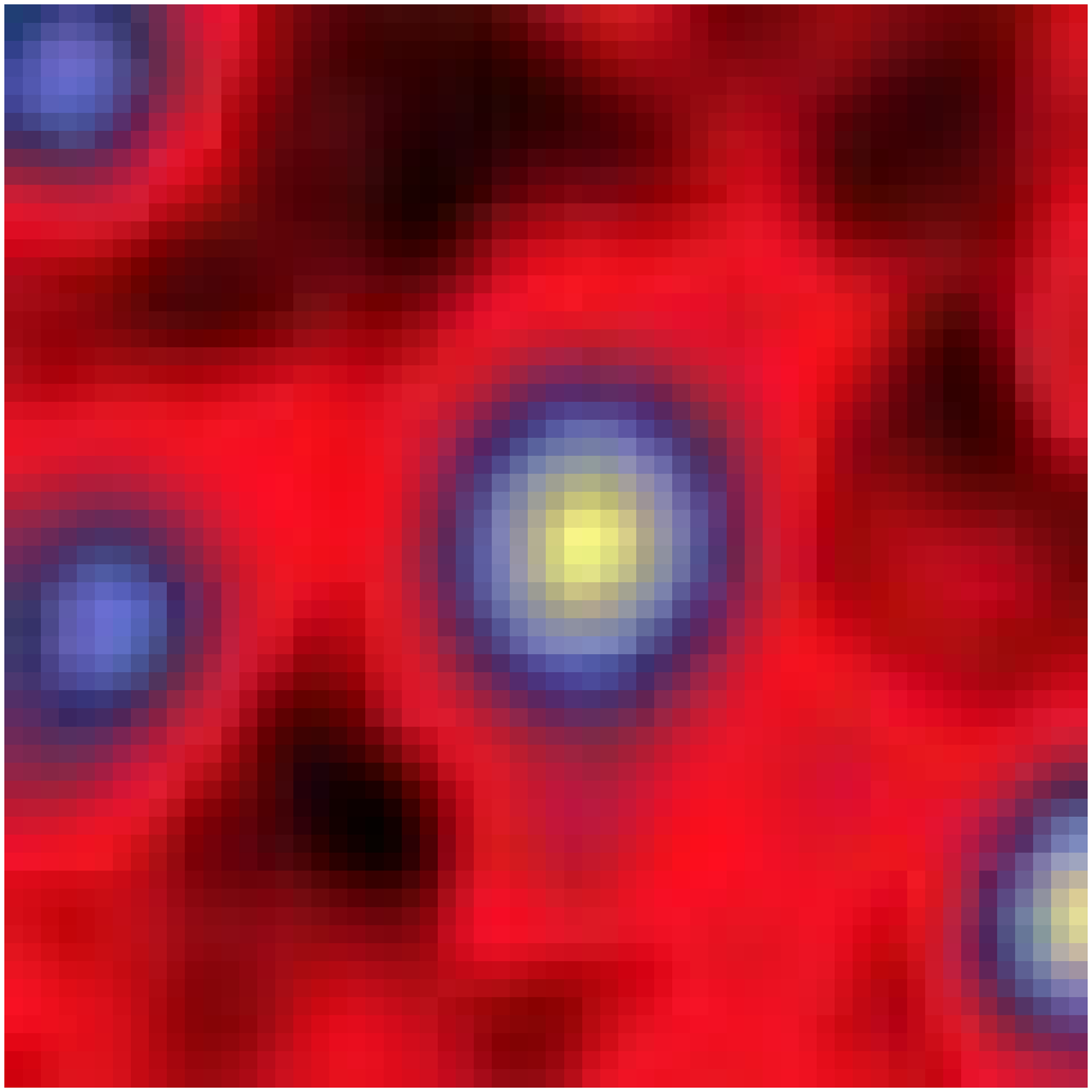}  \FGb{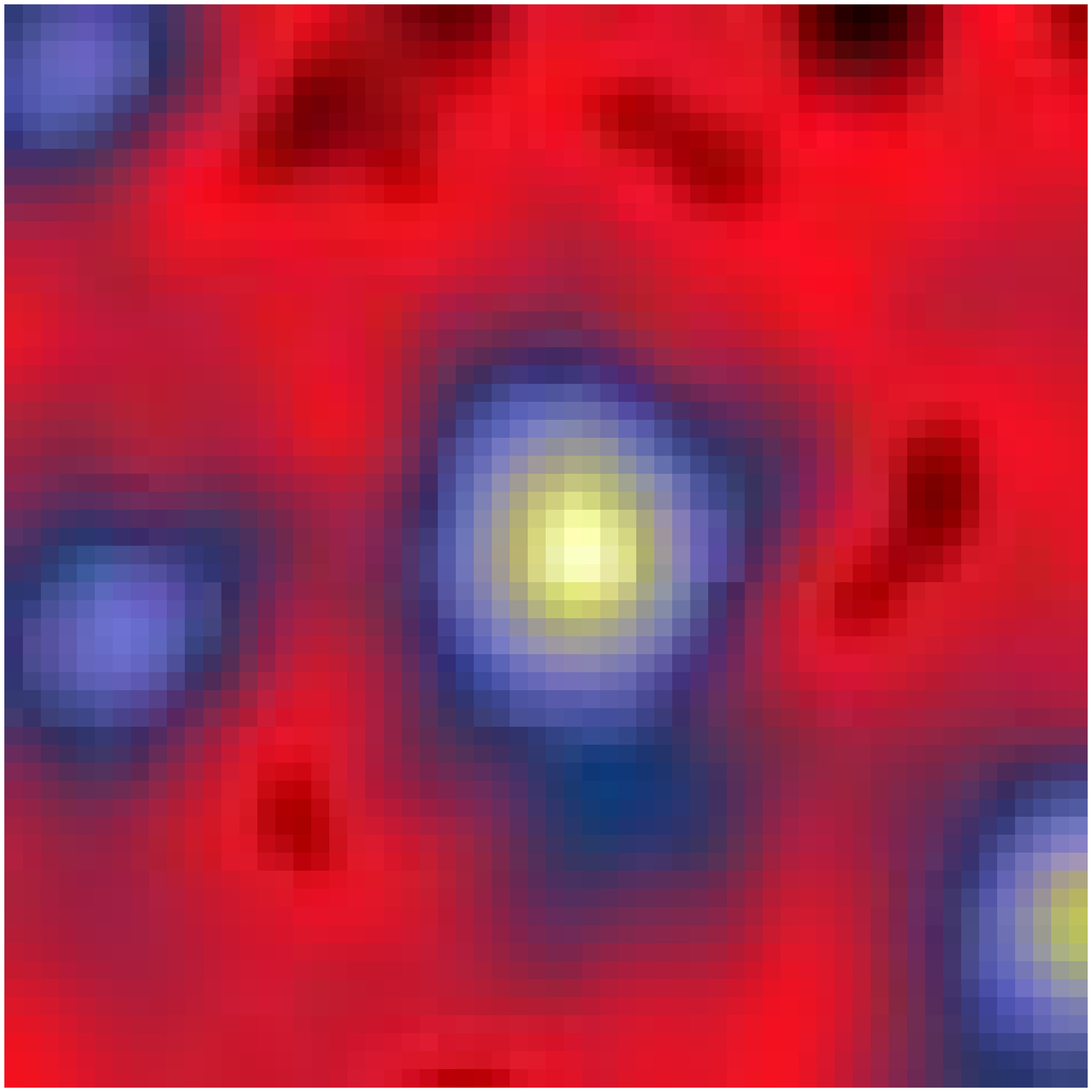}  \FGb{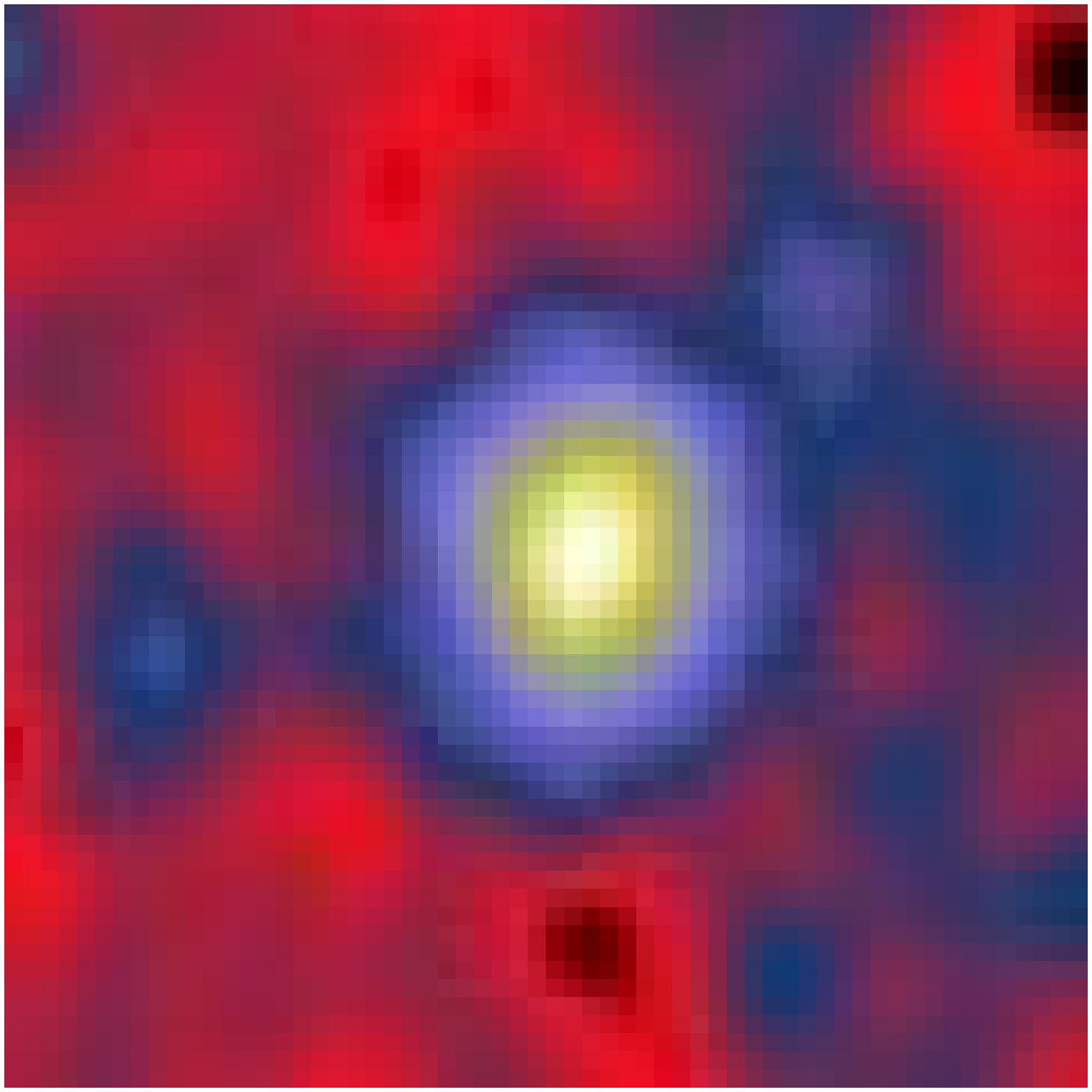}  \FGb{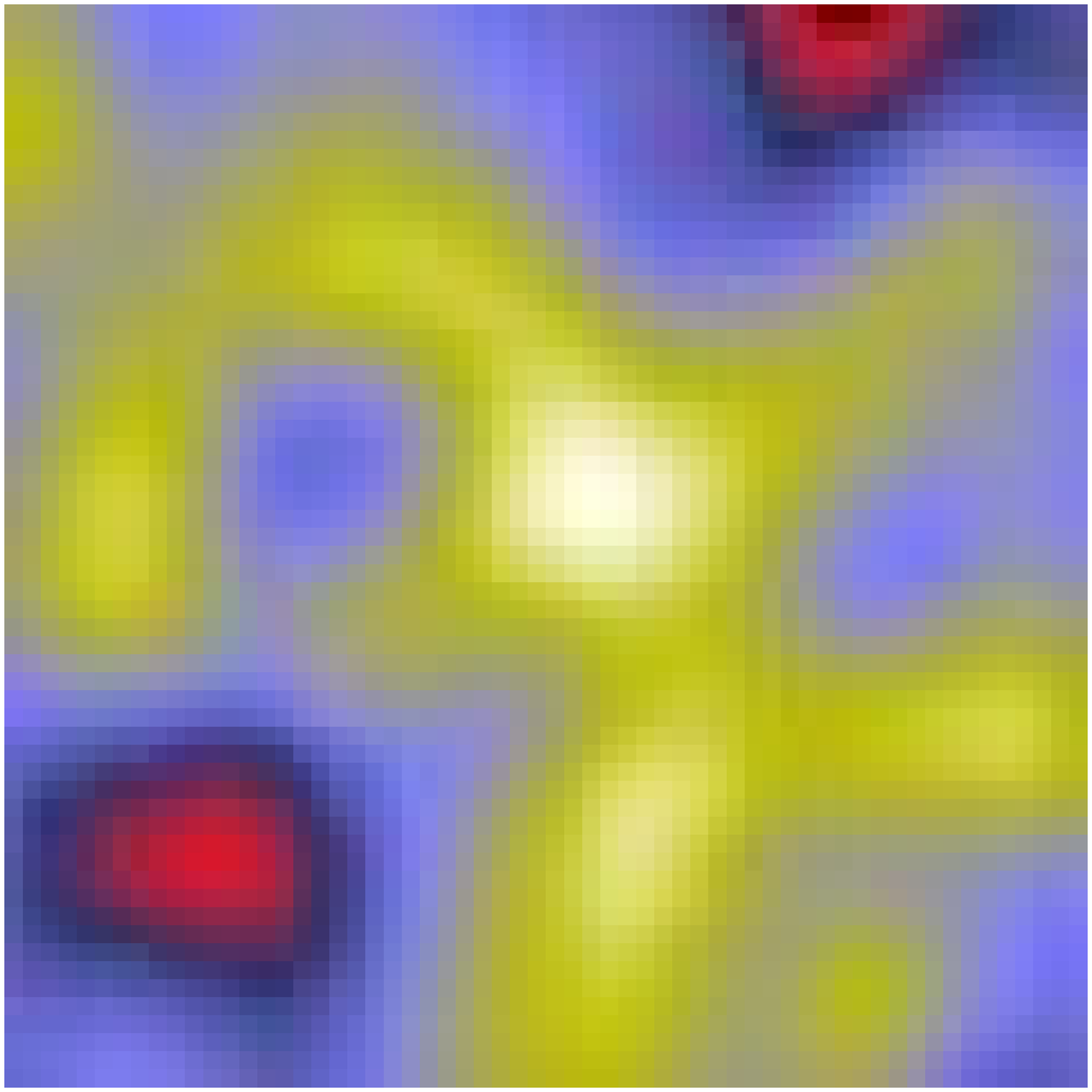} \\  {\#25   NCIA012182}  \\
  \FGb{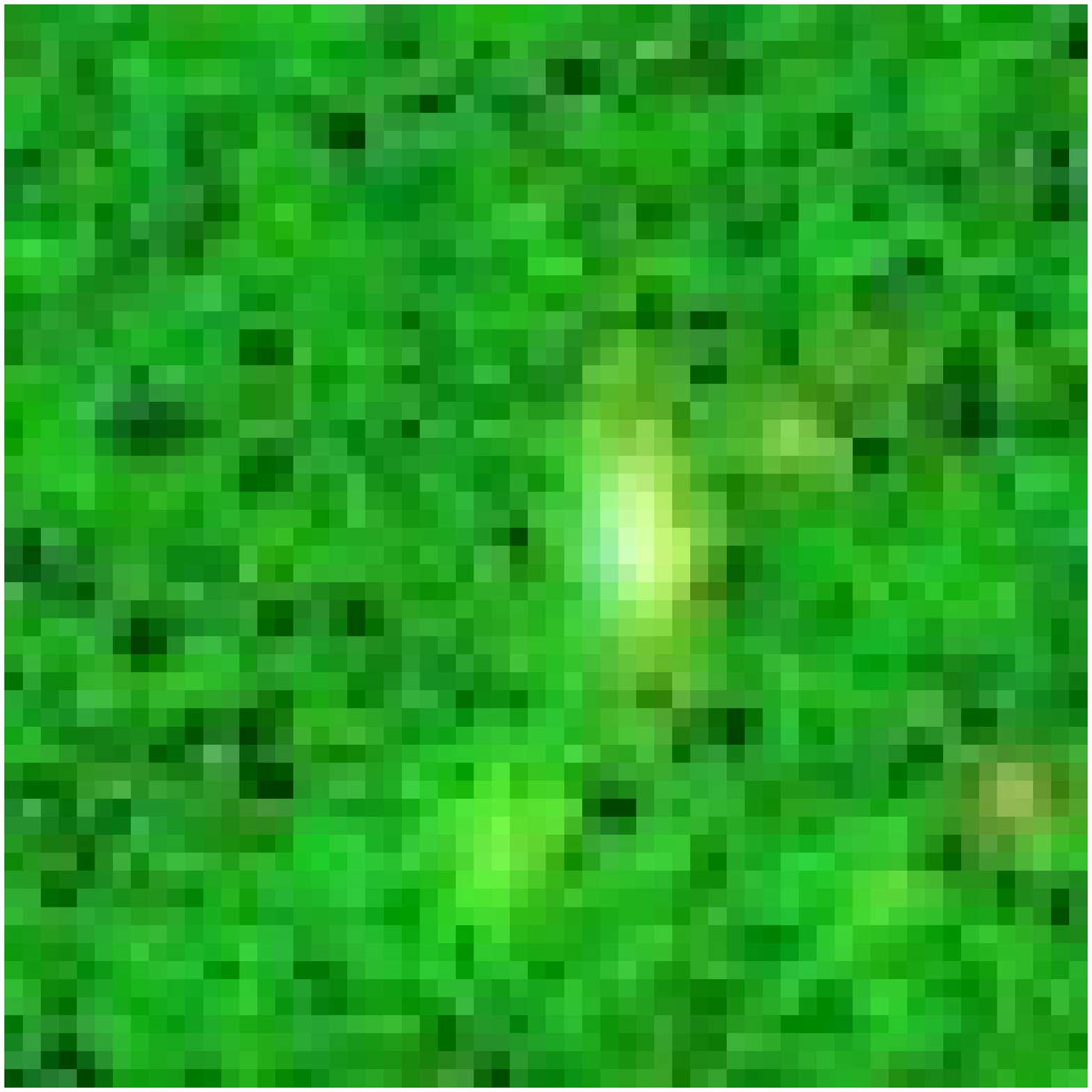}  \FGb{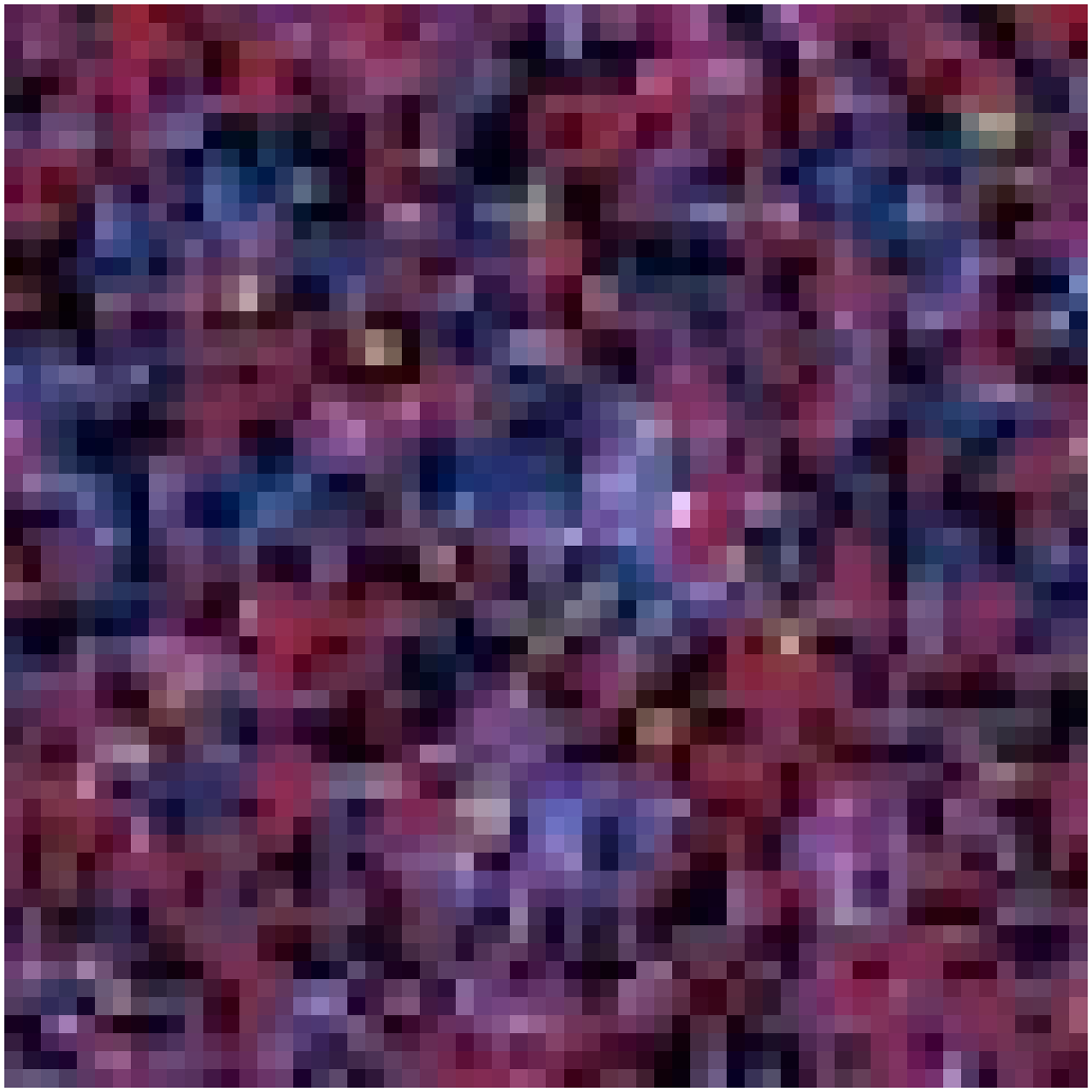}  \FGb{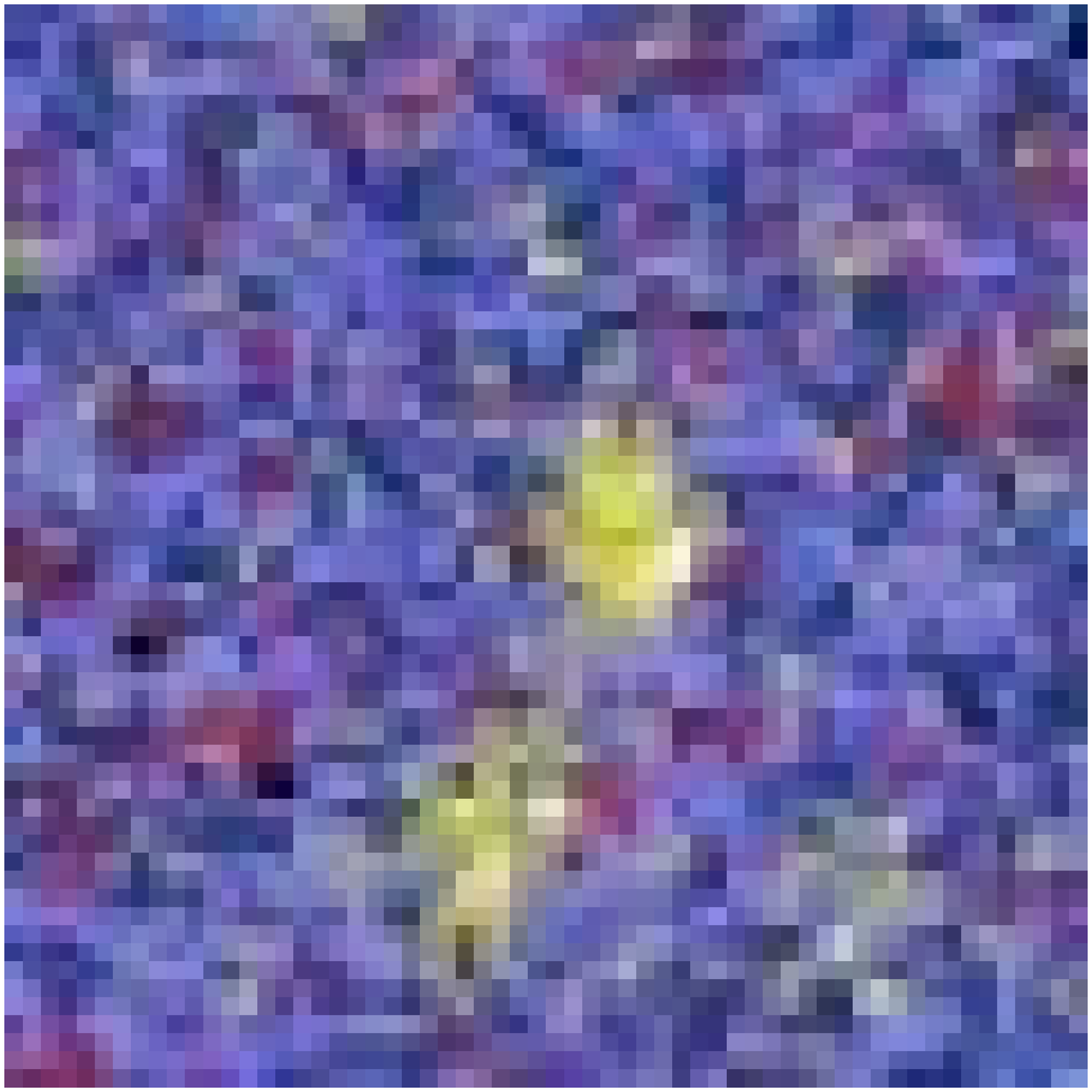}  \FGb{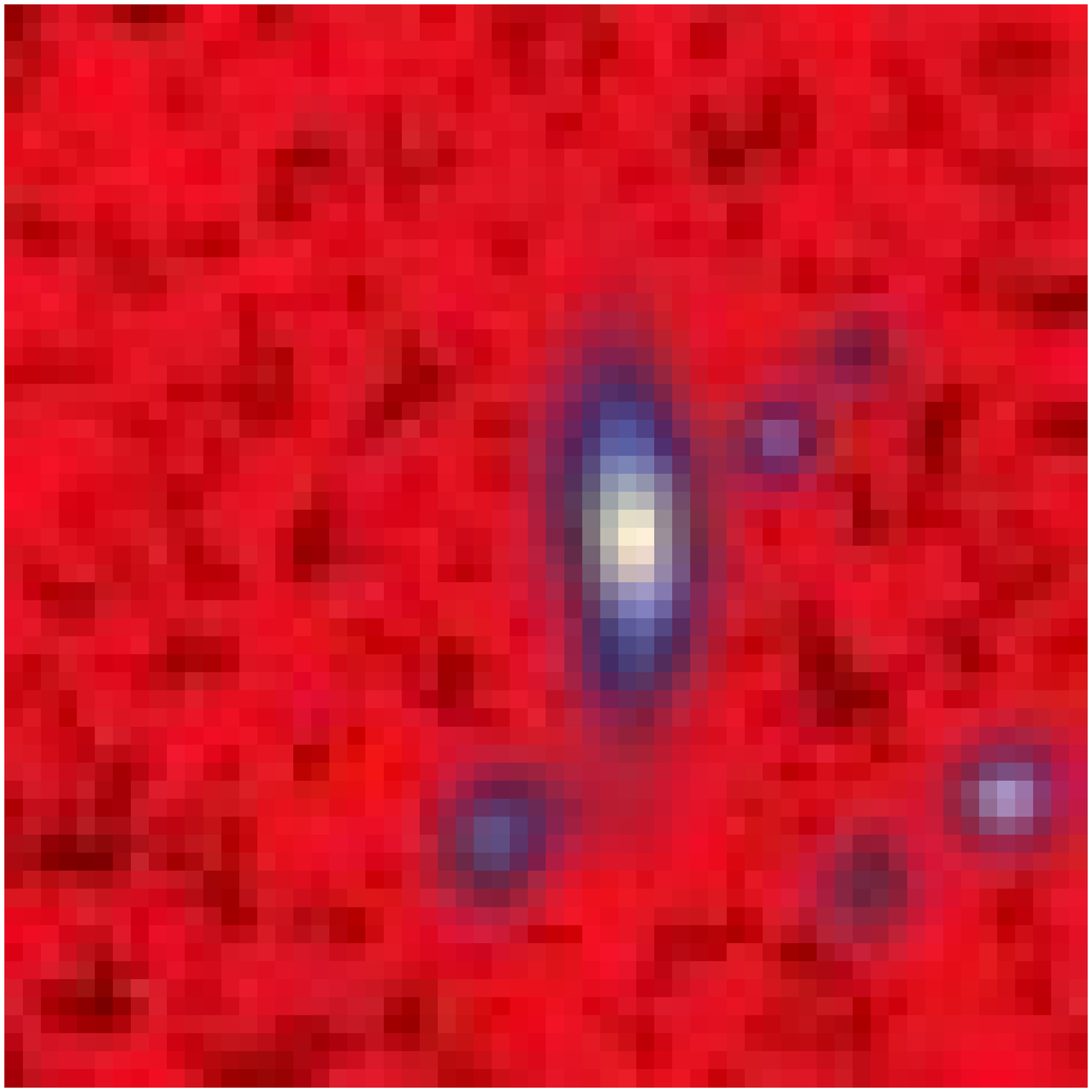}  \FGb{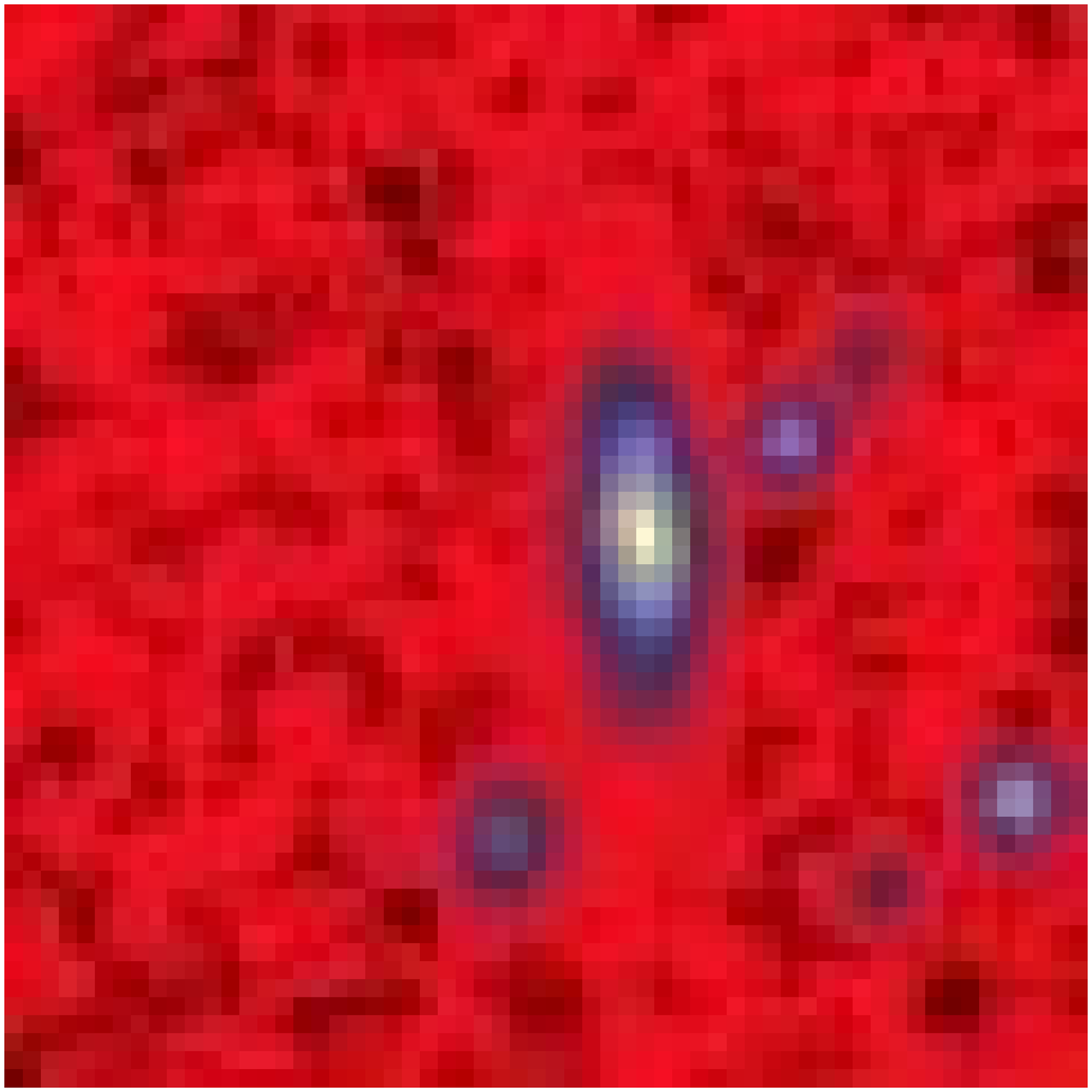}  \FGb{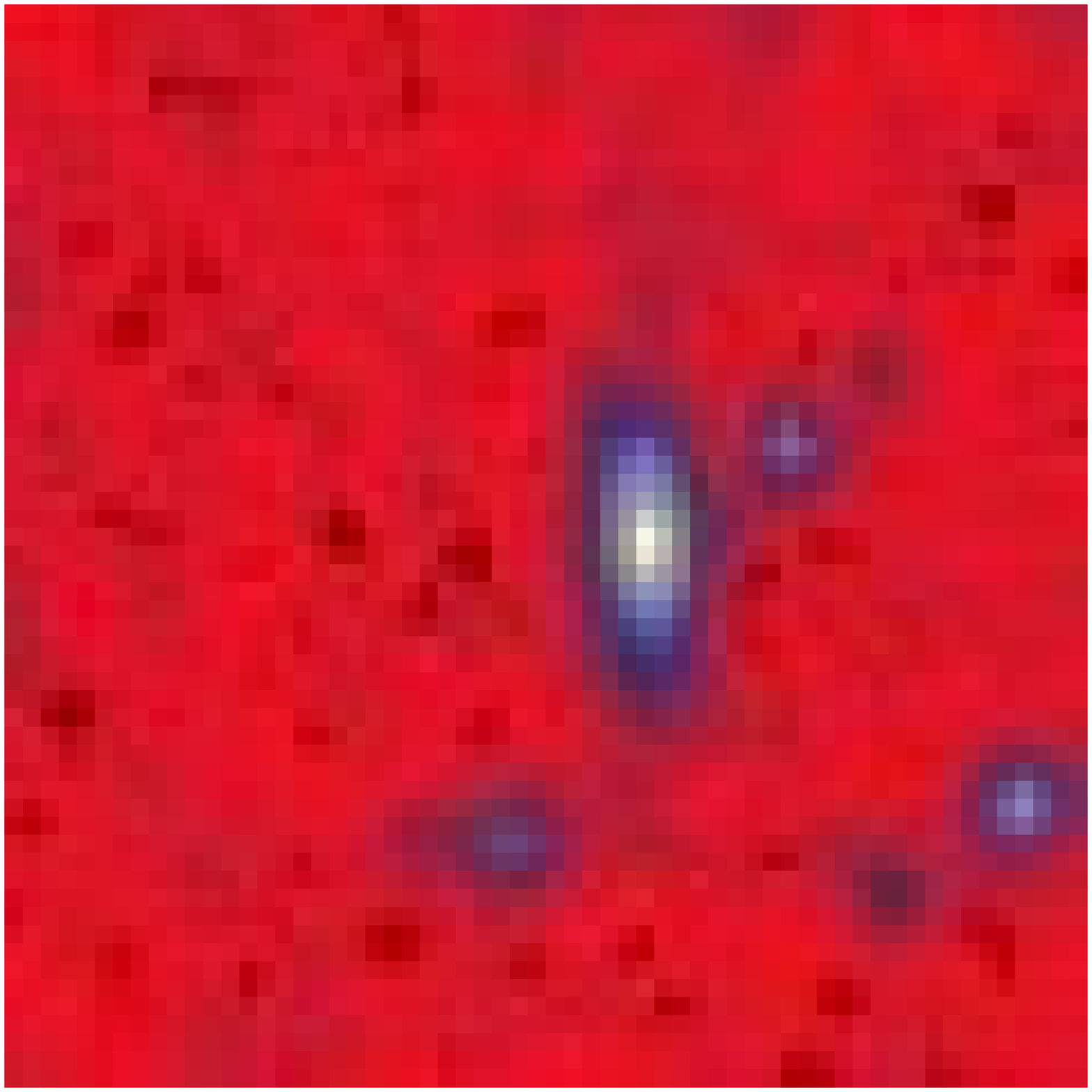}  \FGb{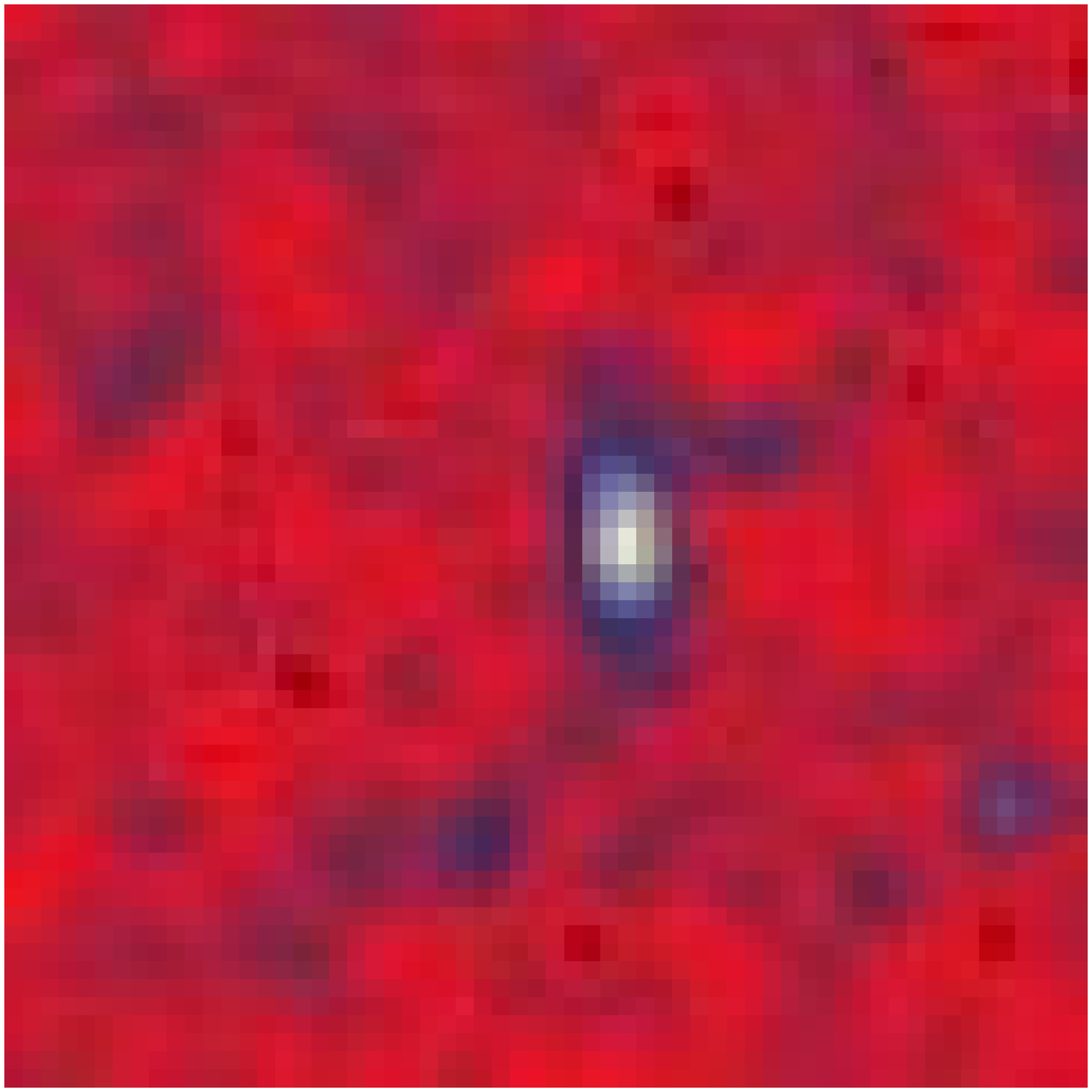}  \FGb{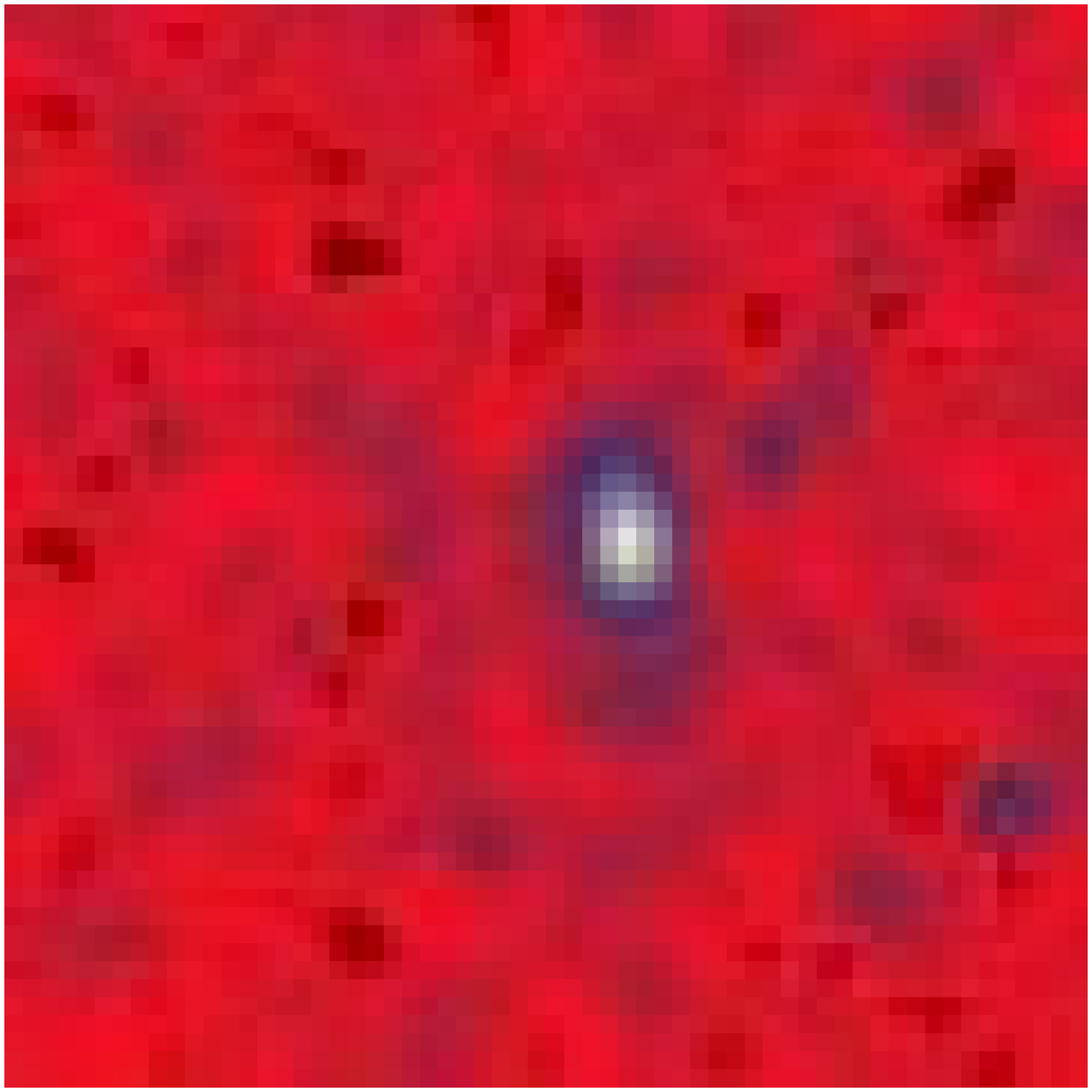}  \FGb{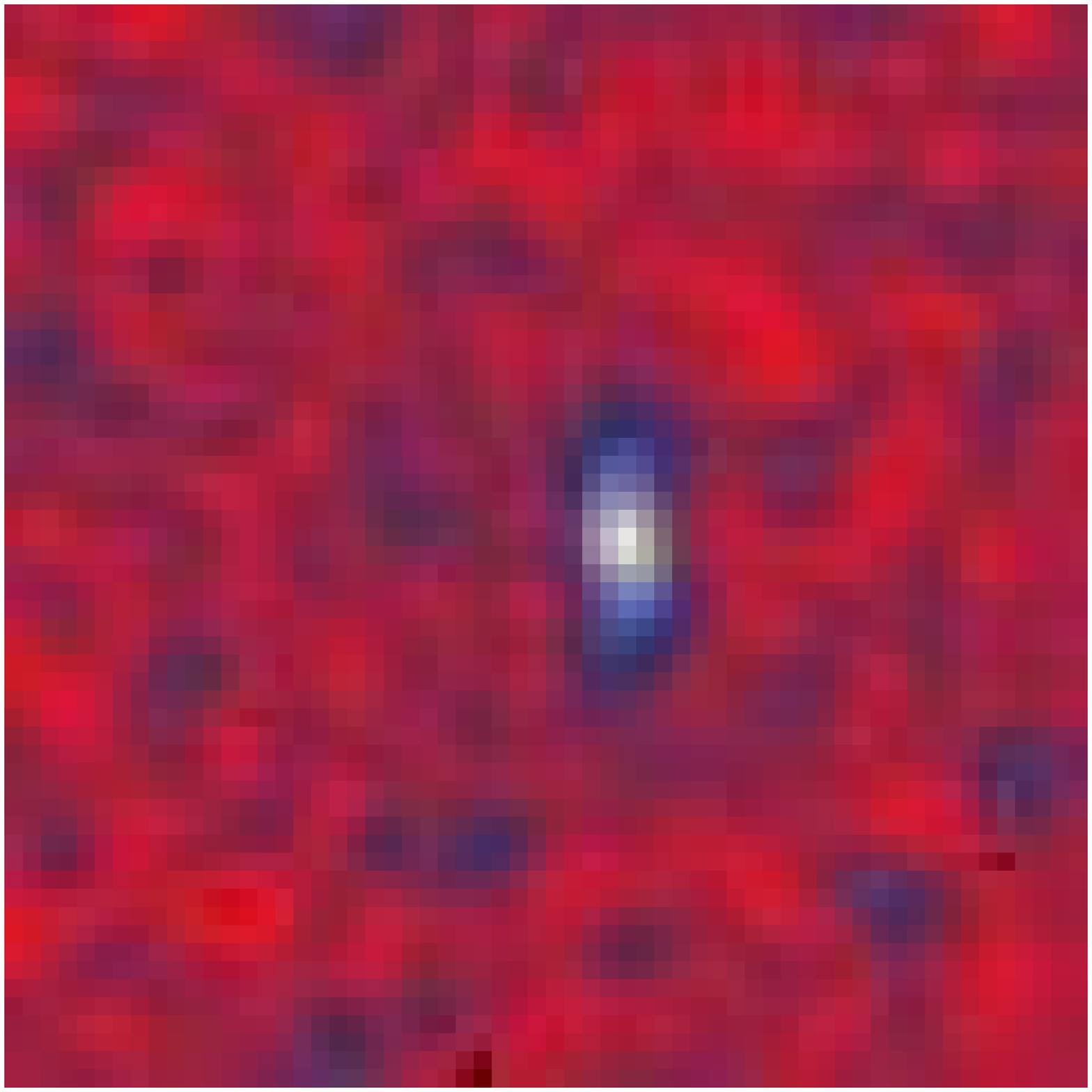}  \FGb{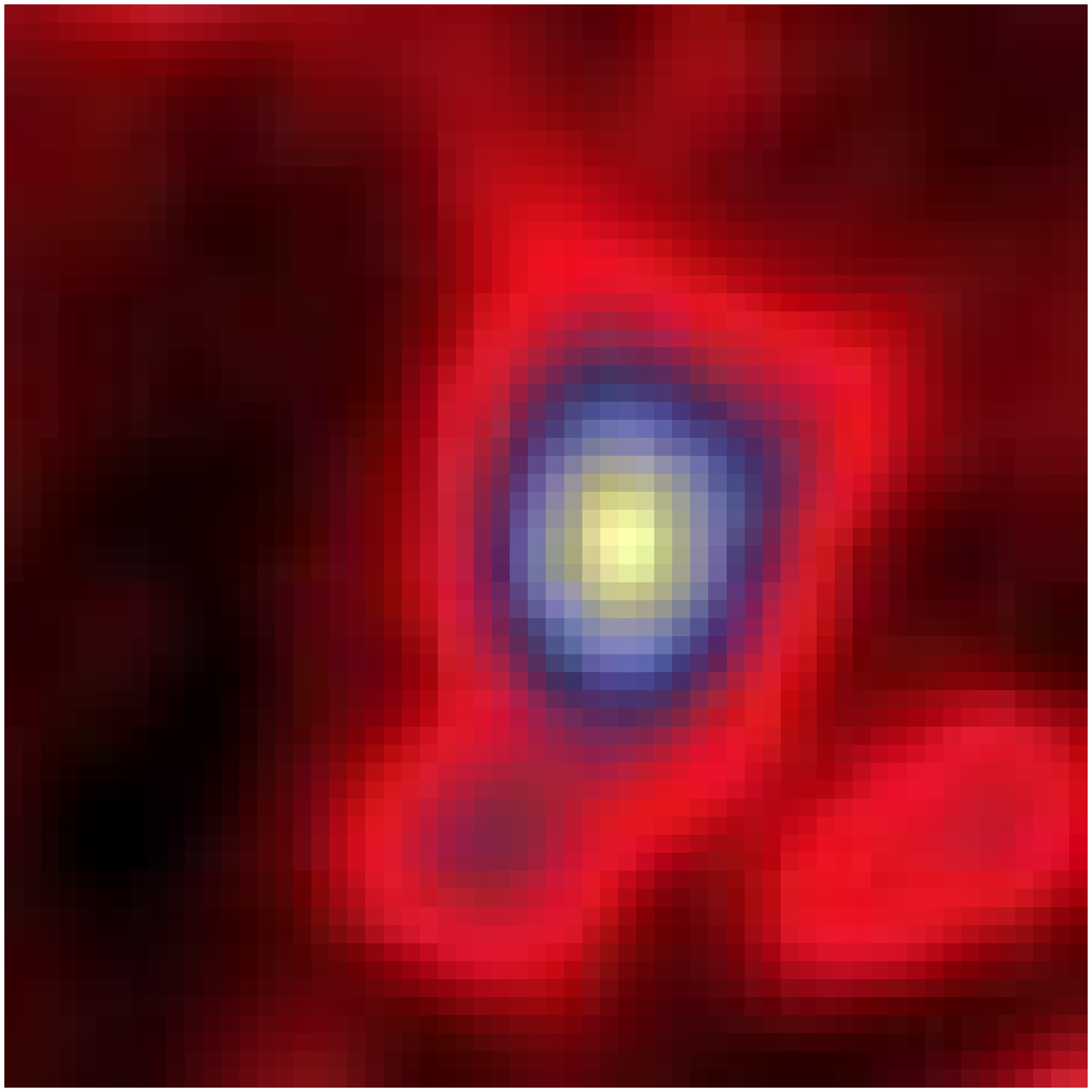}  \FGb{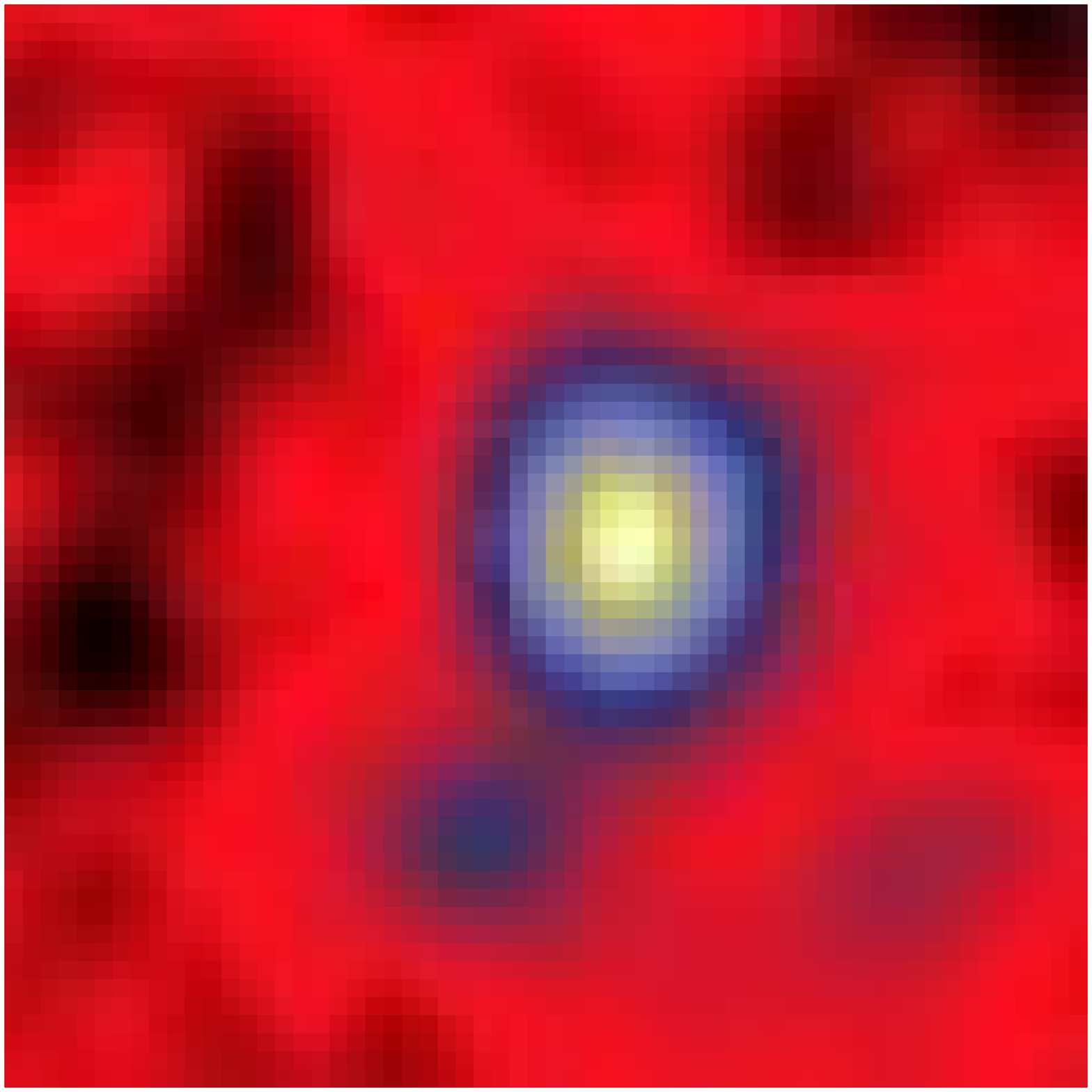}  \FGb{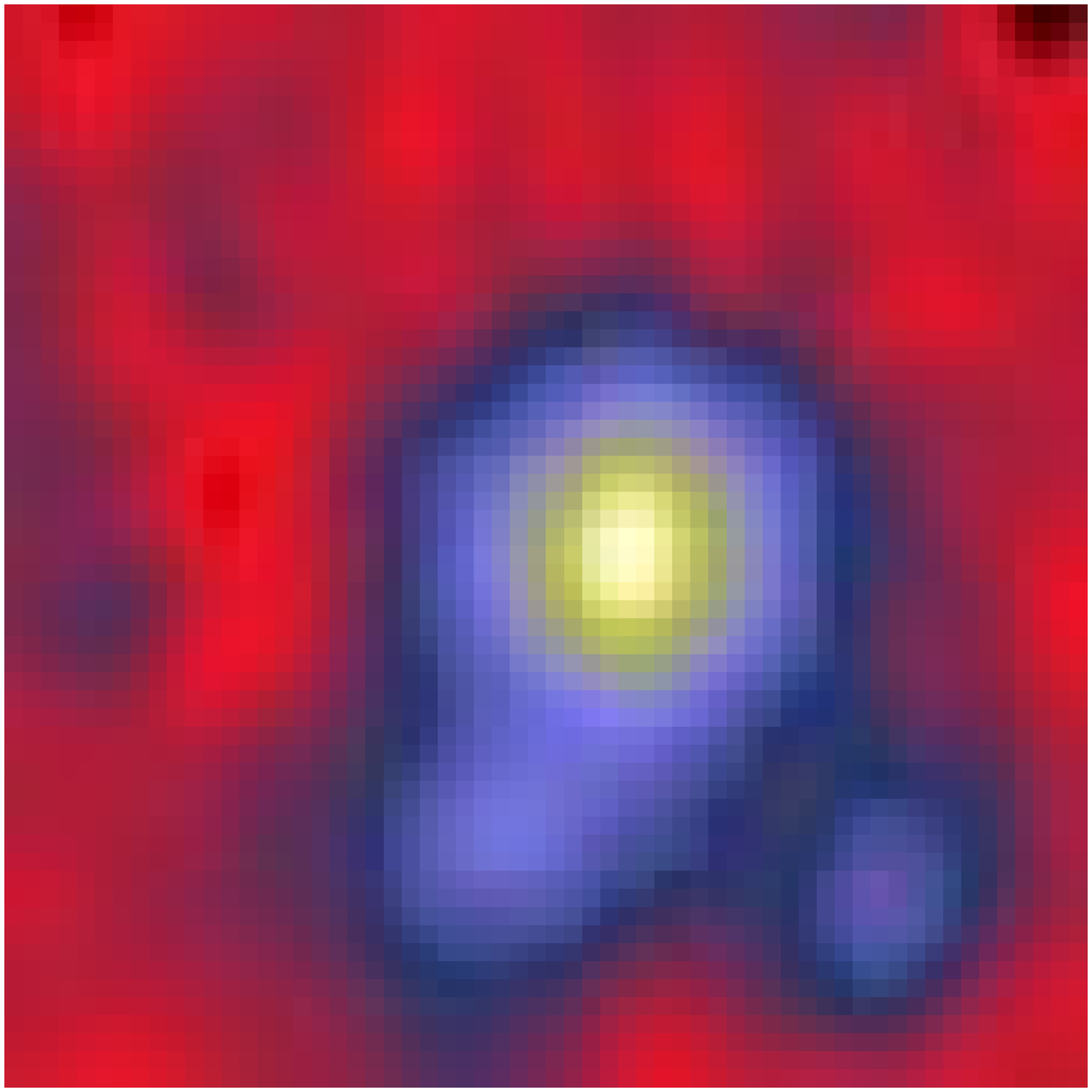}  \FGb{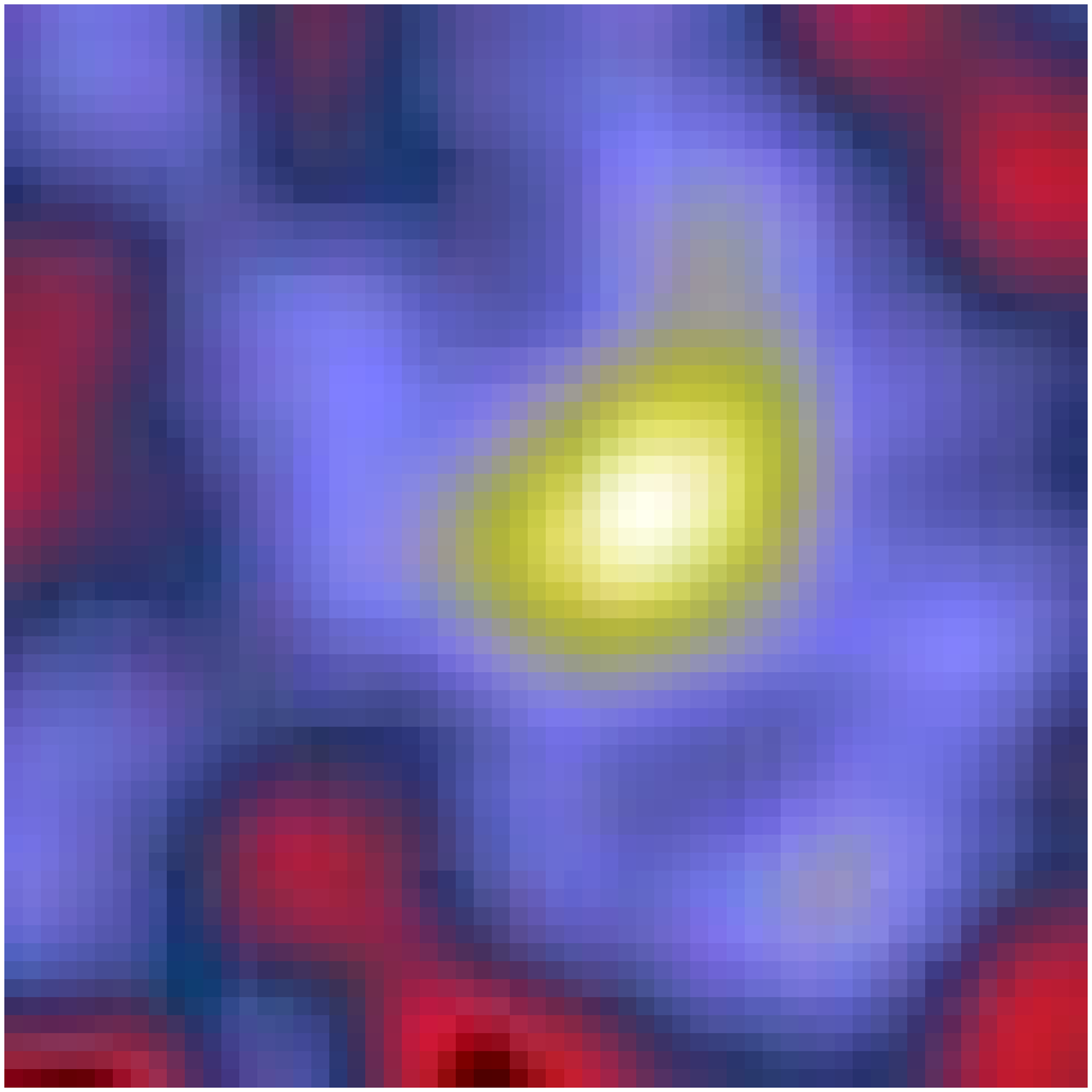} \\  {\#26   NCIA012187}  \\
{ \\ \textbf{Figure \ref{fig:allgal}.(cont)} ~~Visible Galaxies within 40\min\ of NGC~891 } \\
\end{figure}

\clearpage
\begin{figure}[ht]
  \HDb{rgb}  \HDb{FUV}  \HDb{NUV}   \HDb{B}     \HDb{R}    \HDb{IR}   \HDb{J}    \HDb{H}     \HDb{K}     \HDb{W1}    \HDb{W2}    \HDb{W3}    \HDb{W4}  \\
  \FGb{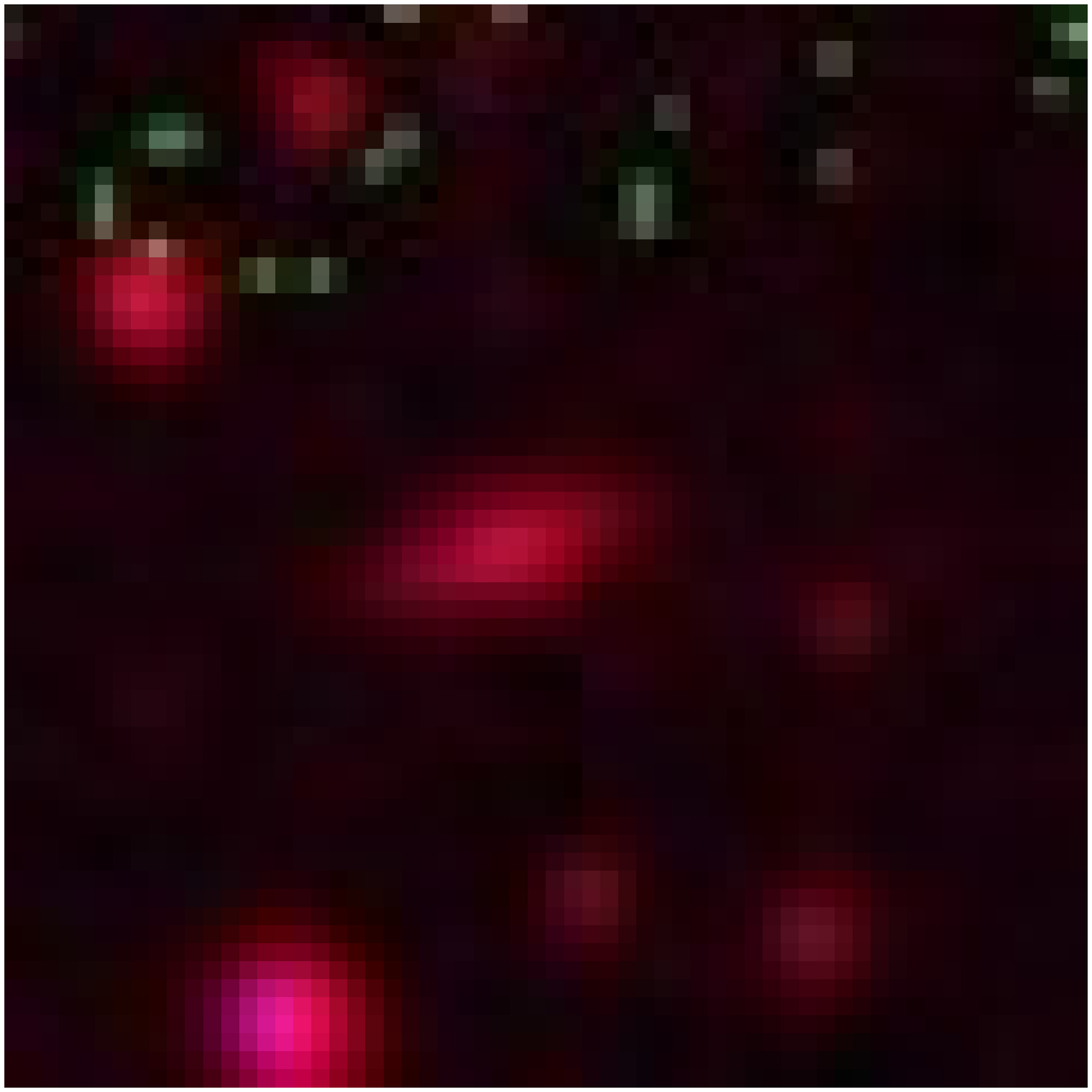}  \FGb{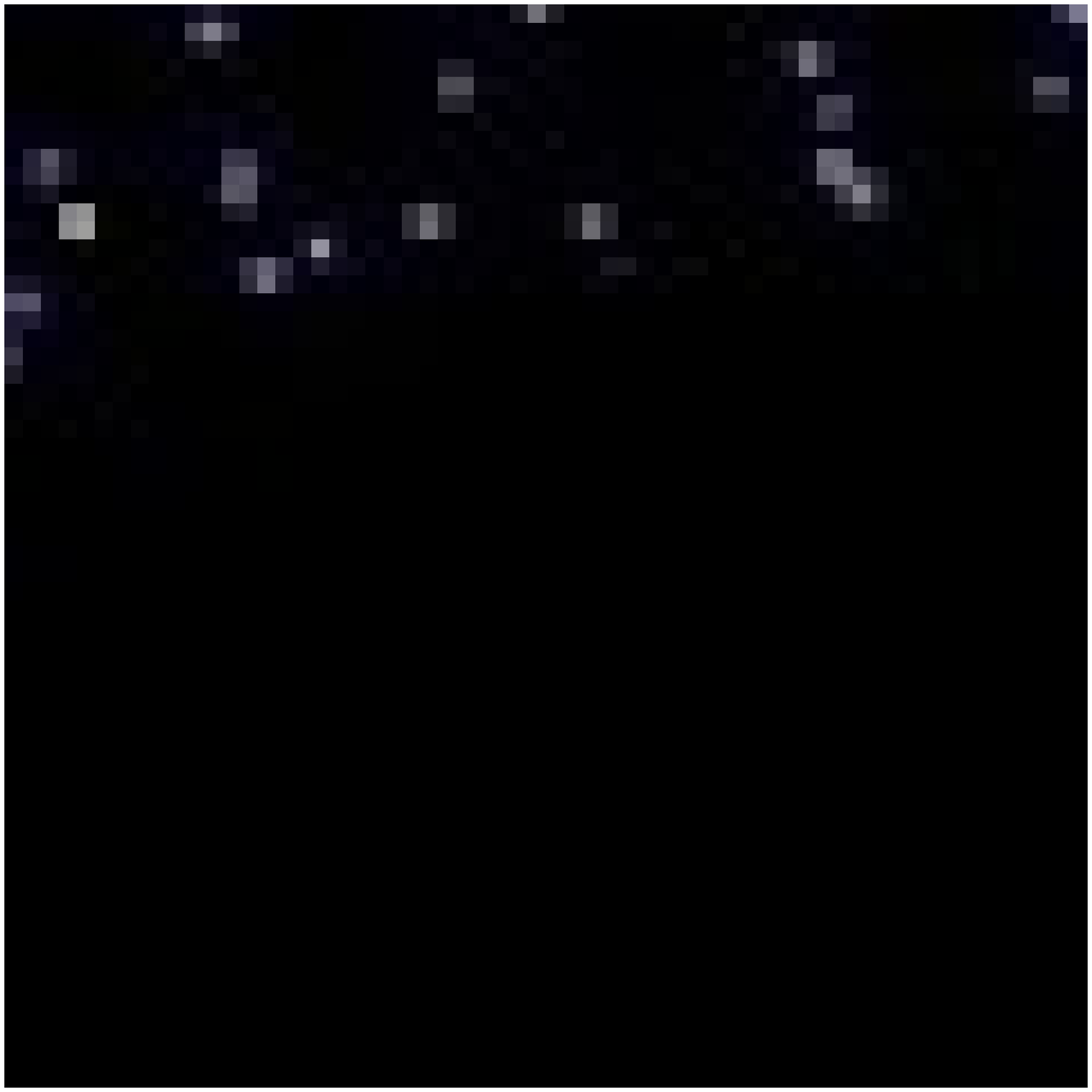}  \FGb{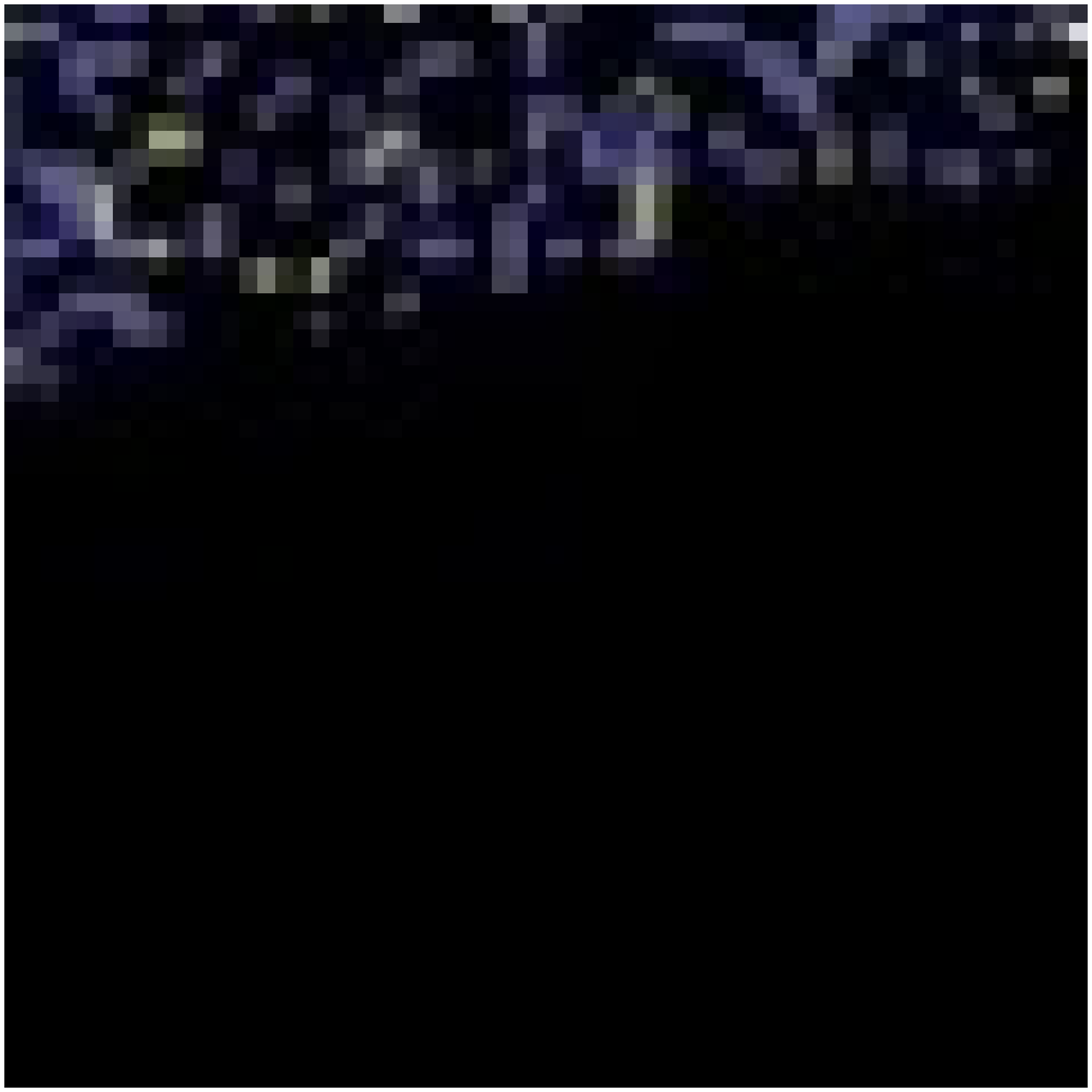}  \FGb{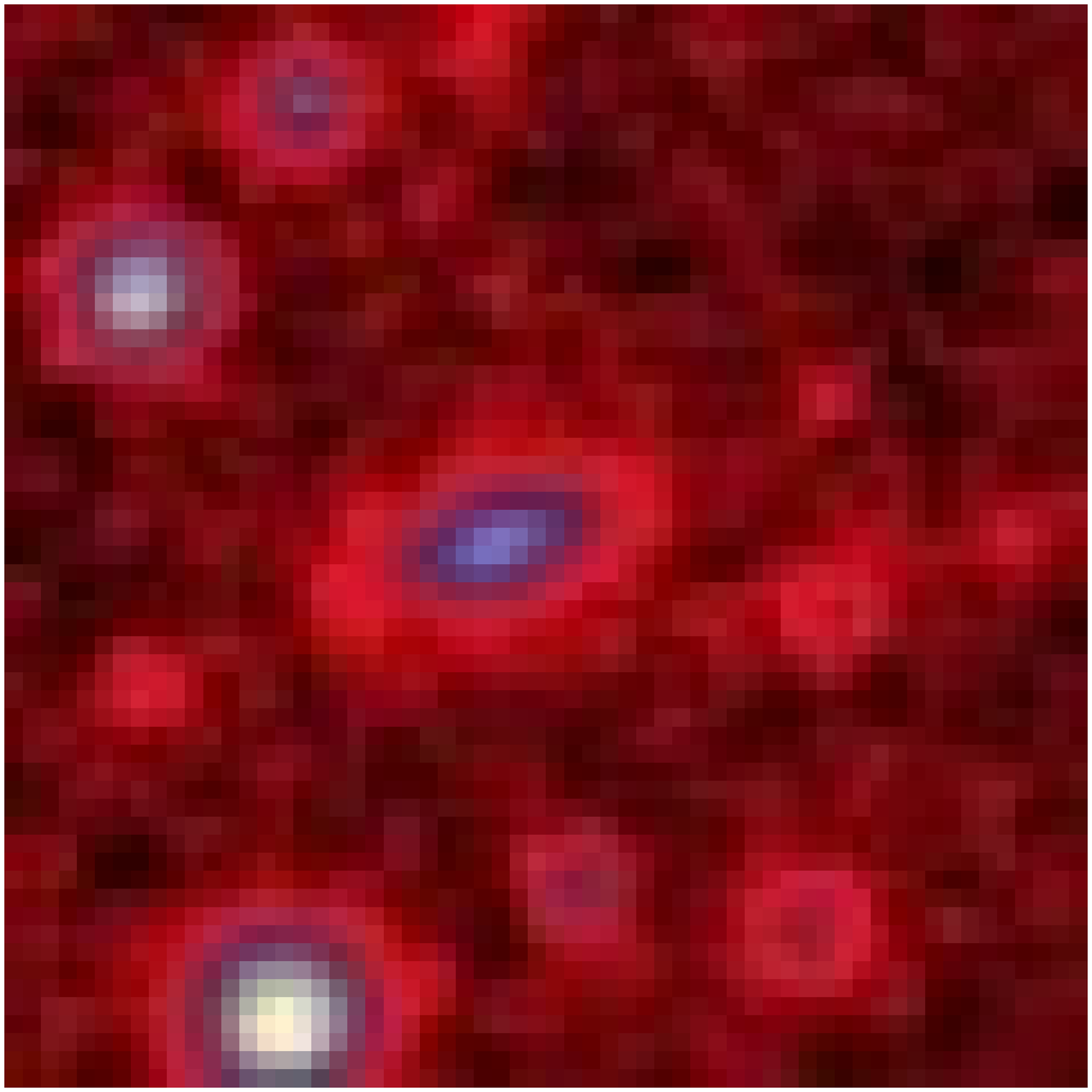}  \FGb{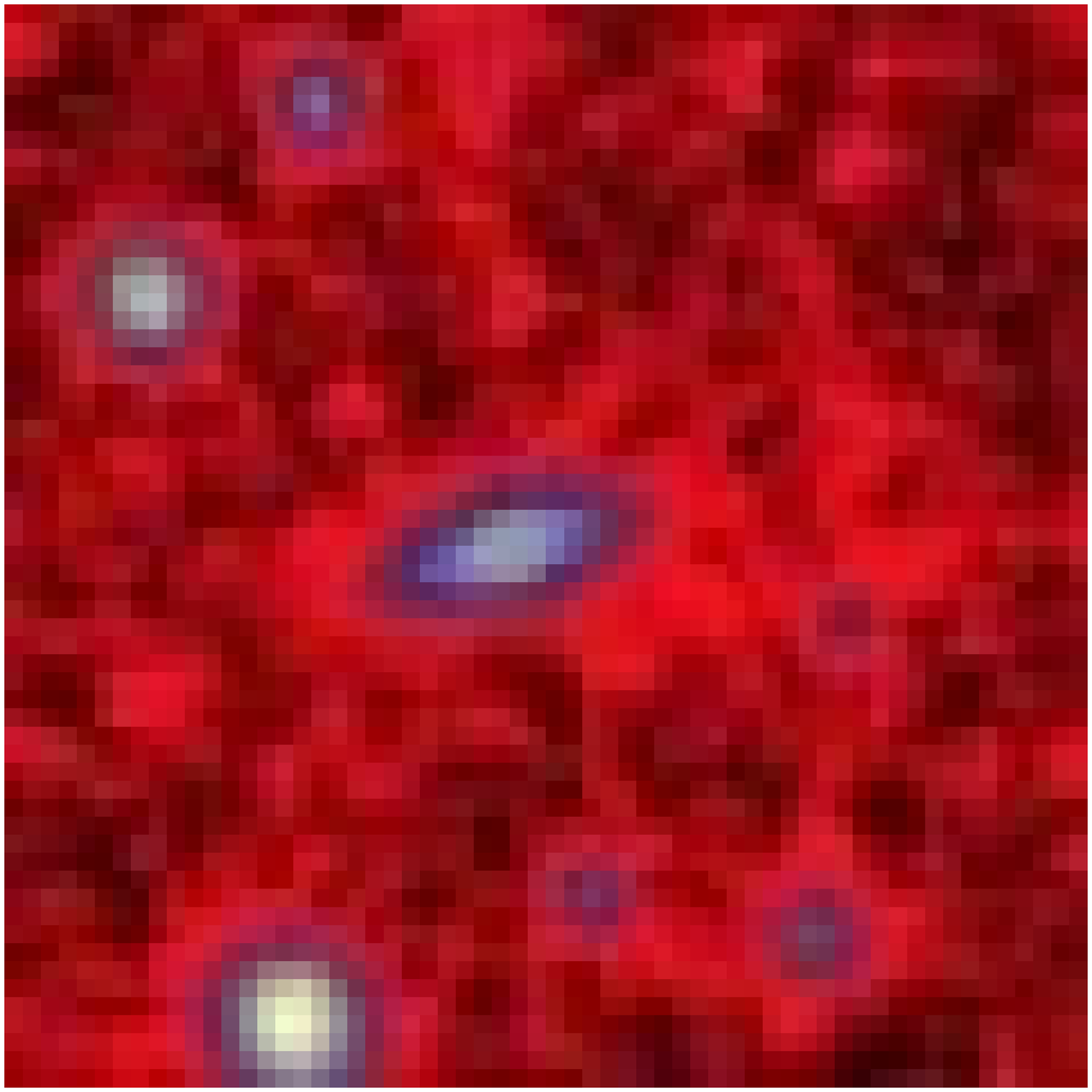}  \FGb{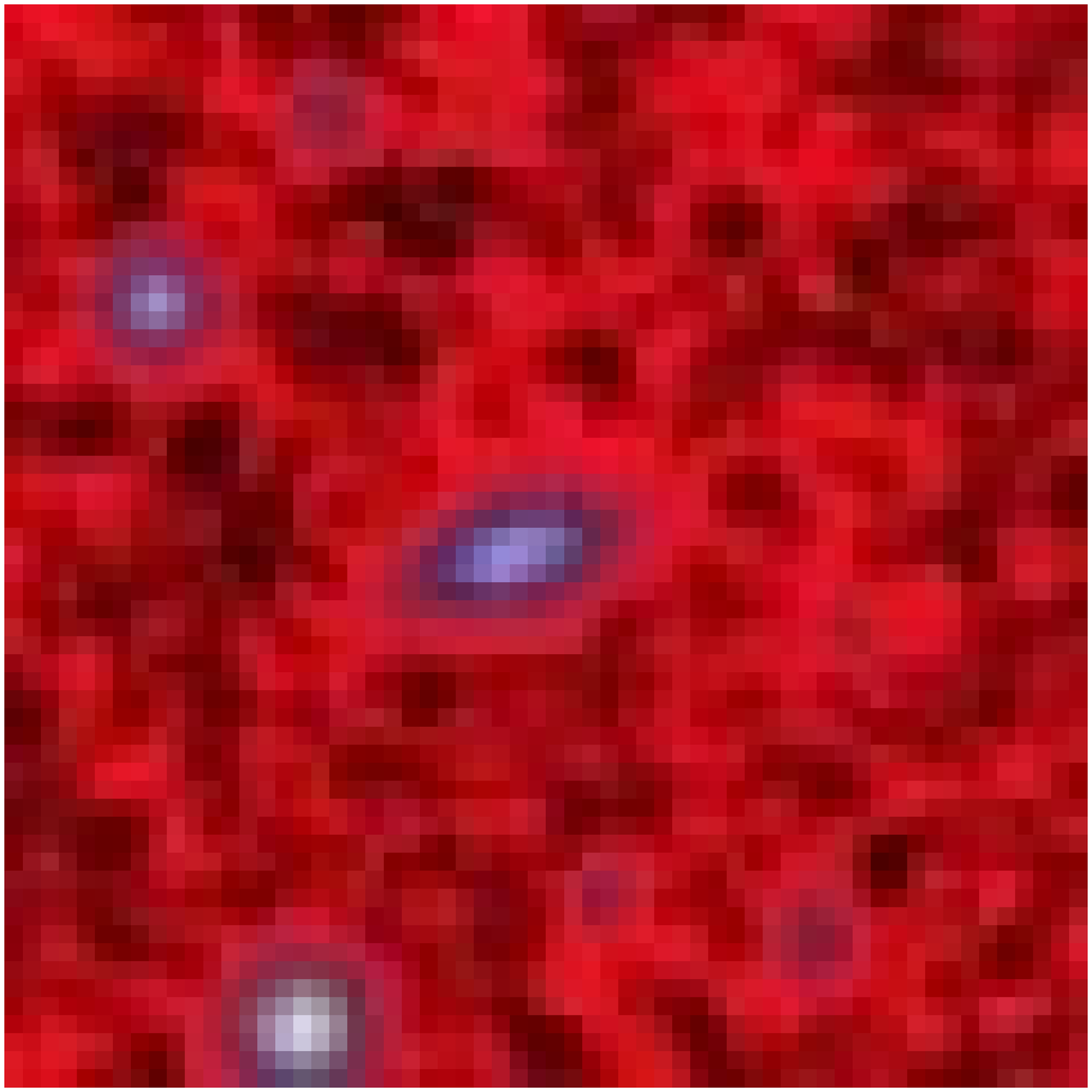}  \FGb{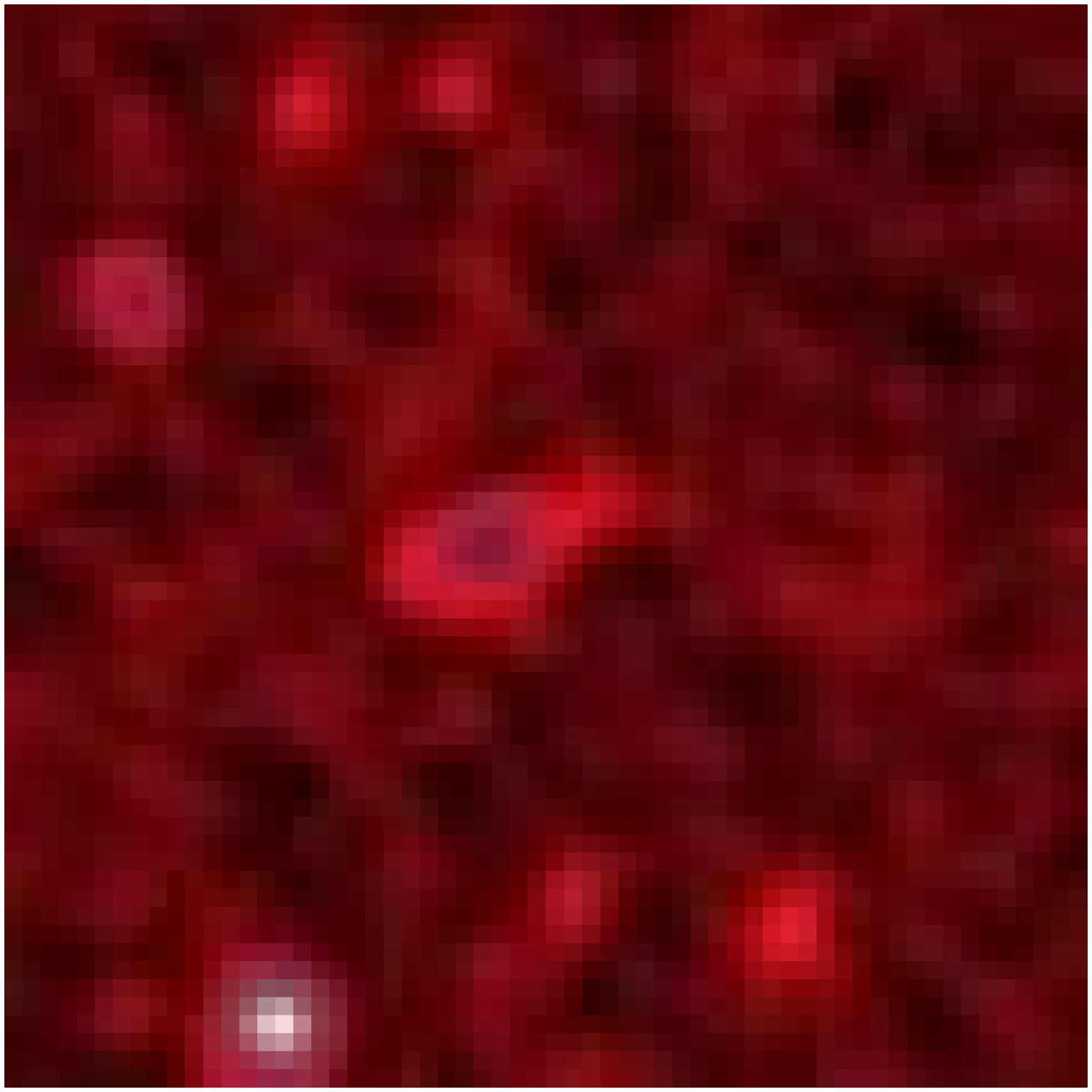}  \FGb{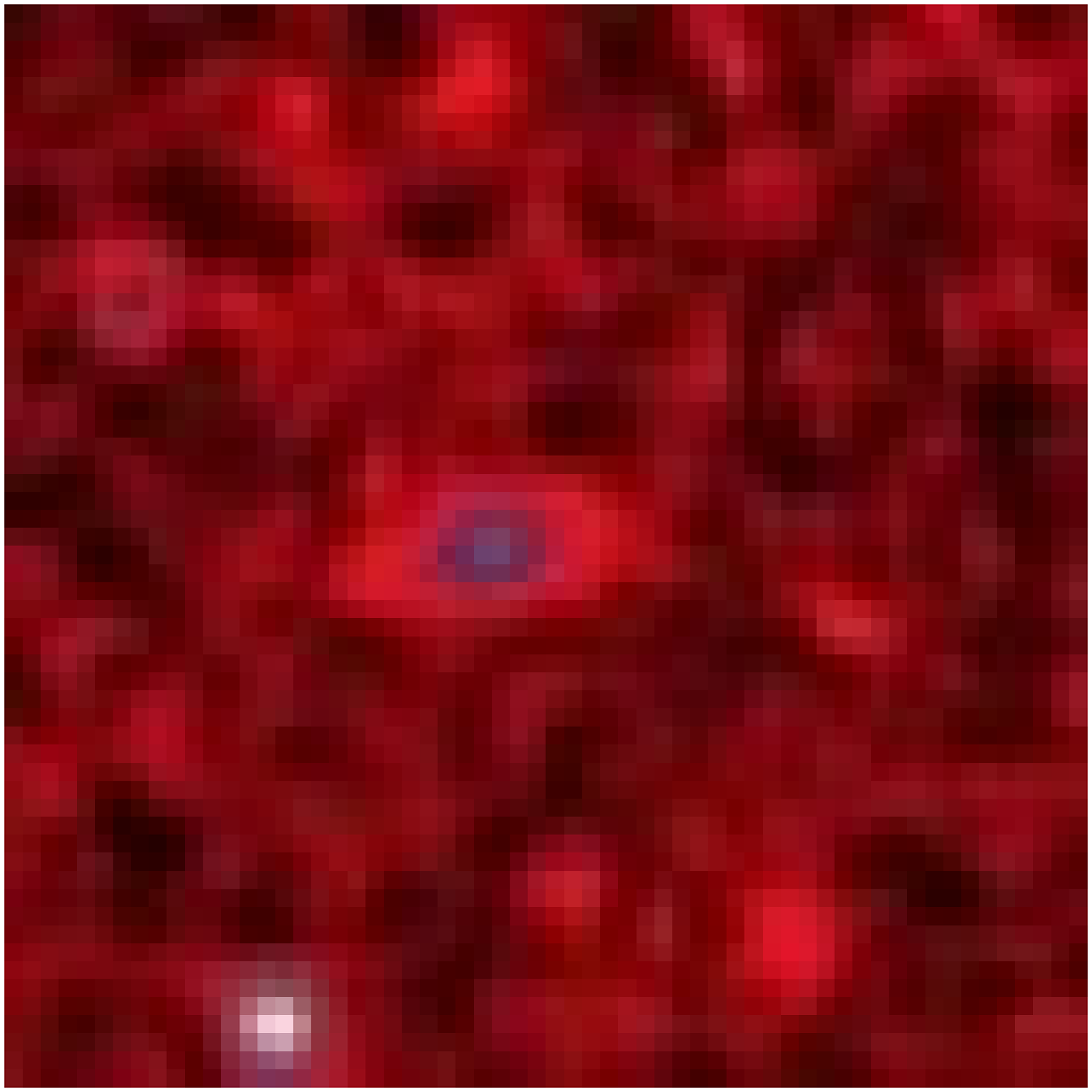}  \FGb{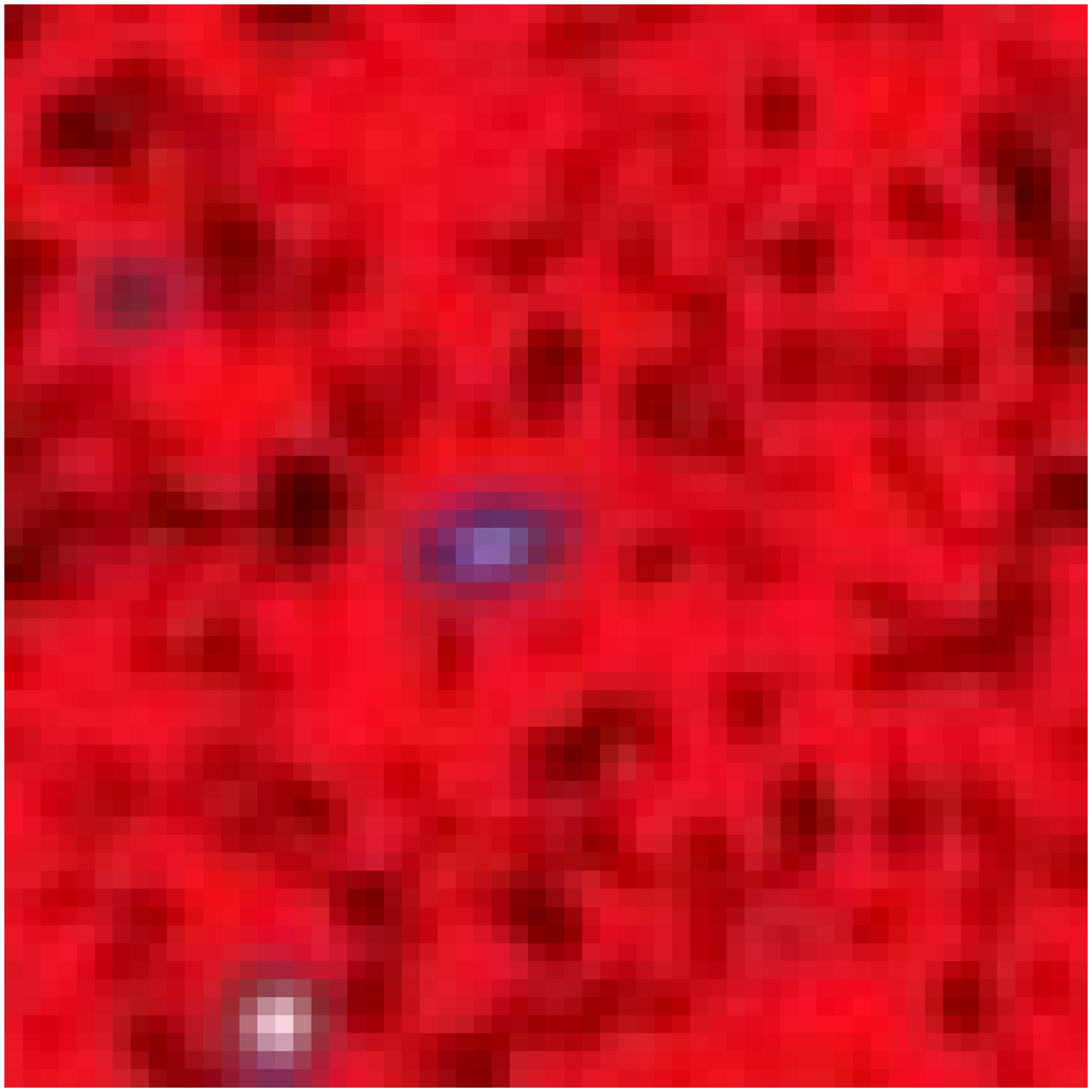}  \FGb{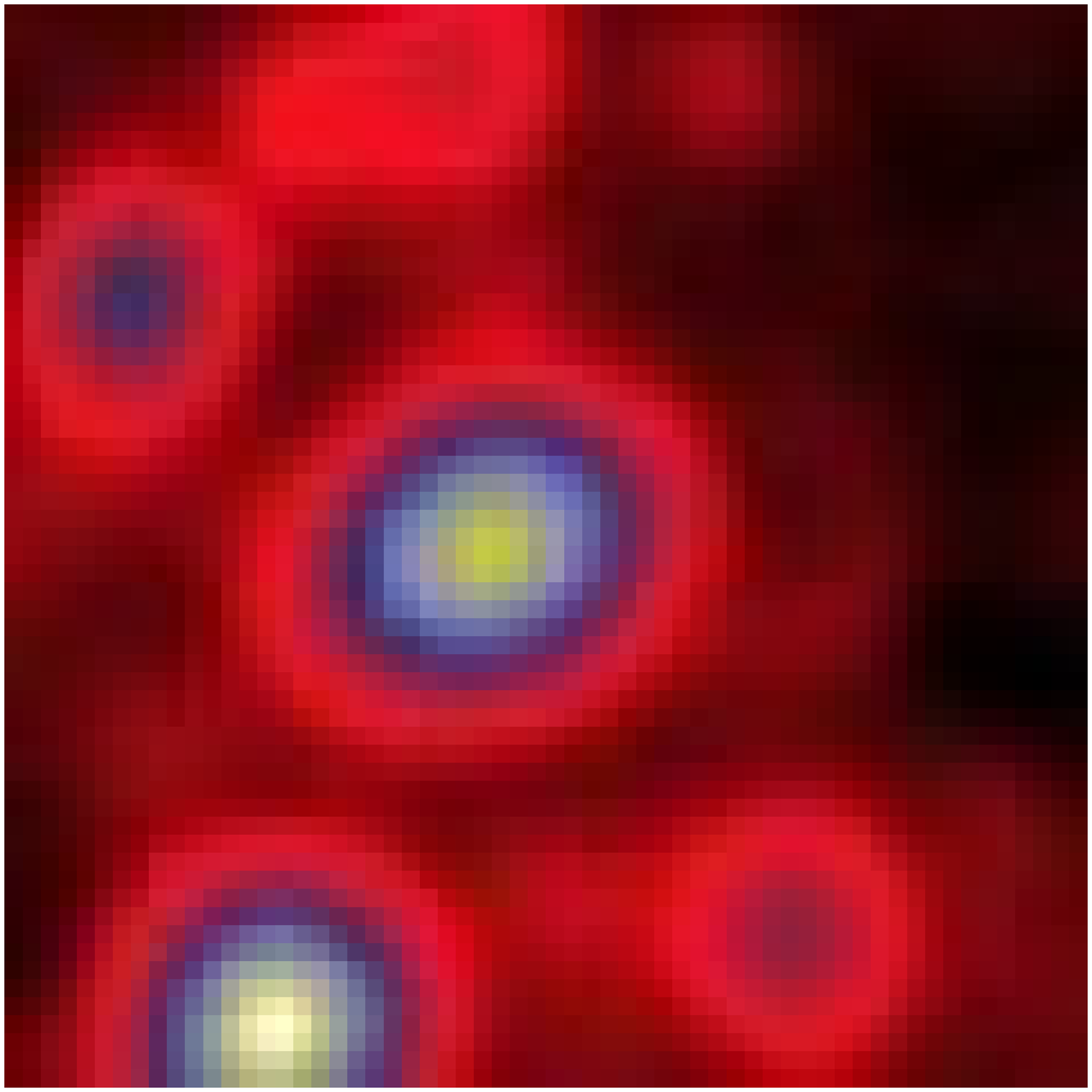}  \FGb{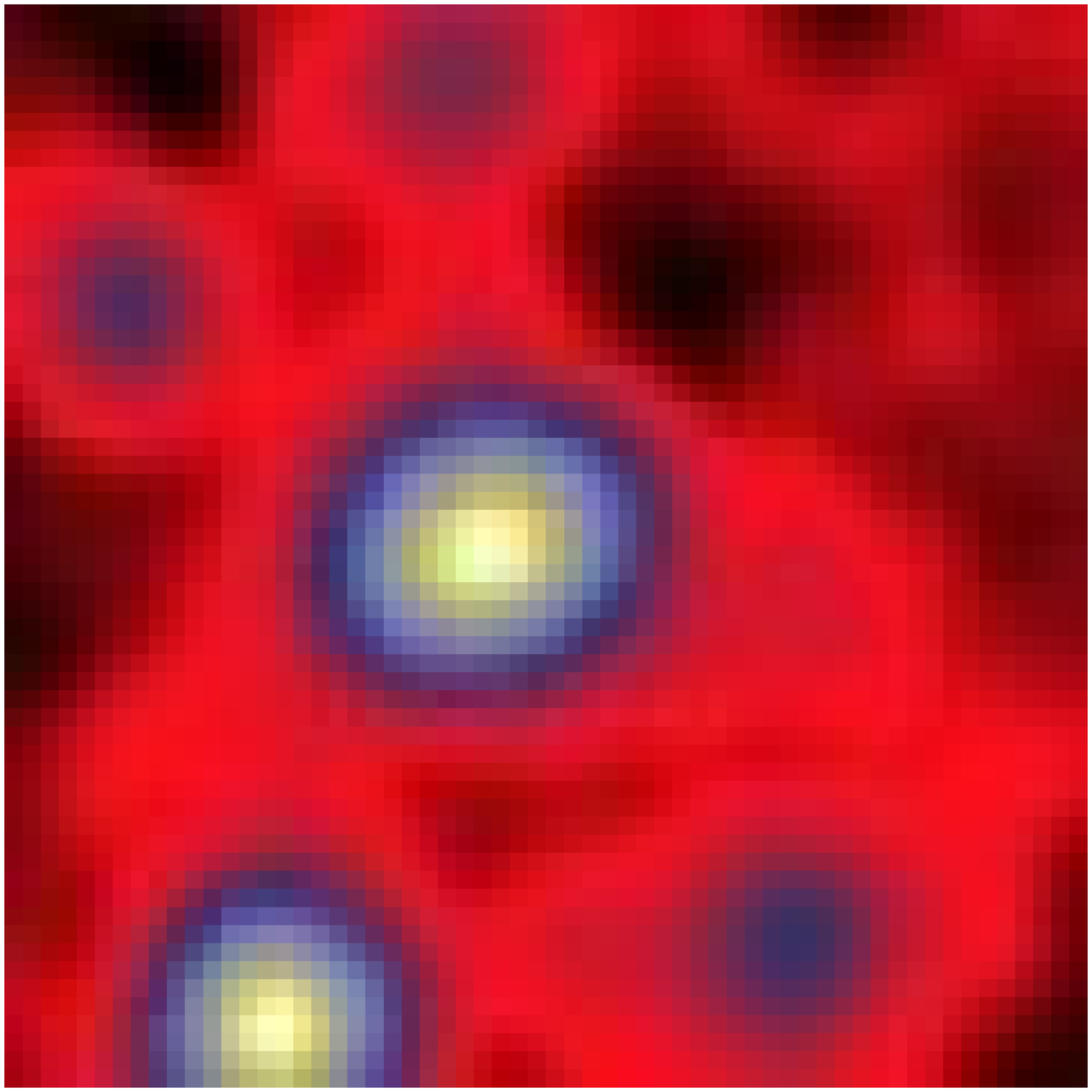}  \FGb{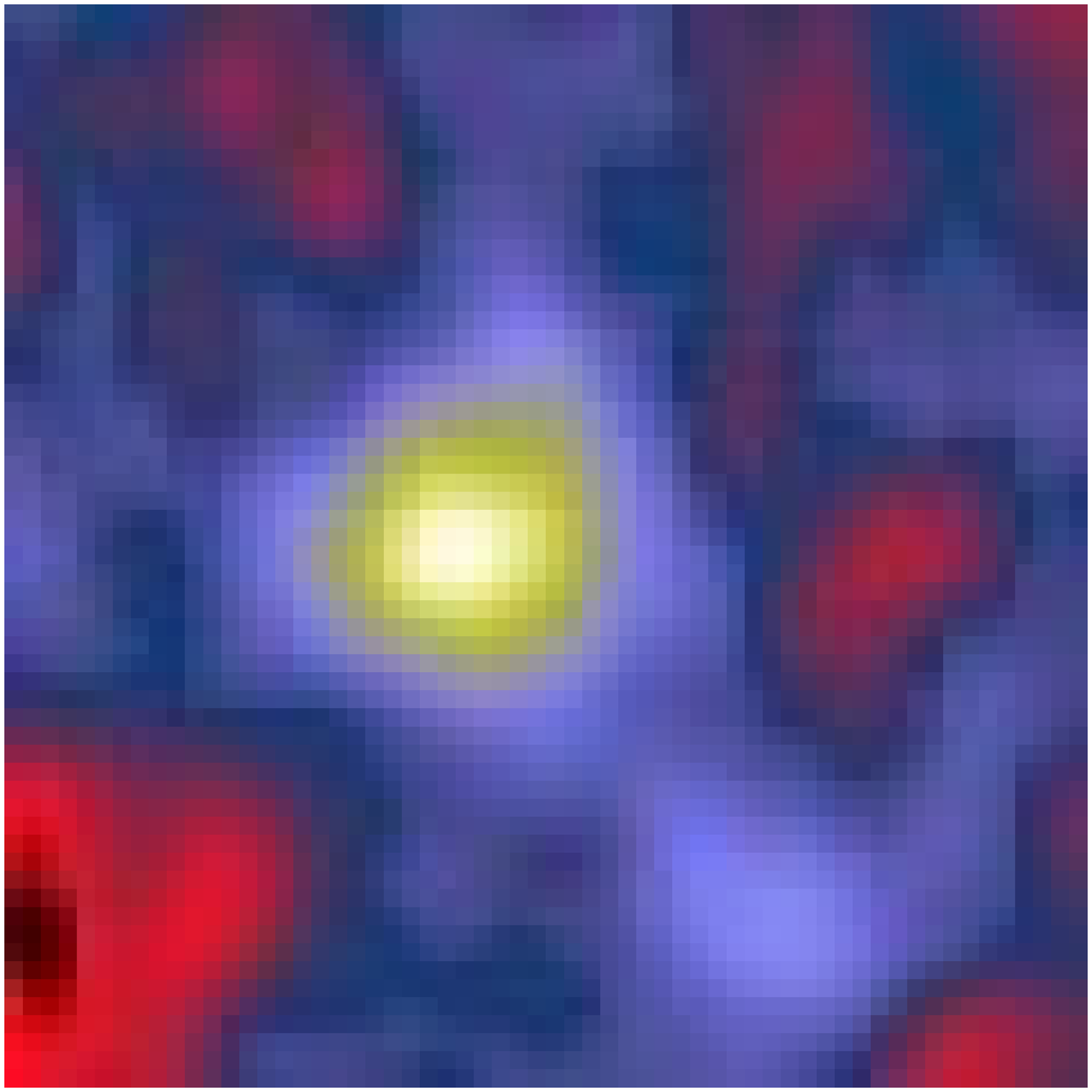}  \FGb{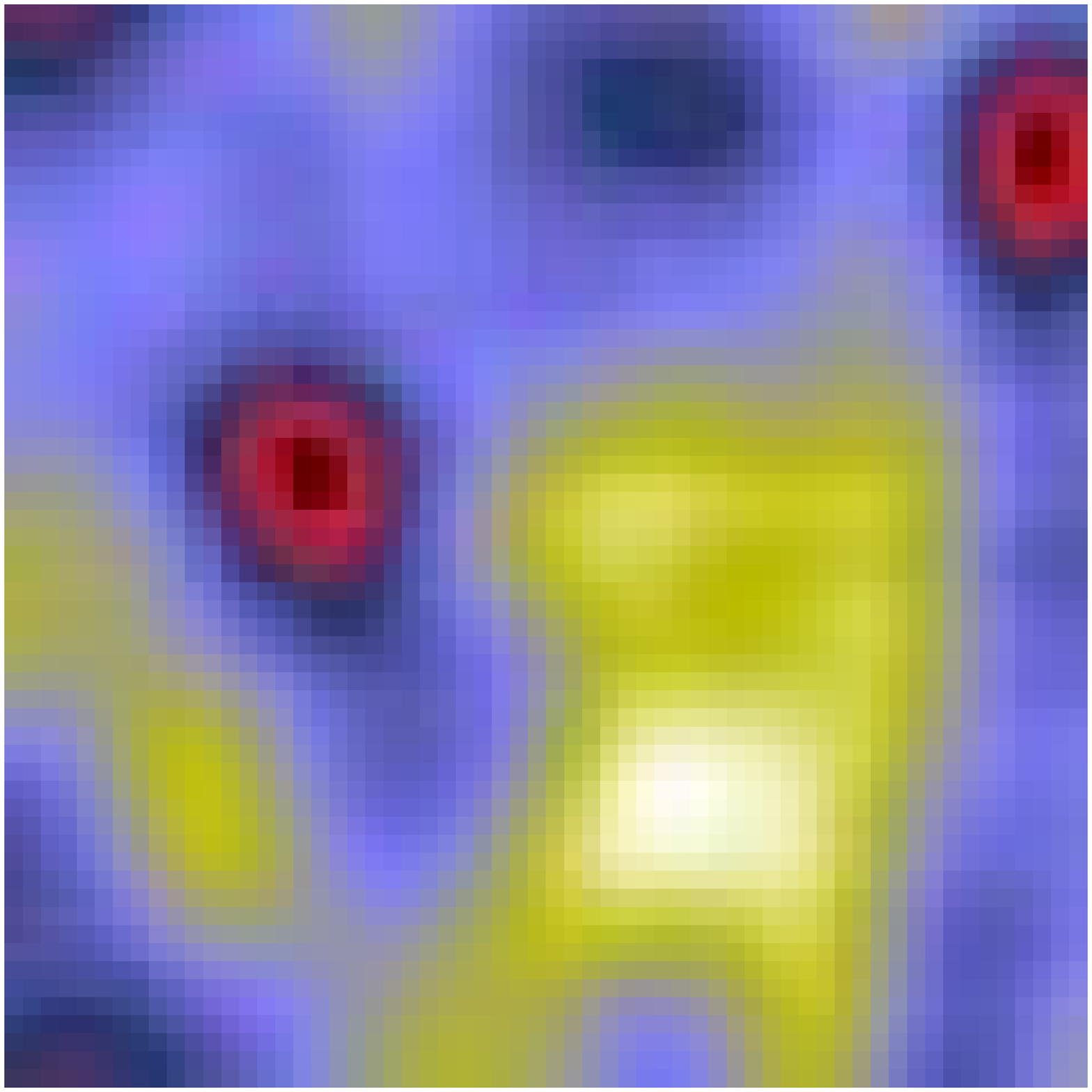} \\  {\#27   NCIA019867}  \\
  \FGb{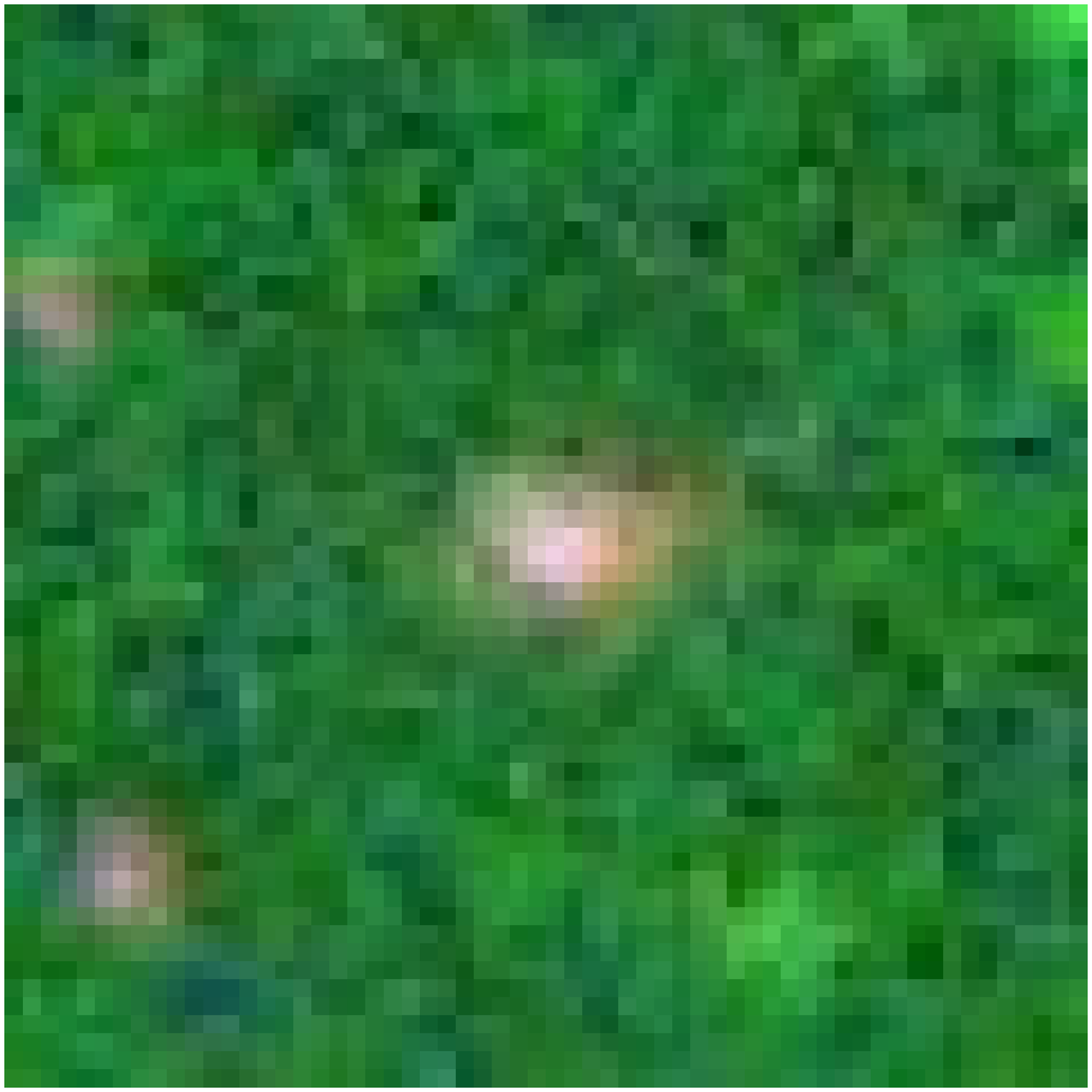}  \FGb{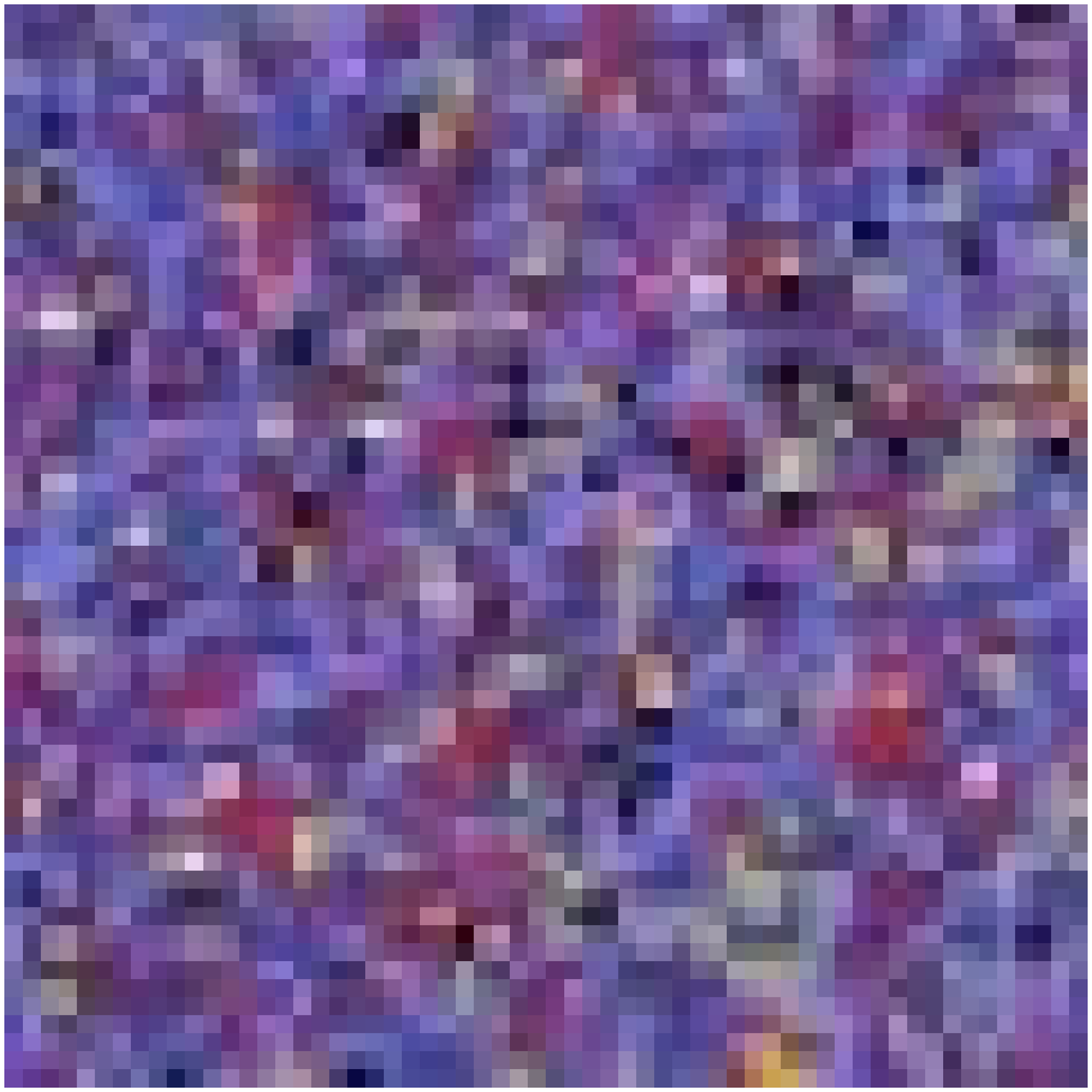}  \FGb{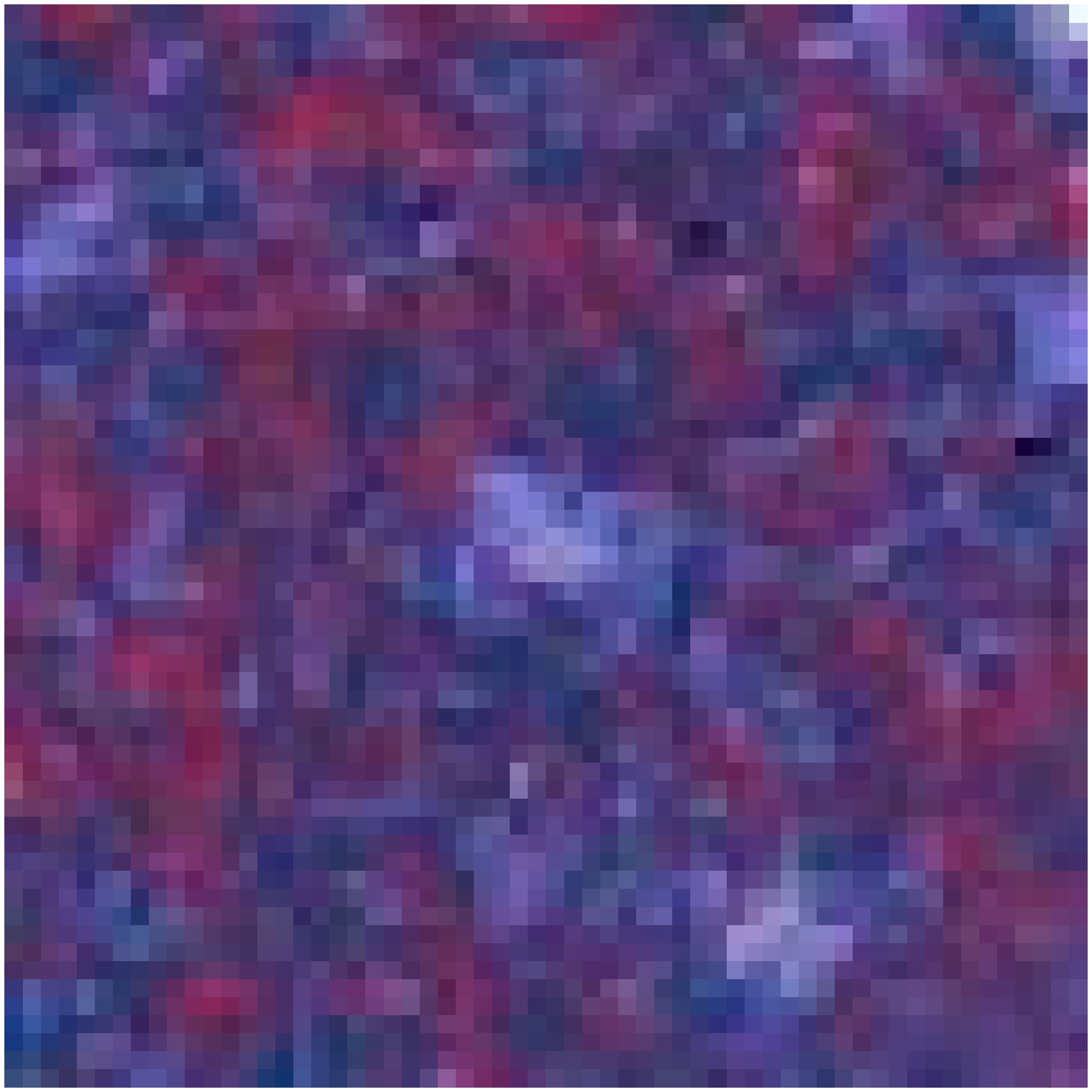}  \FGb{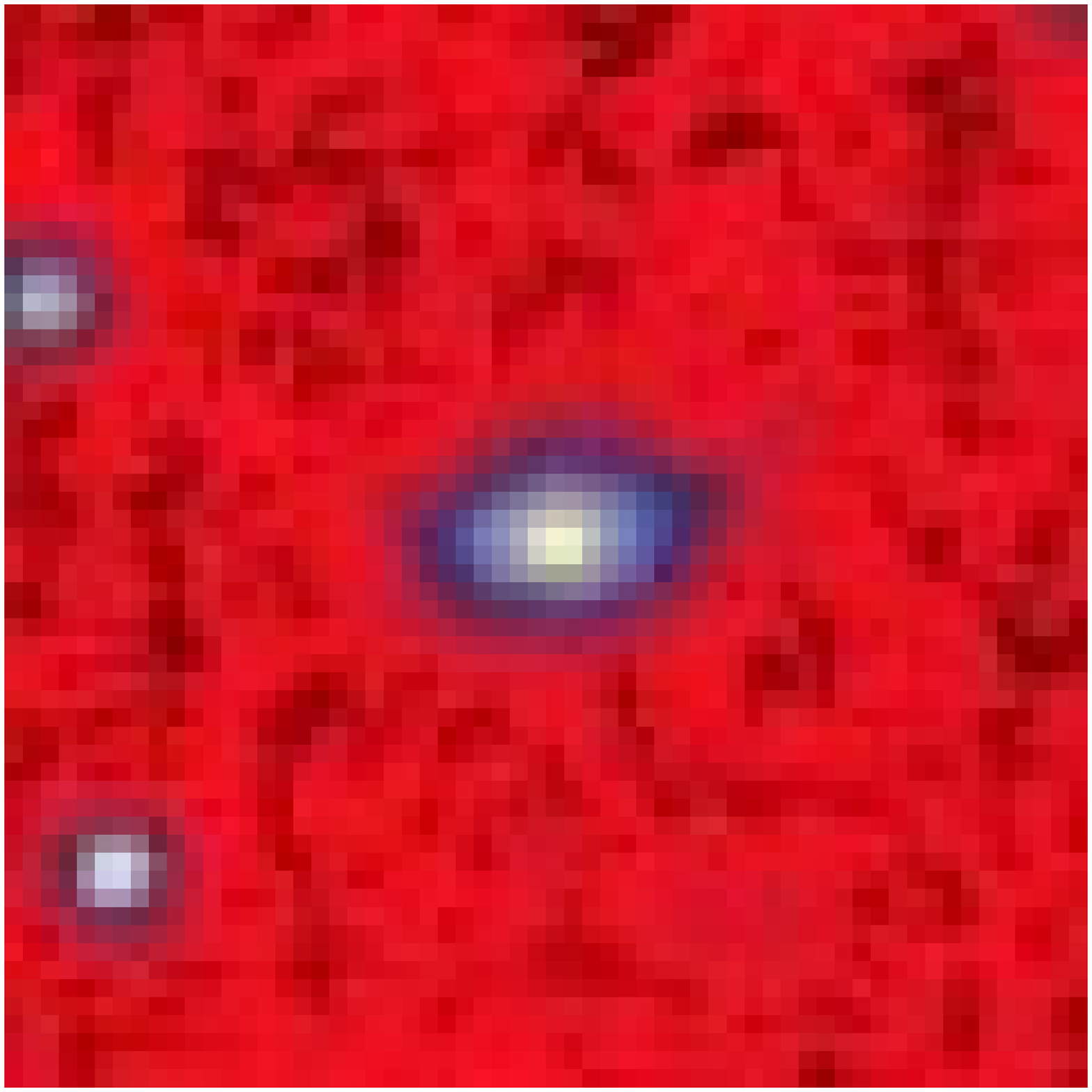}  \FGb{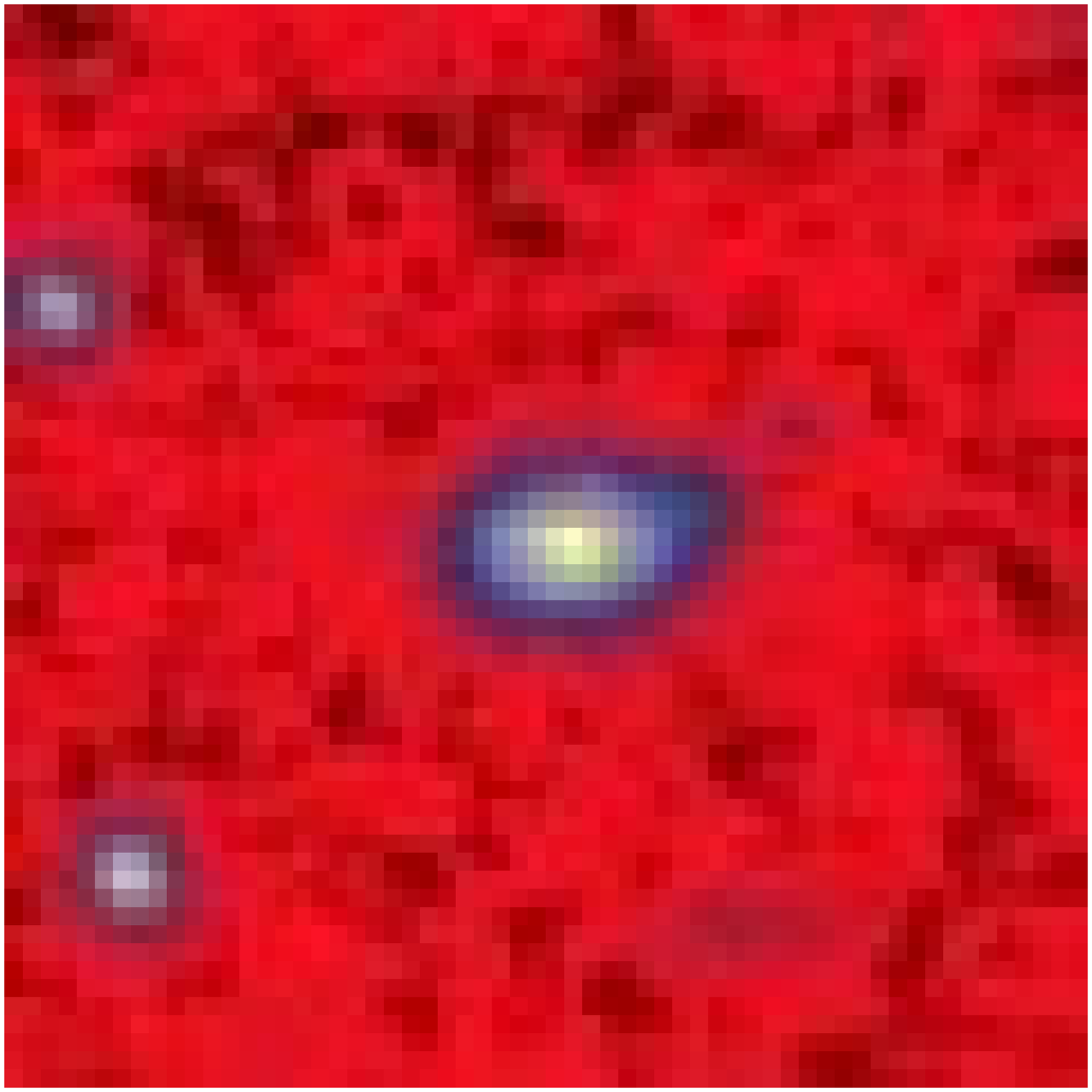}  \FGb{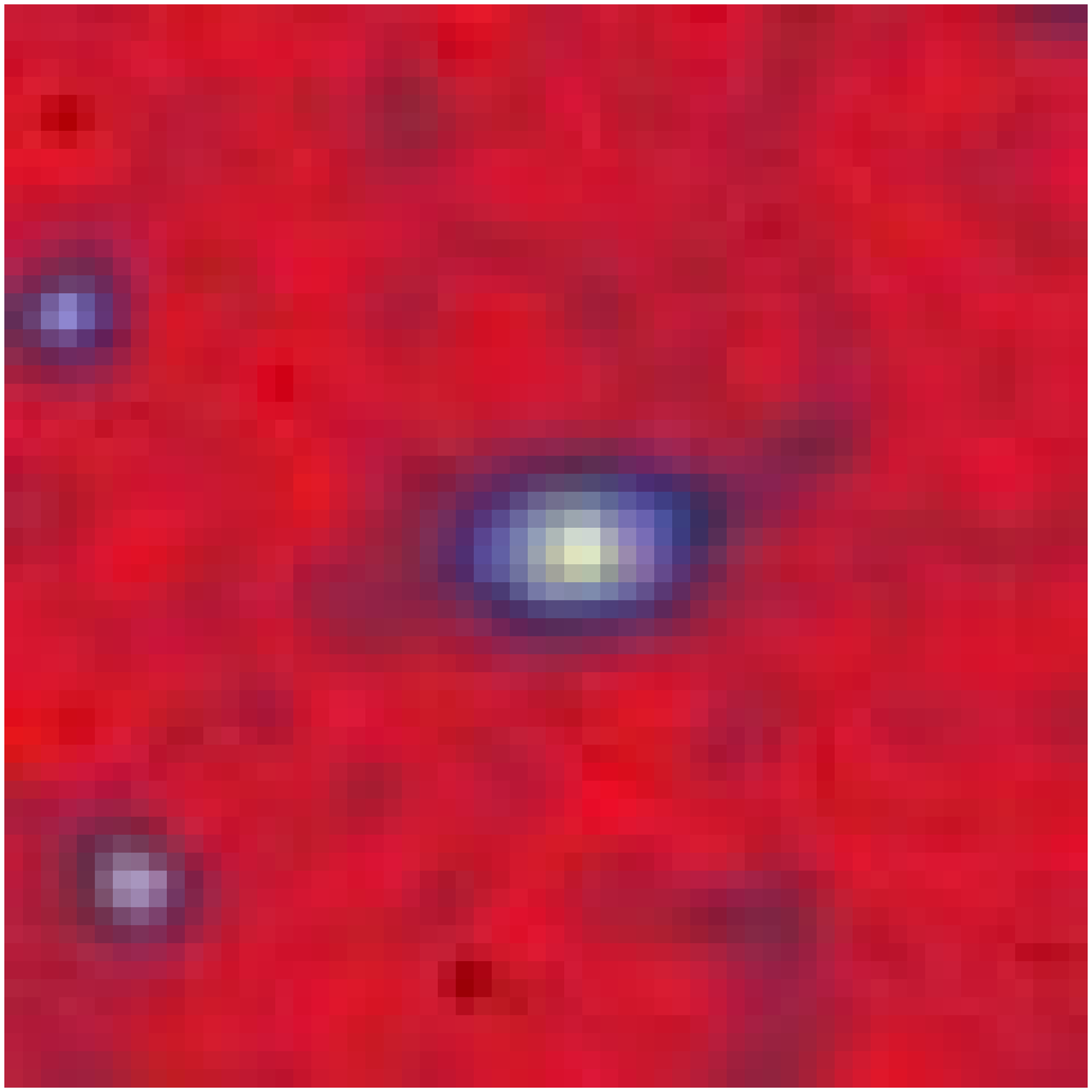}  \FGb{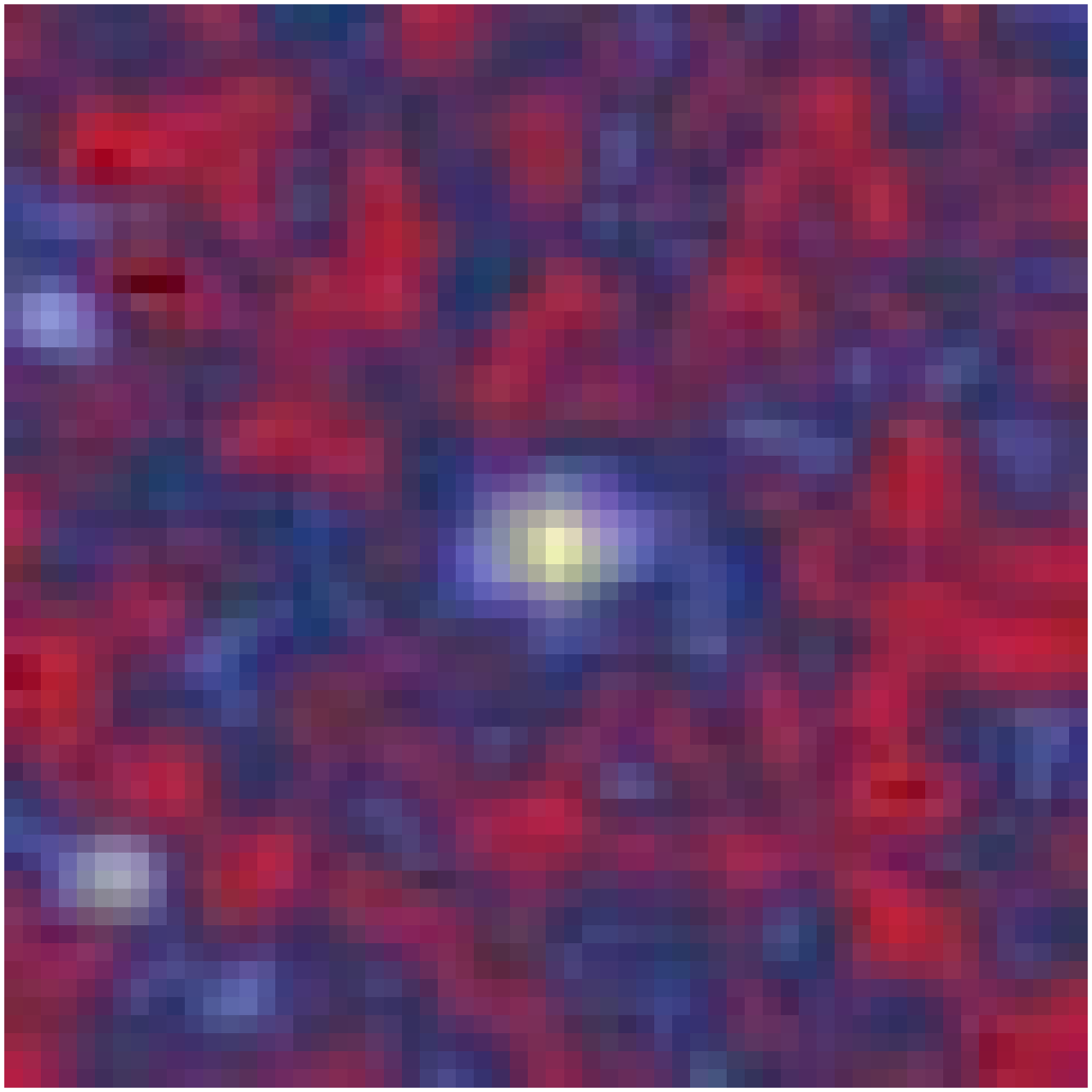}  \FGb{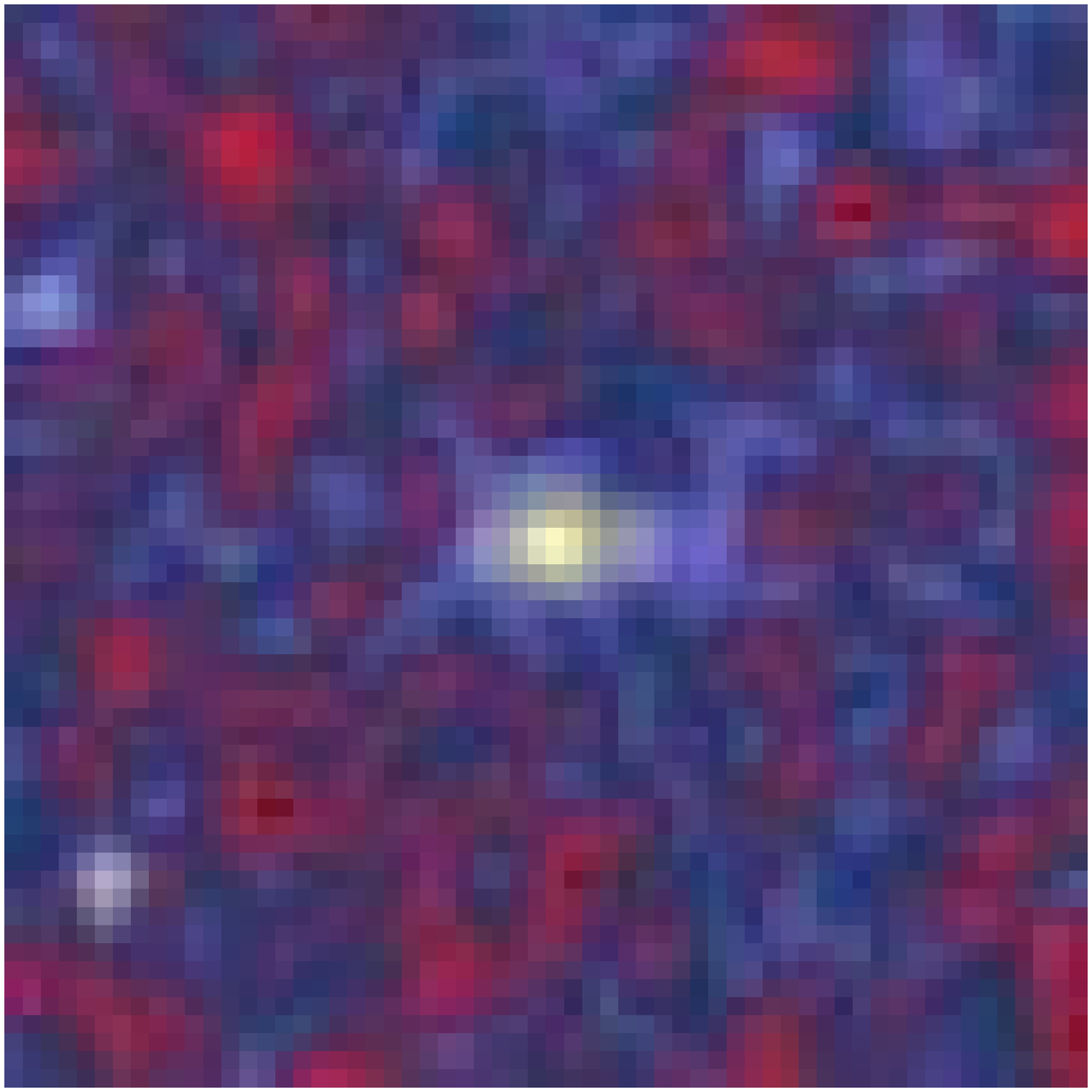}  \FGb{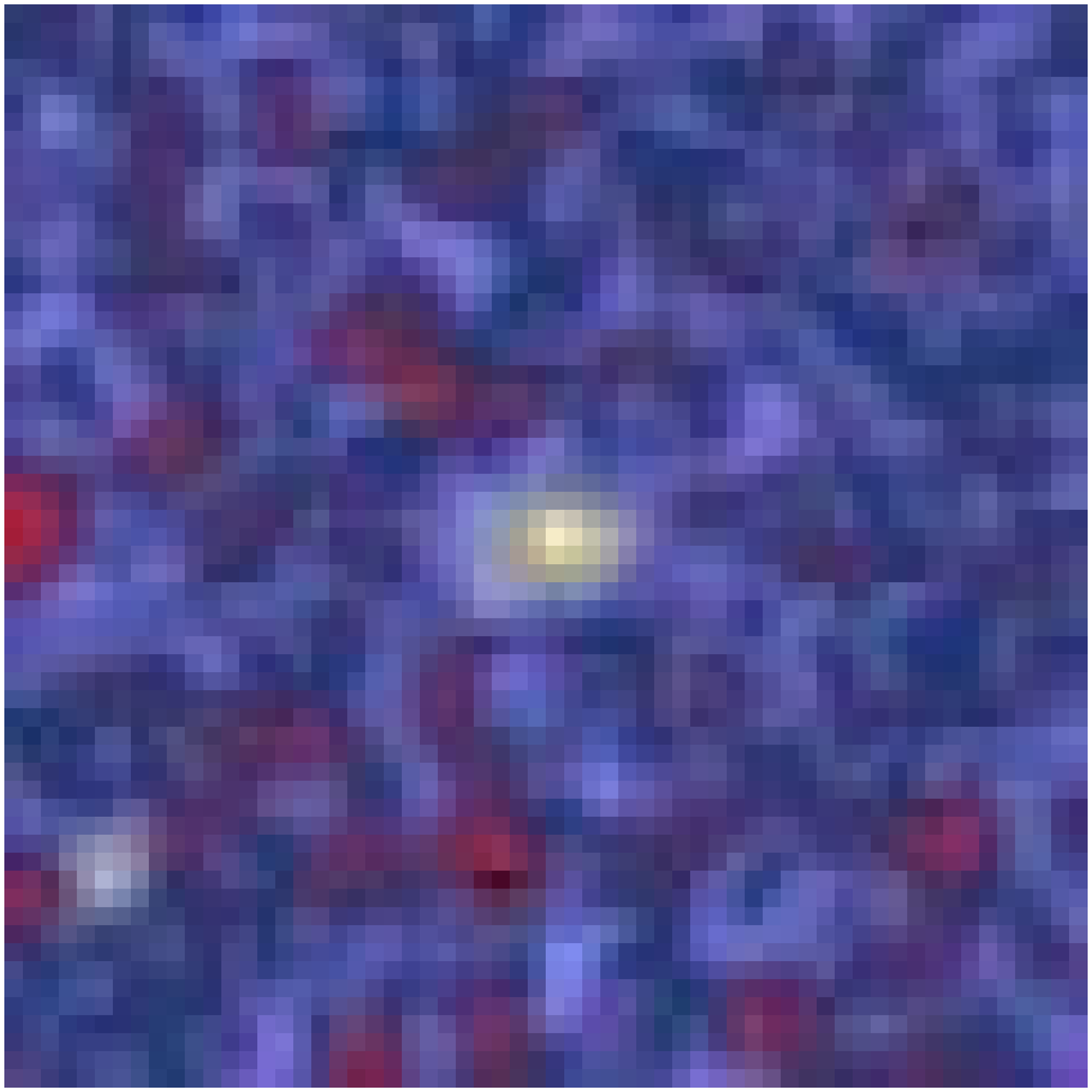}  \FGb{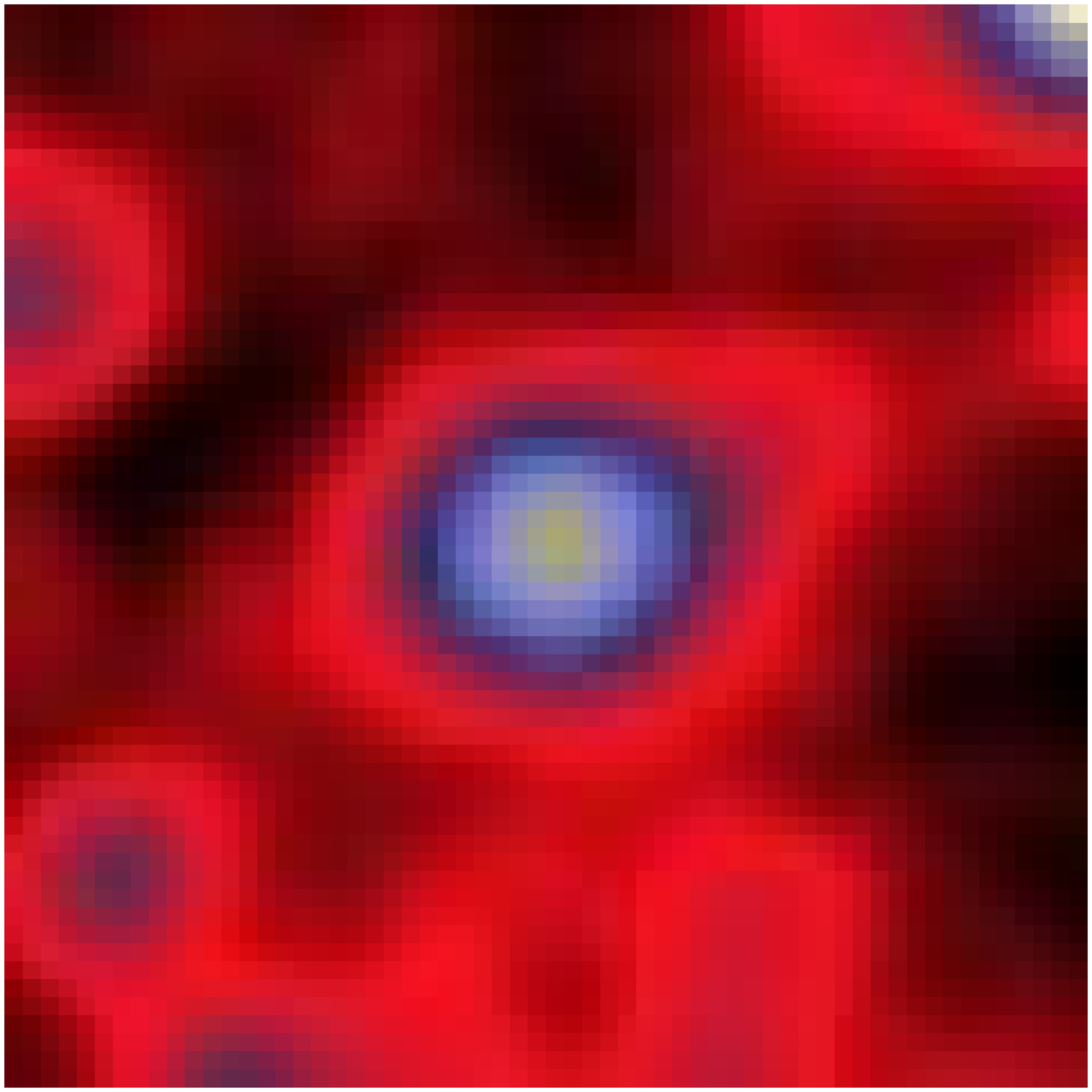}  \FGb{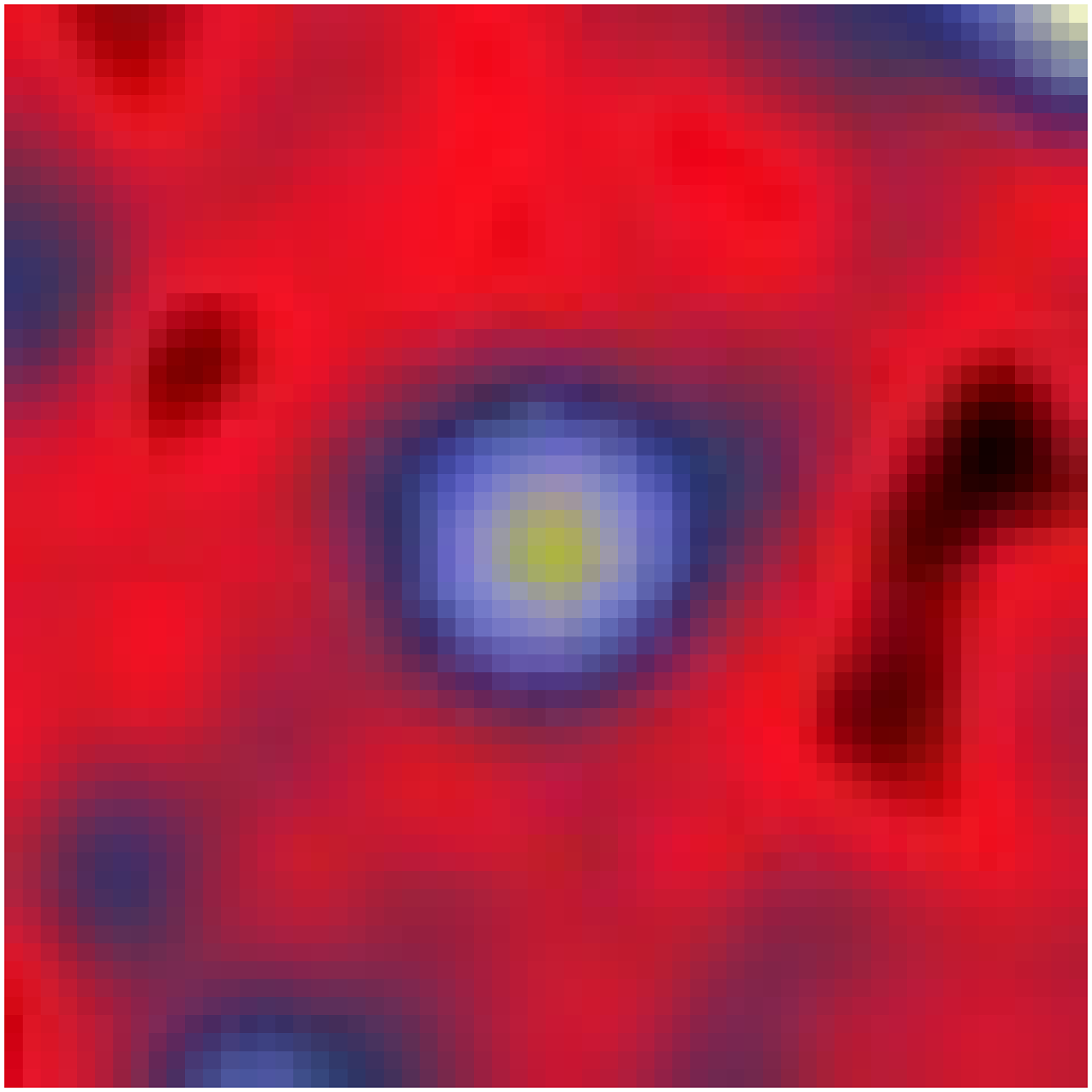}  \FGb{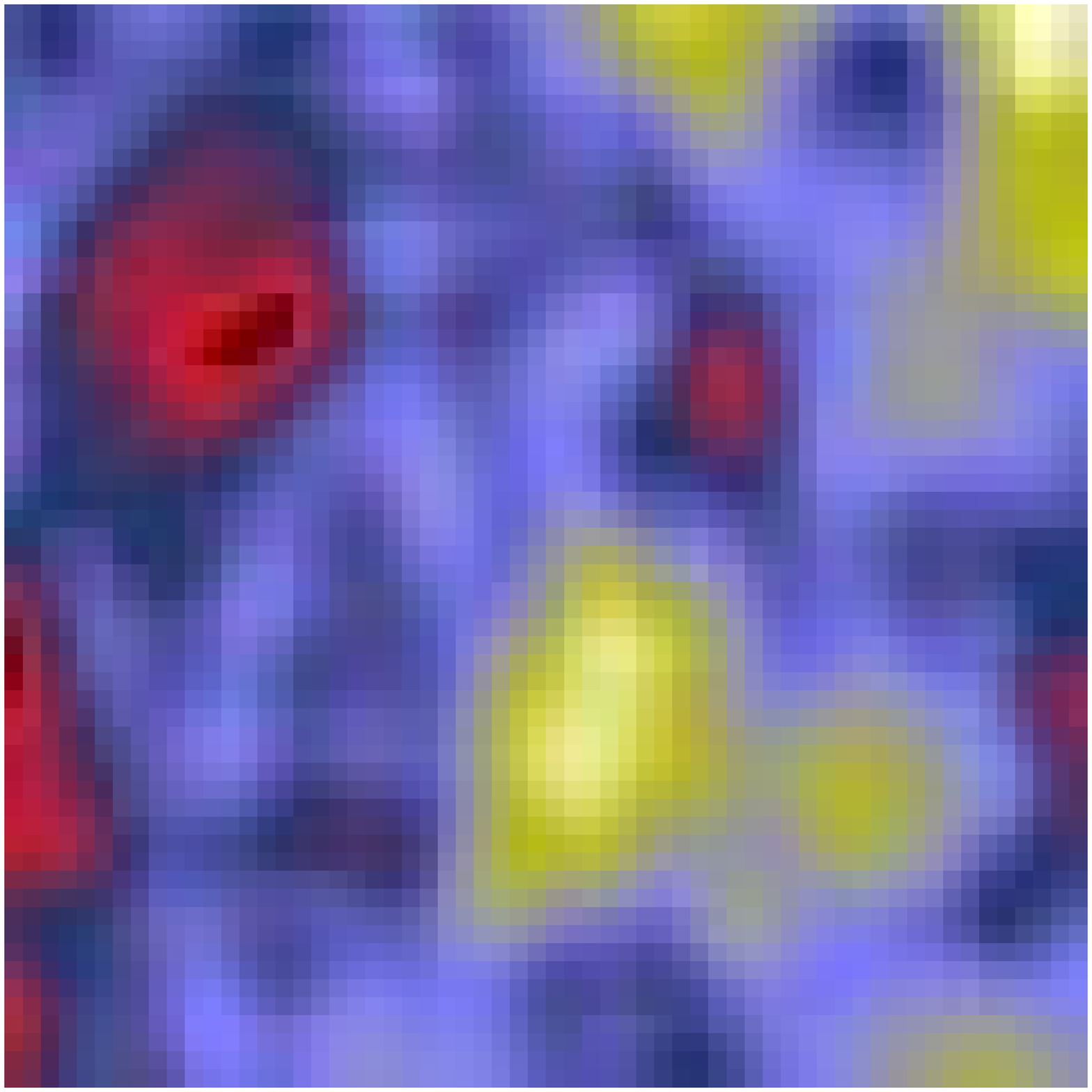}  \FGb{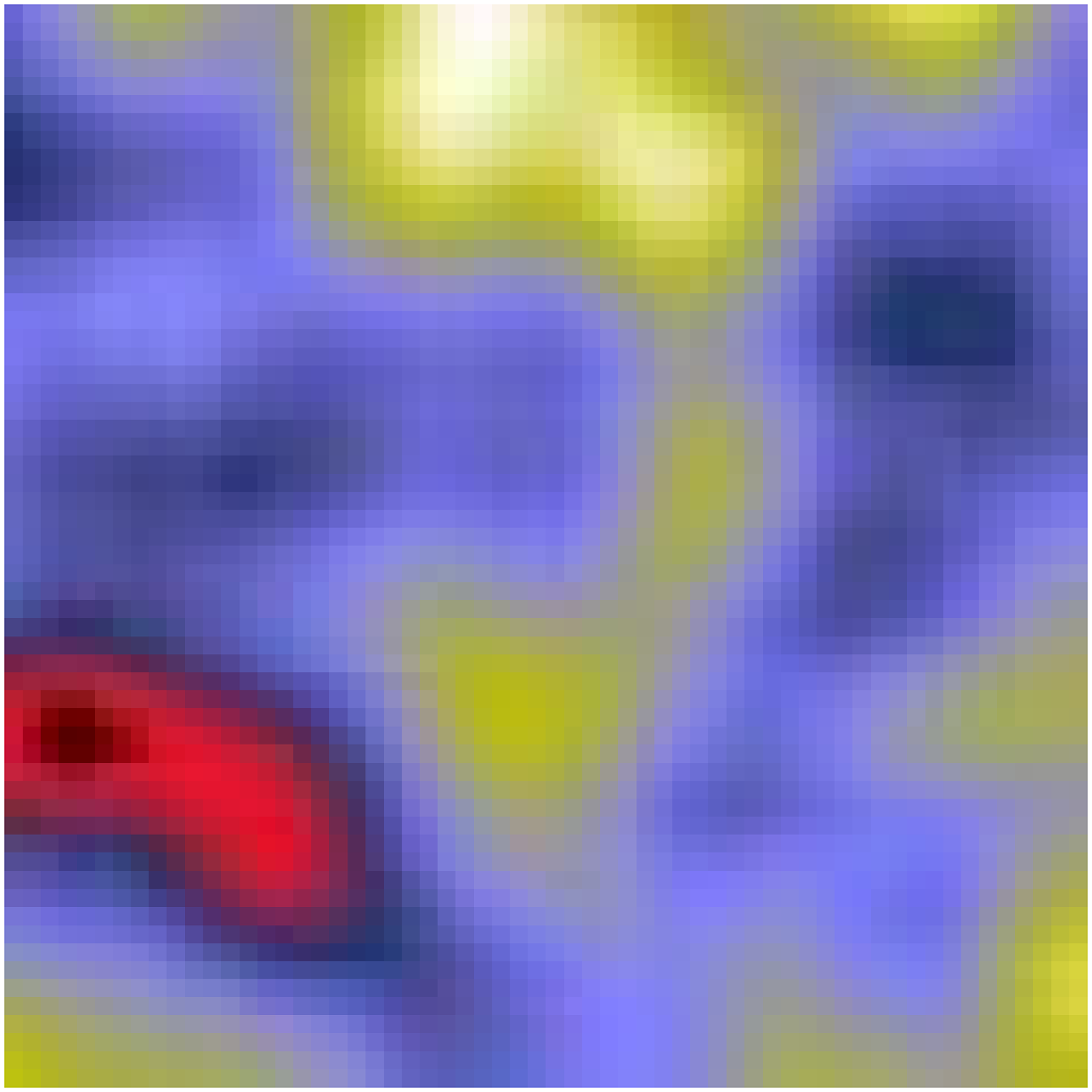} \\  {\#28   NCIA026109}  \\
  \FGb{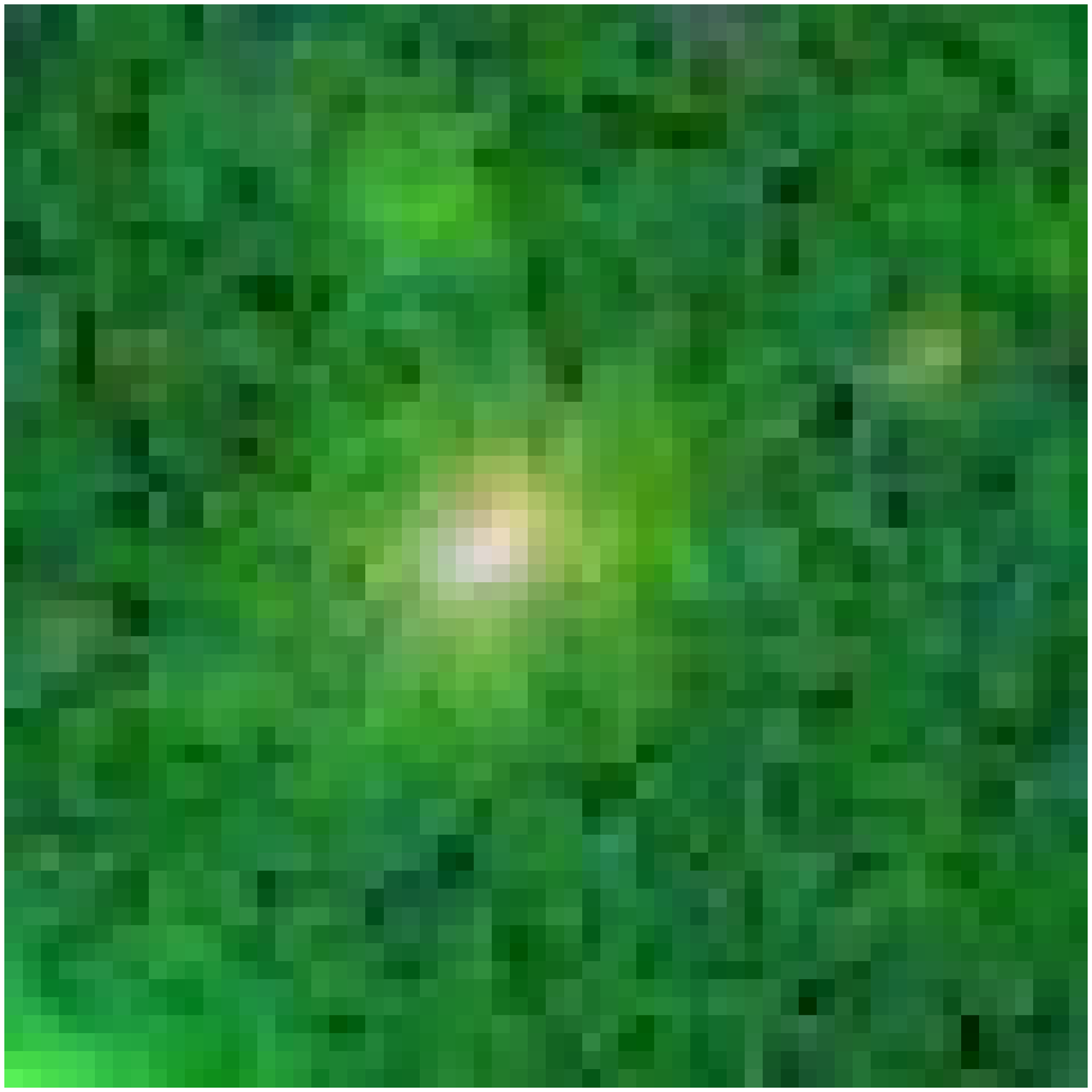}  \FGb{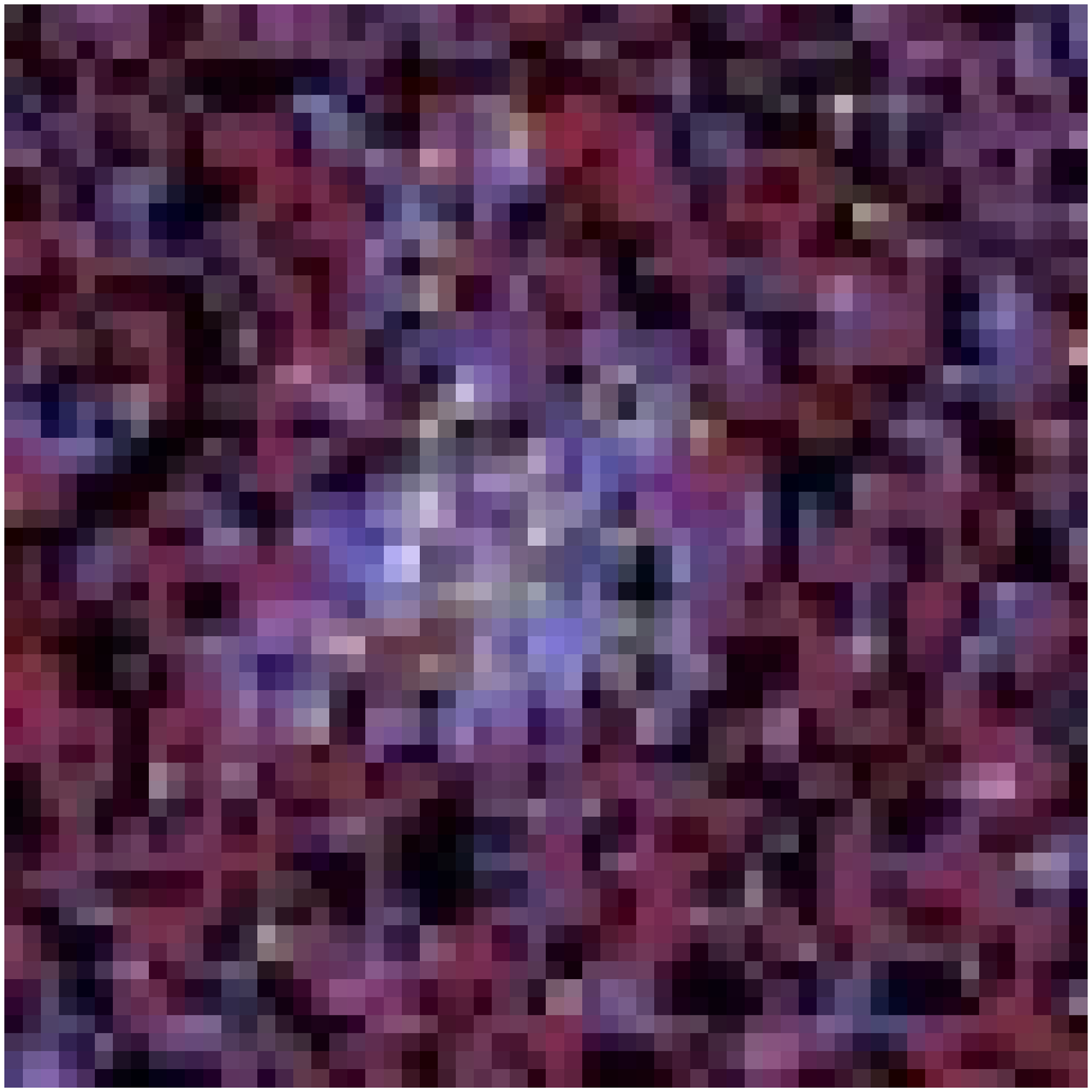}  \FGb{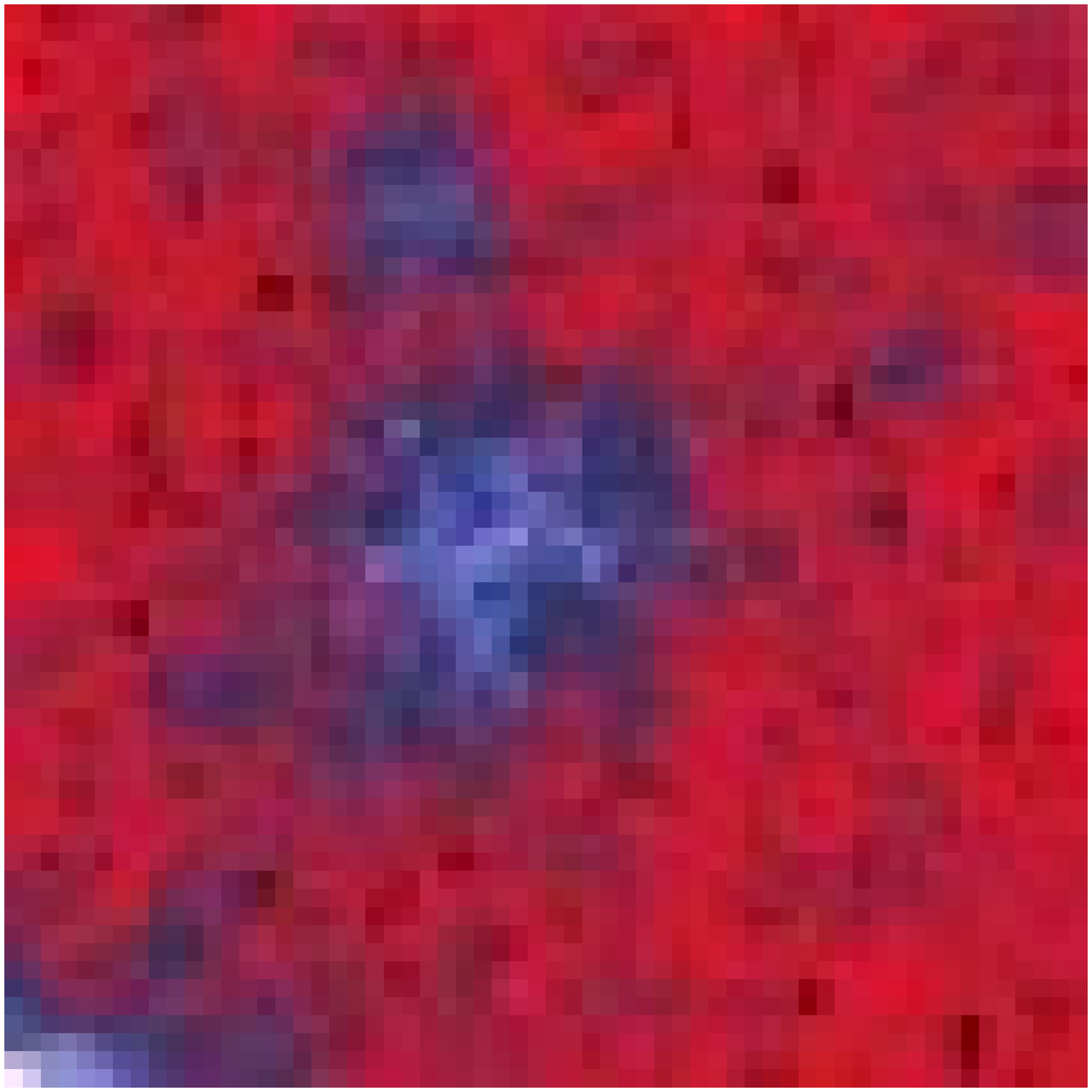}  \FGb{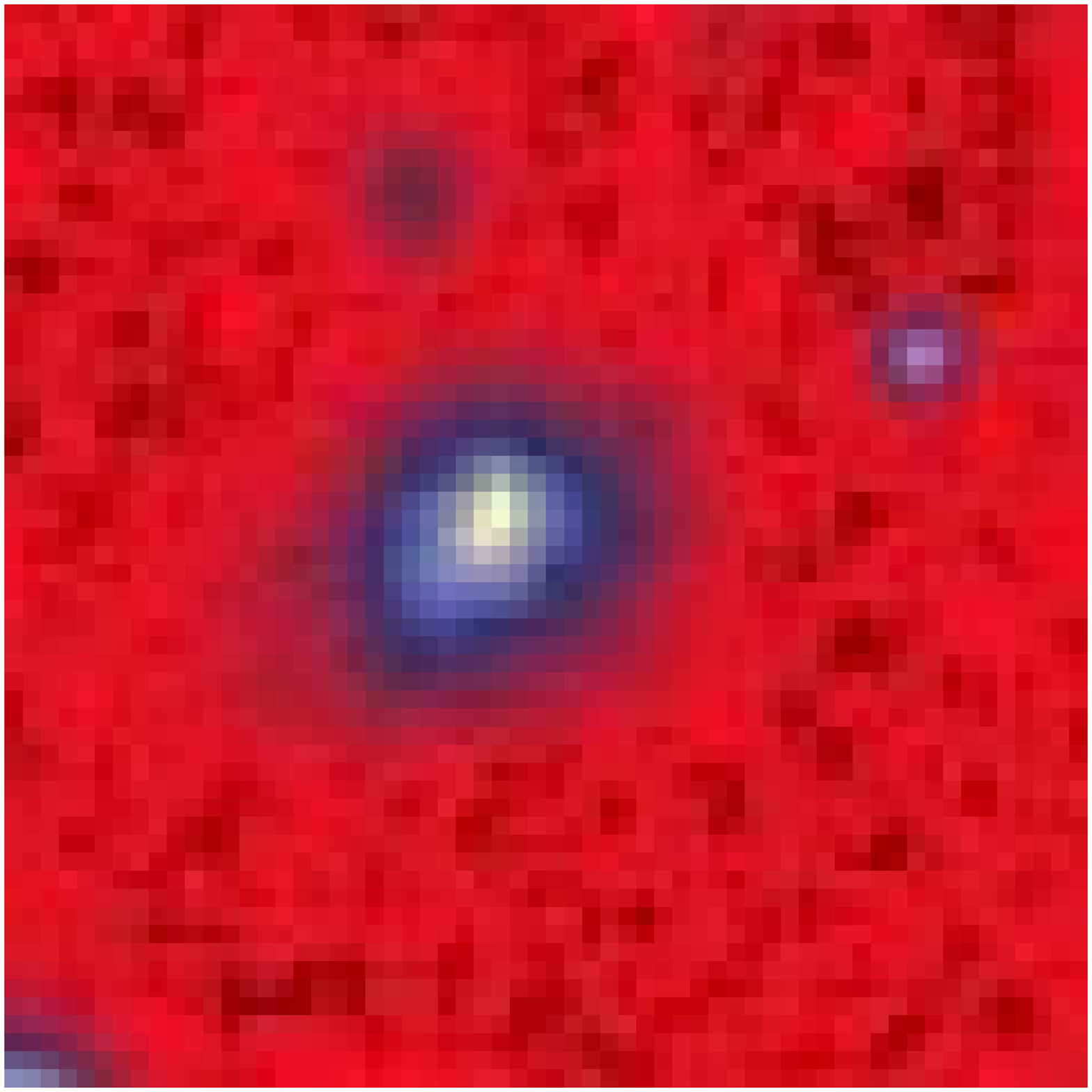}  \FGb{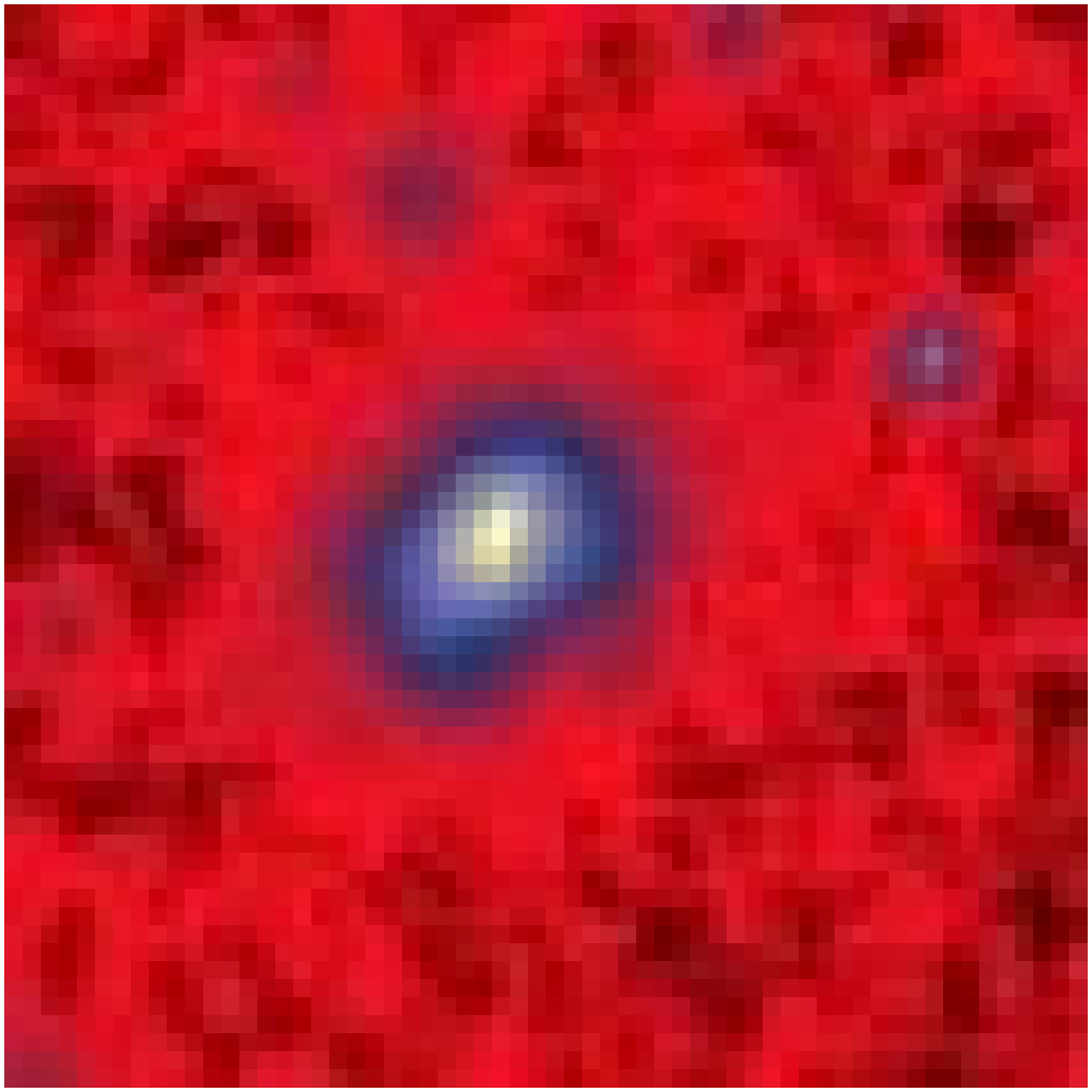}  \FGb{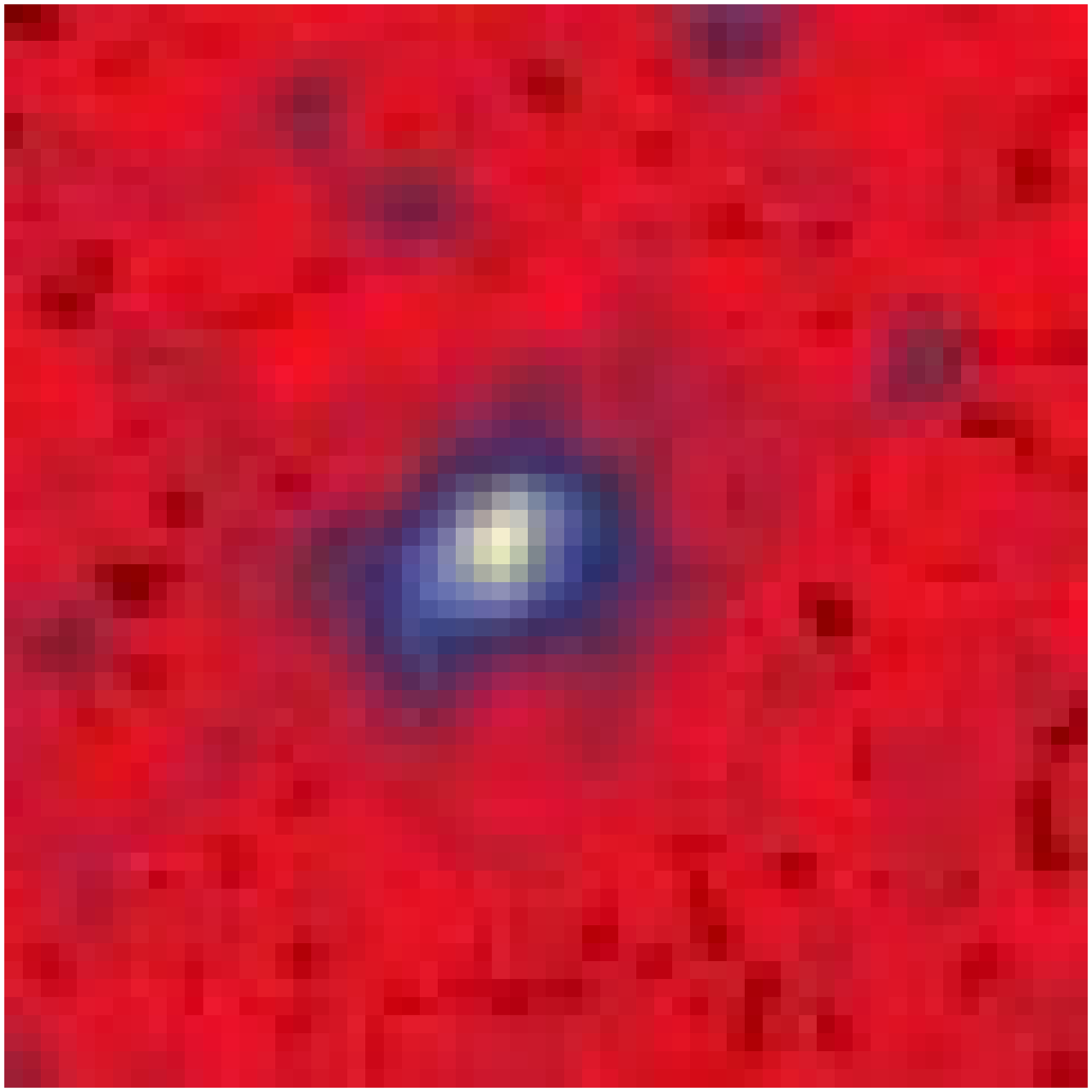}  \FGb{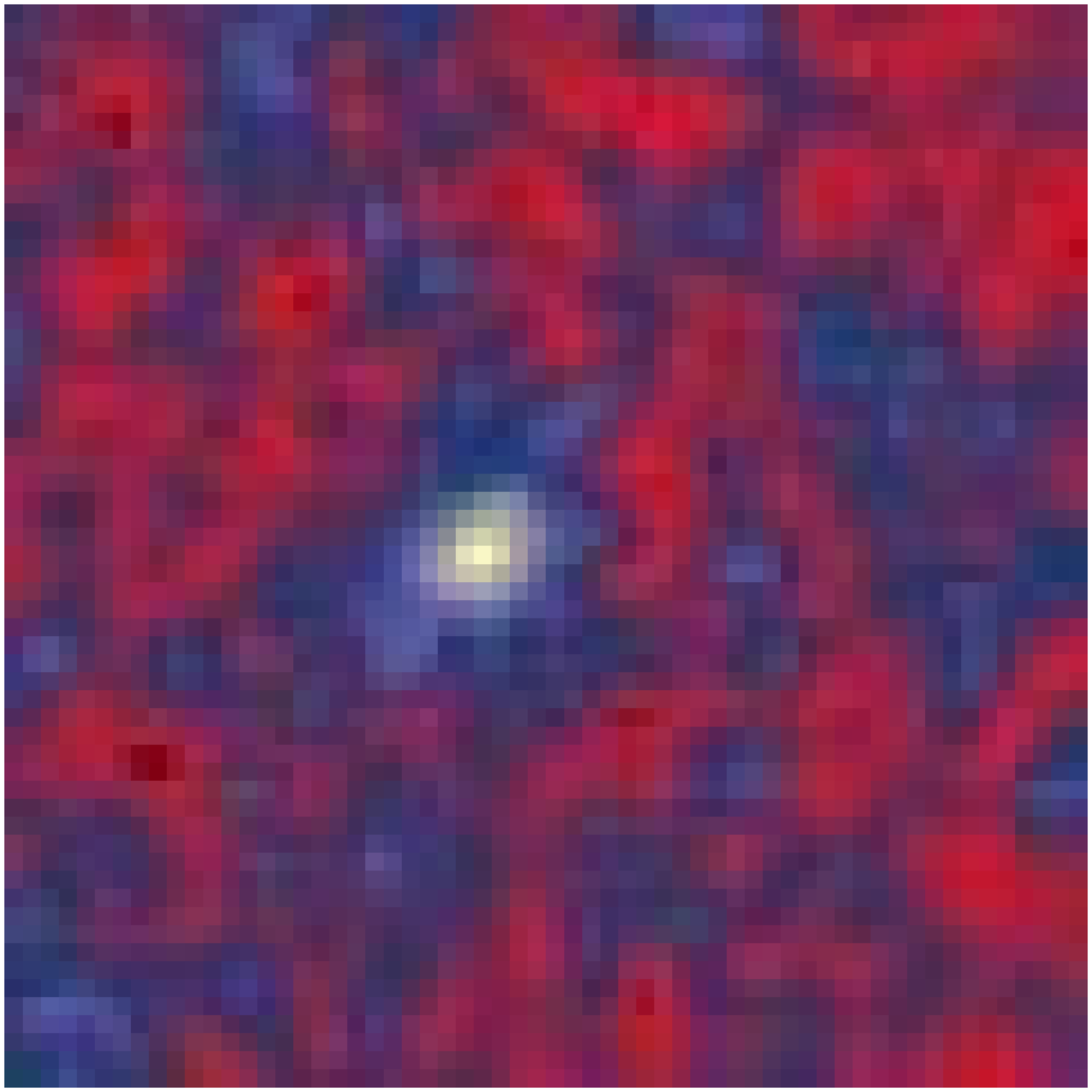}  \FGb{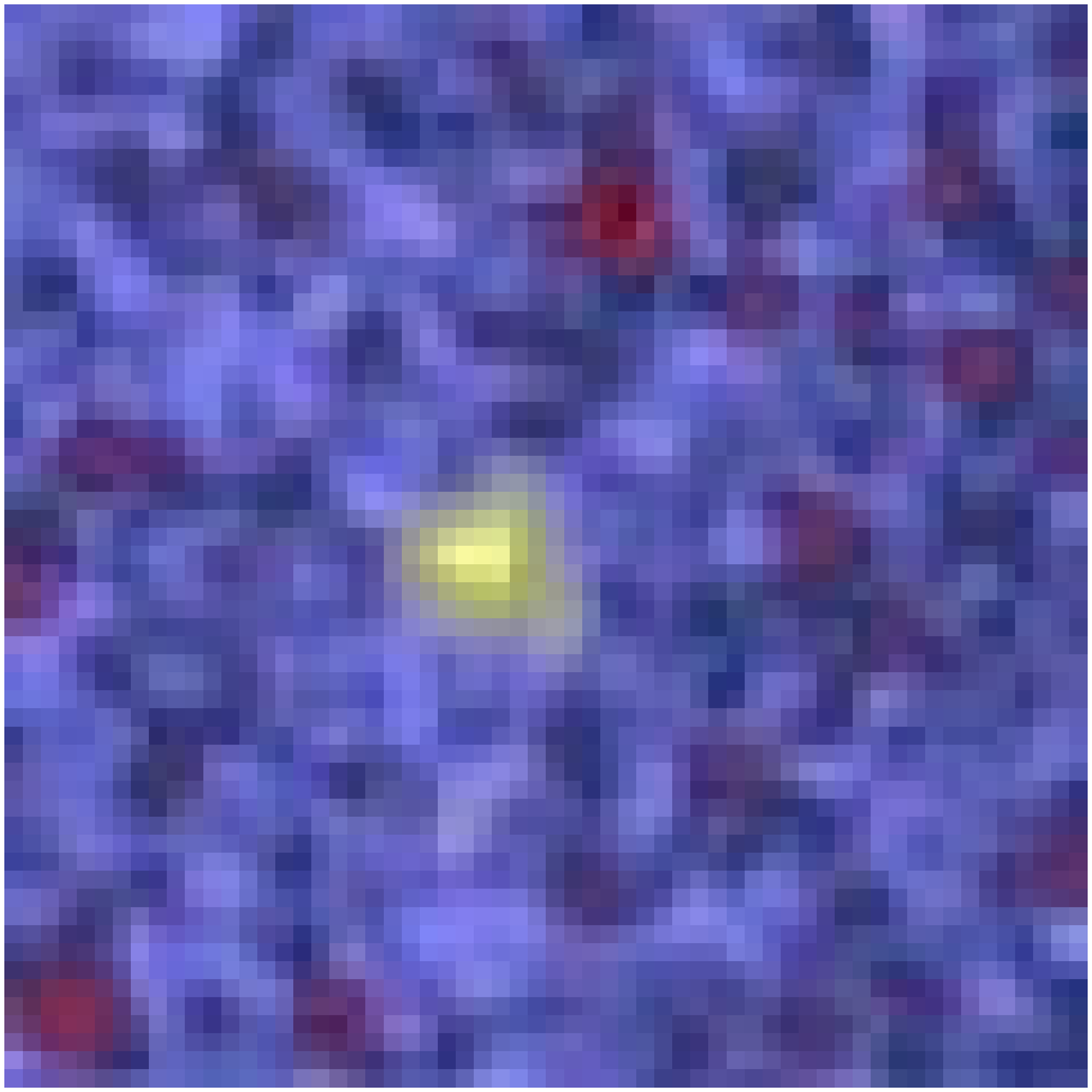}  \FGb{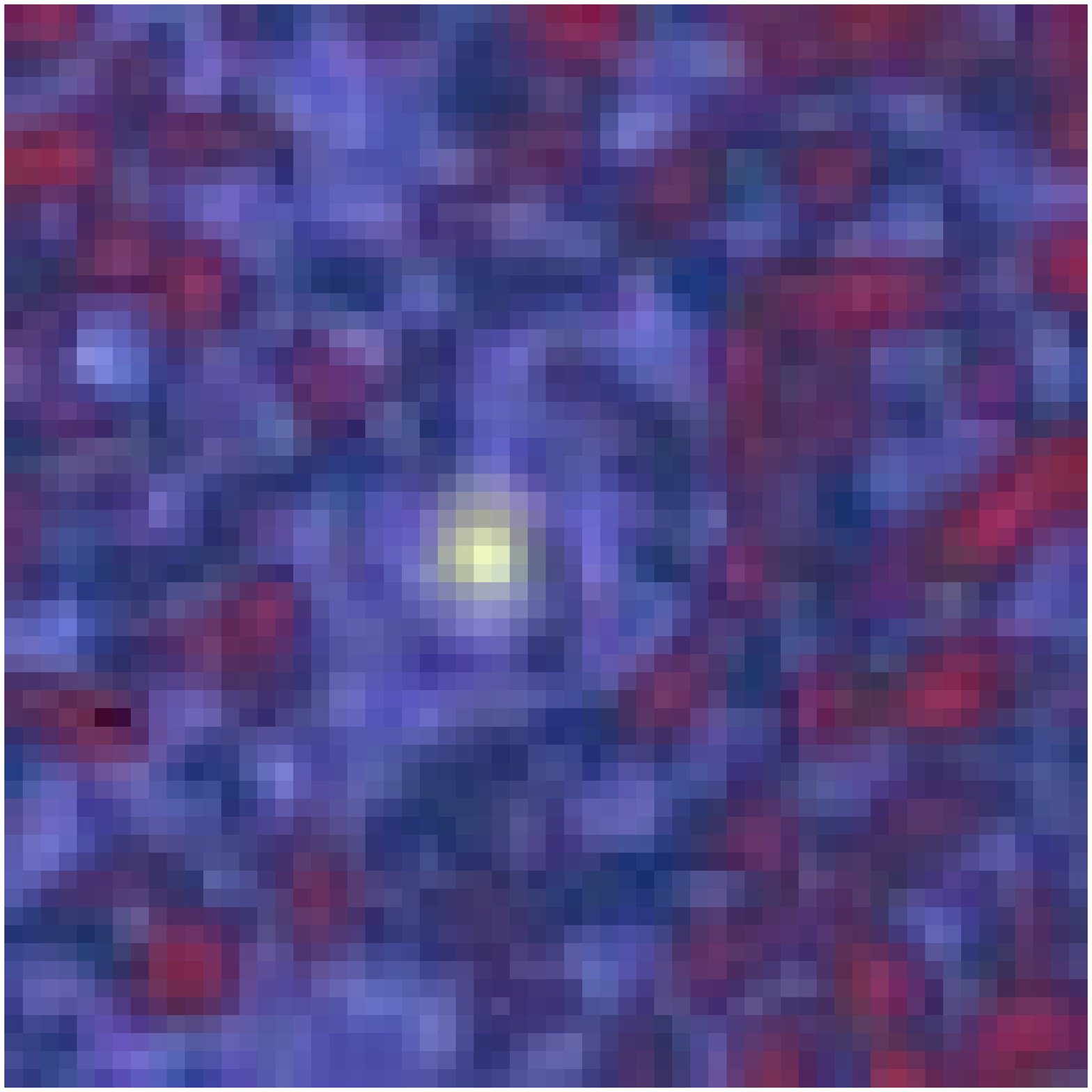}  \FGb{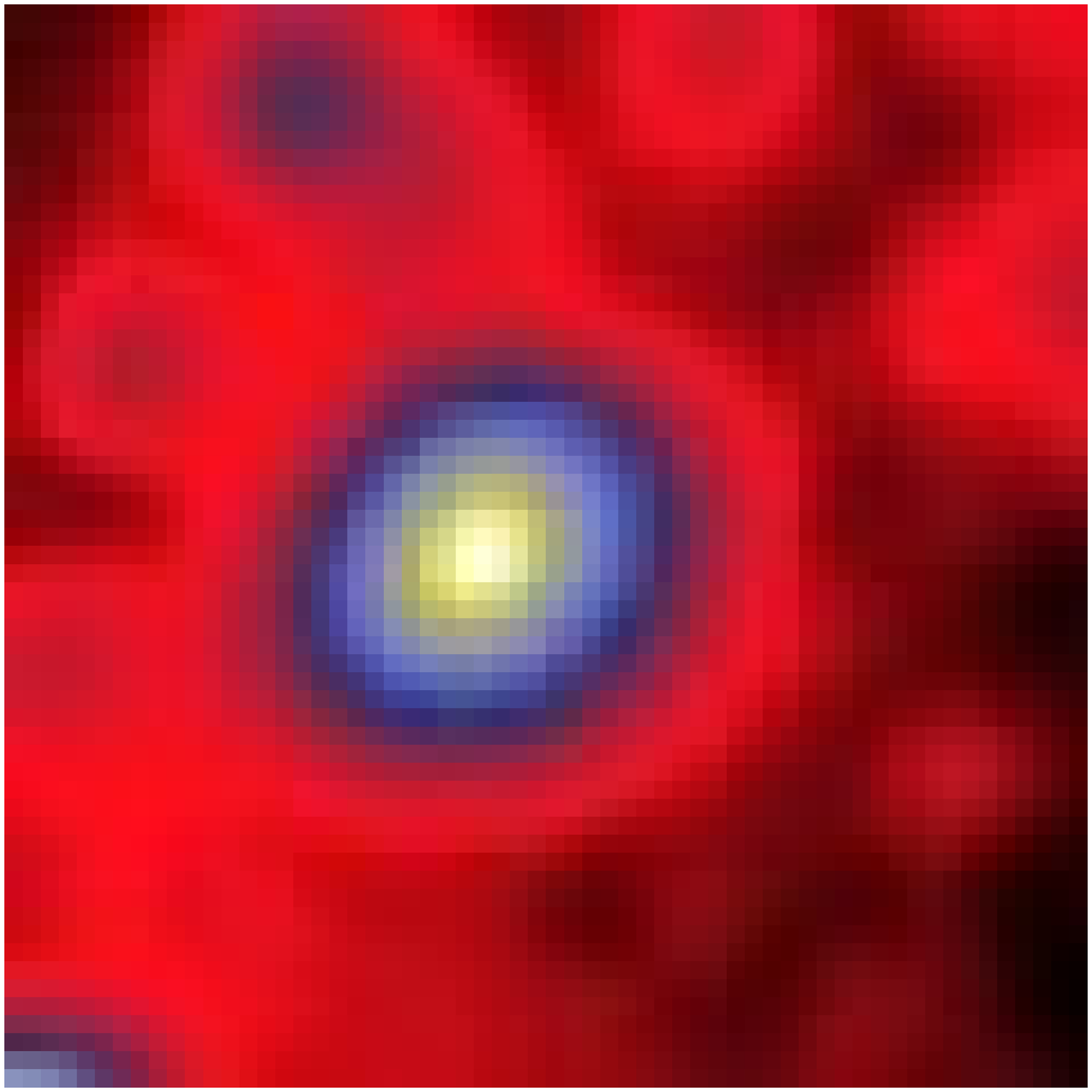}  \FGb{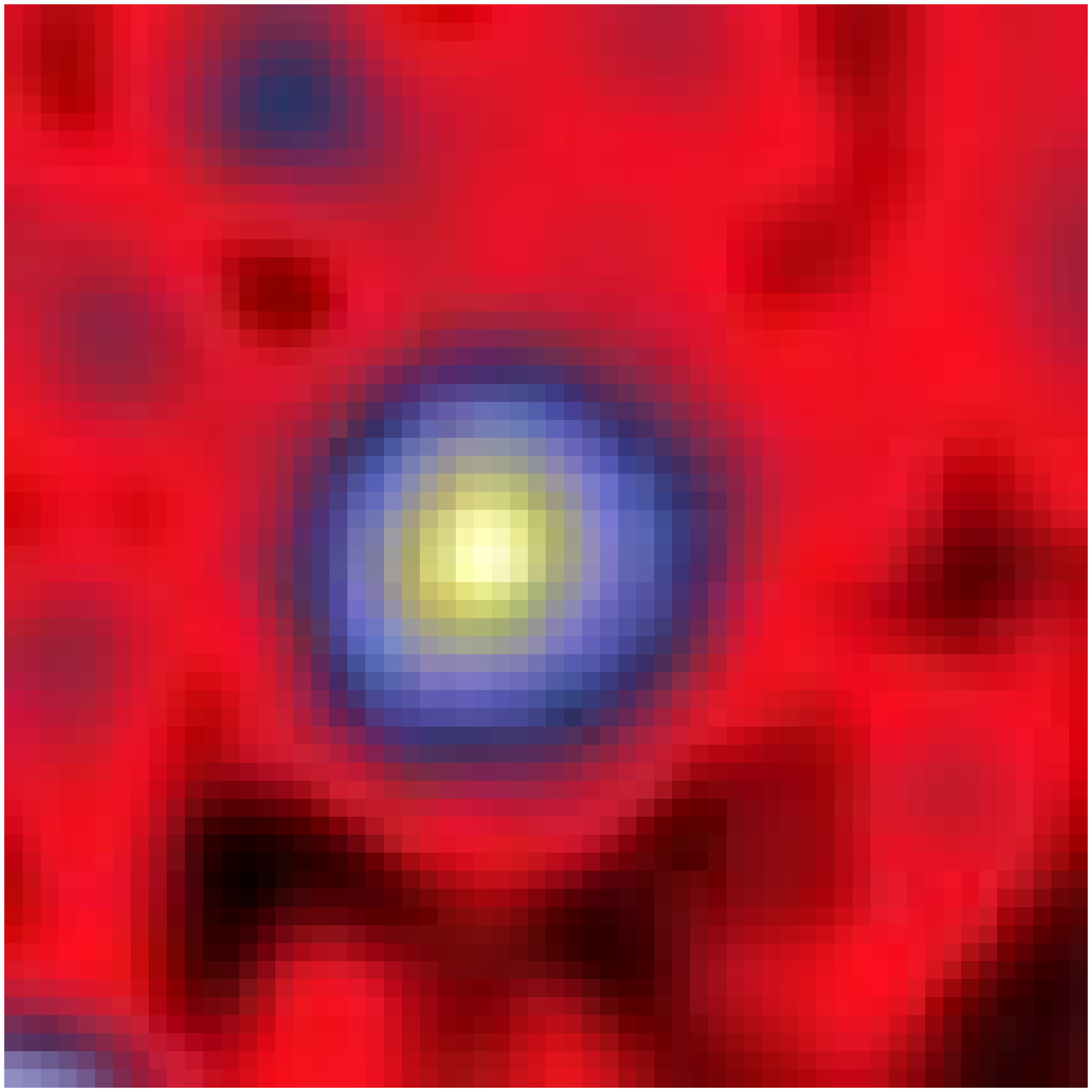}  \FGb{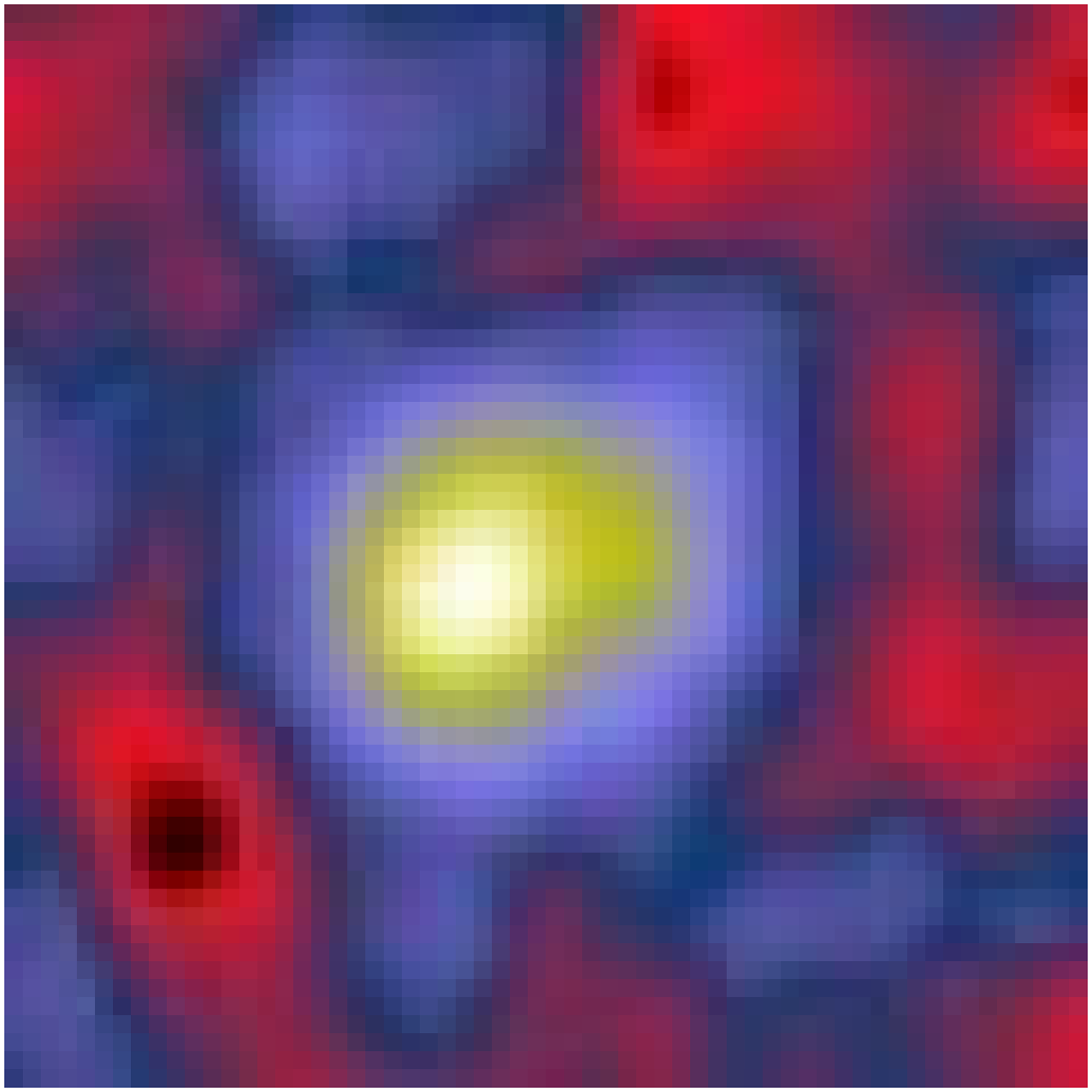}  \FGb{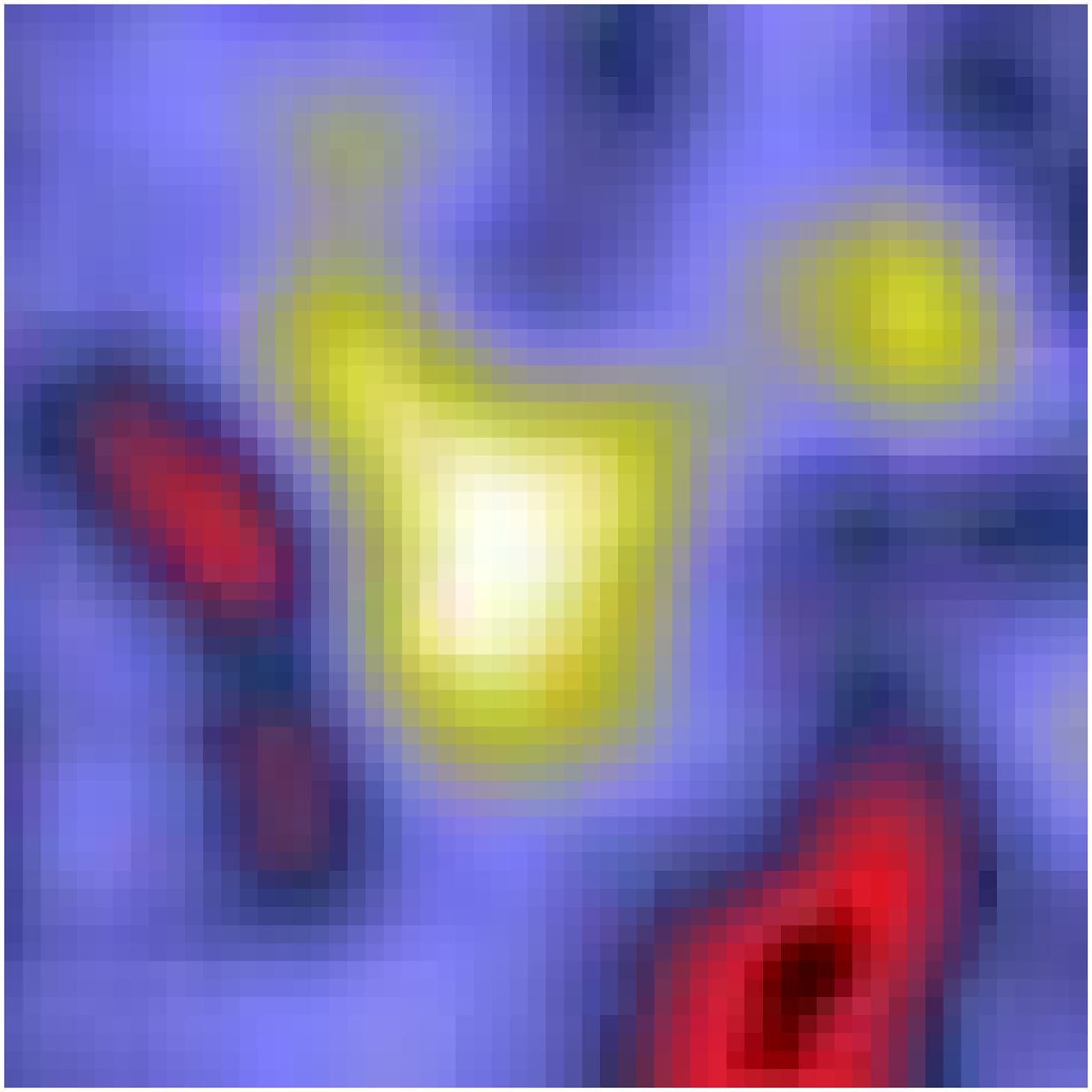} \\  {\#29   NCIA016387}  \\
  \FGb{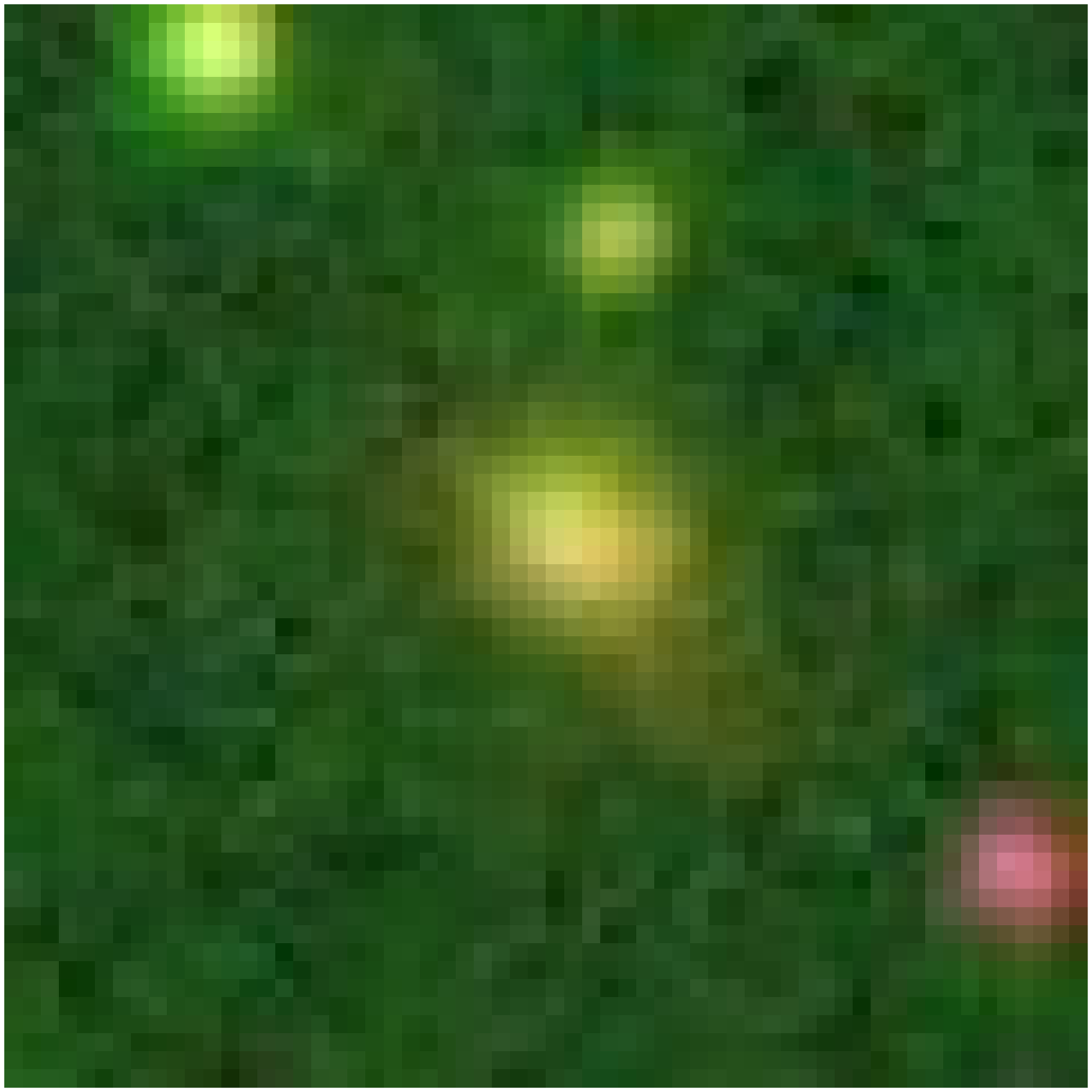}  \FGb{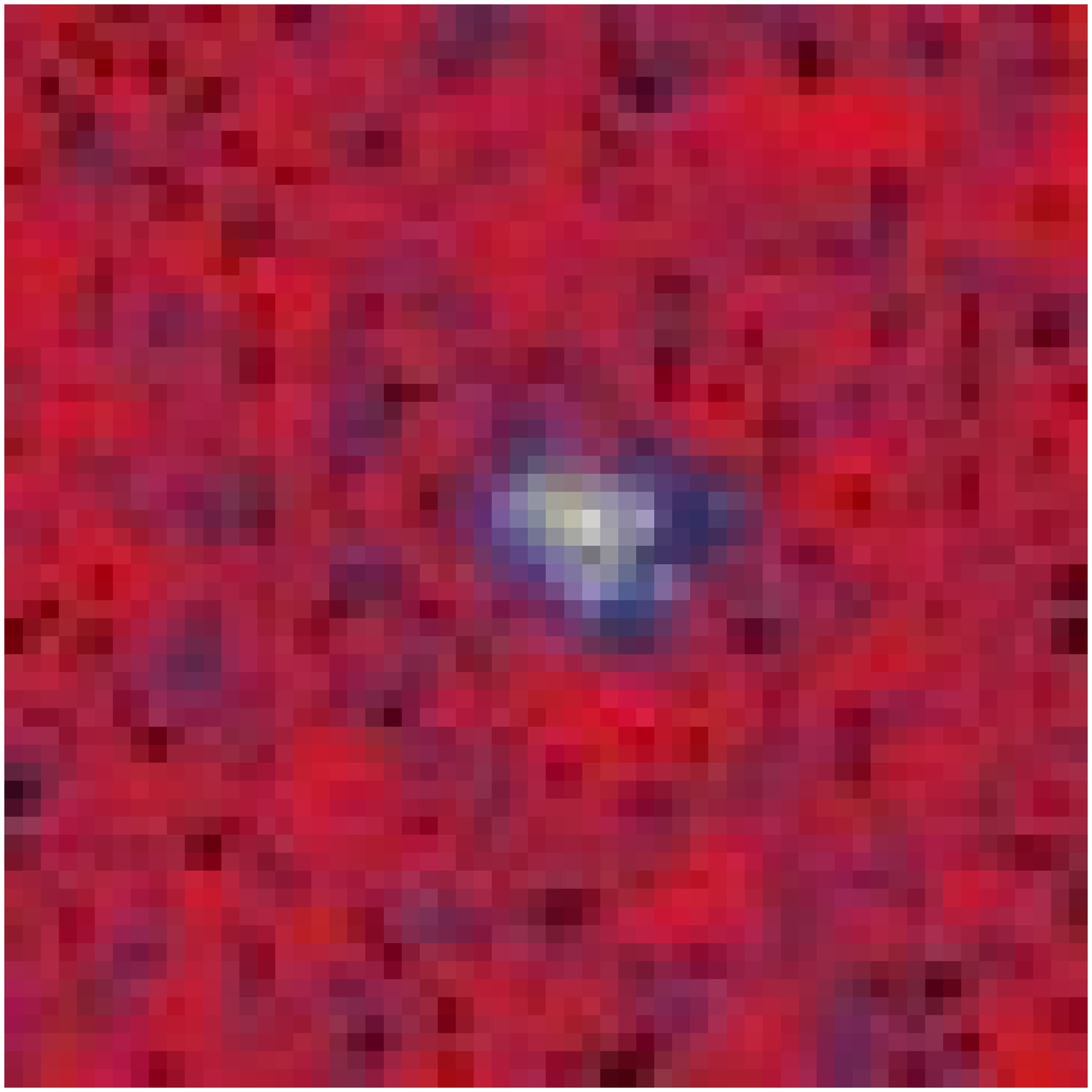}  \FGb{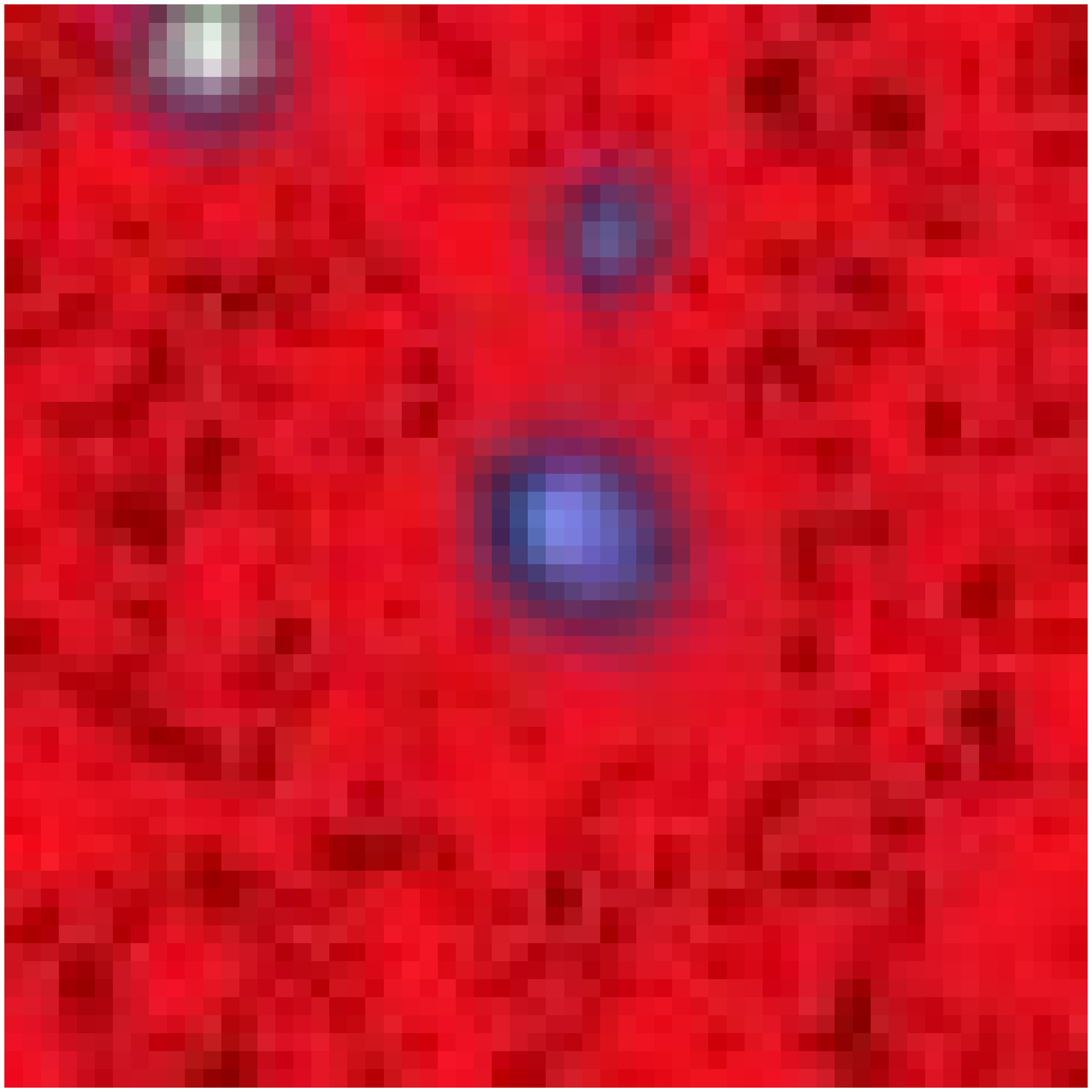}  \FGb{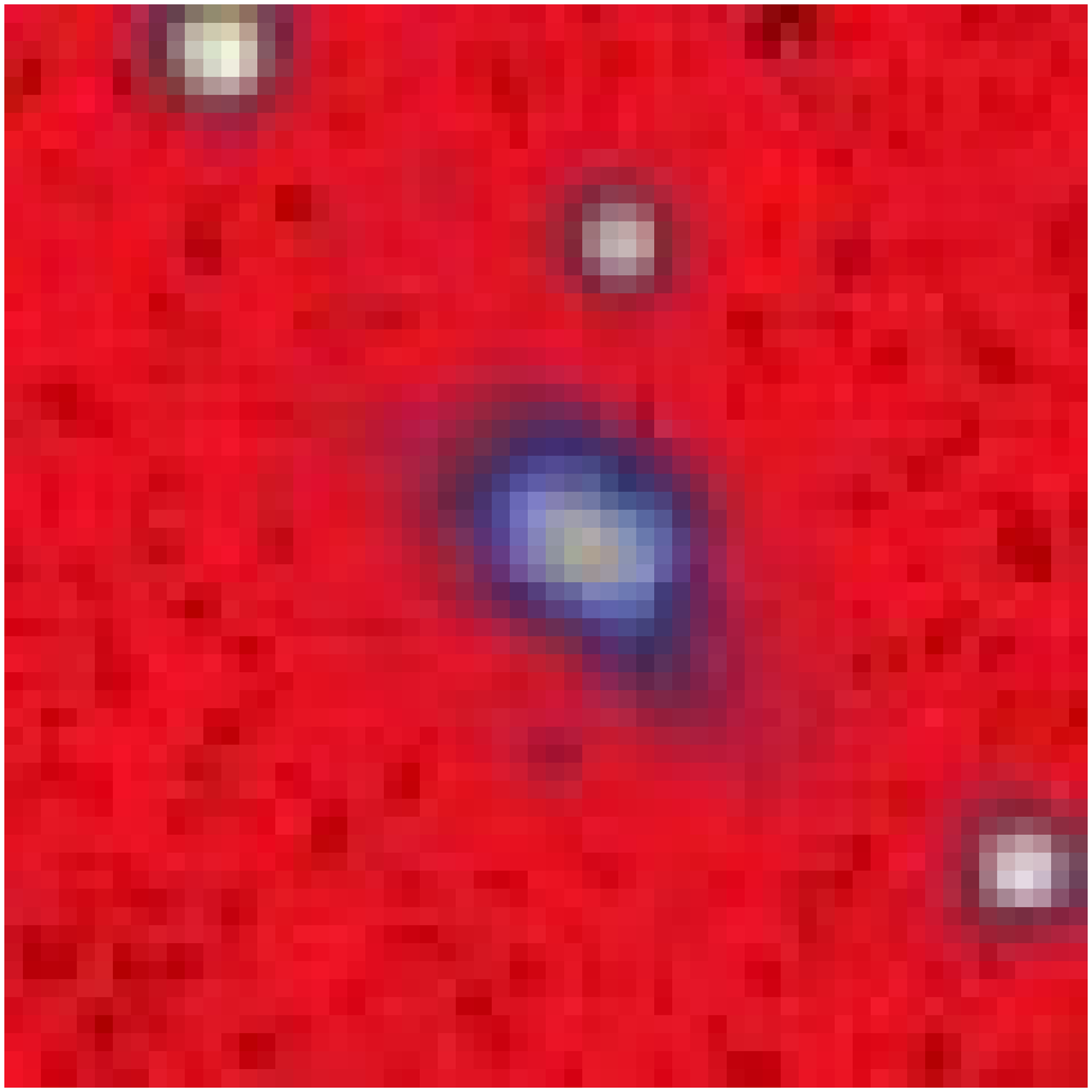}  \FGb{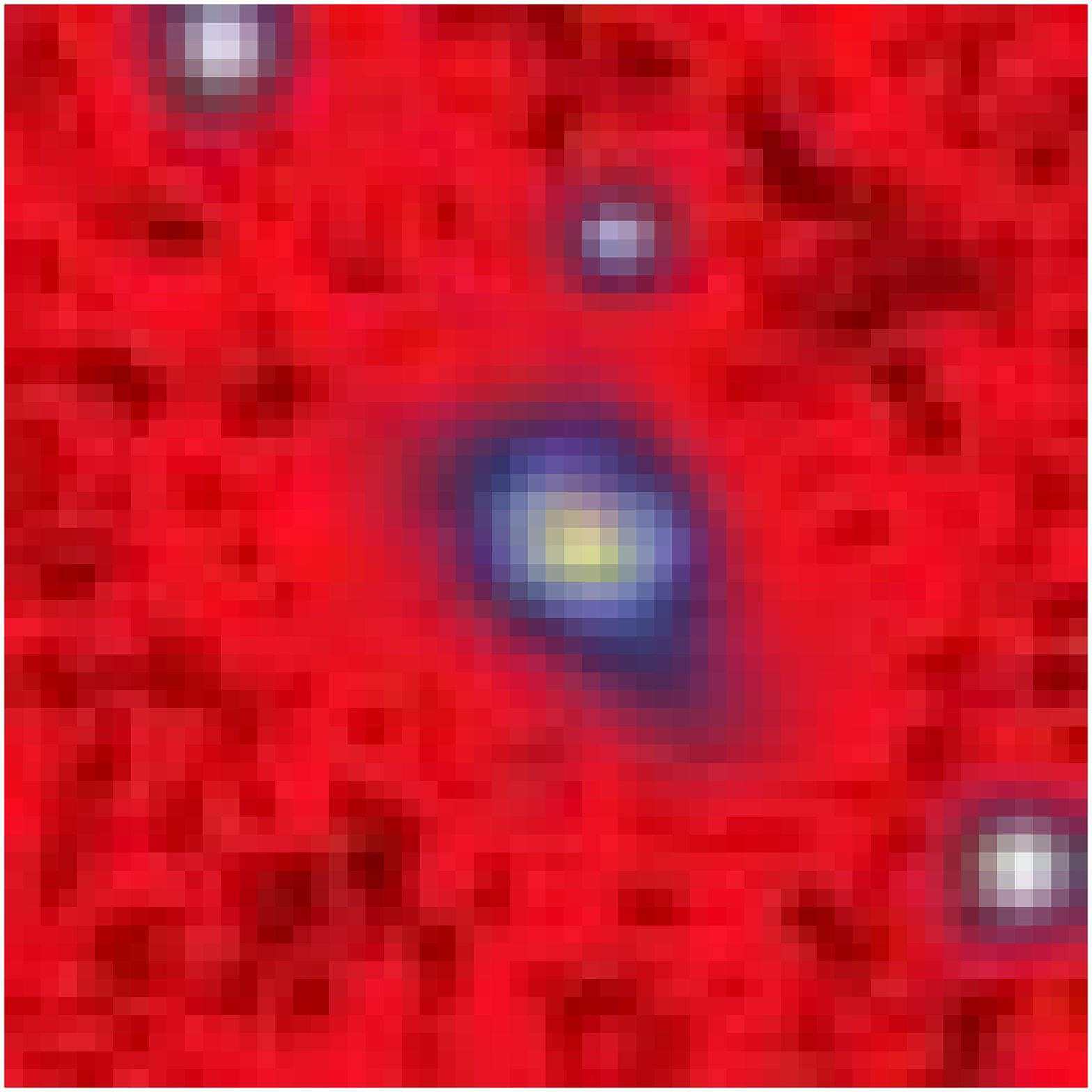}  \FGb{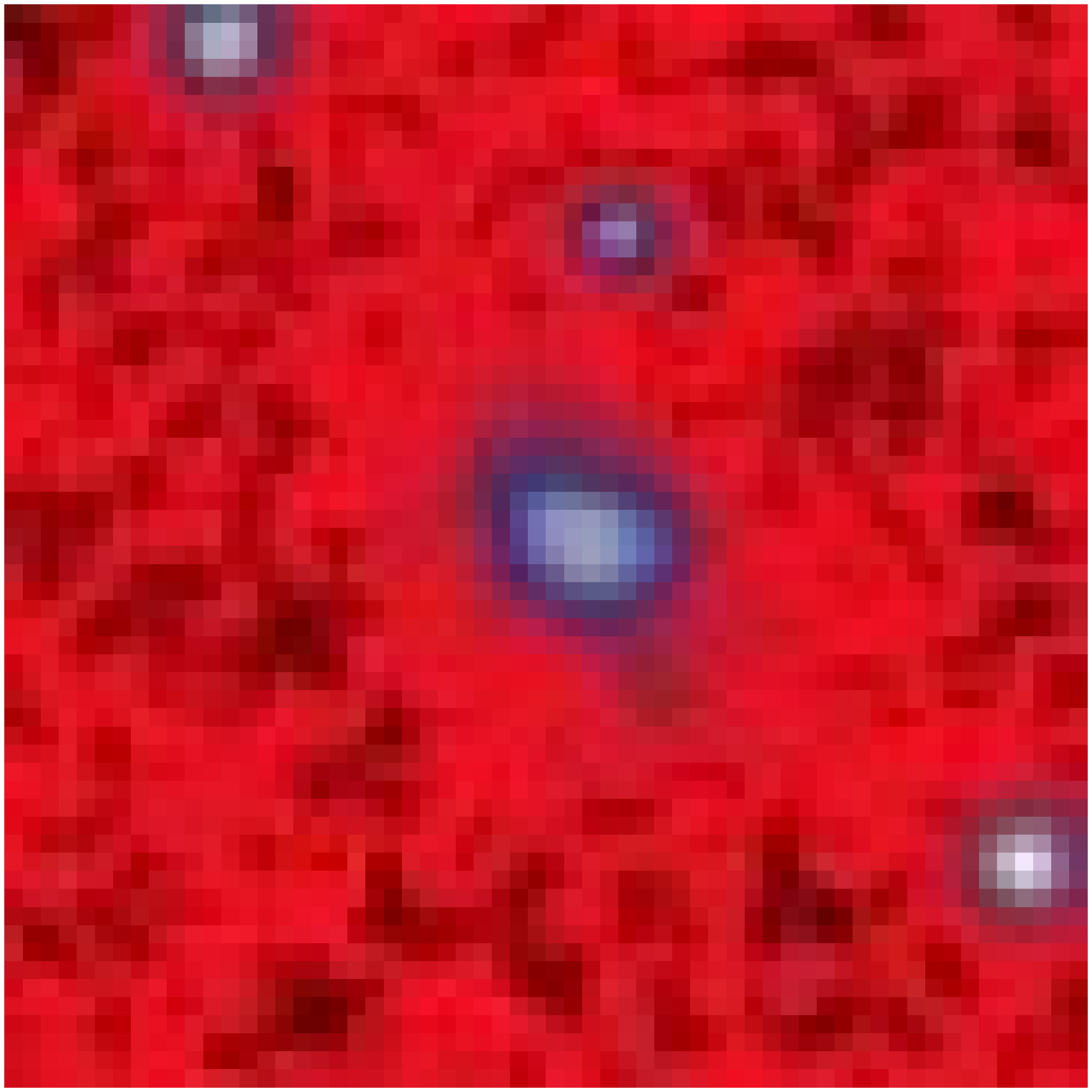}  \FGb{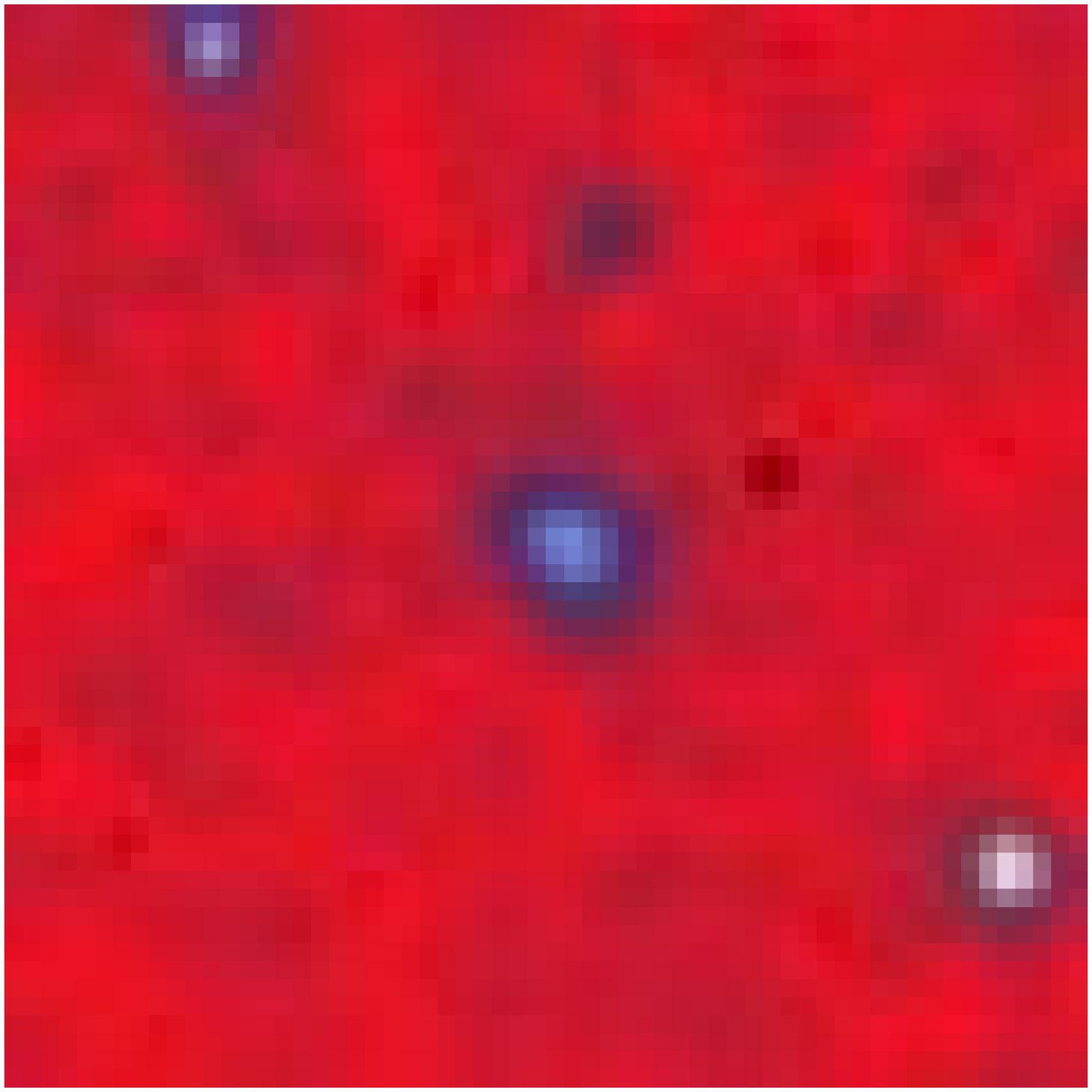}  \FGb{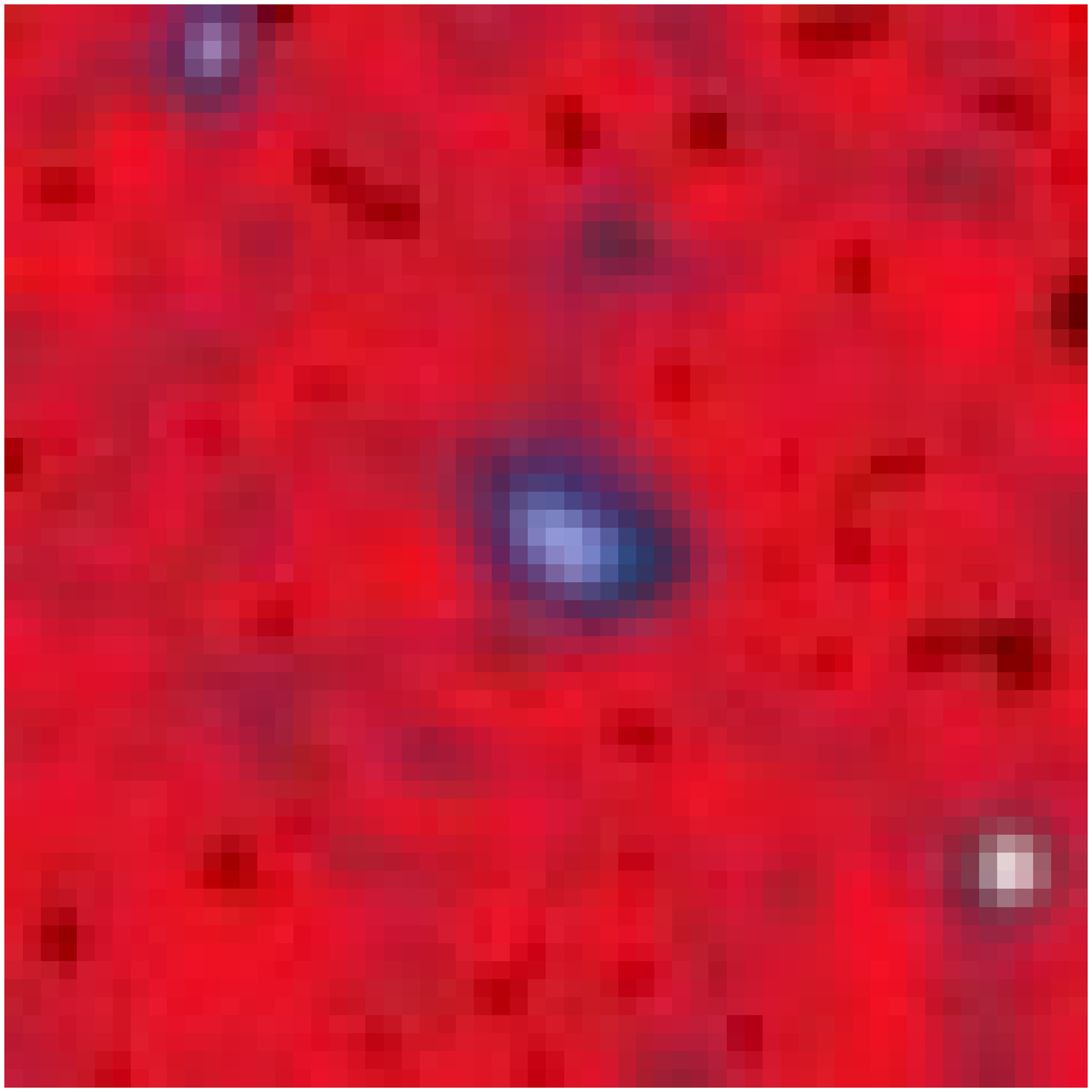}  \FGb{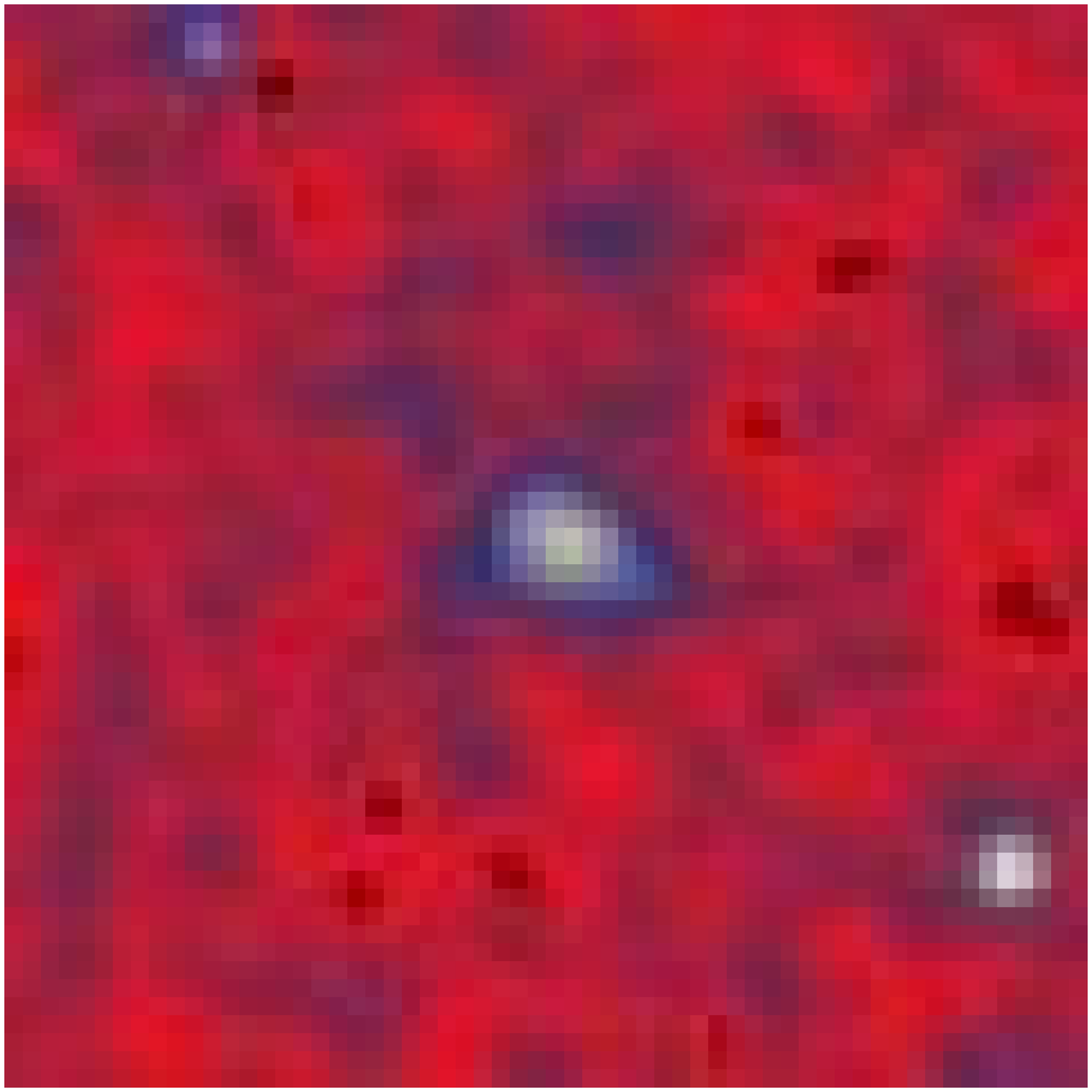}  \FGb{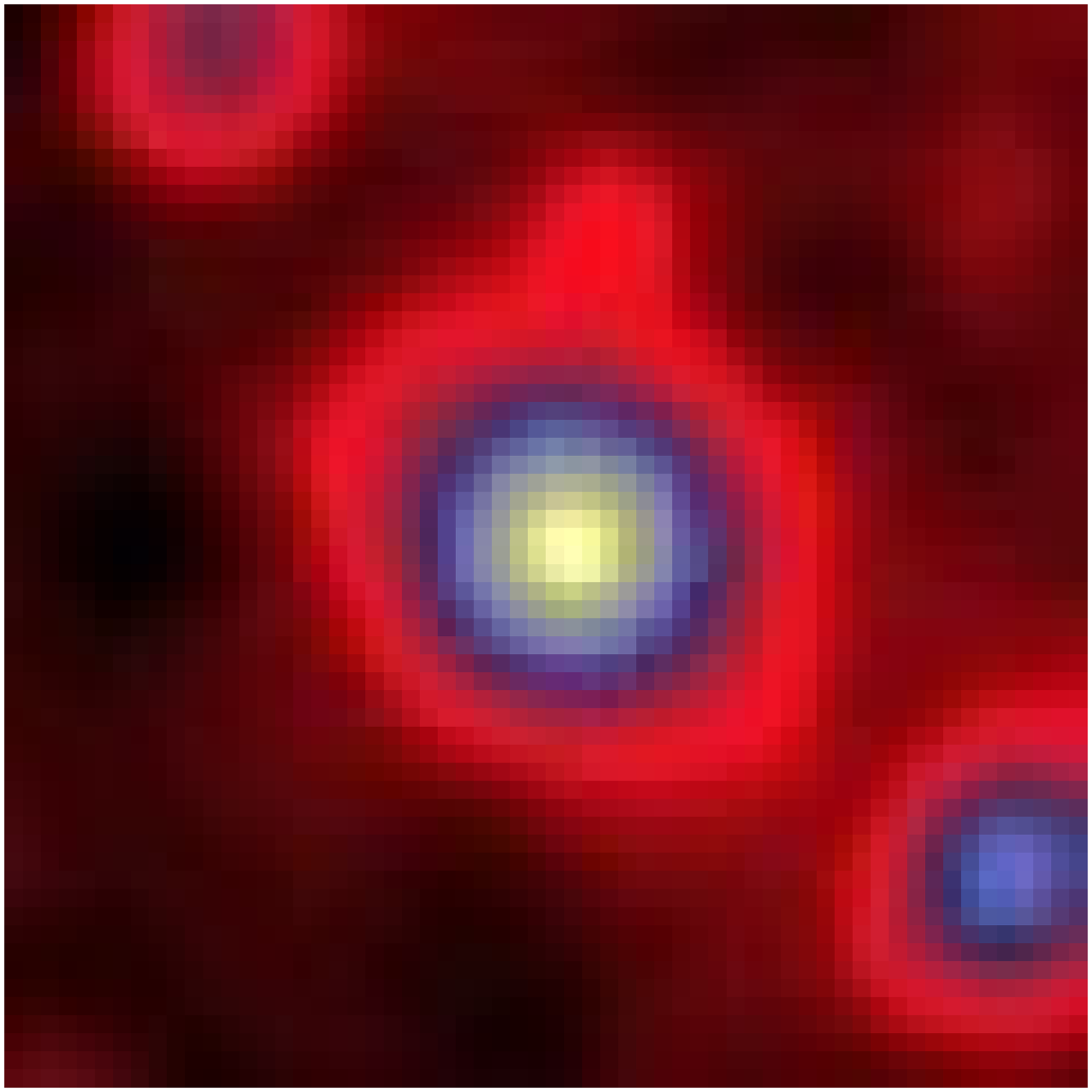}  \FGb{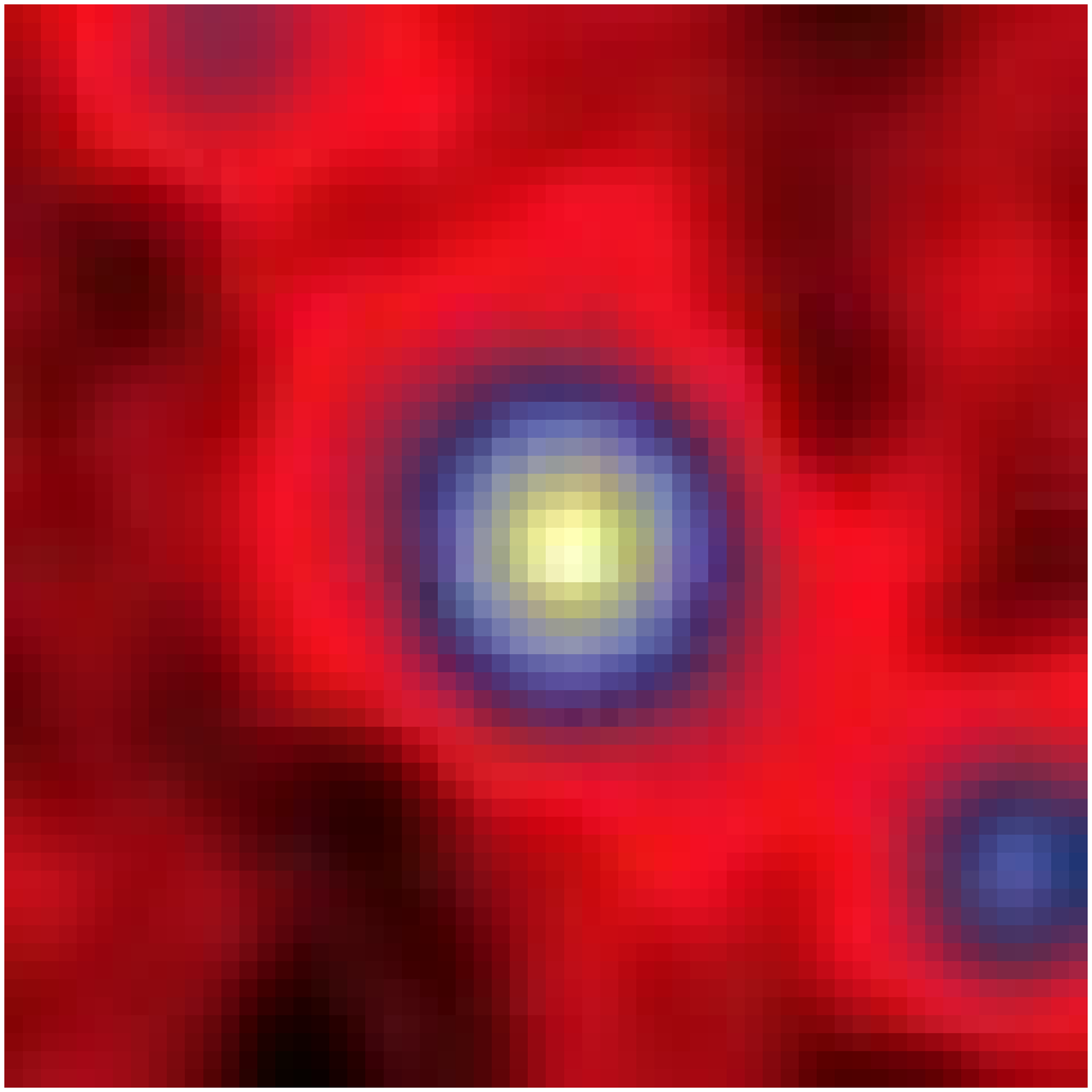}  \FGb{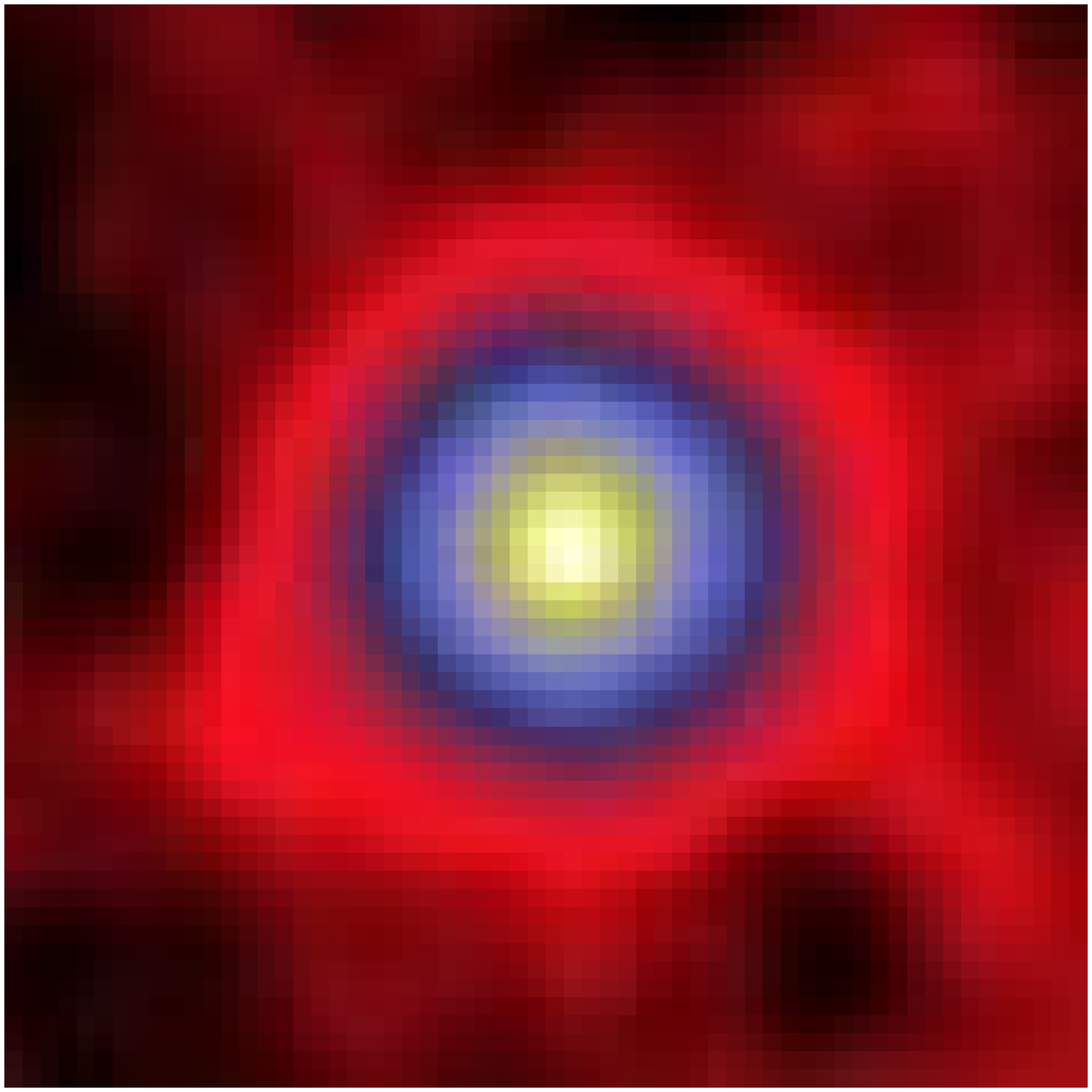}  \FGb{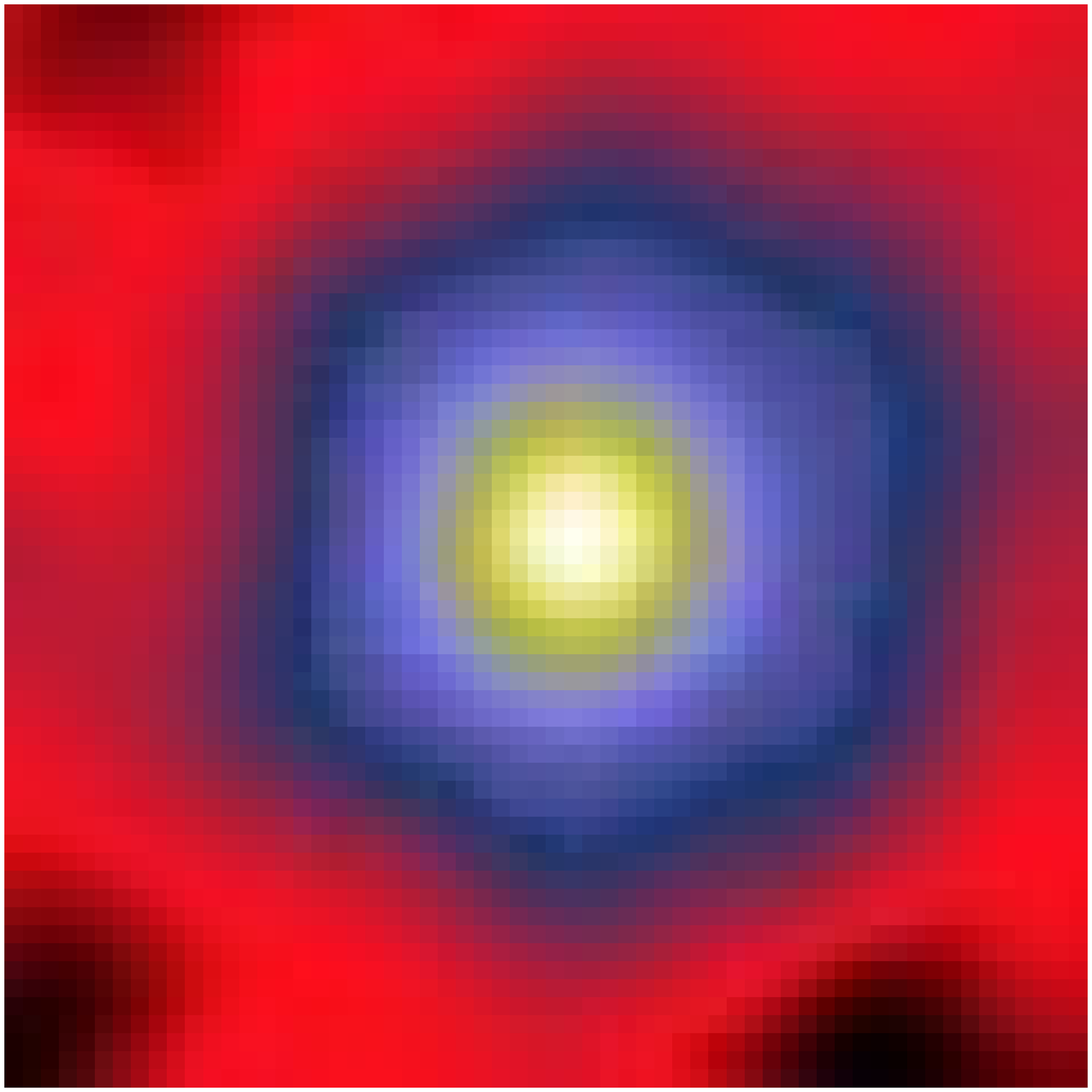} \\  {\#30   NCIA031935}  \\
  \FGb{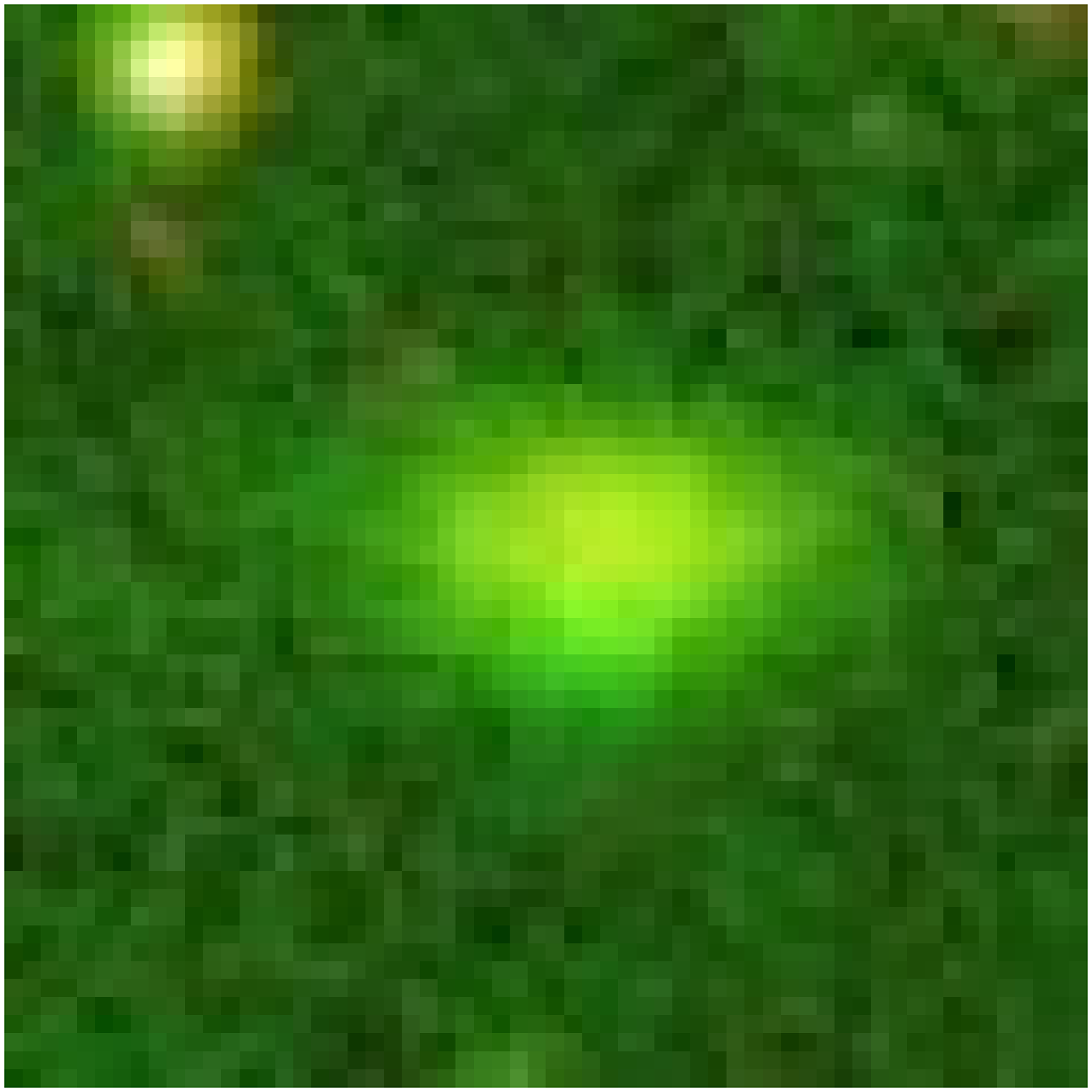}  \FGb{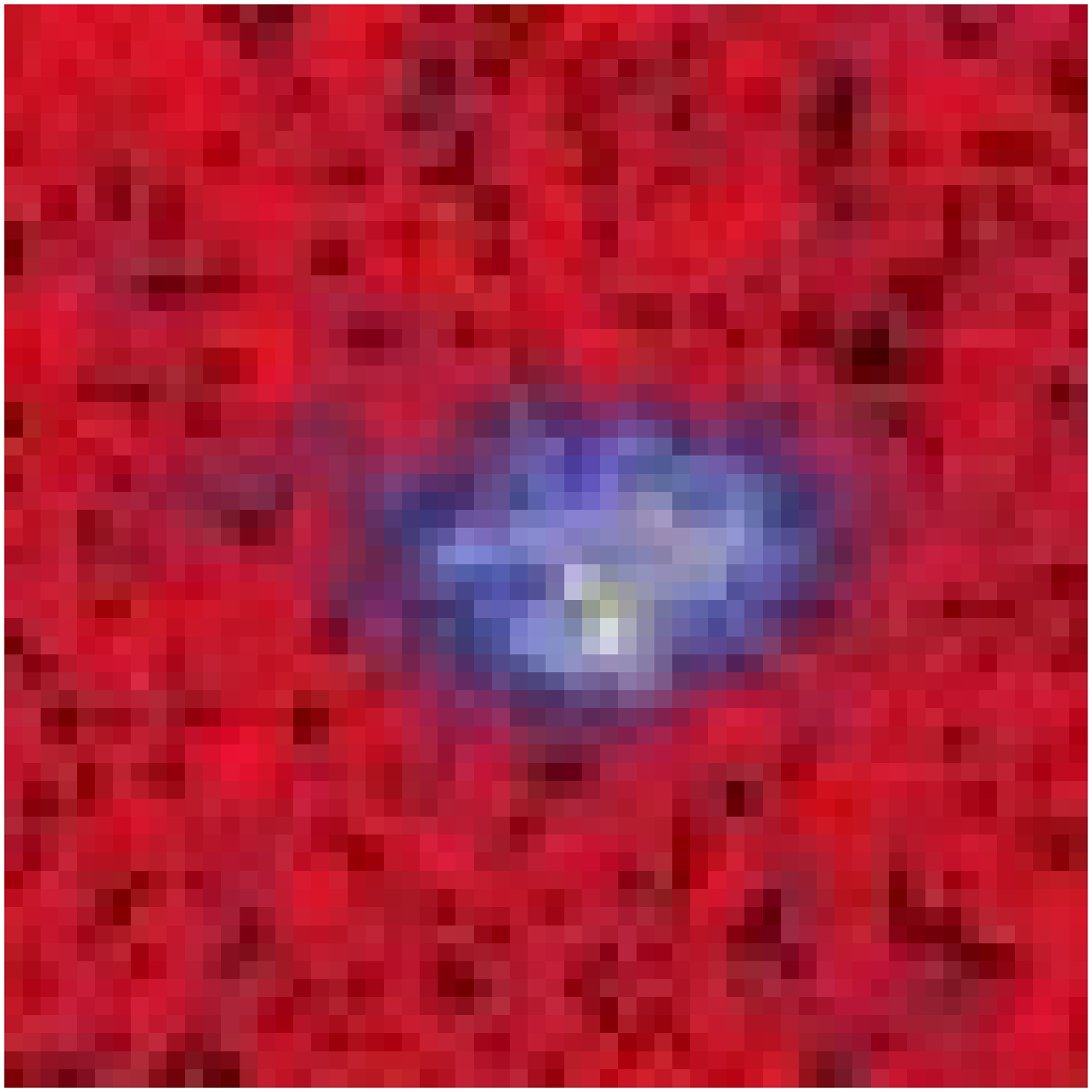}  \FGb{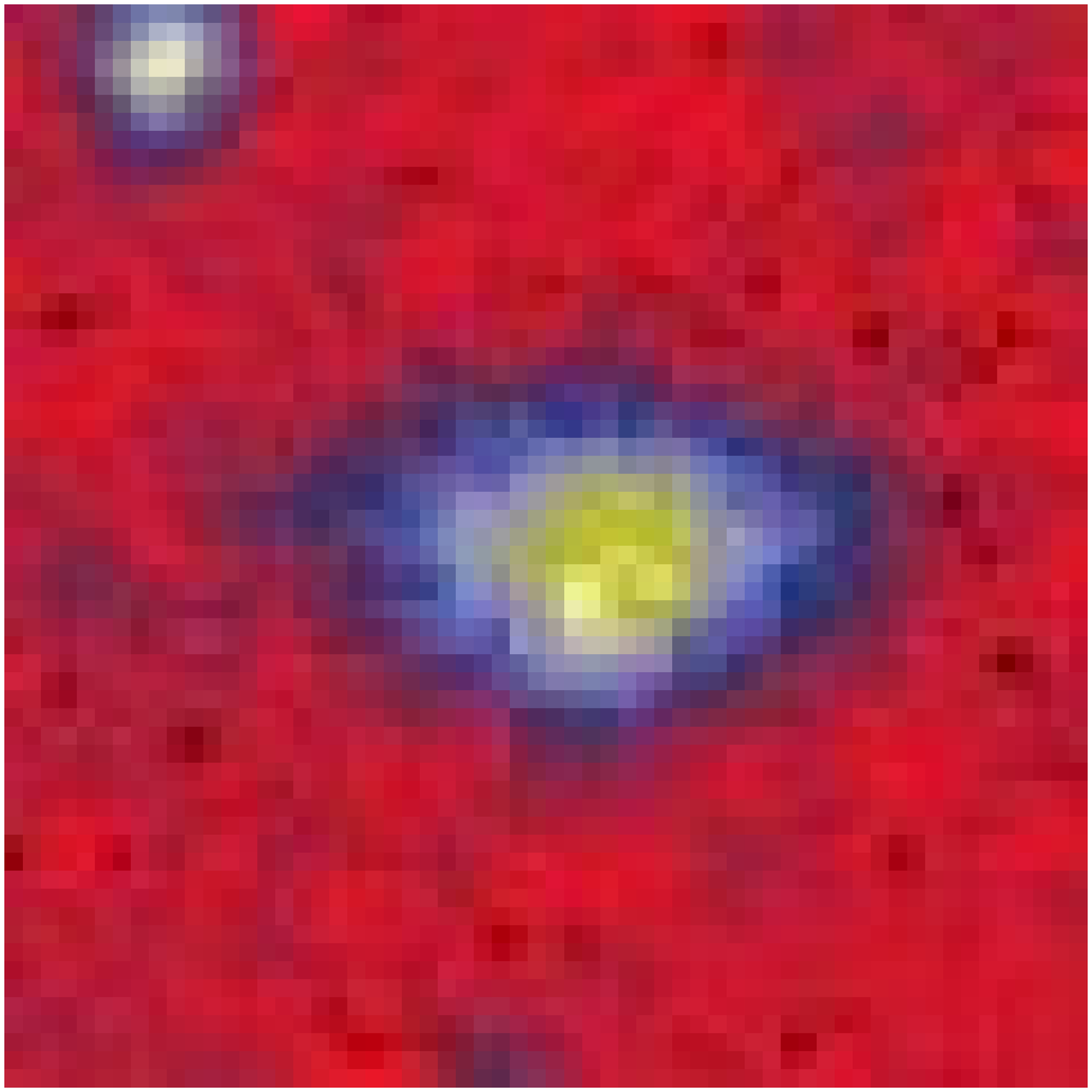}  \FGb{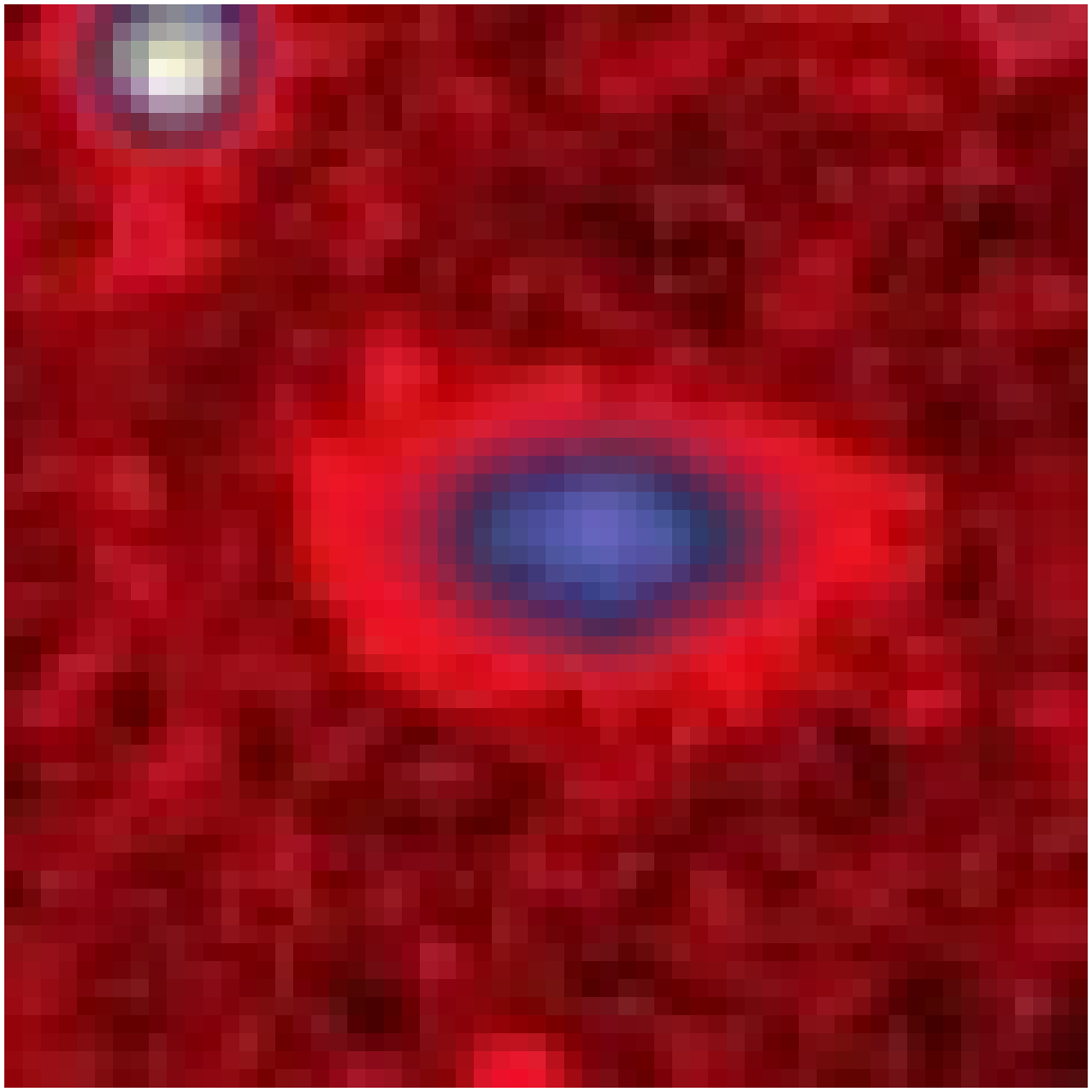}  \FGb{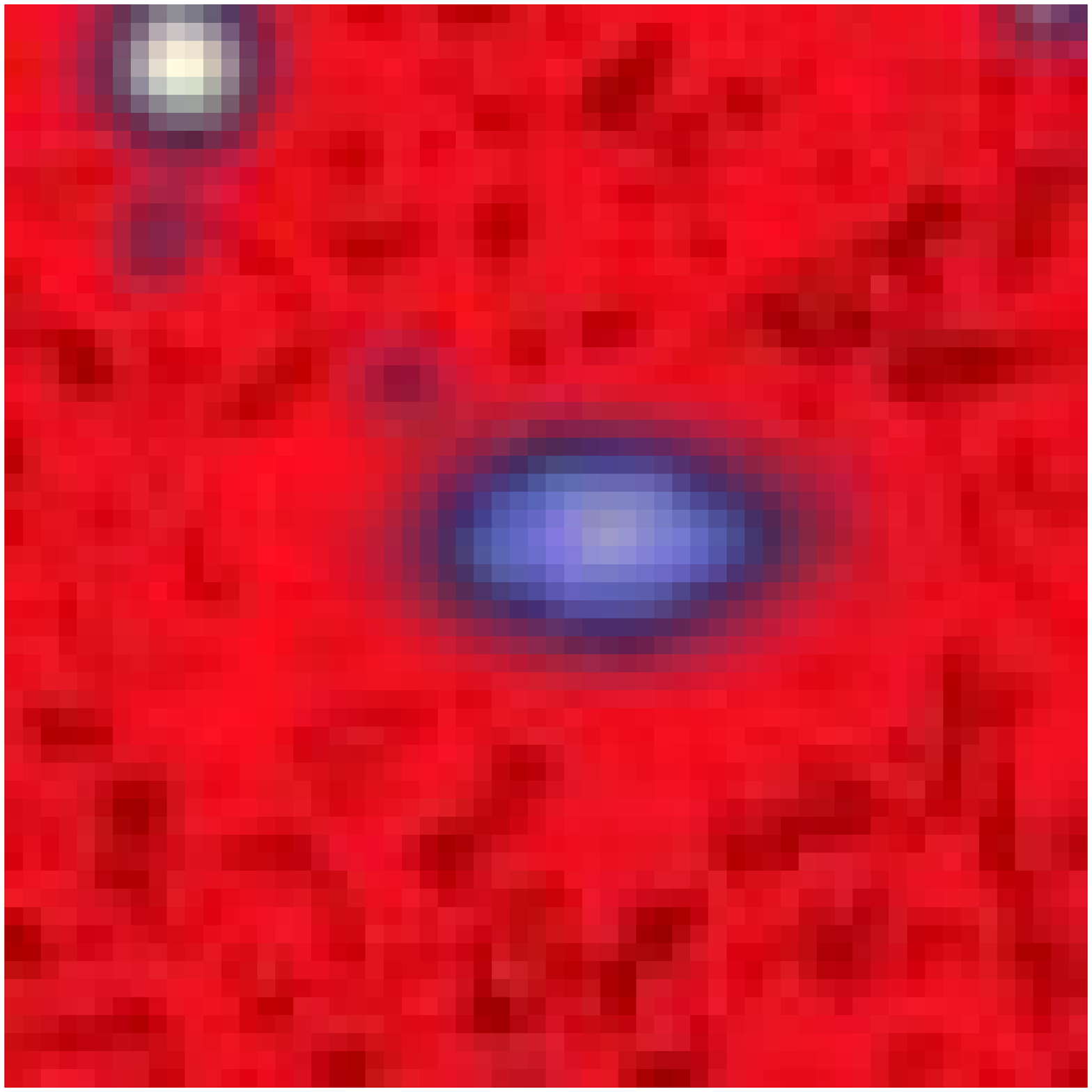}  \FGb{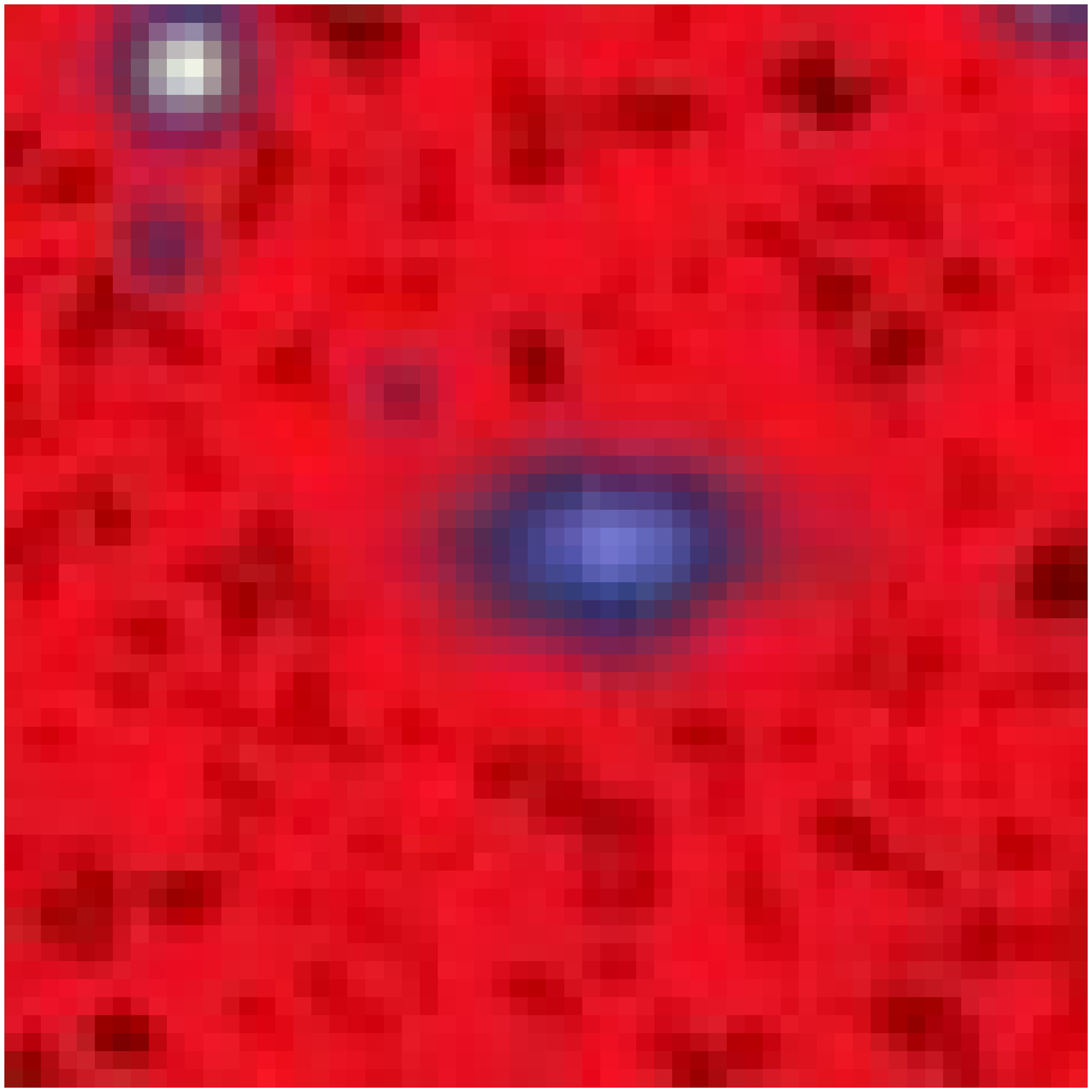}  \FGb{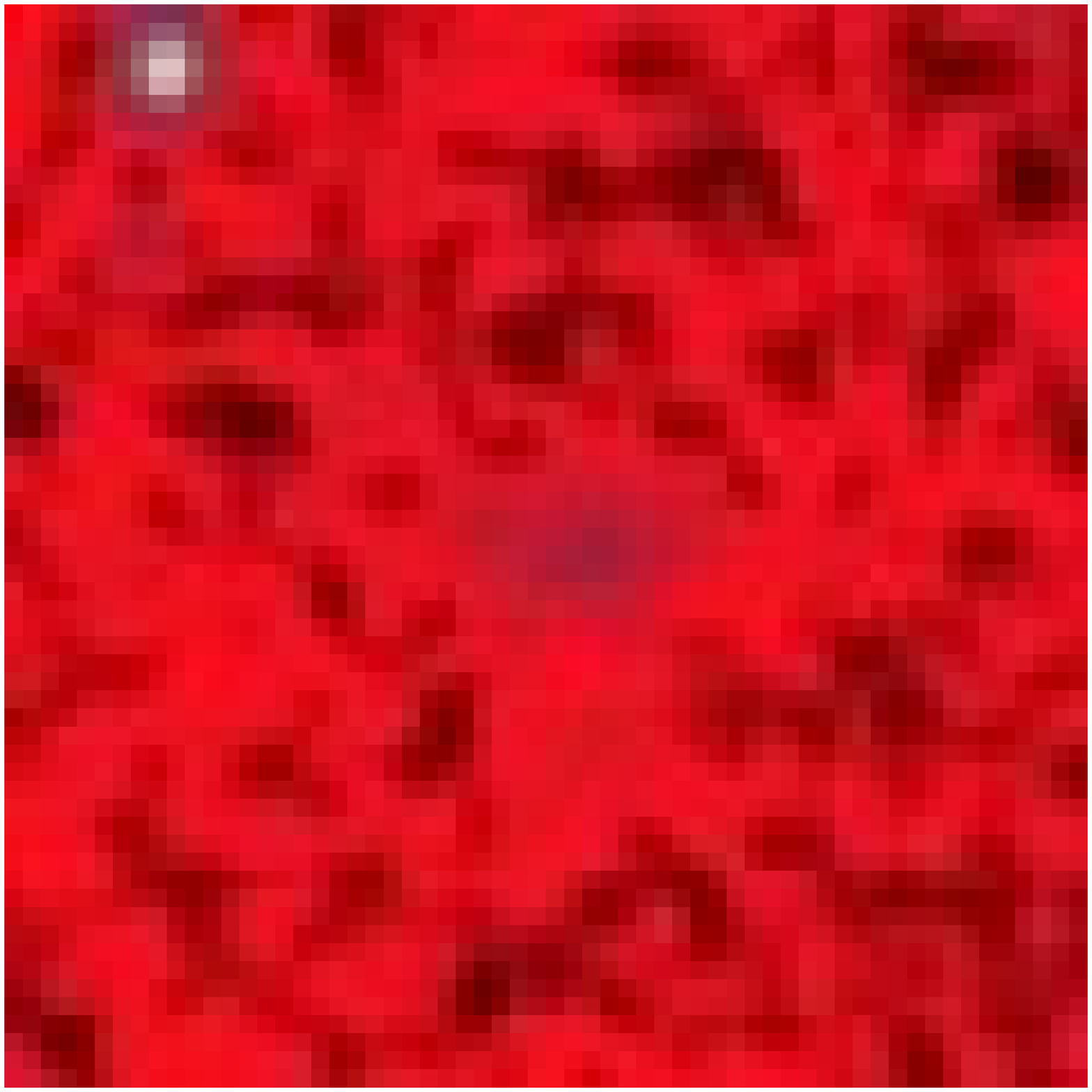}  \FGb{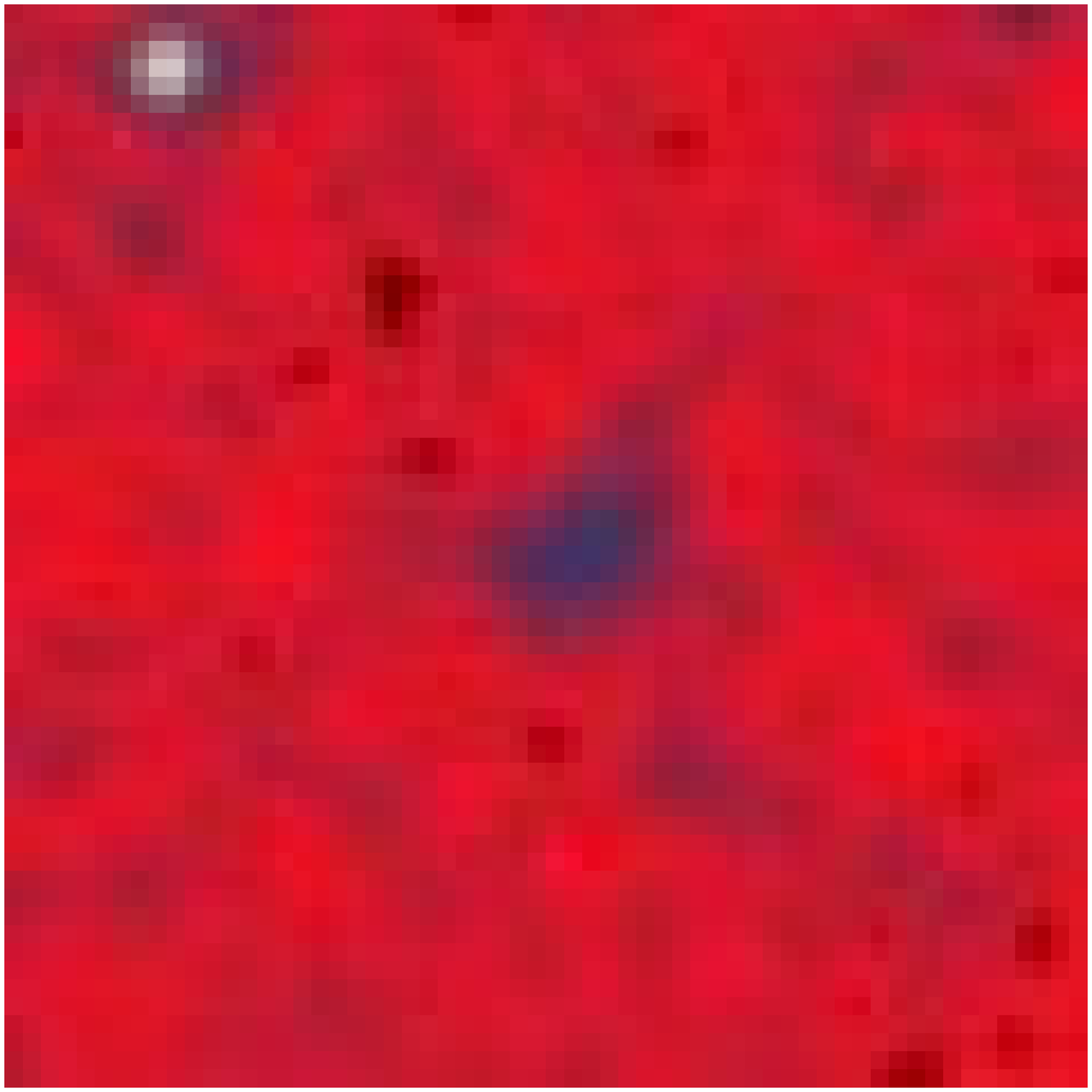}  \FGb{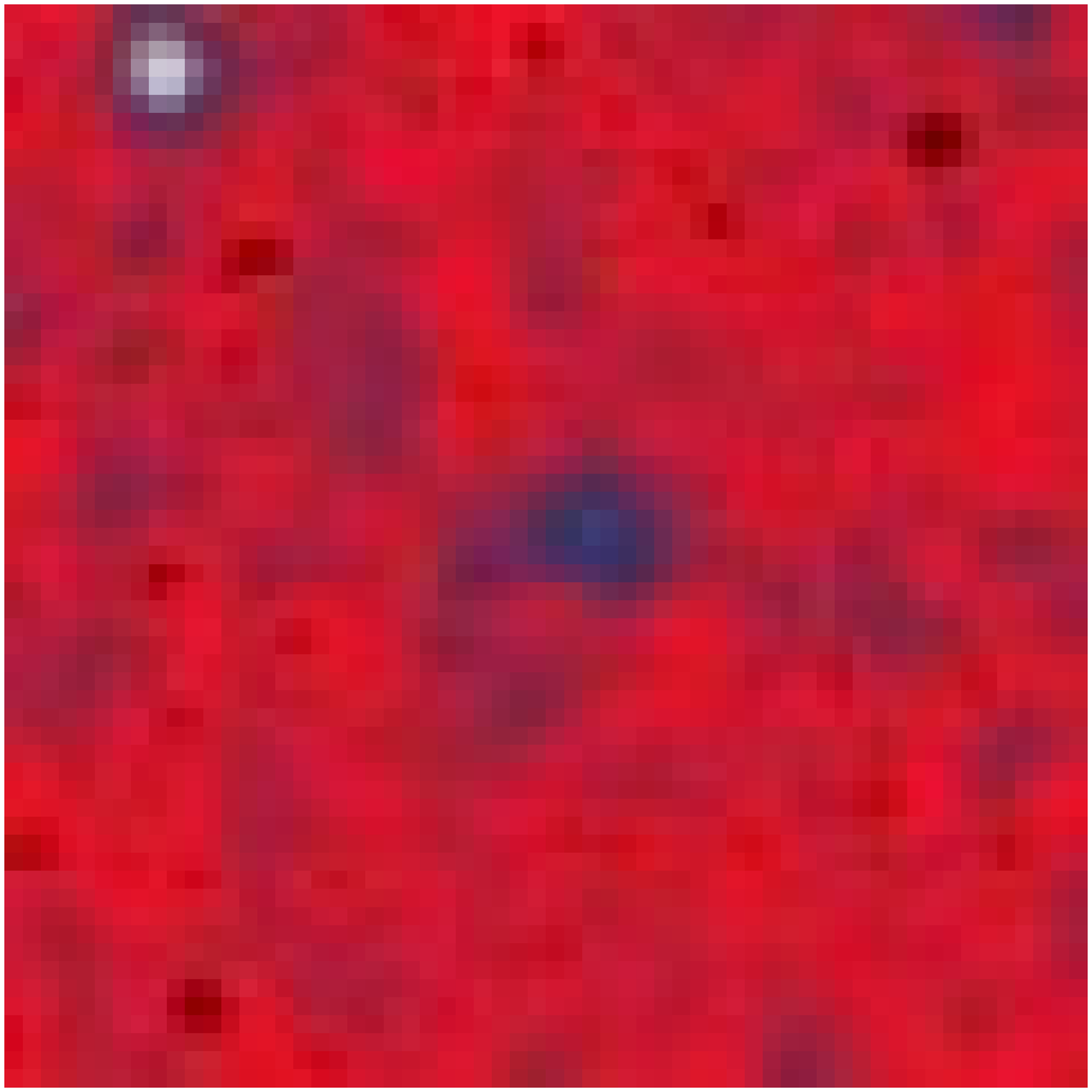}  \FGb{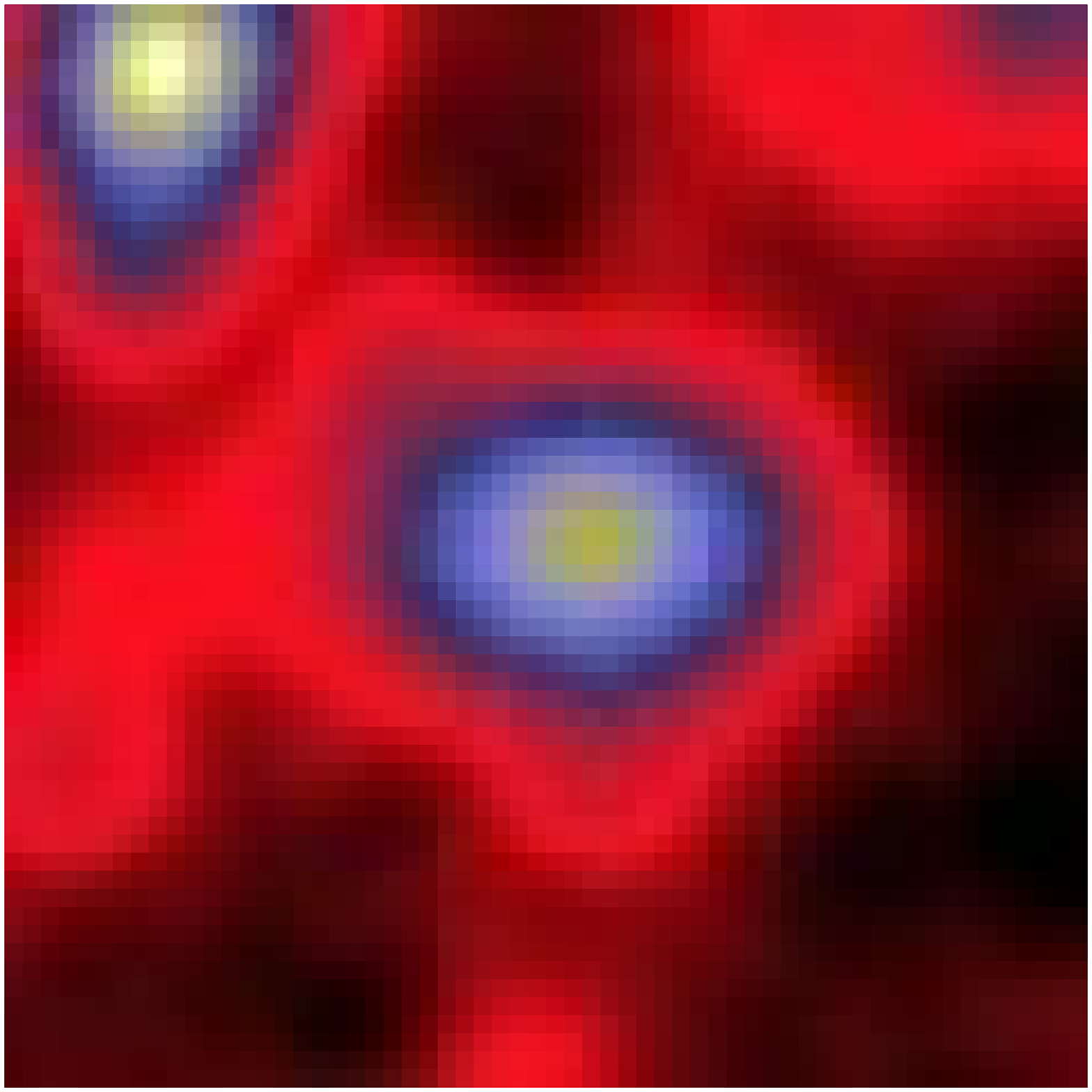}  \FGb{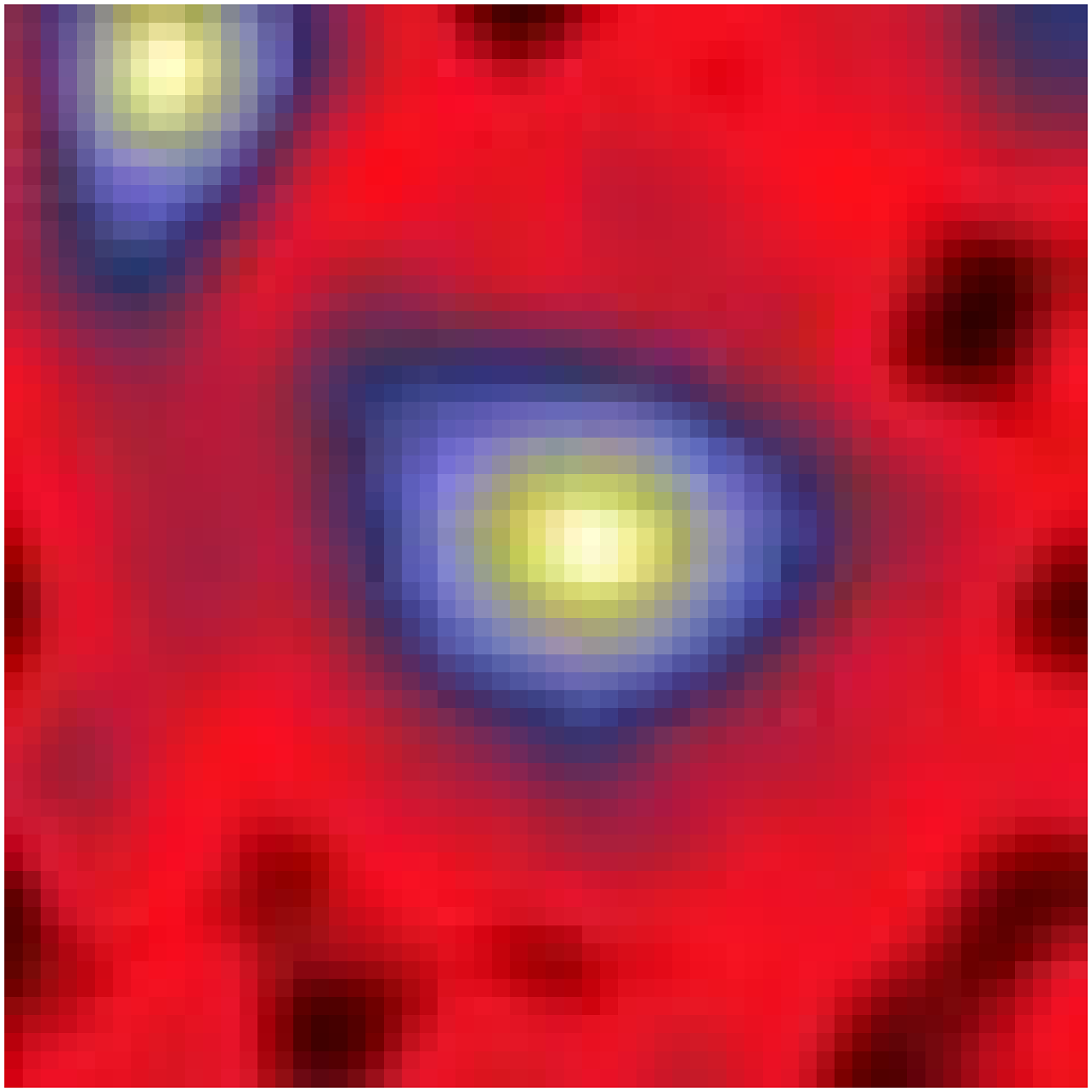}  \FGb{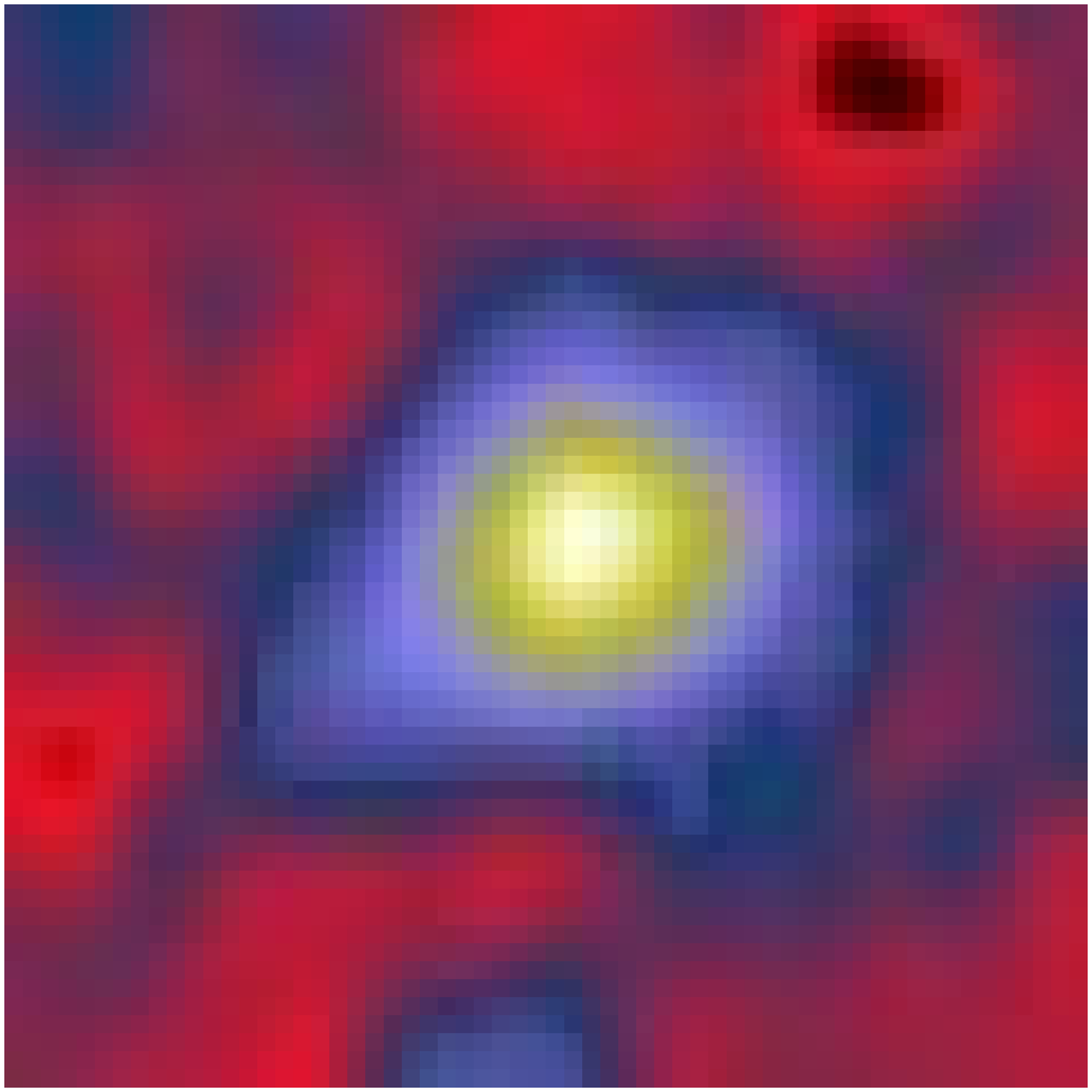}  \FGb{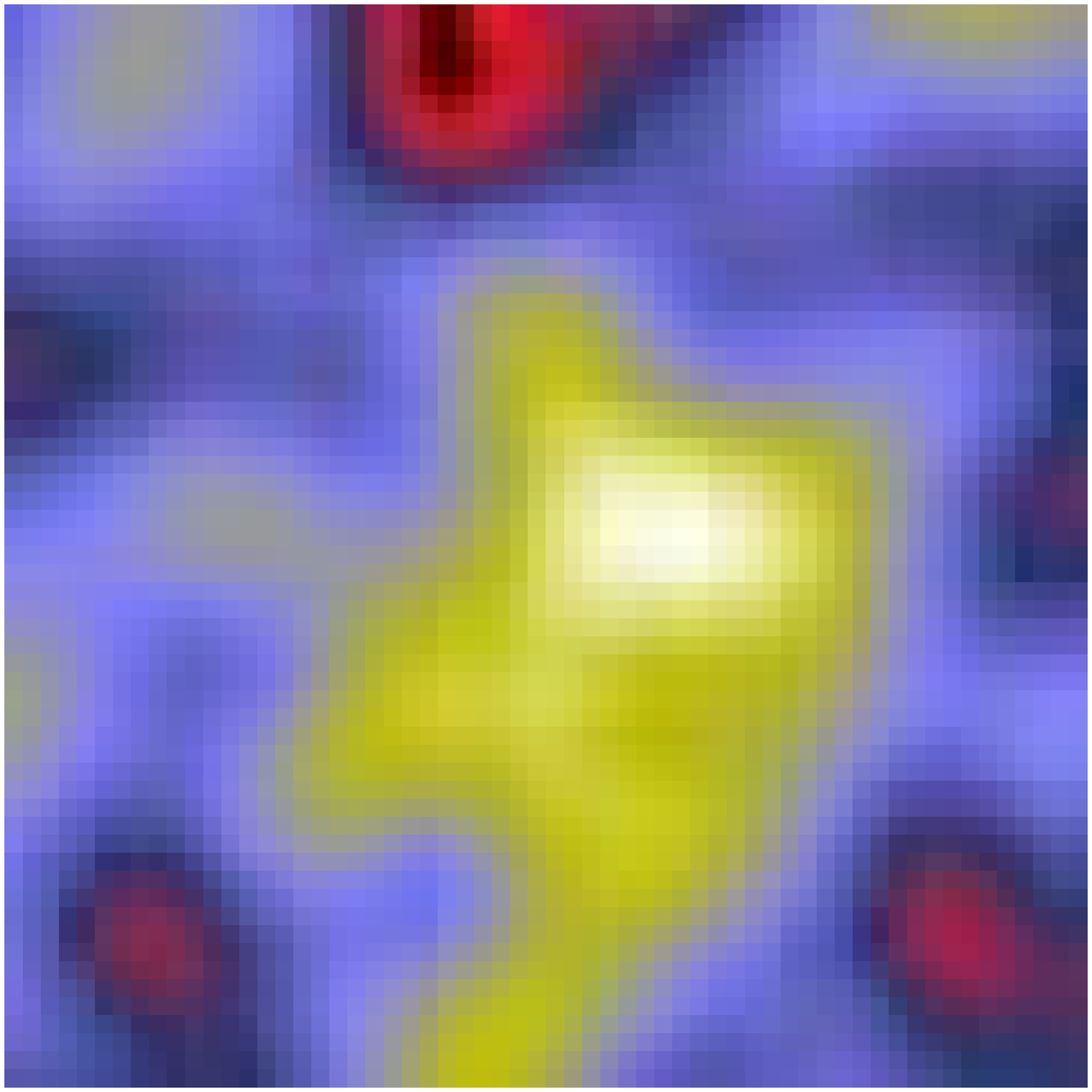} \\  {\#31   NCIA016766}  \\
  \FGb{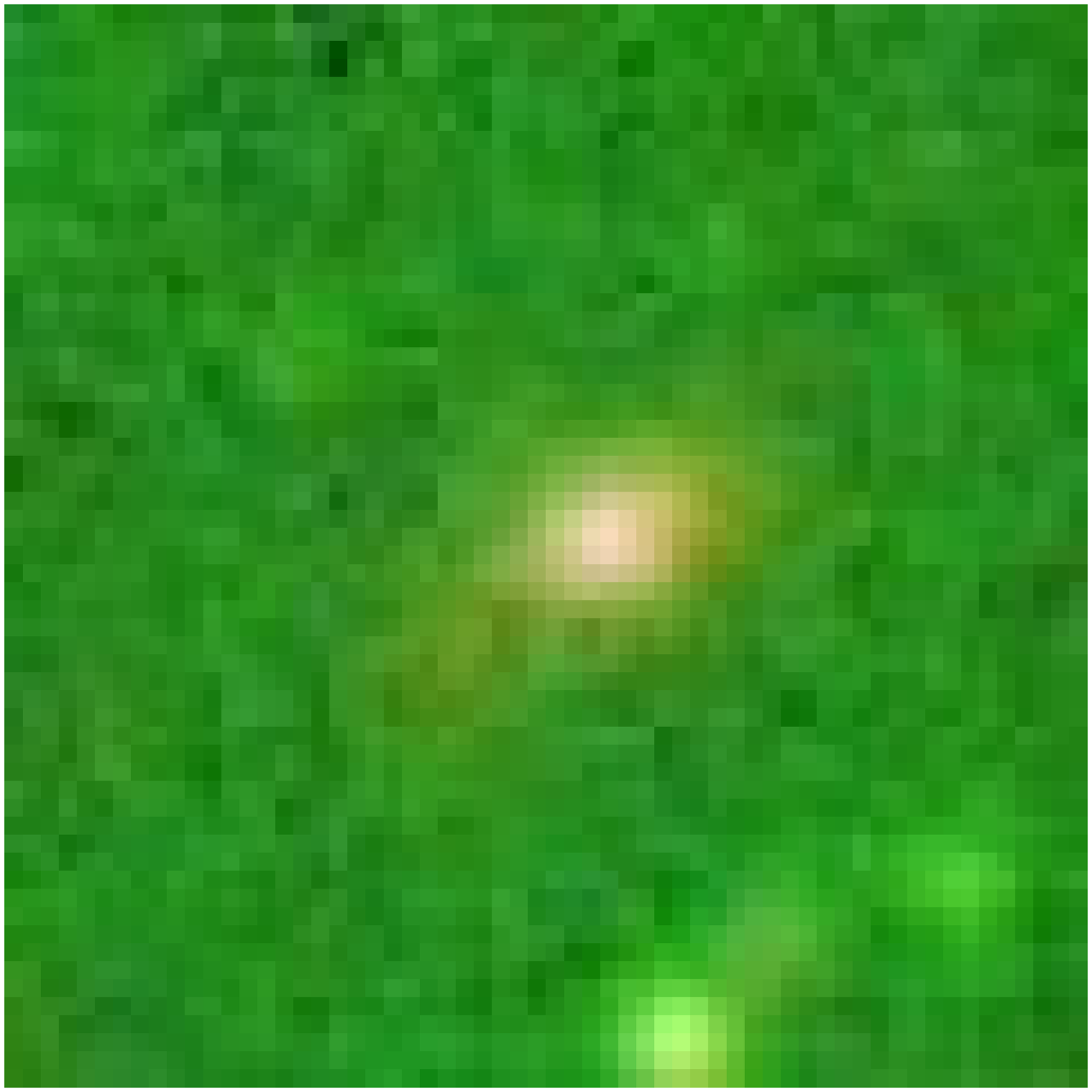}  \FGb{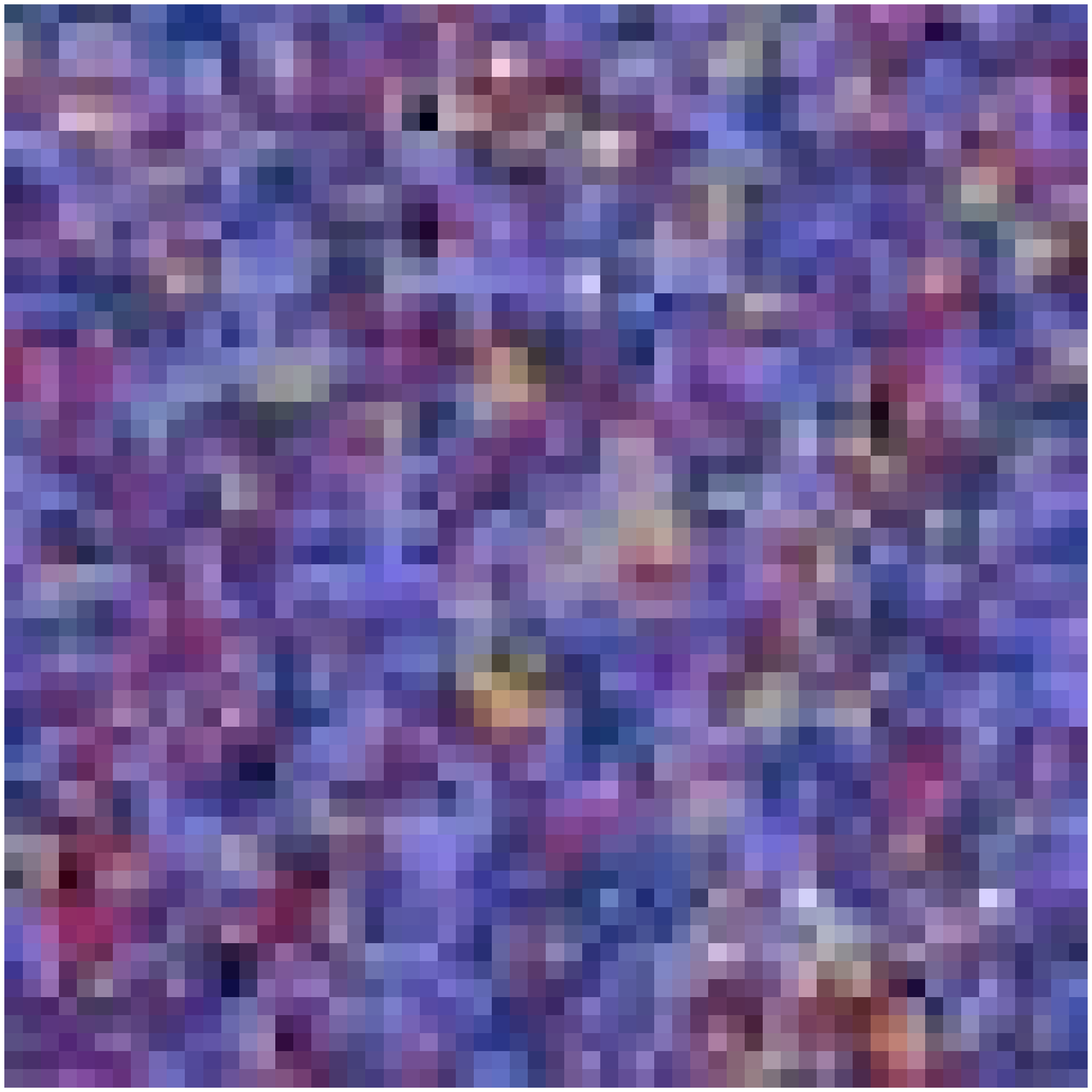}  \FGb{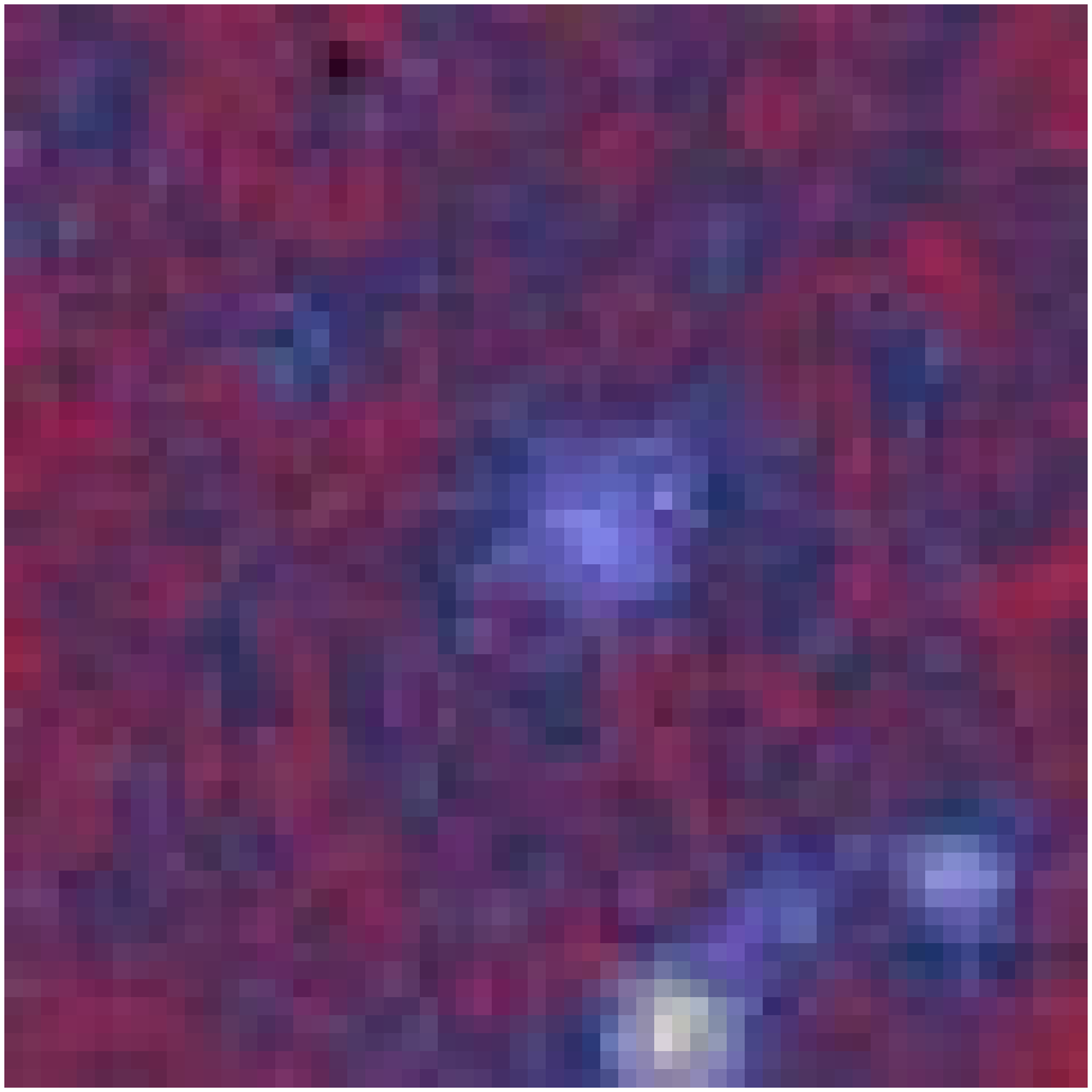}  \FGb{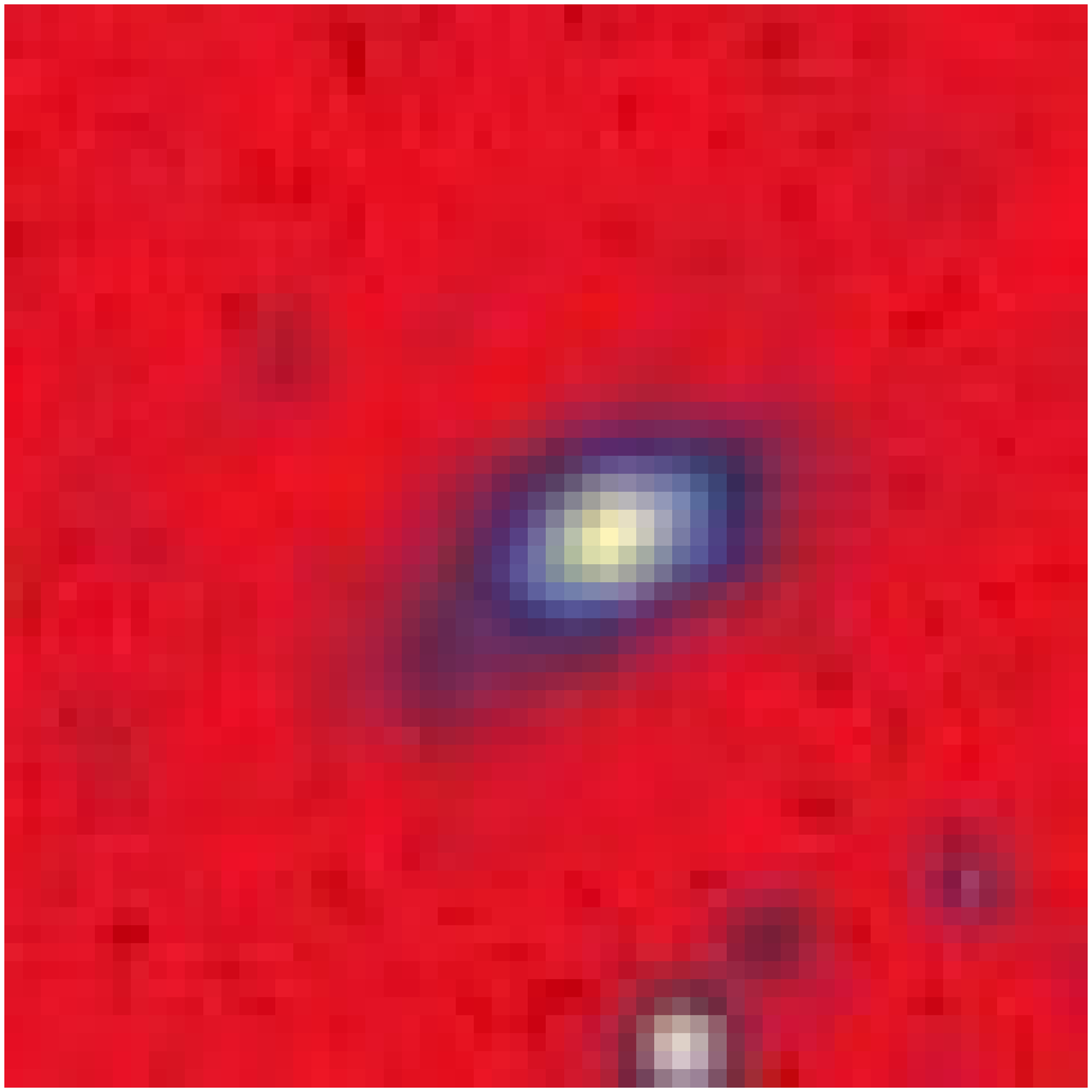}  \FGb{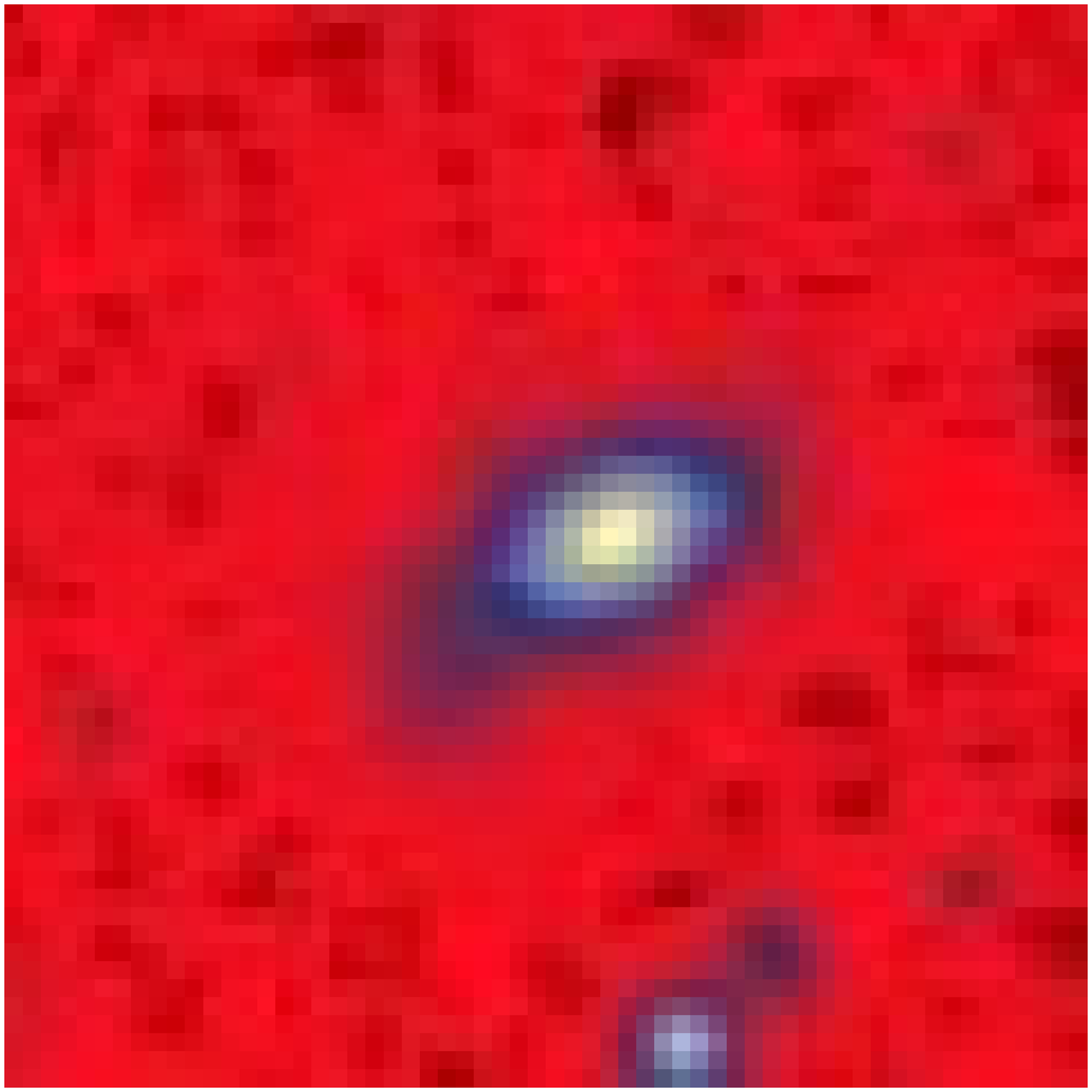}  \FGb{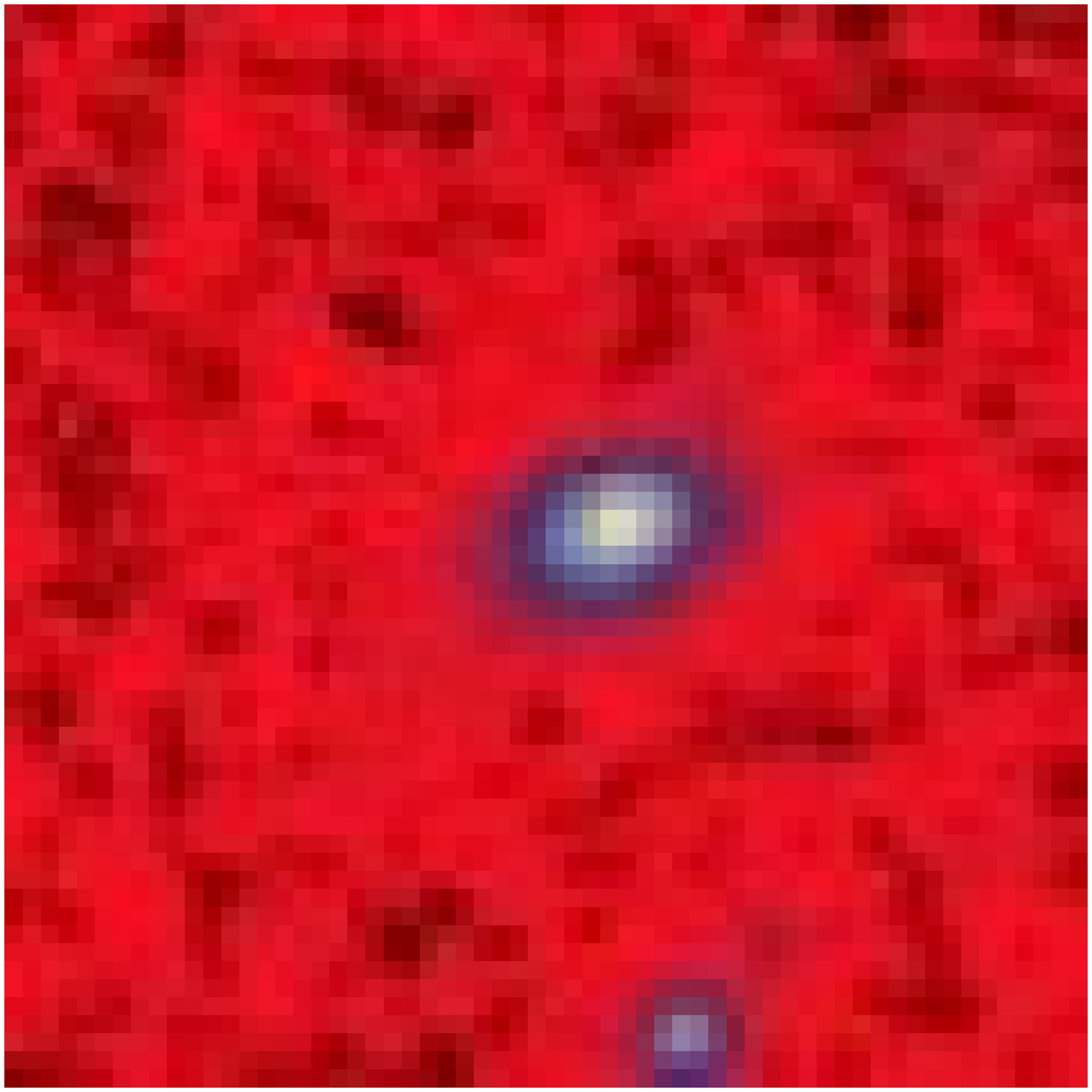}  \FGb{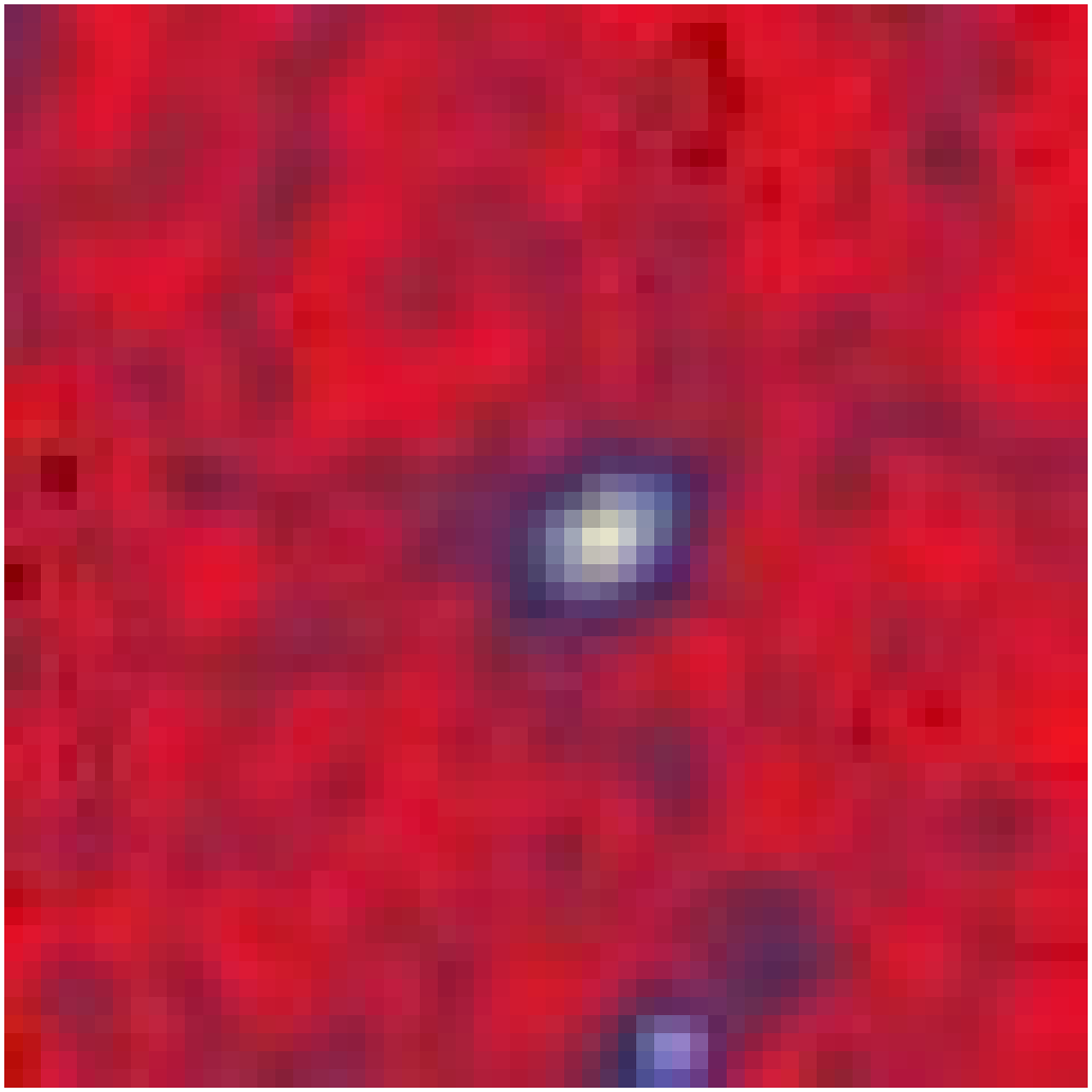}  \FGb{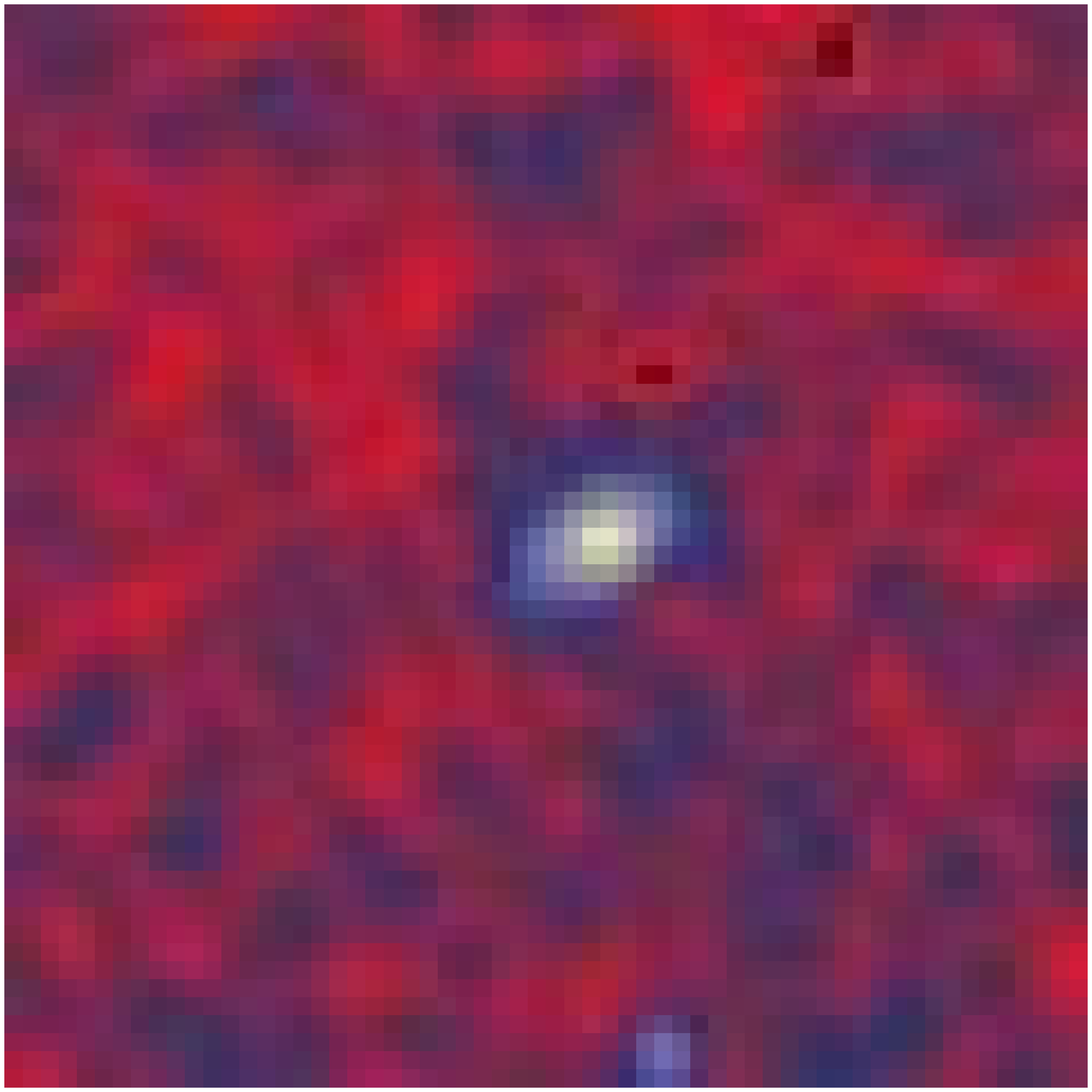}  \FGb{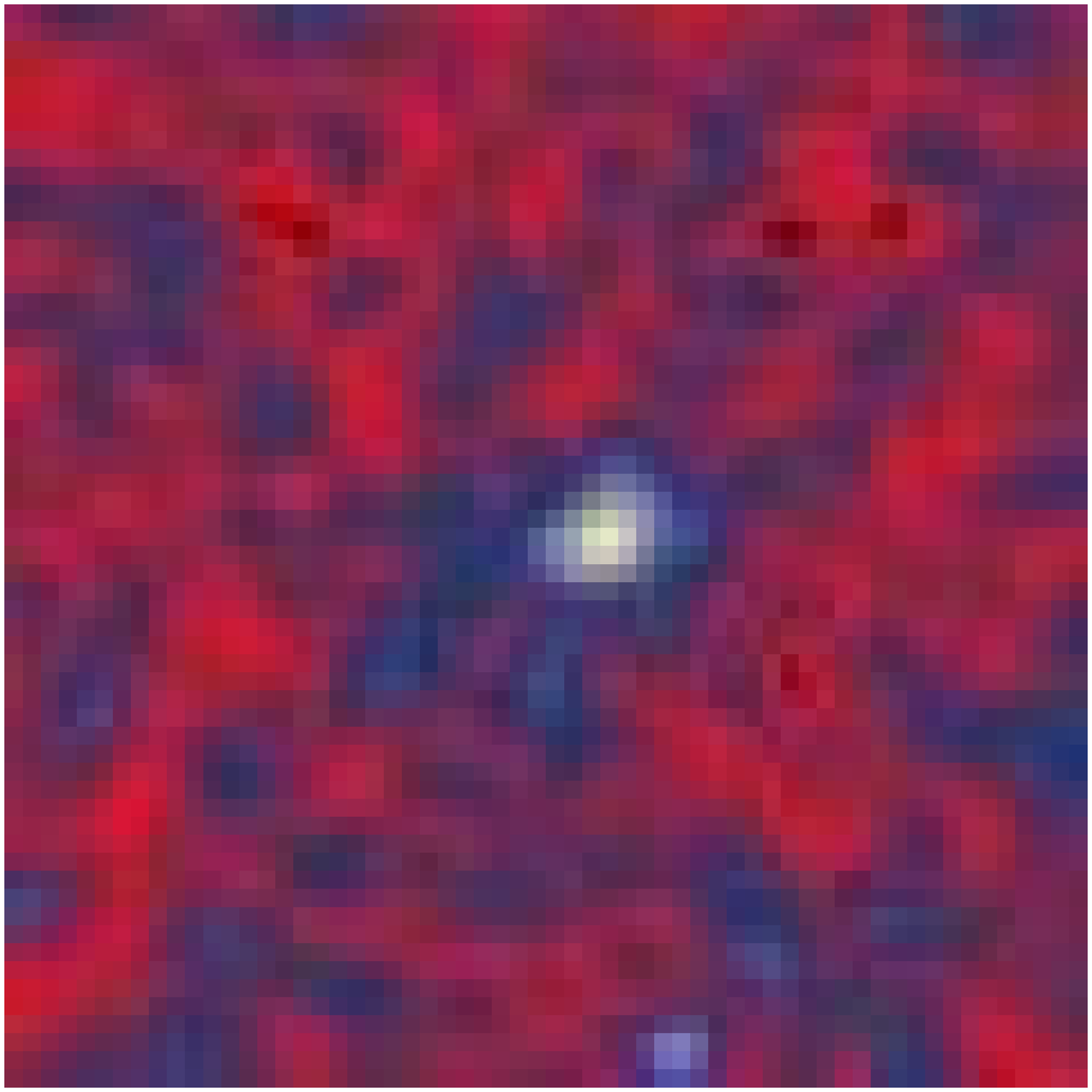}  \FGb{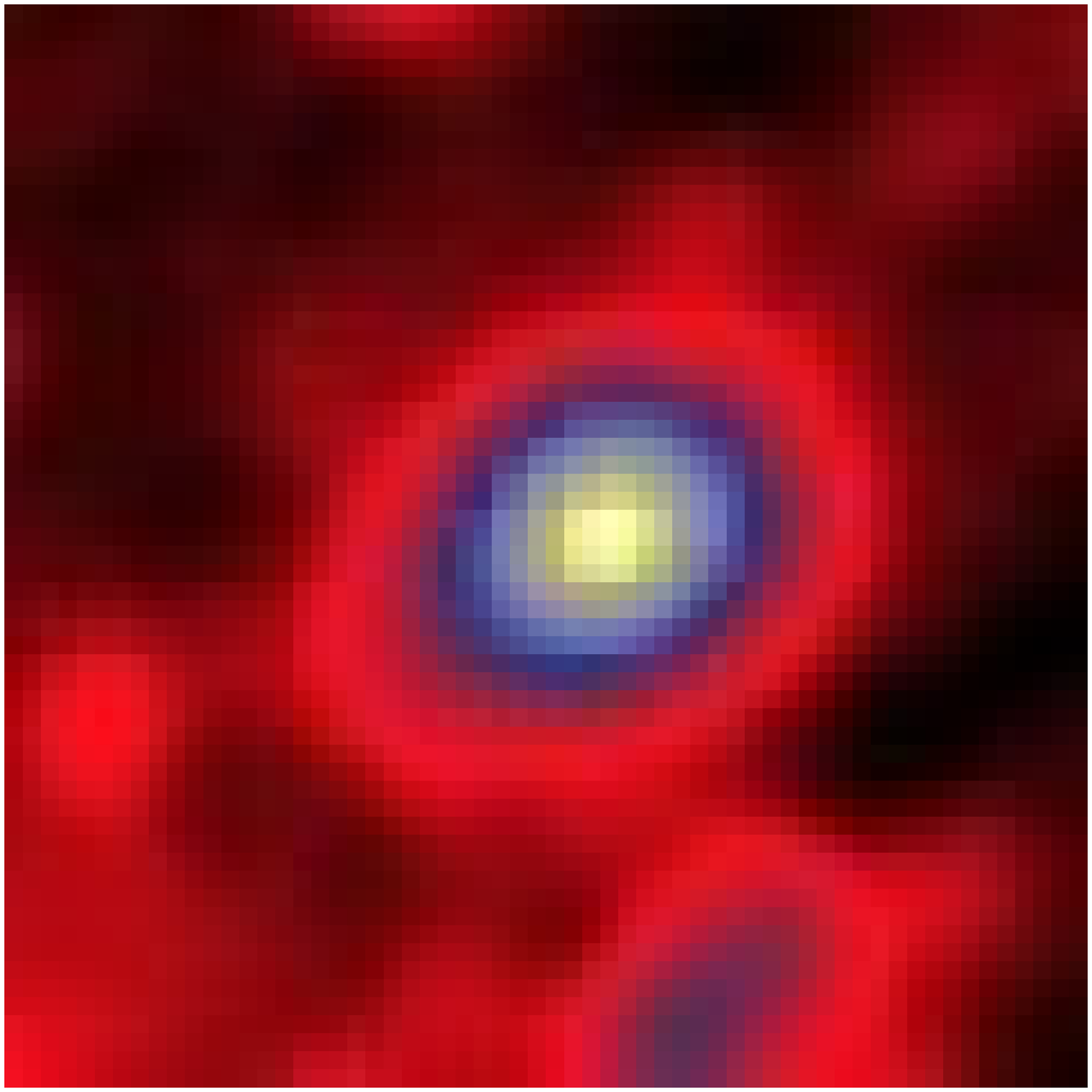}  \FGb{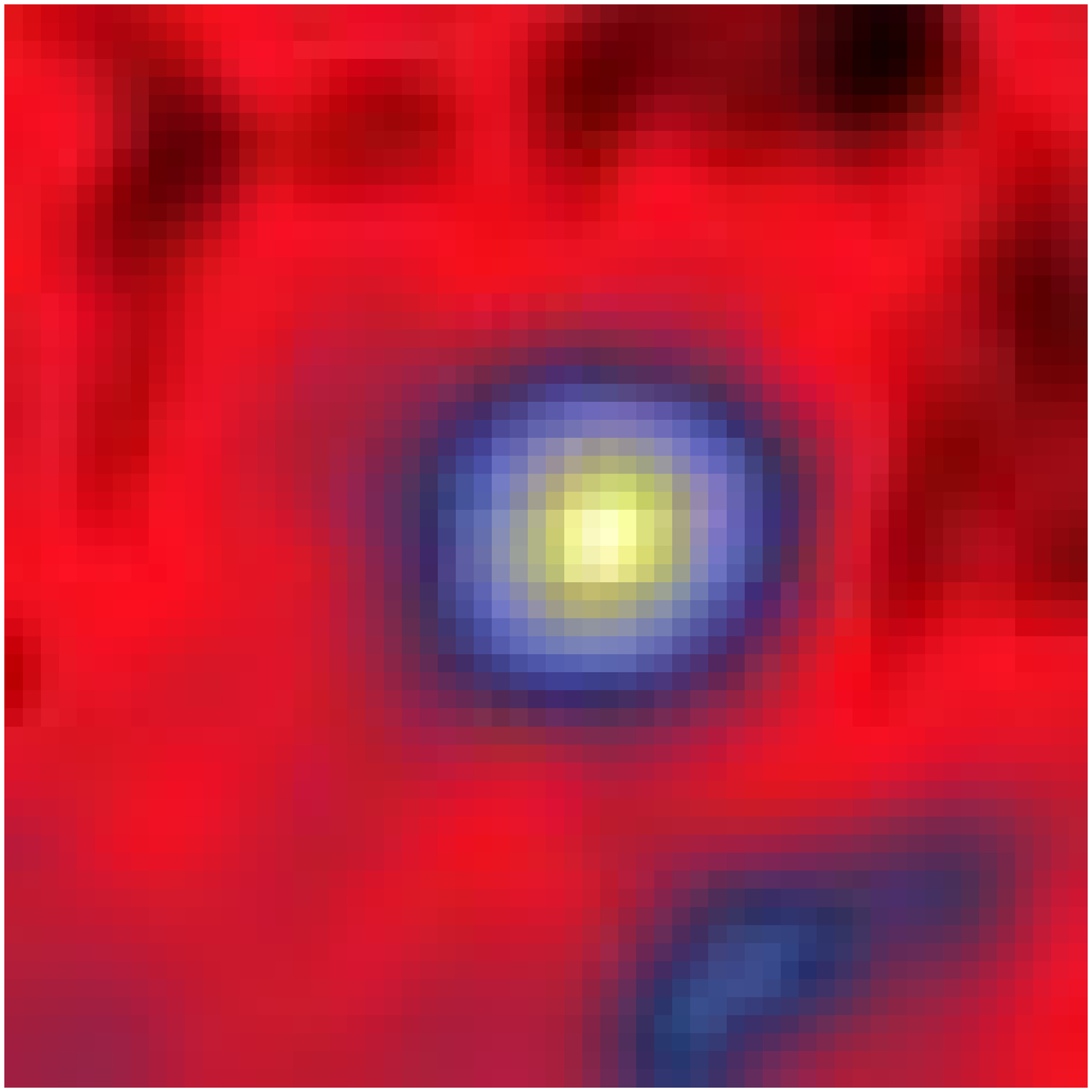}  \FGb{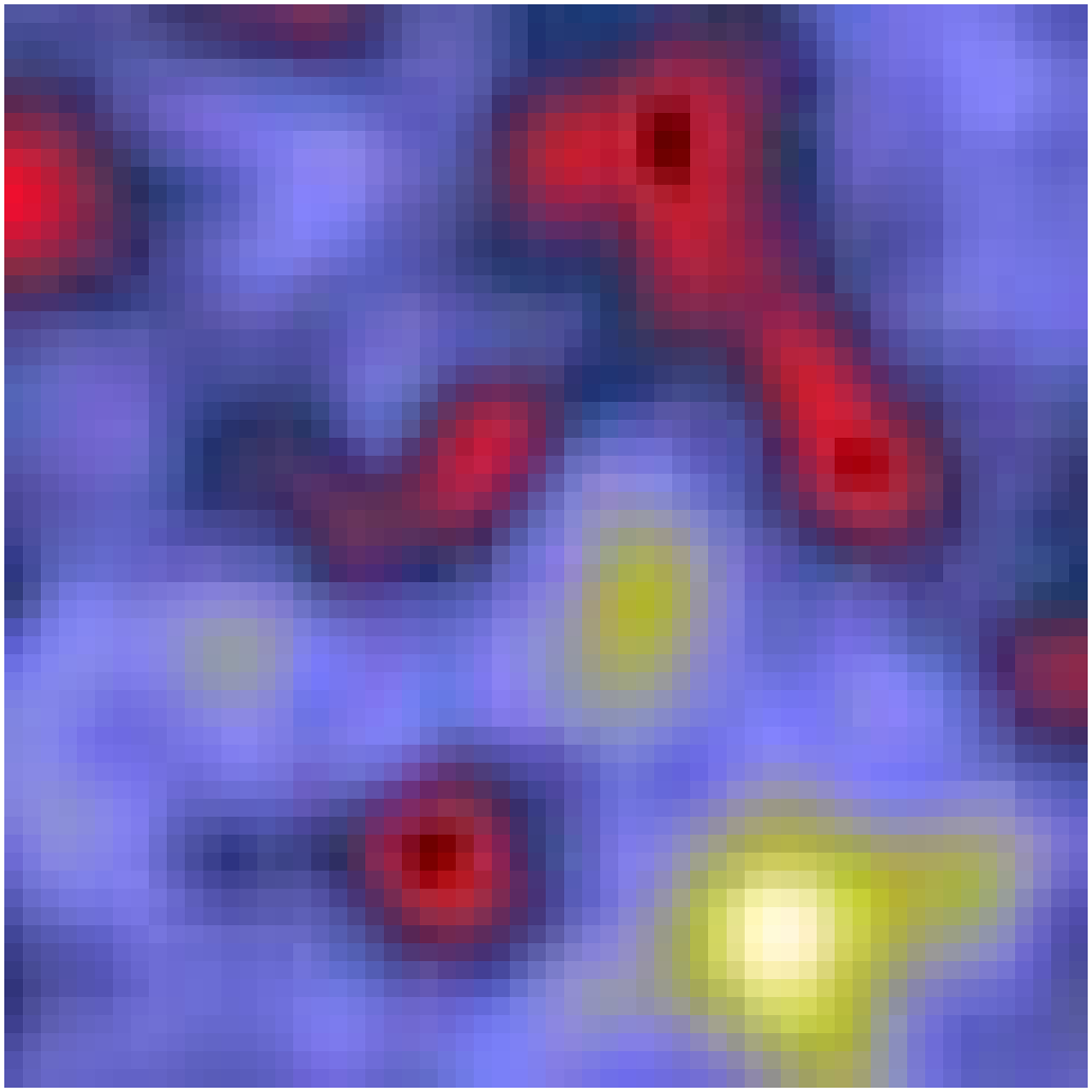}  \FGb{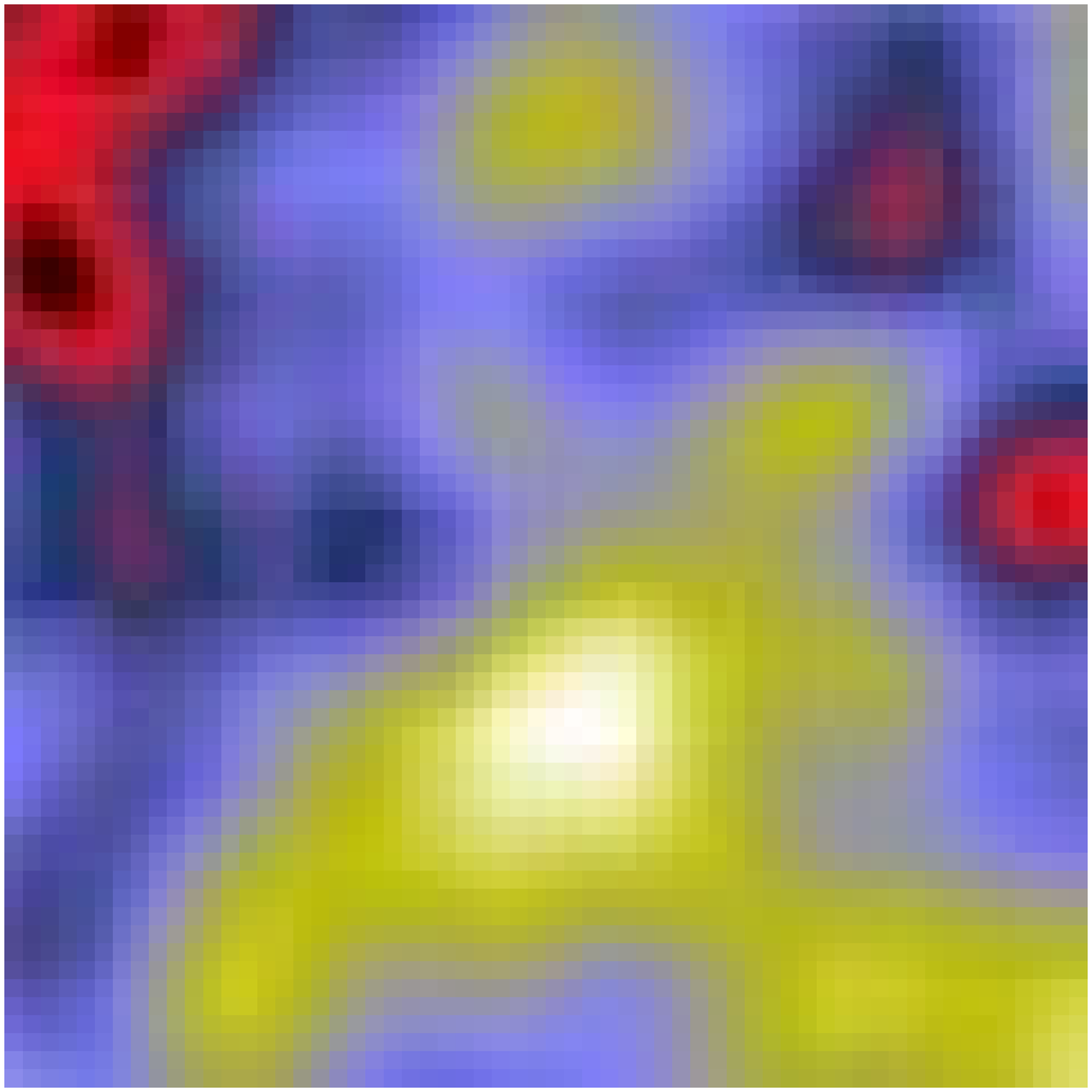} \\  {\#32   NCIA029870}  \\
  \FGb{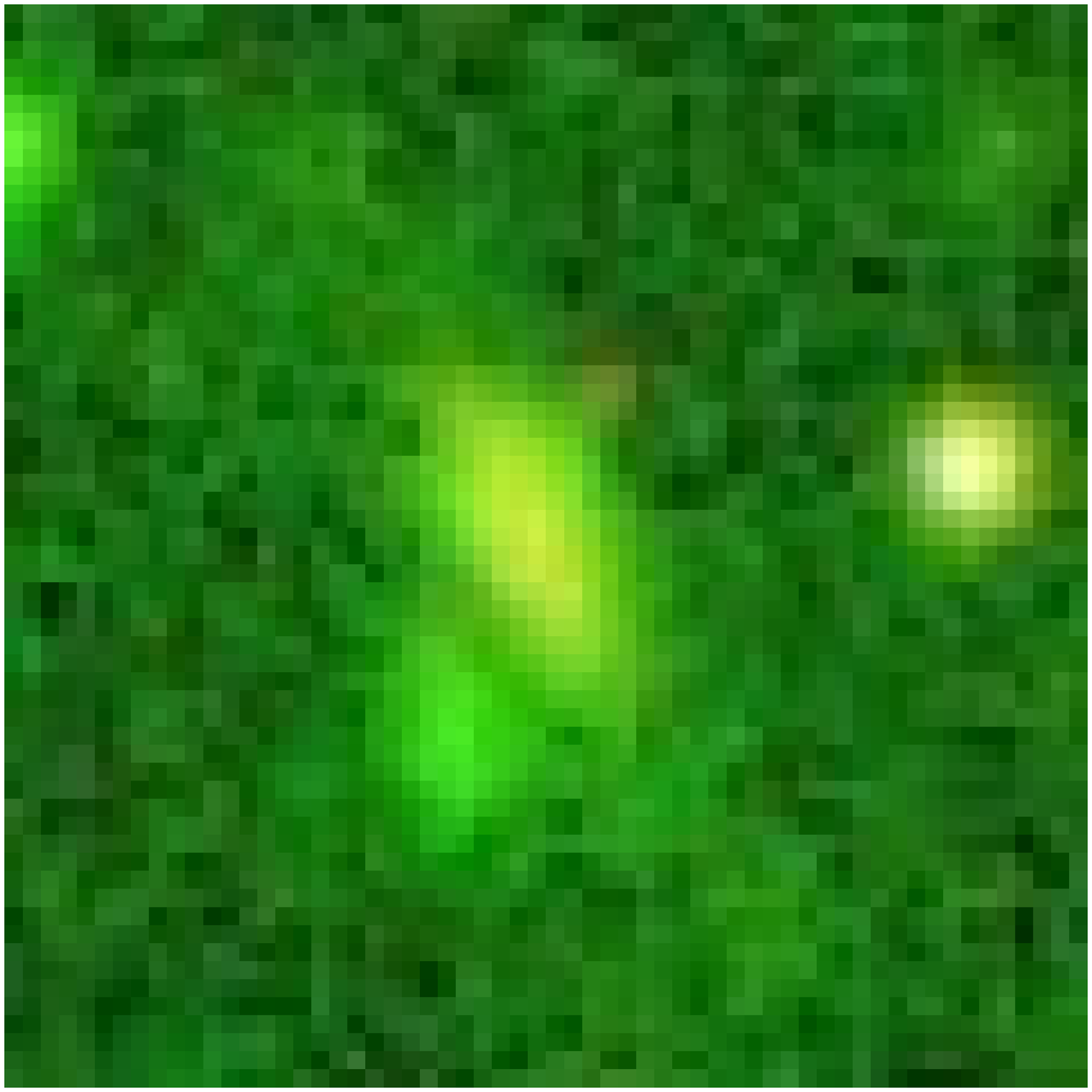}  \FGb{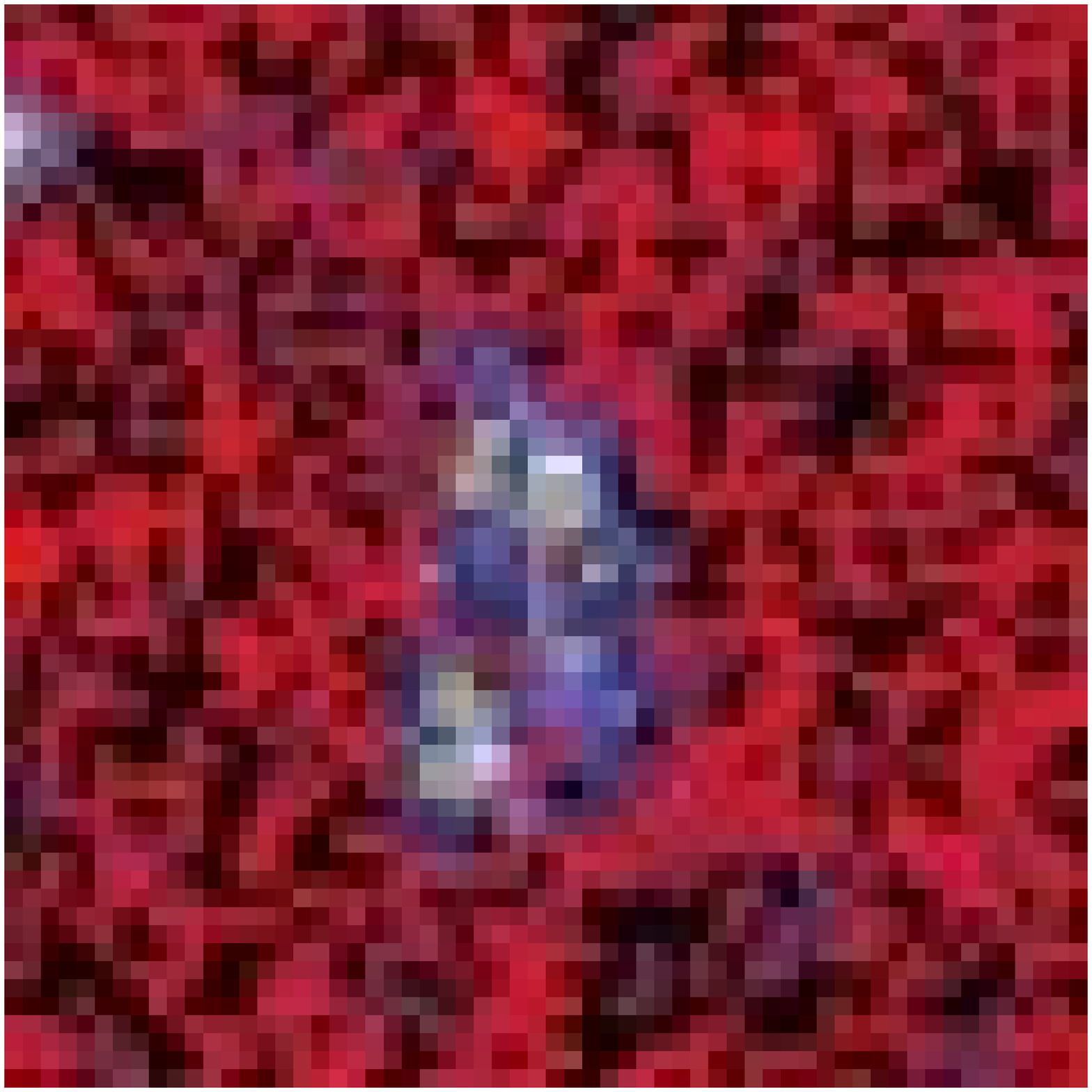}  \FGb{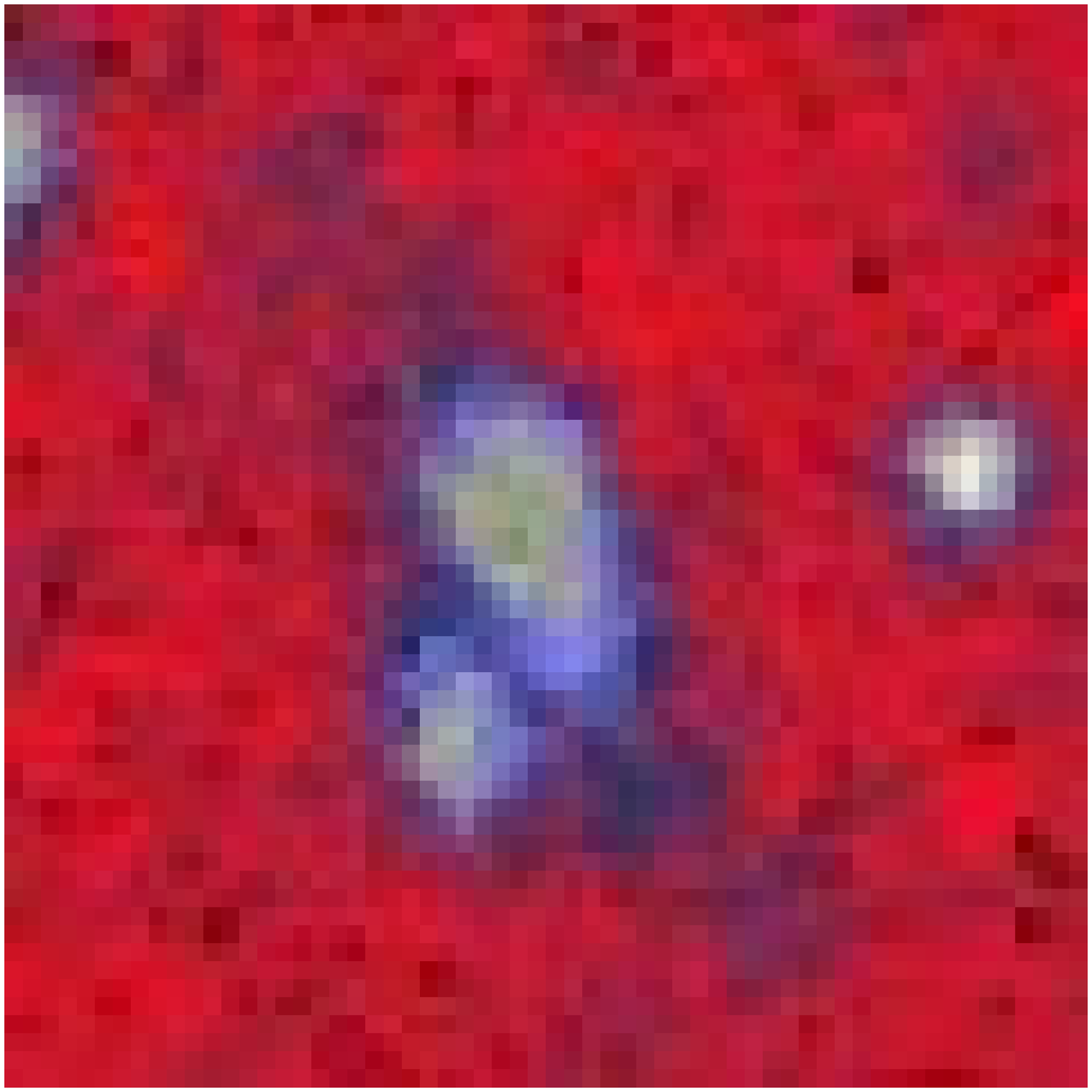}  \FGb{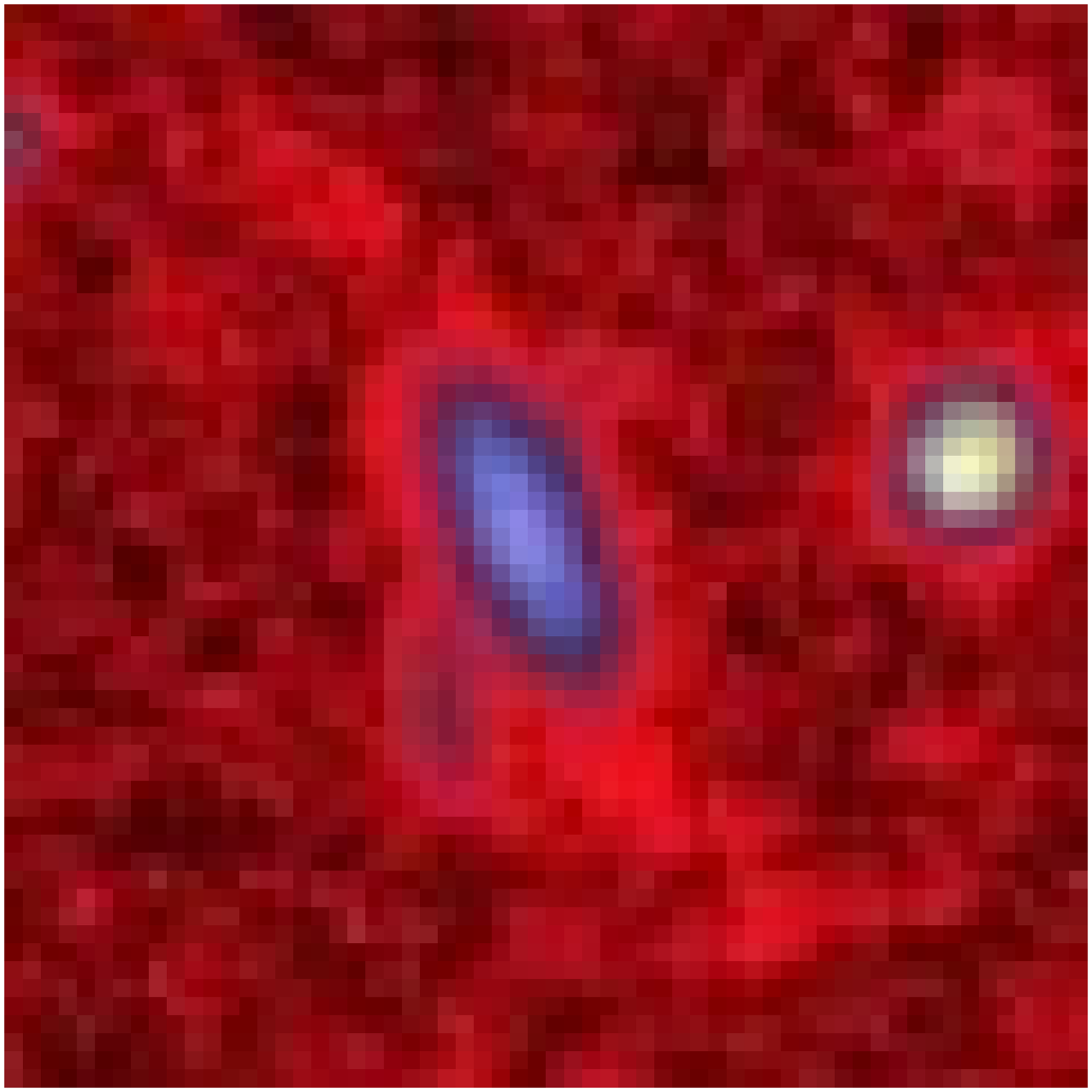}  \FGb{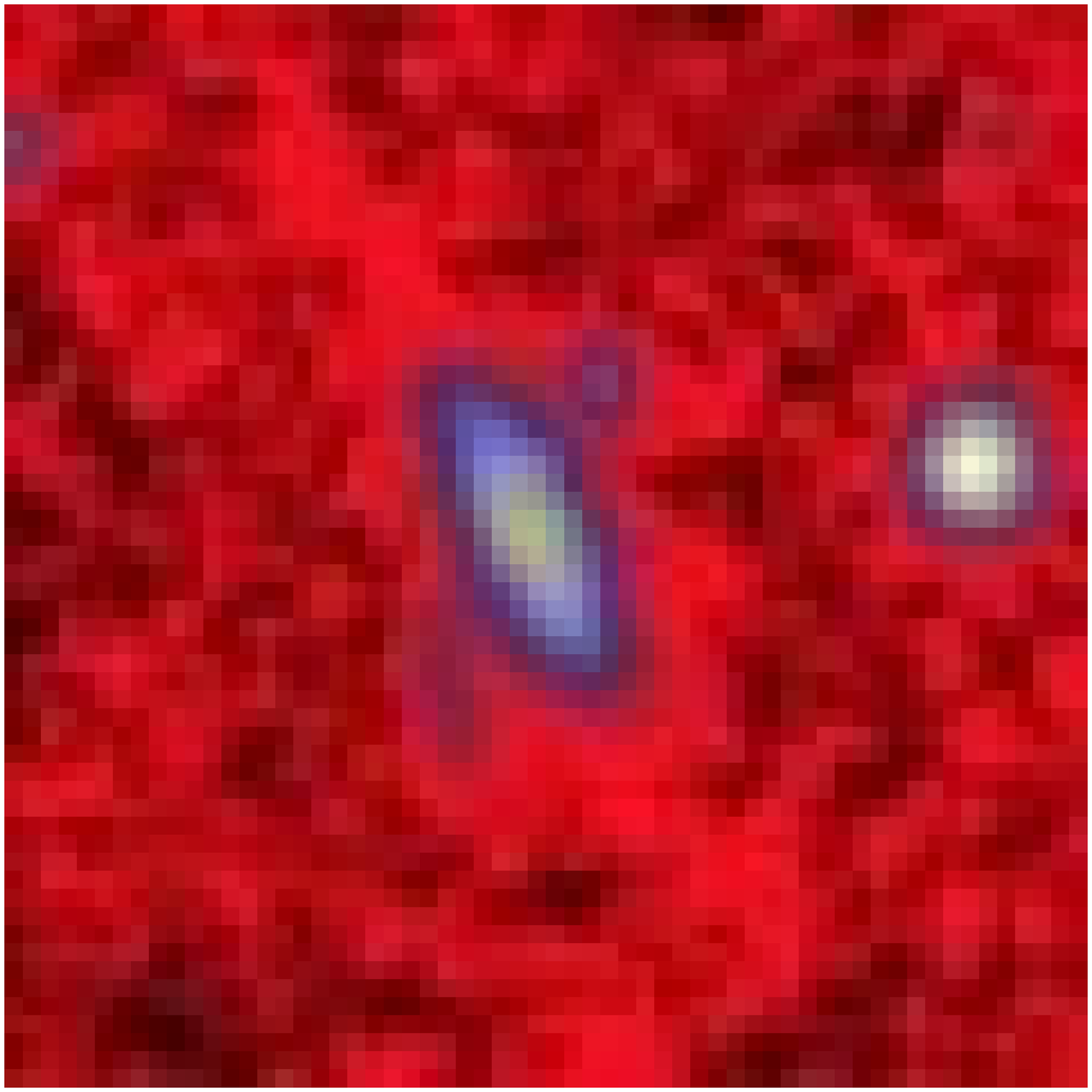}  \FGb{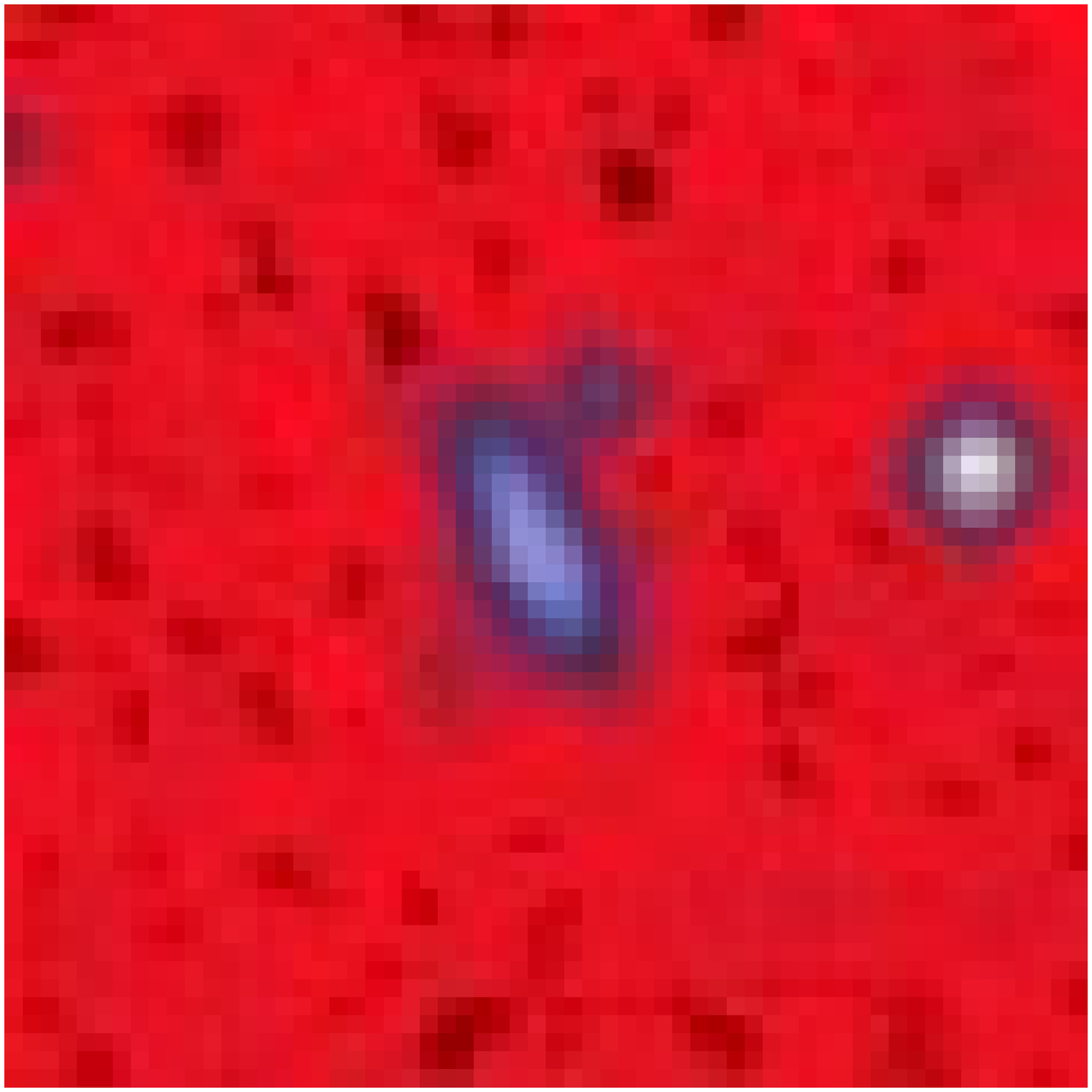}  \FGb{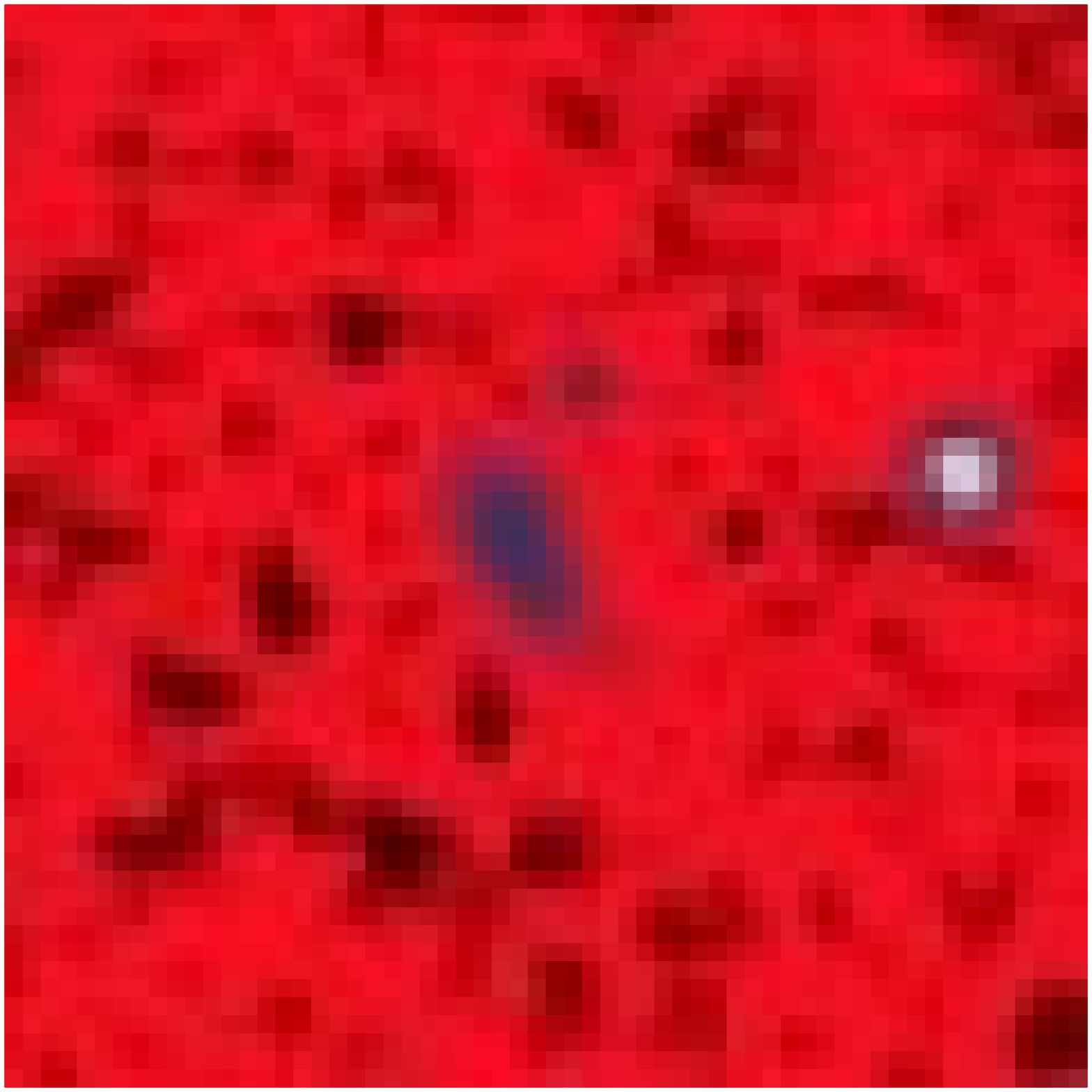}  \FGb{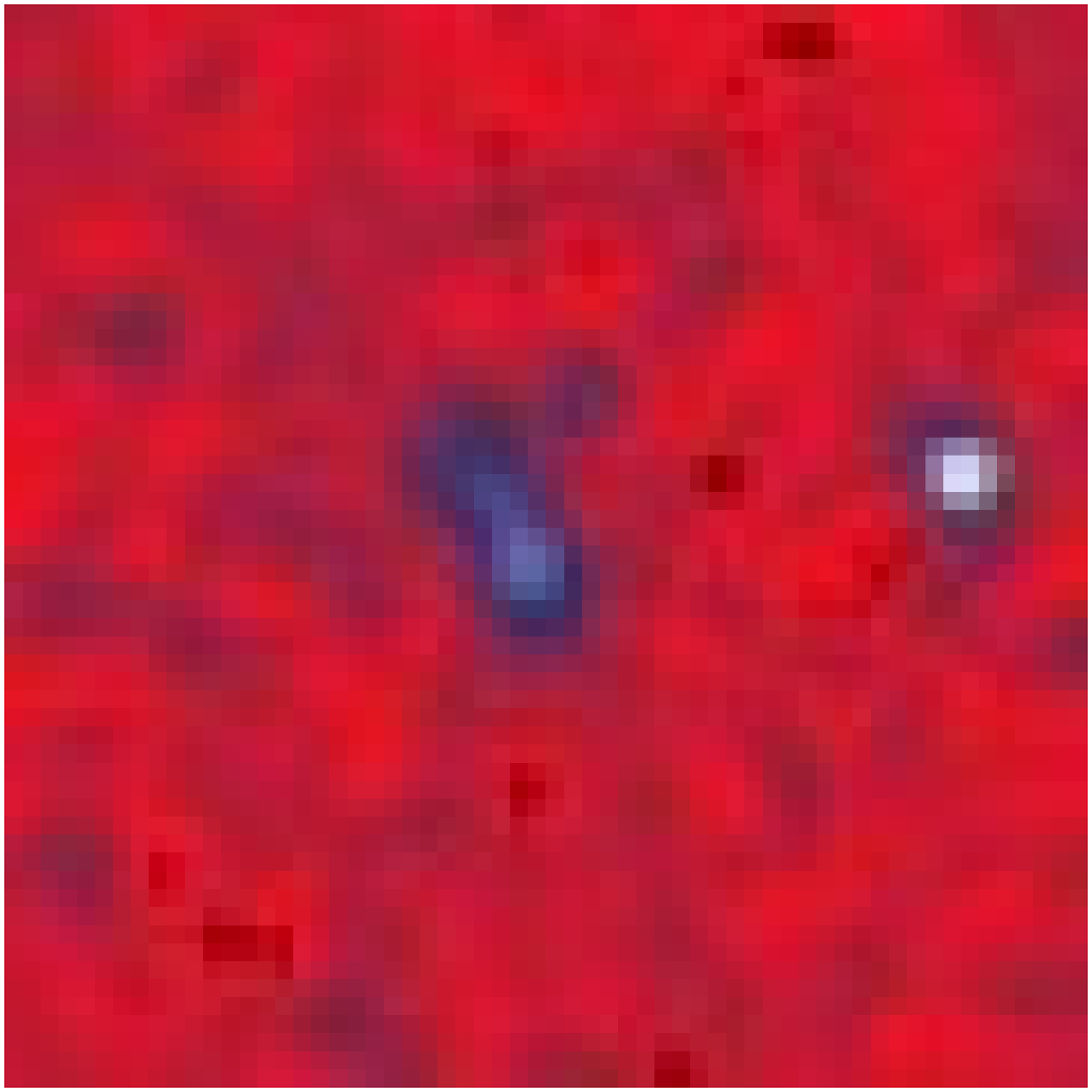}  \FGb{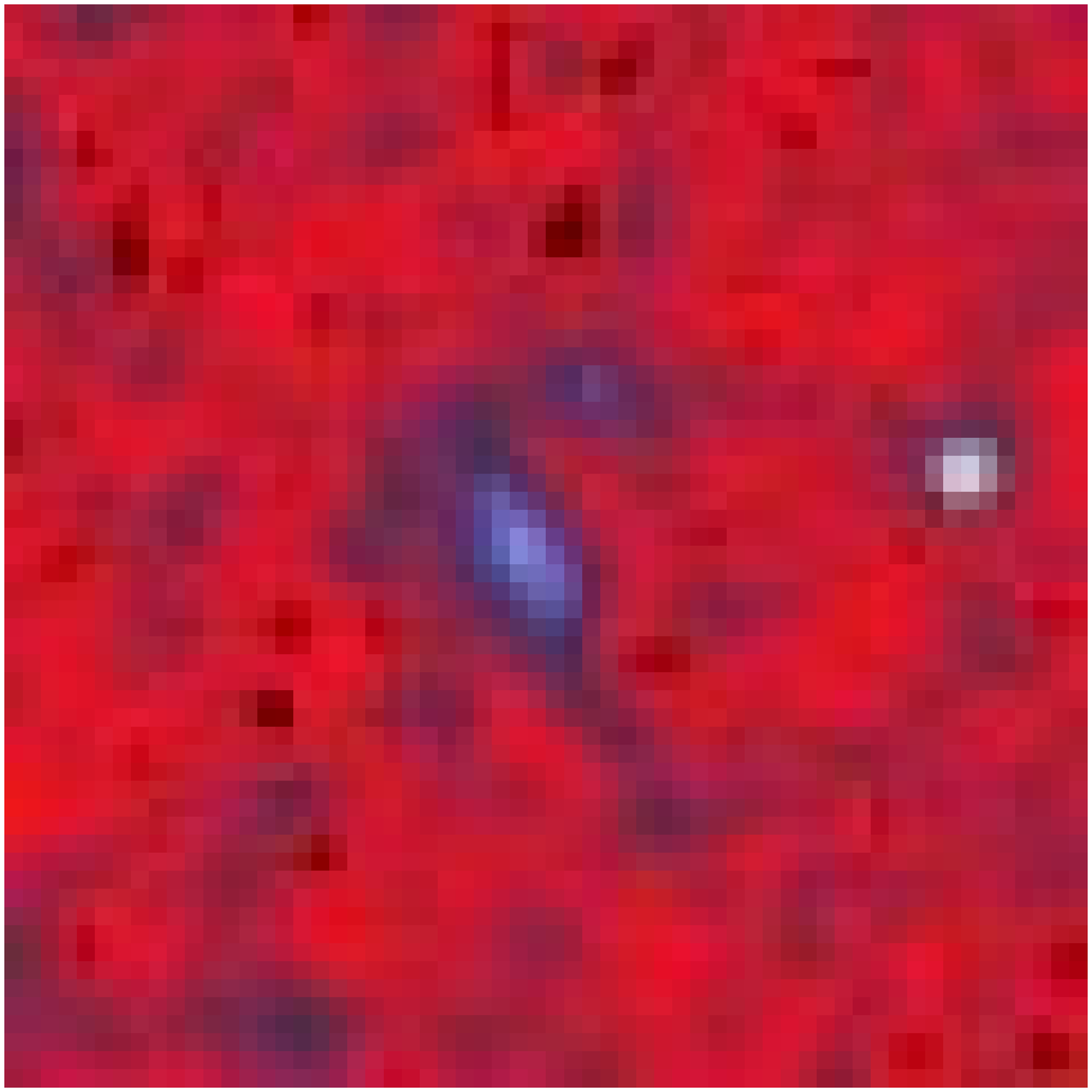}  \FGb{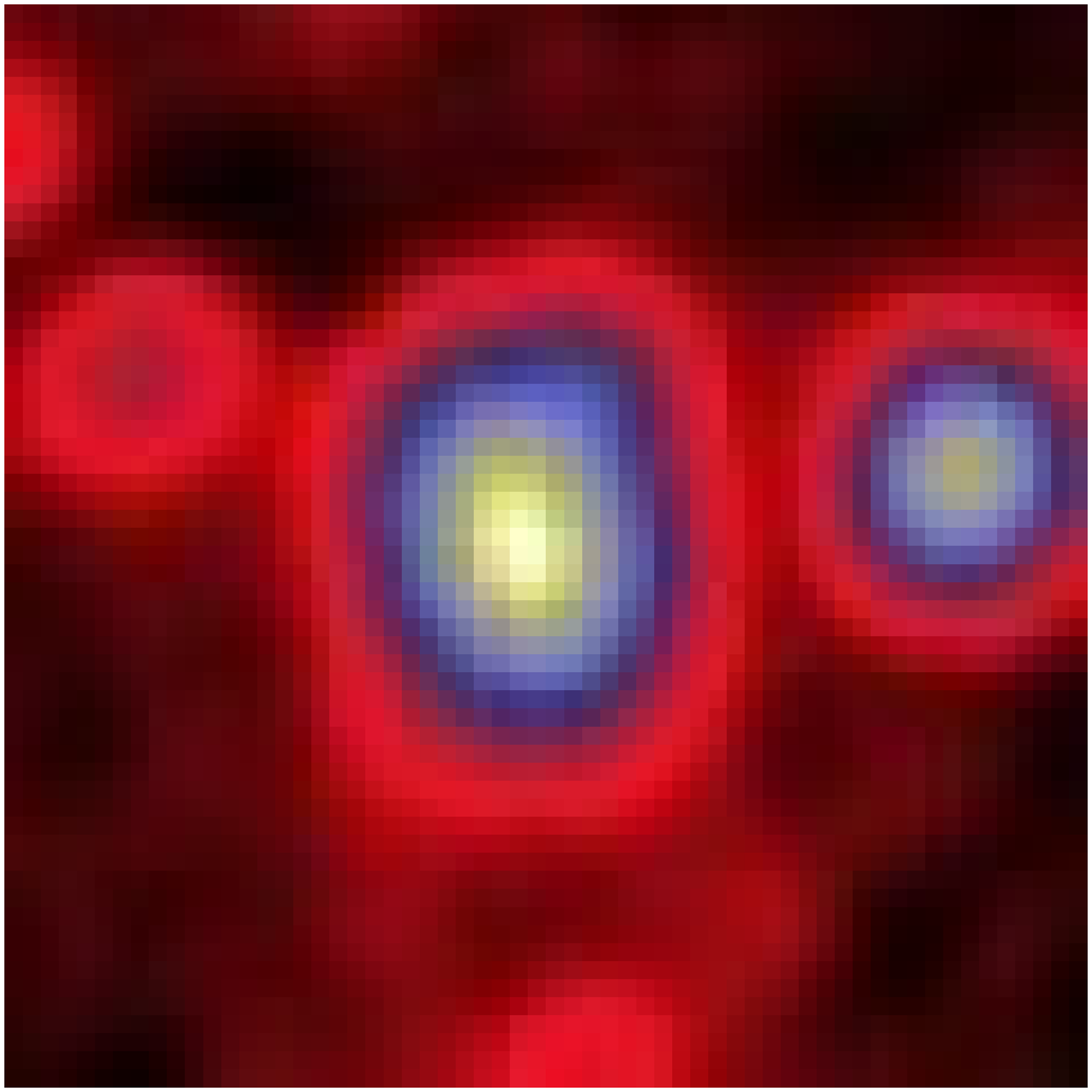}  \FGb{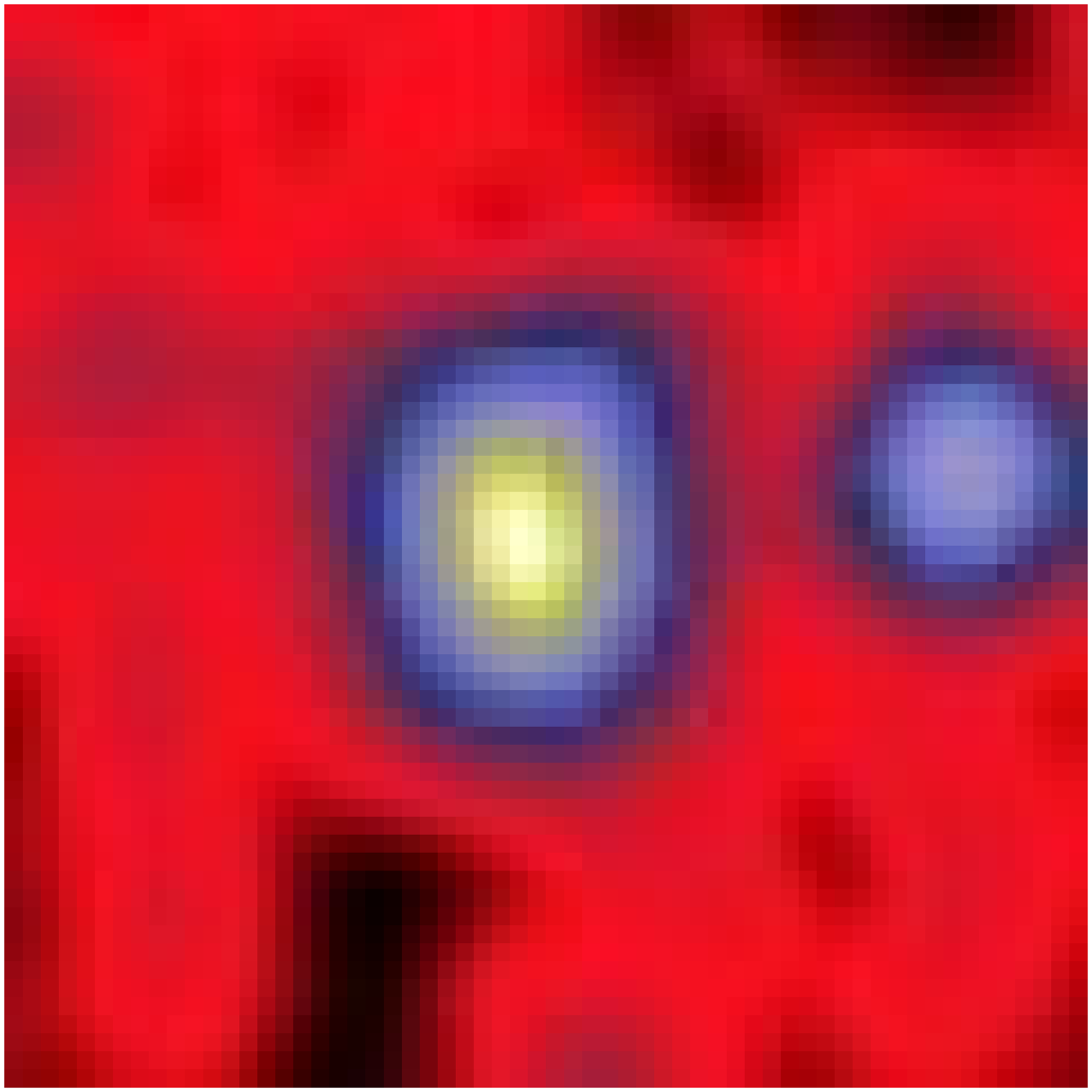}  \FGb{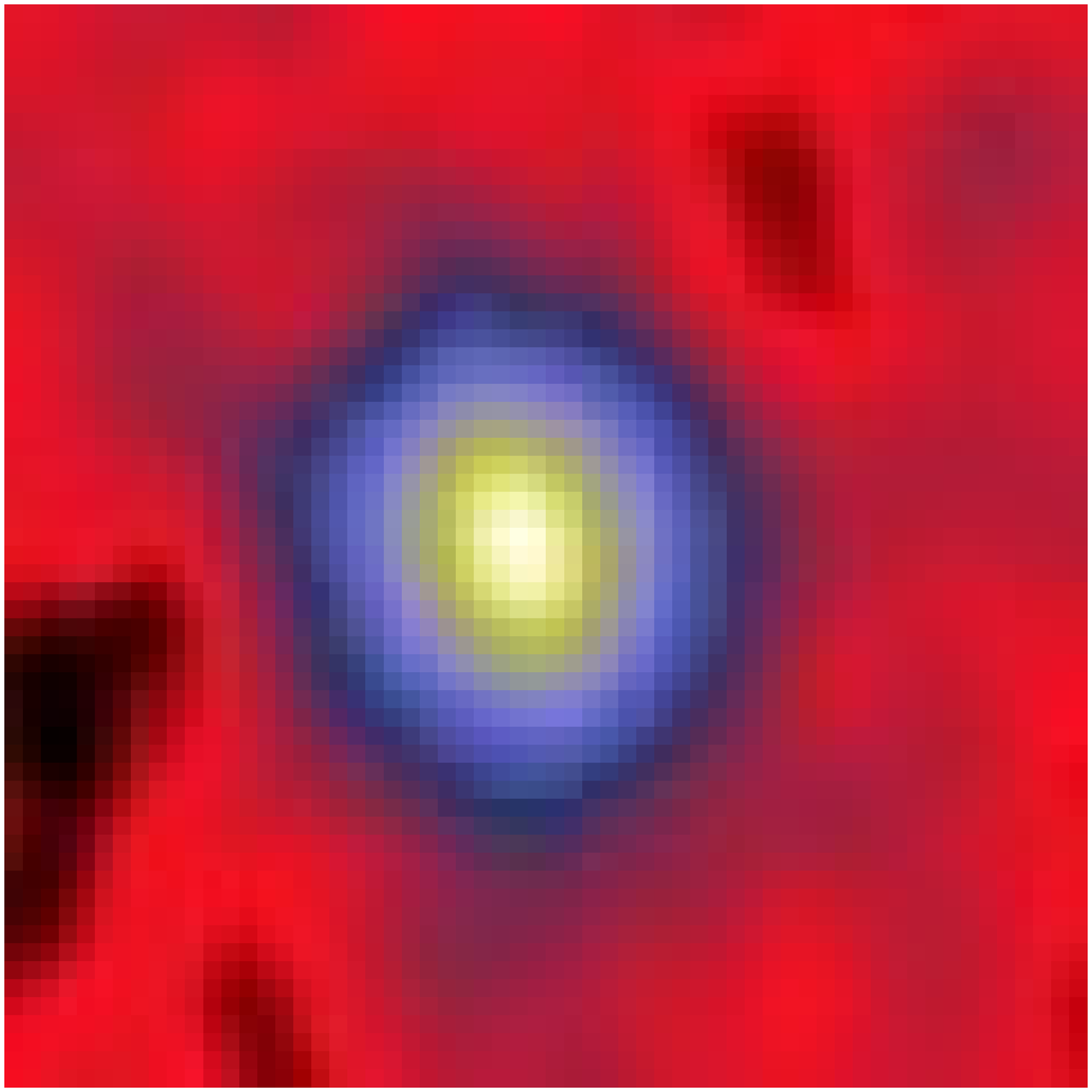}  \FGb{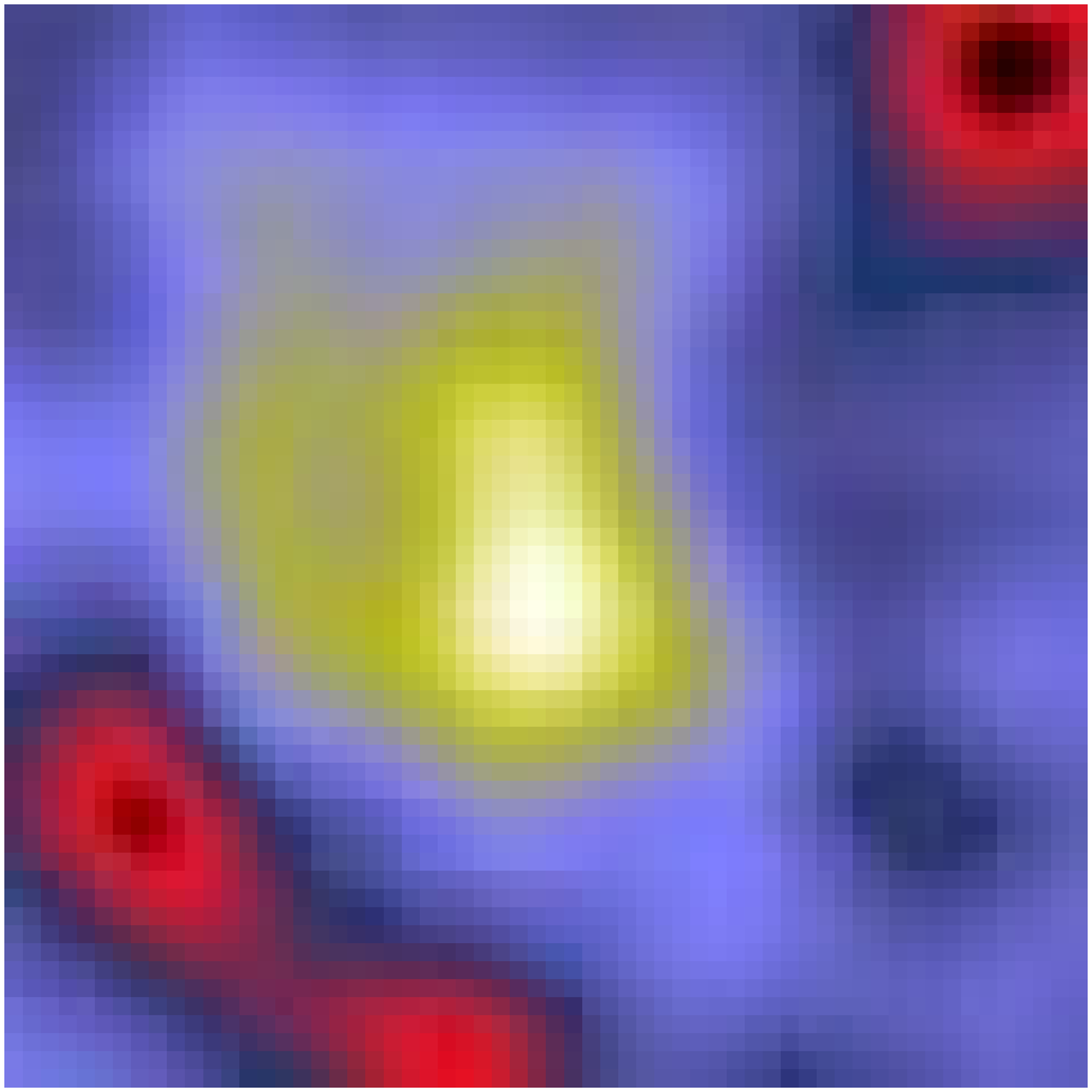} \\  {\#33   NCIA010841}  \\
  \FGb{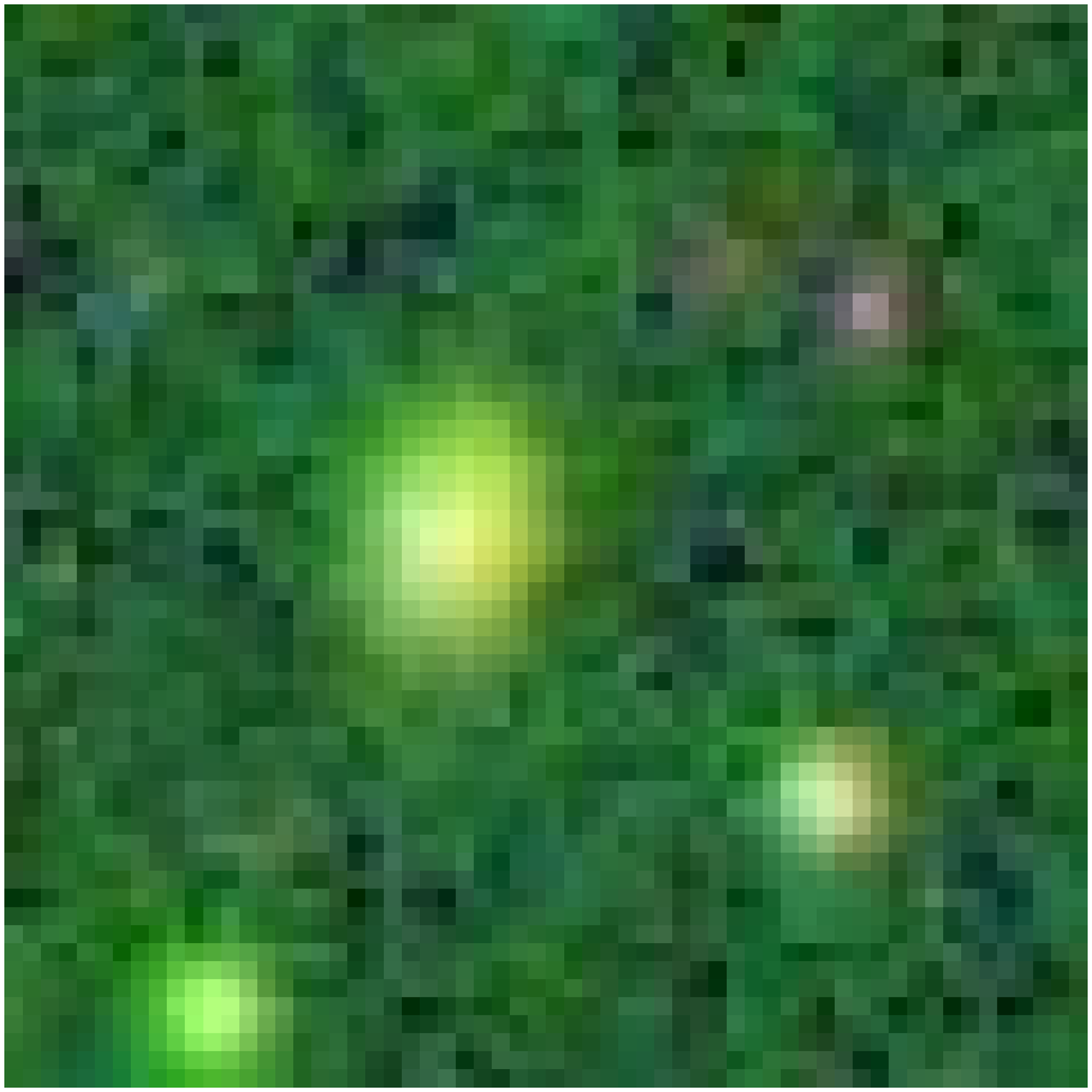}  \FGb{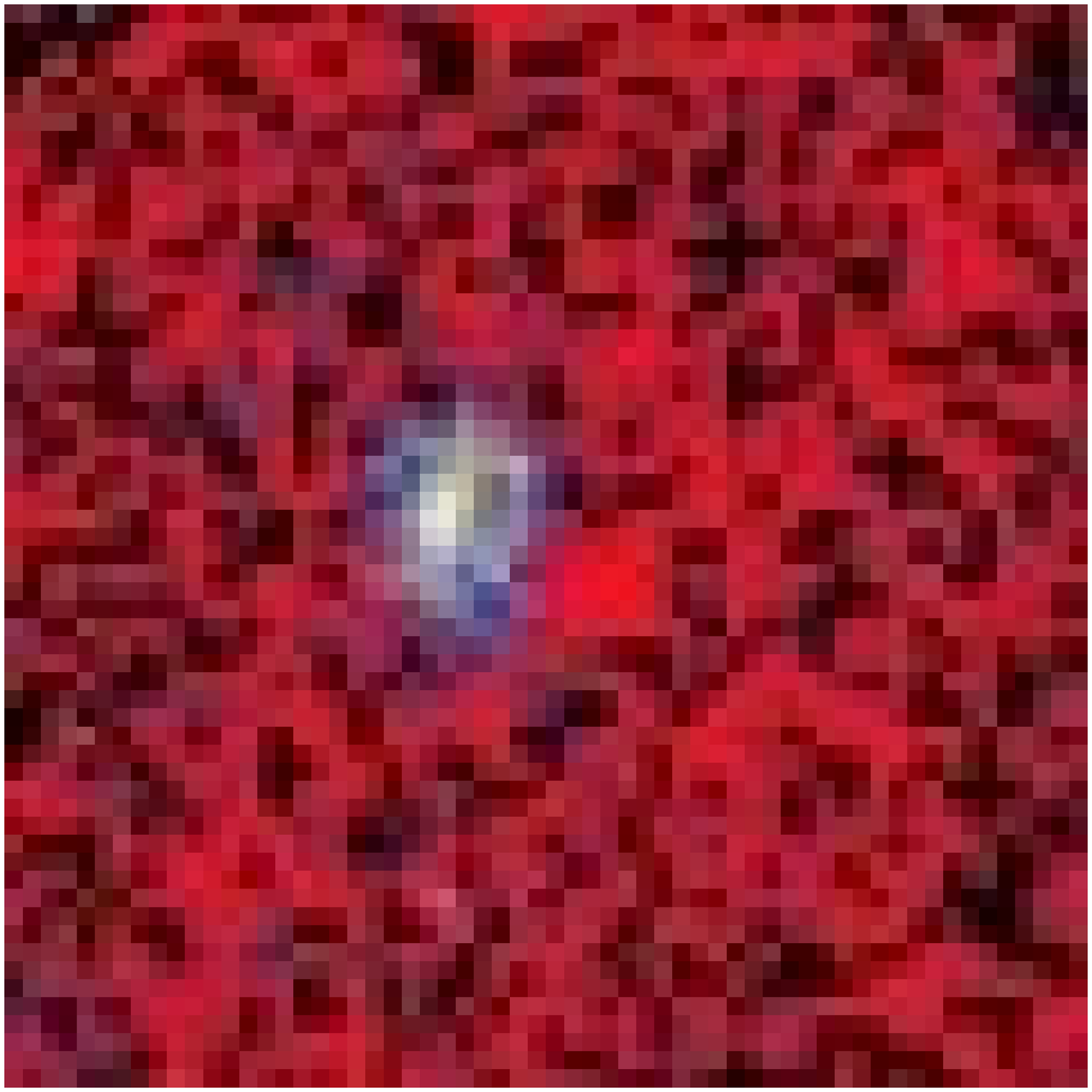}  \FGb{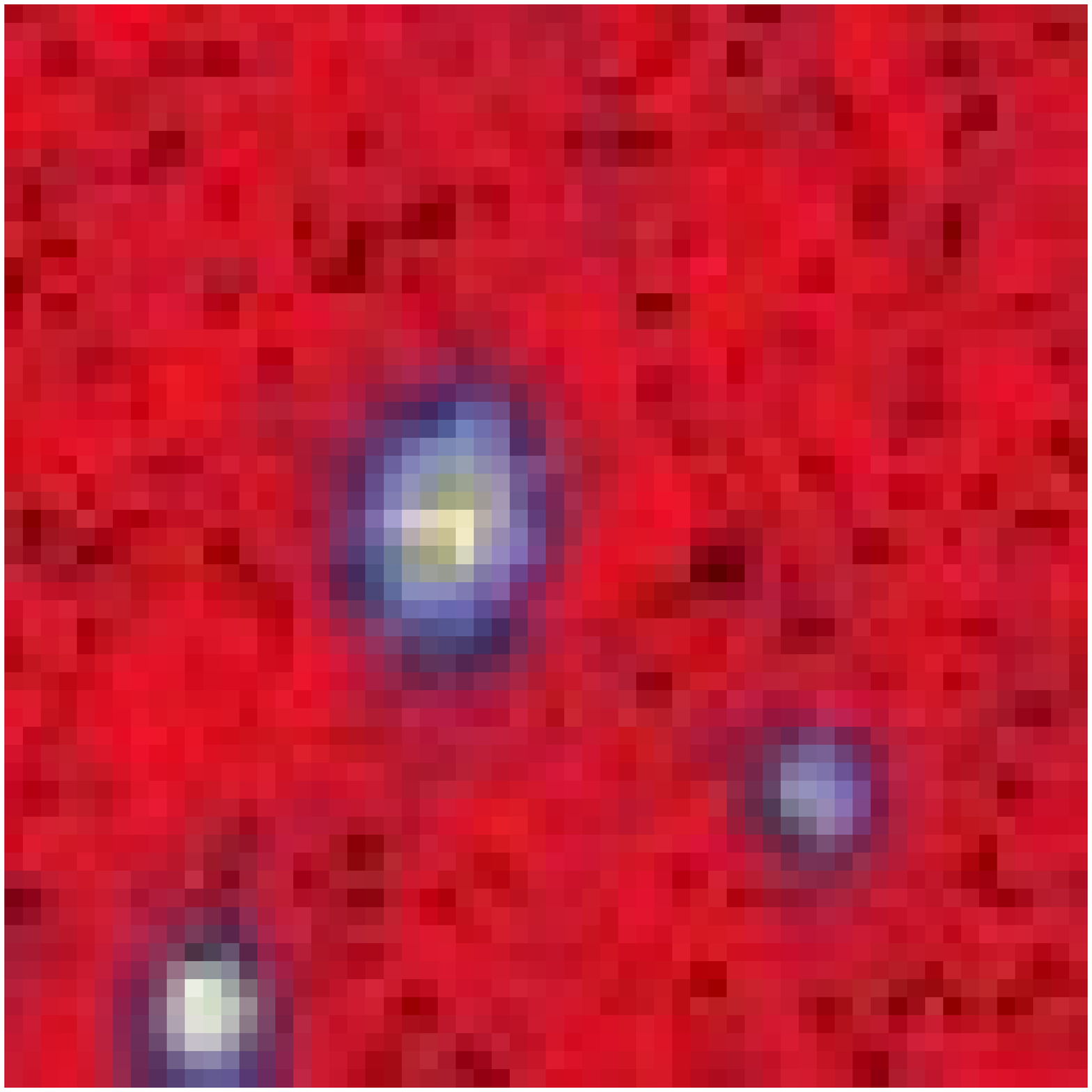}  \FGb{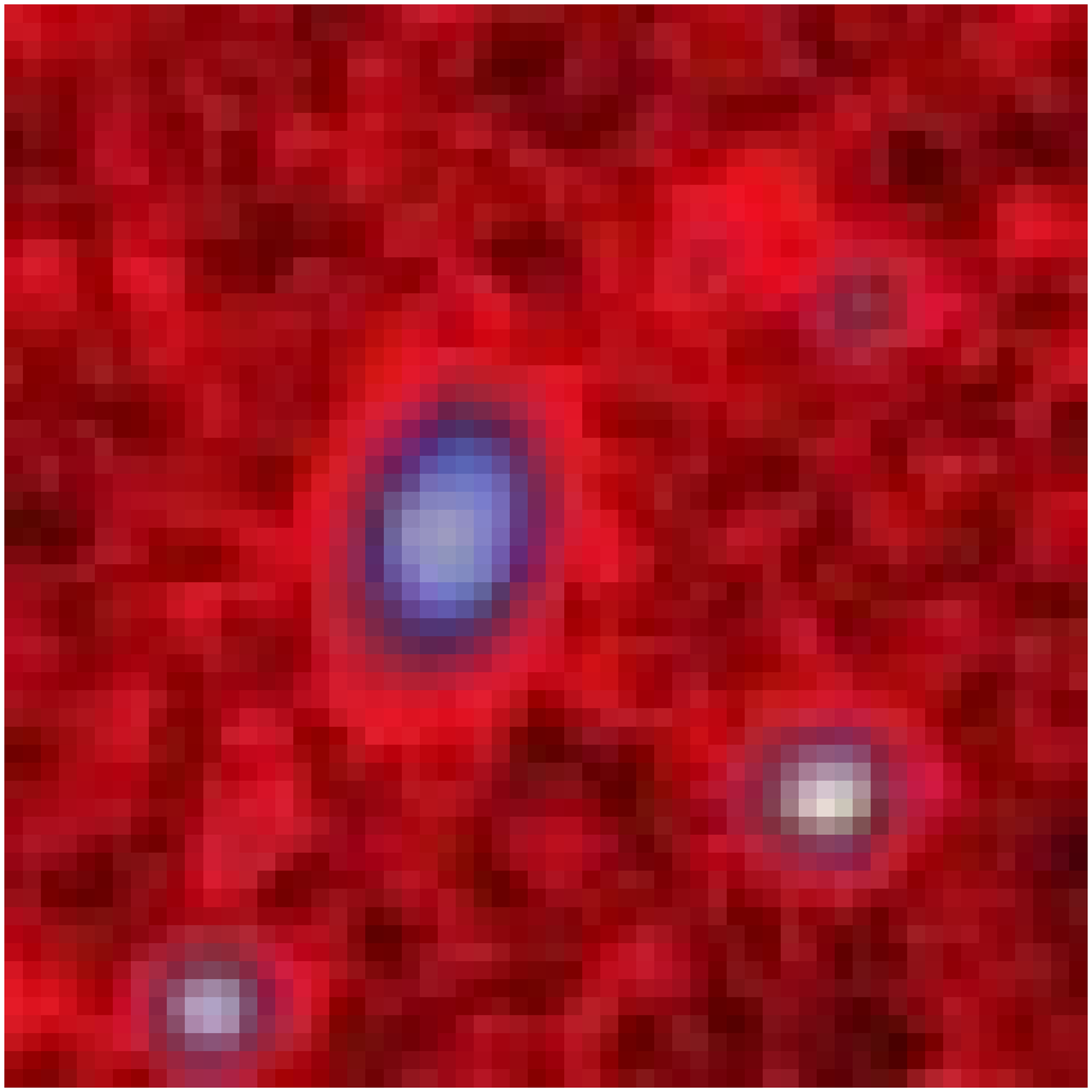}  \FGb{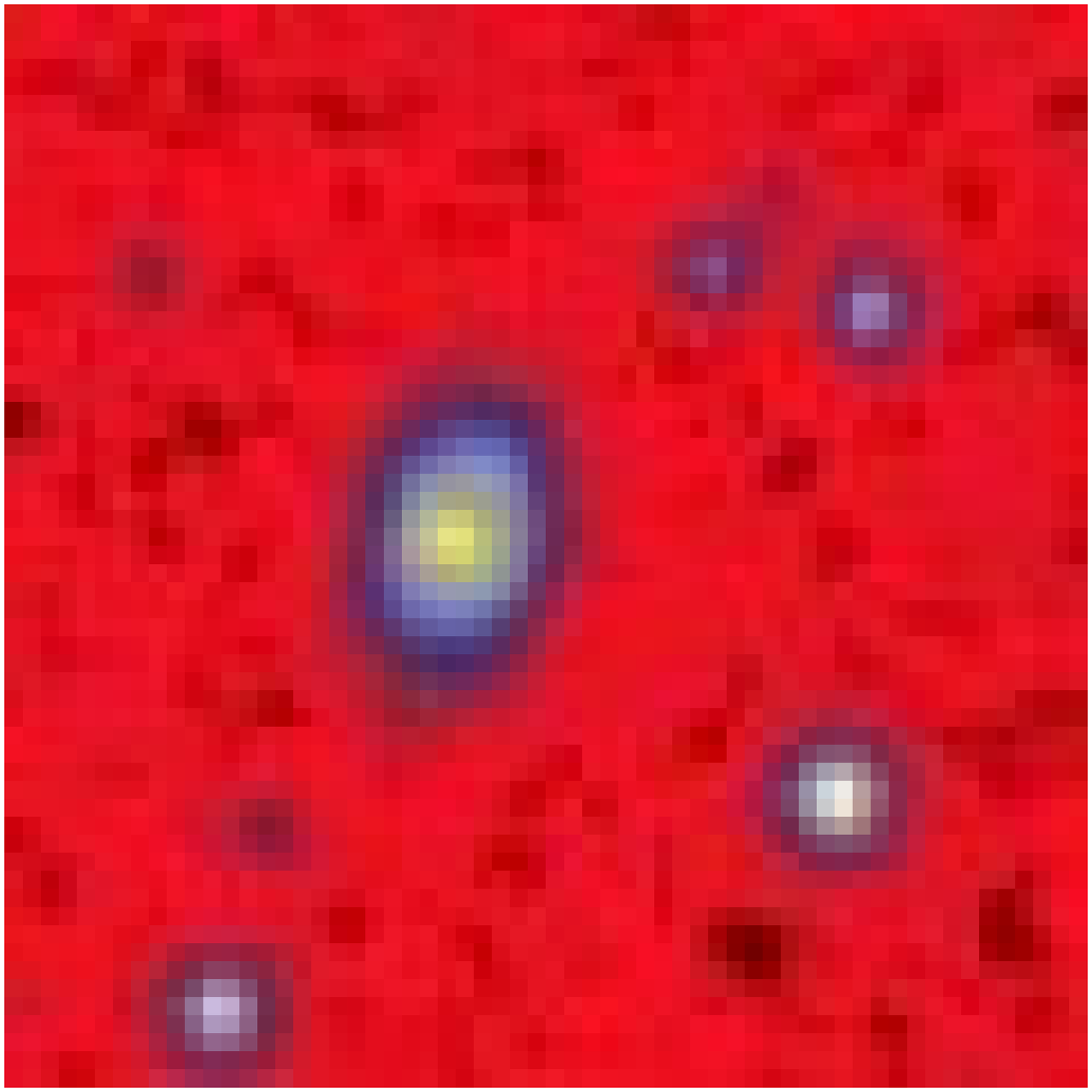}  \FGb{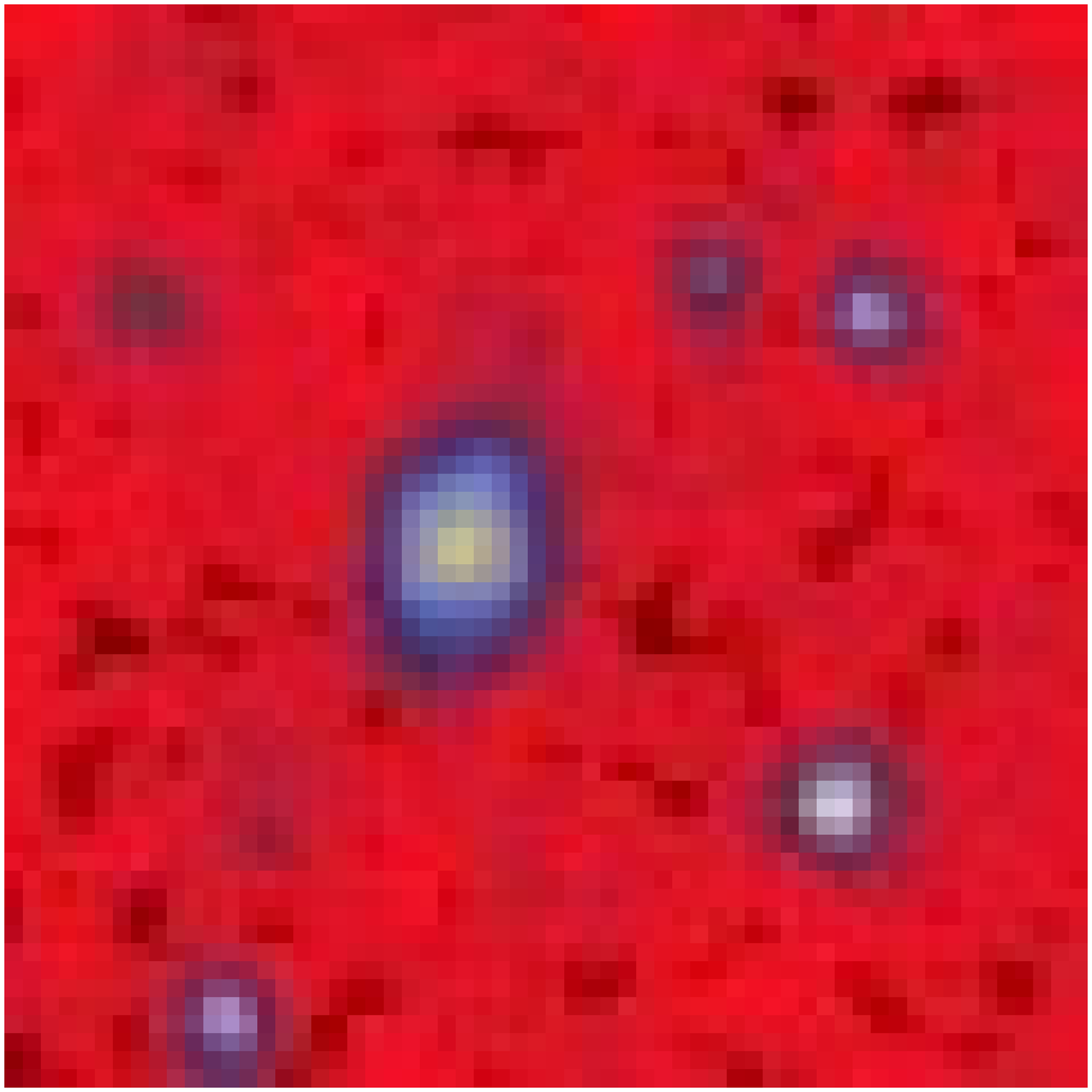}  \FGb{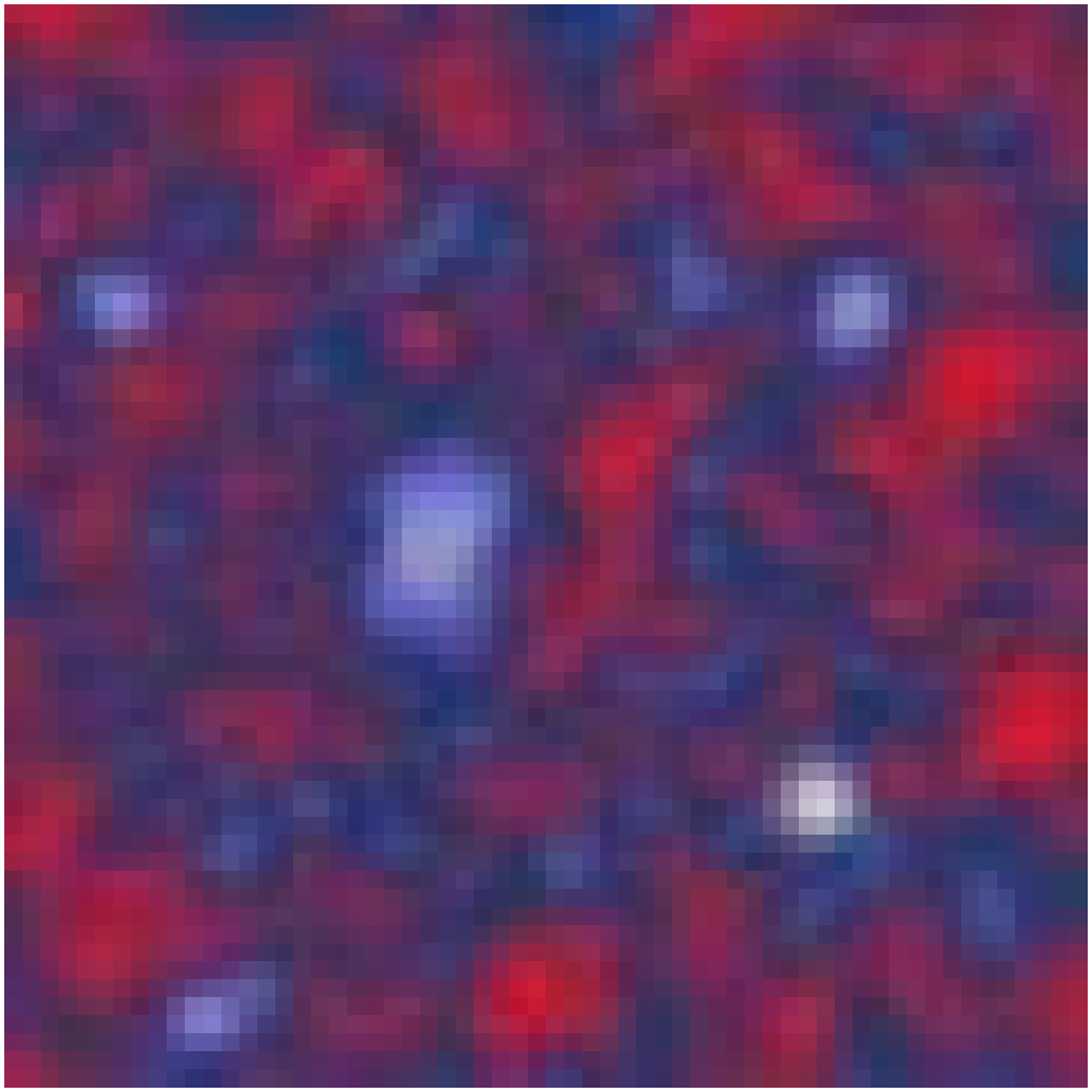}  \FGb{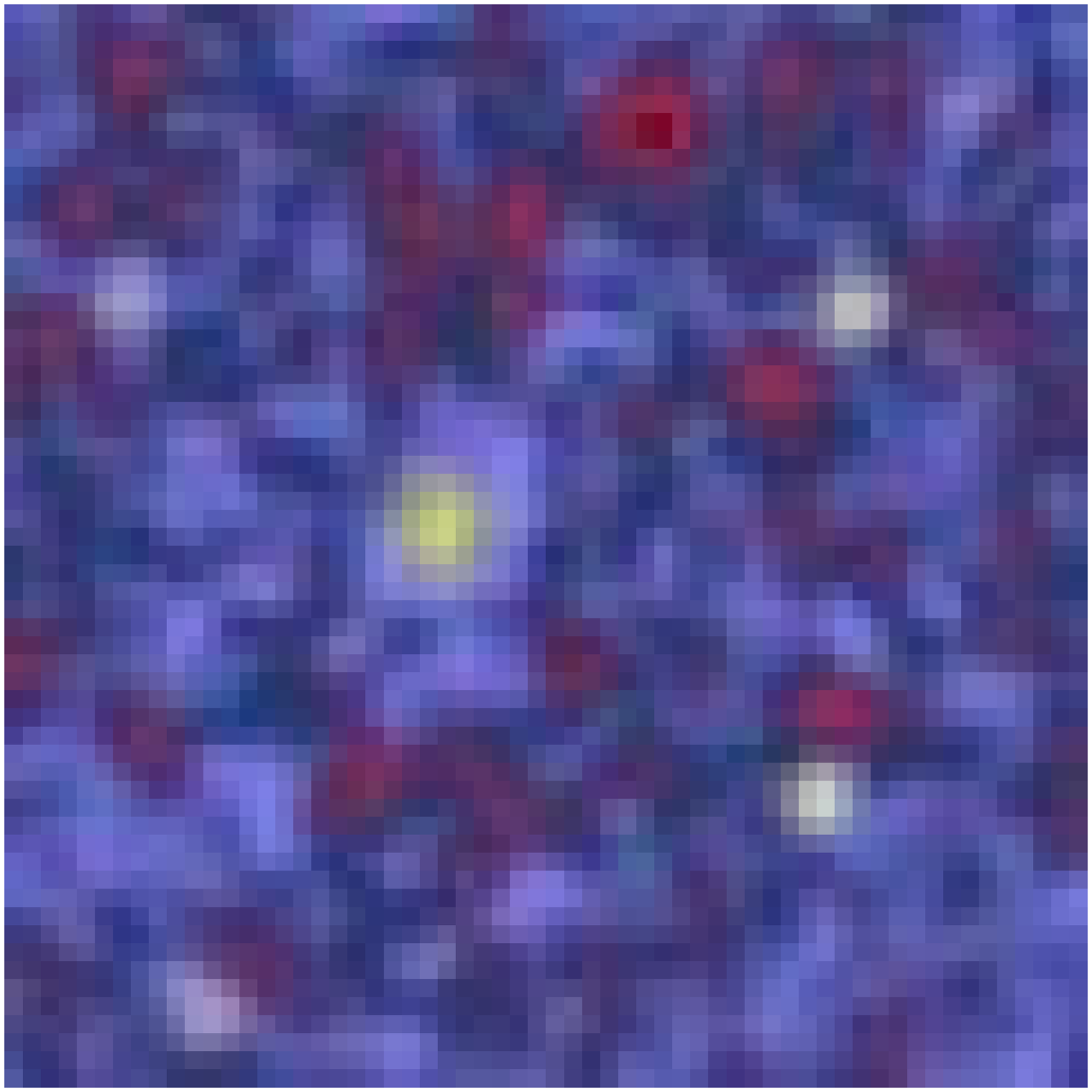}  \FGb{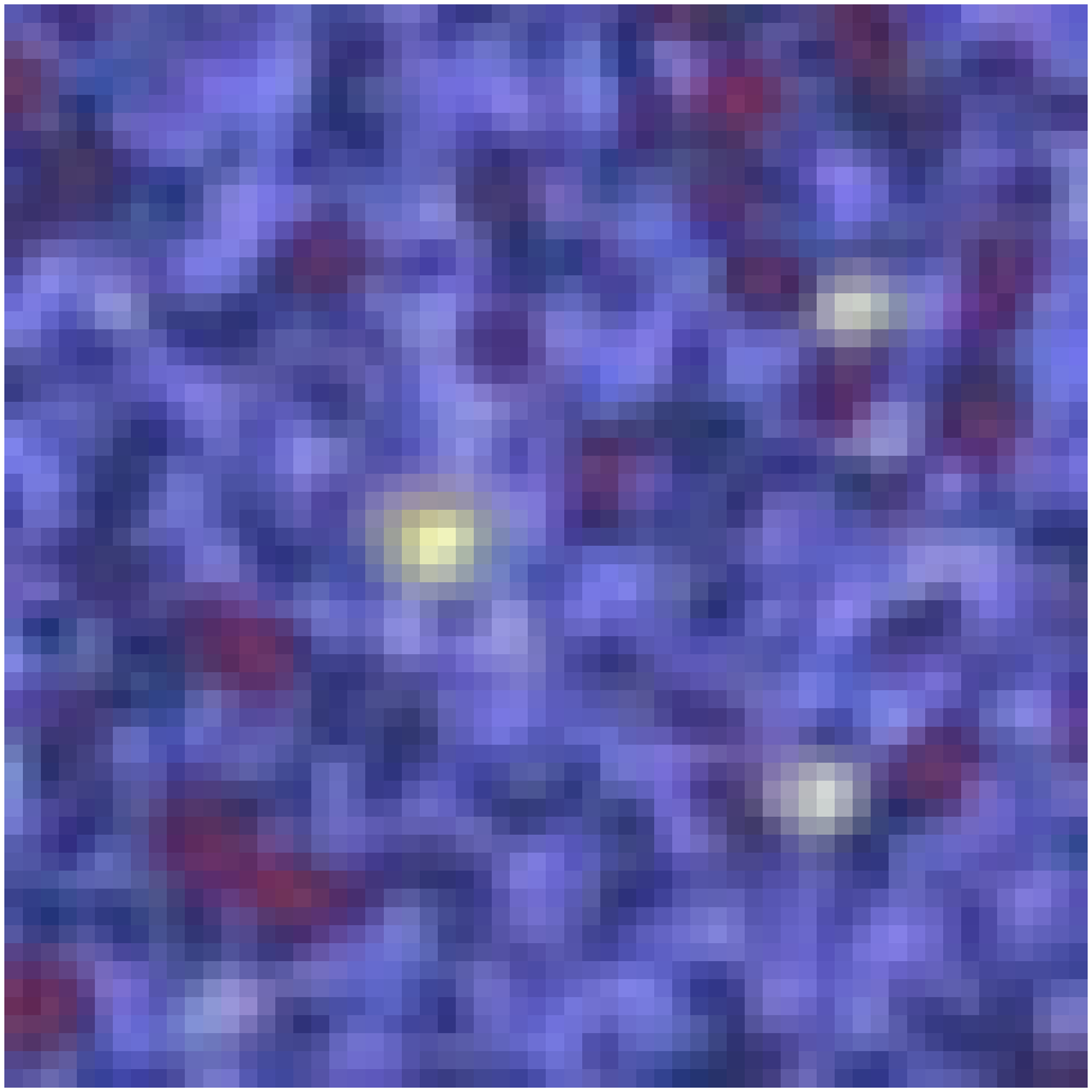}  \FGb{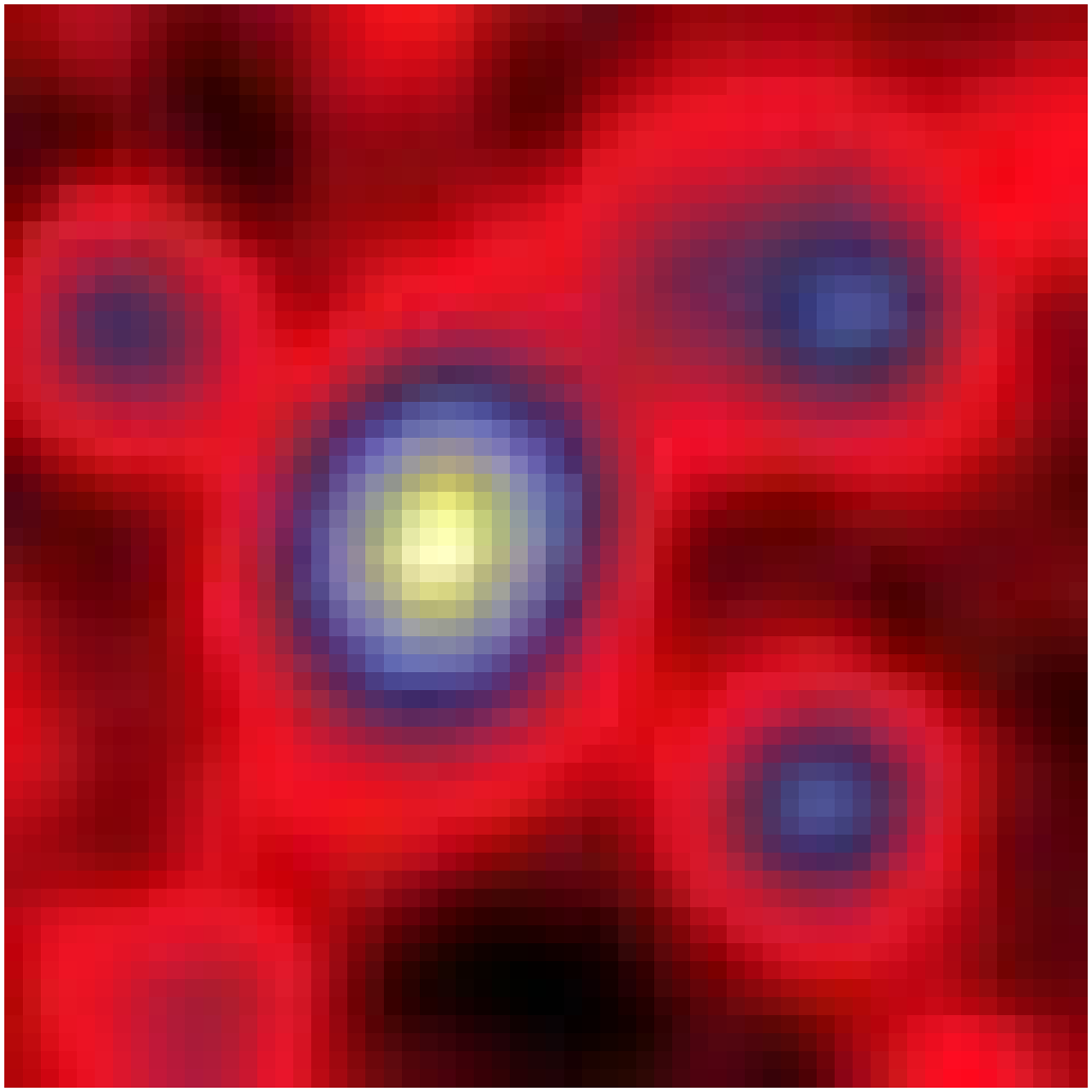}  \FGb{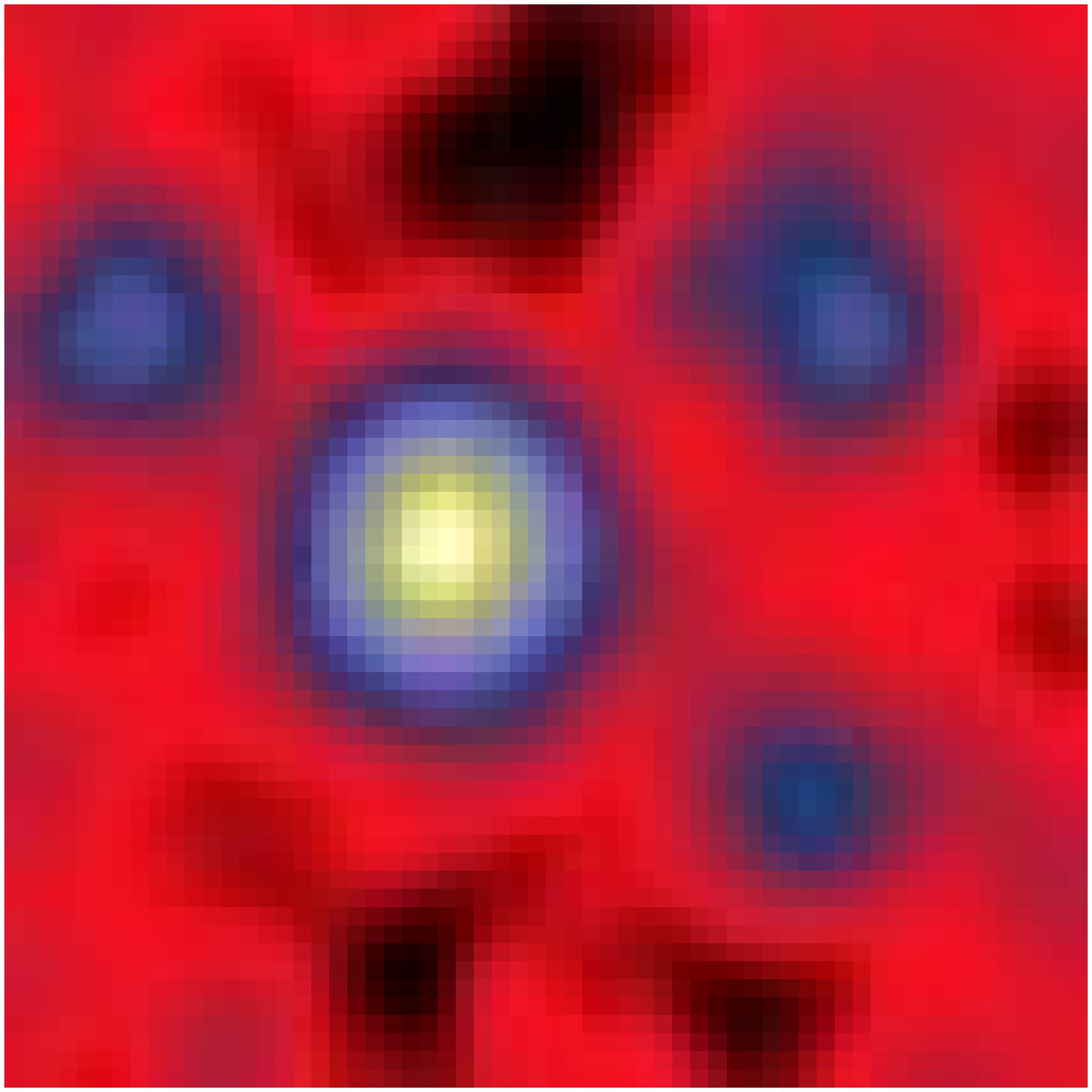}  \FGb{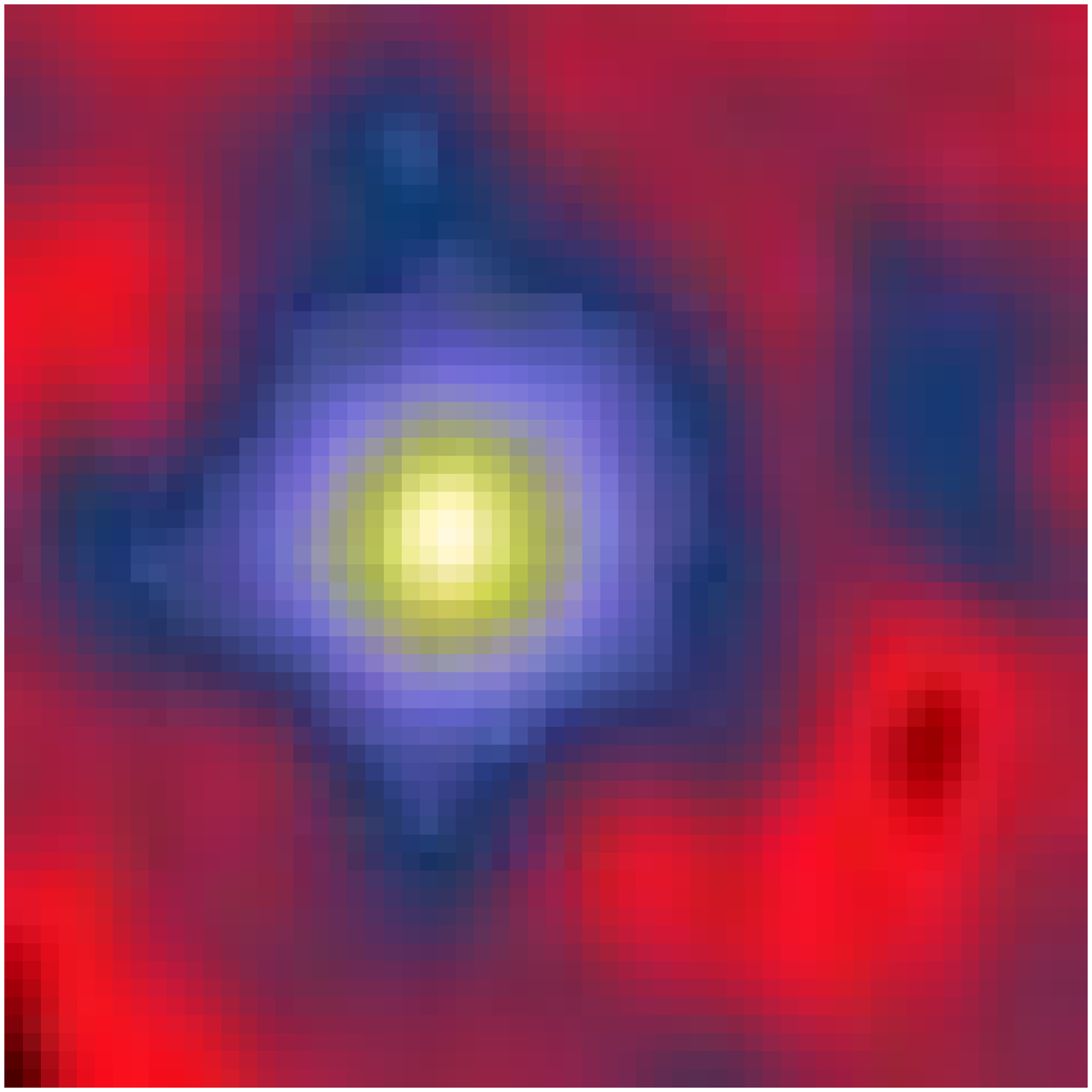}  \FGb{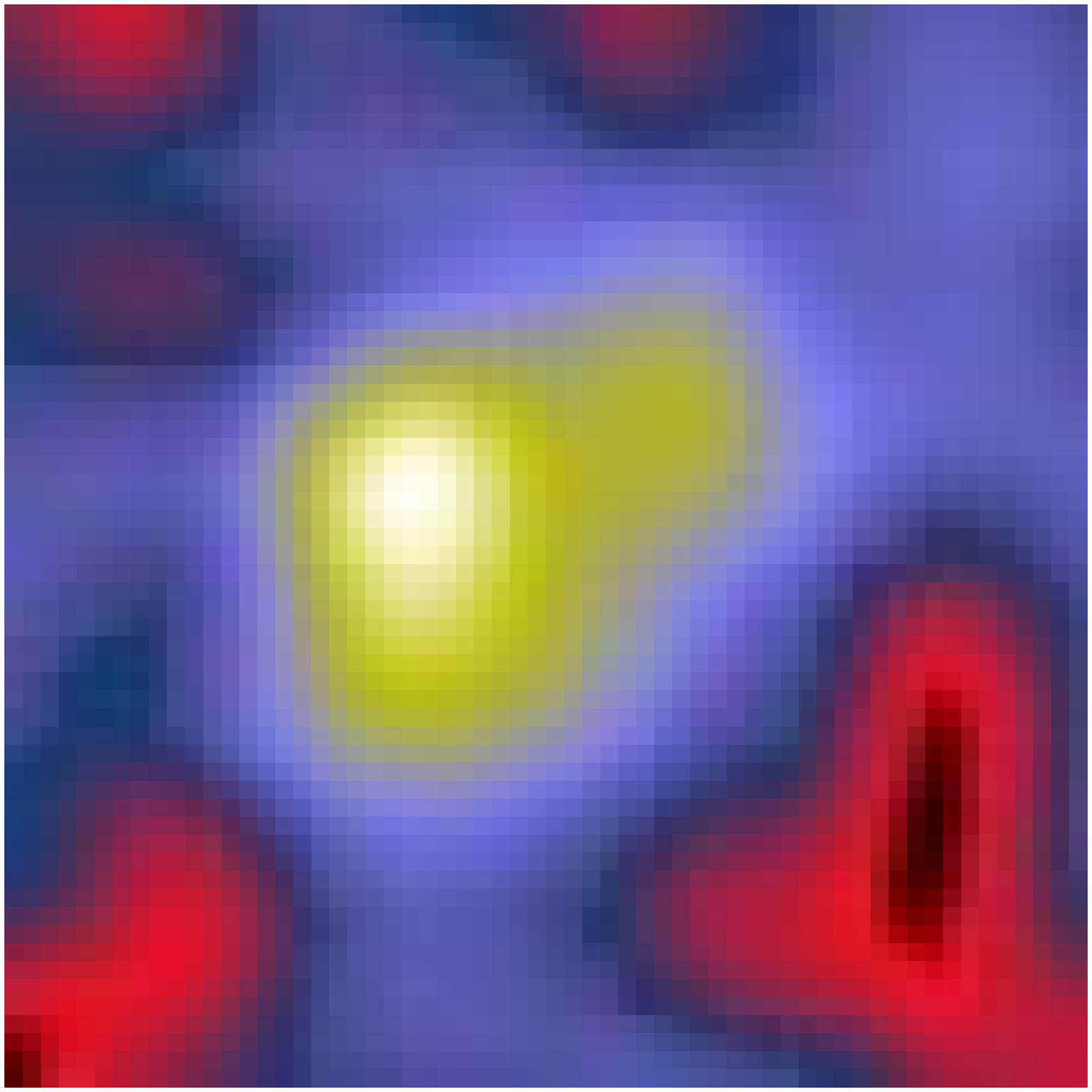} \\  {\#34   NCIA008246}  \\
  \FGb{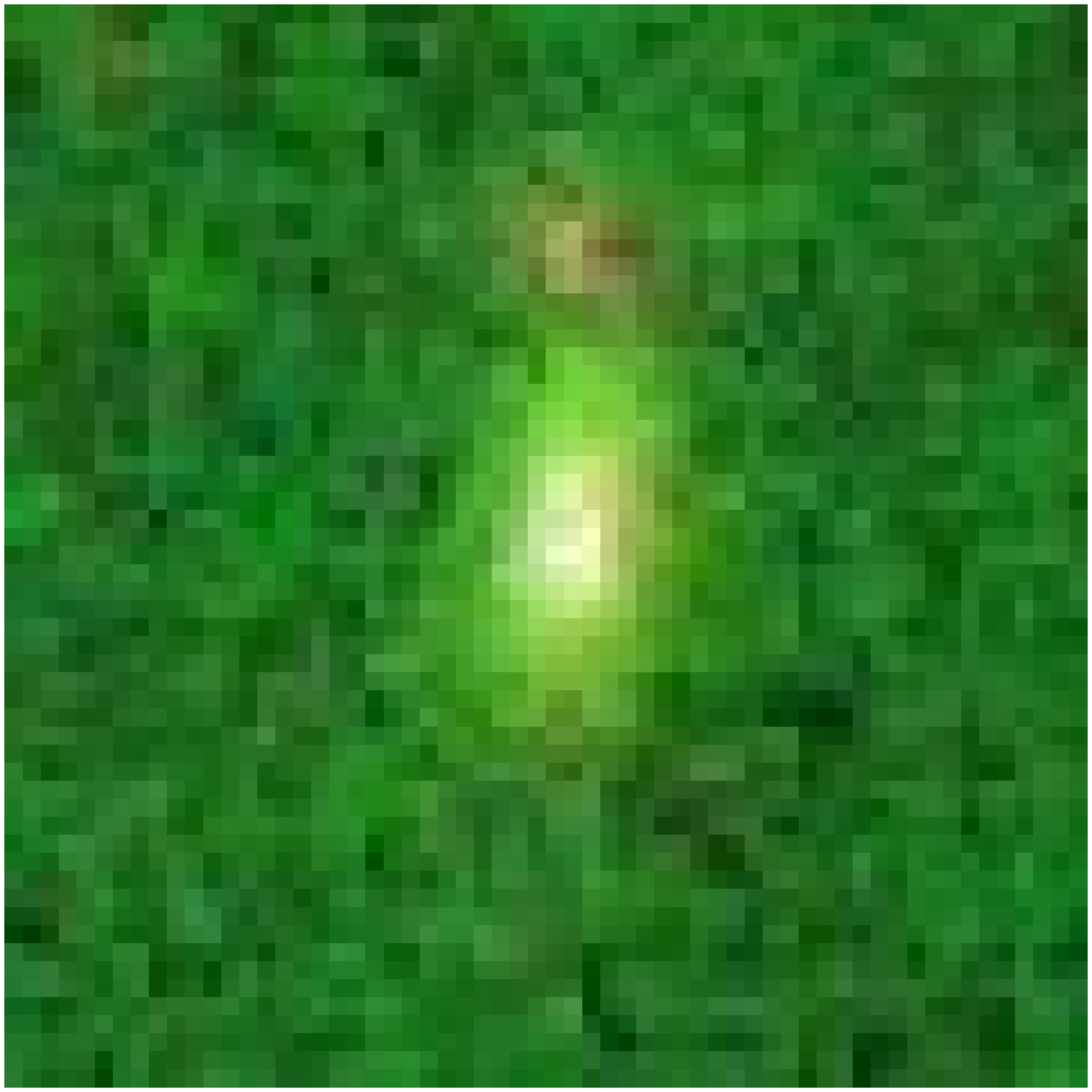}  \FGb{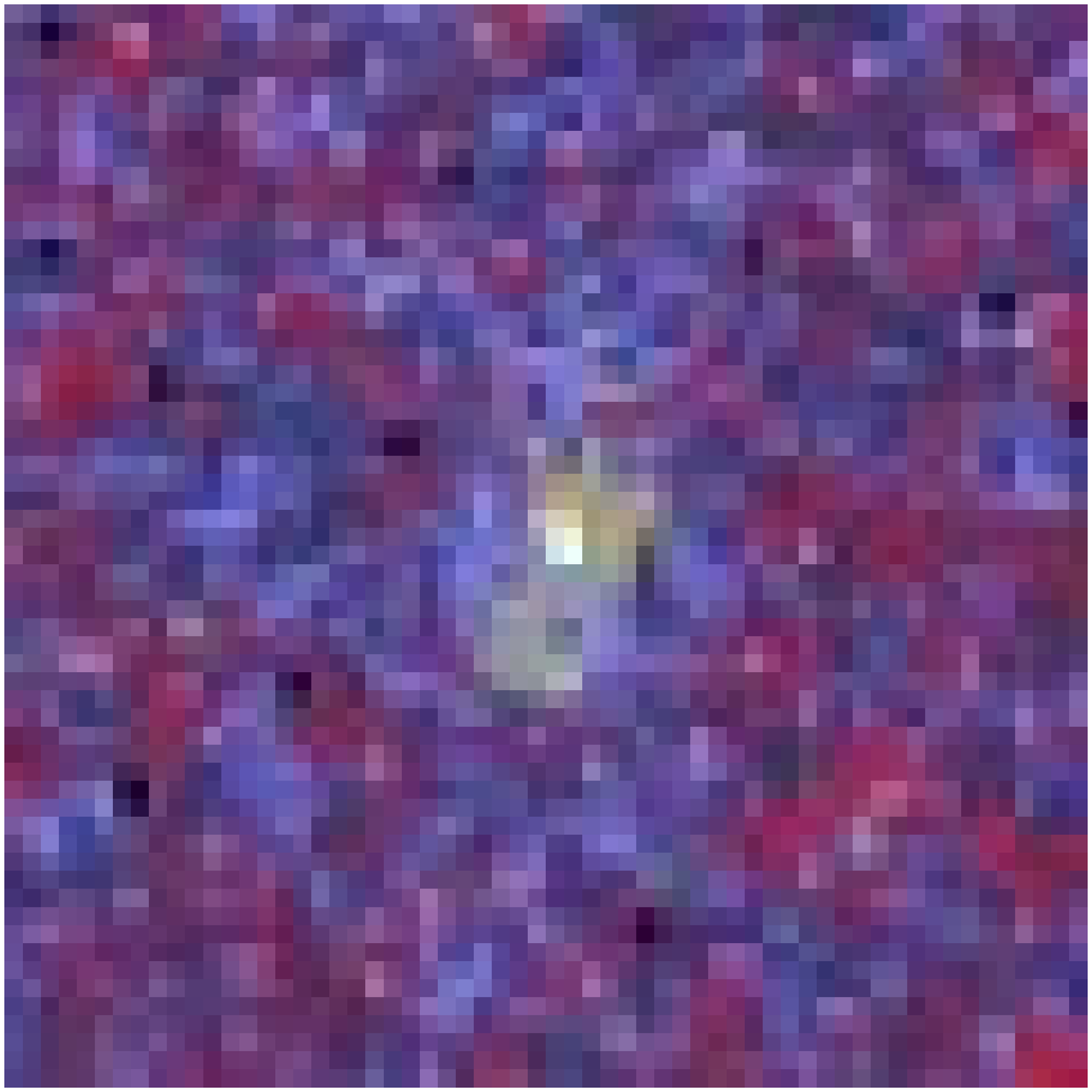}  \FGb{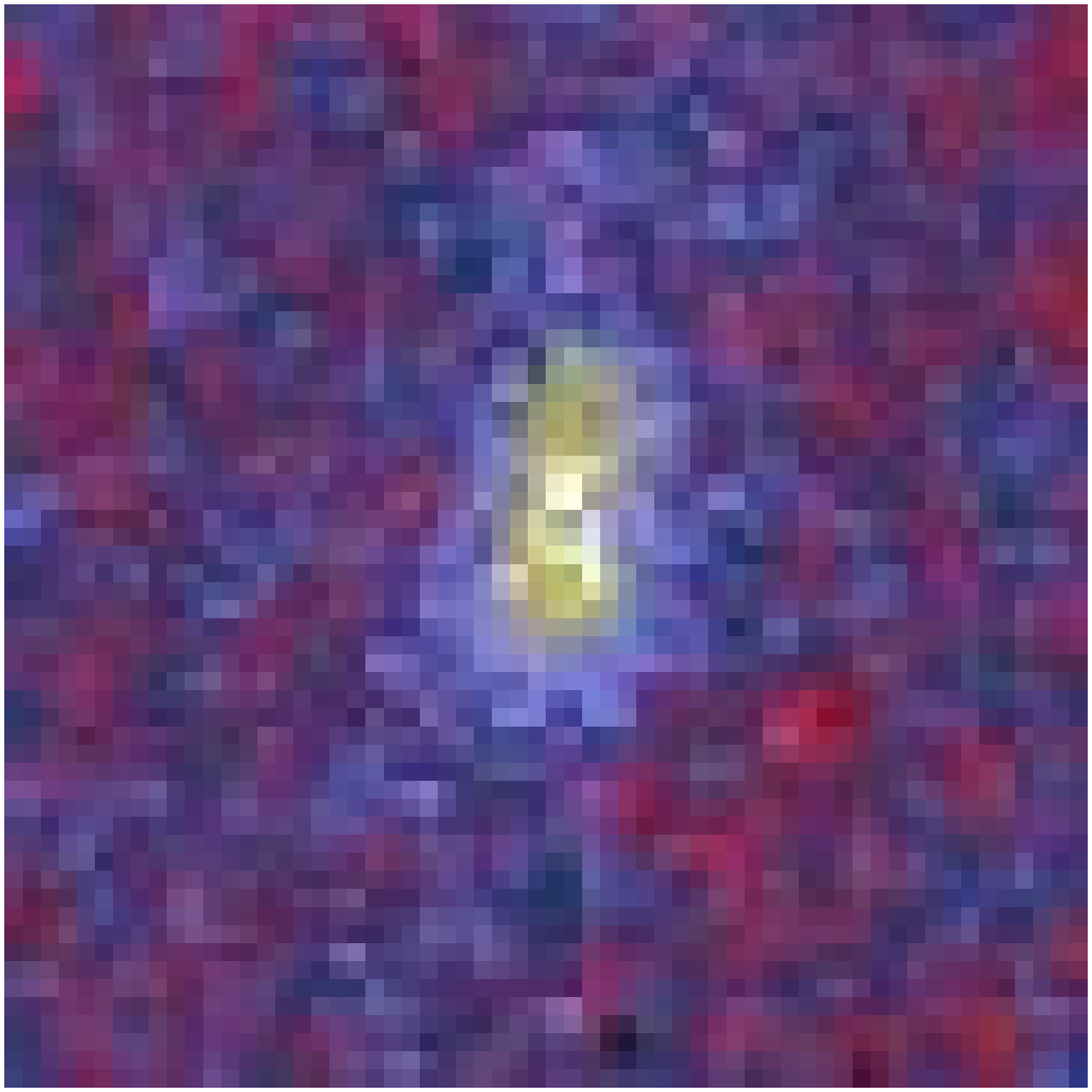}  \FGb{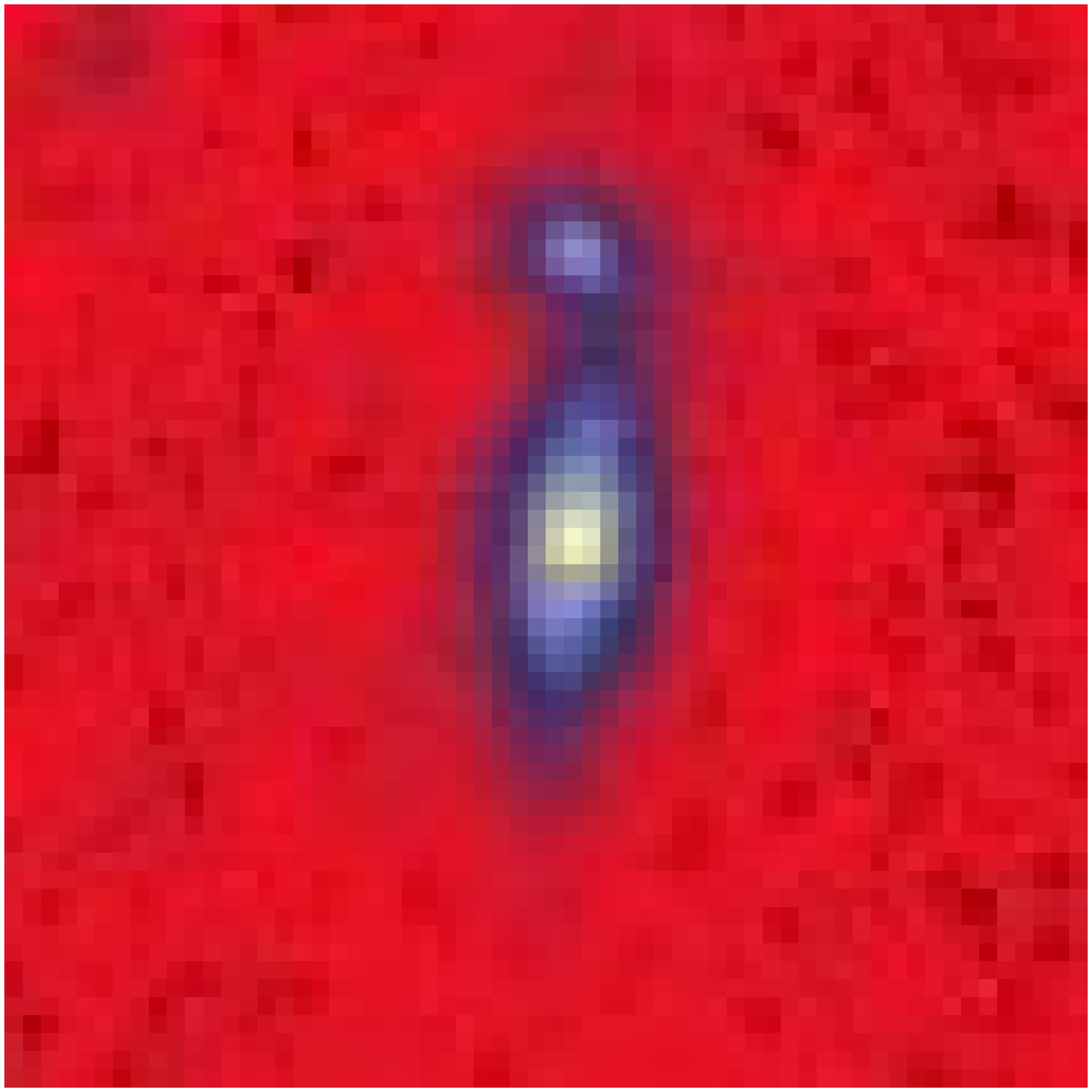}  \FGb{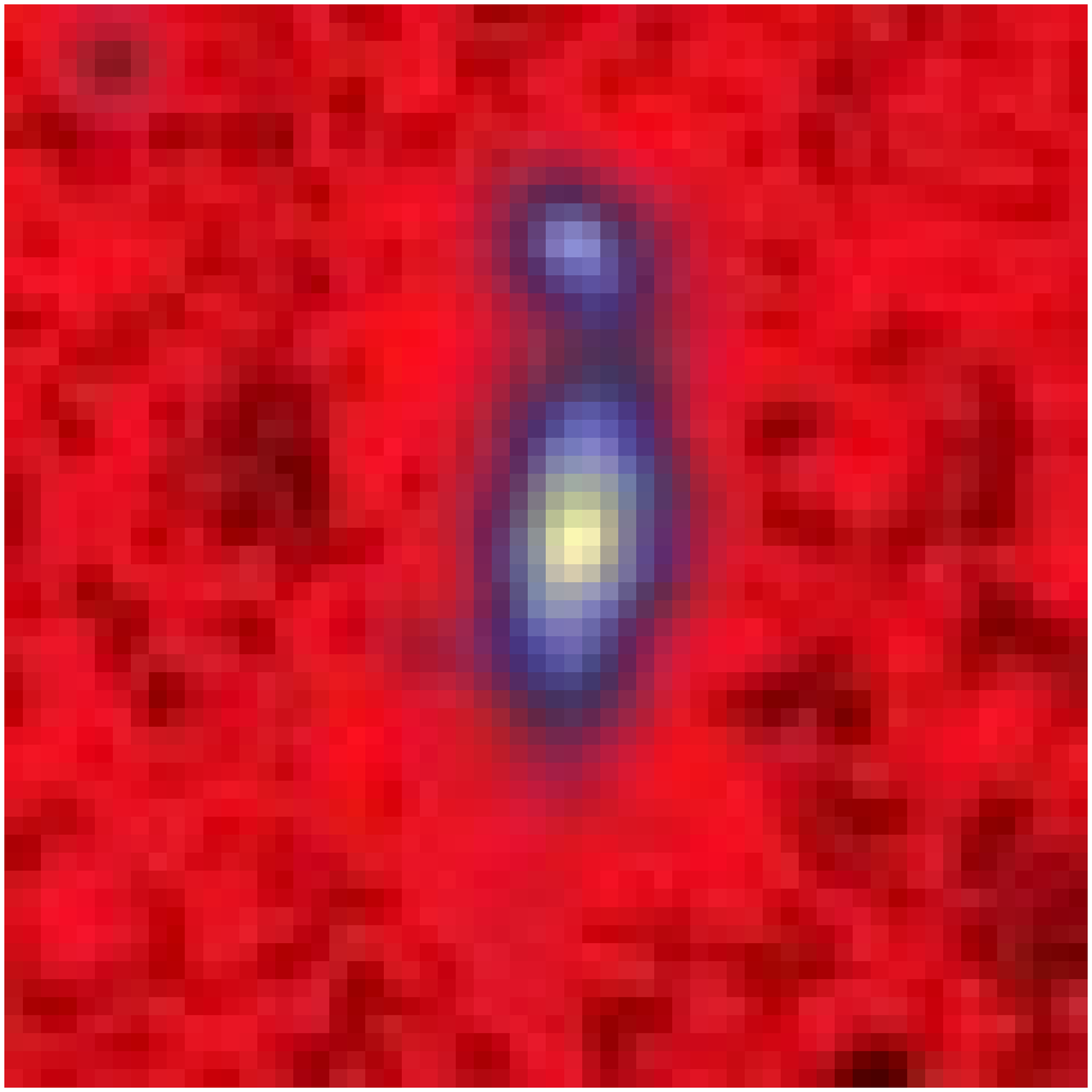}  \FGb{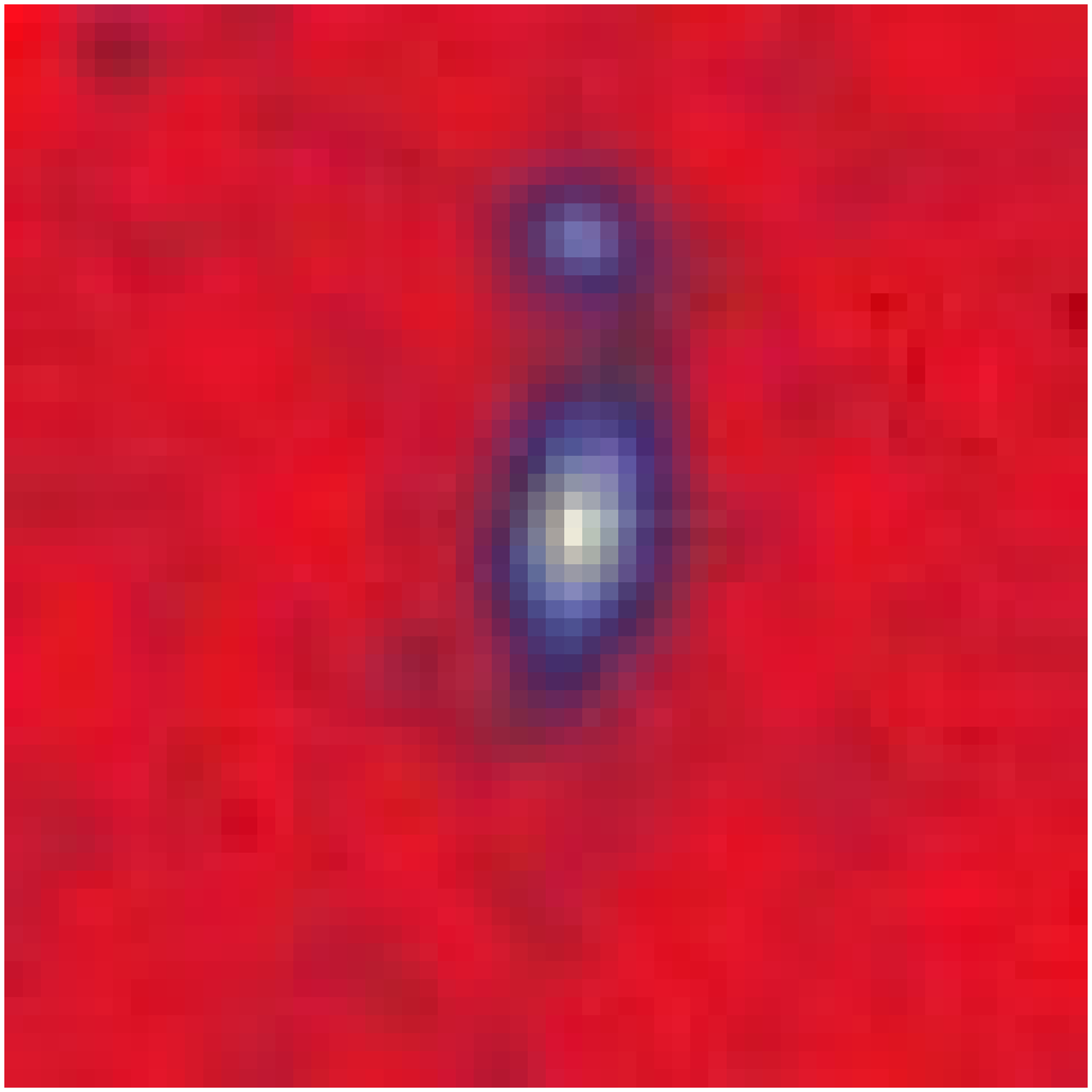}  \FGb{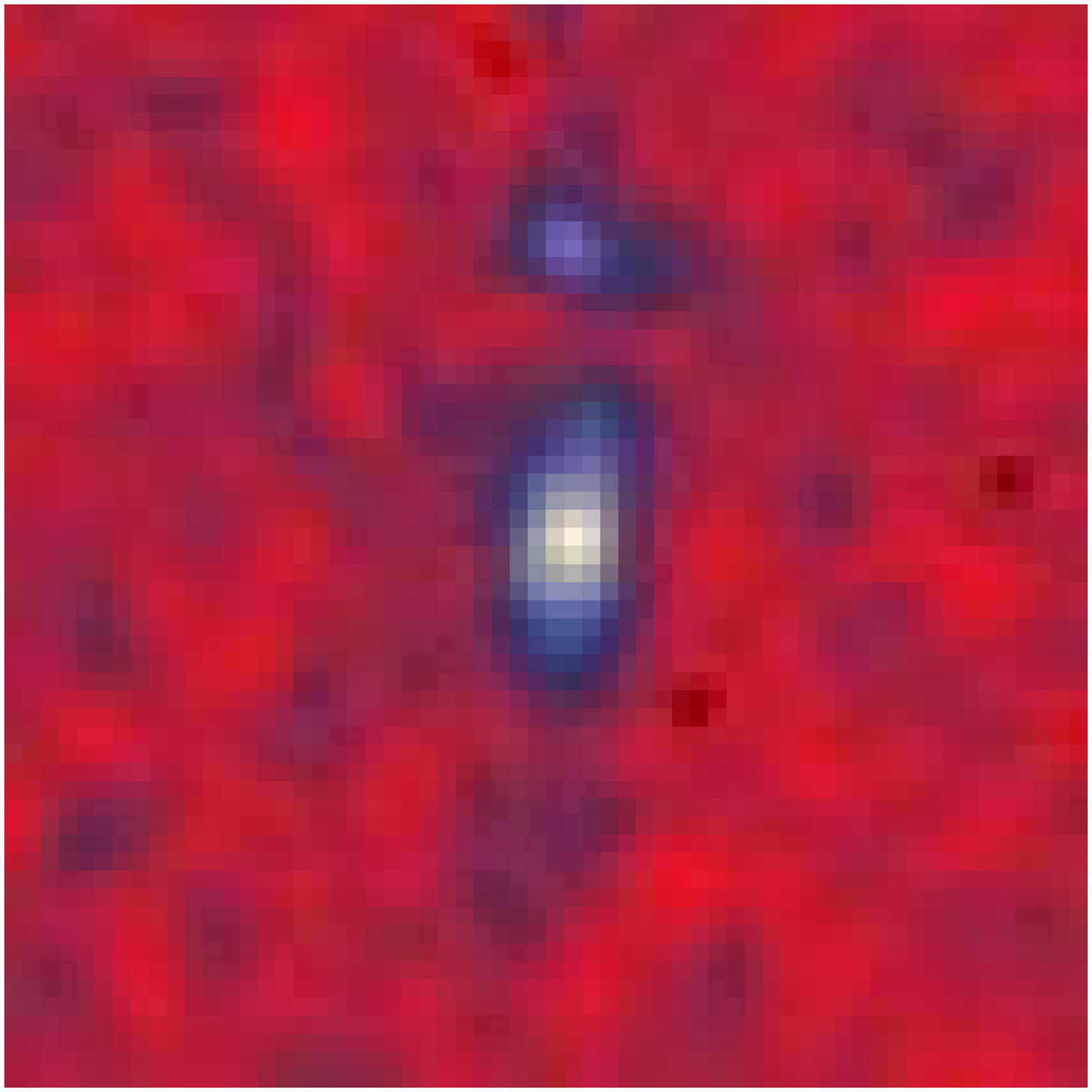}  \FGb{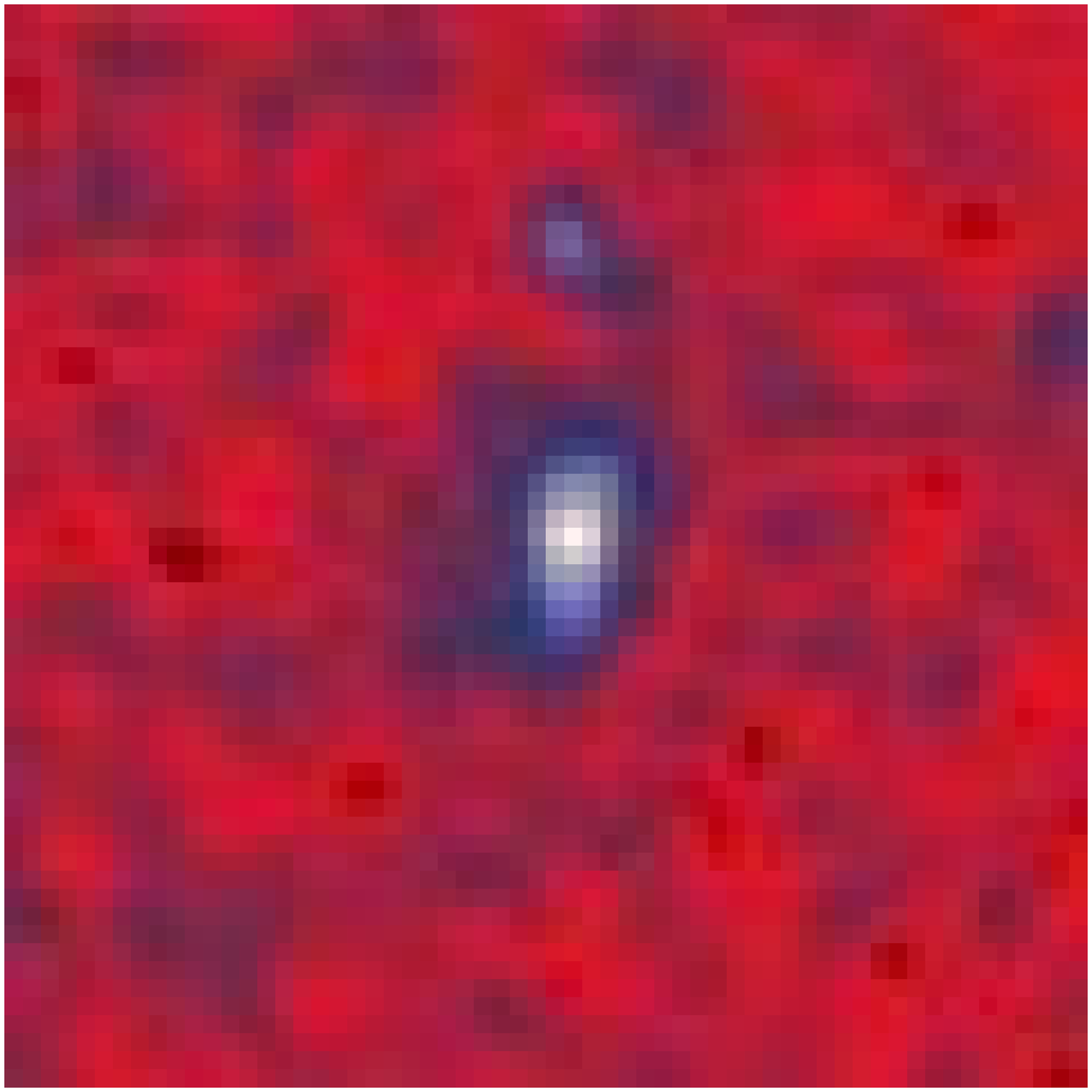}  \FGb{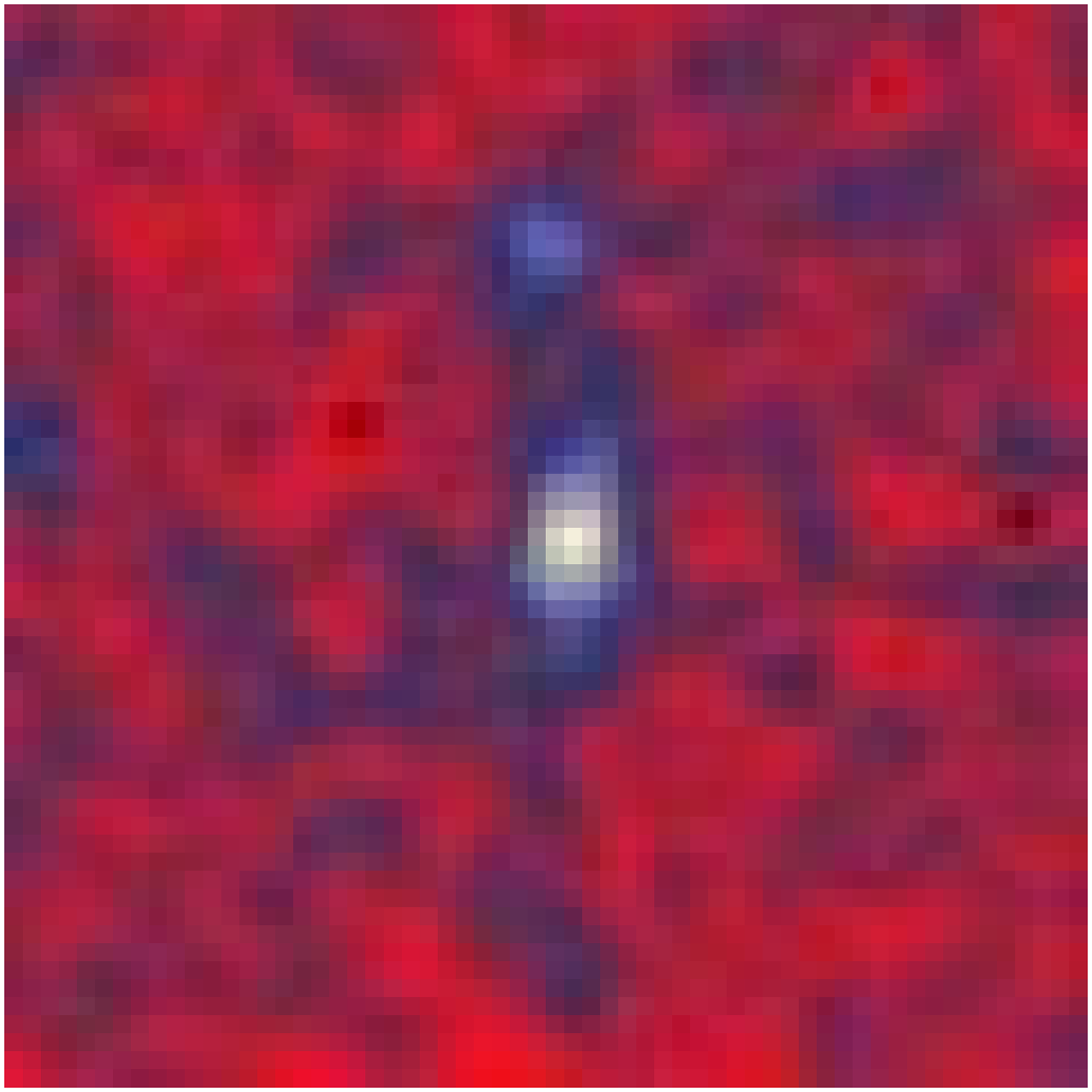}  \FGb{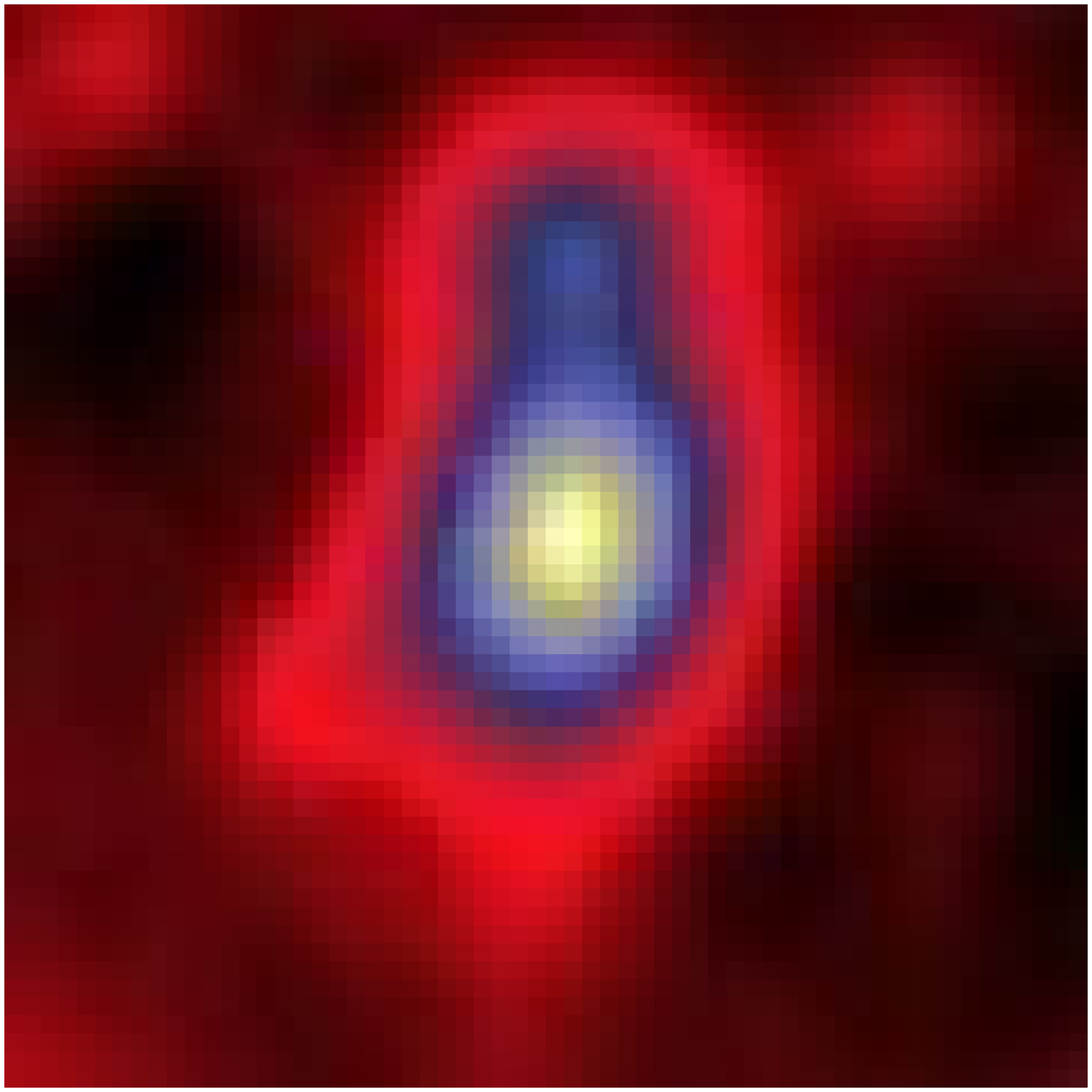}  \FGb{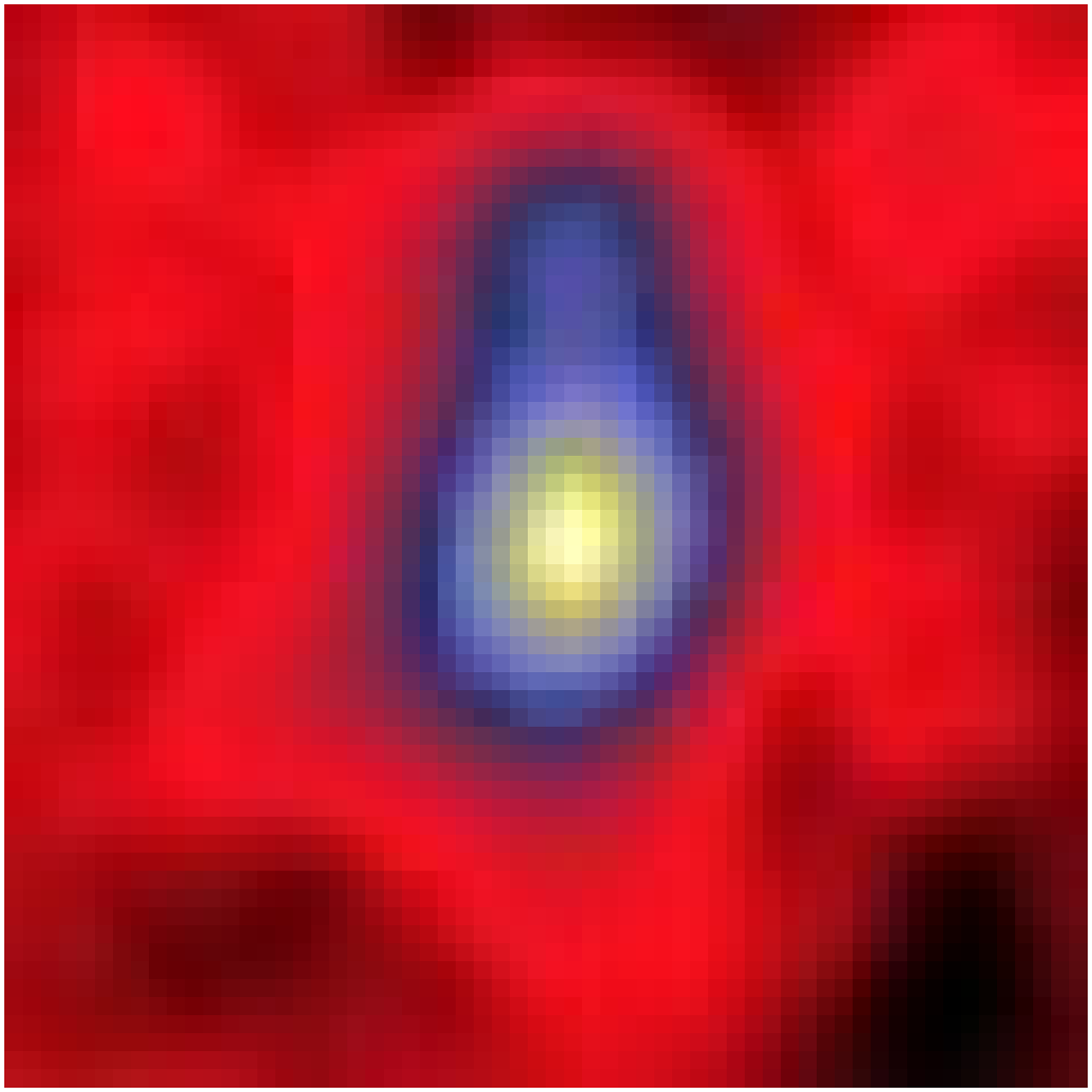}  \FGb{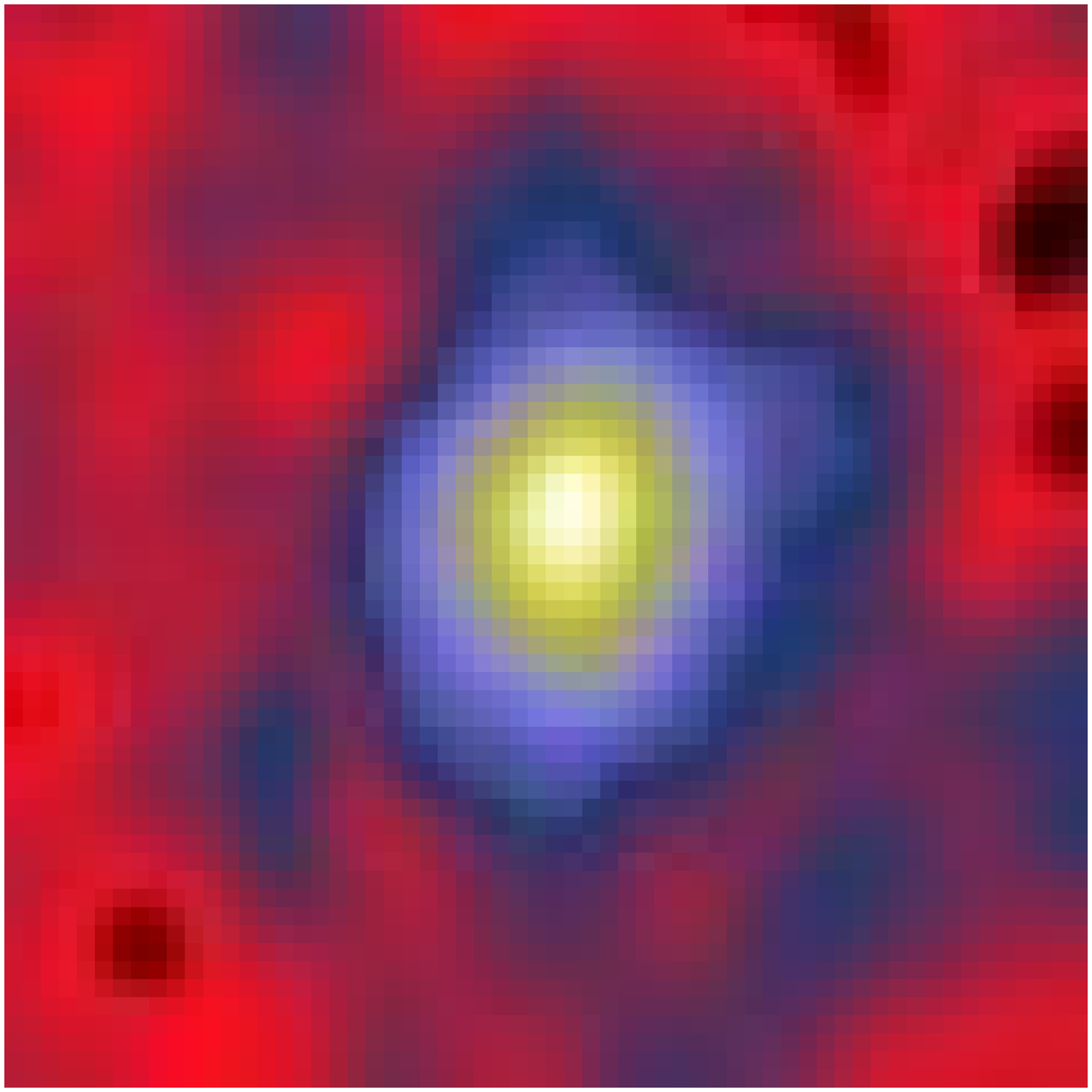}  \FGb{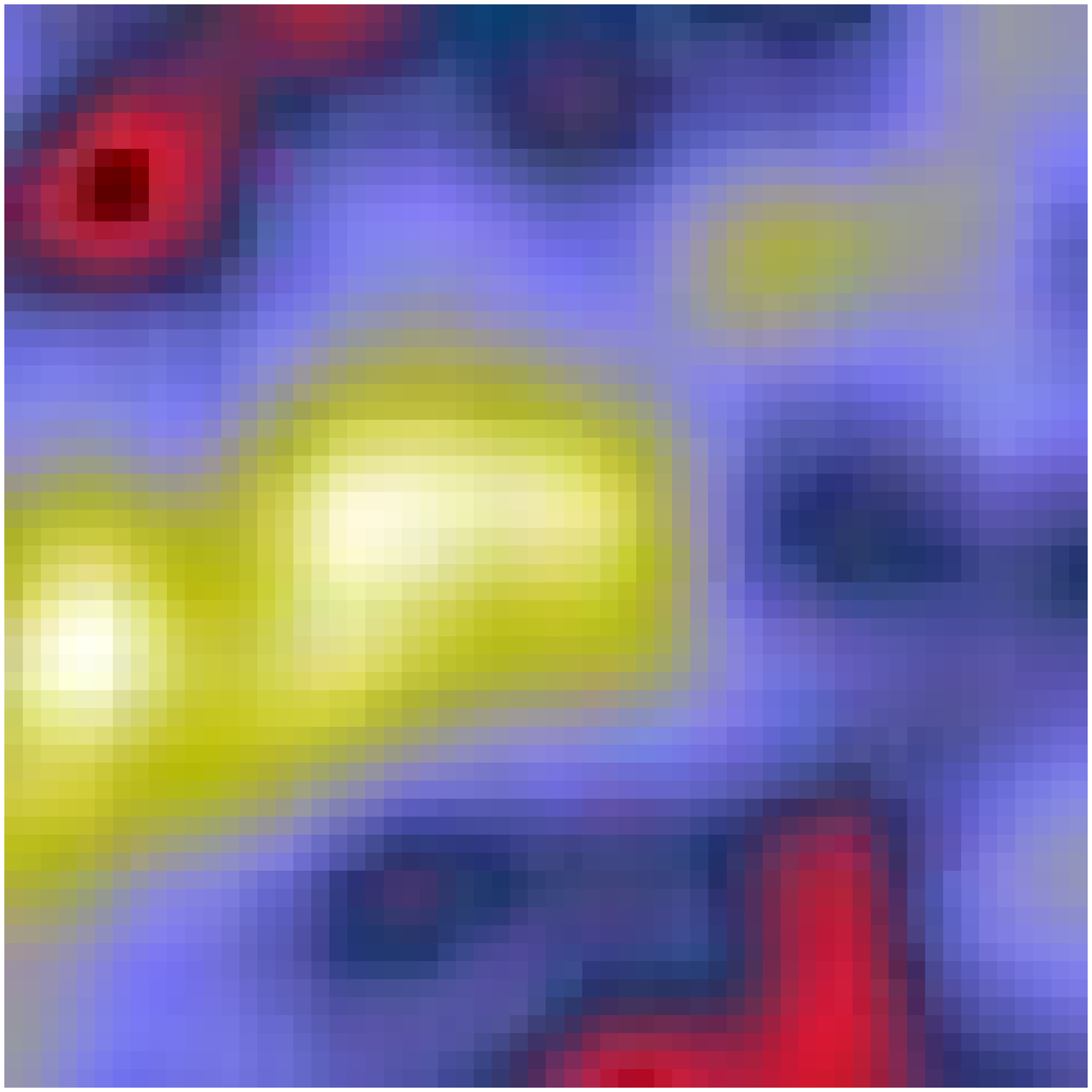} \\  {\#35   NCIA031681}  \\
  \FGb{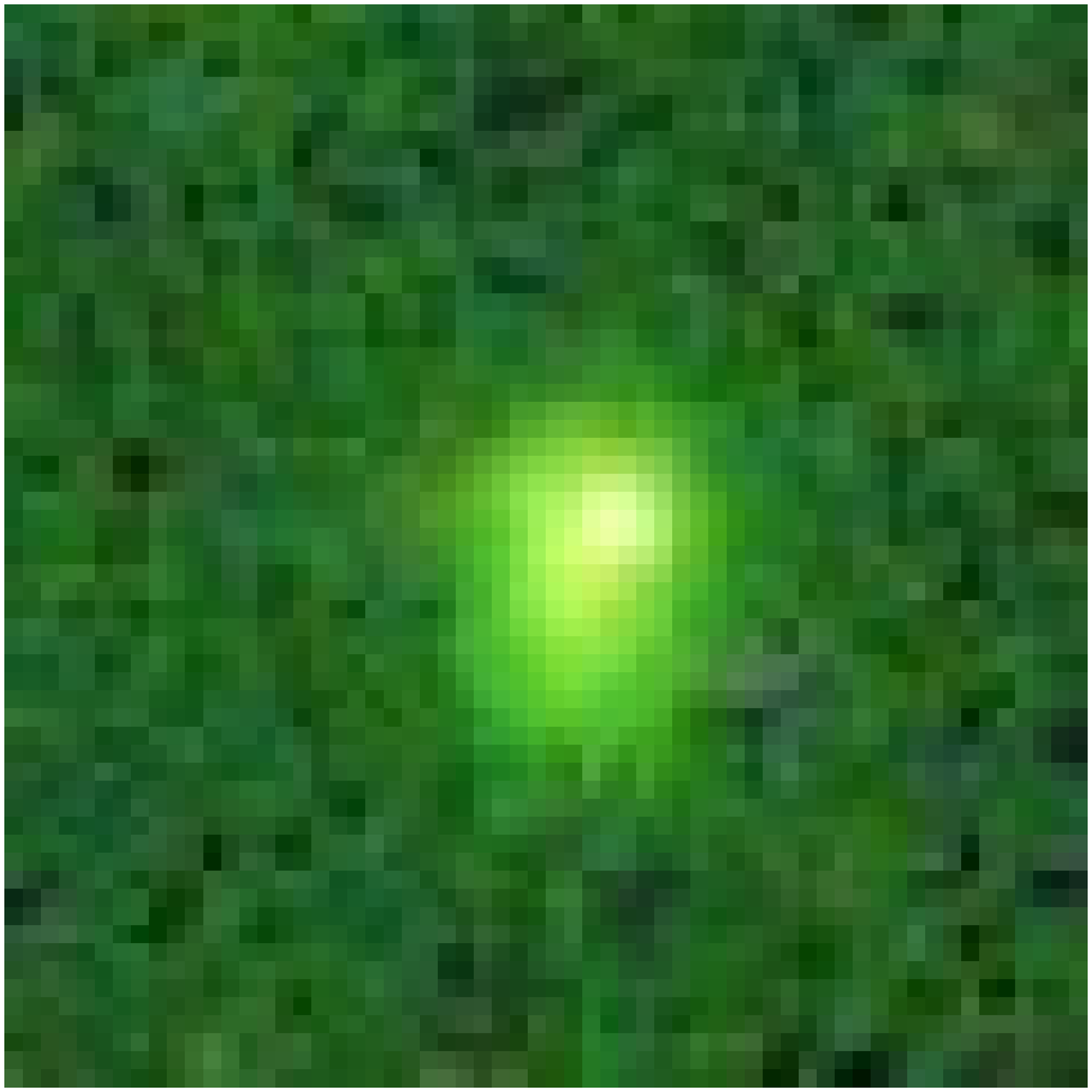}  \FGb{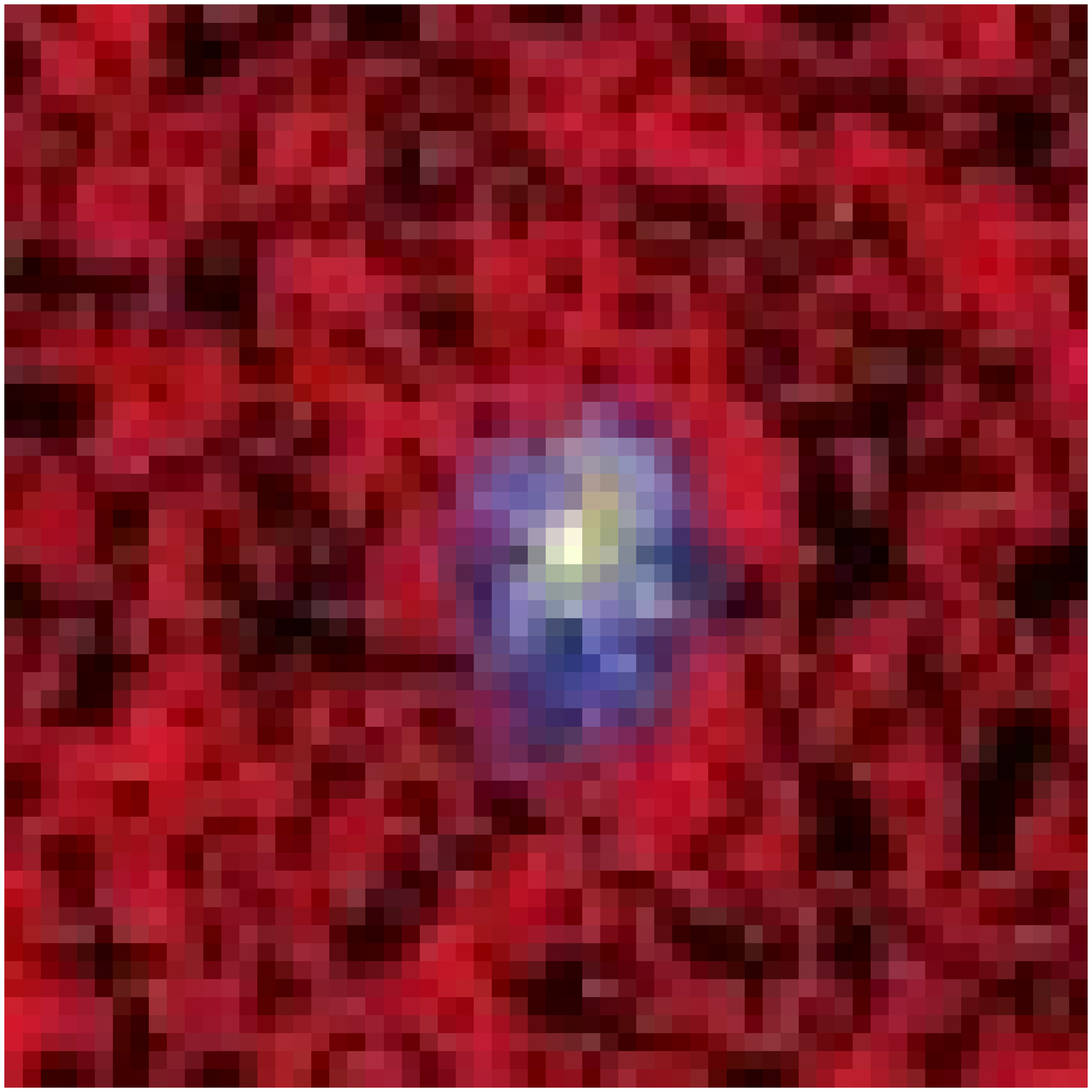}  \FGb{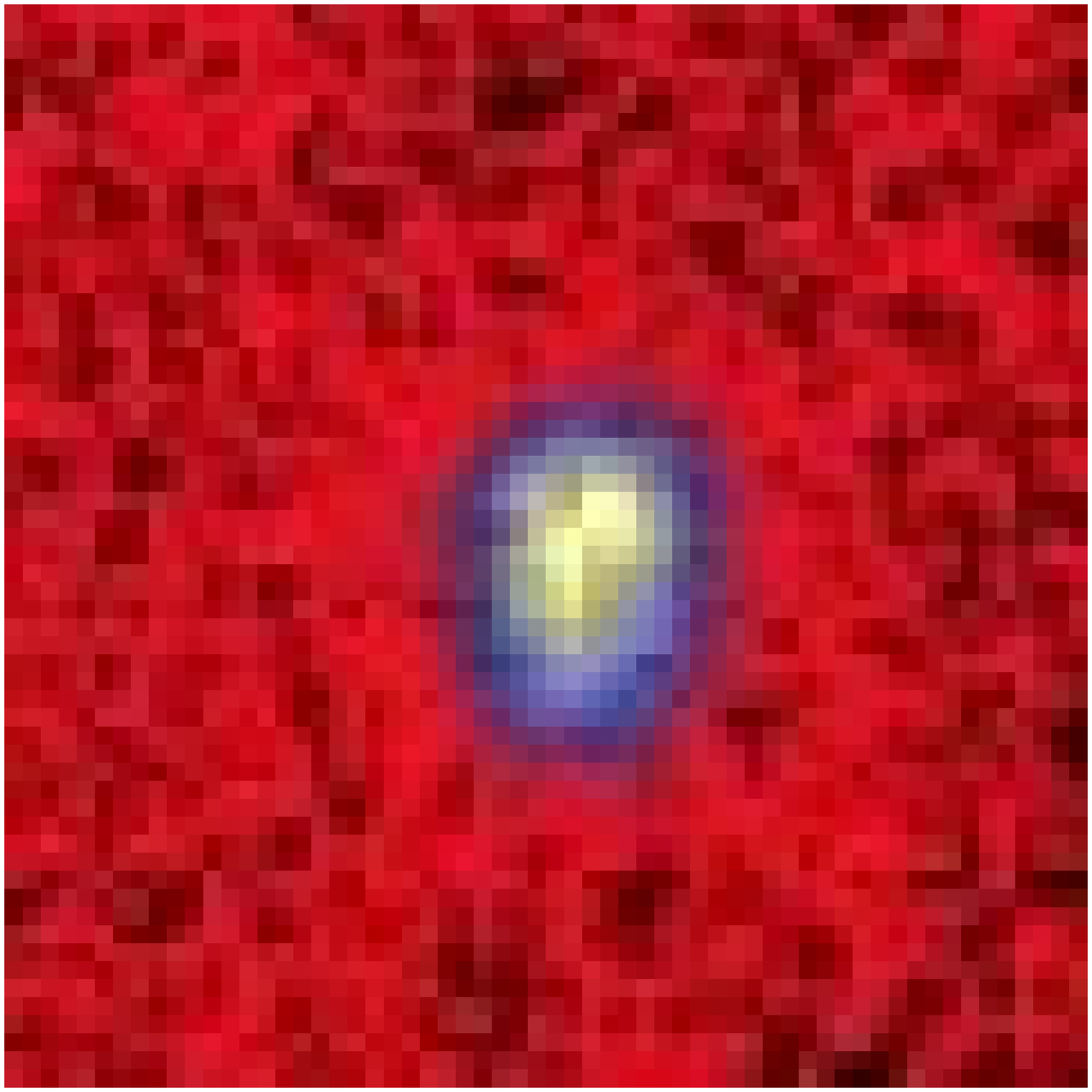}  \FGb{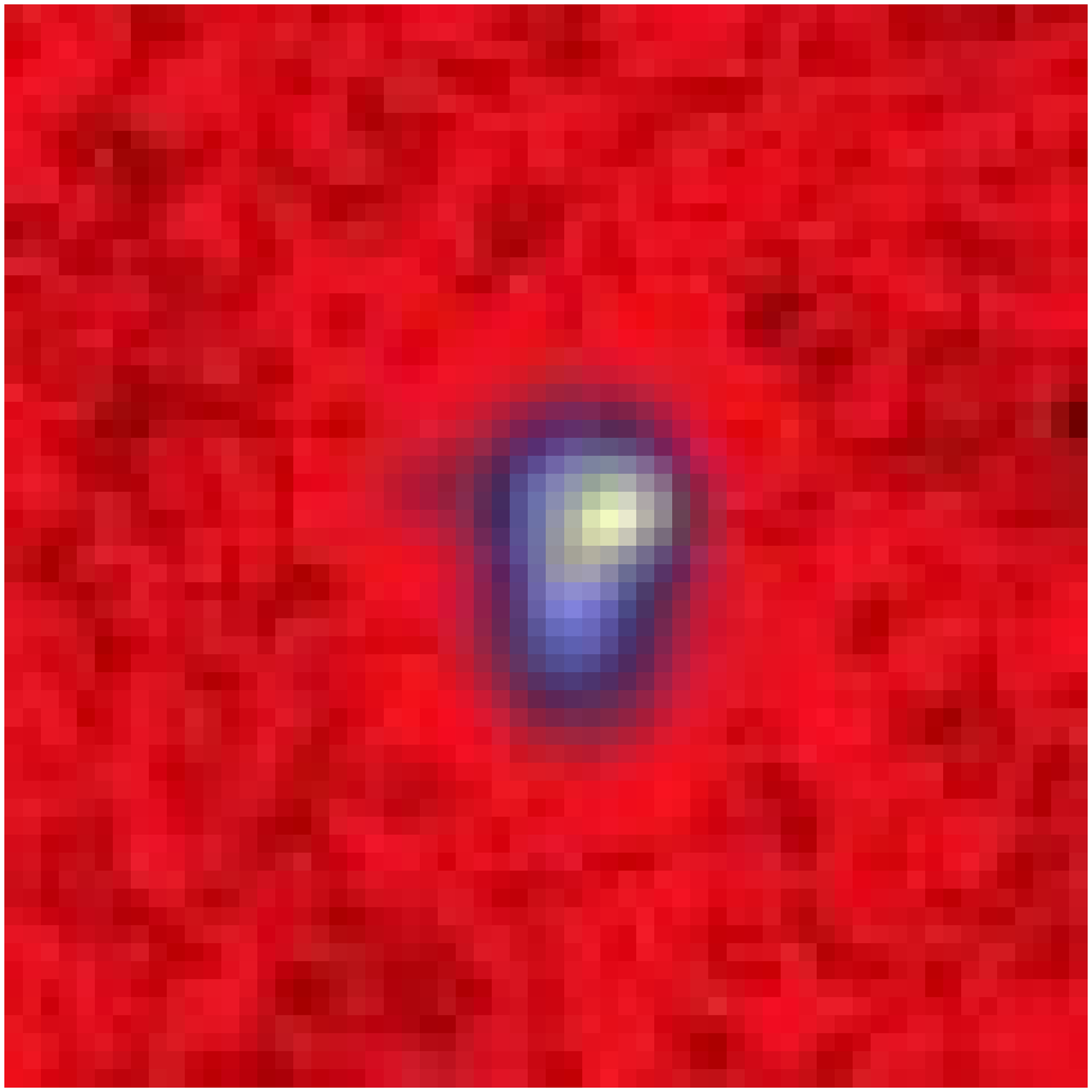}  \FGb{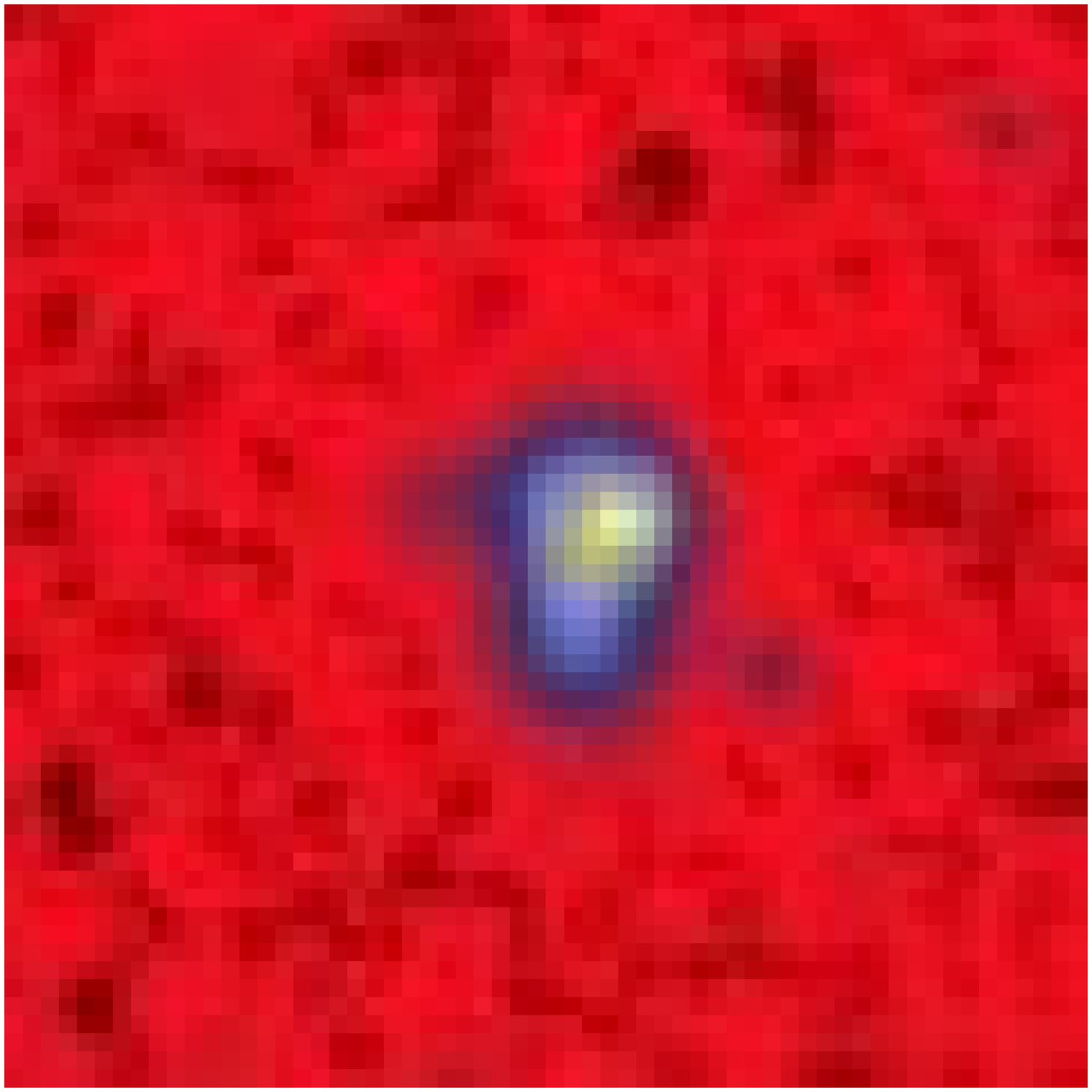}  \FGb{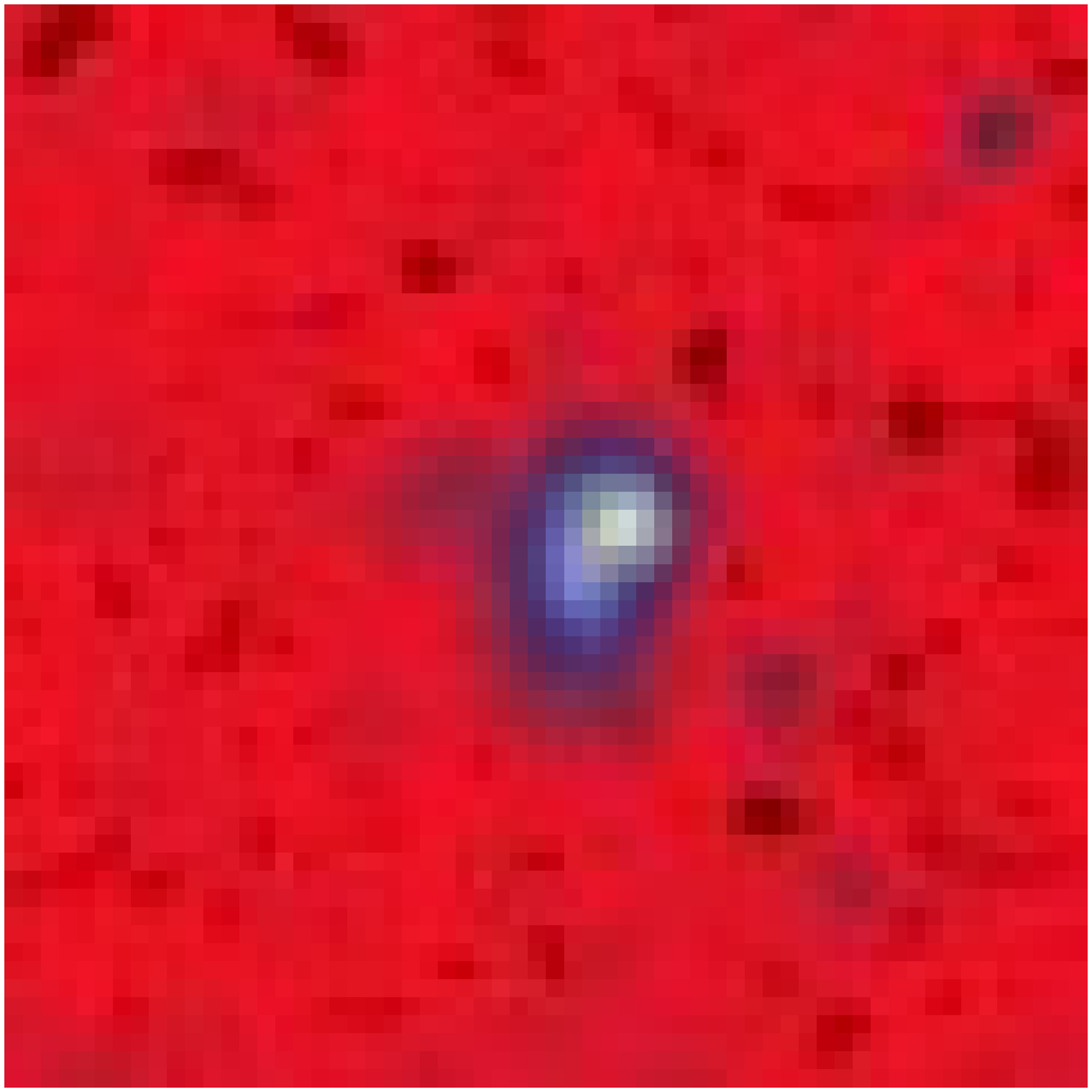}  \FGb{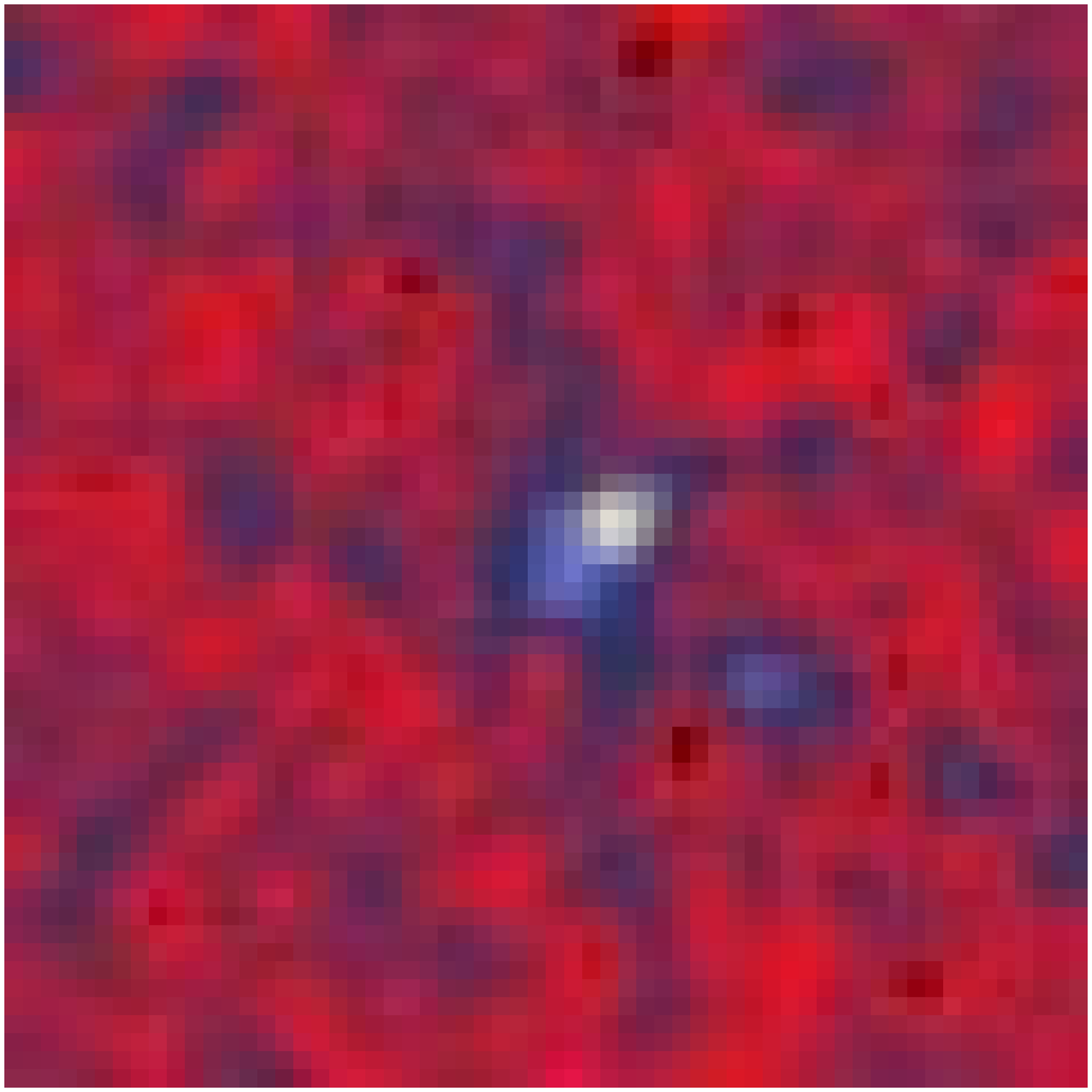}  \FGb{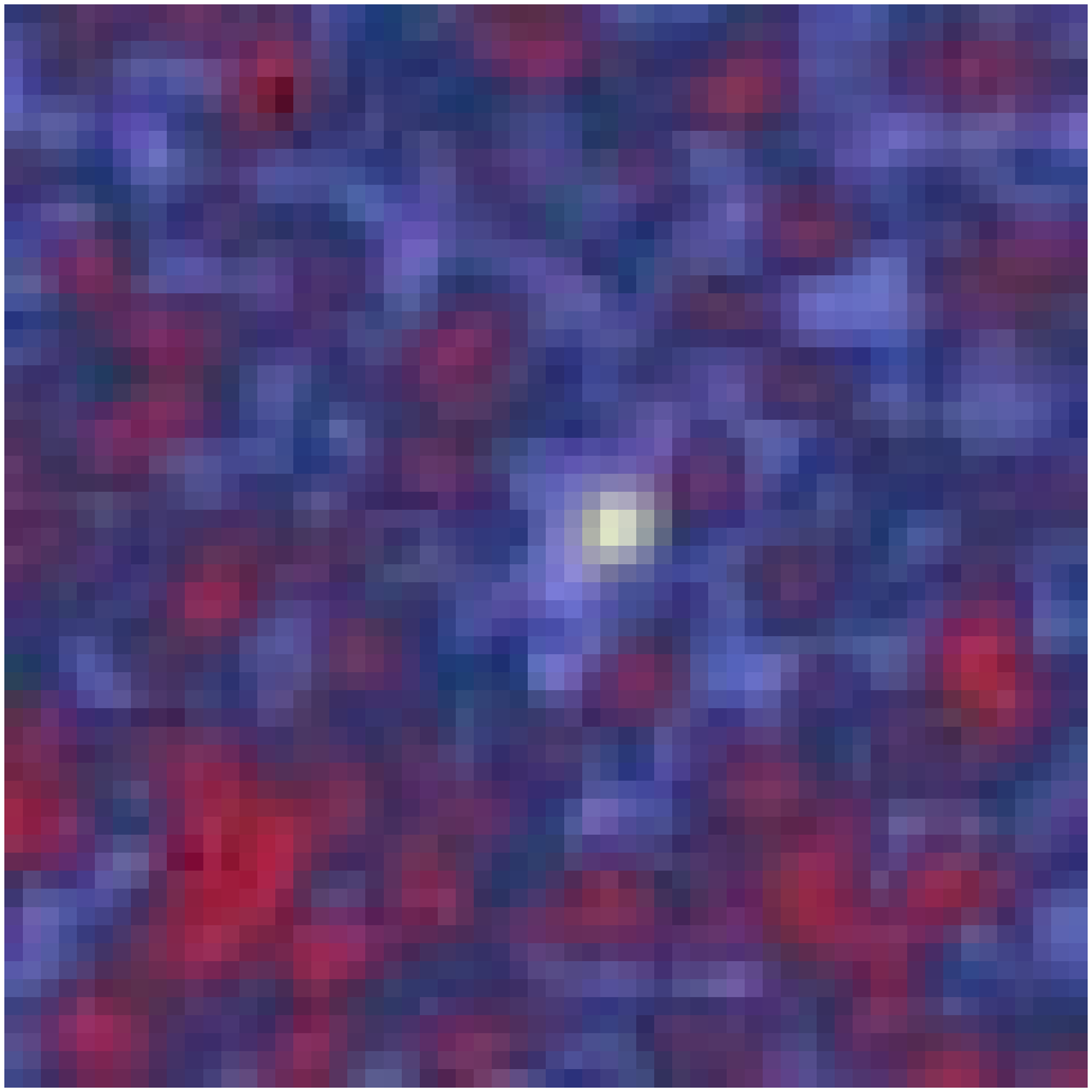}  \FGb{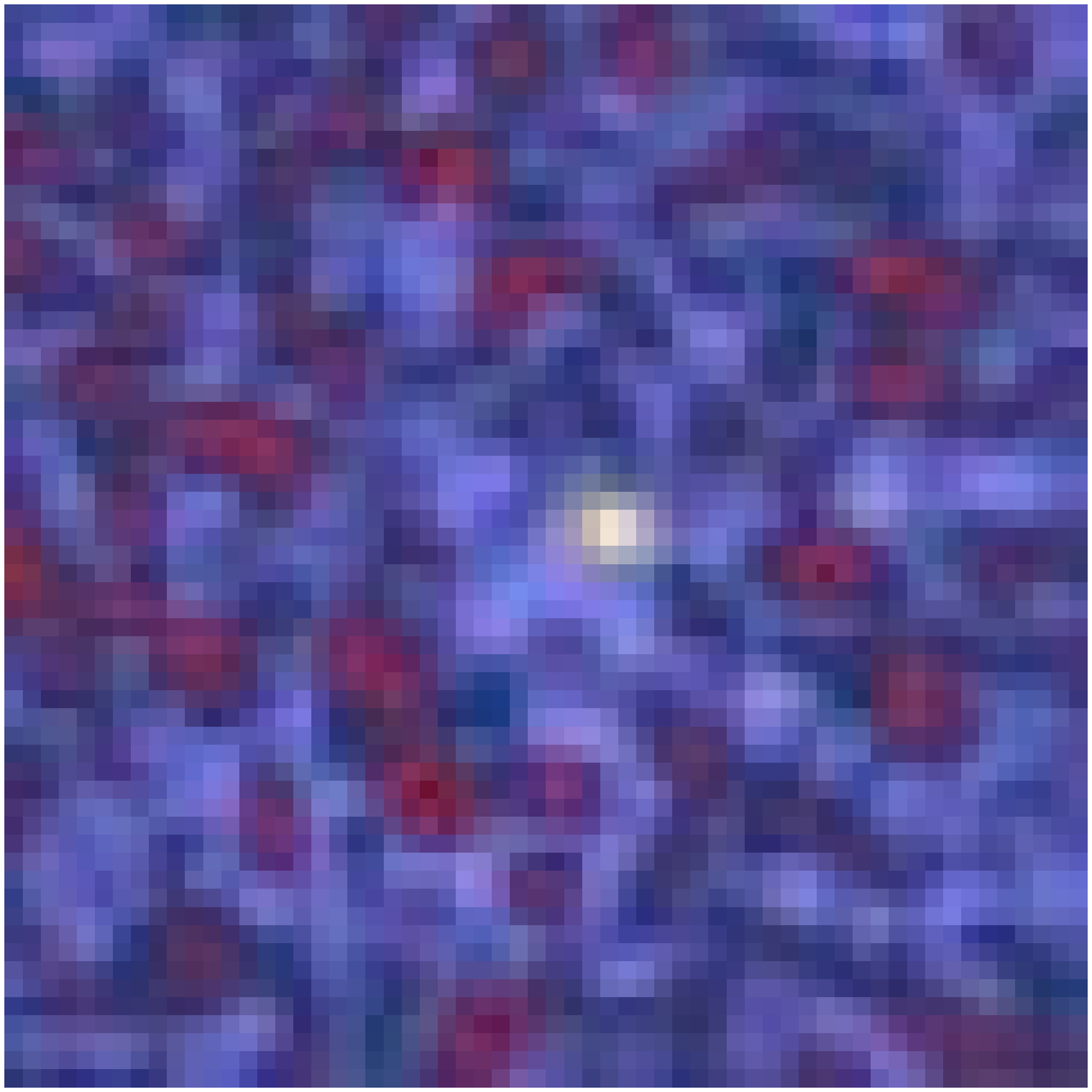}  \FGb{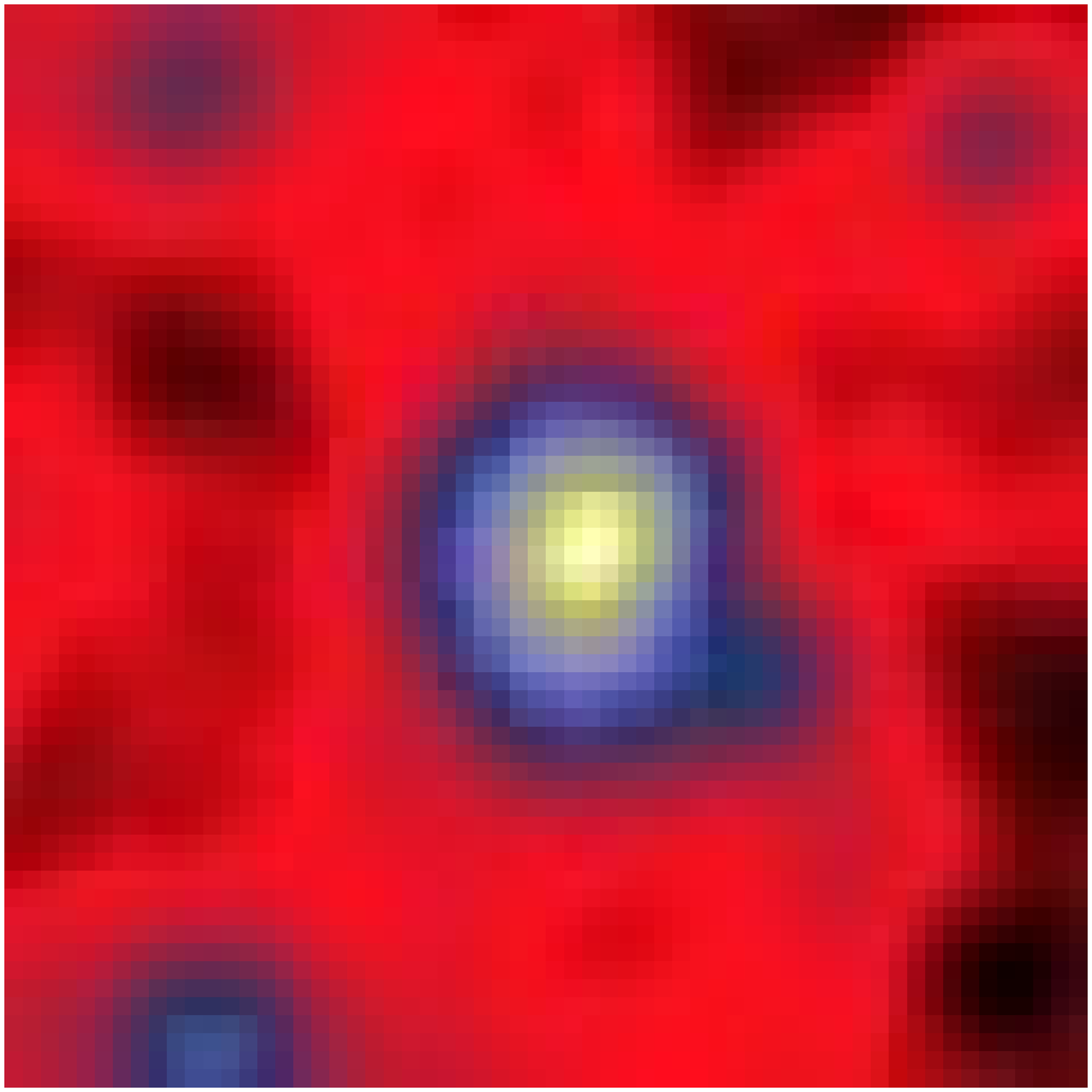}  \FGb{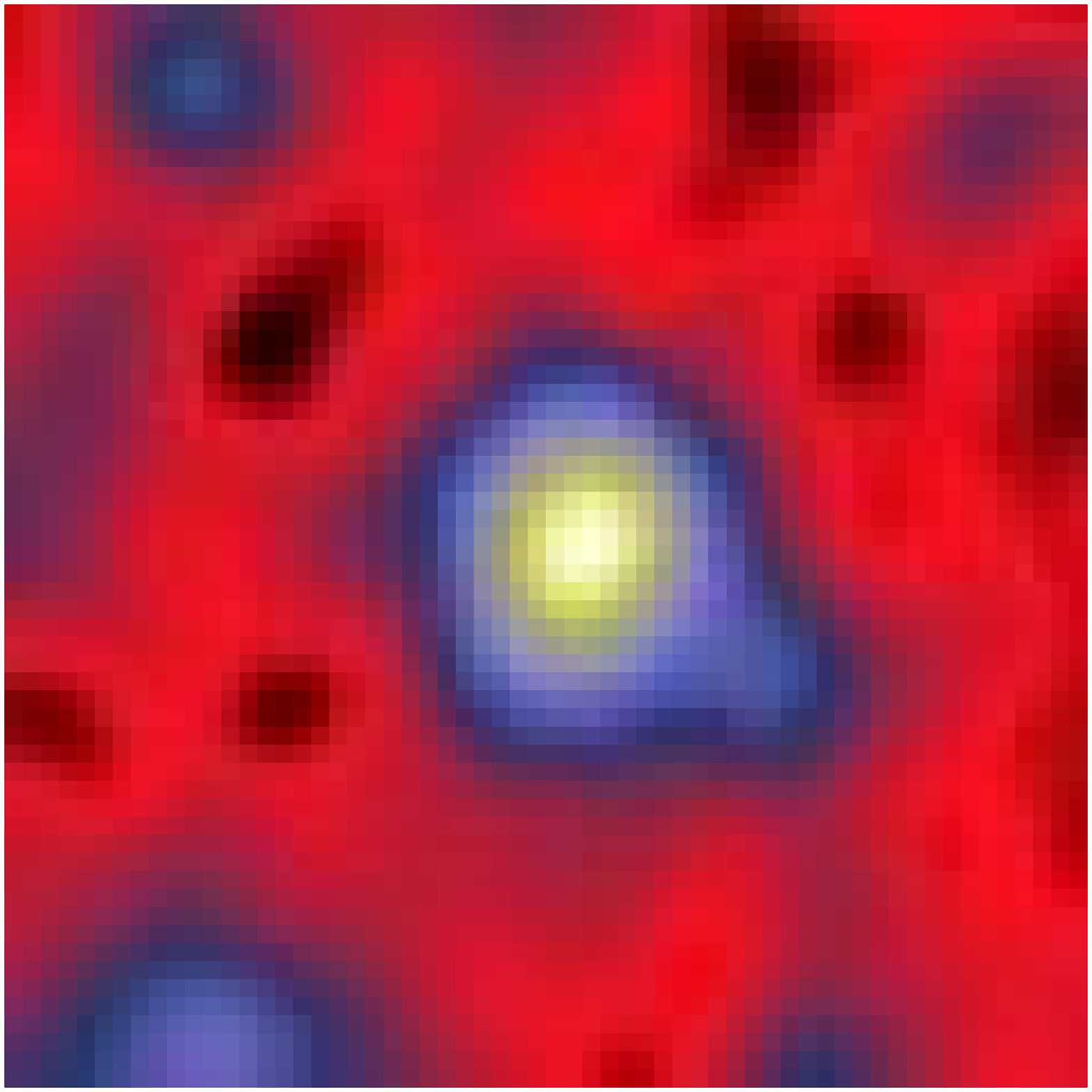}  \FGb{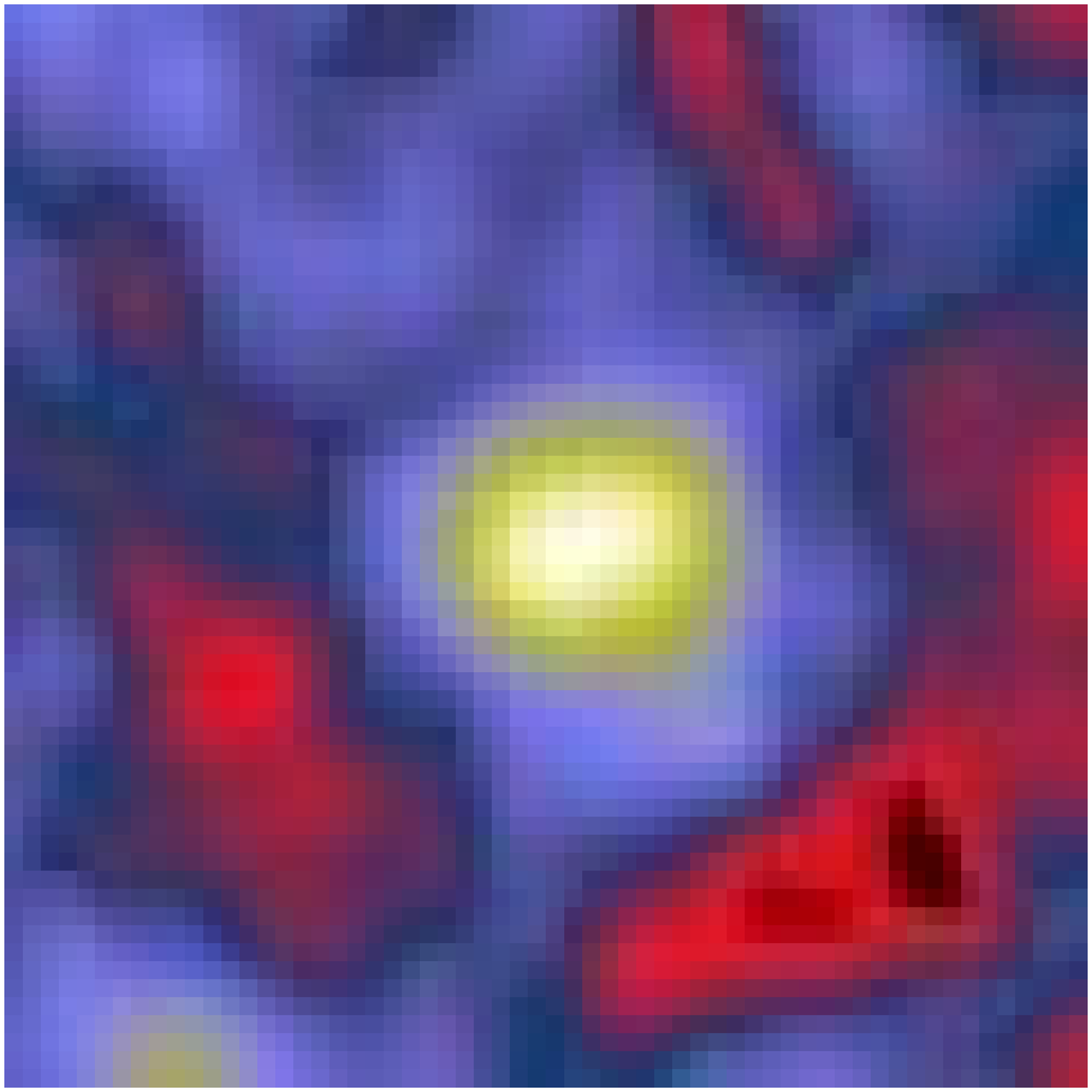}  \FGb{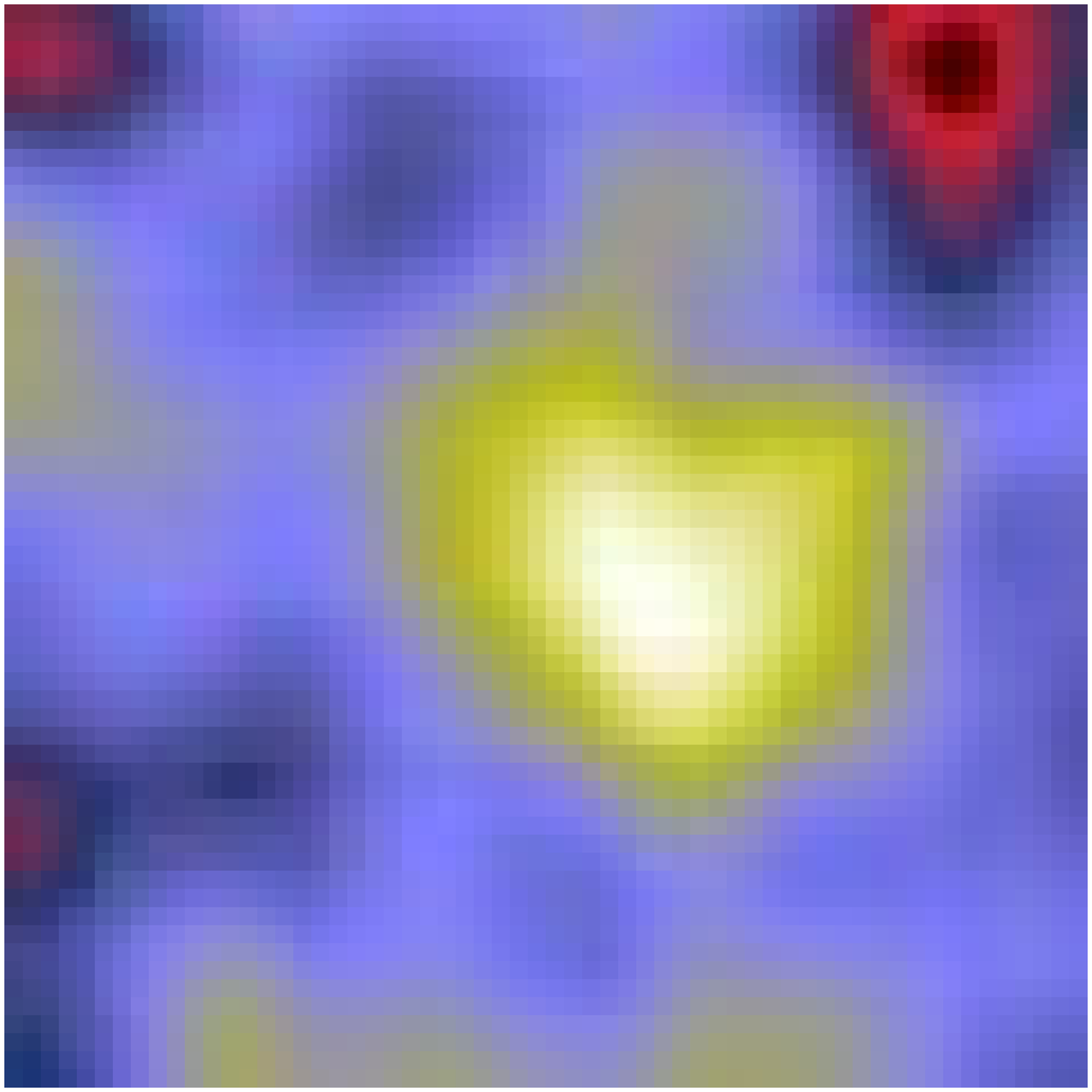} \\  {\#36   NCIA010315}  \\
  \FGb{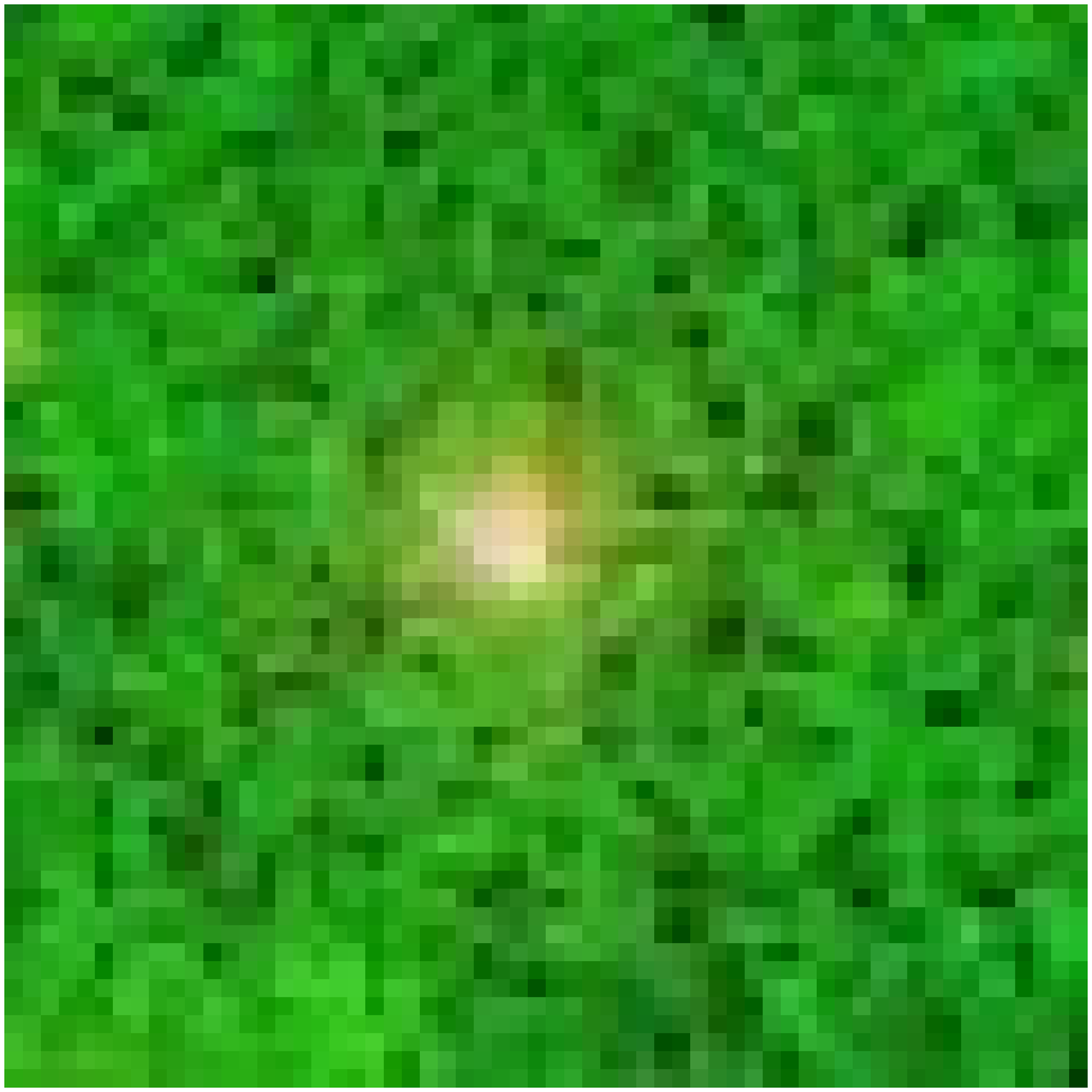}  \FGb{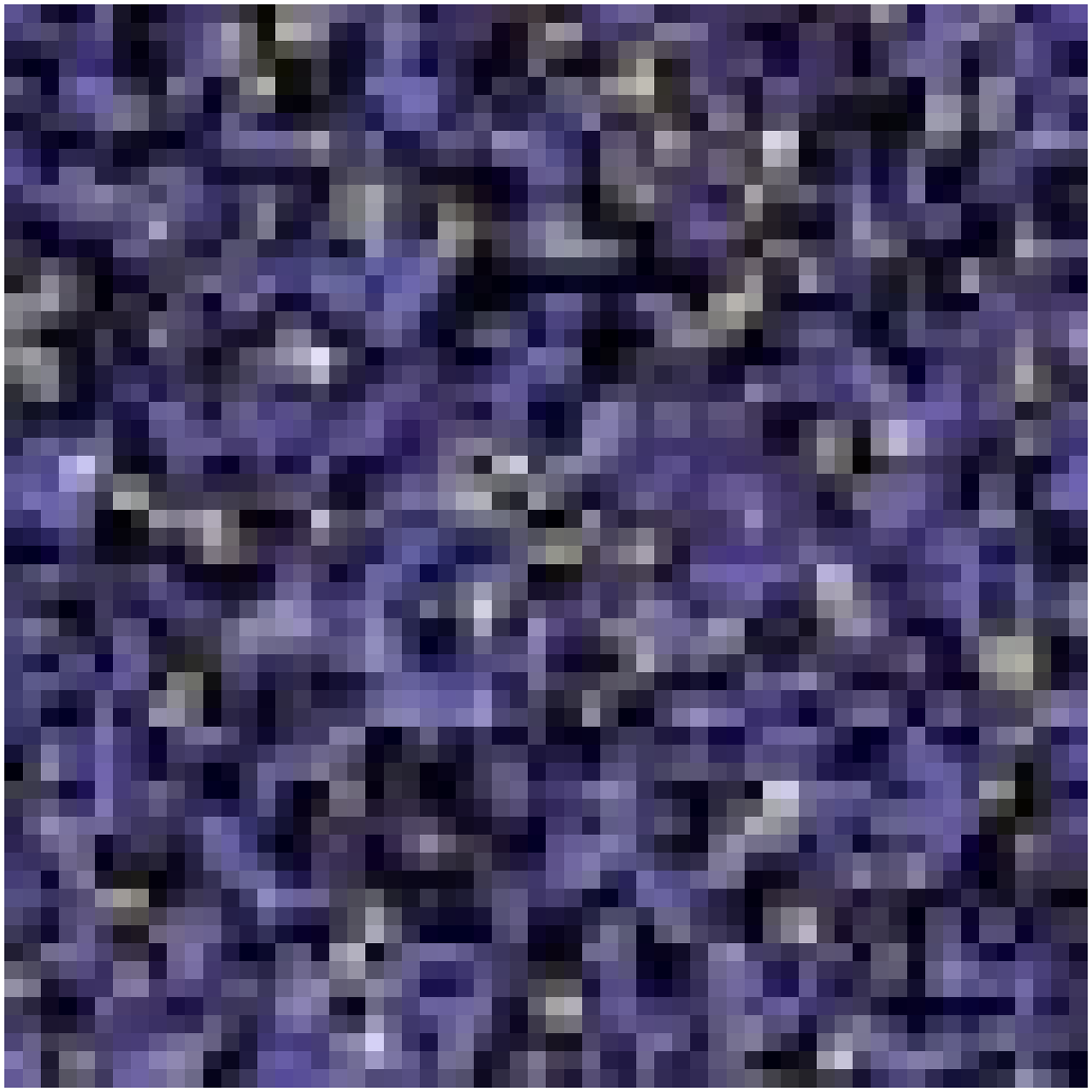}  \FGb{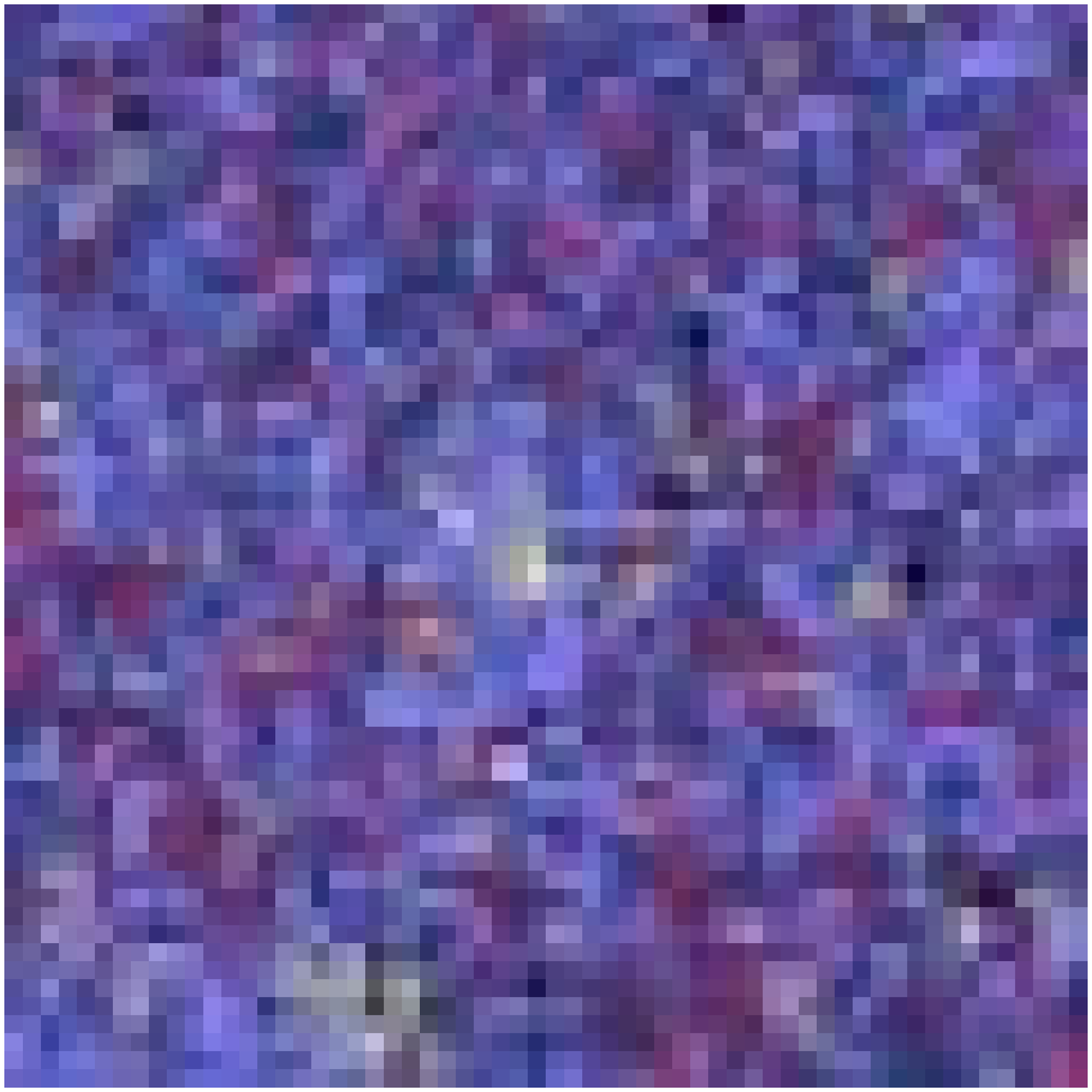}  \FGb{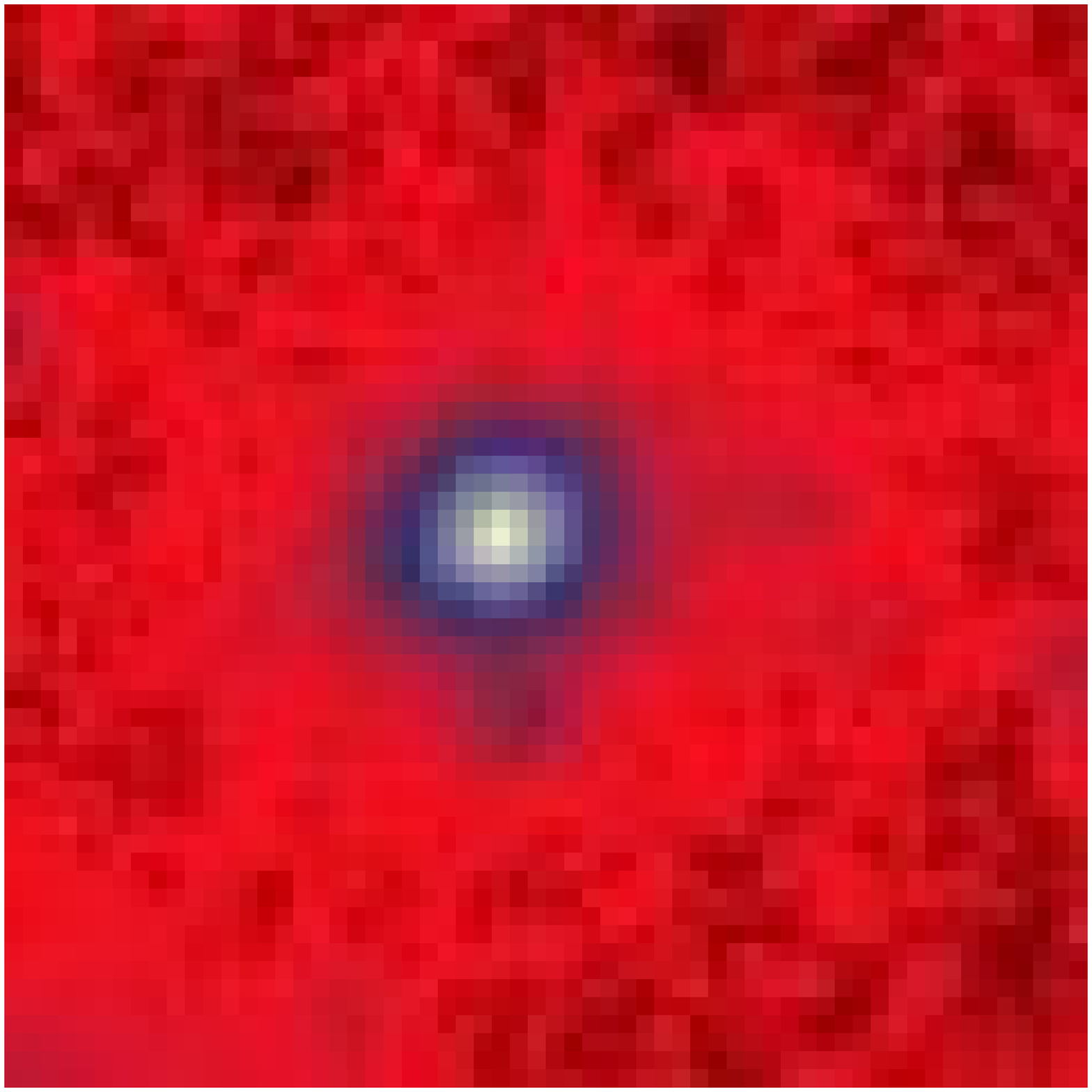}  \FGb{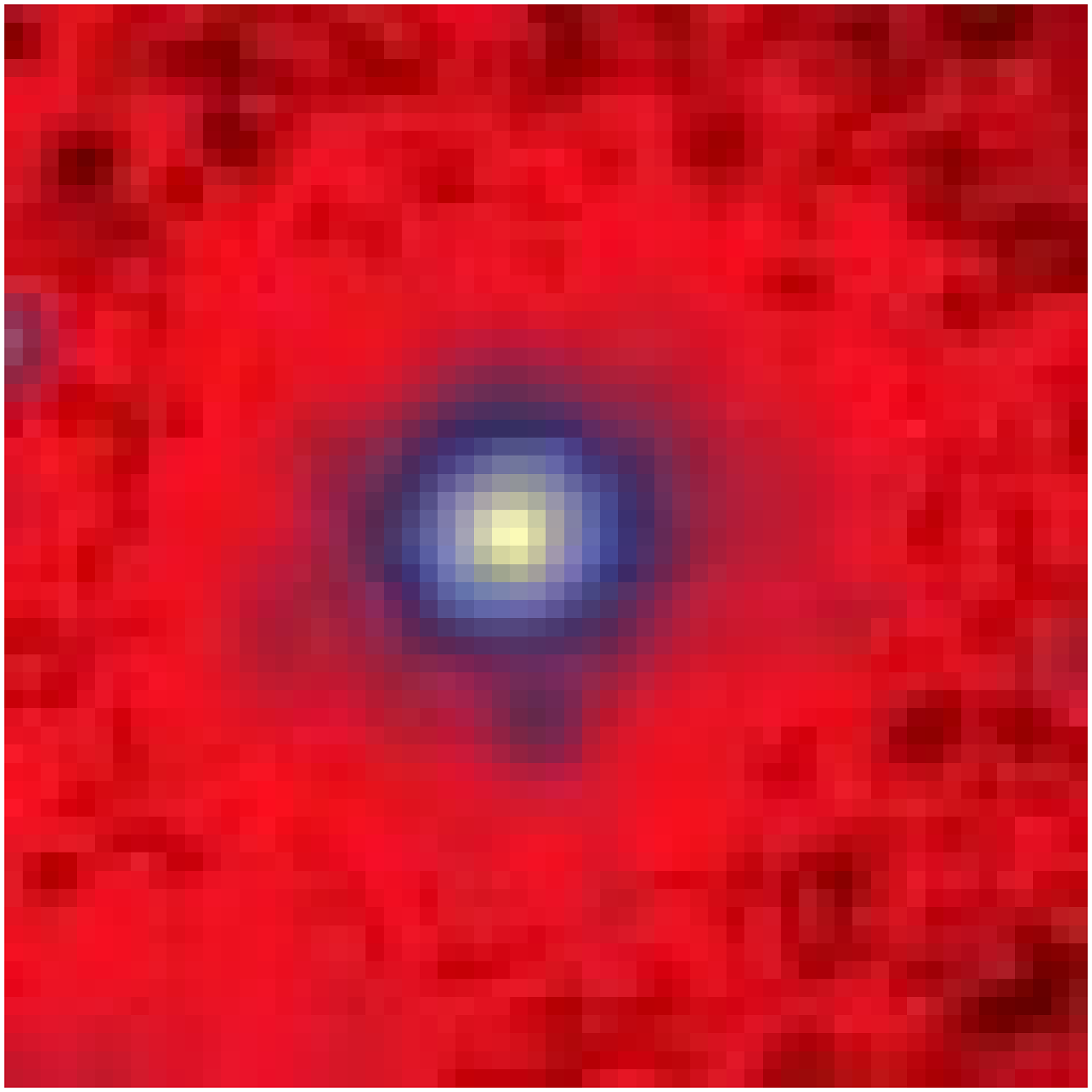}  \FGb{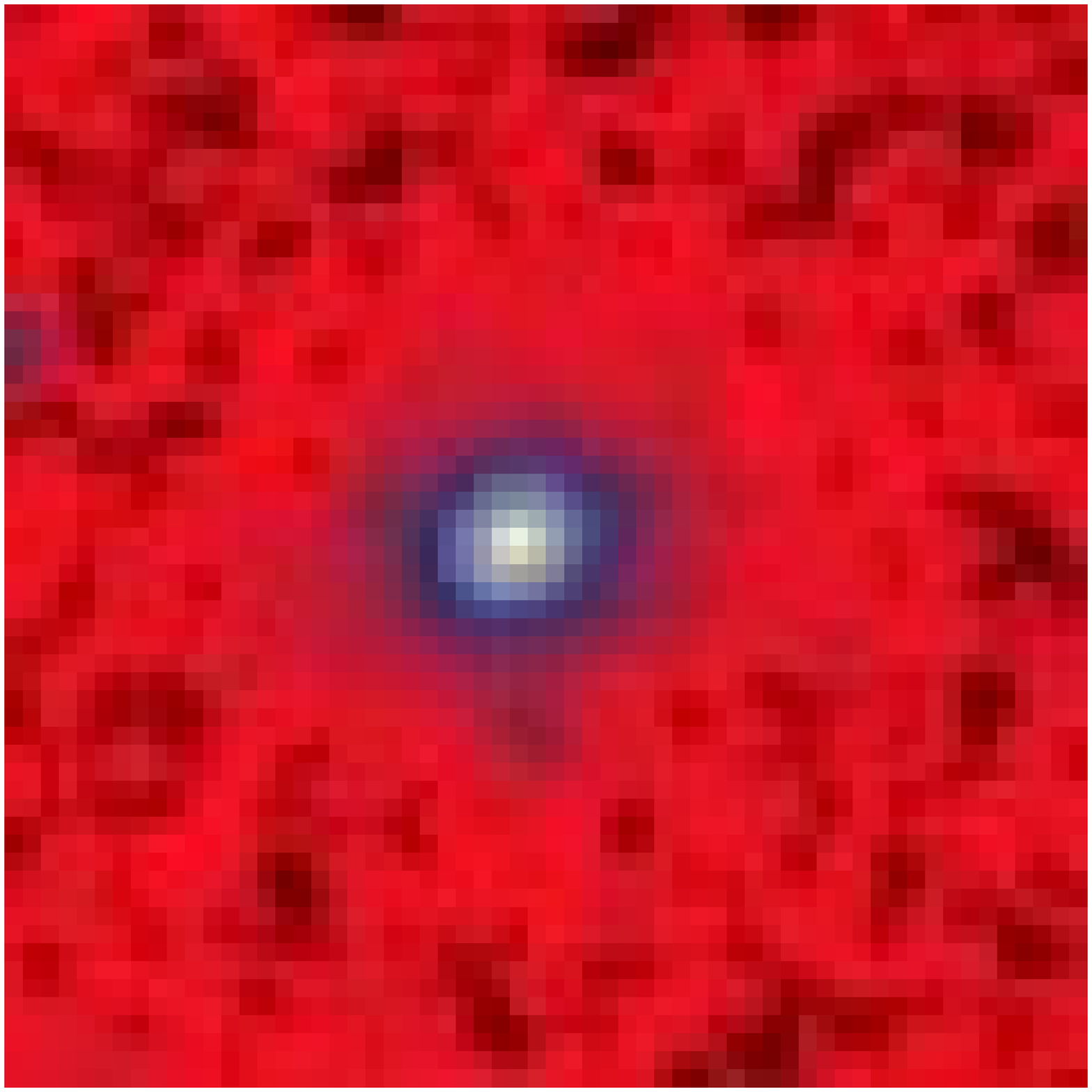}  \FGb{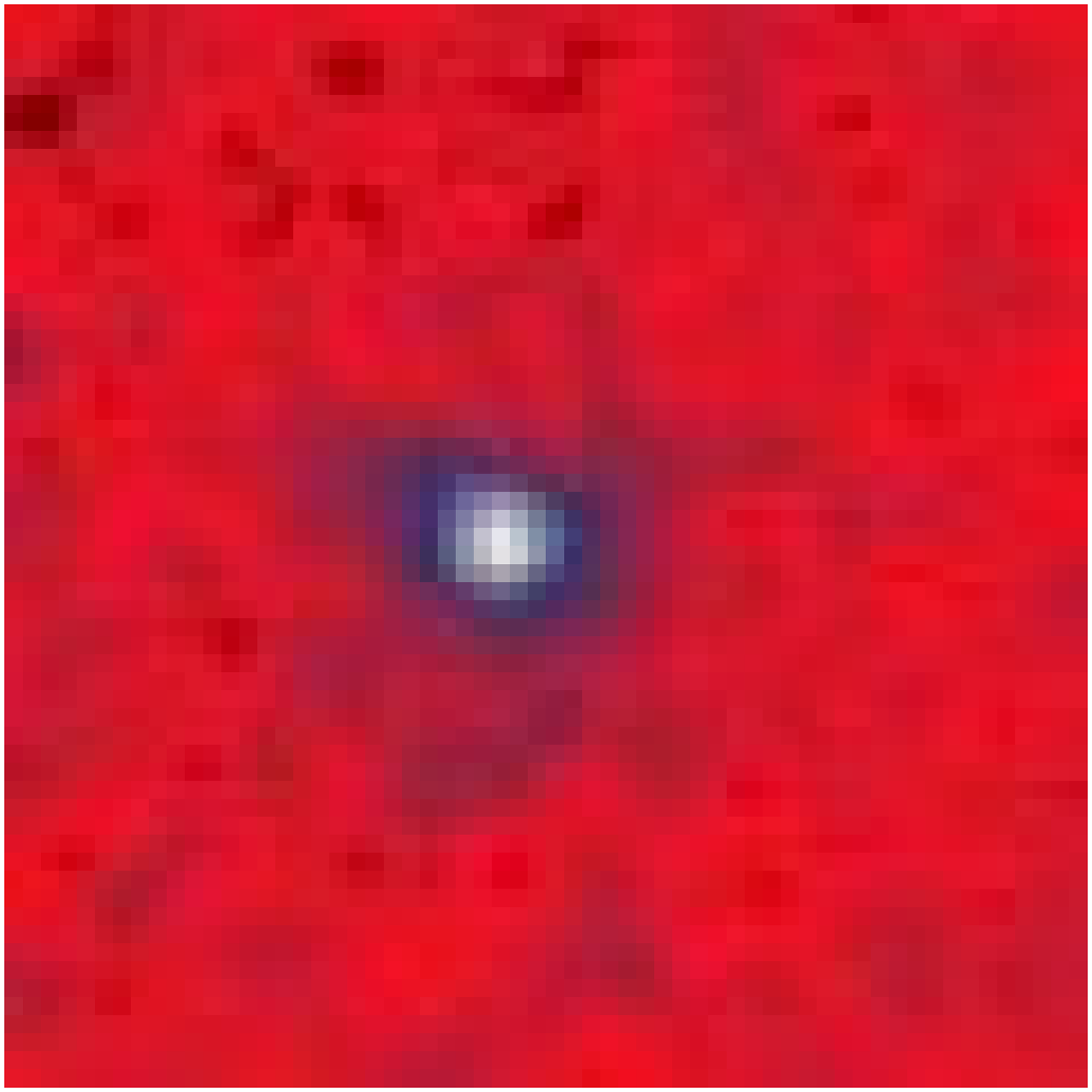}  \FGb{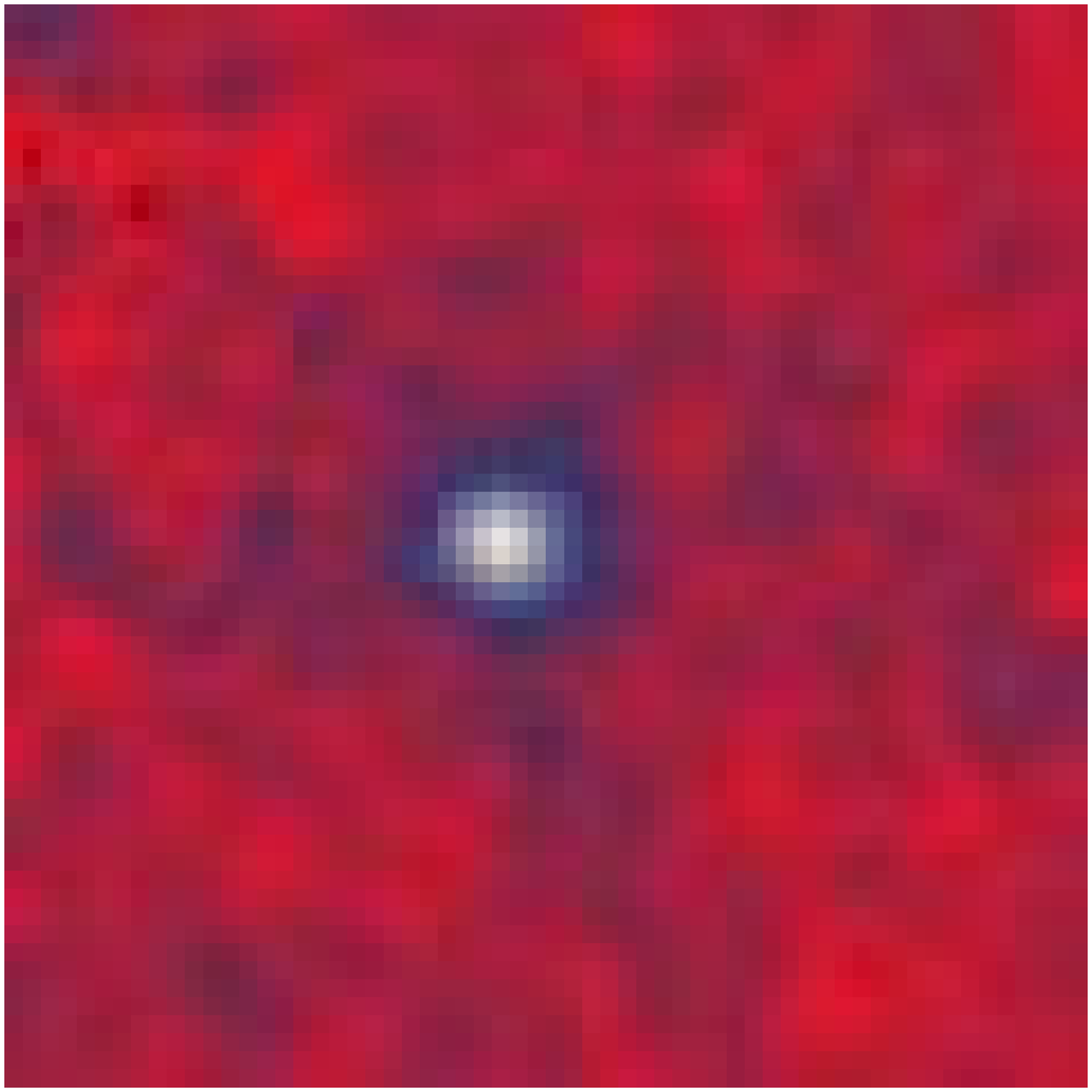}  \FGb{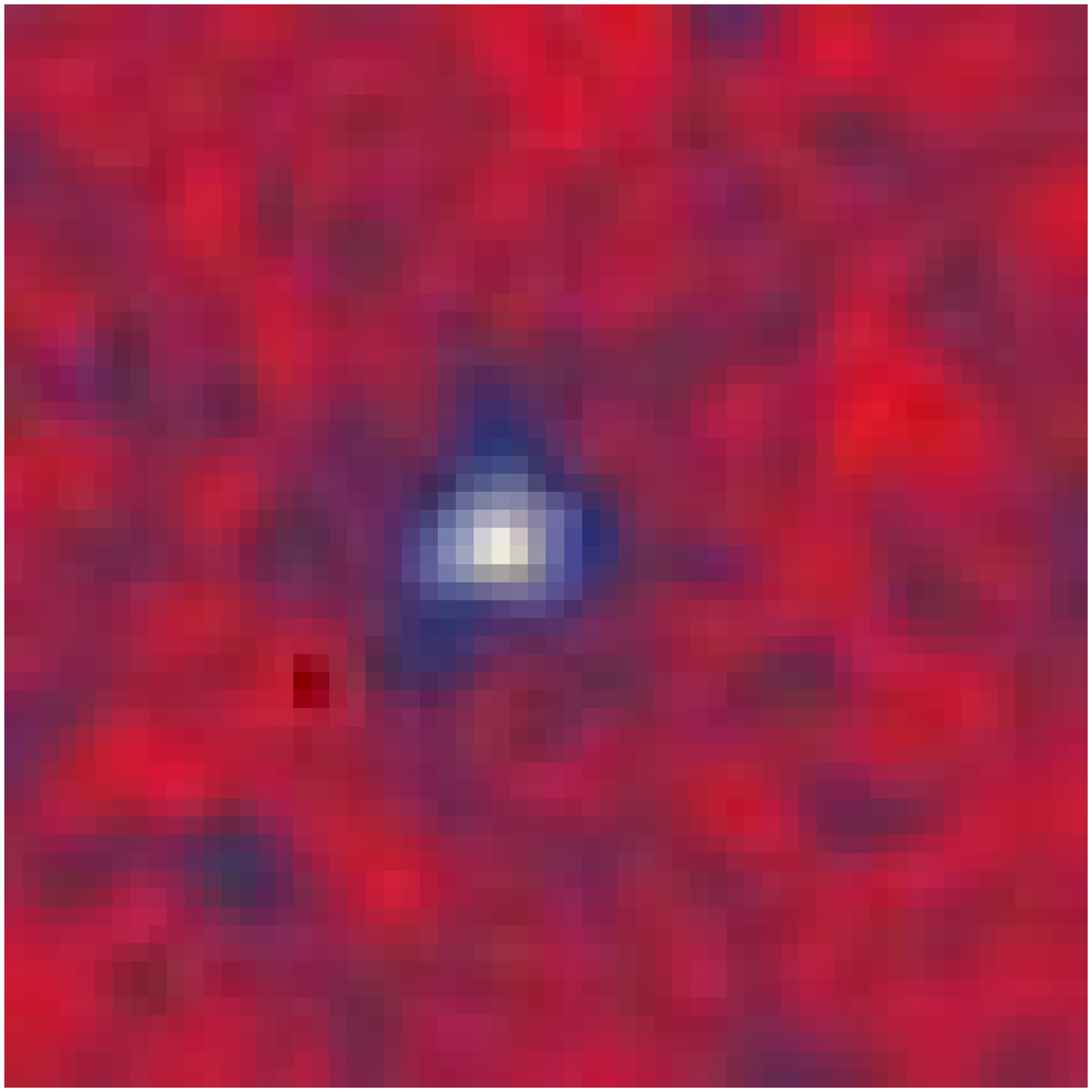}  \FGb{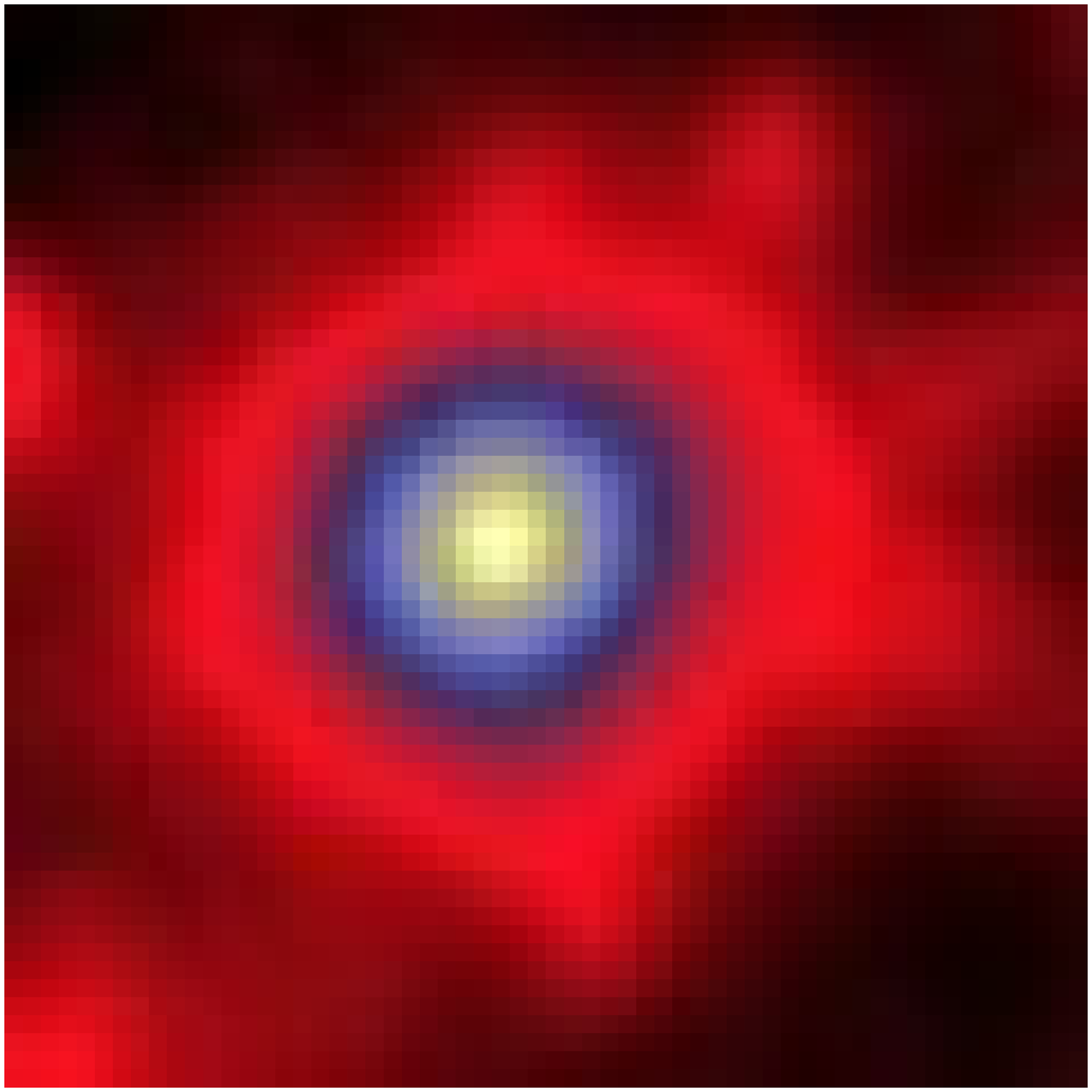}  \FGb{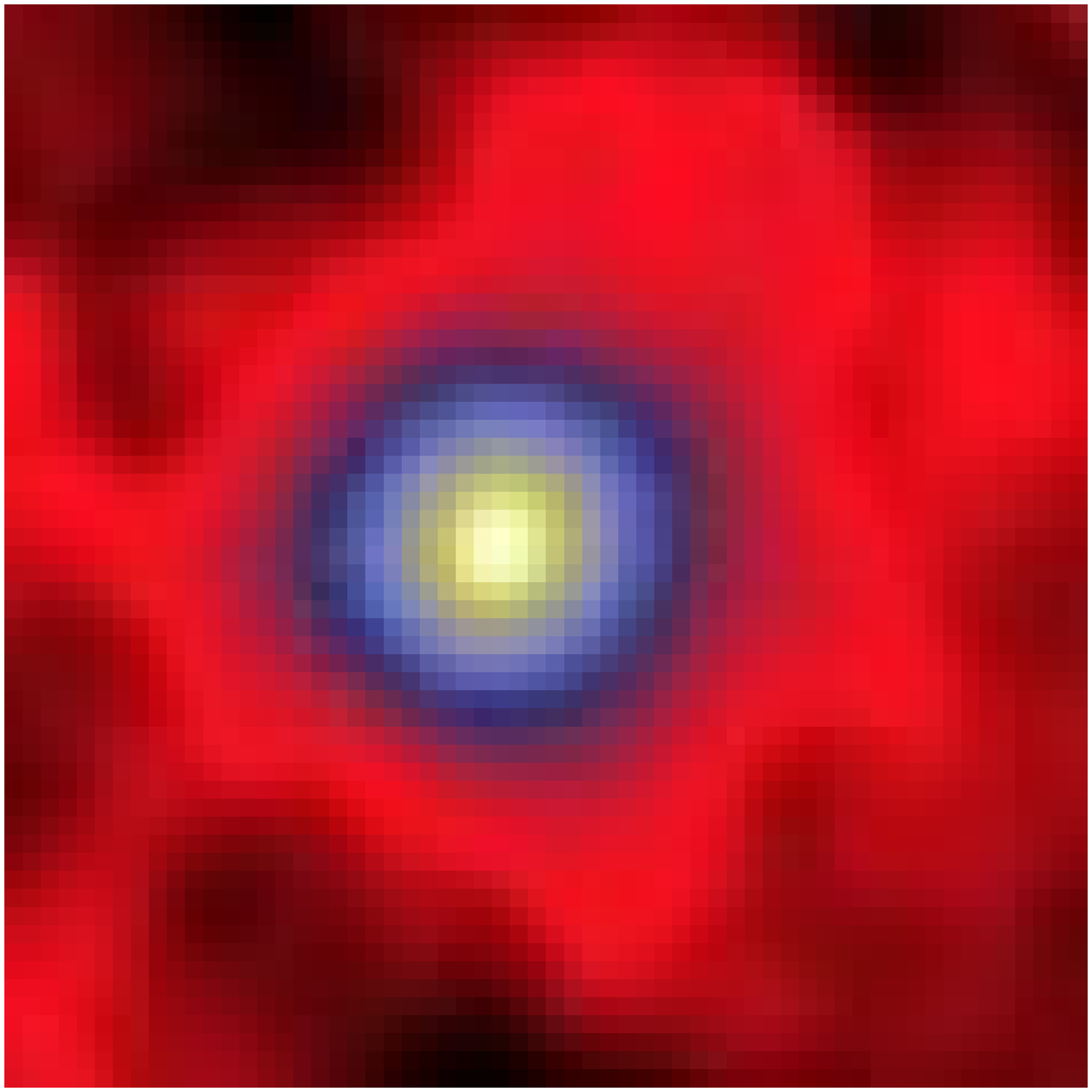}  \FGb{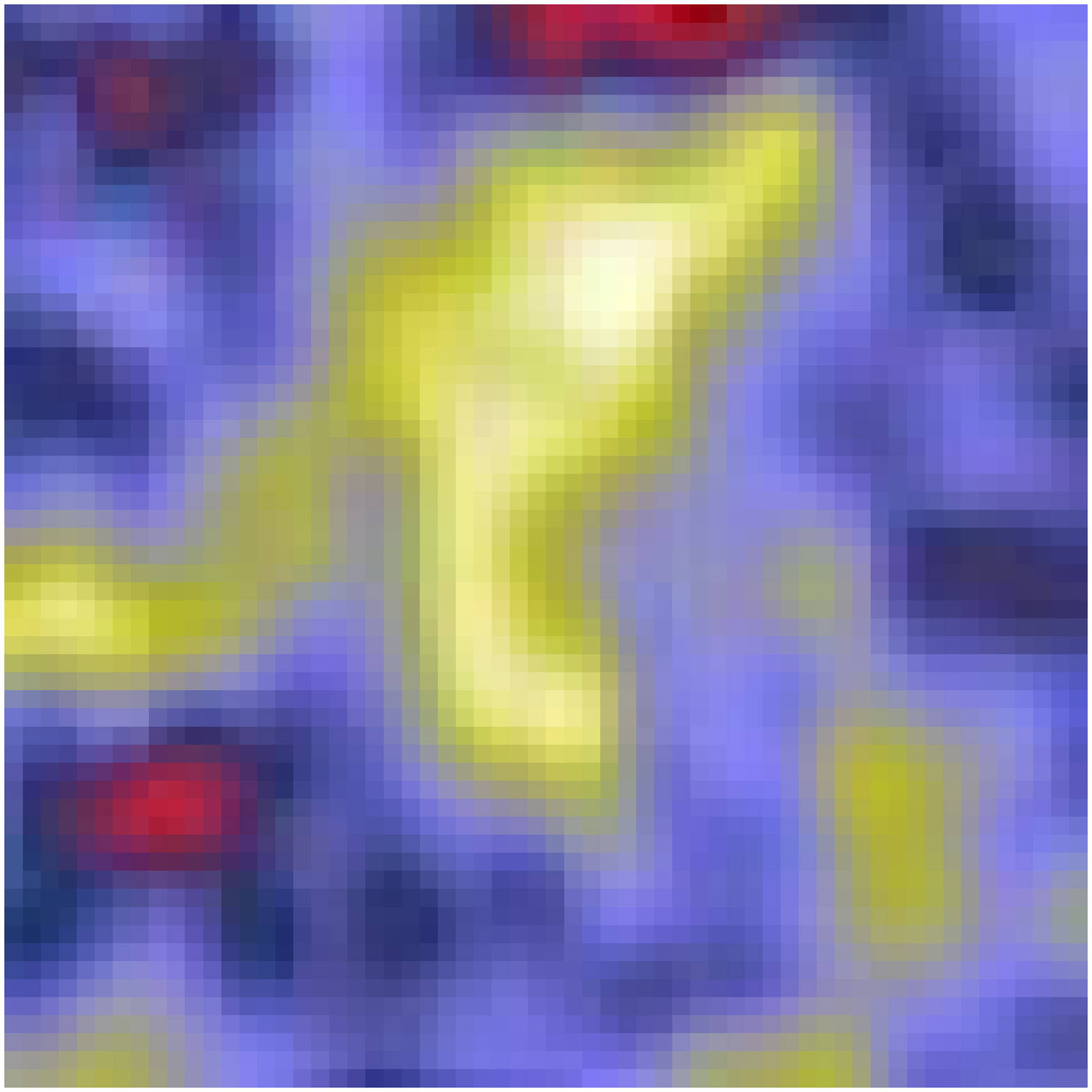}  \FGb{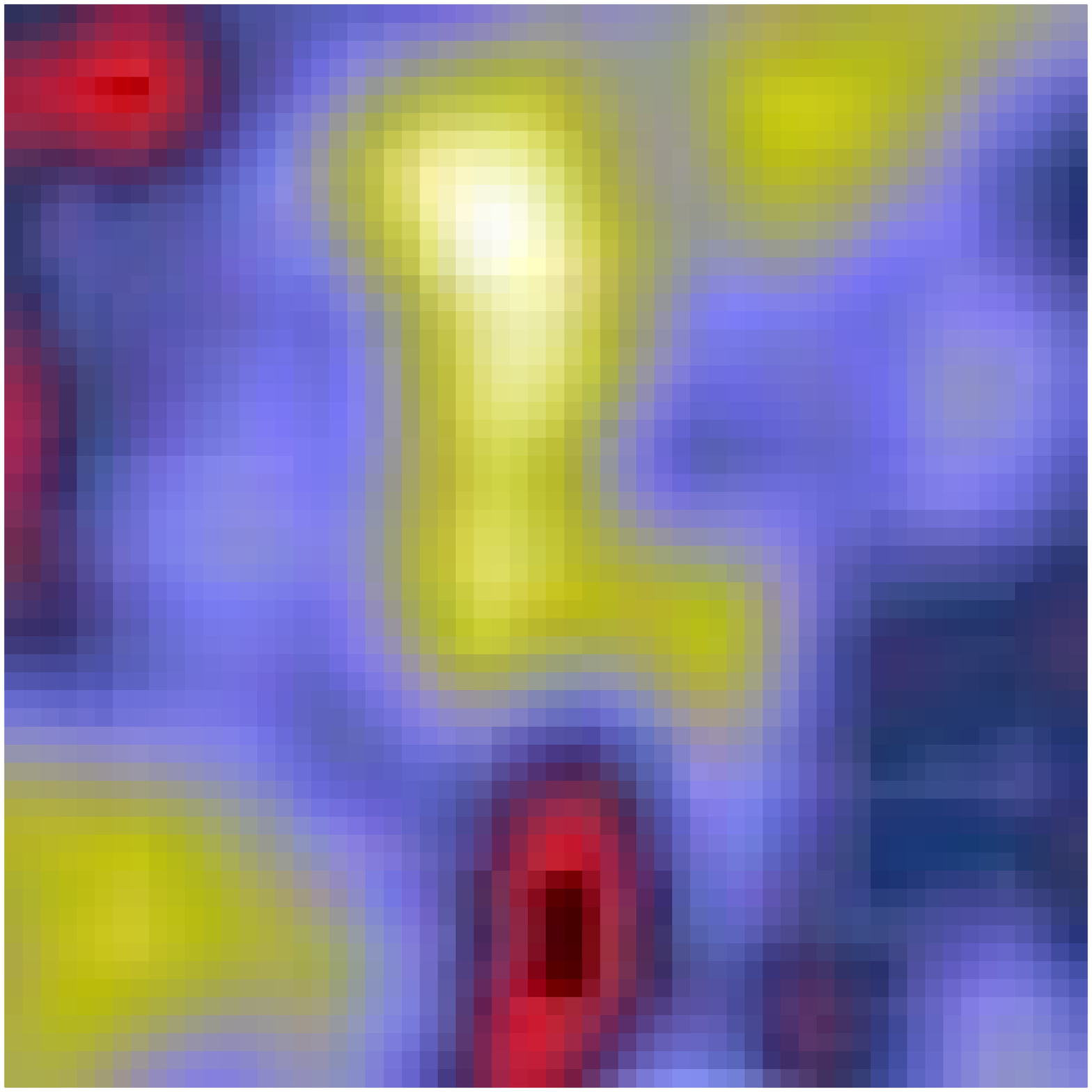} \\  {\#37   NCIA004881}  \\
  \FGb{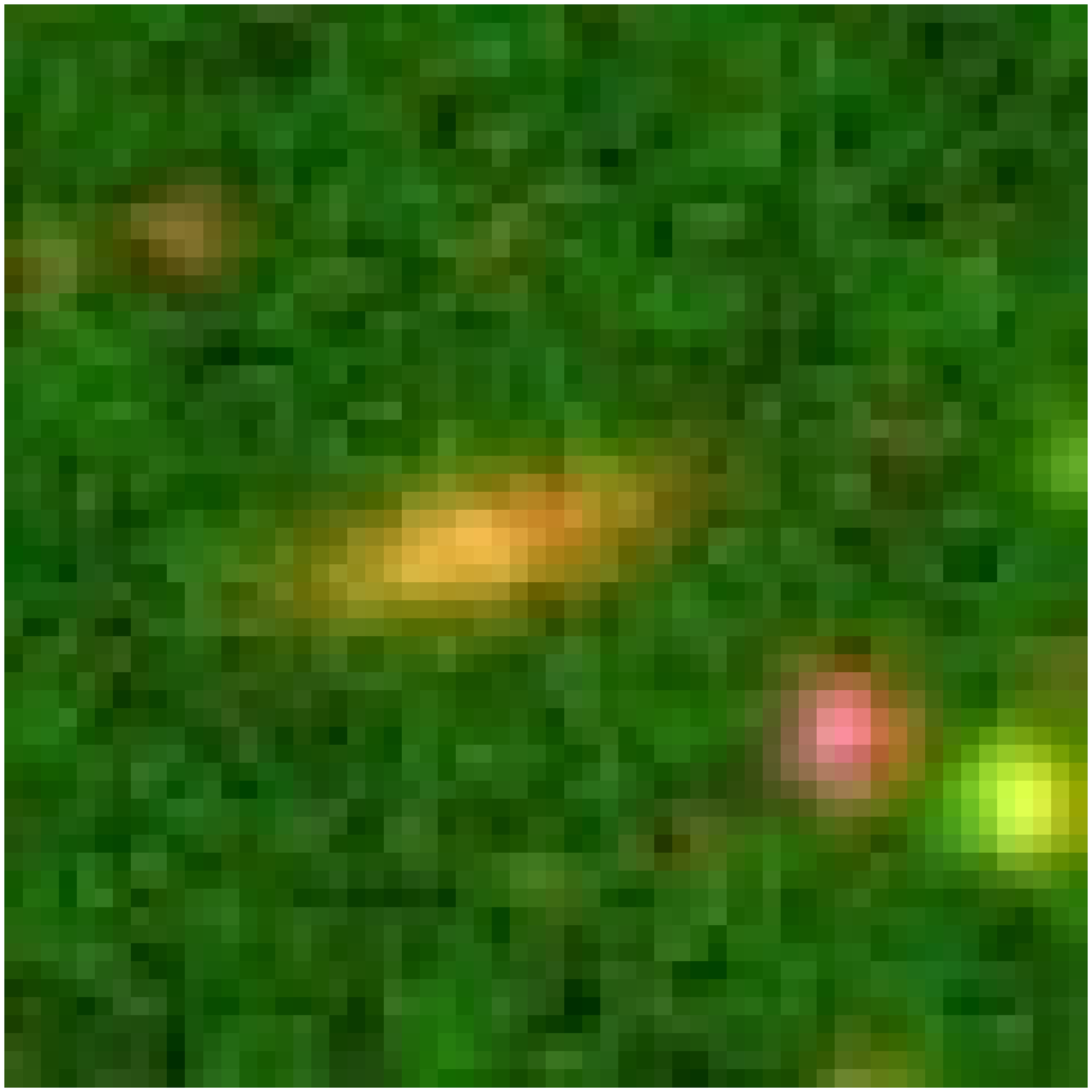}  \FGb{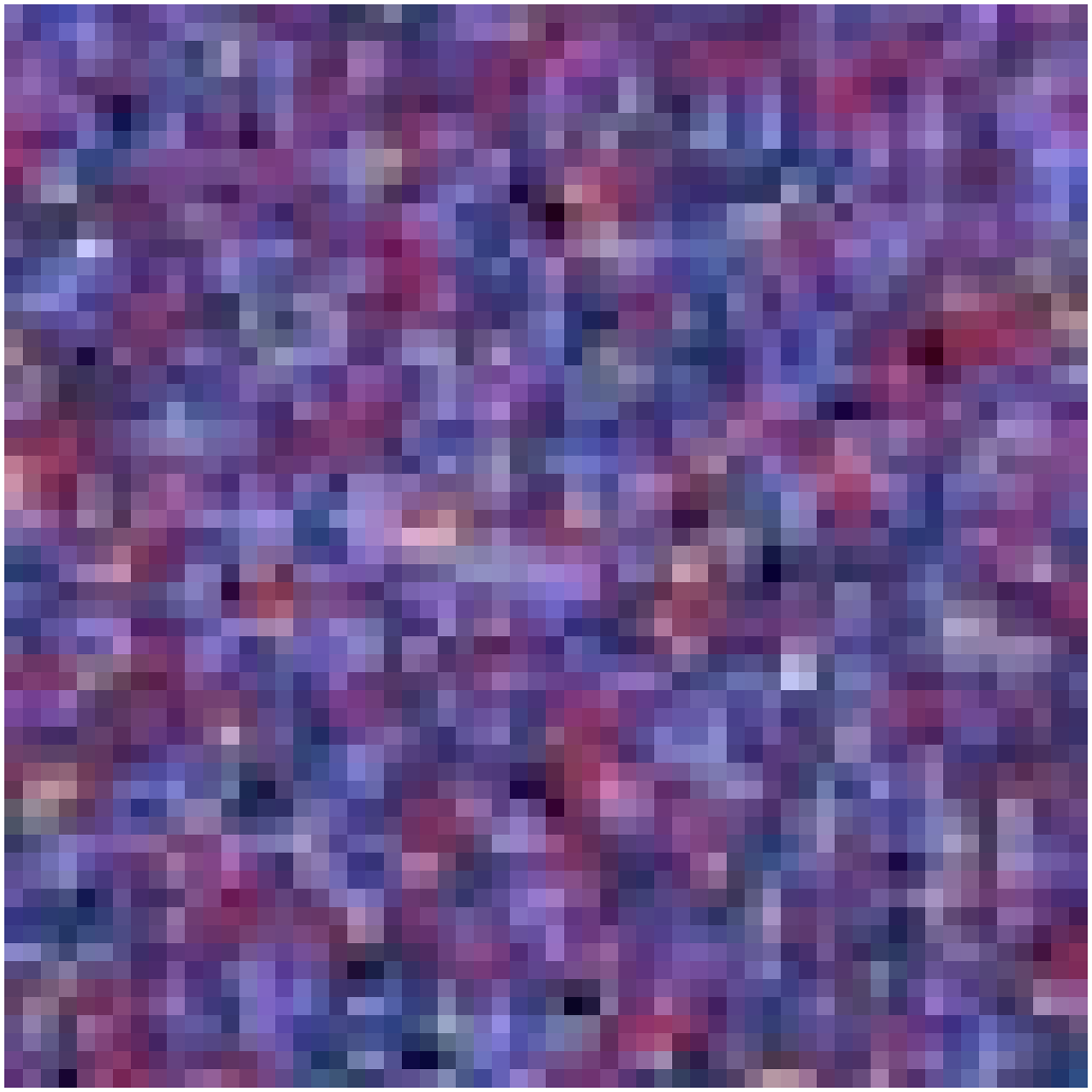}  \FGb{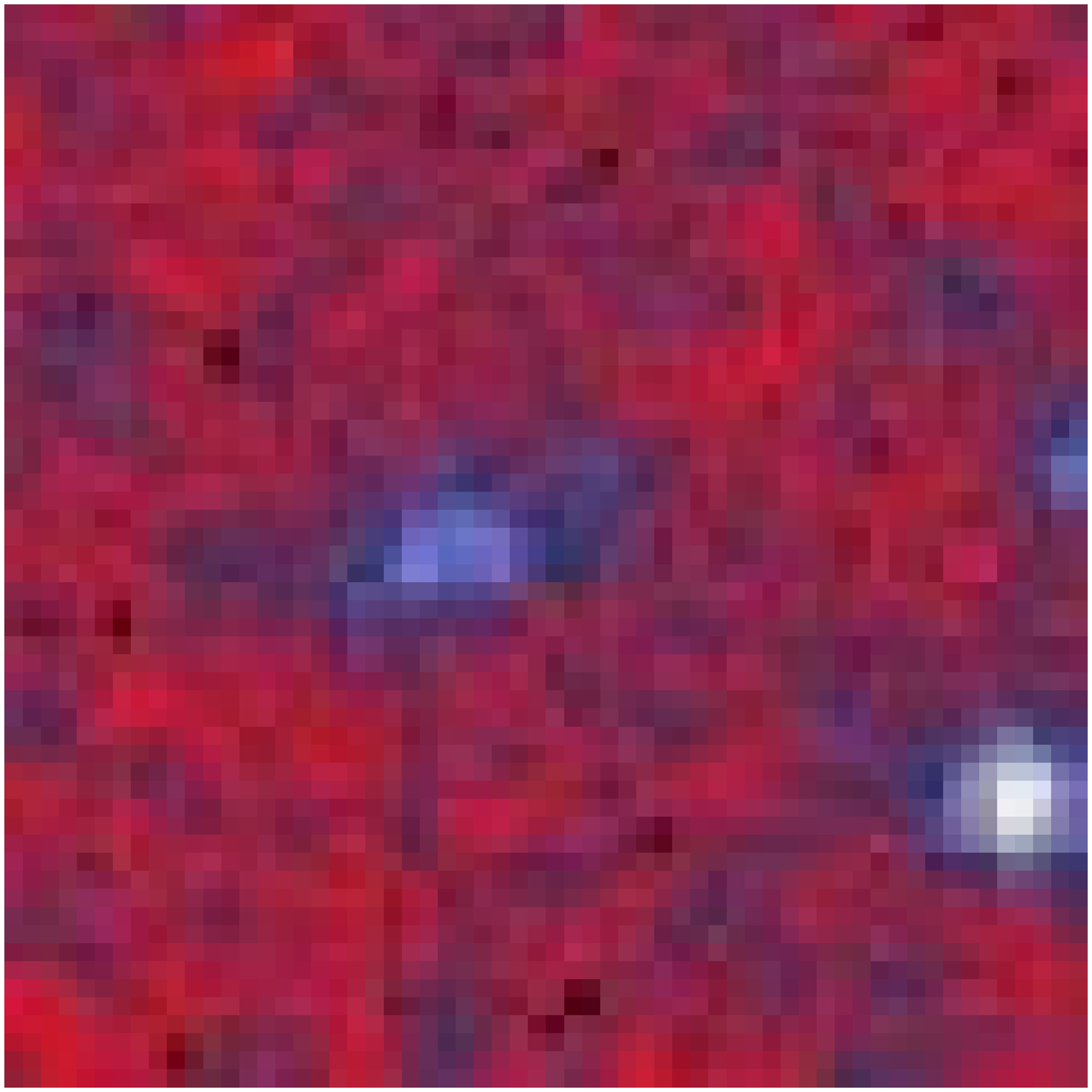}  \FGb{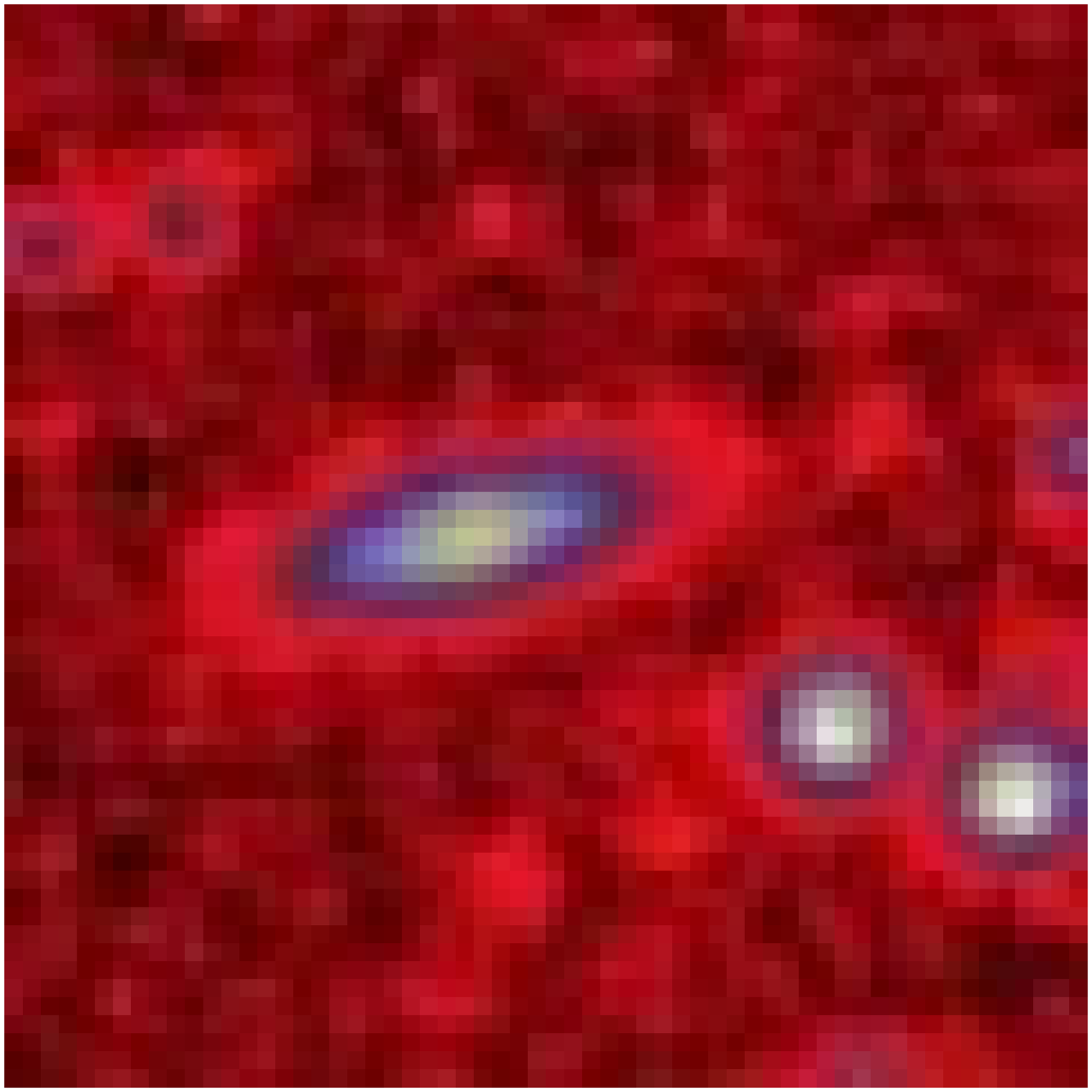}  \FGb{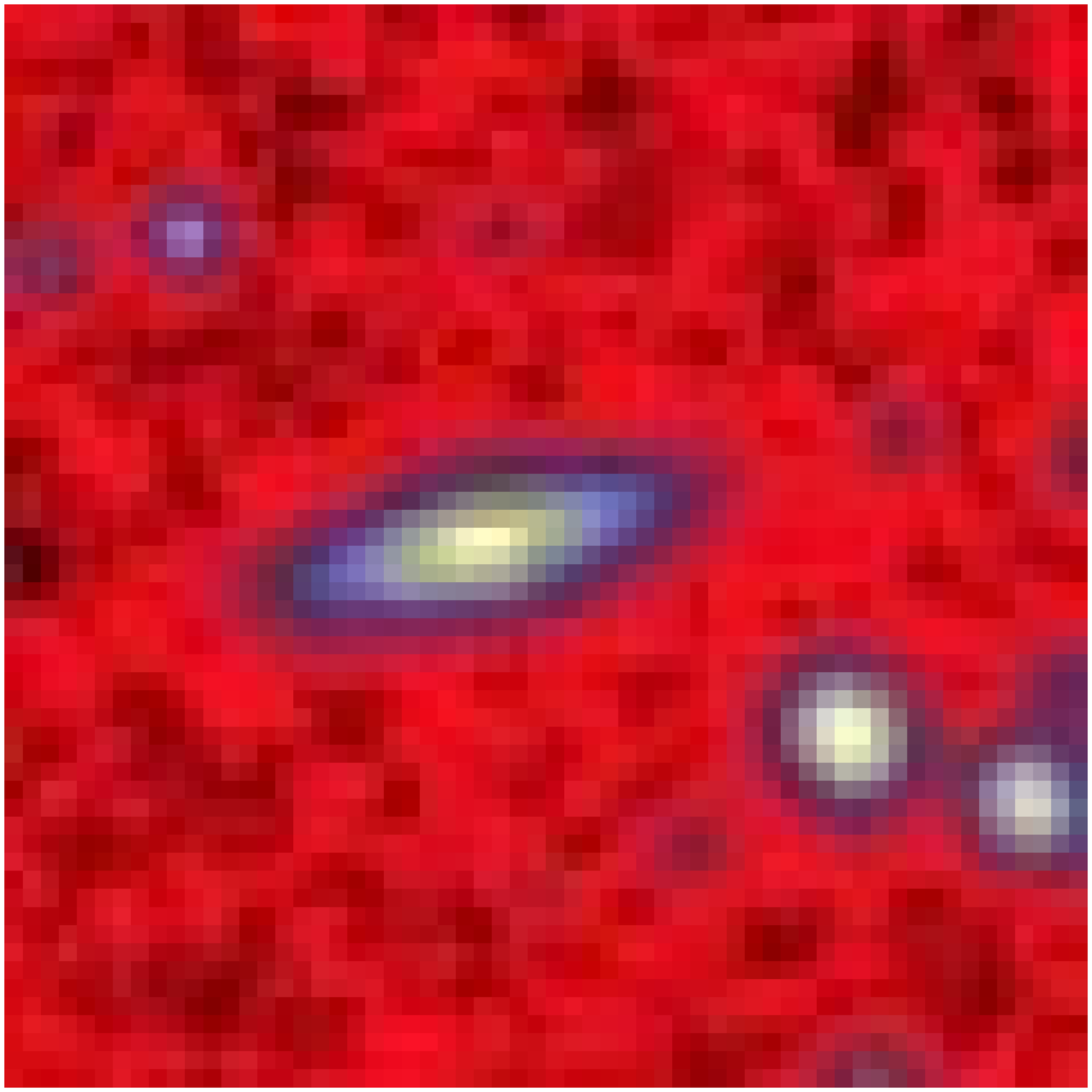}  \FGb{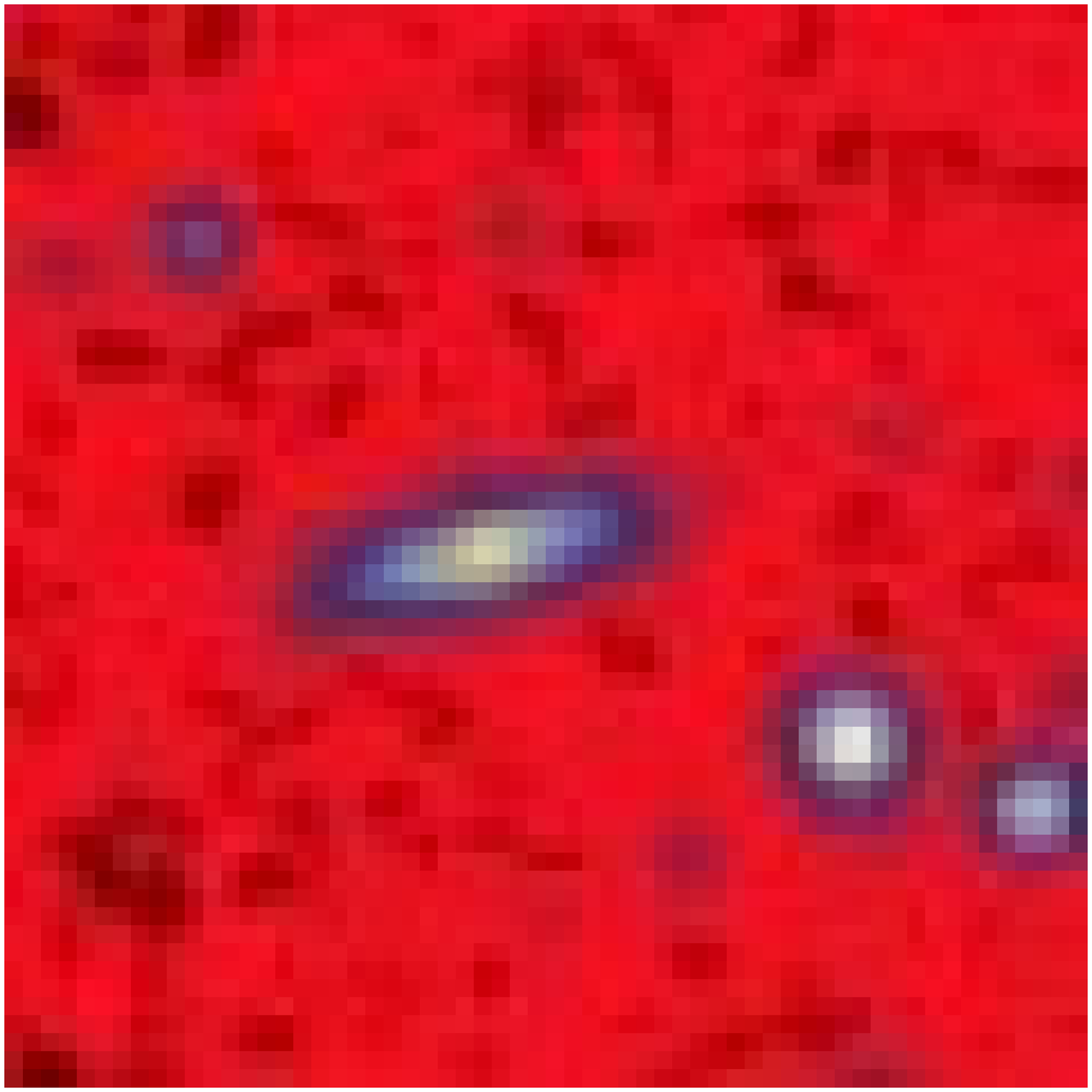}  \FGb{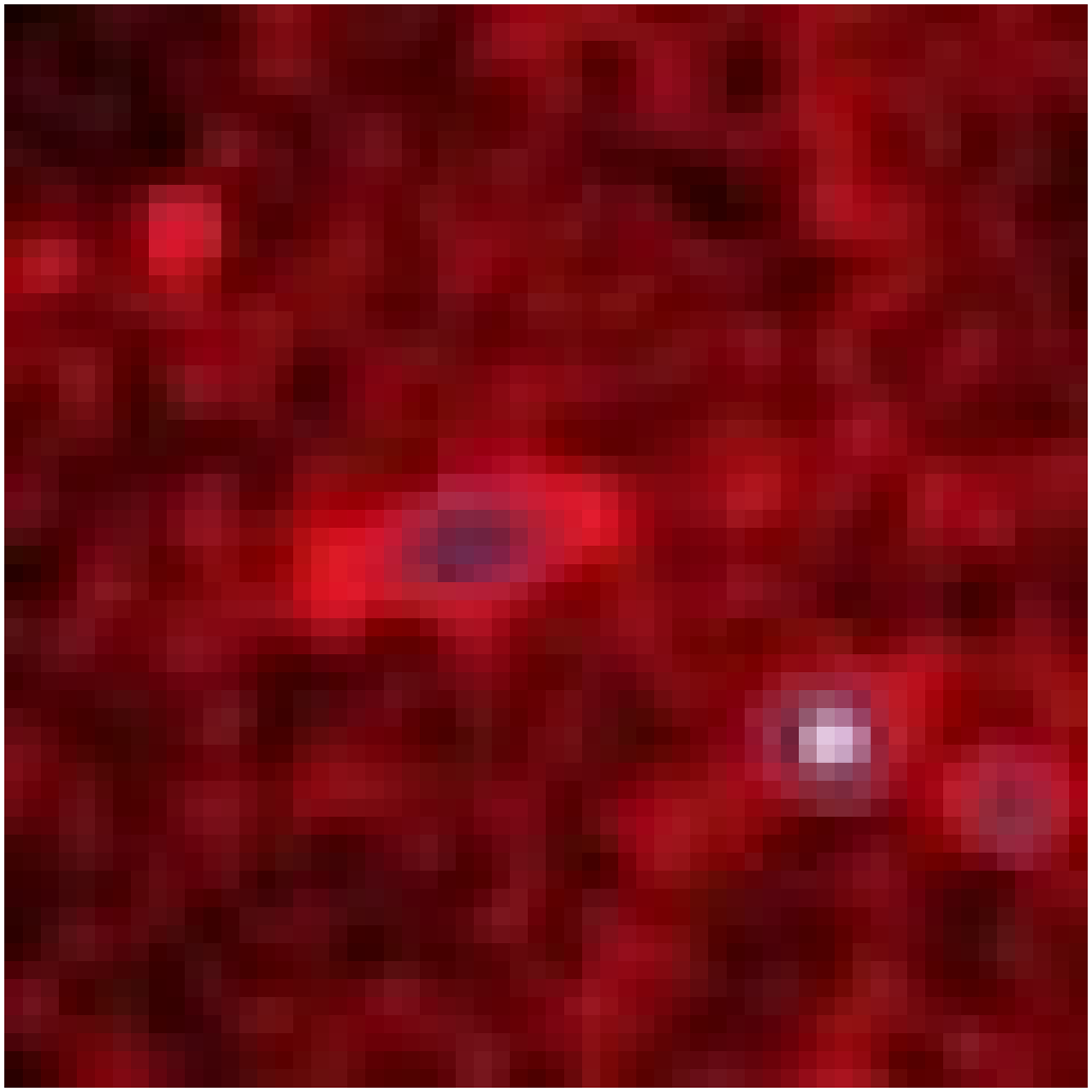}  \FGb{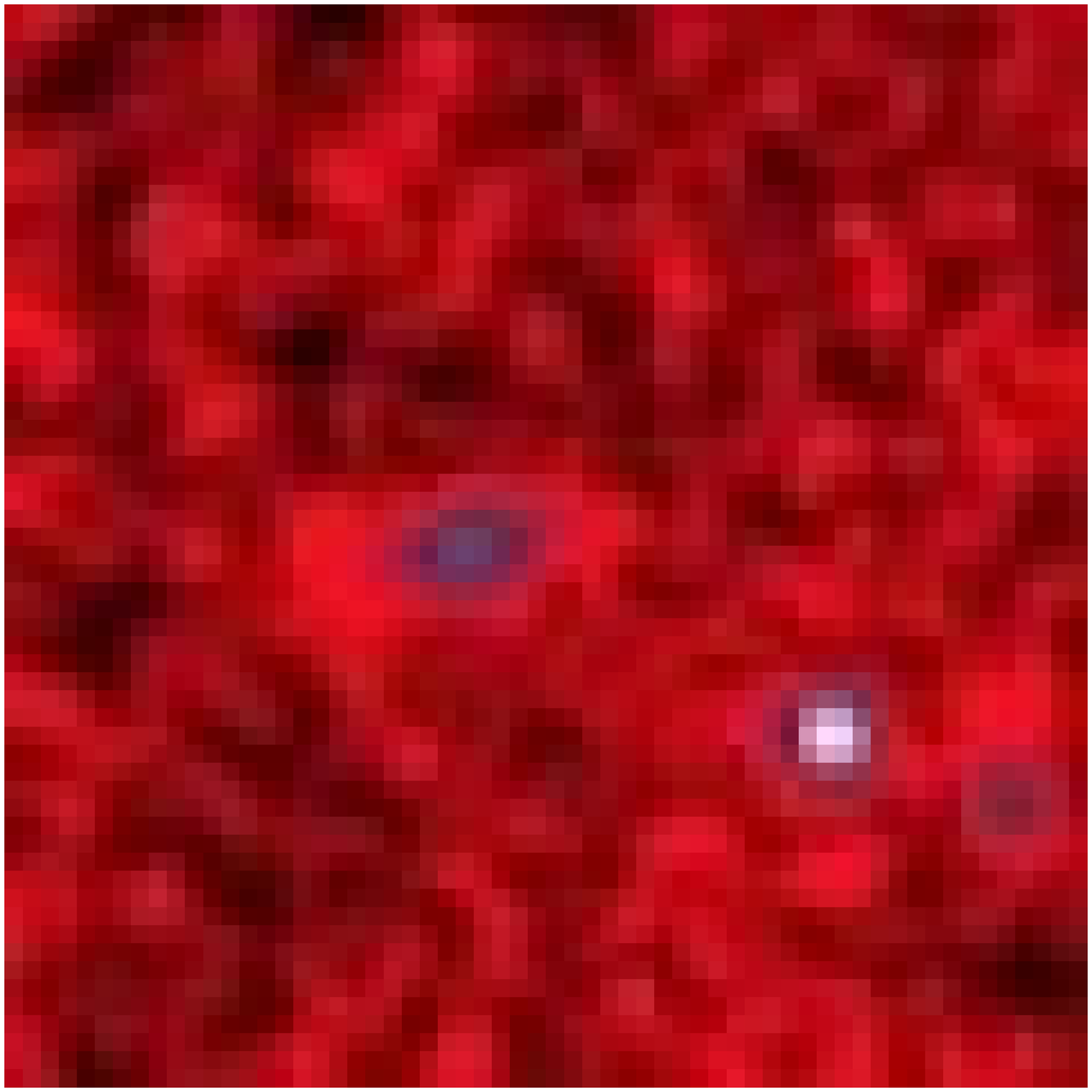}  \FGb{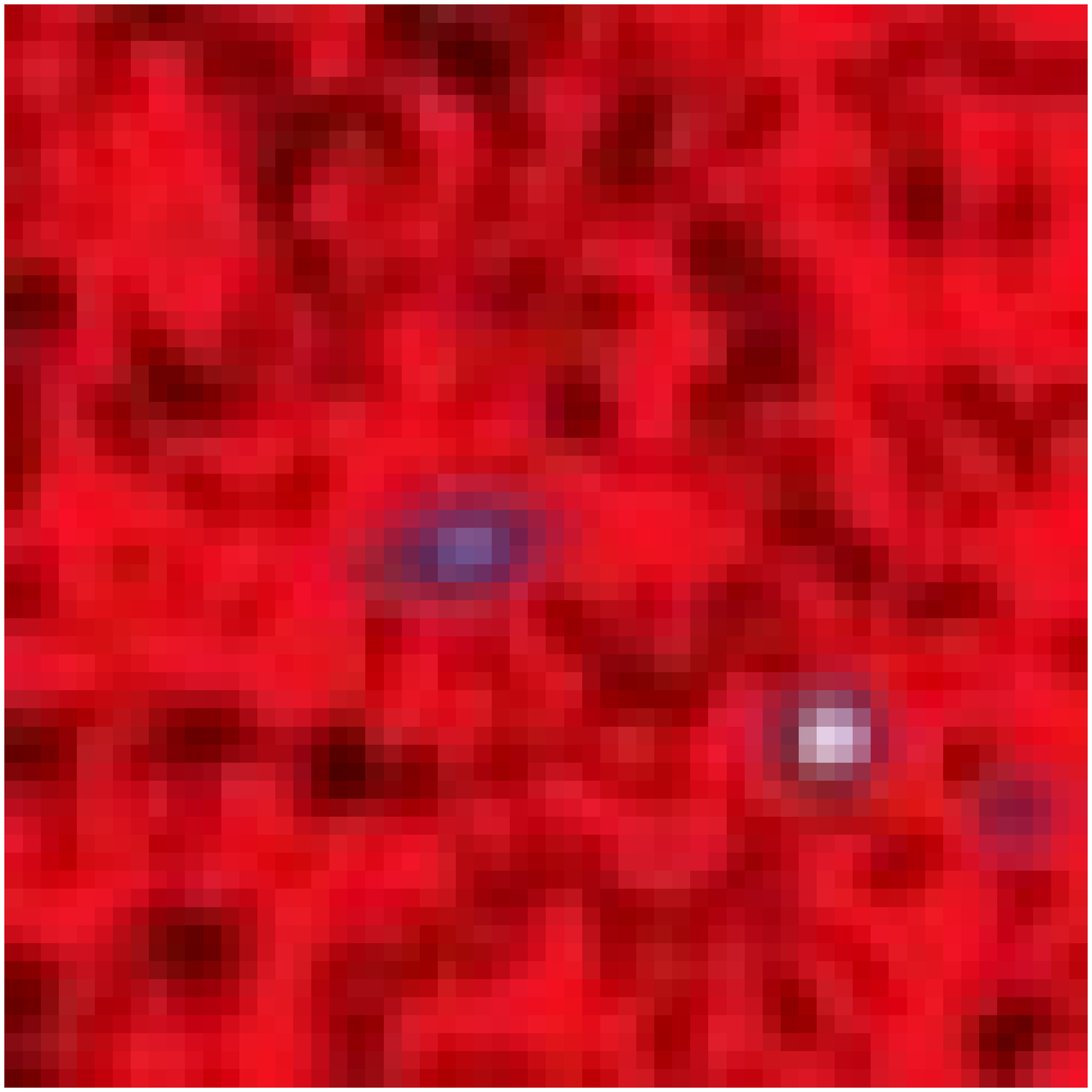}  \FGb{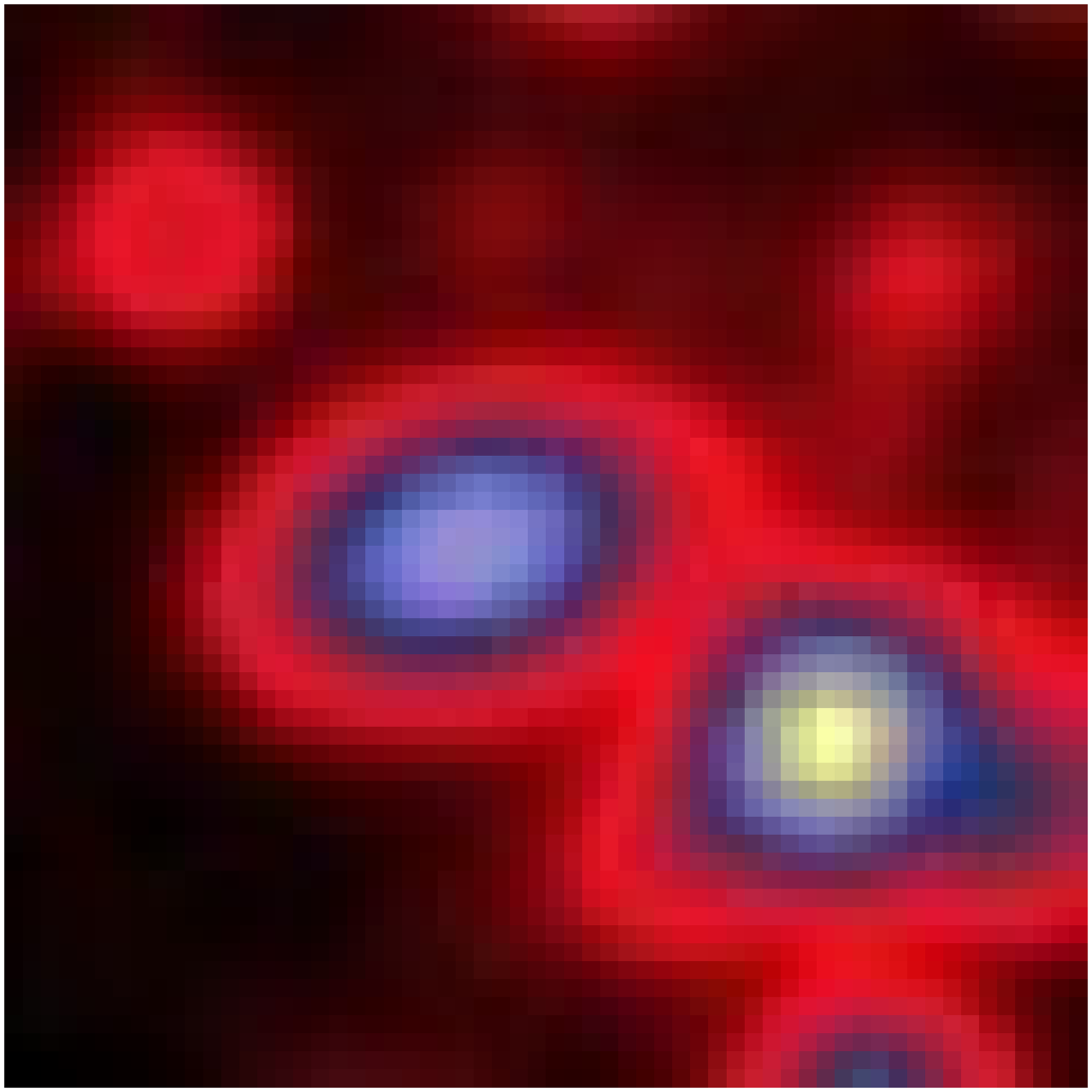}  \FGb{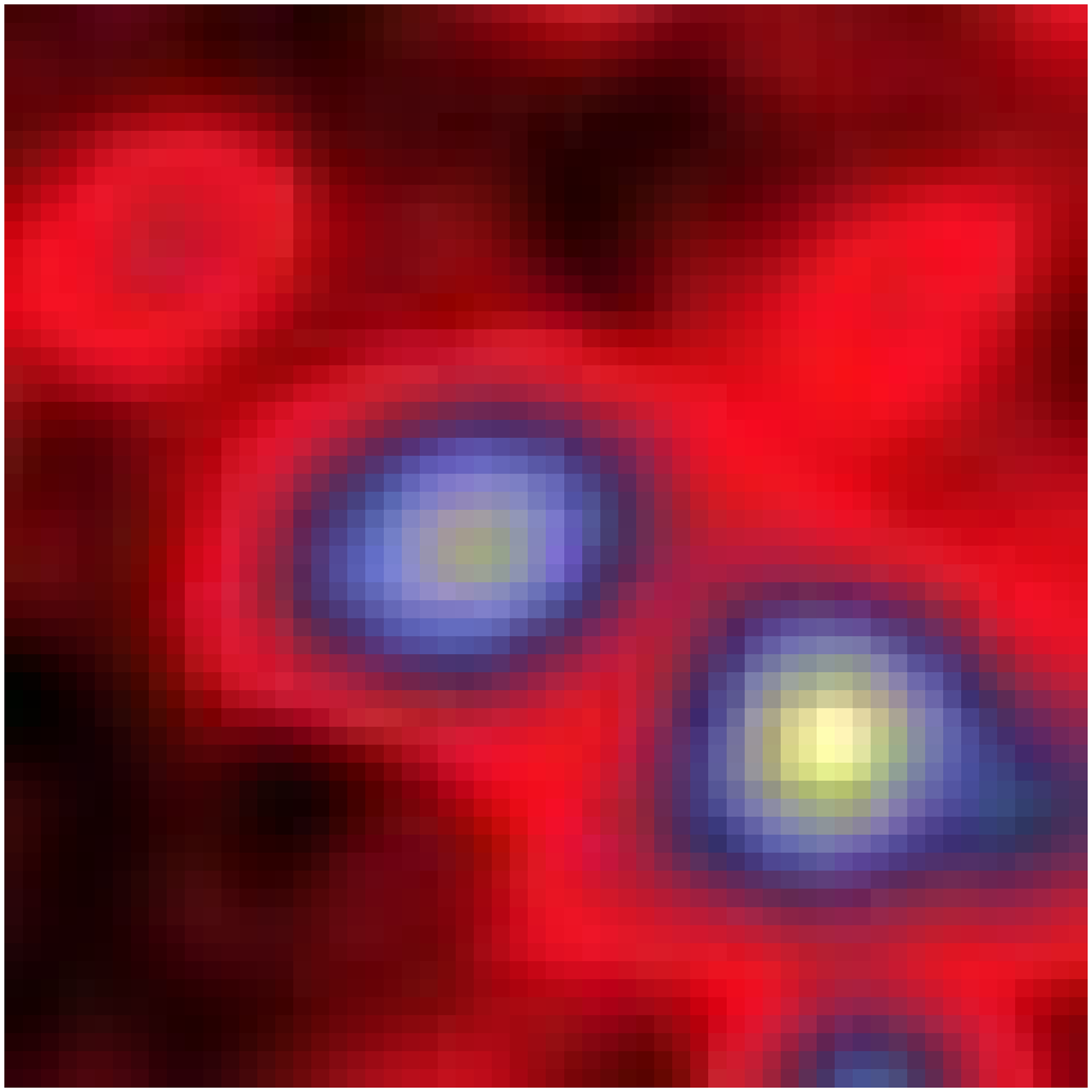}  \FGb{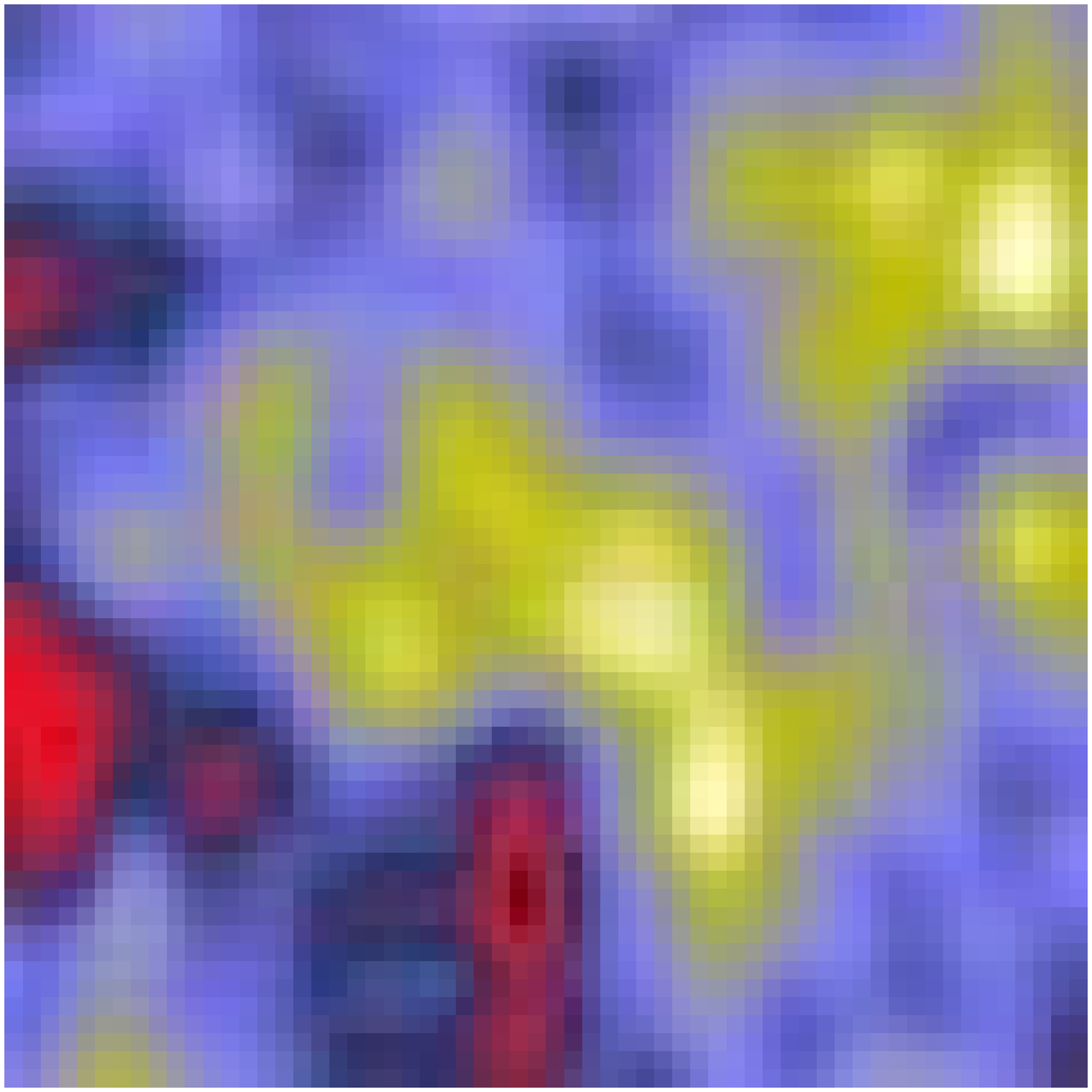}  \FGb{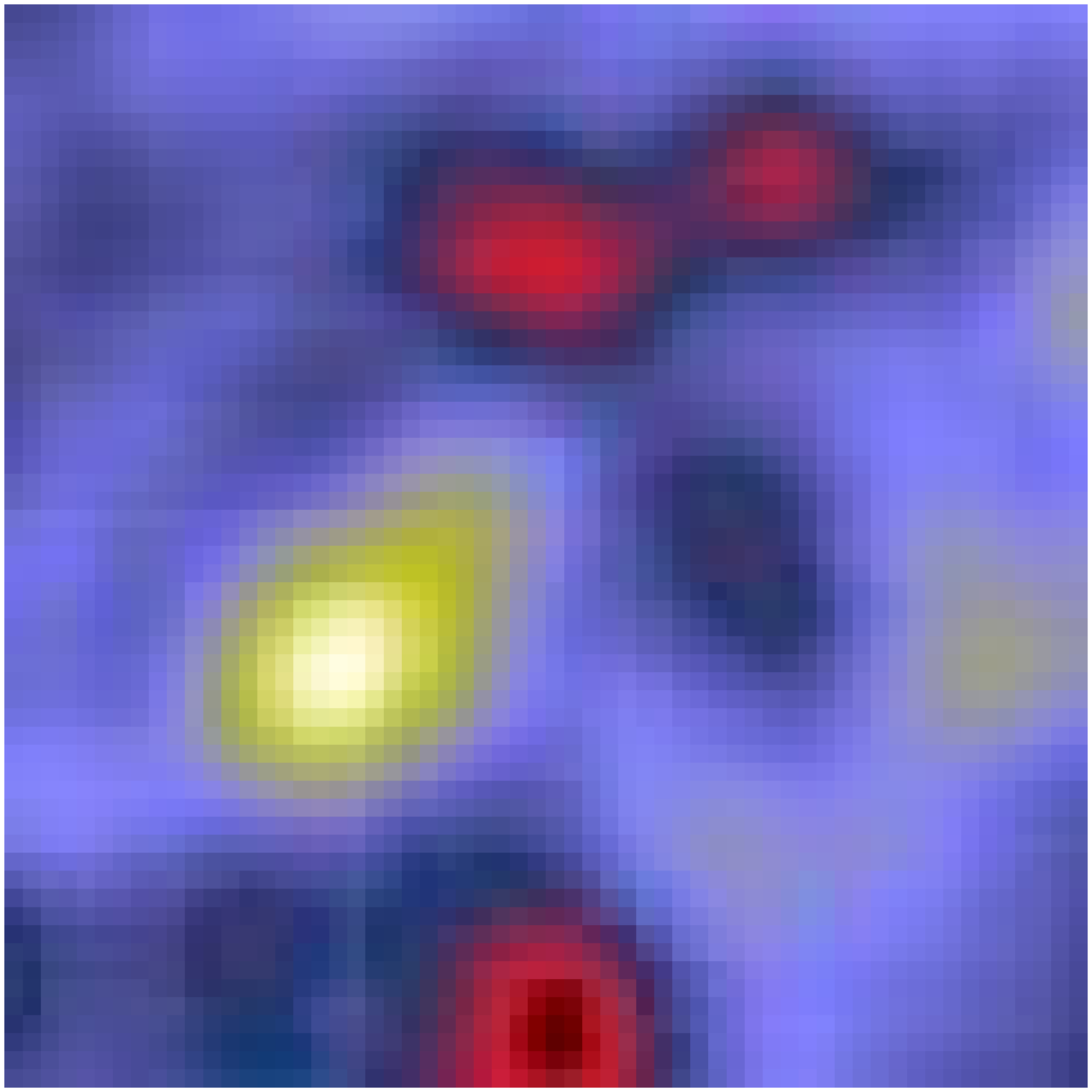} \\  {\#38   NCIA027059}  \\
  \FGb{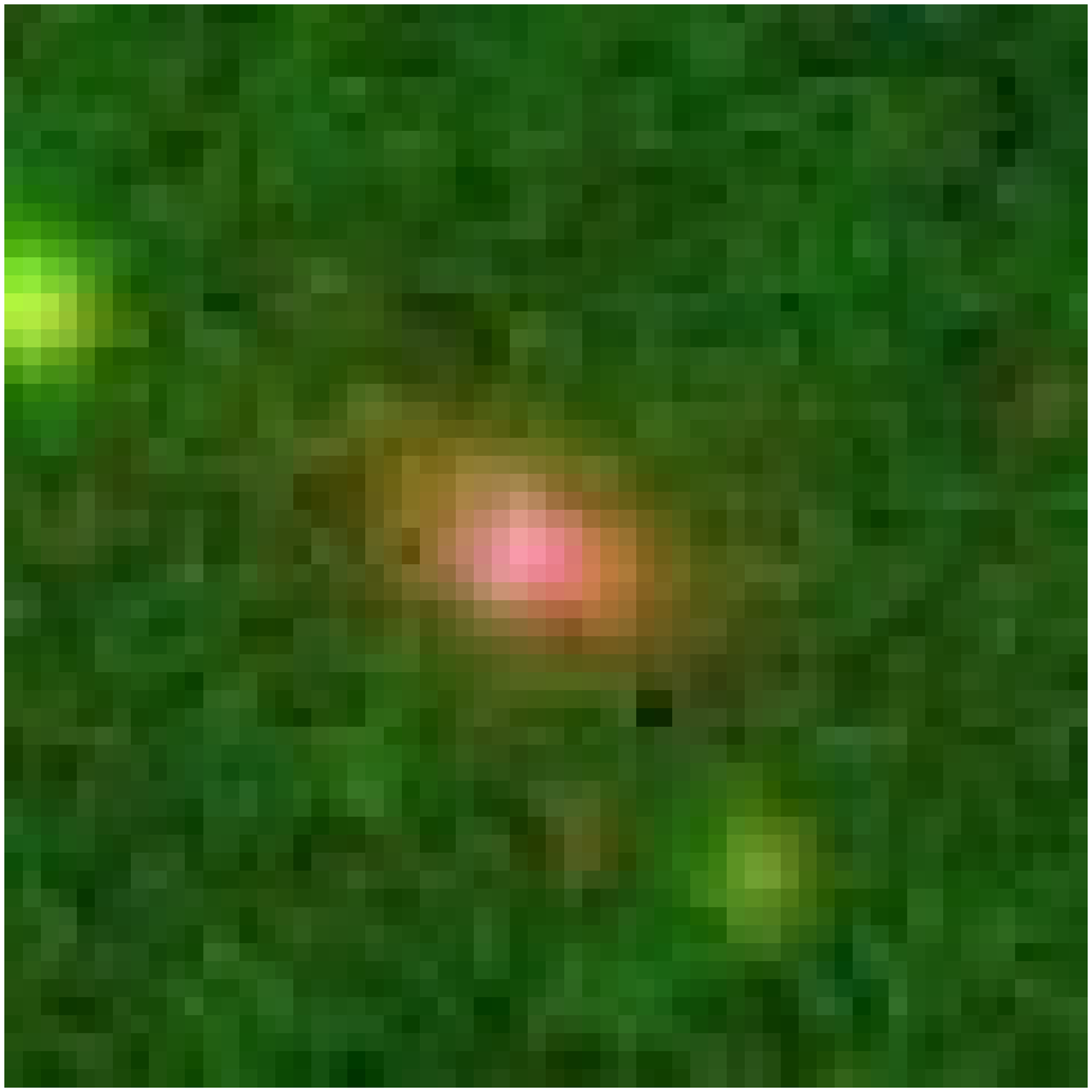}  \FGb{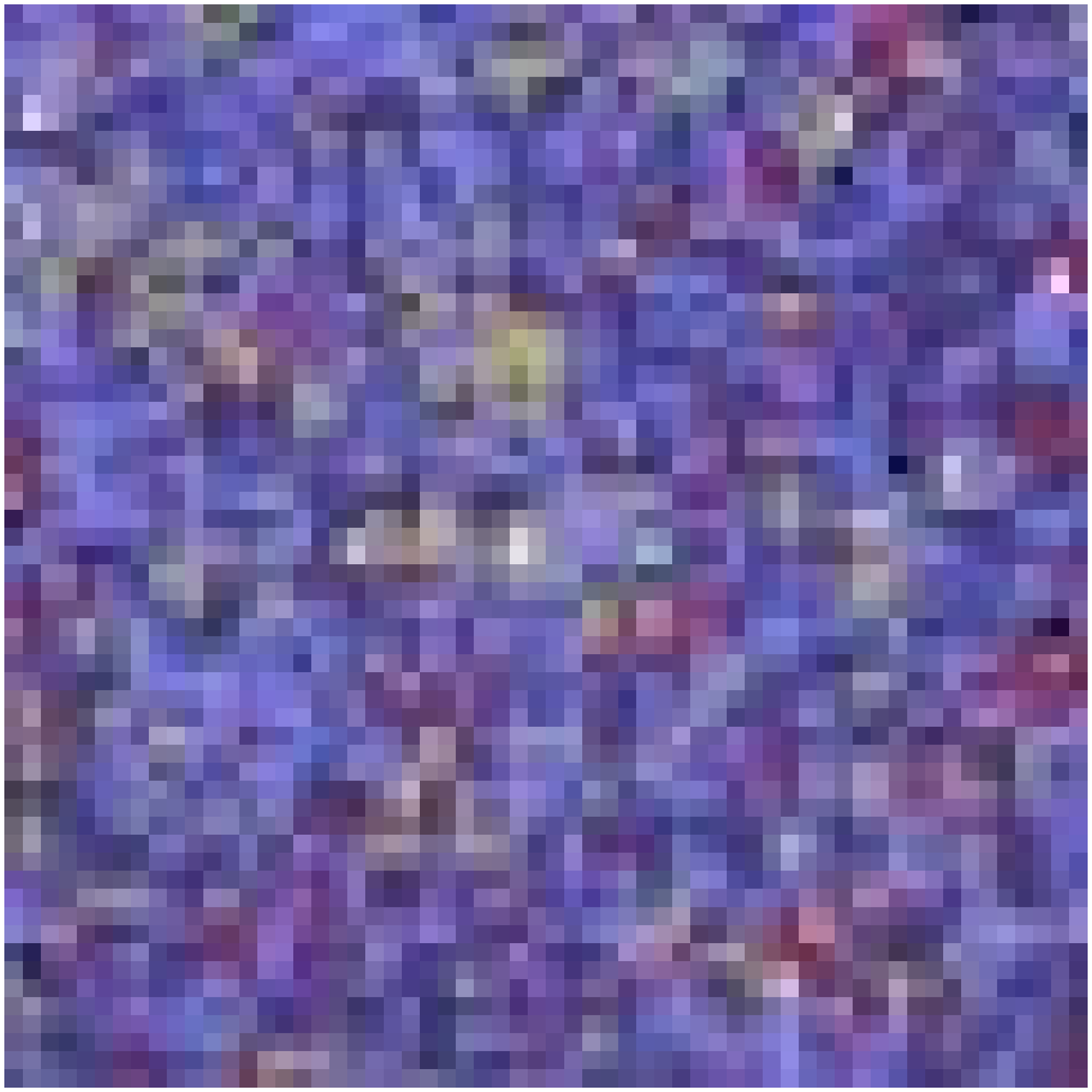}  \FGb{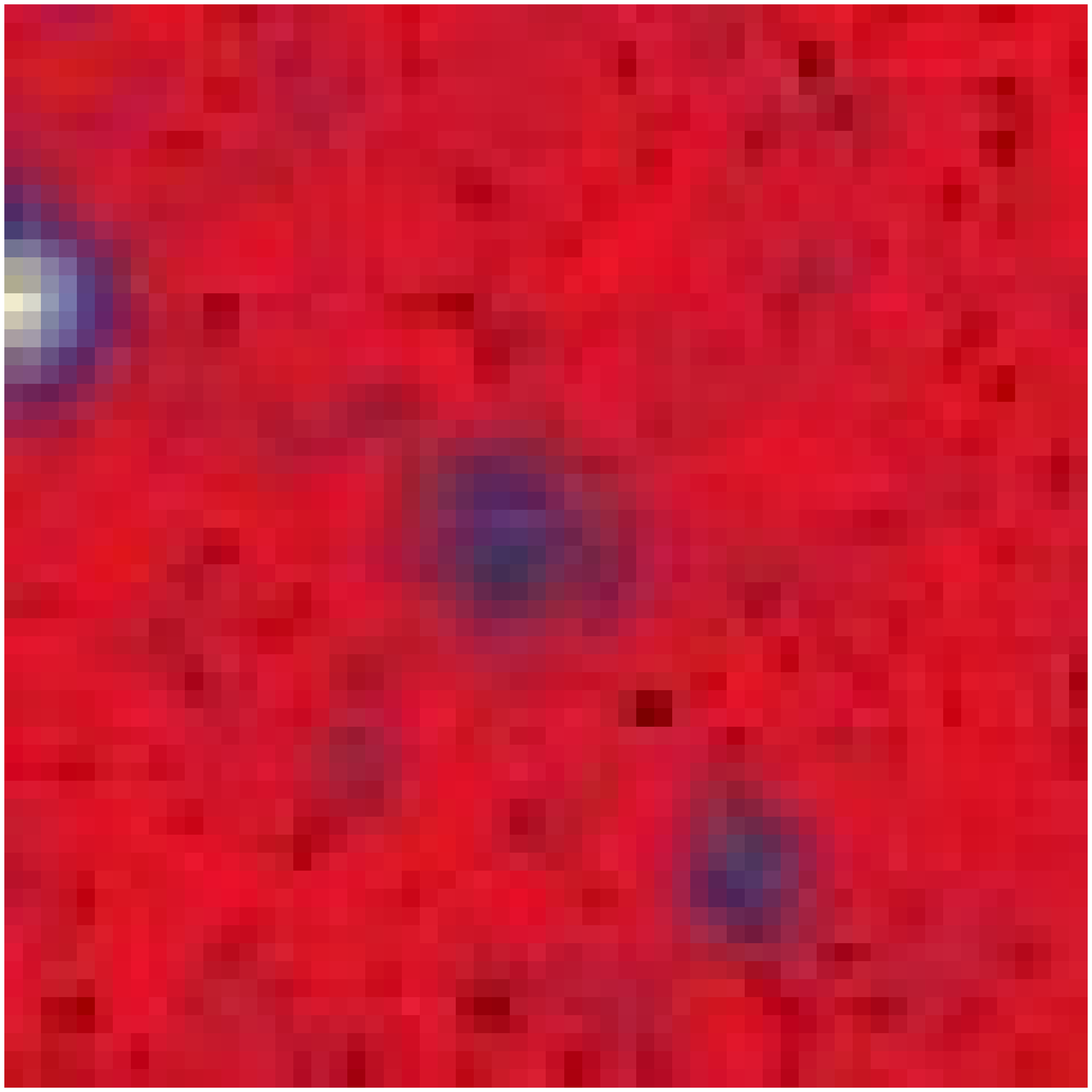}  \FGb{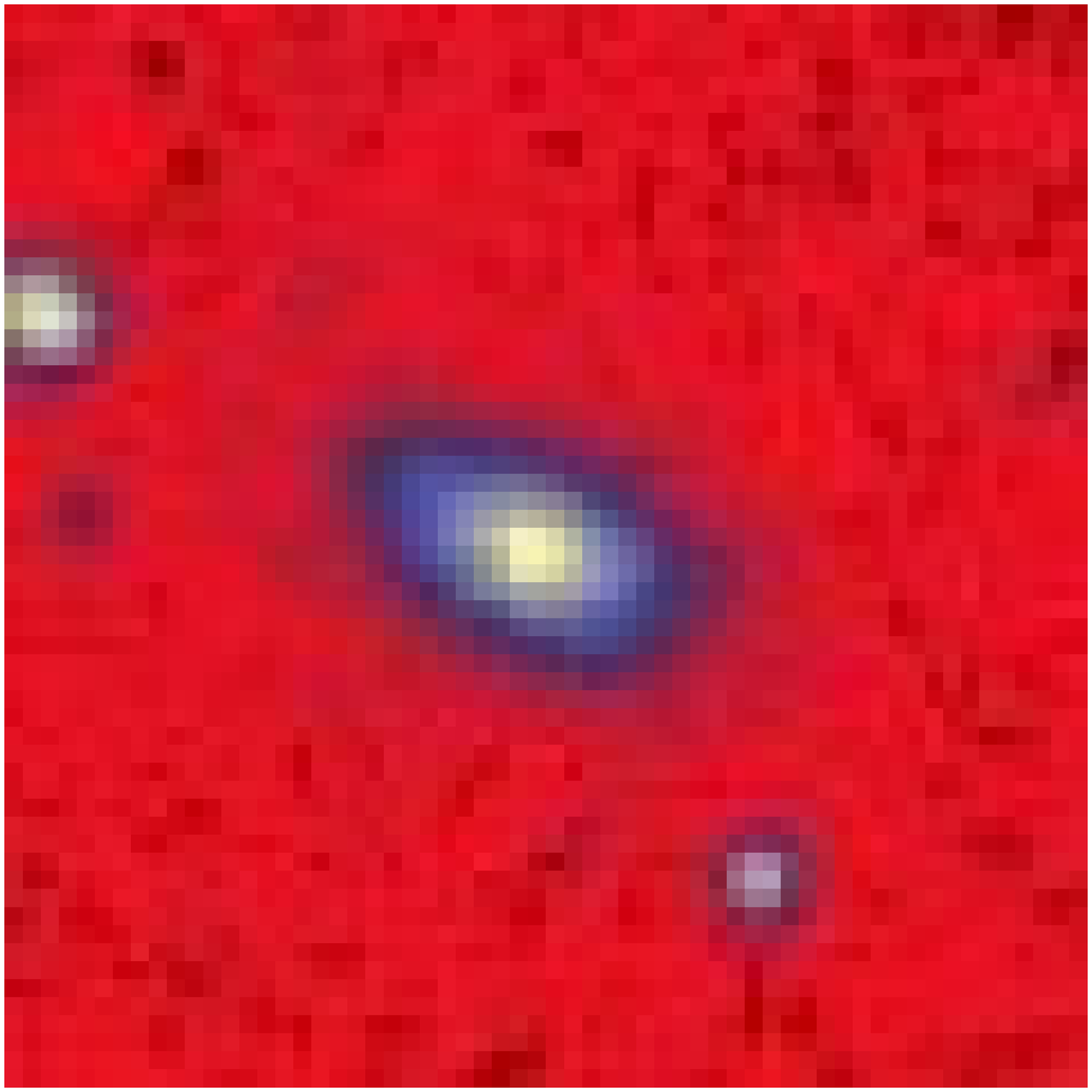}  \FGb{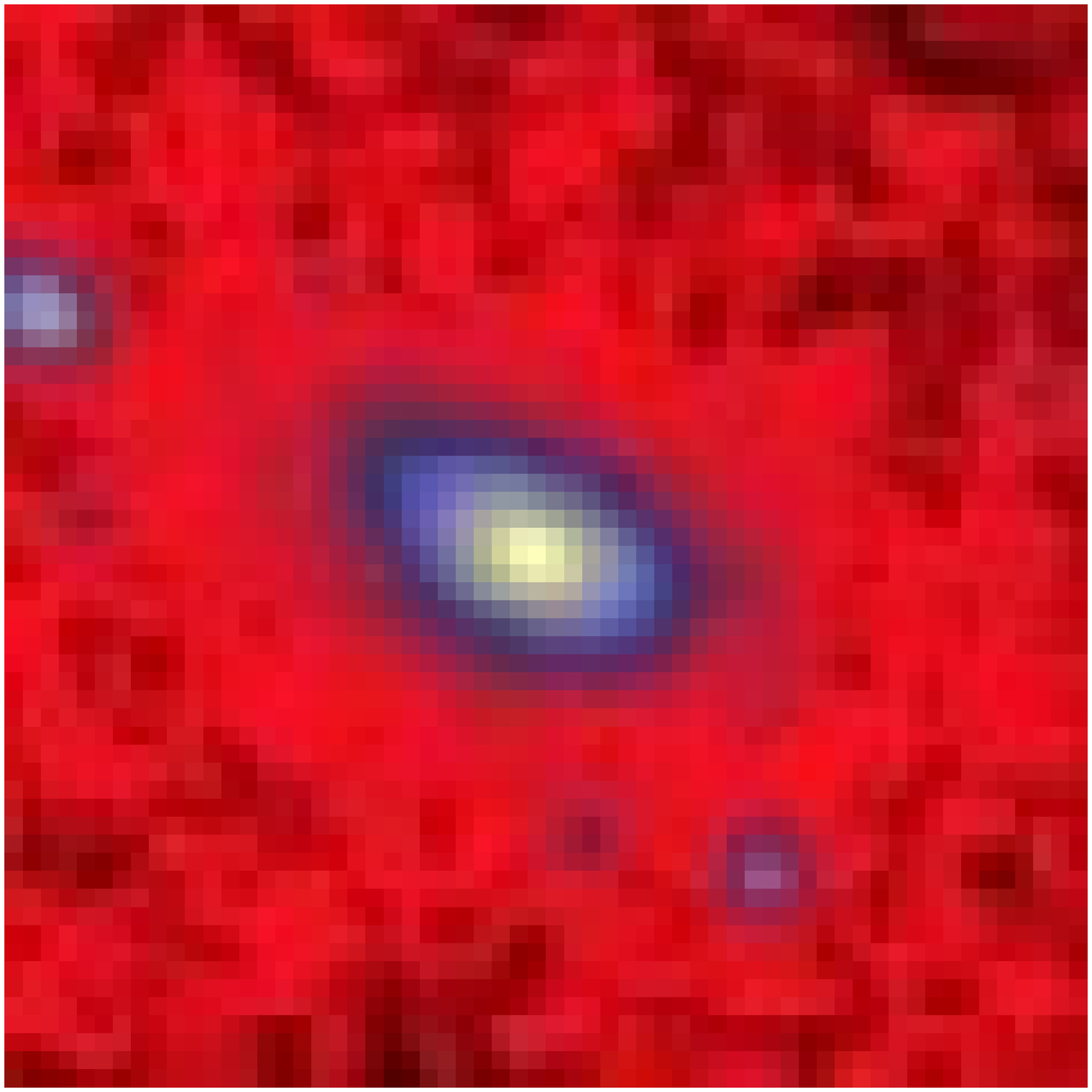}  \FGb{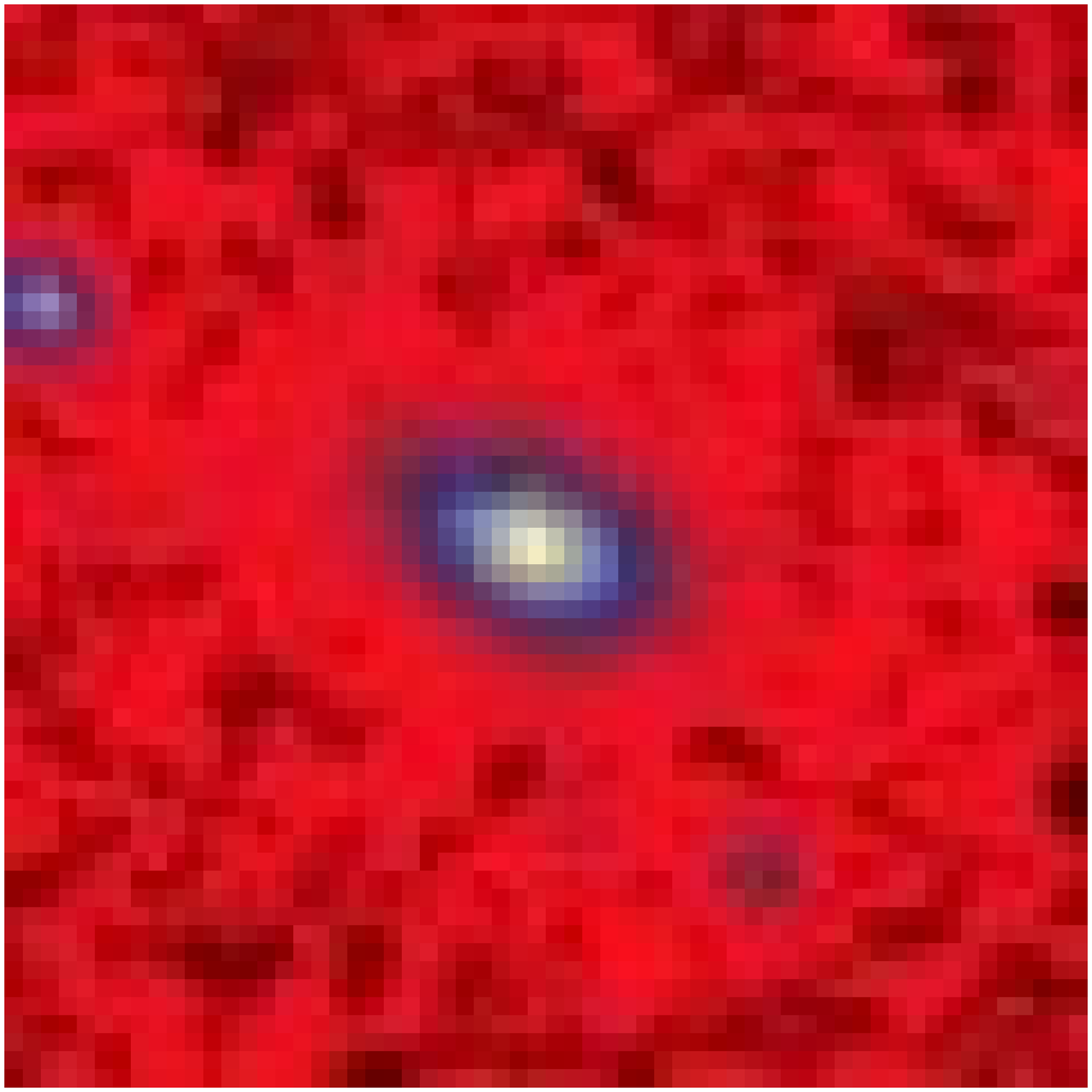}  \FGb{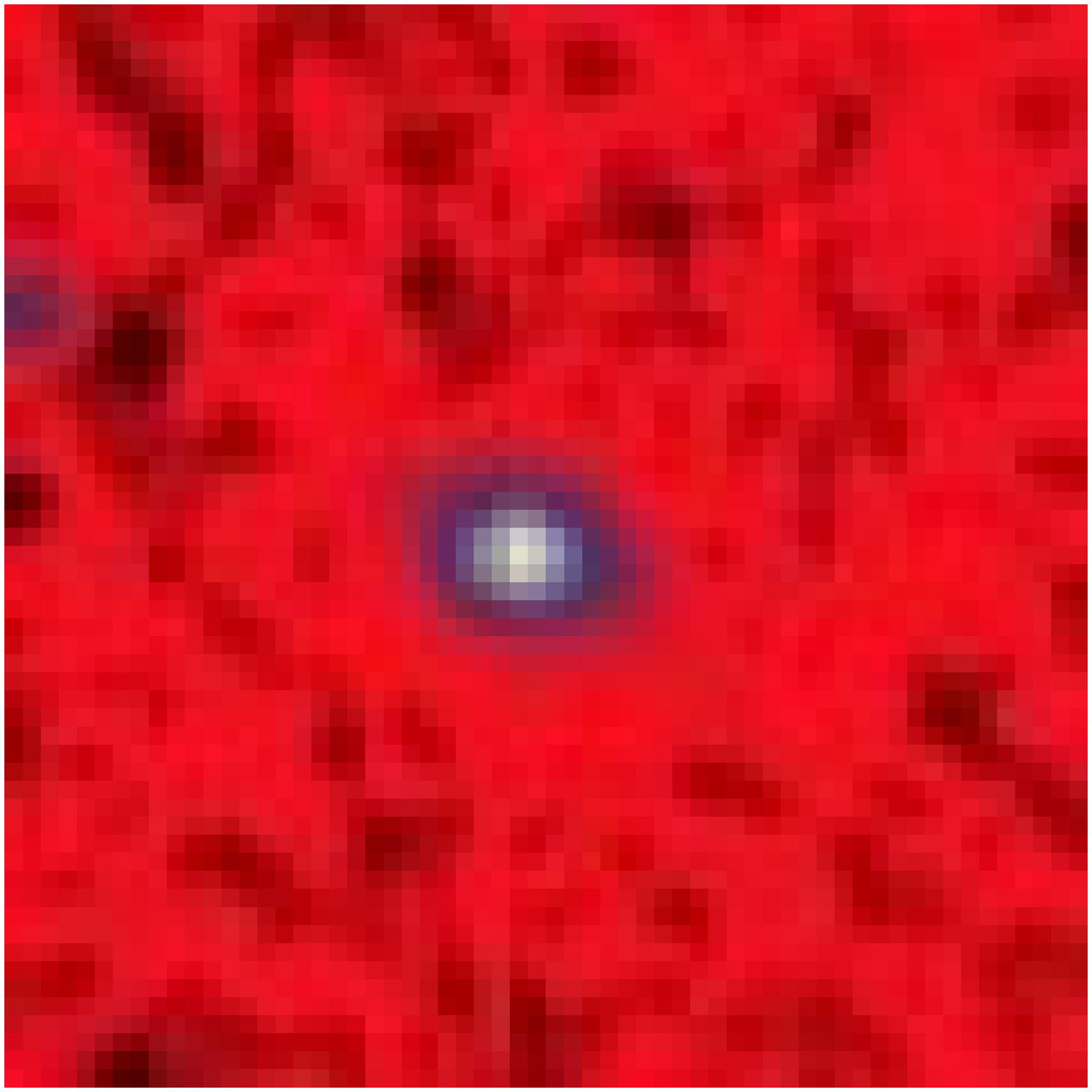}  \FGb{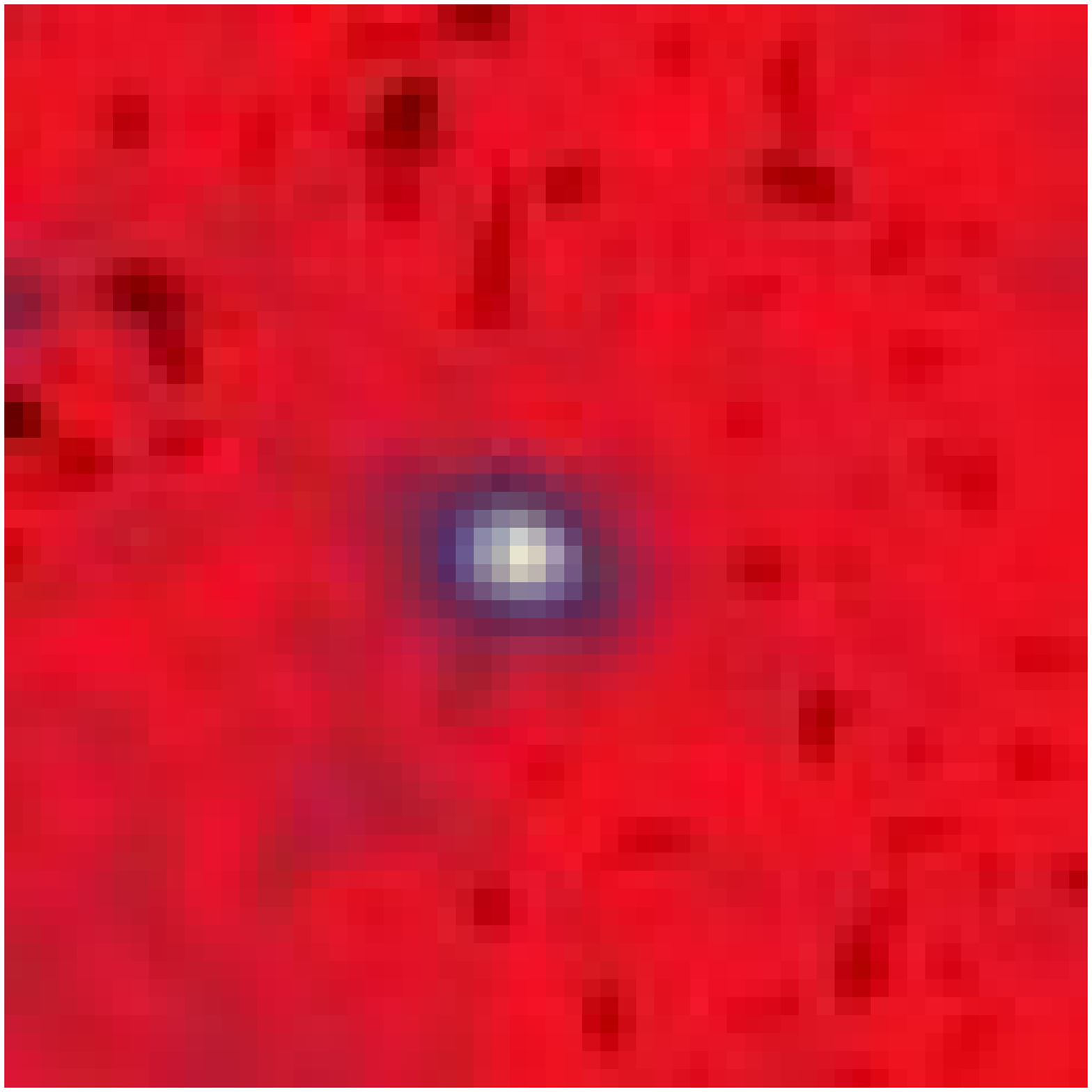}  \FGb{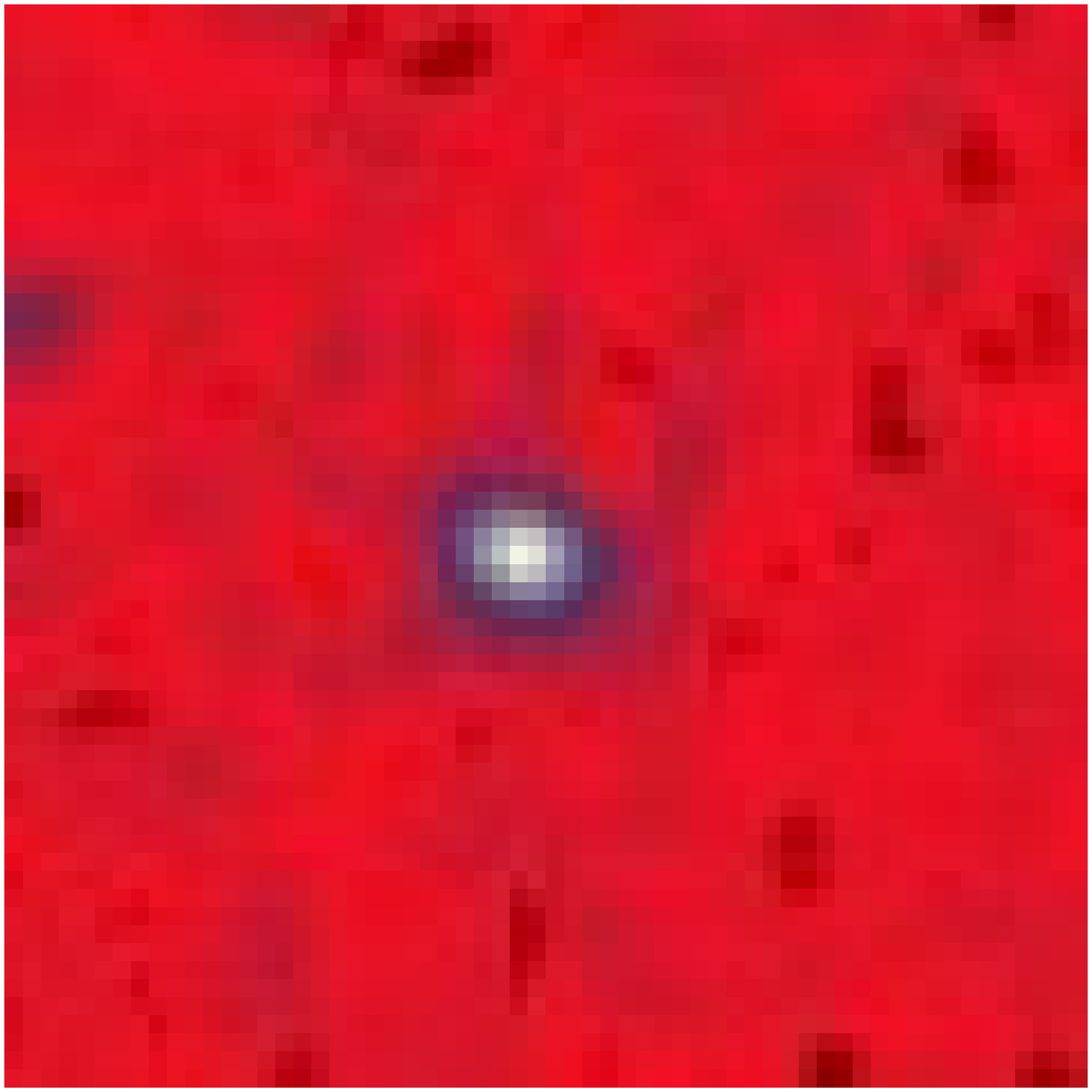}  \FGb{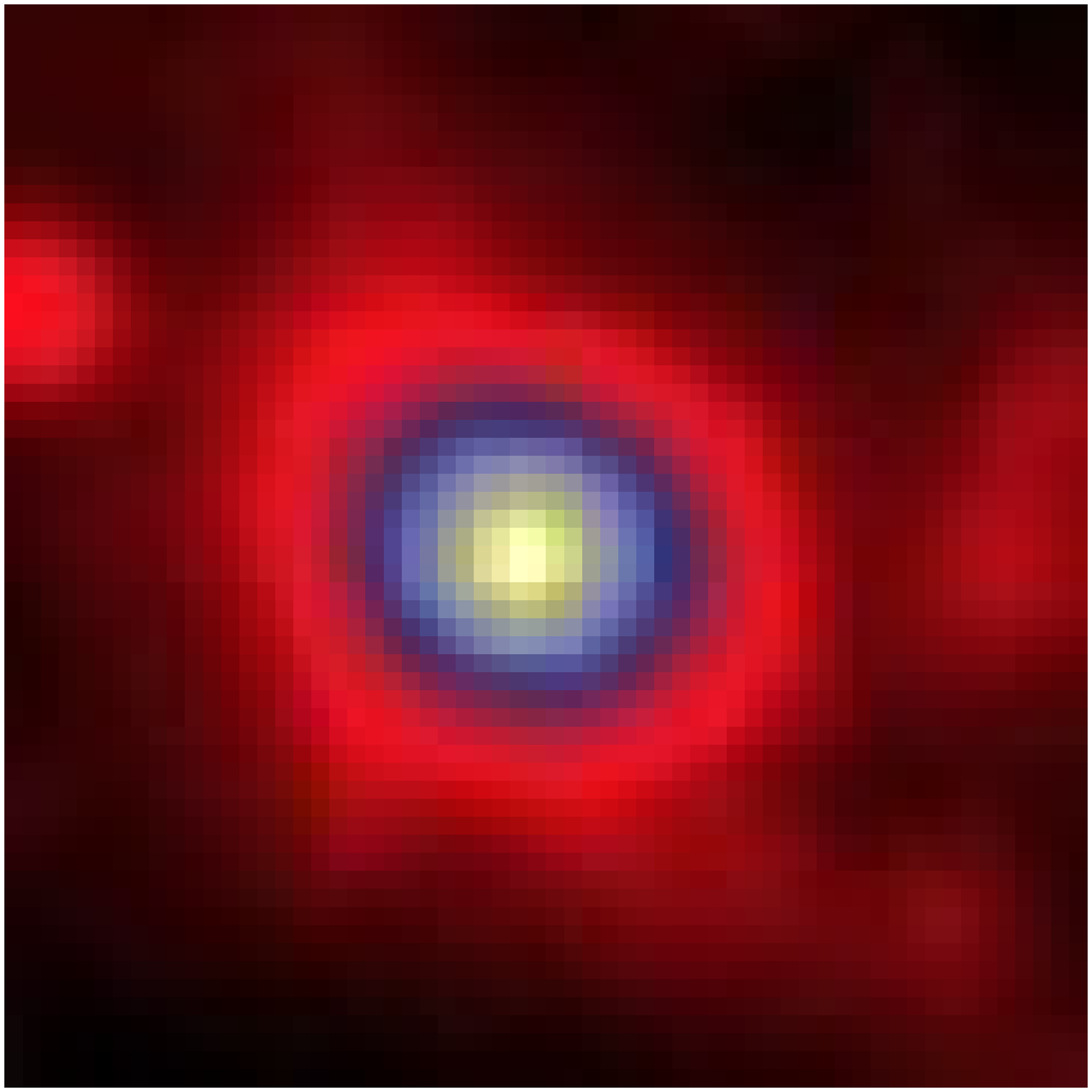}  \FGb{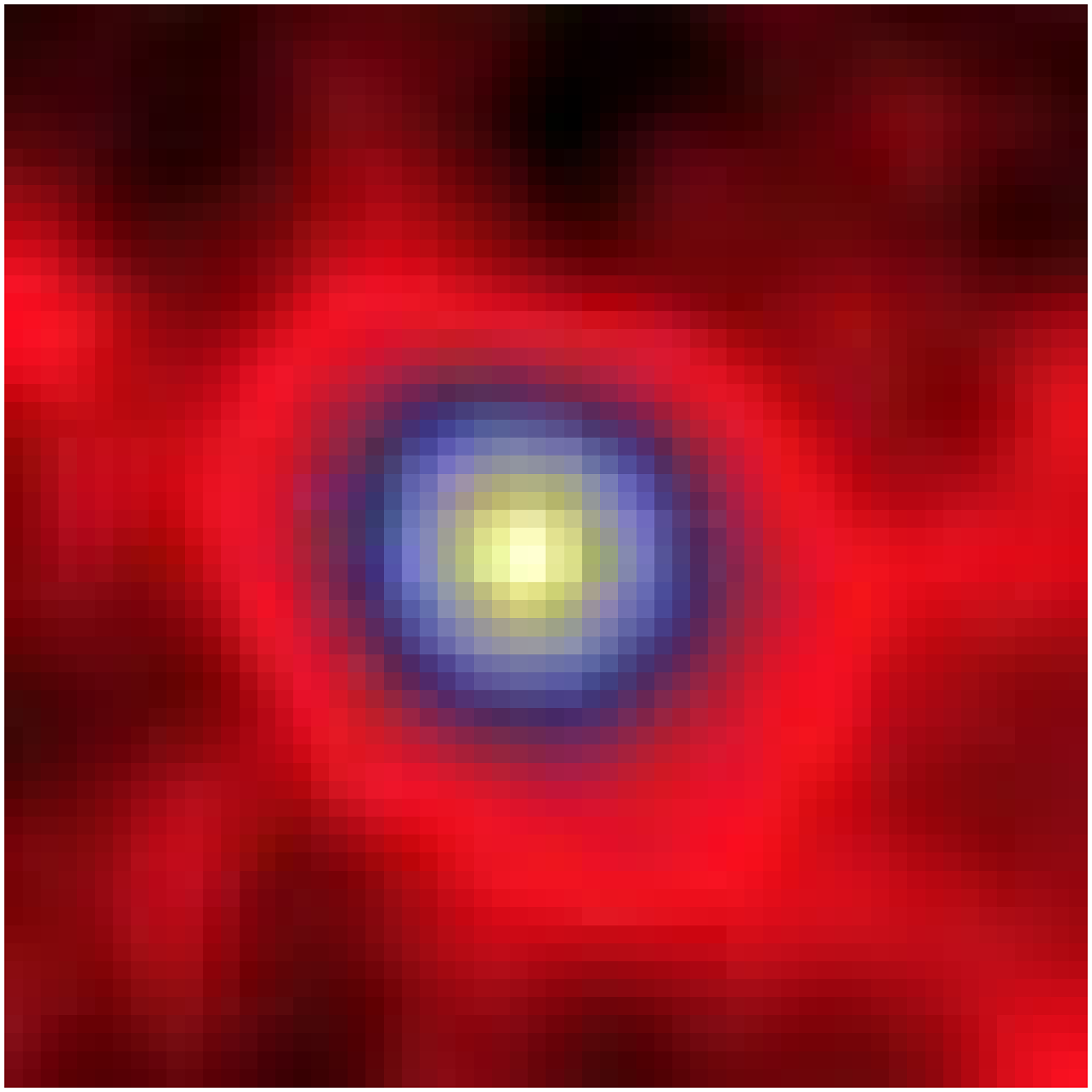}  \FGb{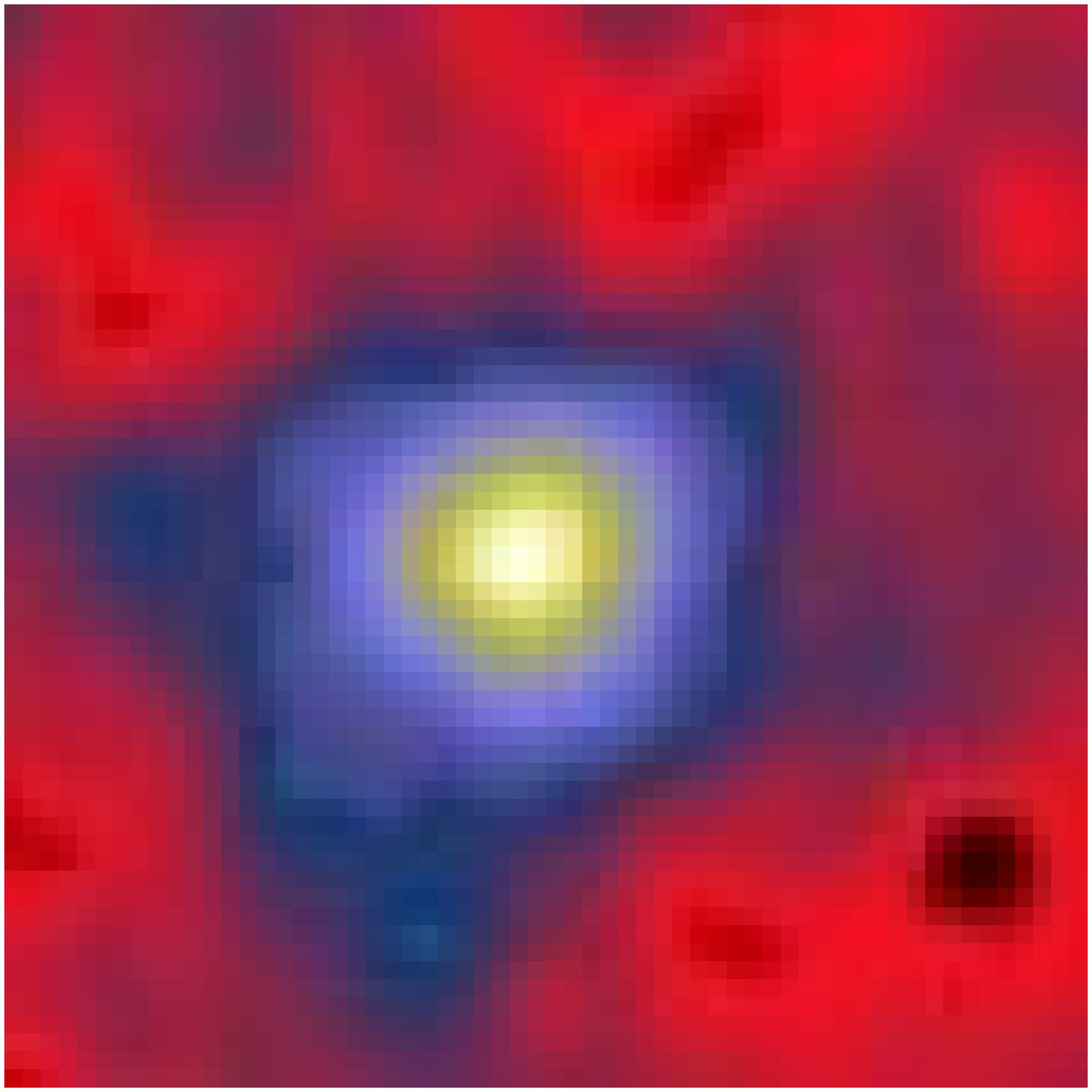}  \FGb{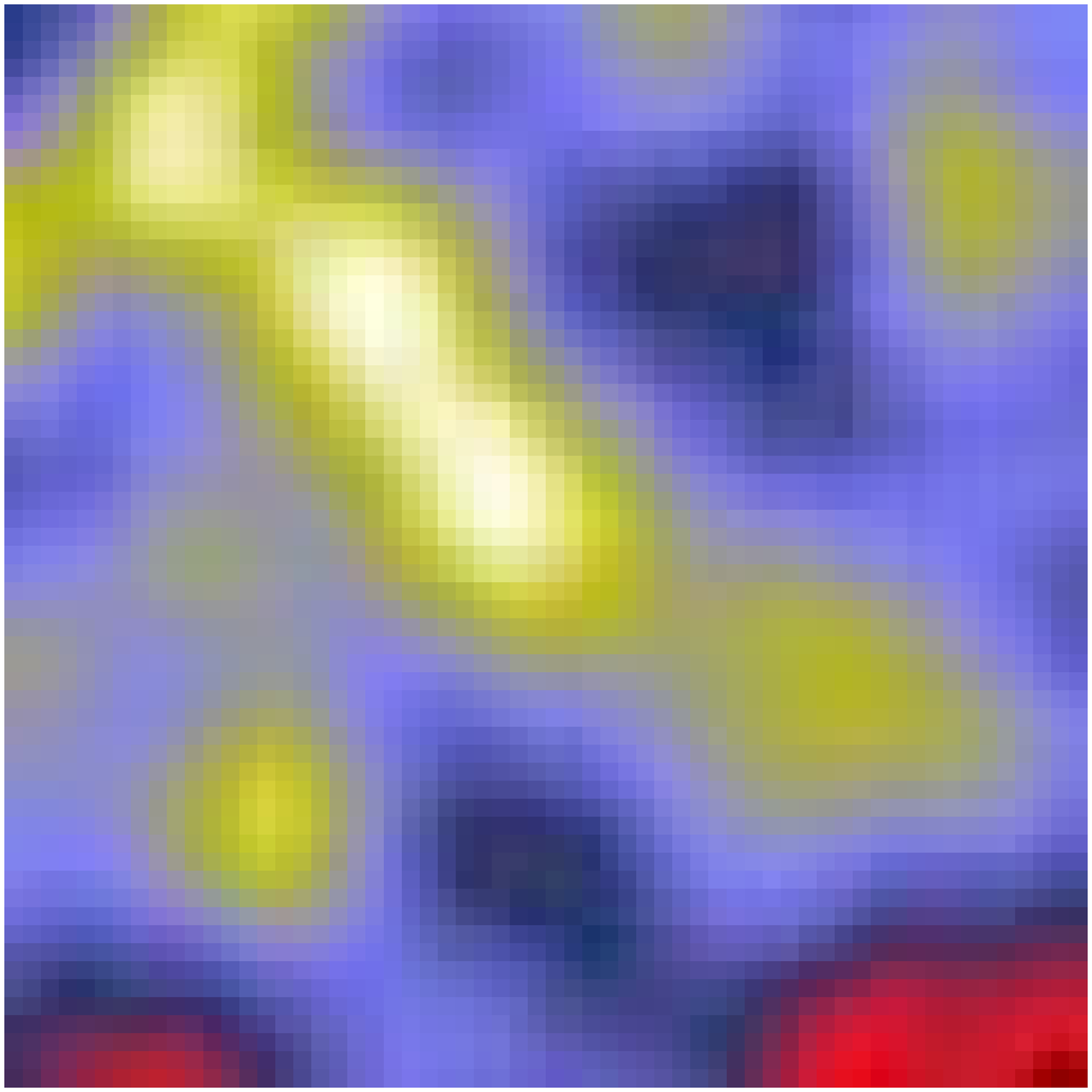} \\  {\#39   NCIA031605}  \\
{ \\ \textbf{Figure \ref{fig:allgal}.(cont)} ~~Visible Galaxies within 40\min\ of NGC~891 } \\
\end{figure}

\clearpage
\begin{figure}[ht]
  \HDb{rgb}  \HDb{FUV}  \HDb{NUV}   \HDb{B}     \HDb{R}    \HDb{IR}   \HDb{J}    \HDb{H}     \HDb{K}     \HDb{W1}    \HDb{W2}    \HDb{W3}    \HDb{W4}  \\
  \FGb{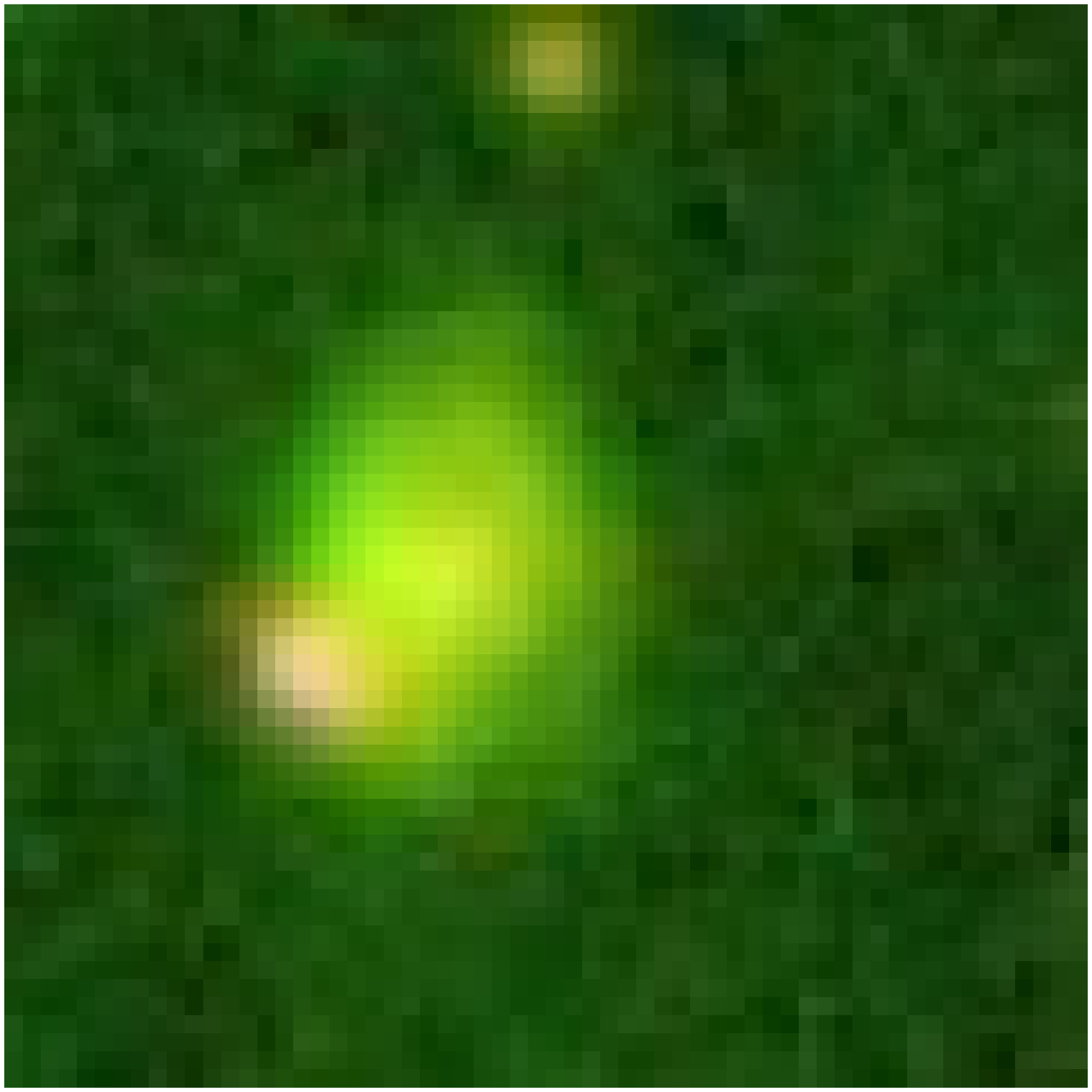}  \FGb{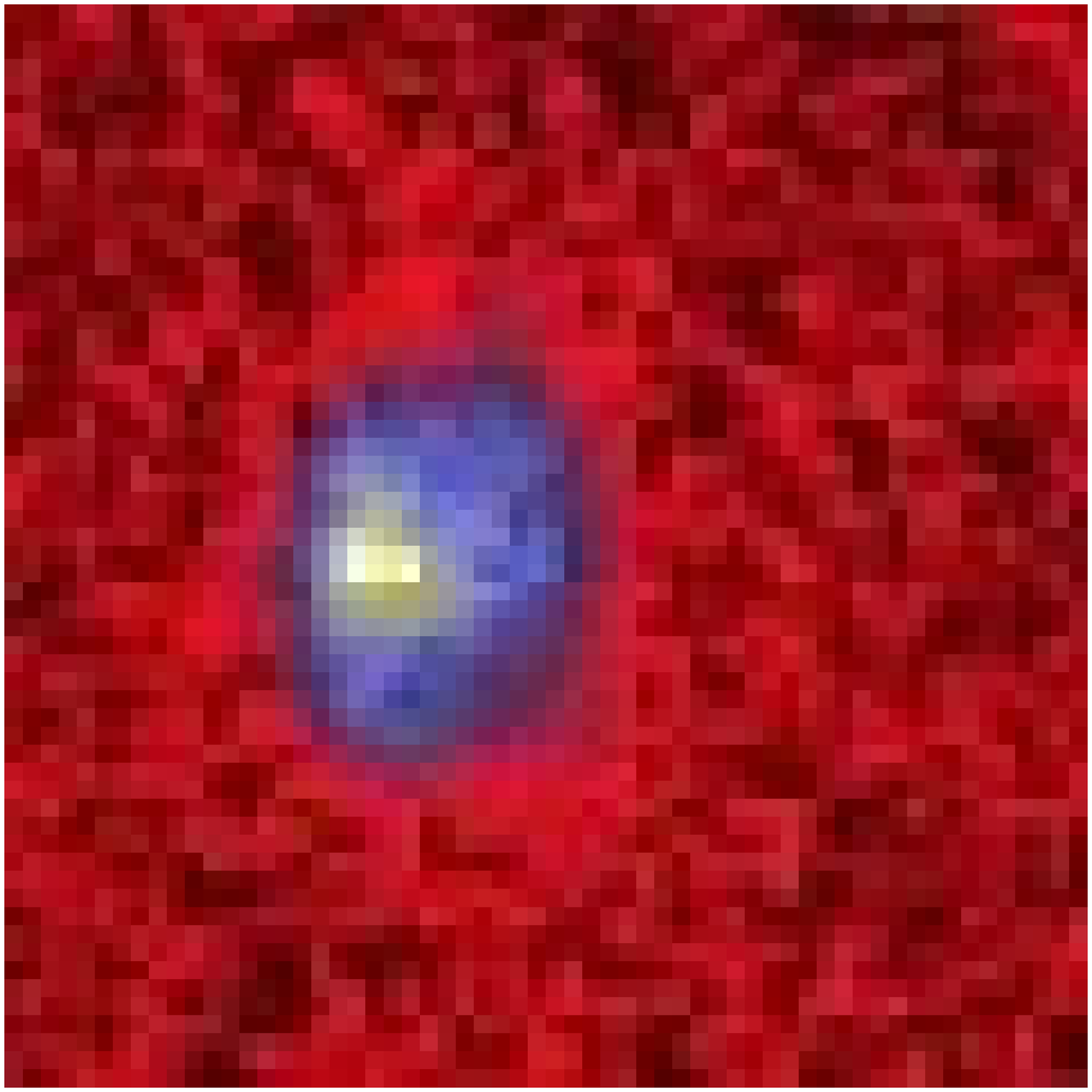}  \FGb{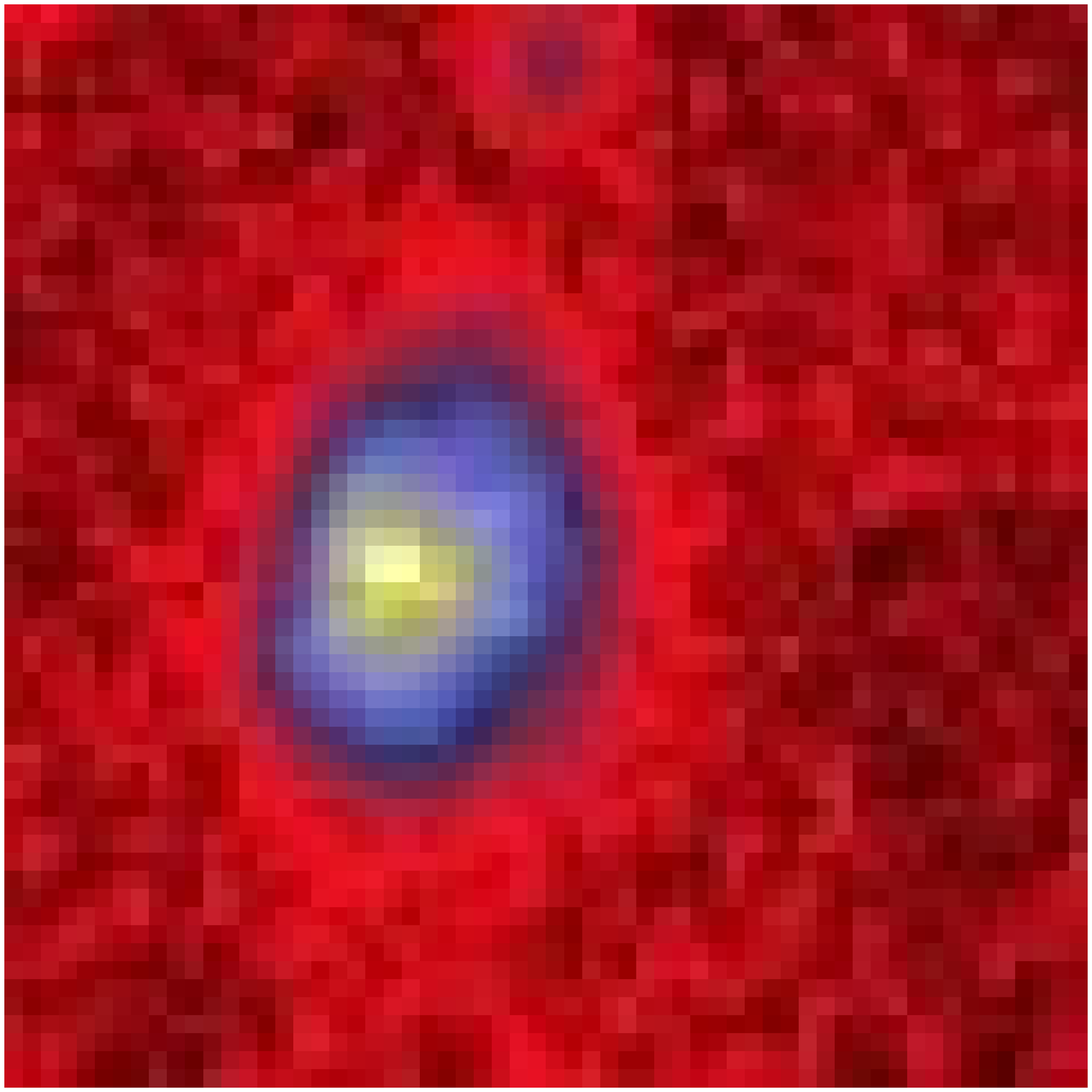}  \FGb{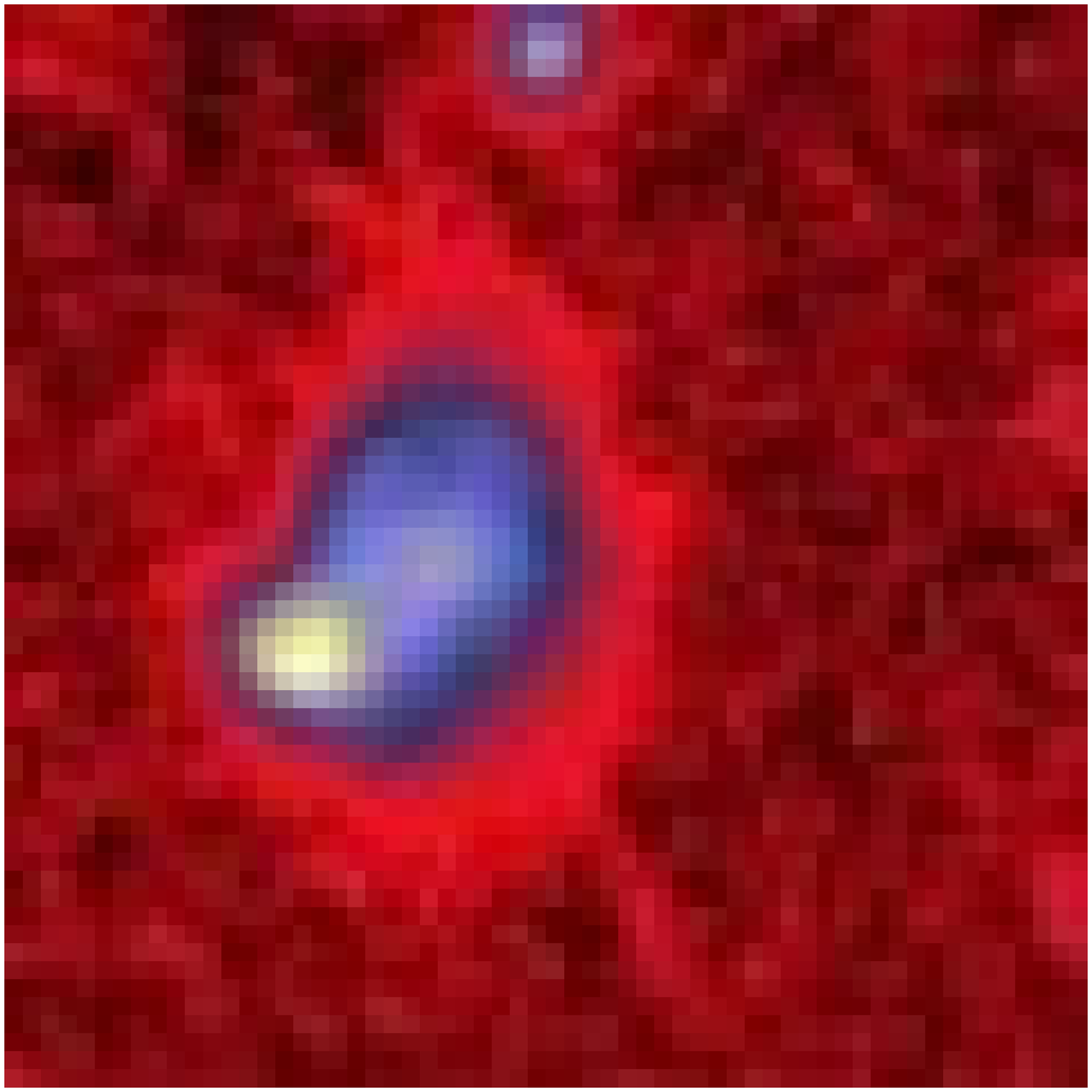}  \FGb{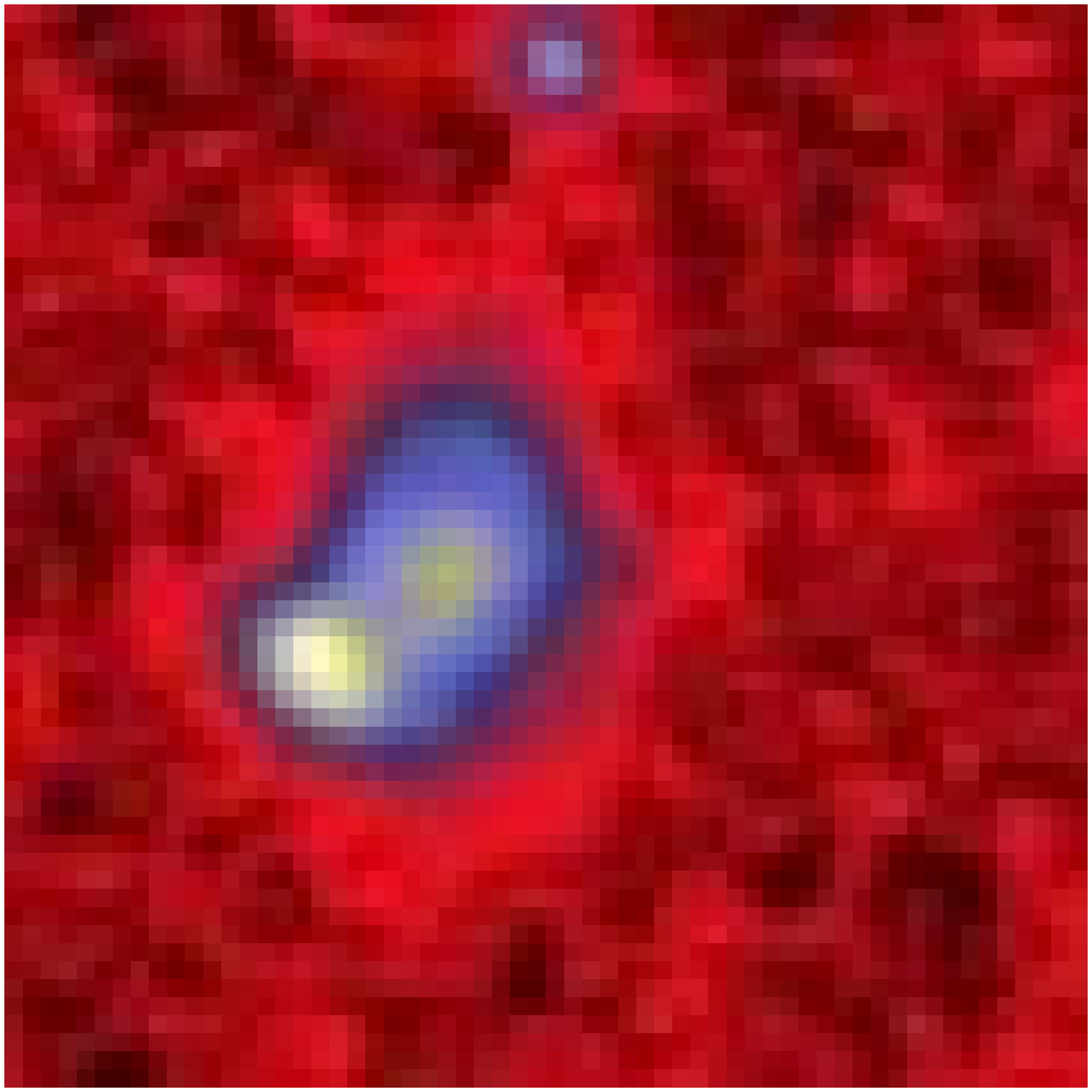}  \FGb{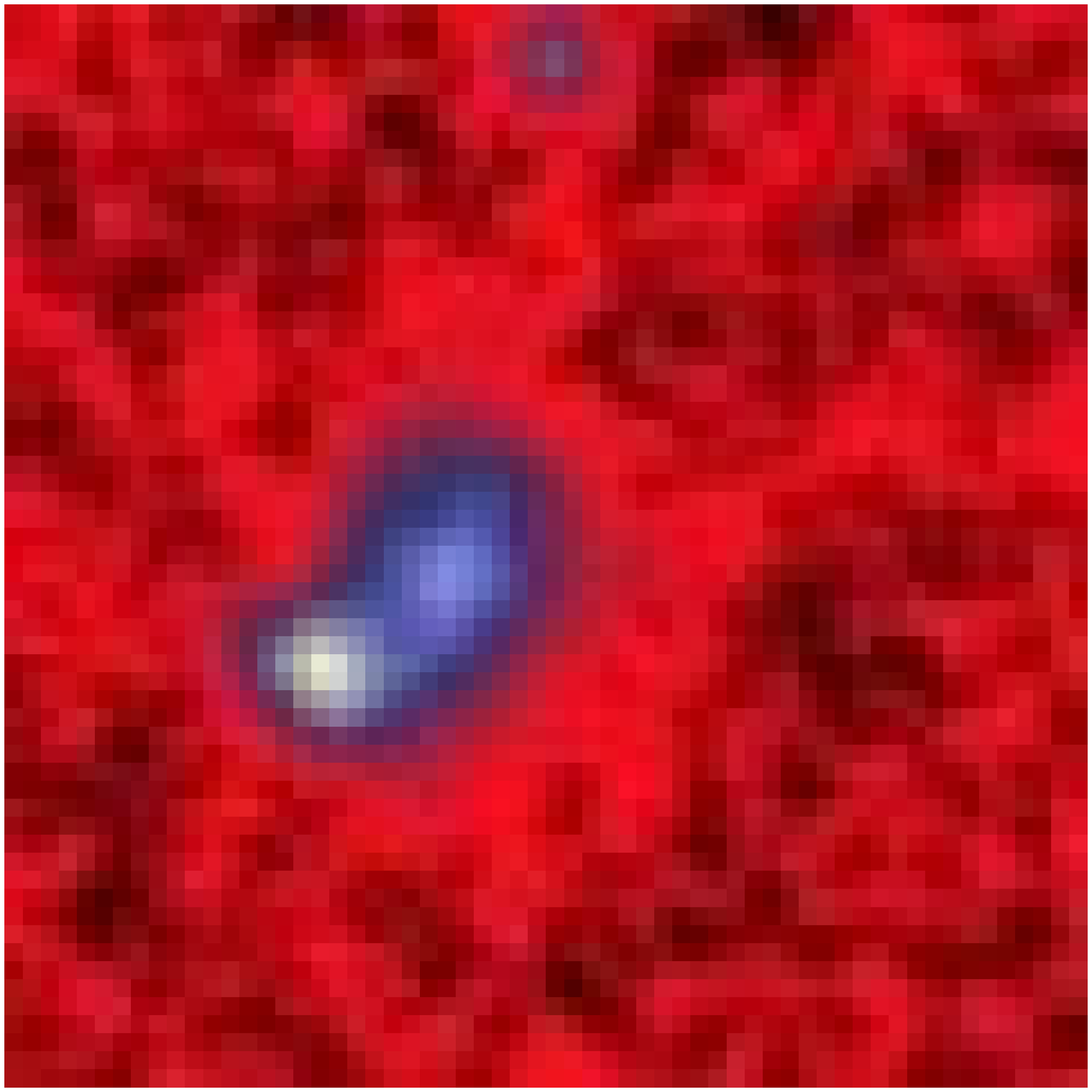}  \FGb{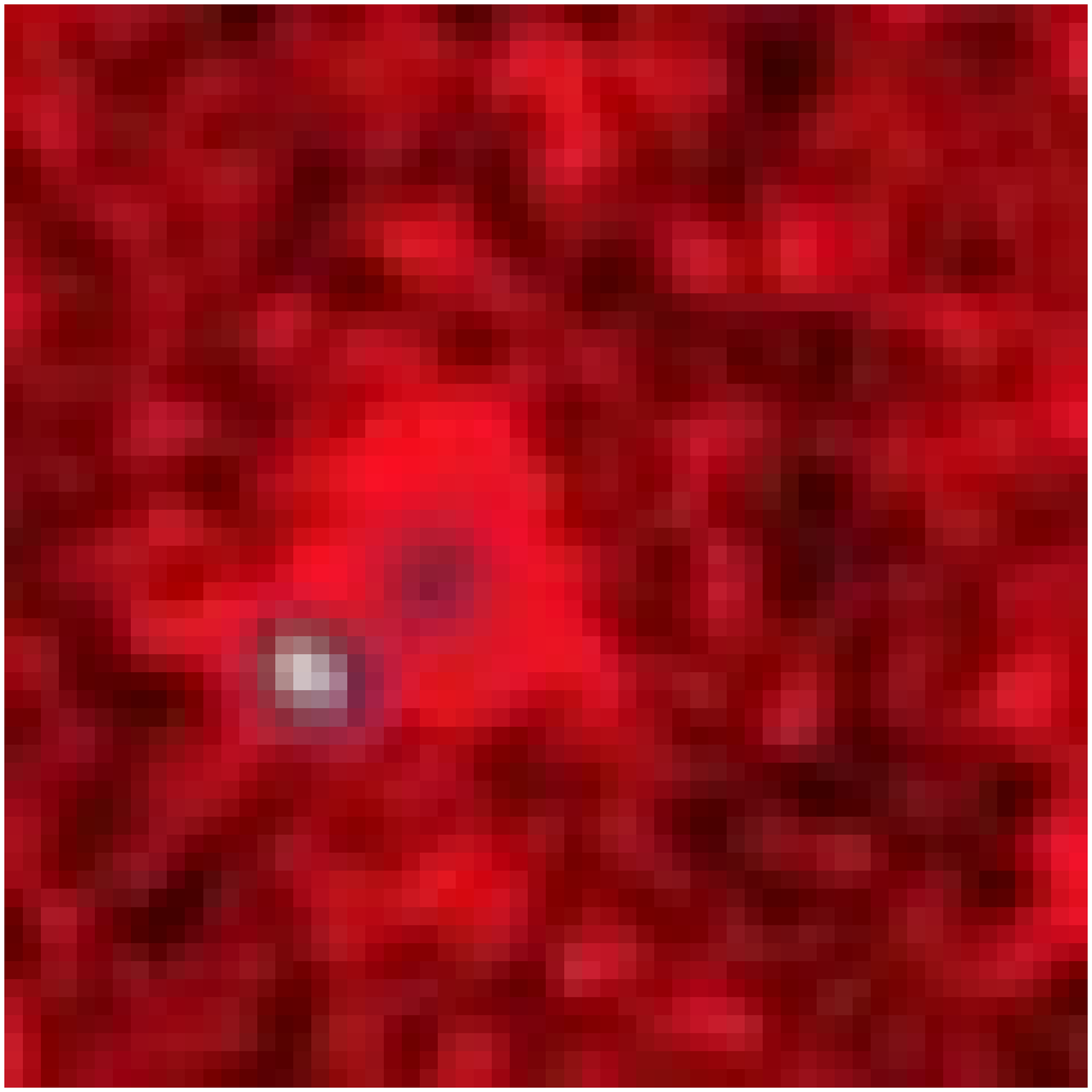}  \FGb{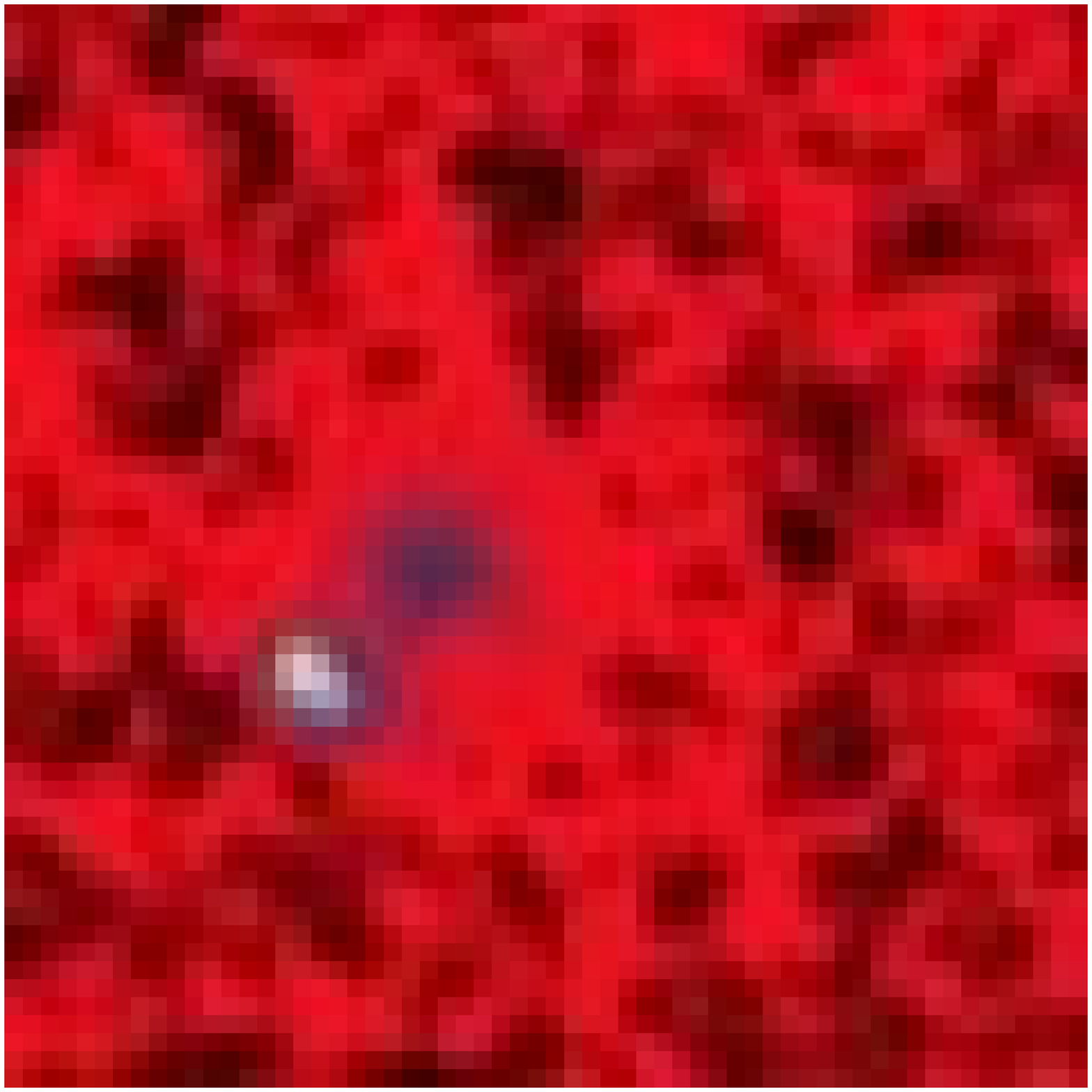}  \FGb{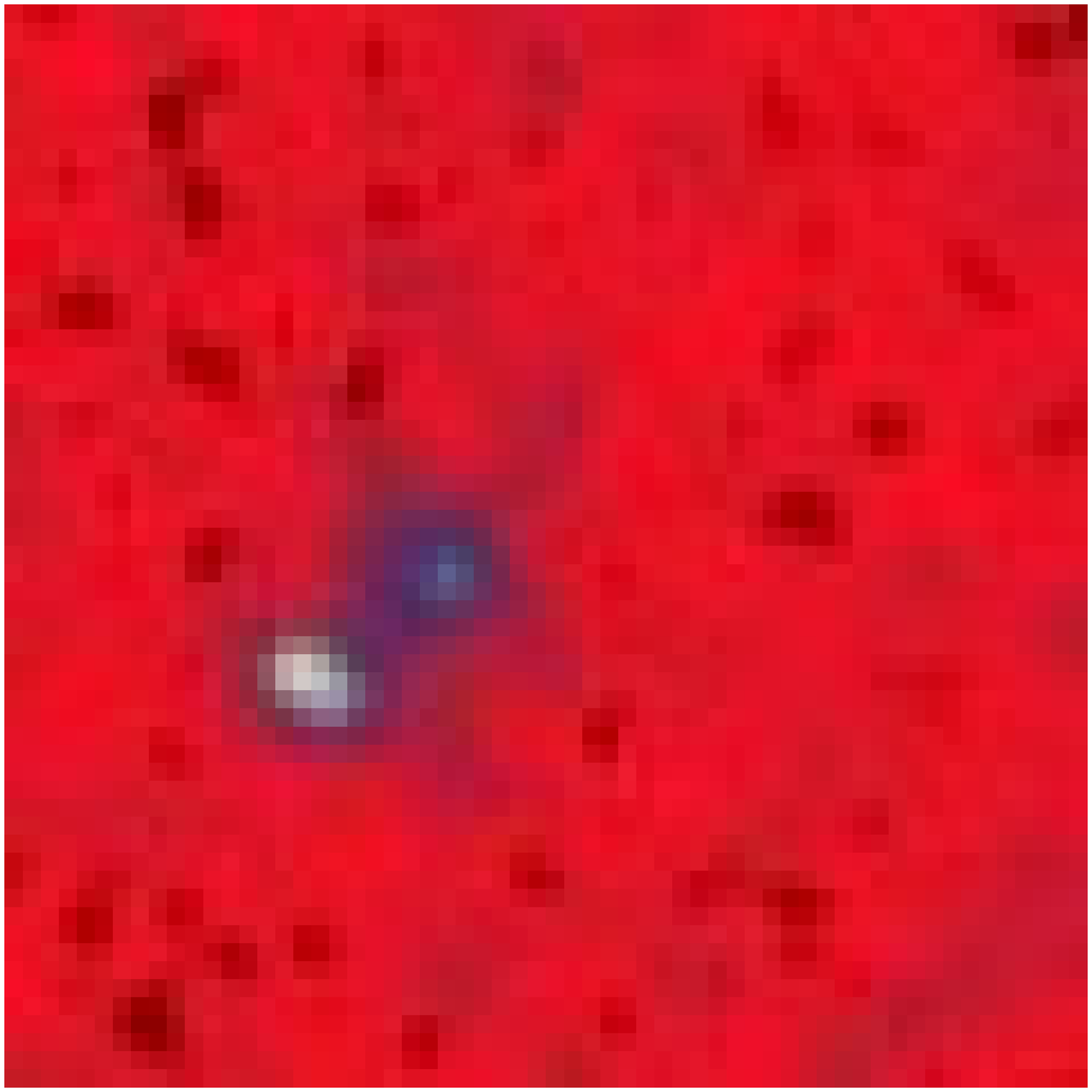}  \FGb{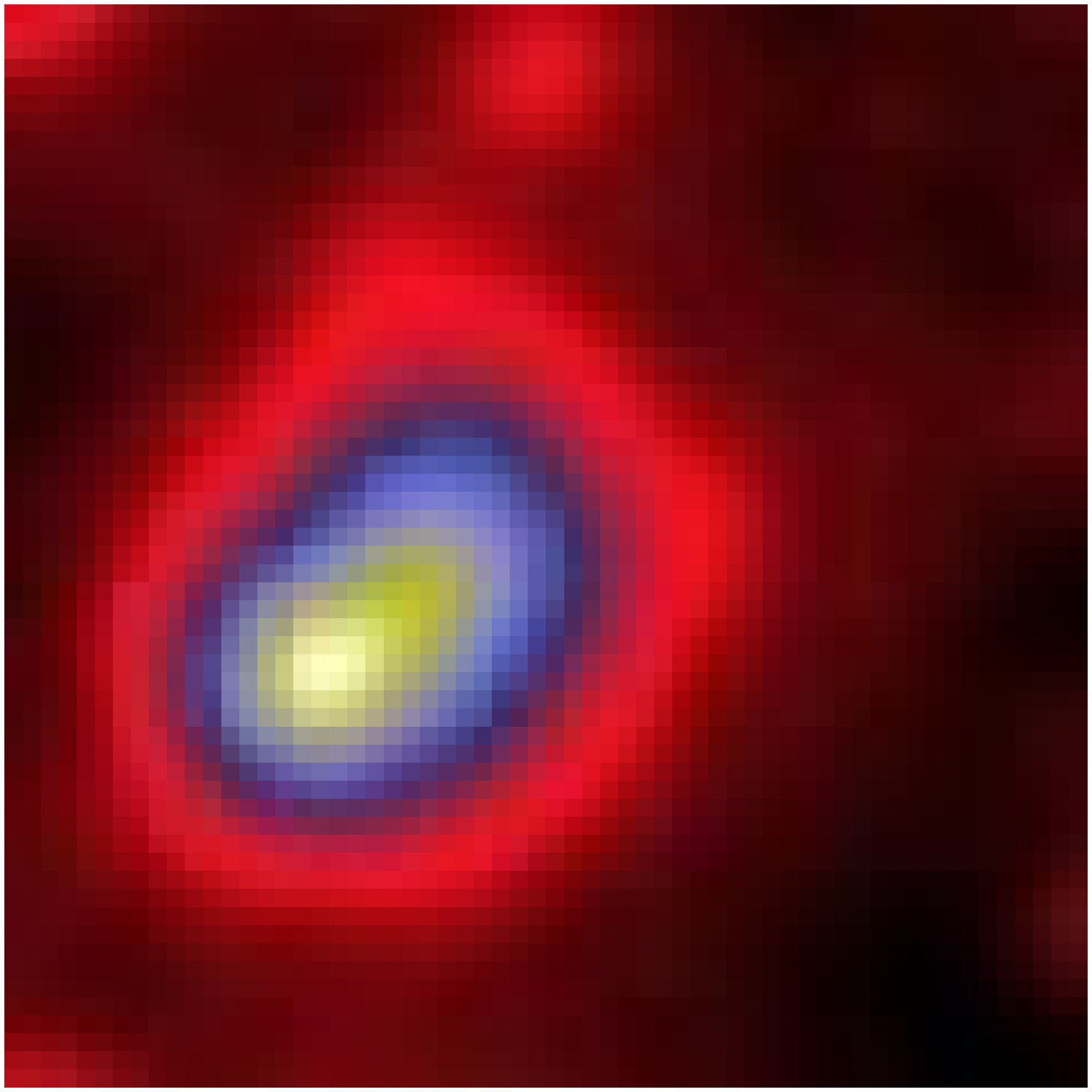}  \FGb{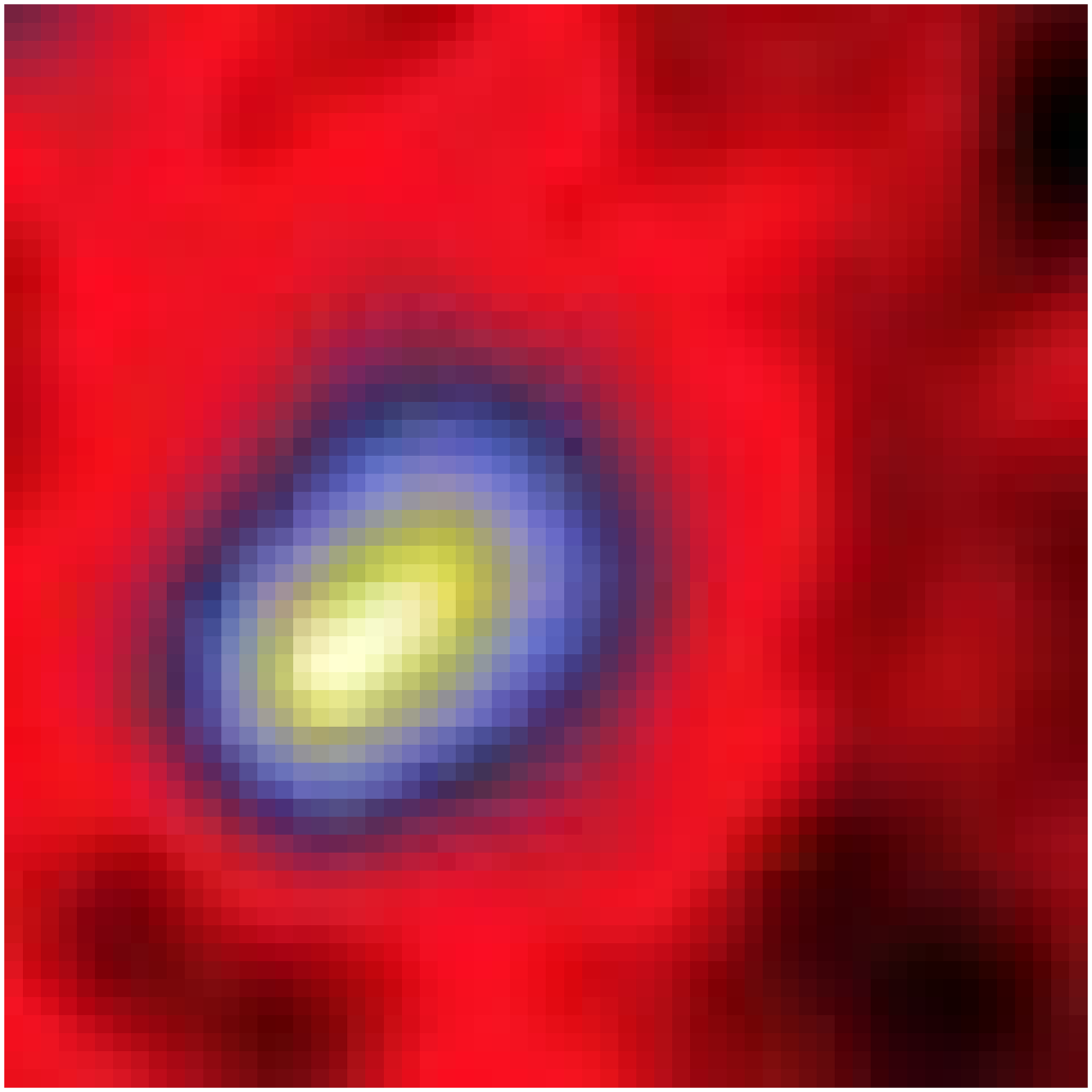}  \FGb{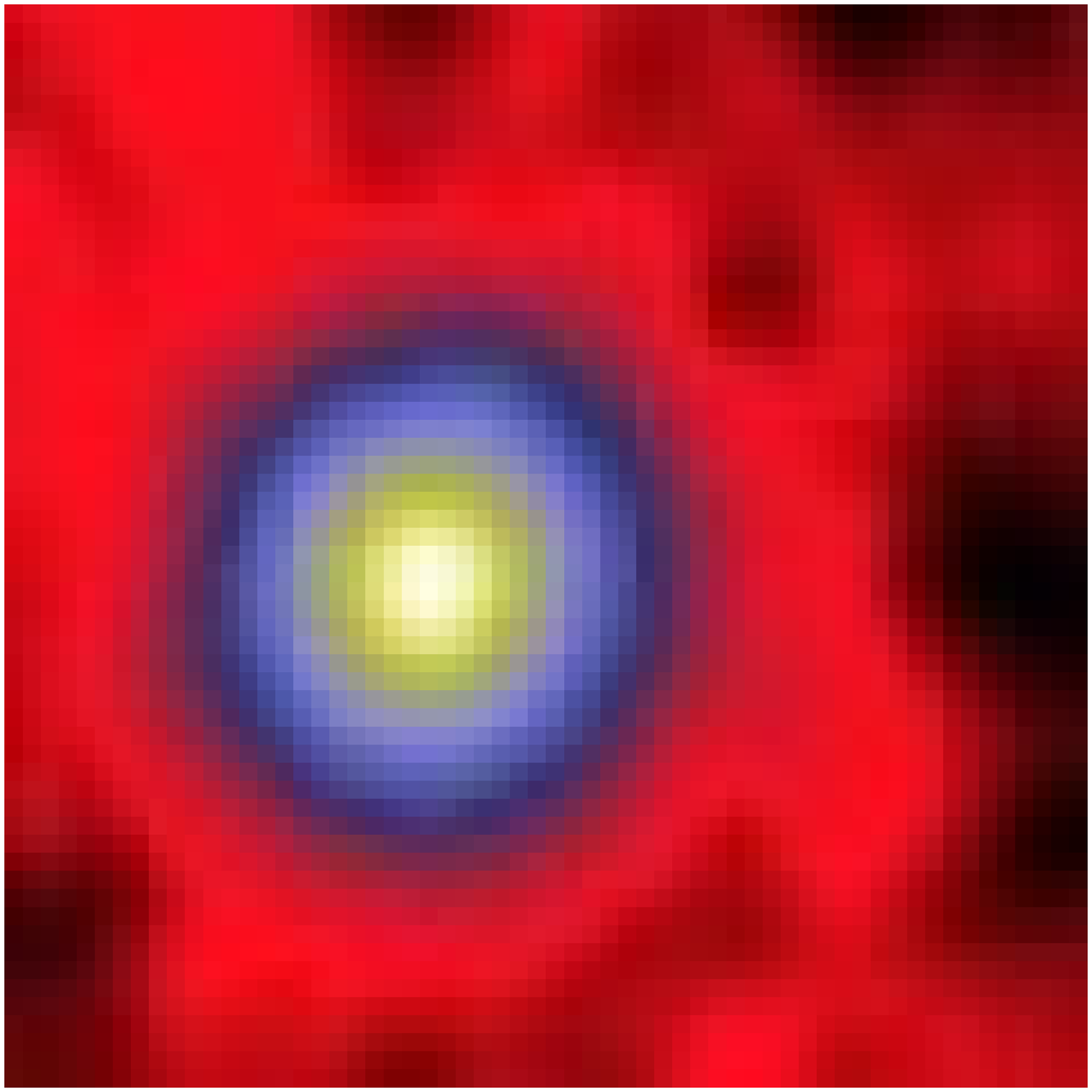}  \FGb{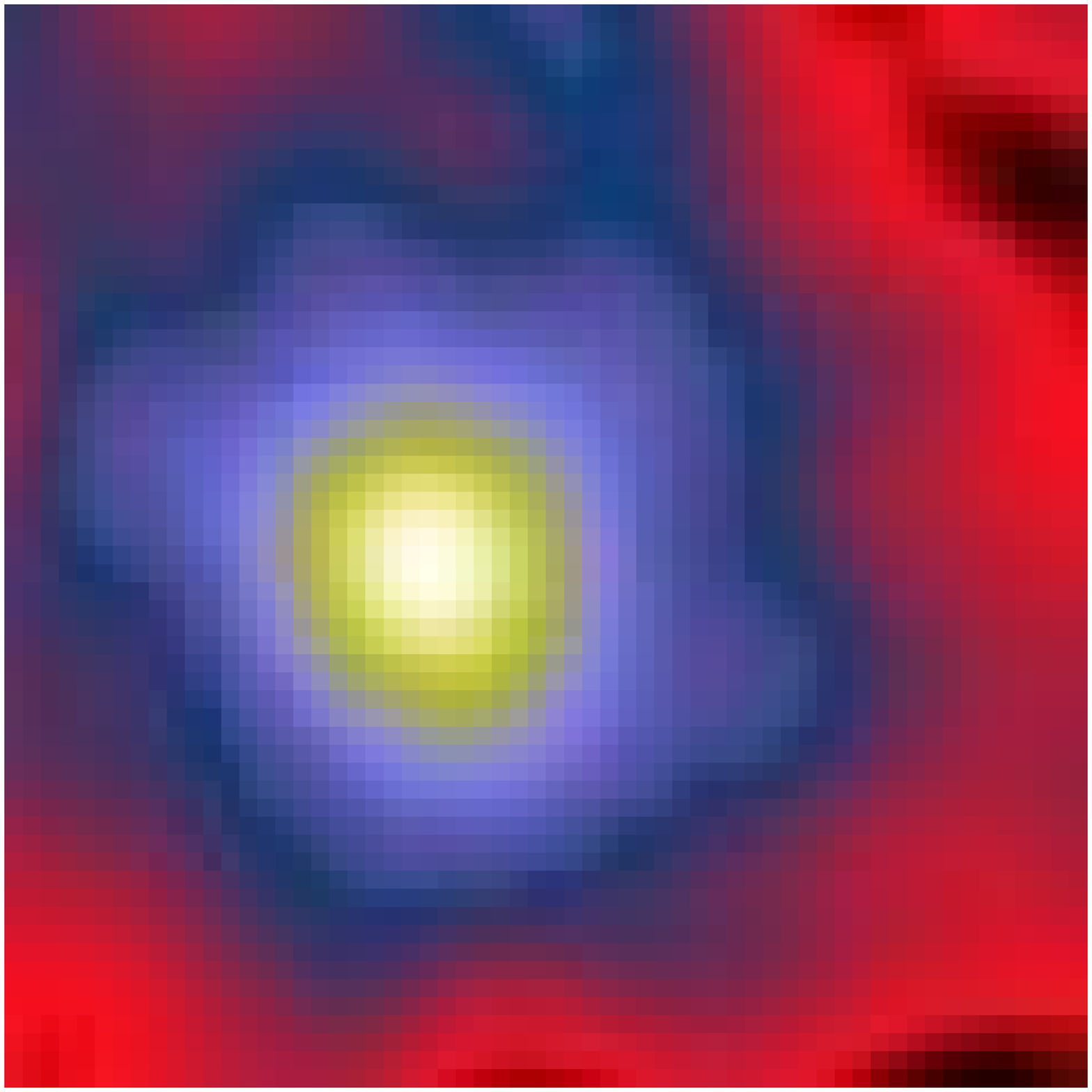} \\  {\#40   NBZ5010876}  \\
  \FGb{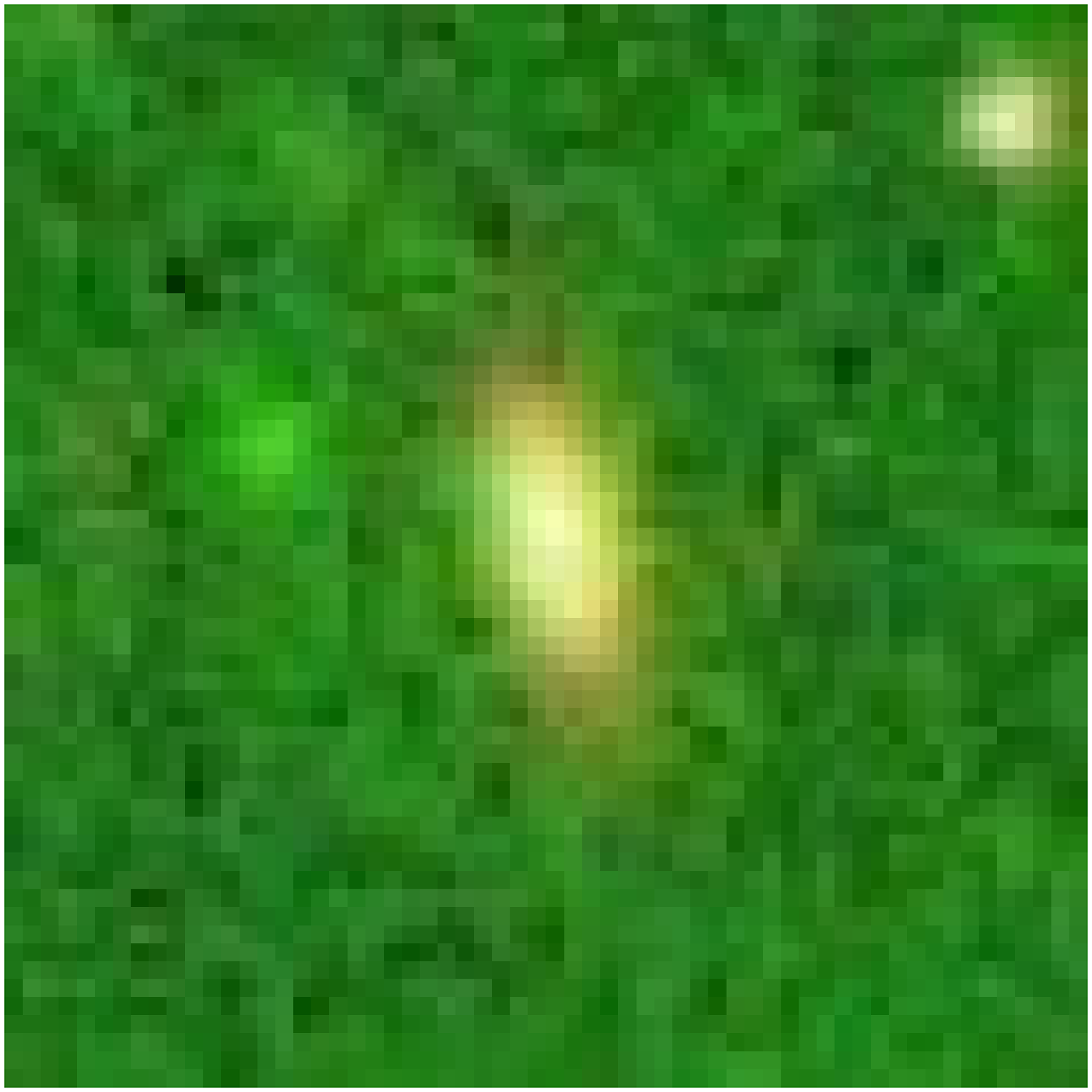}  \FGb{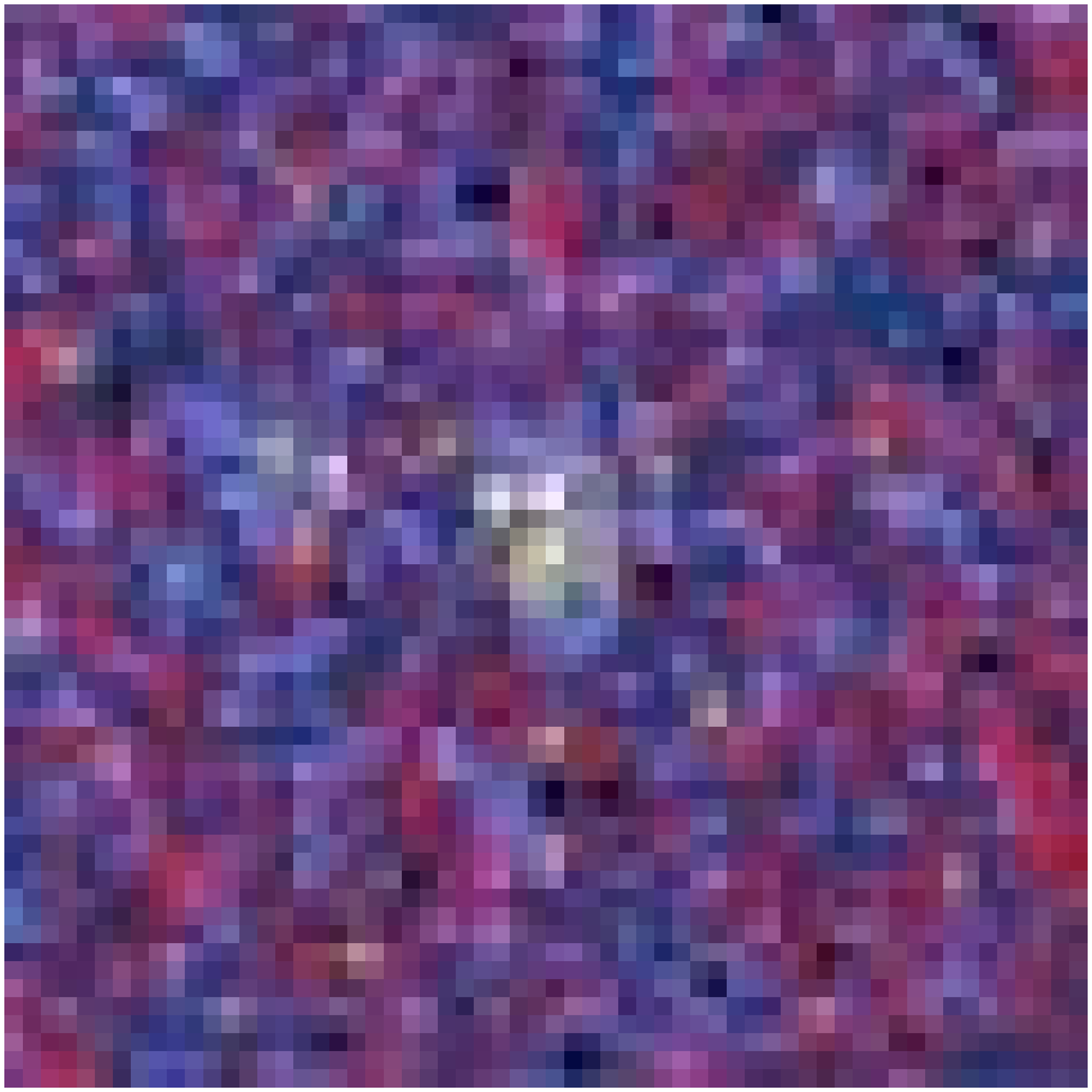}  \FGb{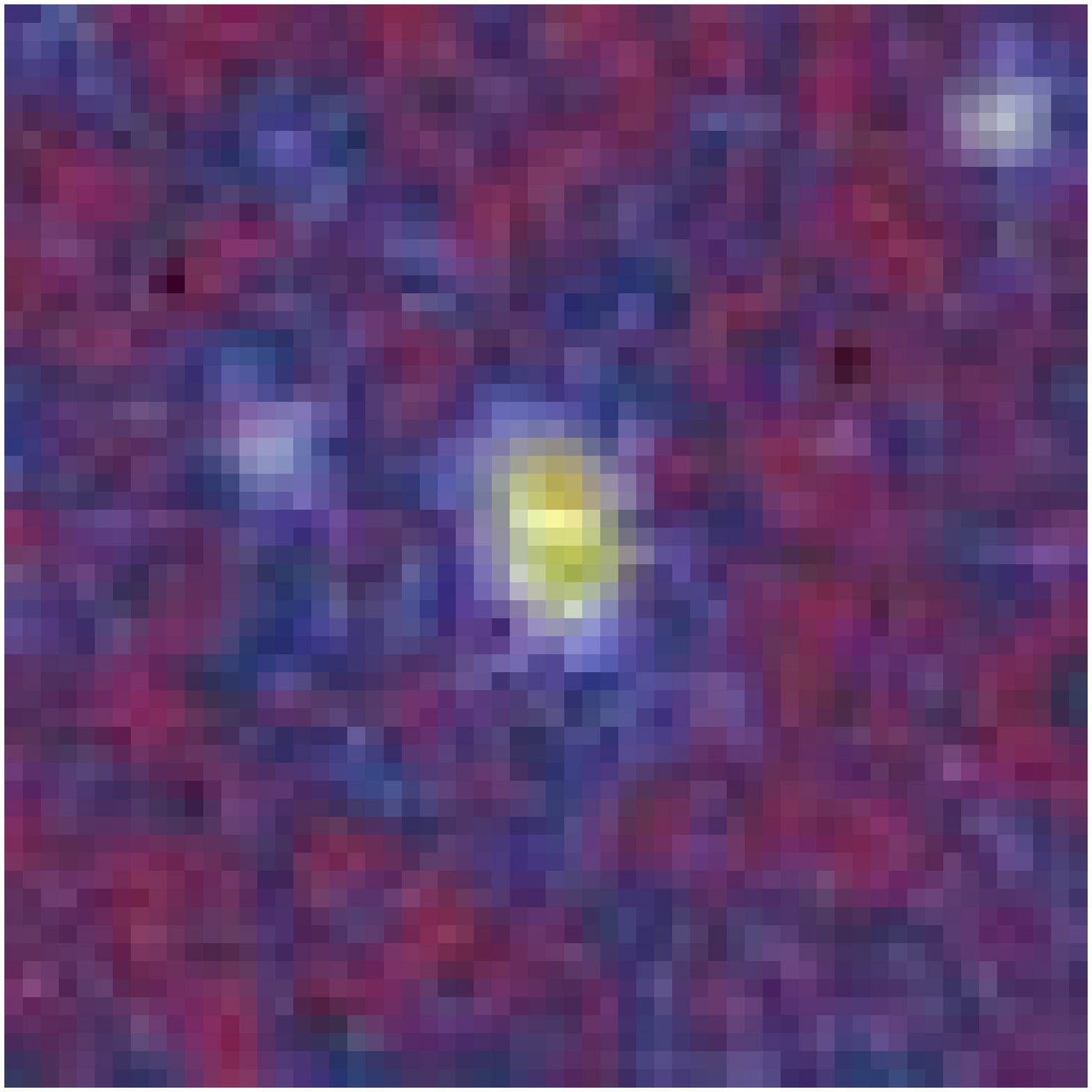}  \FGb{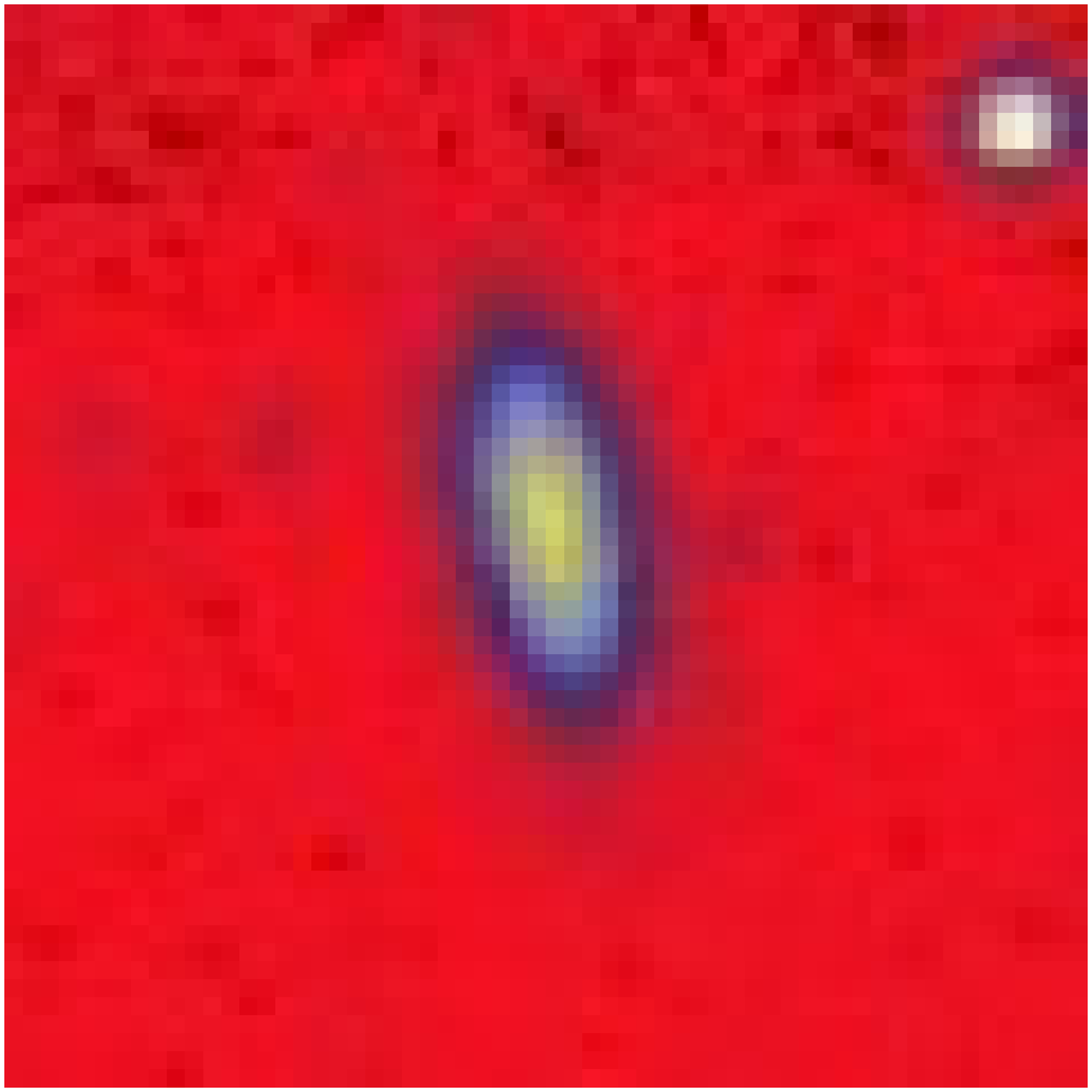}  \FGb{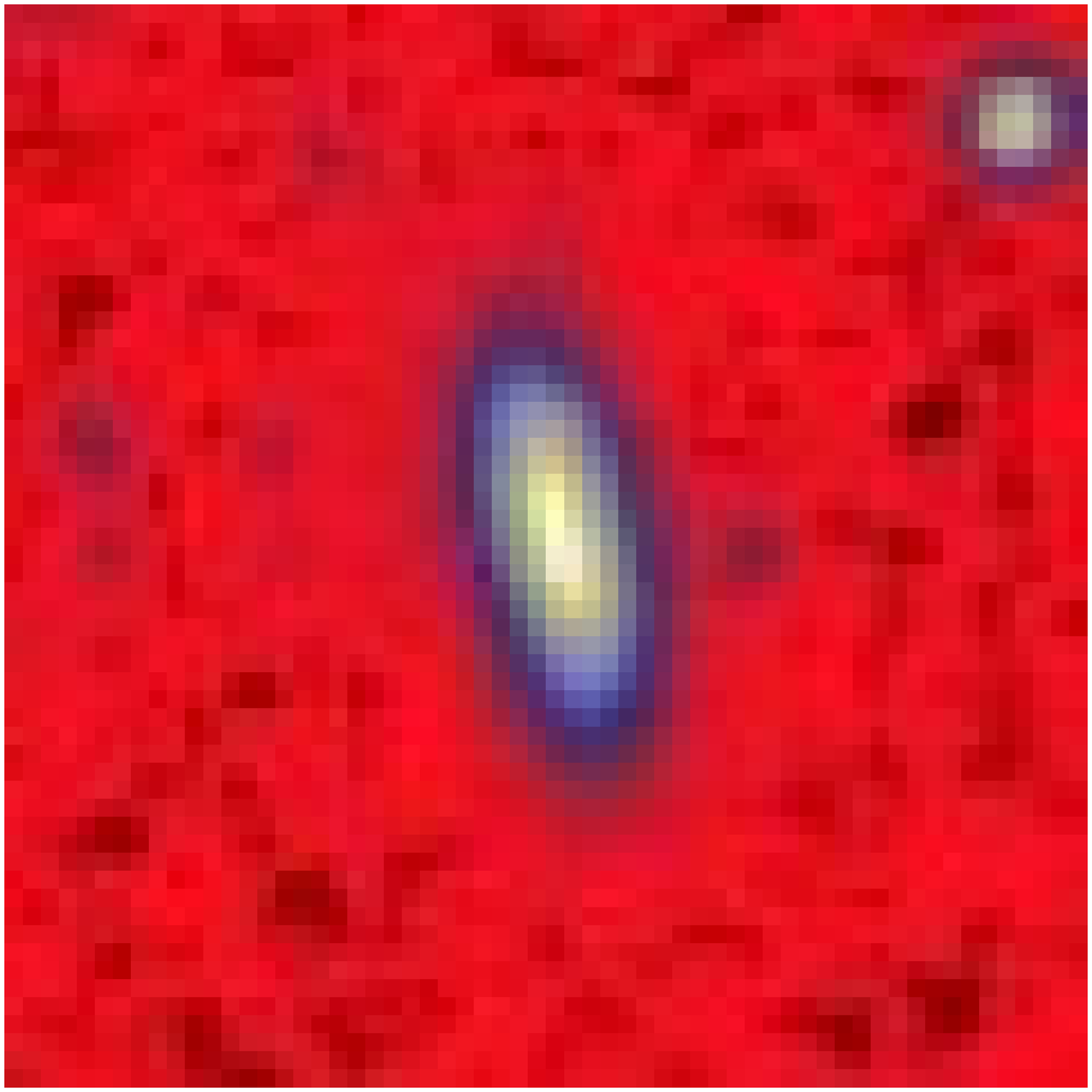}  \FGb{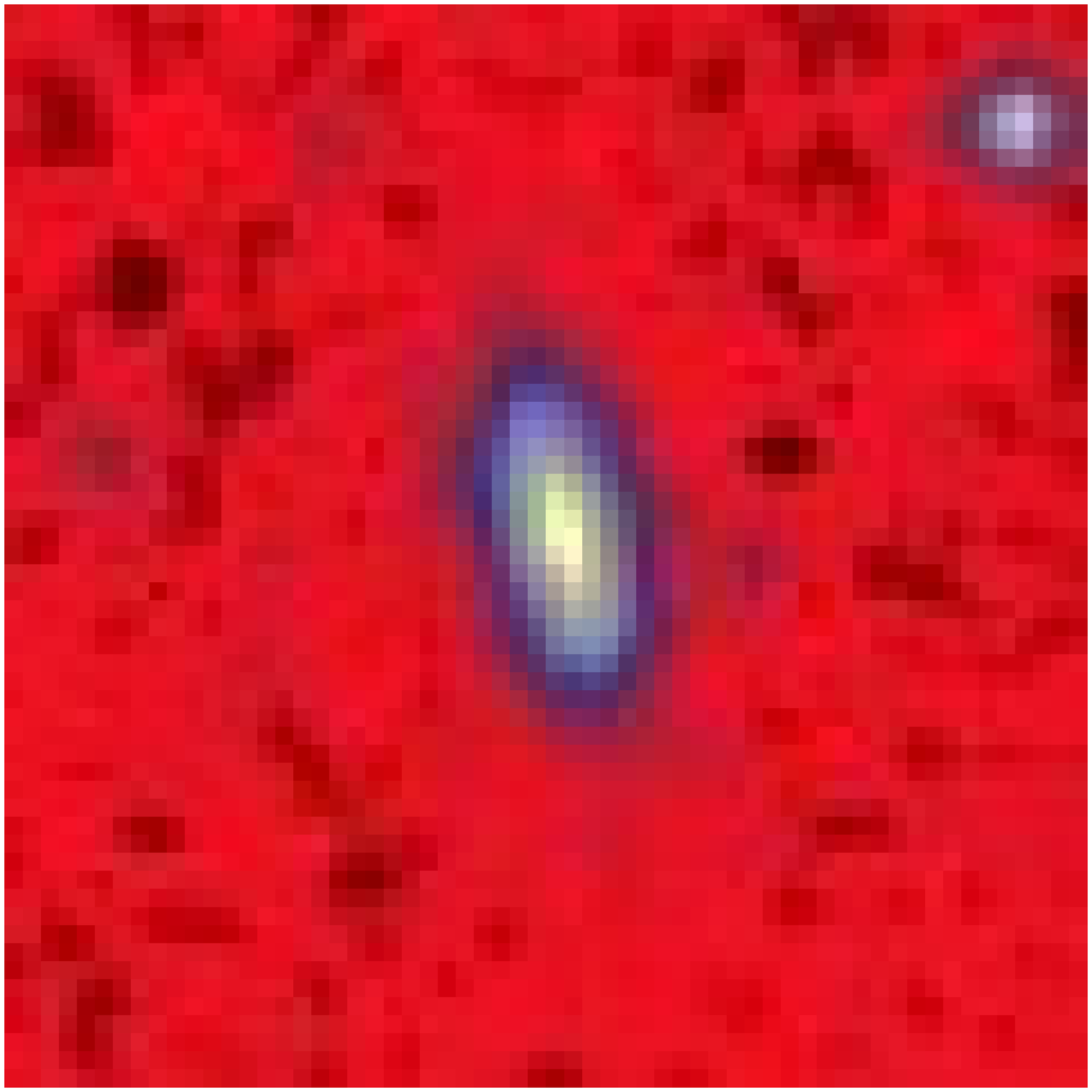}  \FGb{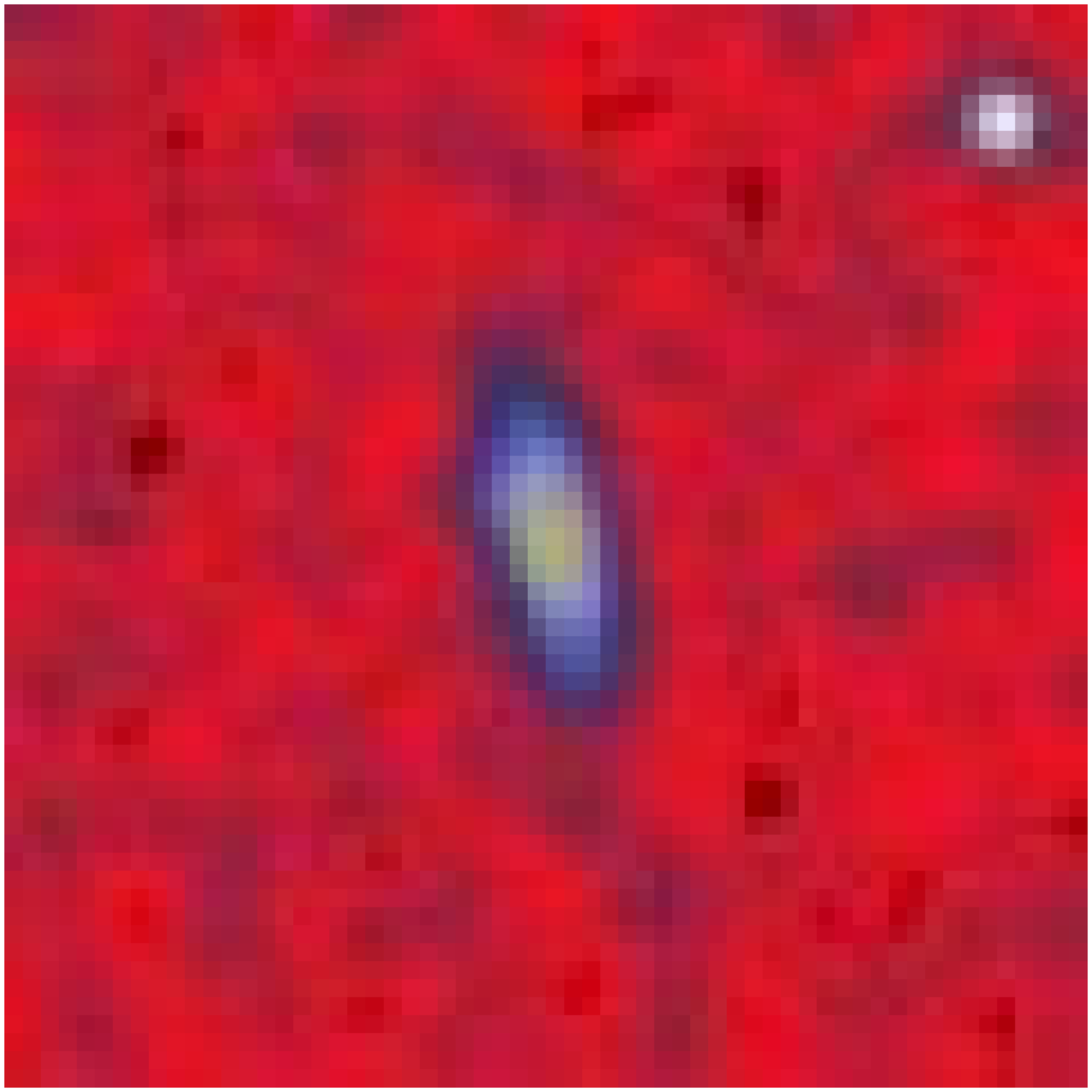}  \FGb{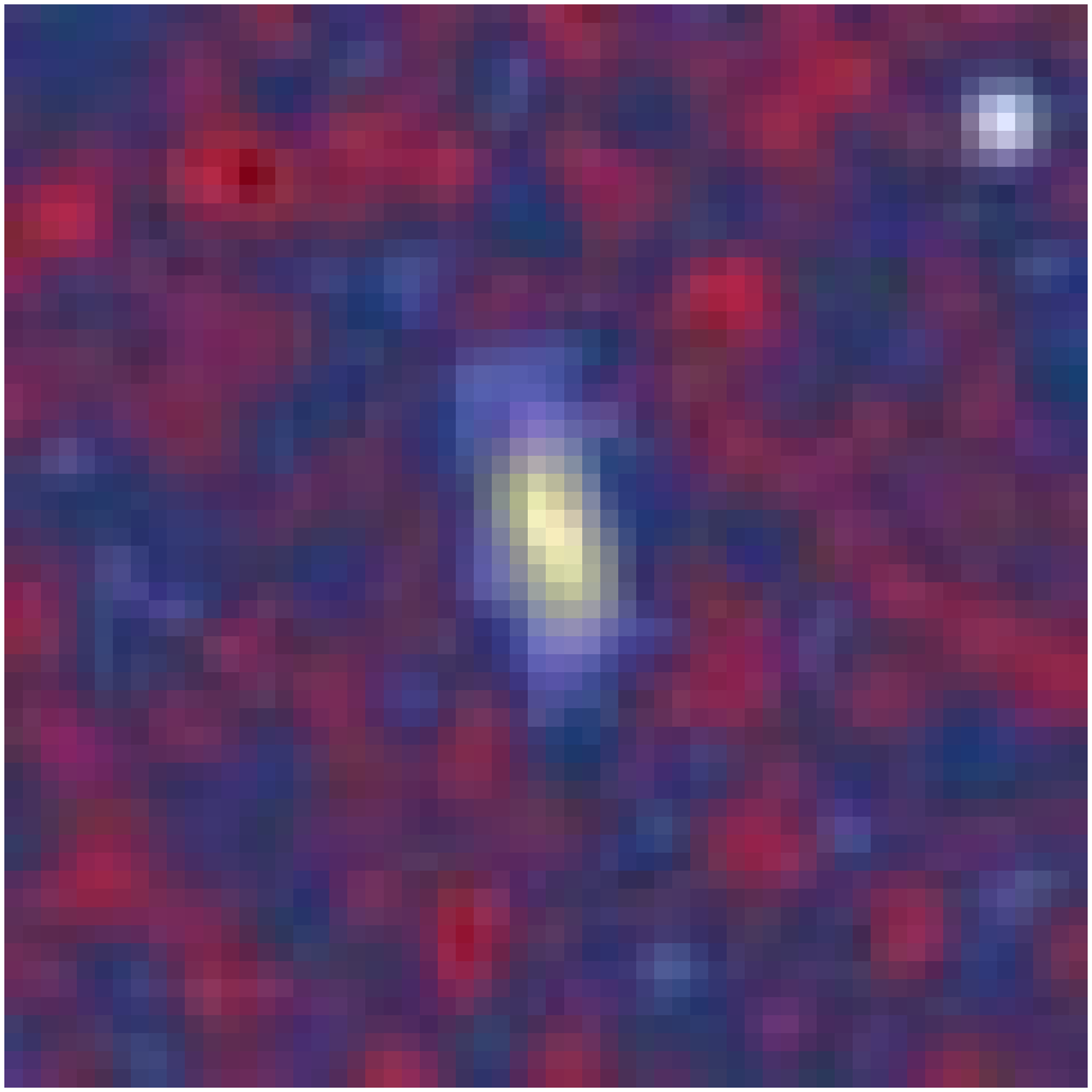}  \FGb{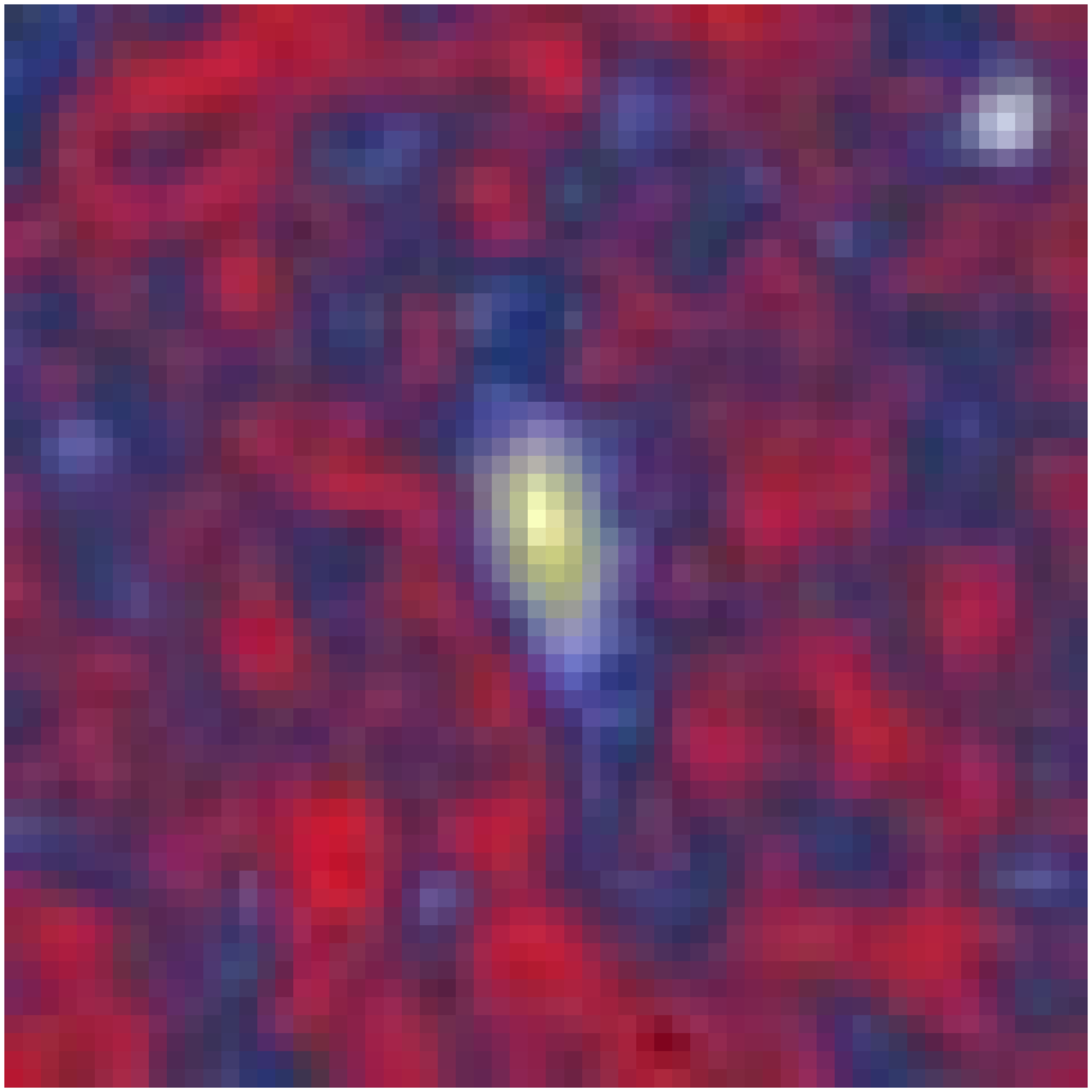}  \FGb{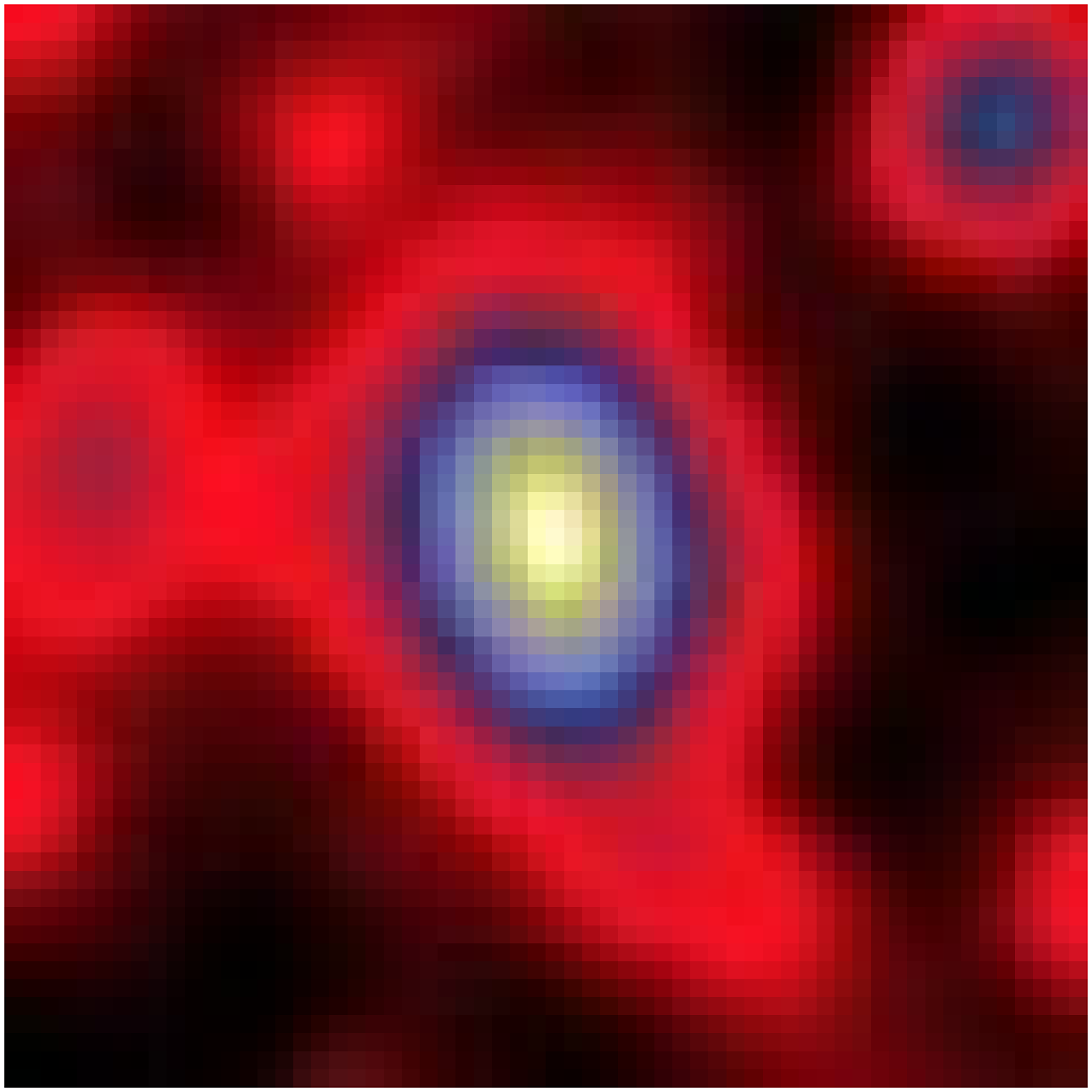}  \FGb{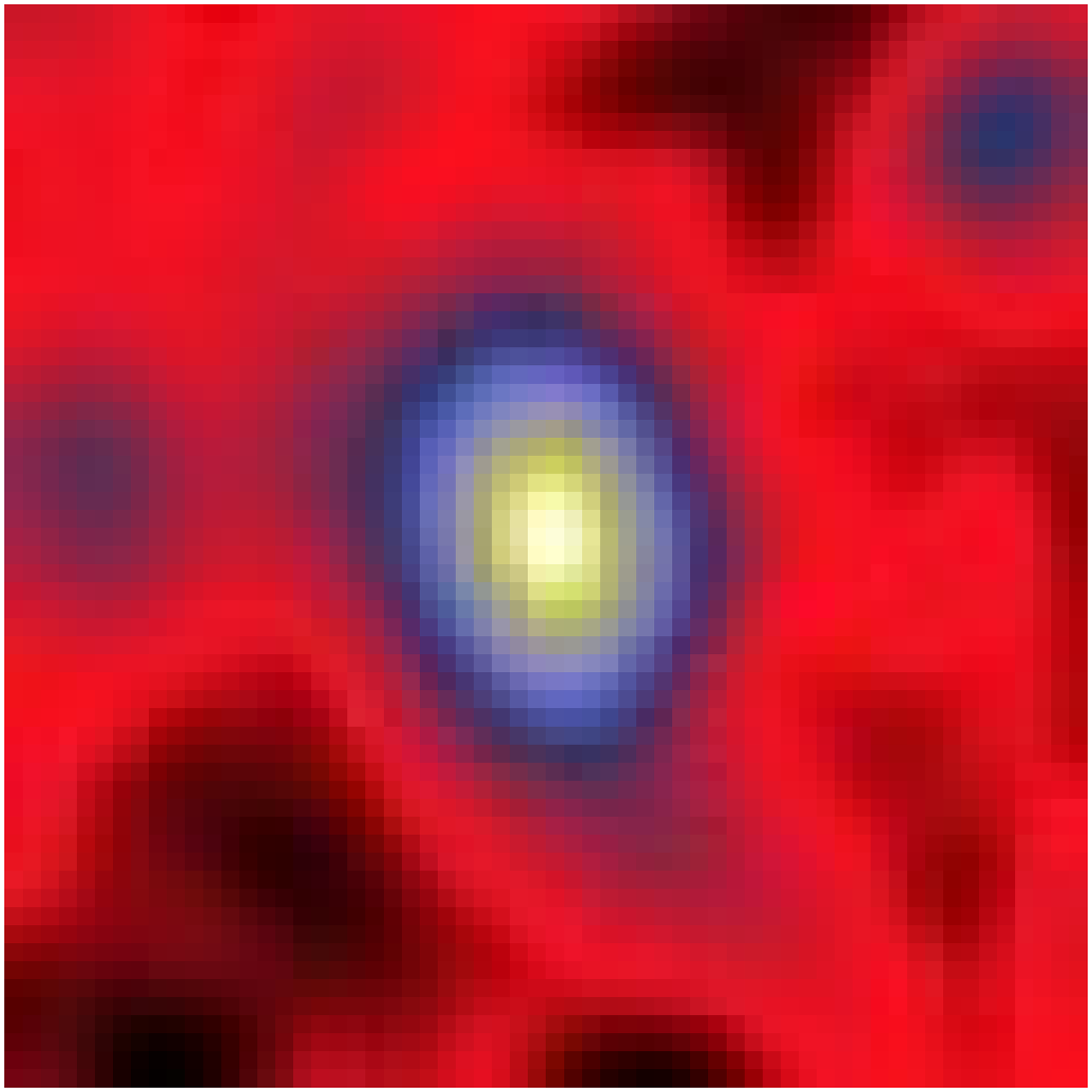}  \FGb{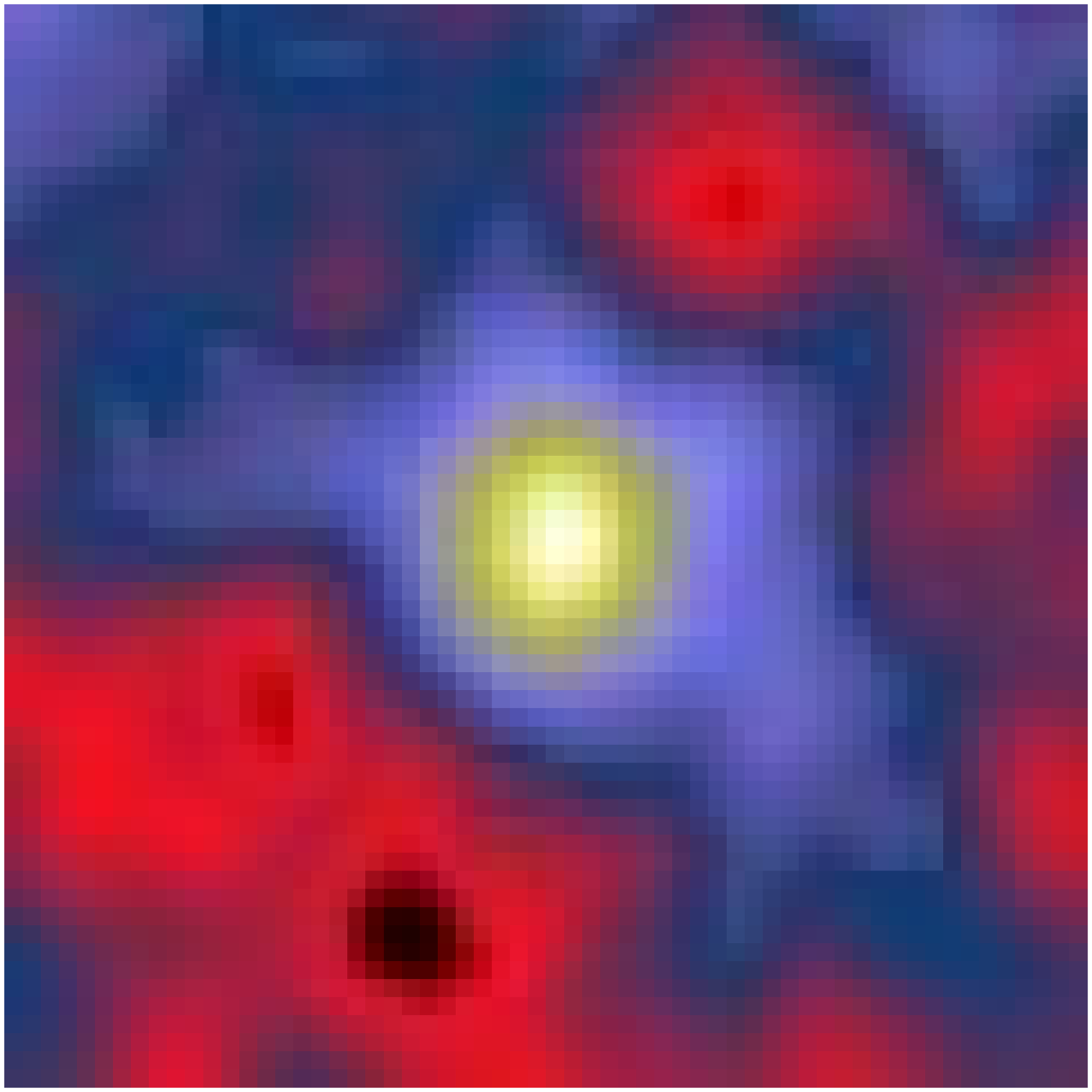}  \FGb{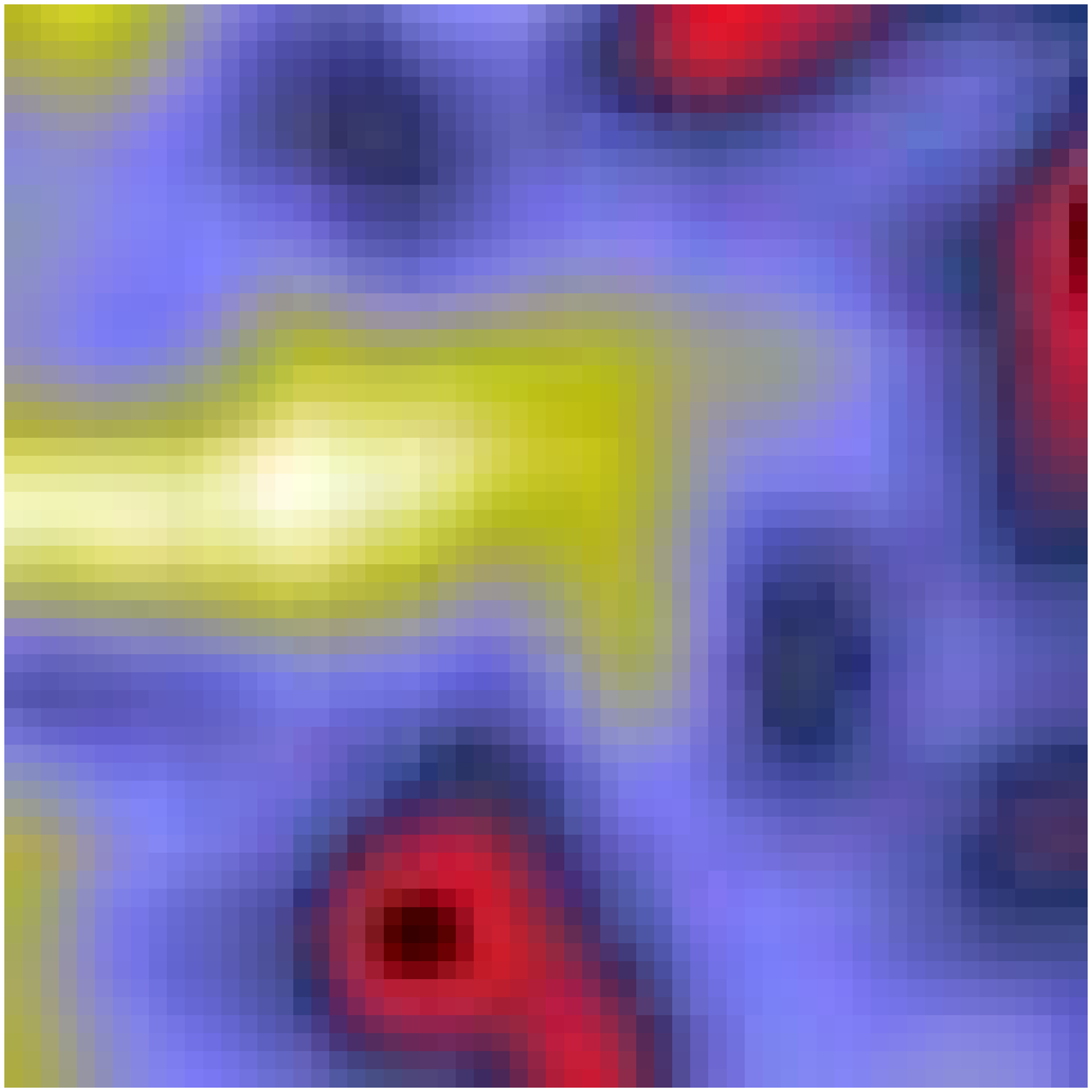} \\  {\#41   NCIA027303}  \\
  \FGb{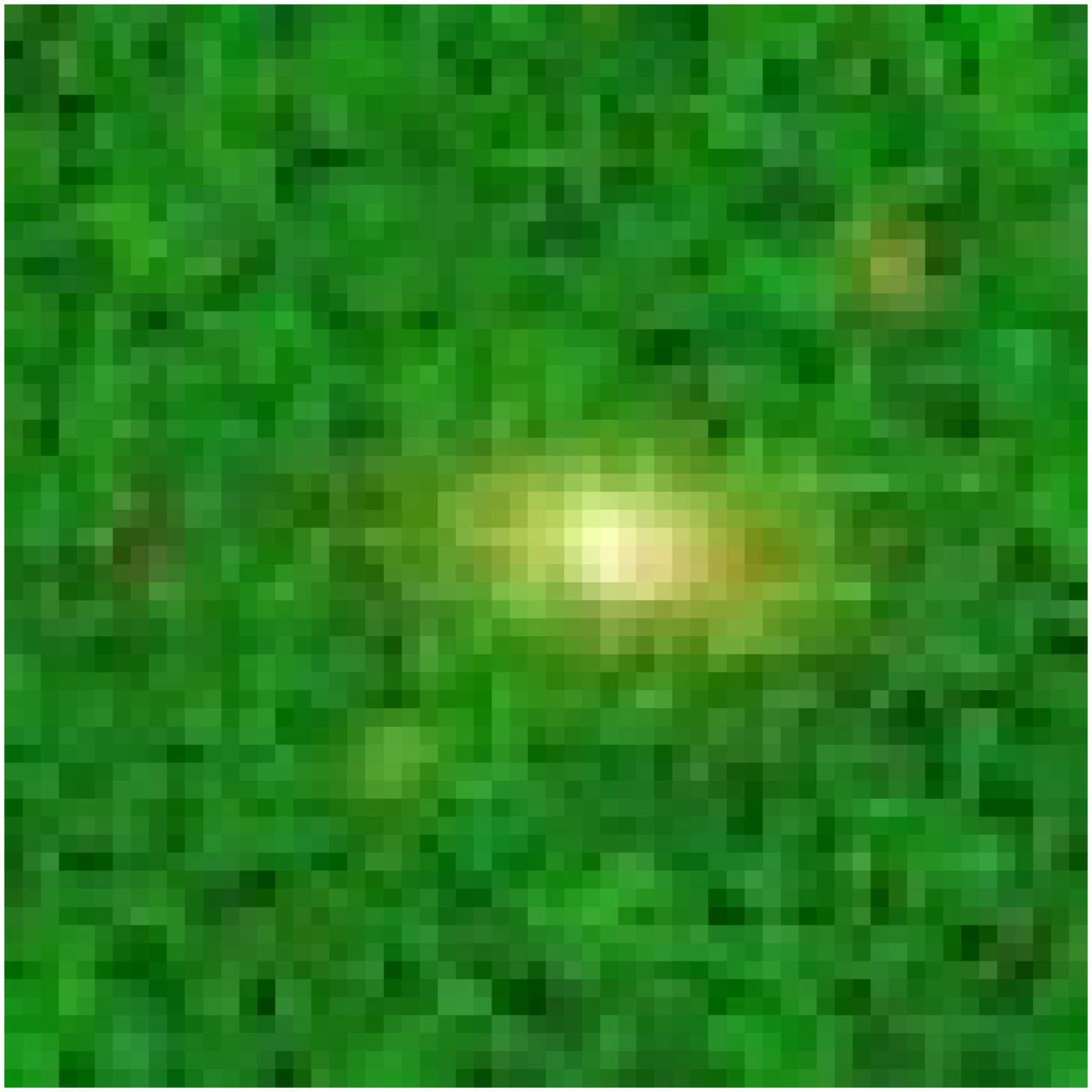}  \FGb{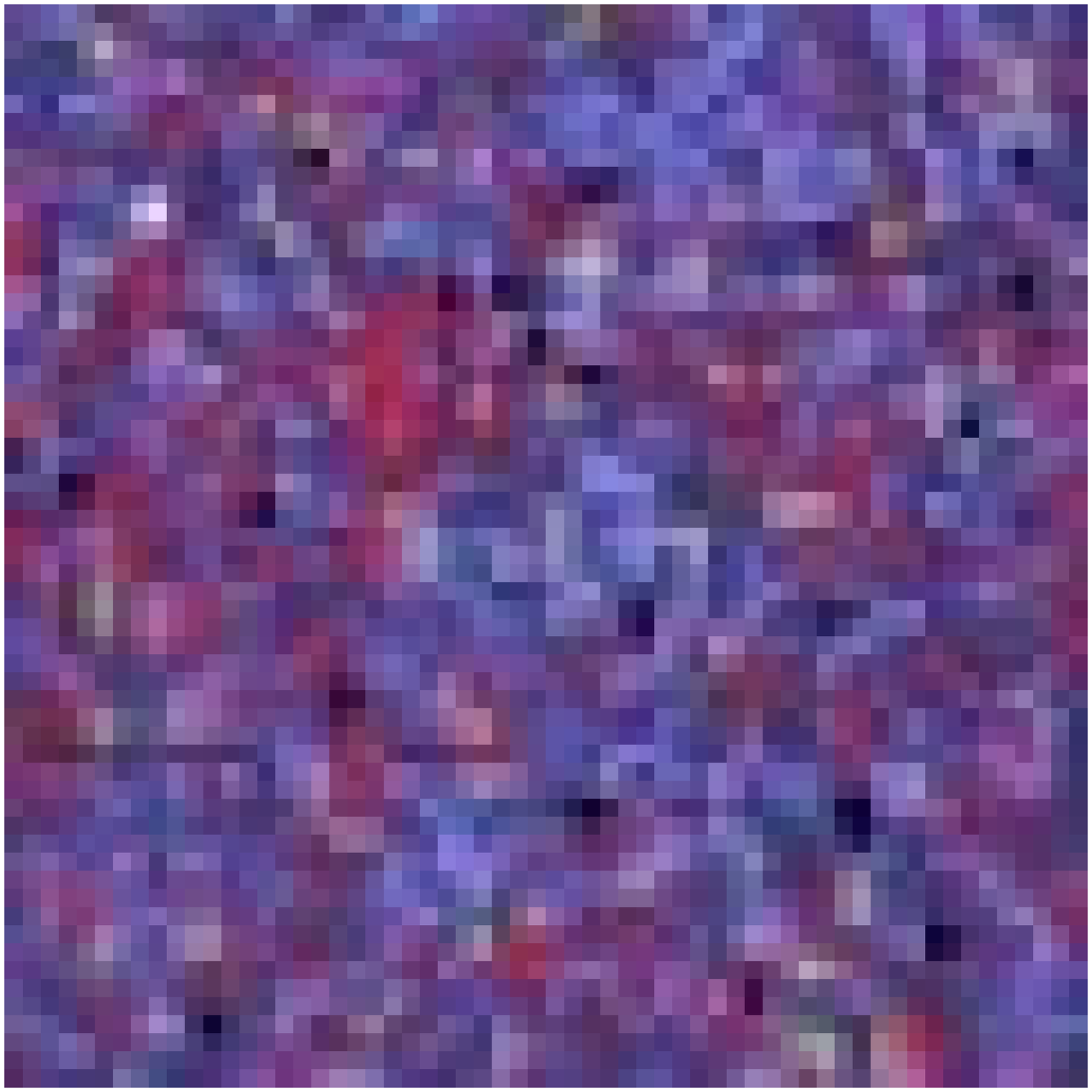}  \FGb{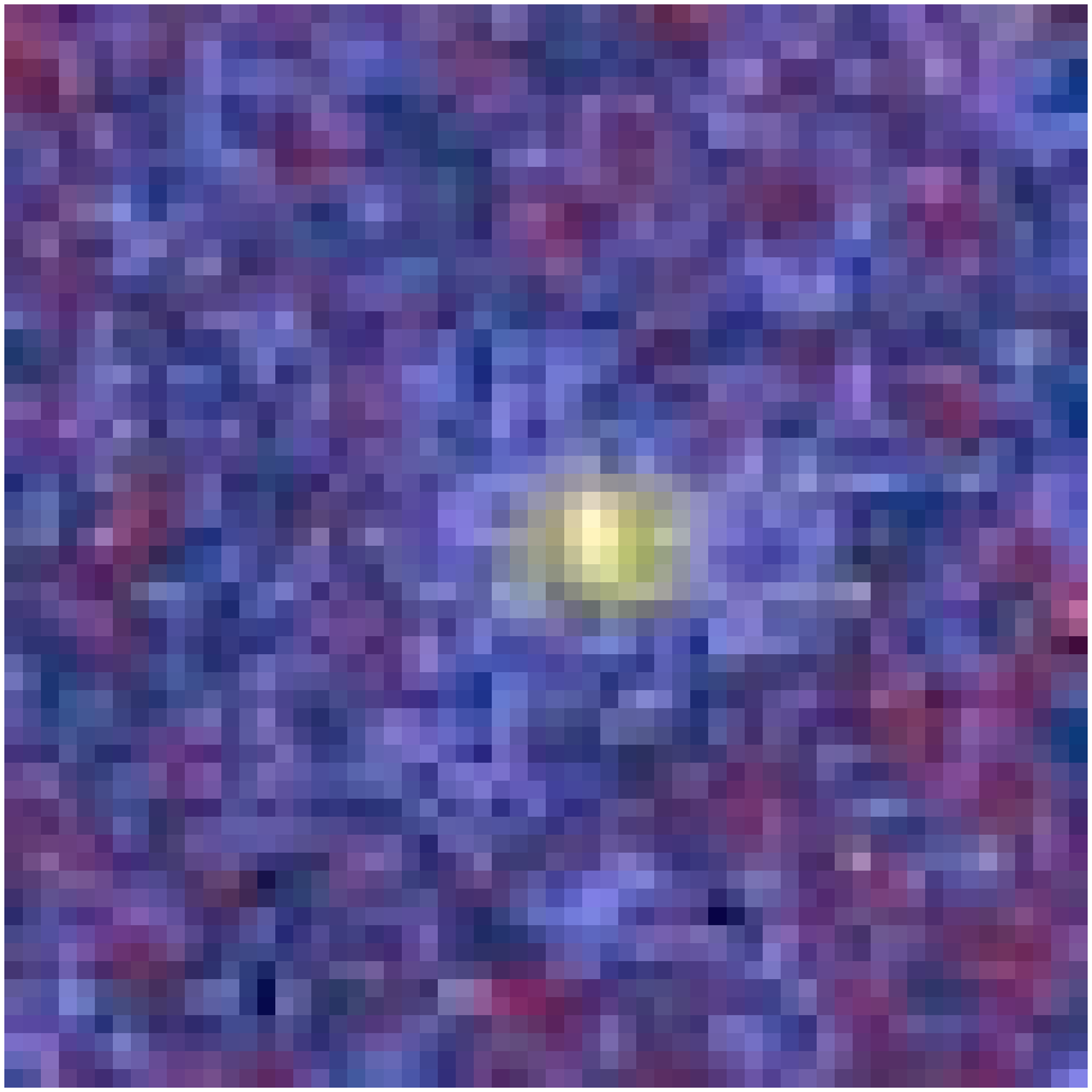}  \FGb{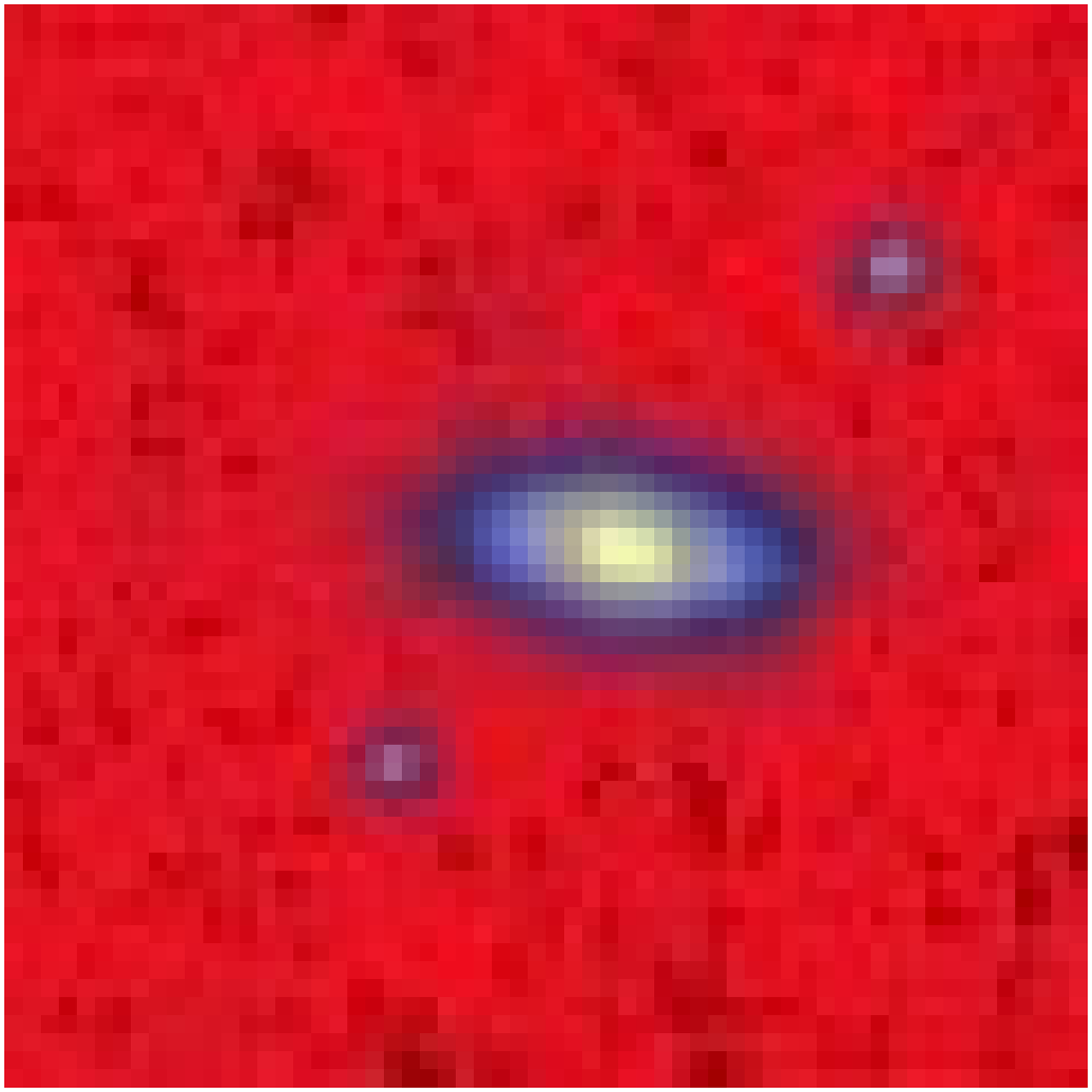}  \FGb{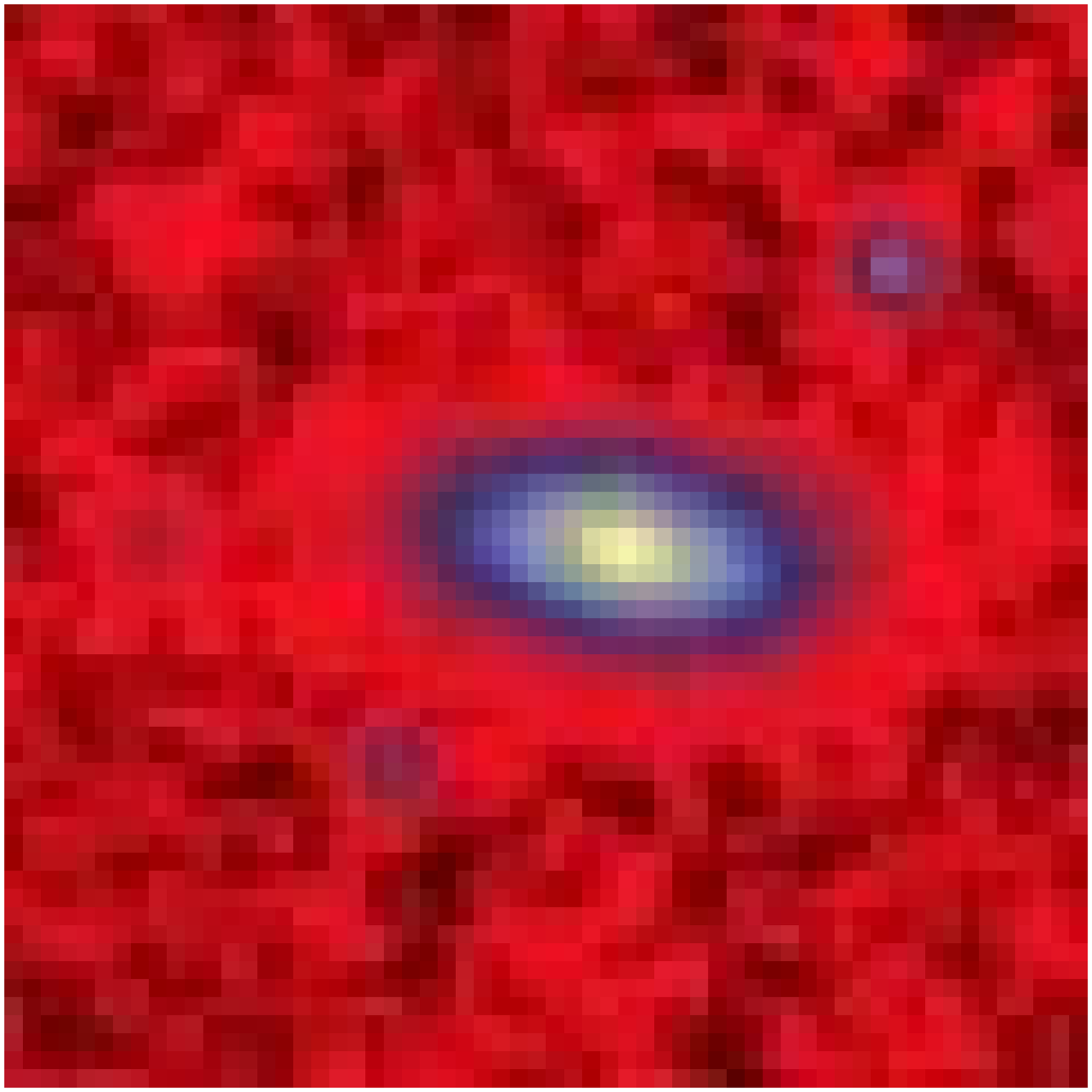}  \FGb{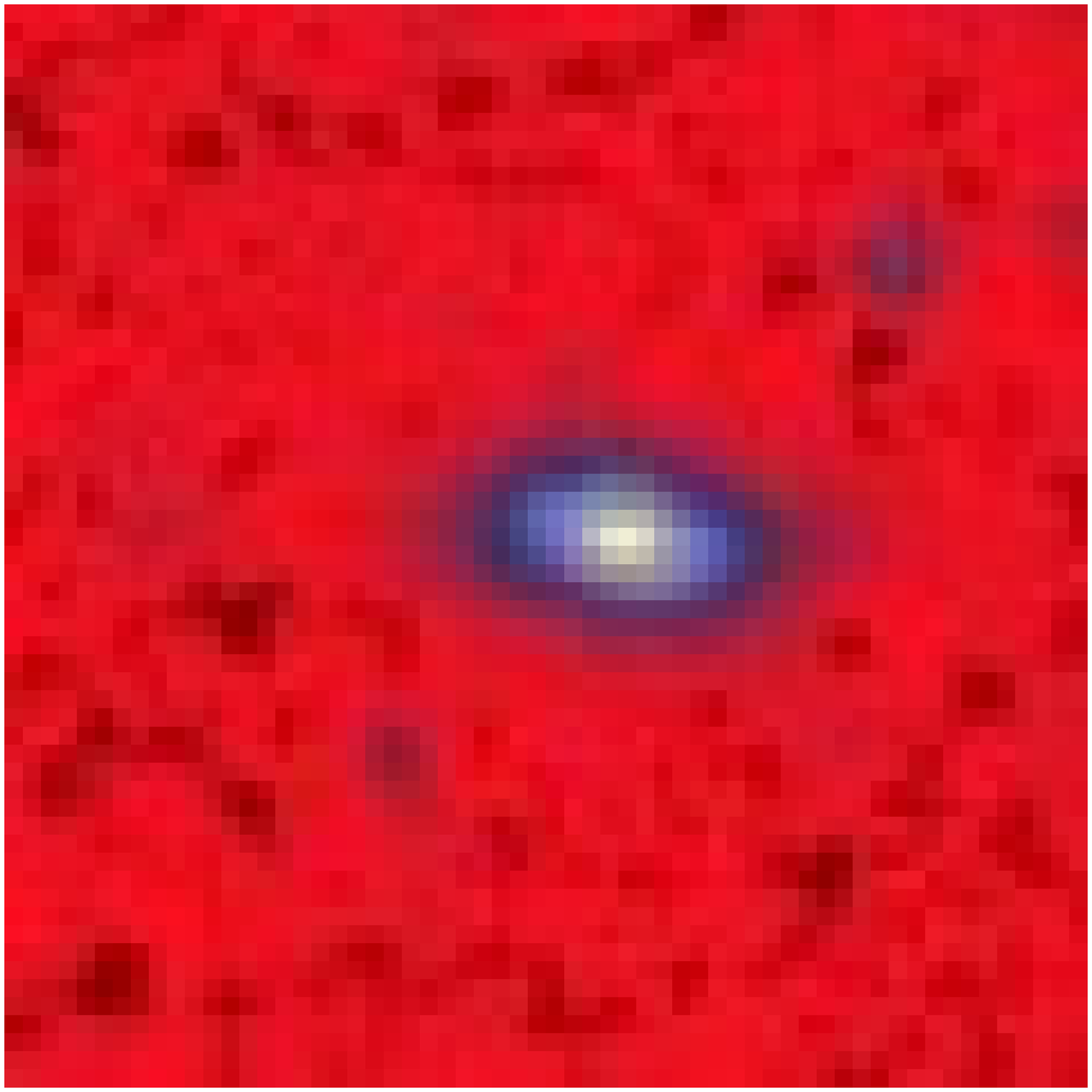}  \FGb{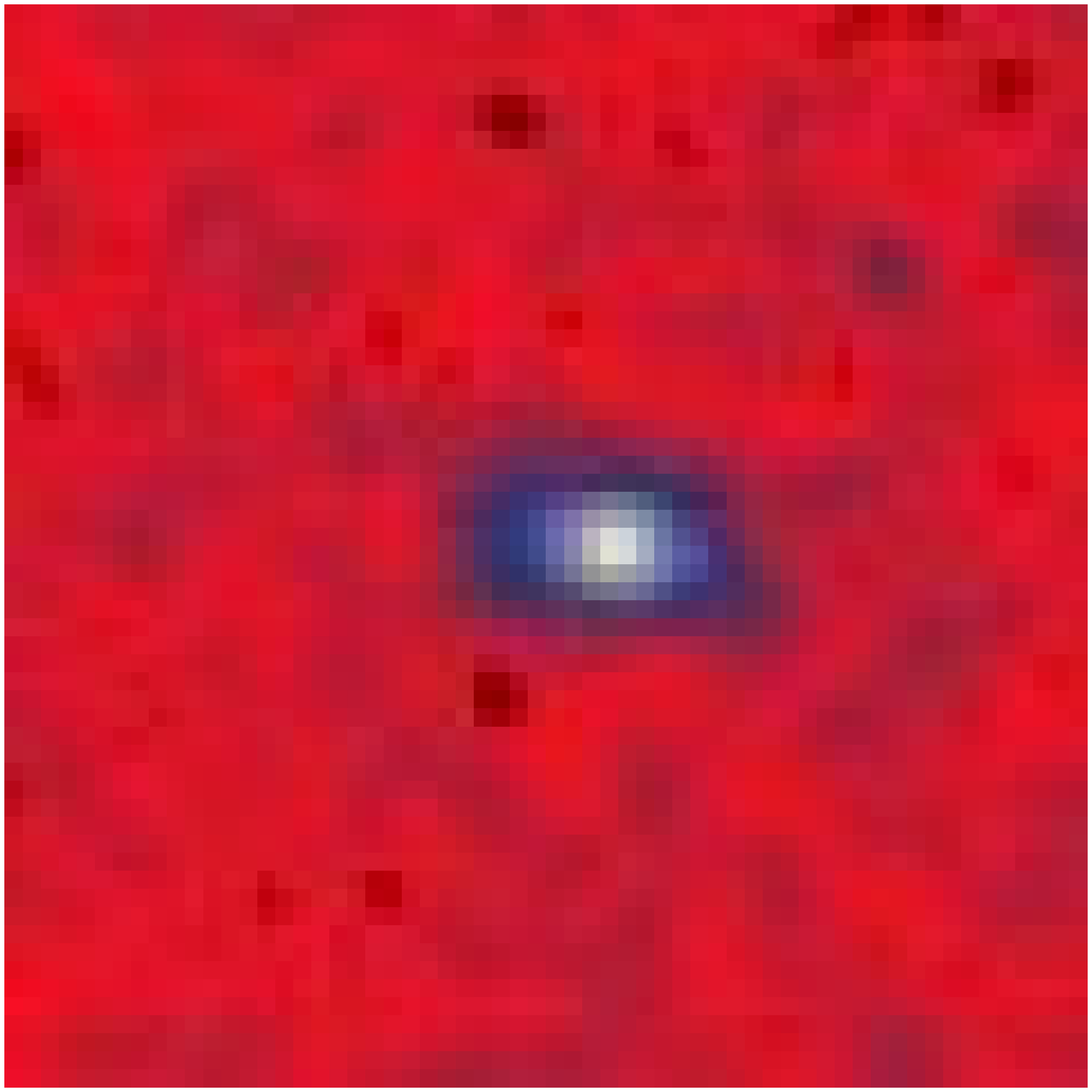}  \FGb{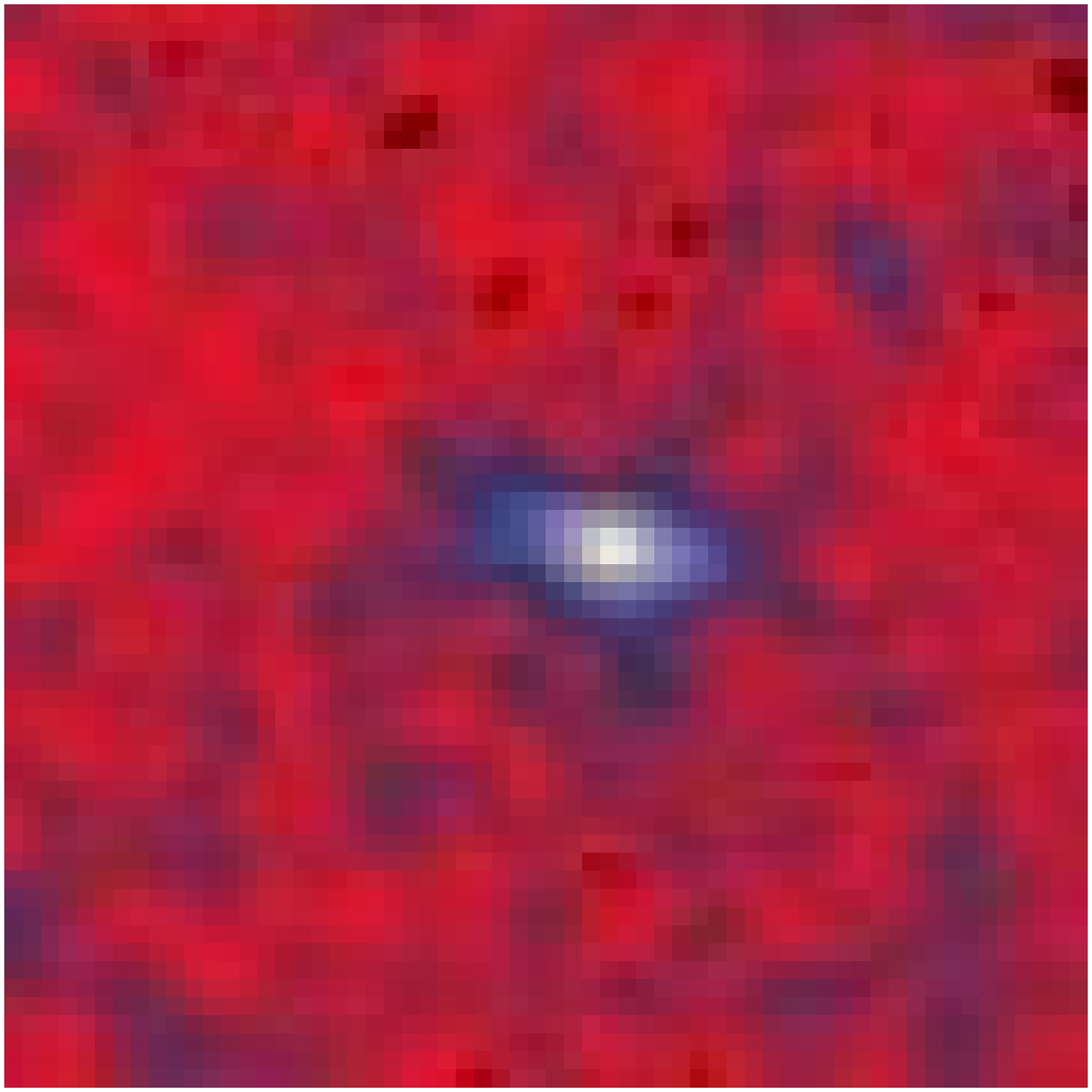}  \FGb{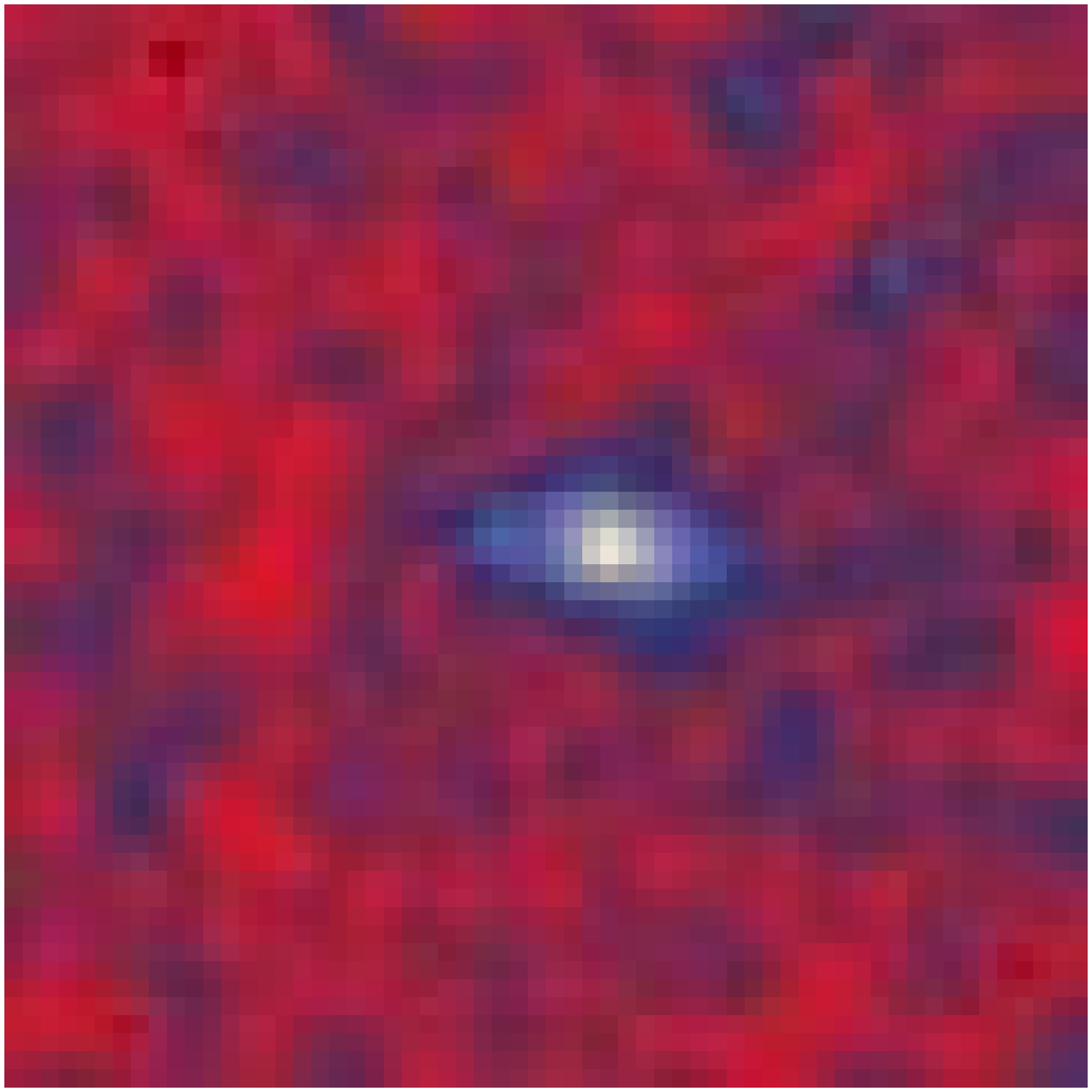}  \FGb{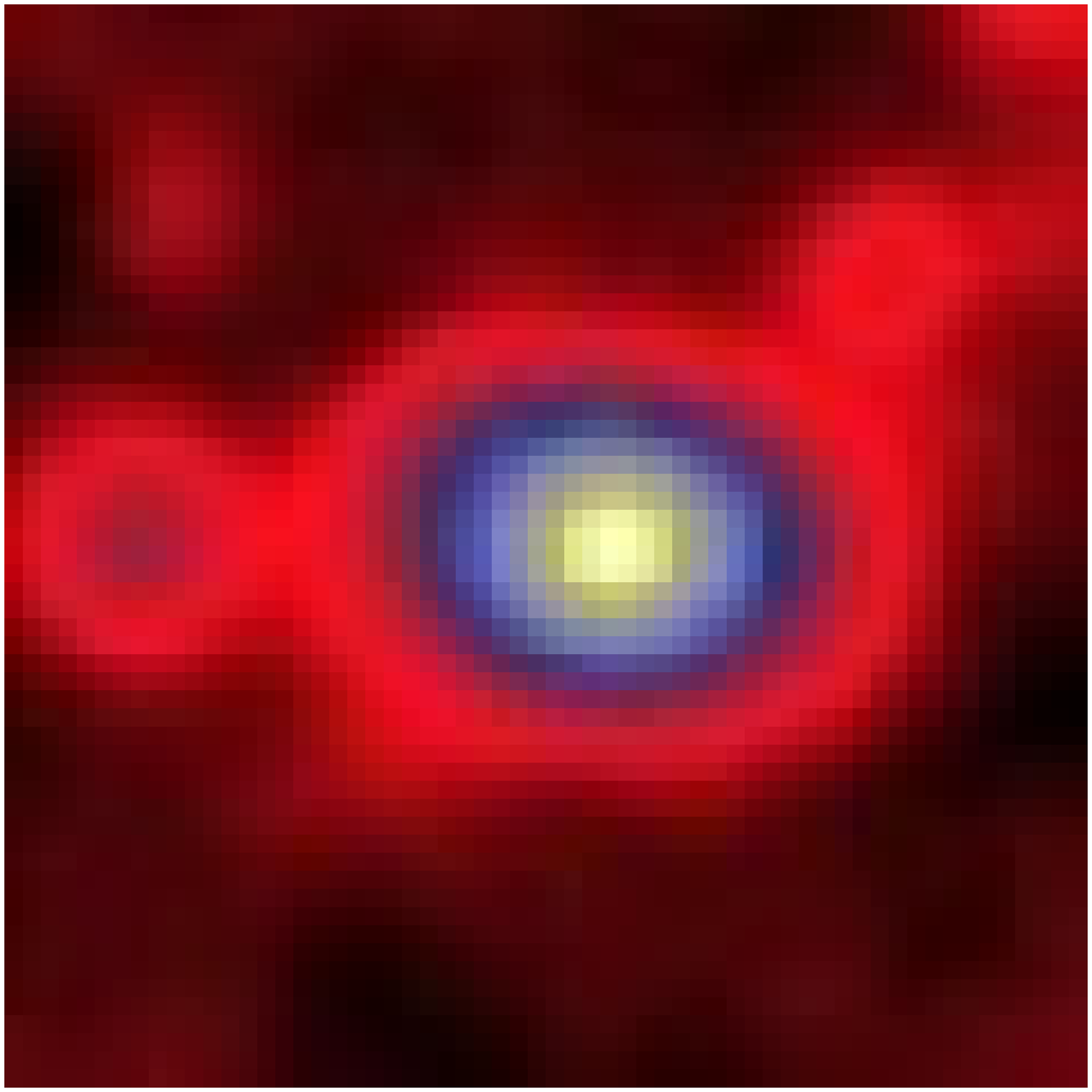}  \FGb{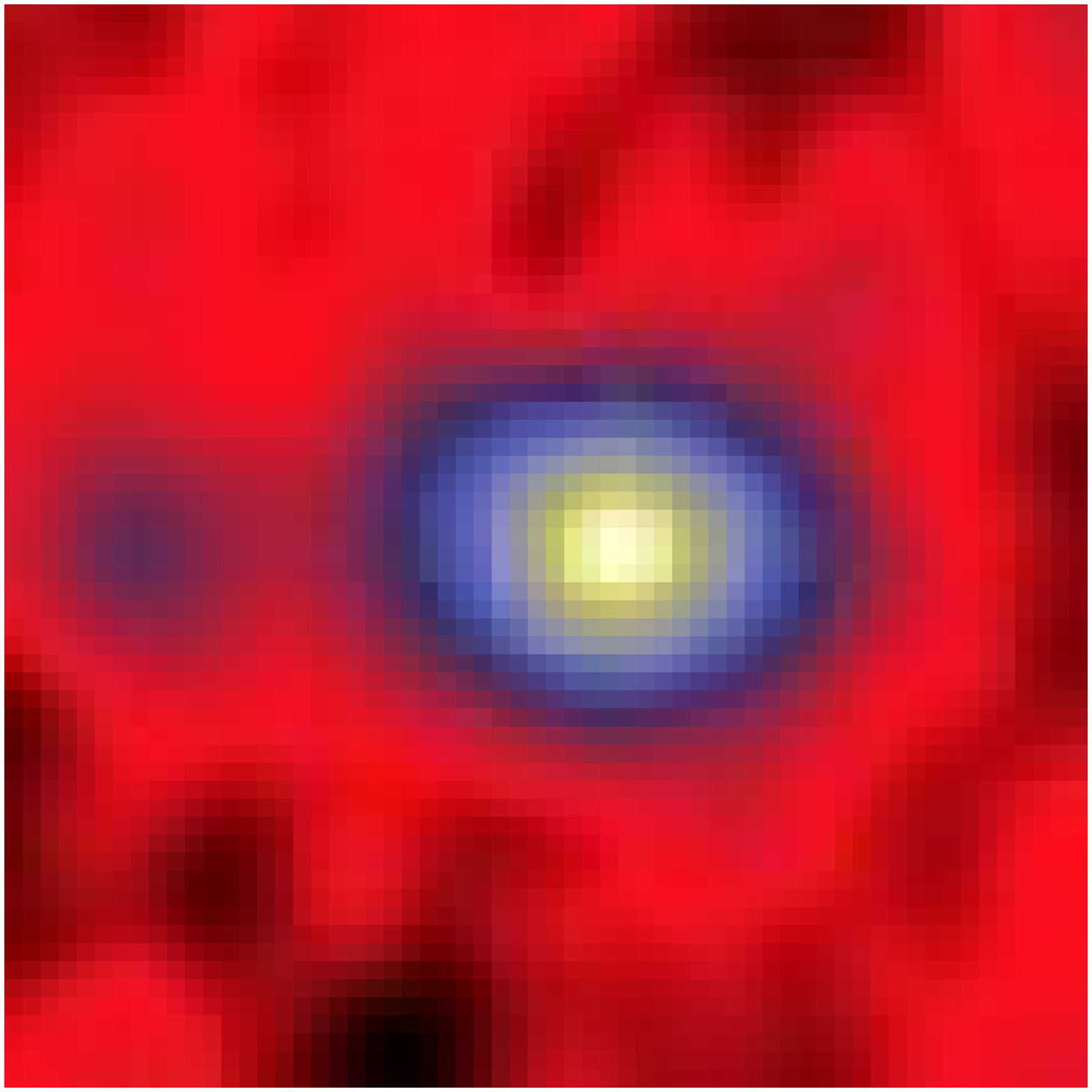}  \FGb{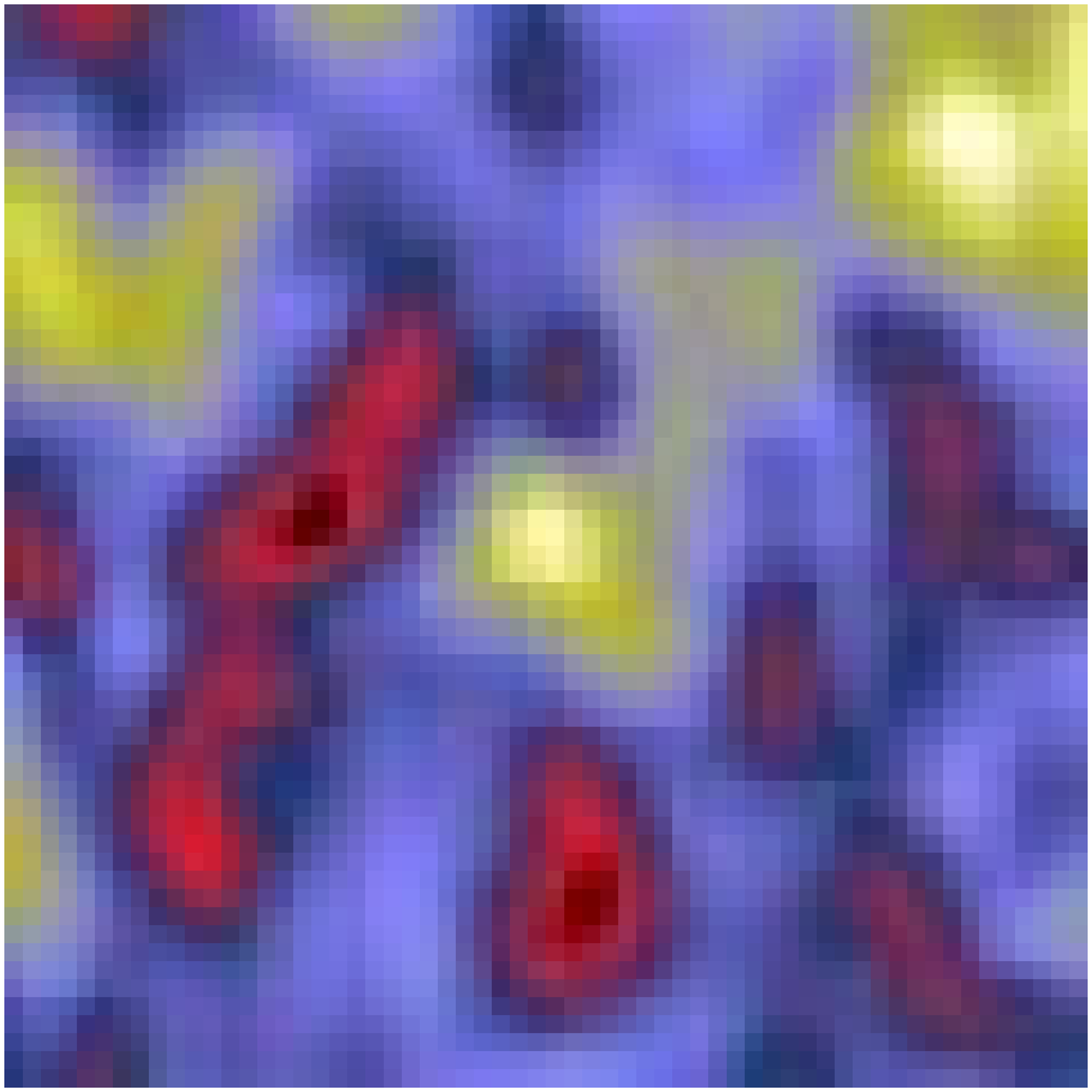}  \FGb{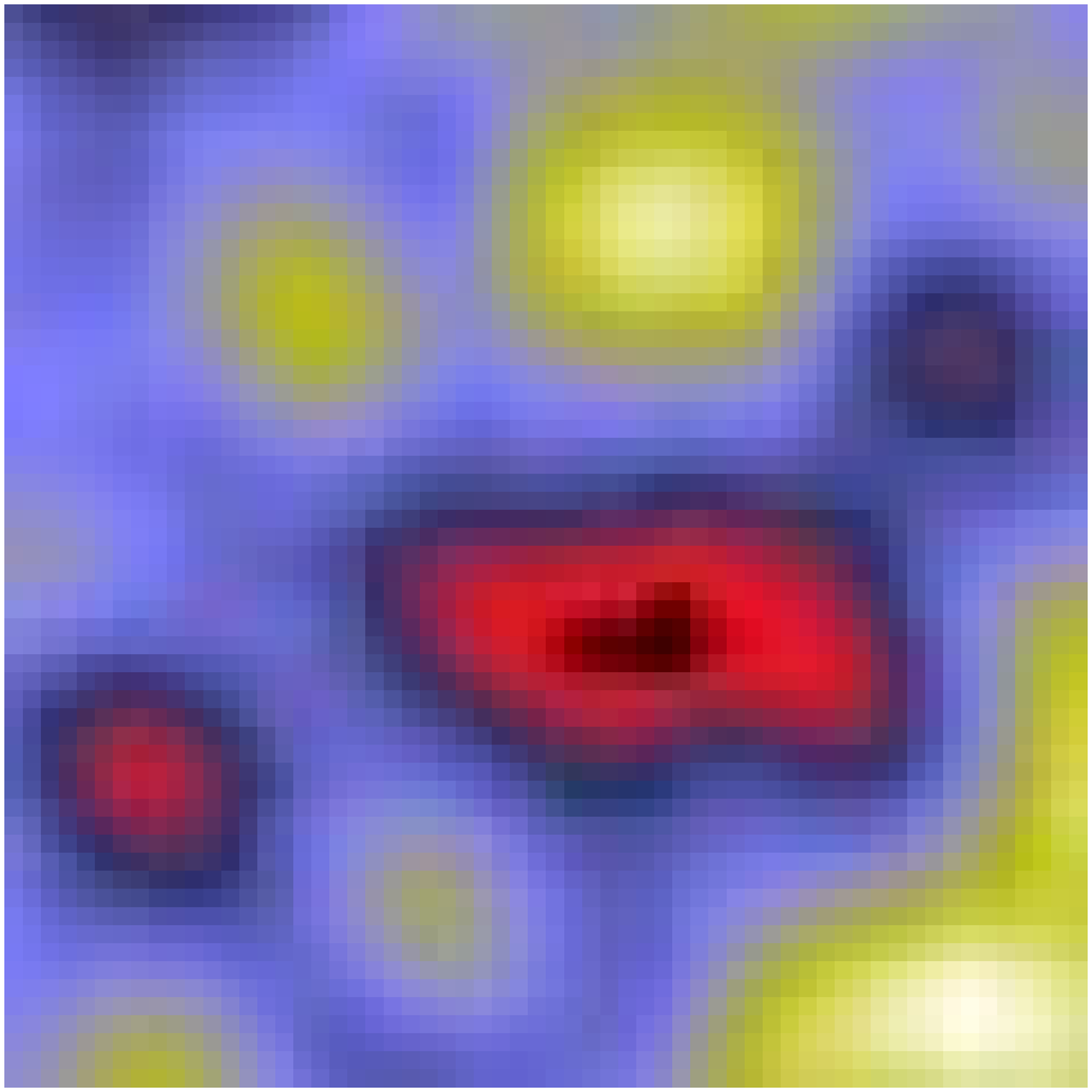} \\  {\#42   NCIA029804}  \\
  \FGb{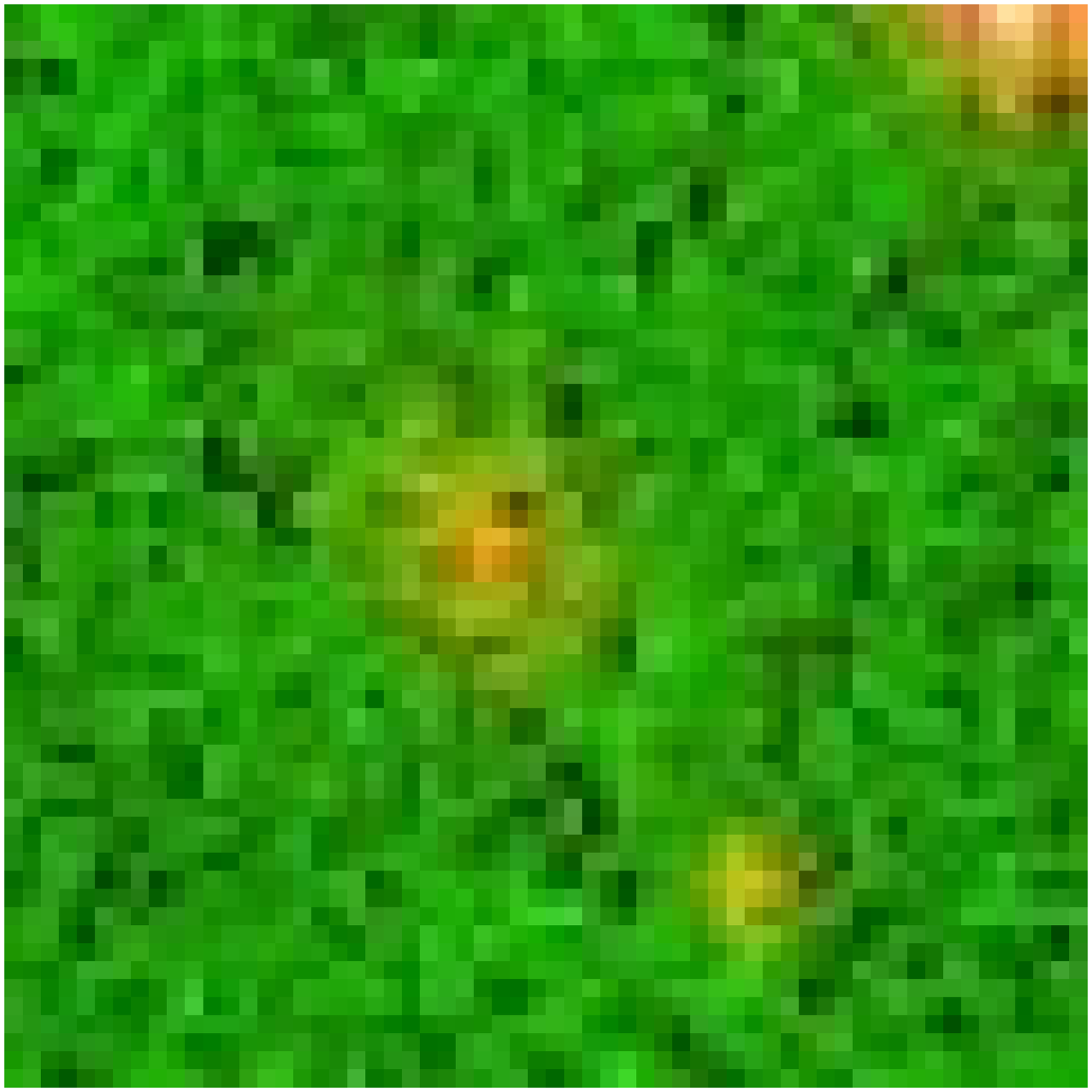}  \FGb{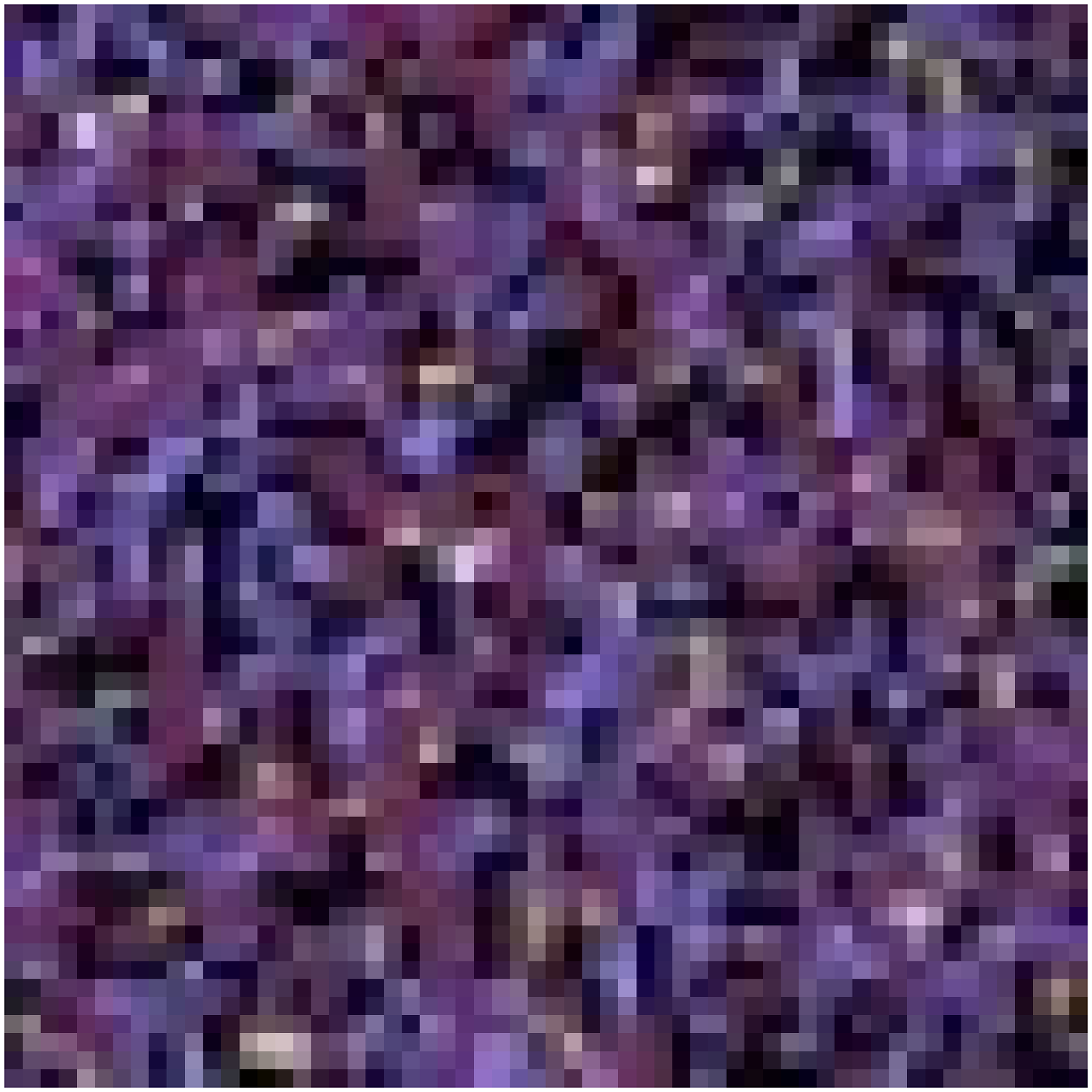}  \FGb{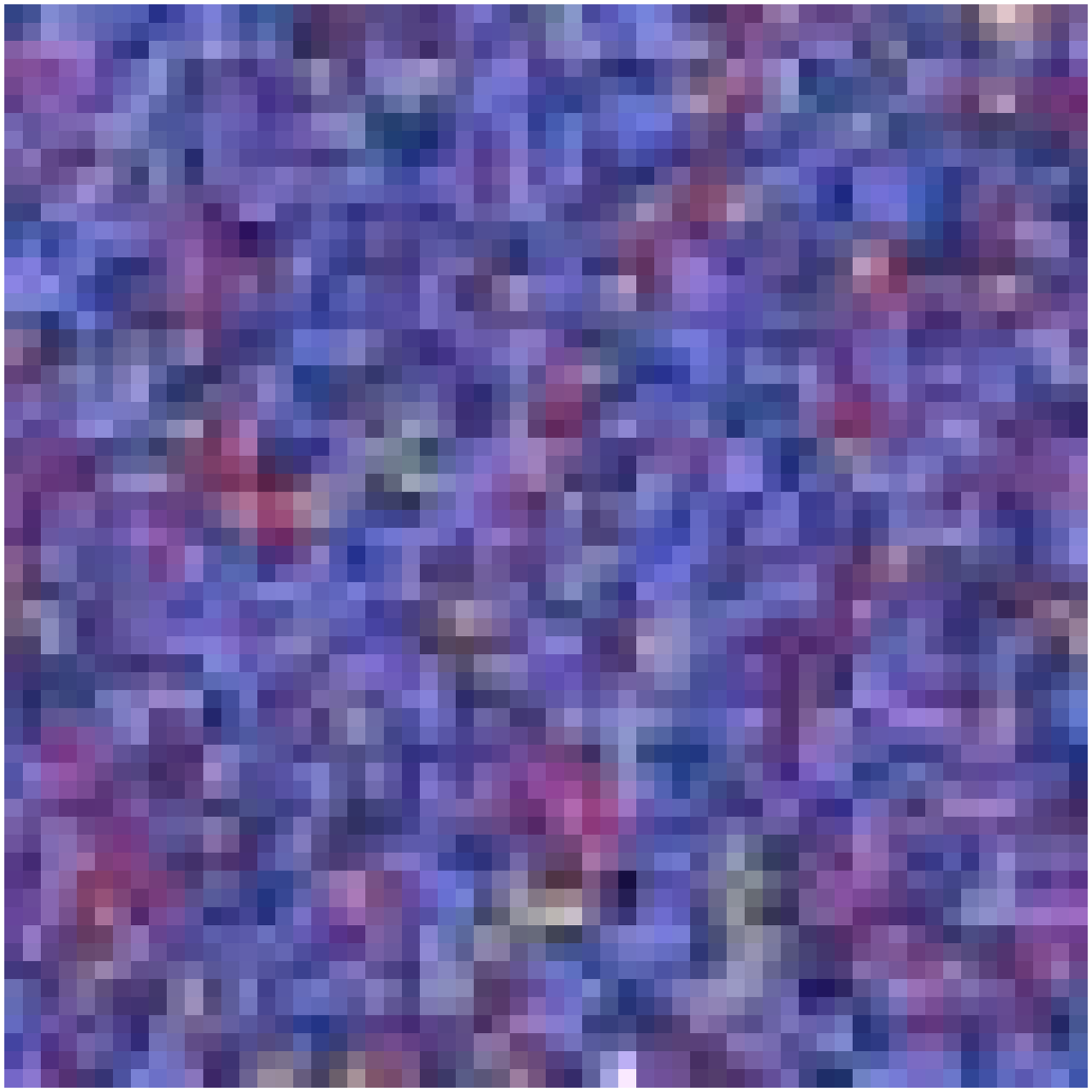}  \FGb{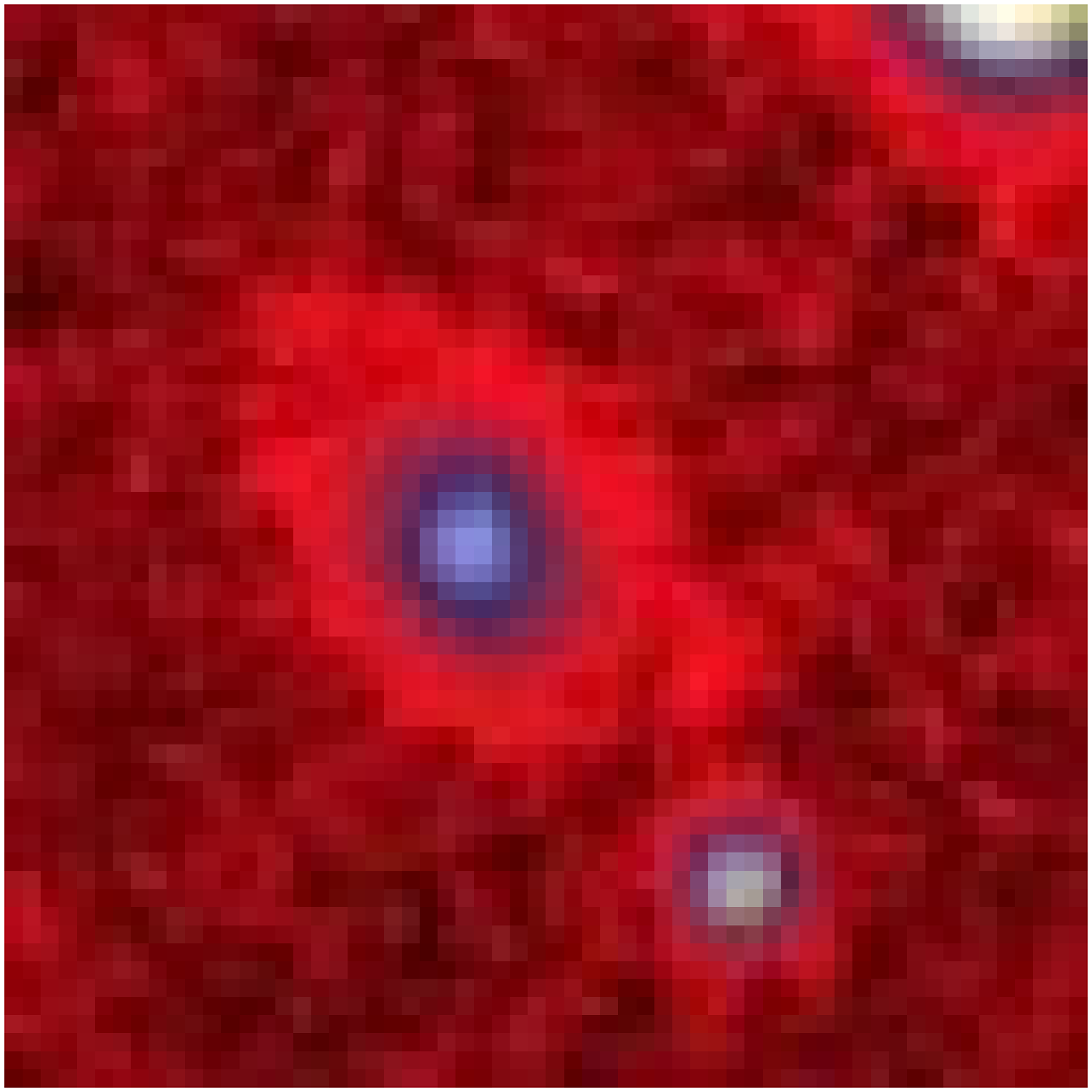}  \FGb{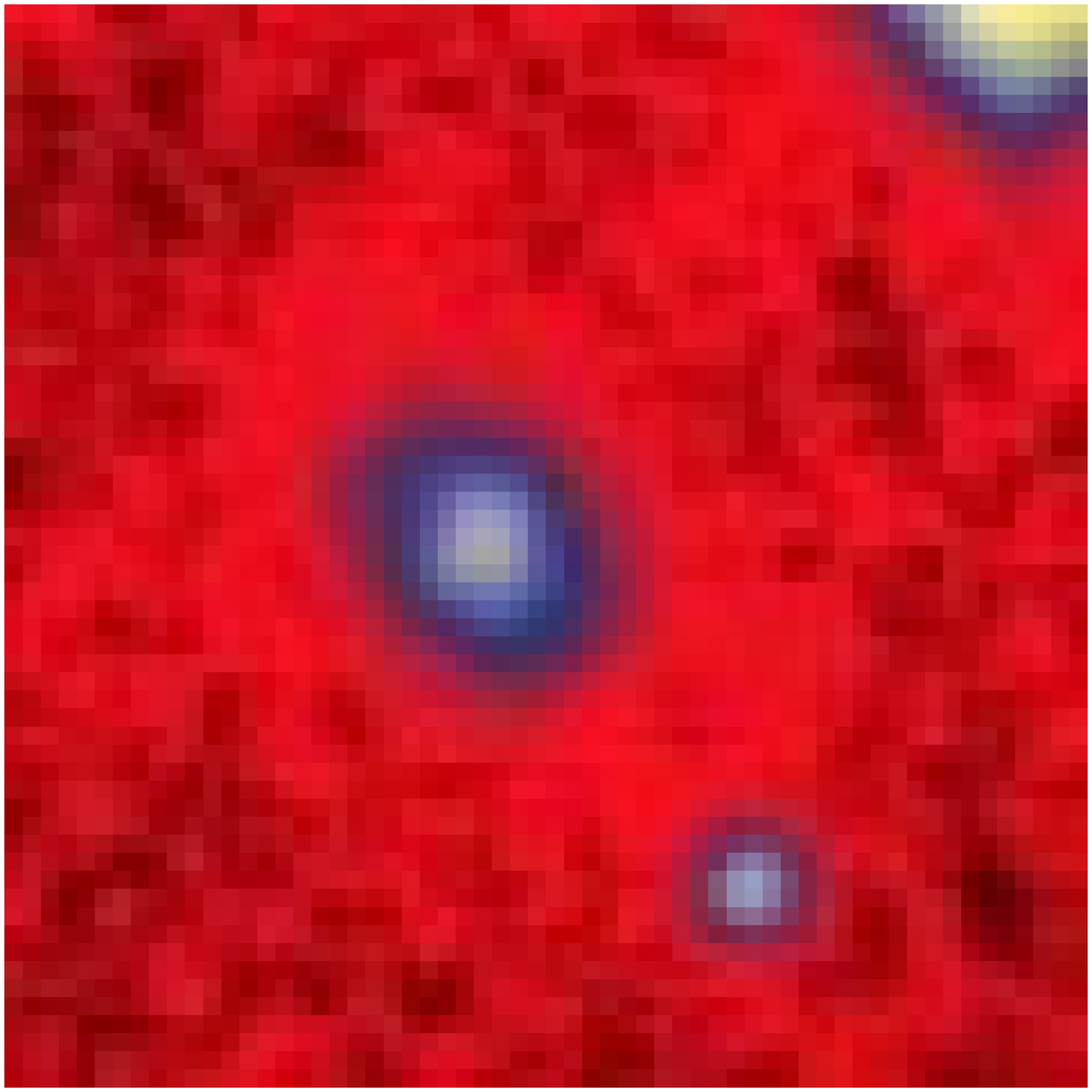}  \FGb{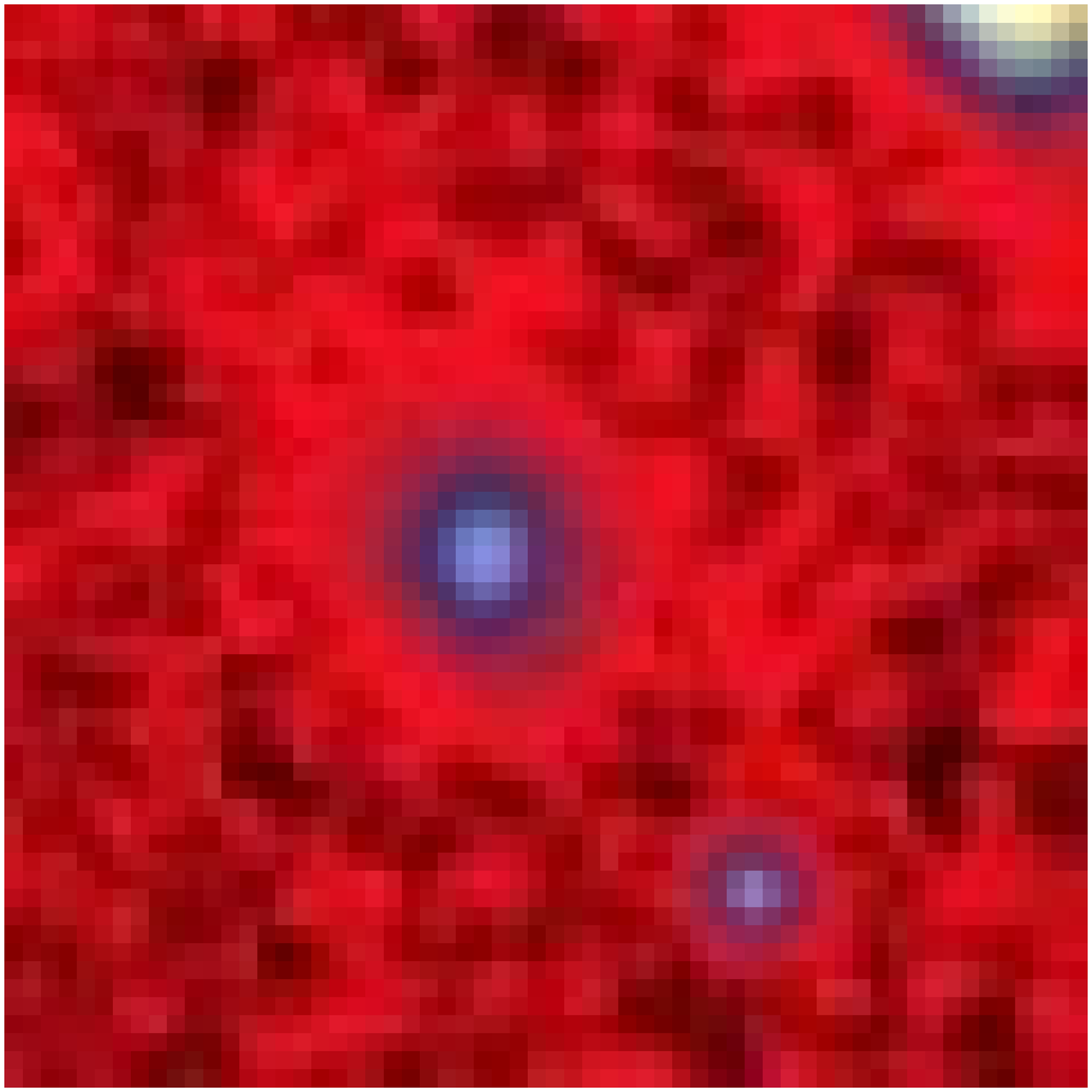}  \FGb{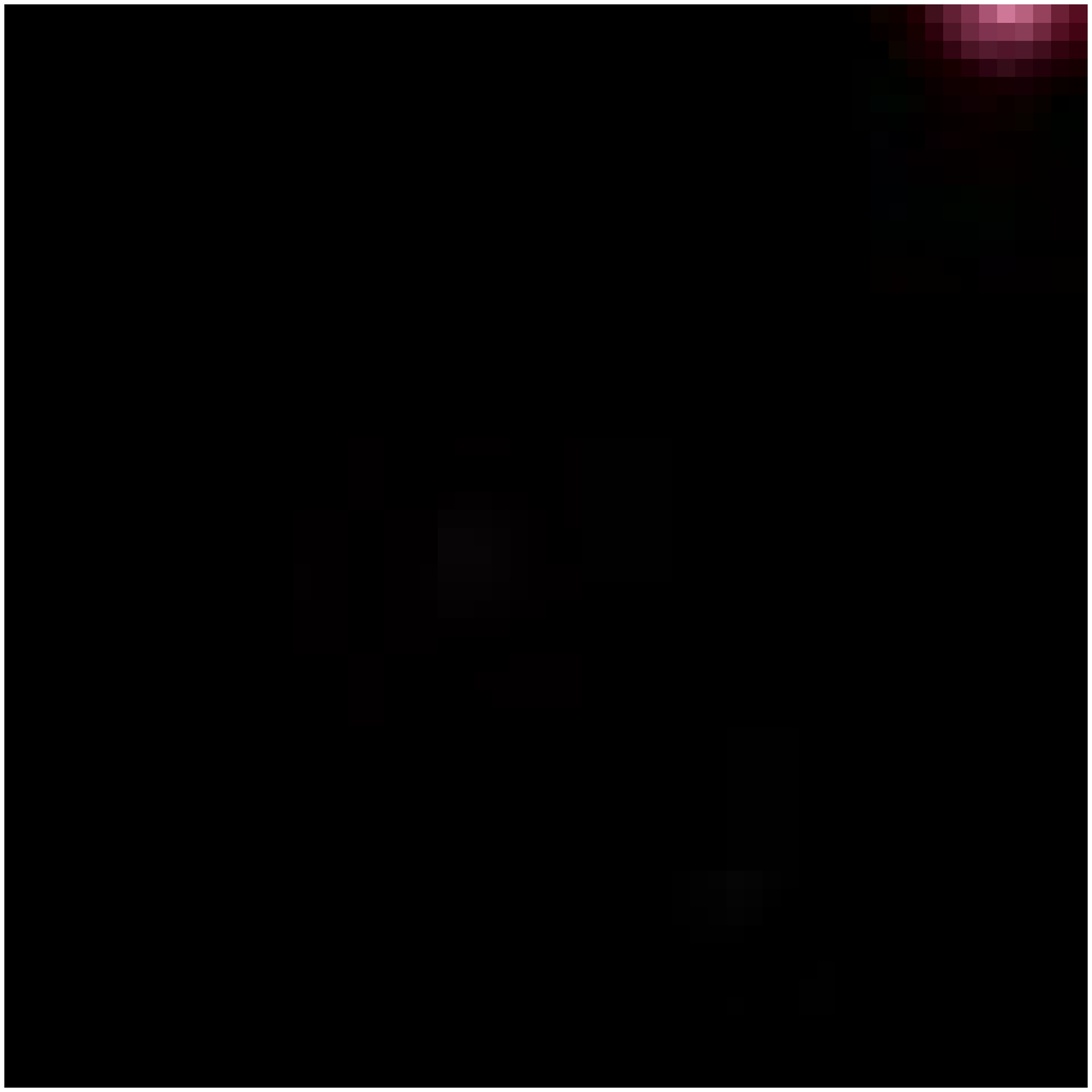}  \FGb{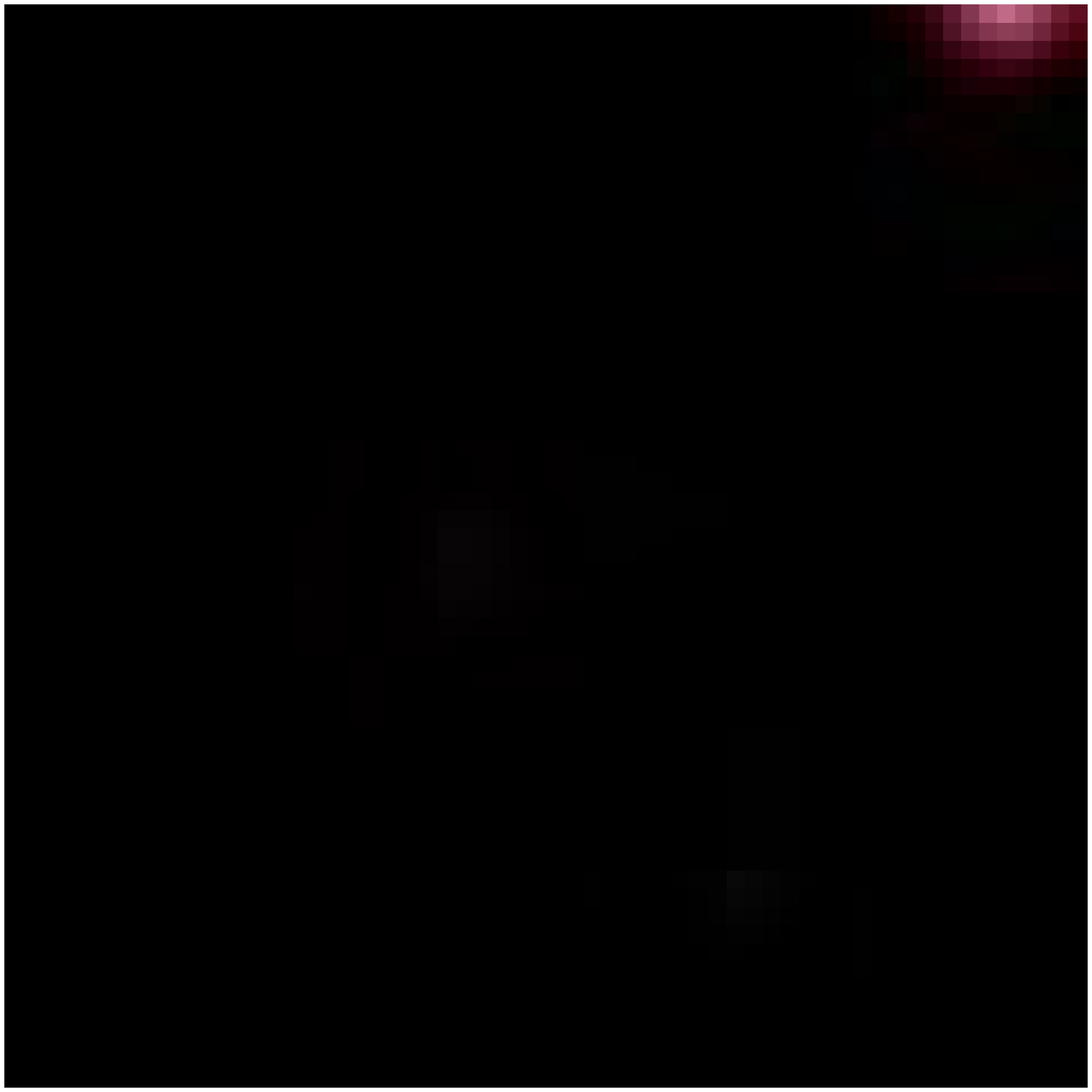}  \FGb{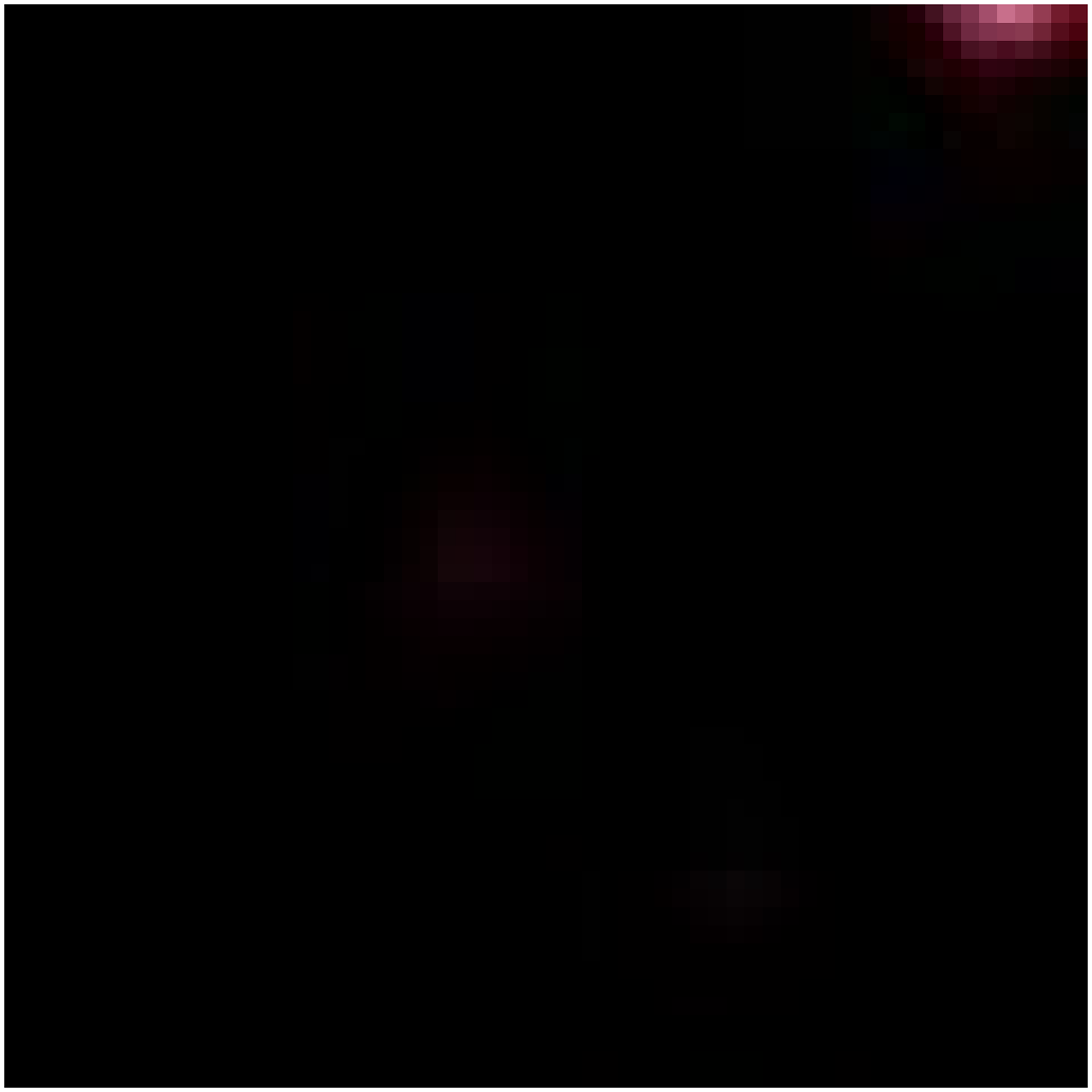}  \FGb{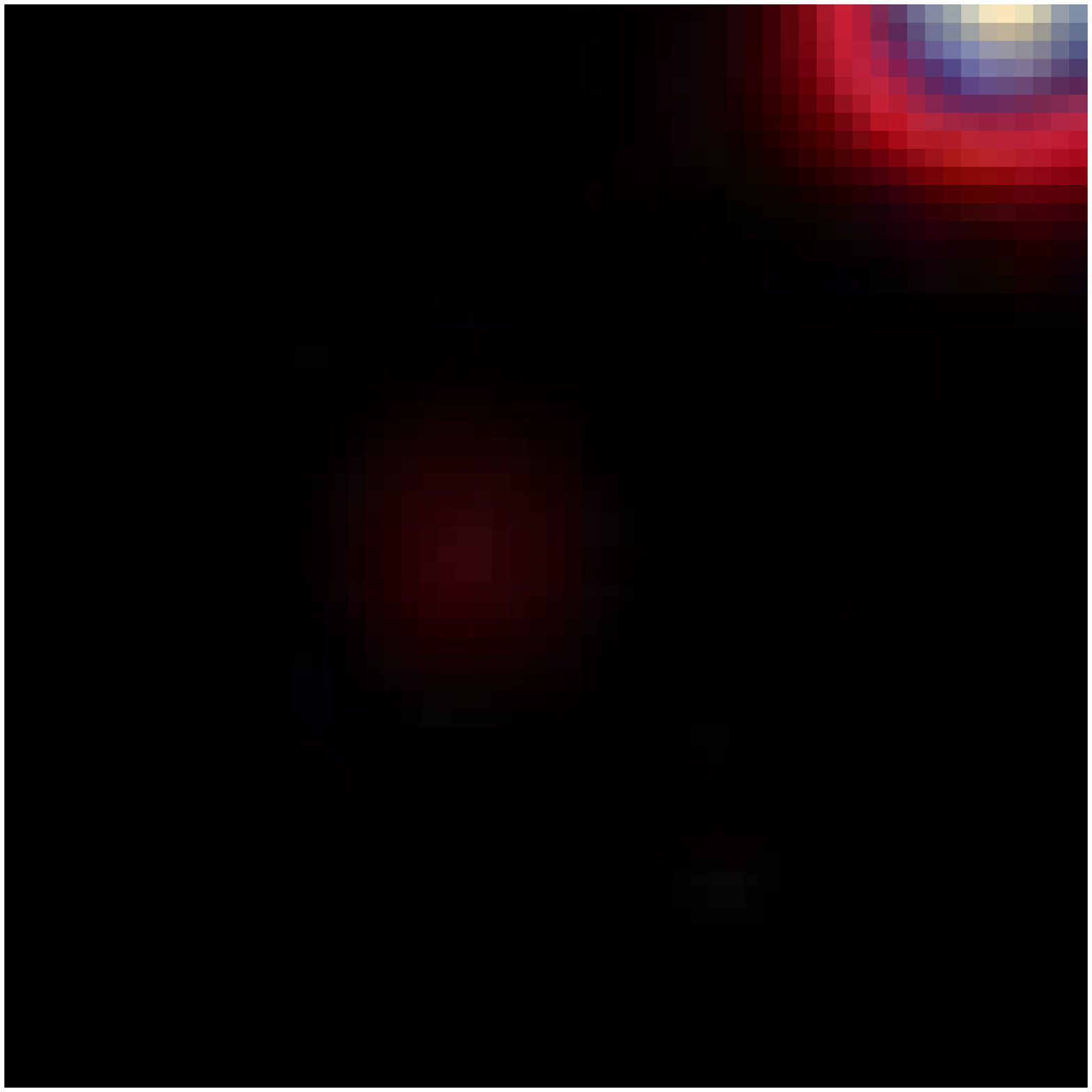}  \FGb{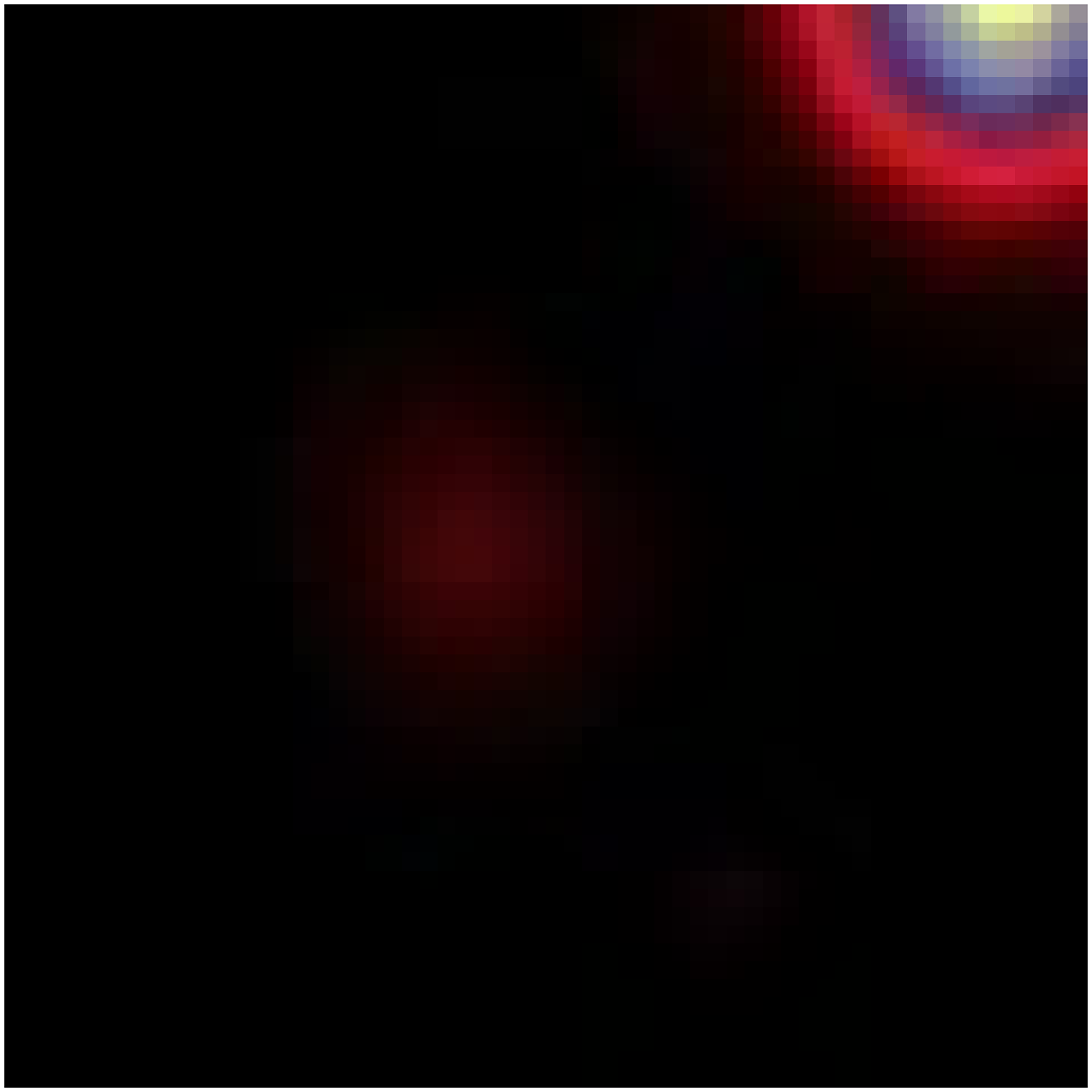}  \FGb{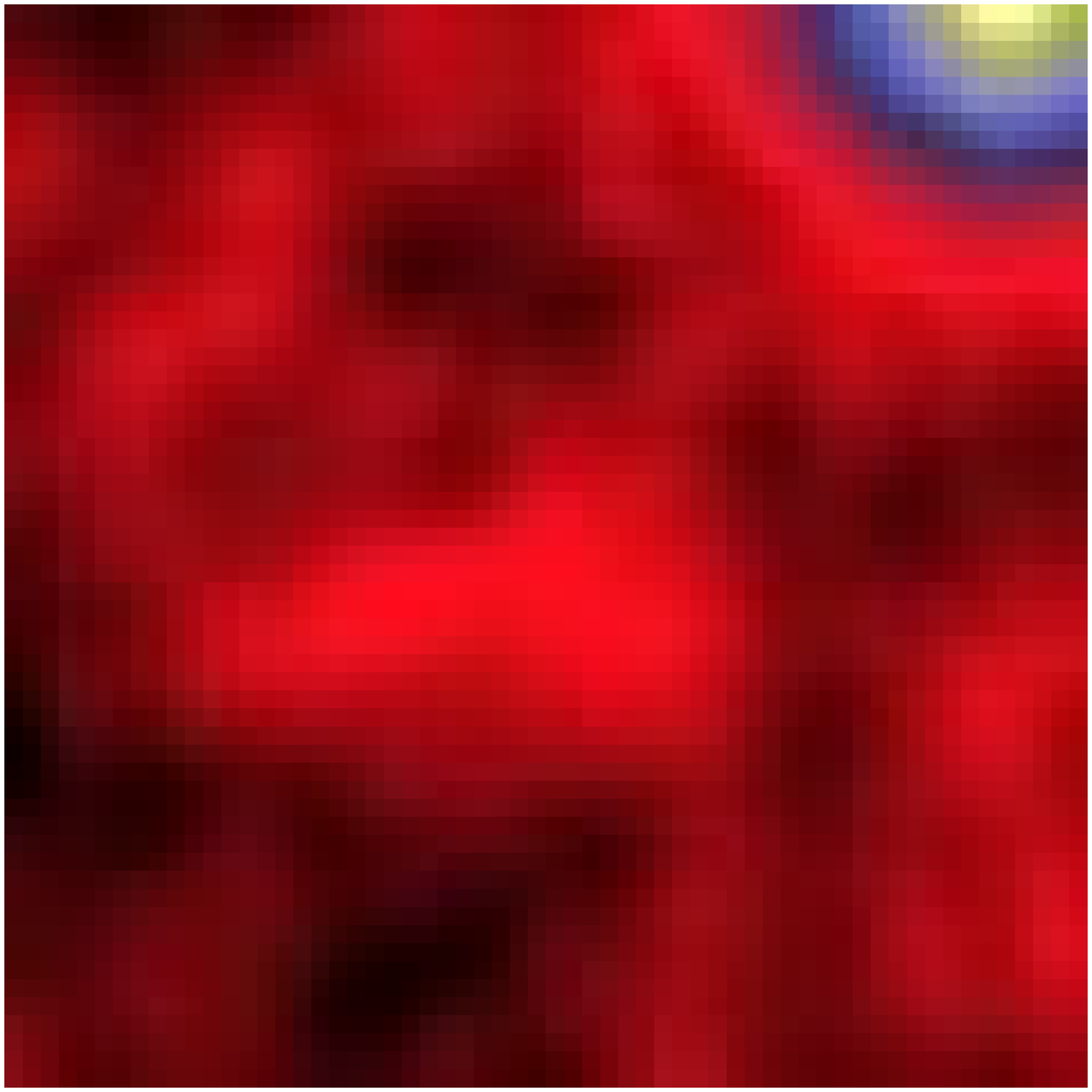}  \FGb{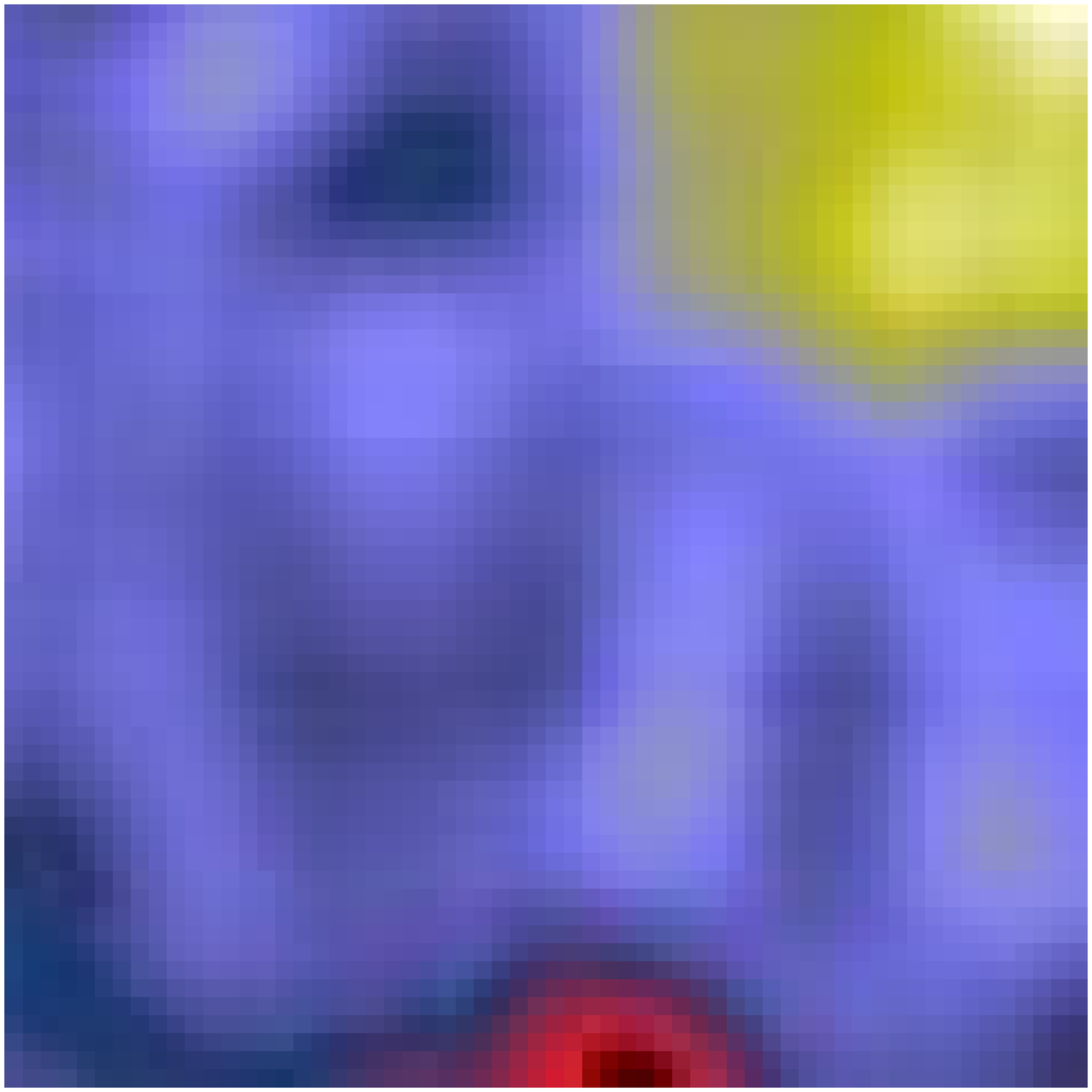} \\  {\#43   NCIA004956}  \\
  \FGb{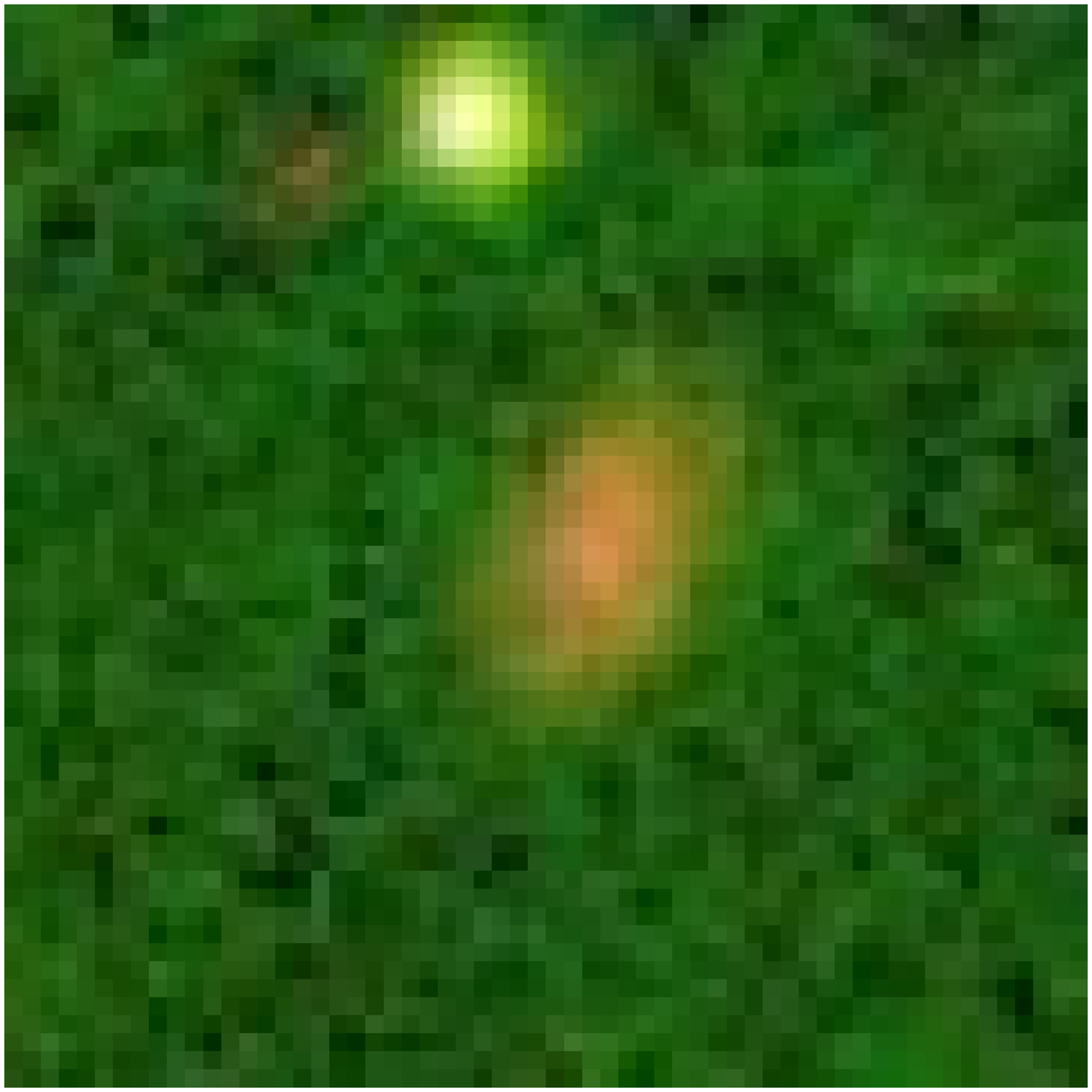}  \FGb{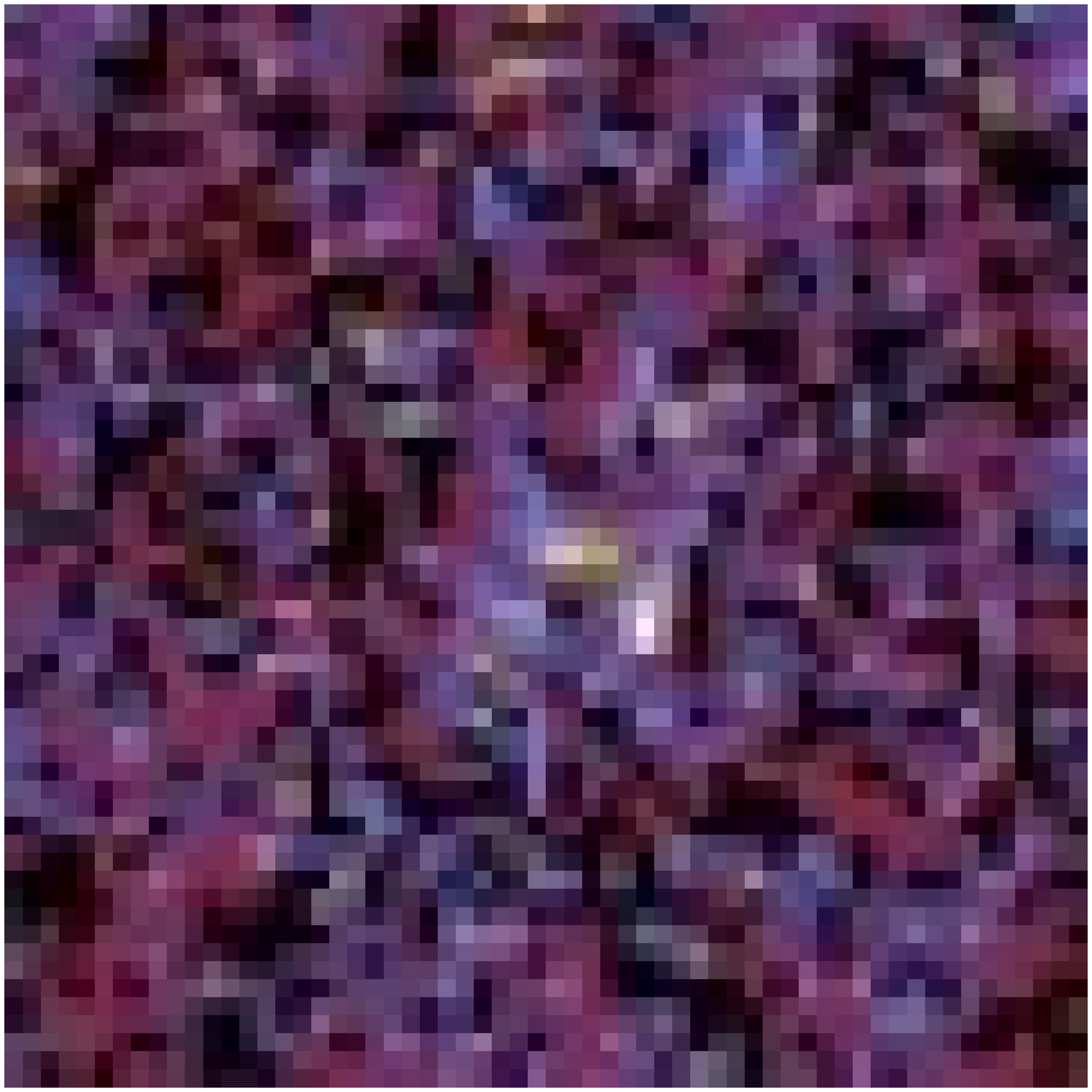}  \FGb{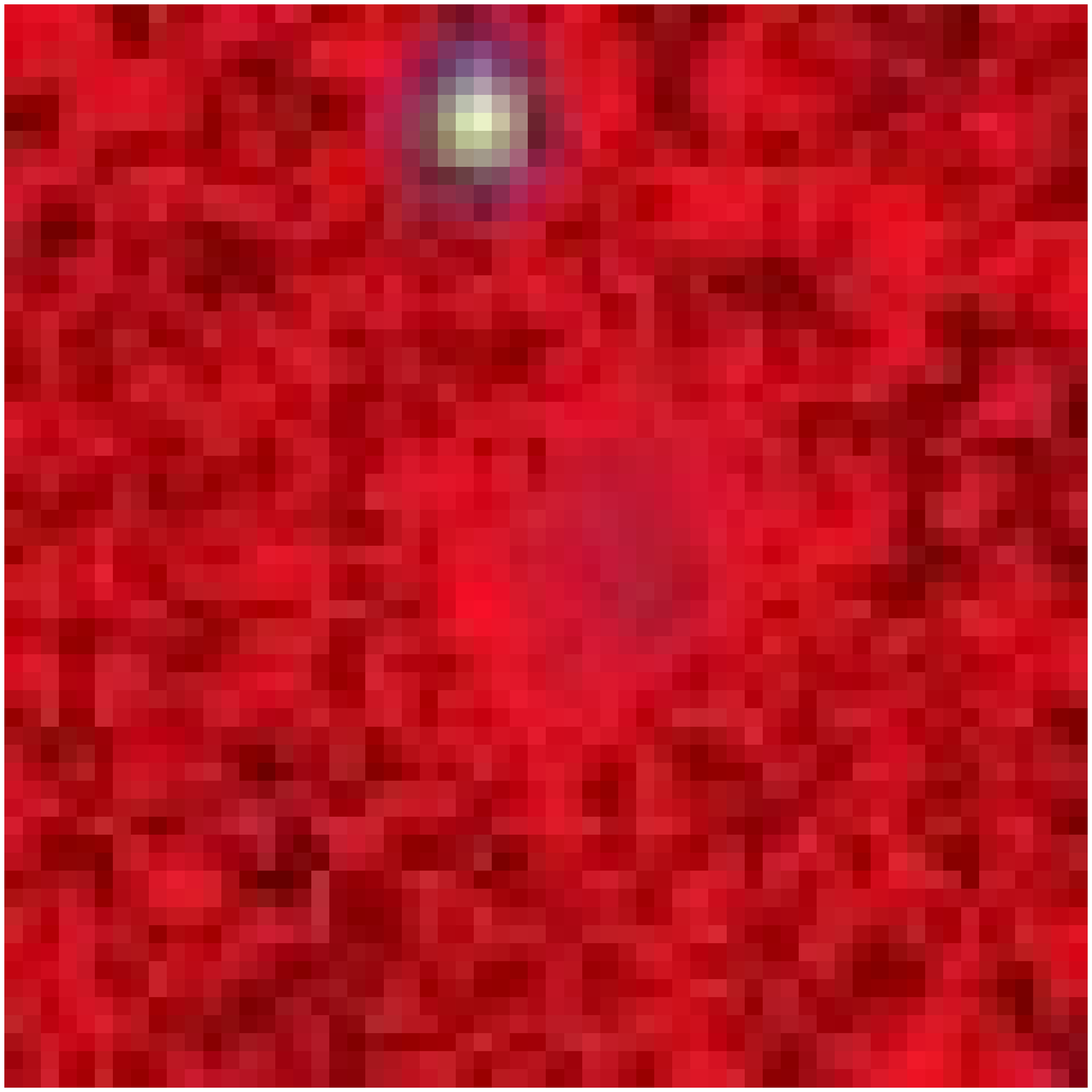}  \FGb{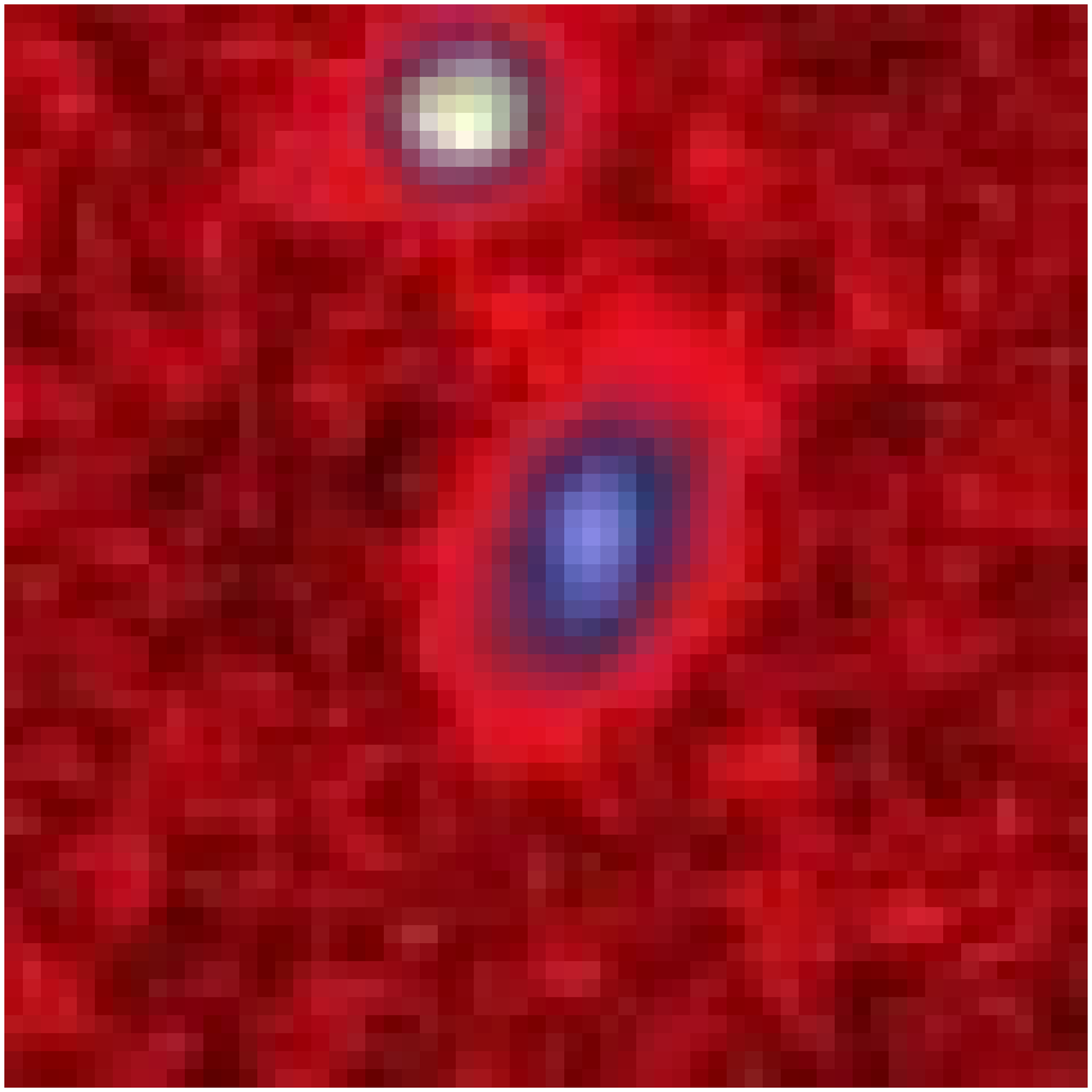}  \FGb{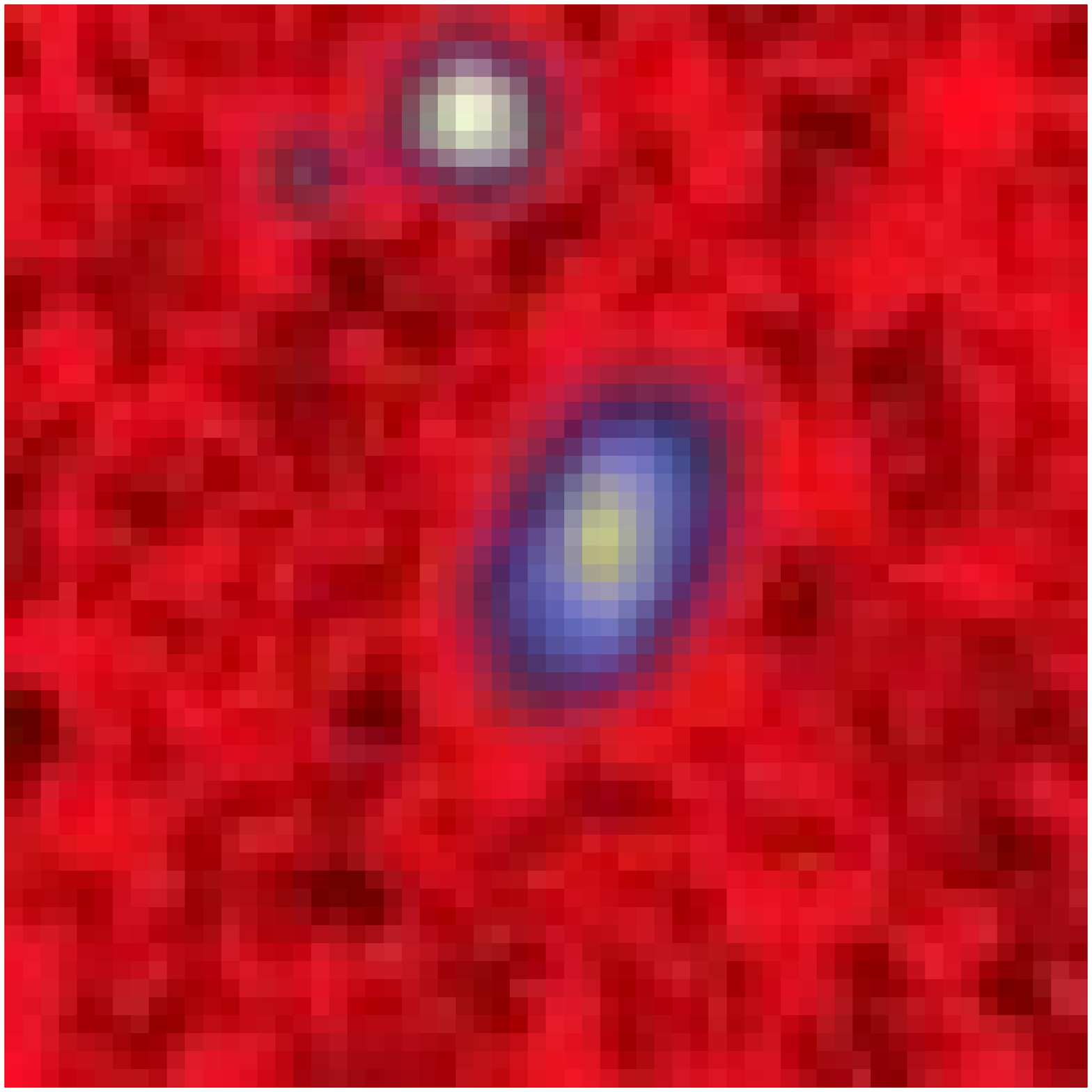}  \FGb{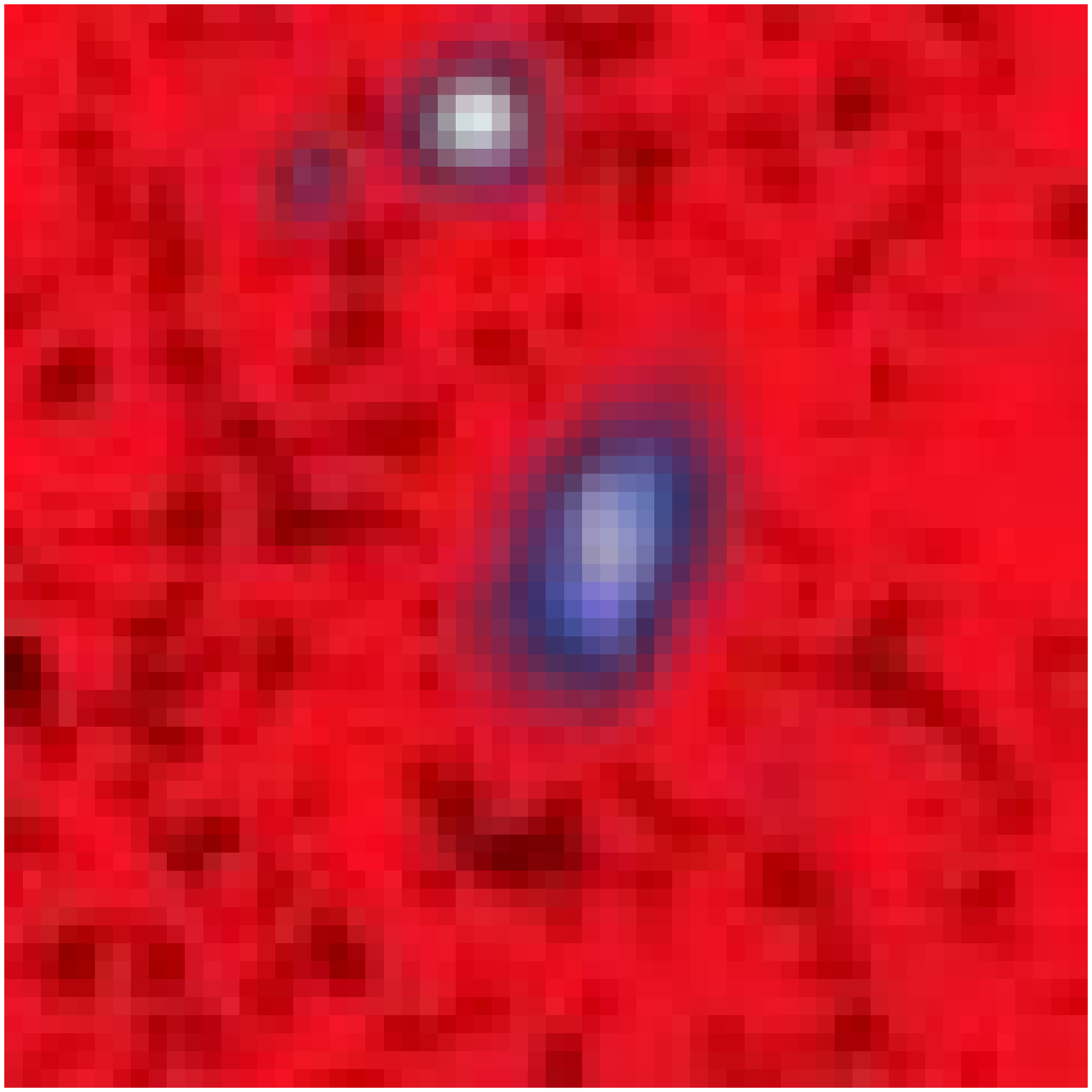}  \FGb{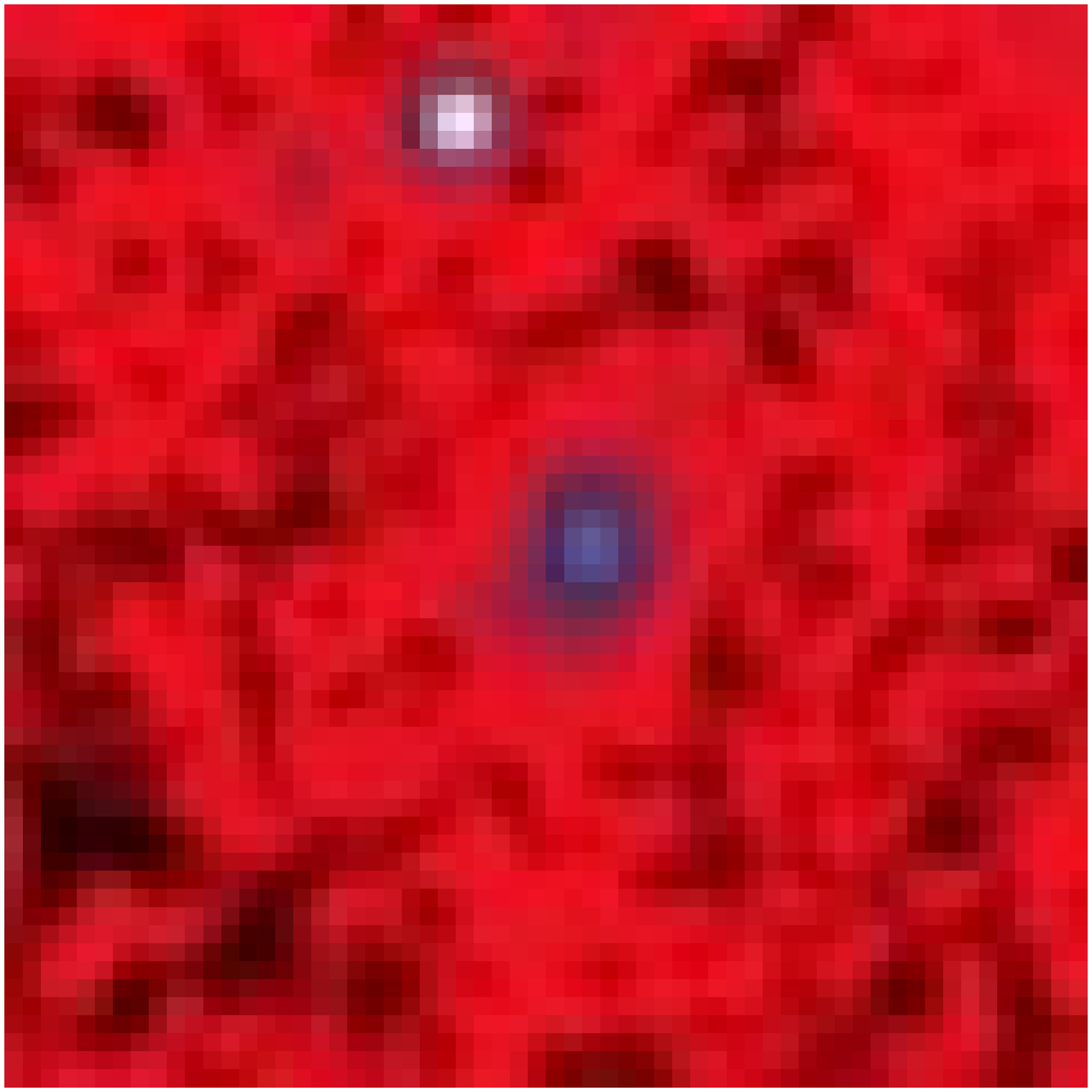}  \FGb{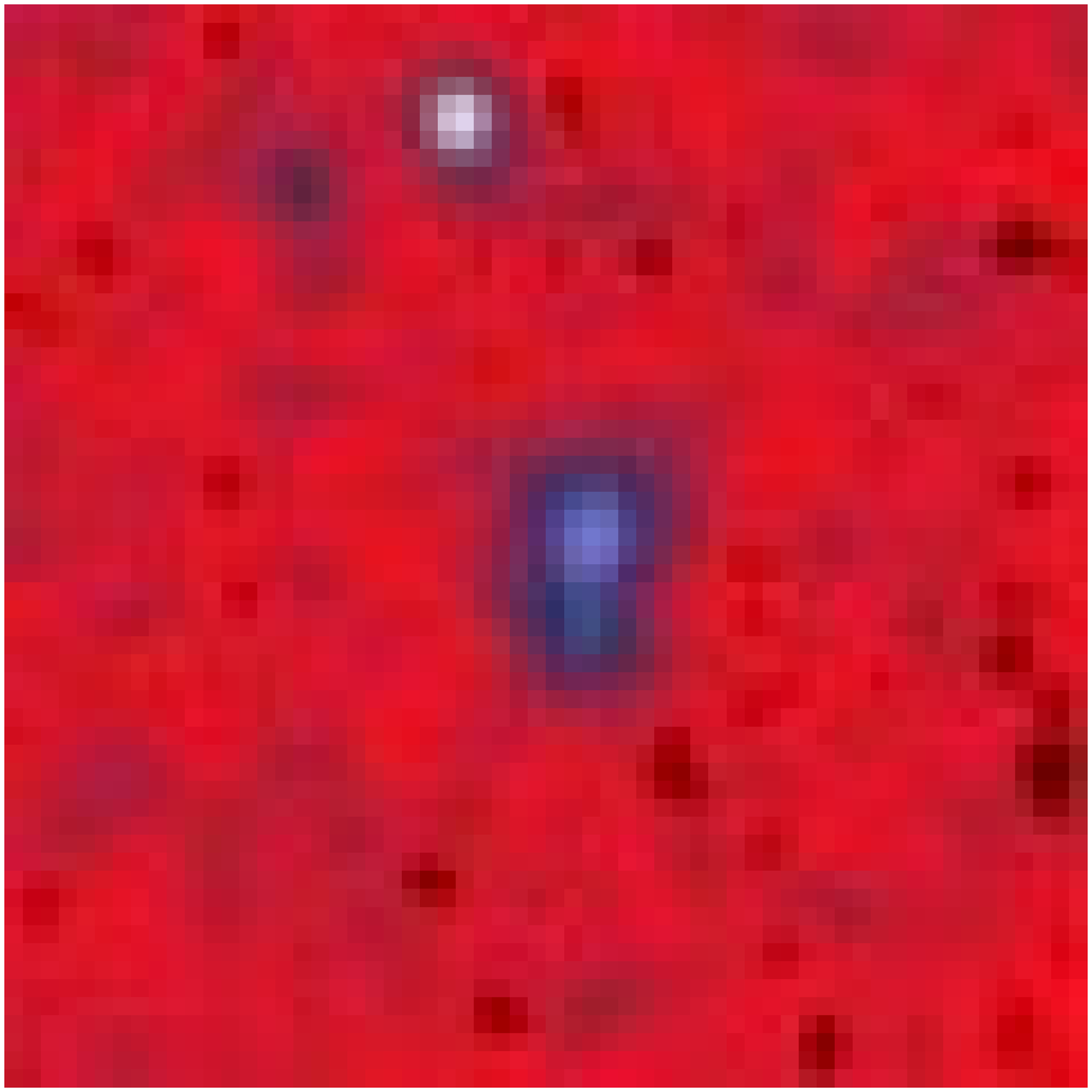}  \FGb{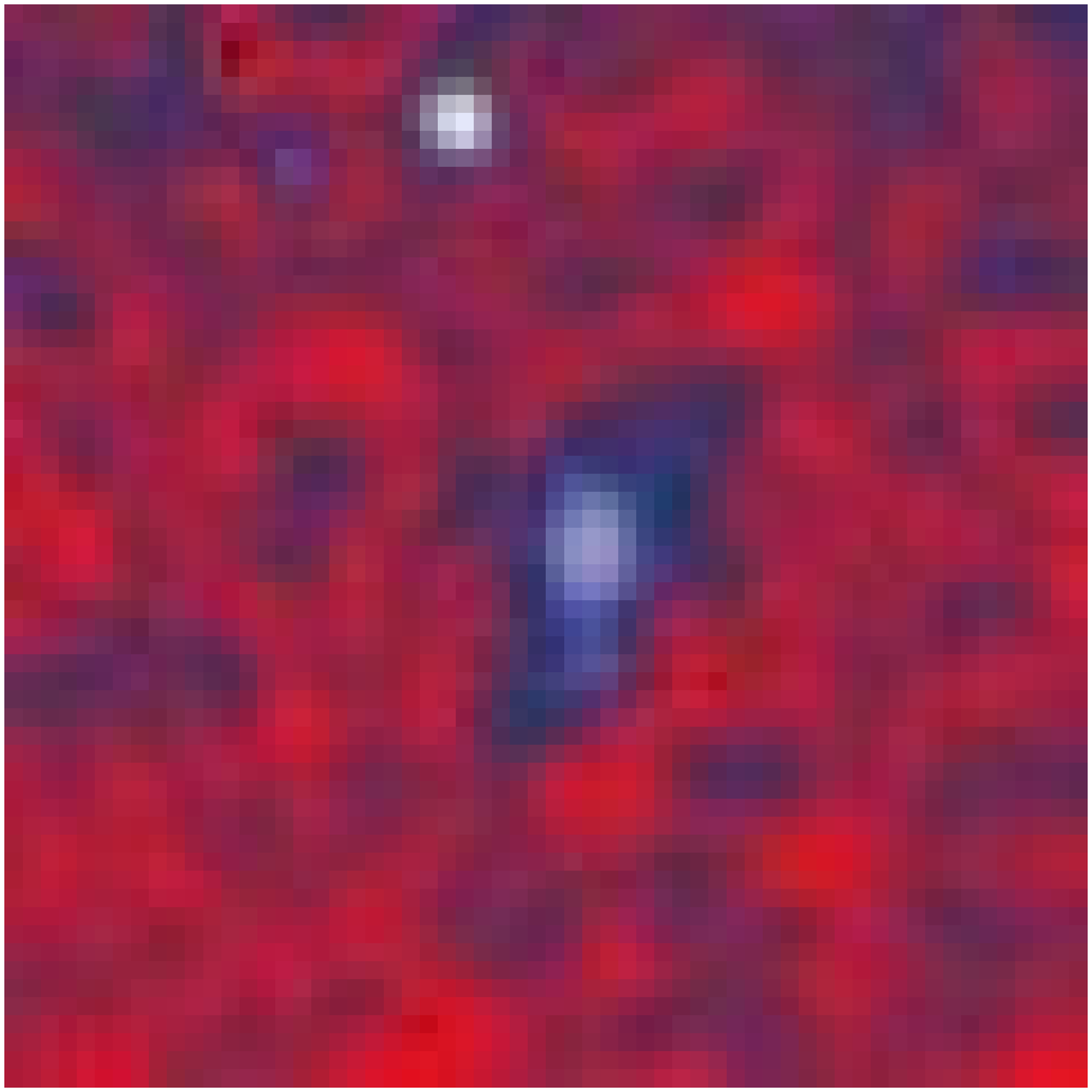}  \FGb{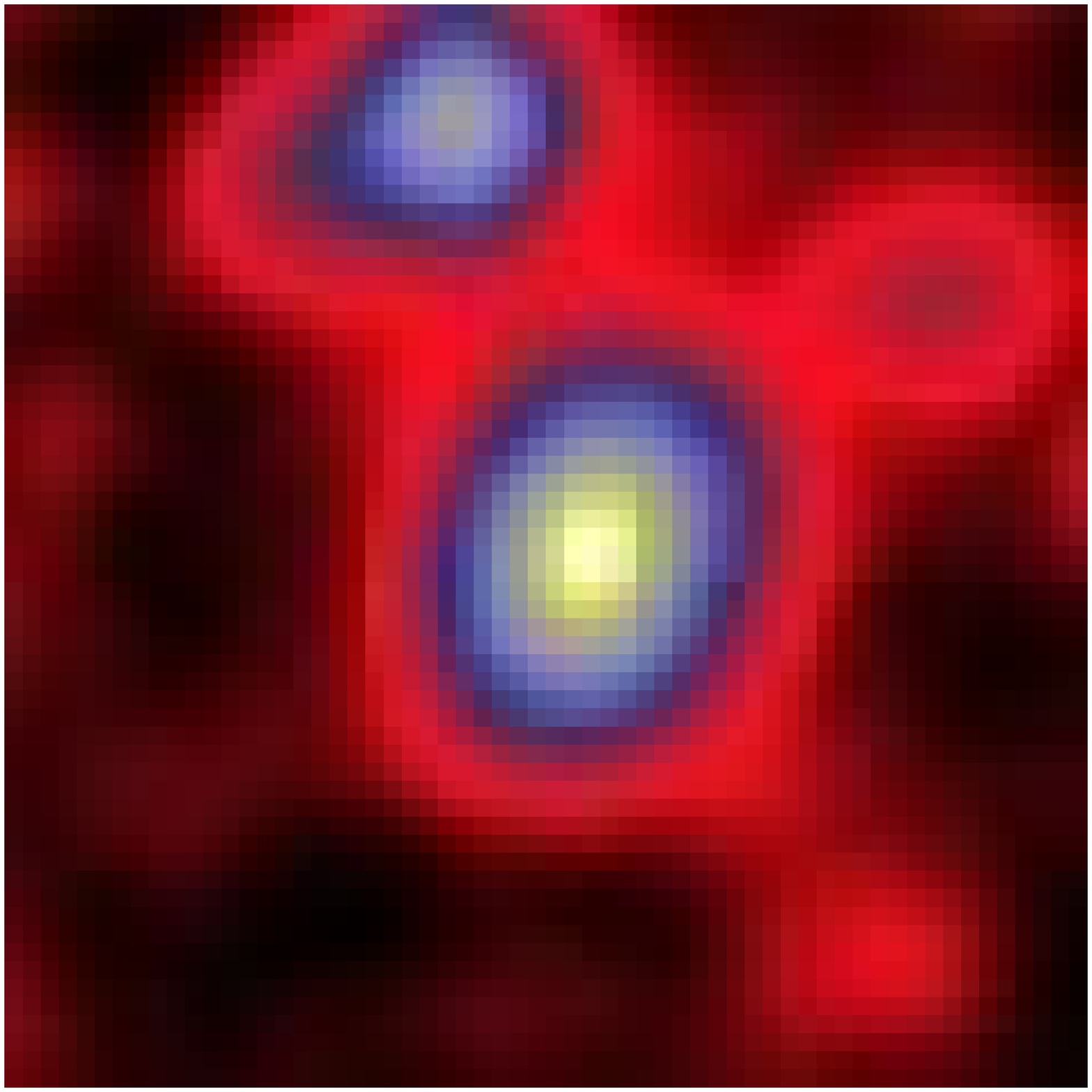}  \FGb{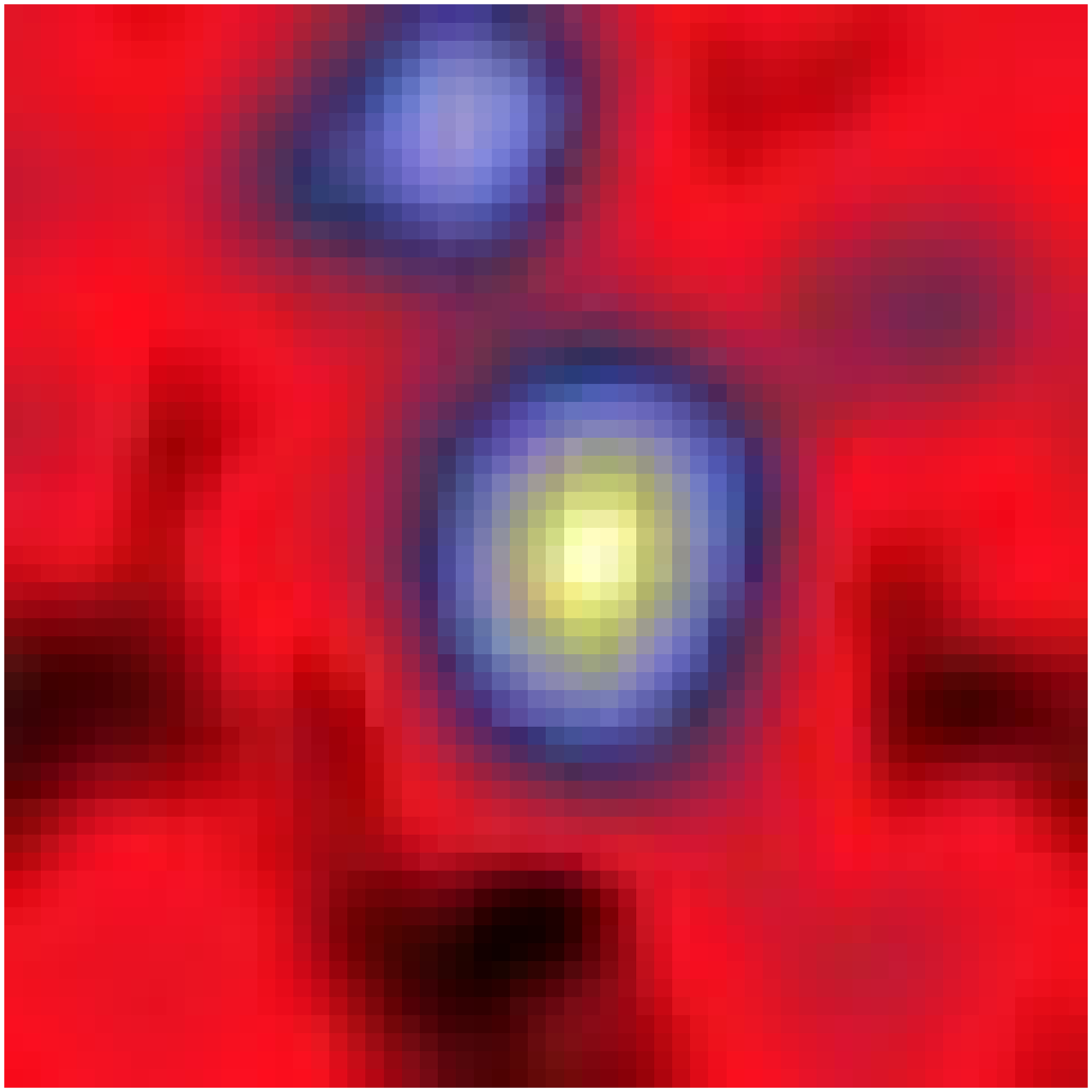}  \FGb{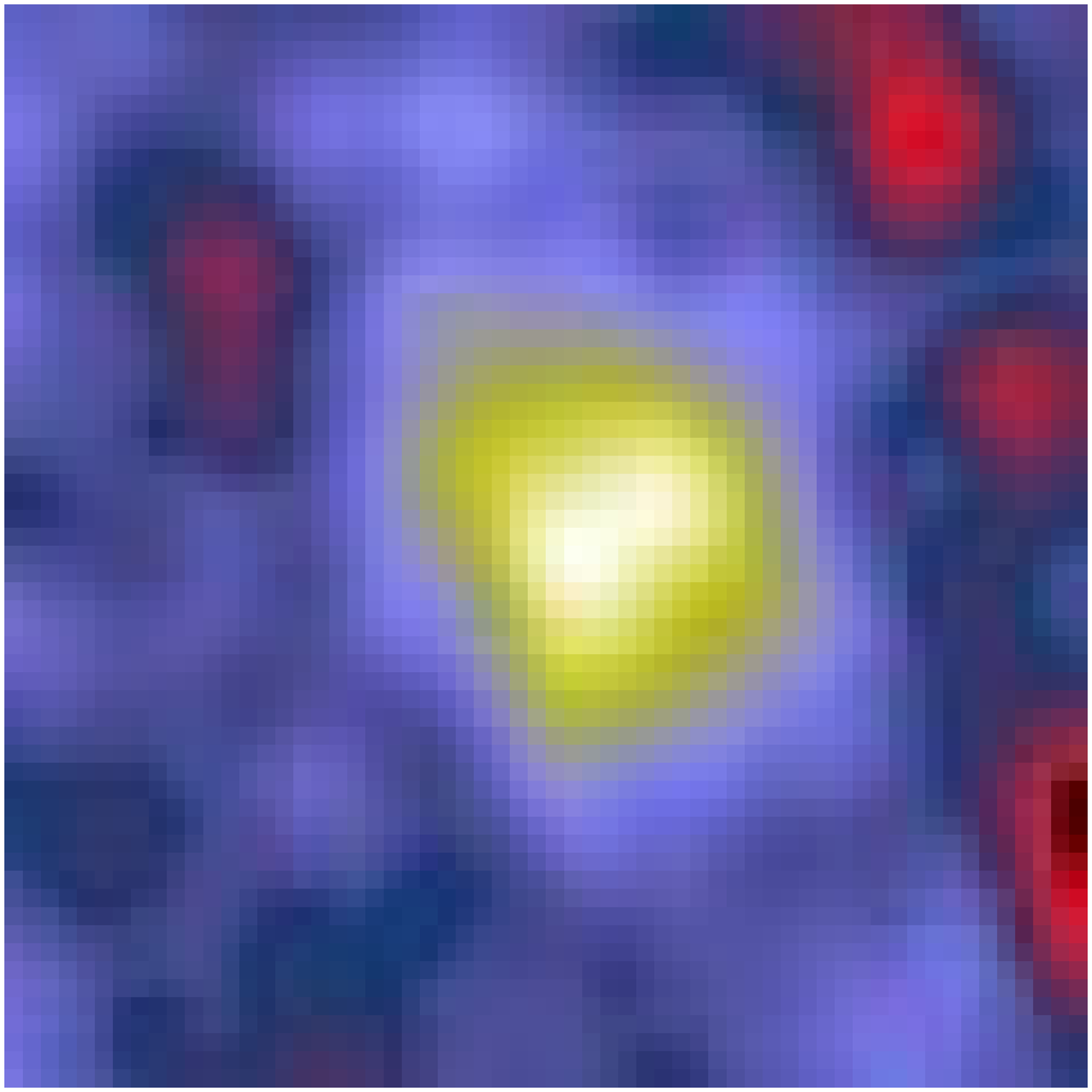}  \FGb{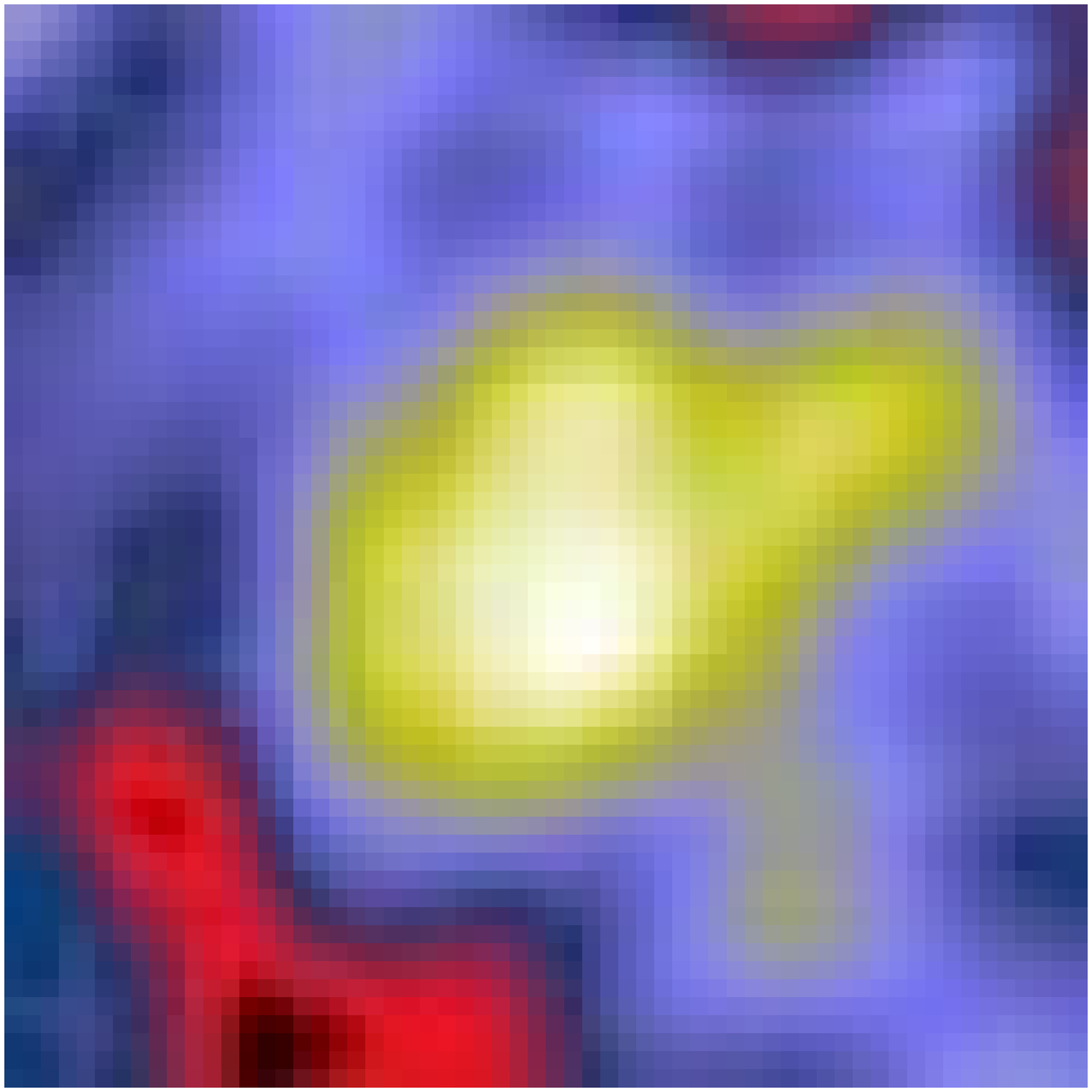} \\  {\#44   NCIA006145}  \\
  \FGb{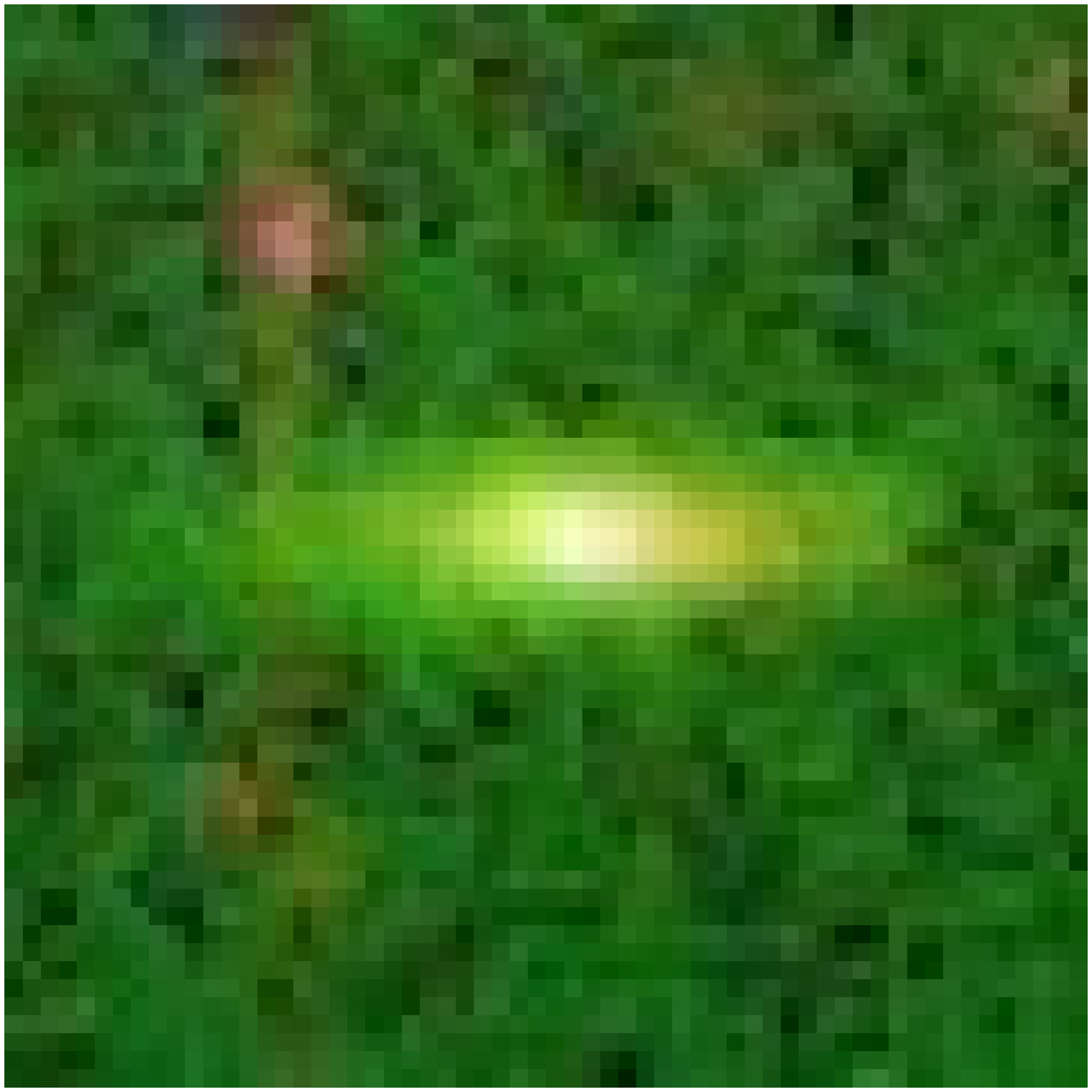}  \FGb{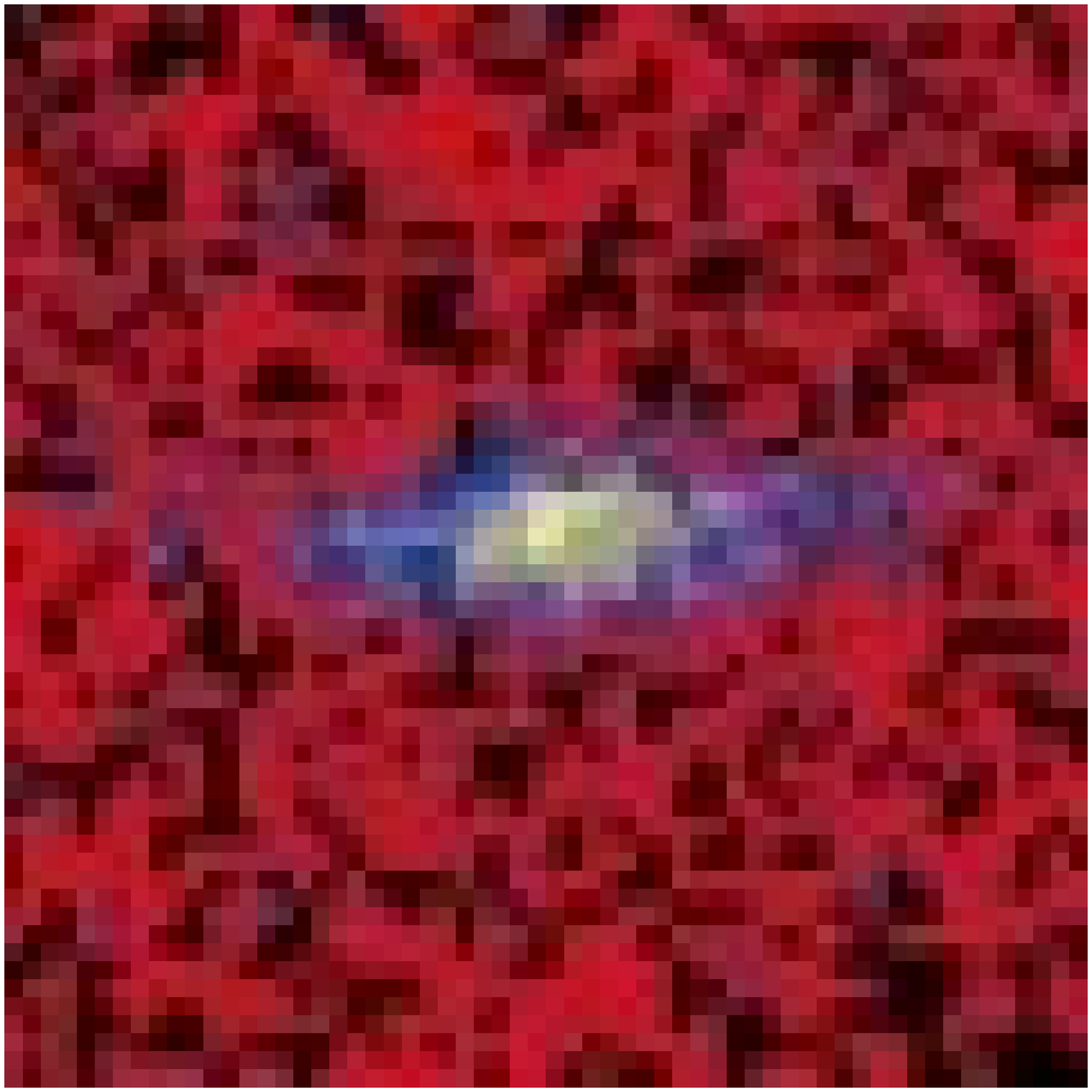}  \FGb{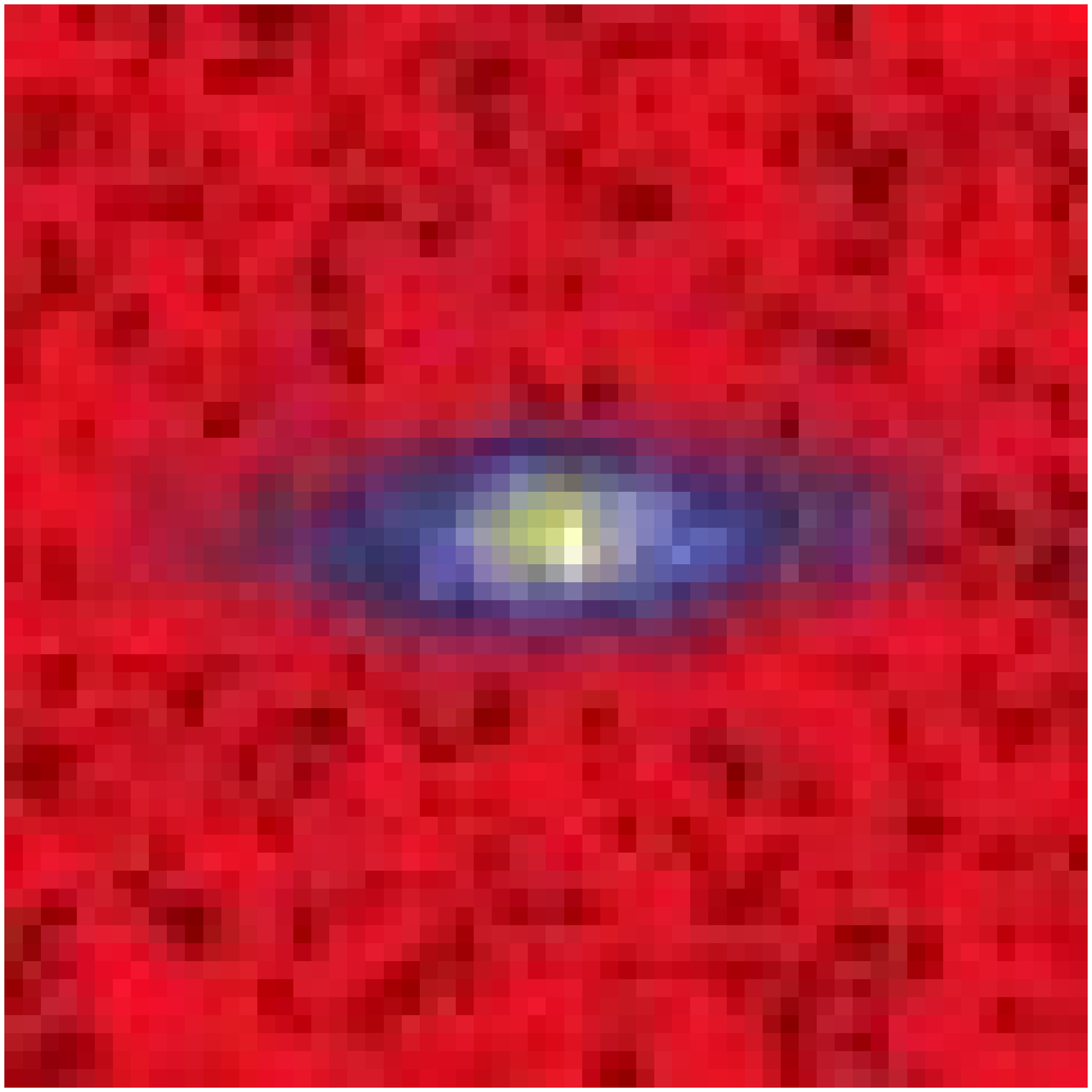}  \FGb{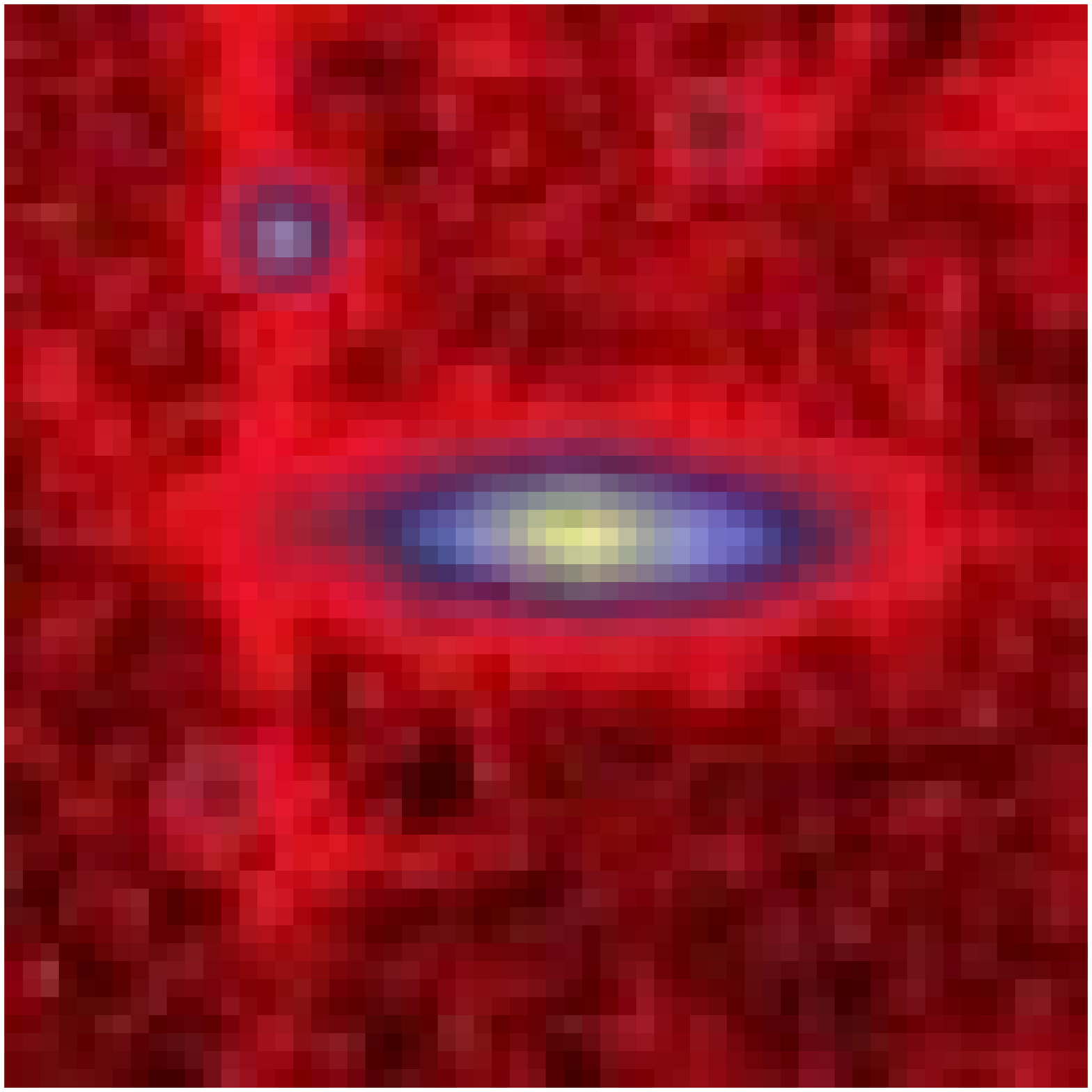}  \FGb{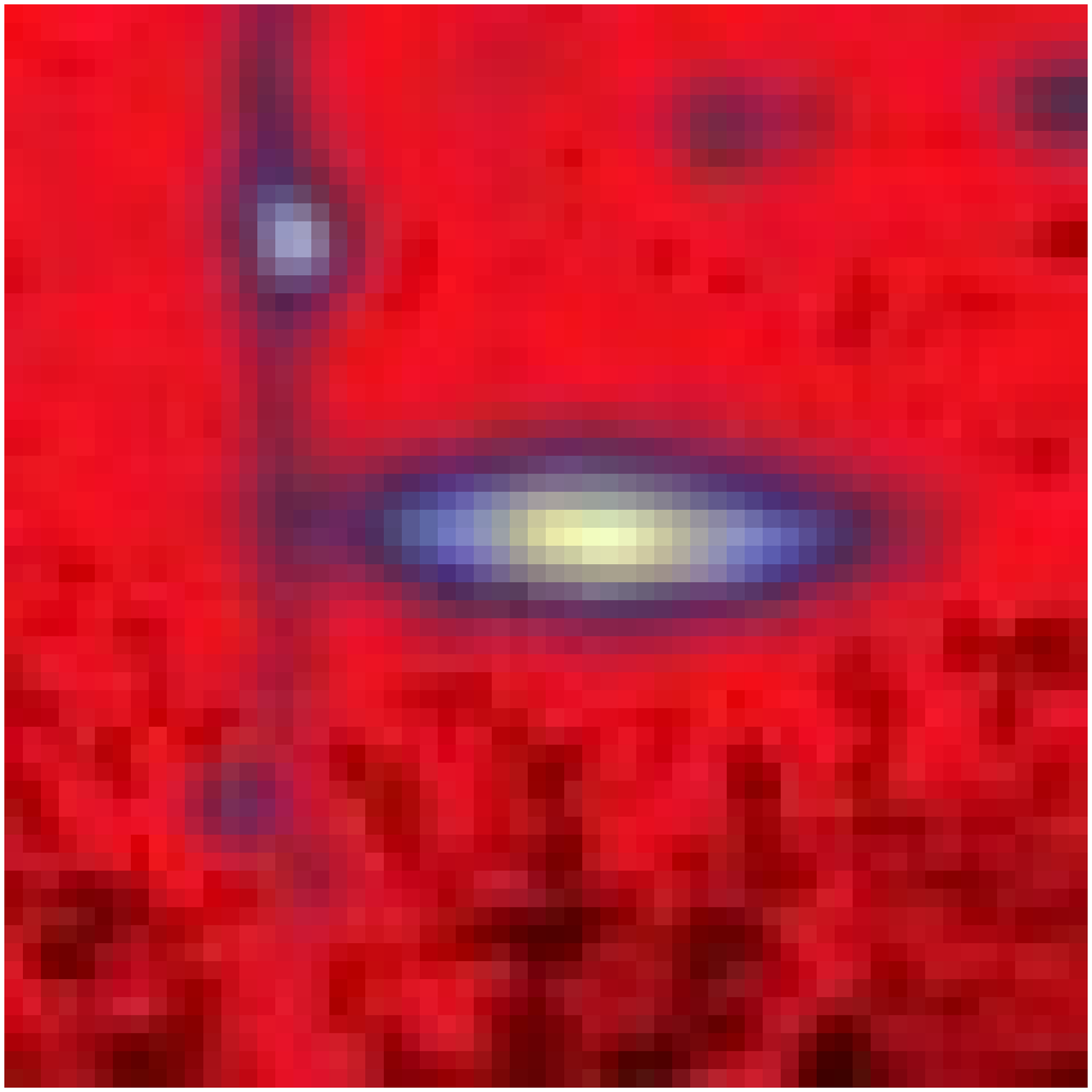}  \FGb{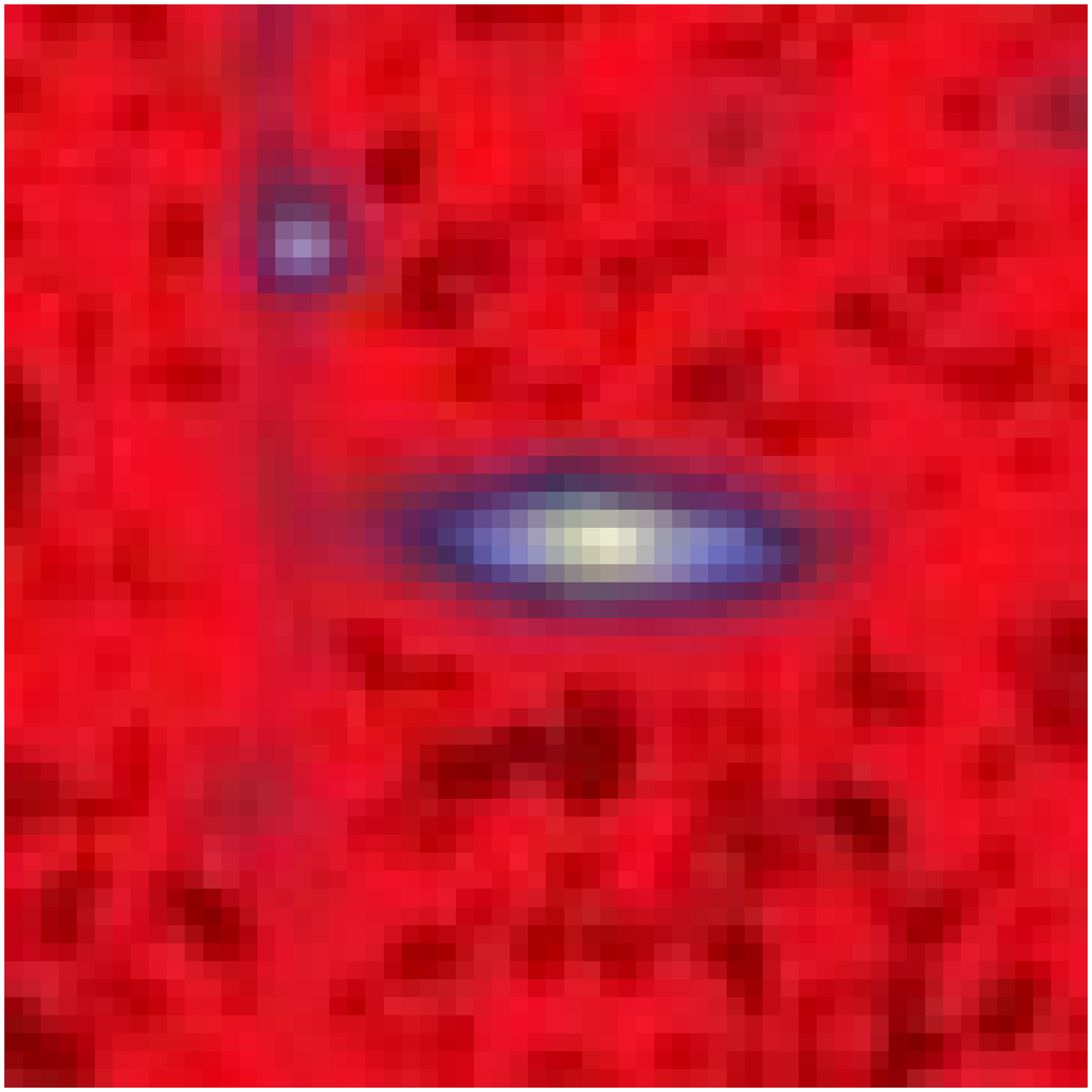}  \FGb{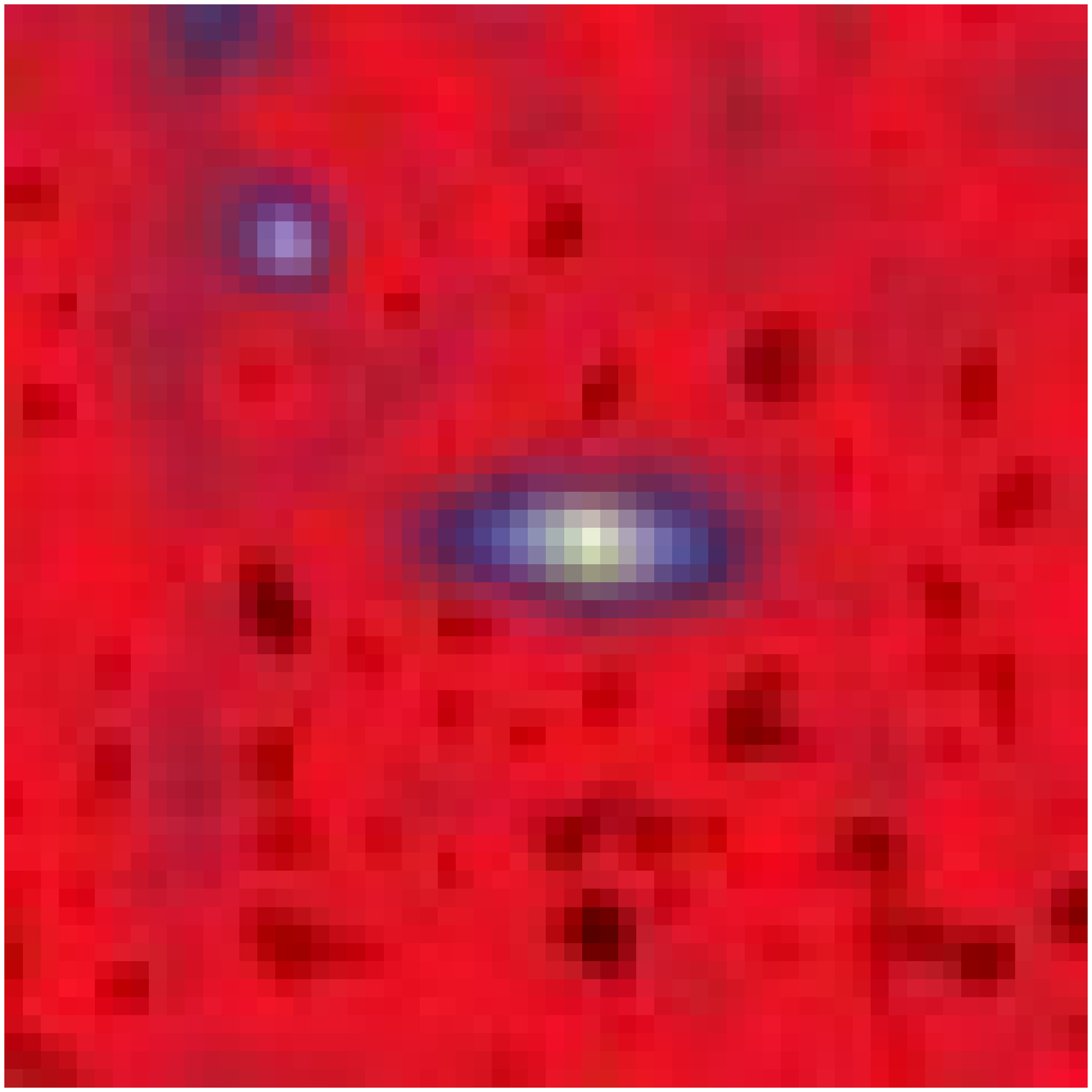}  \FGb{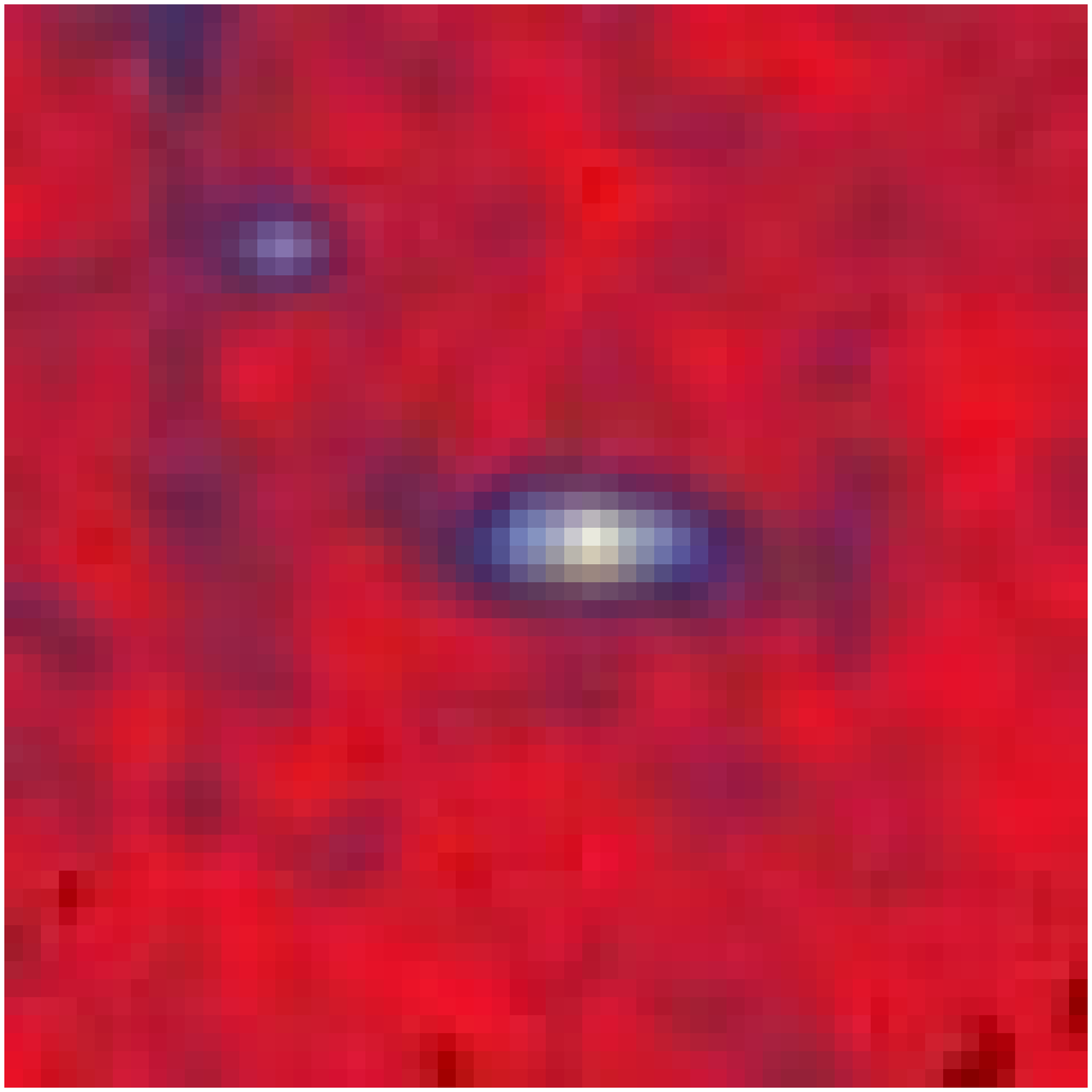}  \FGb{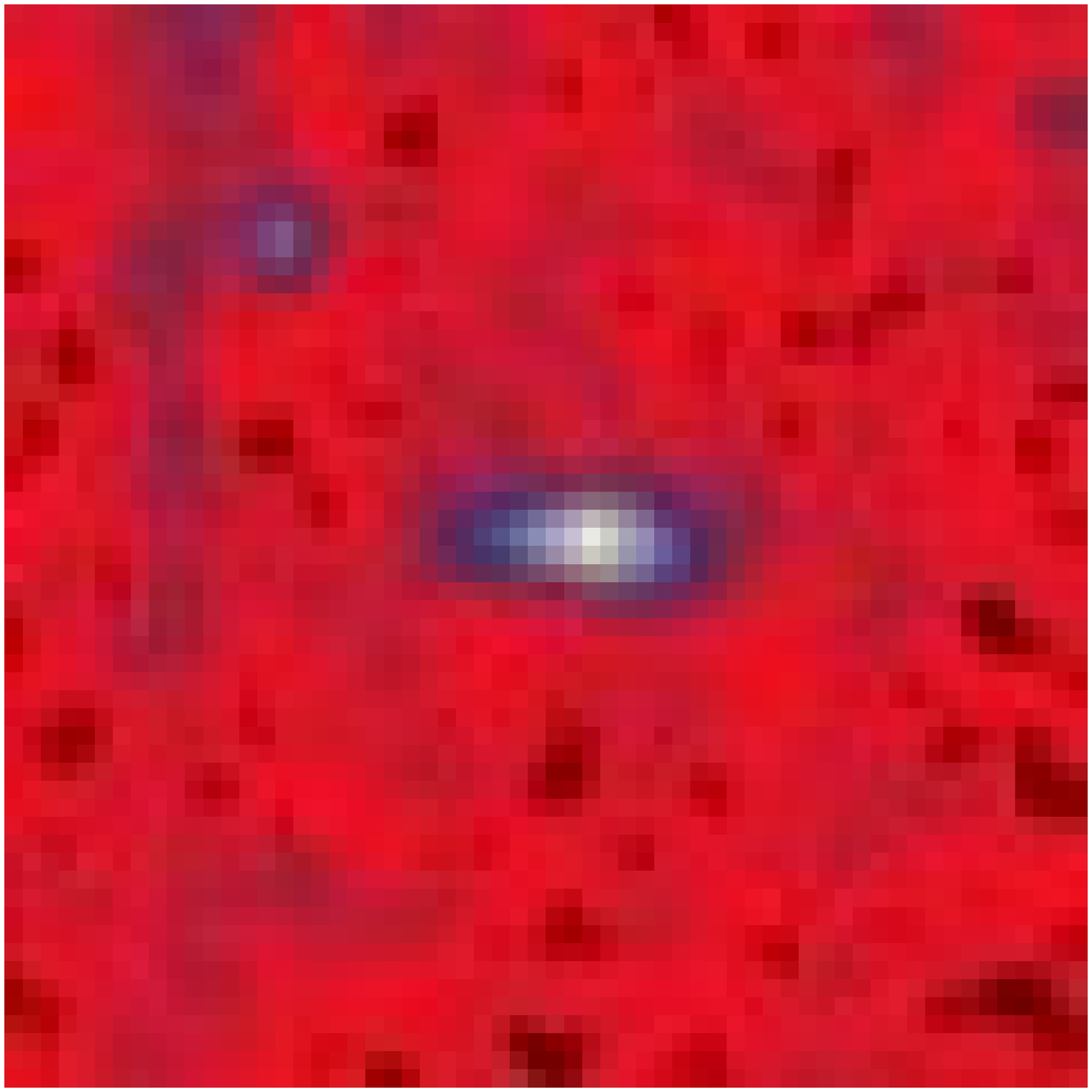}  \FGb{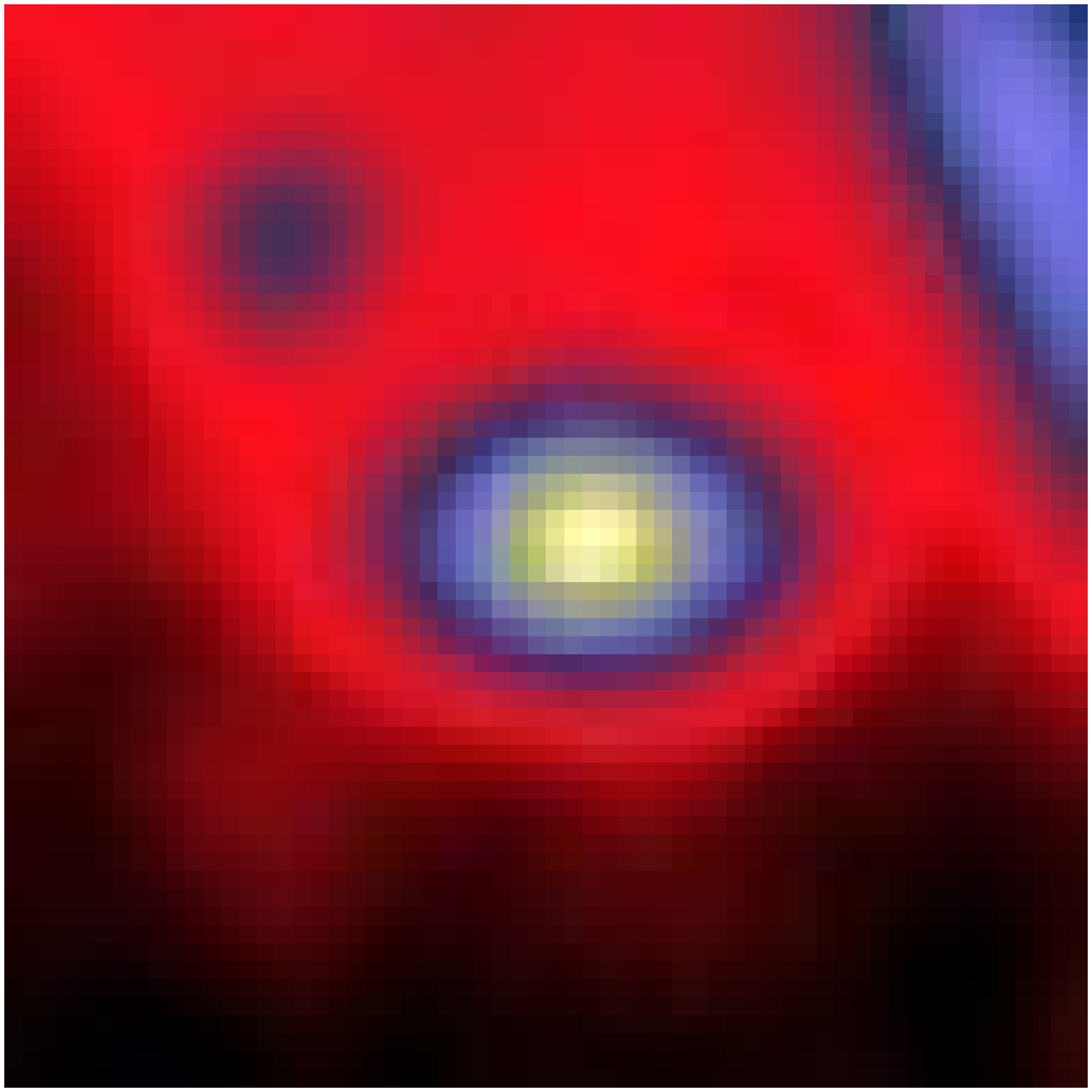}  \FGb{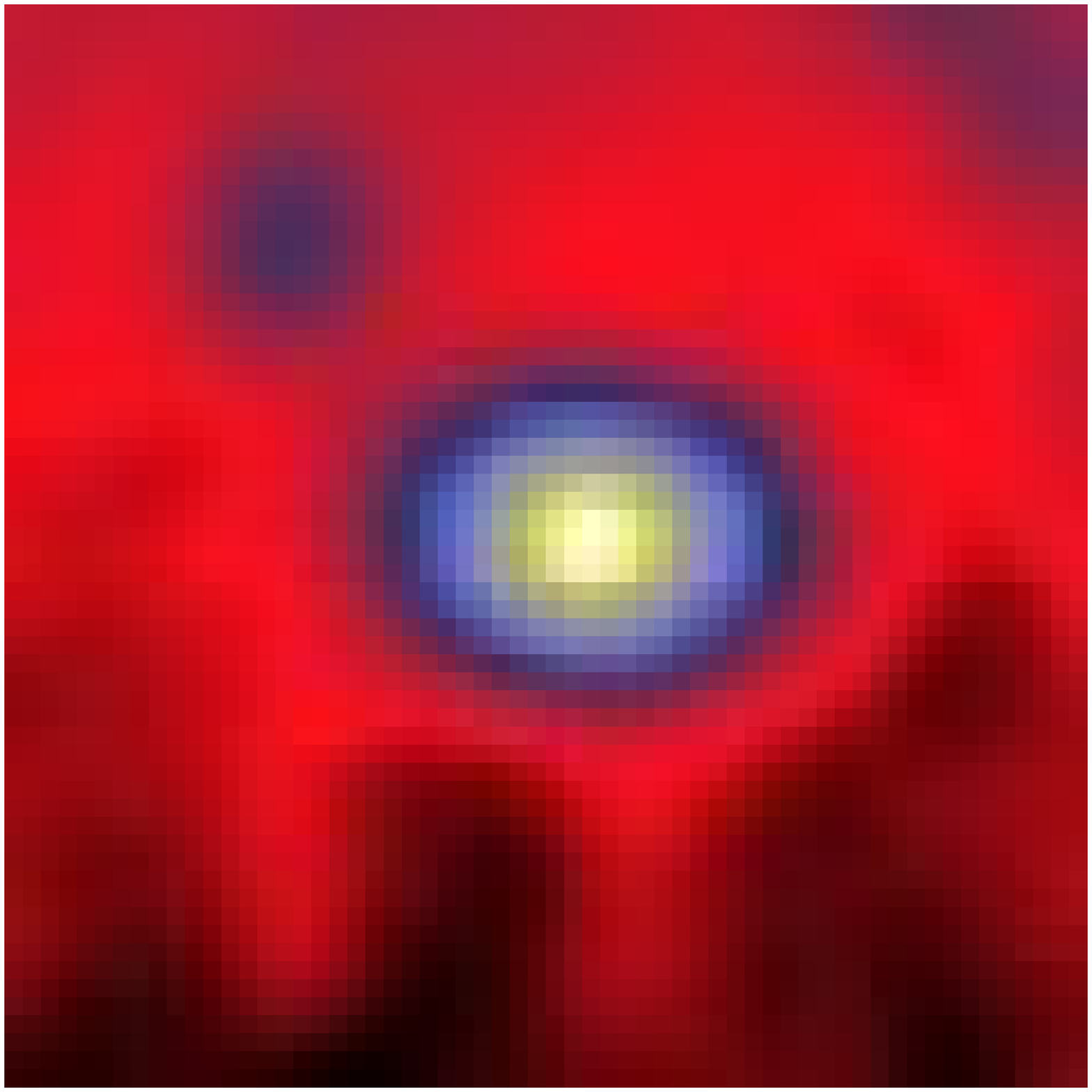}  \FGb{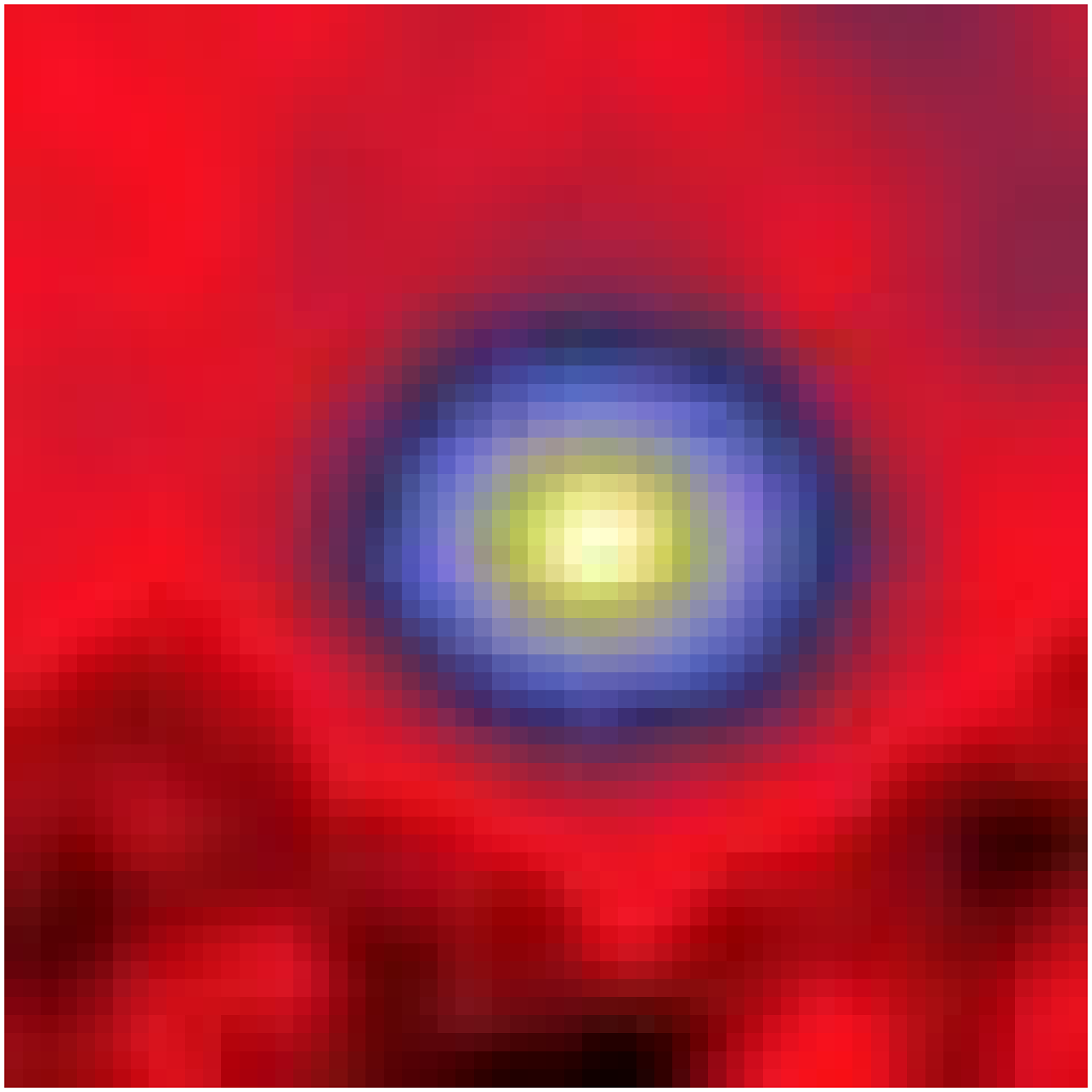}  \FGb{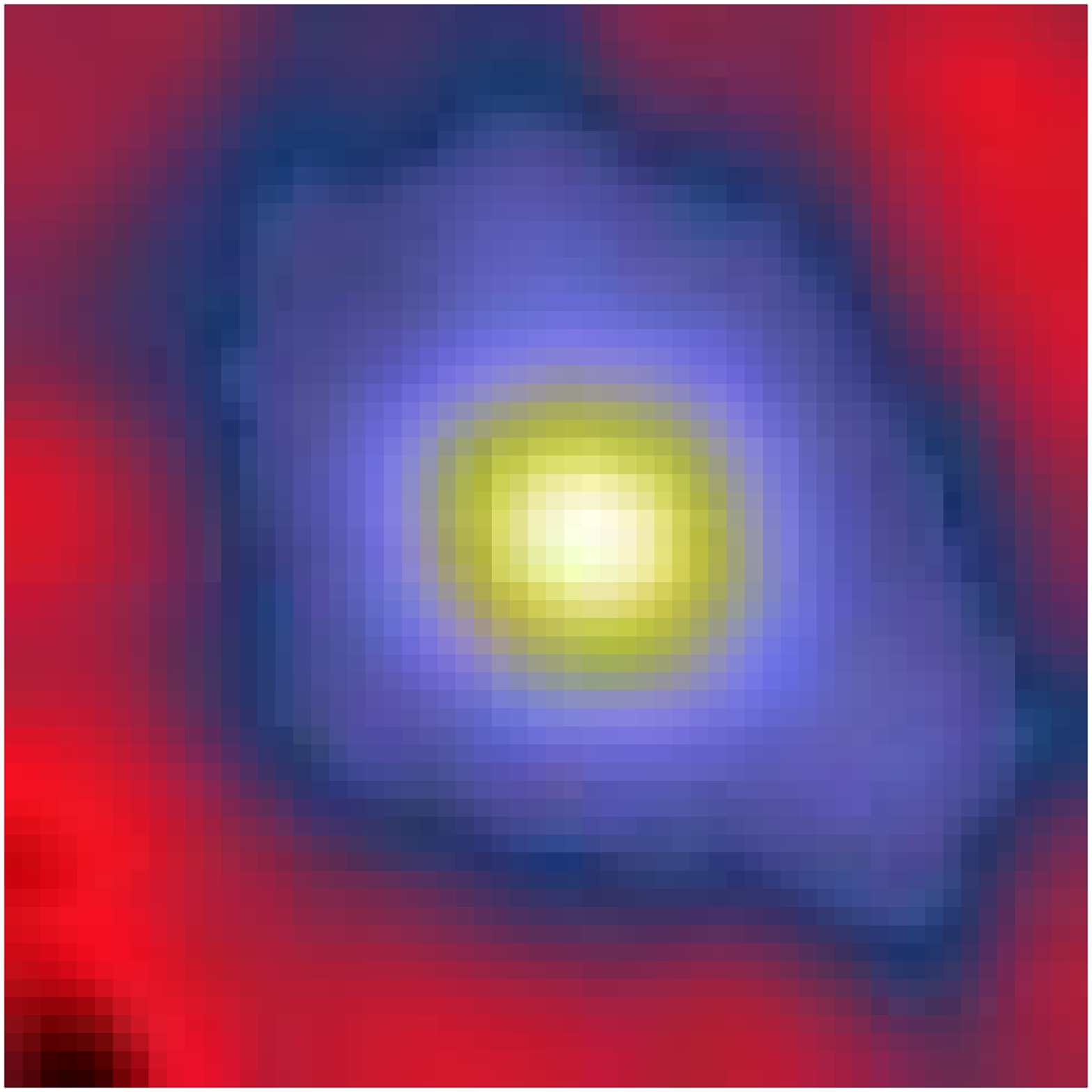} \\  {\#45   NCIA001071}  \\
  \FGb{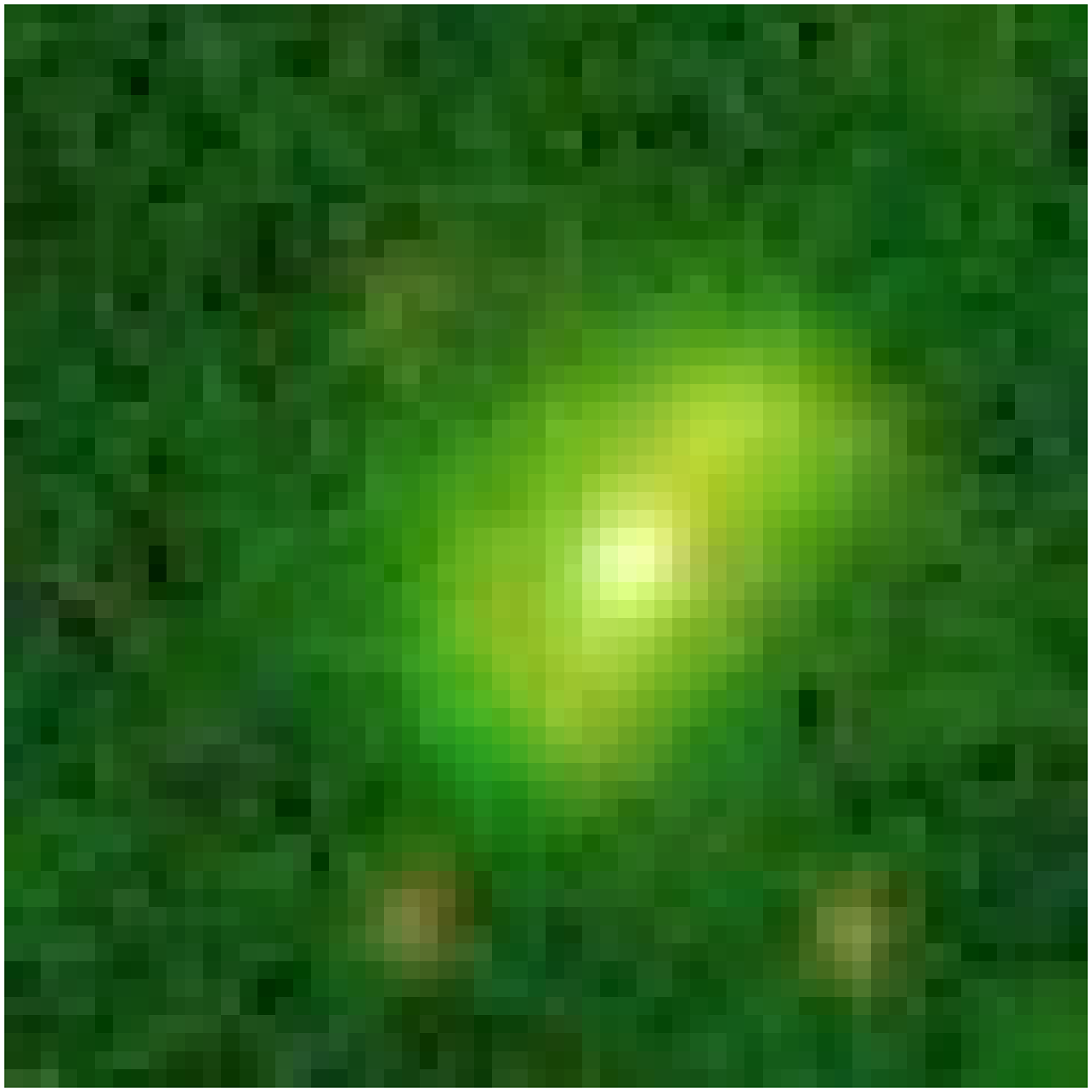}  \FGb{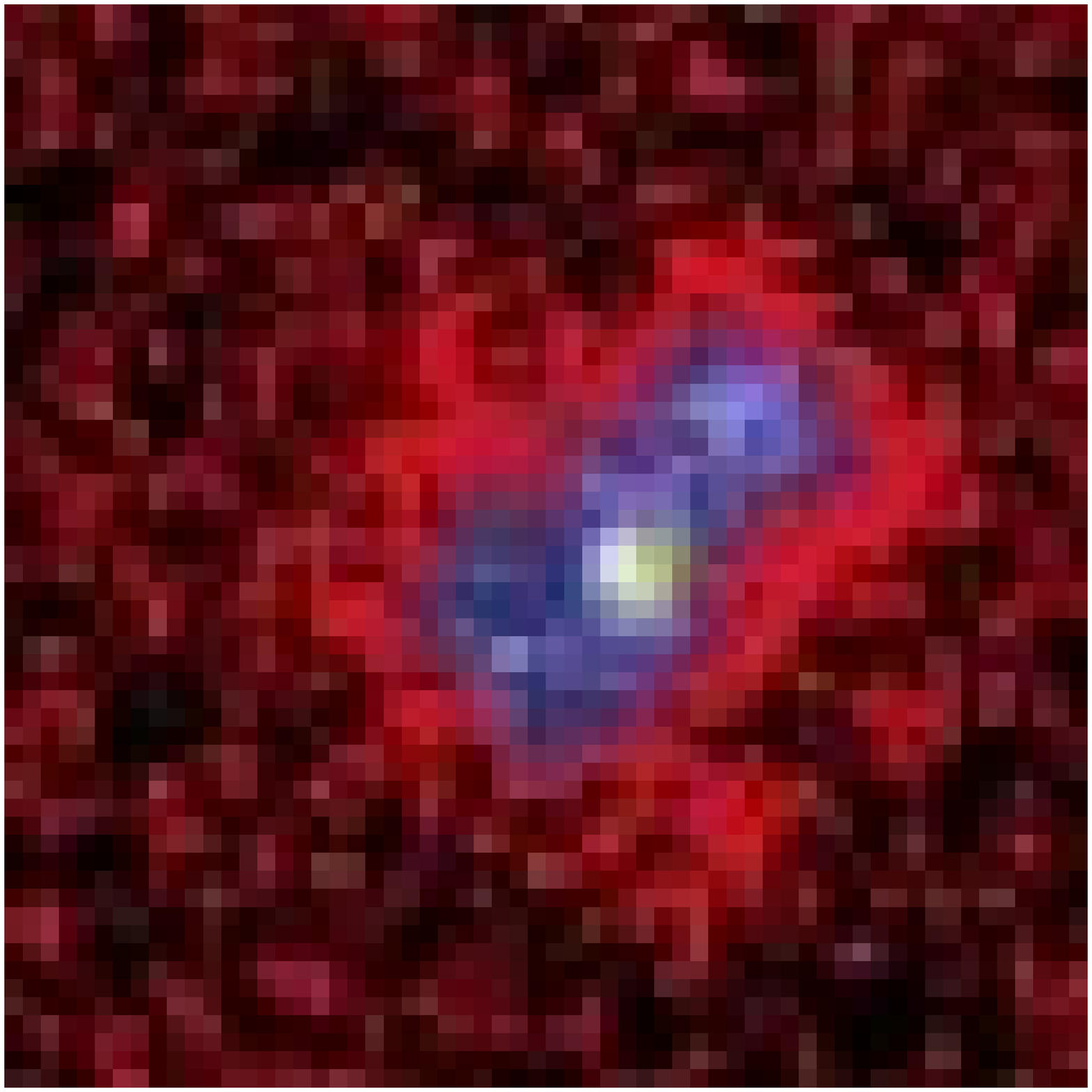}  \FGb{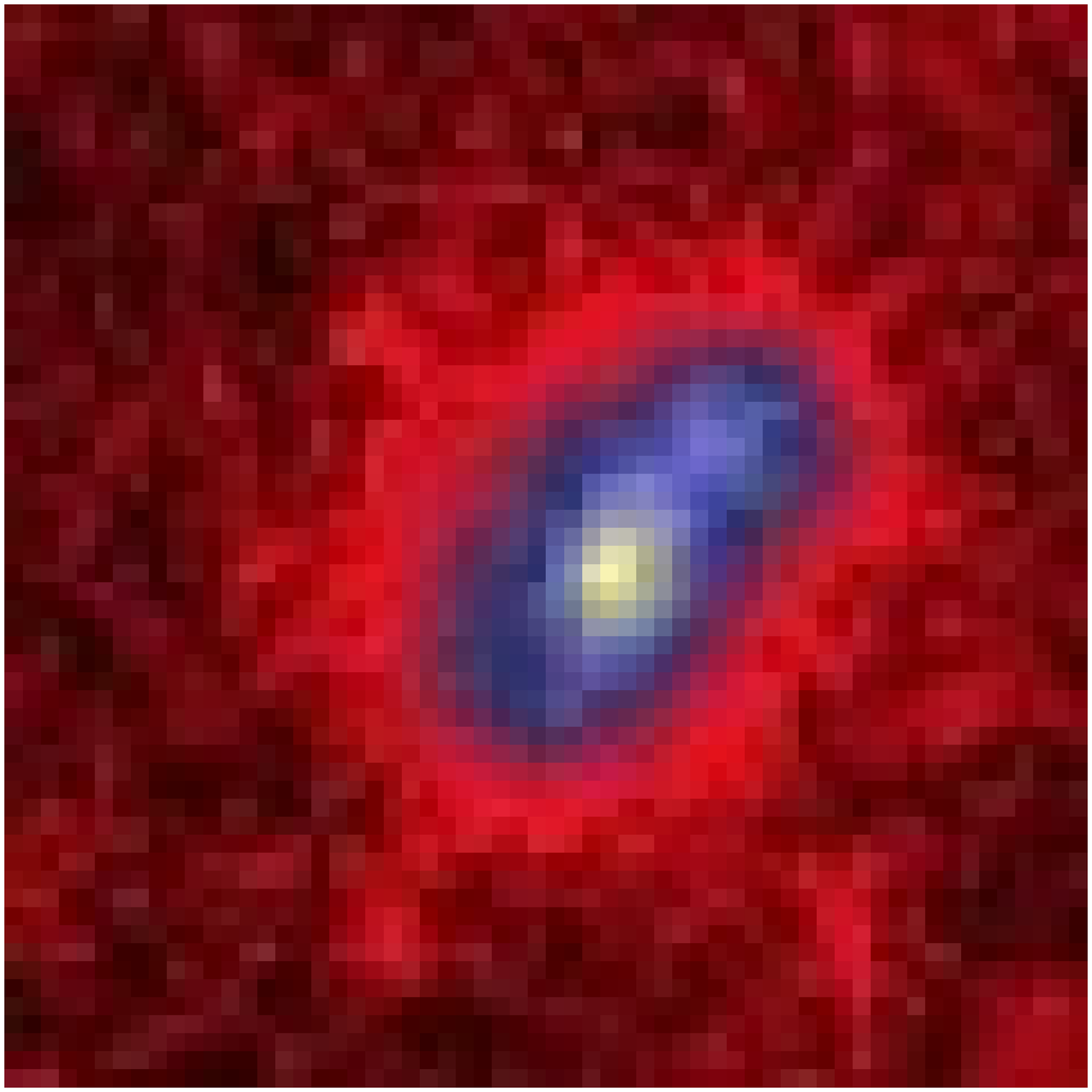}  \FGb{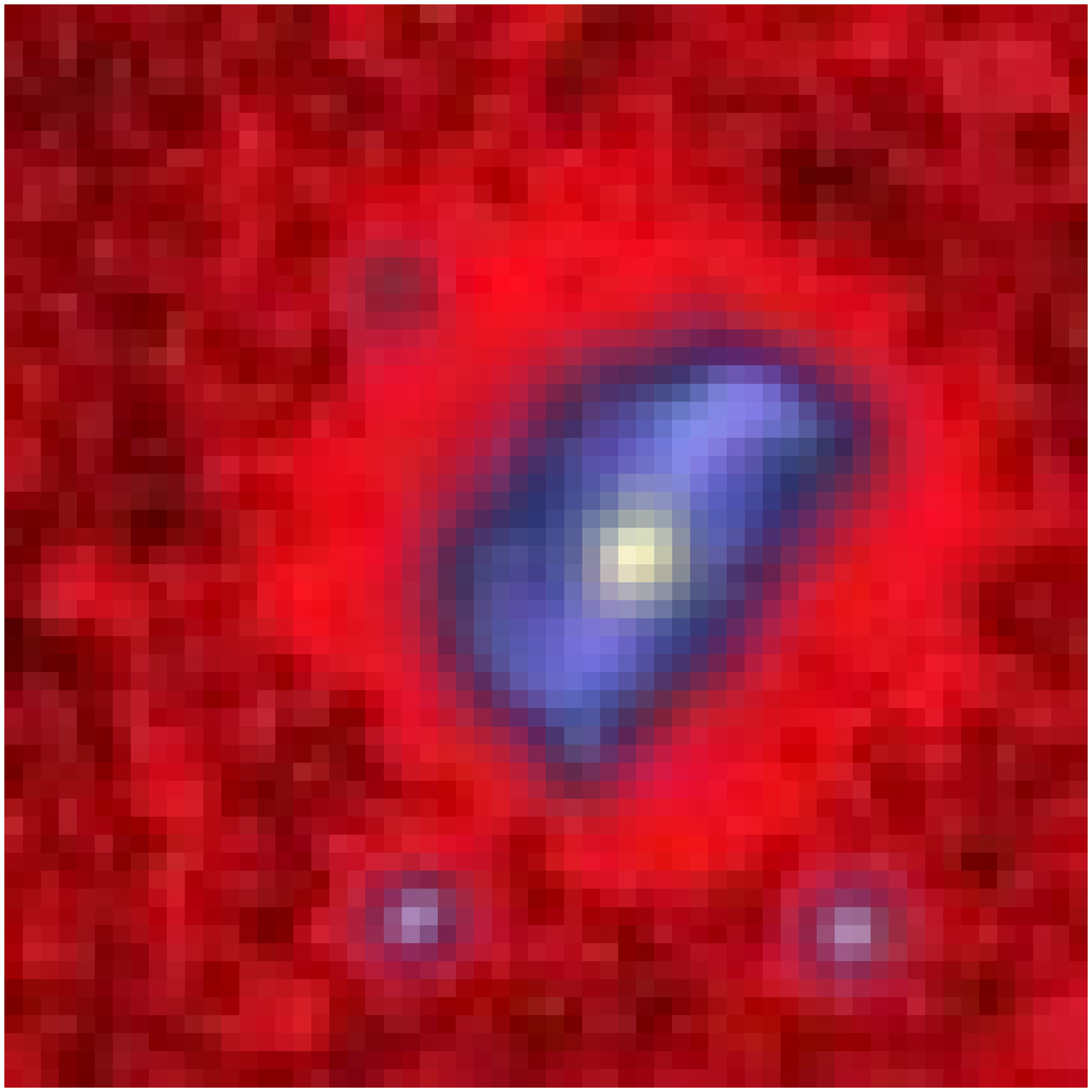}  \FGb{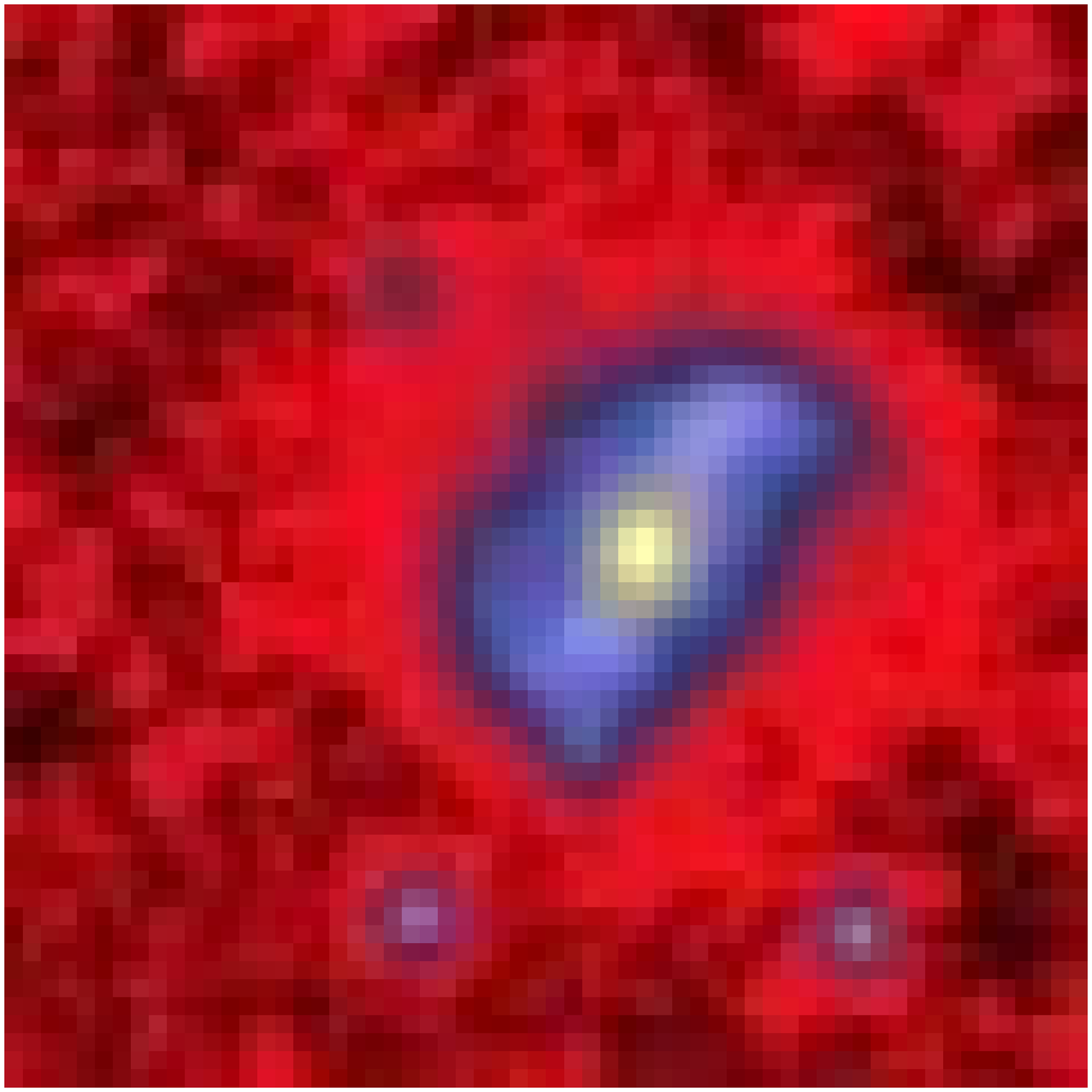}  \FGb{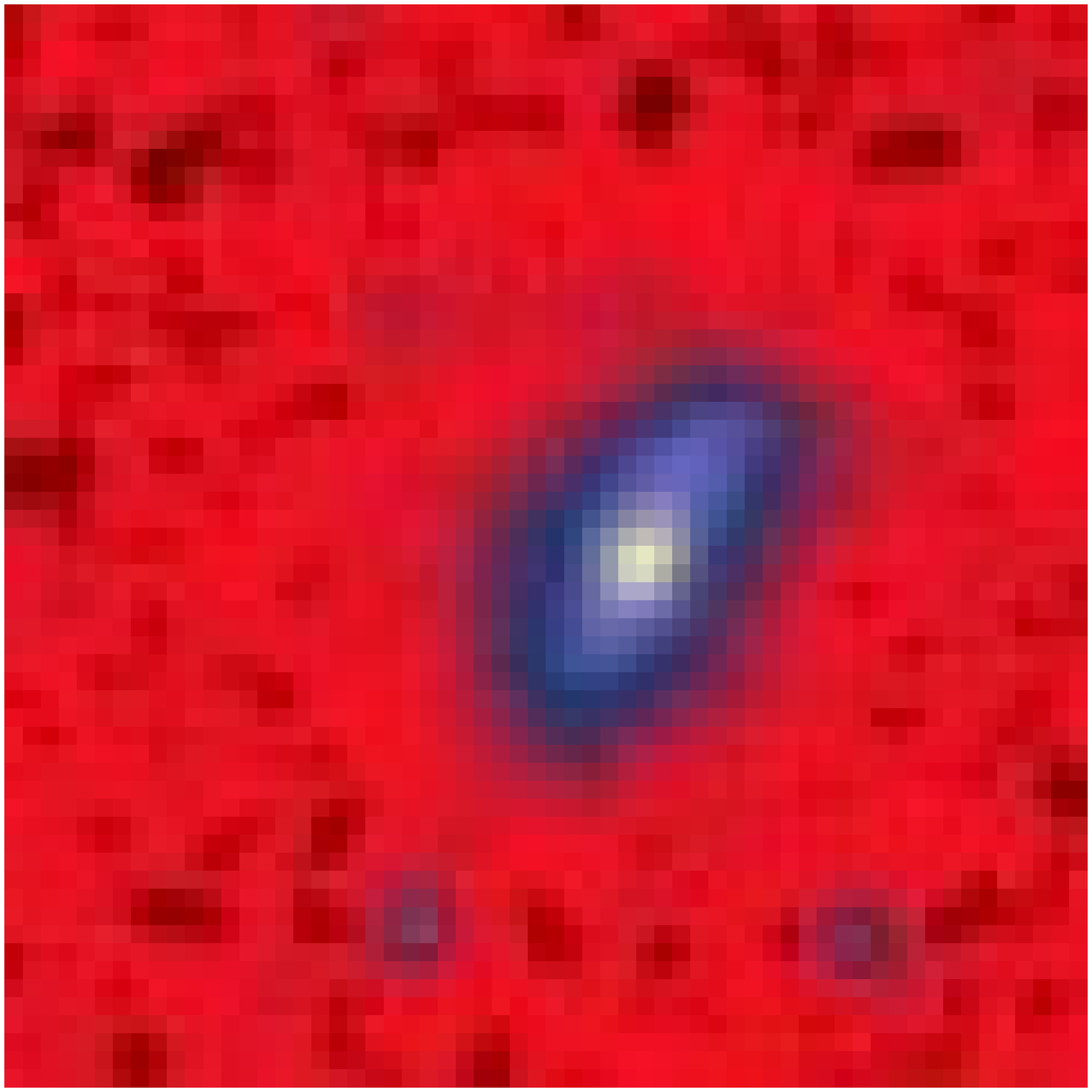}  \FGb{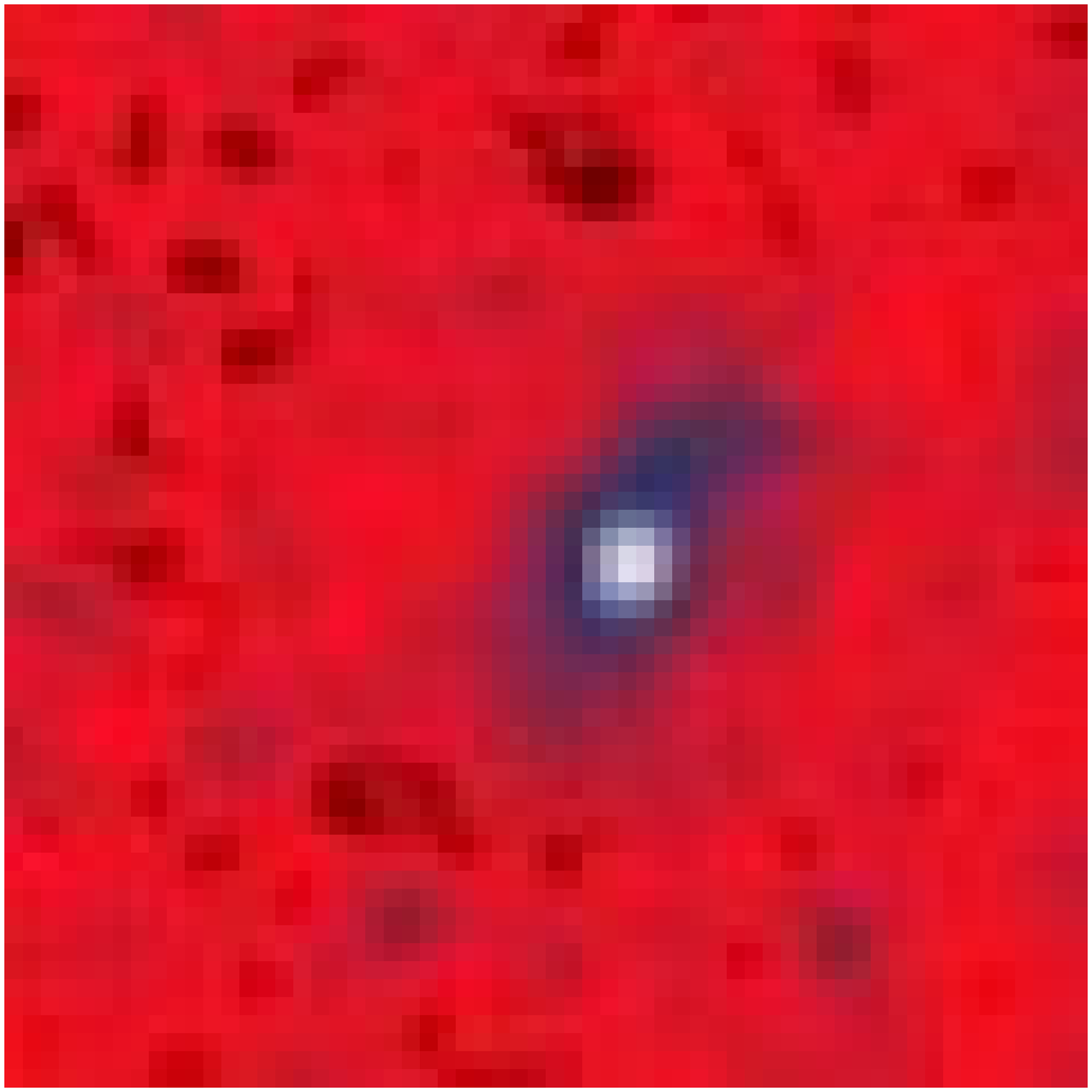}  \FGb{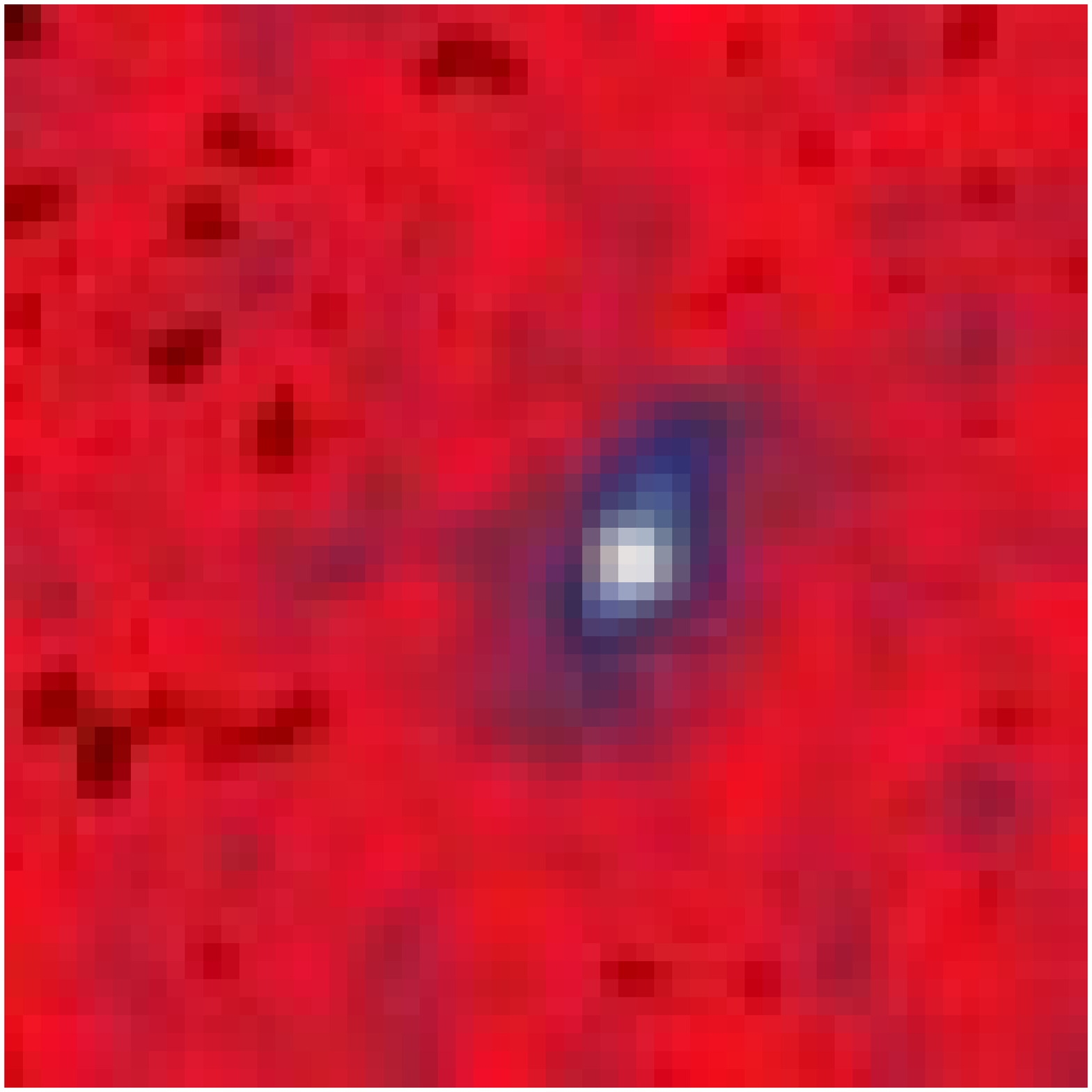}  \FGb{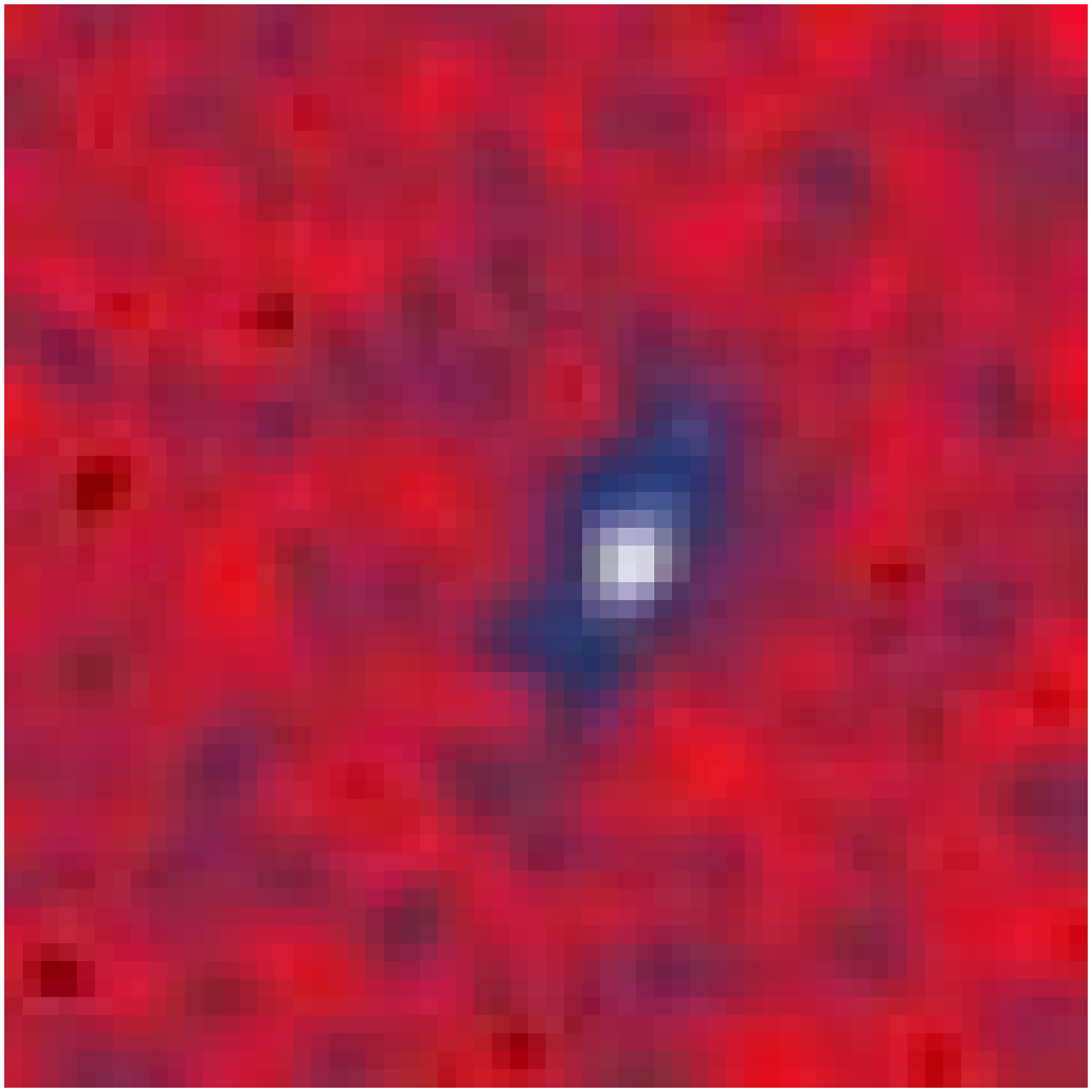}  \FGb{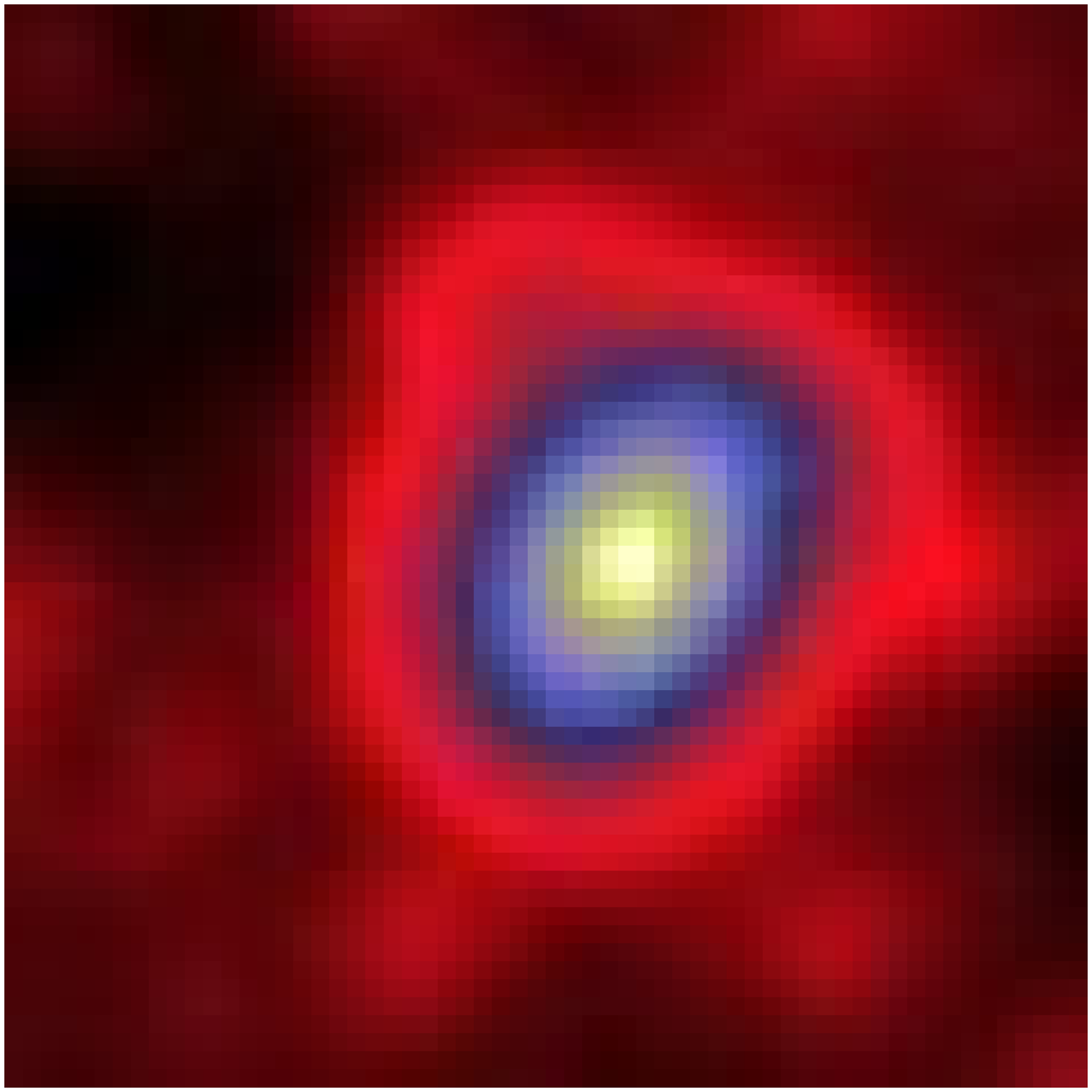}  \FGb{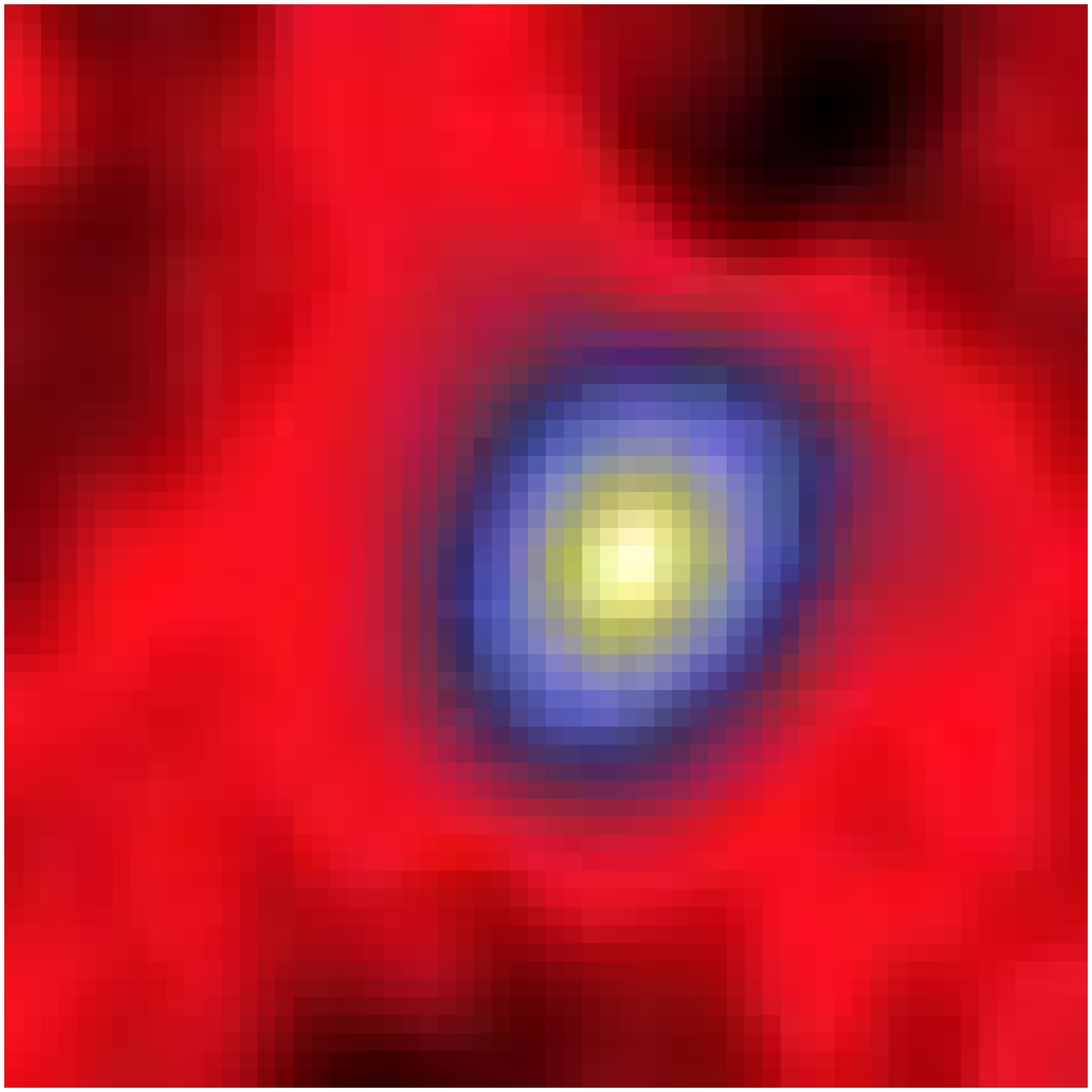}  \FGb{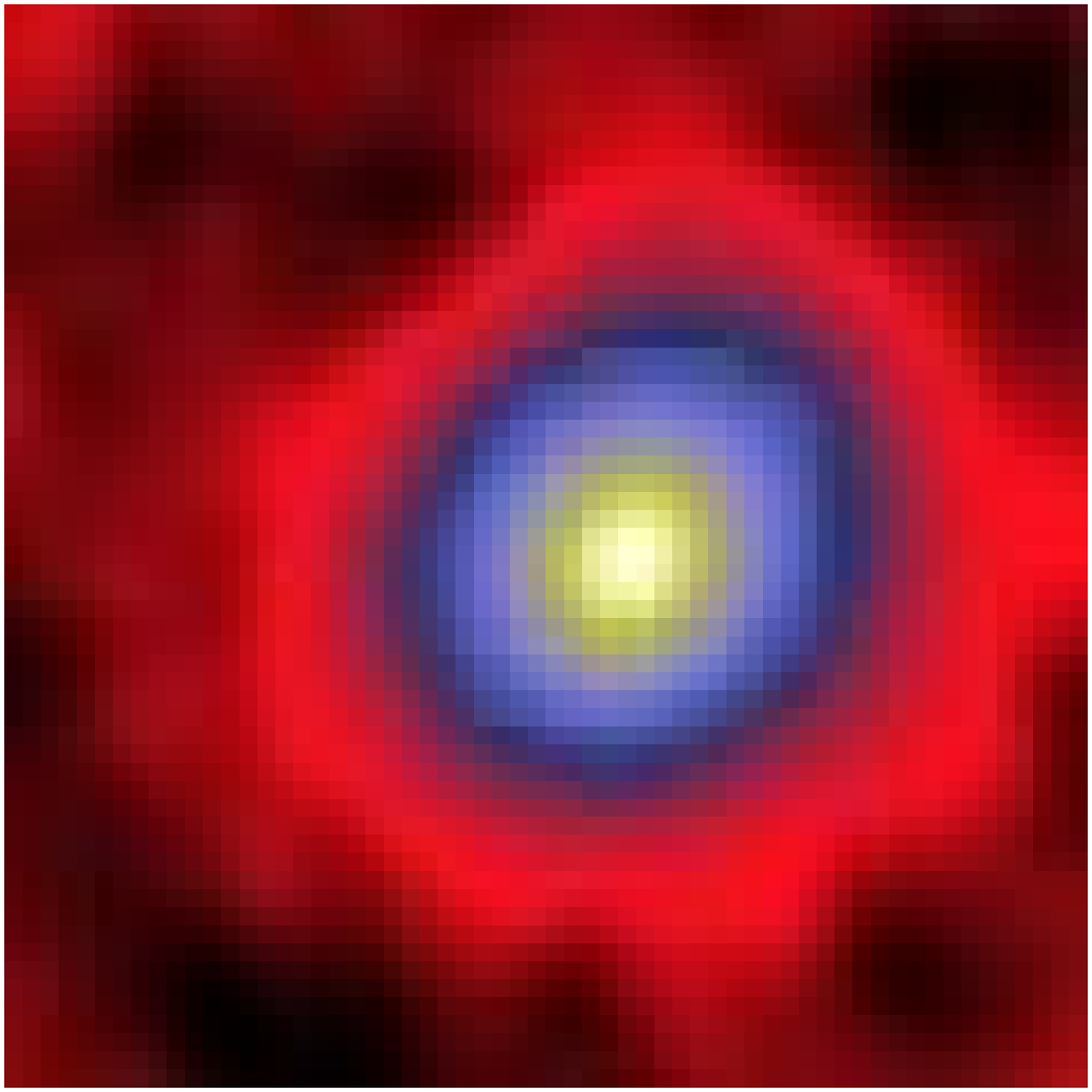}  \FGb{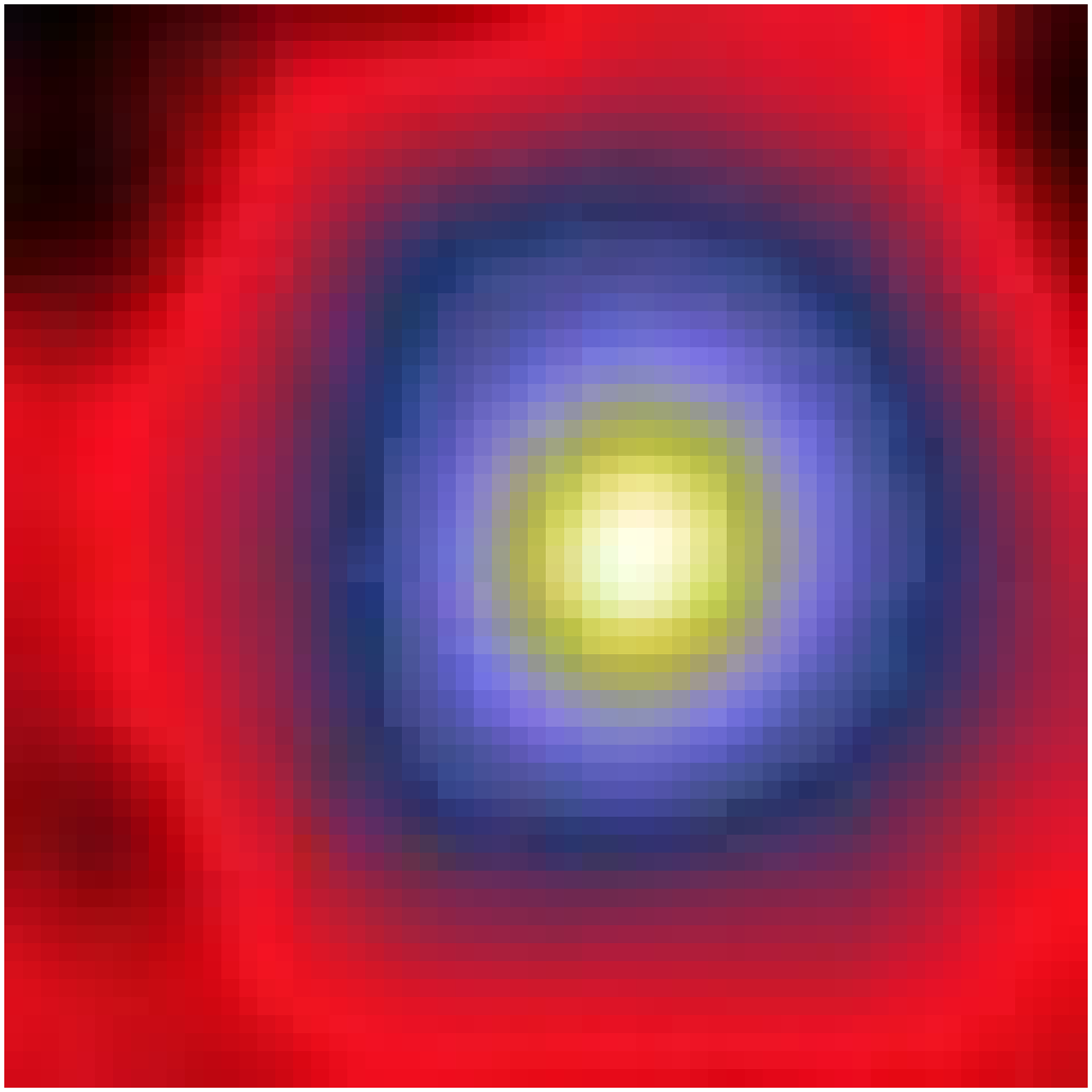} \\  {\#46   NCIA001356}  \\
  \FGb{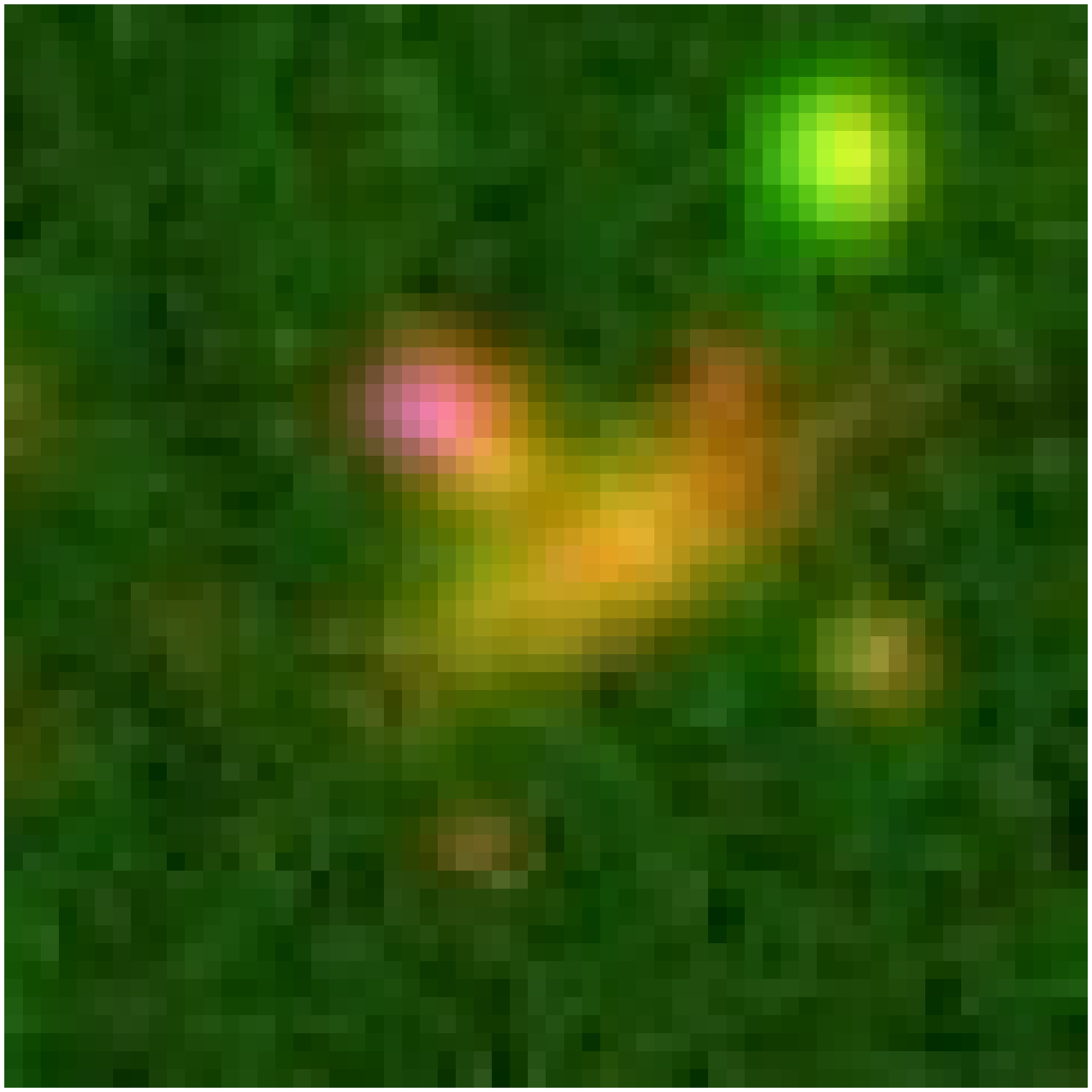}  \FGb{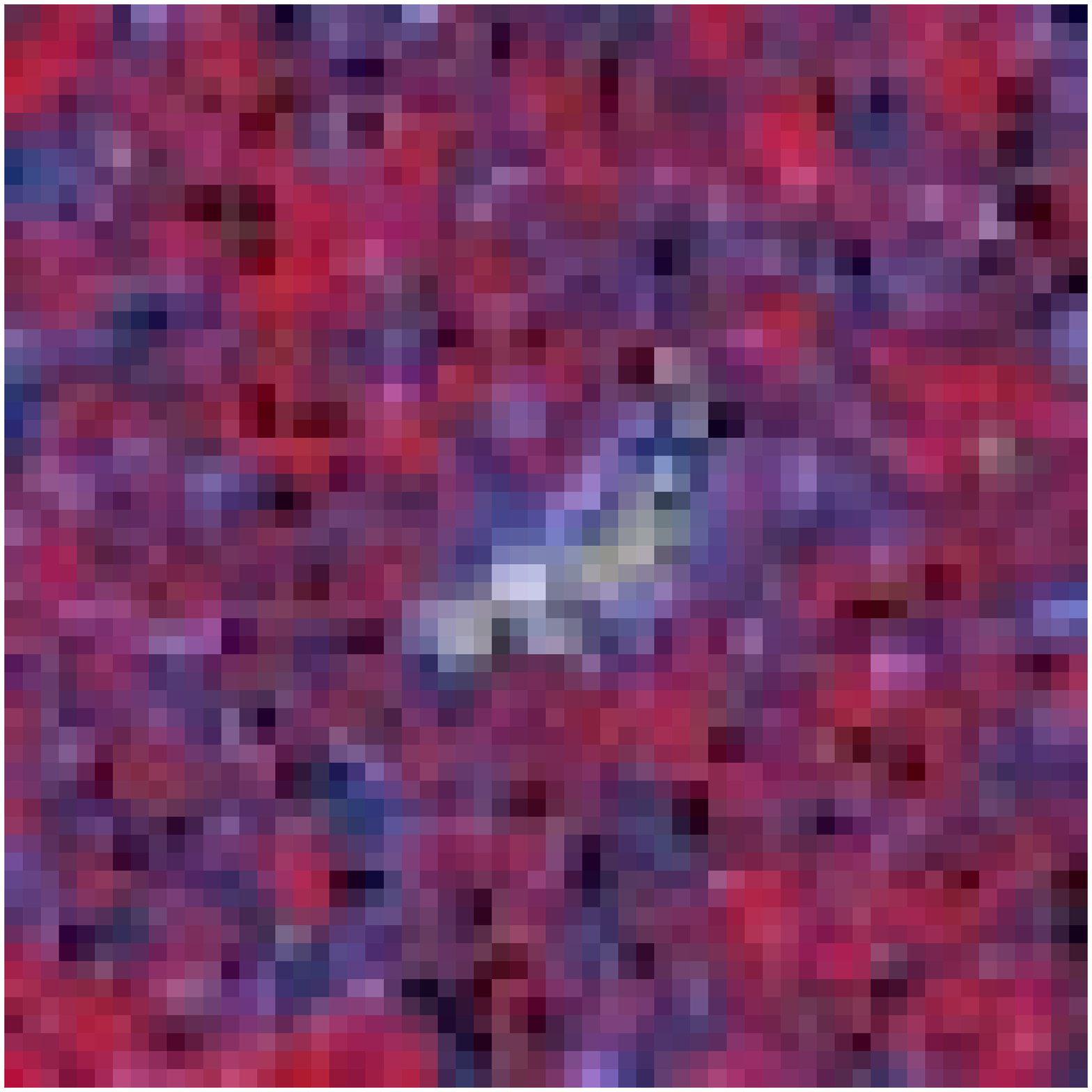}  \FGb{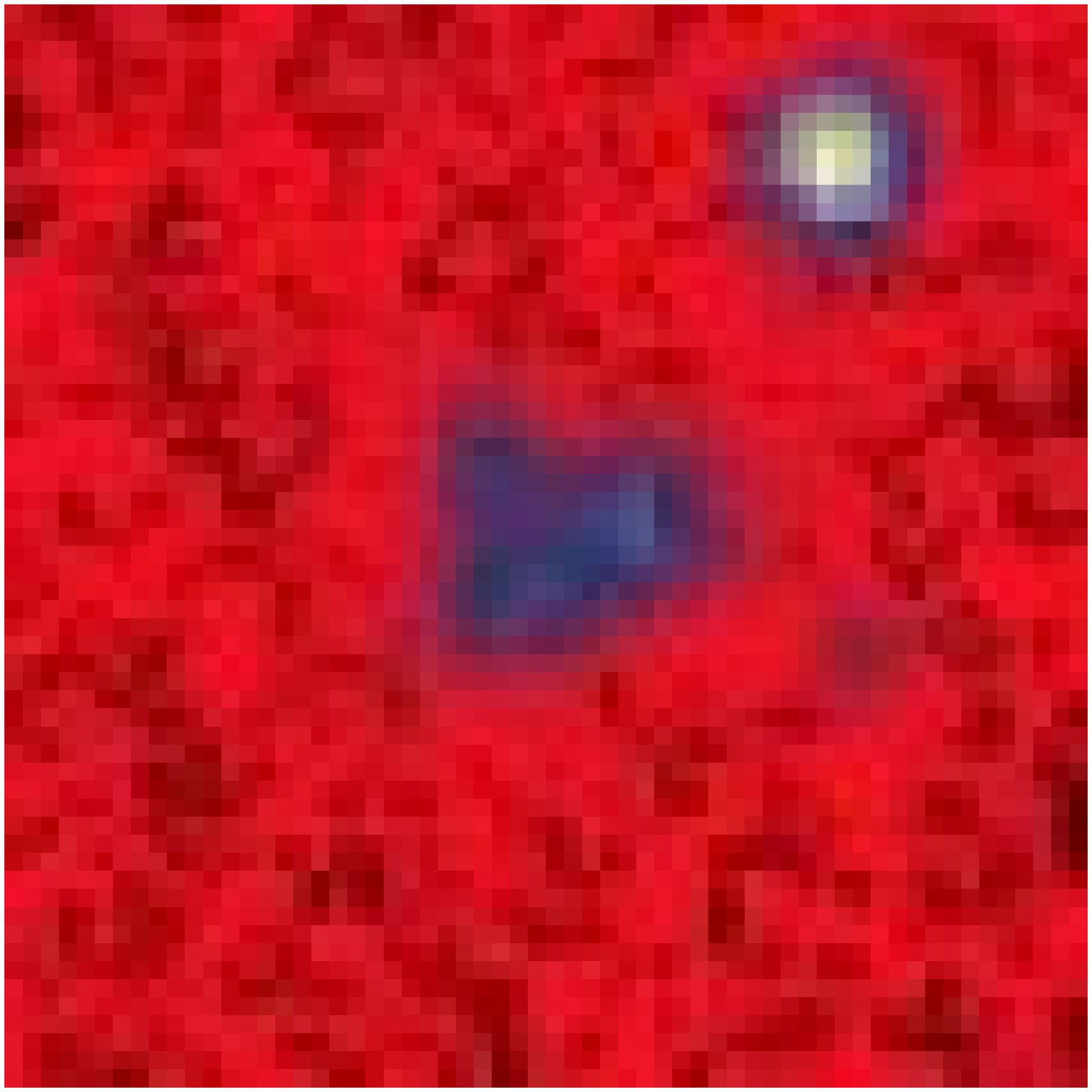}  \FGb{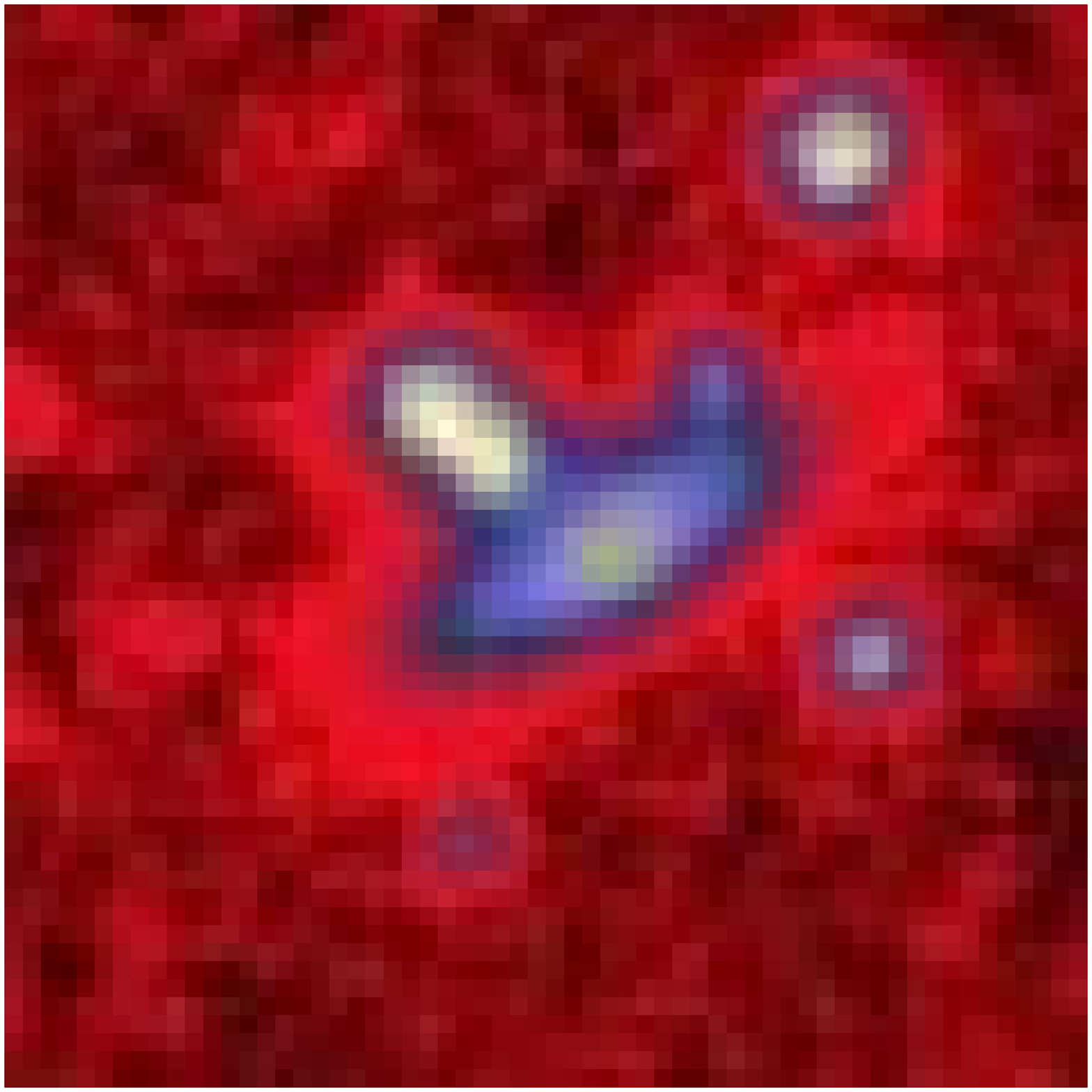}  \FGb{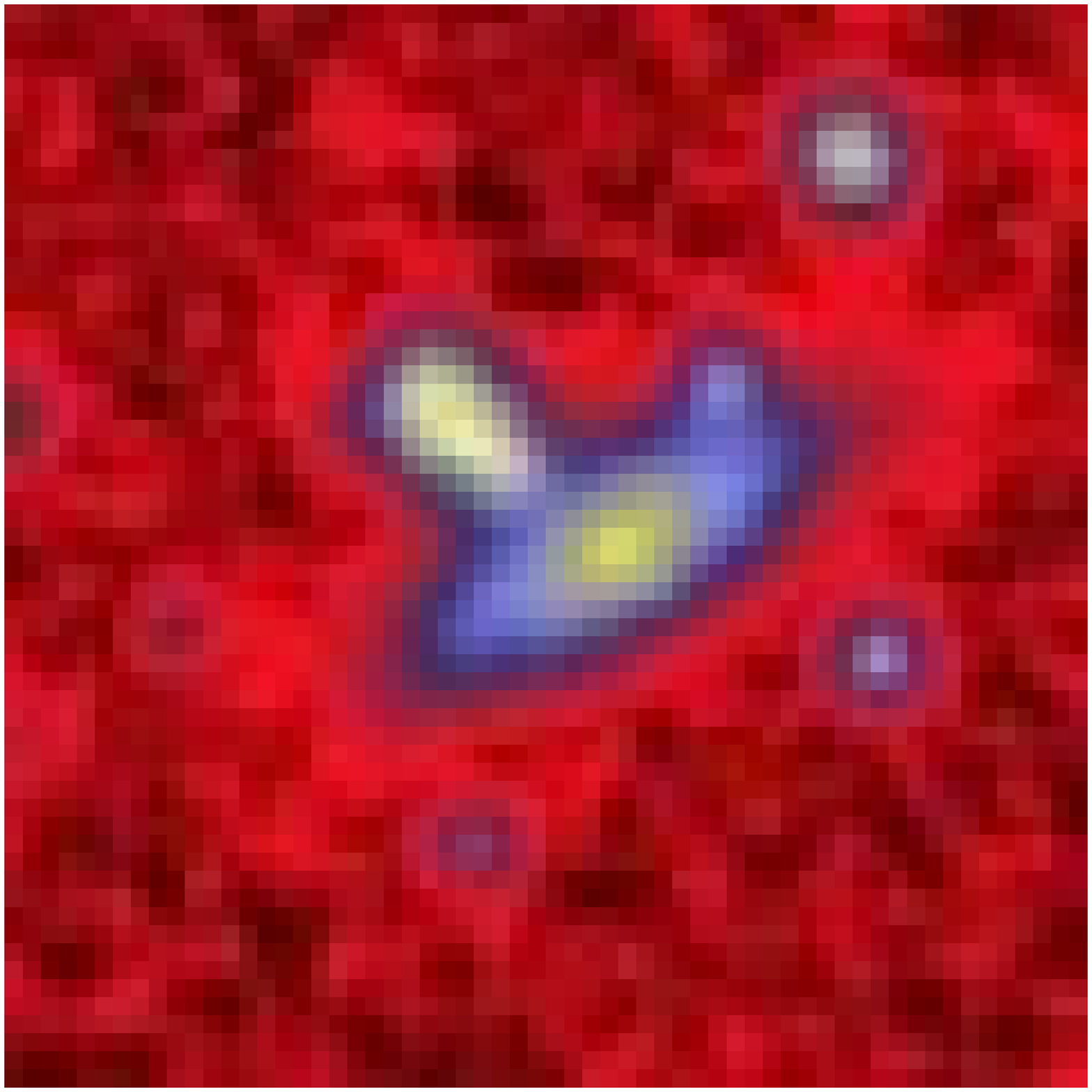}  \FGb{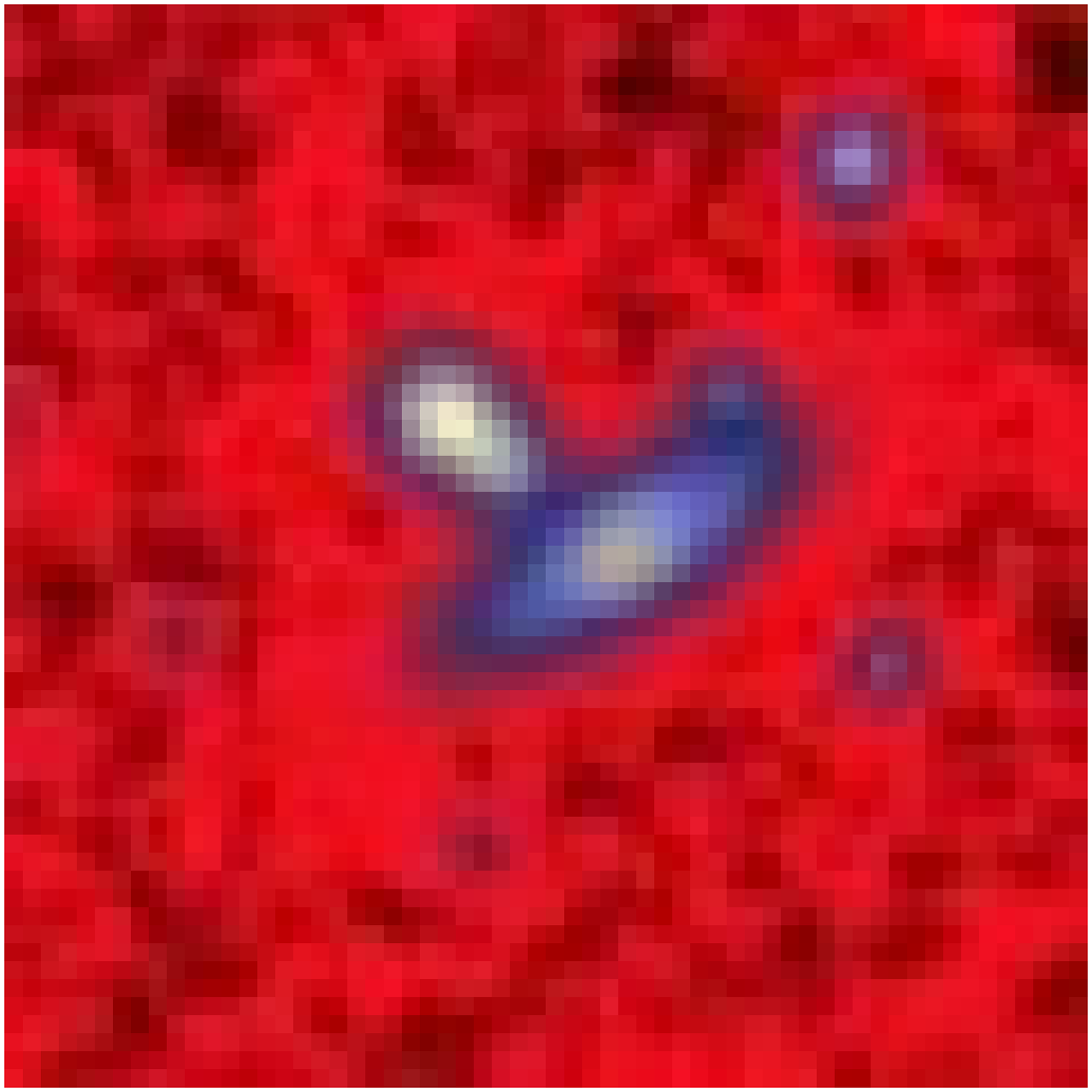}  \FGb{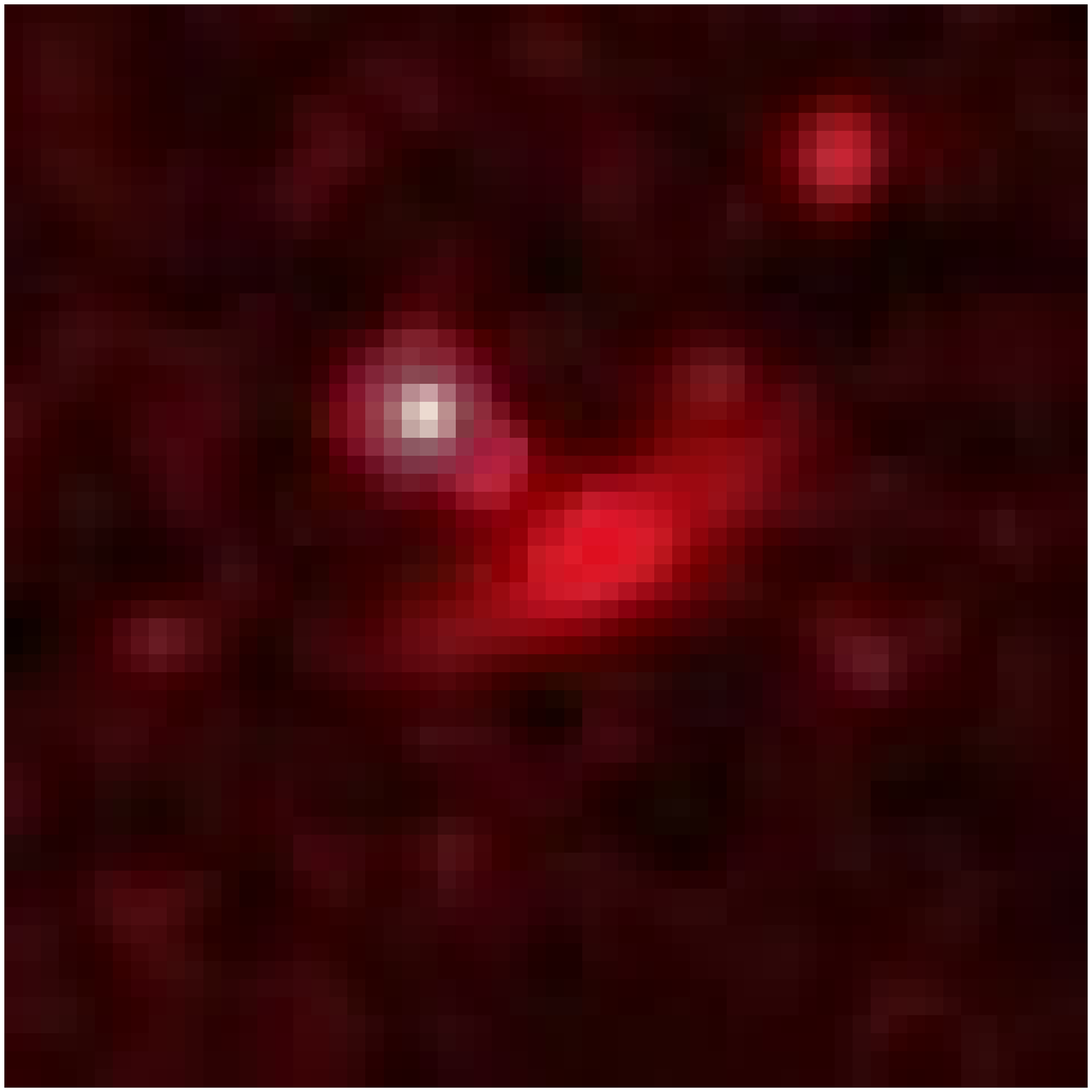}  \FGb{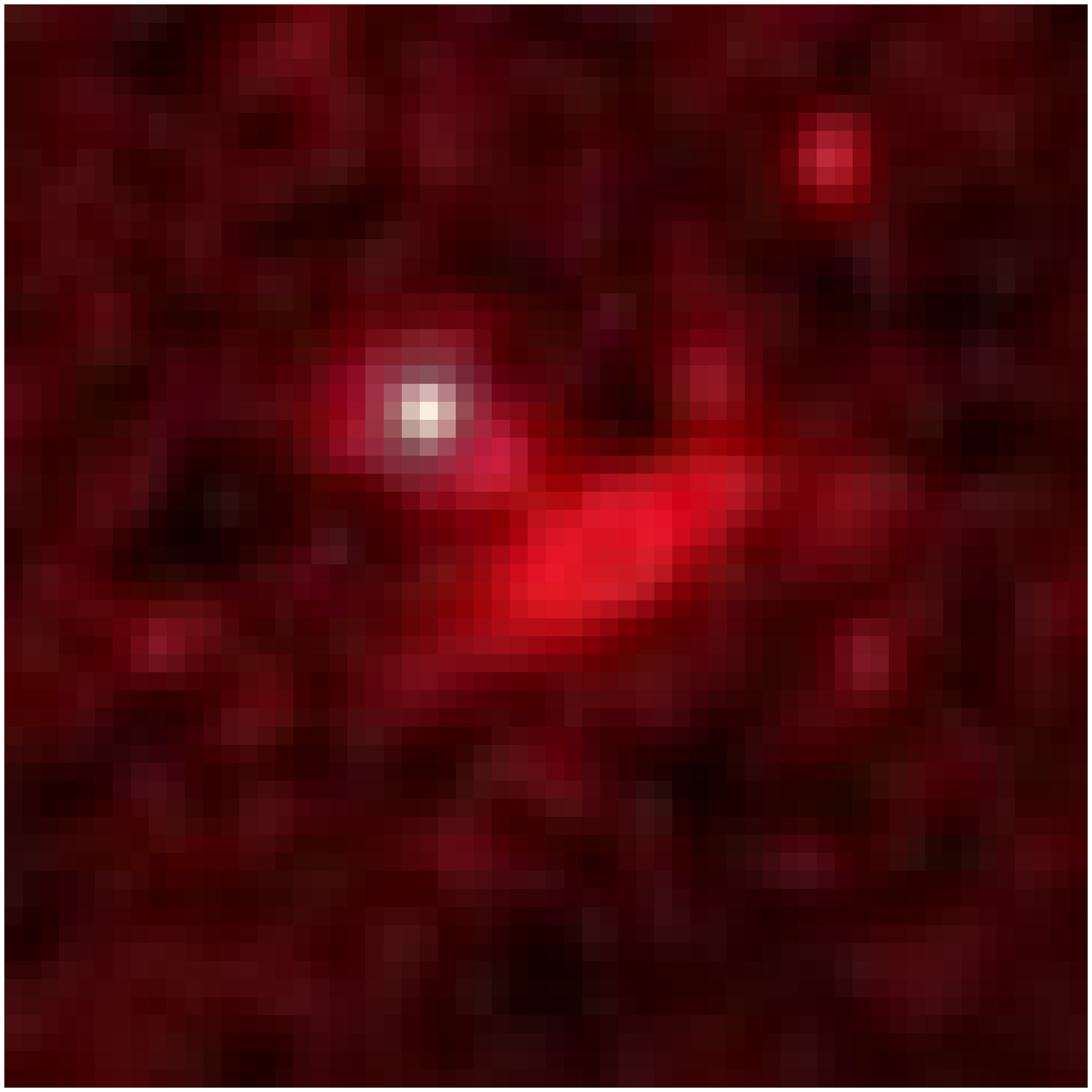}  \FGb{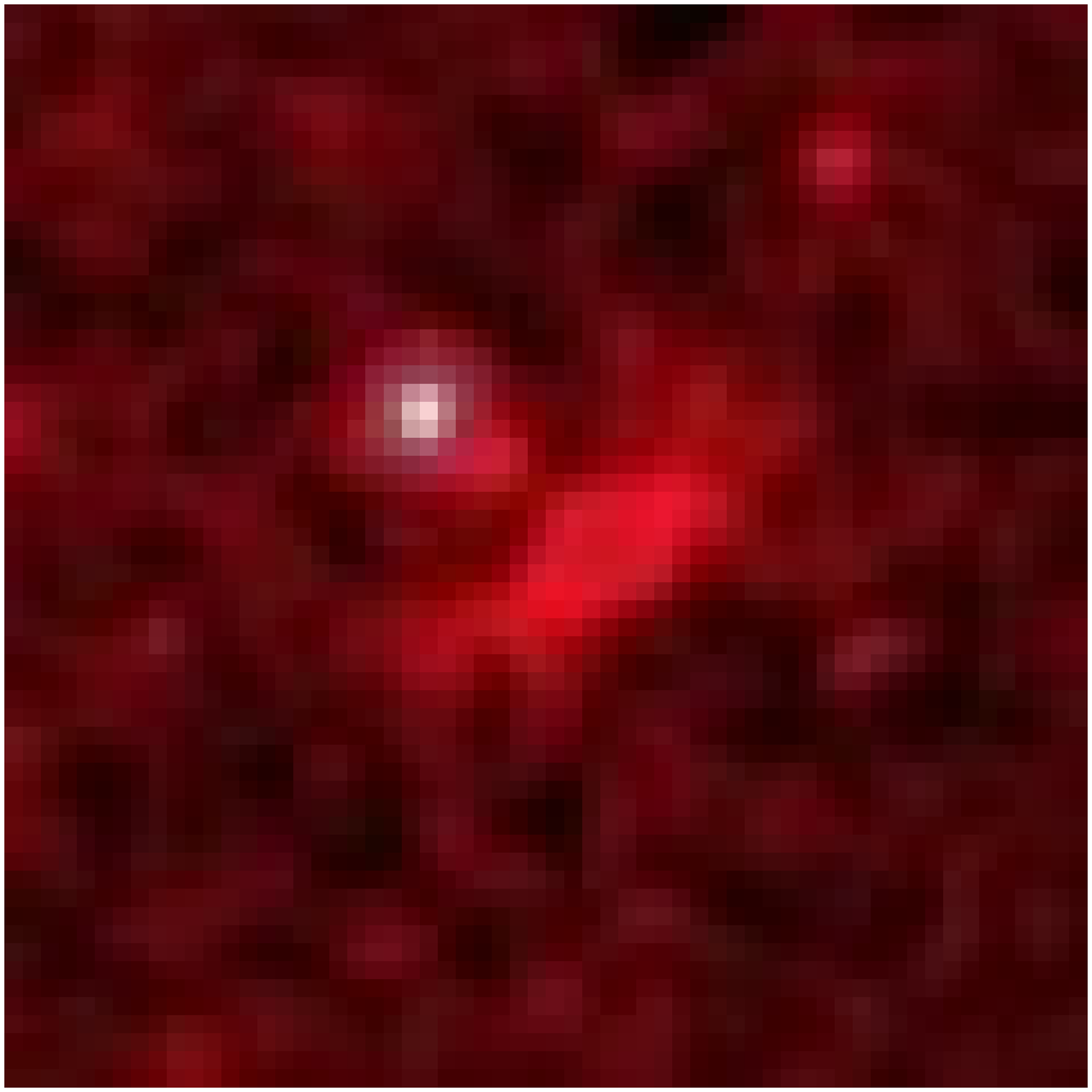}  \FGb{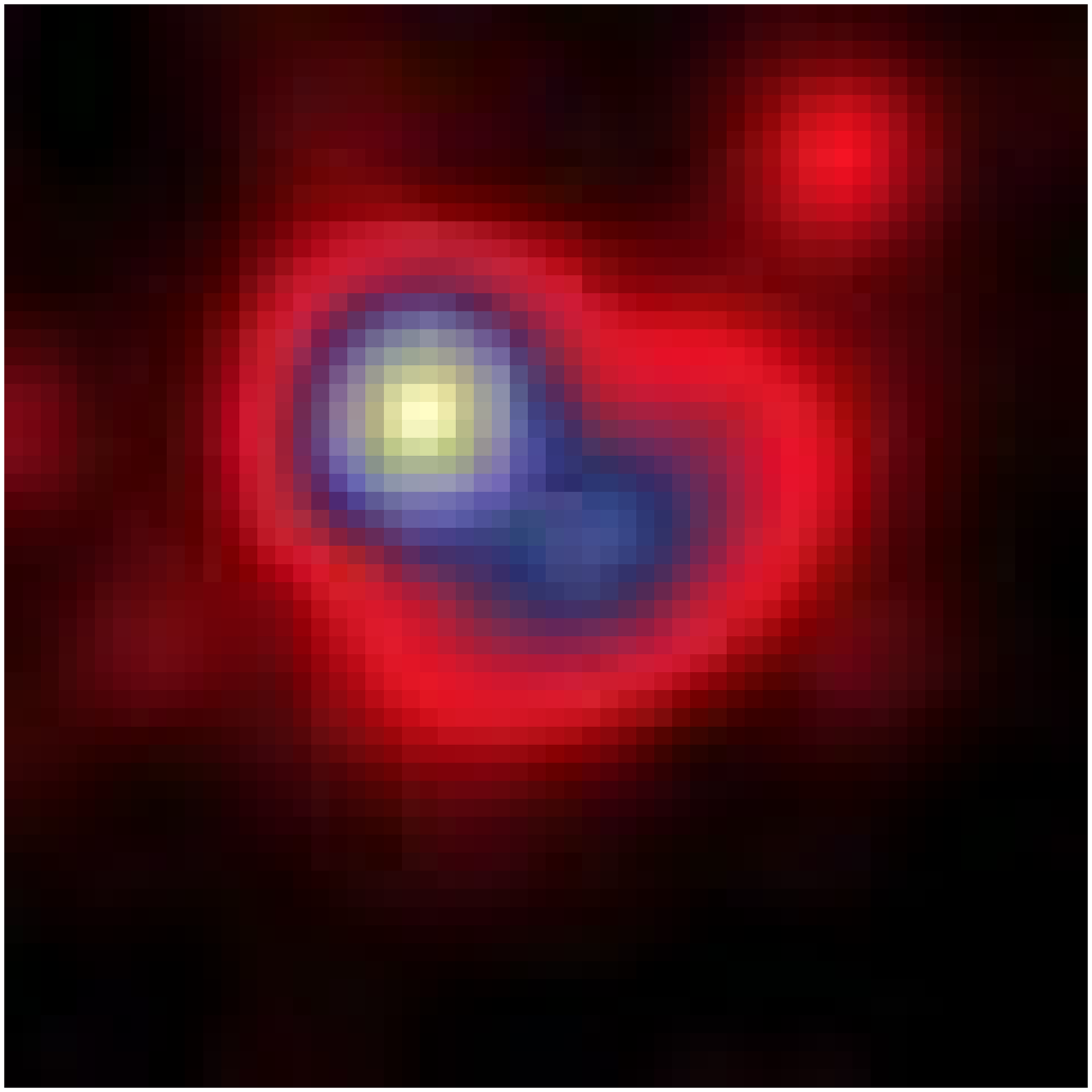}  \FGb{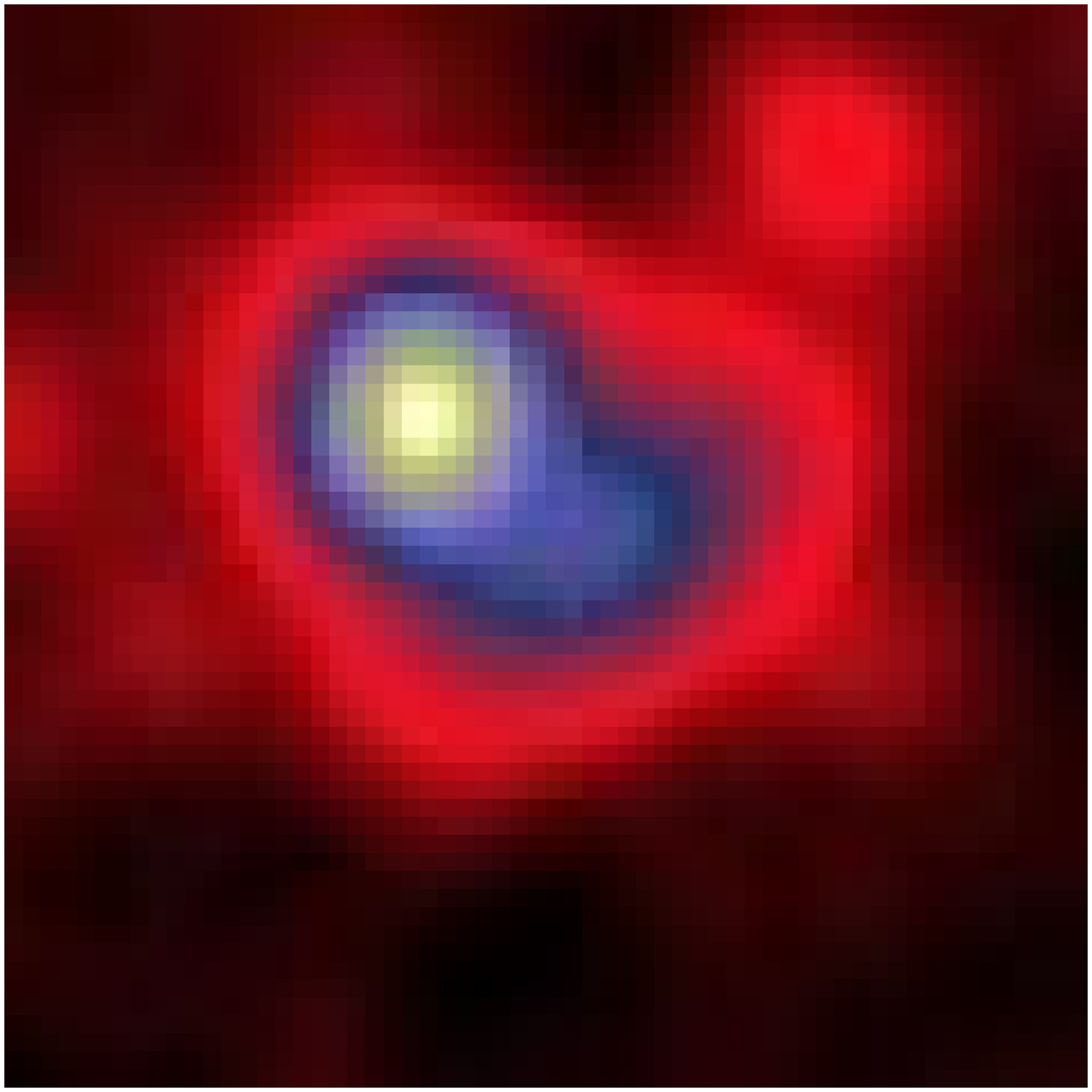}  \FGb{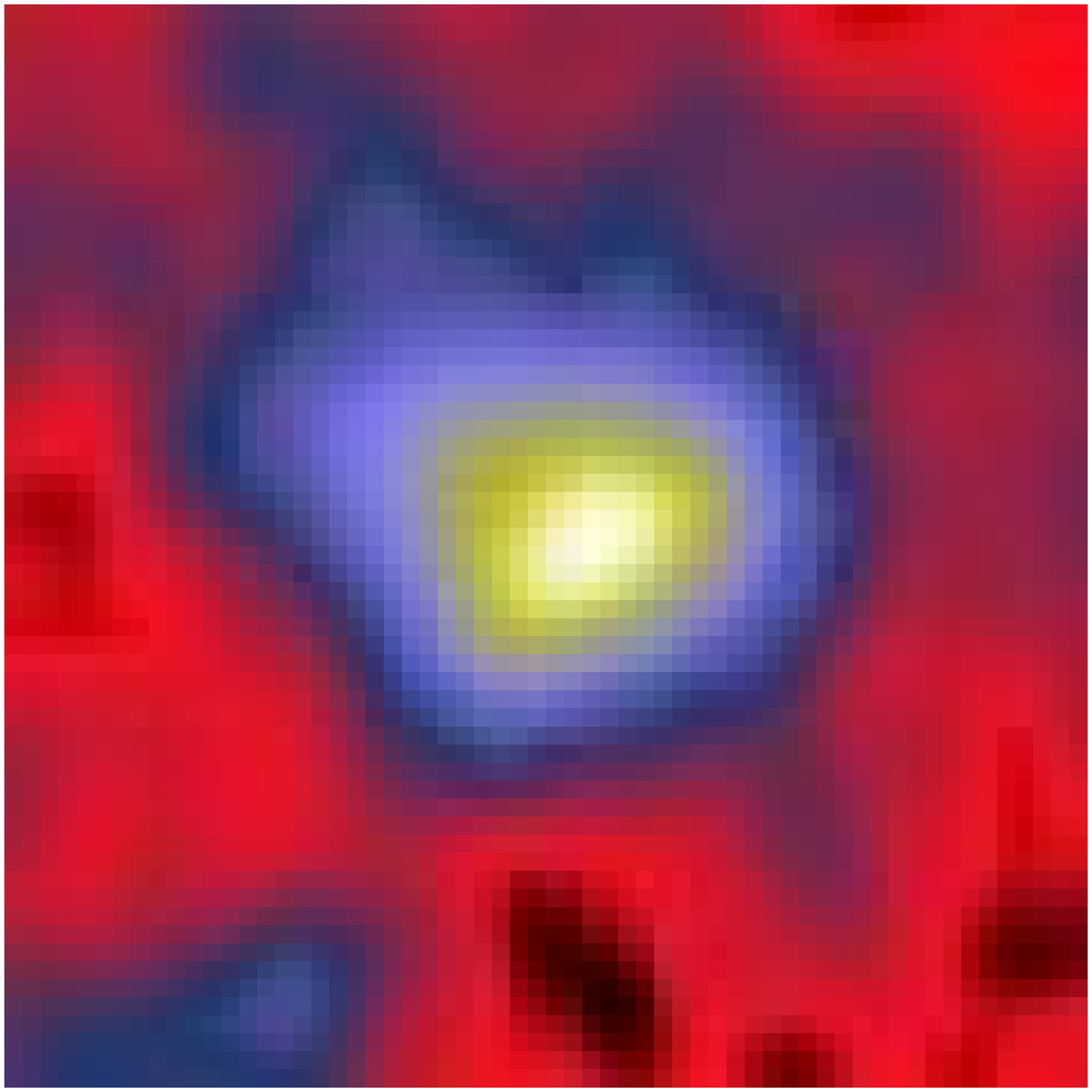}  \FGb{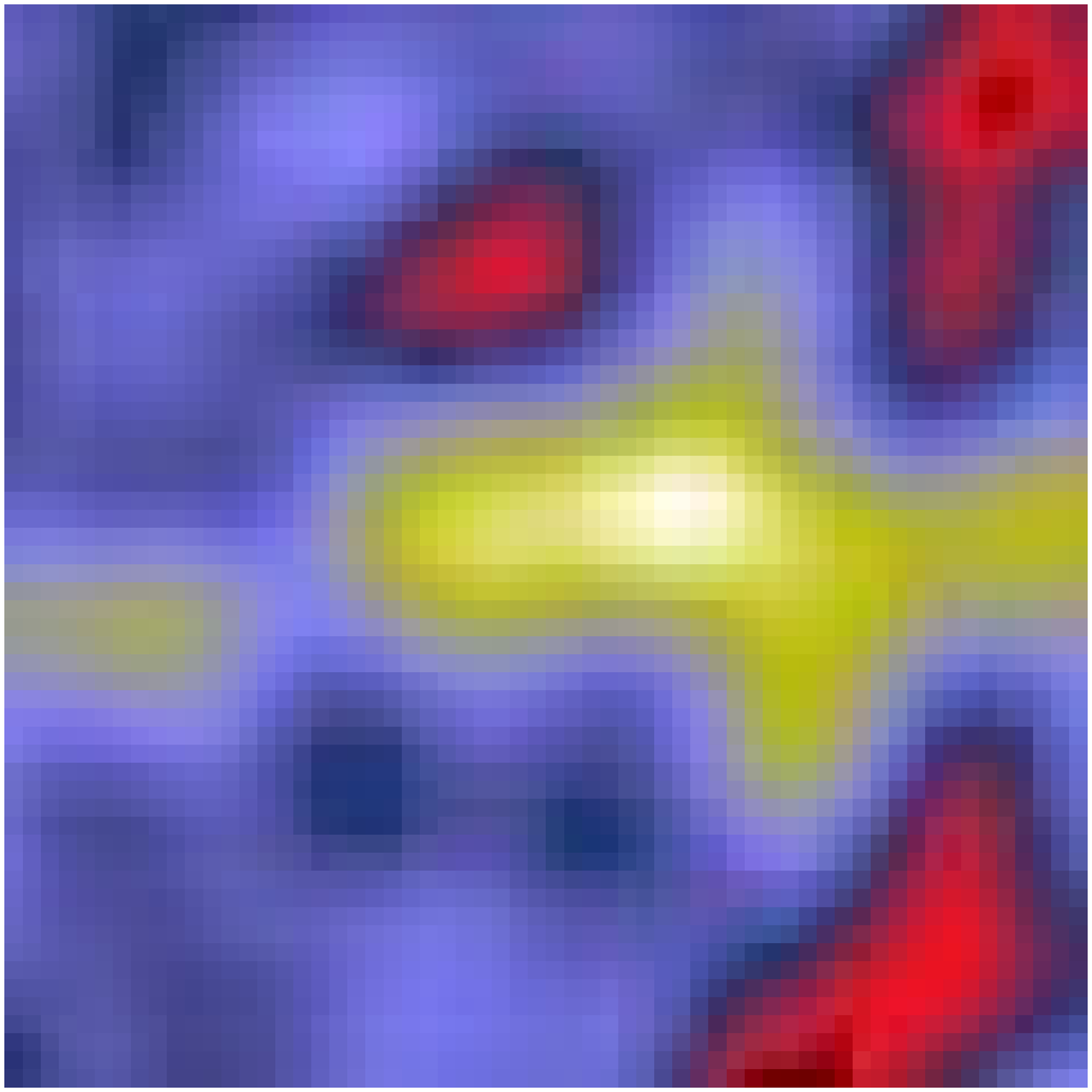} \\  {\#47   NCIA015505}  \\
  \FGb{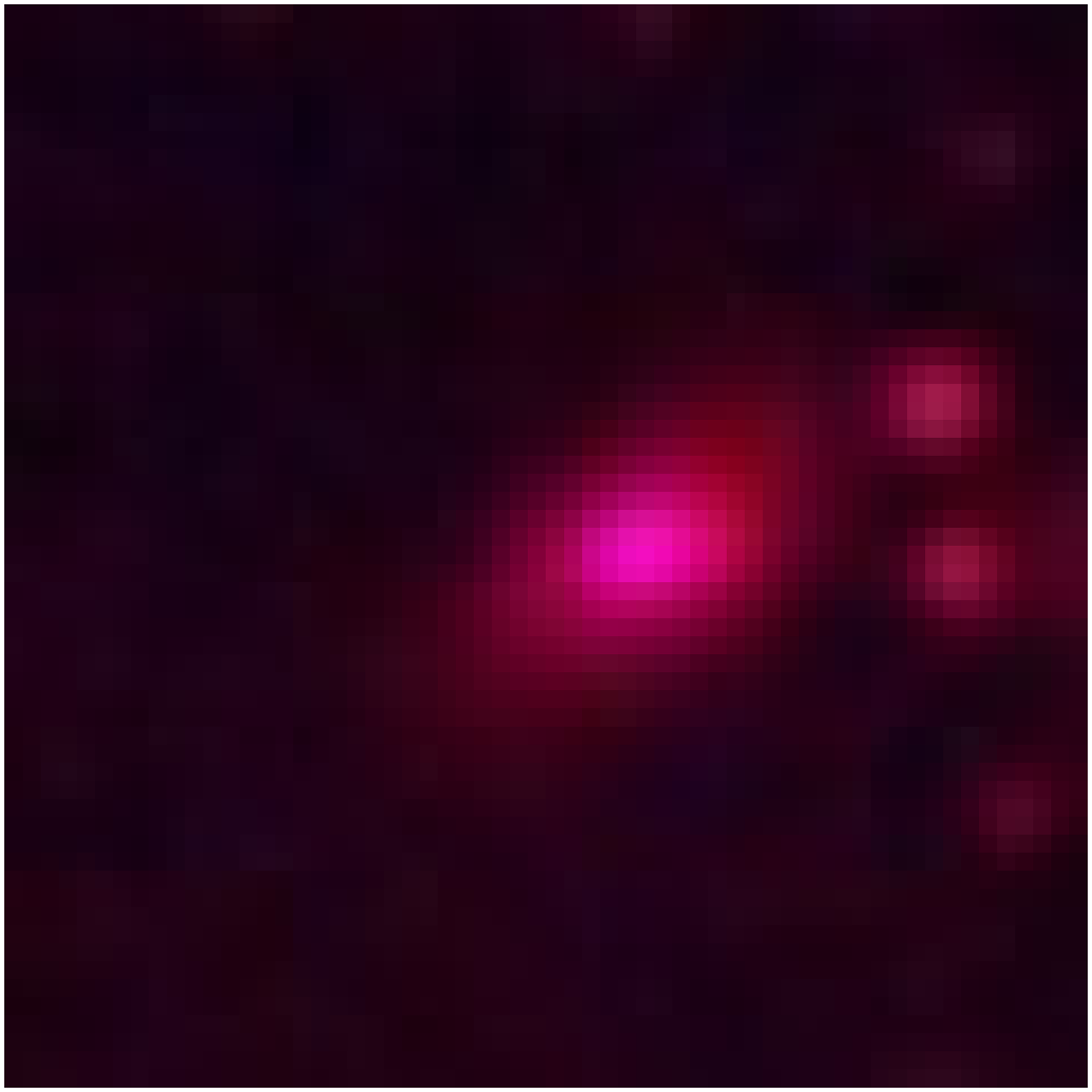}  \FGb{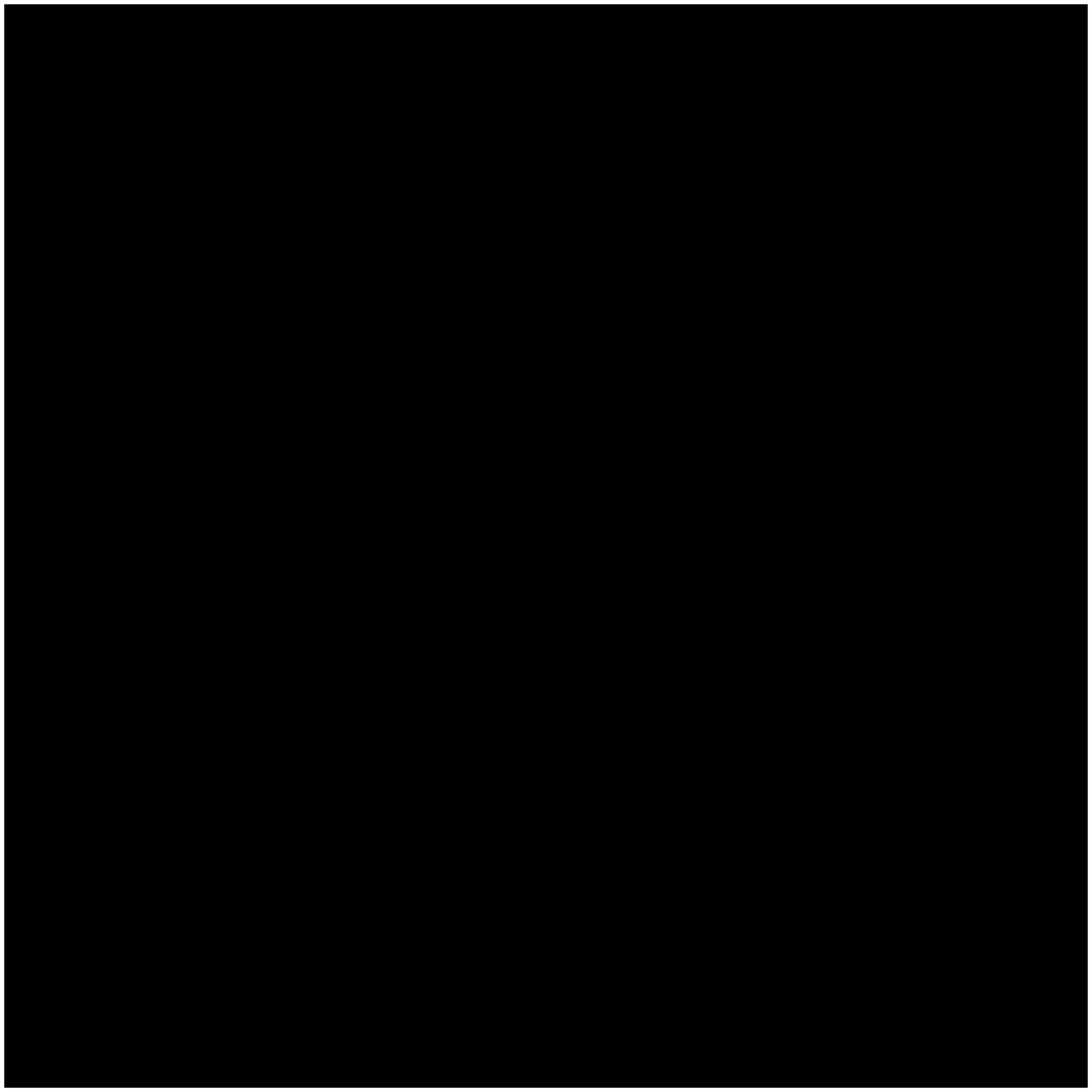}  \FGb{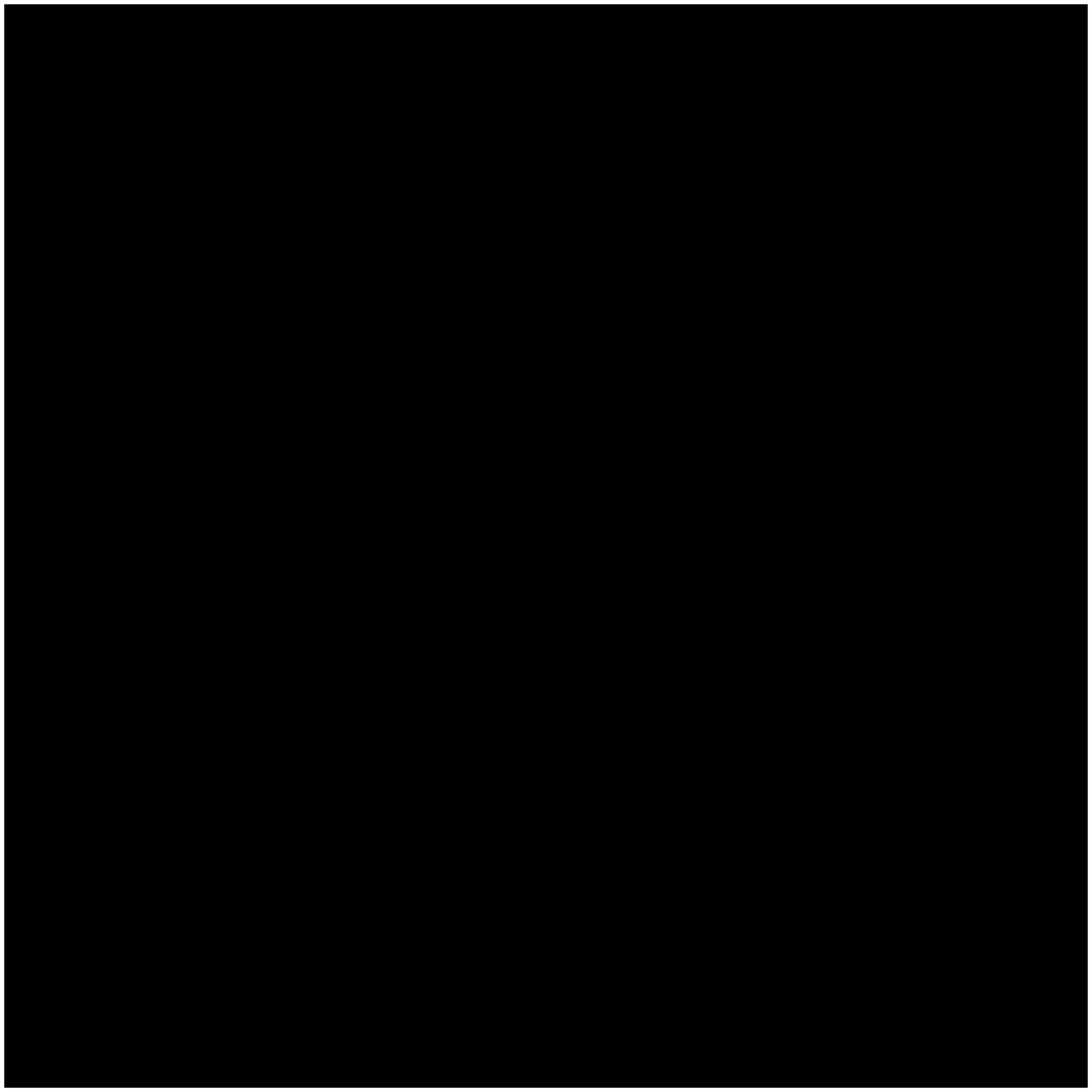}  \FGb{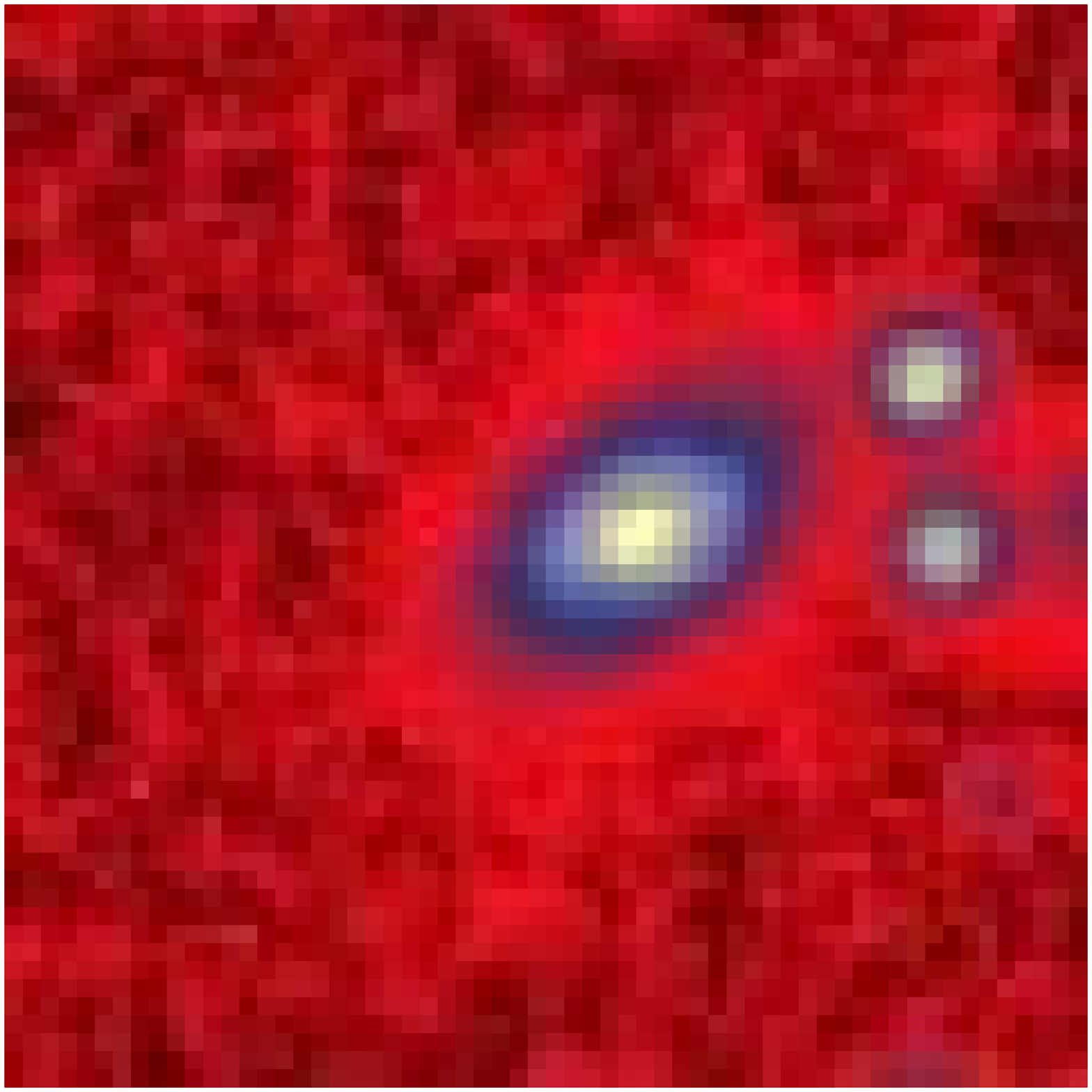}  \FGb{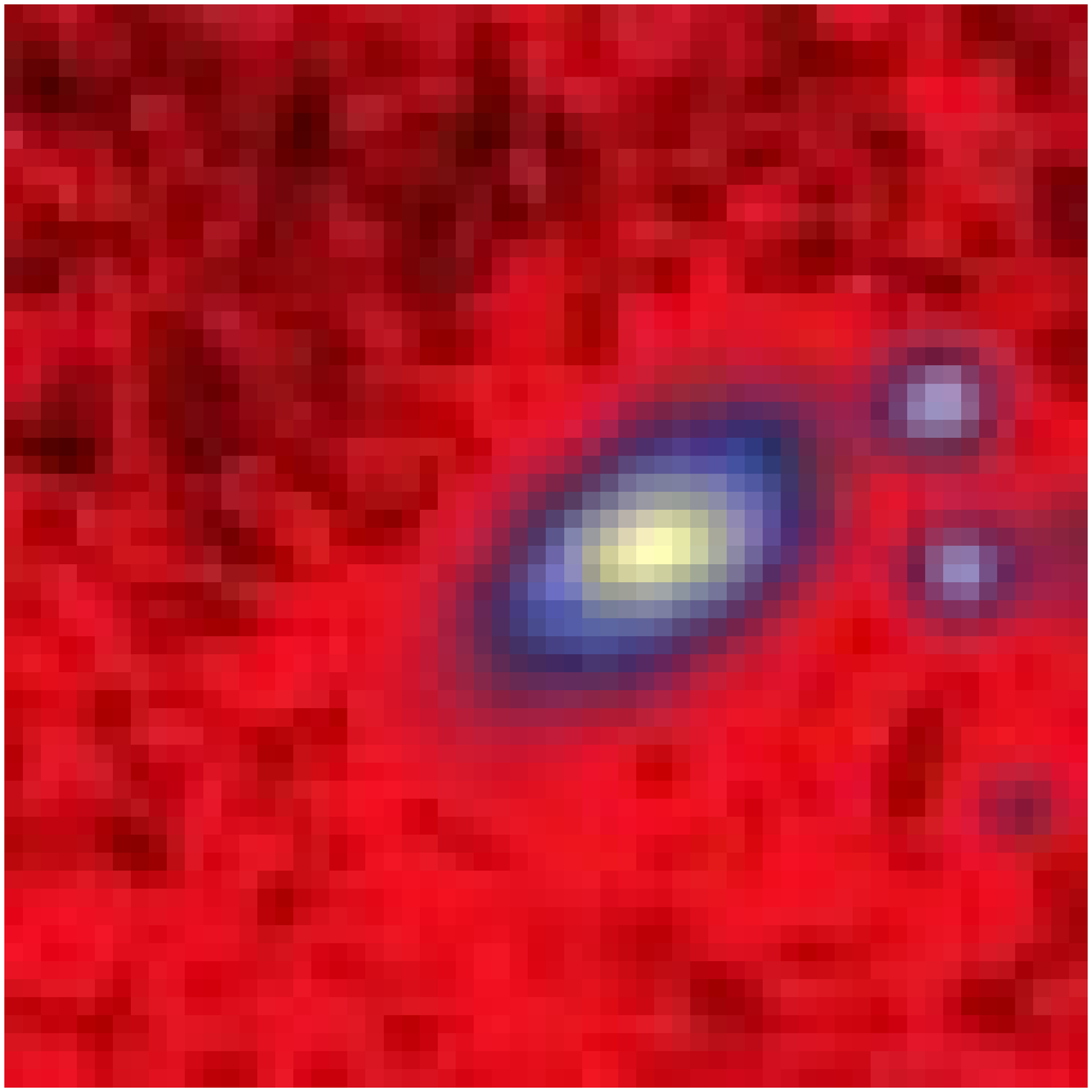}  \FGb{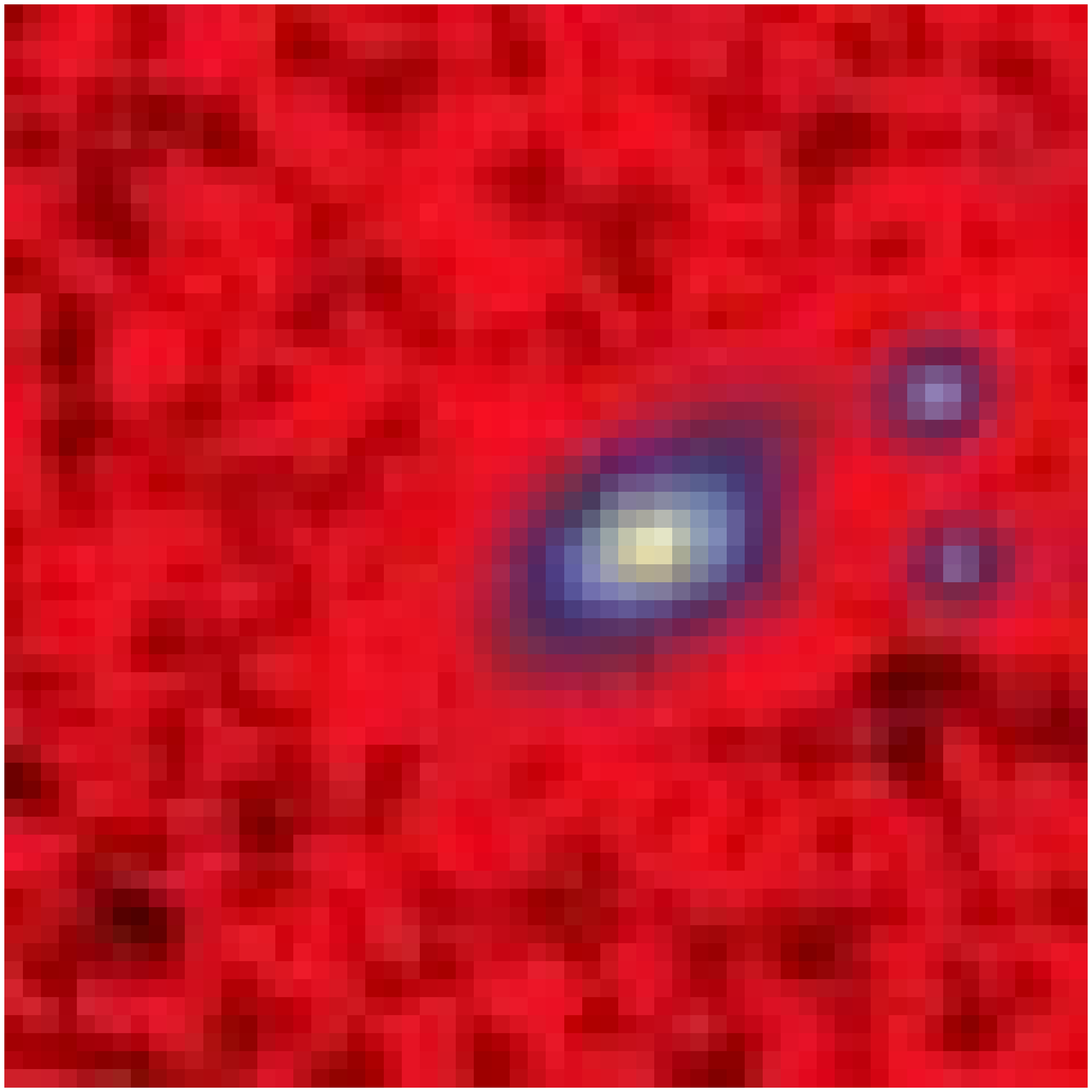}  \FGb{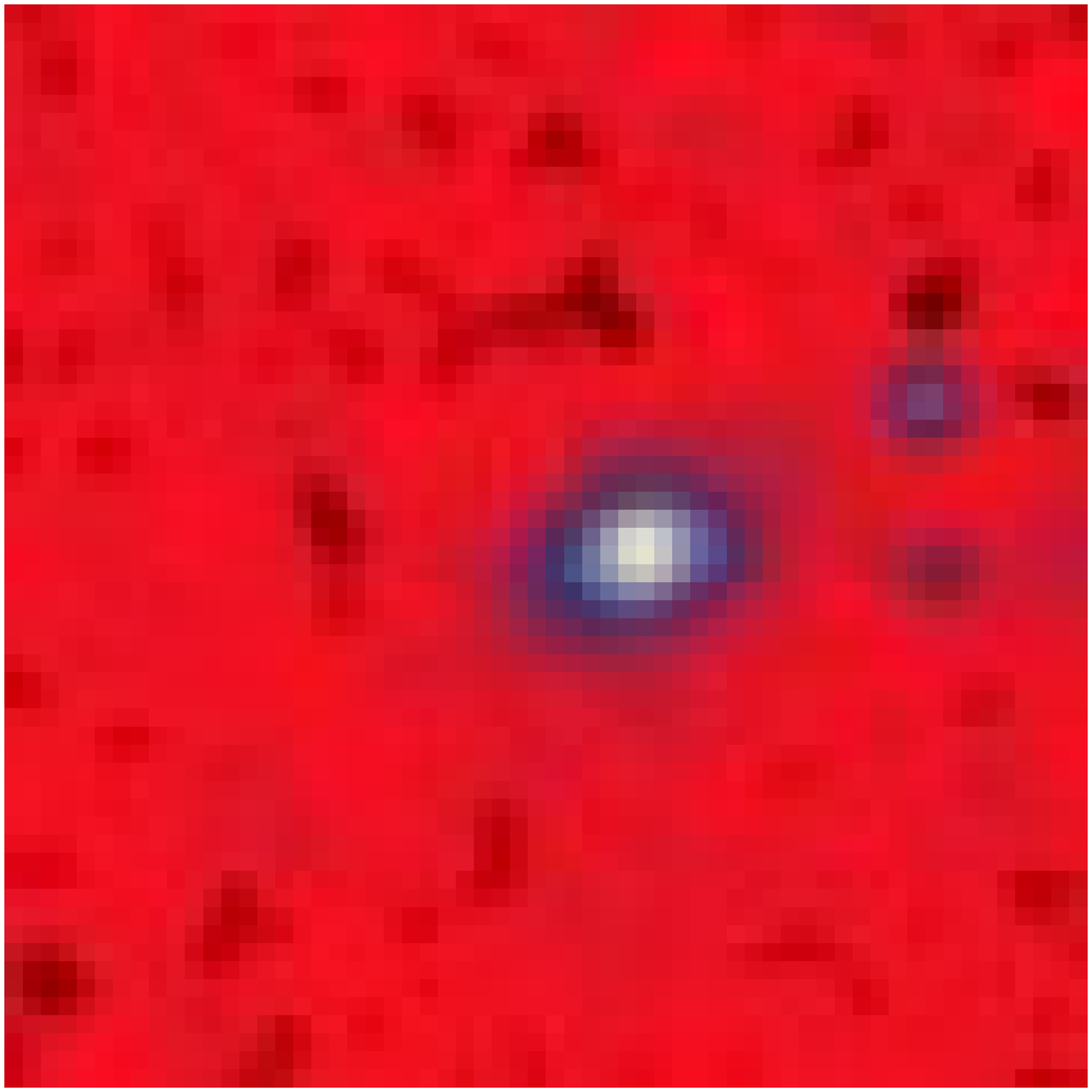}  \FGb{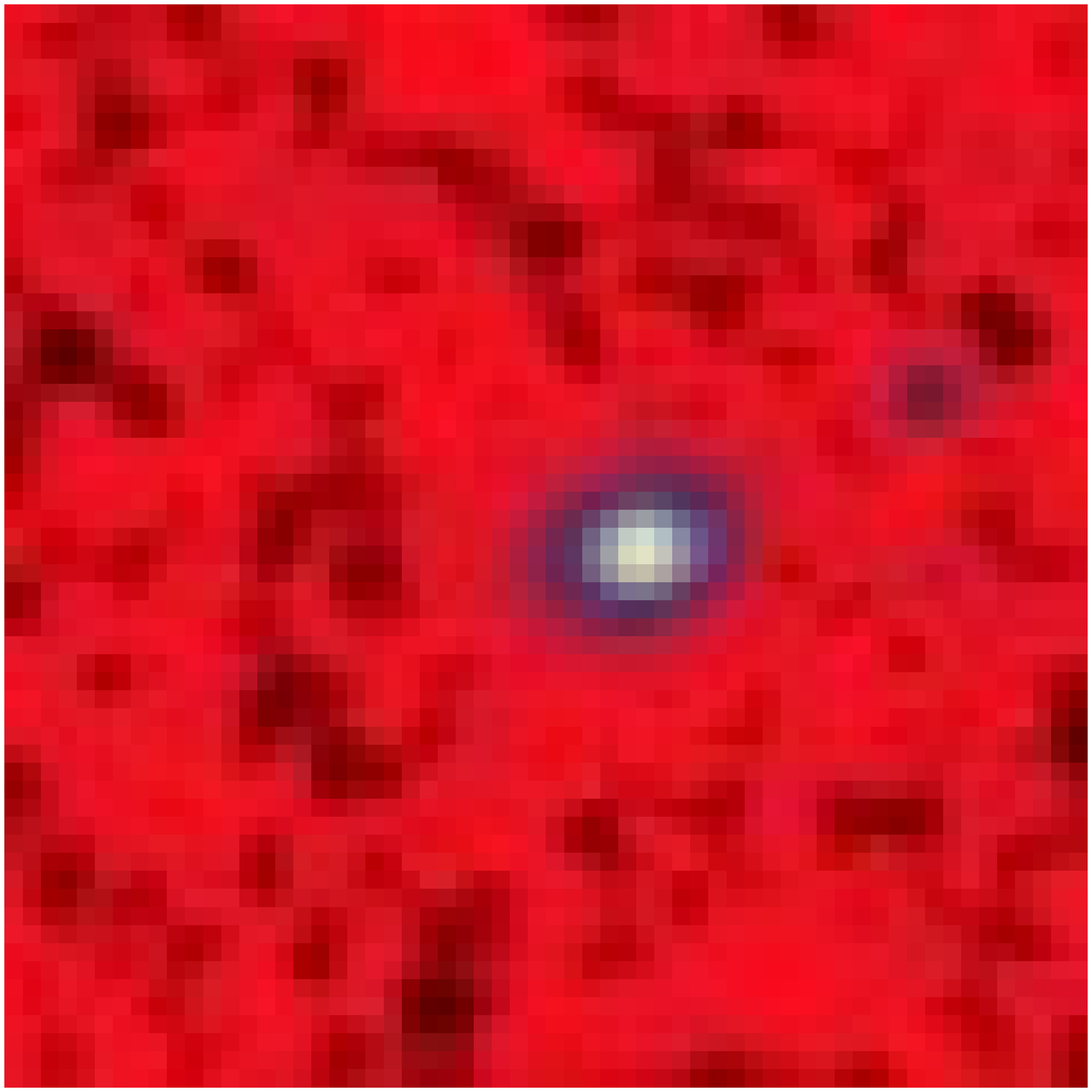}  \FGb{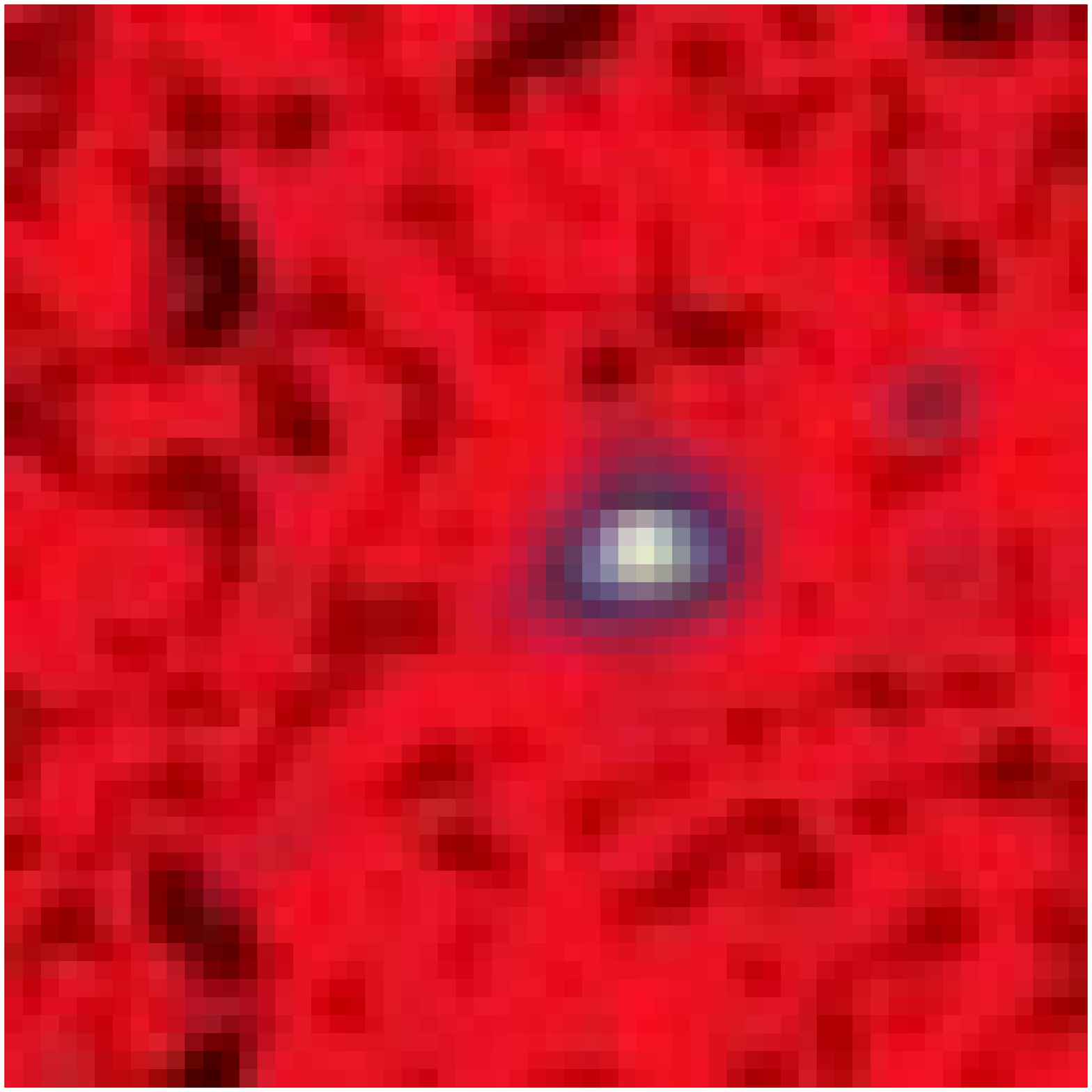}  \FGb{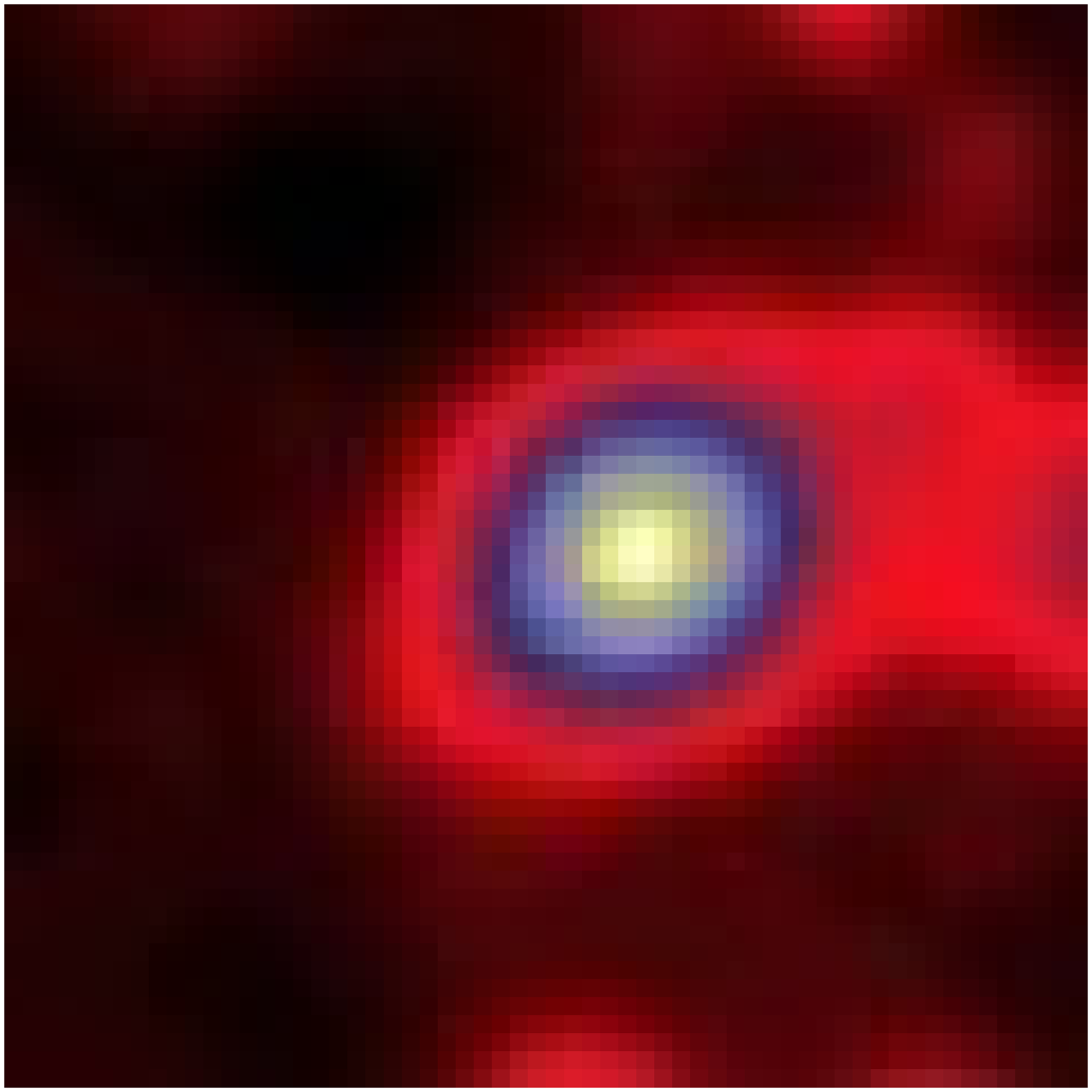}  \FGb{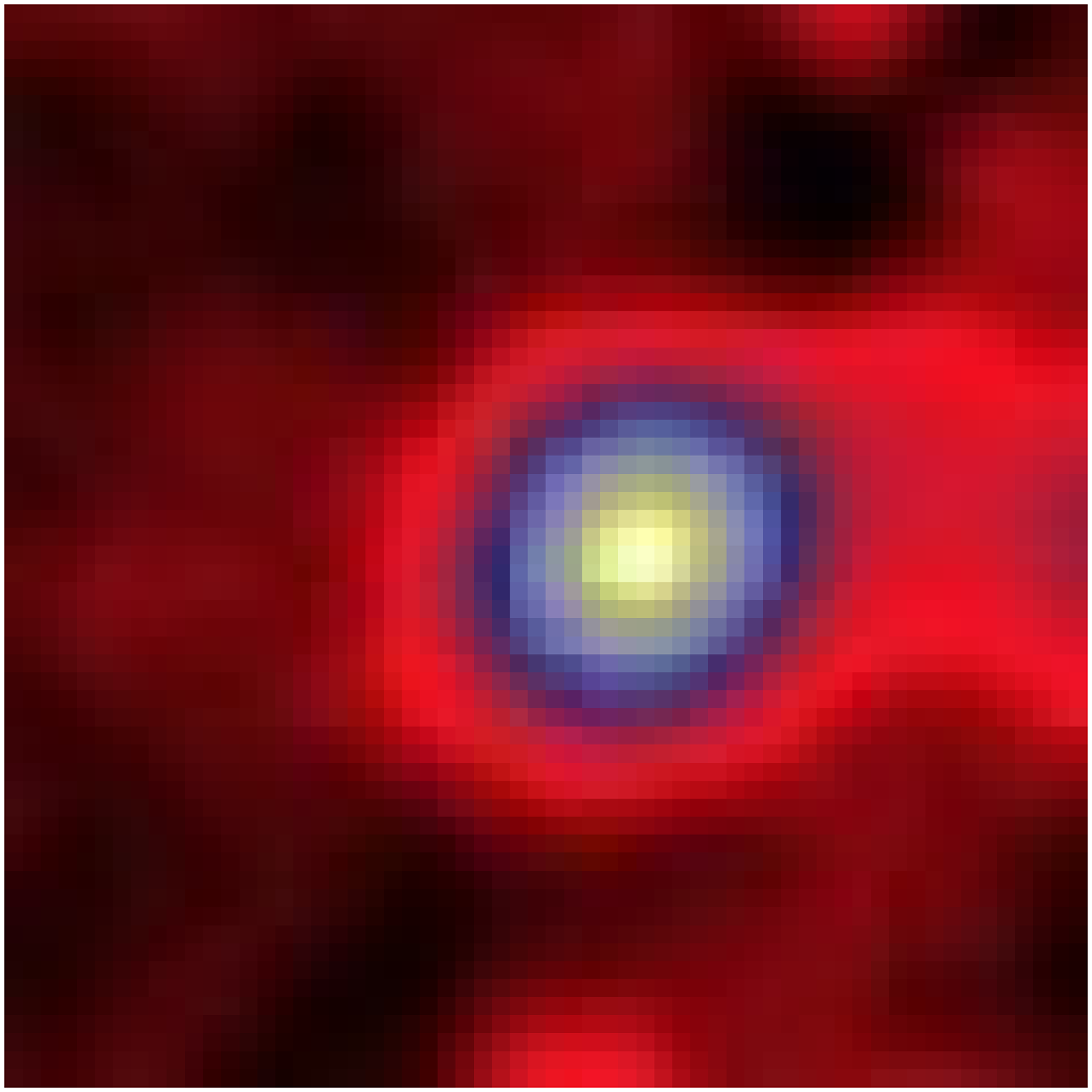}  \FGb{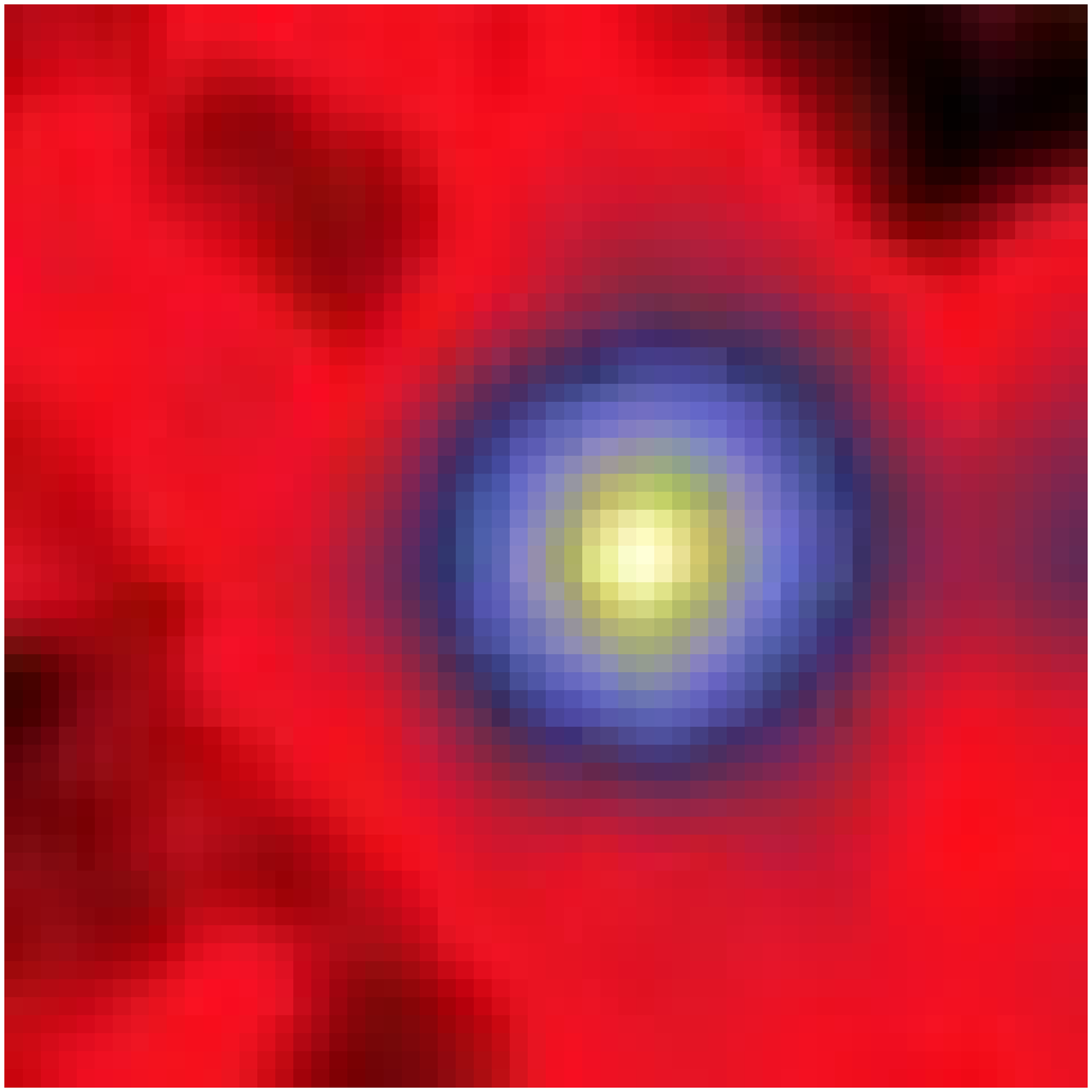}  \FGb{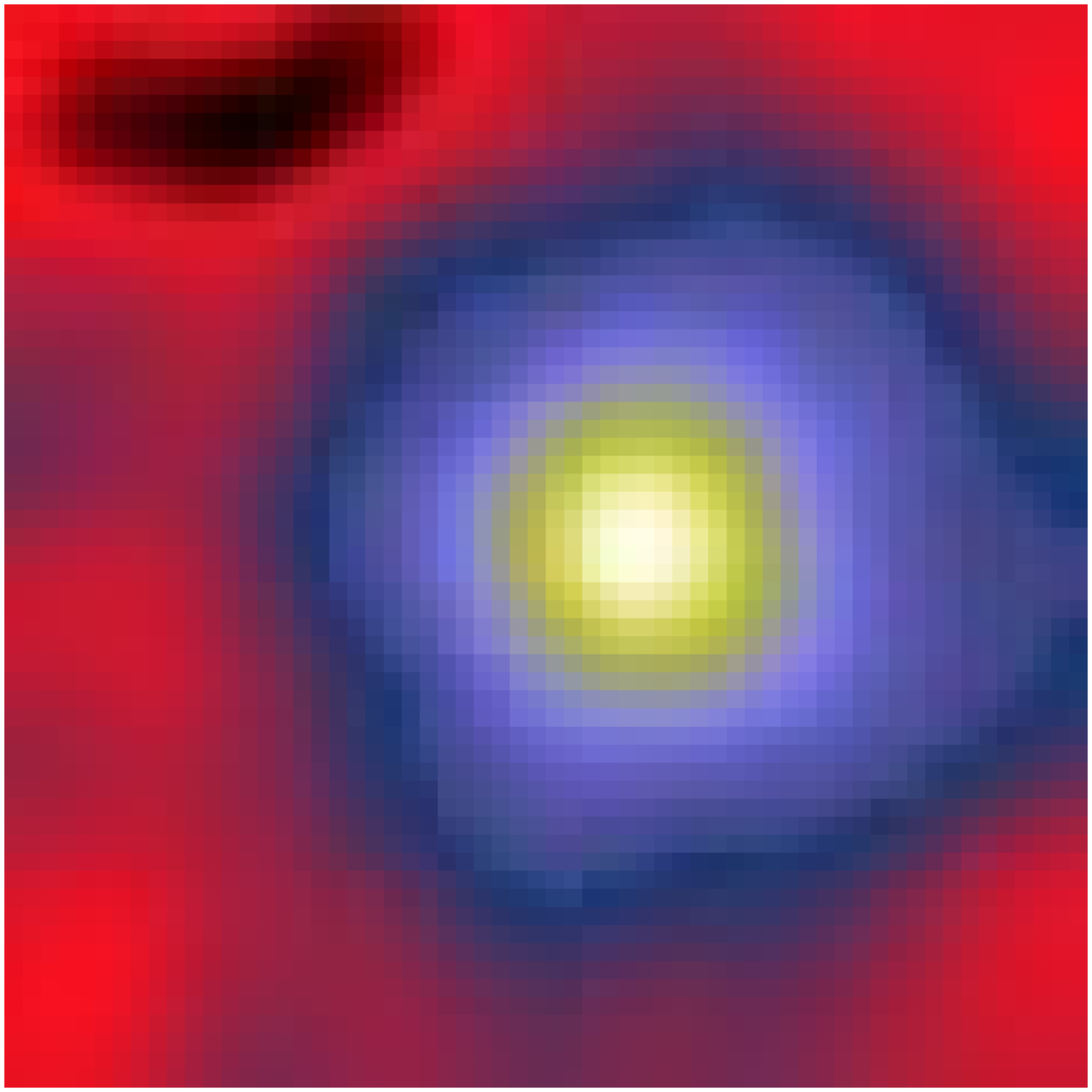} \\  {\#48   NBZ5000254}  \\
  \FGb{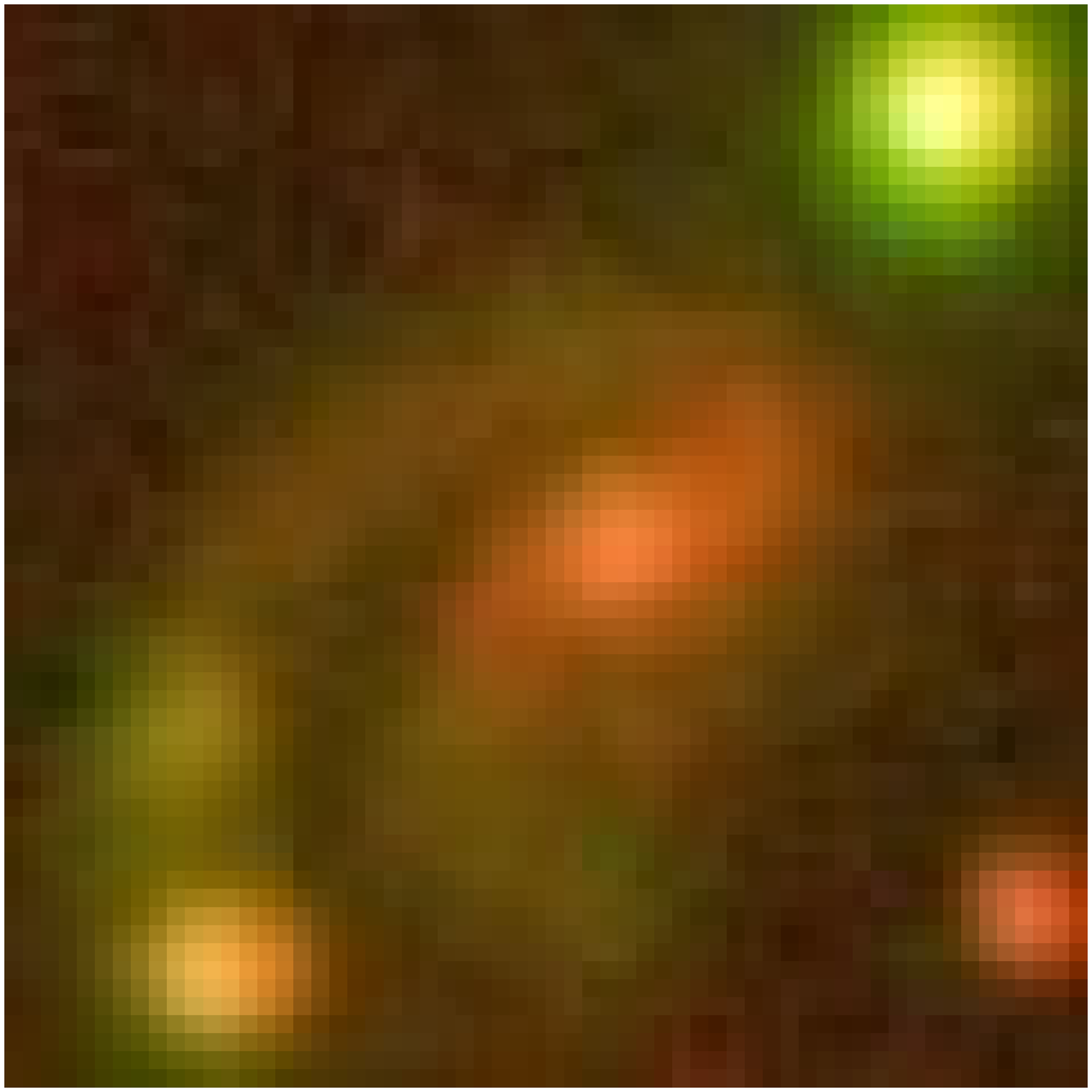}  \FGb{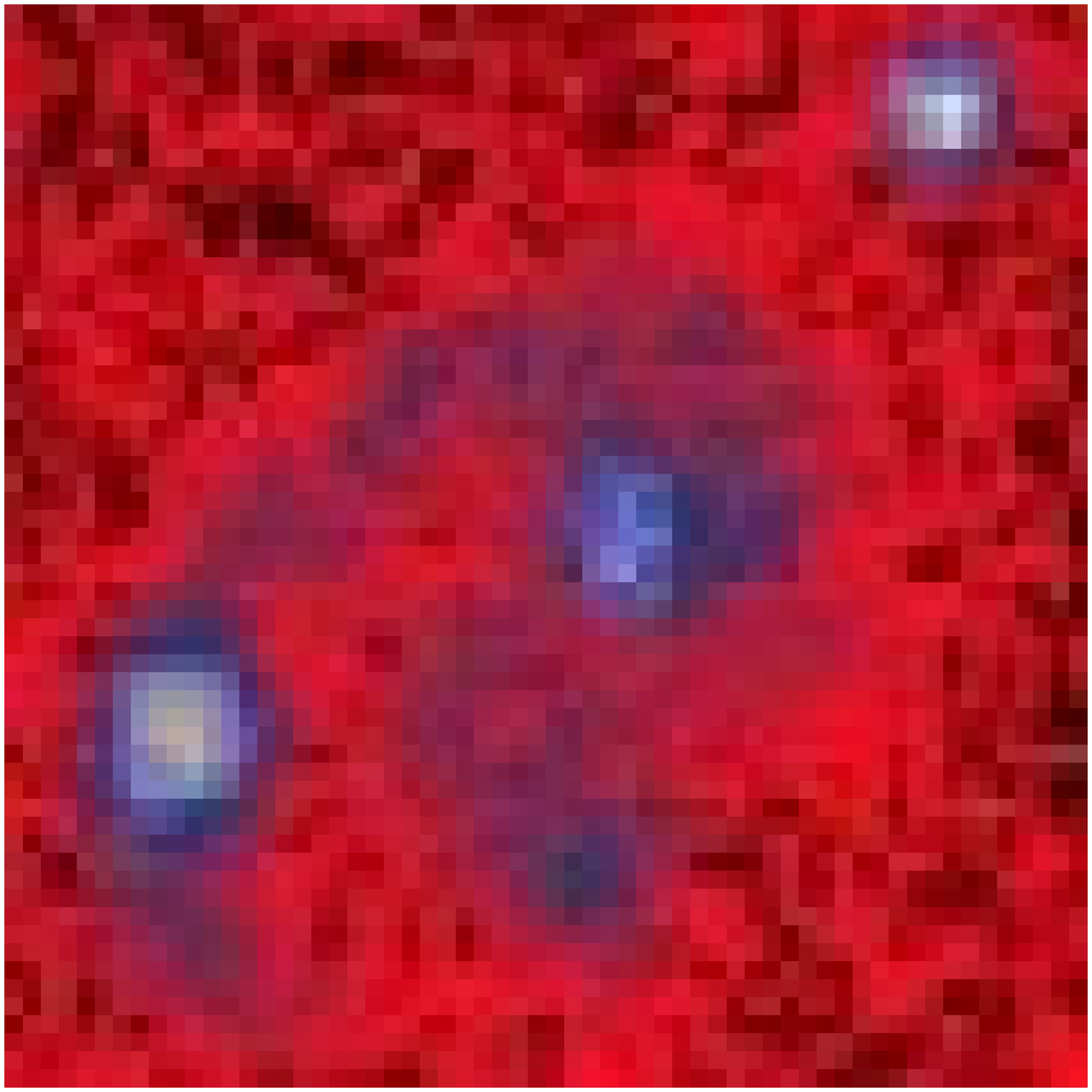}  \FGb{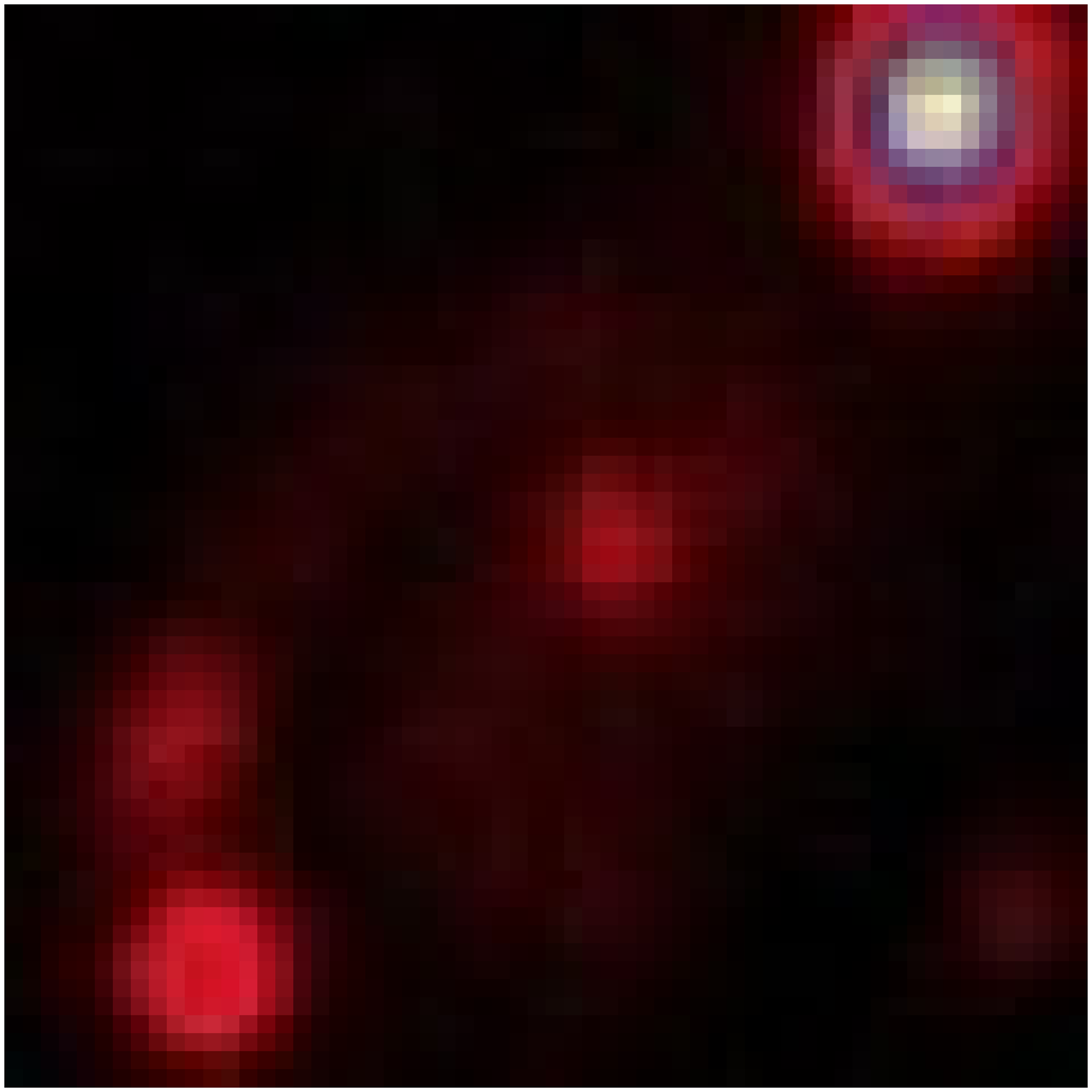}  \FGb{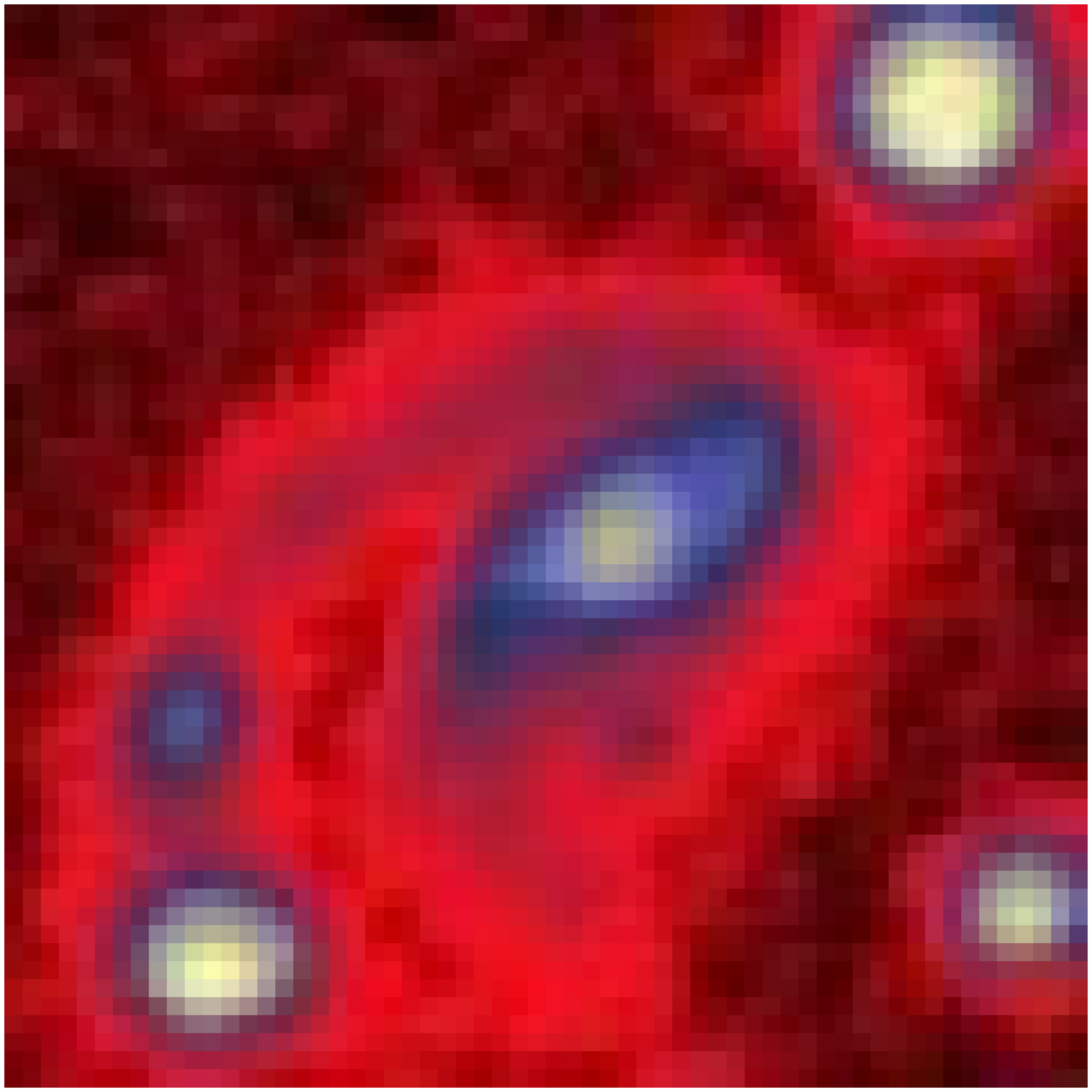}  \FGb{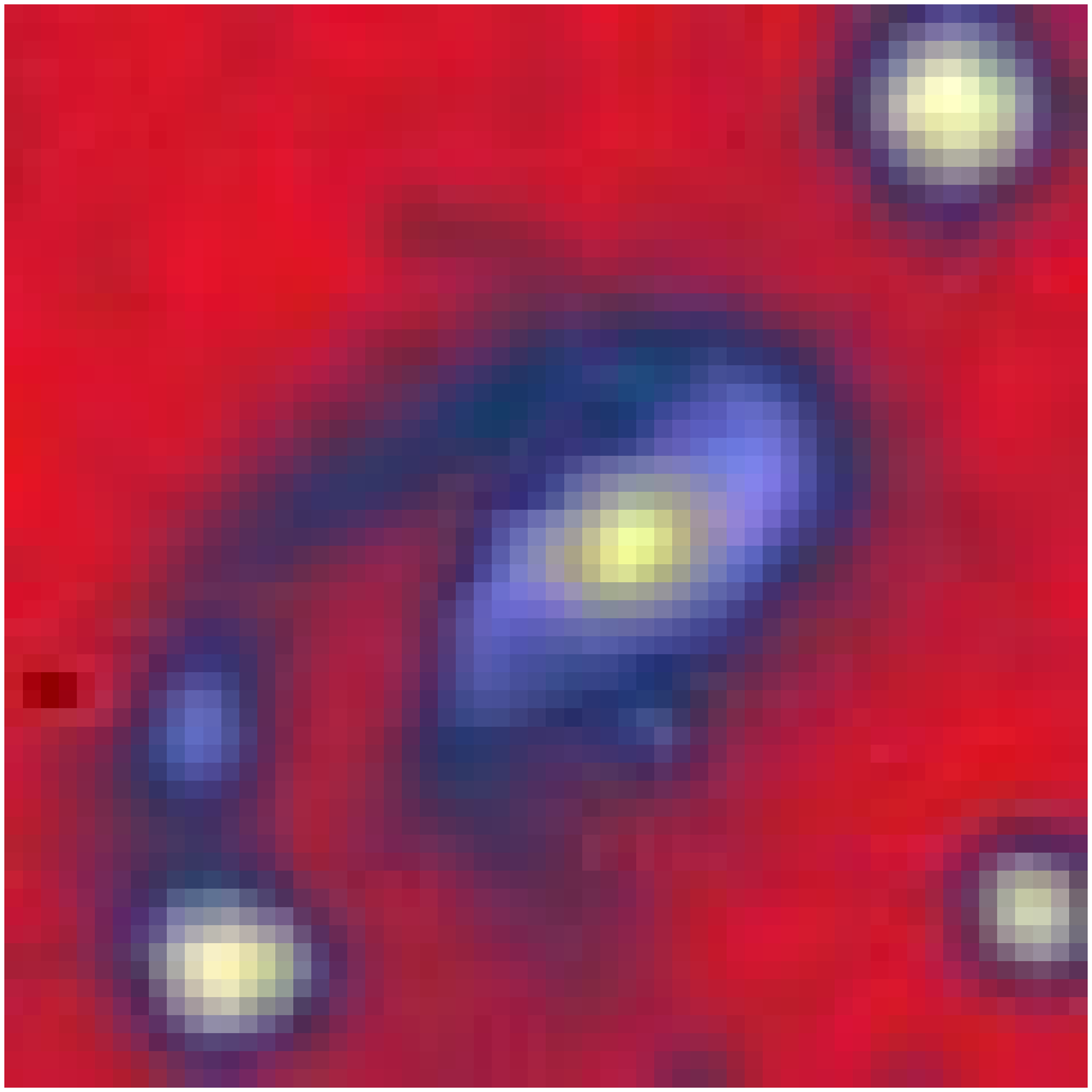}  \FGb{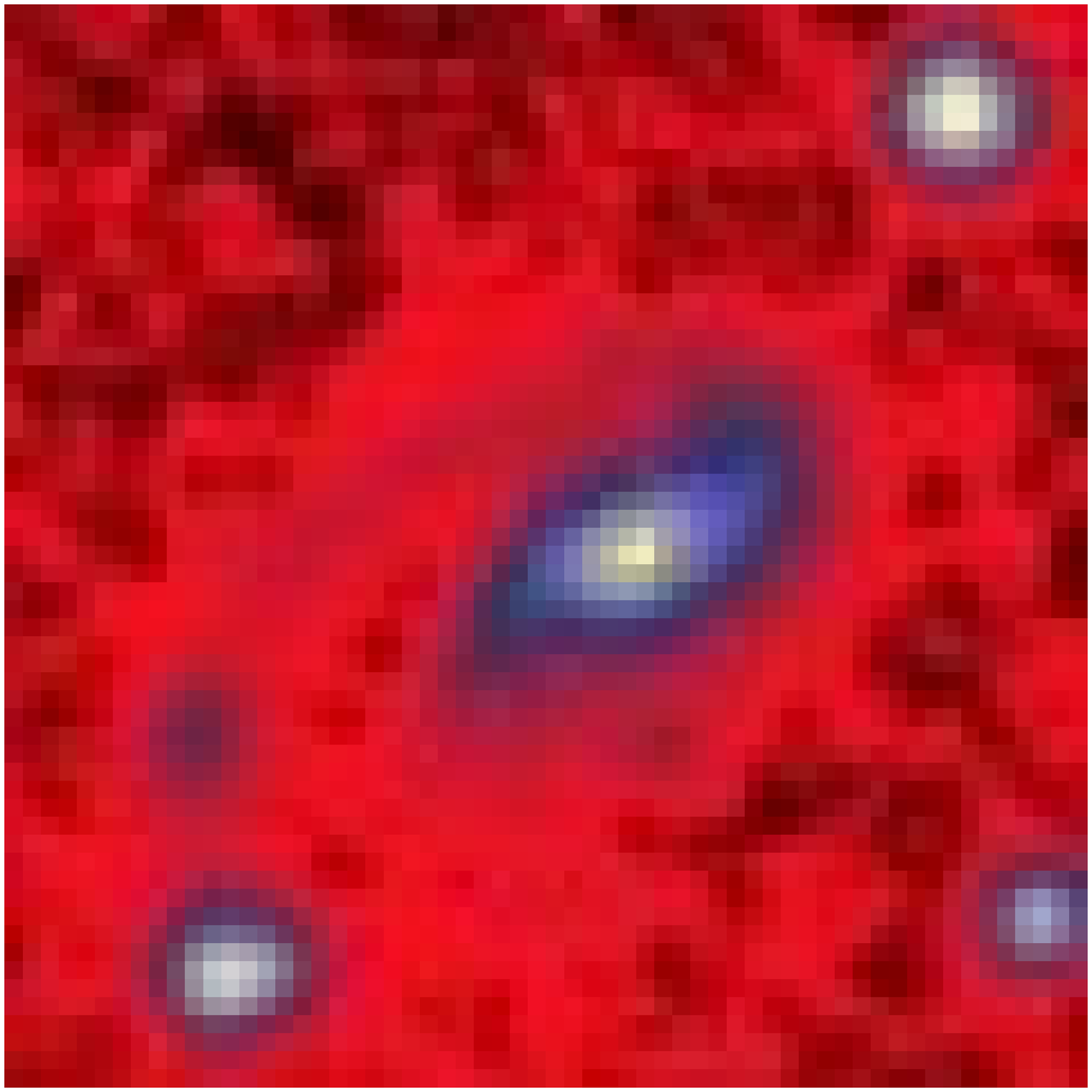}  \FGb{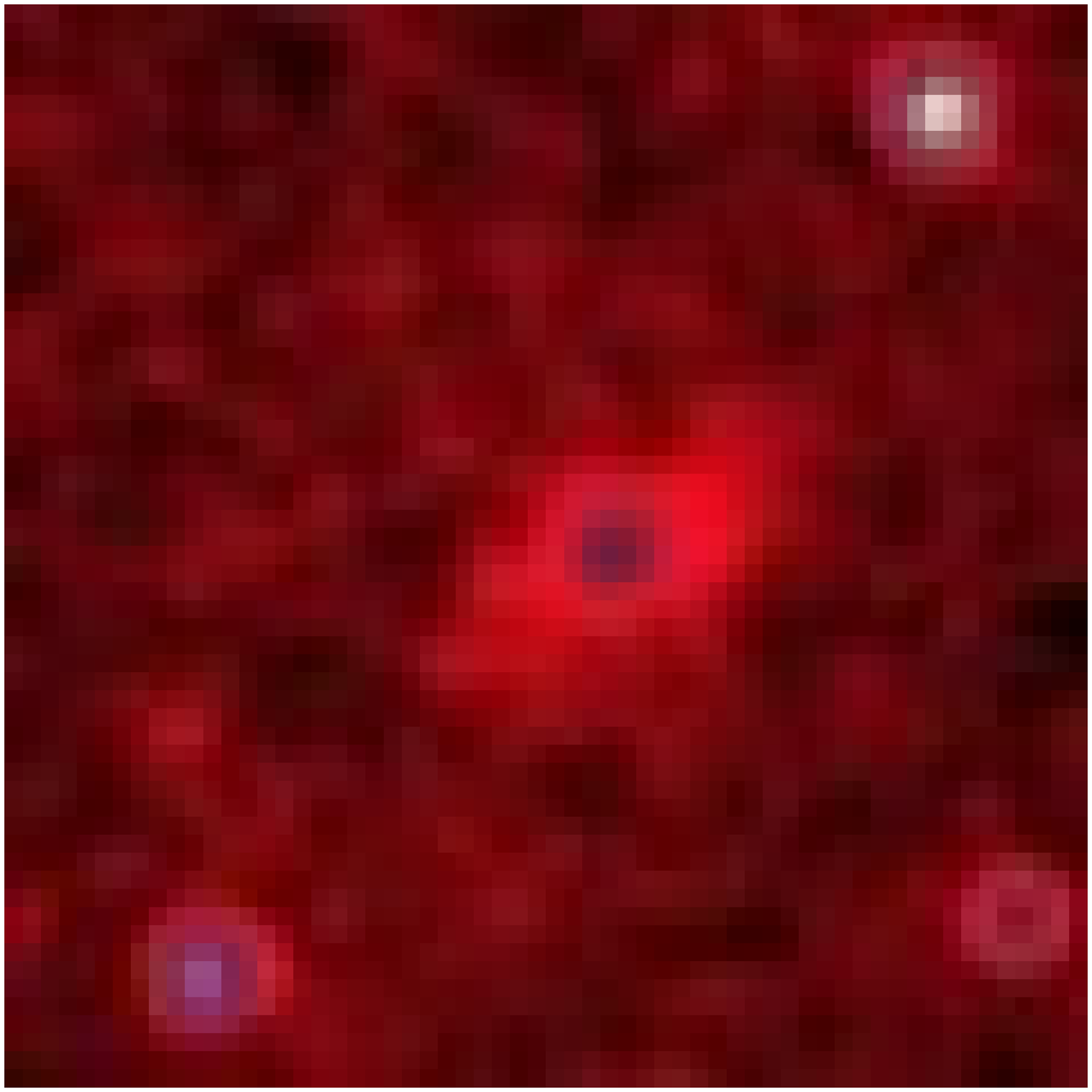}  \FGb{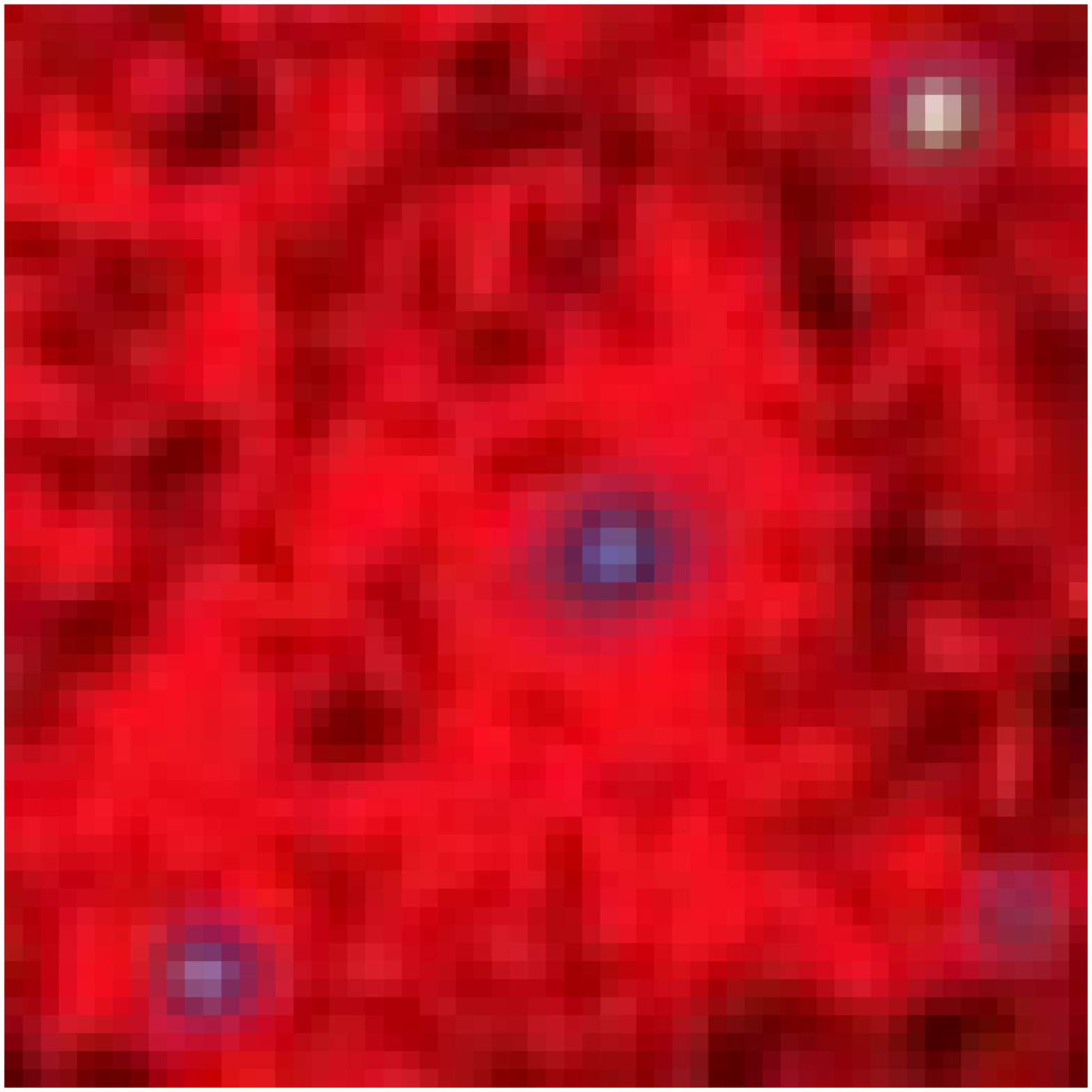}  \FGb{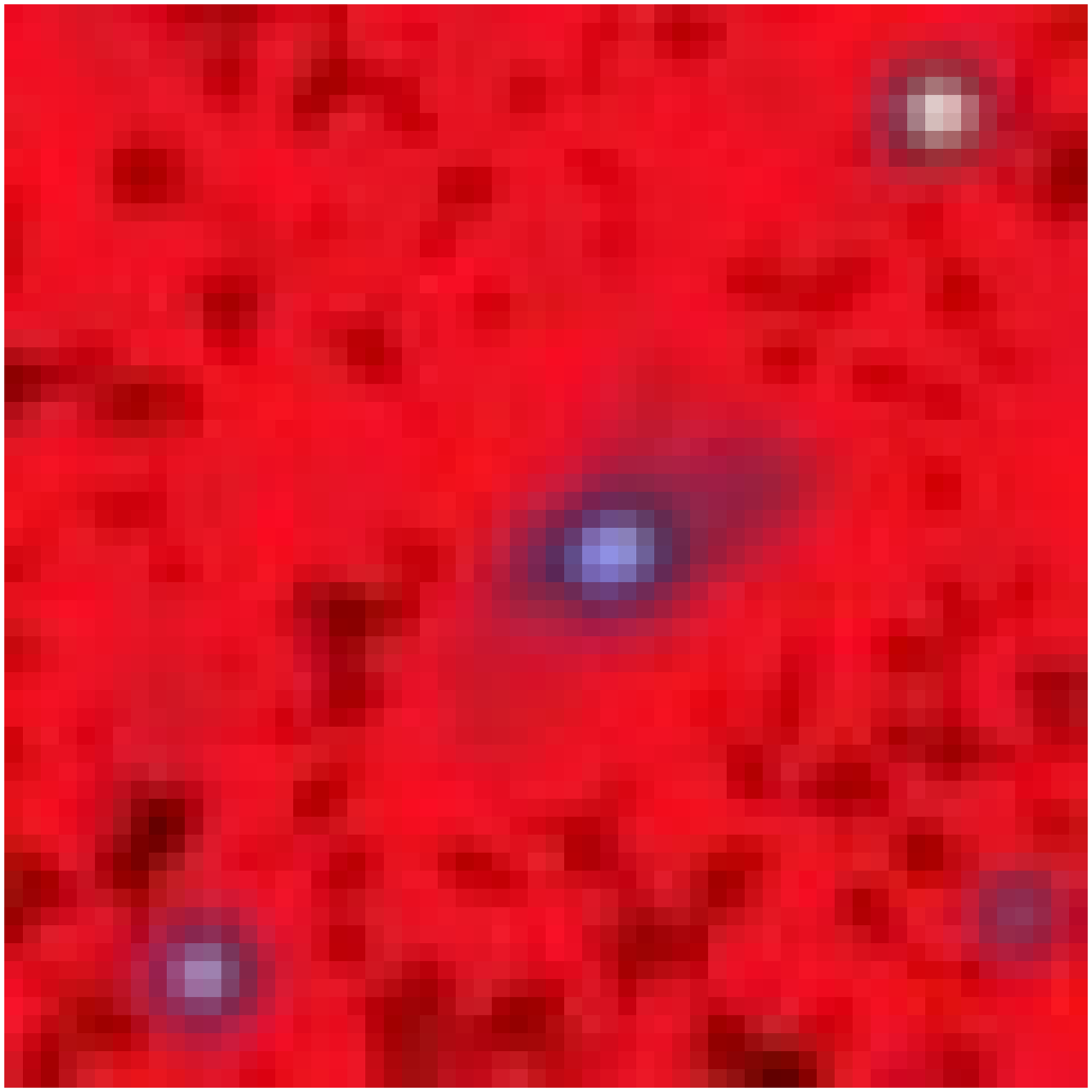}  \FGb{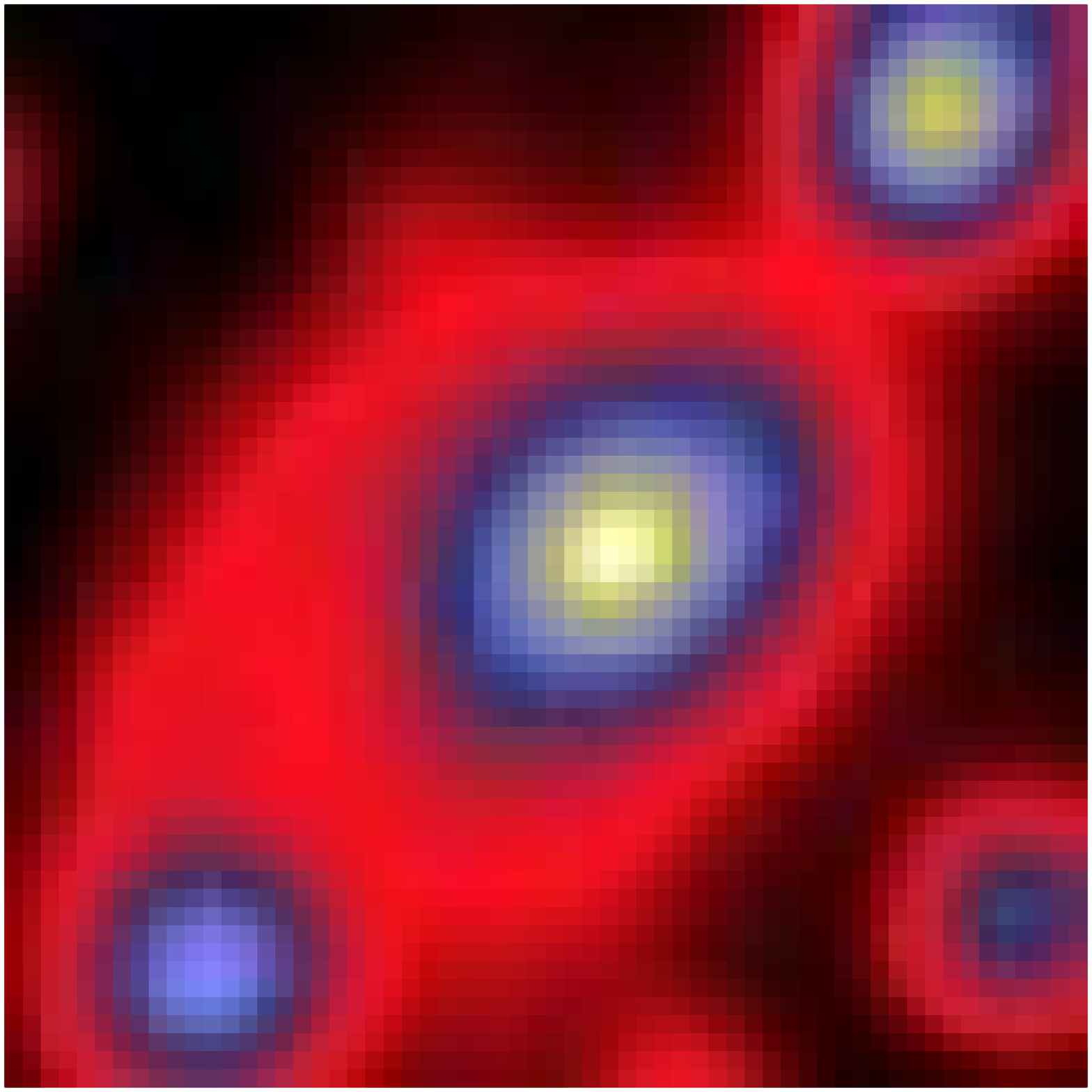}  \FGb{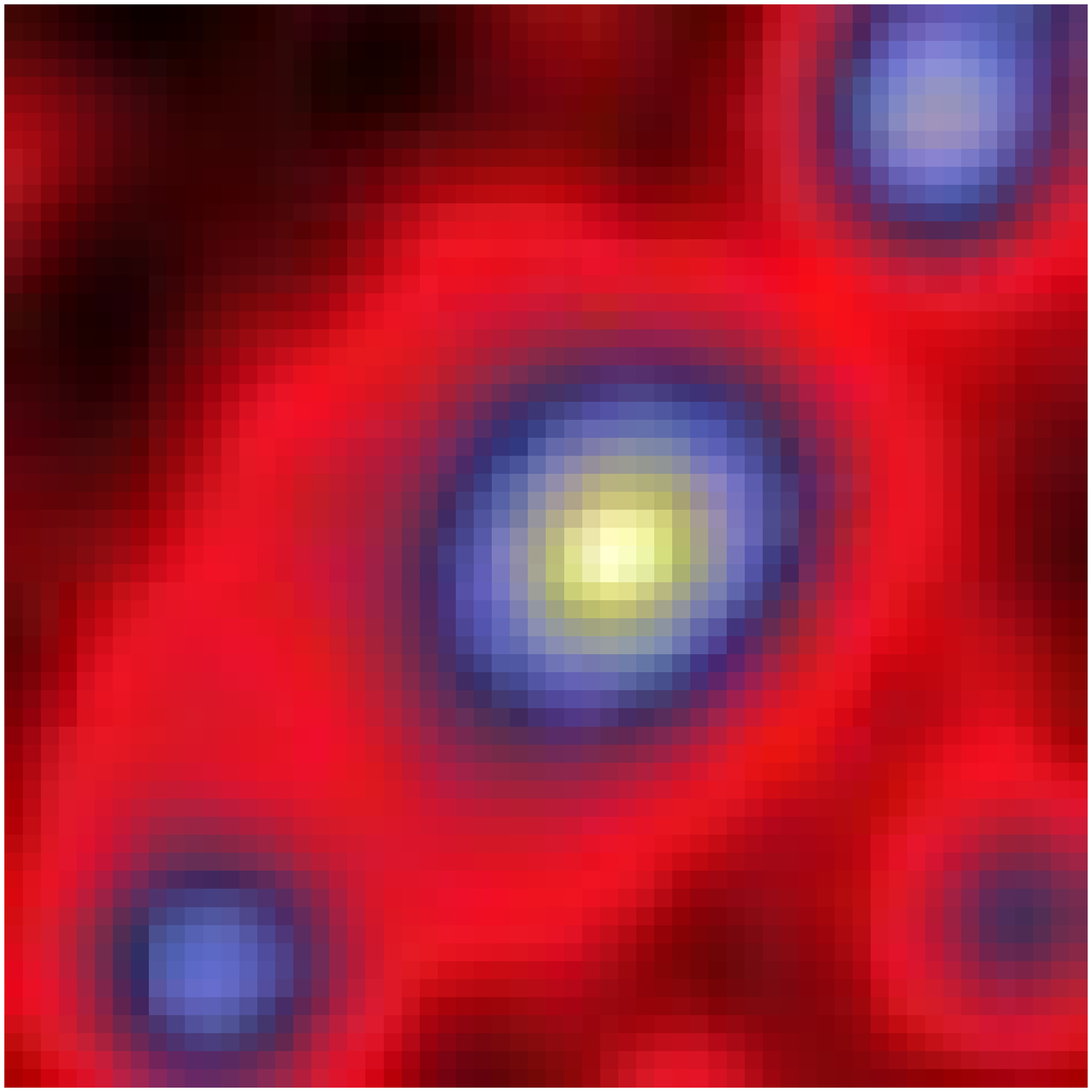}  \FGb{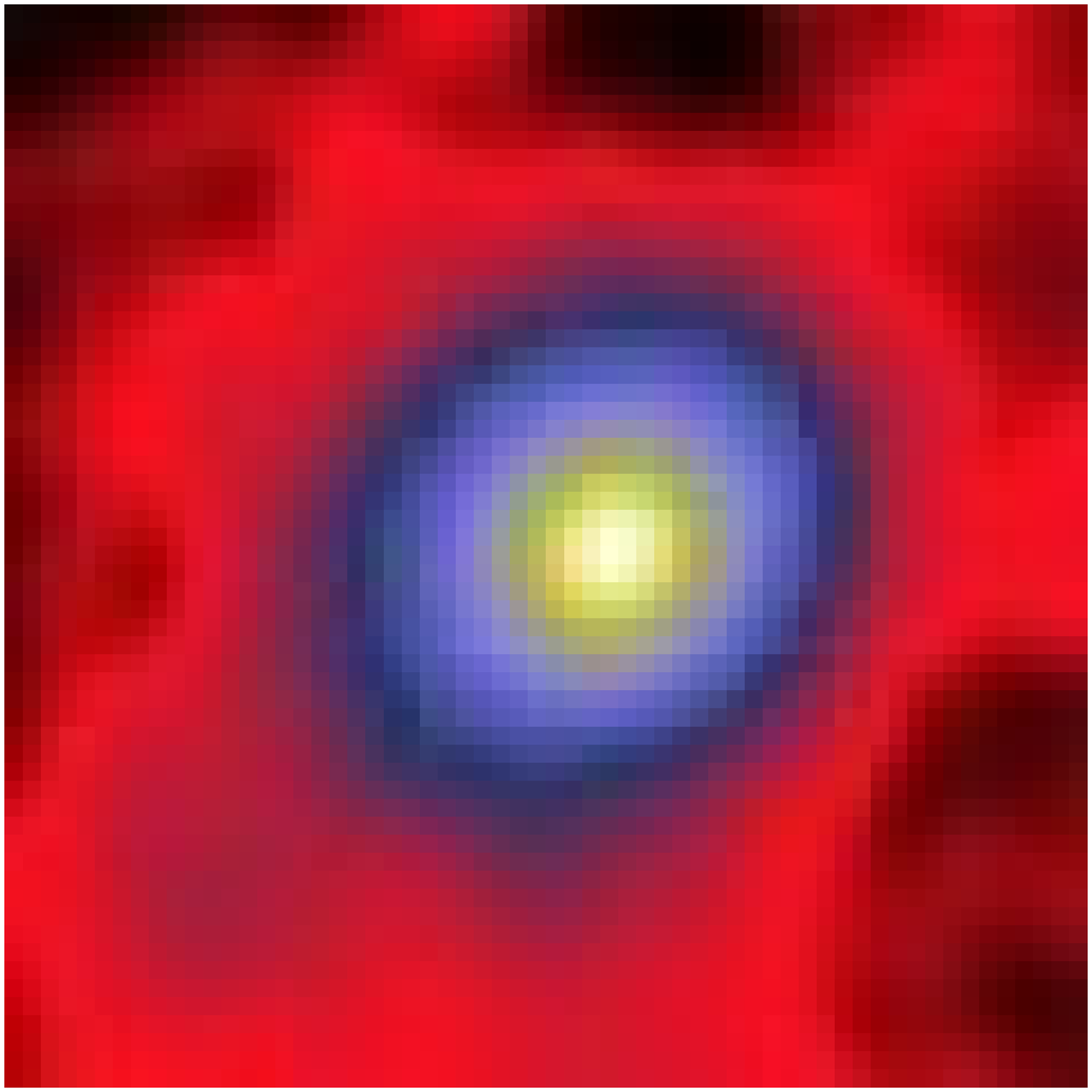}  \FGb{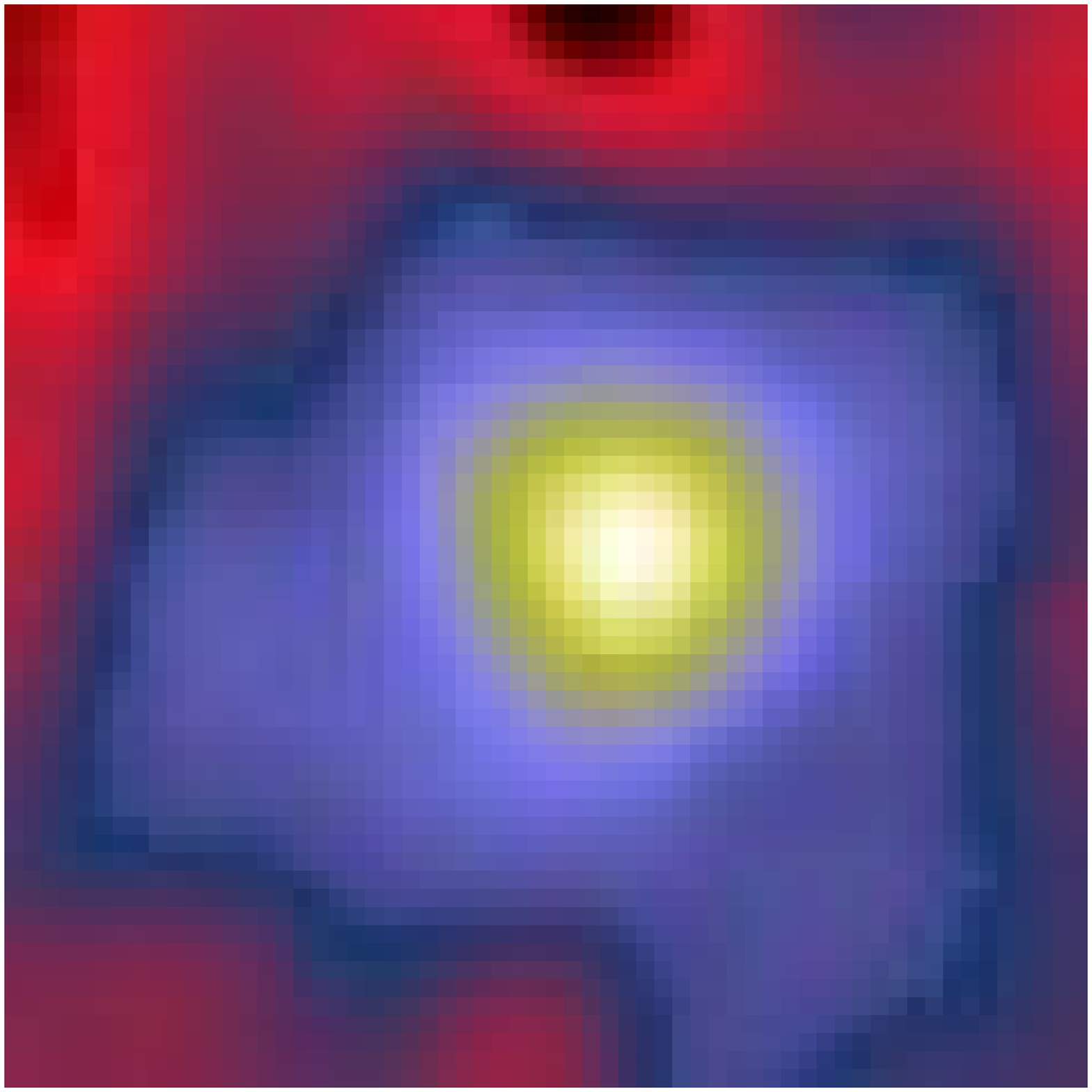} \\  {\#49   NCIA000887}  \\
  \FGb{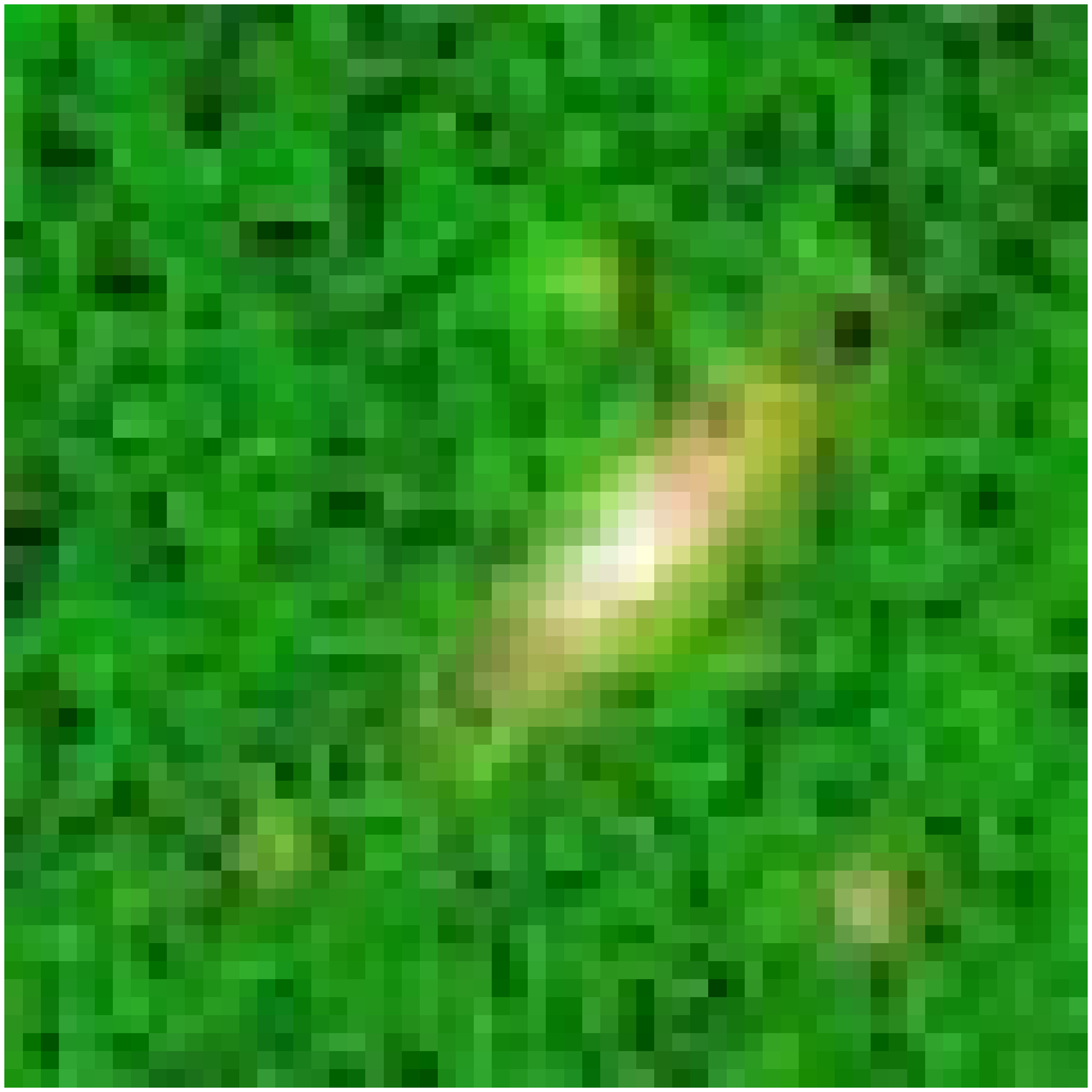}  \FGb{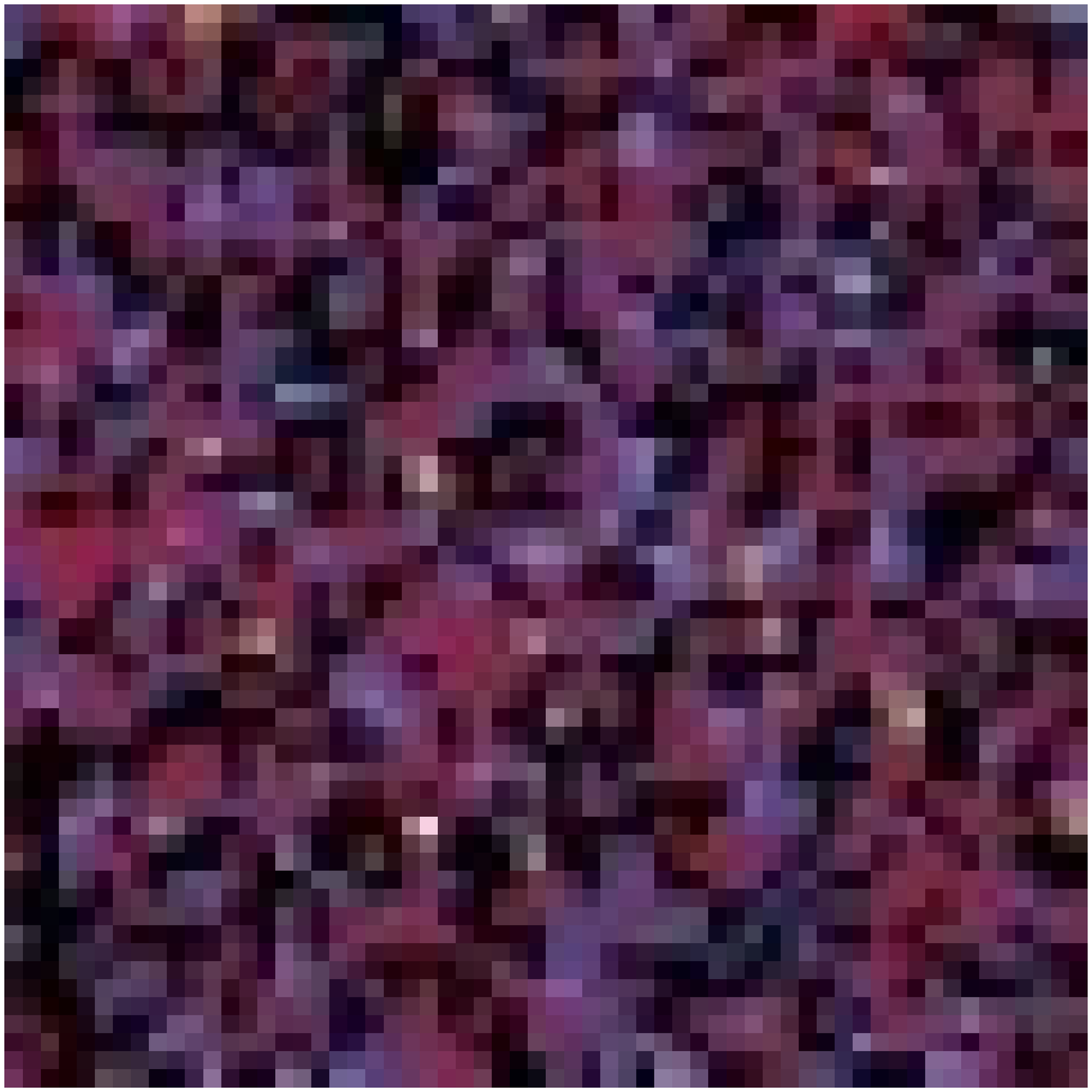}  \FGb{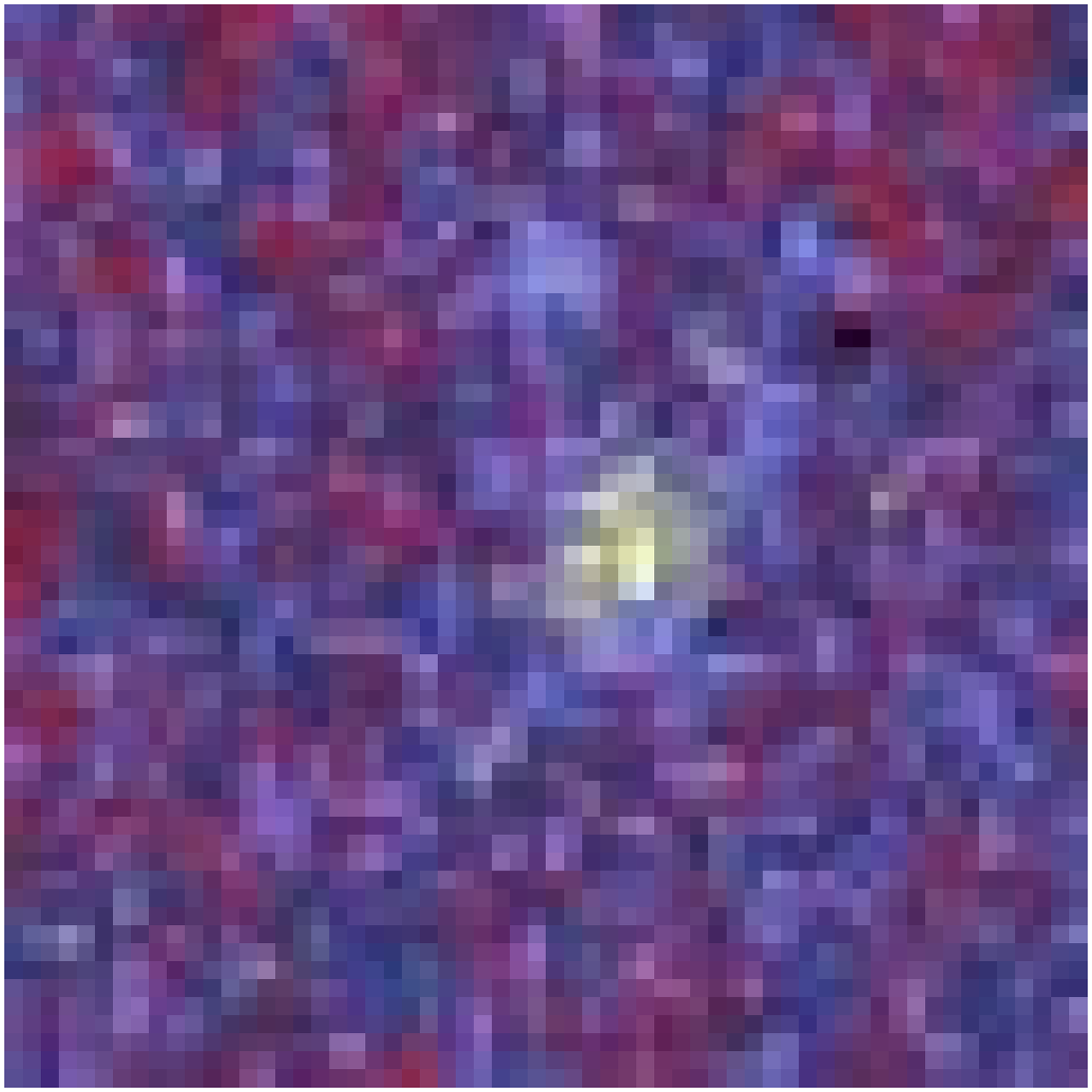}  \FGb{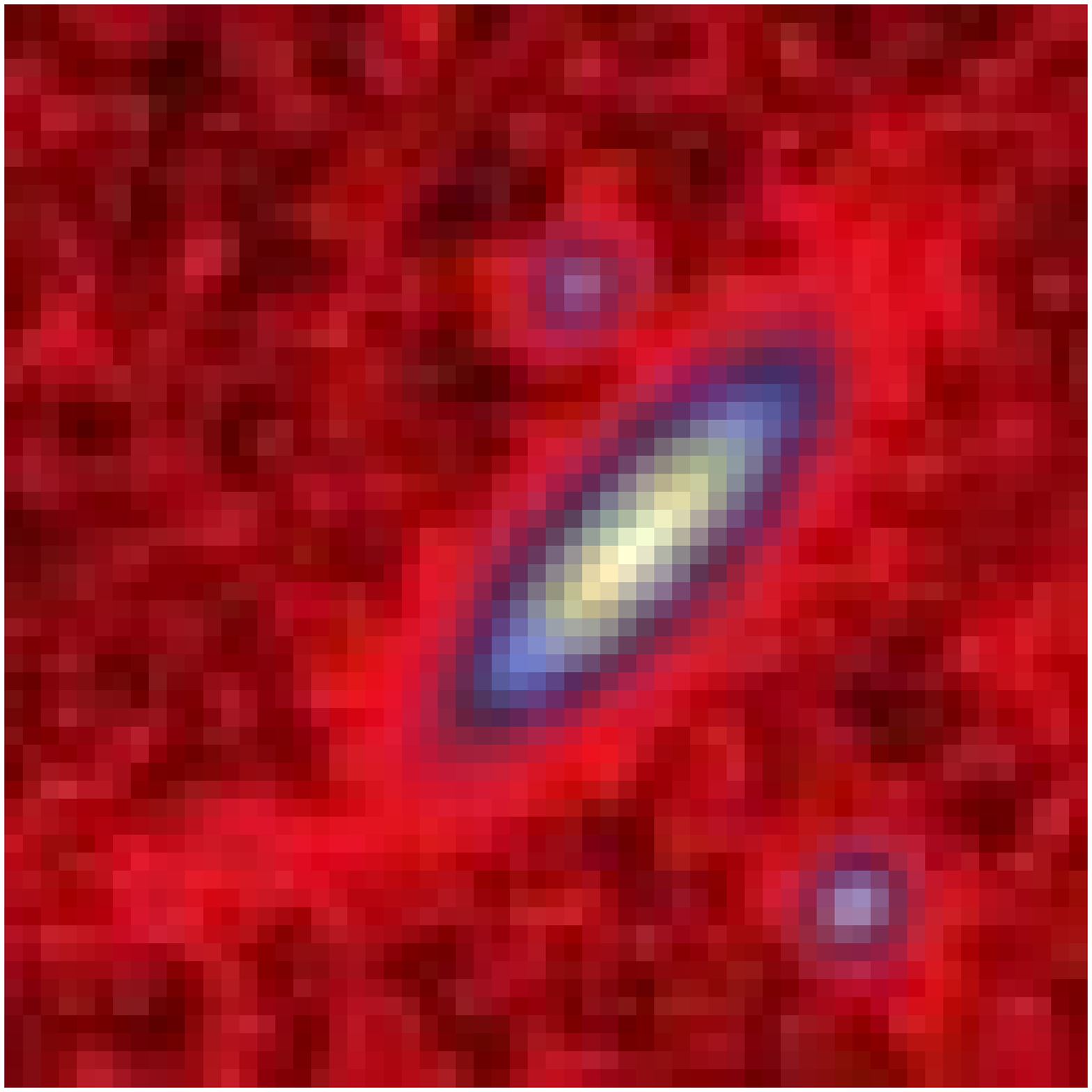}  \FGb{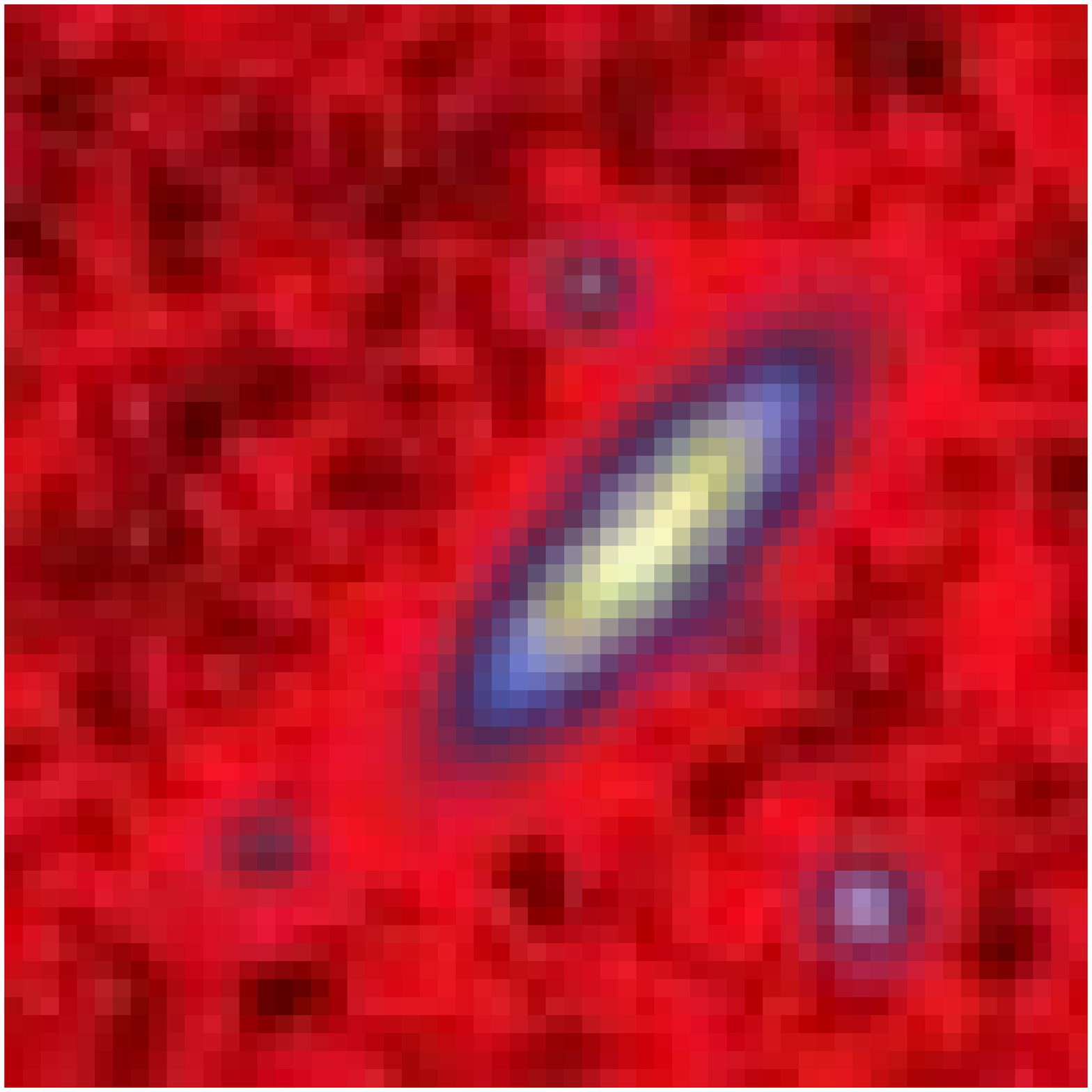}  \FGb{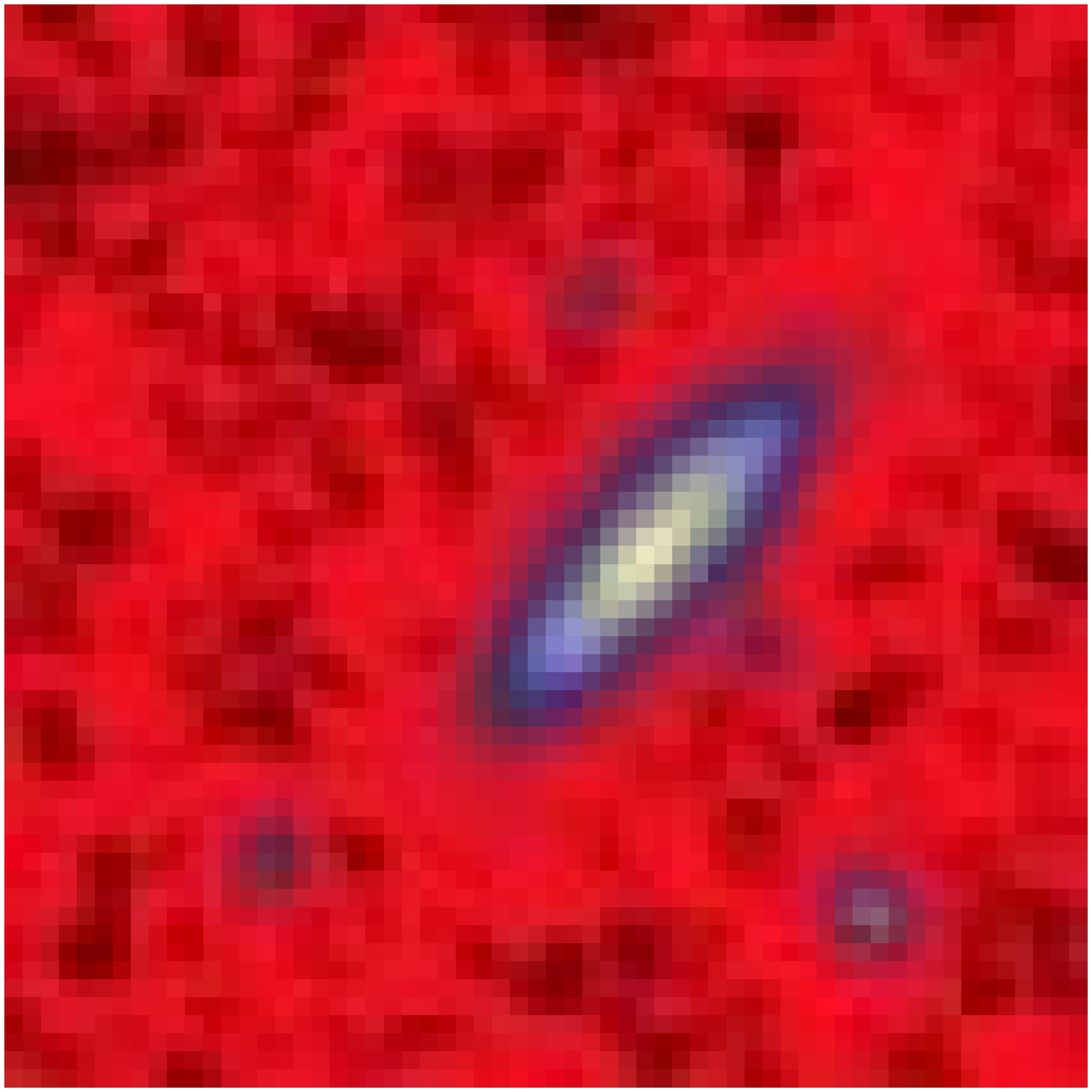}  \FGb{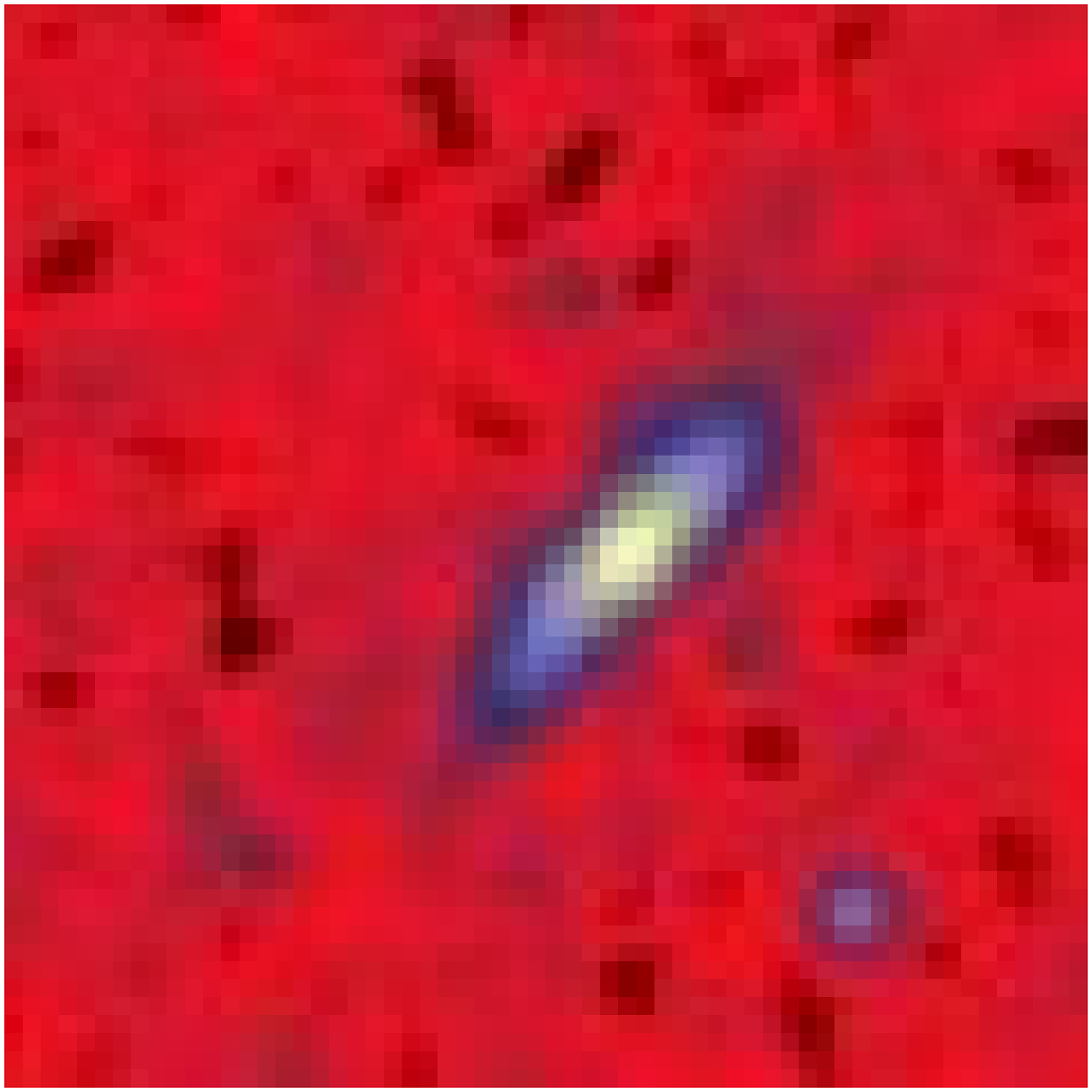}  \FGb{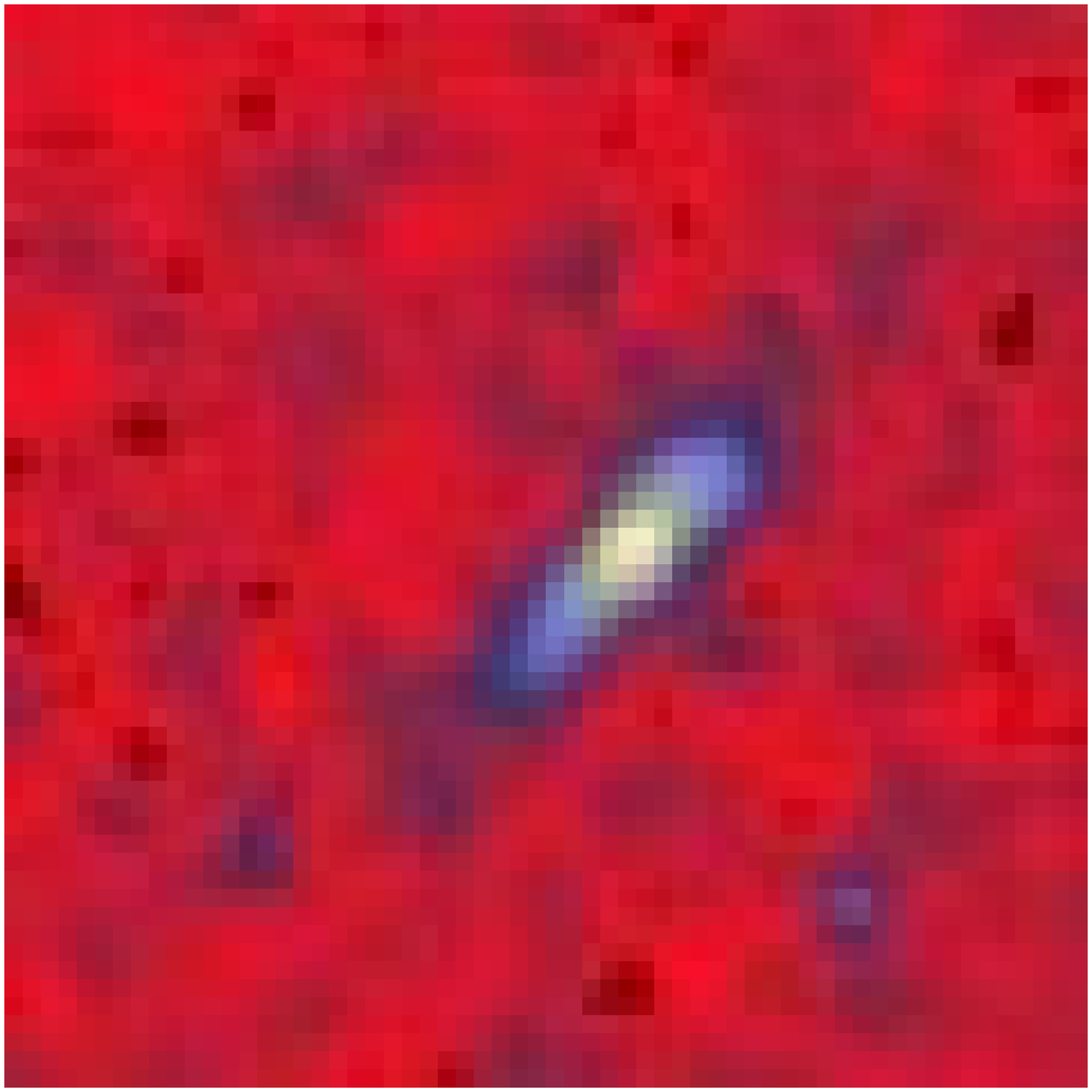}  \FGb{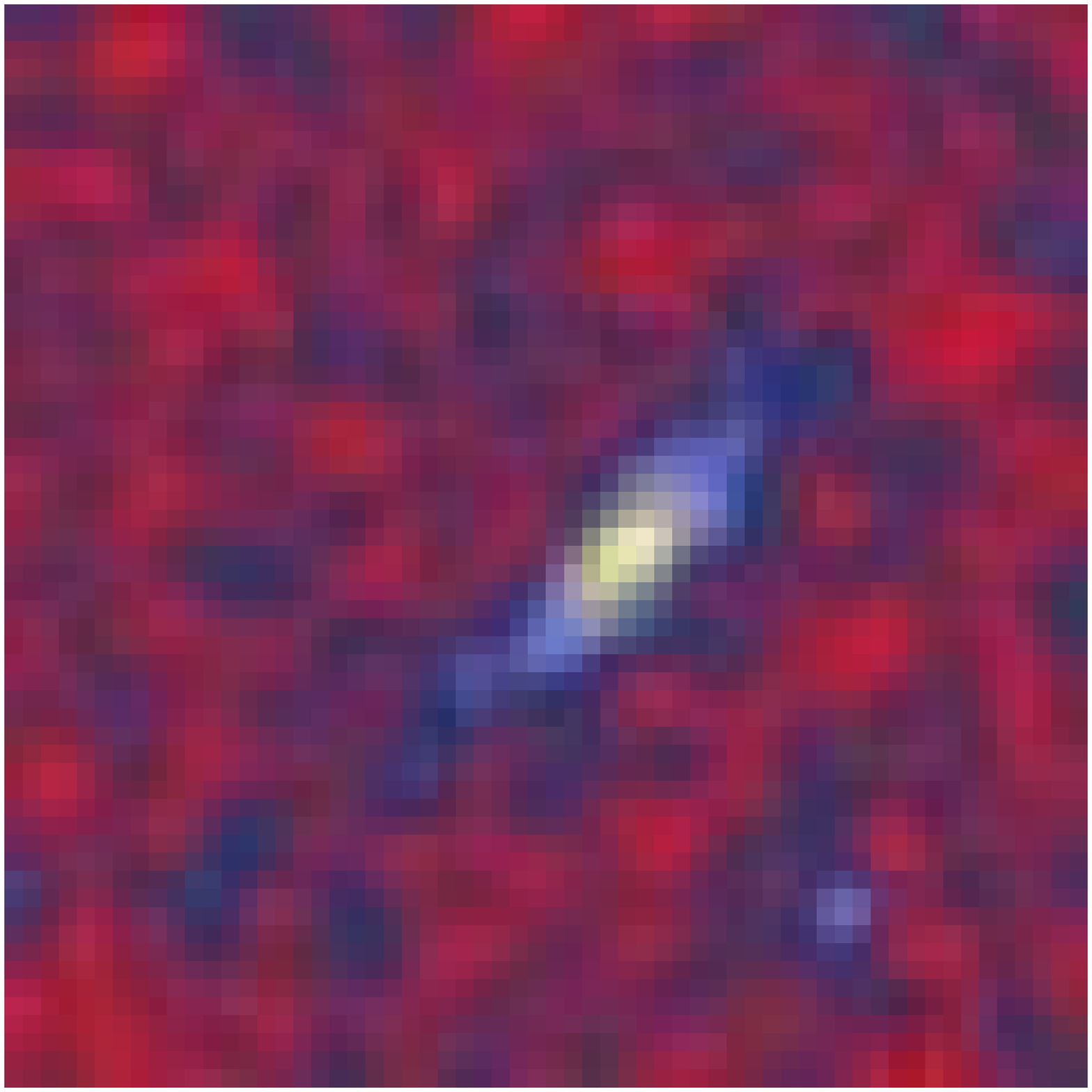}  \FGb{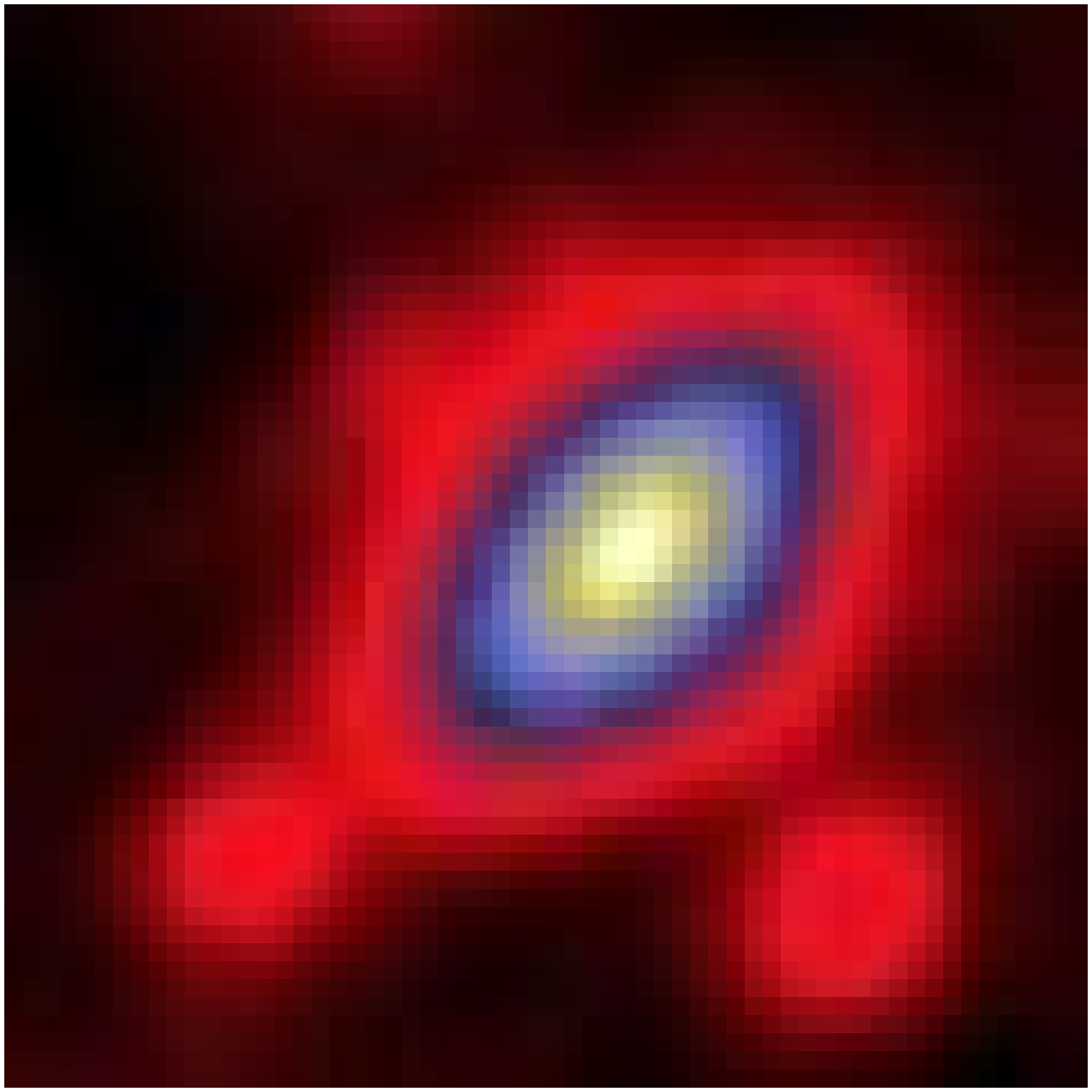}  \FGb{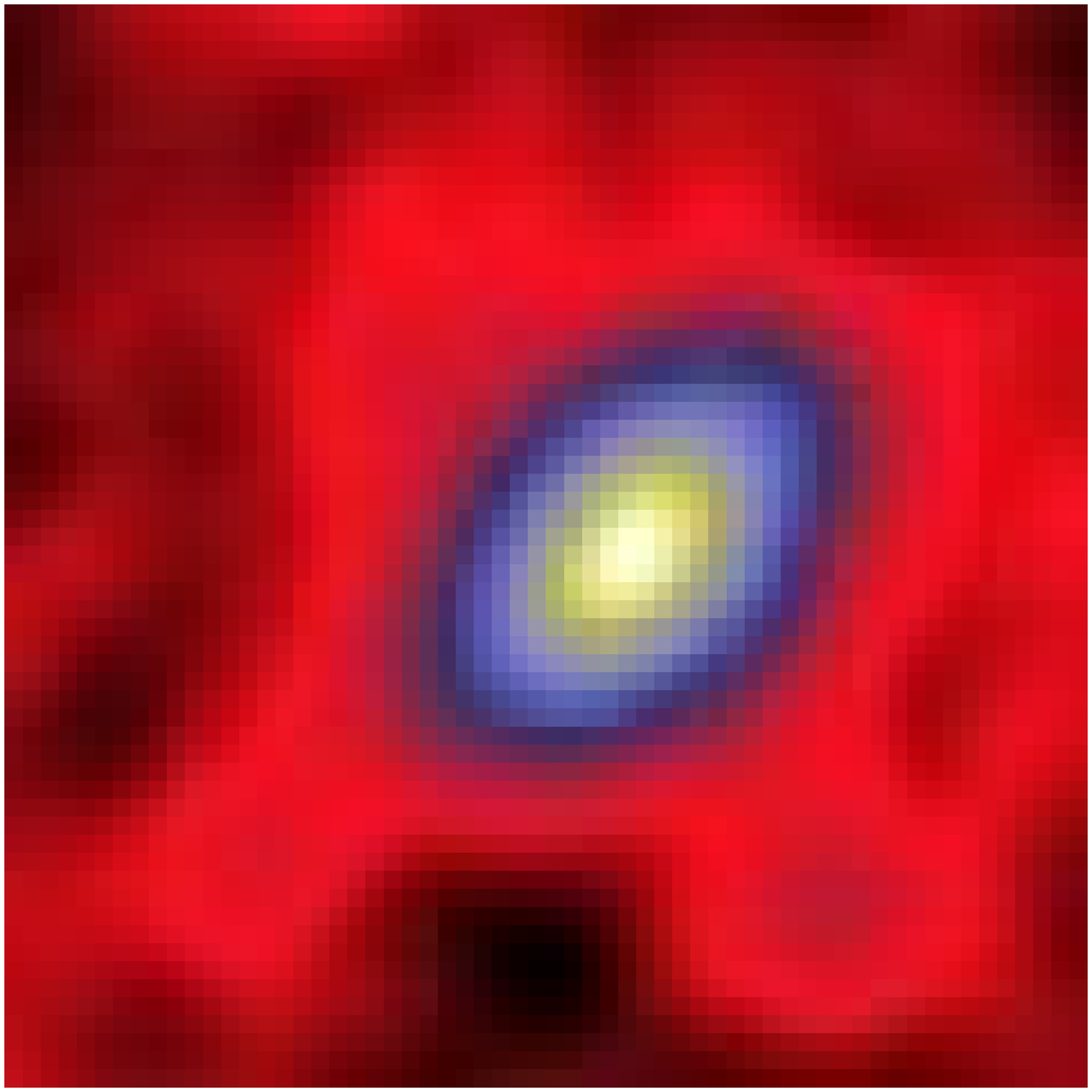}  \FGb{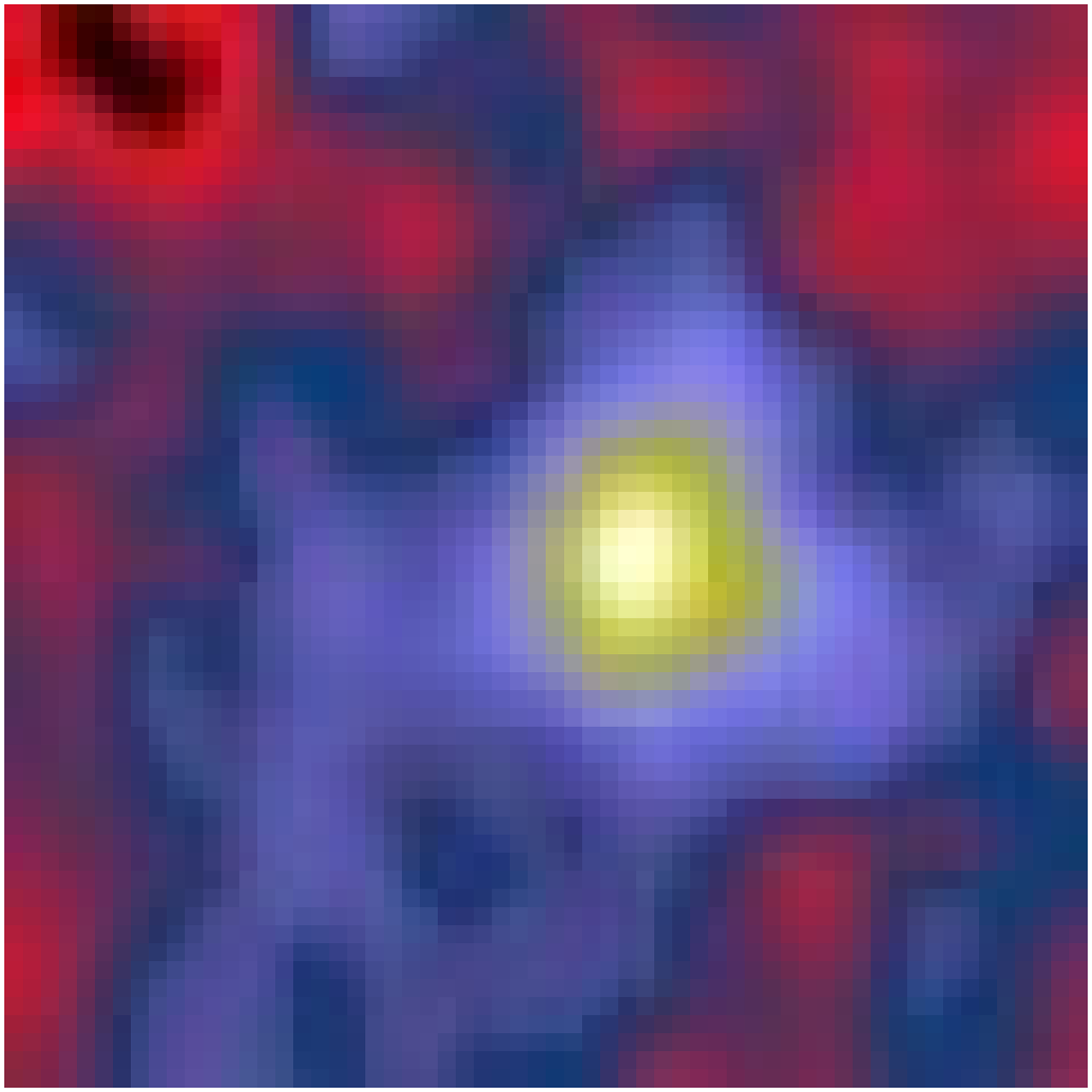}  \FGb{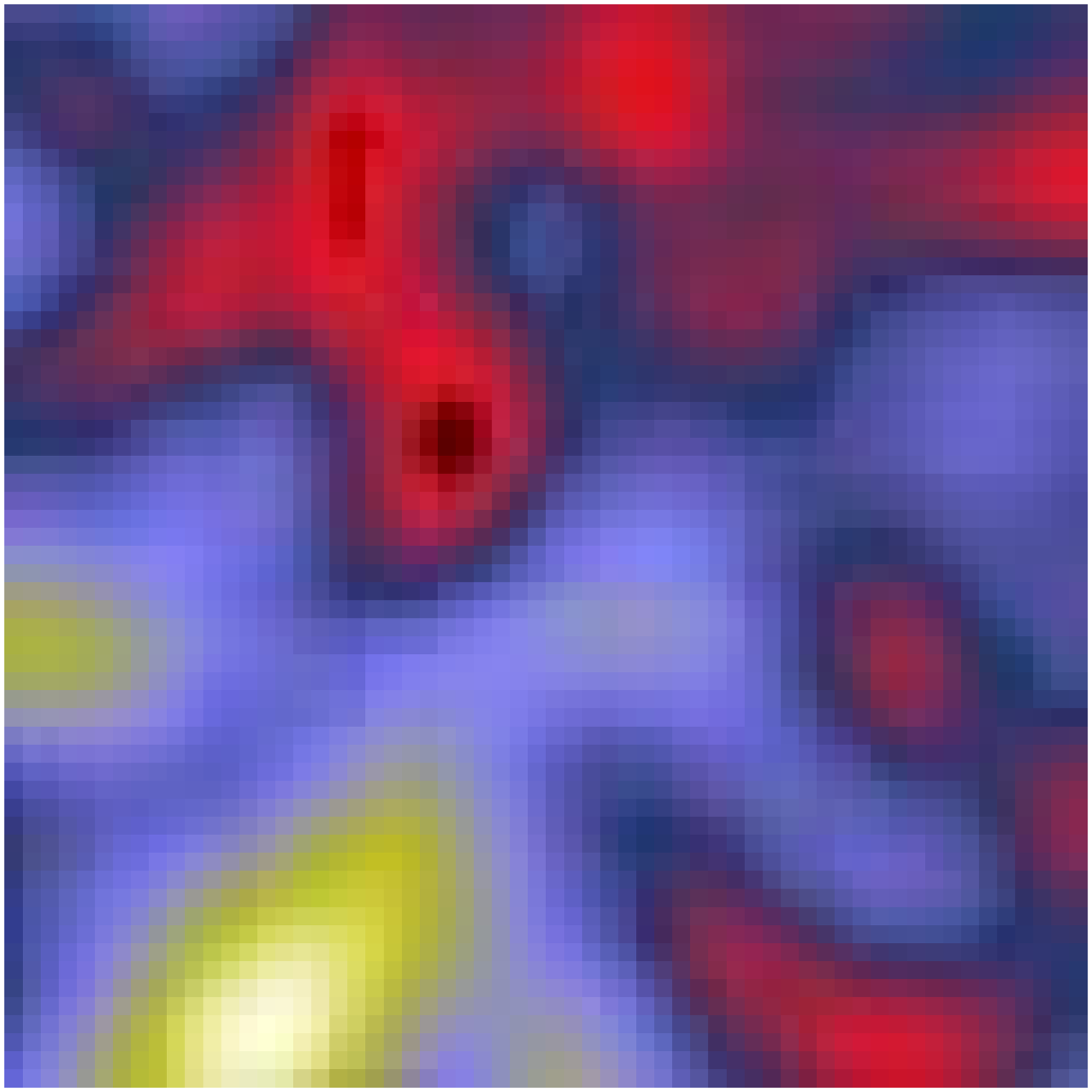} \\  {\#50   NCIA001268}  \\
  \FGb{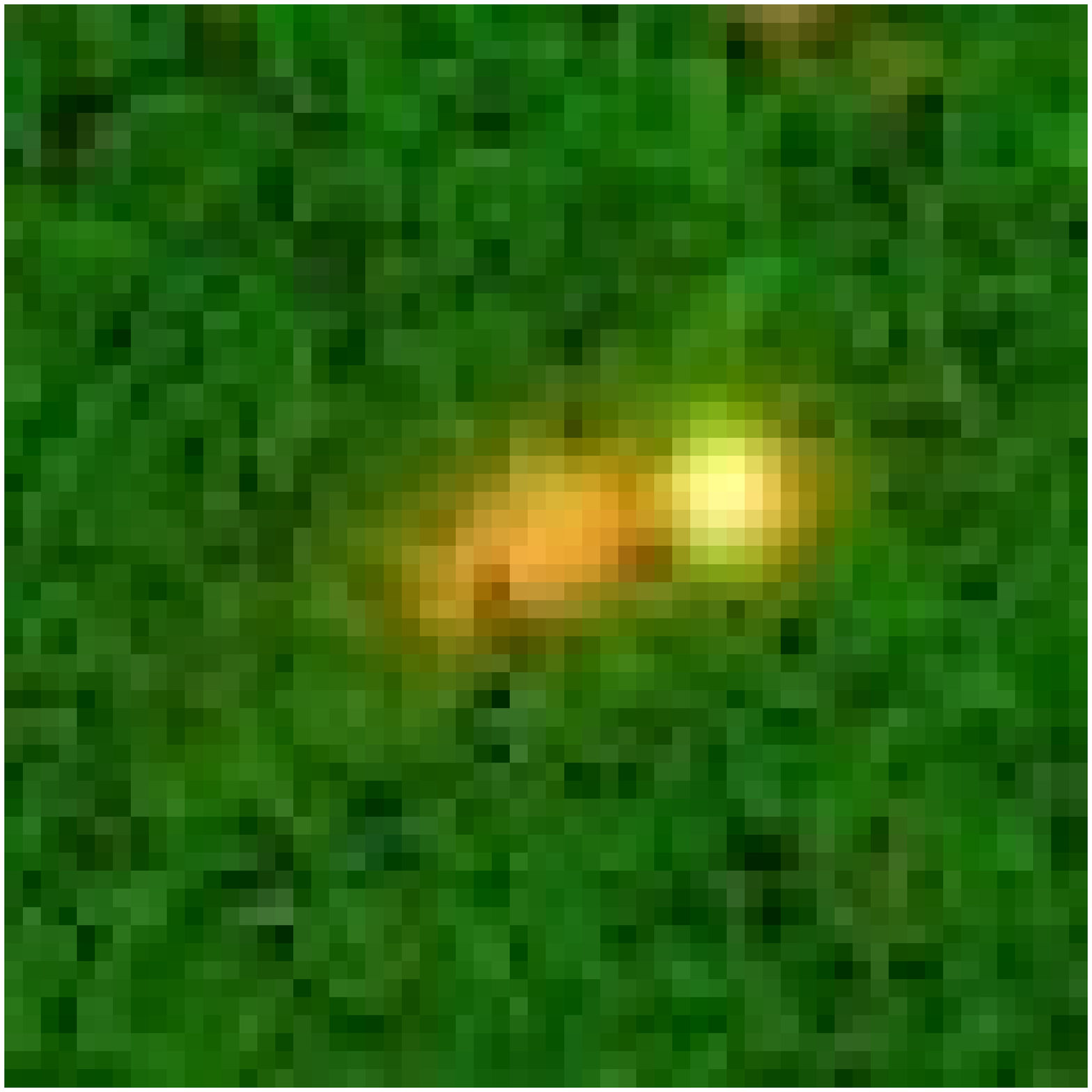}  \FGb{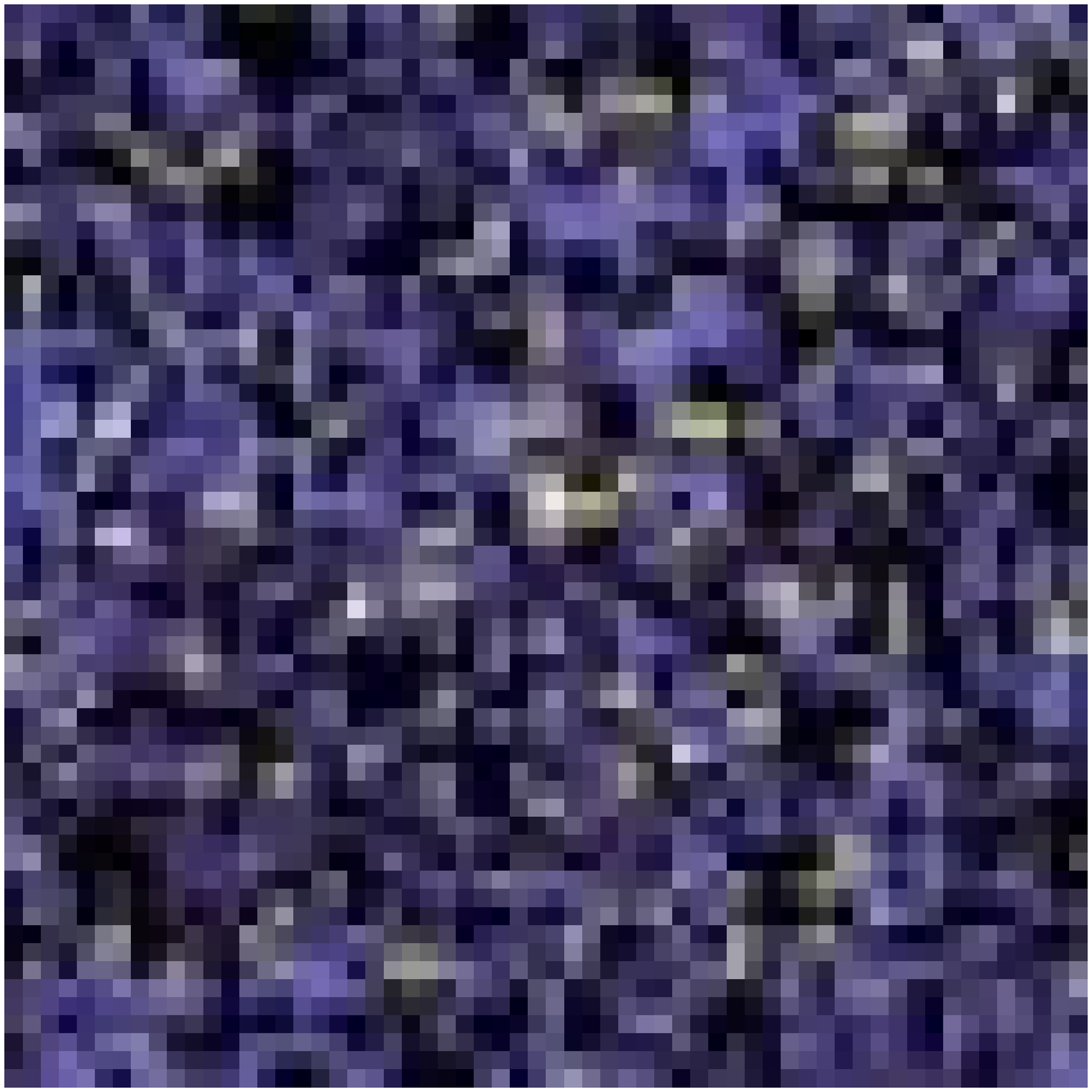}  \FGb{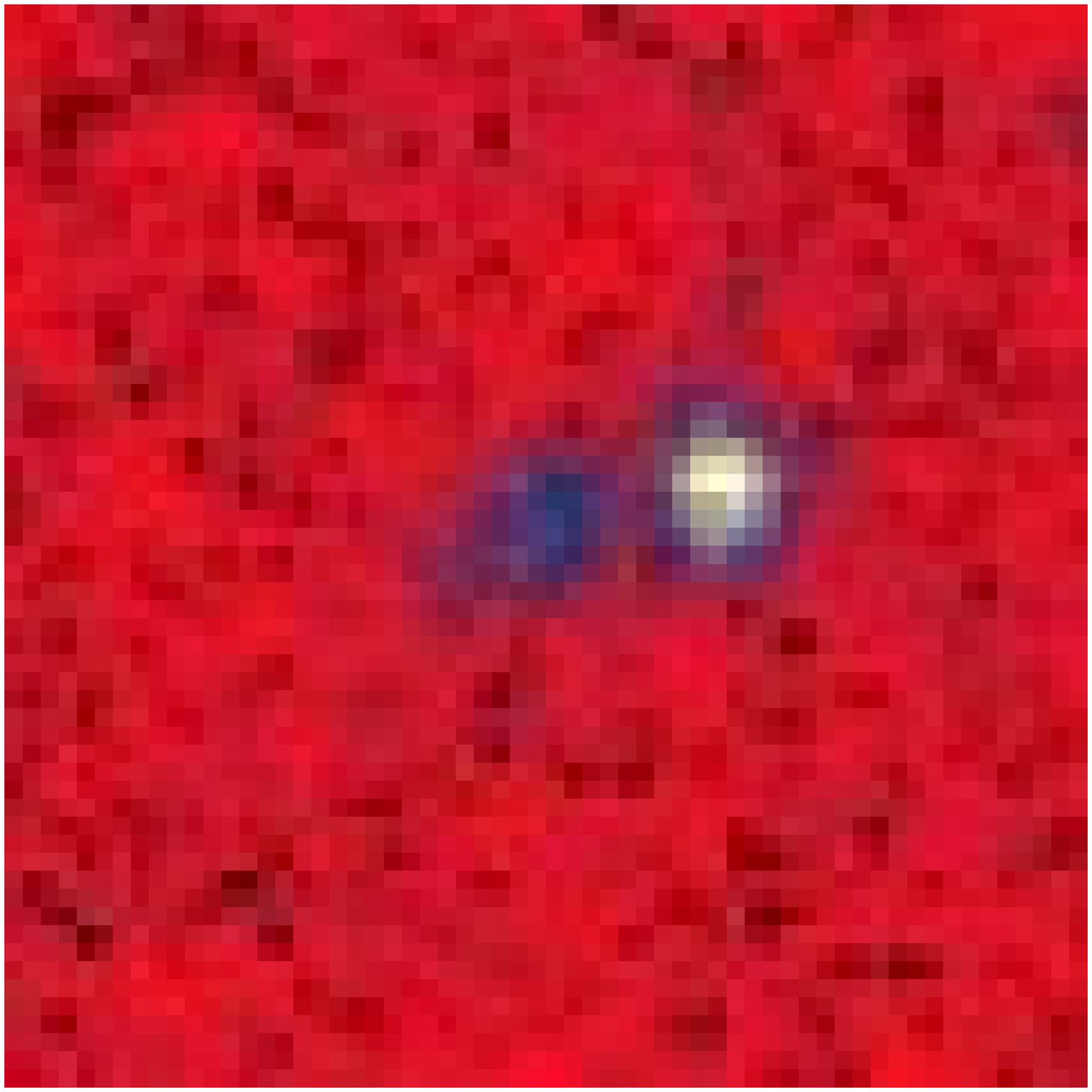}  \FGb{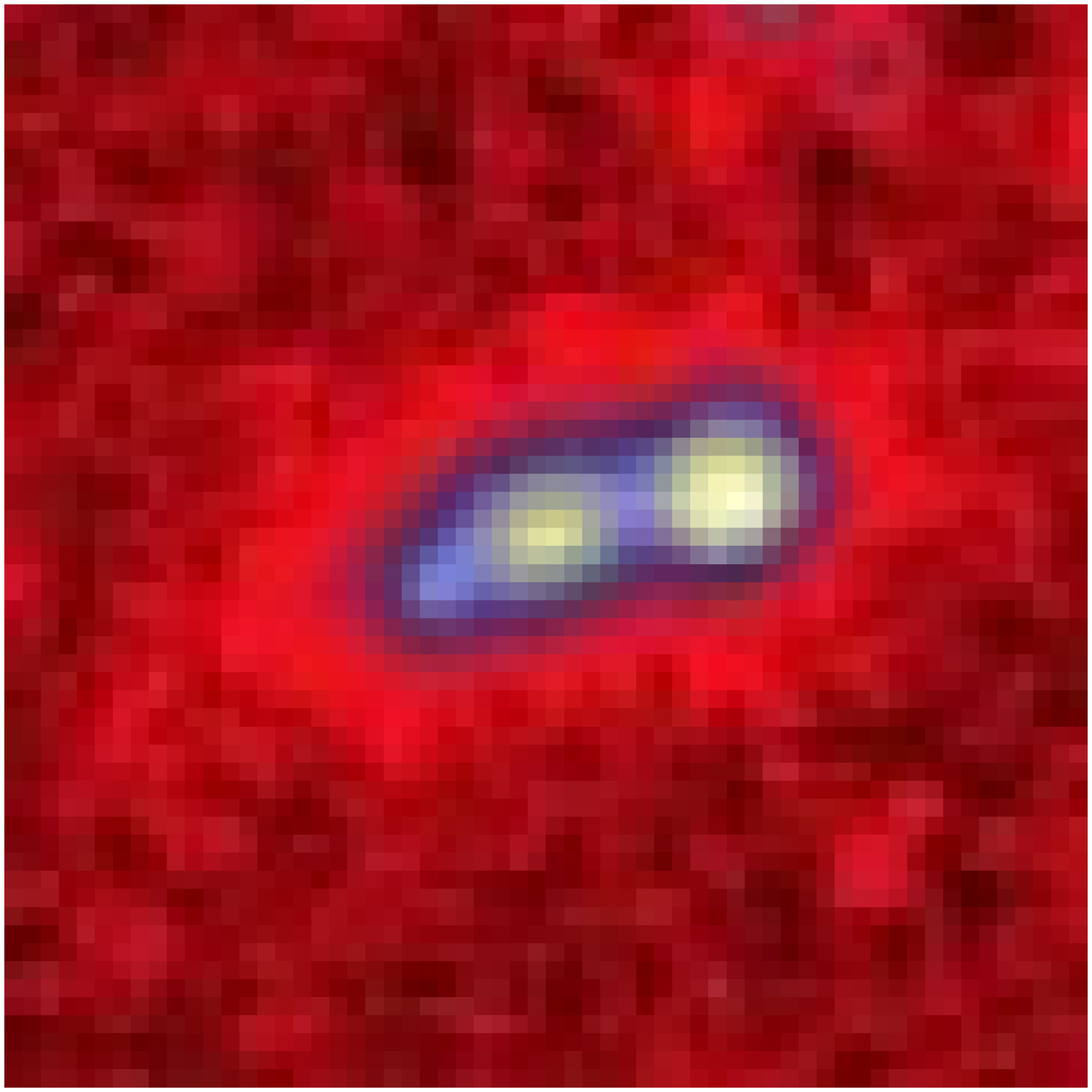}  \FGb{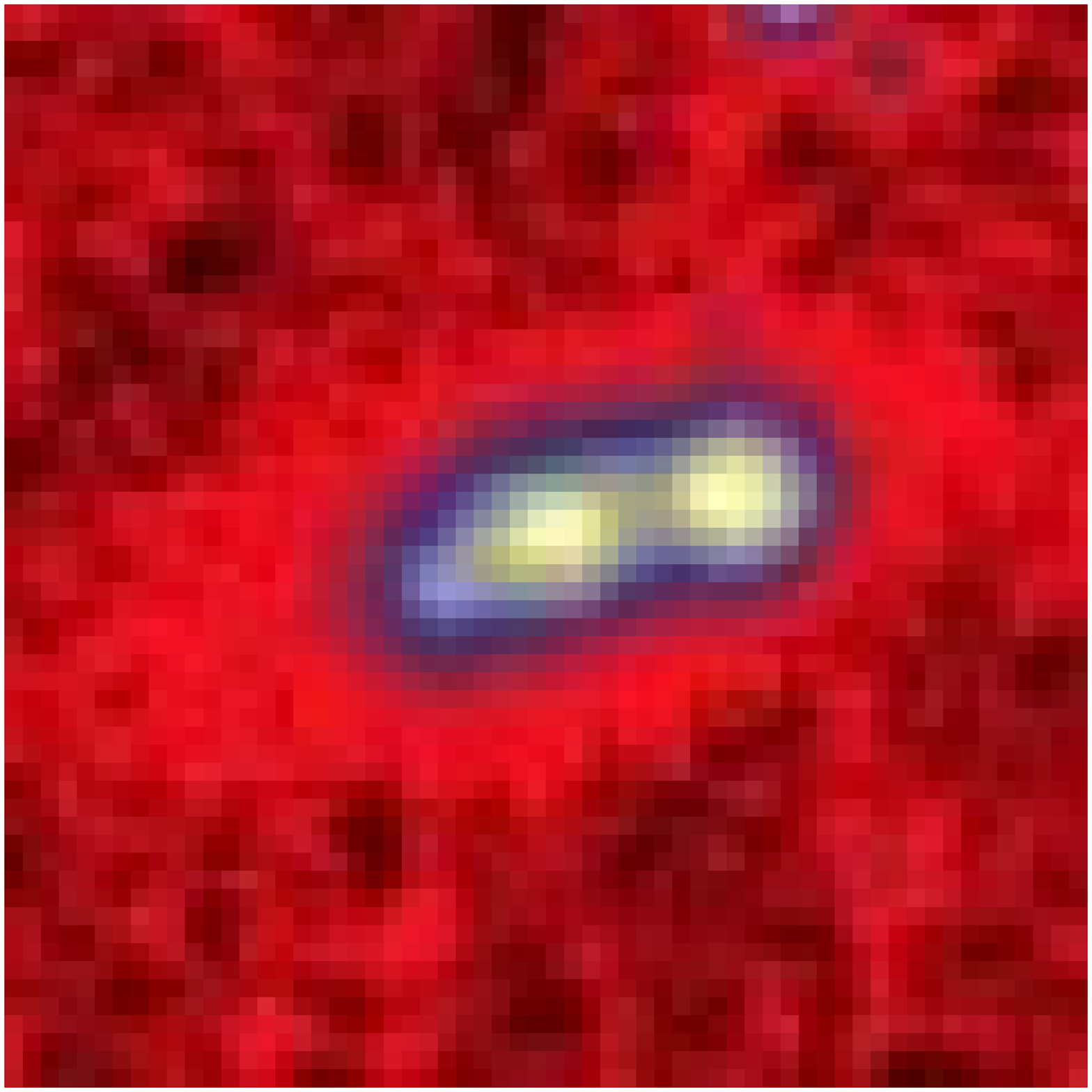}  \FGb{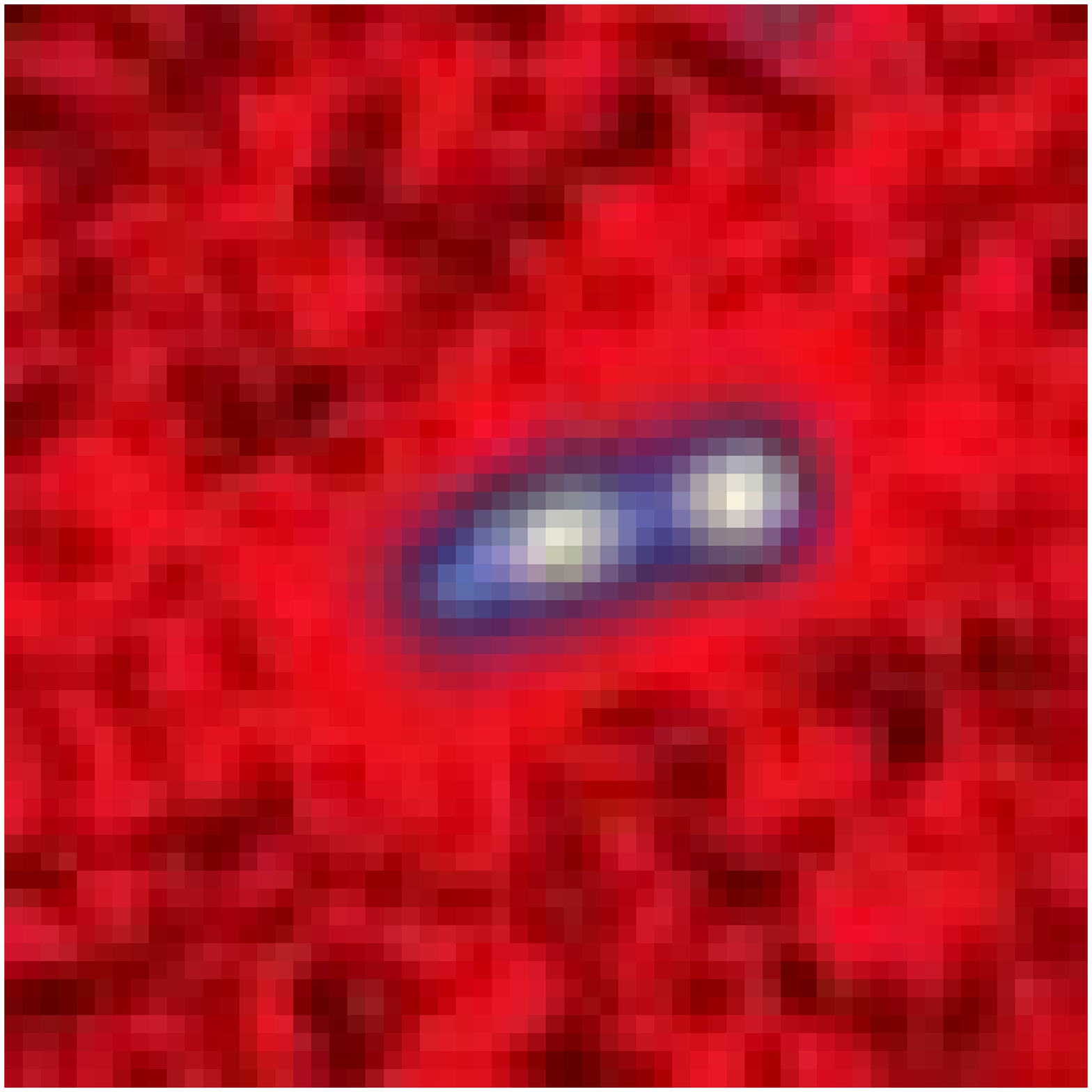}  \FGb{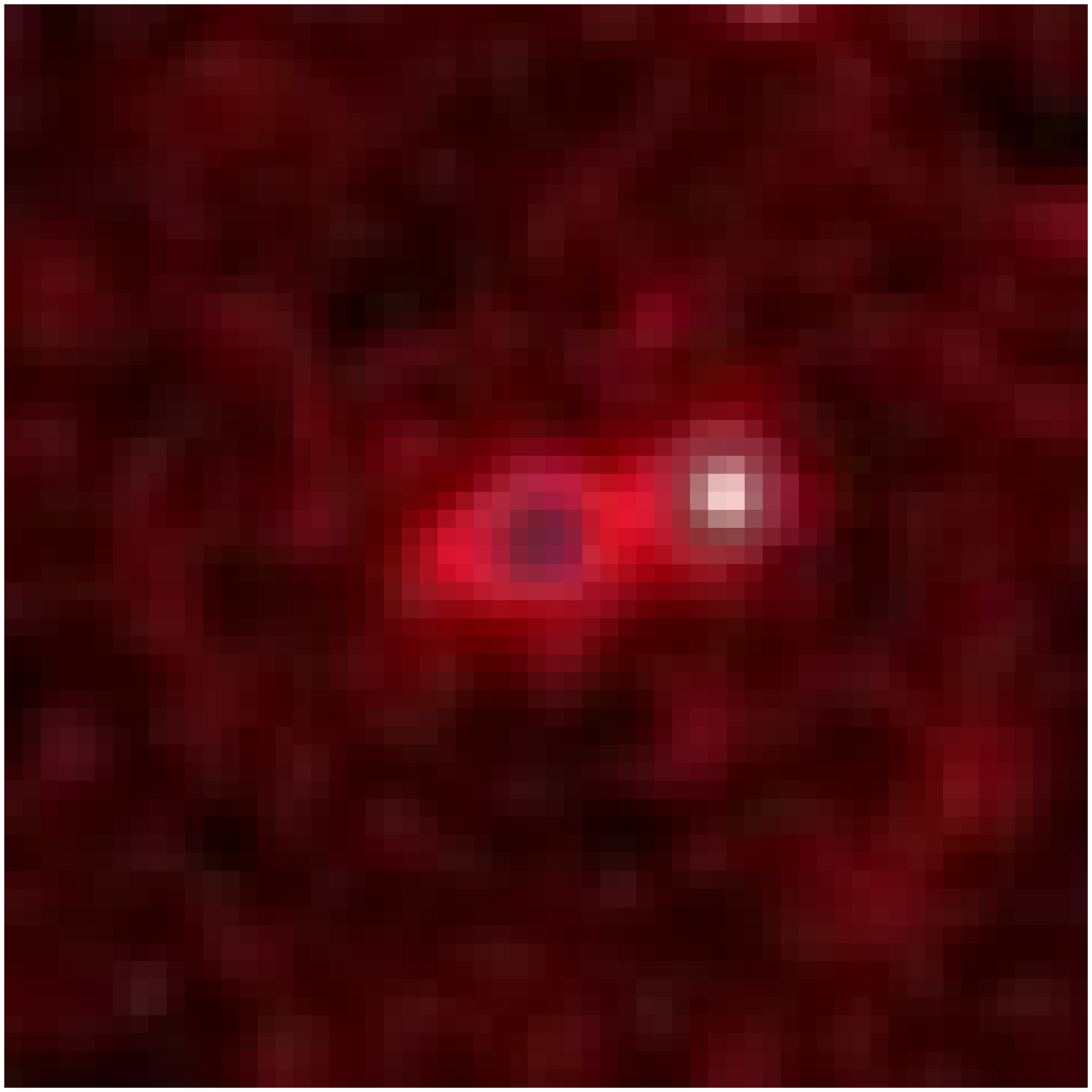}  \FGb{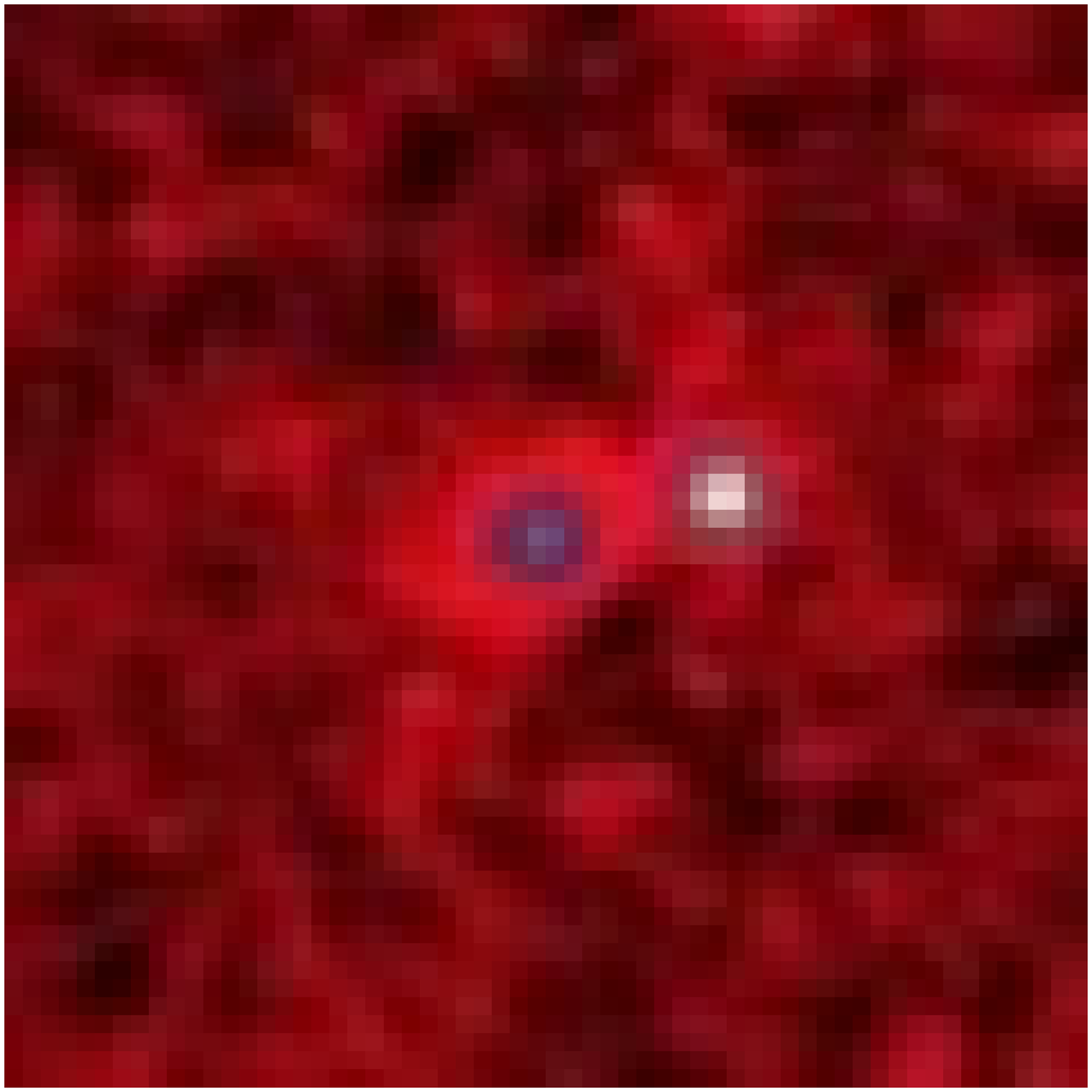}  \FGb{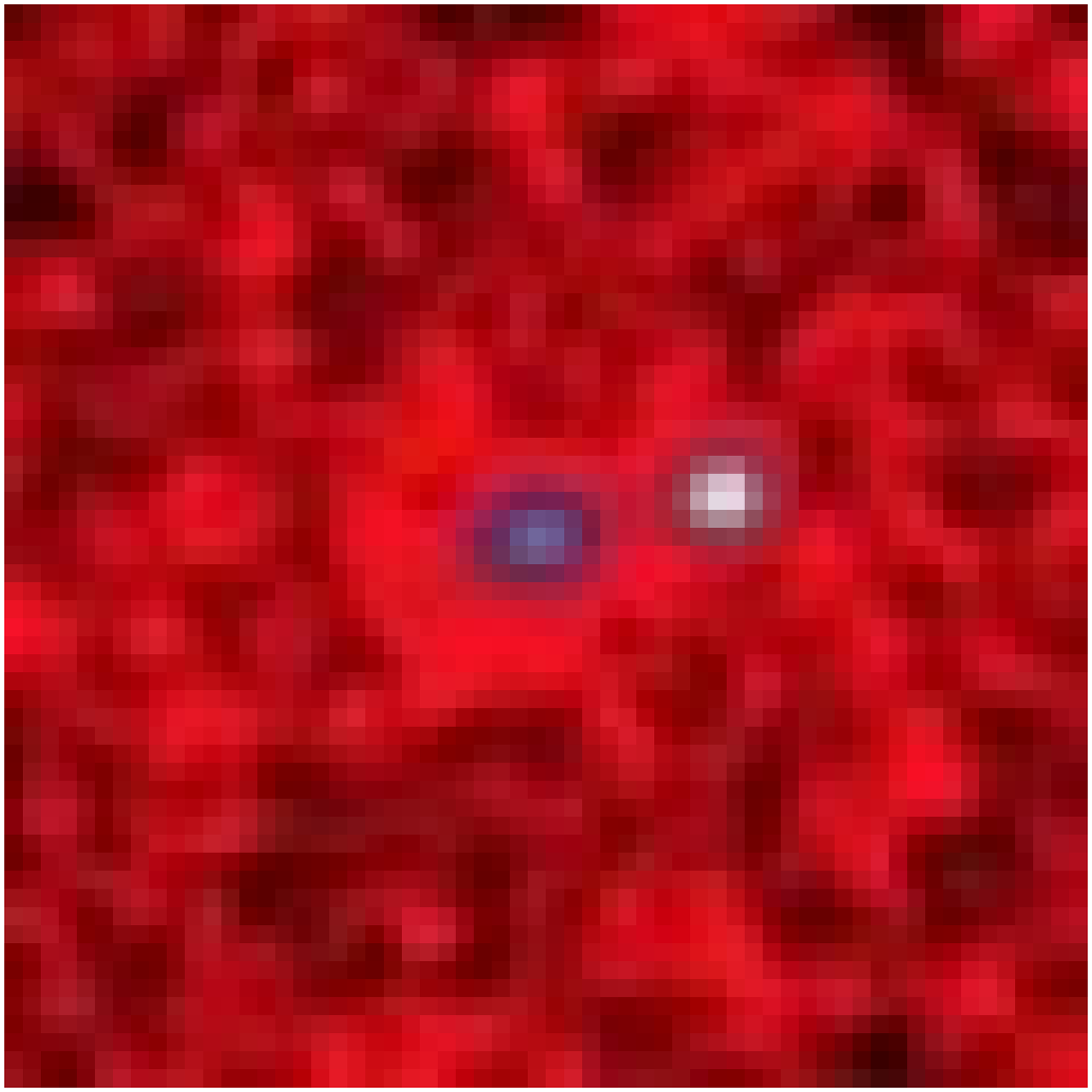}  \FGb{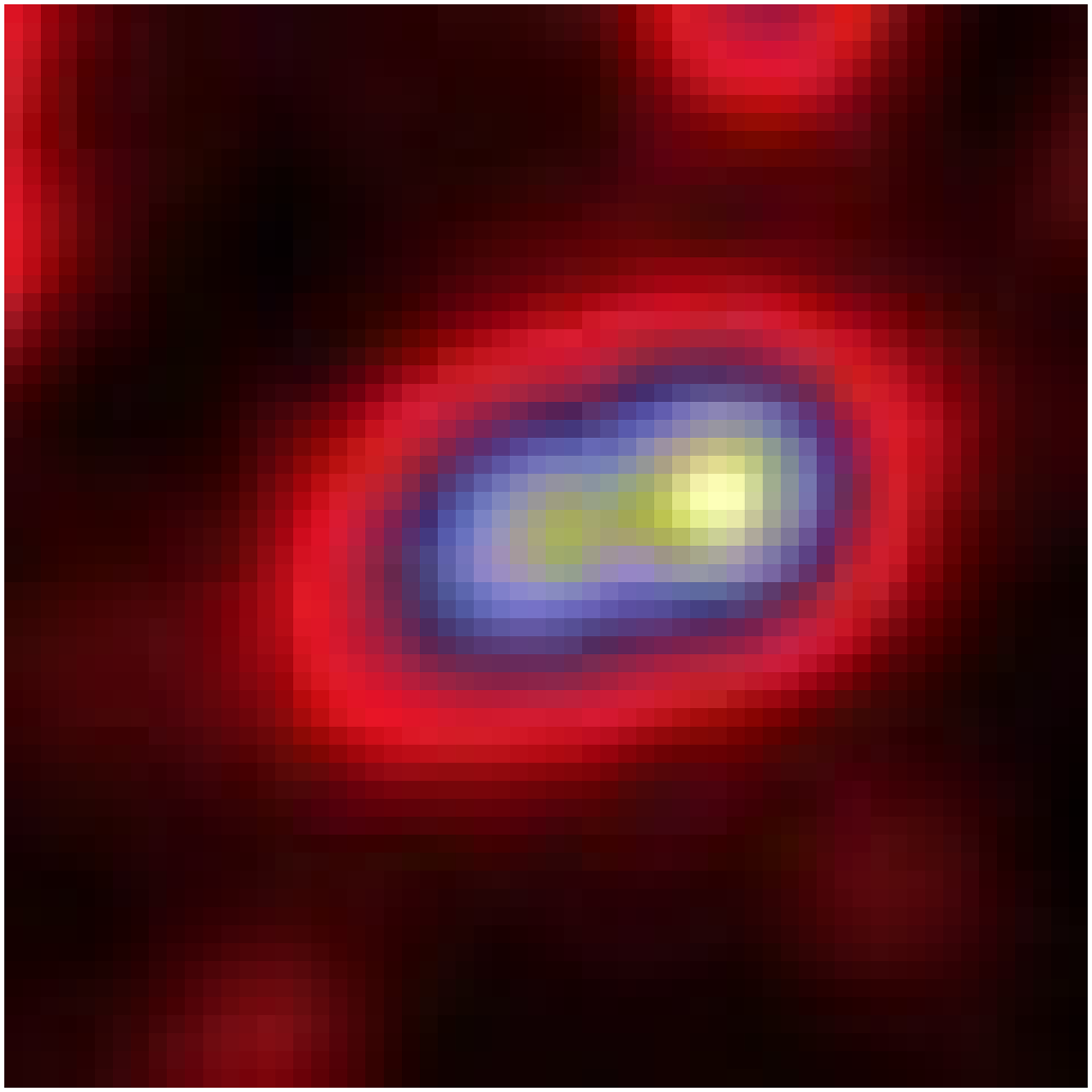}  \FGb{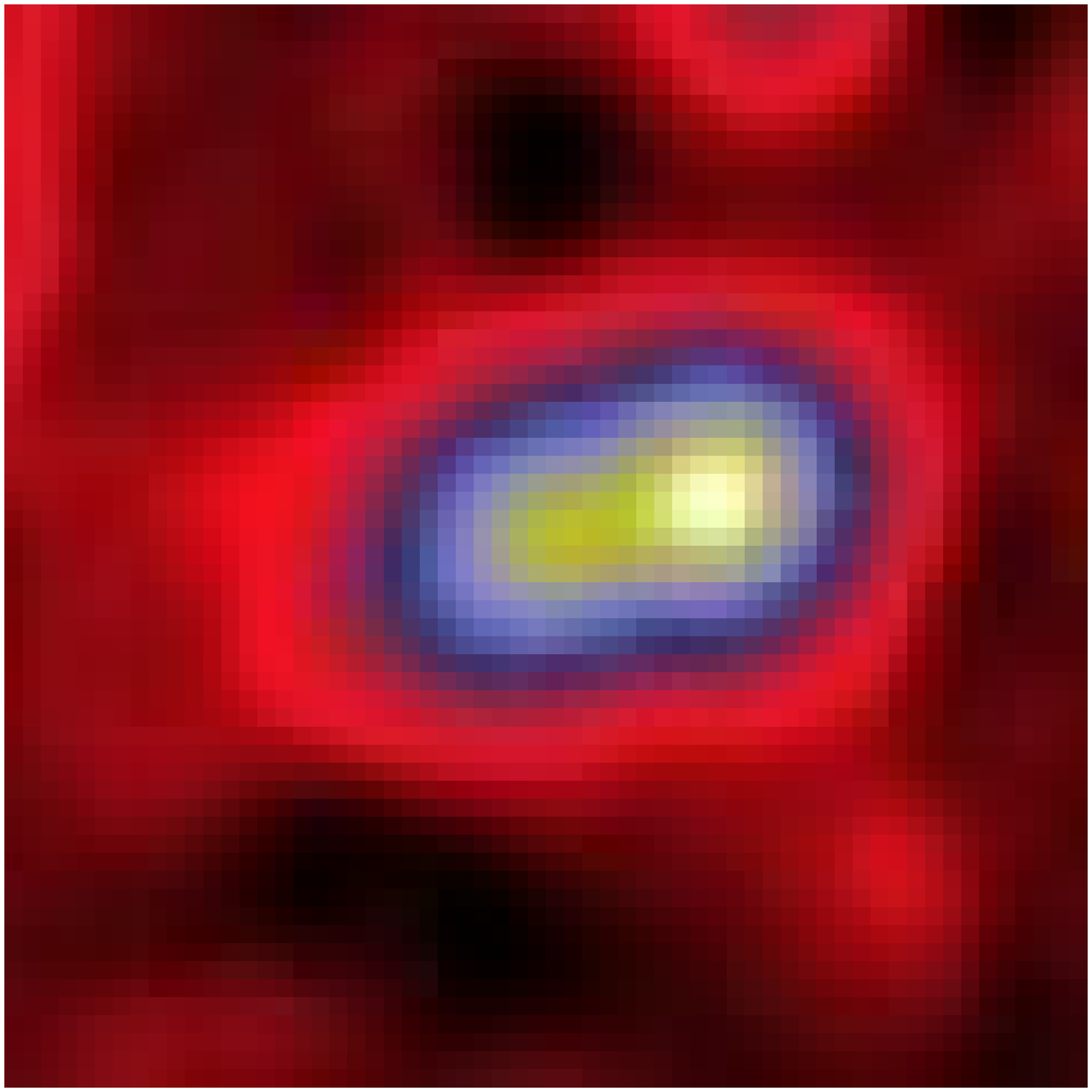}  \FGb{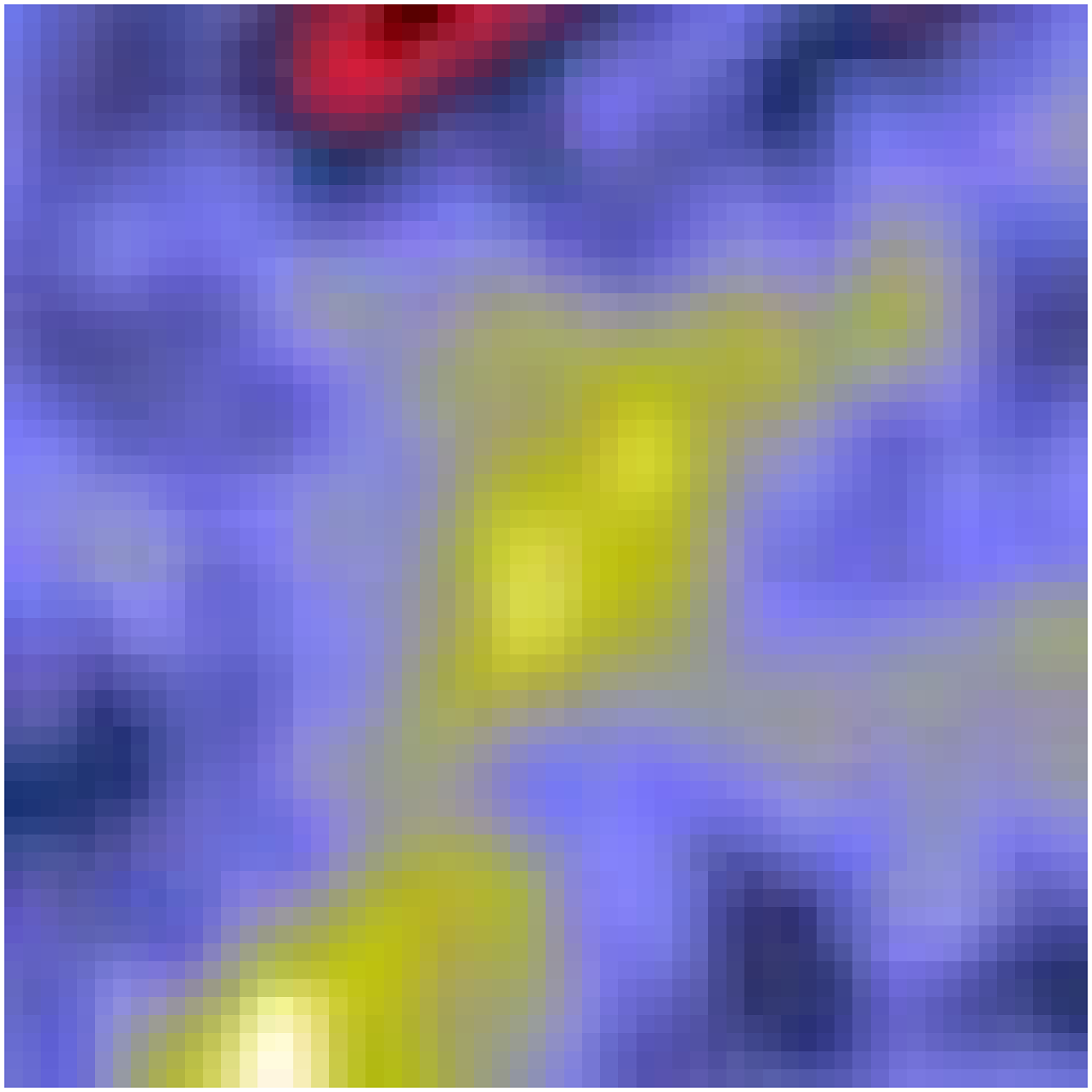}  \FGb{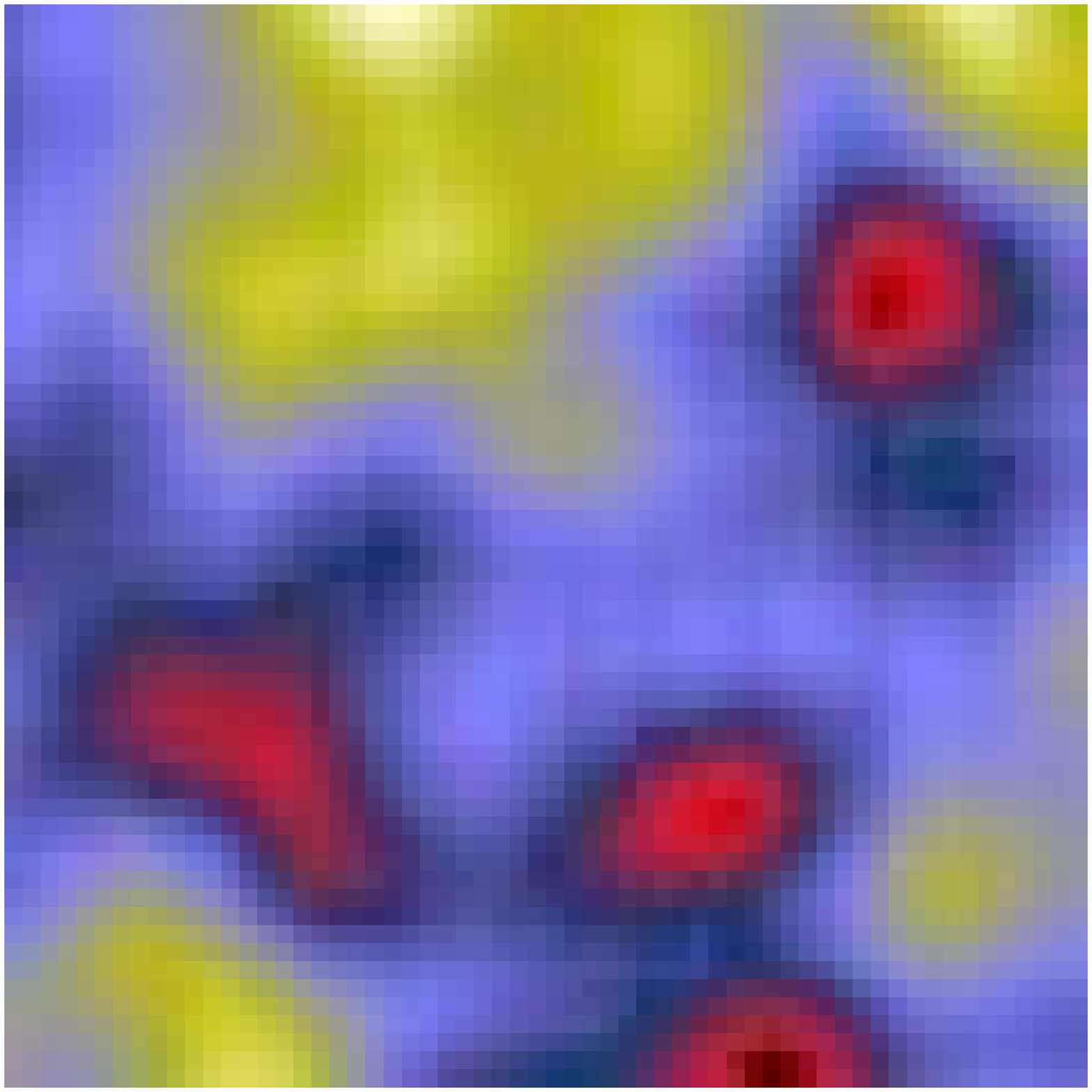} \\  {\#51   NCIA001160}  \\
  \FGb{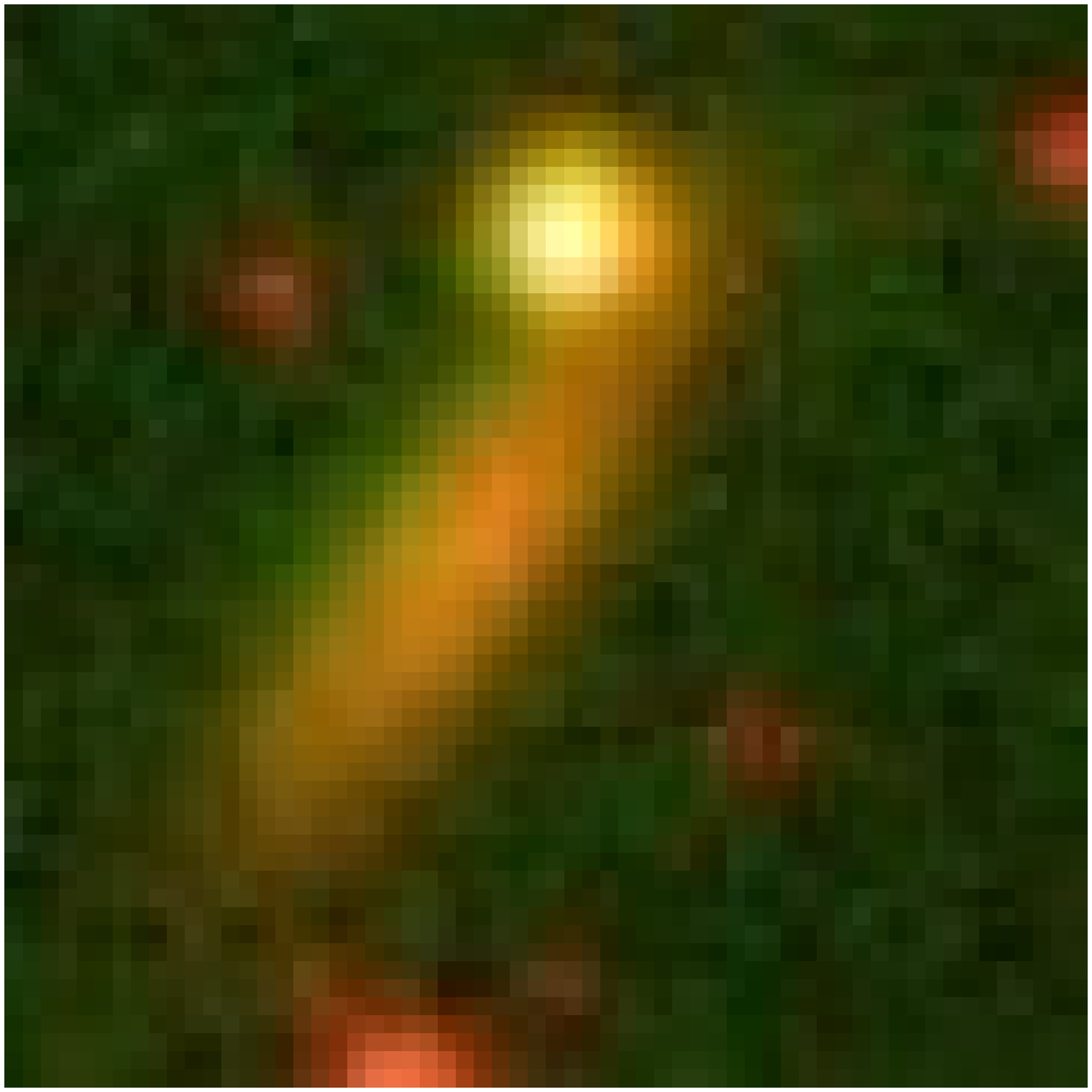}  \FGb{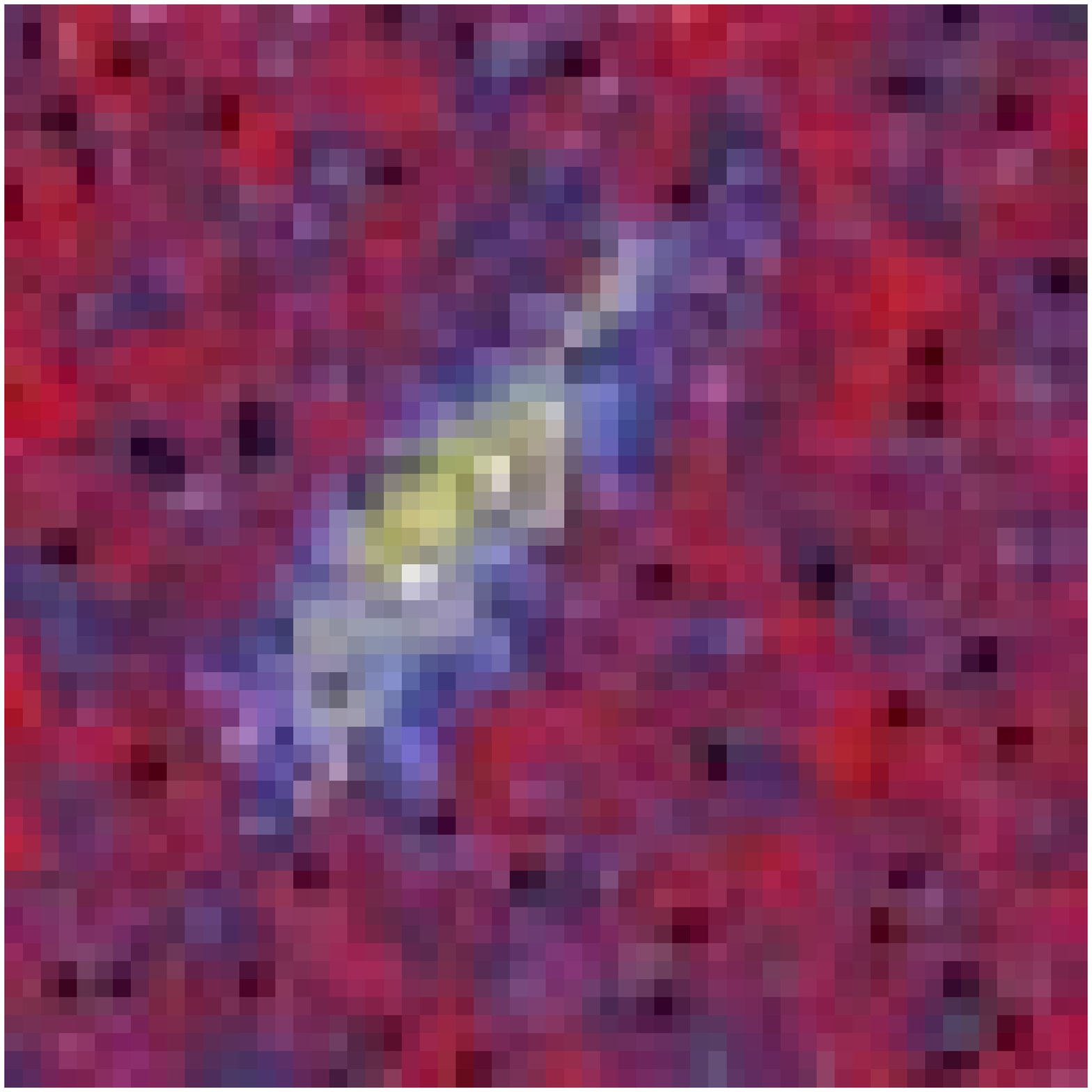}  \FGb{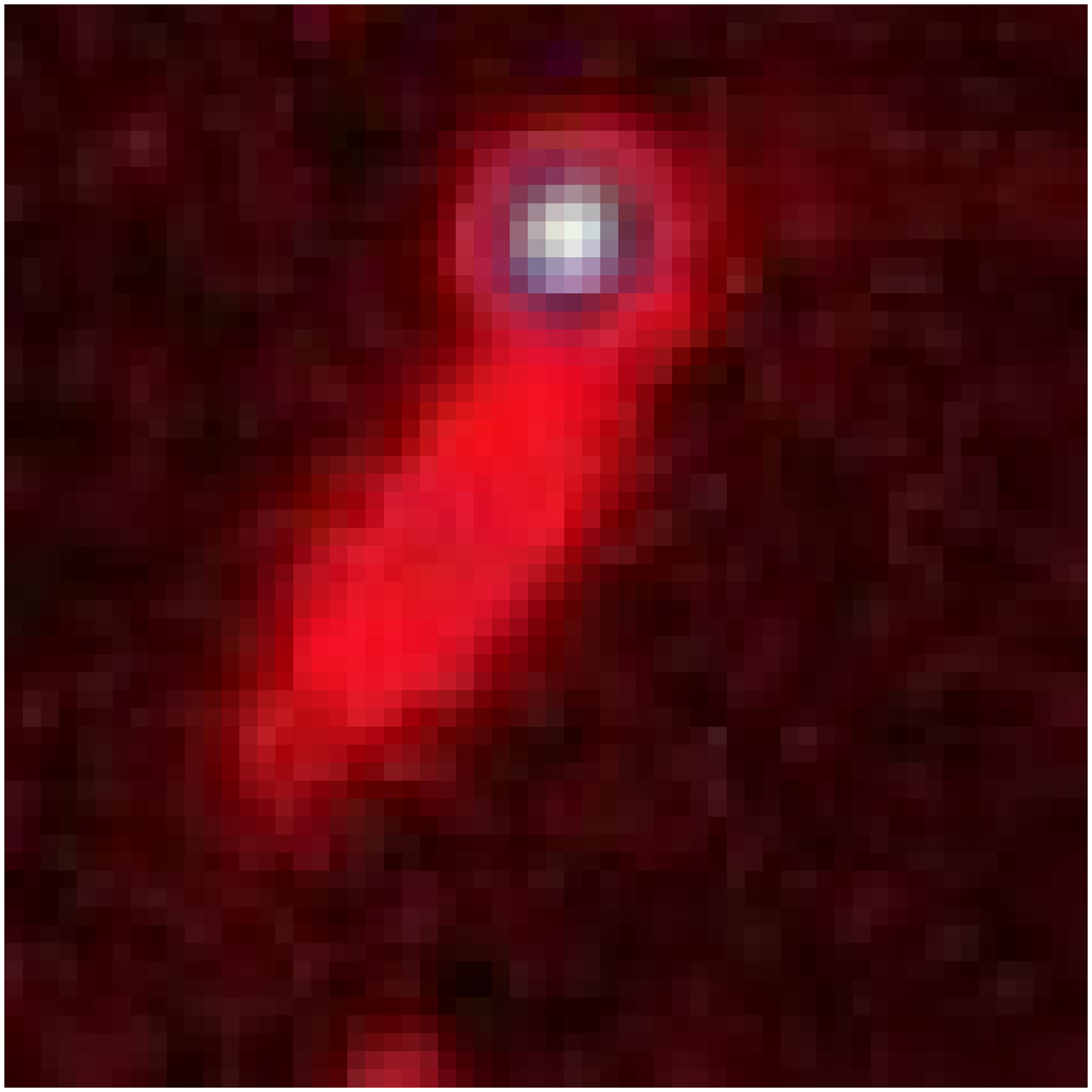}  \FGb{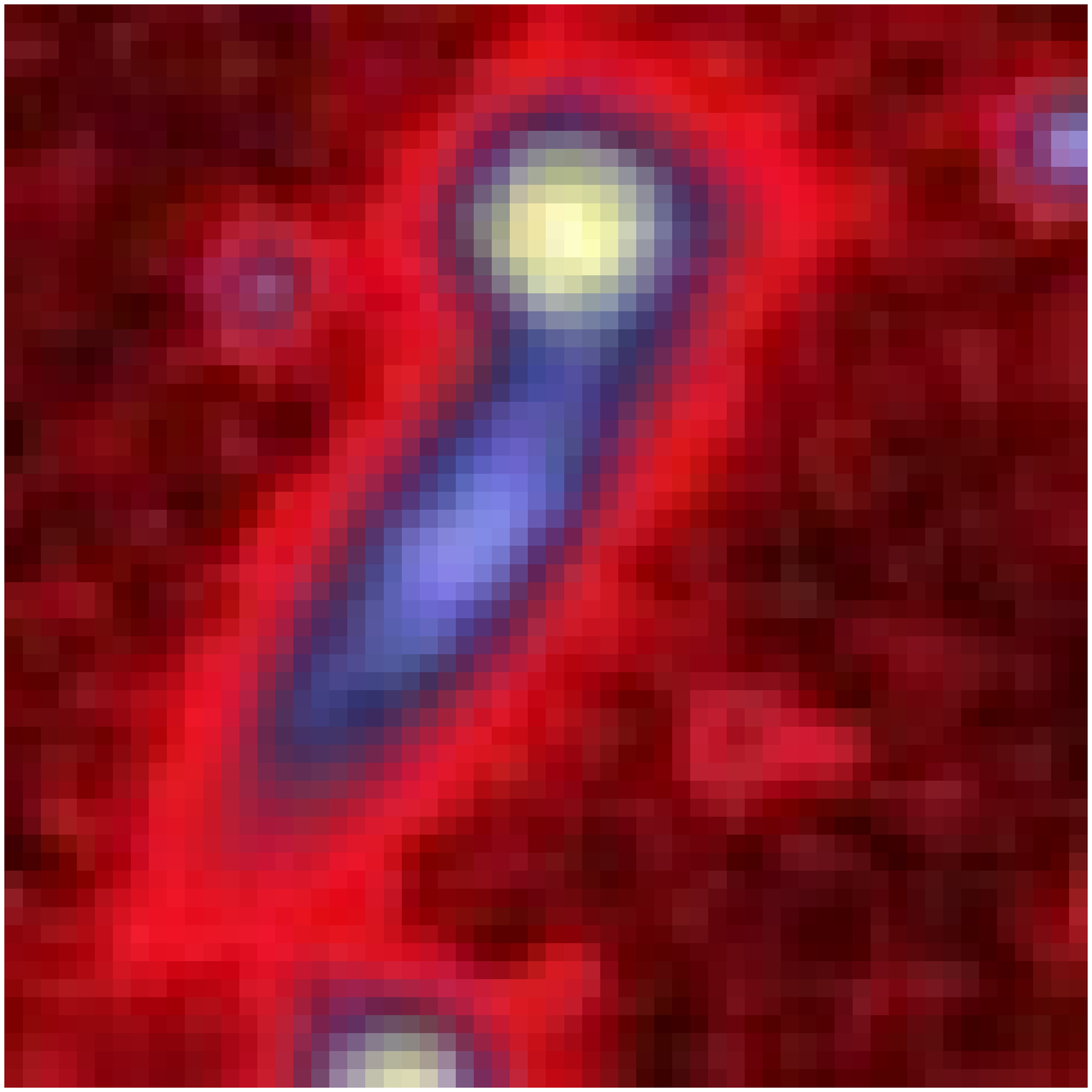}  \FGb{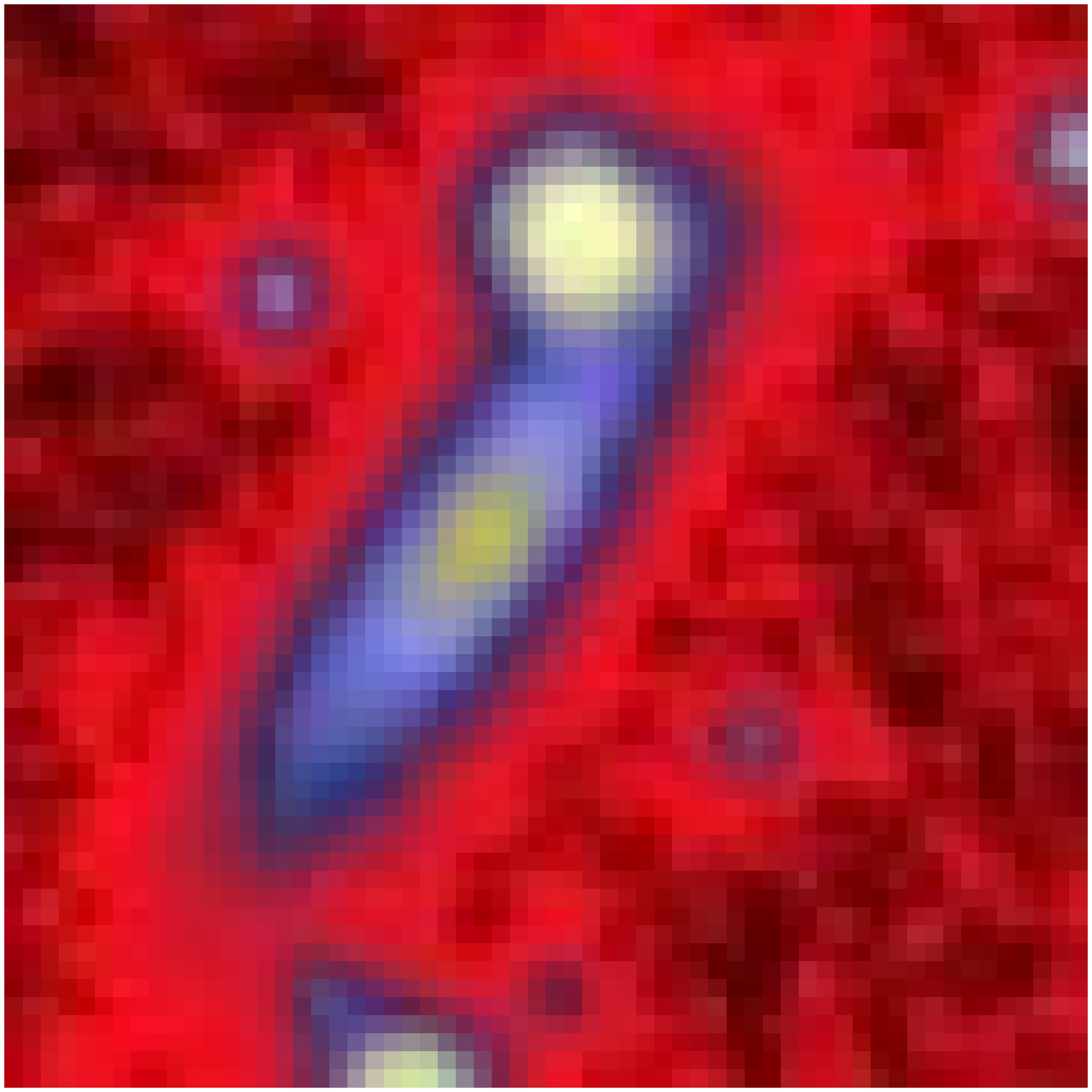}  \FGb{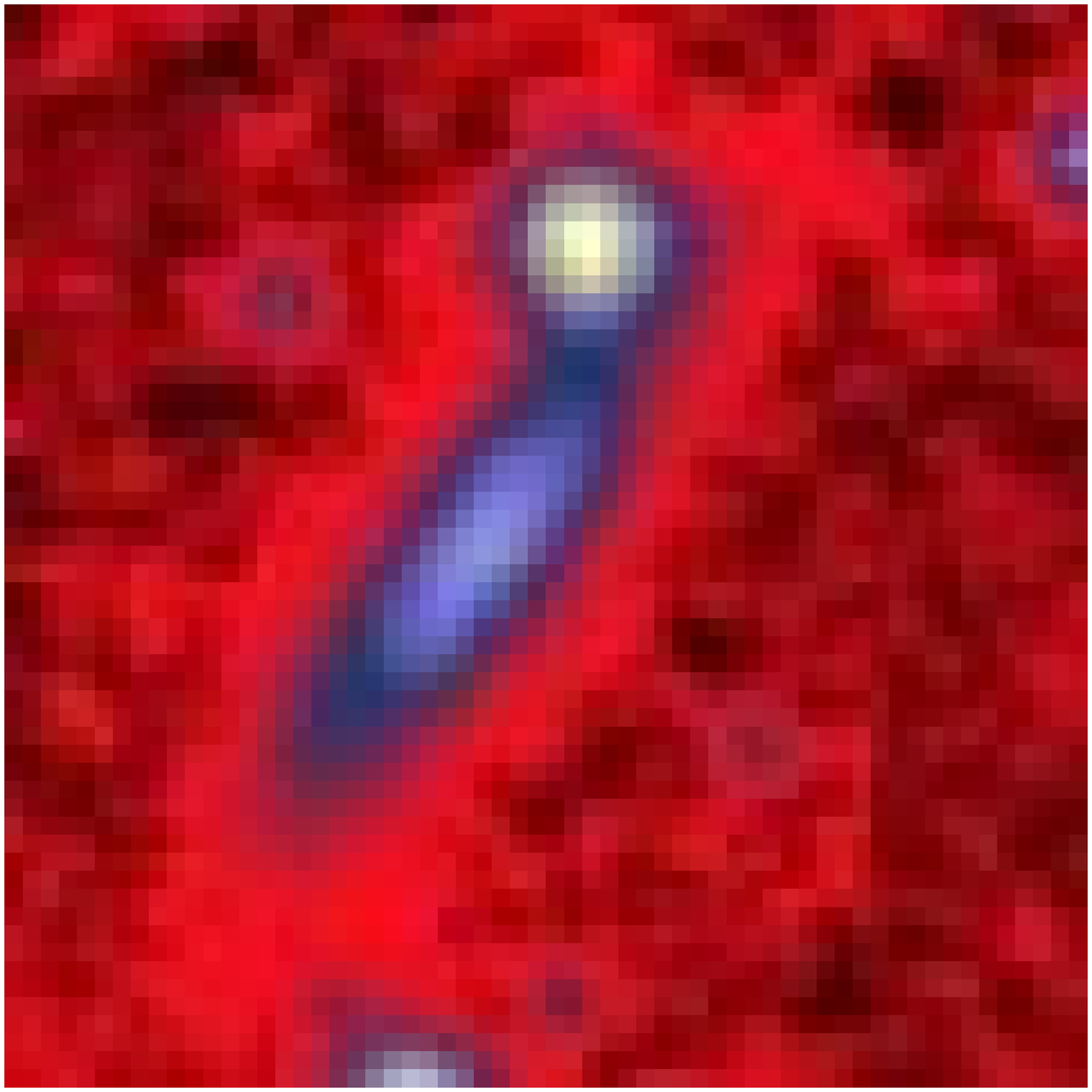}  \FGb{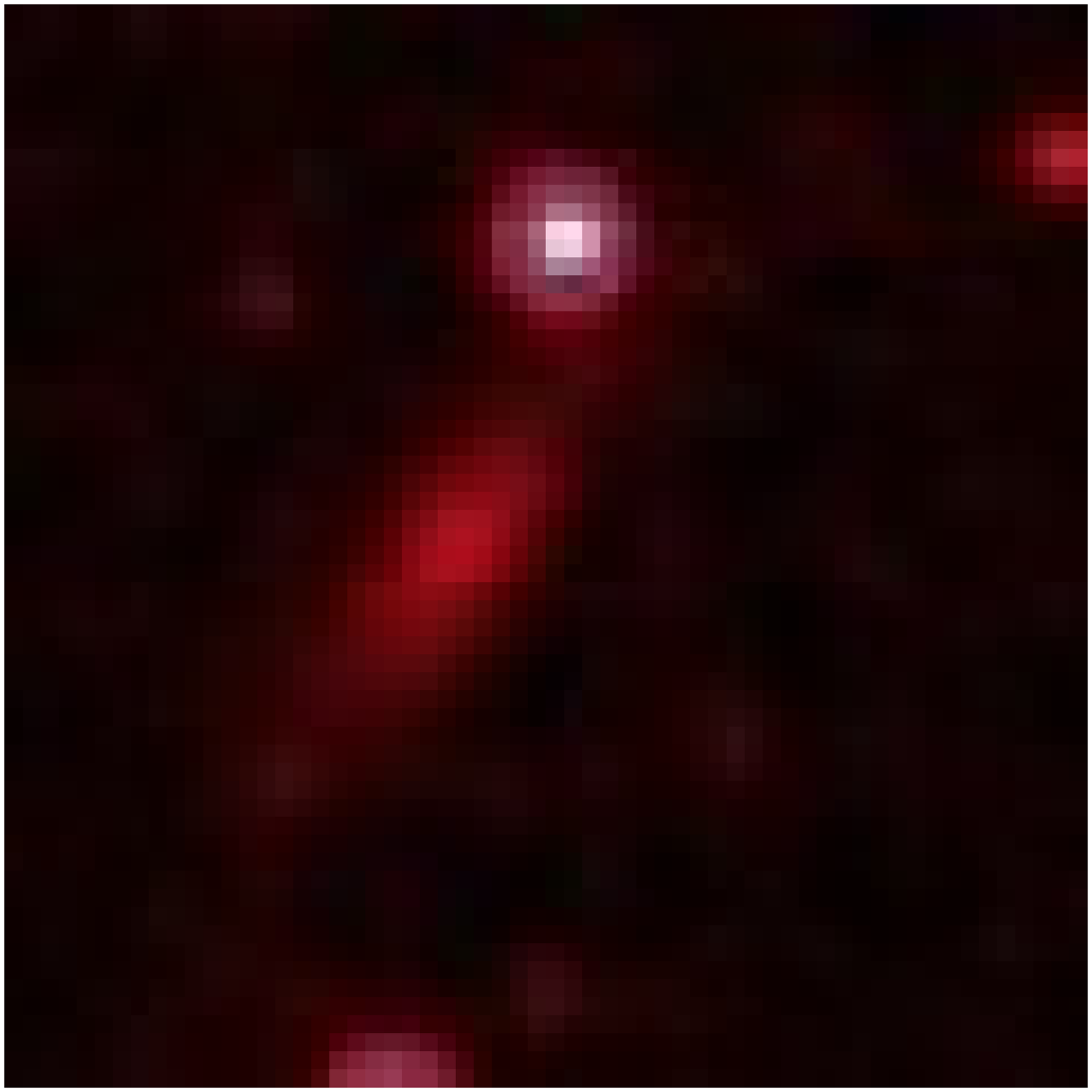}  \FGb{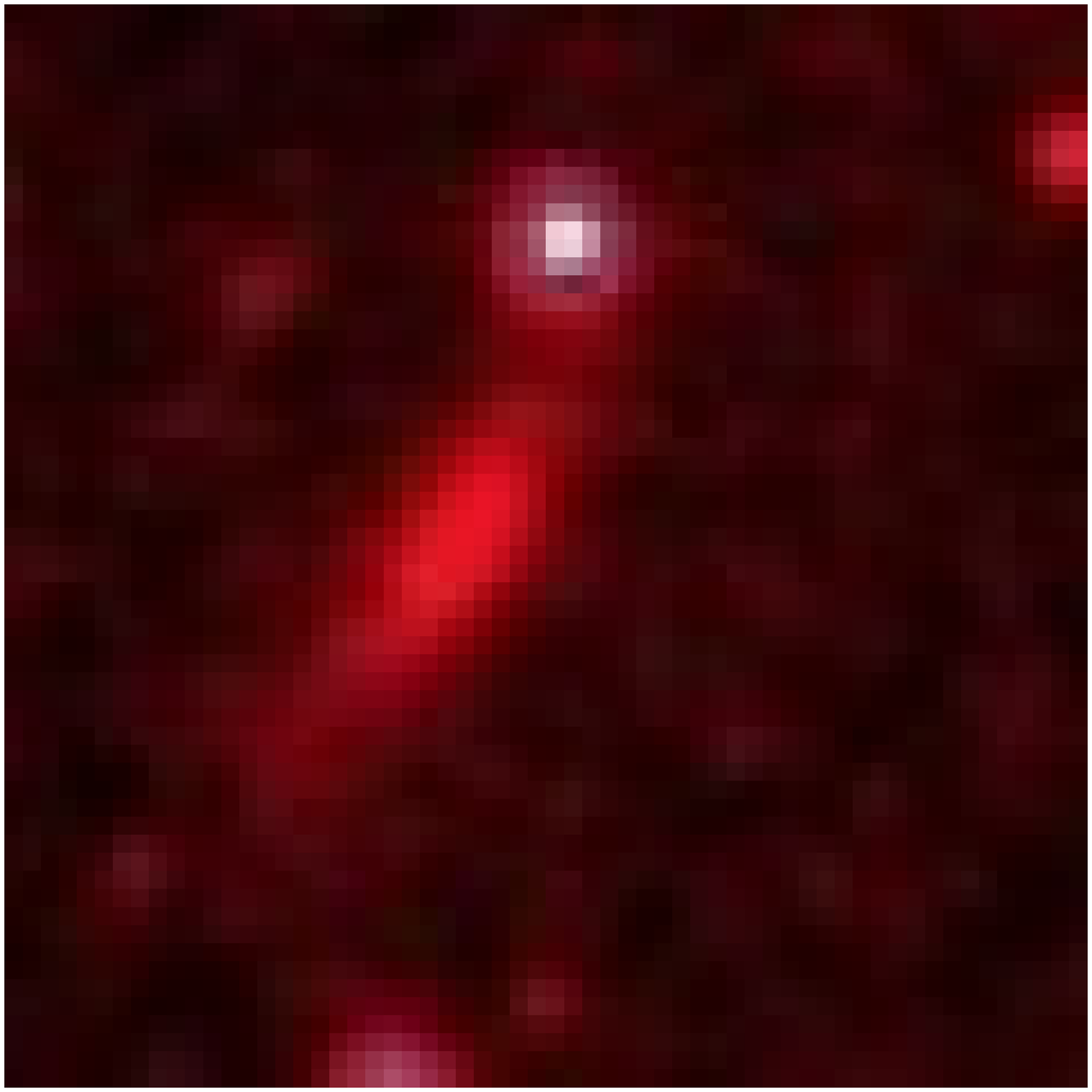}  \FGb{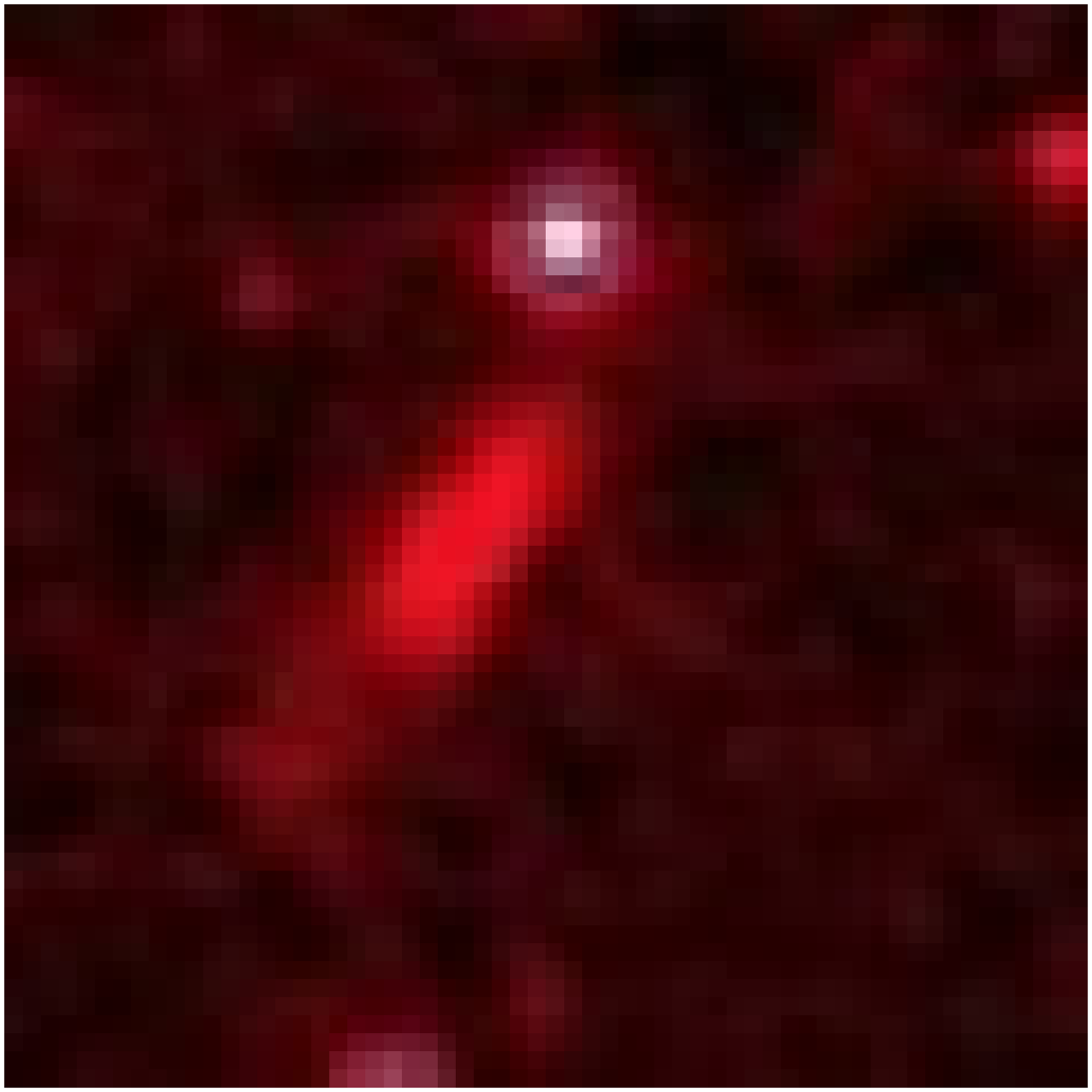}  \FGb{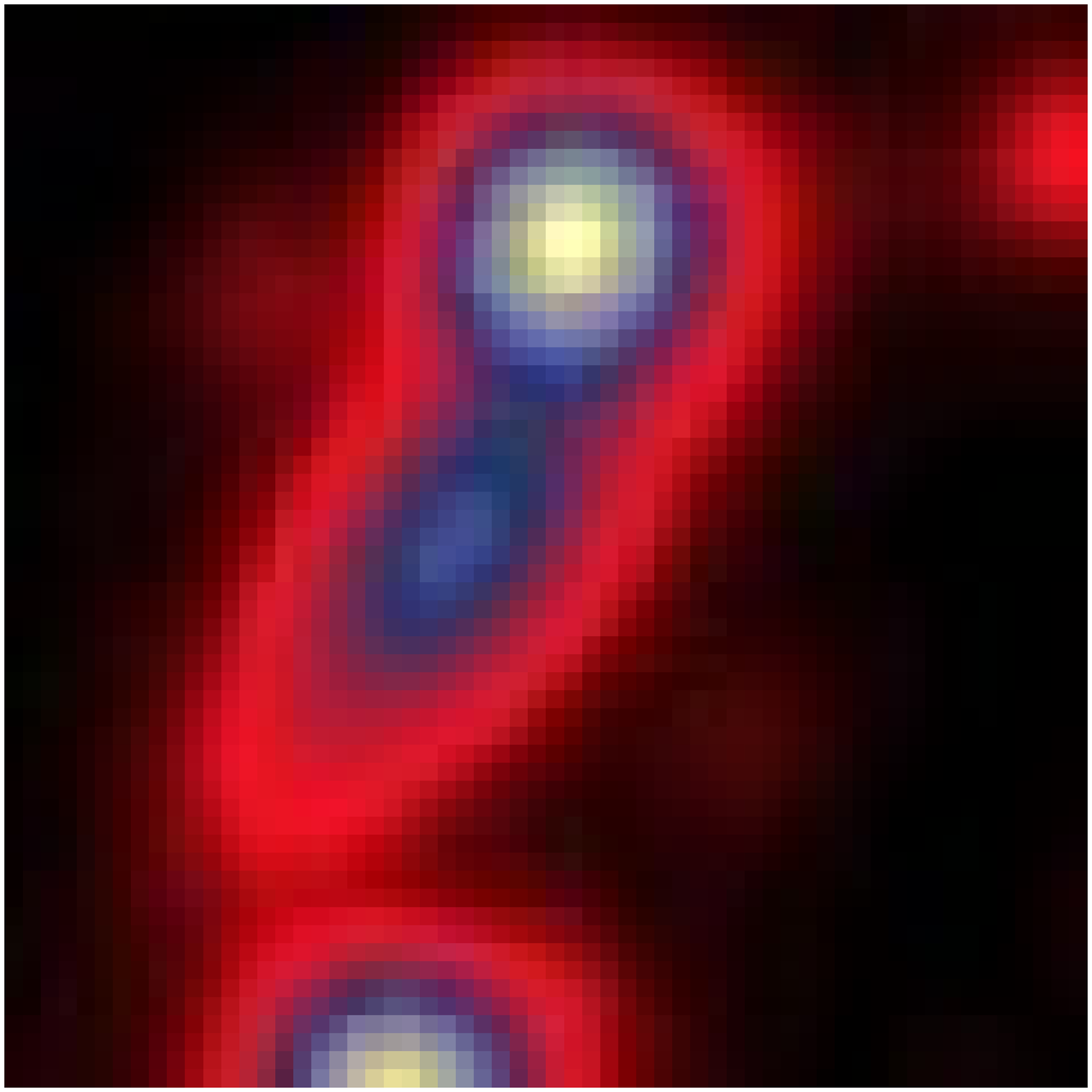}  \FGb{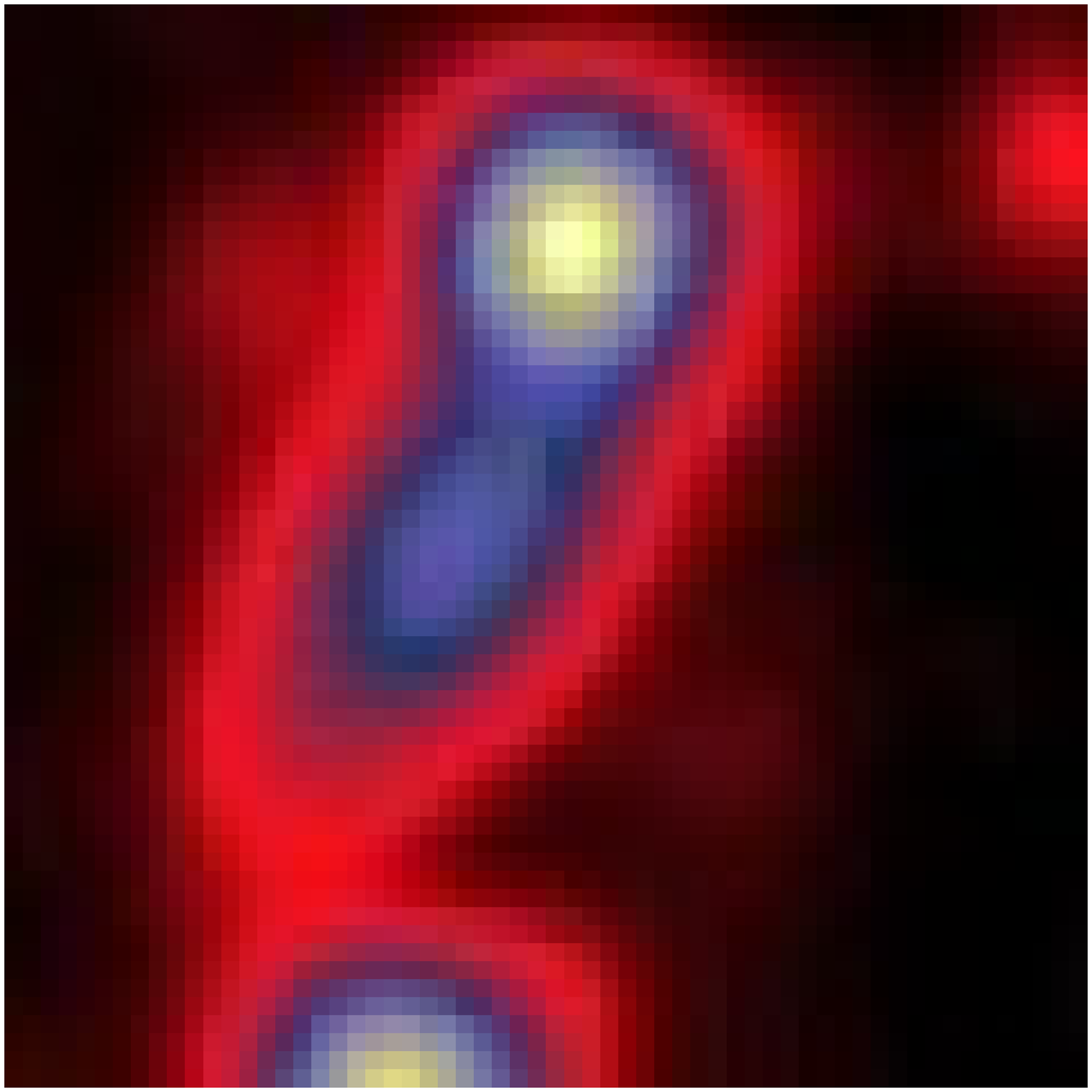}  \FGb{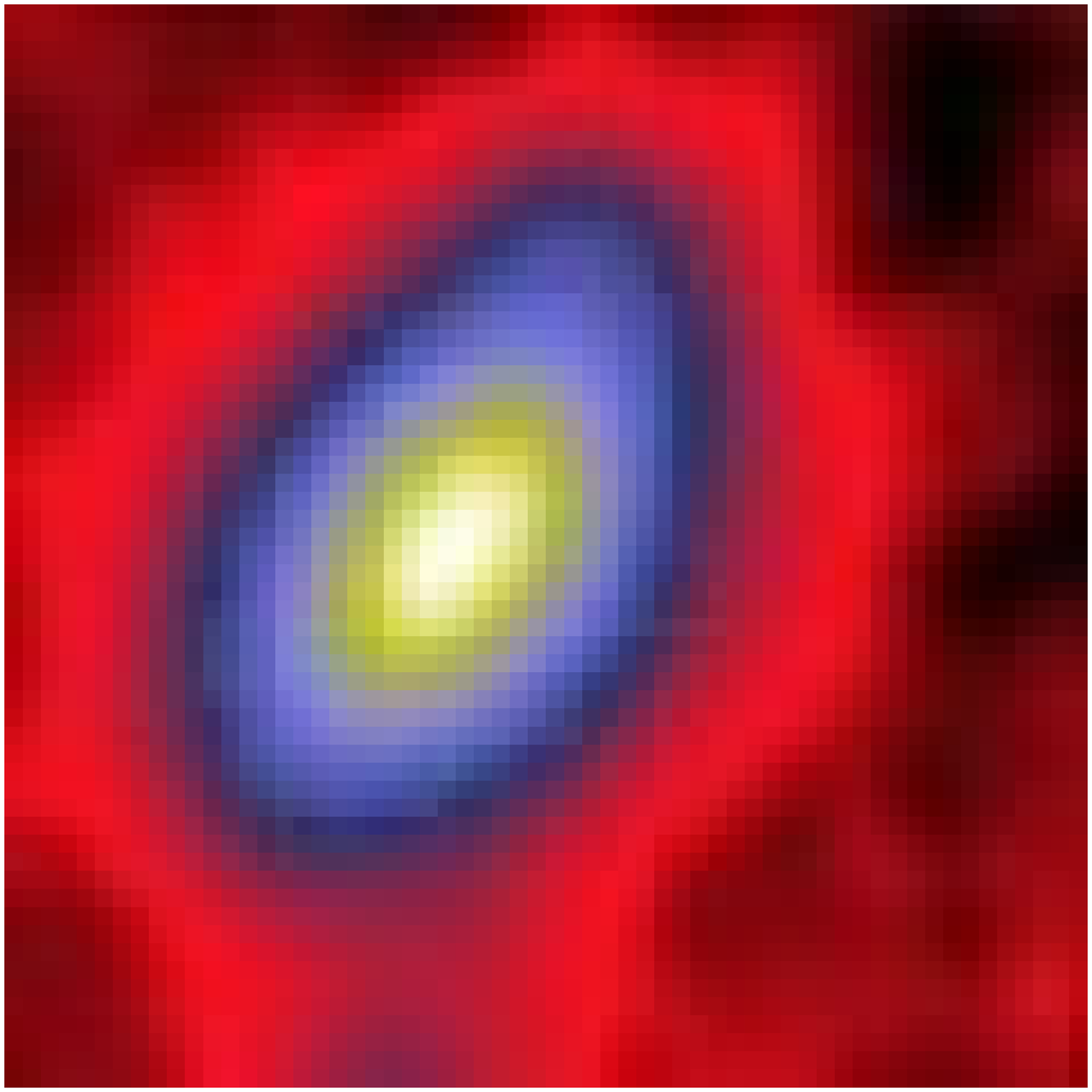}  \FGb{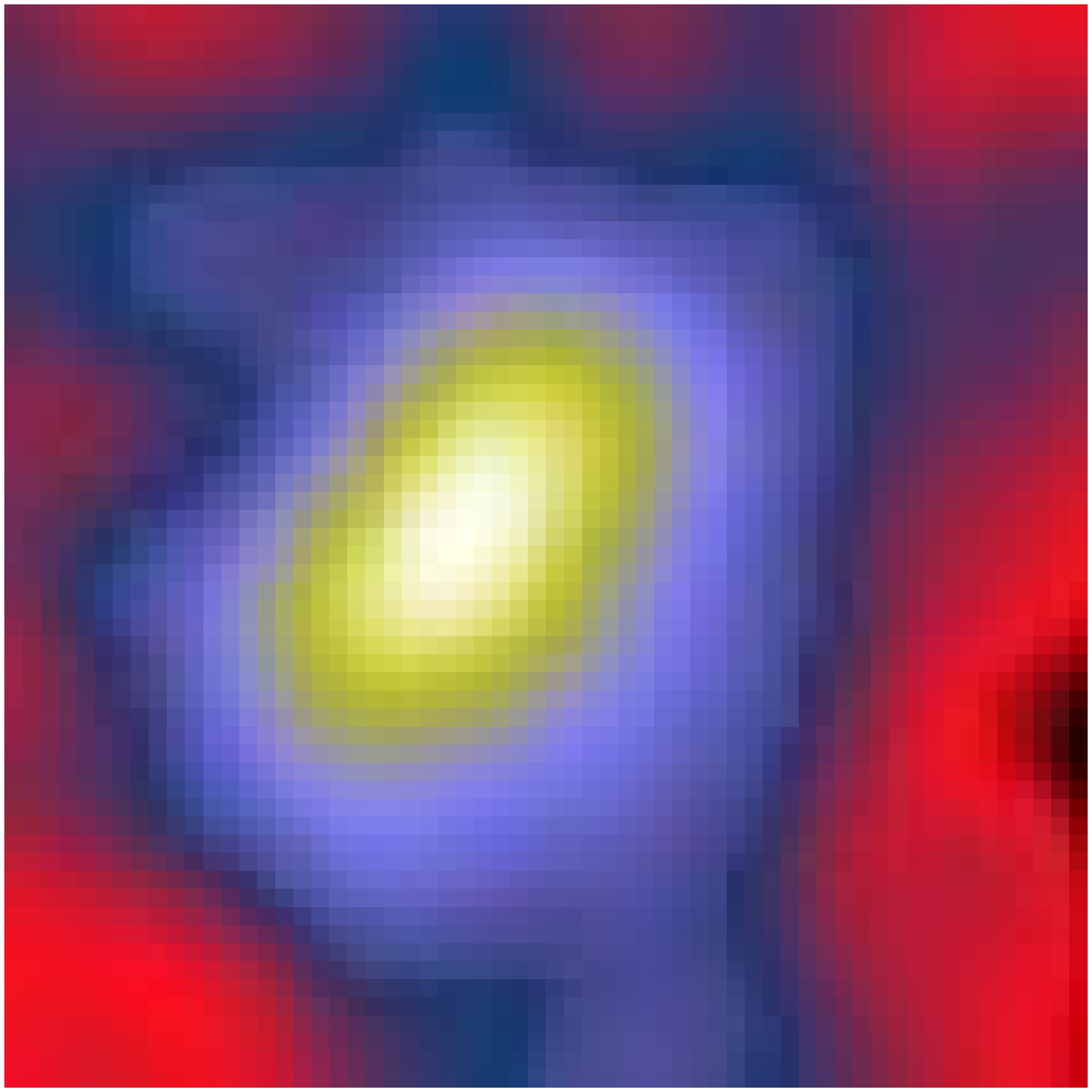} \\  {\#52   NCIA000452}  \\
{ \\ \textbf{Figure \ref{fig:allgal}.(cont)} ~~Visible Galaxies within 40\min\ of NGC~891 } \\
\end{figure}

\clearpage
\begin{figure}[ht]
  \HDb{rgb}  \HDb{FUV}  \HDb{NUV}   \HDb{B}     \HDb{R}    \HDb{IR}   \HDb{J}    \HDb{H}     \HDb{K}     \HDb{W1}    \HDb{W2}    \HDb{W3}    \HDb{W4}  \\
  \FGb{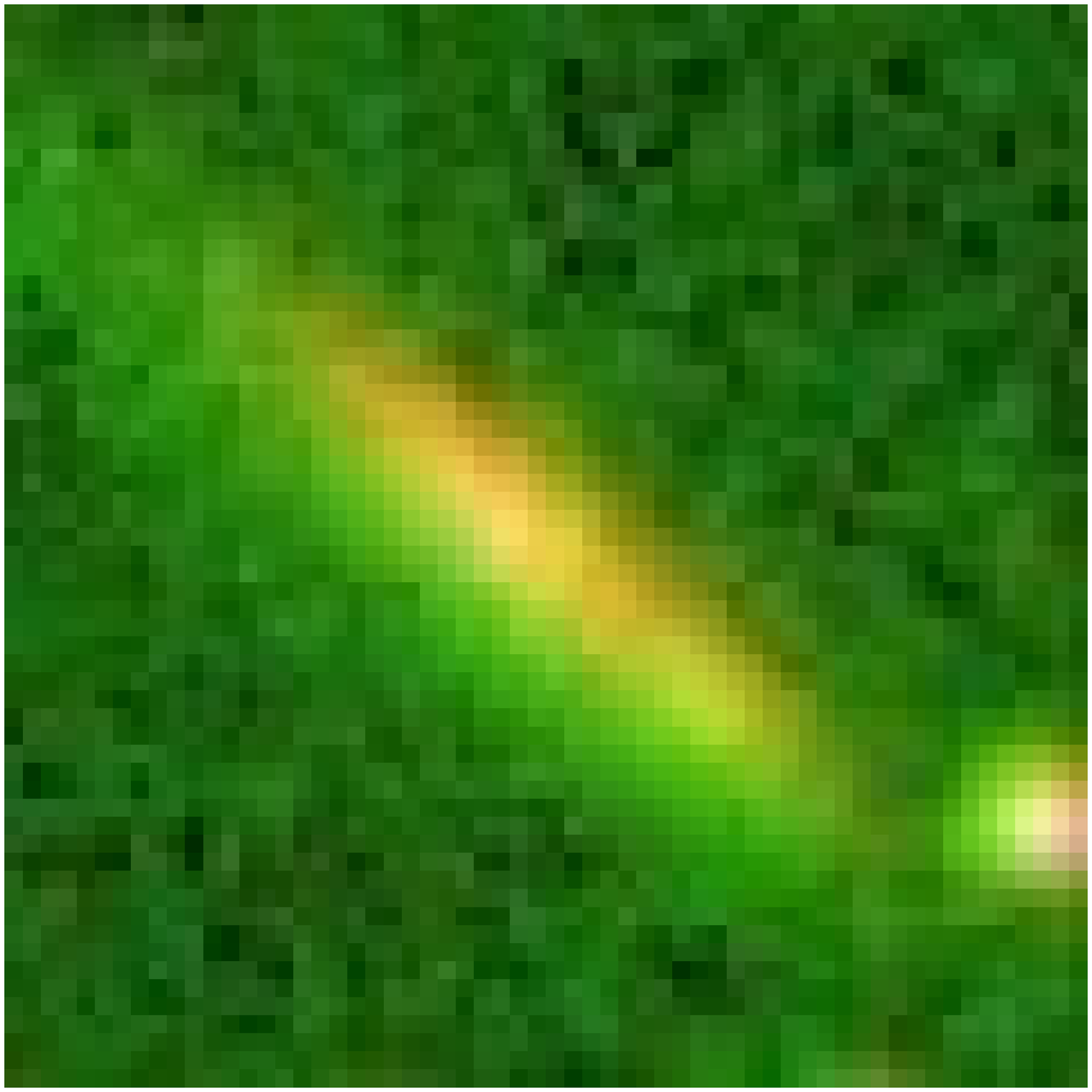}  \FGb{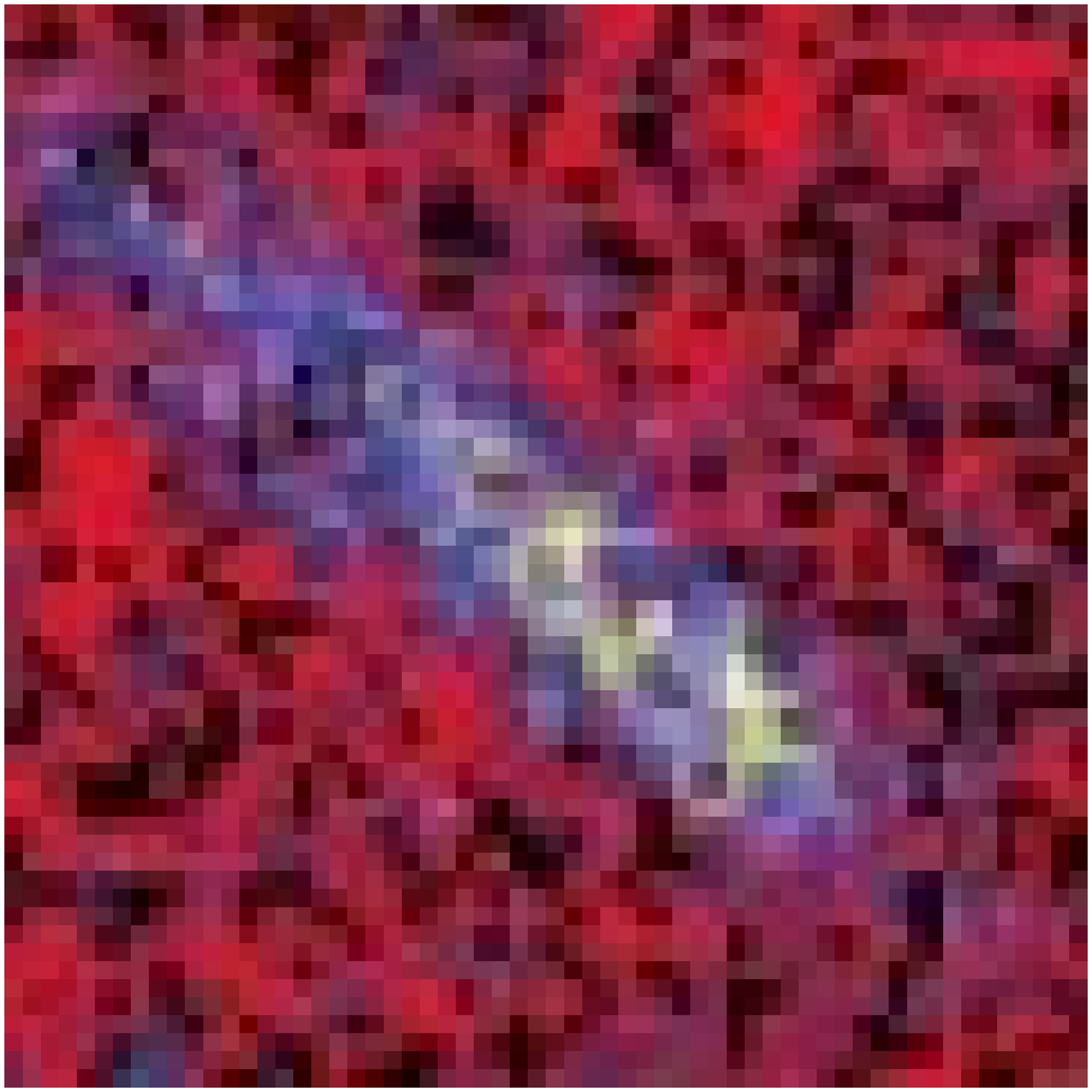}  \FGb{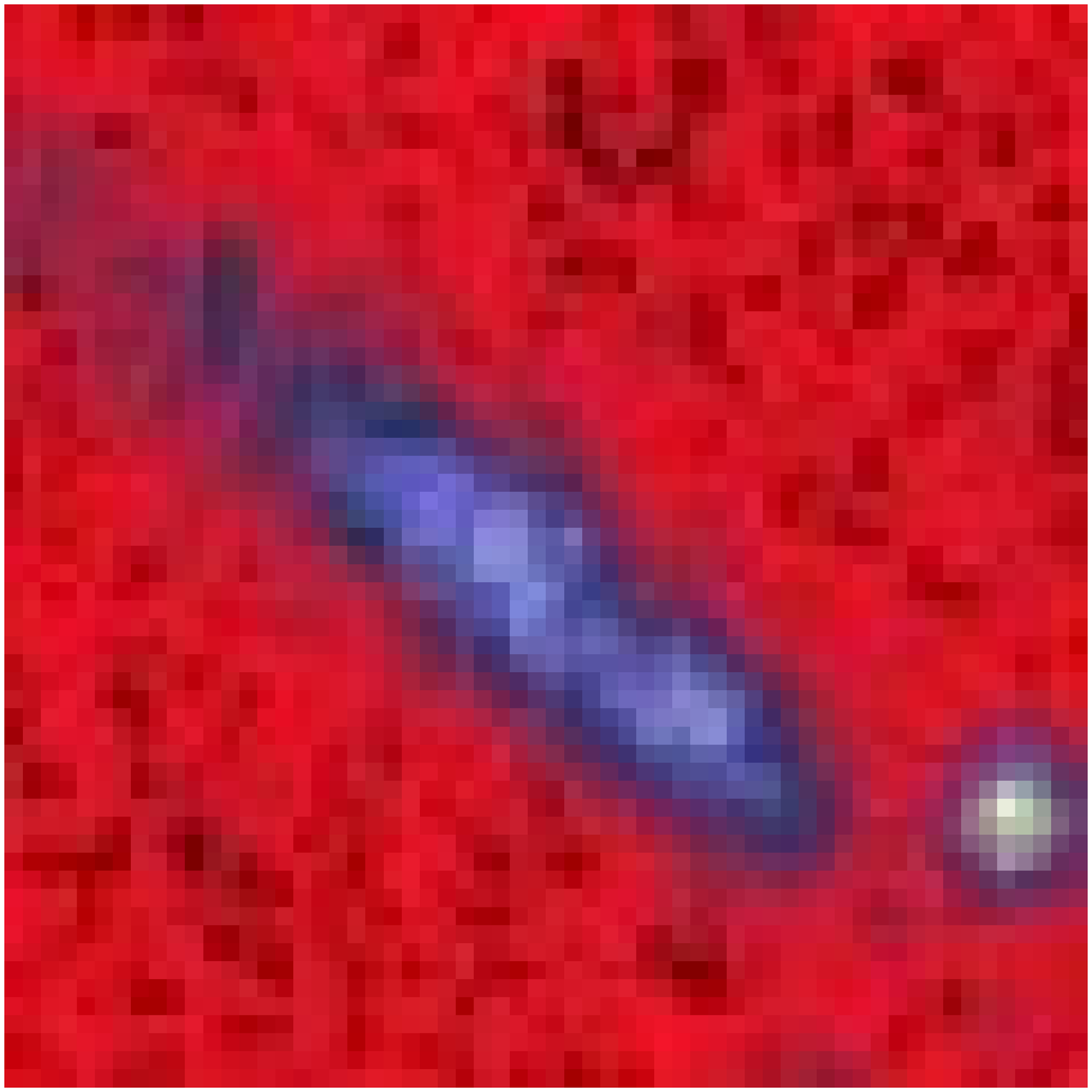}  \FGb{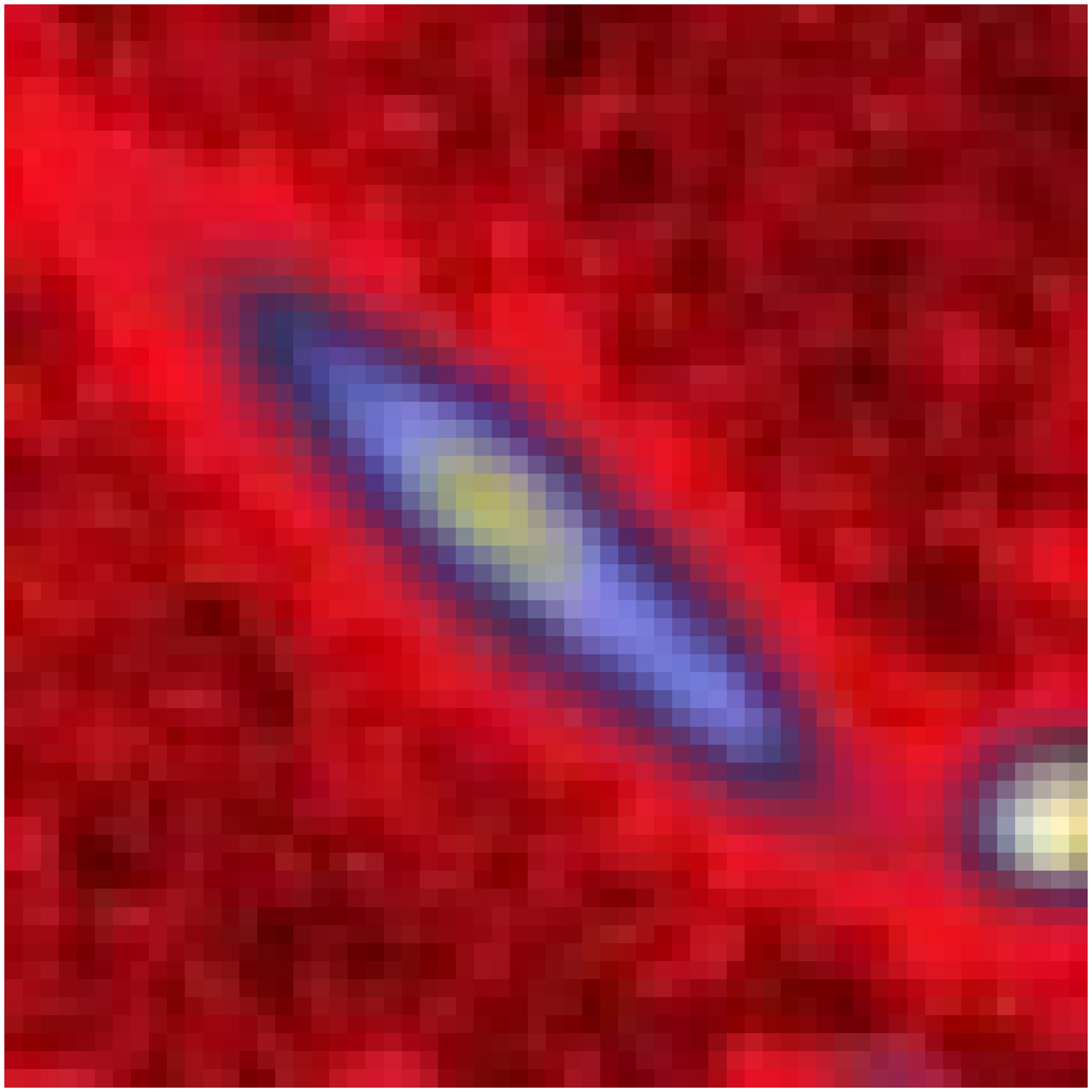}  \FGb{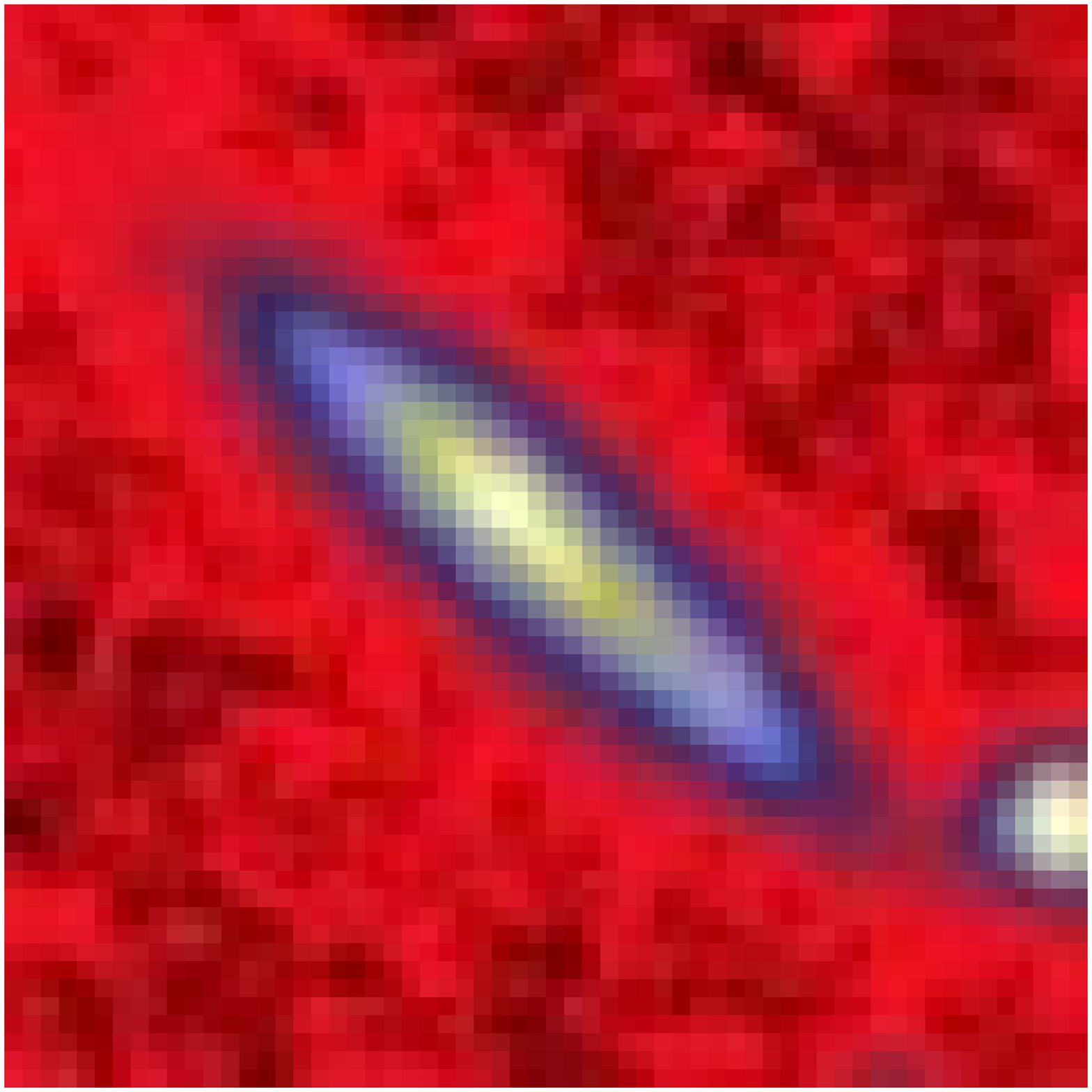}  \FGb{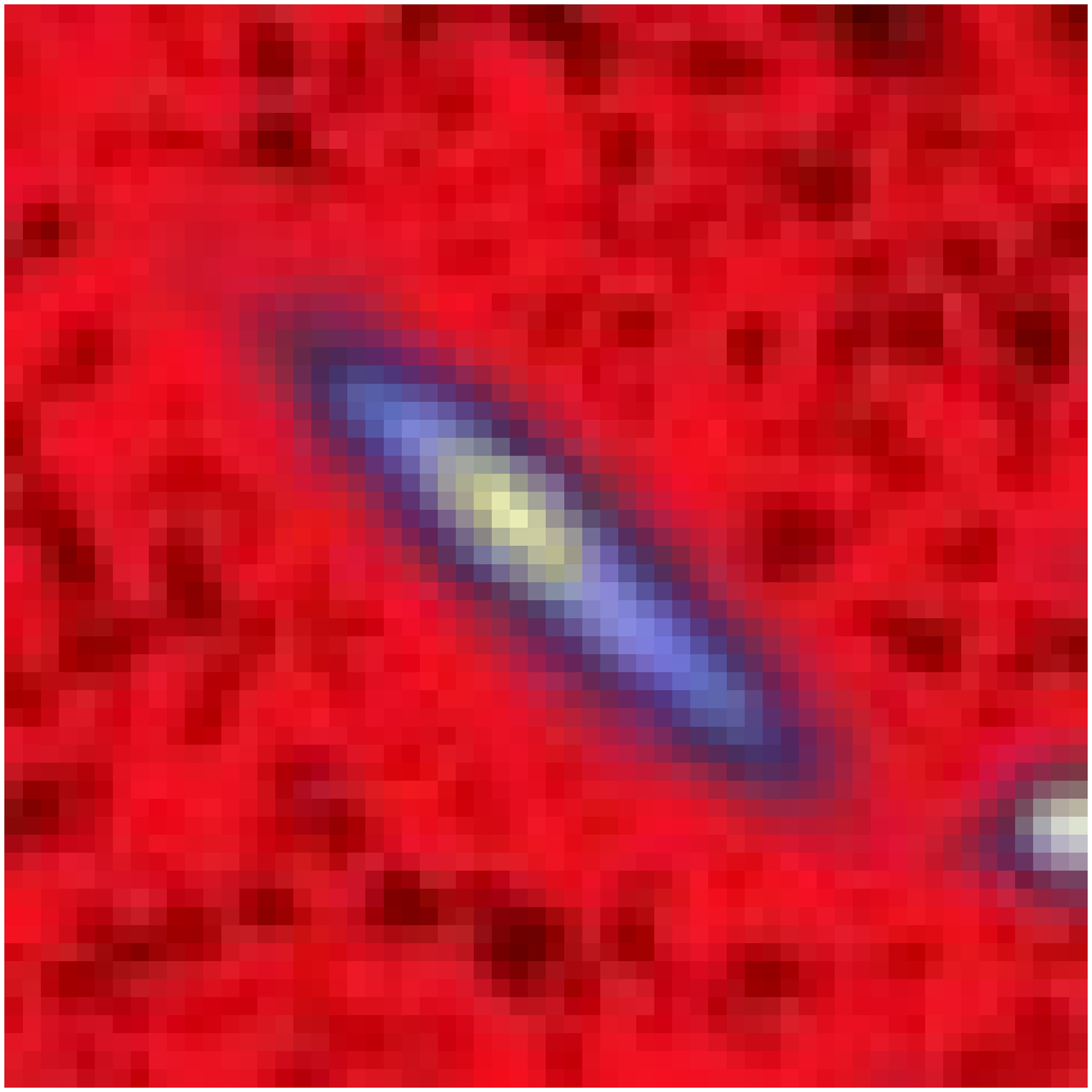}  \FGb{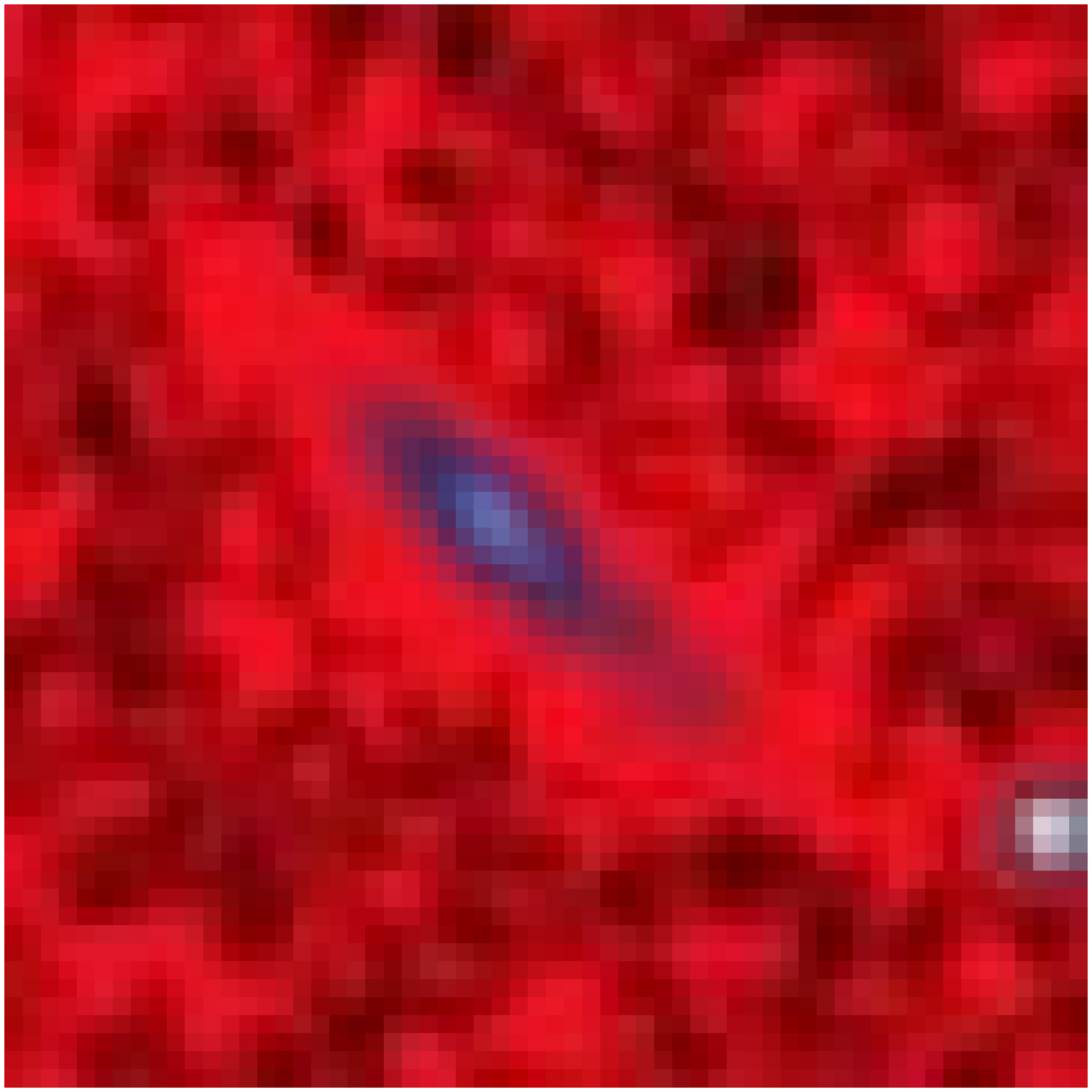}  \FGb{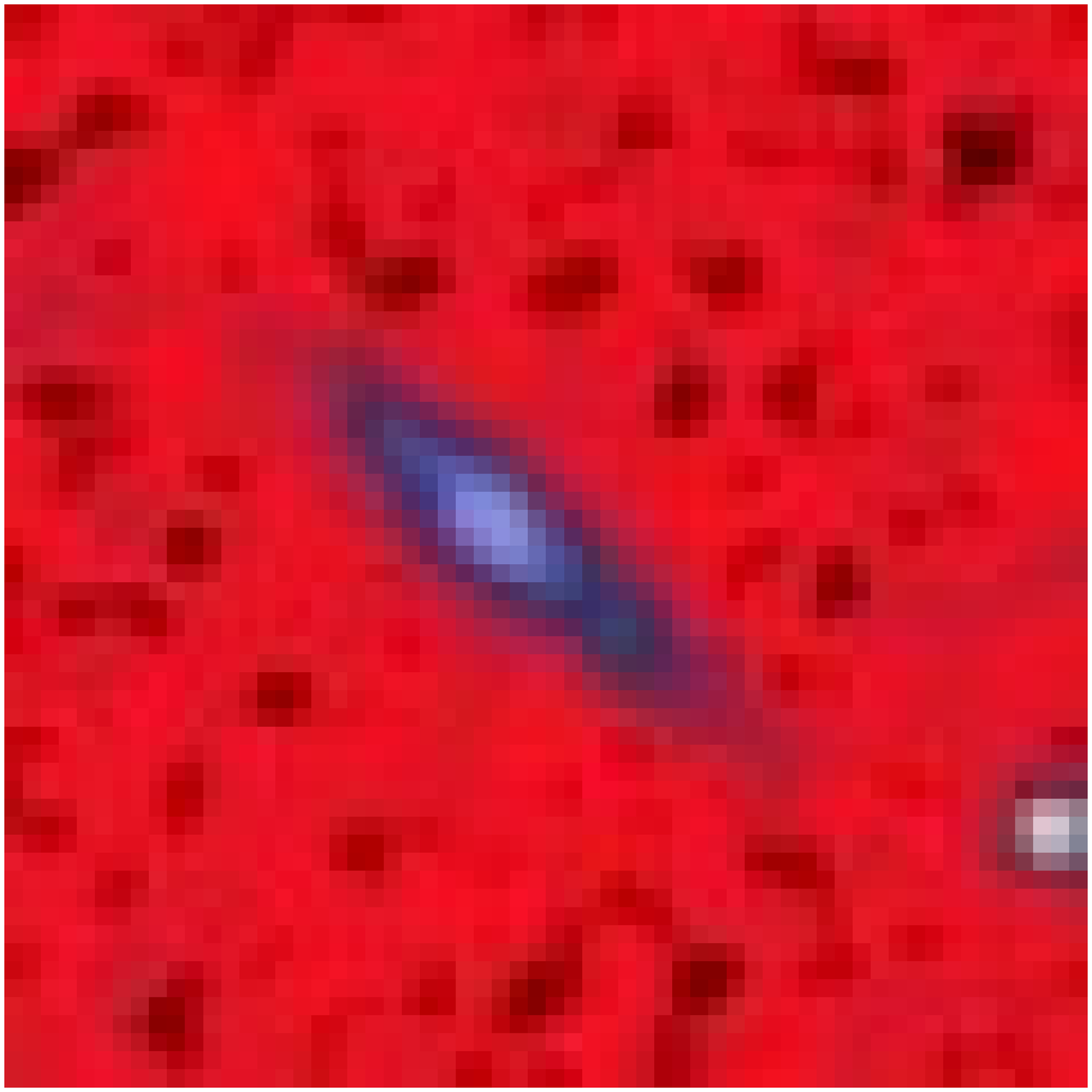}  \FGb{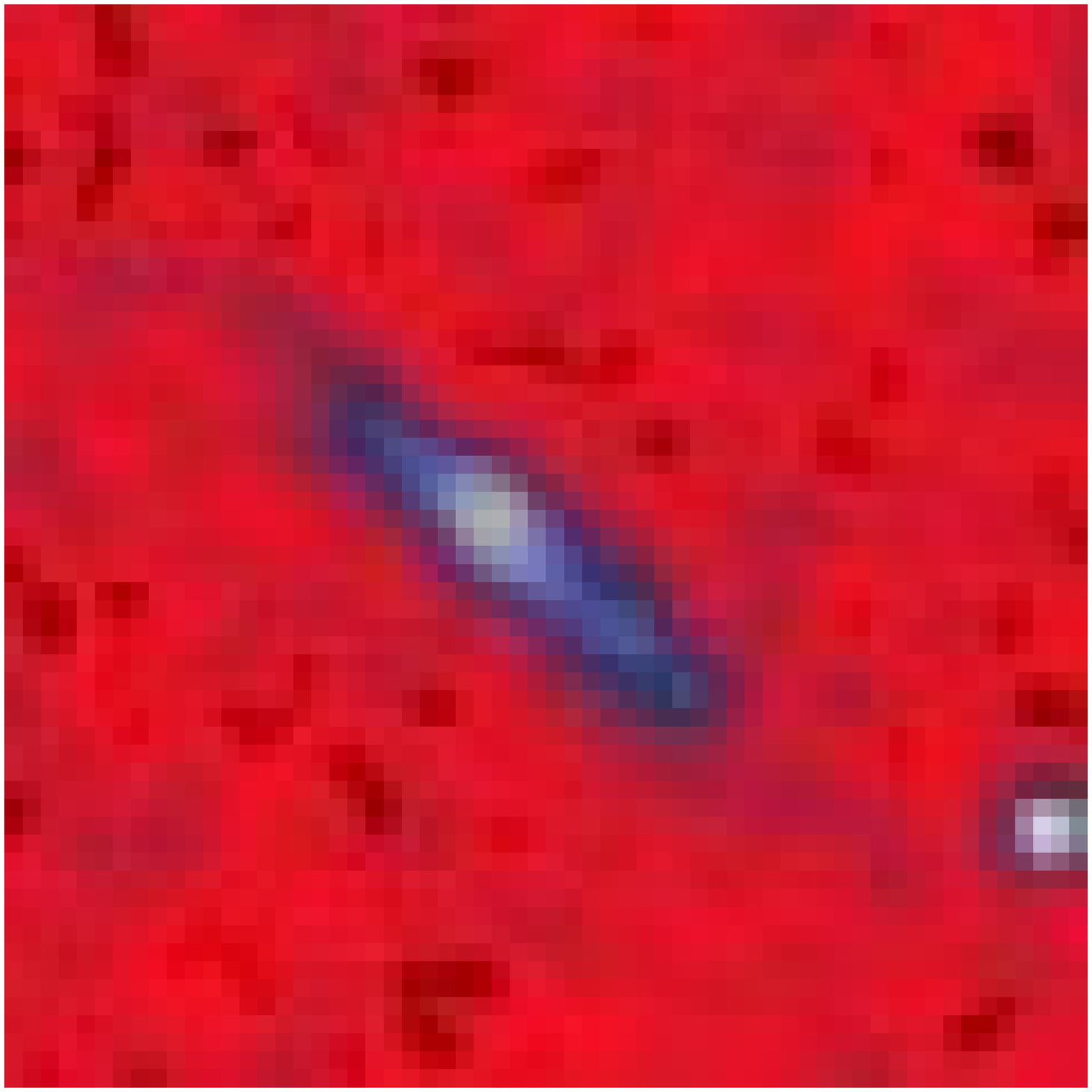}  \FGb{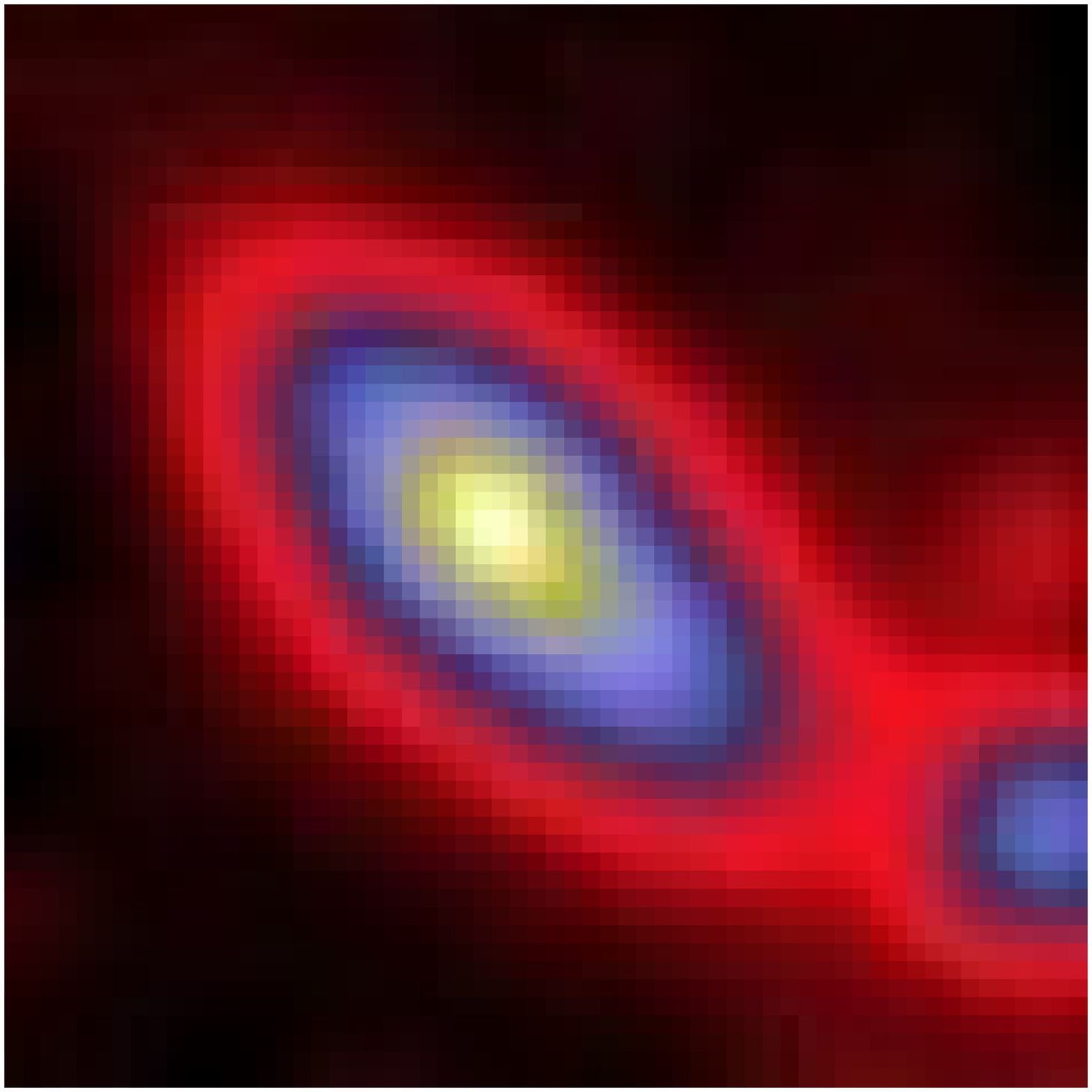}  \FGb{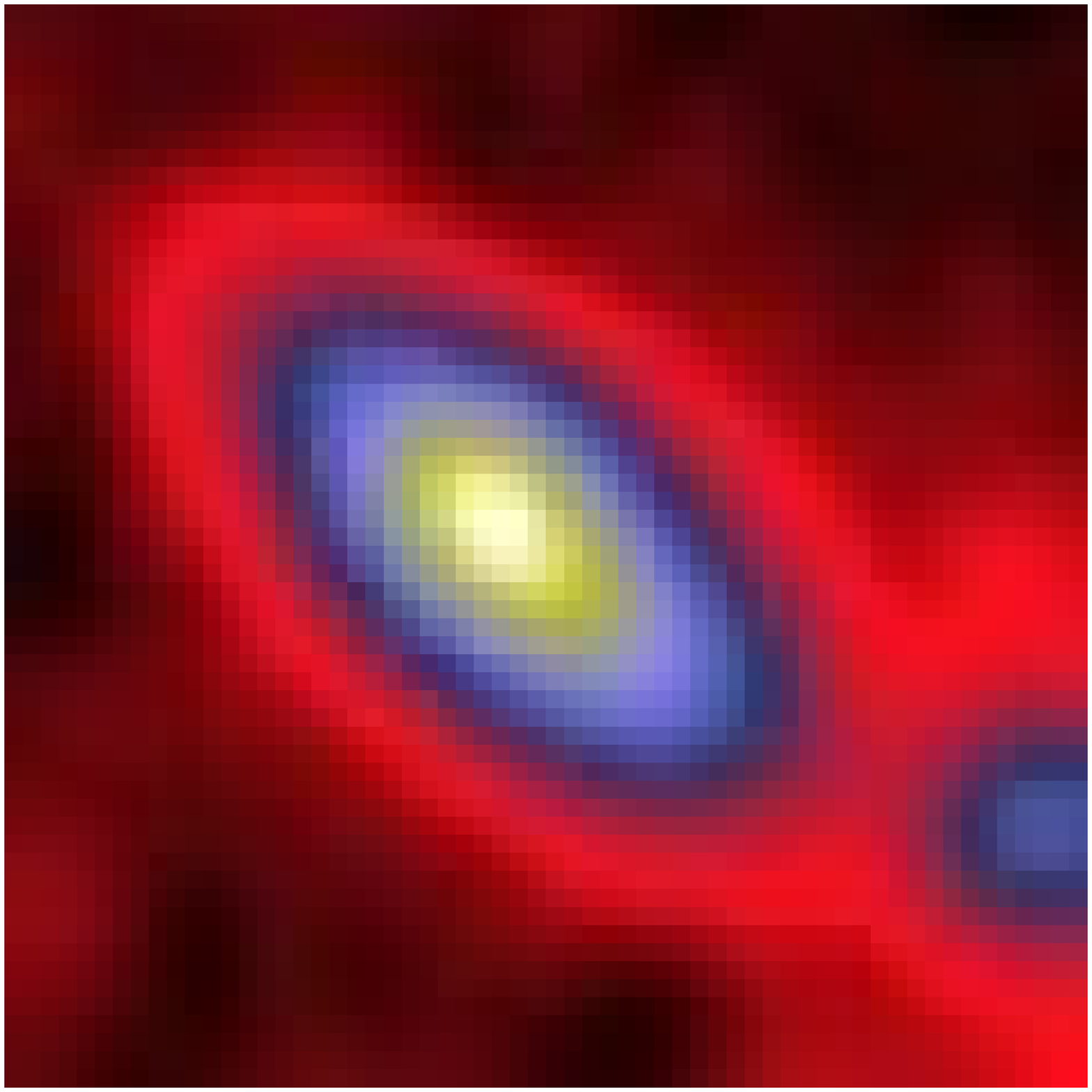}  \FGb{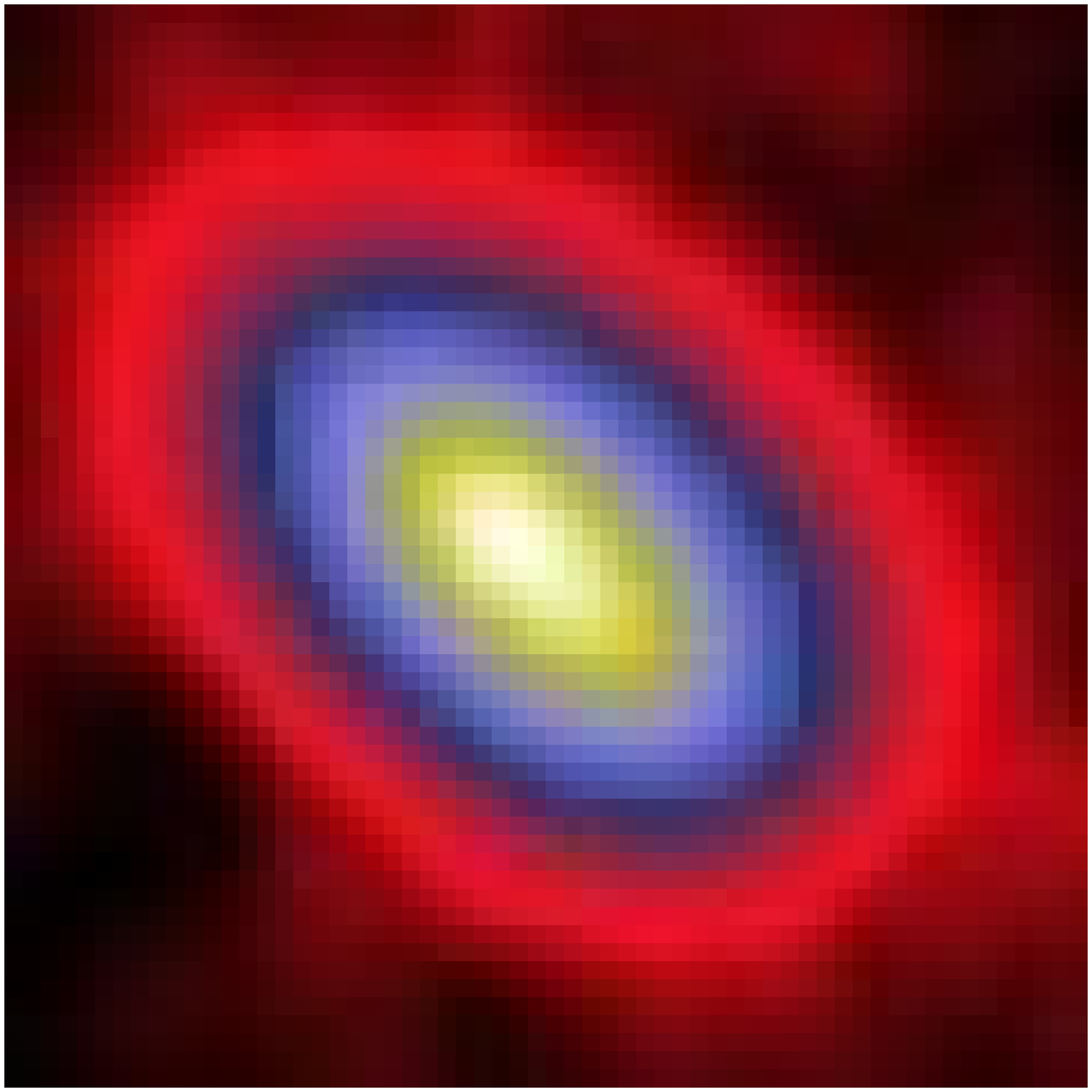}  \FGb{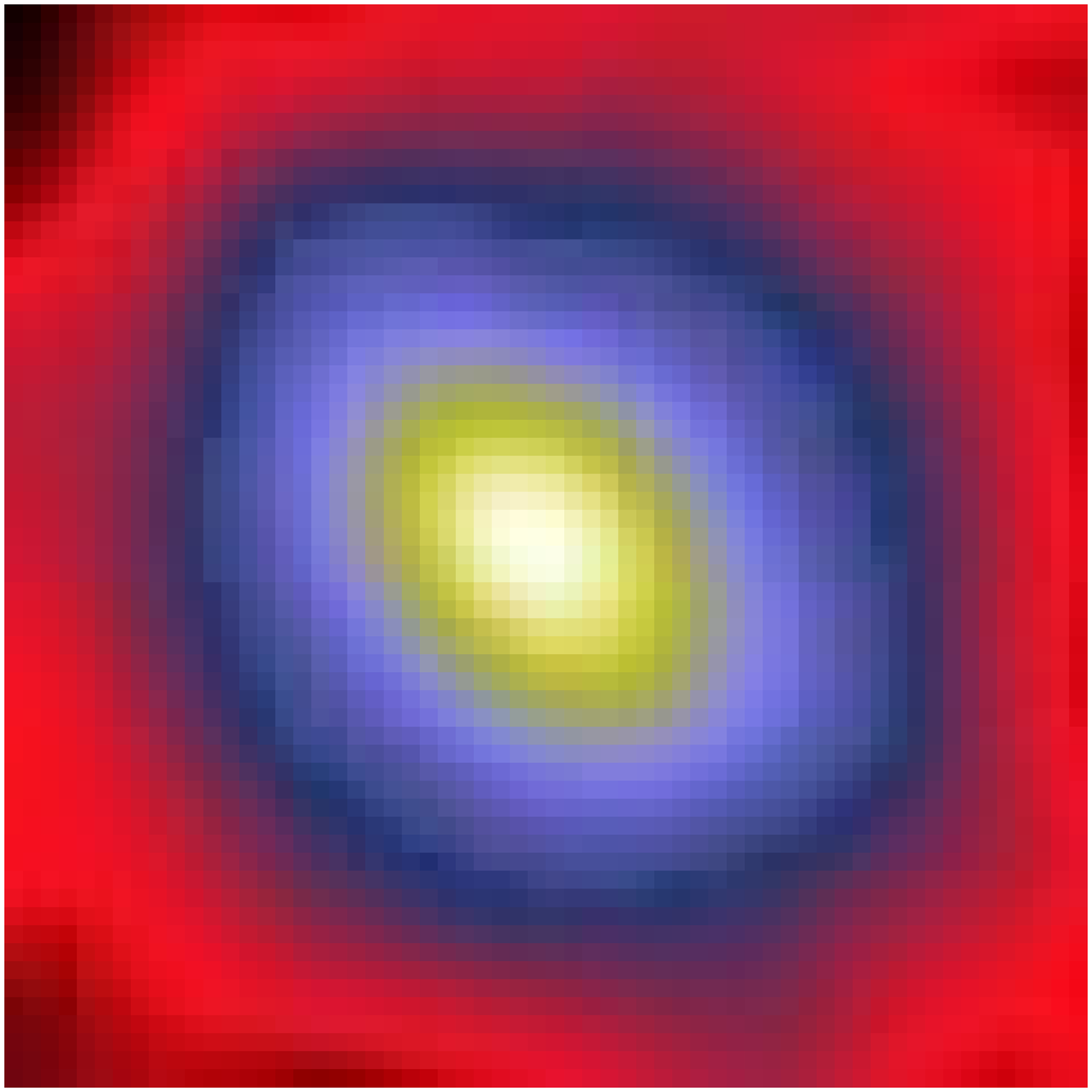} \\  {\#53   NCIA001126}  \\
  \FGb{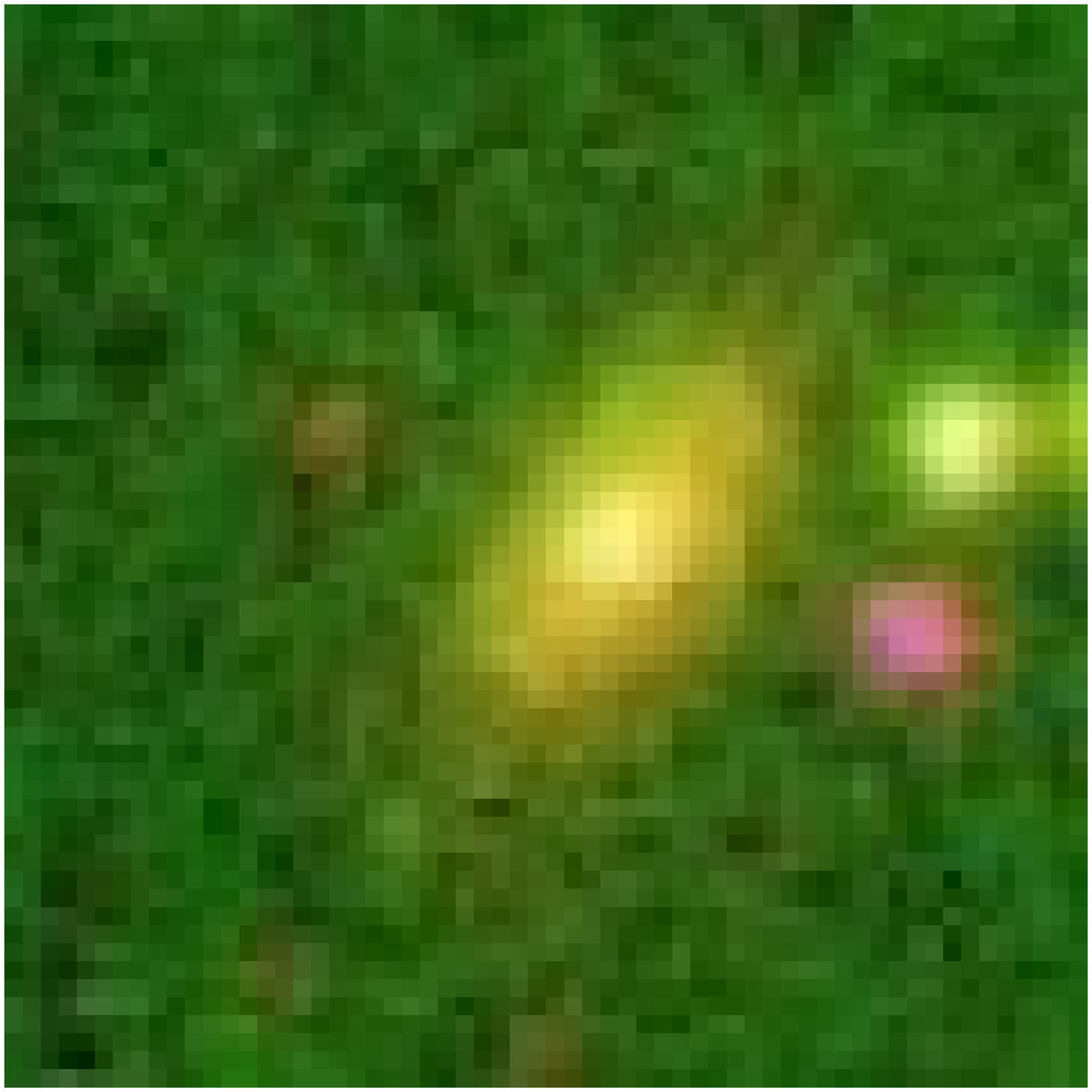}  \FGb{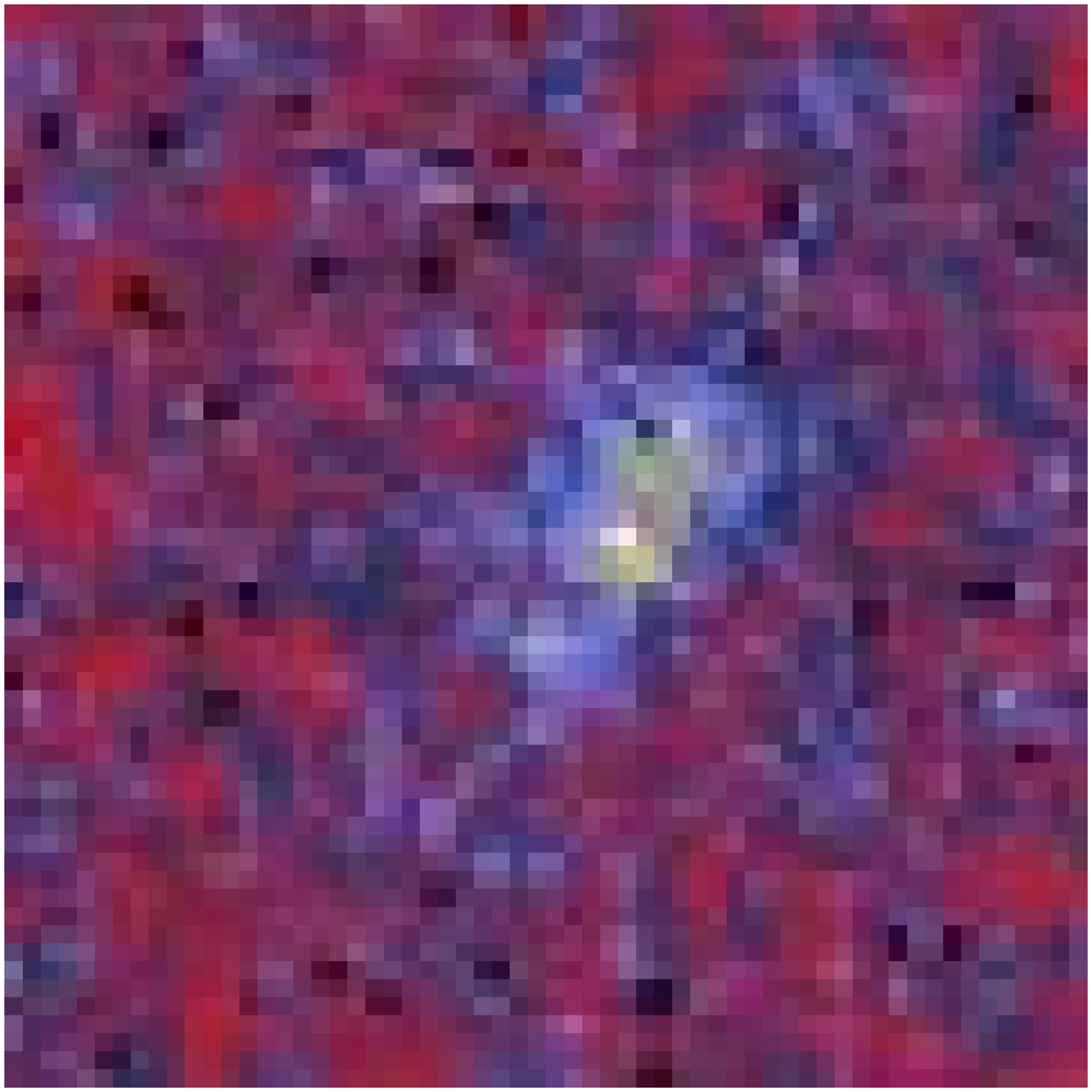}  \FGb{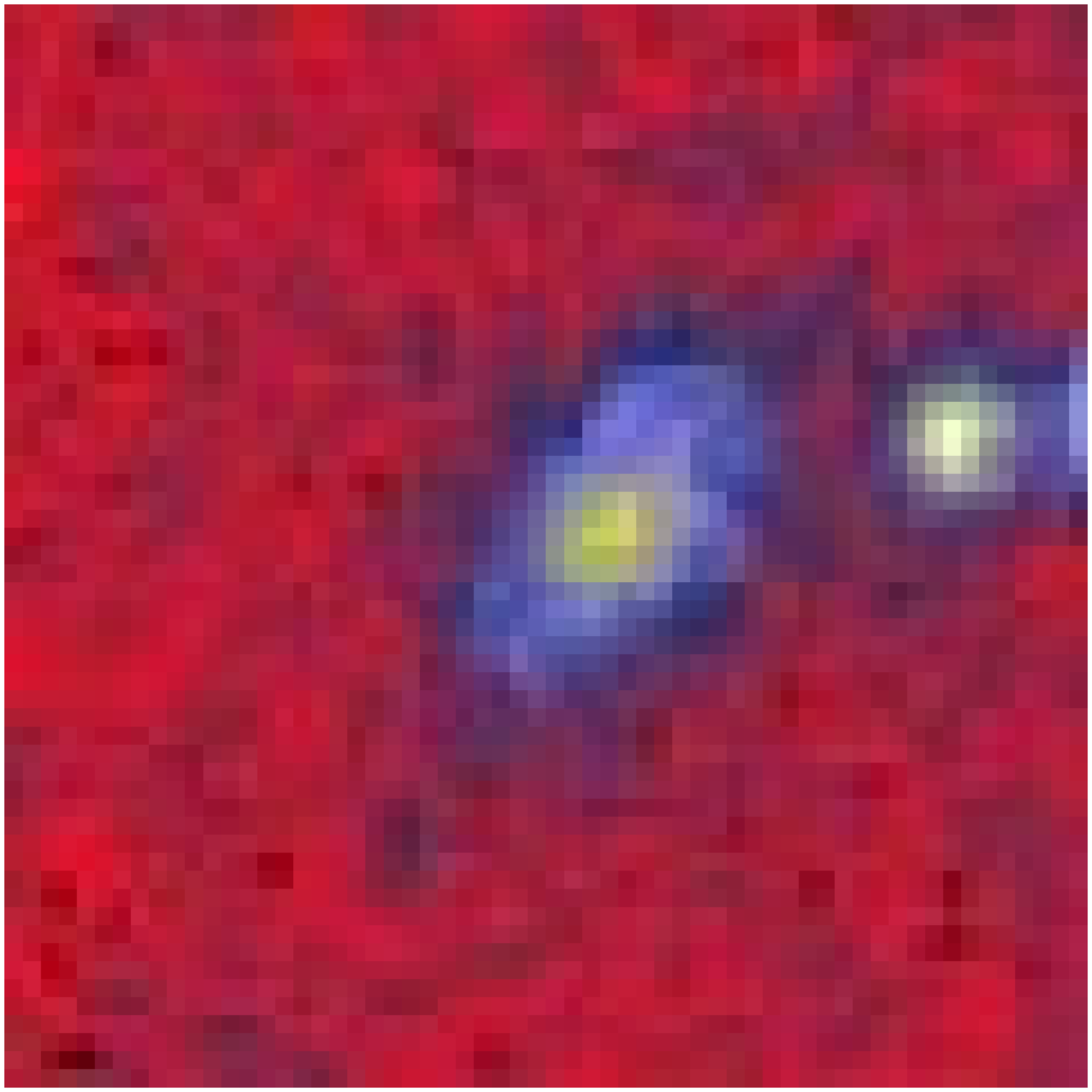}  \FGb{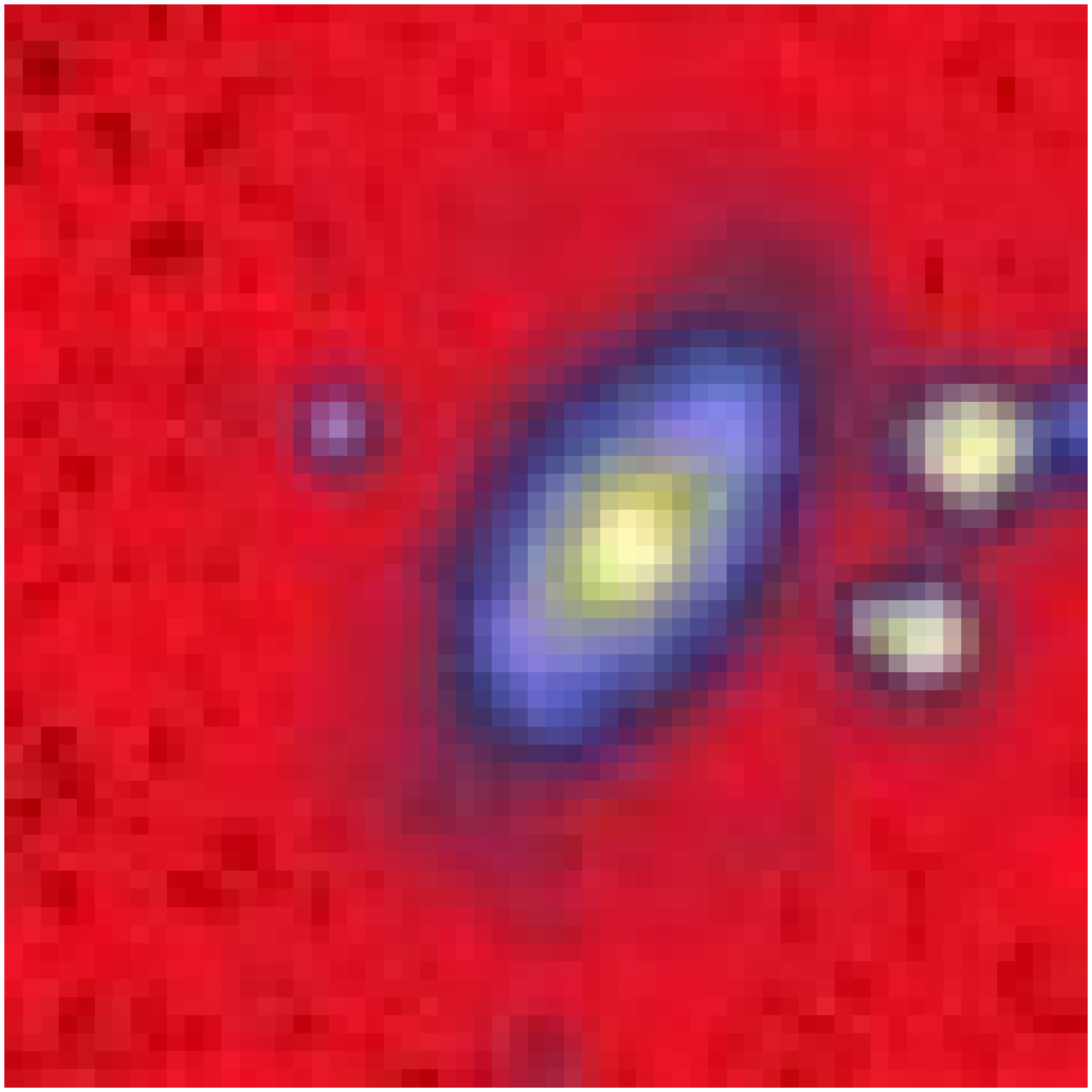}  \FGb{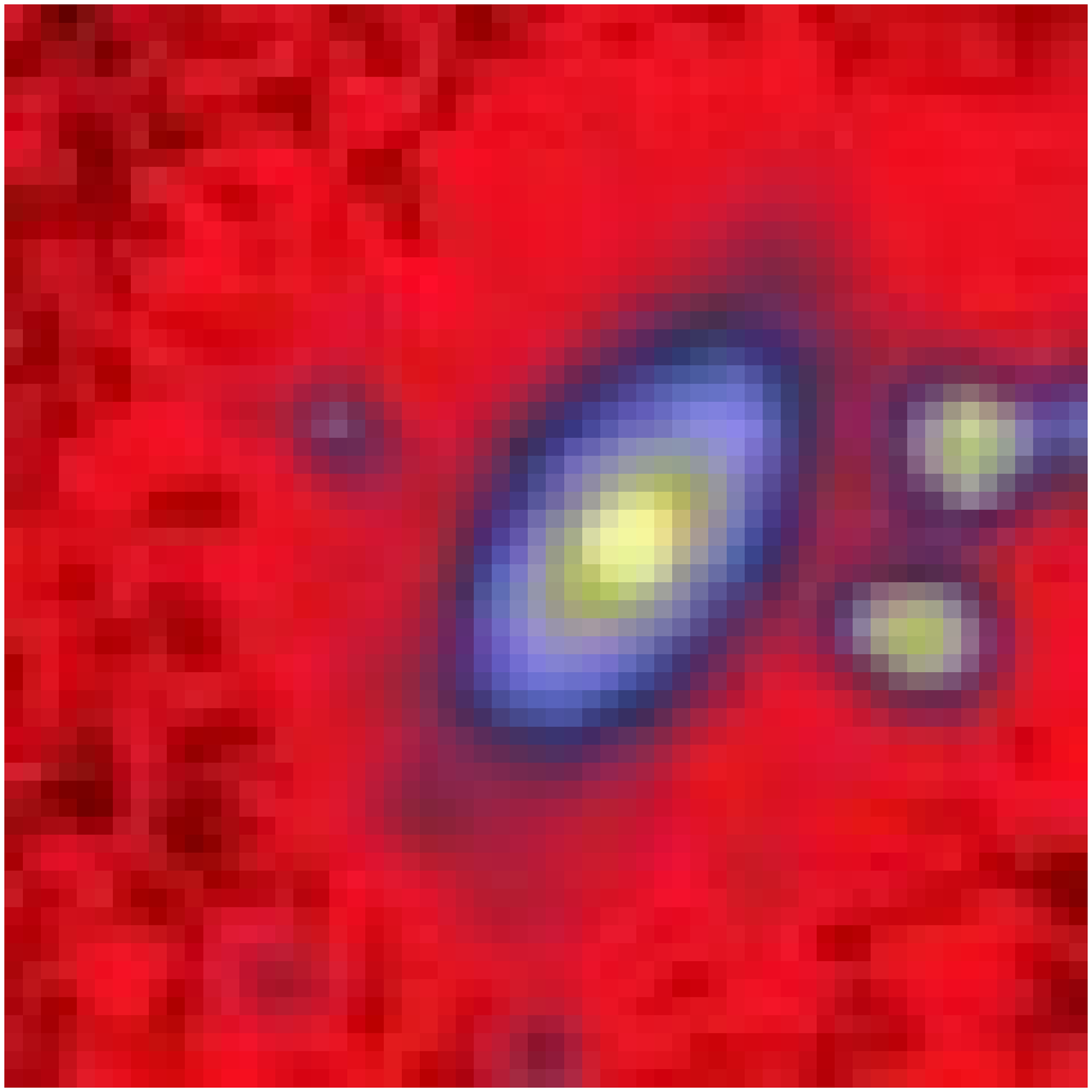}  \FGb{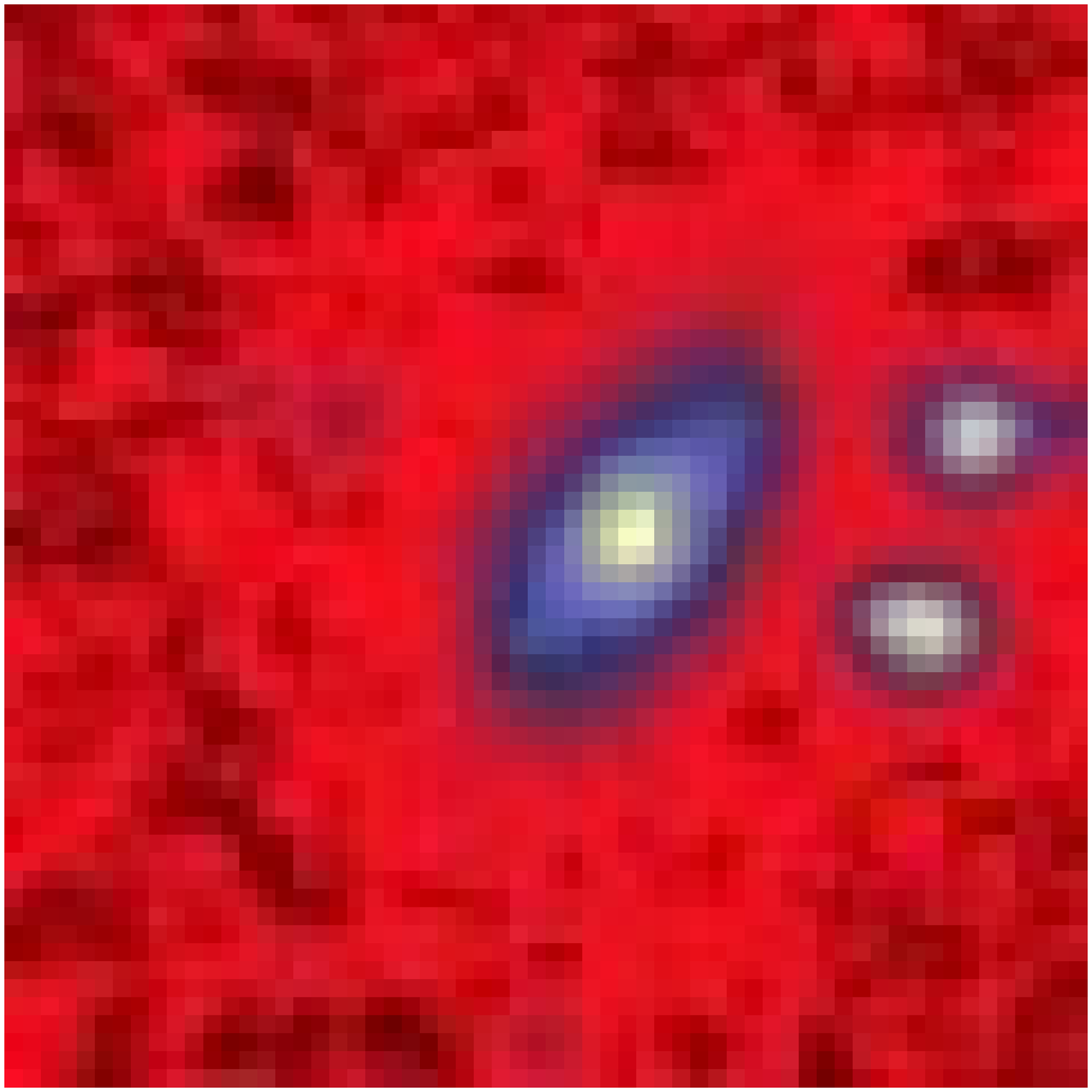}  \FGb{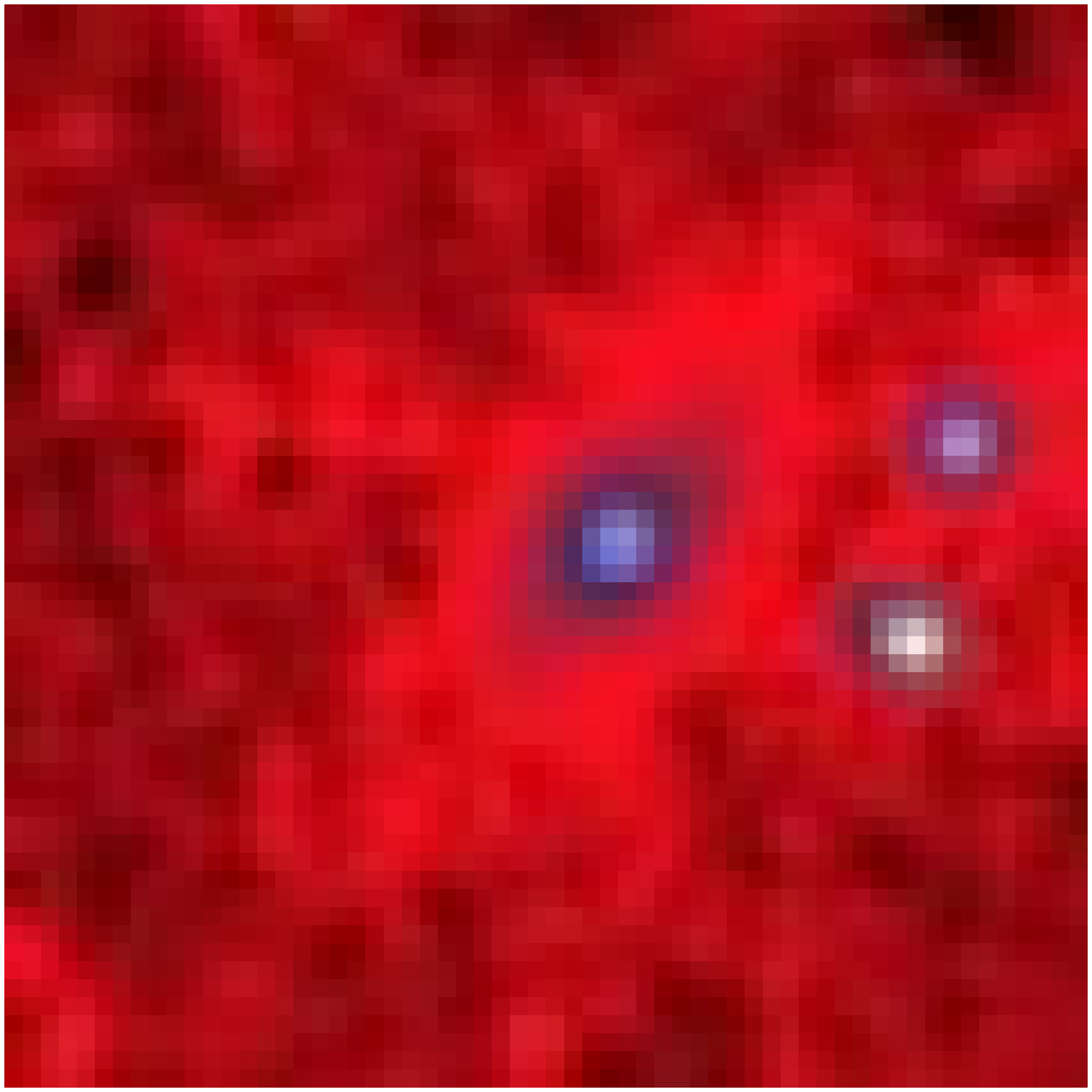}  \FGb{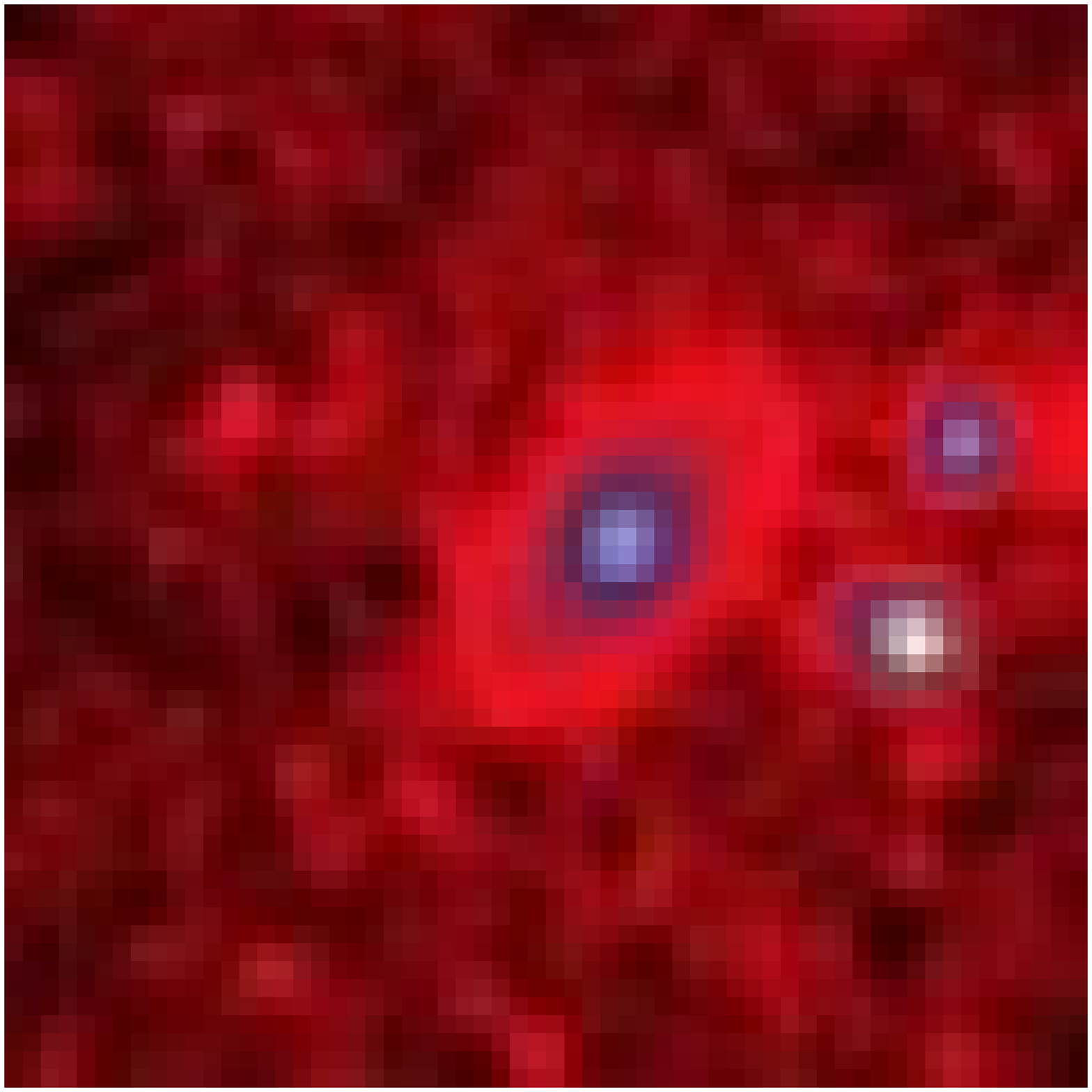}  \FGb{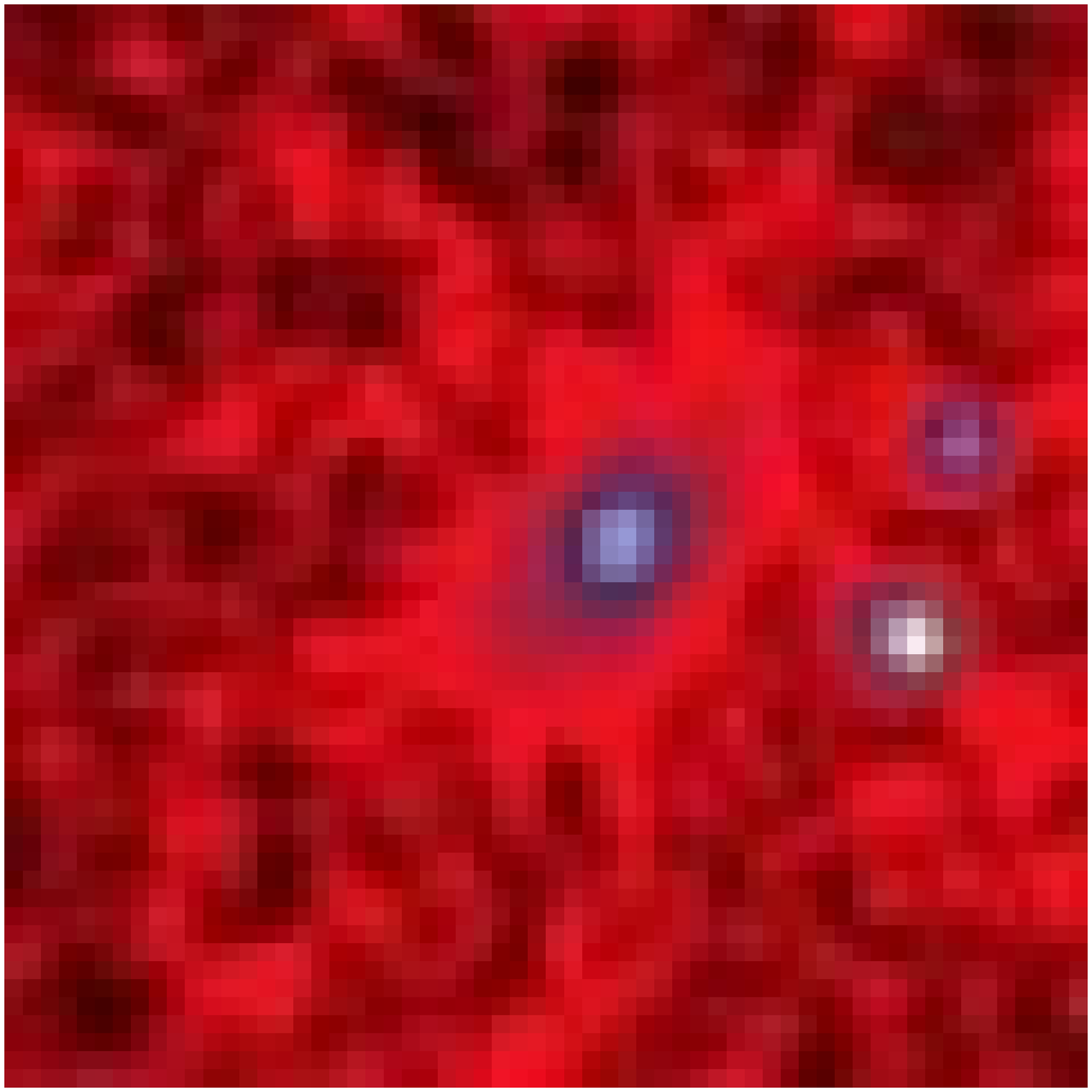}  \FGb{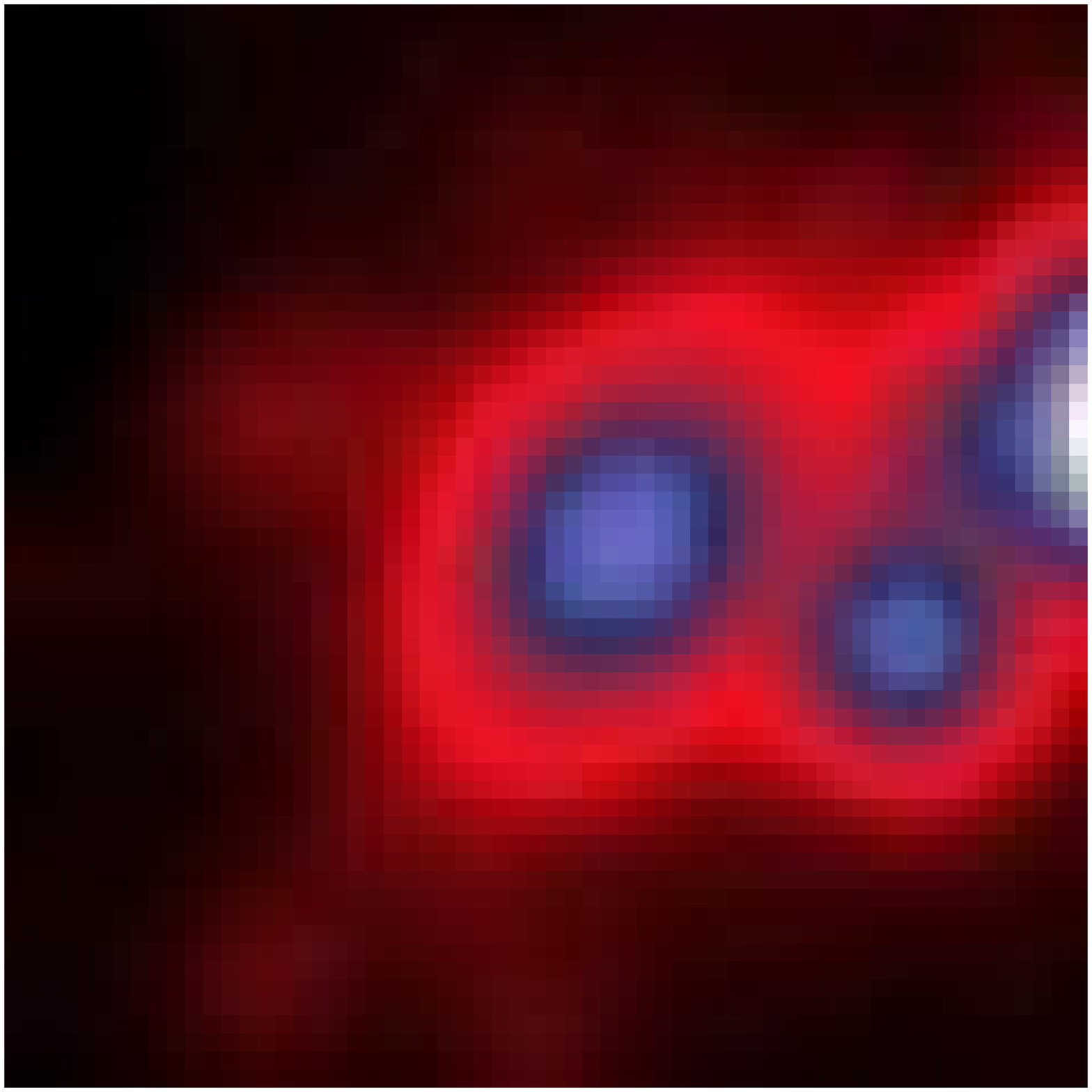}  \FGb{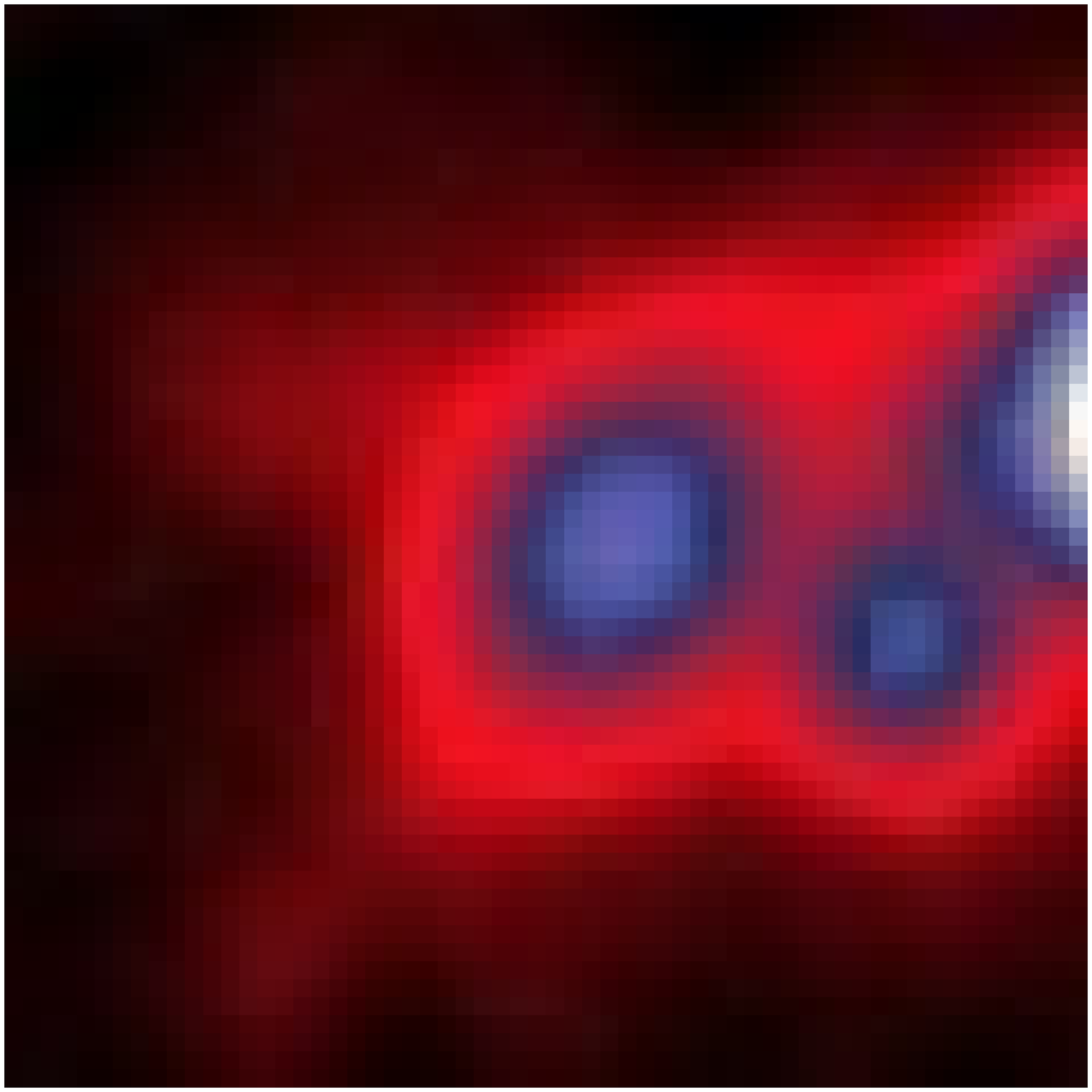}  \FGb{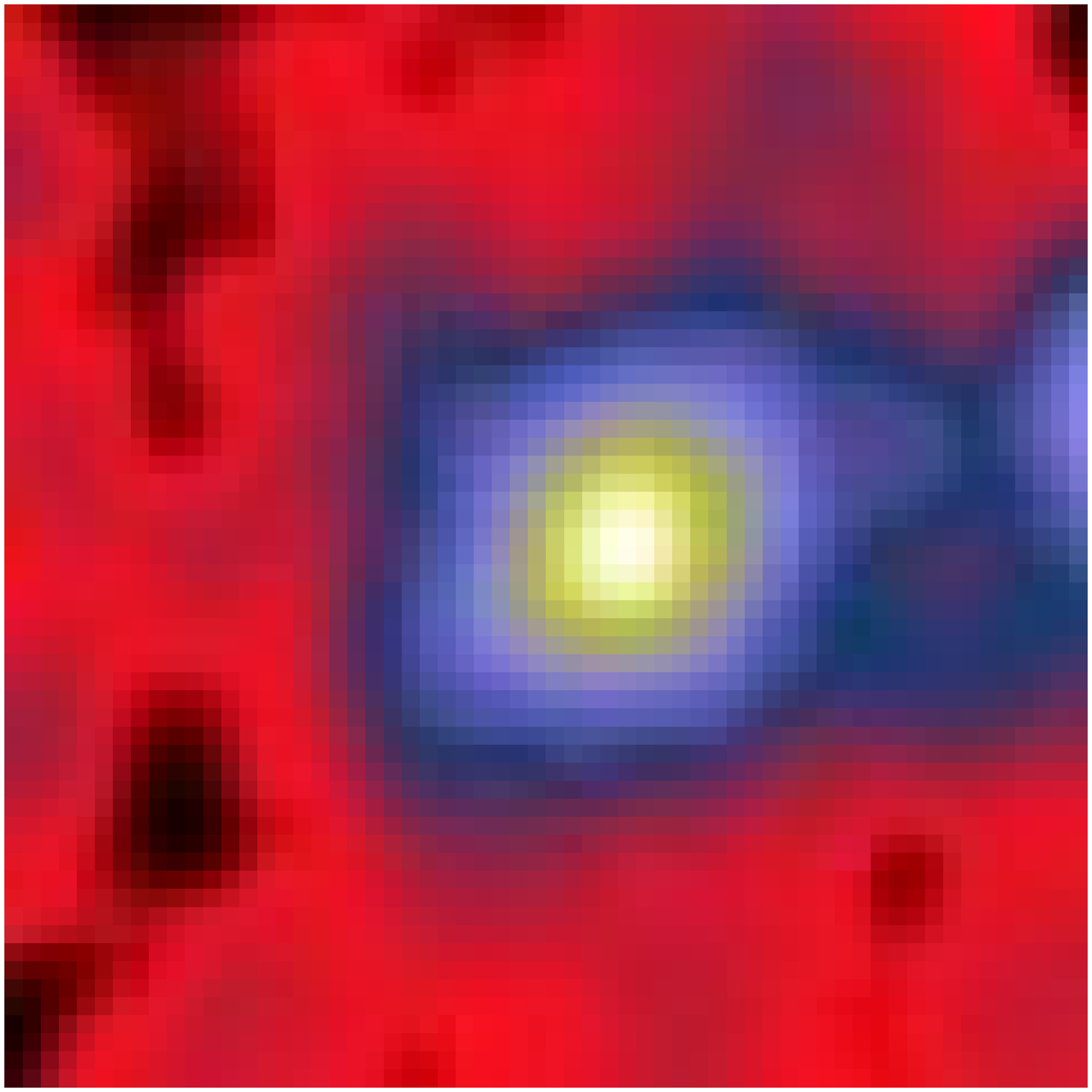}  \FGb{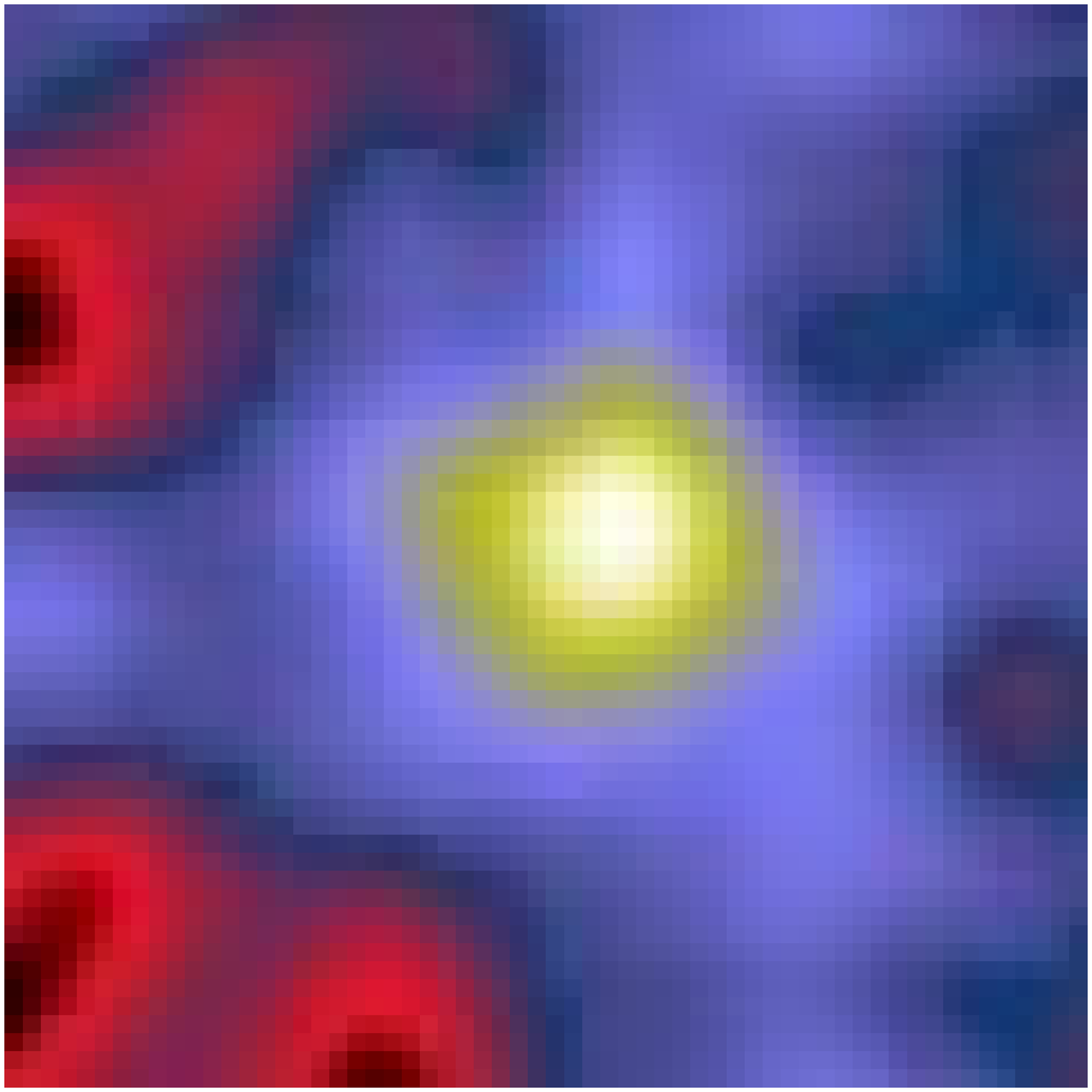} \\  {\#54   NCIA000277}  \\
  \FGb{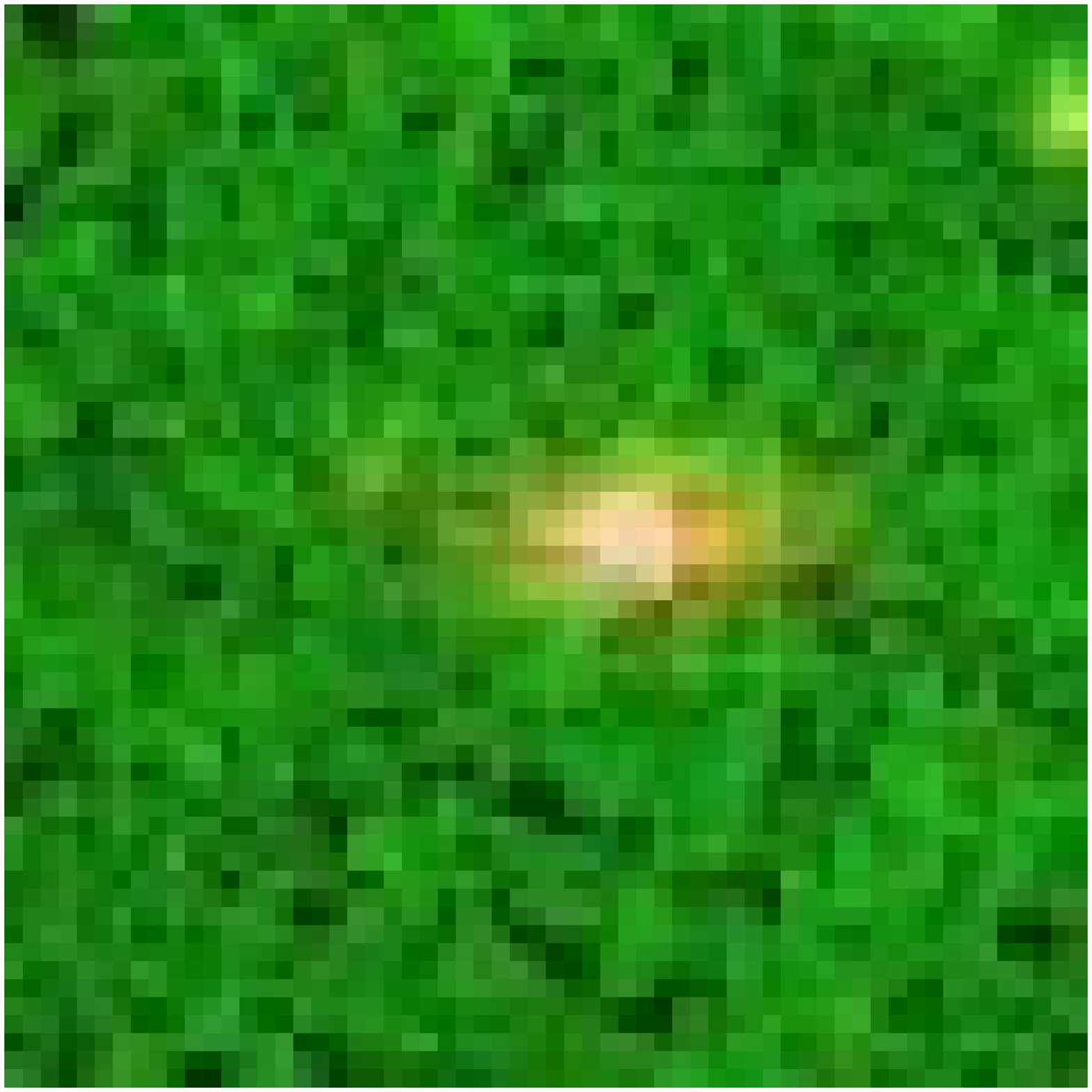}  \FGb{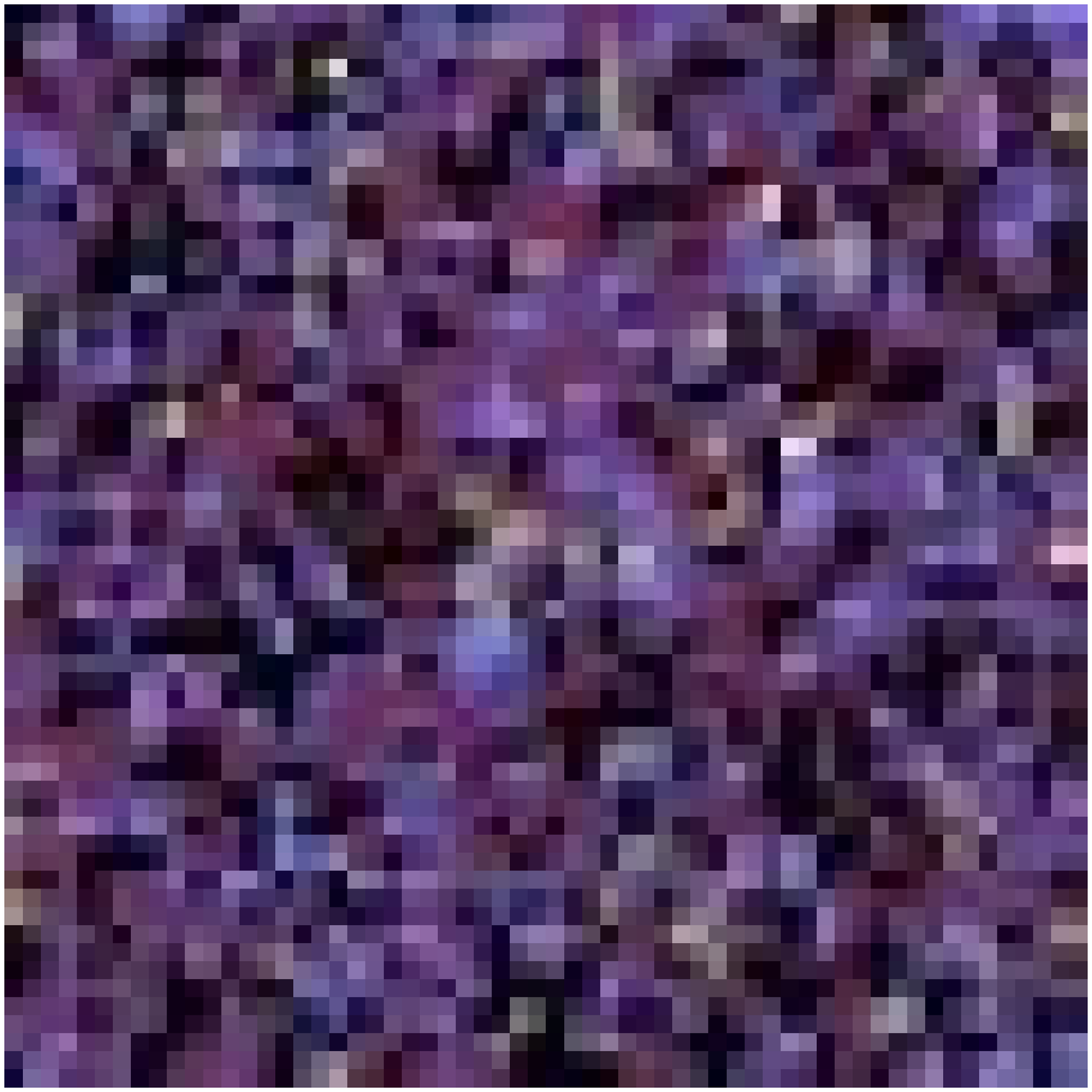}  \FGb{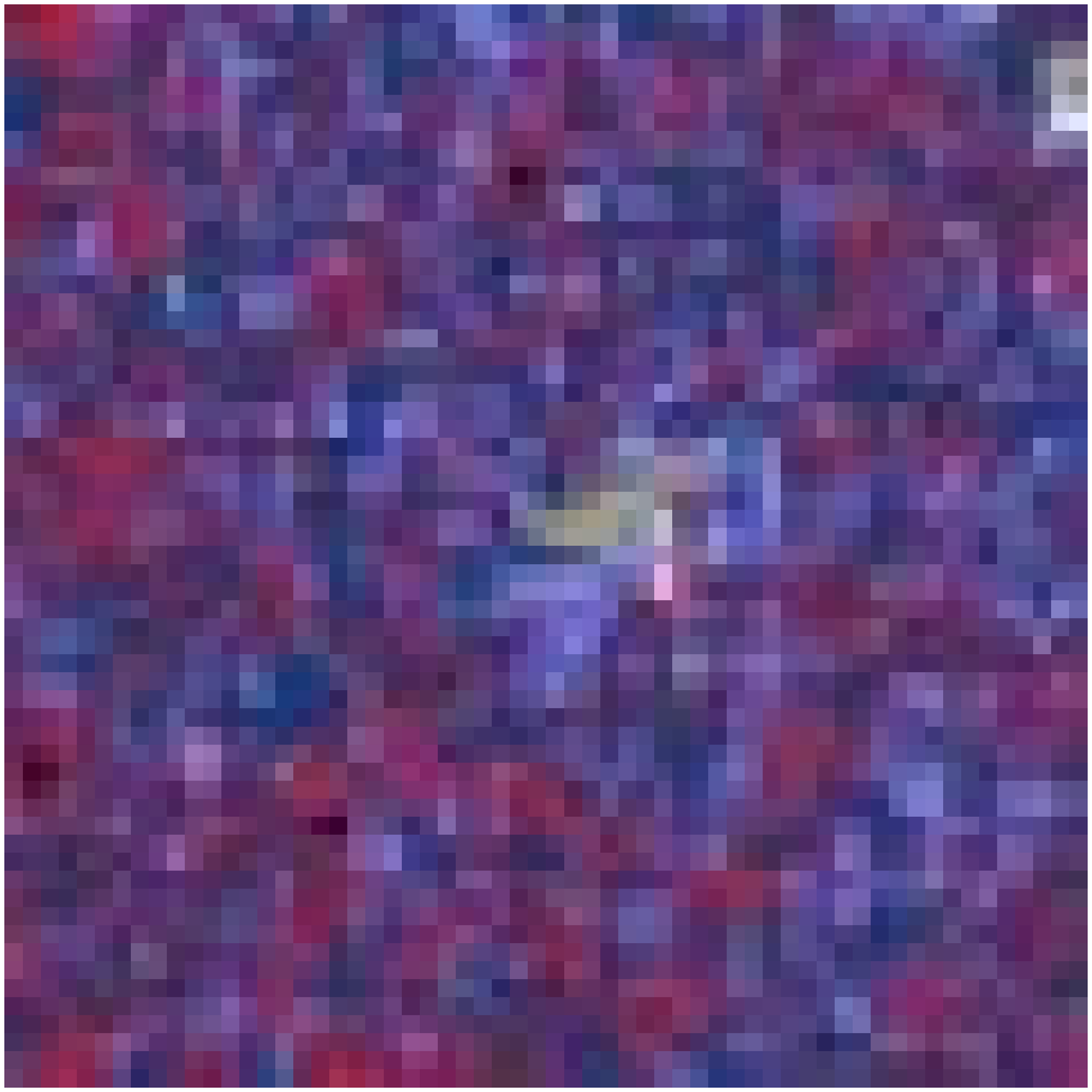}  \FGb{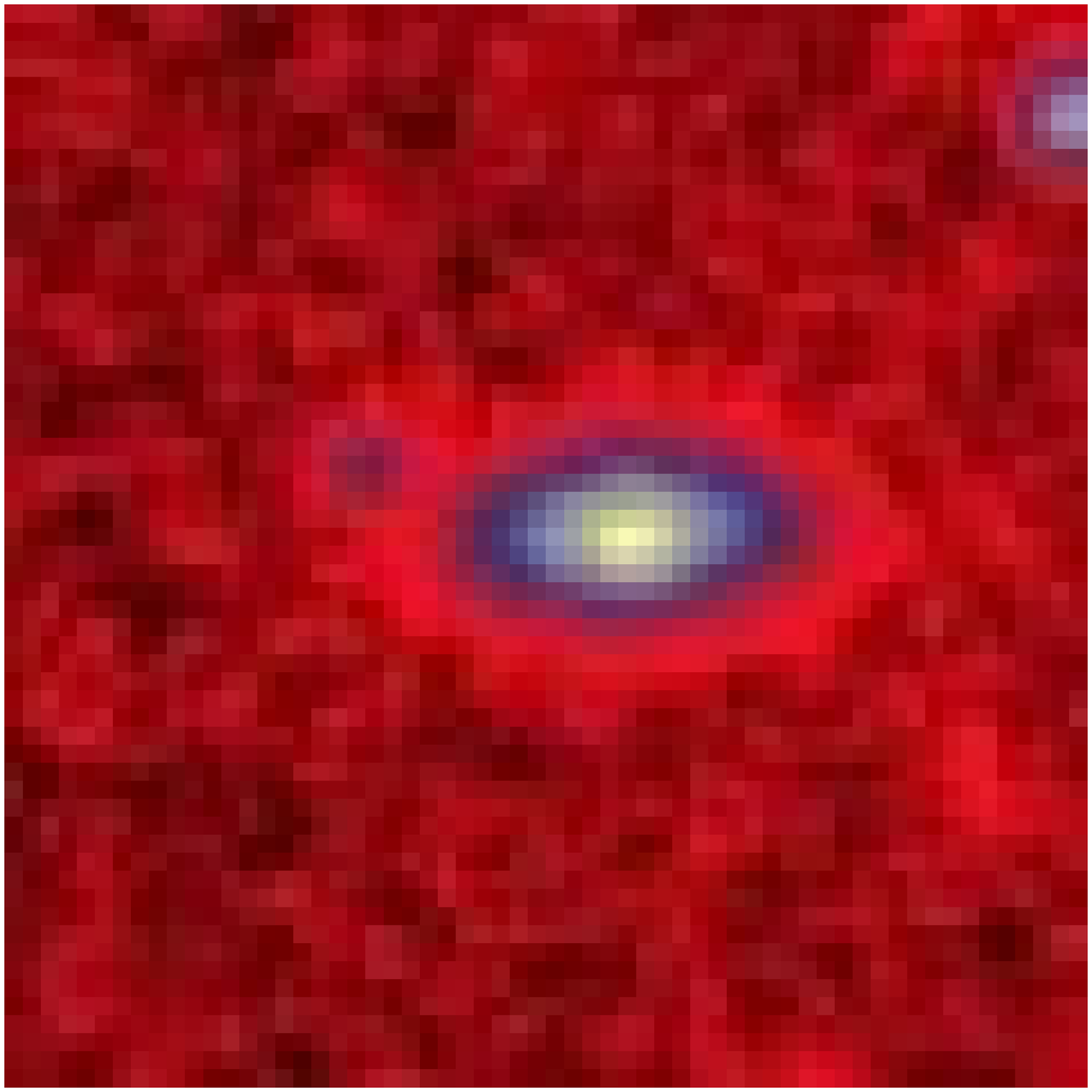}  \FGb{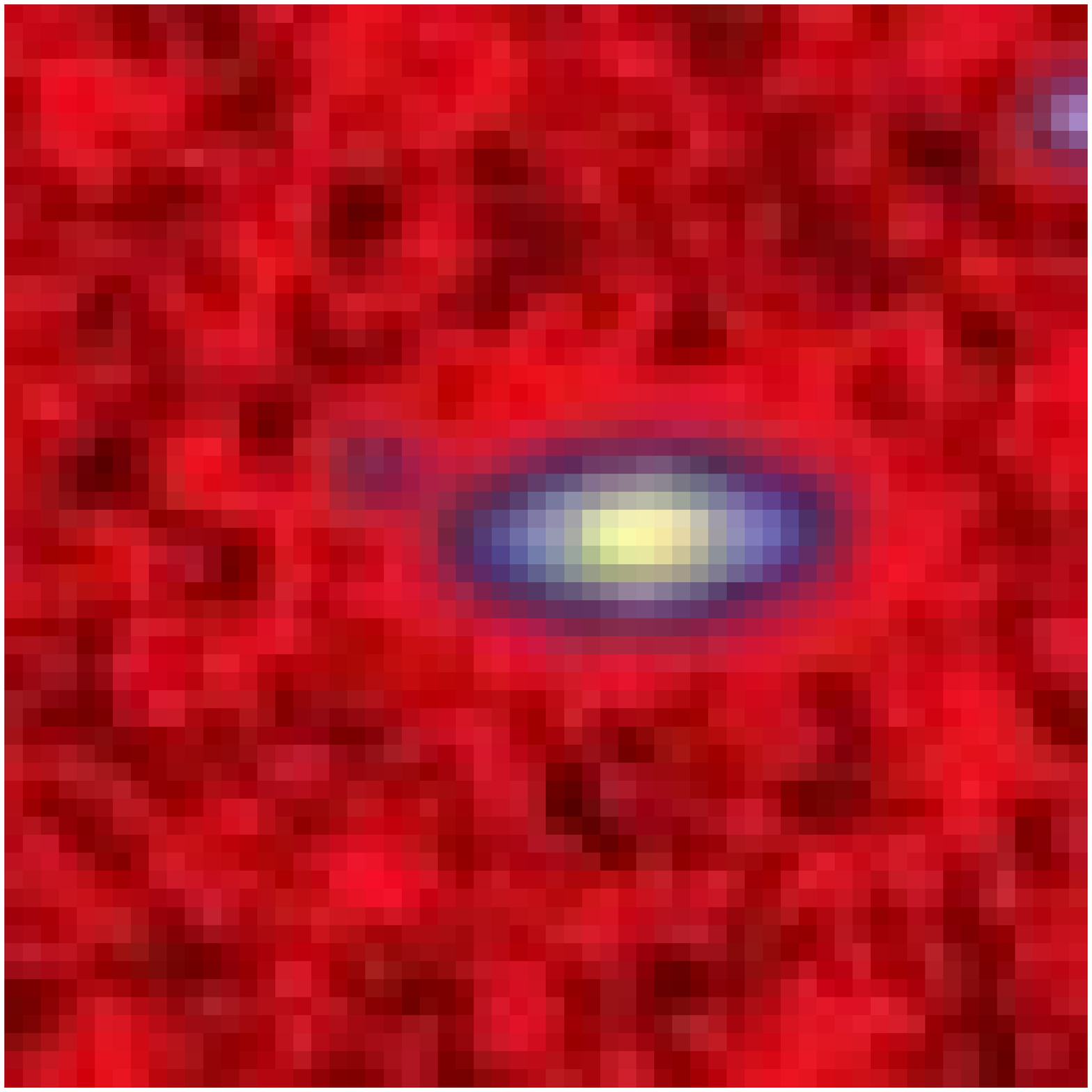}  \FGb{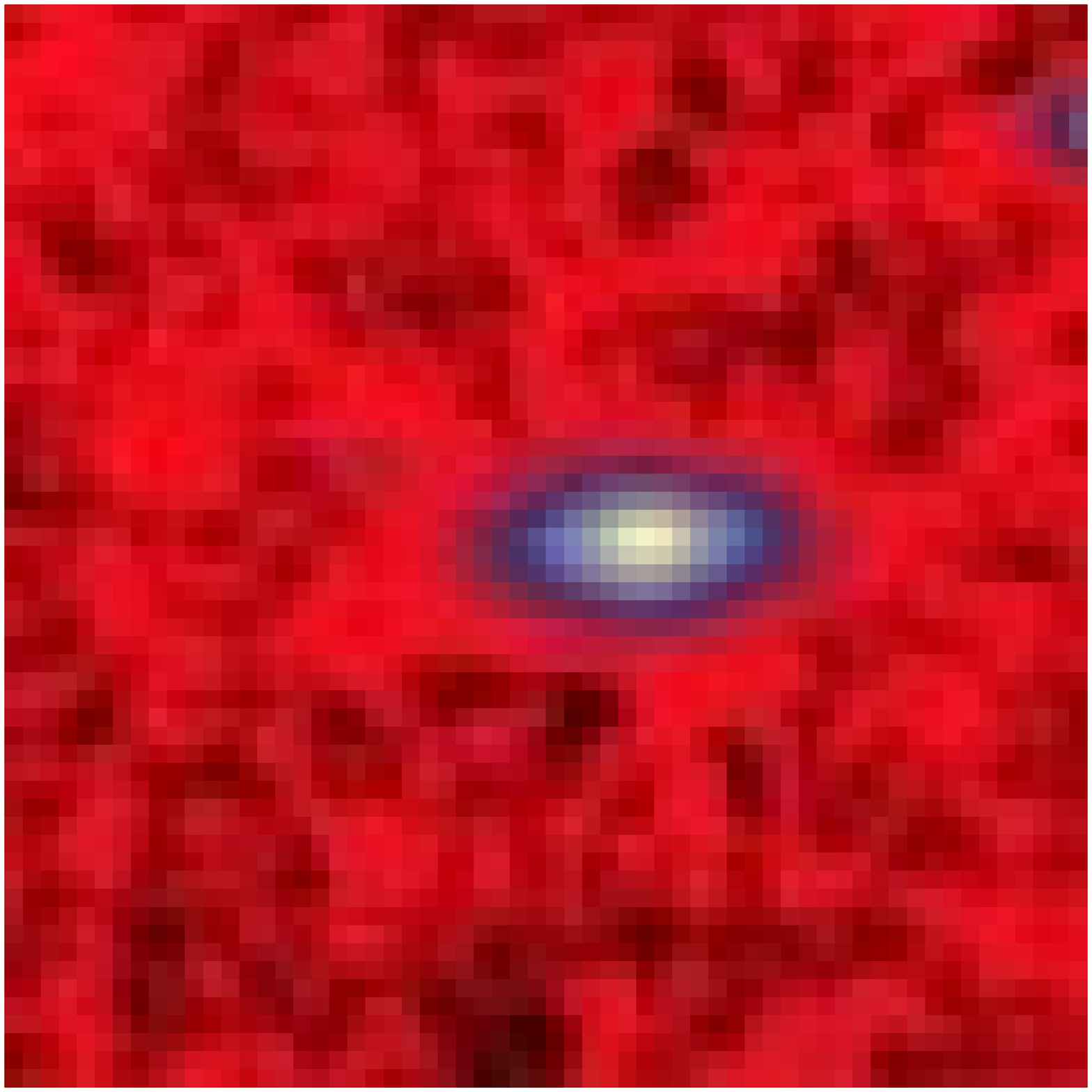}  \FGb{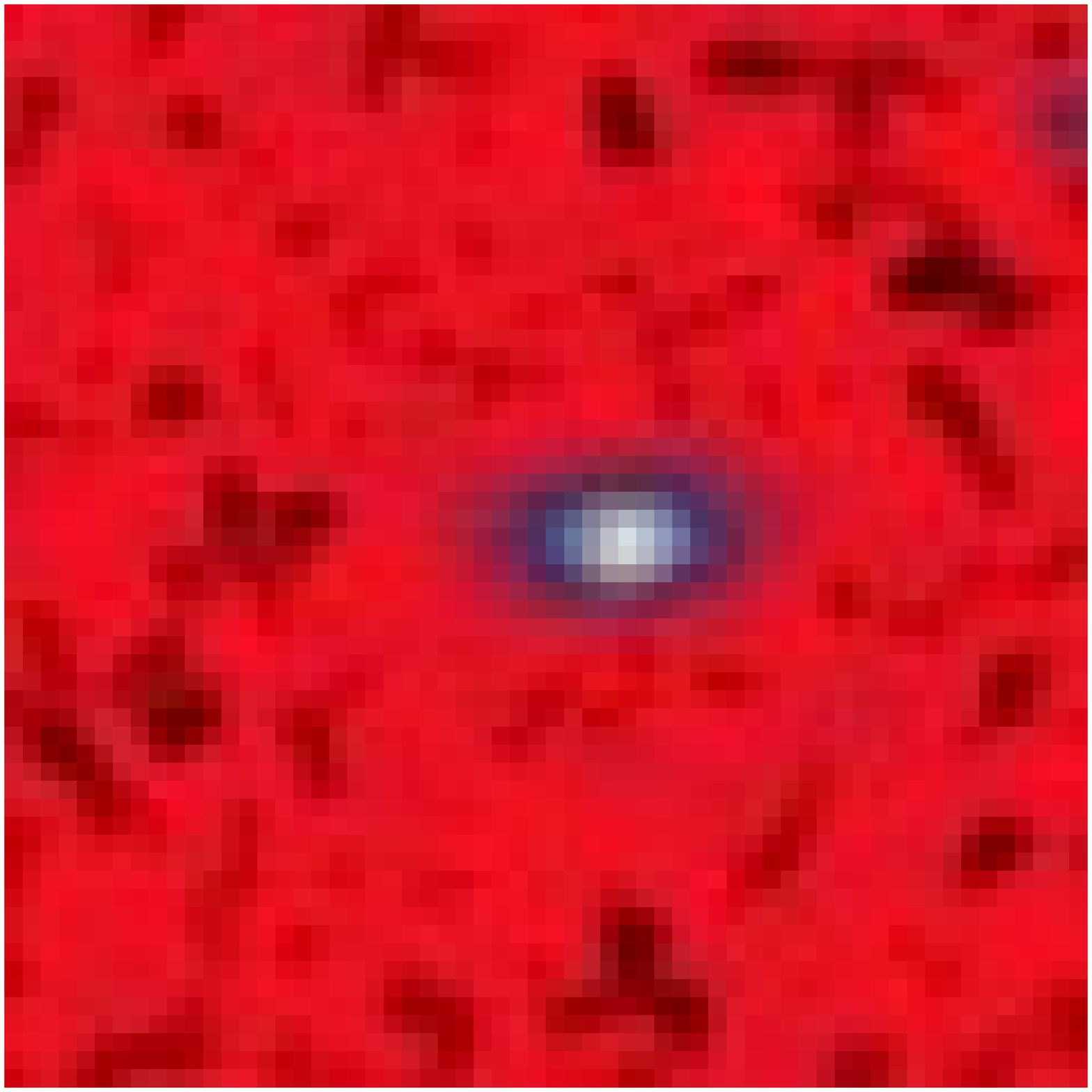}  \FGb{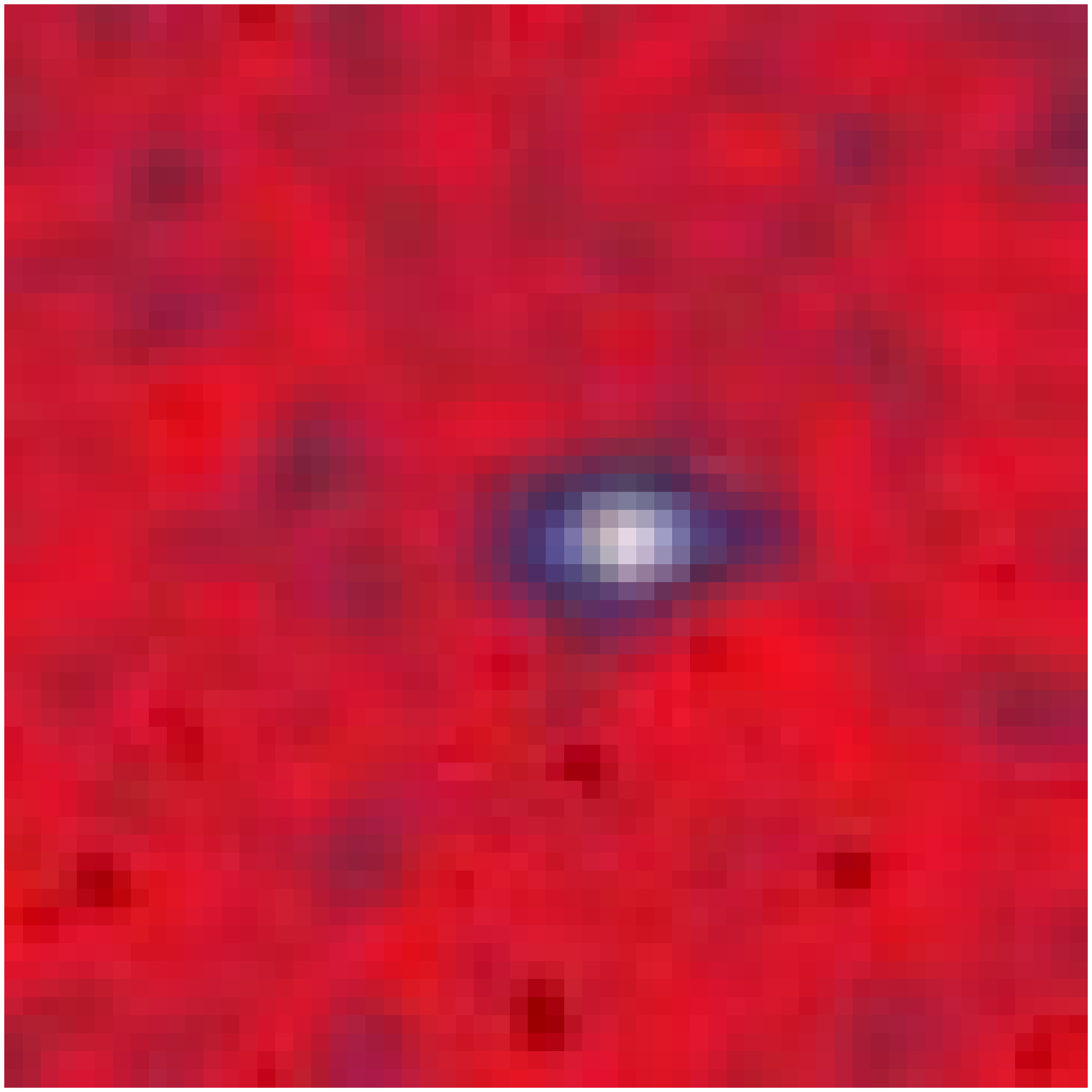}  \FGb{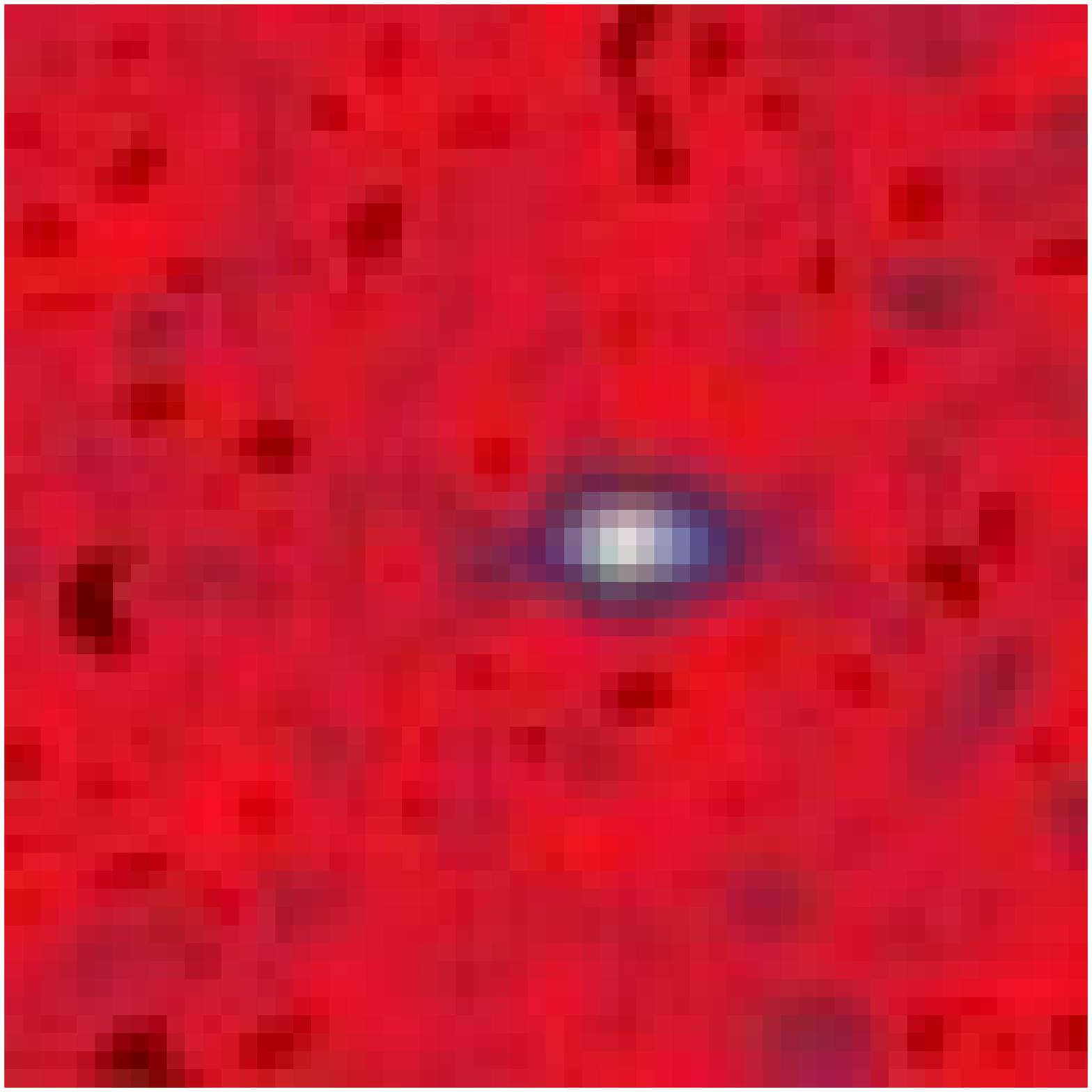}  \FGb{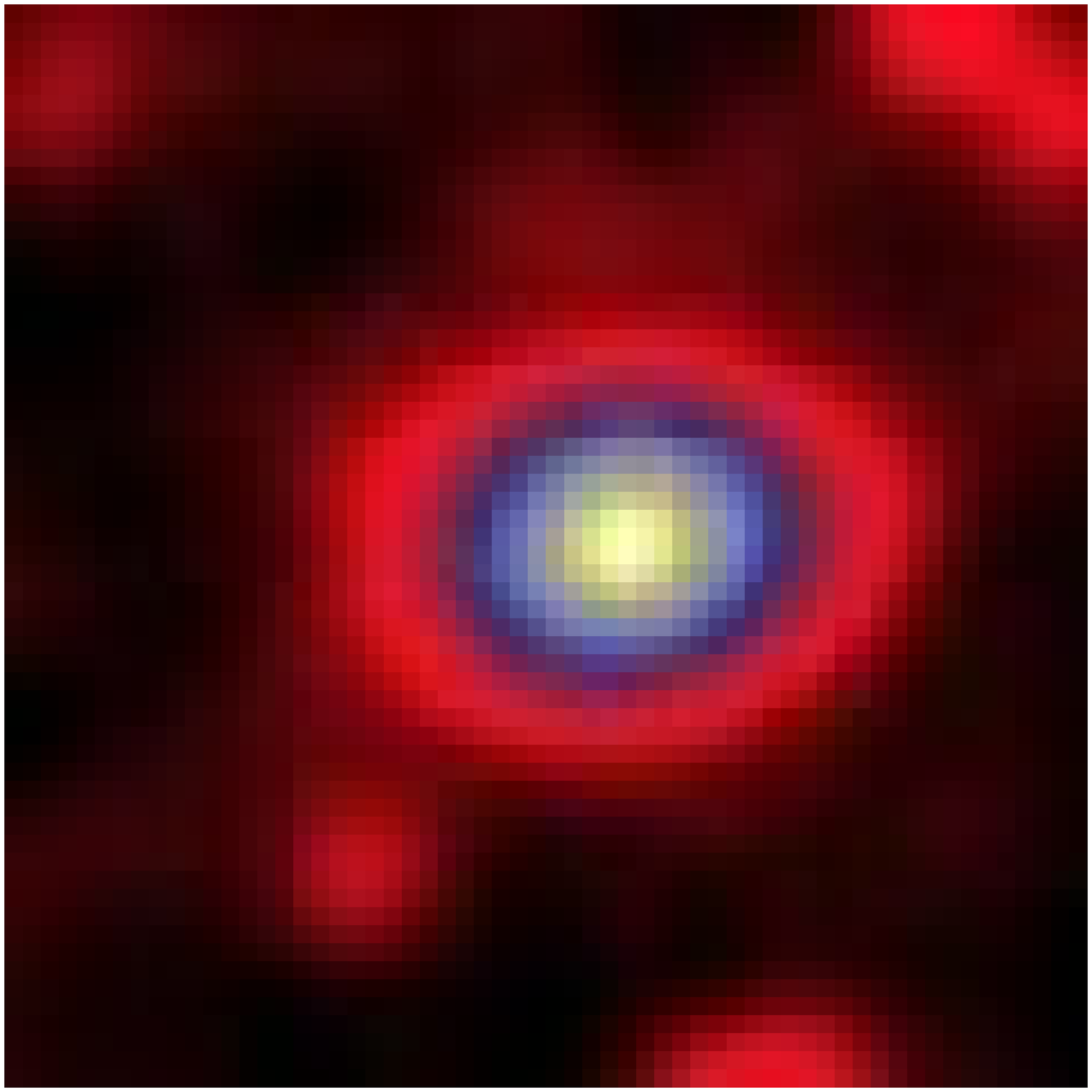}  \FGb{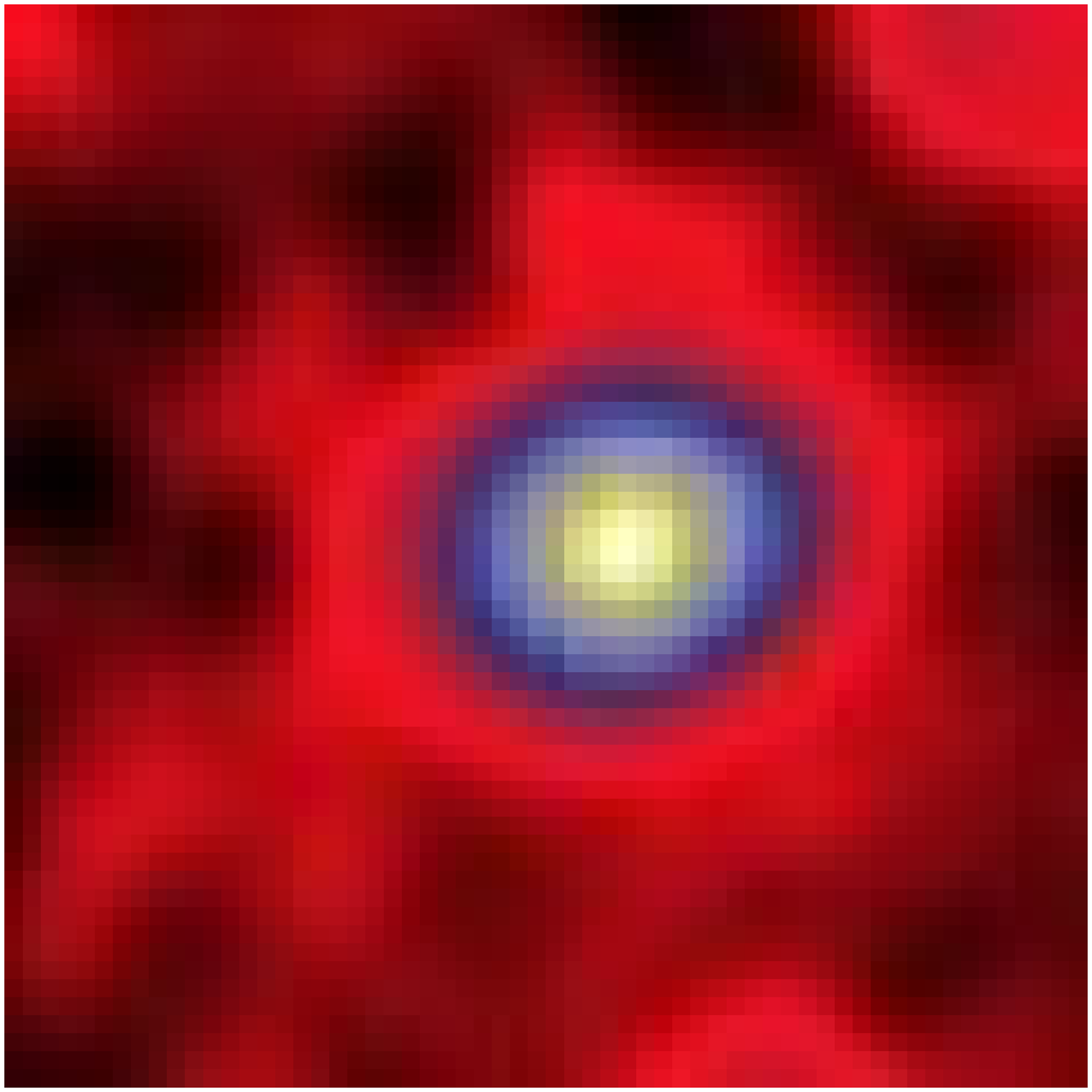}  \FGb{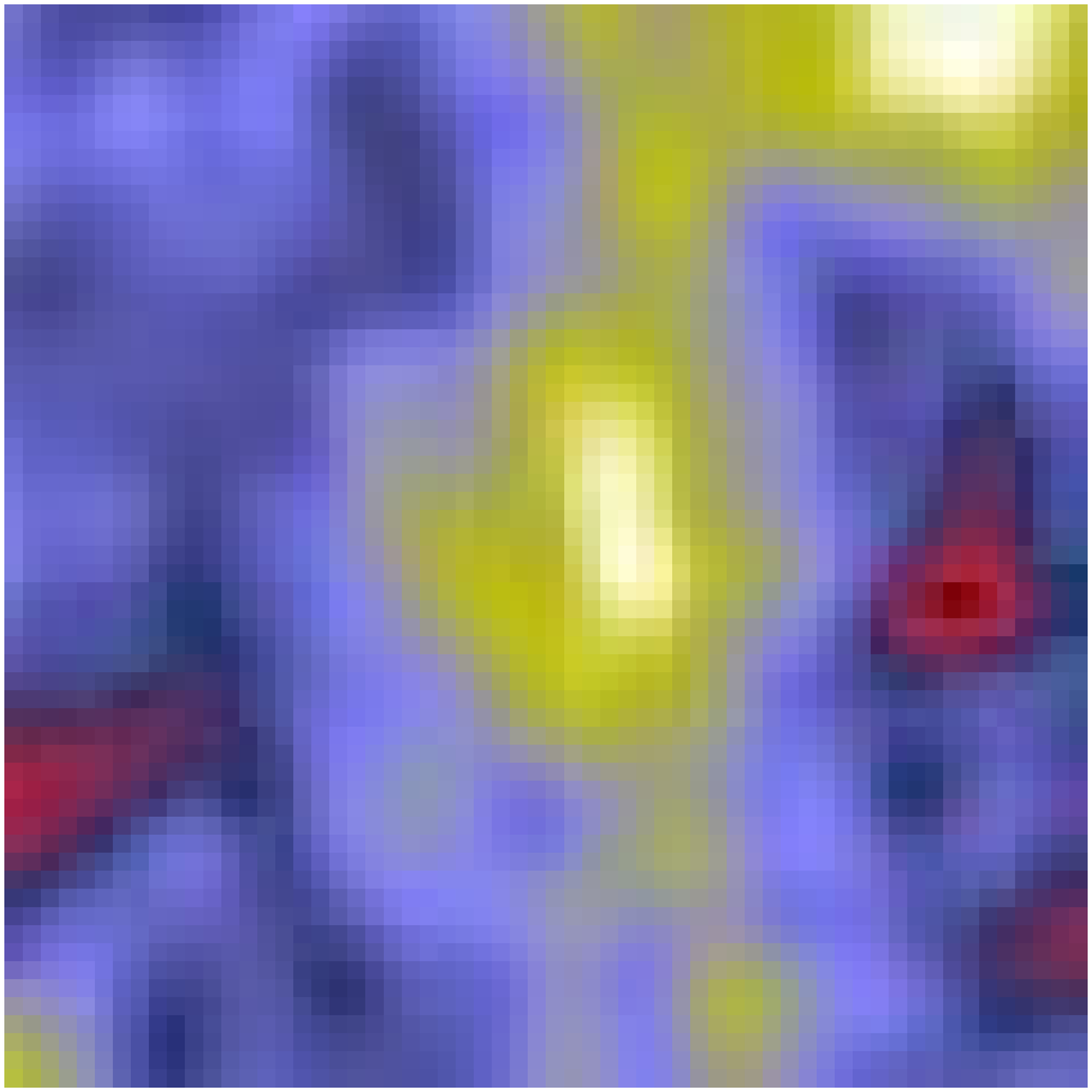}  \FGb{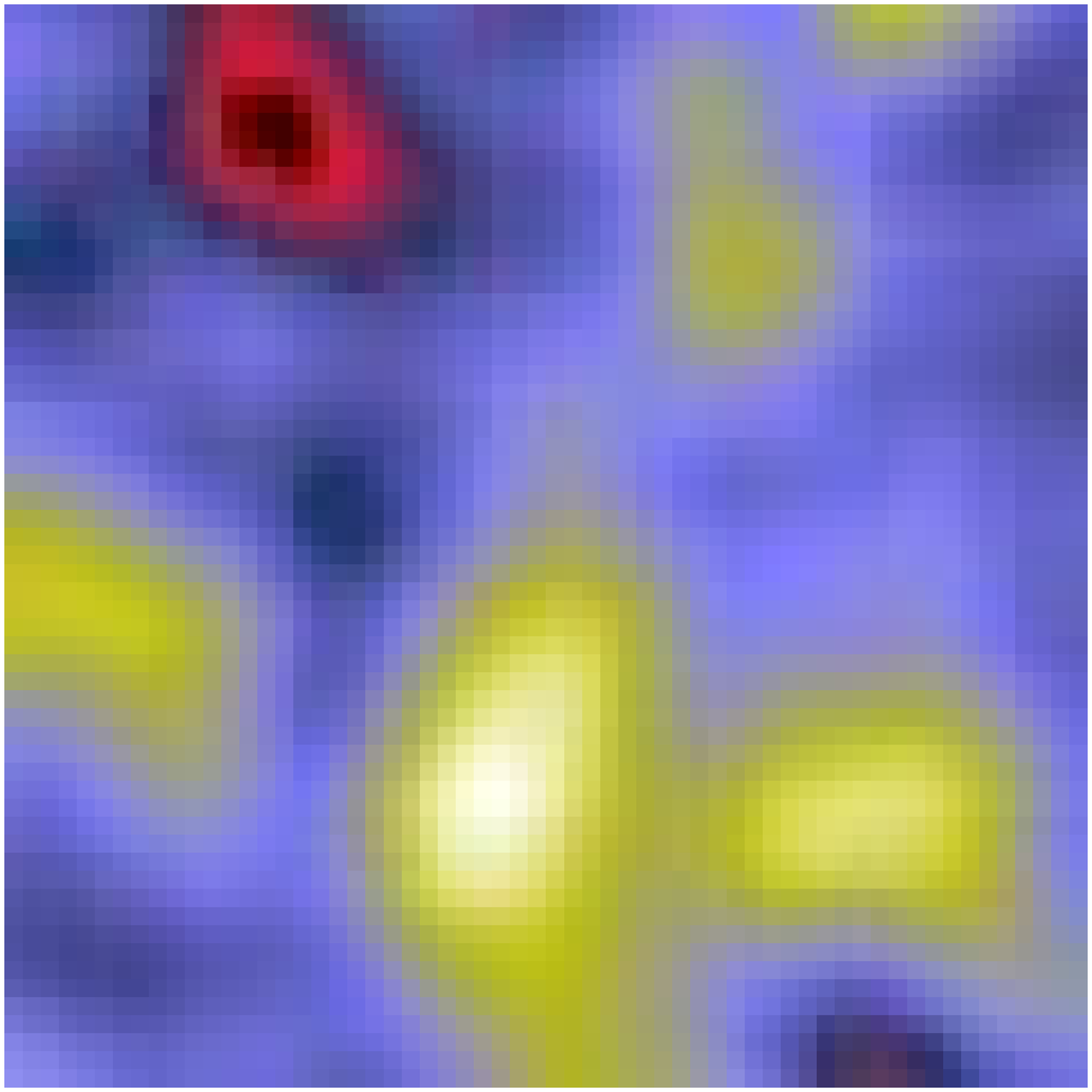} \\  {\#55   NCIA001249}  \\
  \FGb{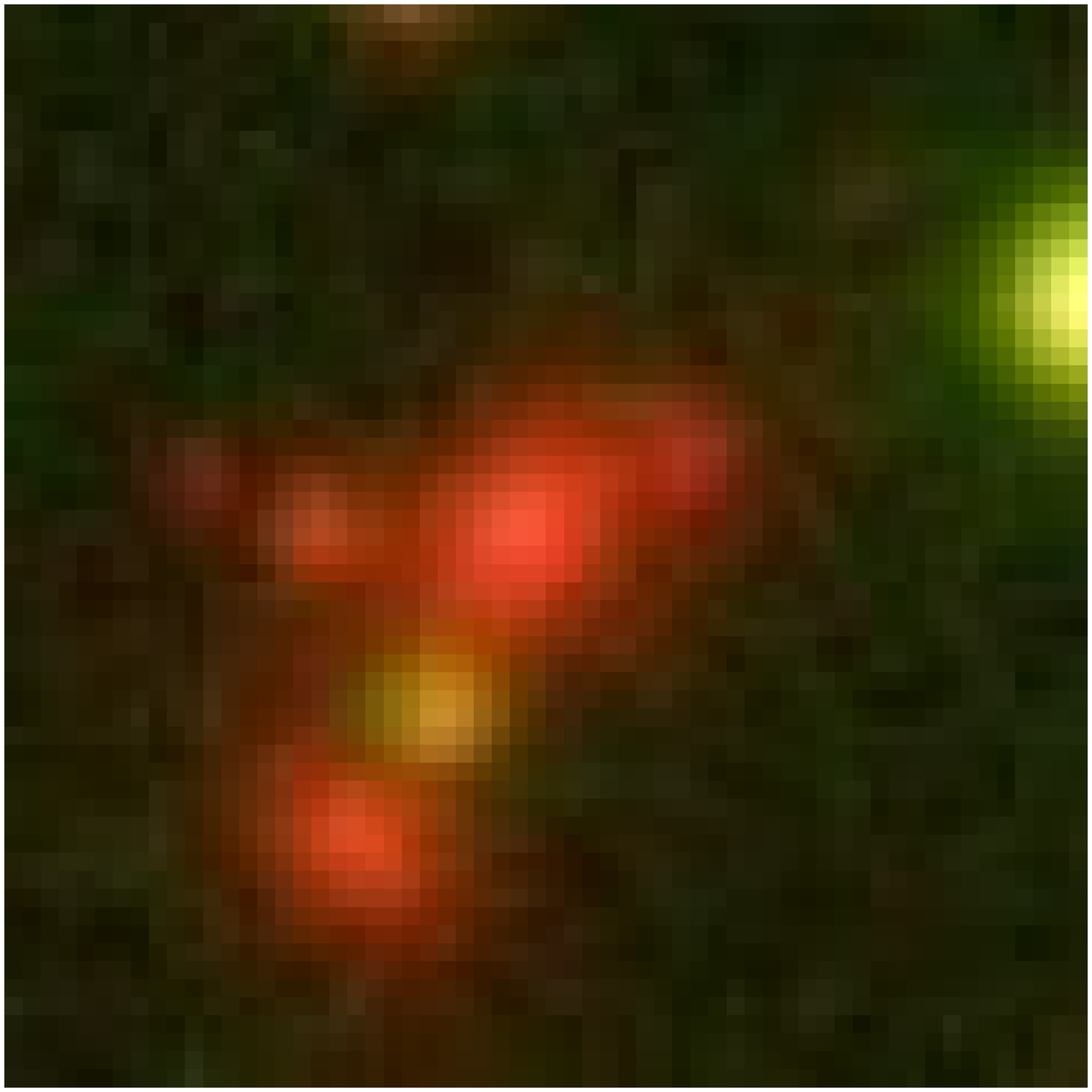}  \FGb{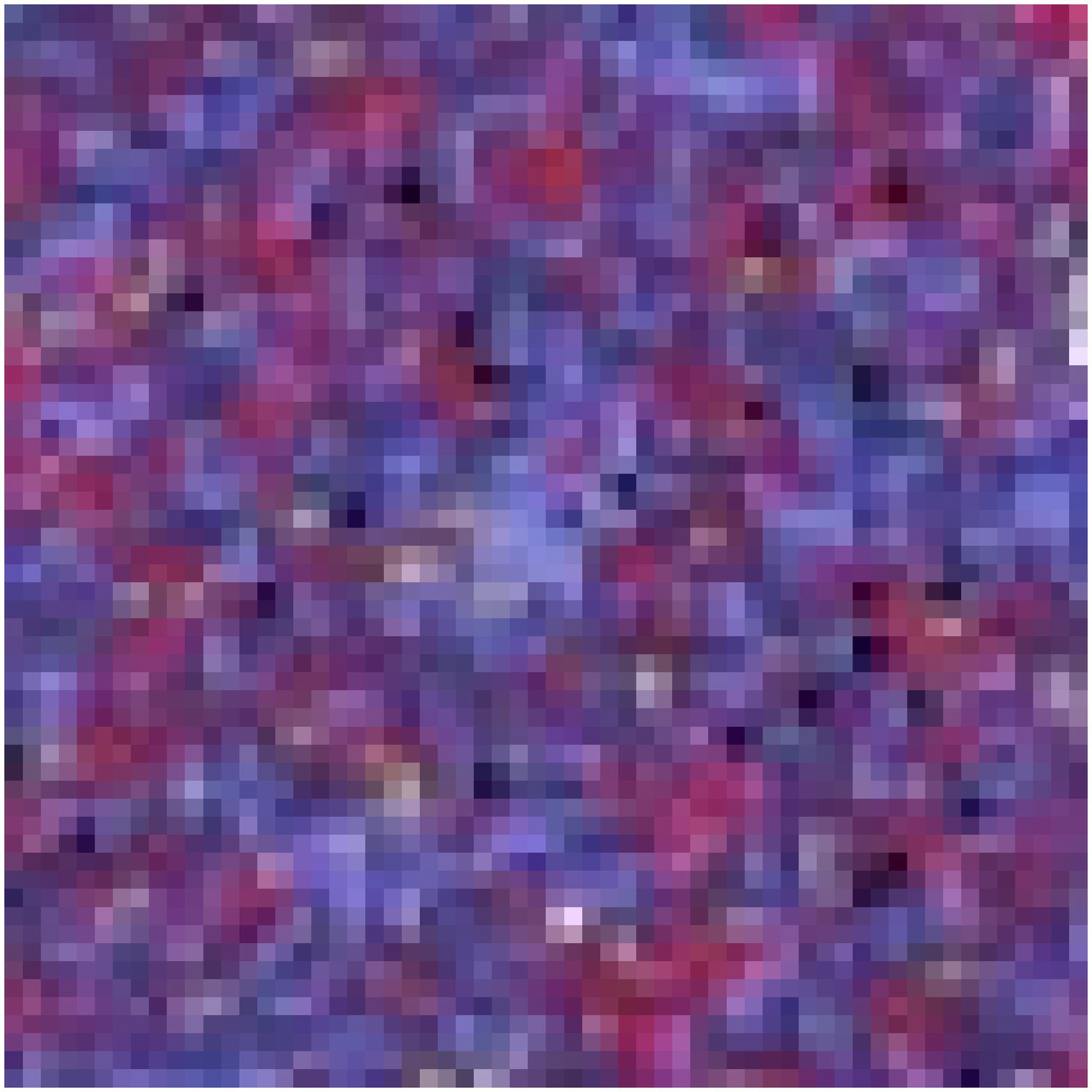}  \FGb{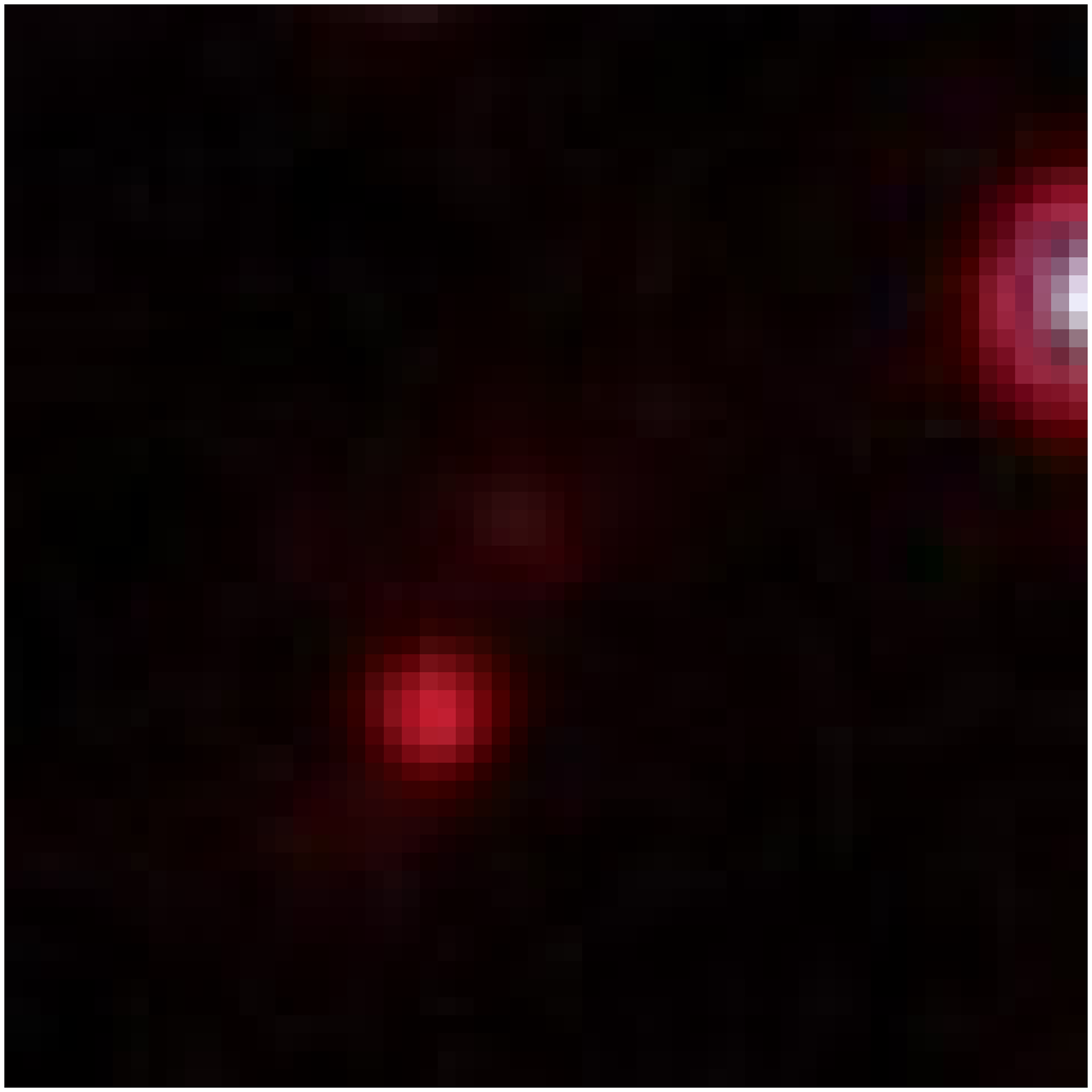}  \FGb{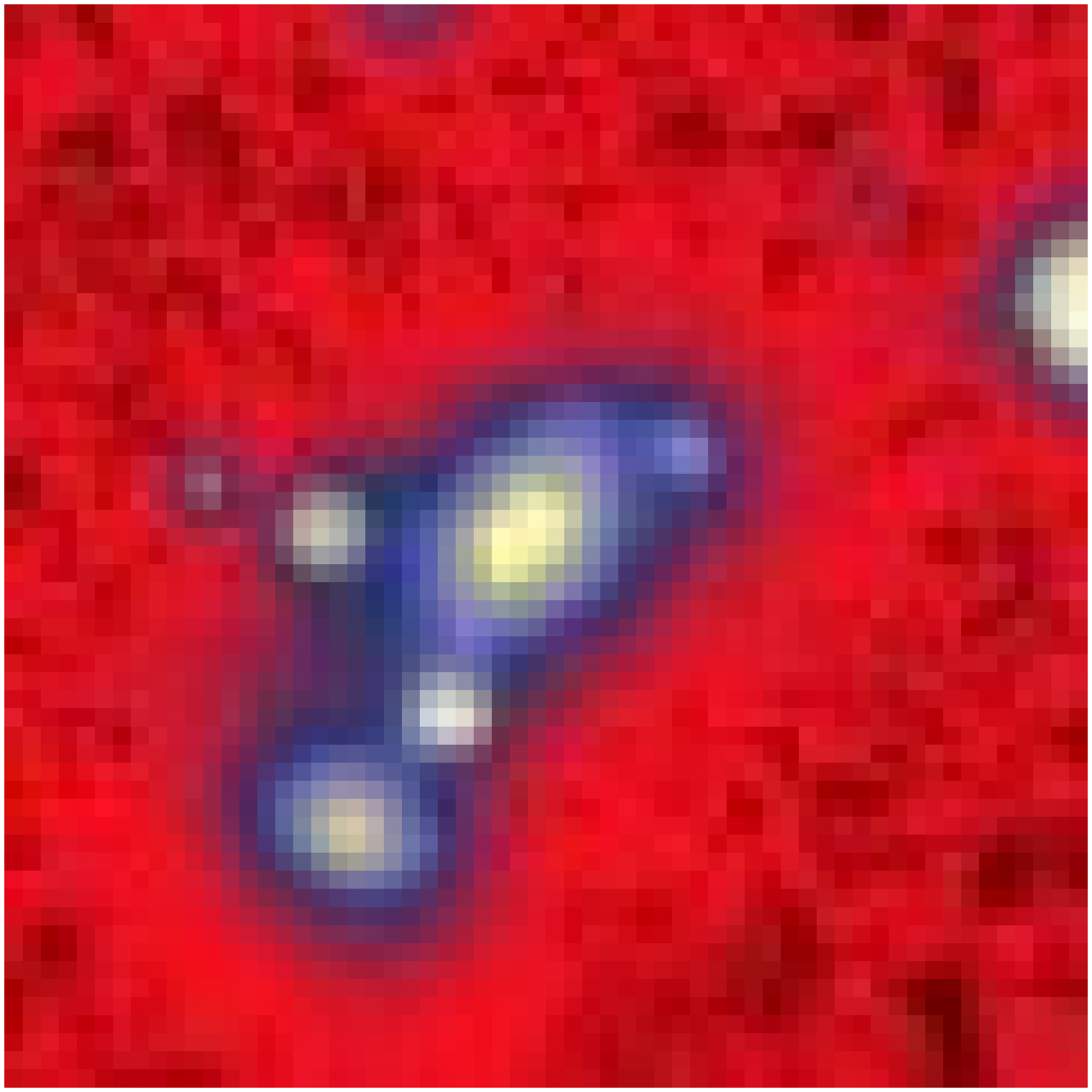}  \FGb{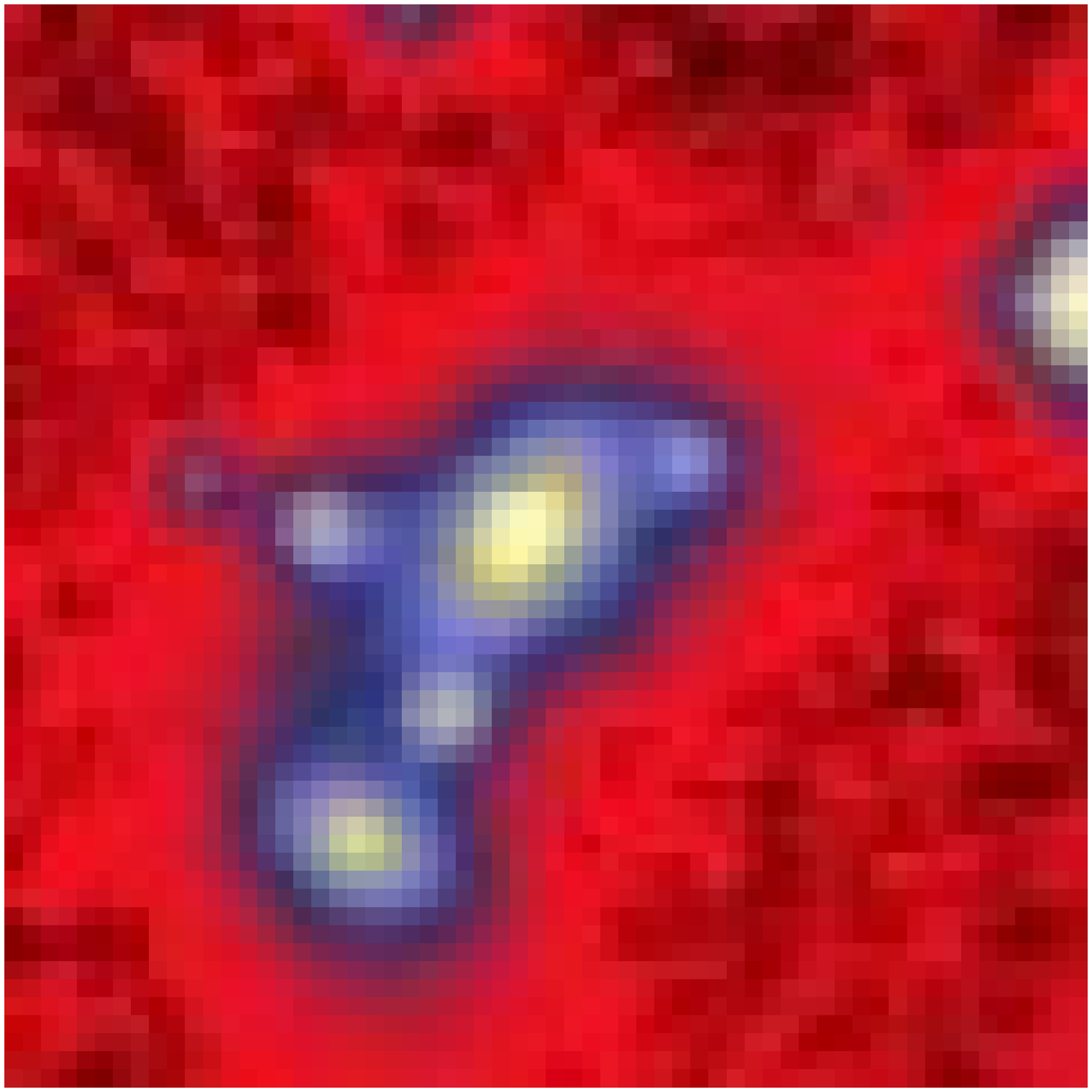}  \FGb{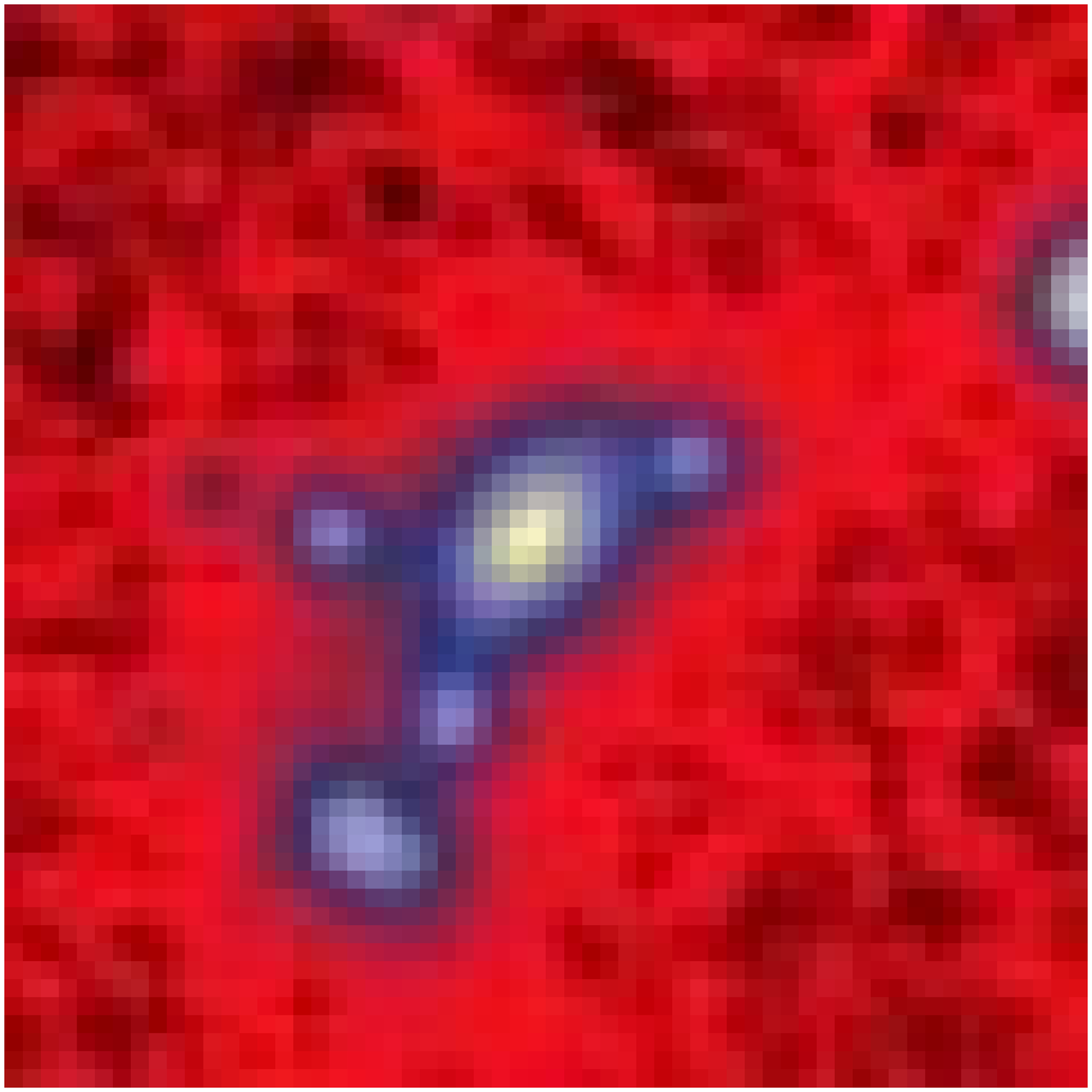}  \FGb{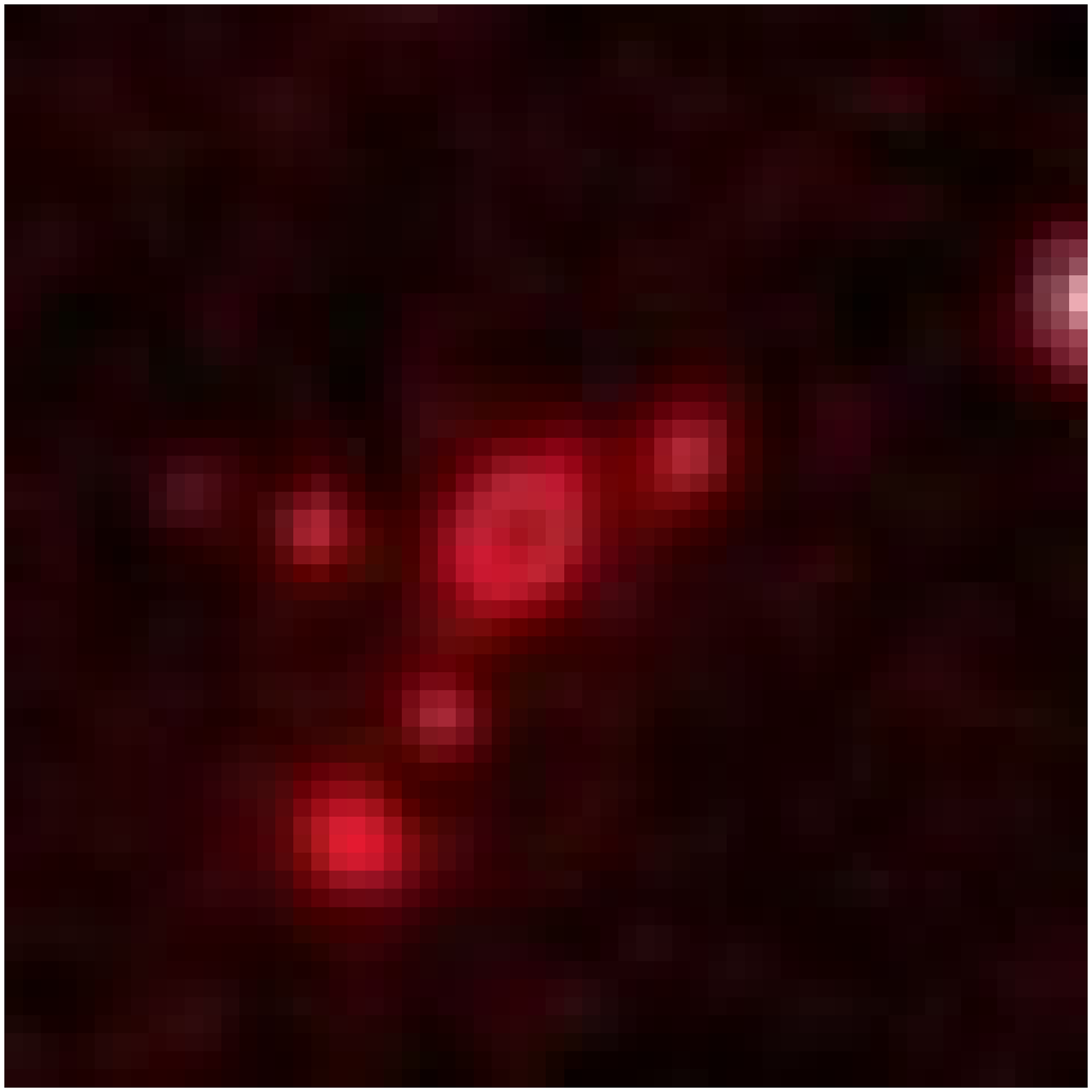}  \FGb{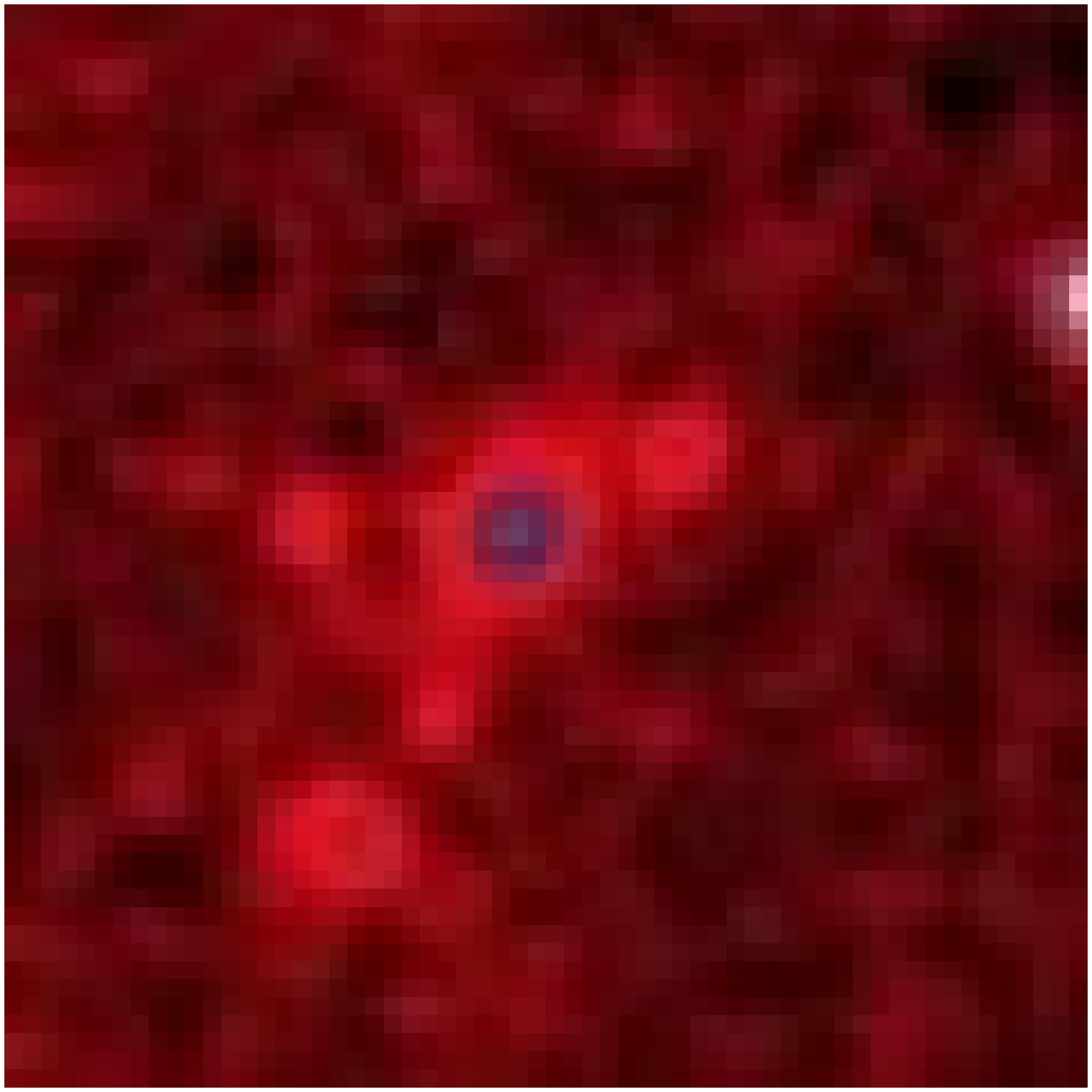}  \FGb{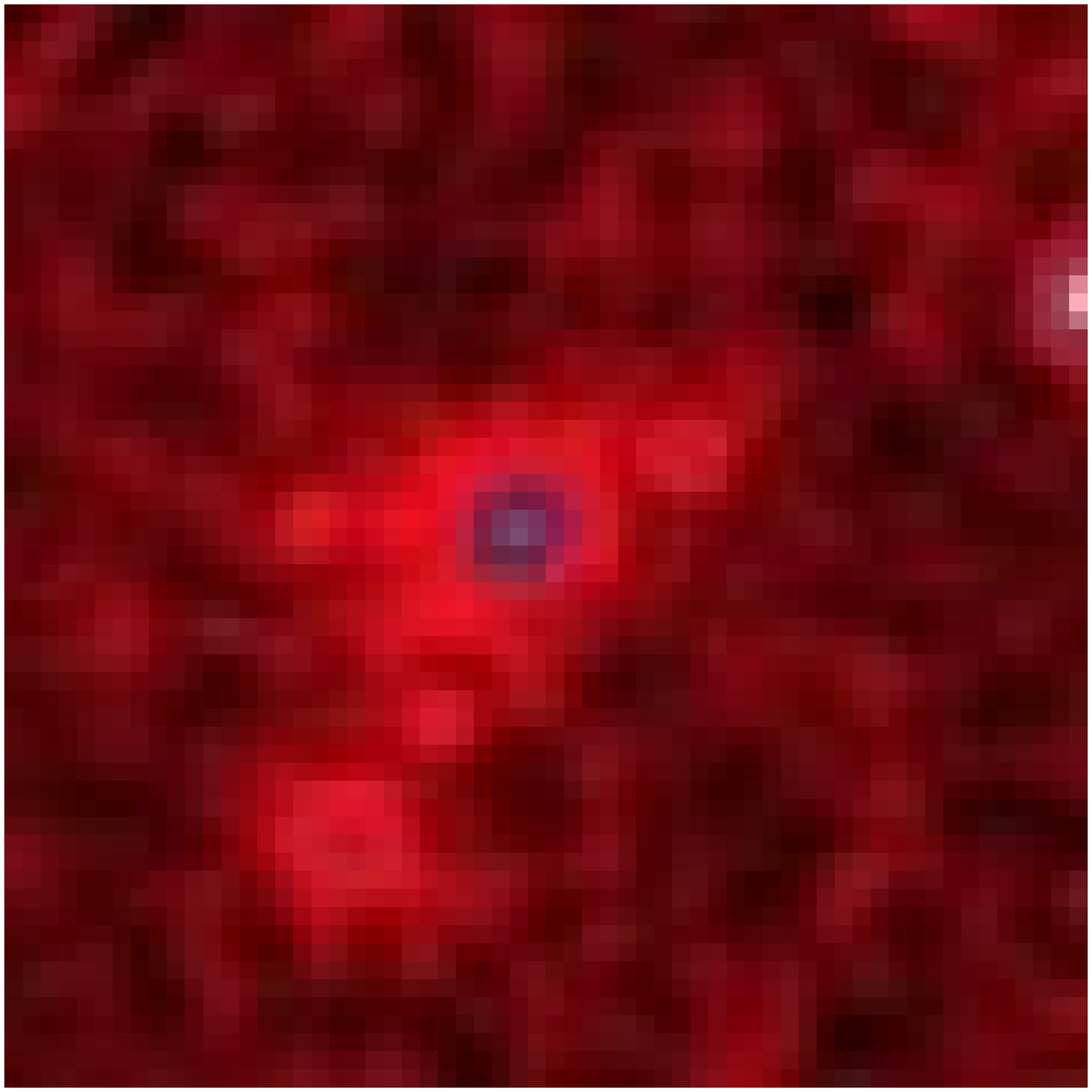}  \FGb{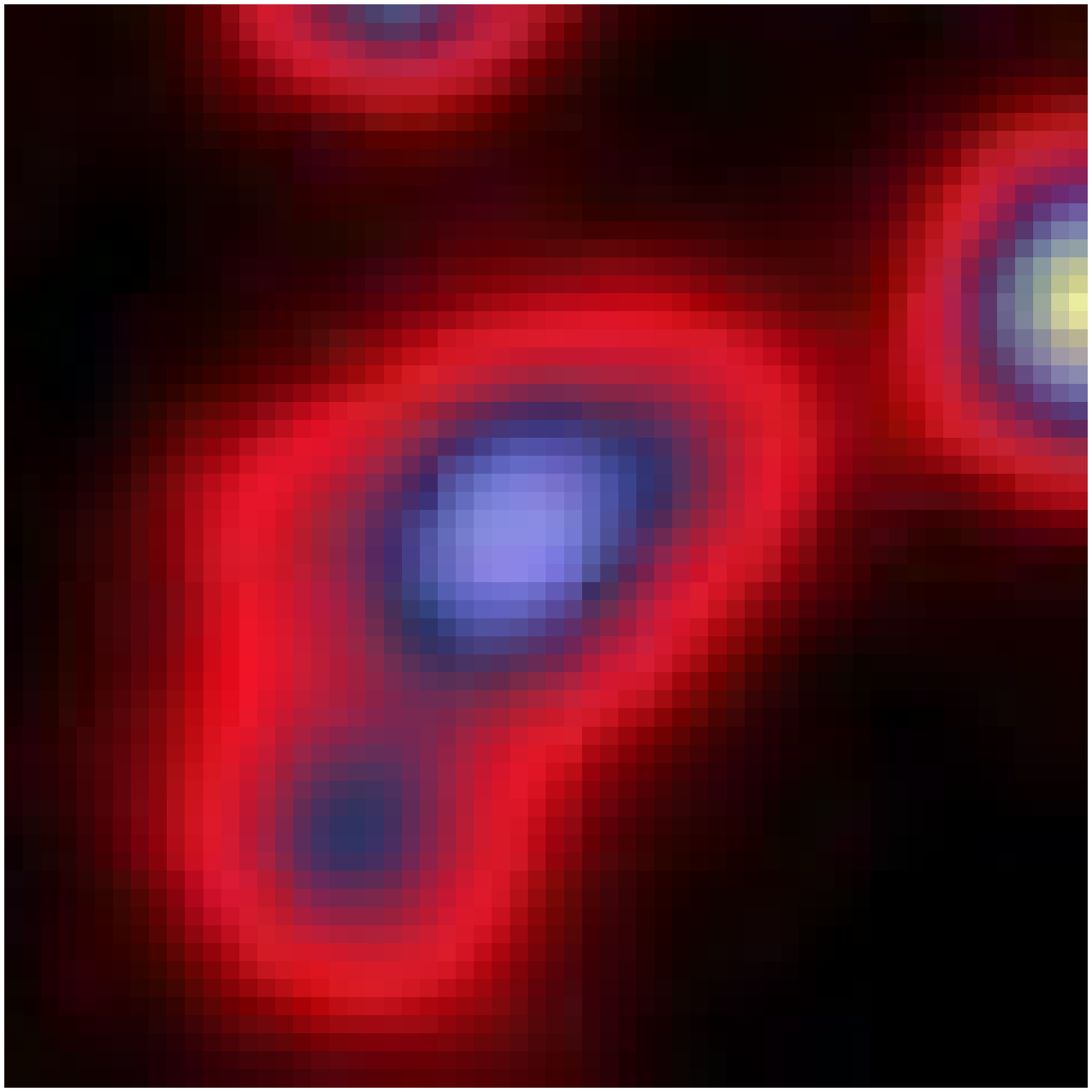}  \FGb{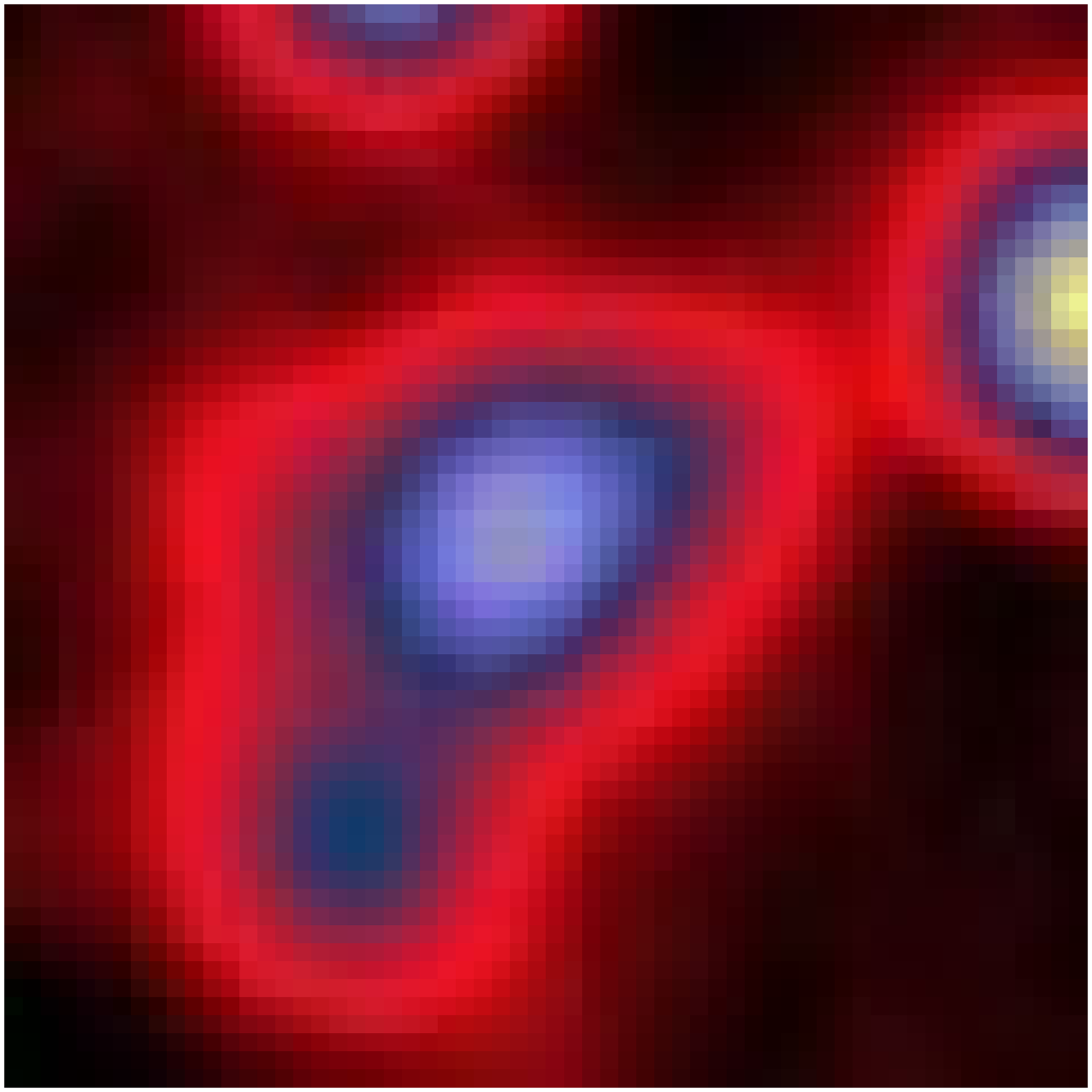}  \FGb{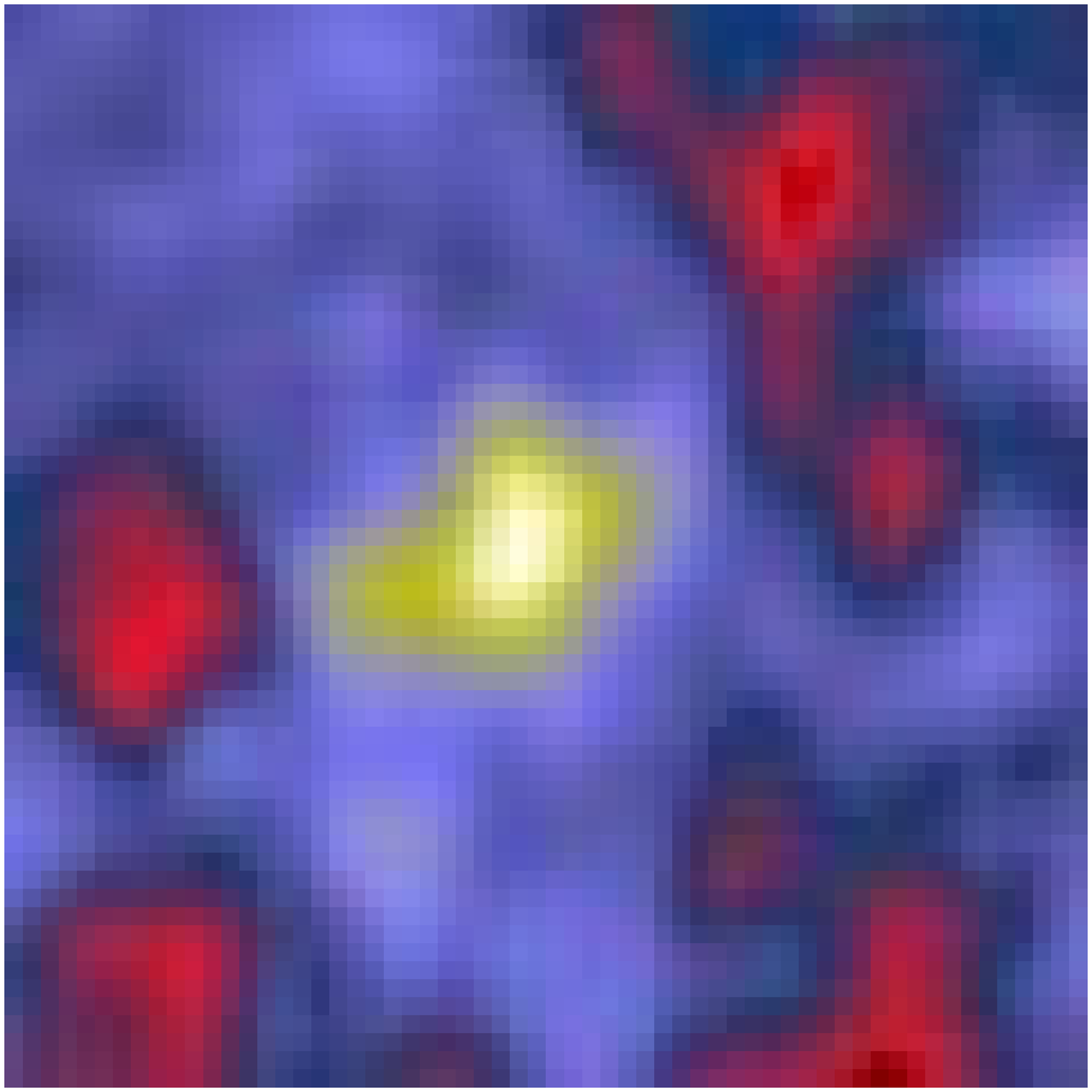}  \FGb{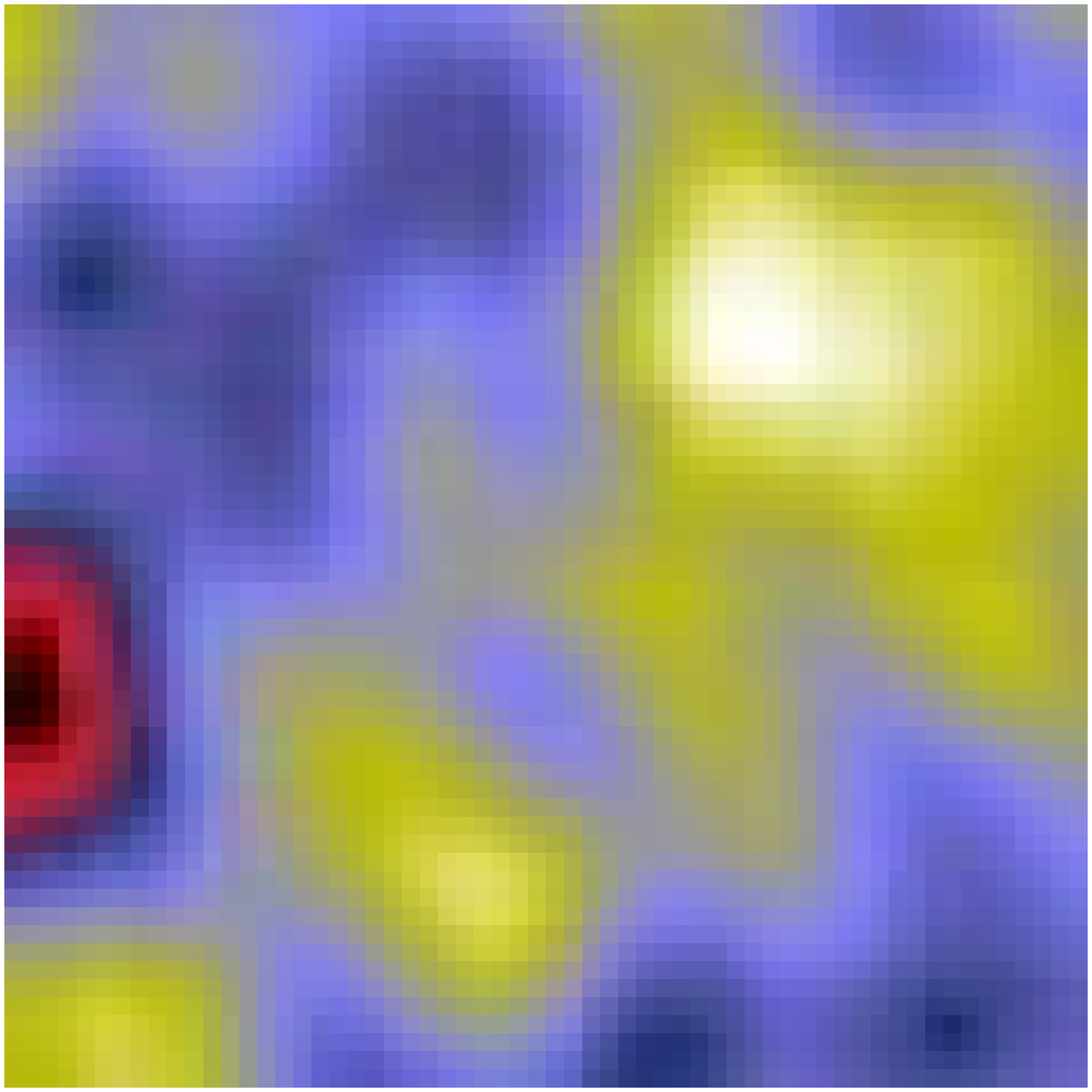} \\  {\#56   NCIA000313}  \\
  \FGb{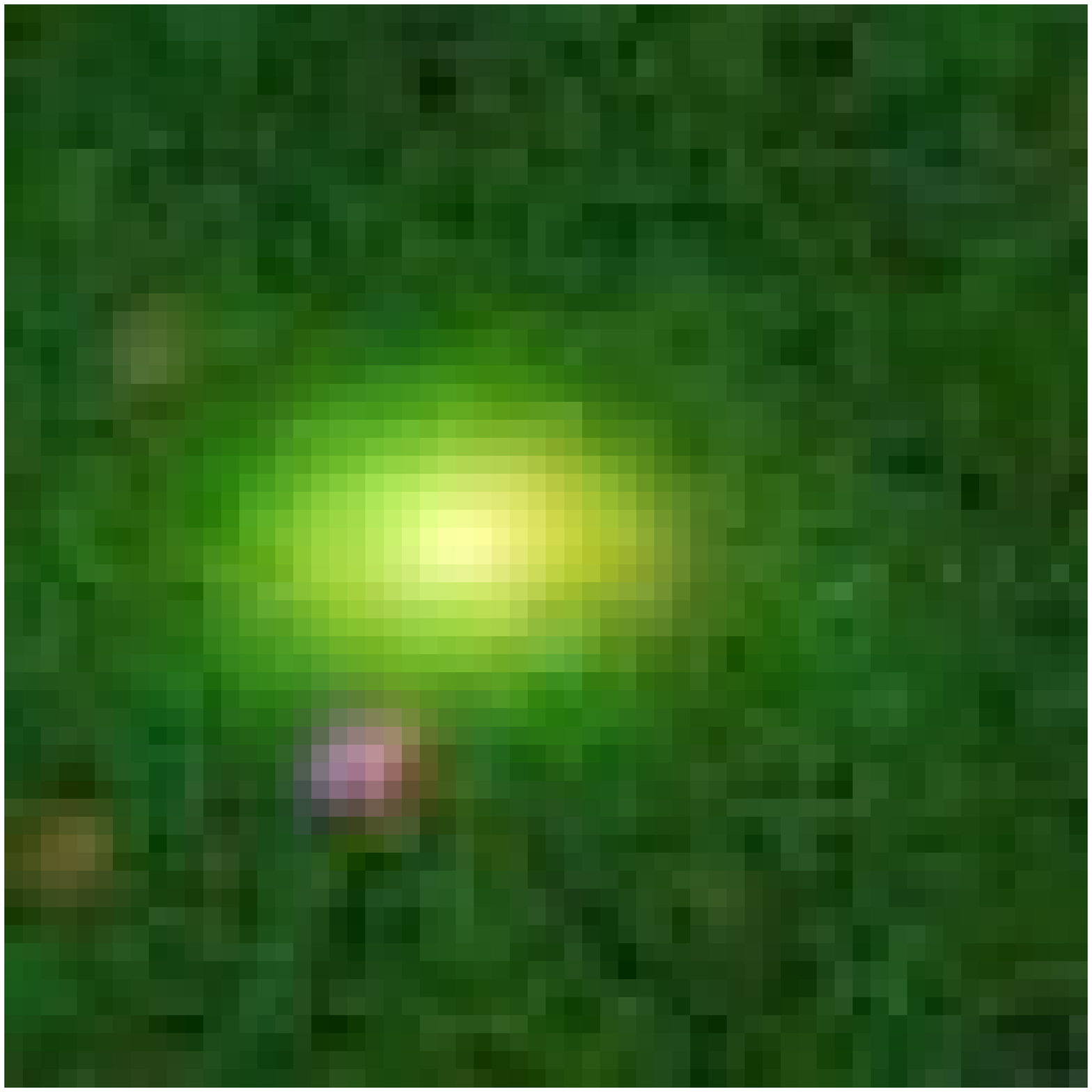}  \FGb{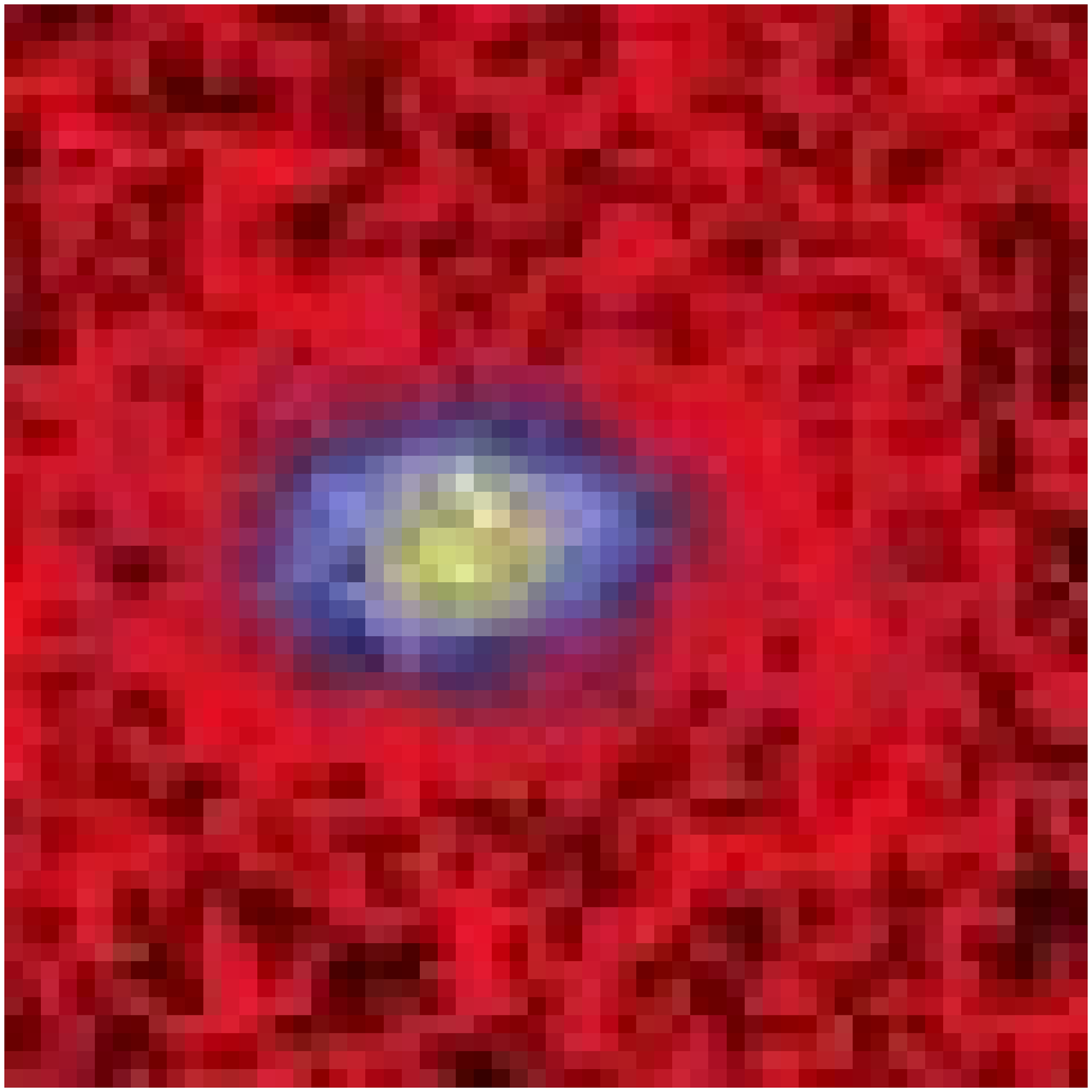}  \FGb{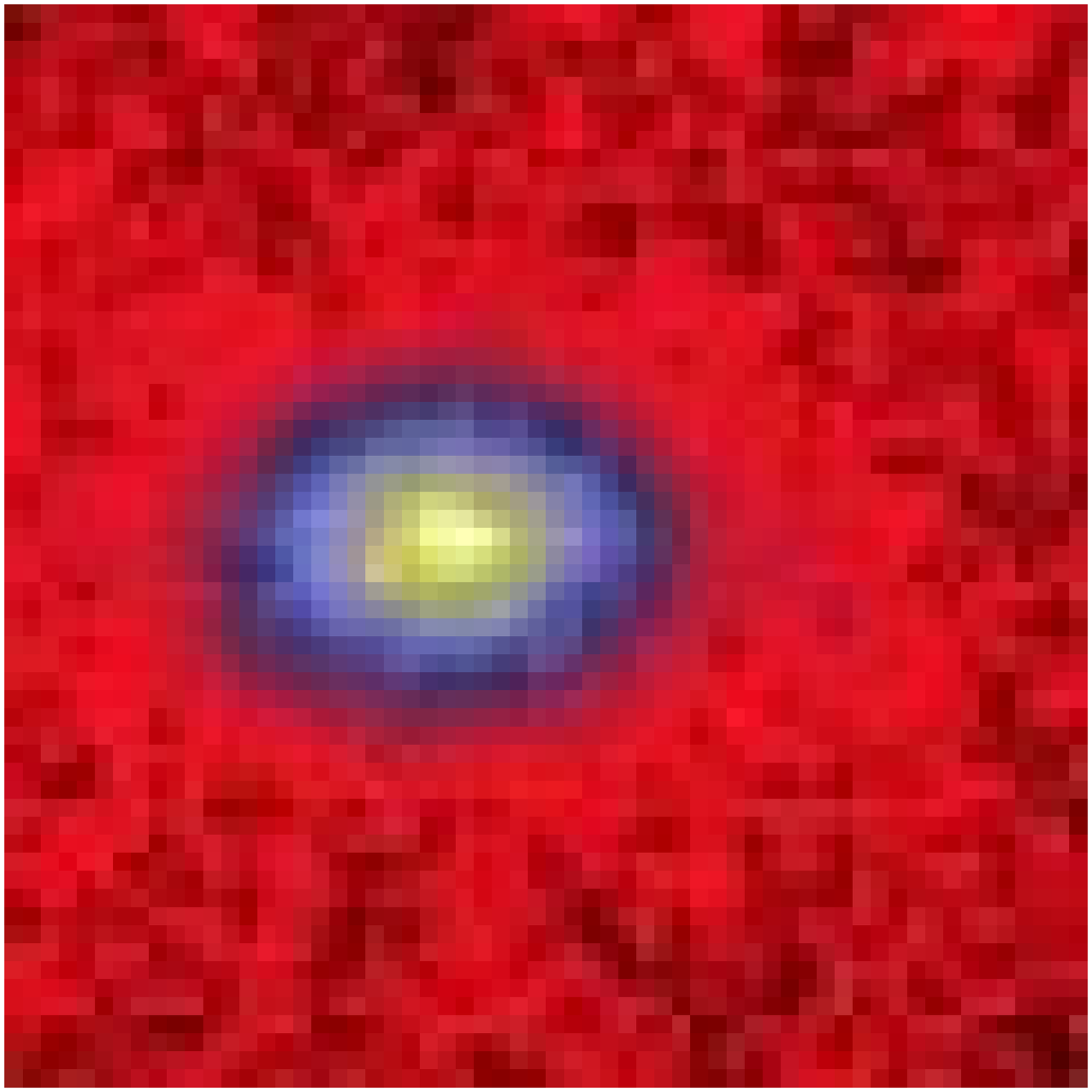}  \FGb{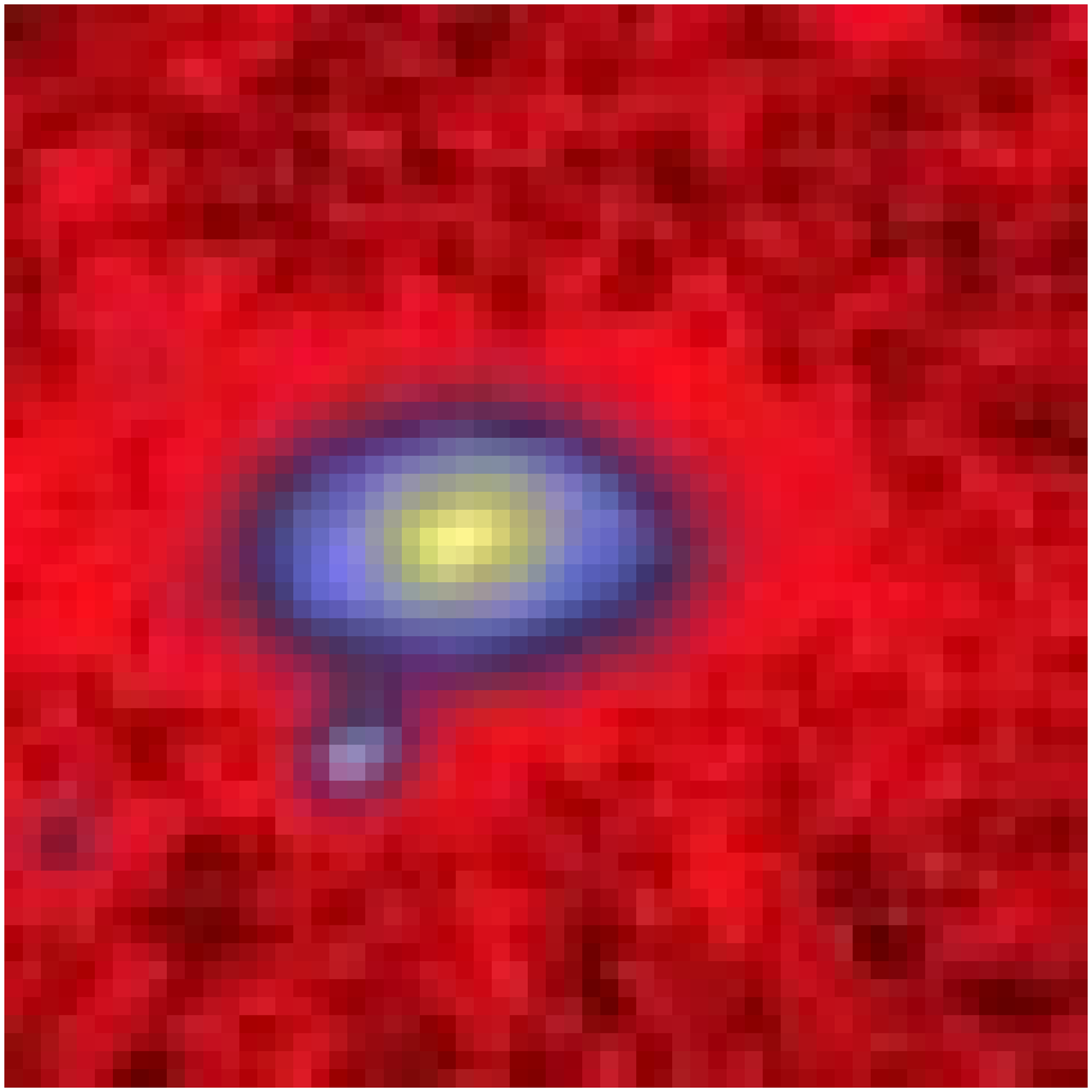}  \FGb{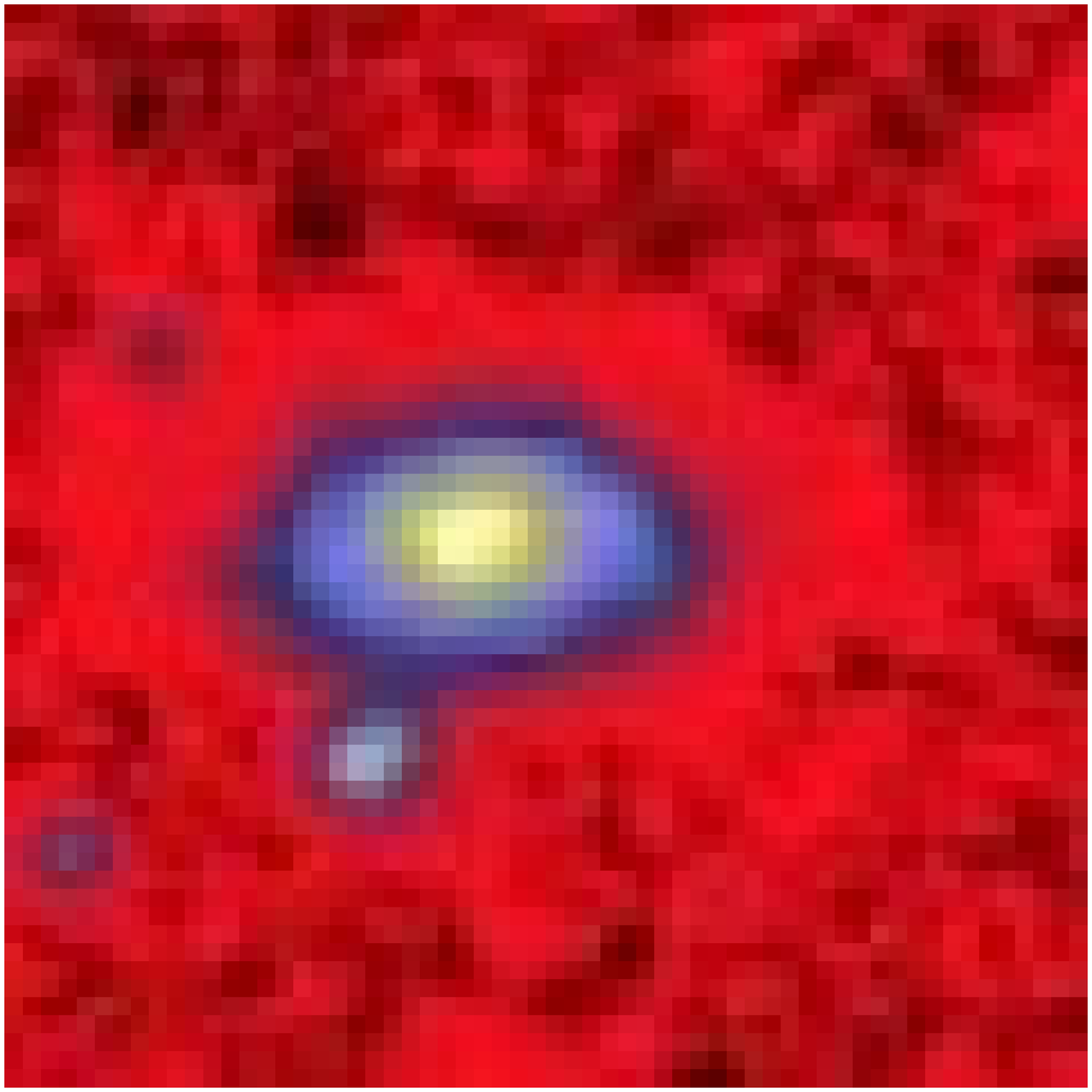}  \FGb{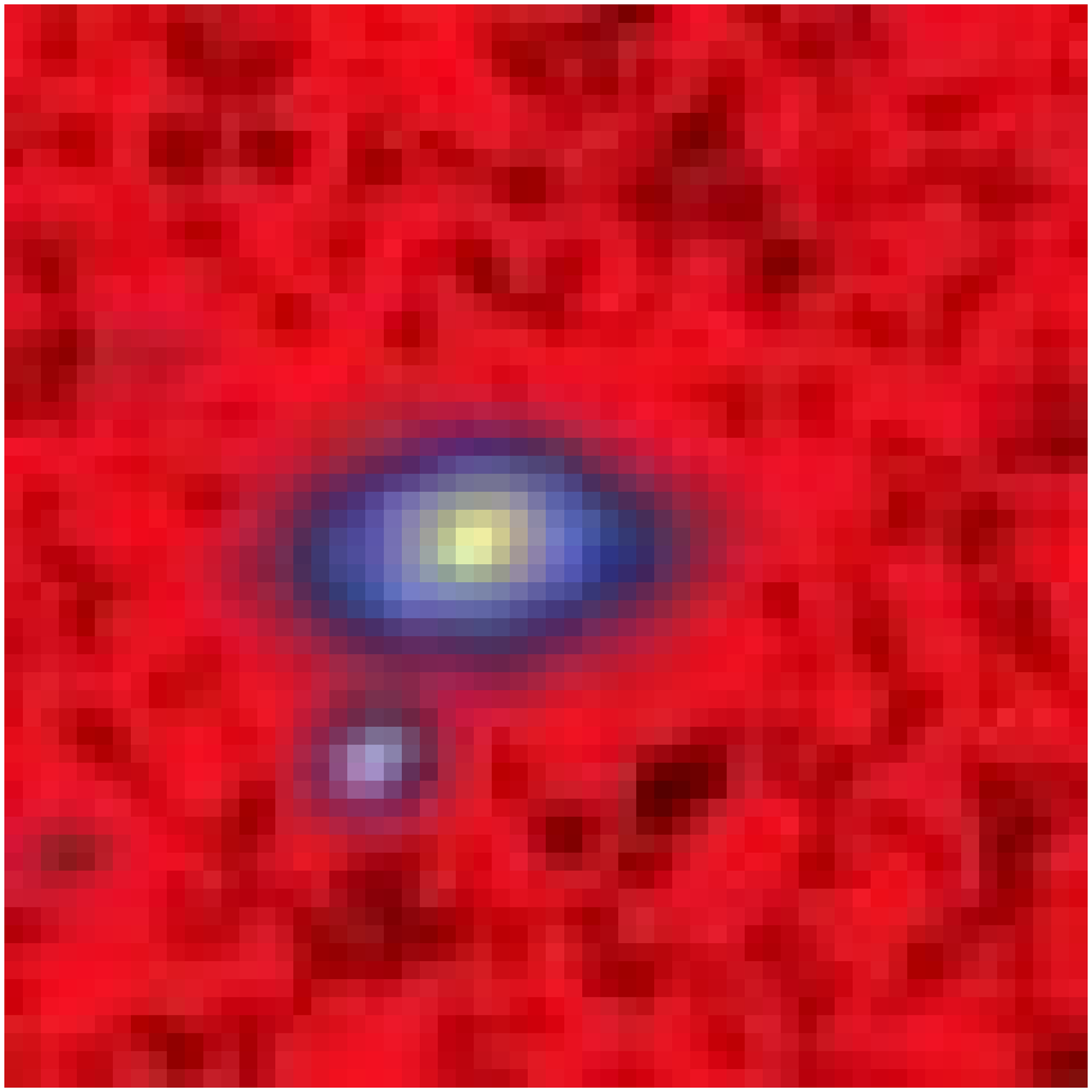}  \FGb{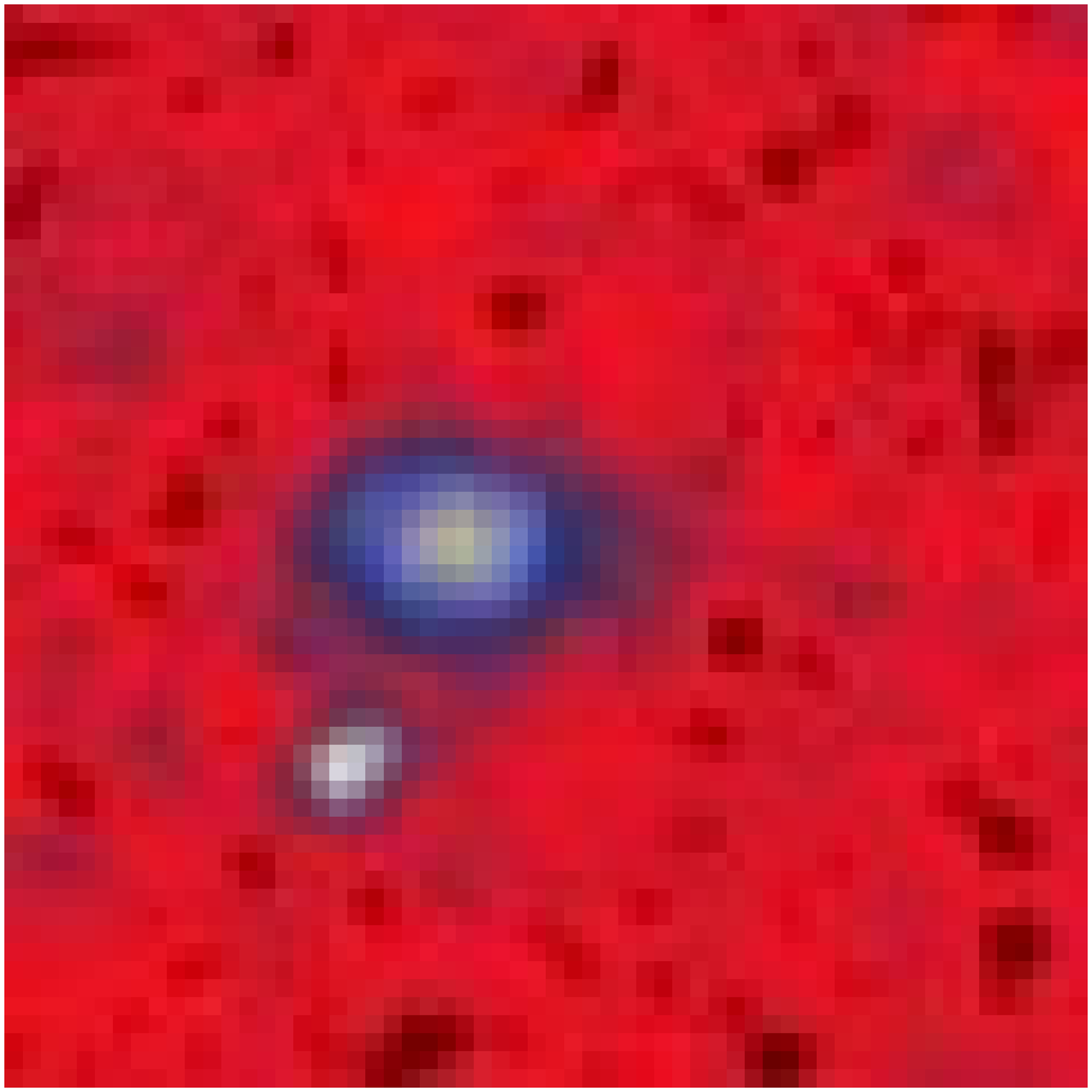}  \FGb{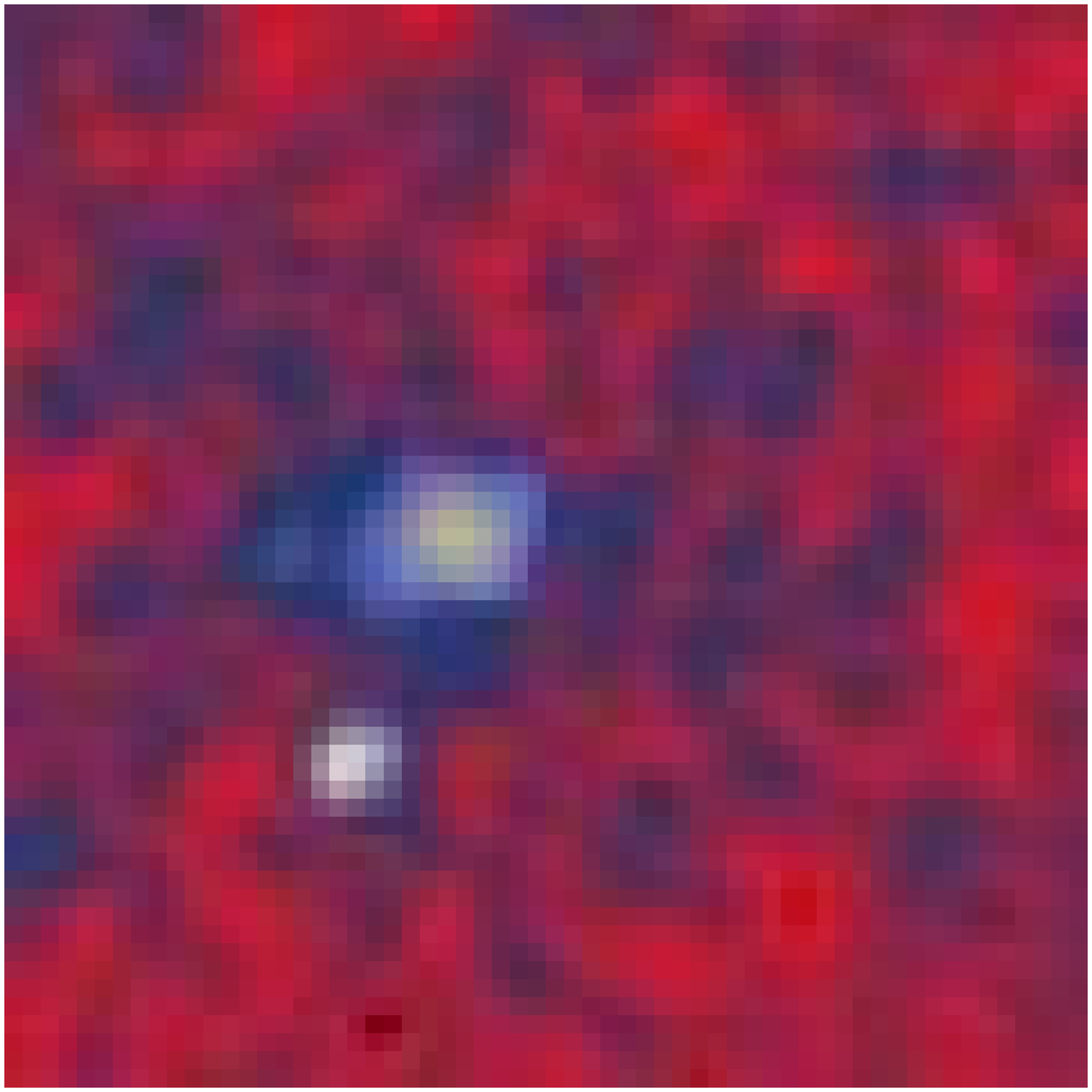}  \FGb{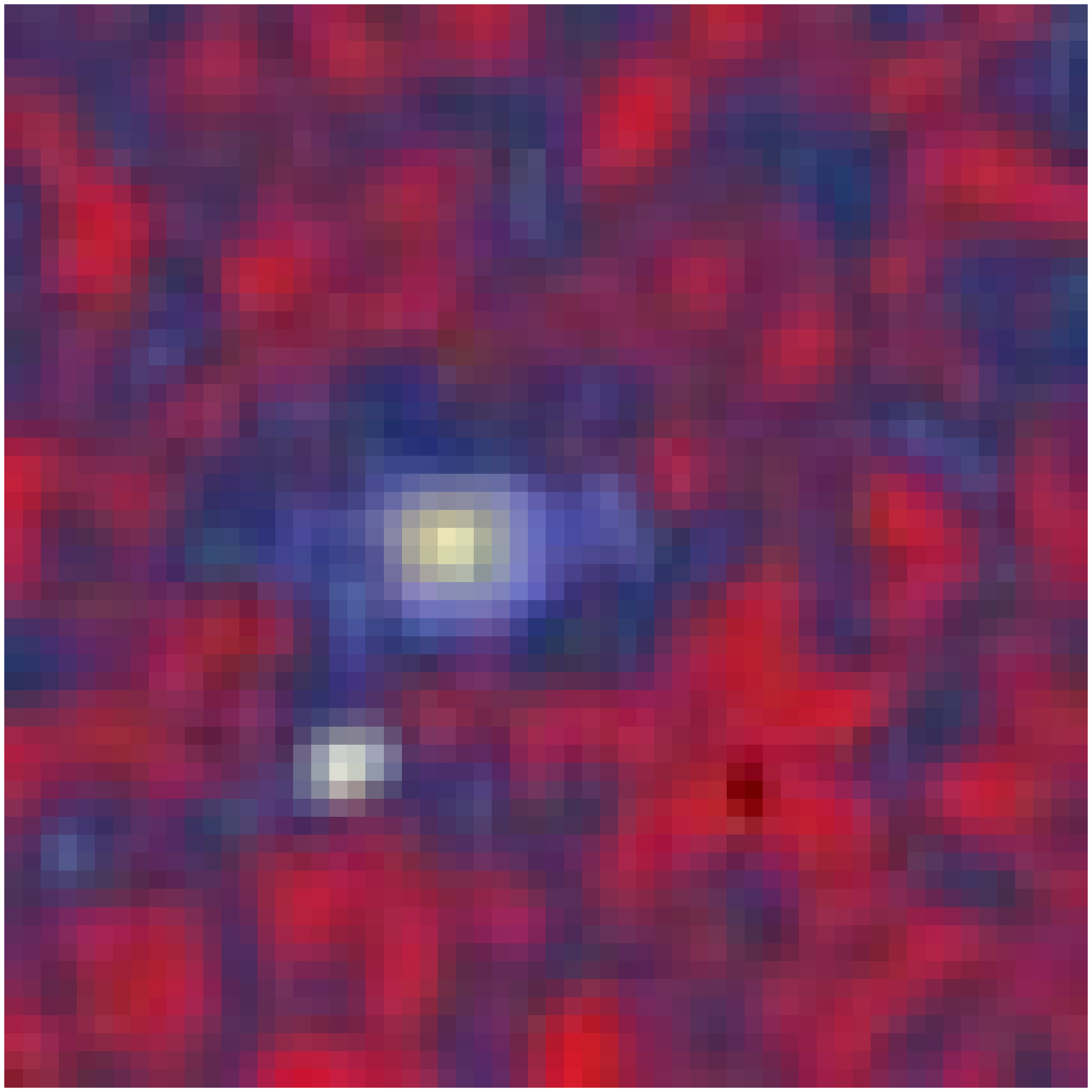}  \FGb{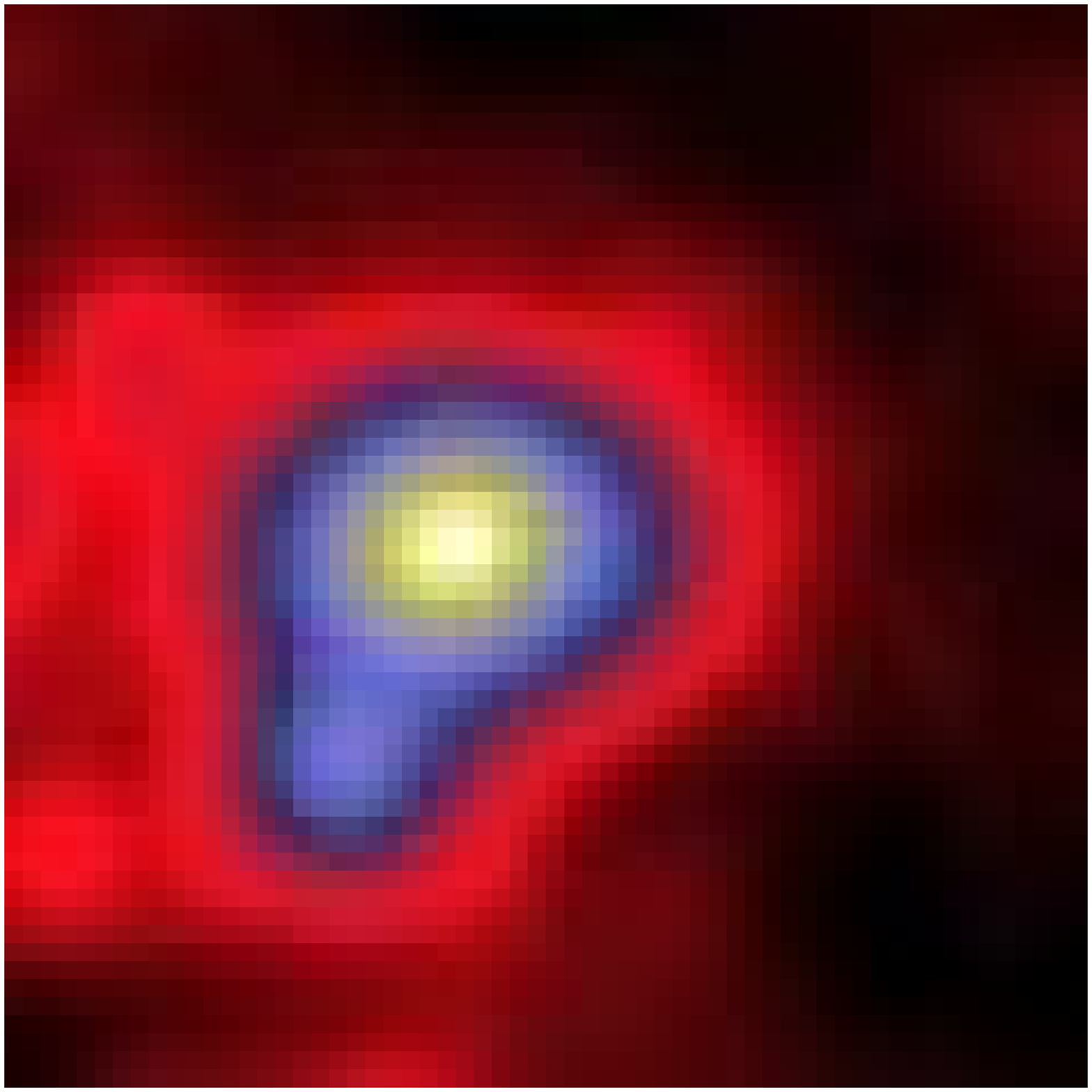}  \FGb{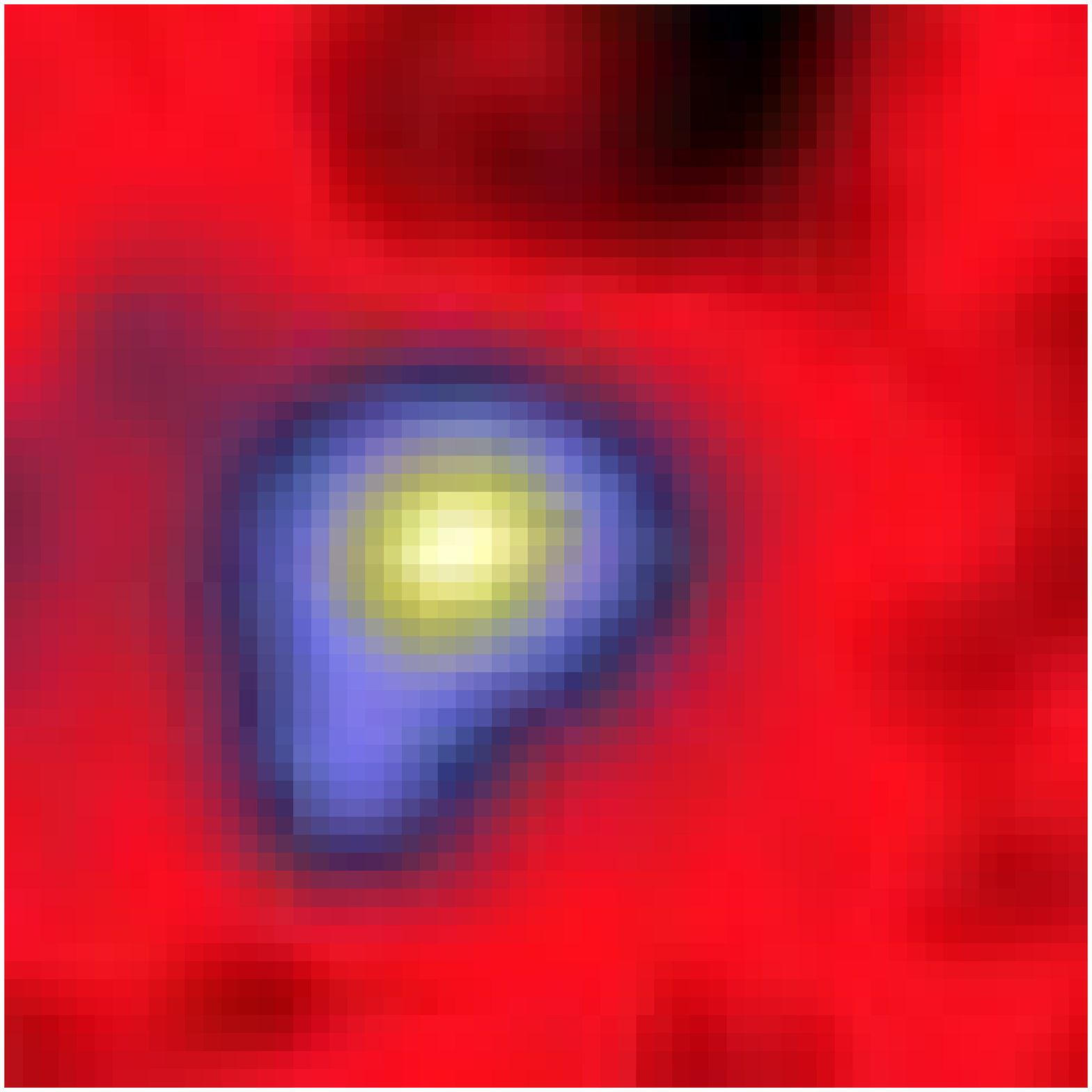}  \FGb{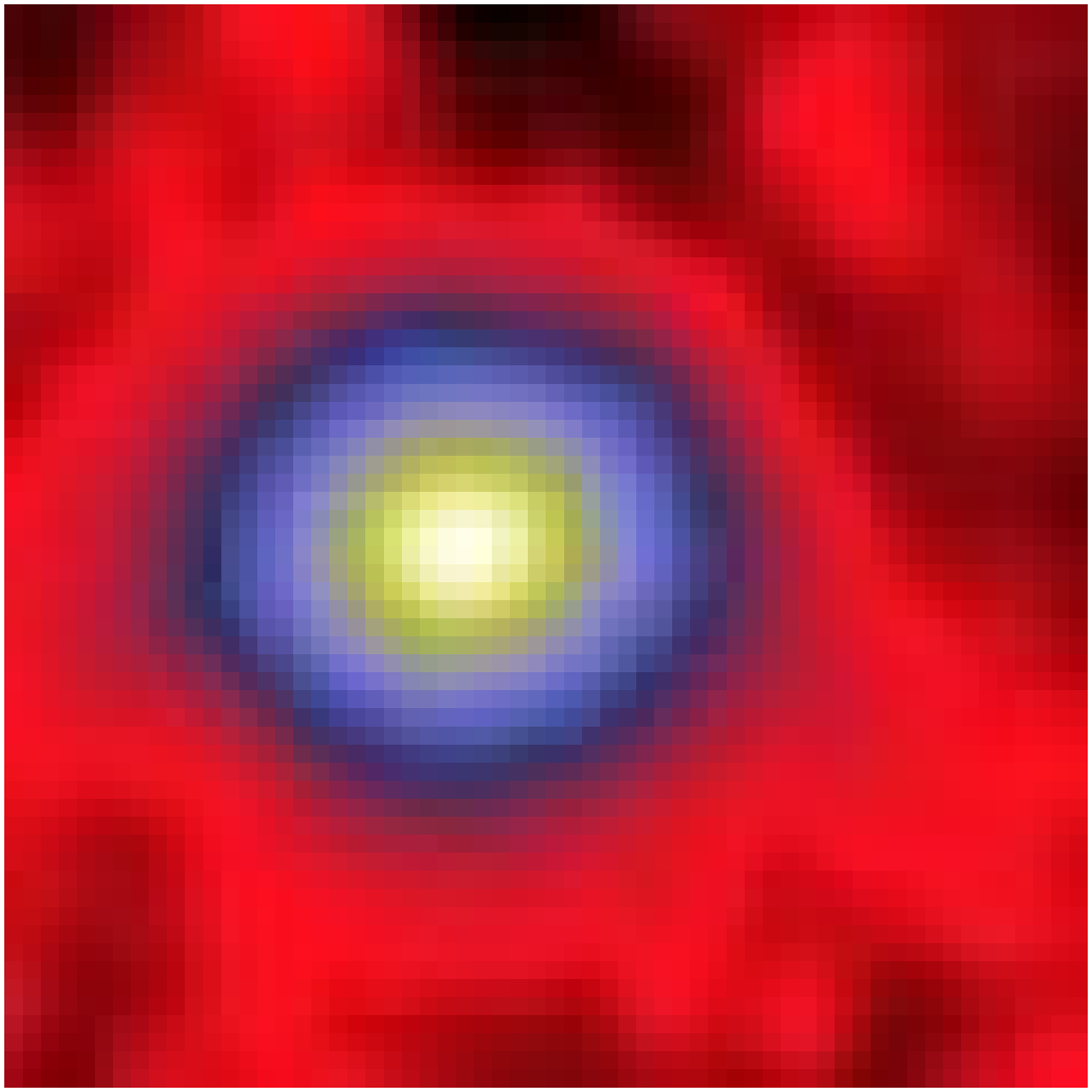}  \FGb{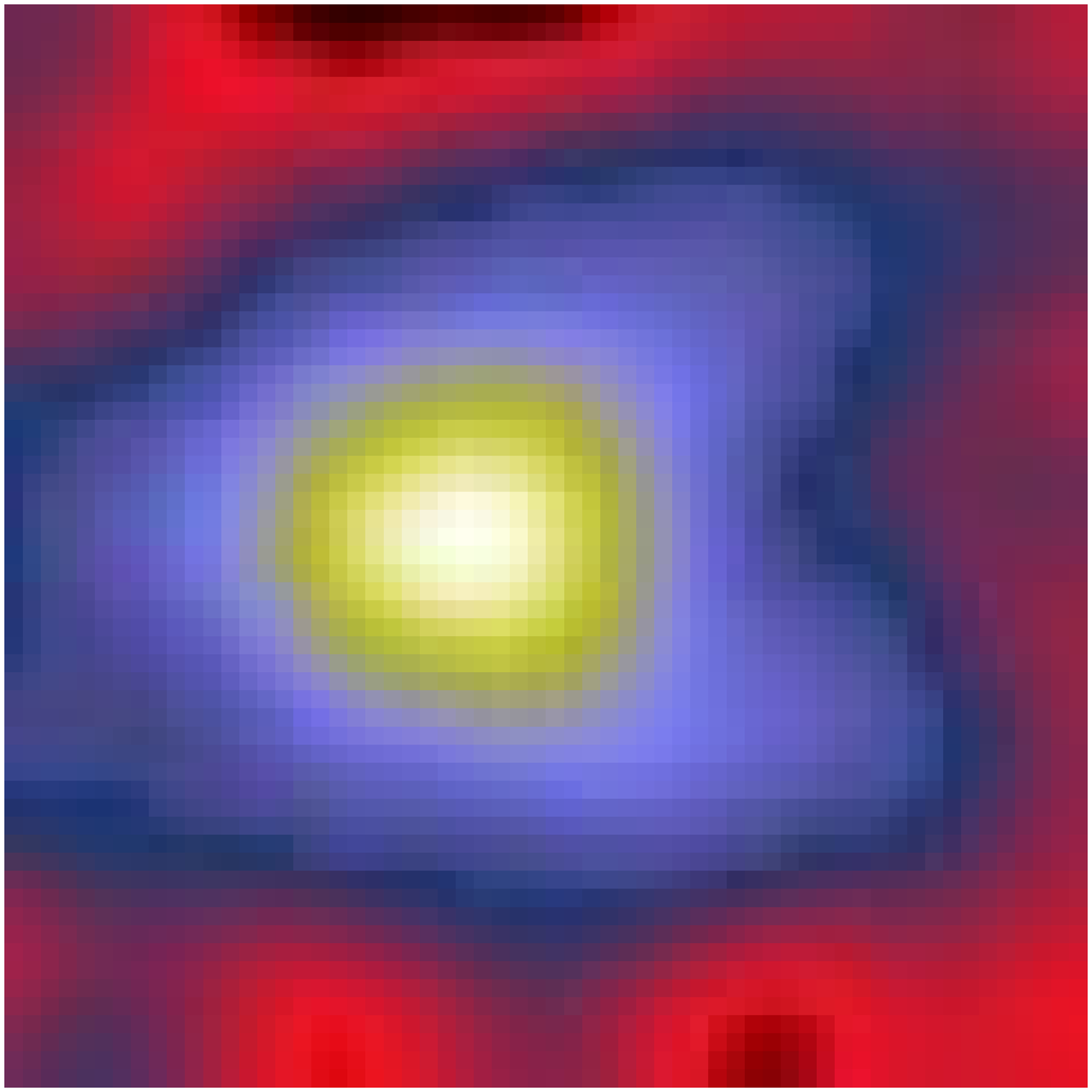} \\  {\#57   NCIA000956}  \\
  \FGb{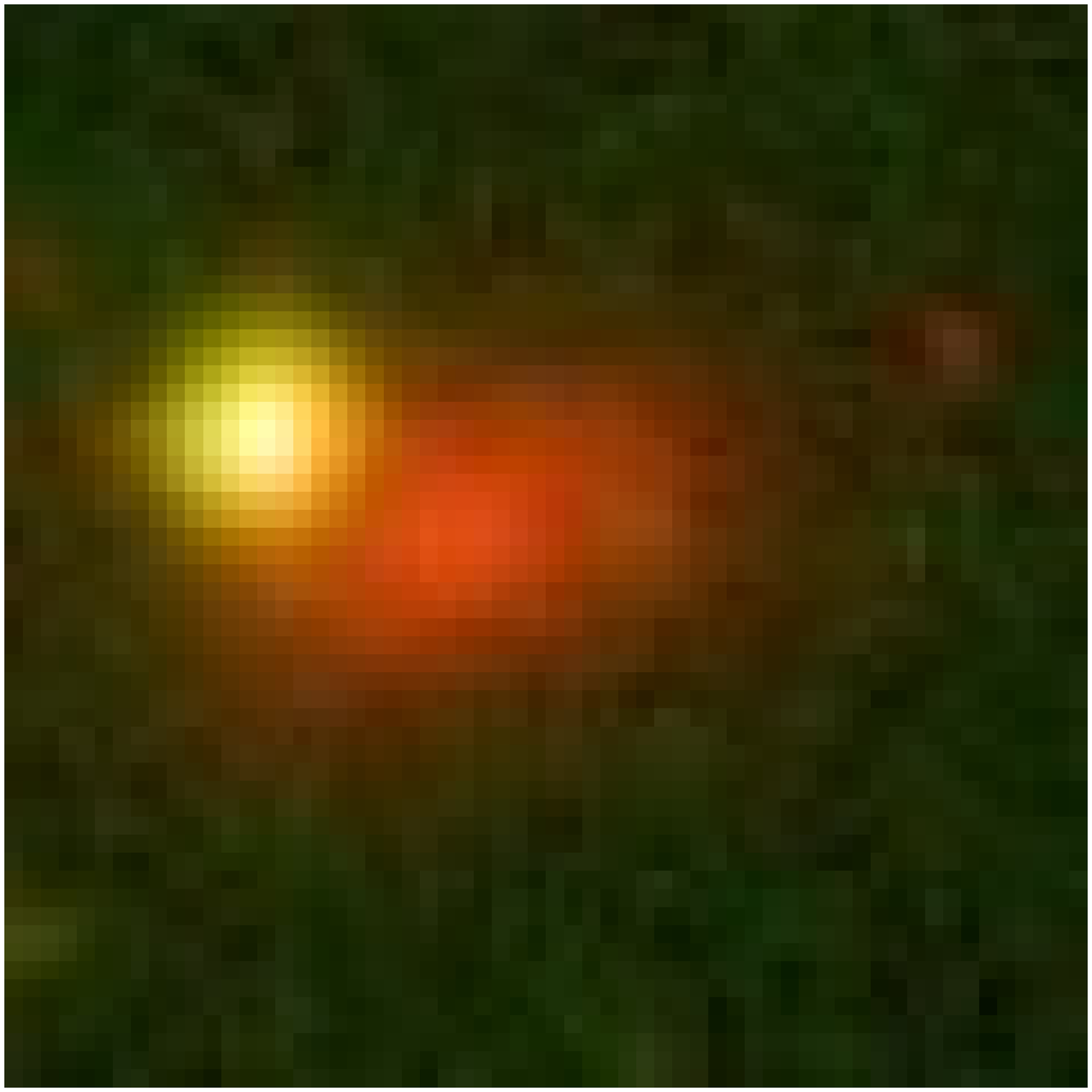}  \FGb{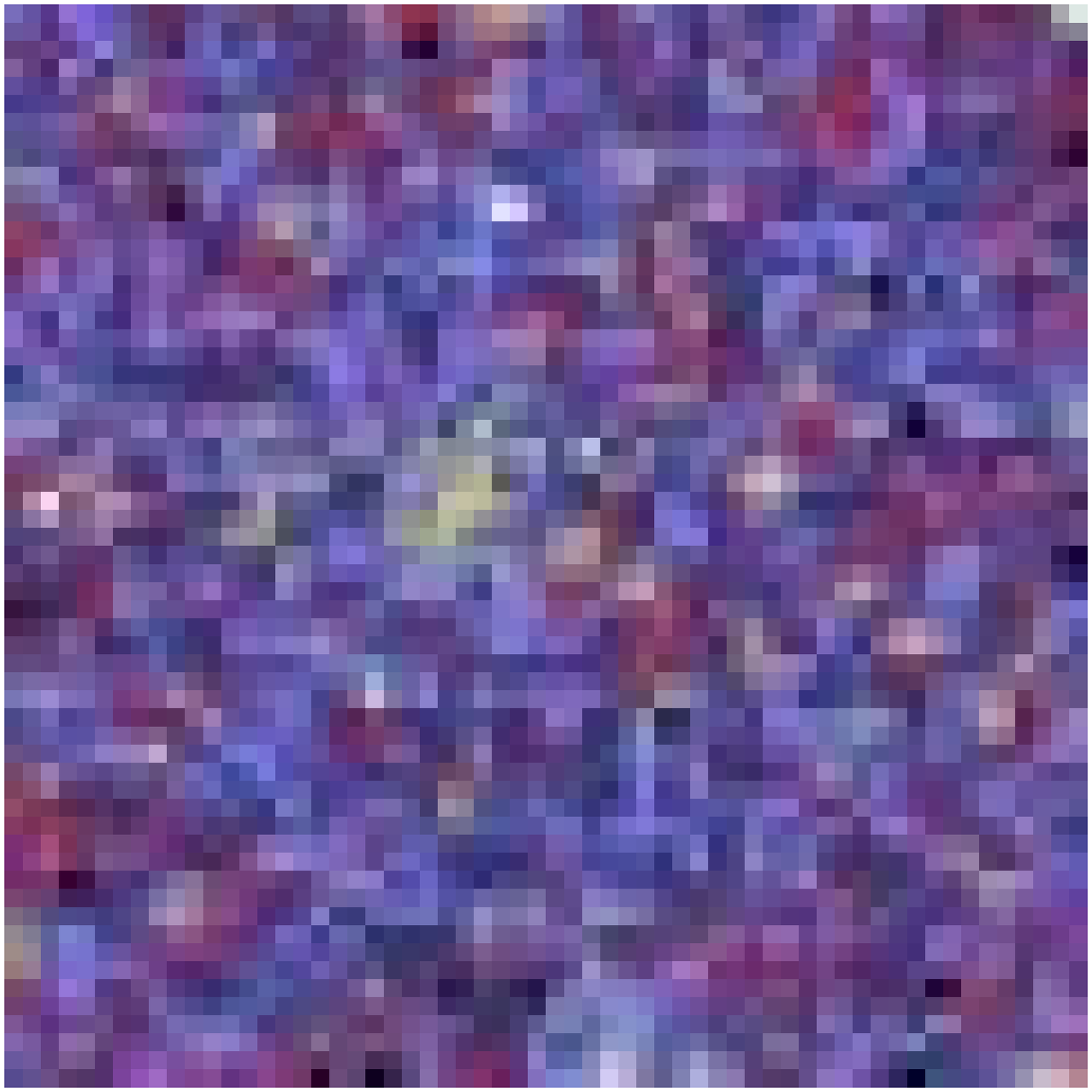}  \FGb{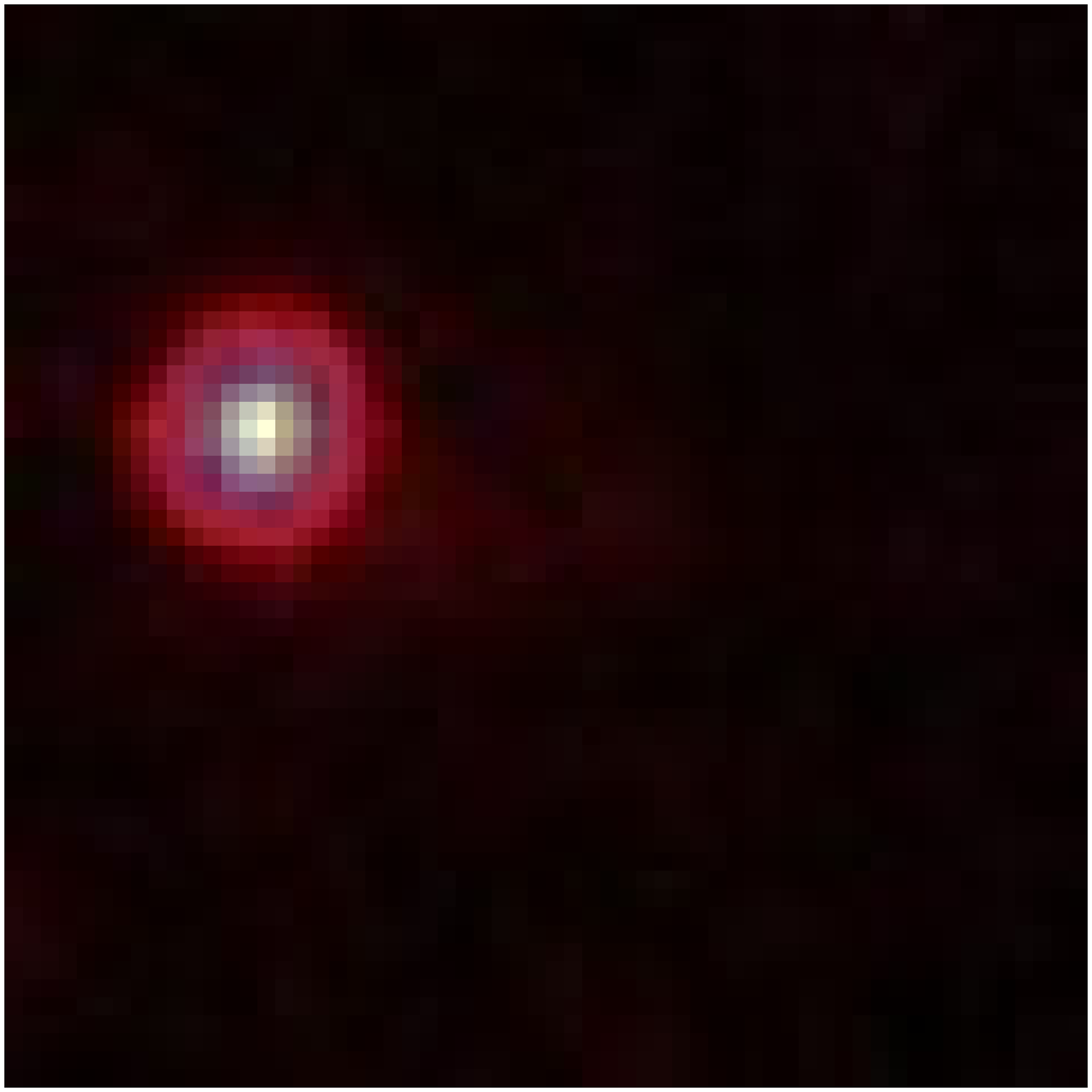}  \FGb{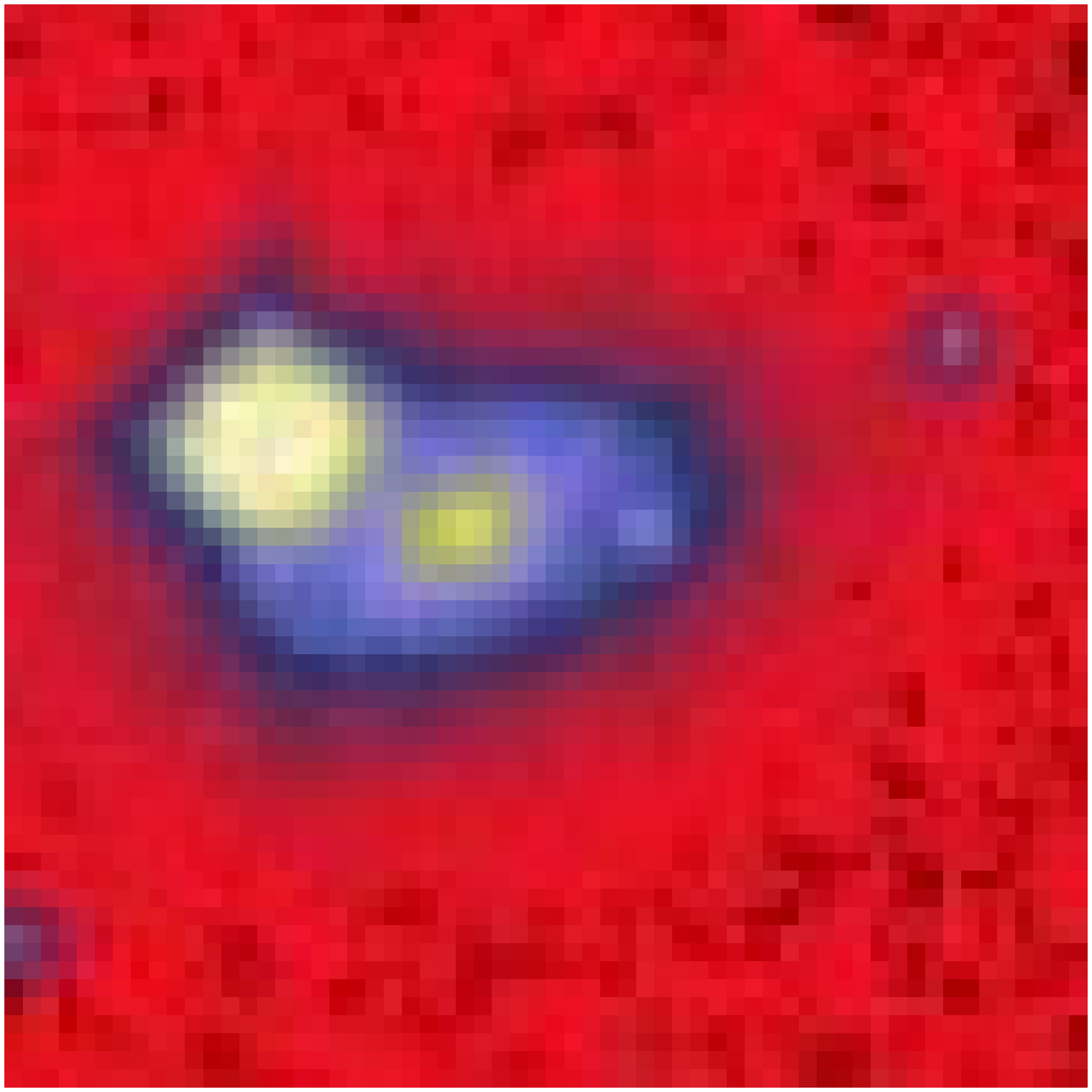}  \FGb{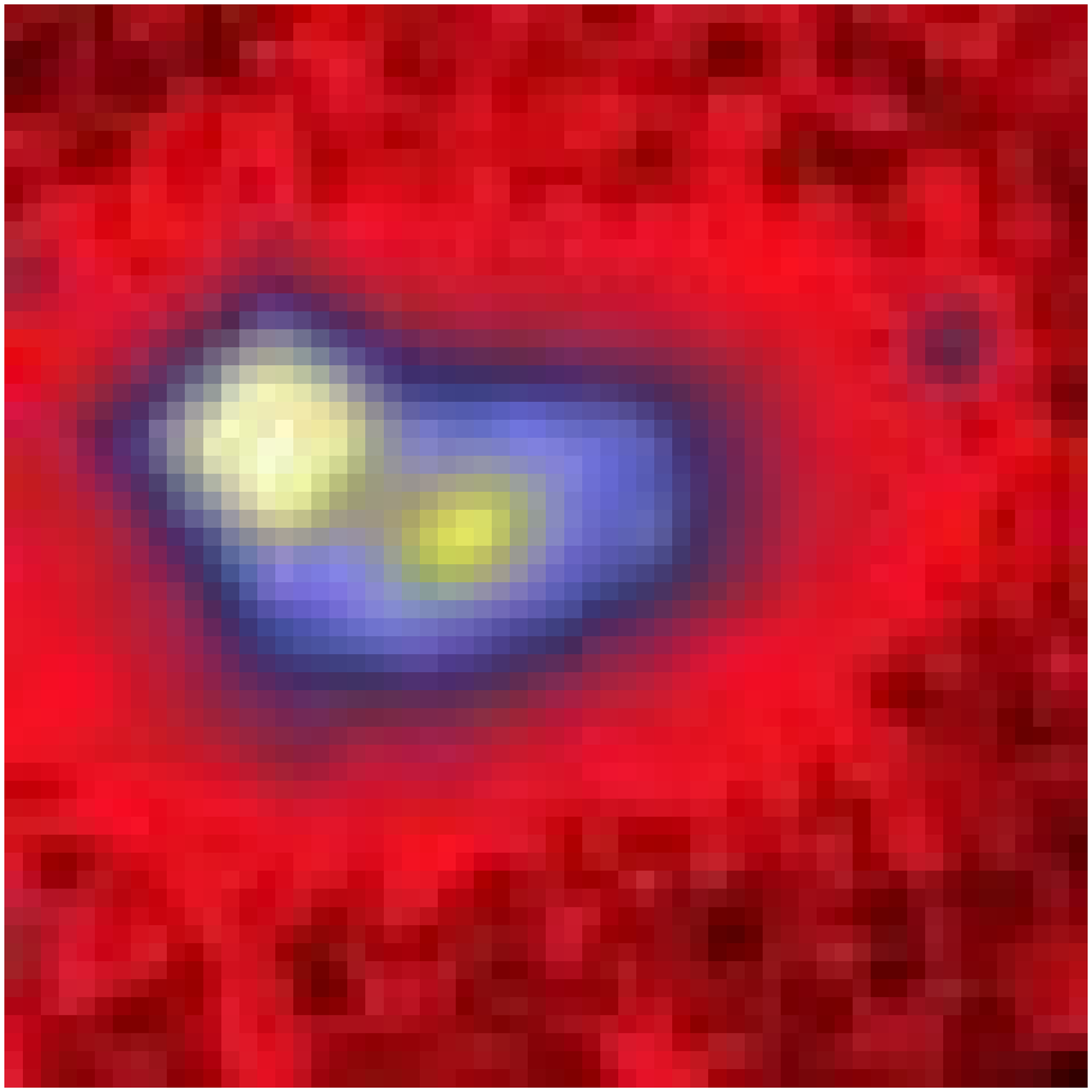}  \FGb{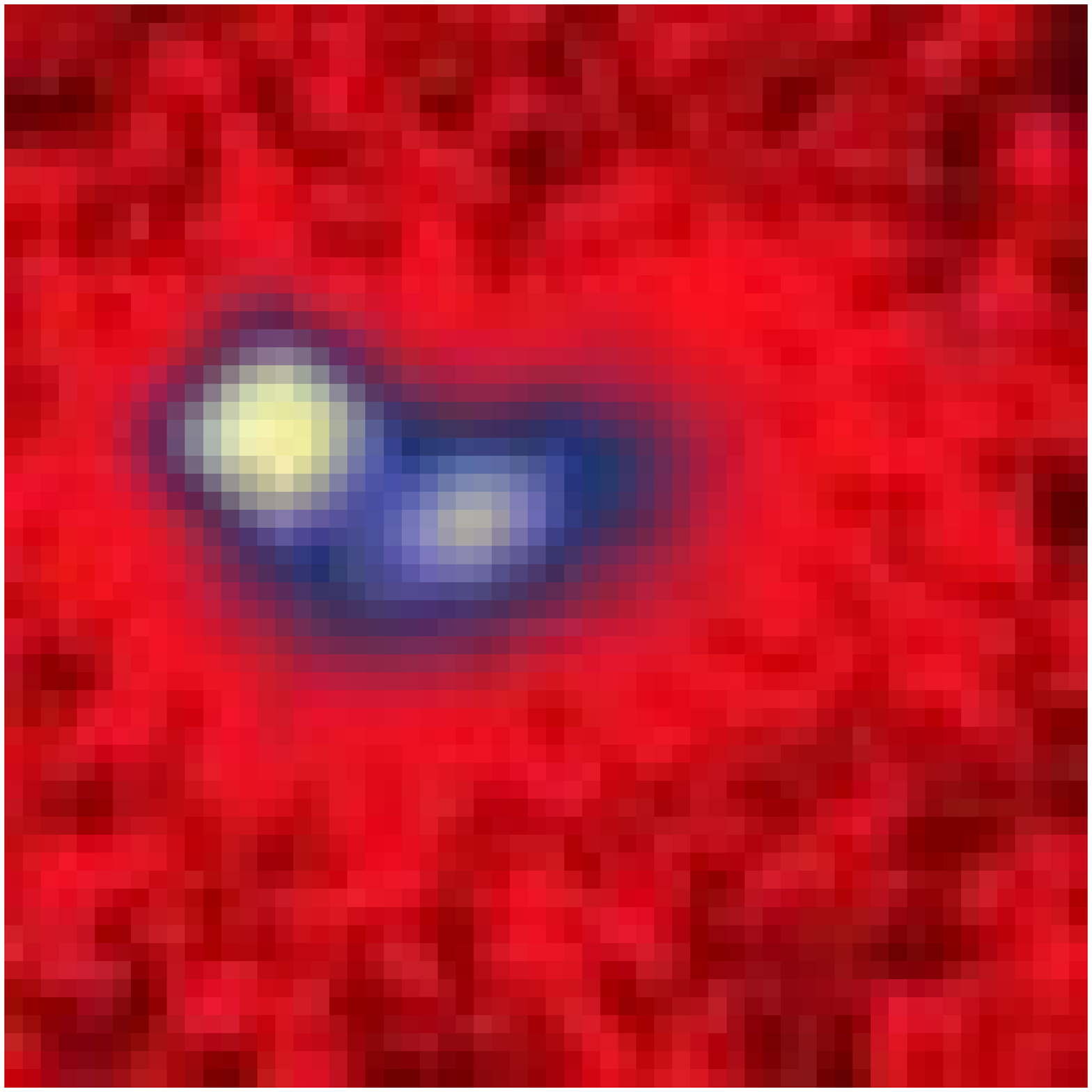}  \FGb{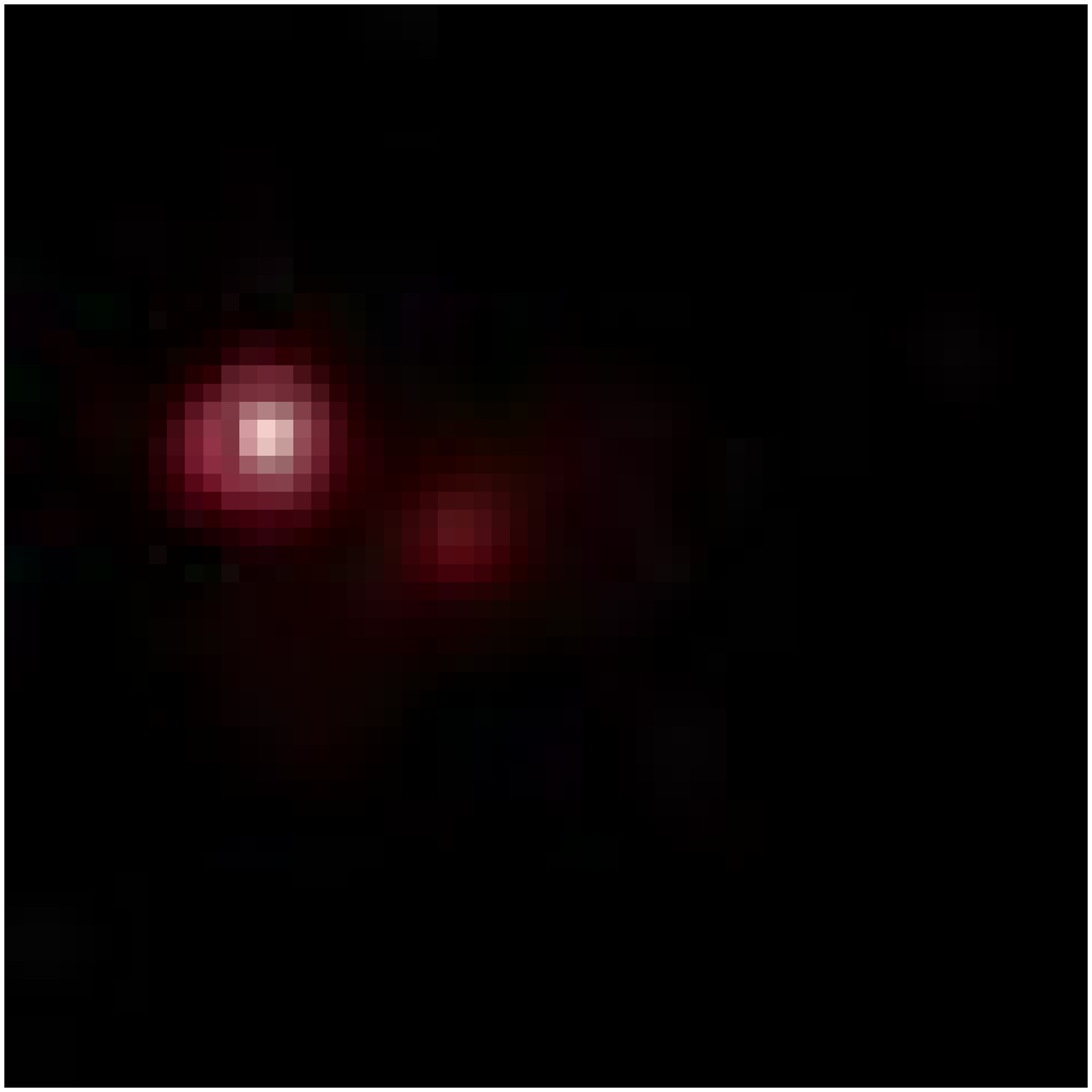}  \FGb{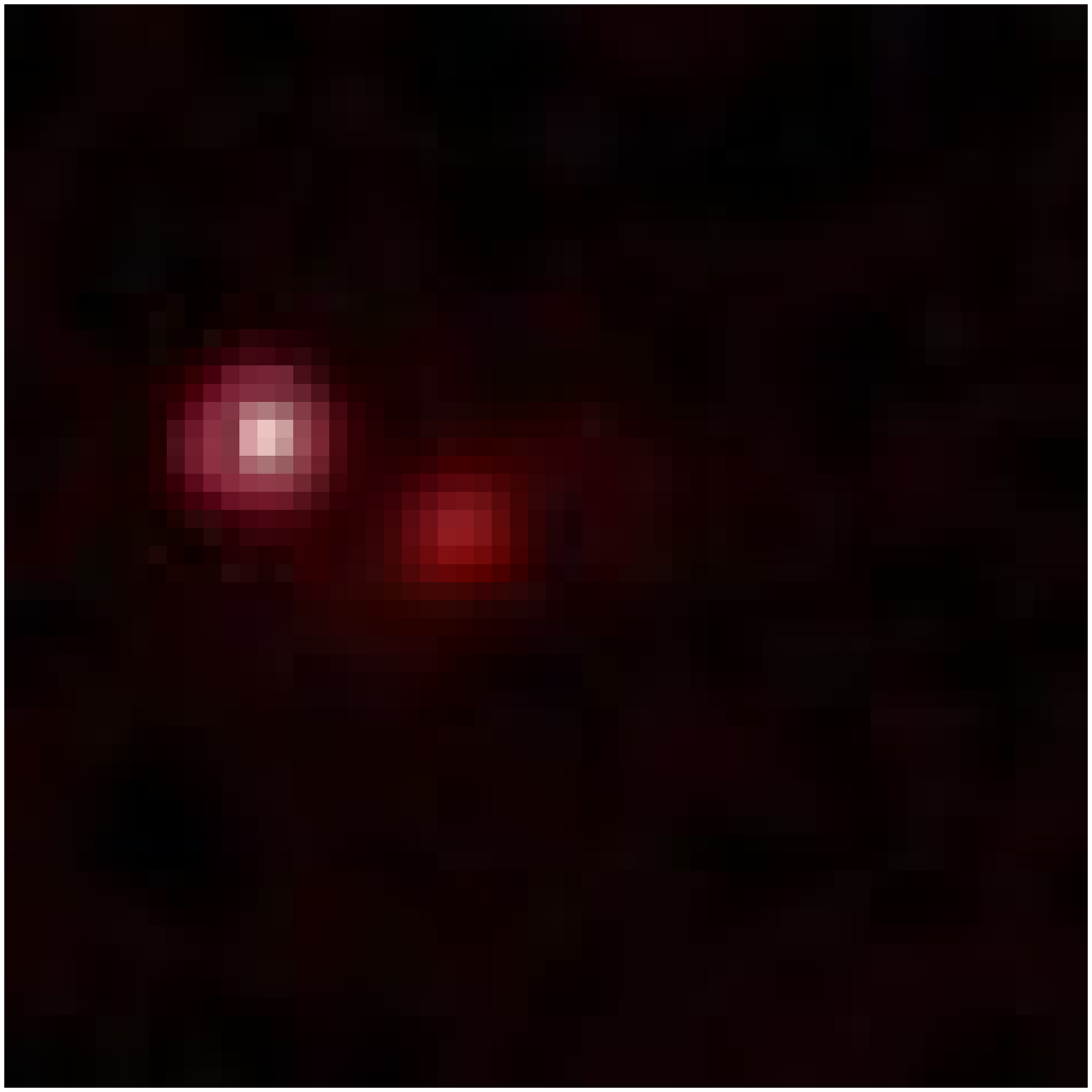}  \FGb{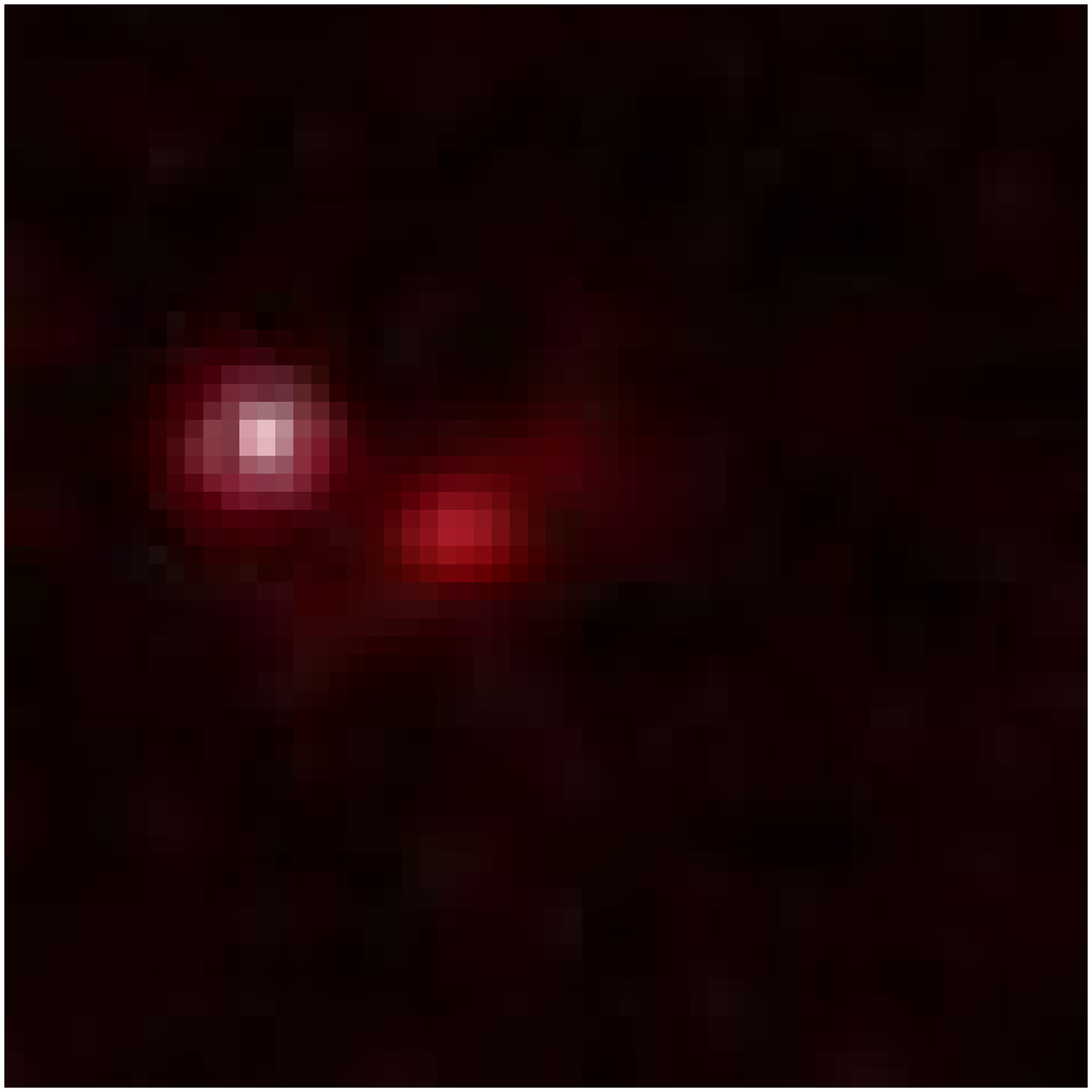}  \FGb{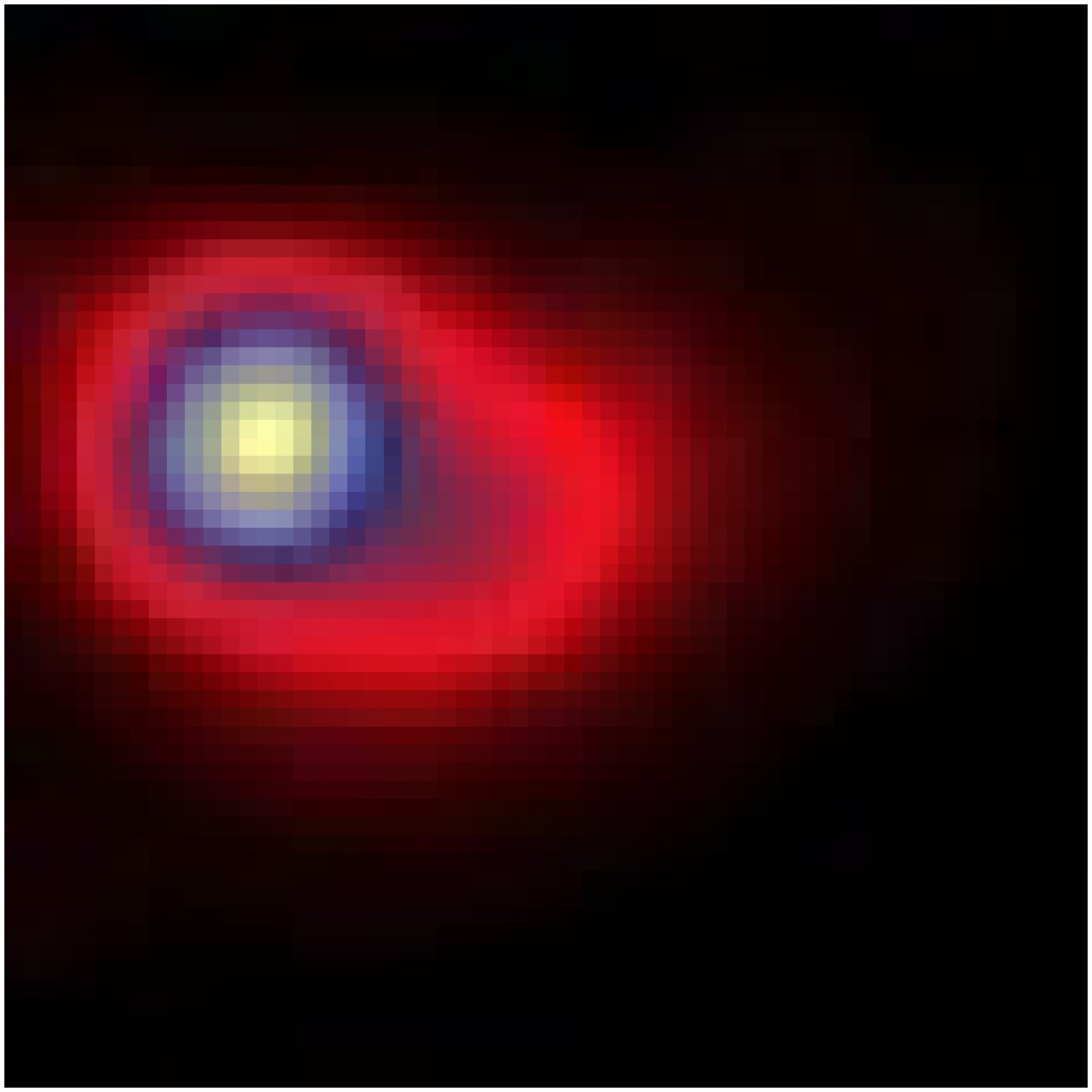}  \FGb{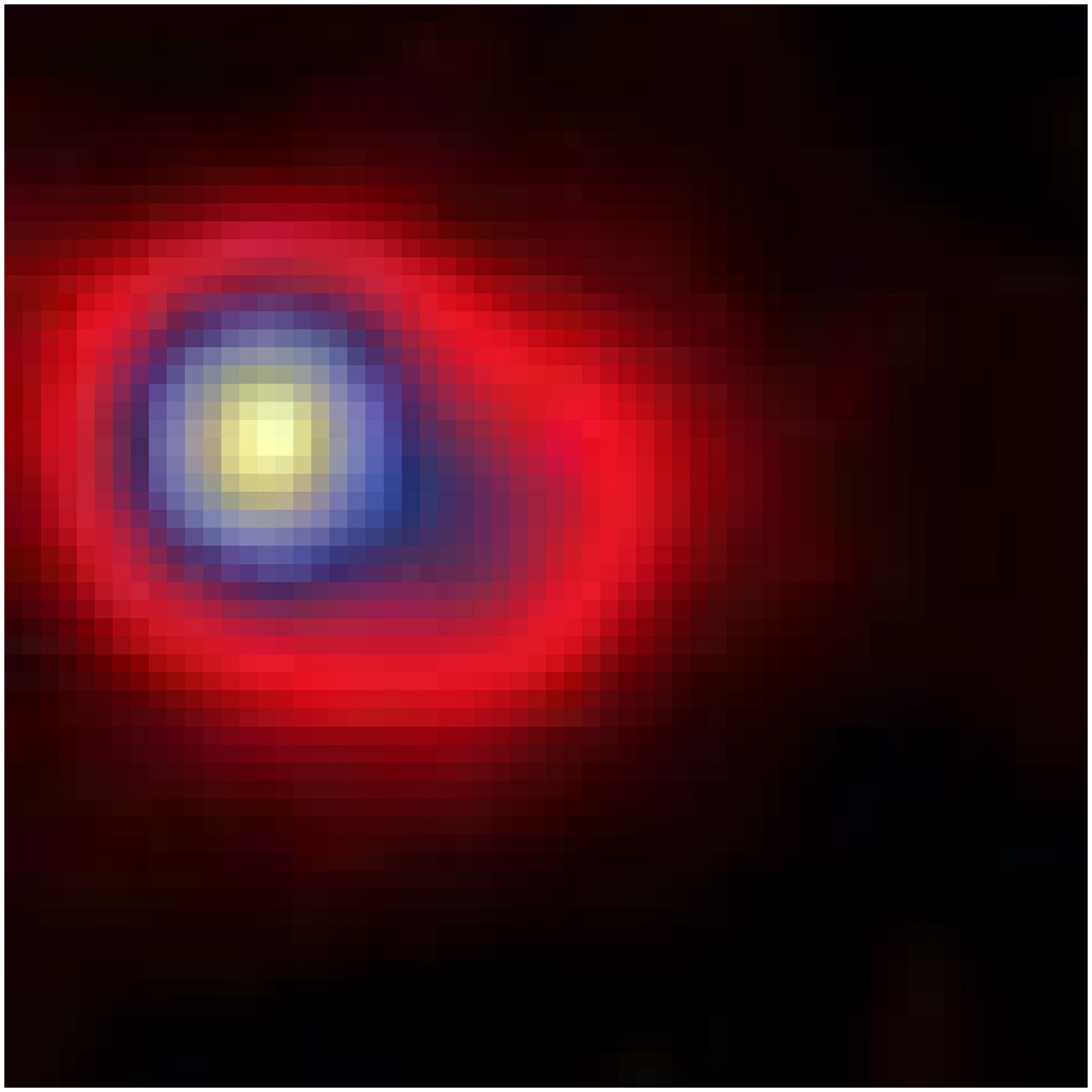}  \FGb{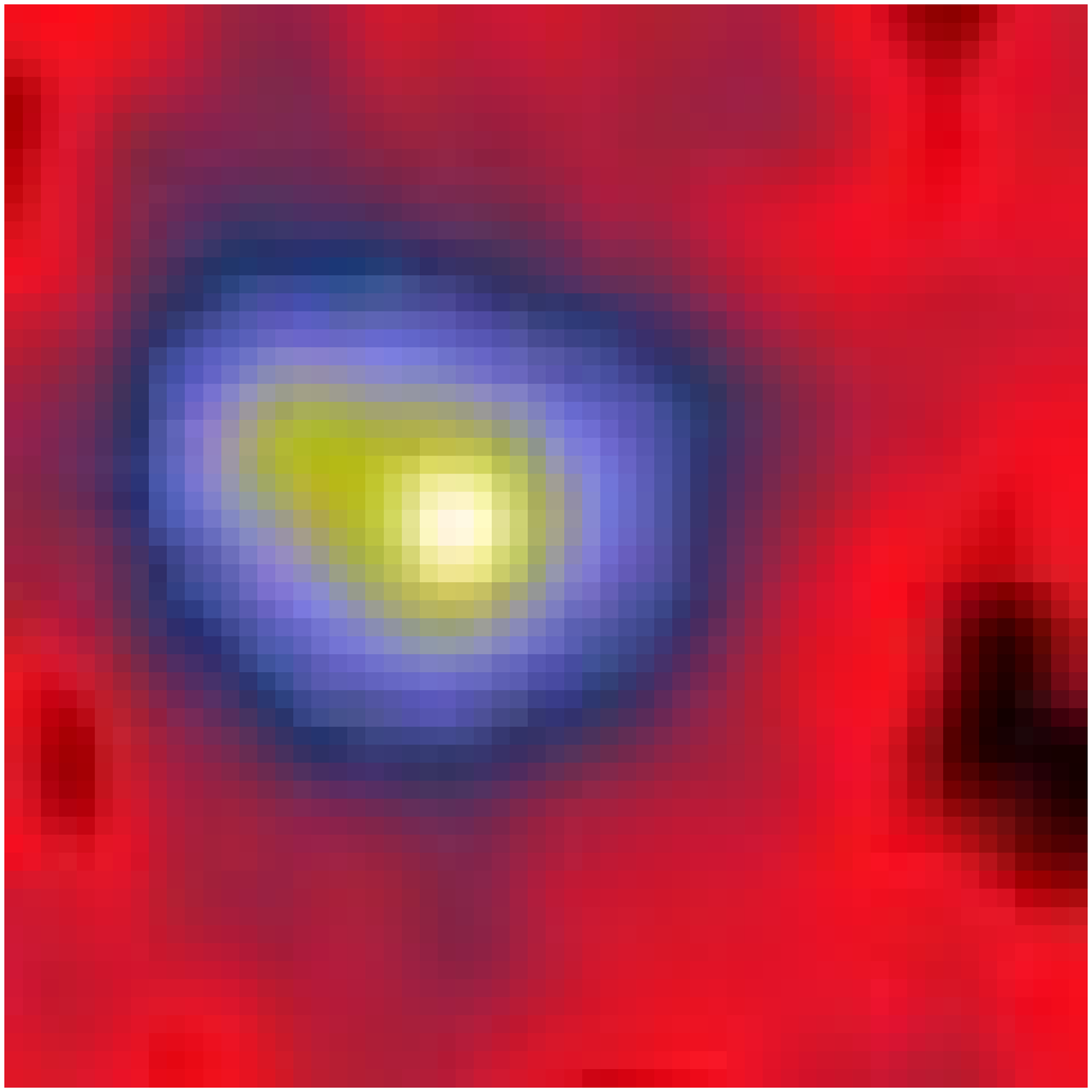}  \FGb{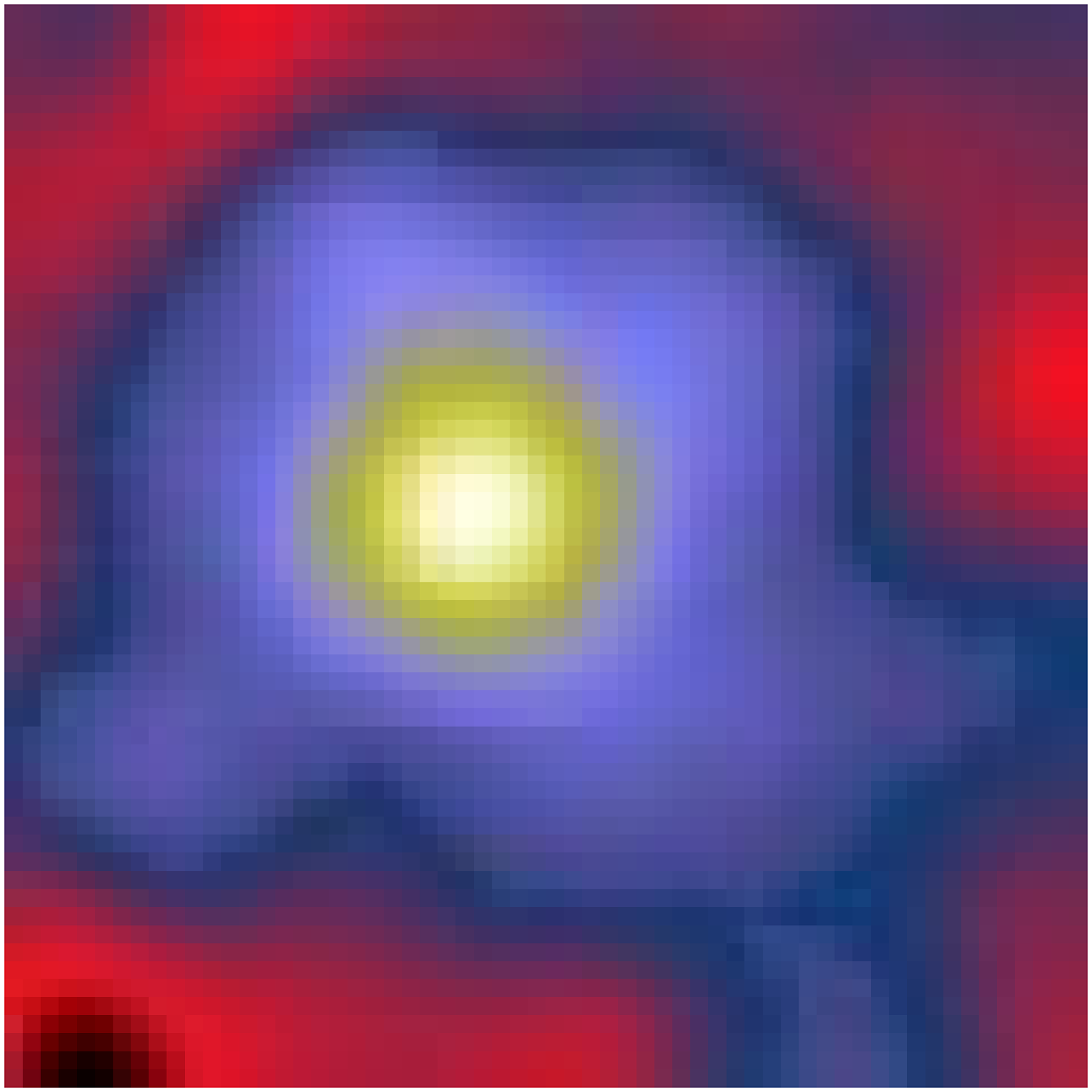} \\  {\#58   NCIA030617}  \\
  \FGb{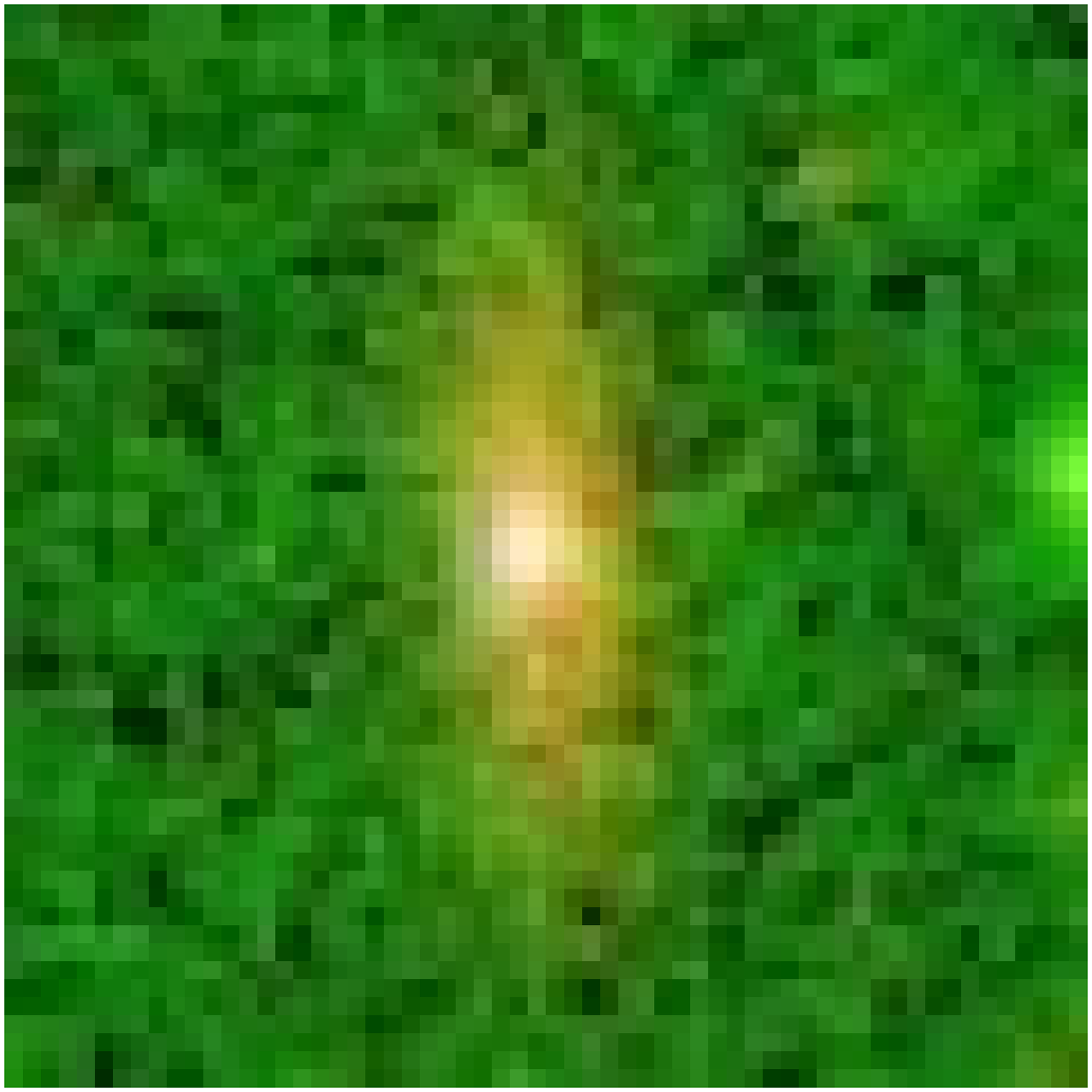}  \FGb{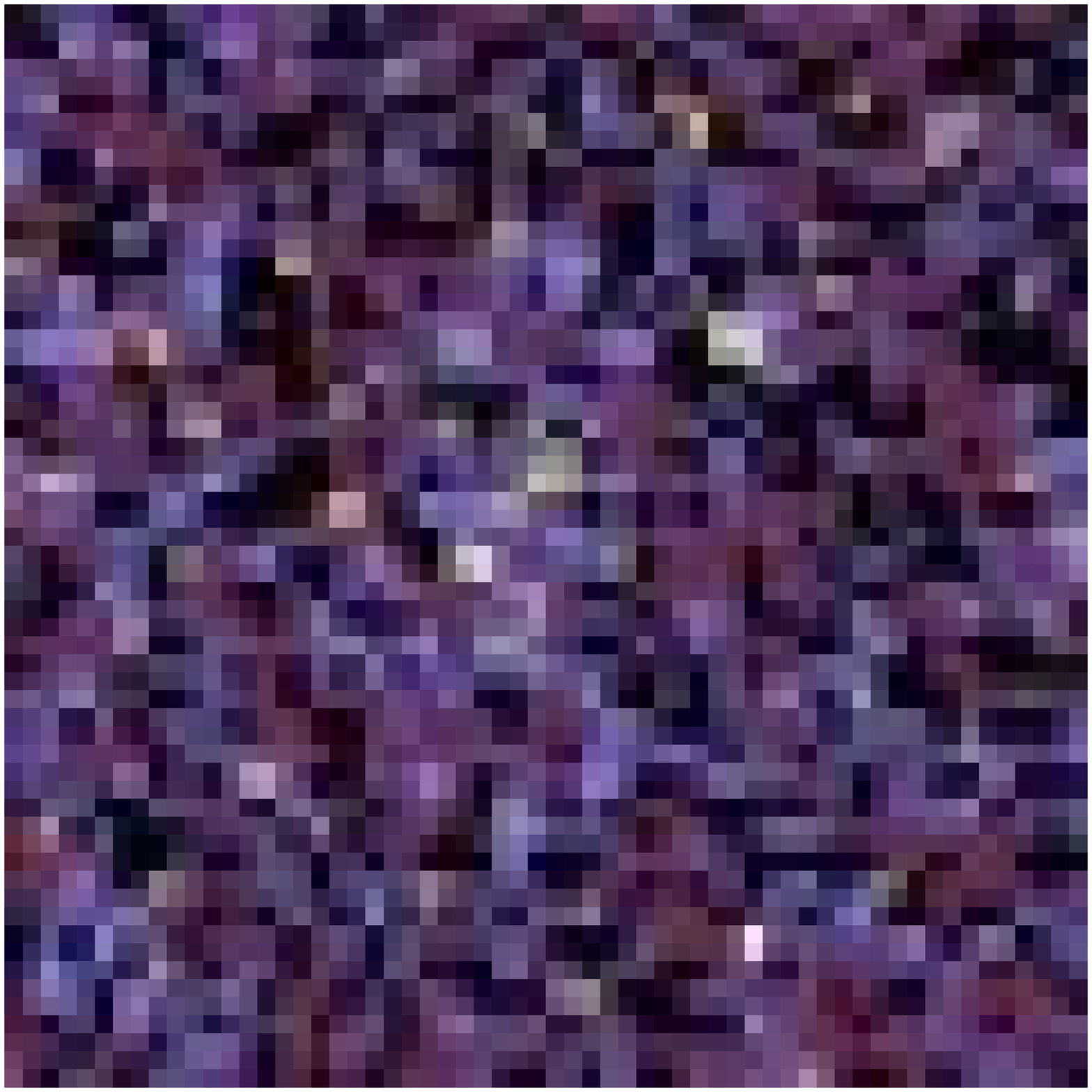}  \FGb{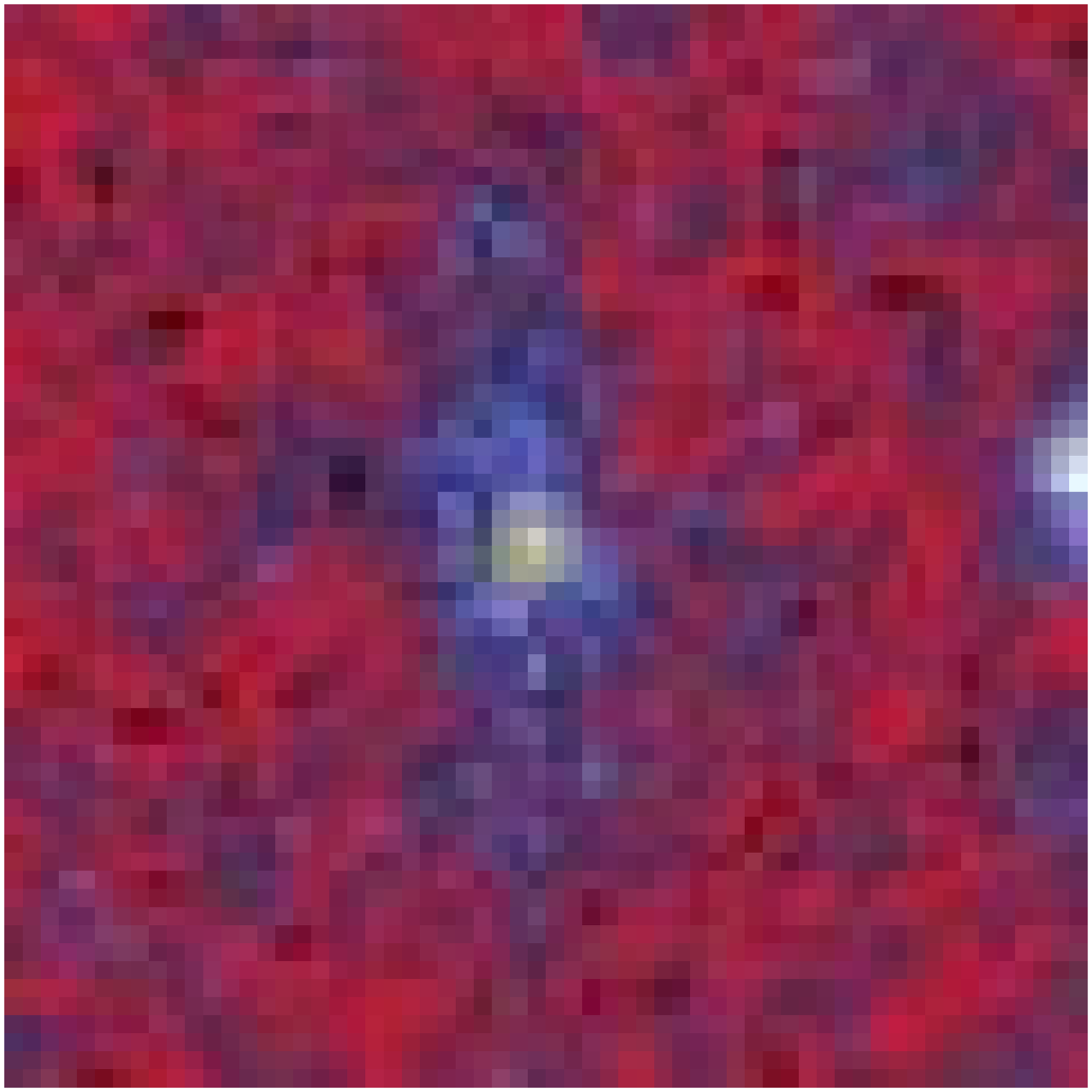}  \FGb{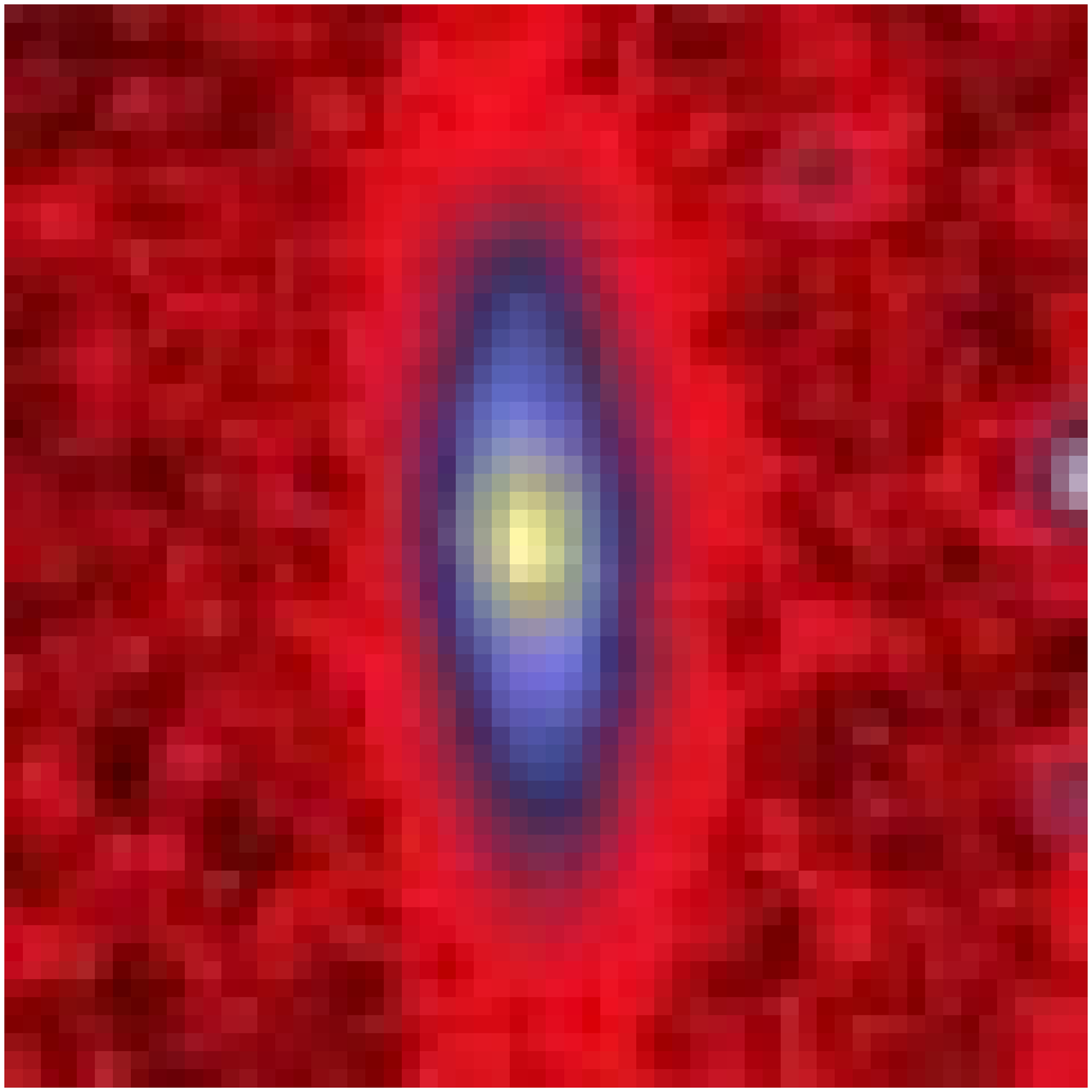}  \FGb{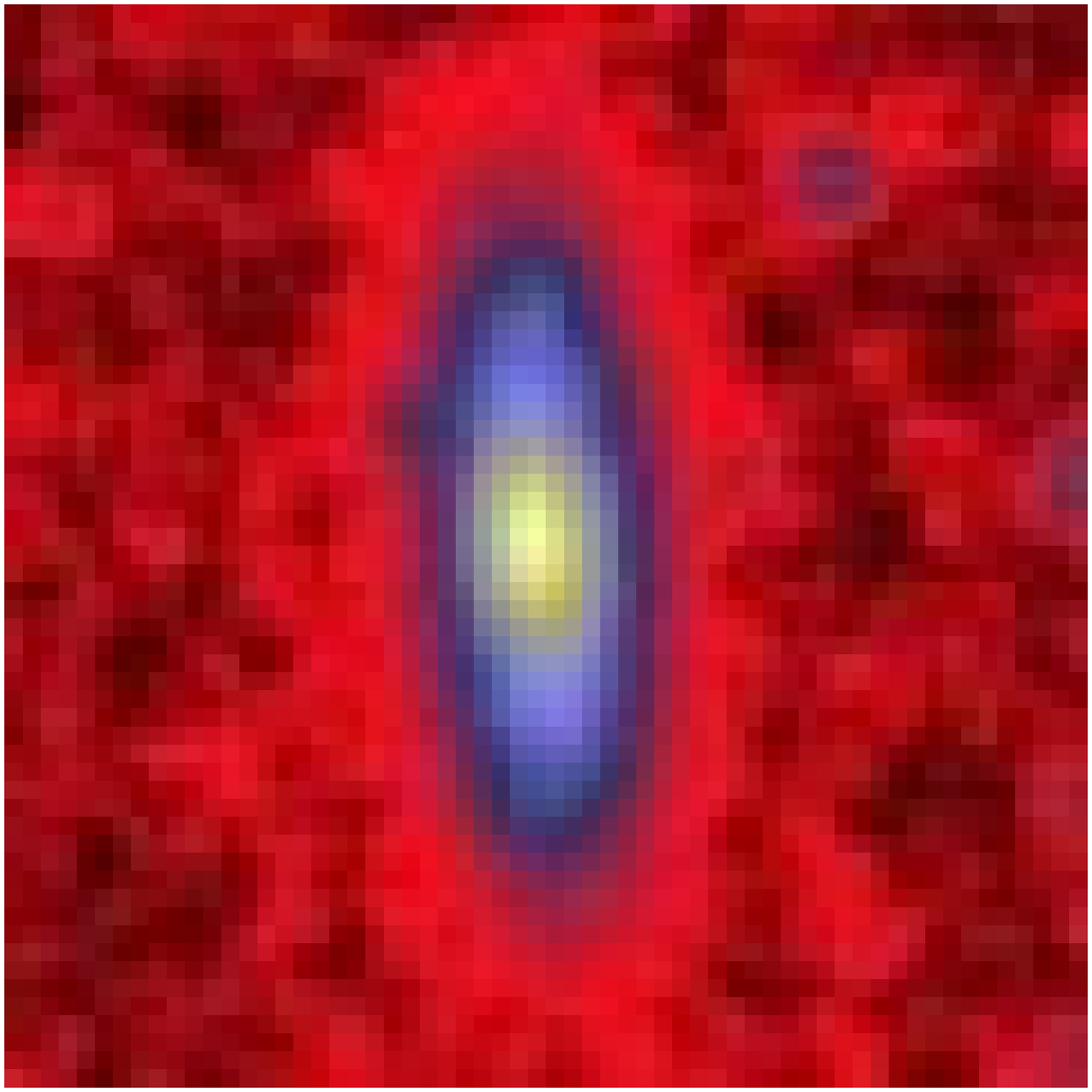}  \FGb{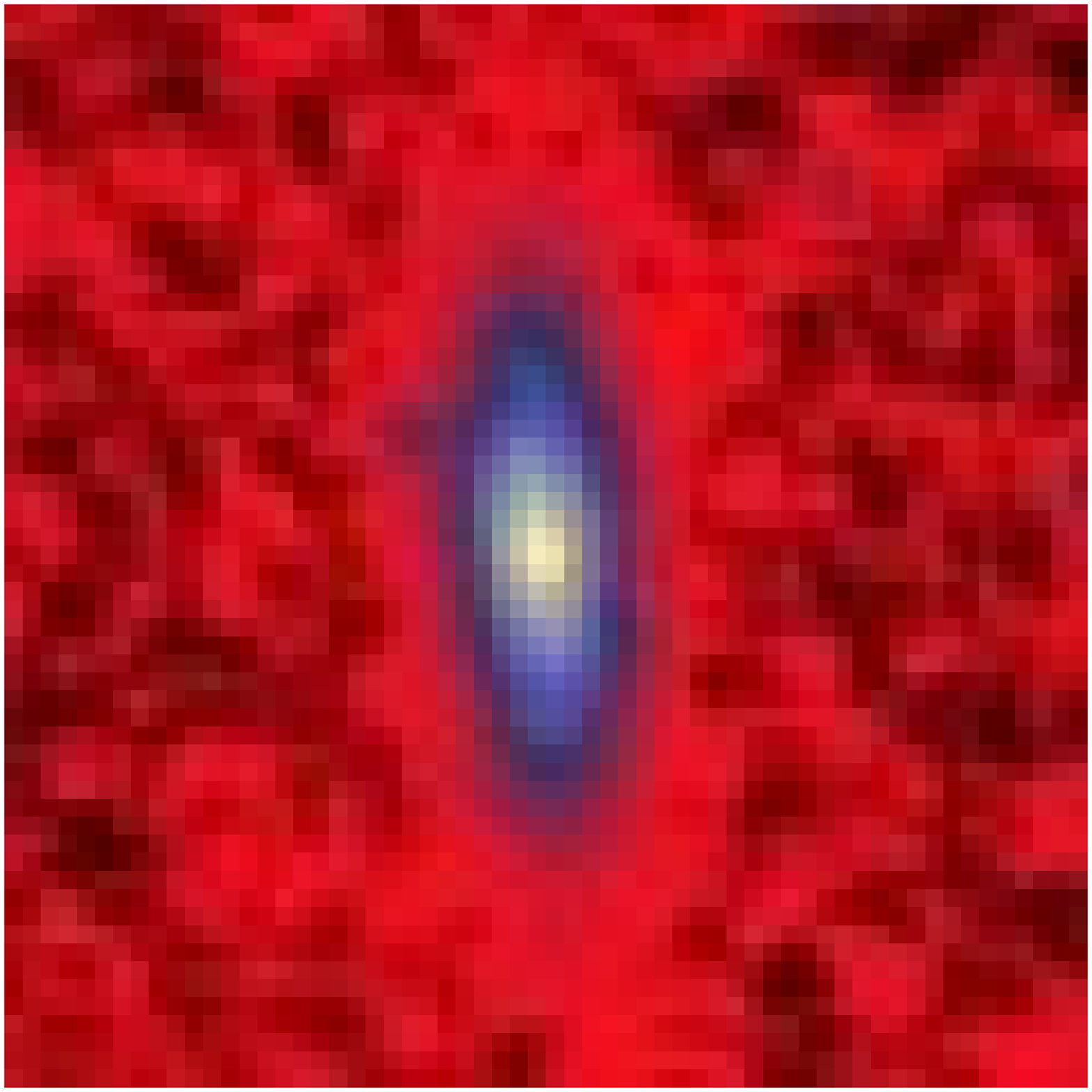}  \FGb{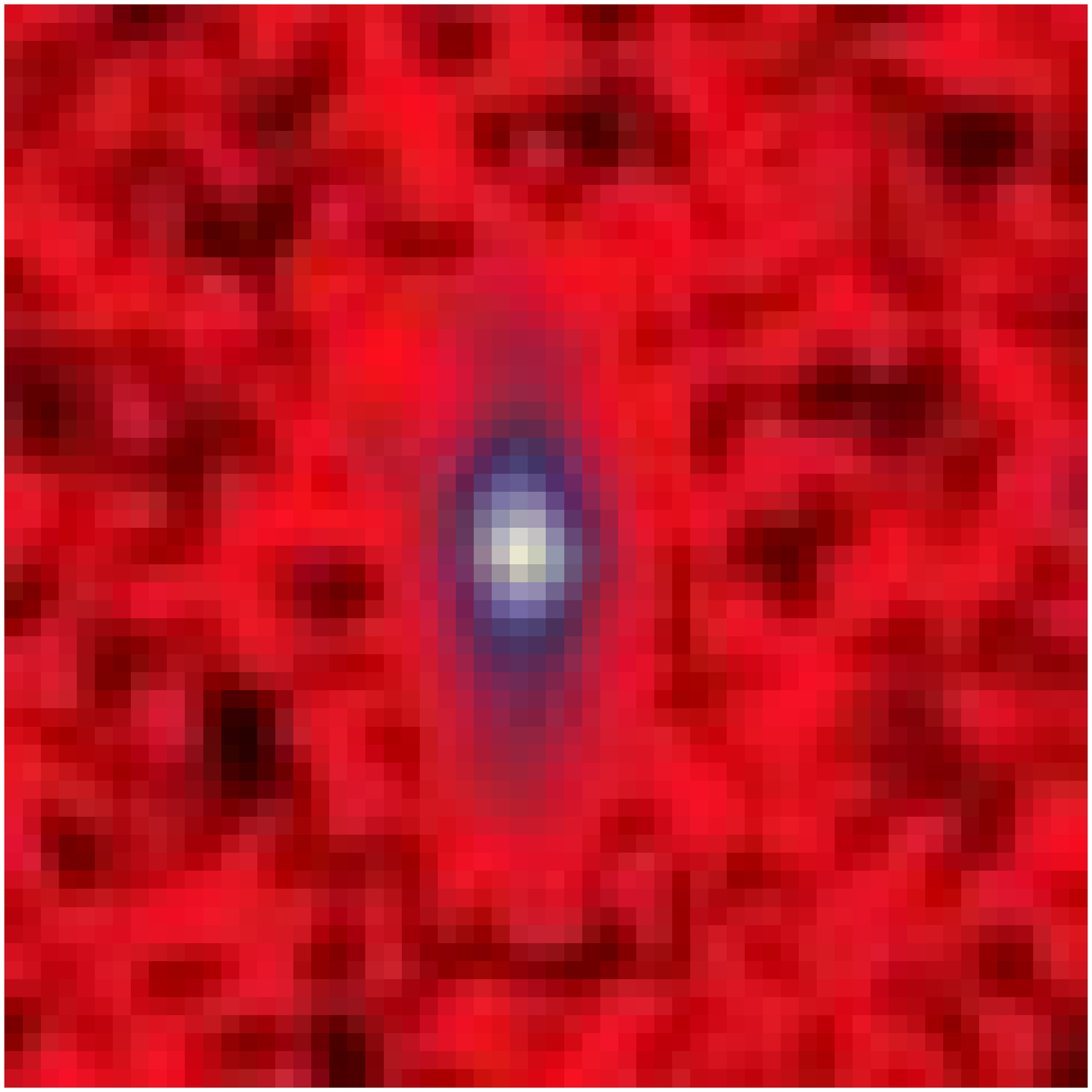}  \FGb{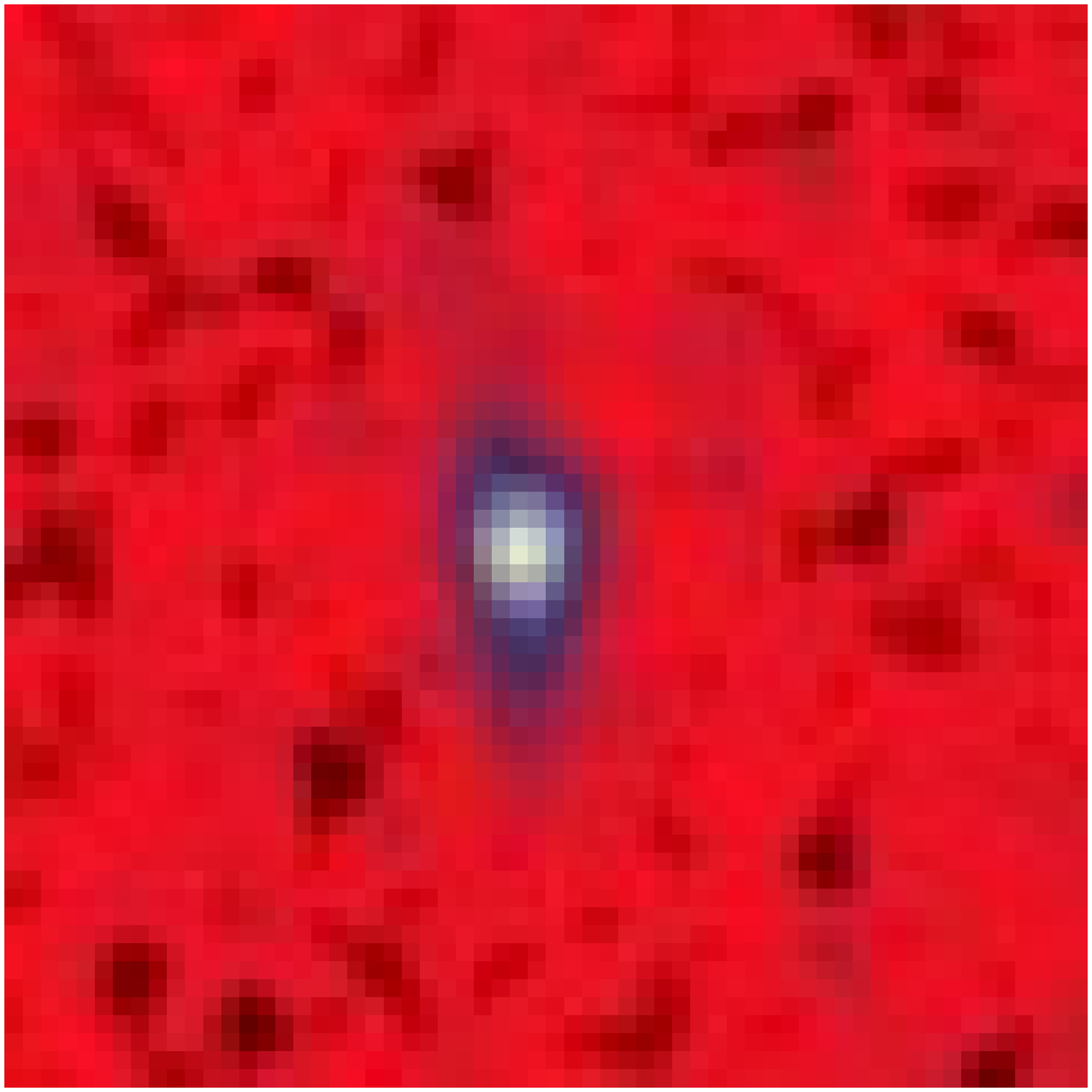}  \FGb{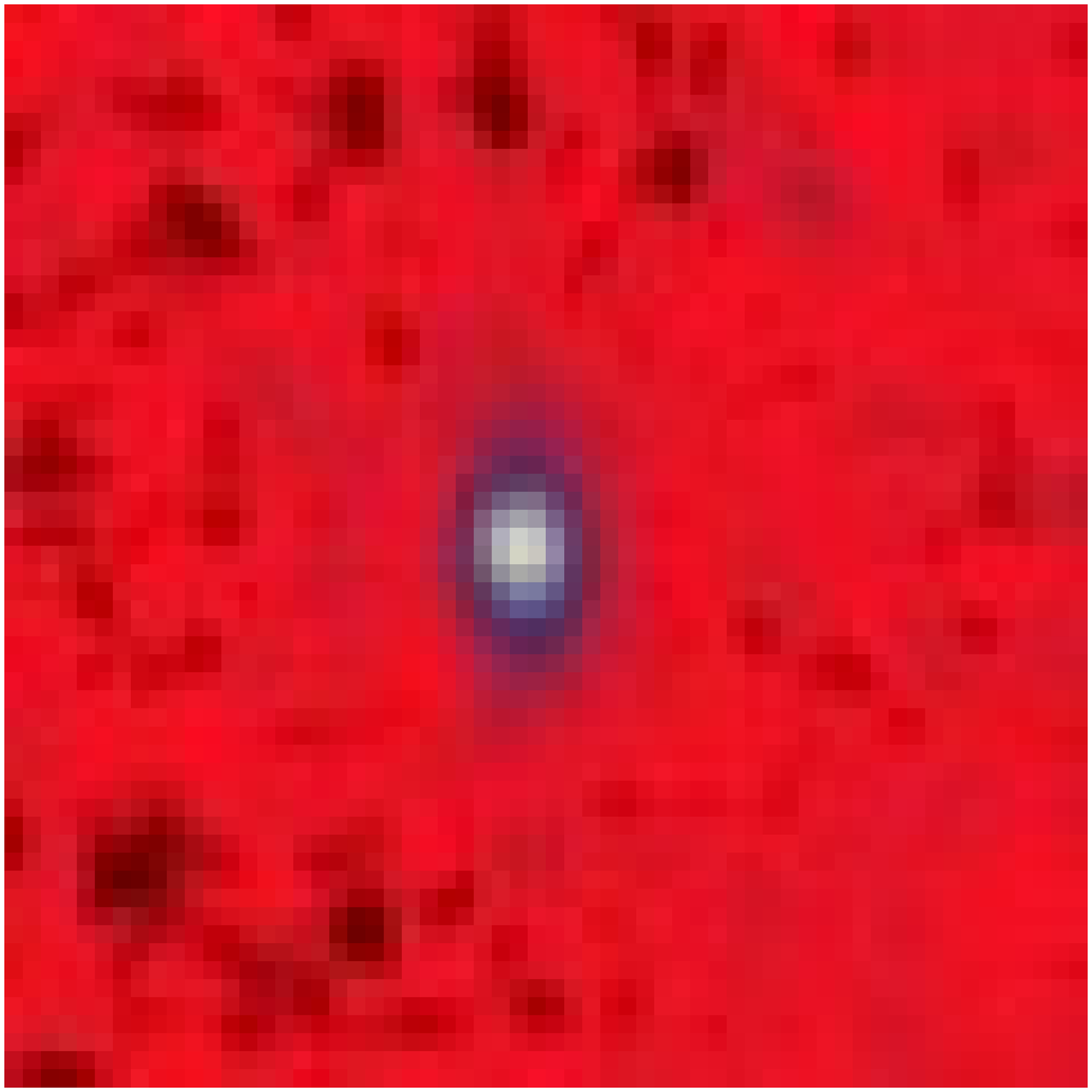}  \FGb{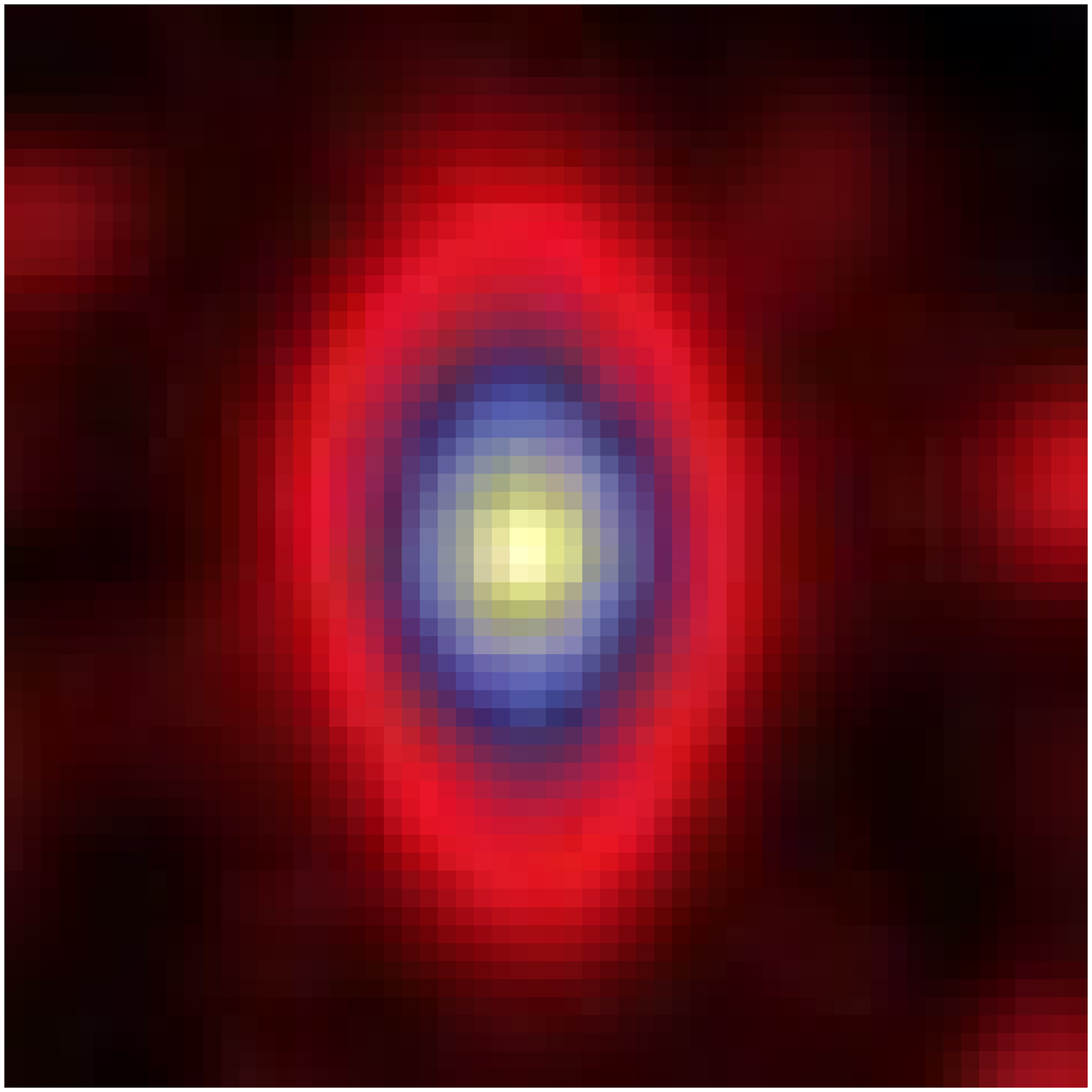}  \FGb{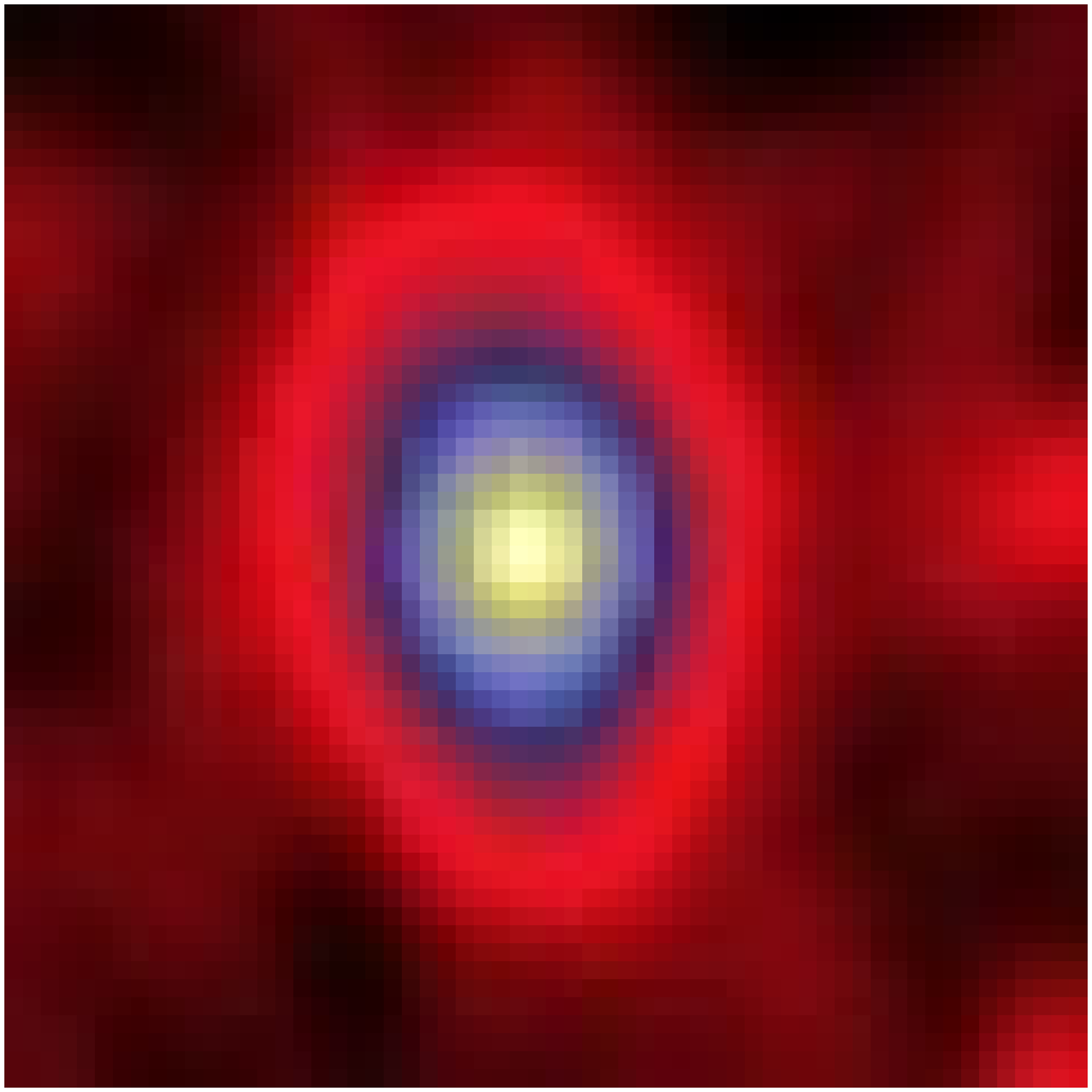}  \FGb{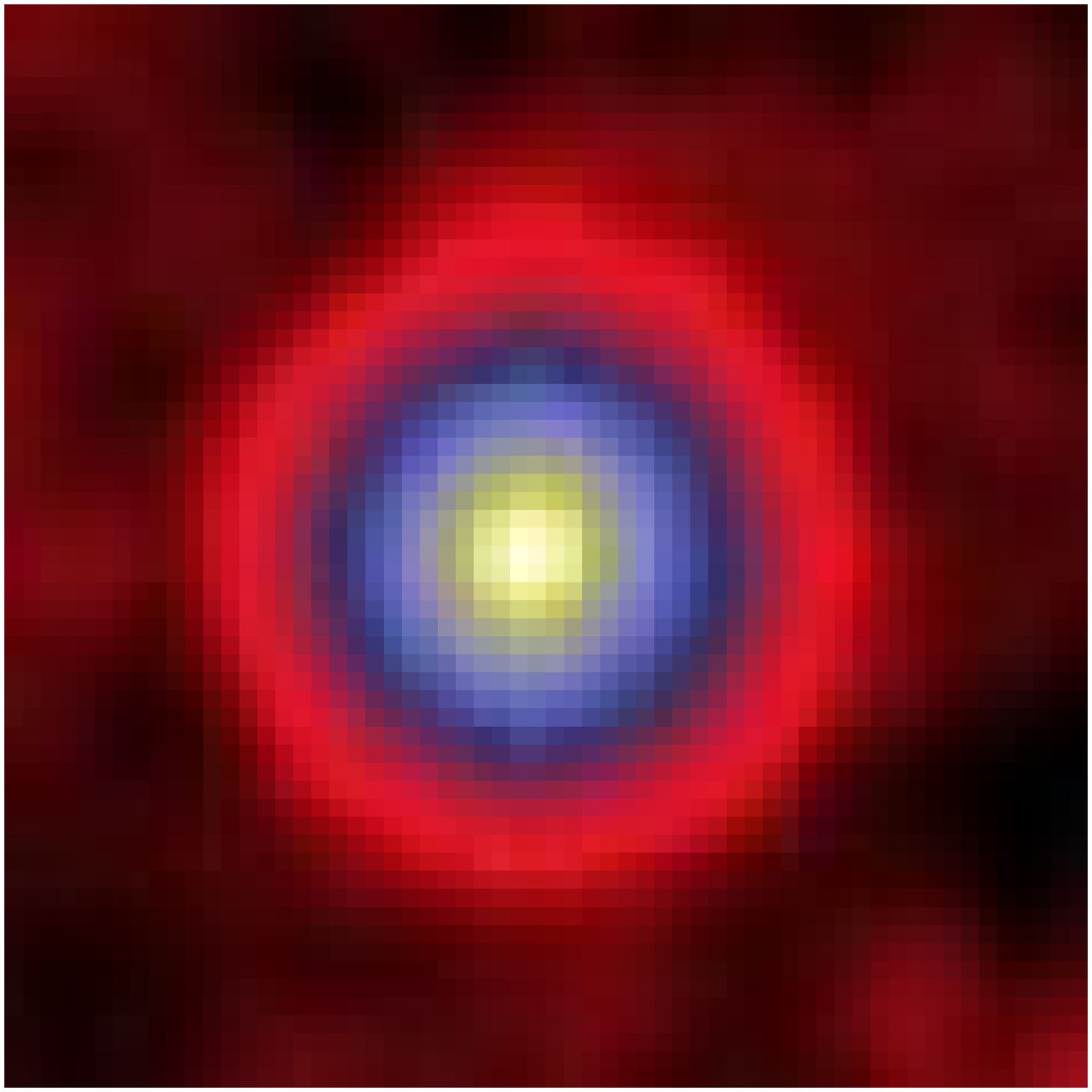}  \FGb{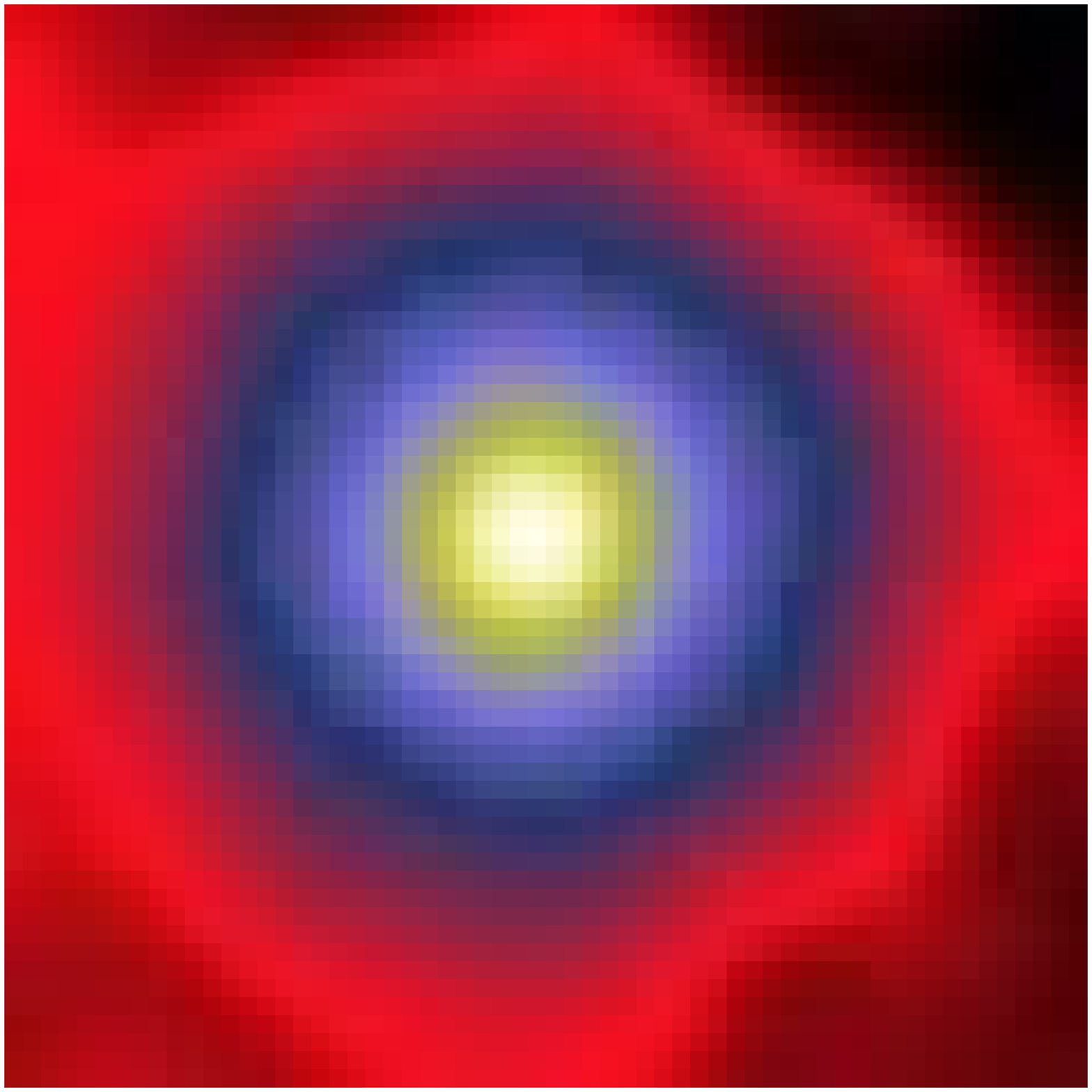} \\  {\#59   NCIA001042}  \\
  \FGb{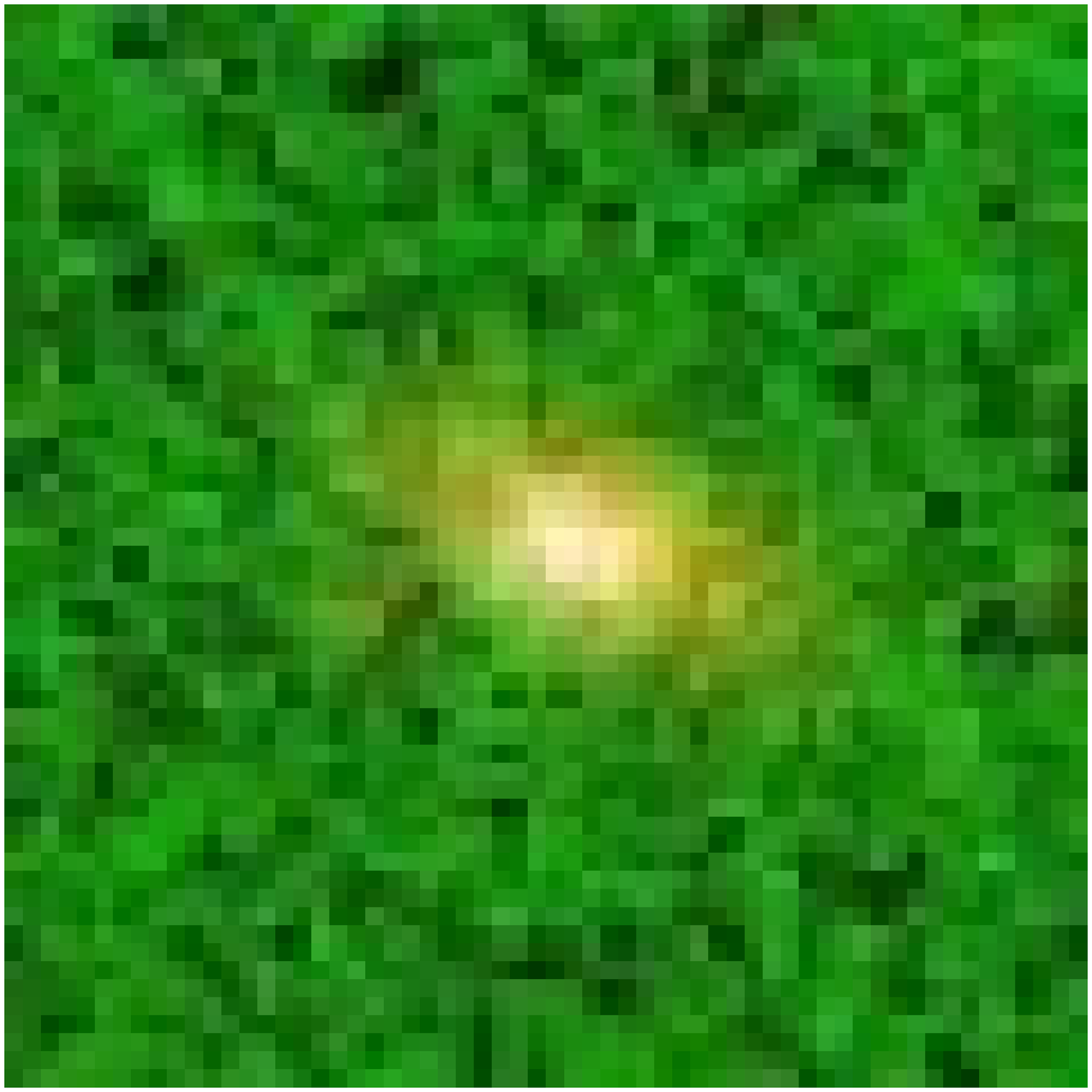}  \FGb{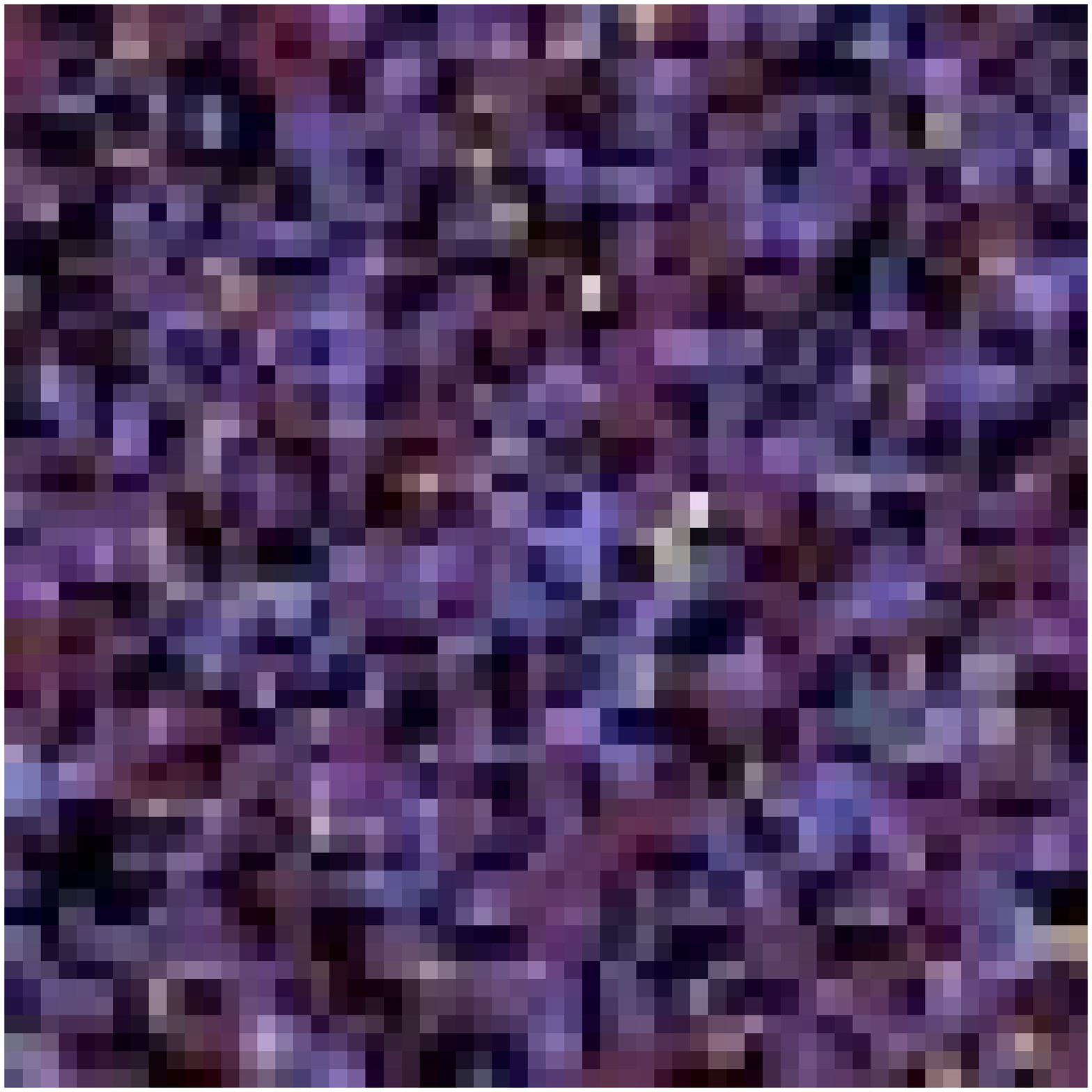}  \FGb{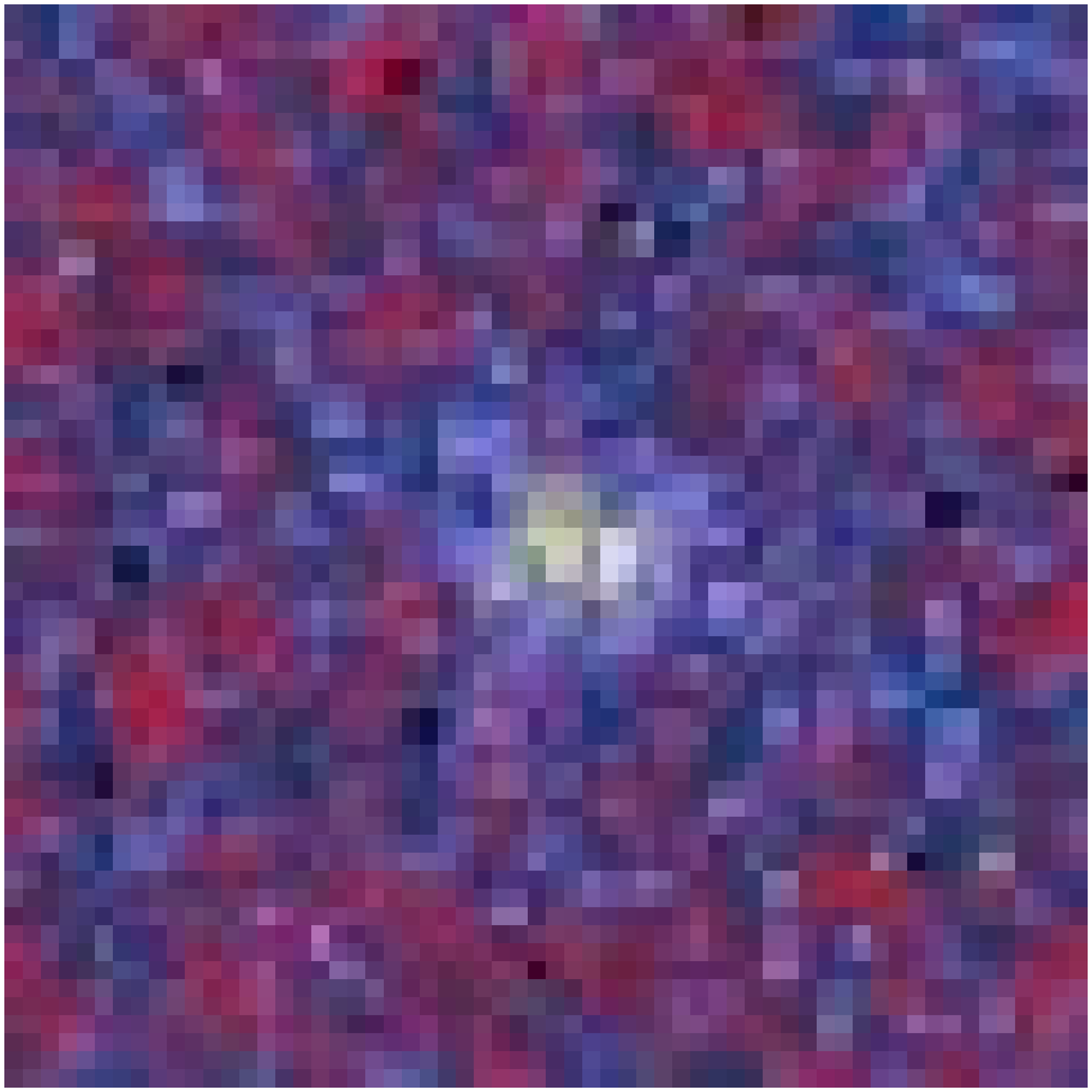}  \FGb{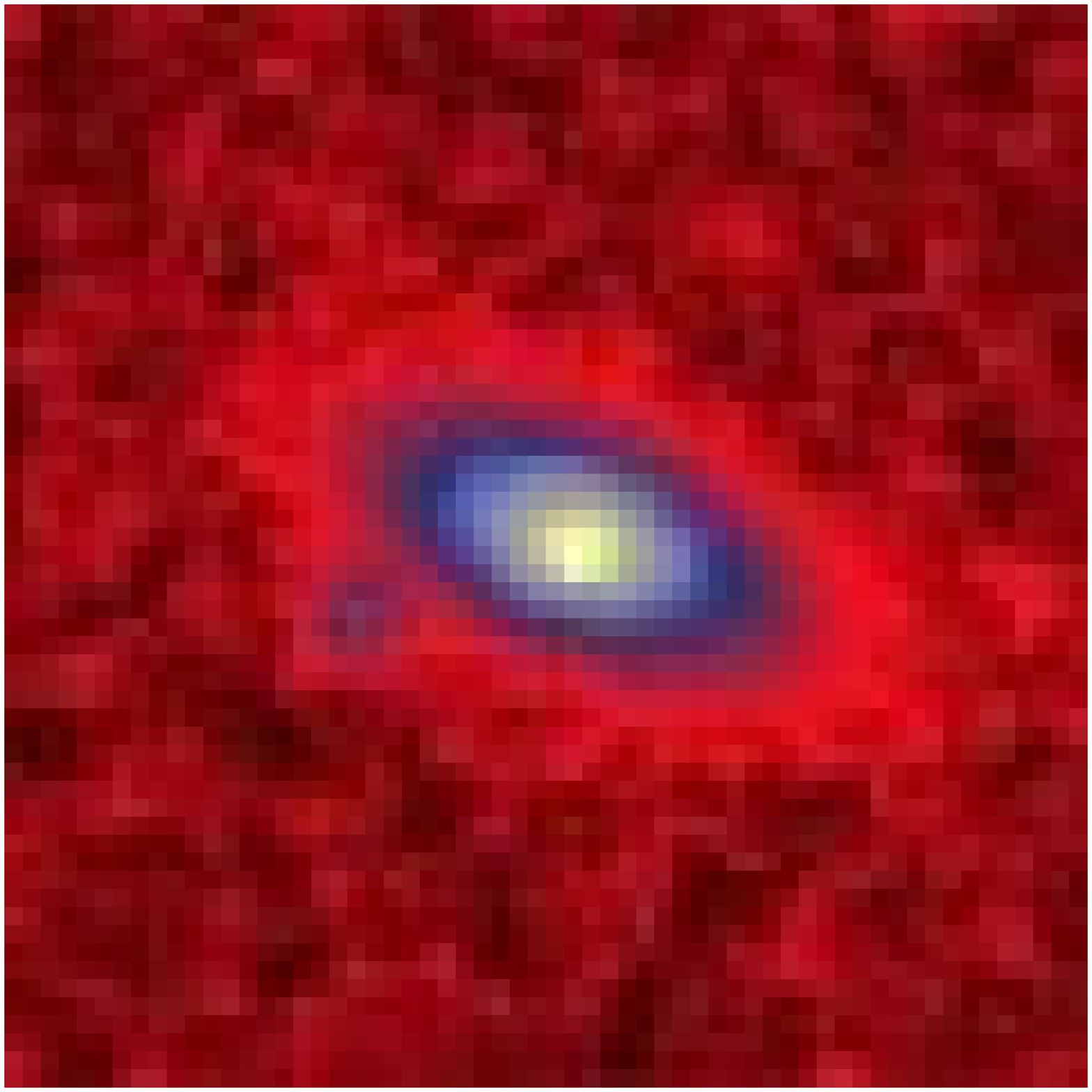}  \FGb{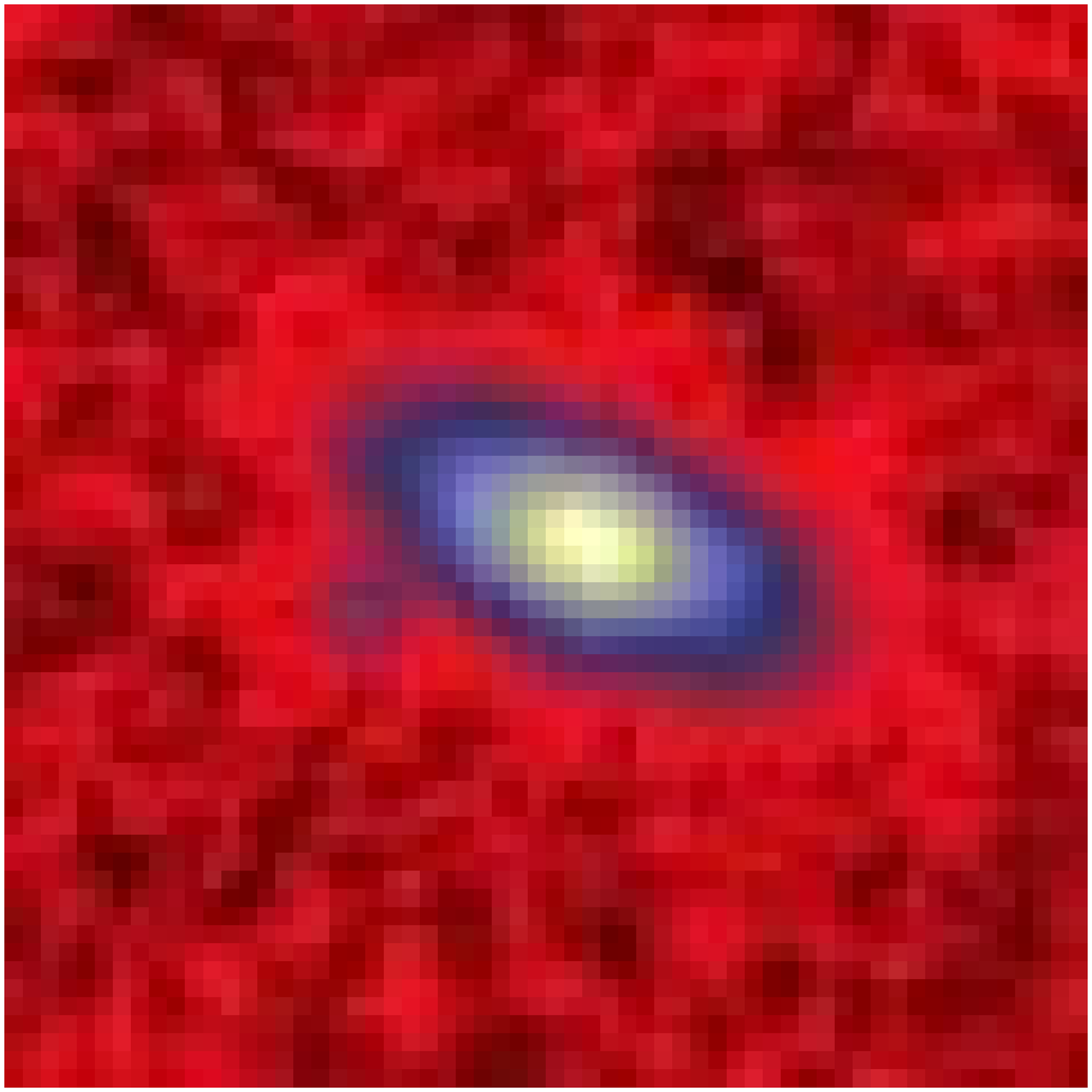}  \FGb{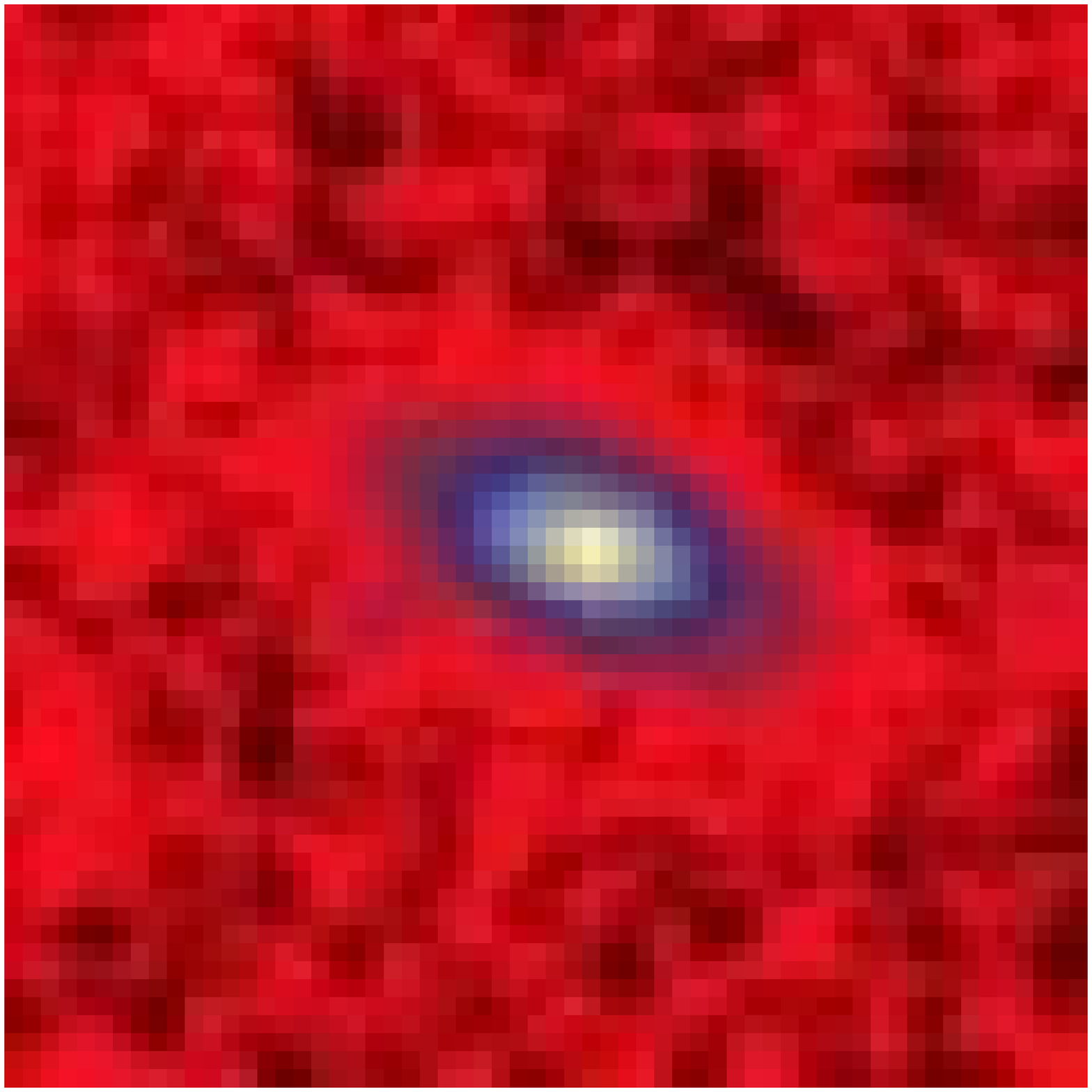}  \FGb{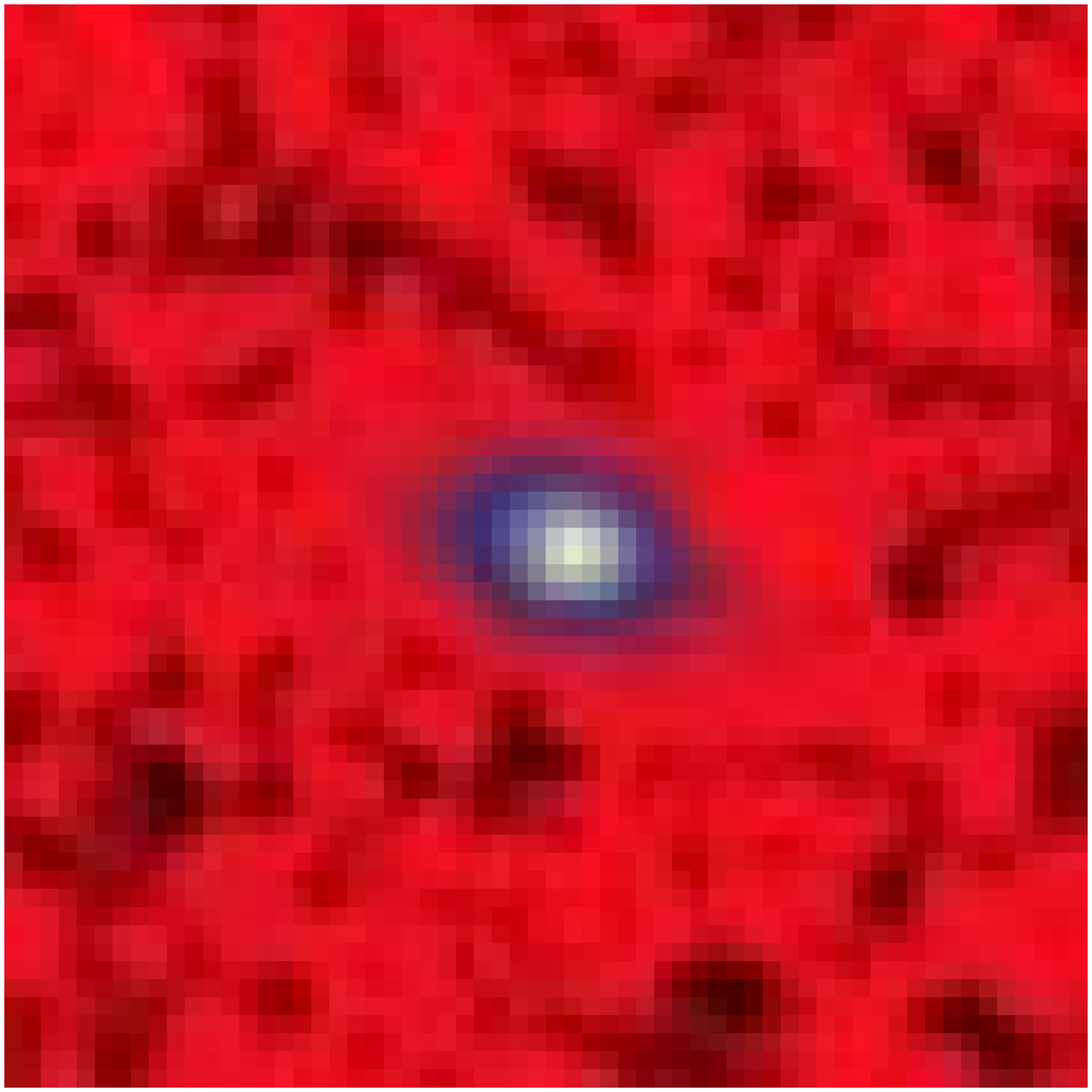}  \FGb{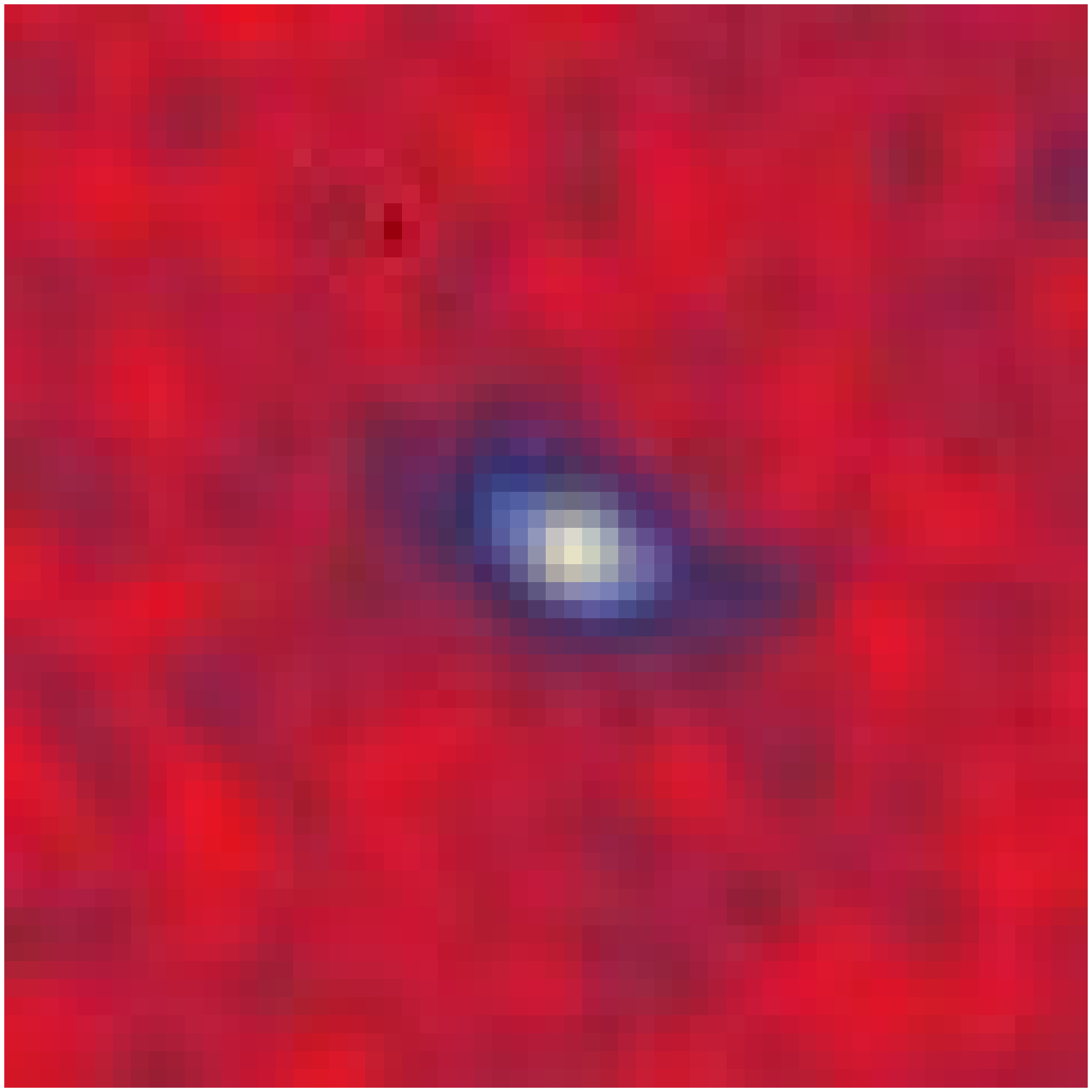}  \FGb{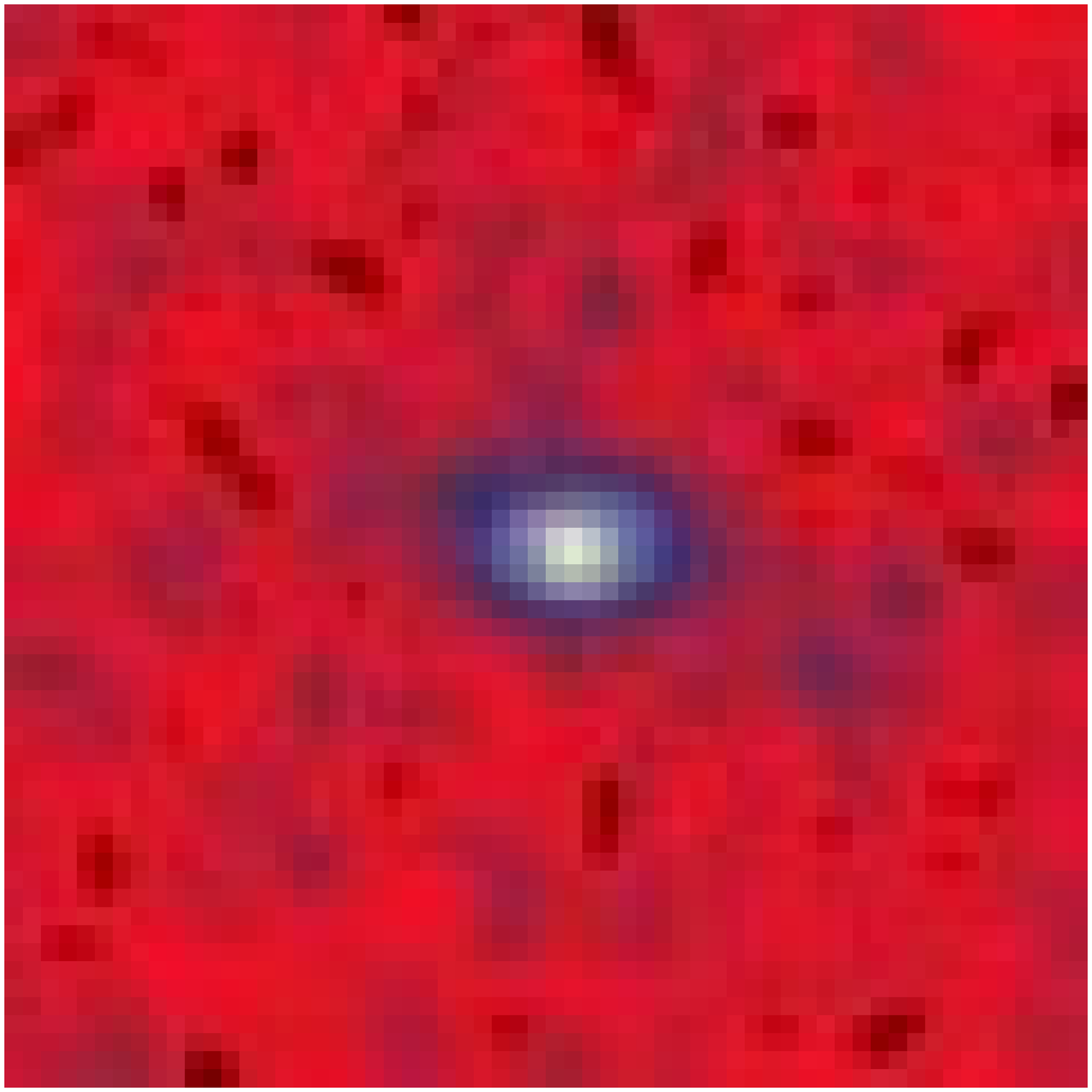}  \FGb{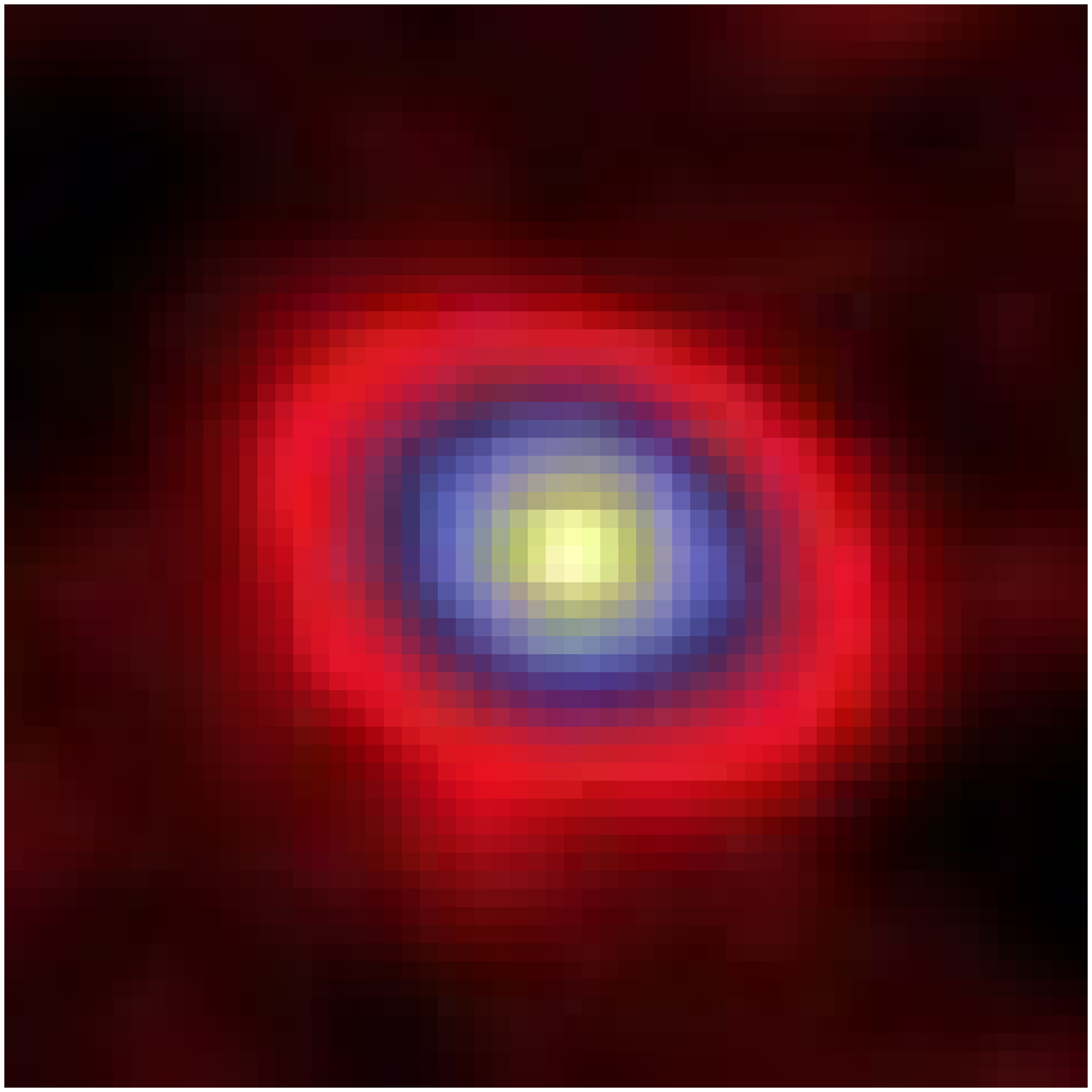}  \FGb{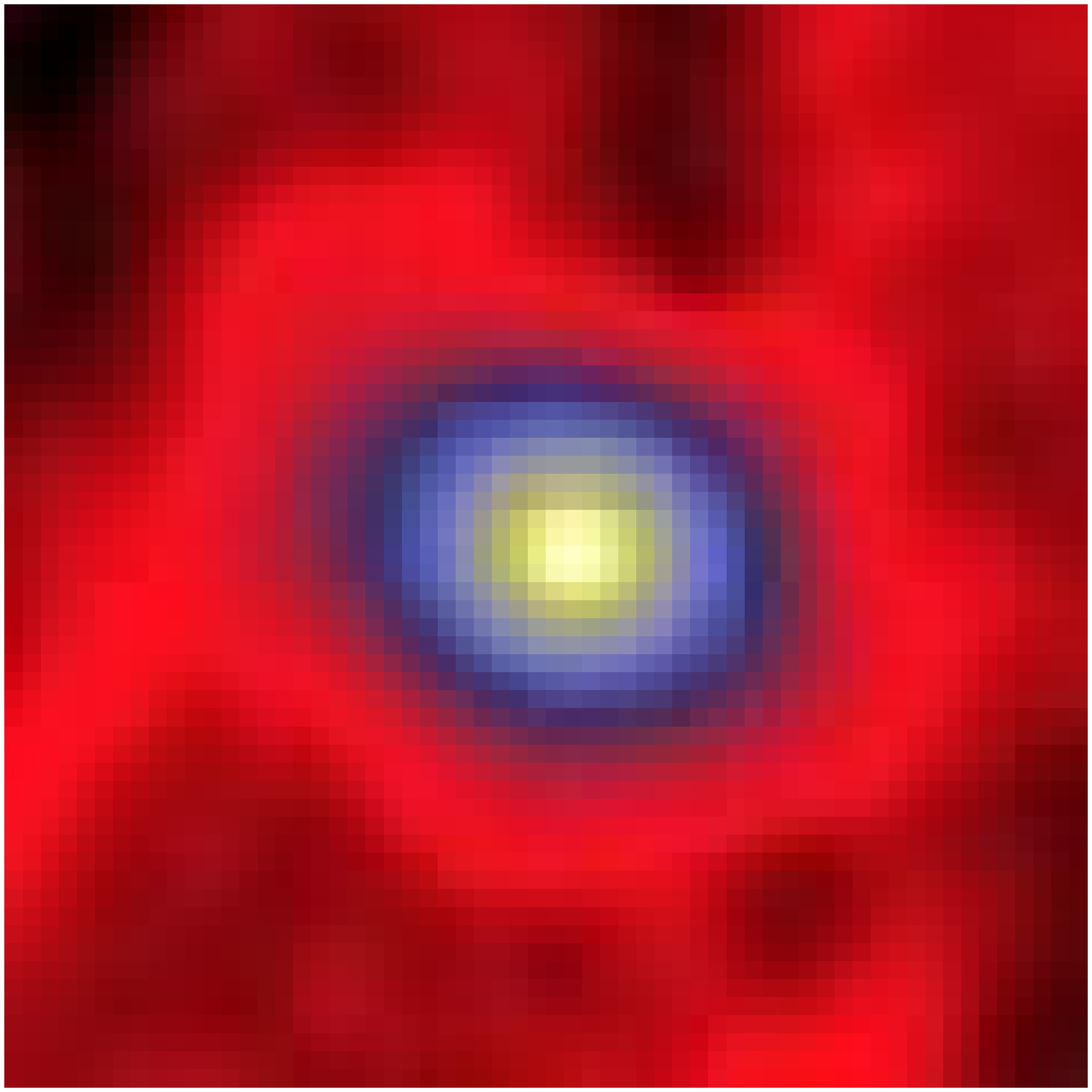}  \FGb{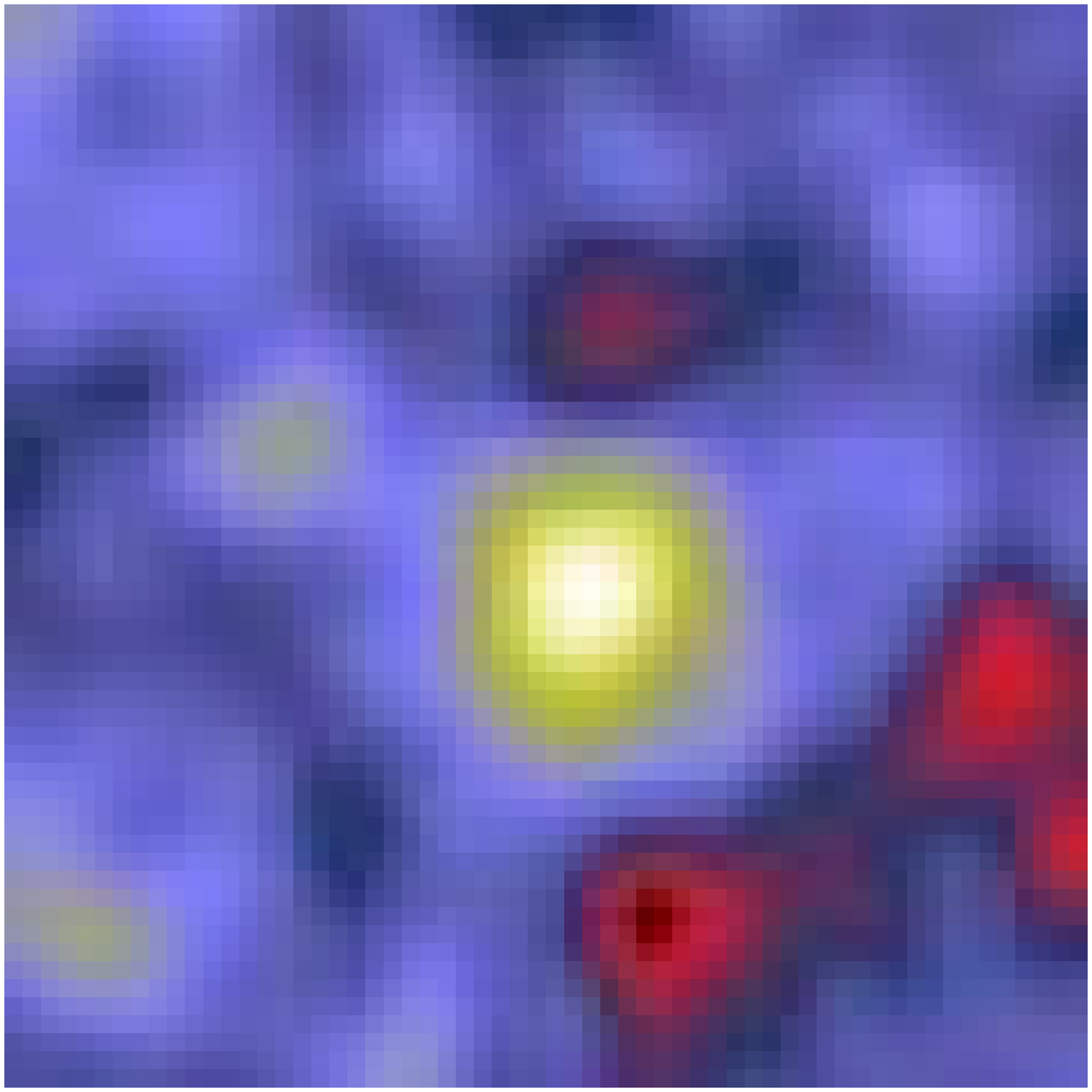}  \FGb{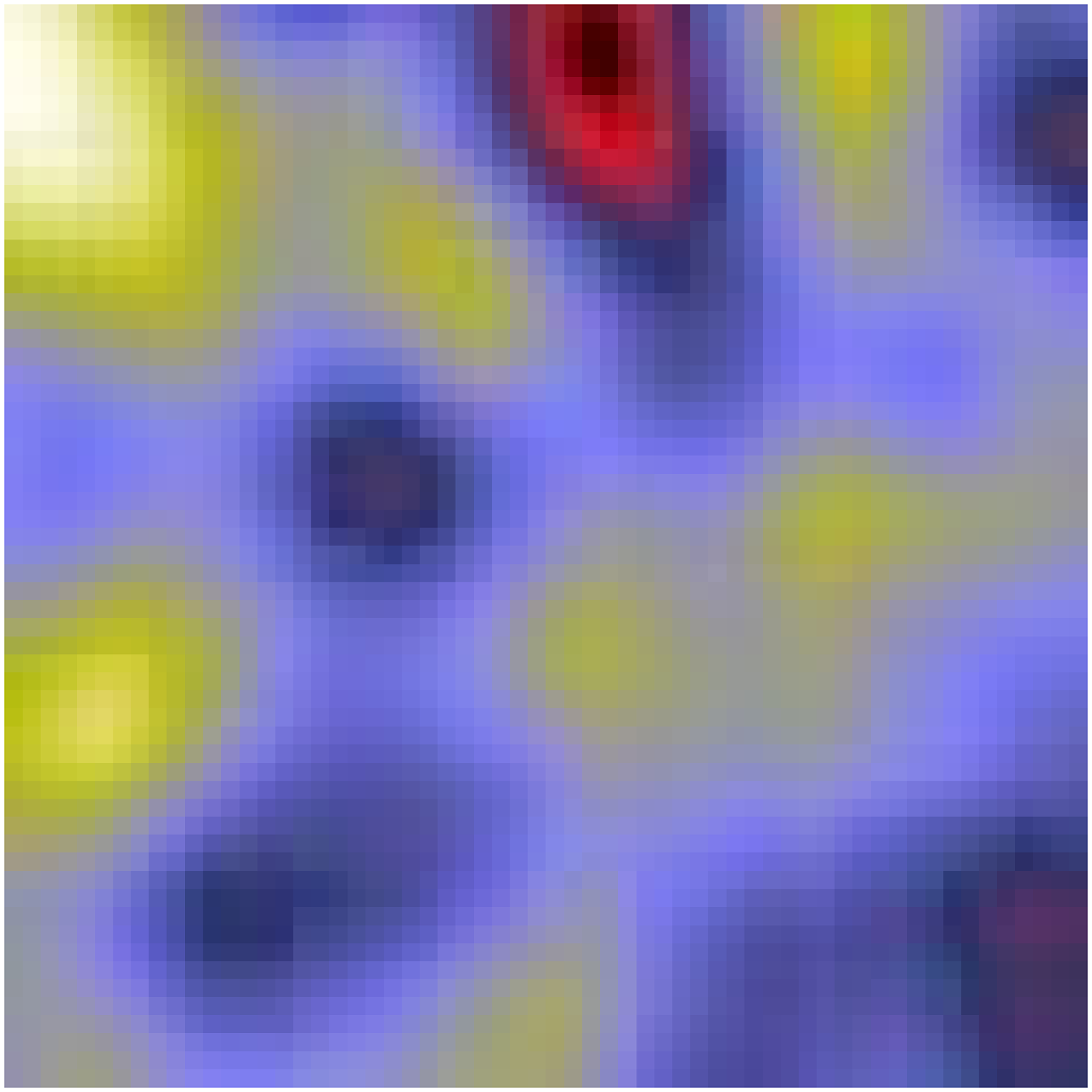} \\  {\#60   NCIA001337}  \\
  \FGb{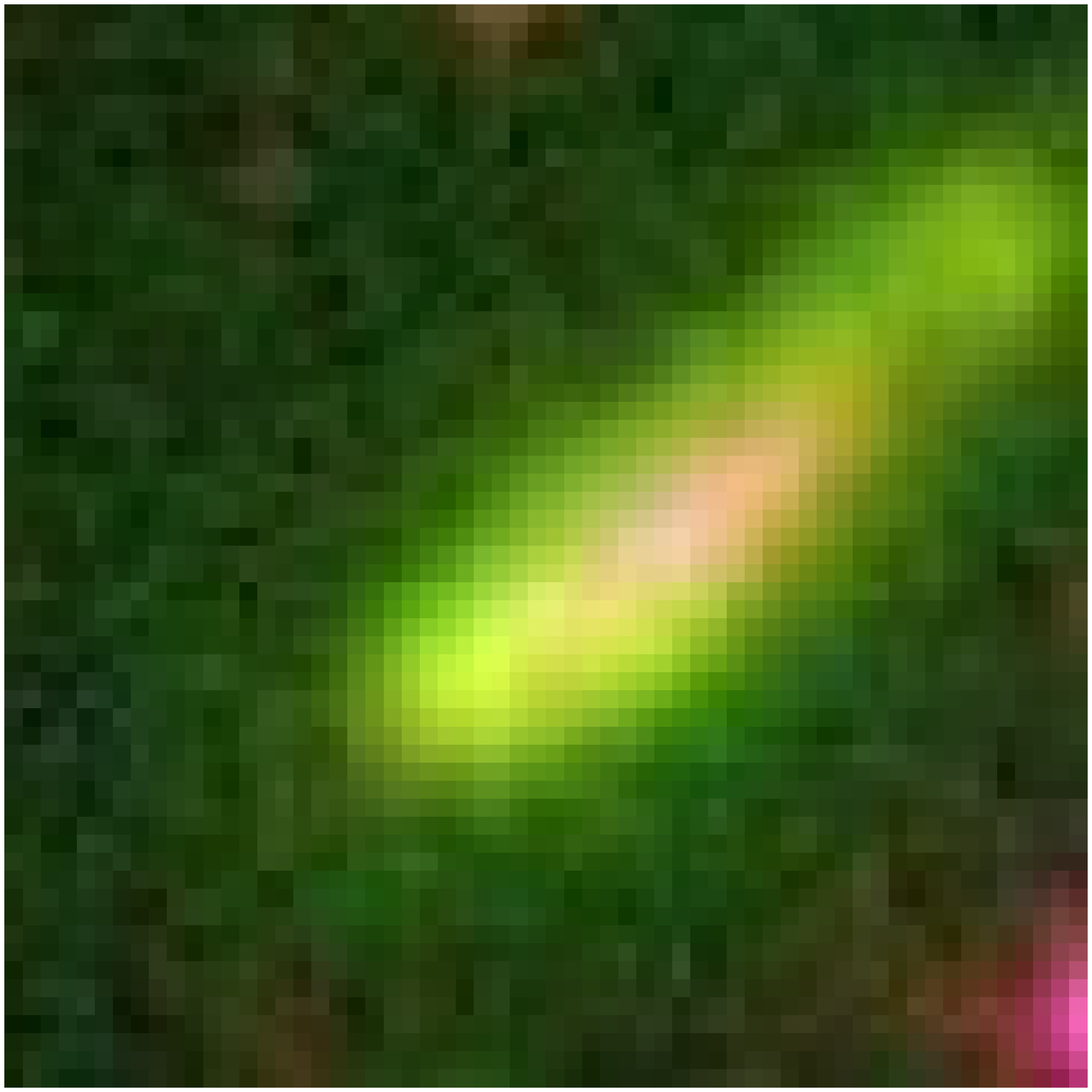}  \FGb{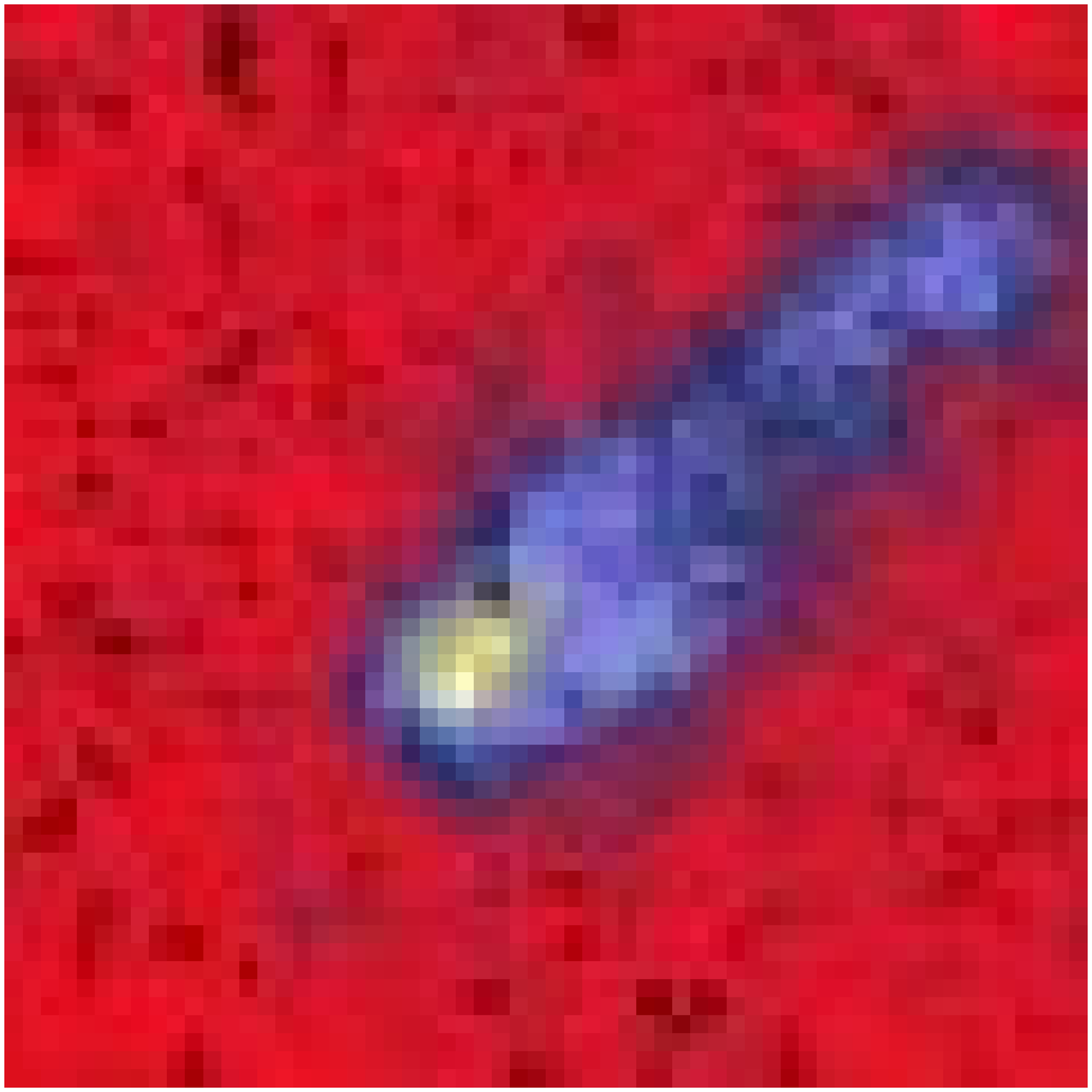}  \FGb{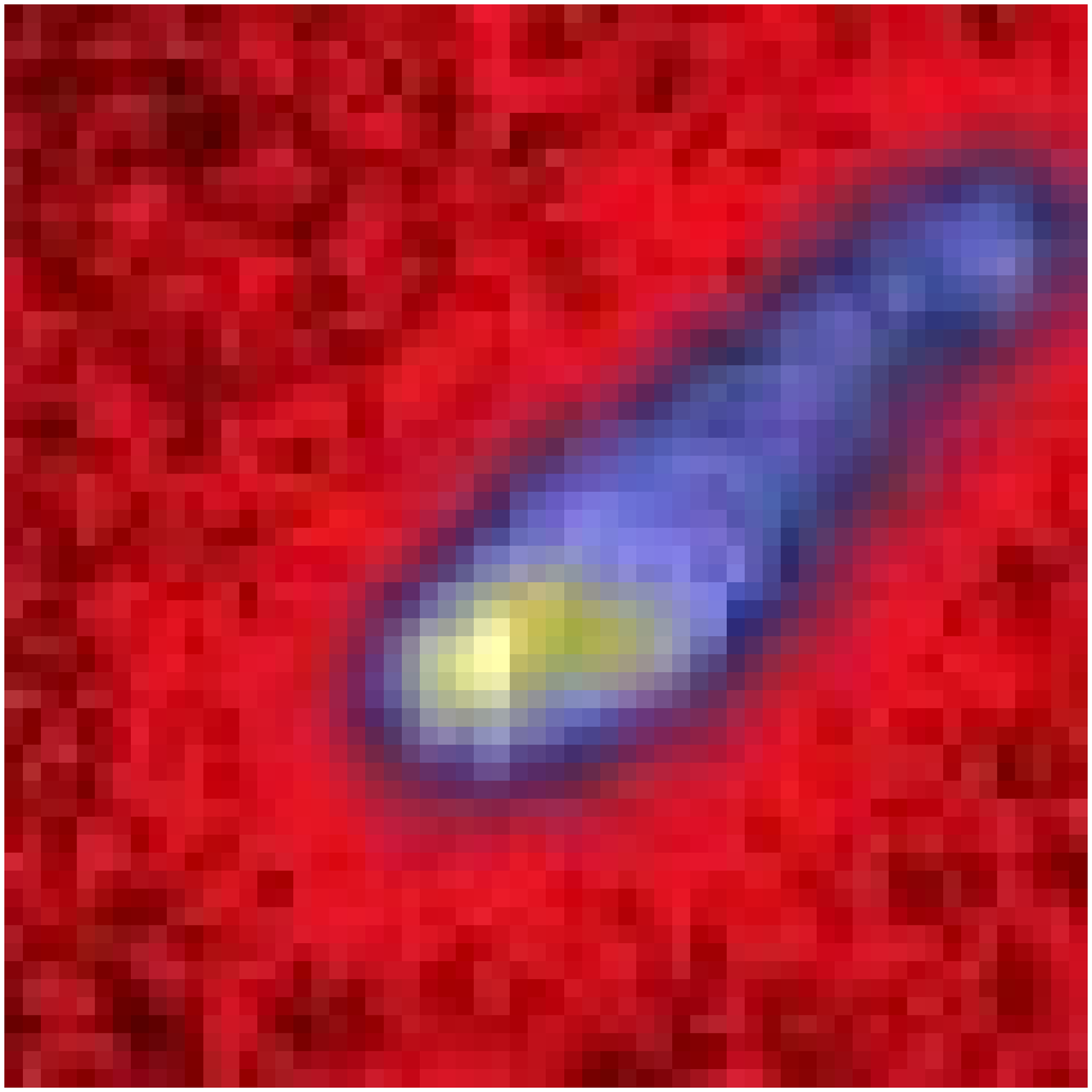}  \FGb{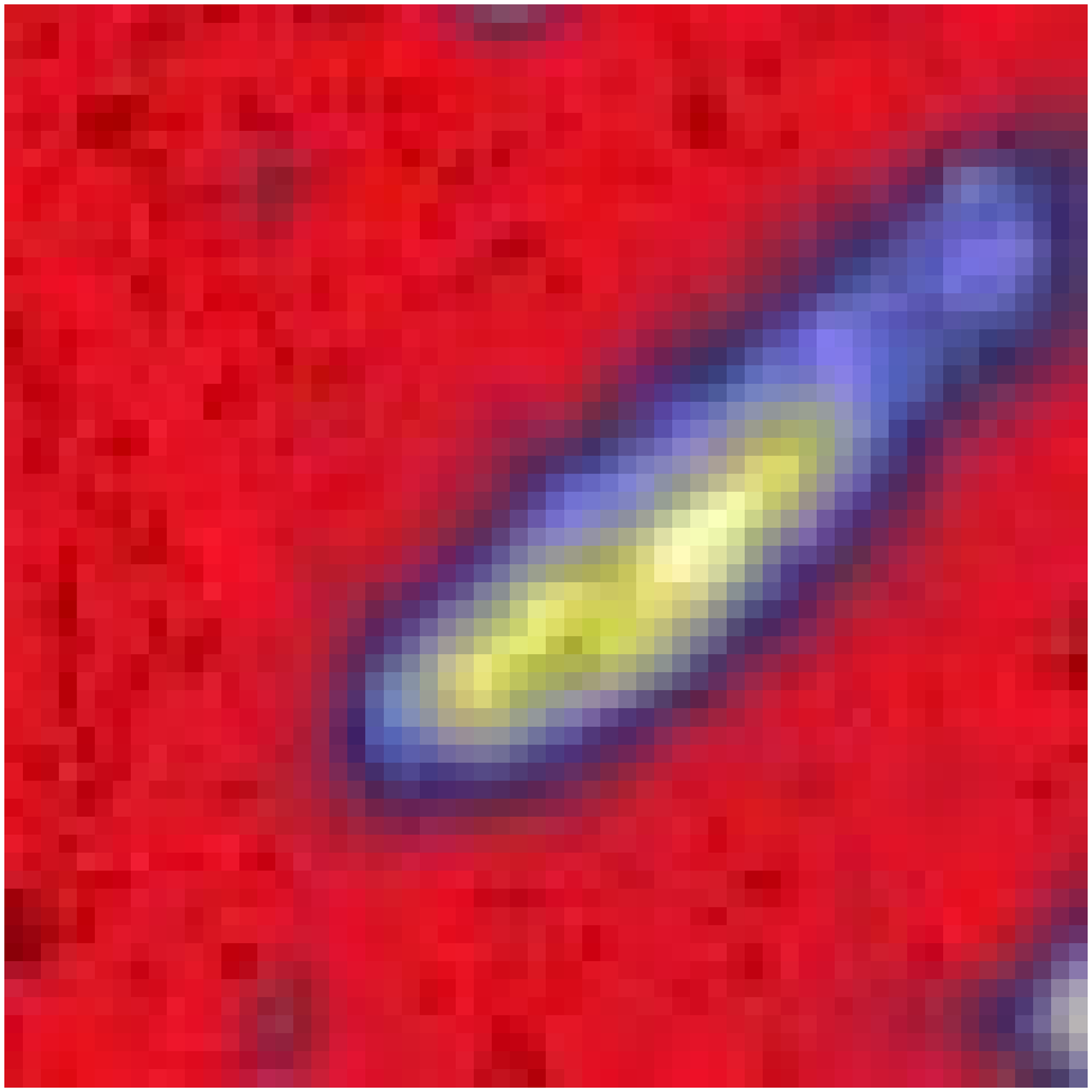}  \FGb{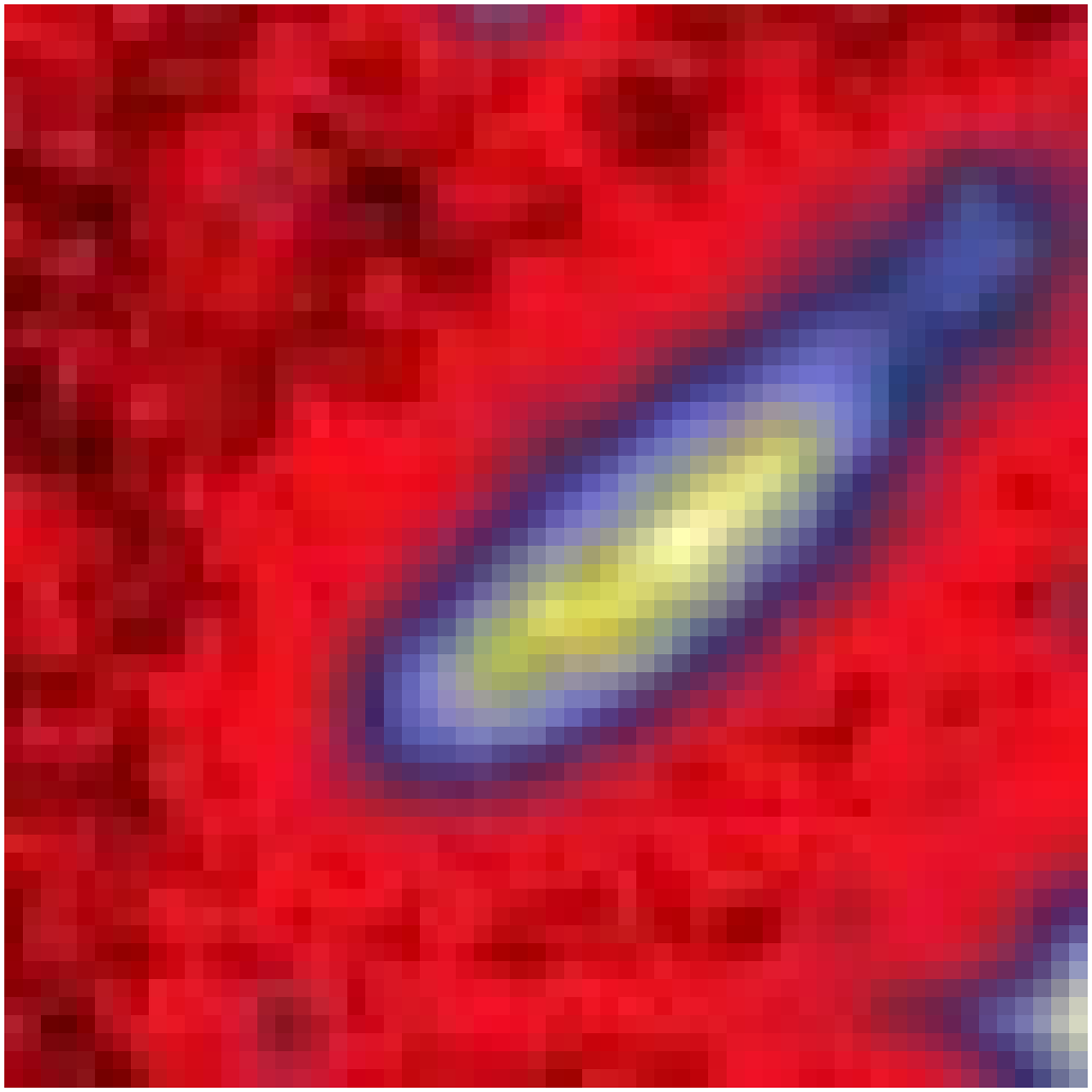}  \FGb{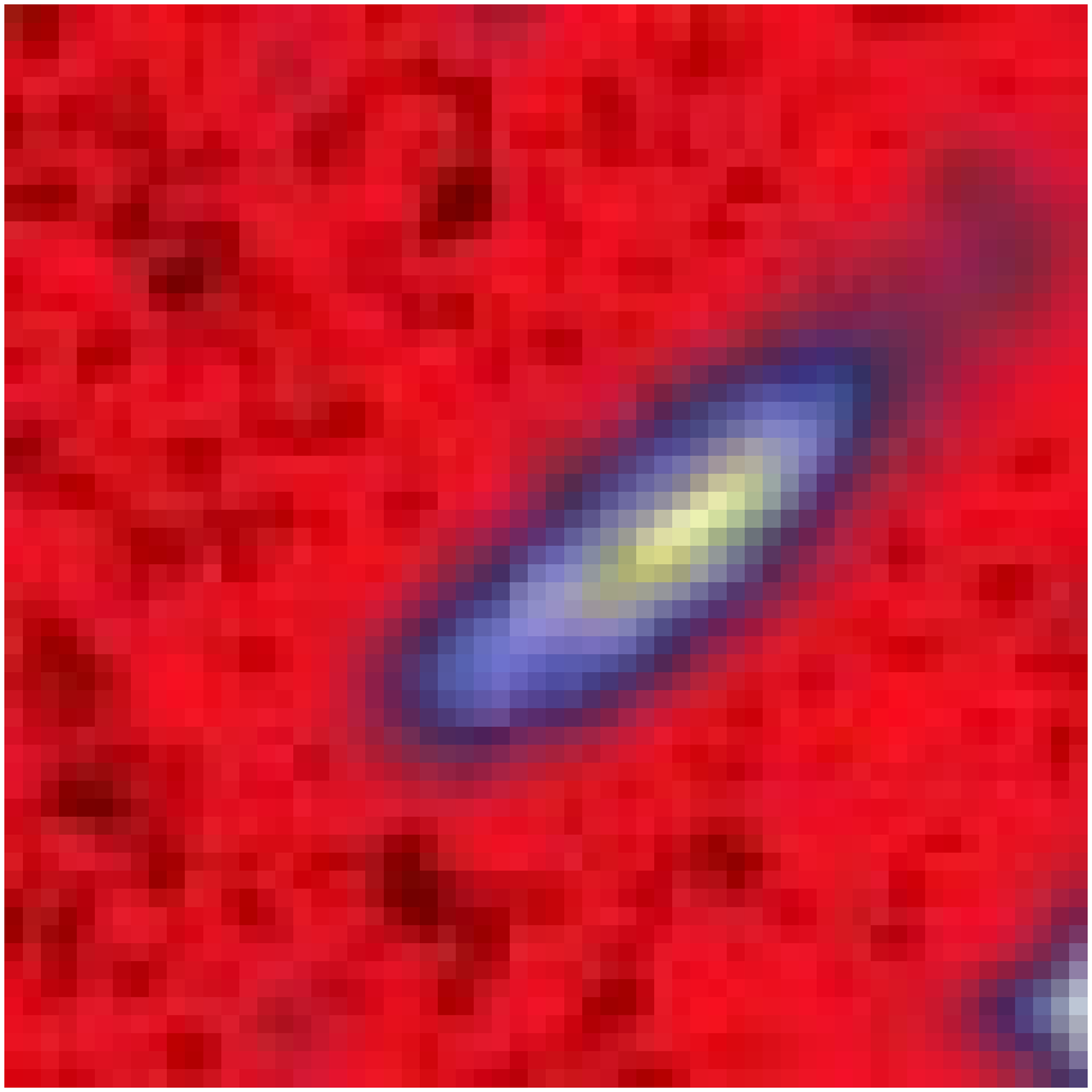}  \FGb{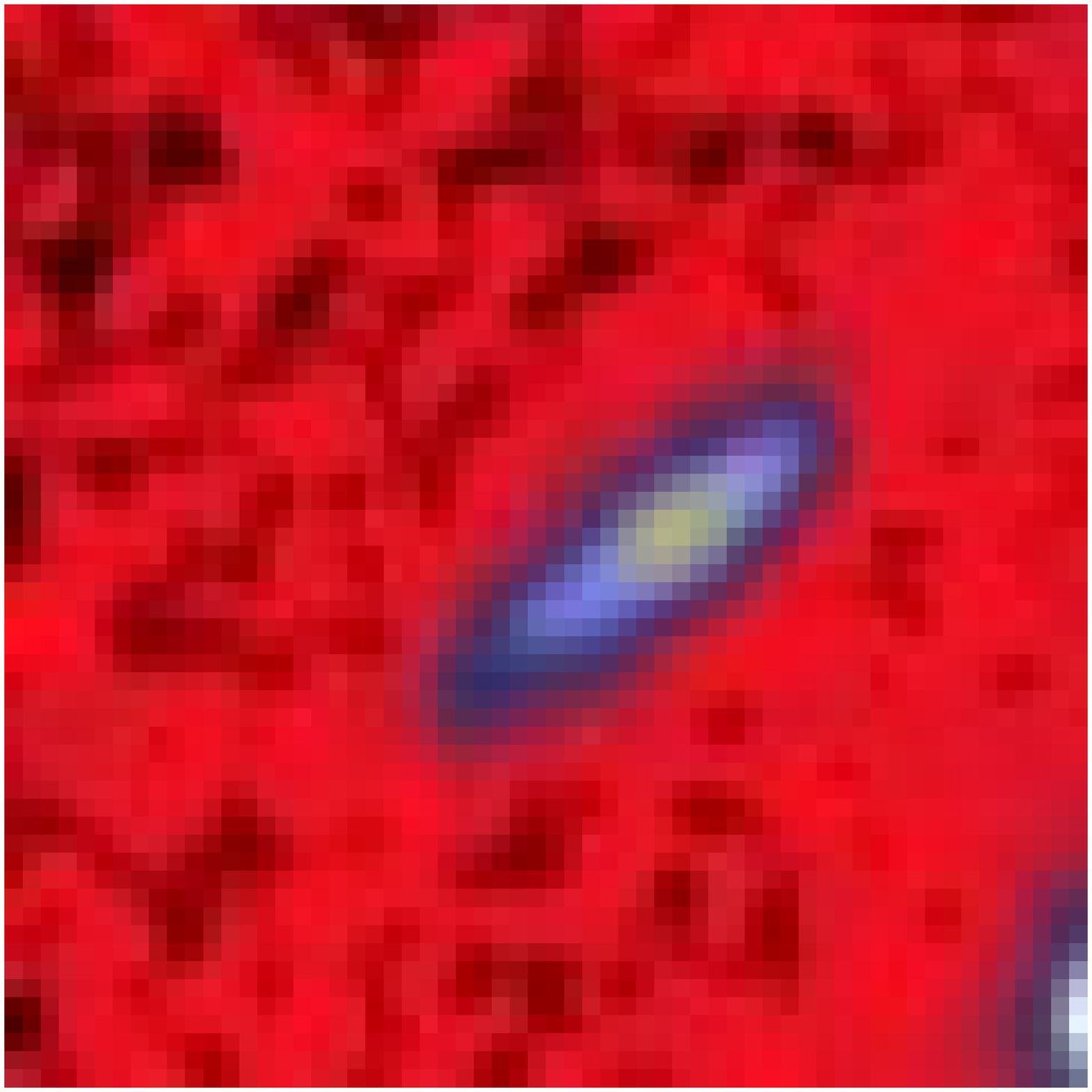}  \FGb{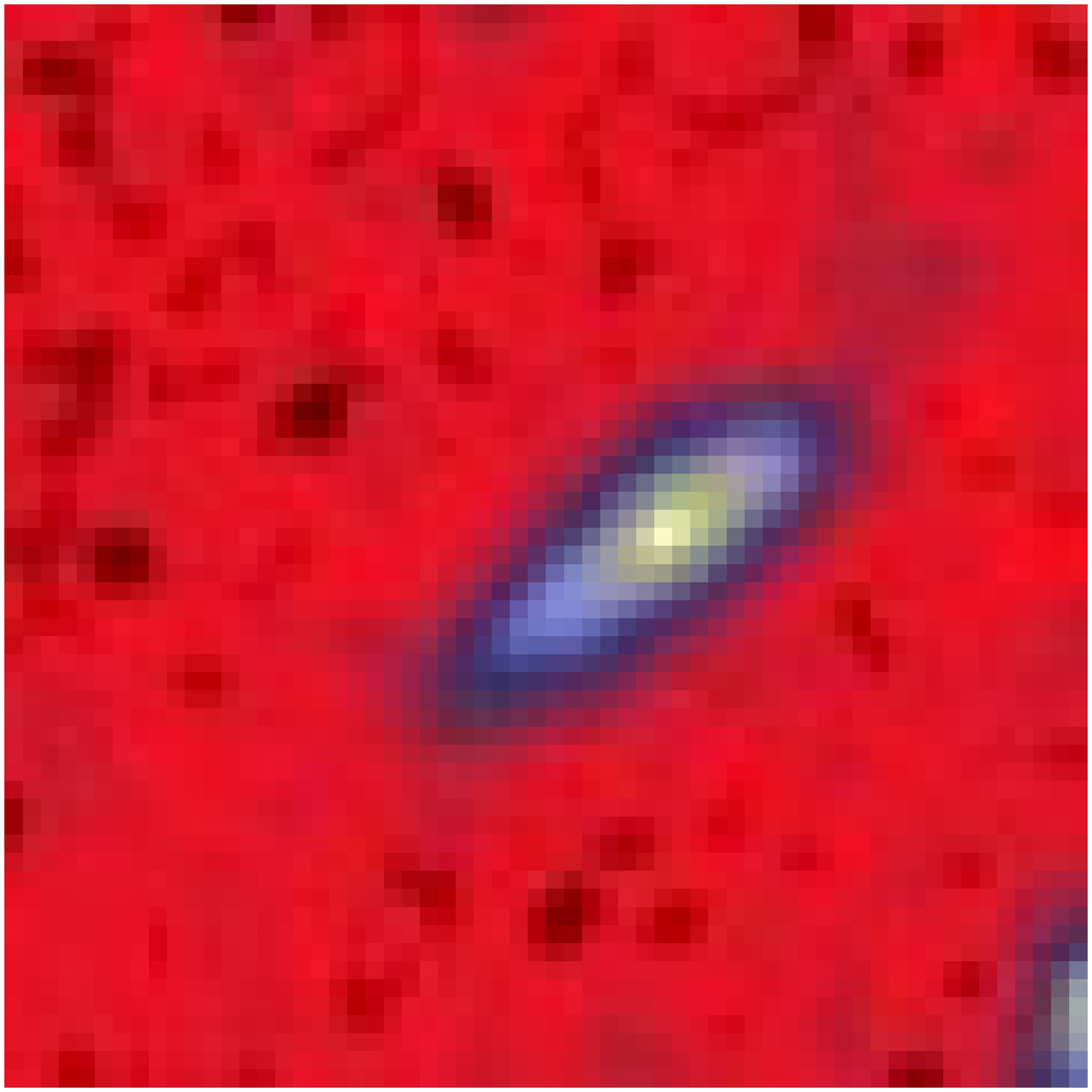}  \FGb{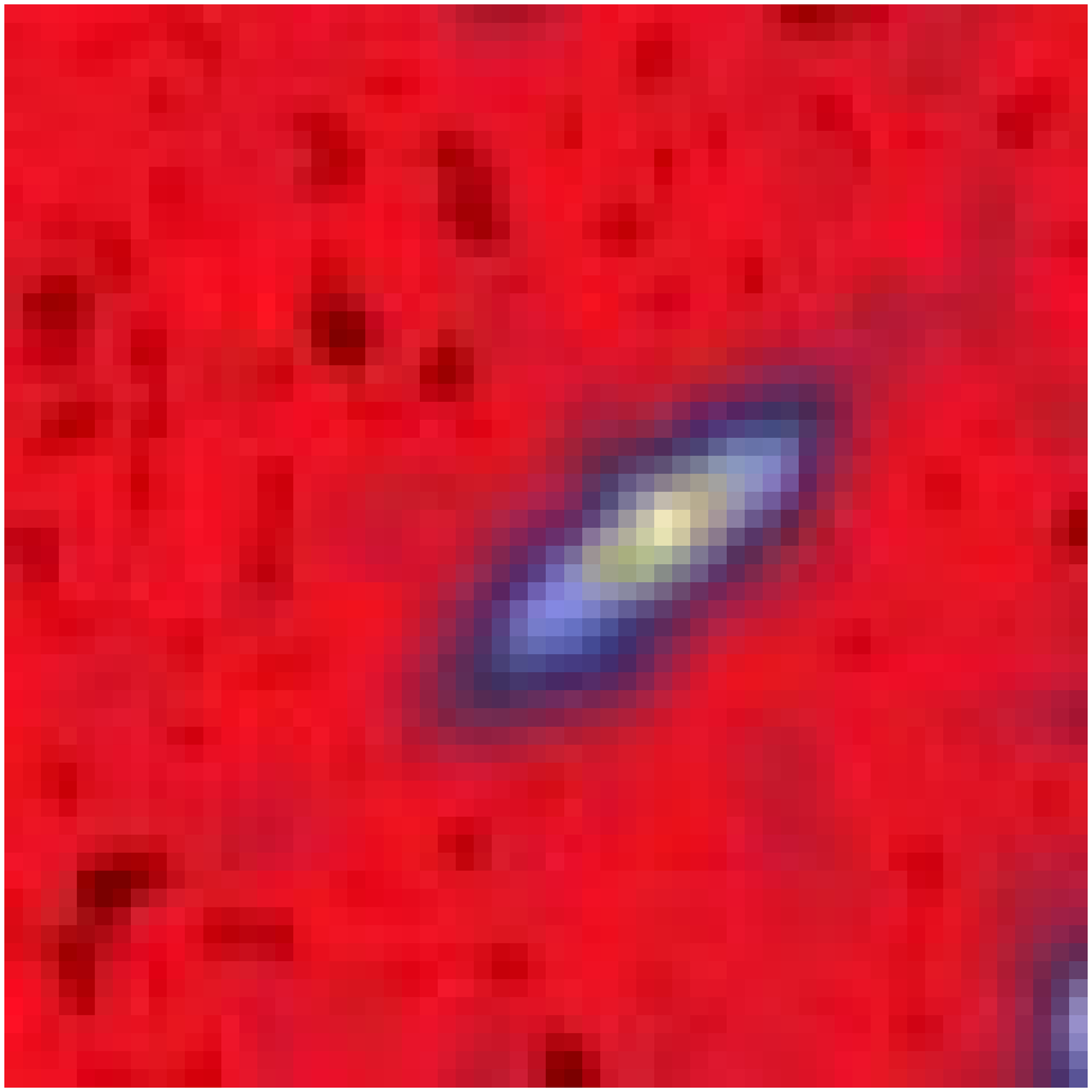}  \FGb{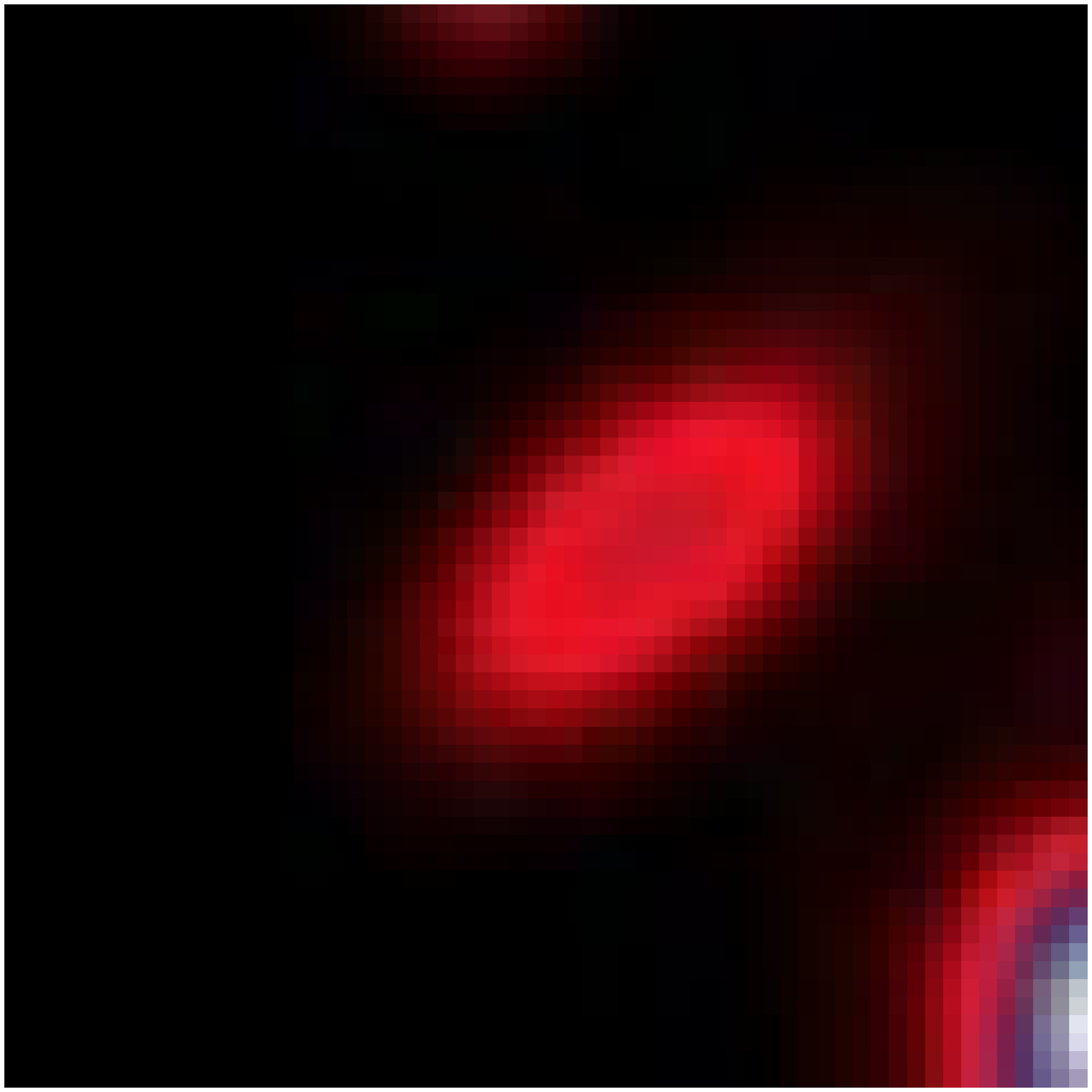}  \FGb{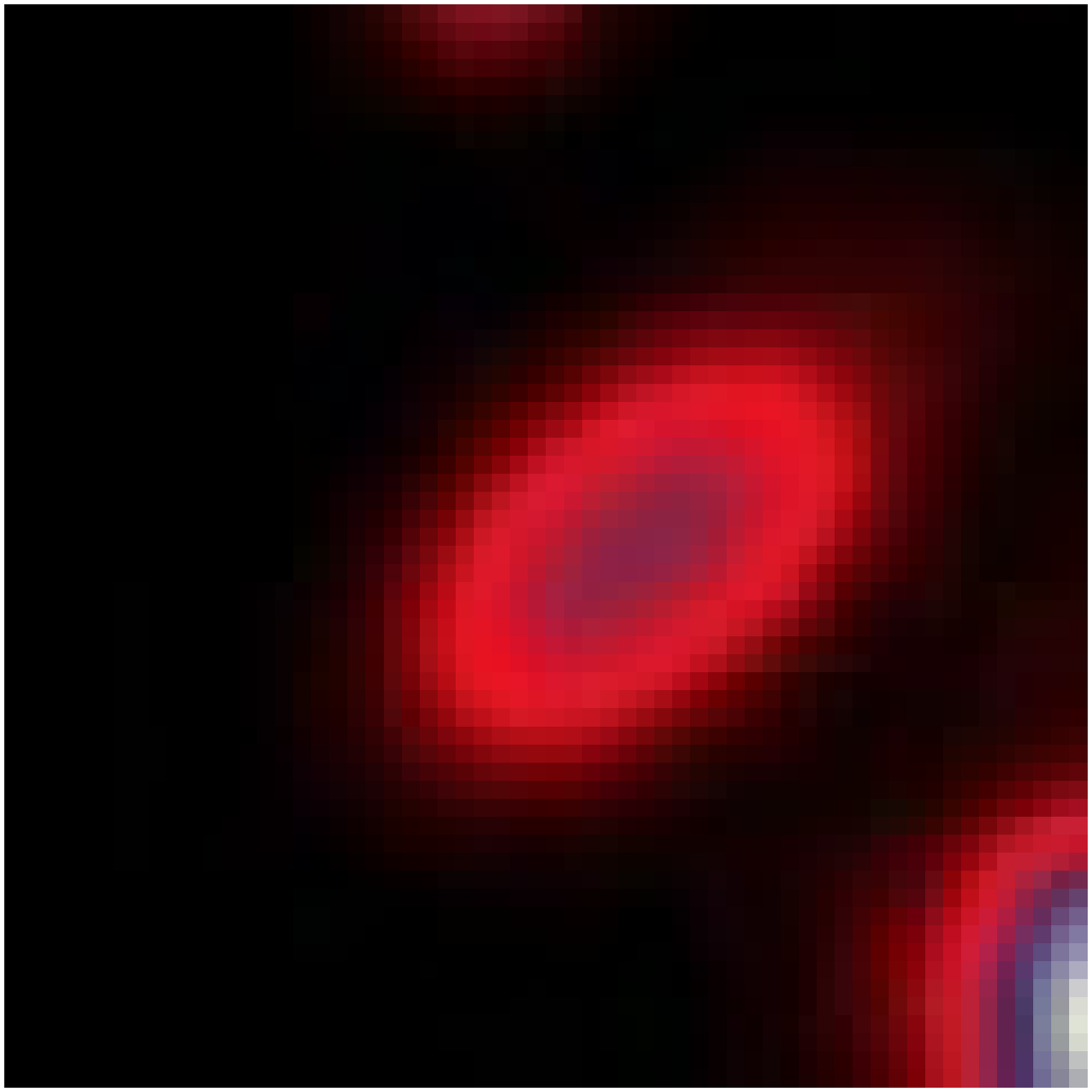}  \FGb{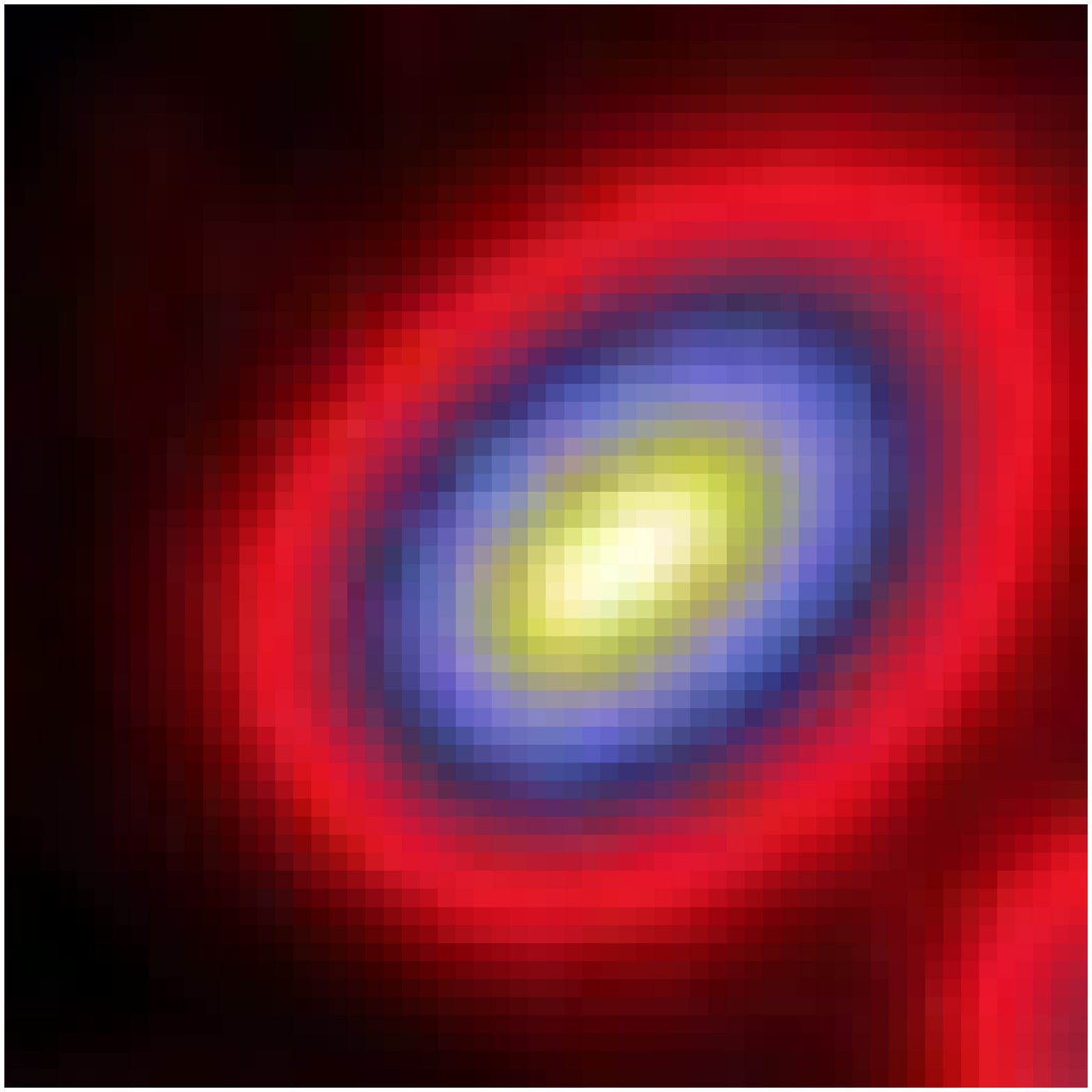}  \FGb{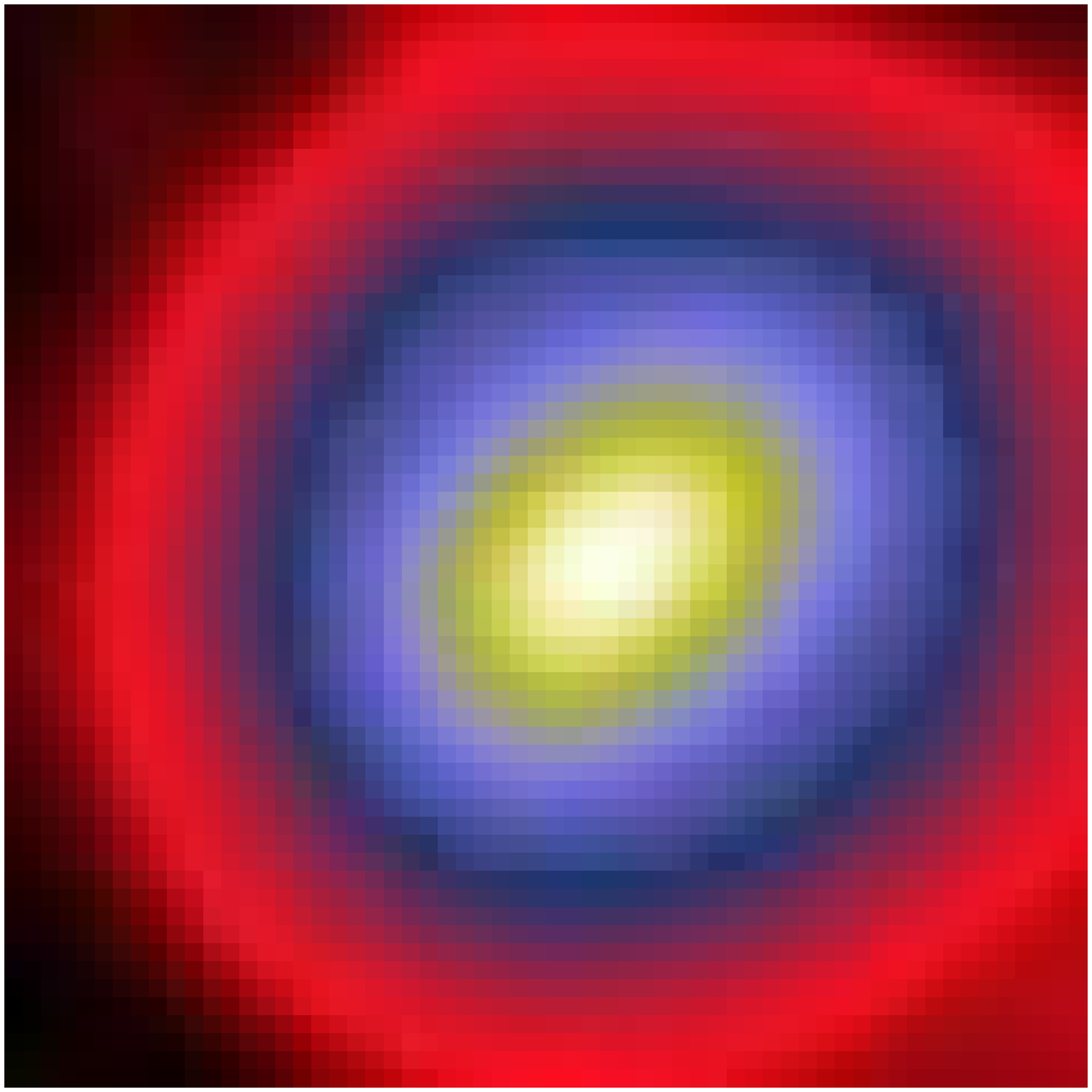} \\  {\#61   NBZ5000678}  \\
  \FGb{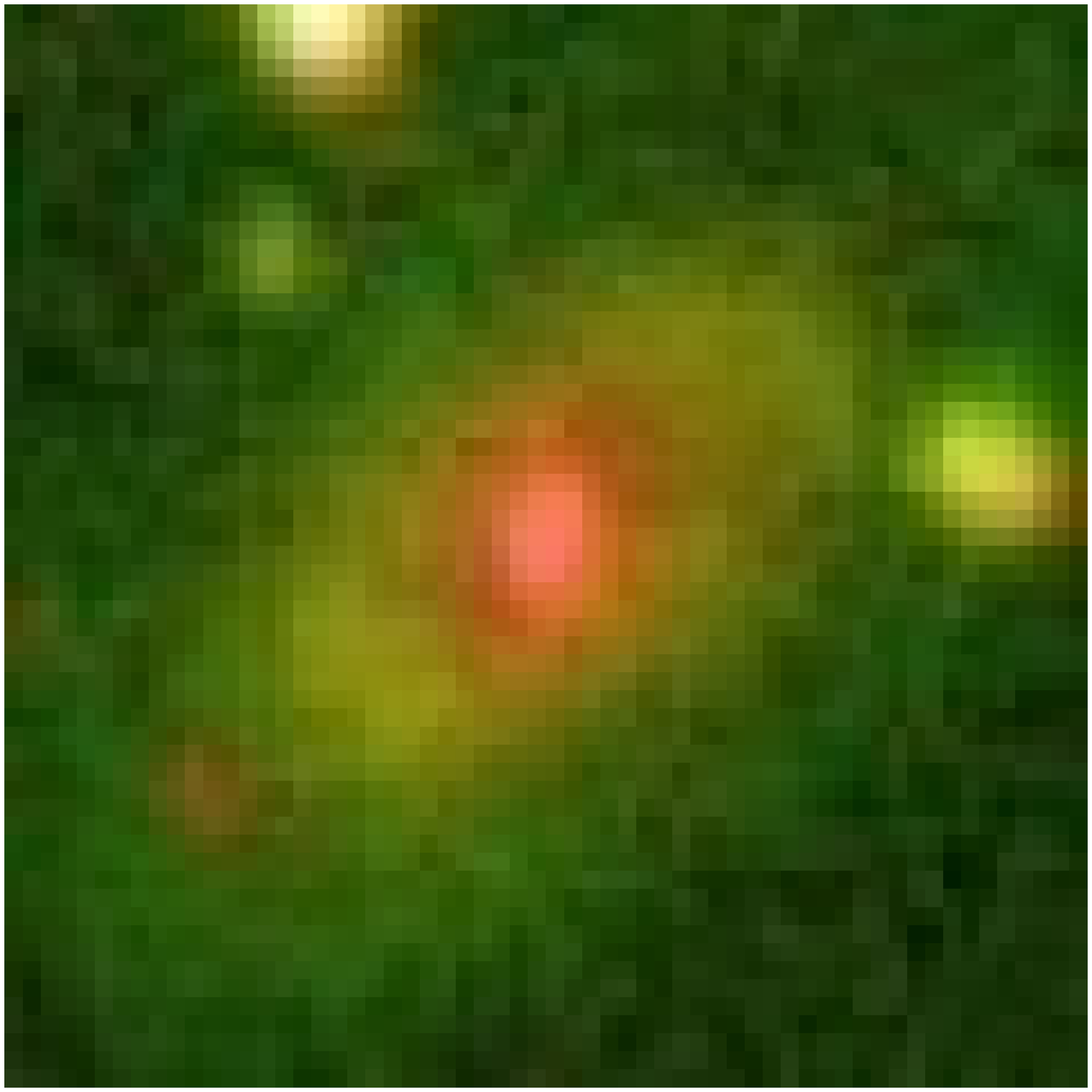}  \FGb{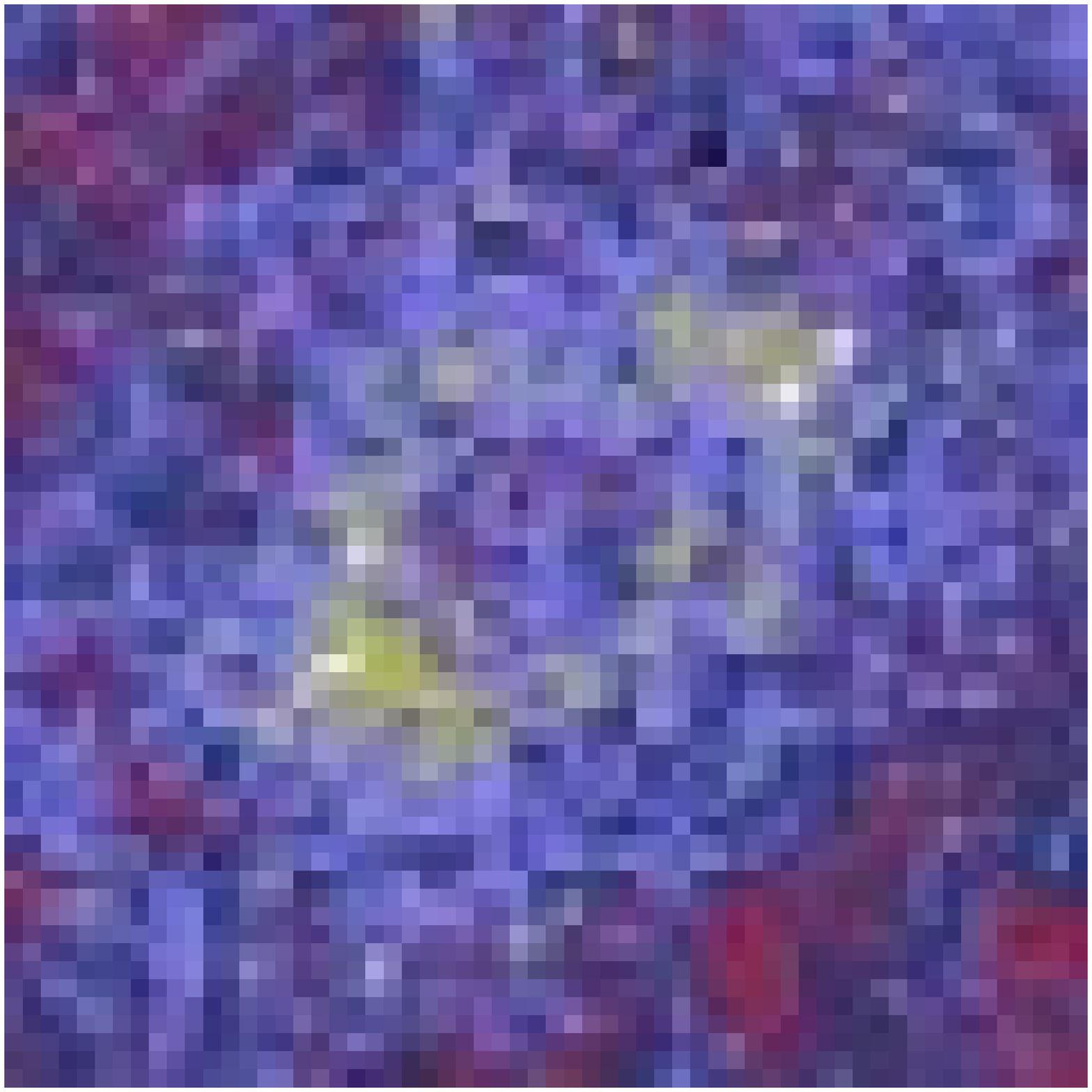}  \FGb{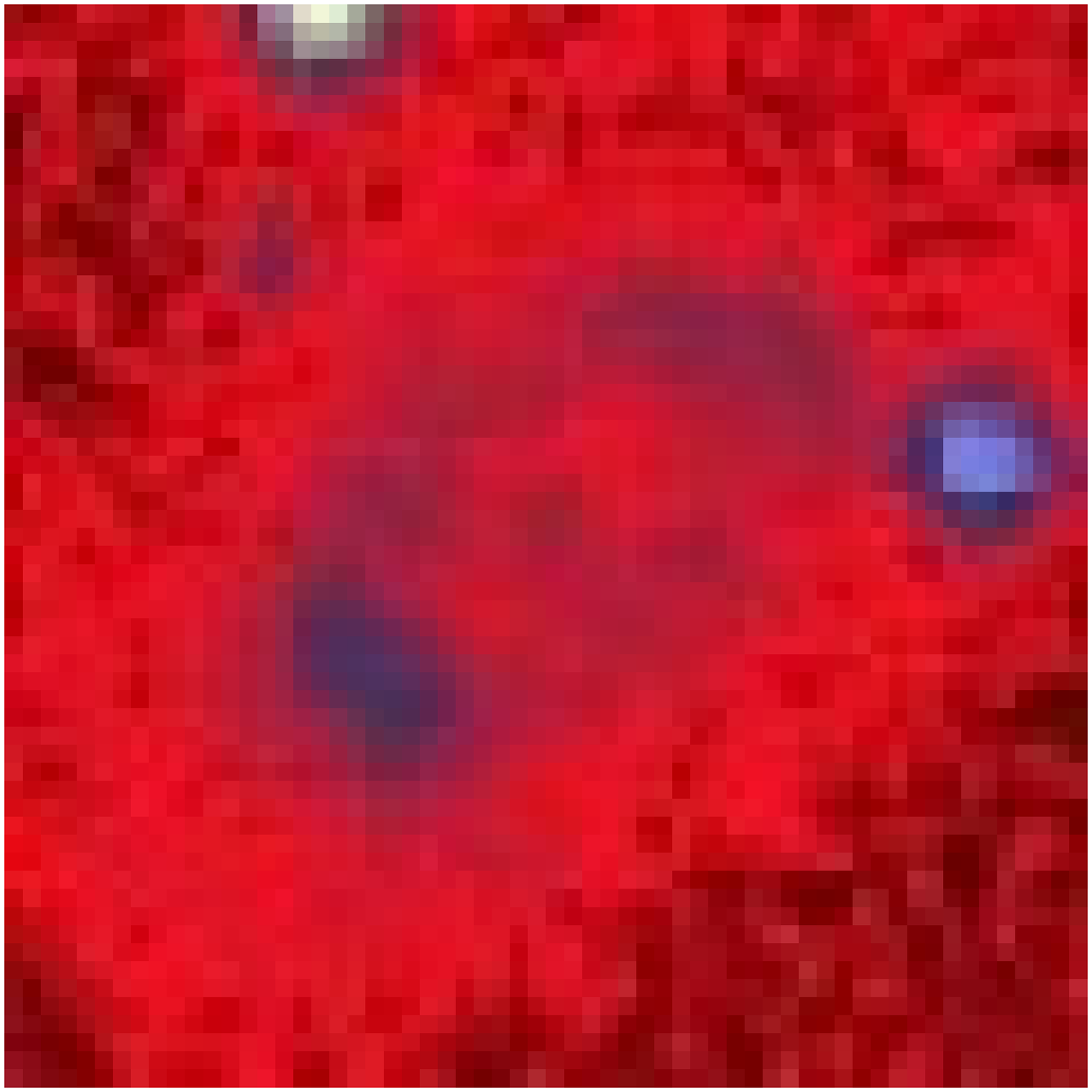}  \FGb{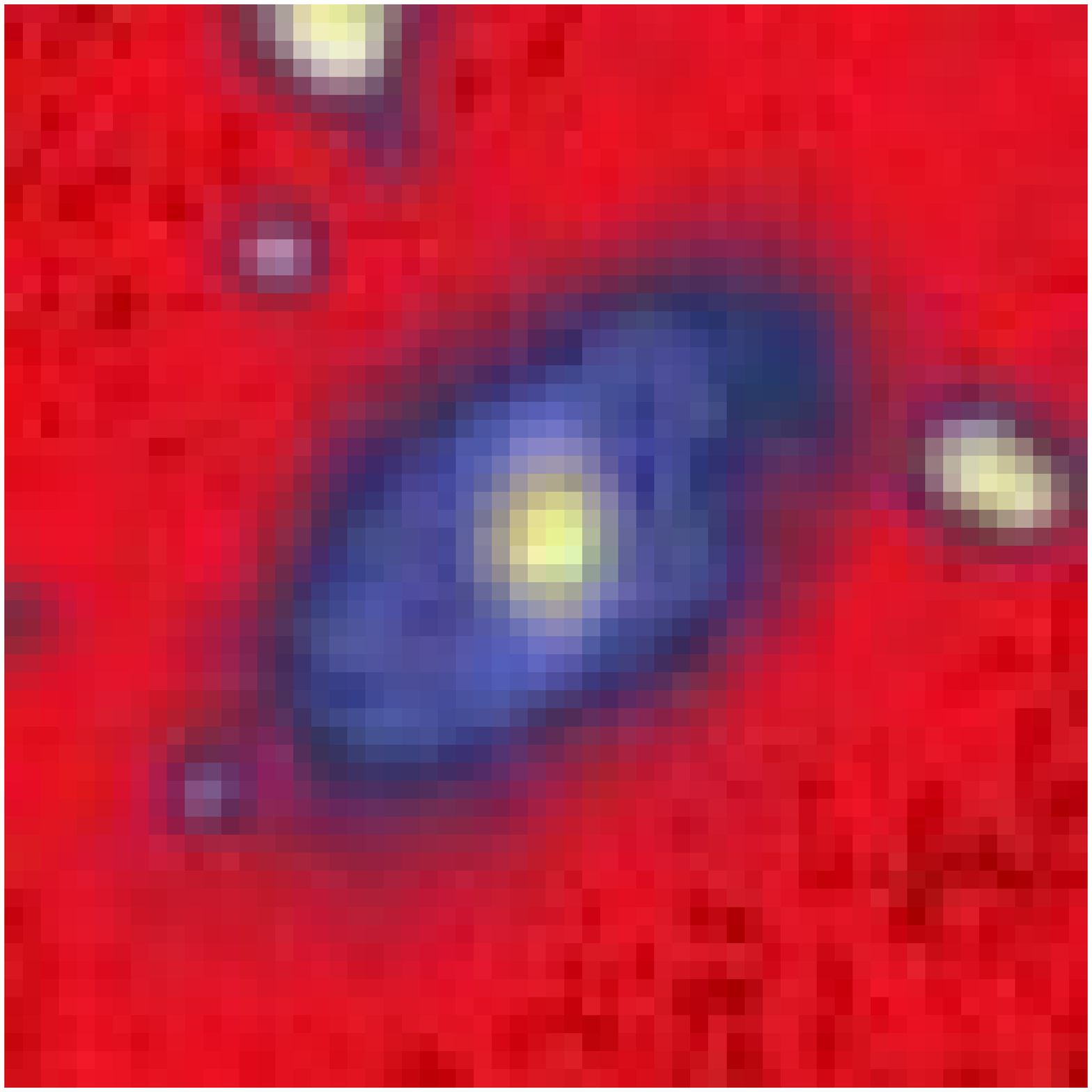}  \FGb{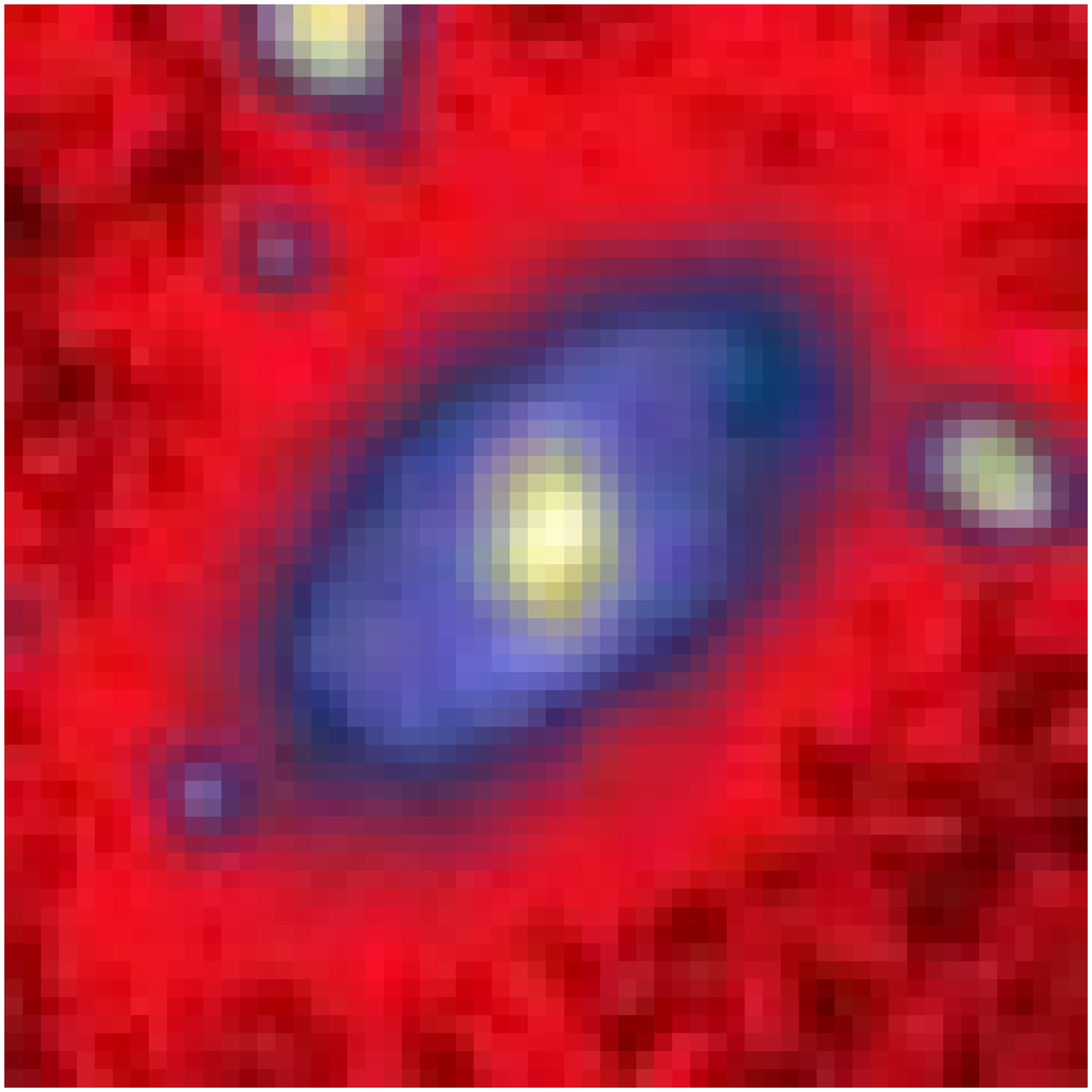}  \FGb{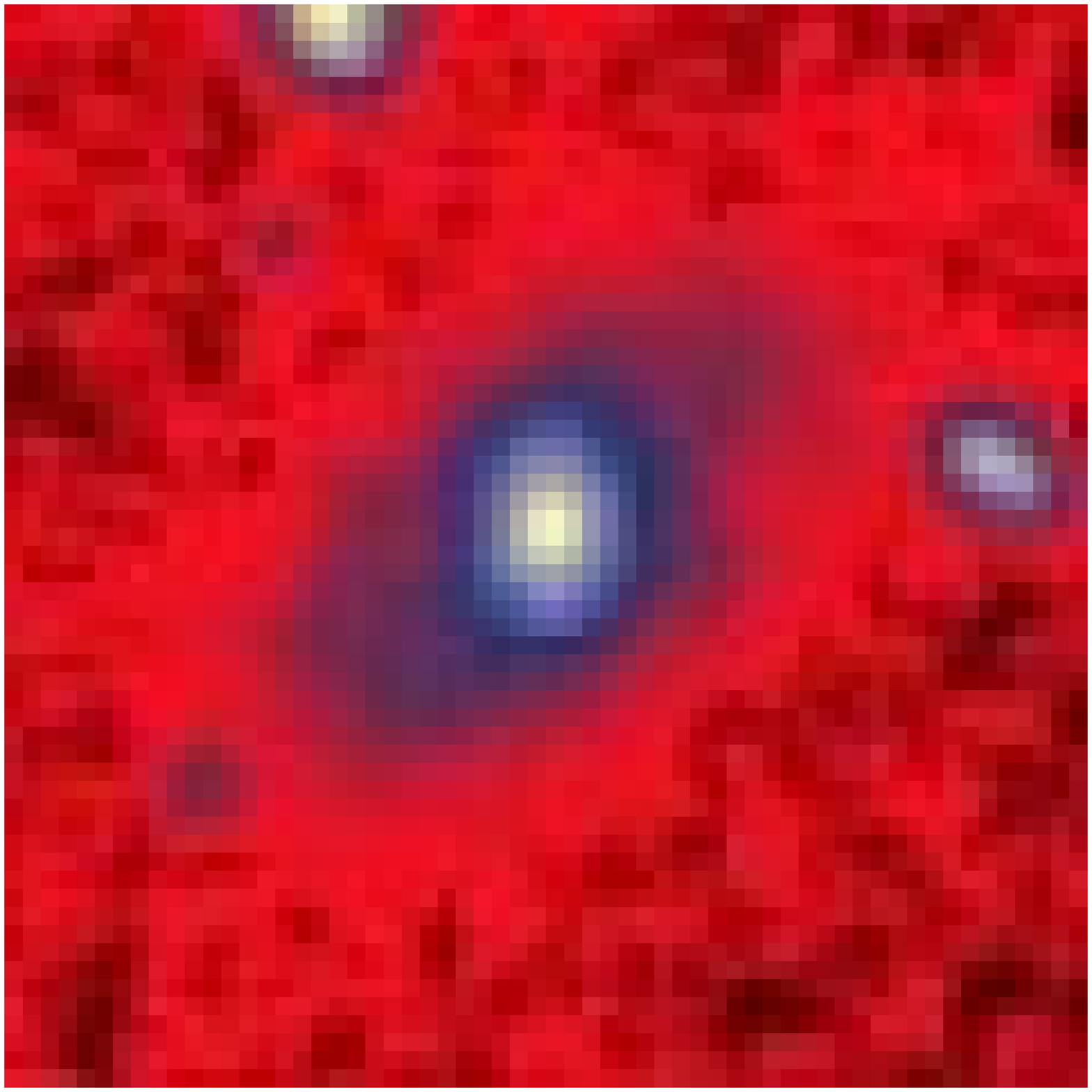}  \FGb{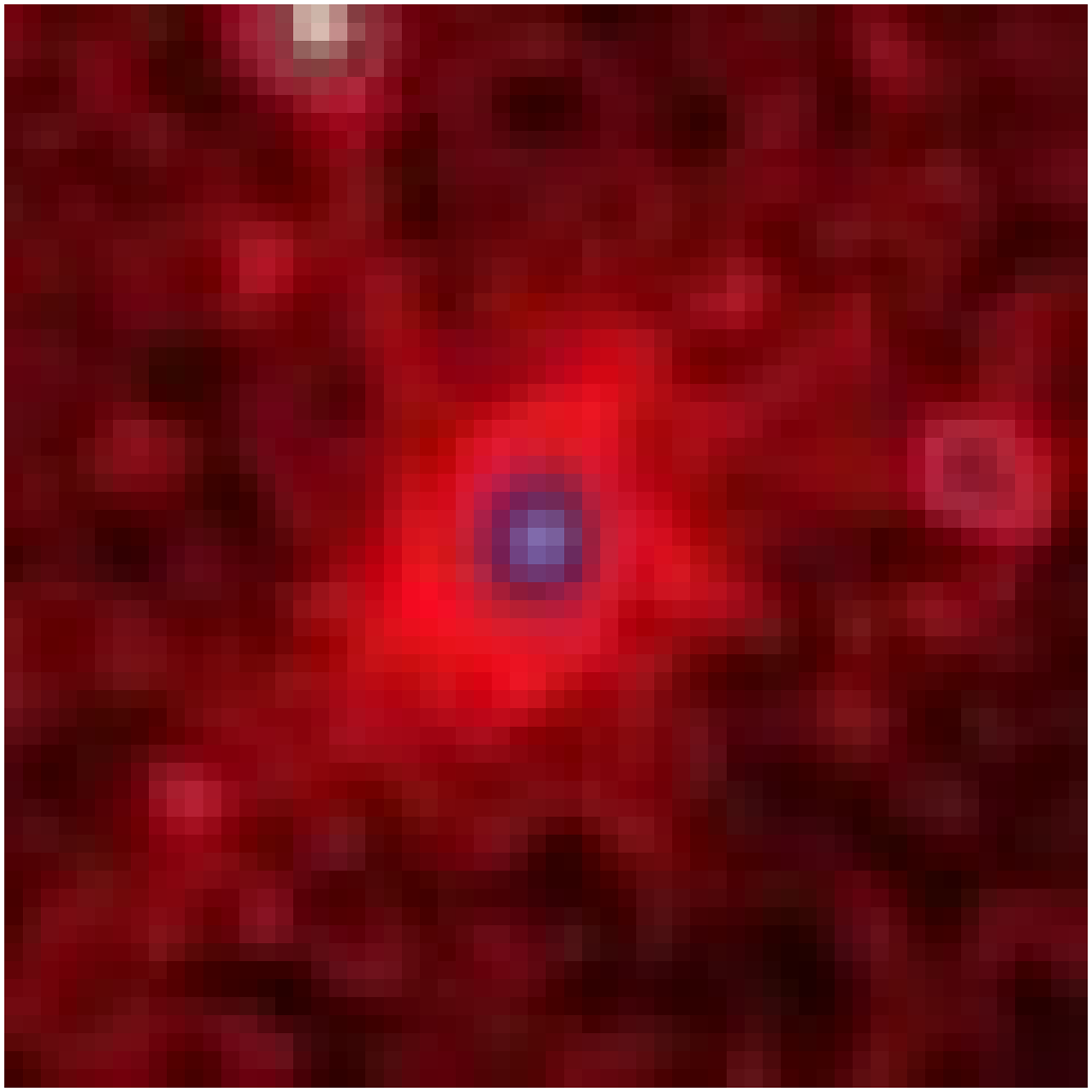}  \FGb{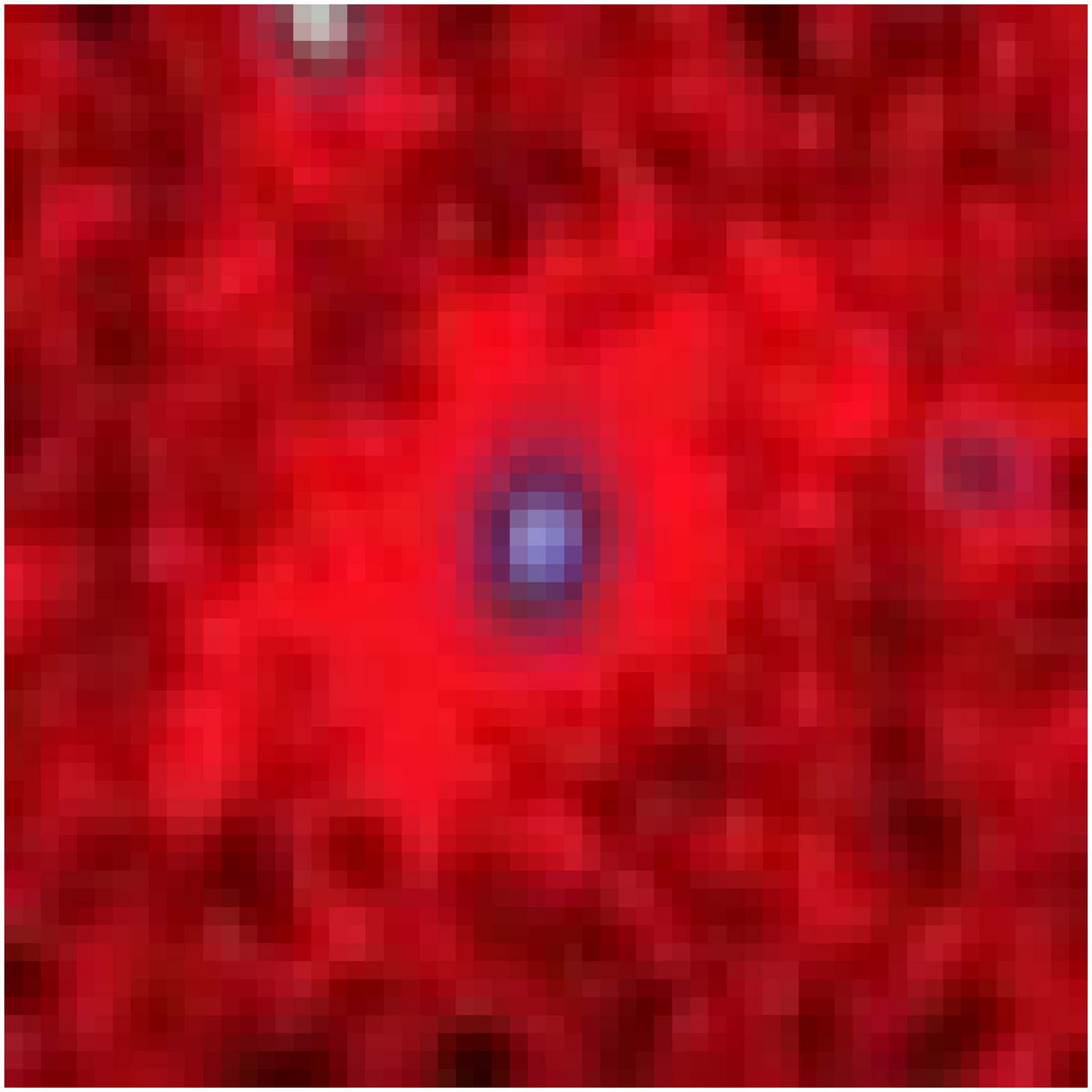}  \FGb{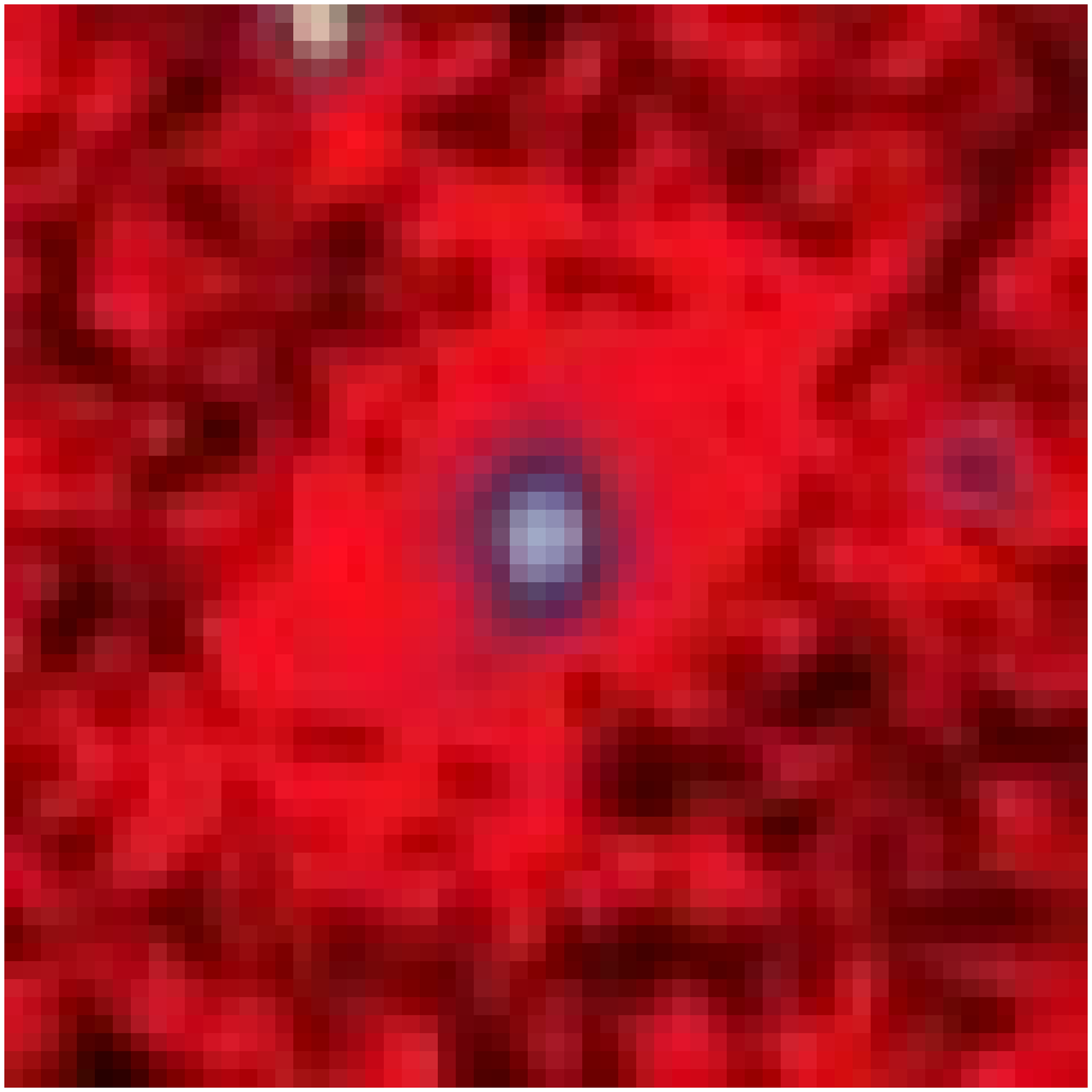}  \FGb{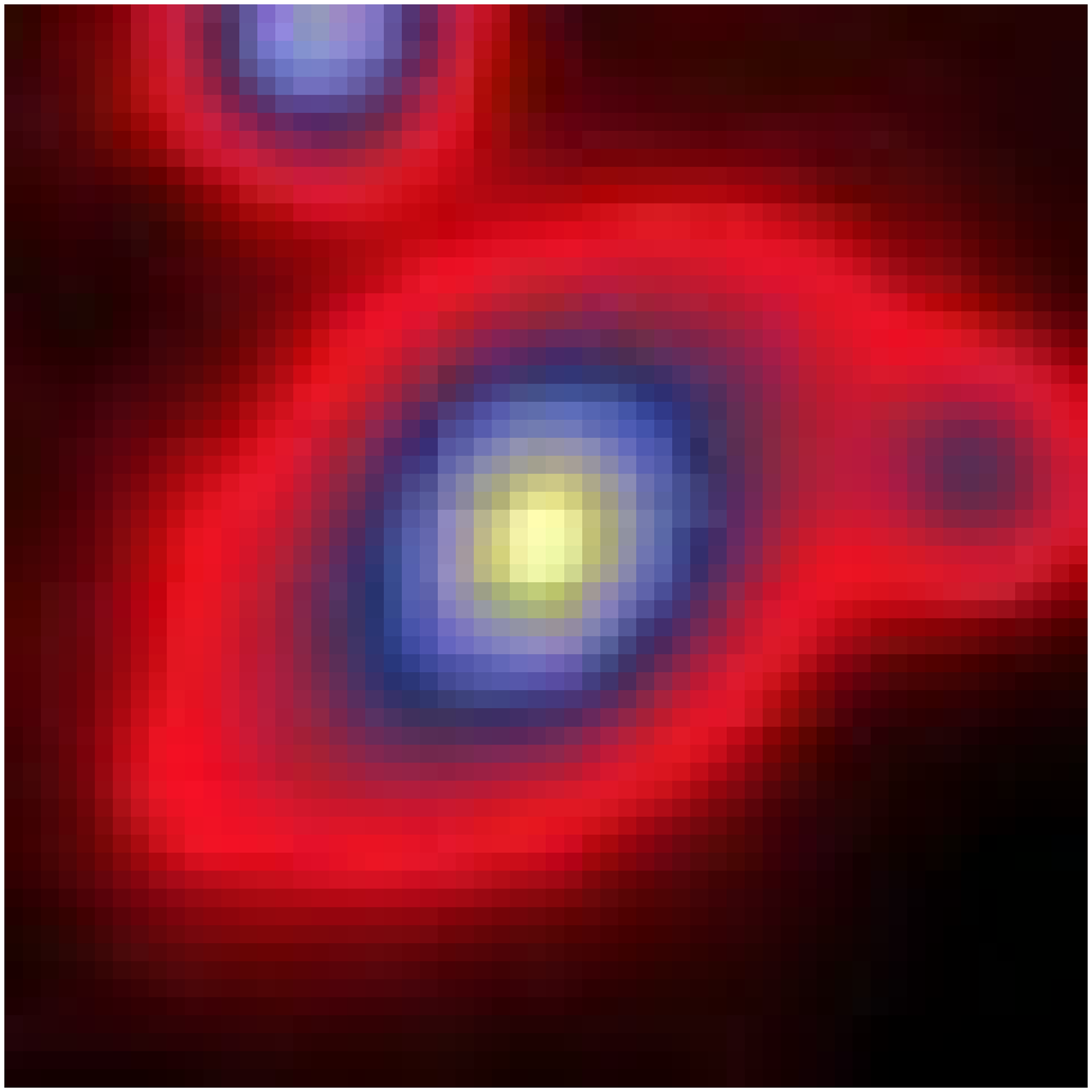}  \FGb{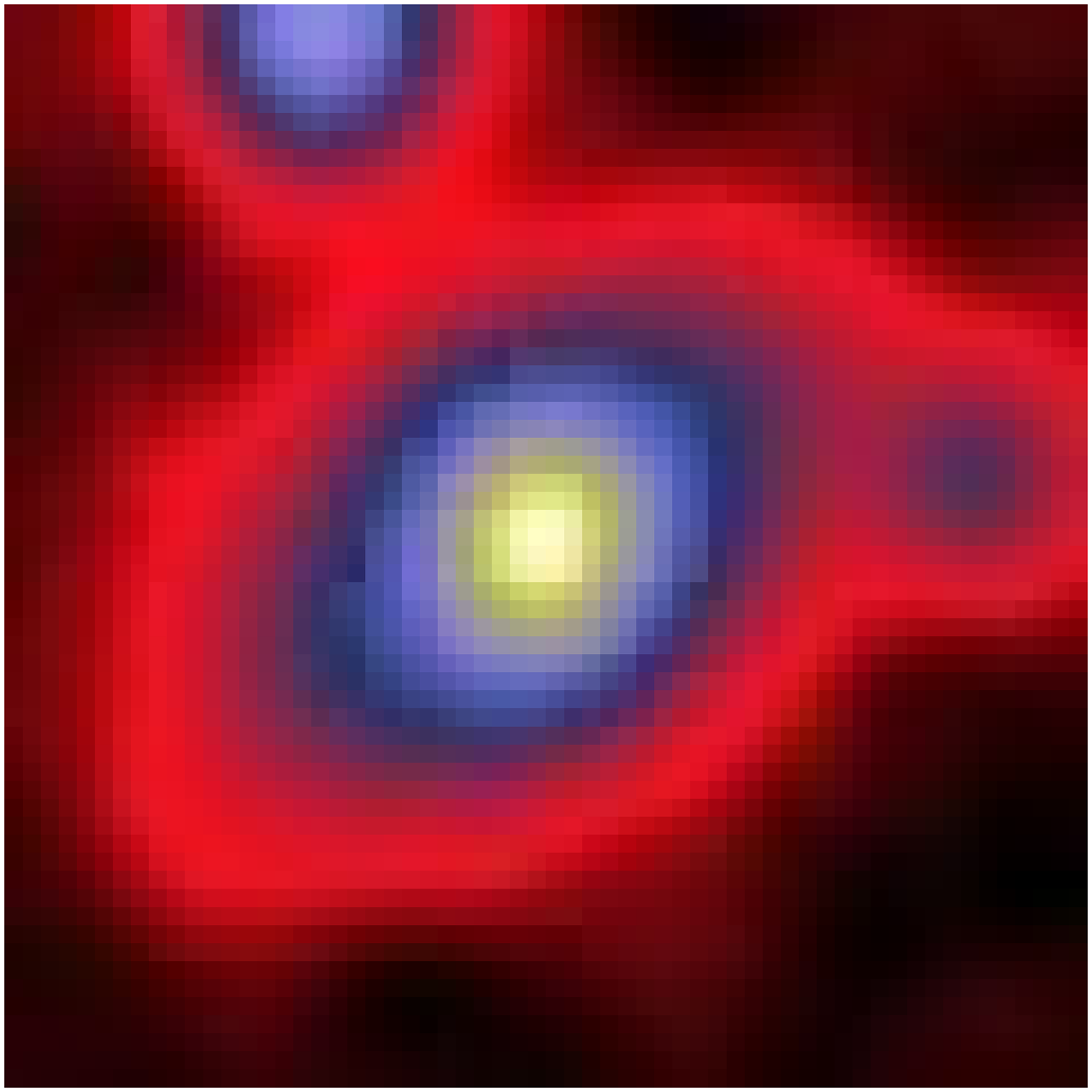}  \FGb{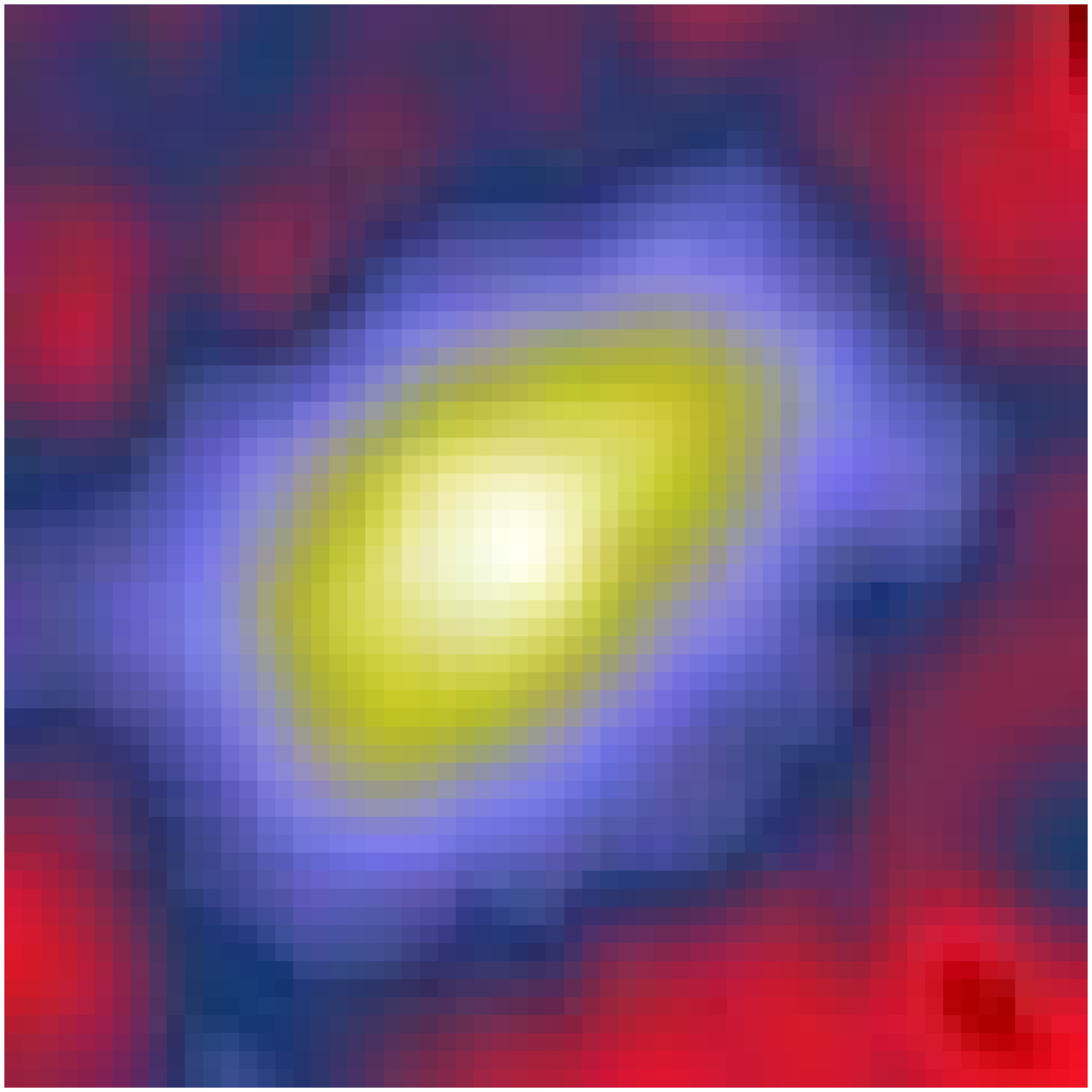}  \FGb{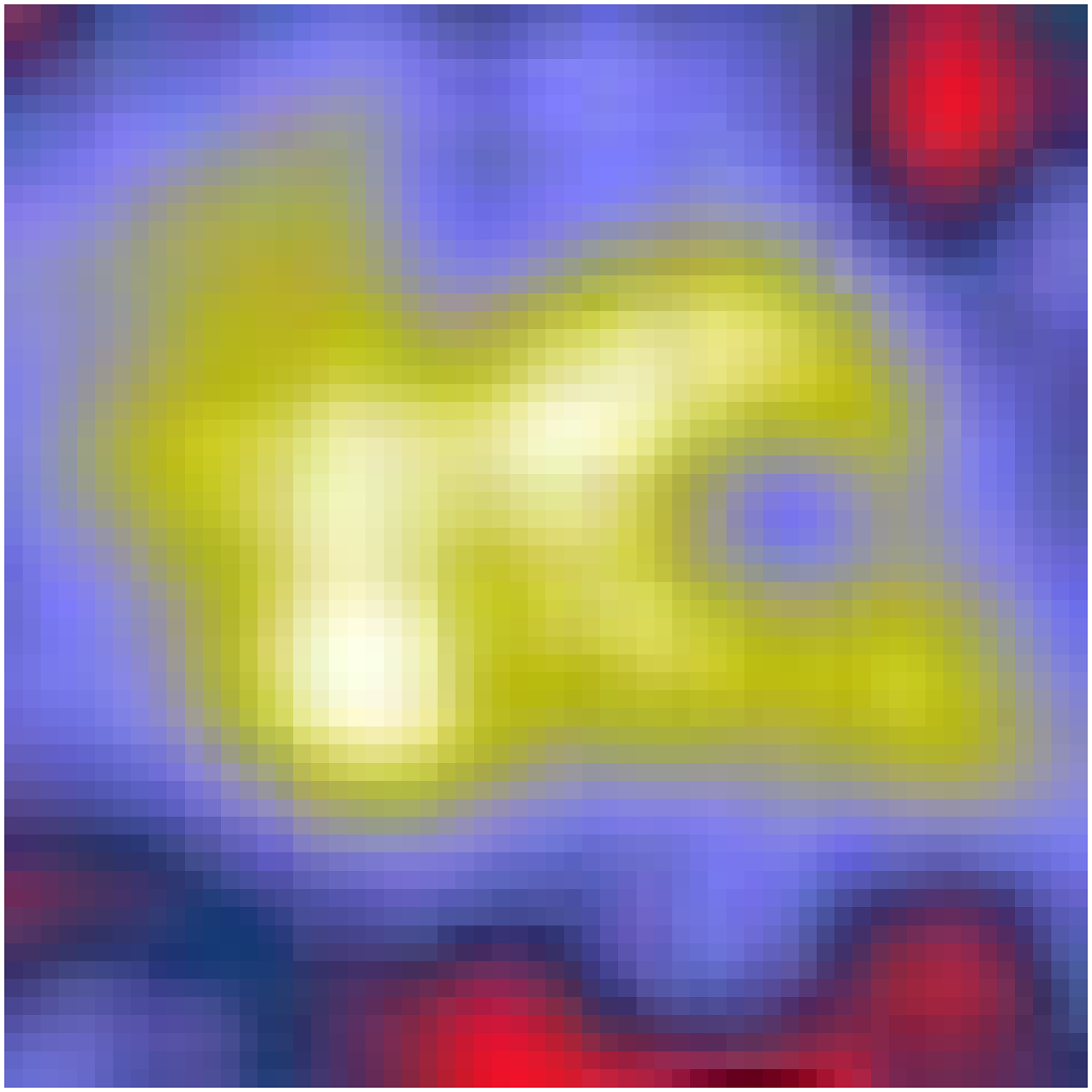} \\  {\#62   NBZ5000685}  \\
  \FGb{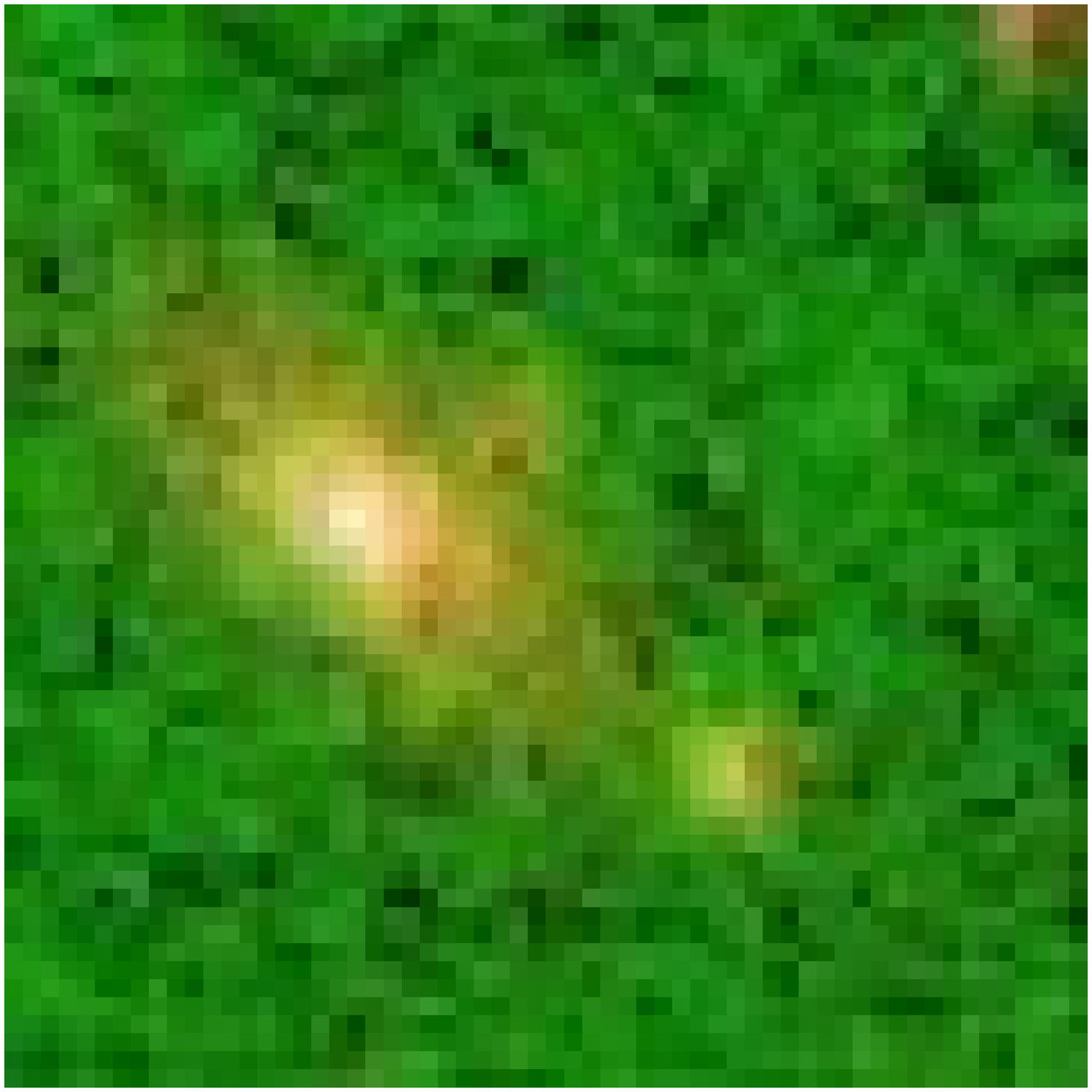}  \FGb{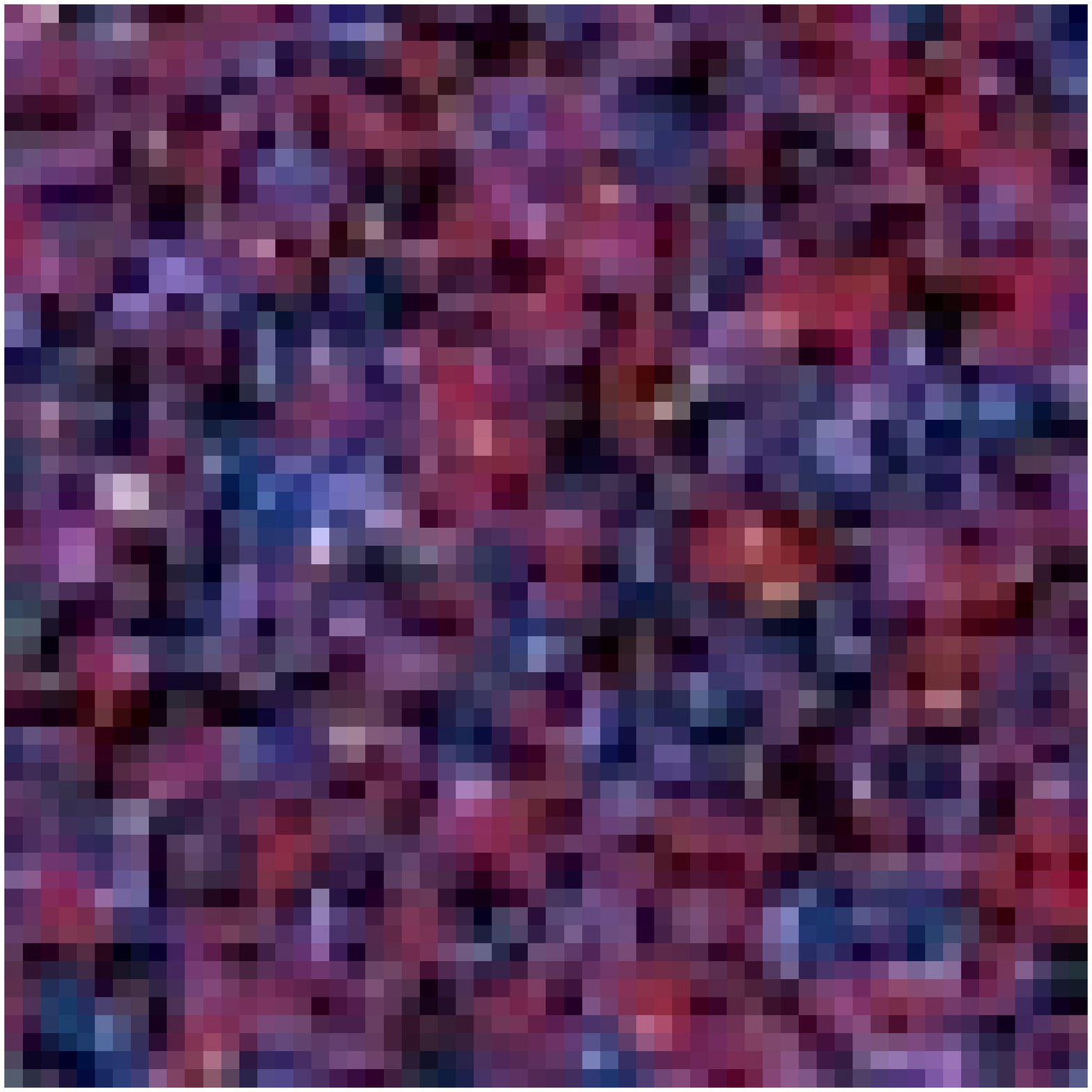}  \FGb{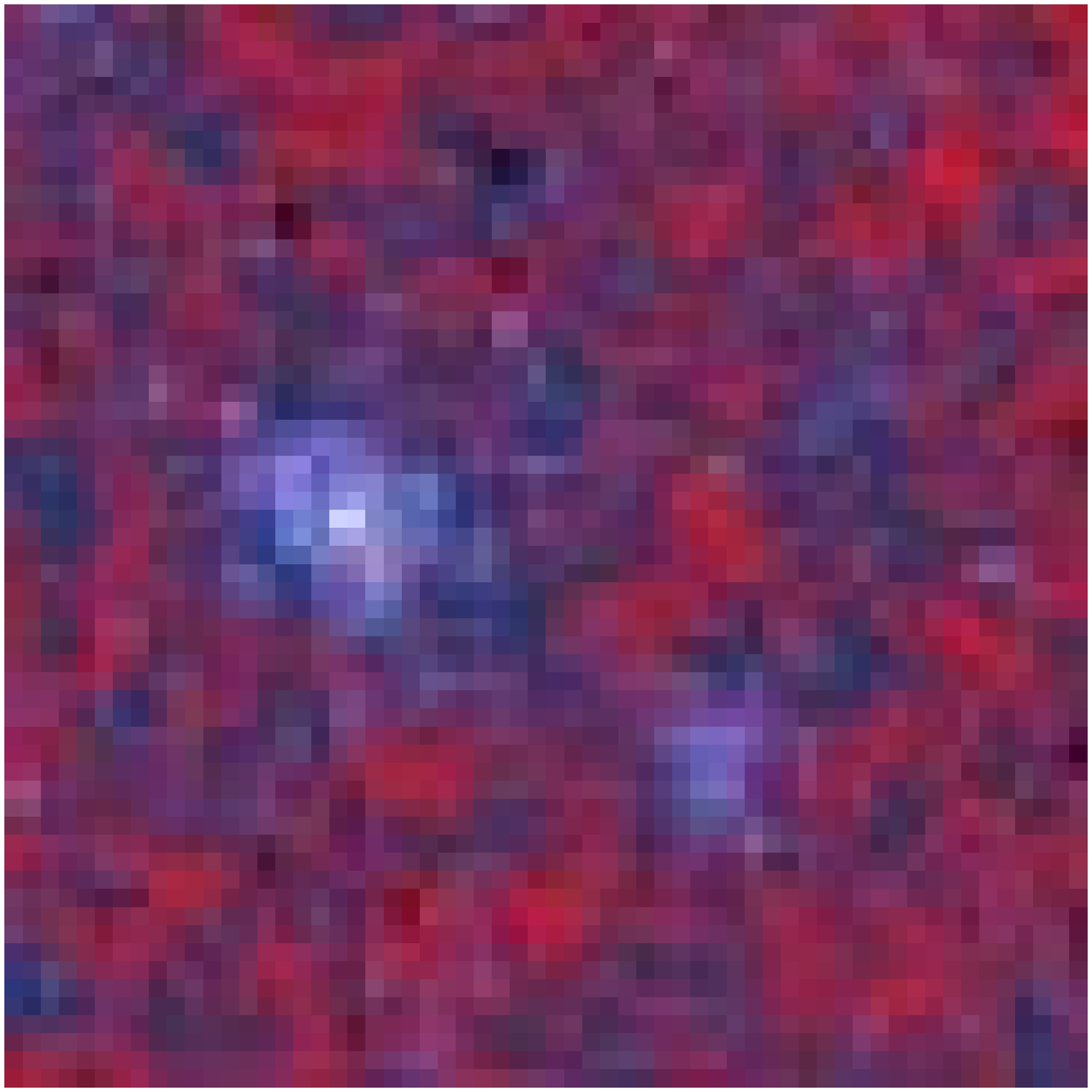}  \FGb{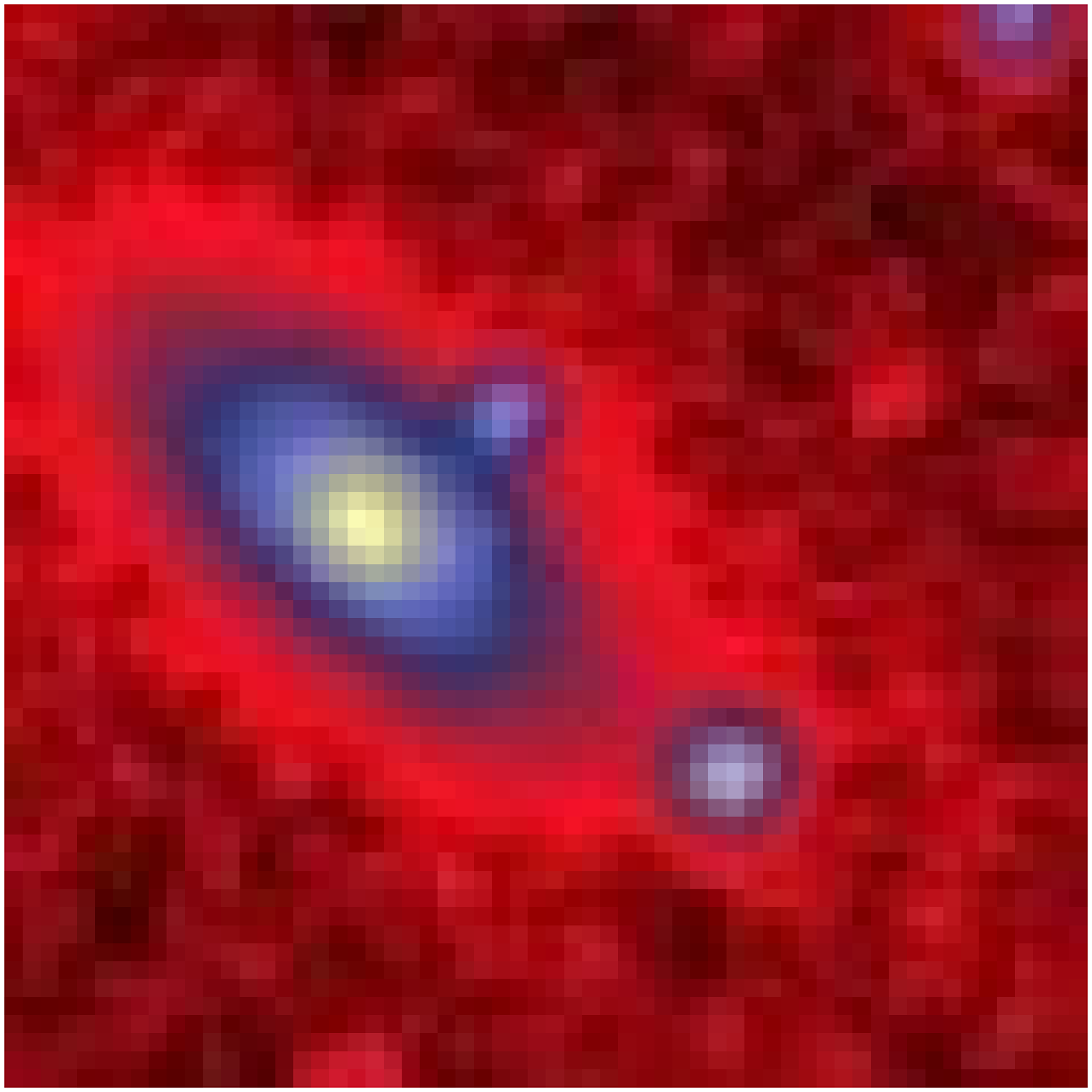}  \FGb{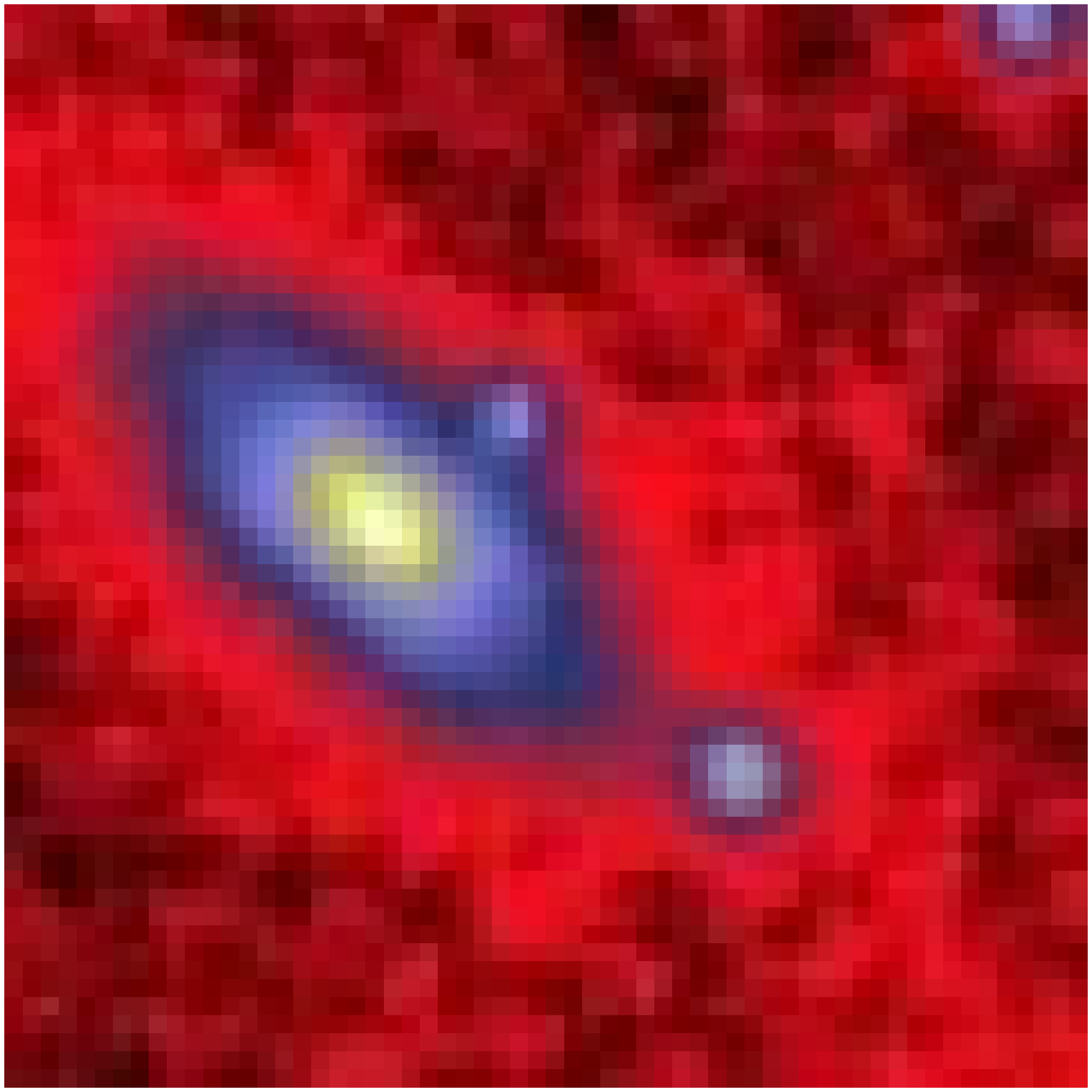}  \FGb{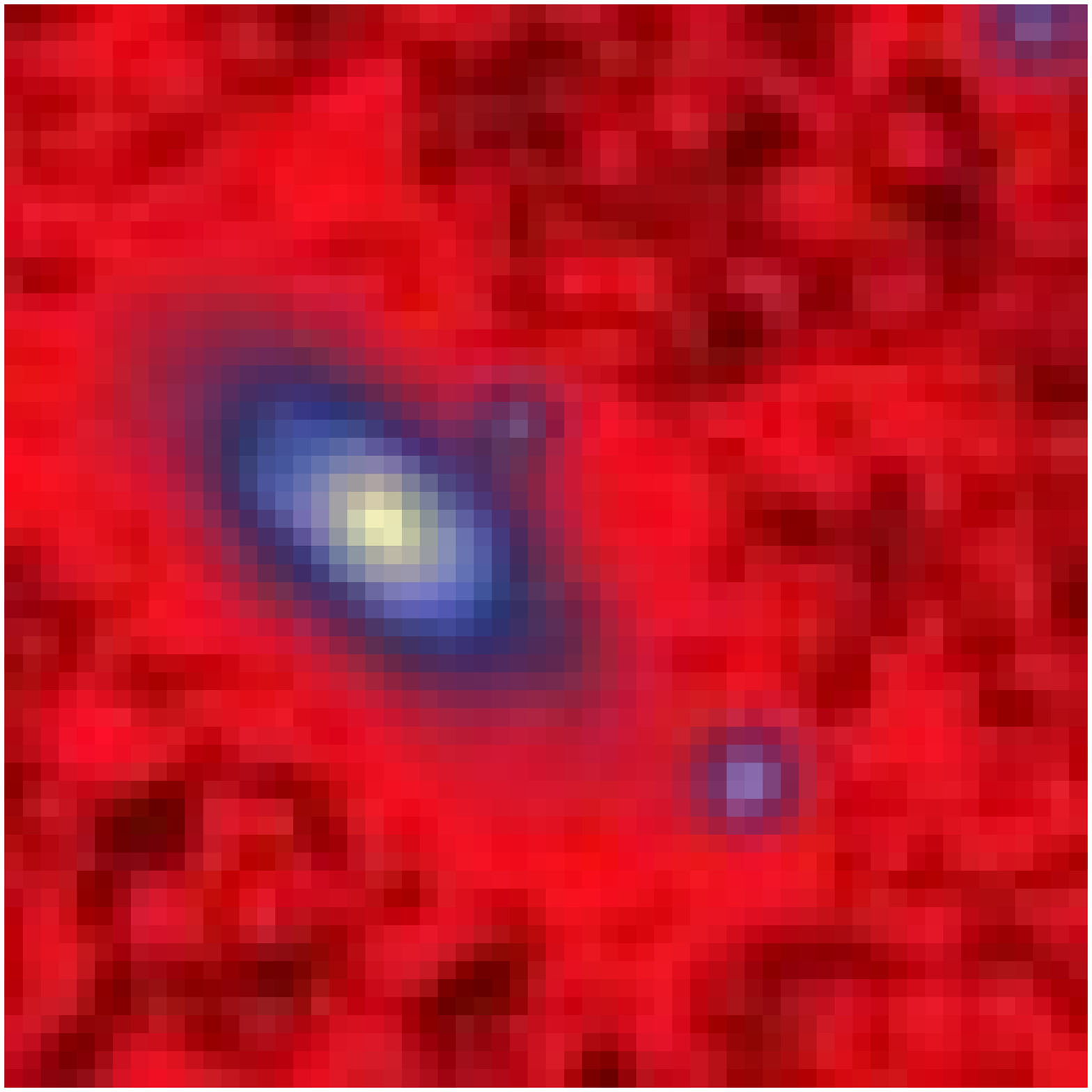}  \FGb{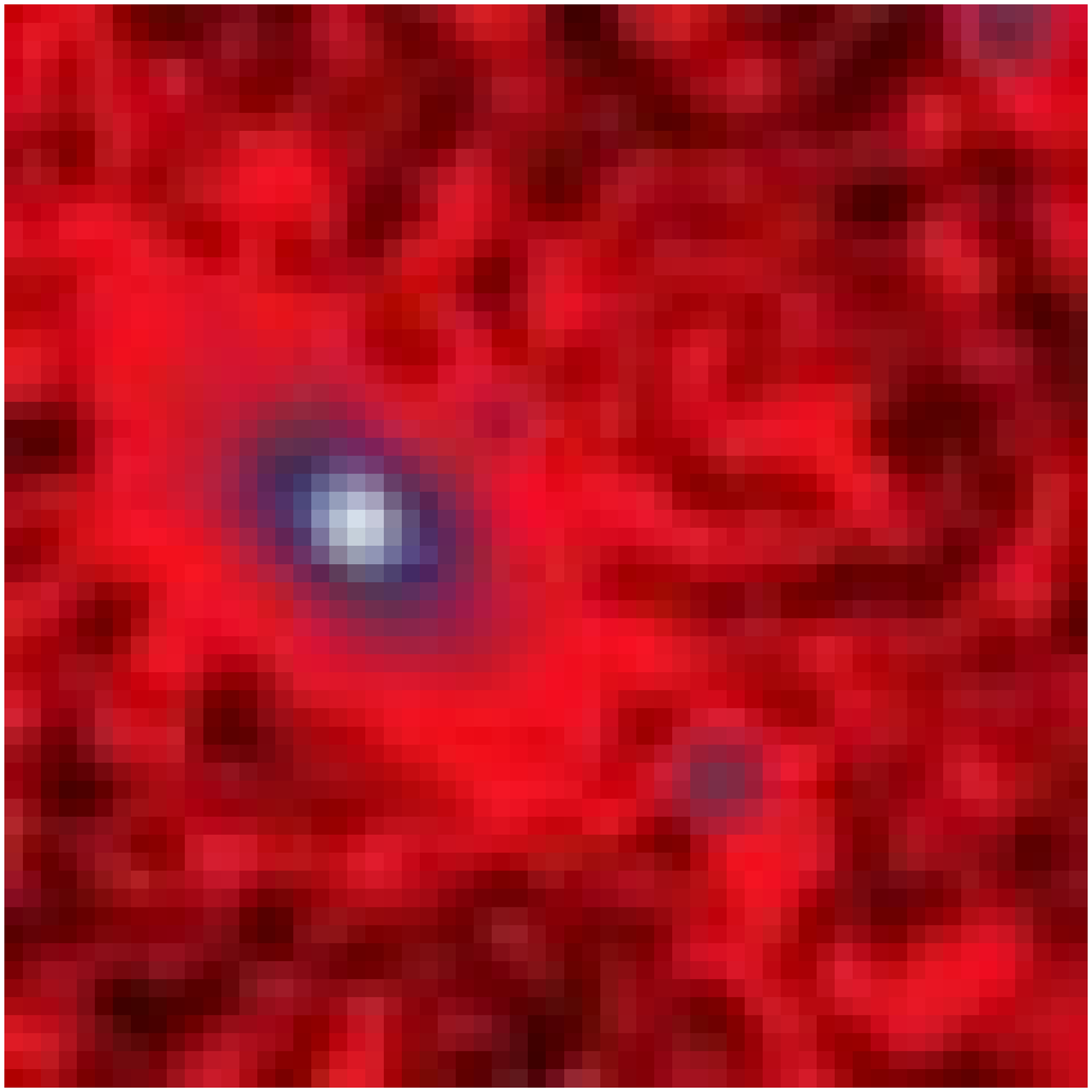}  \FGb{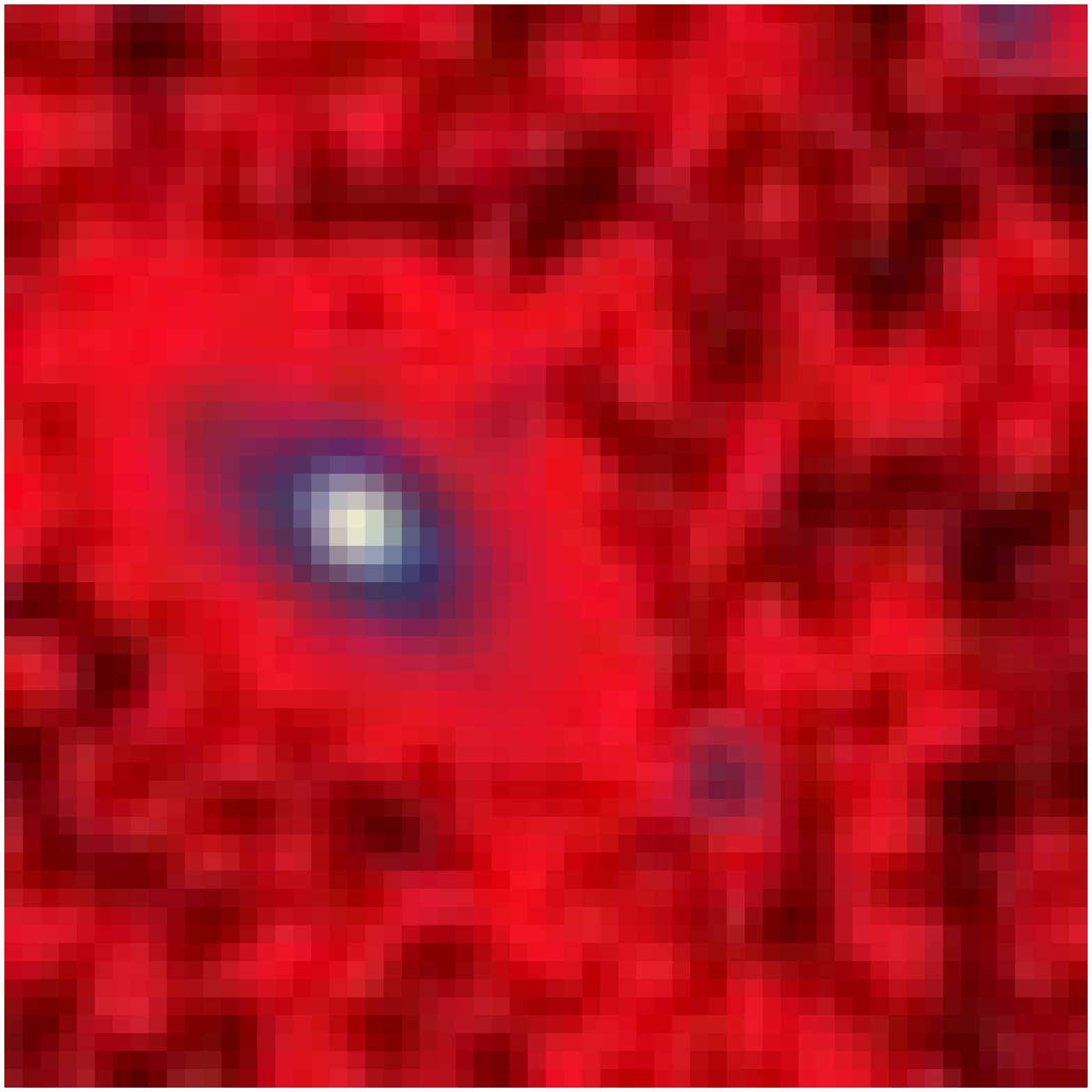}  \FGb{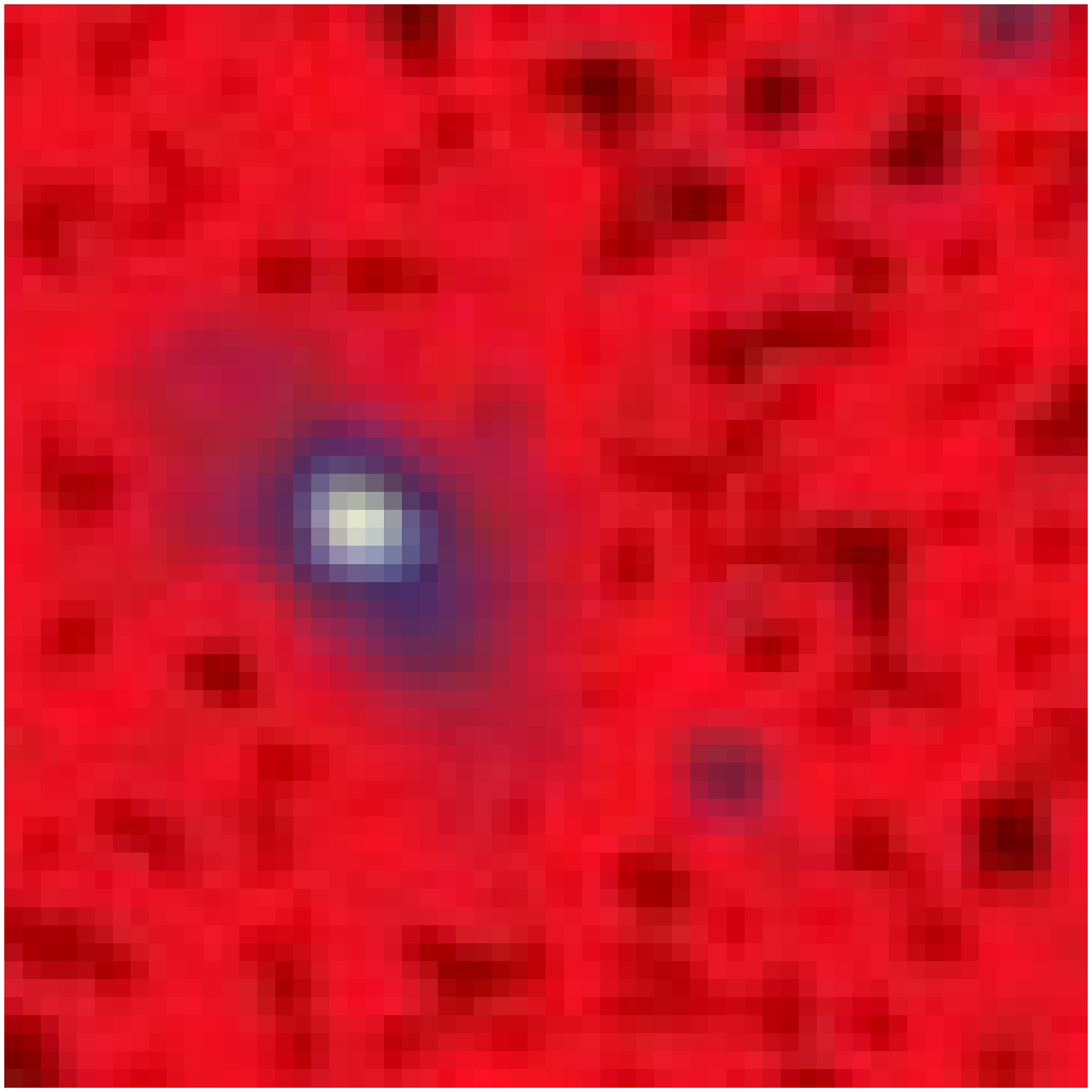}  \FGb{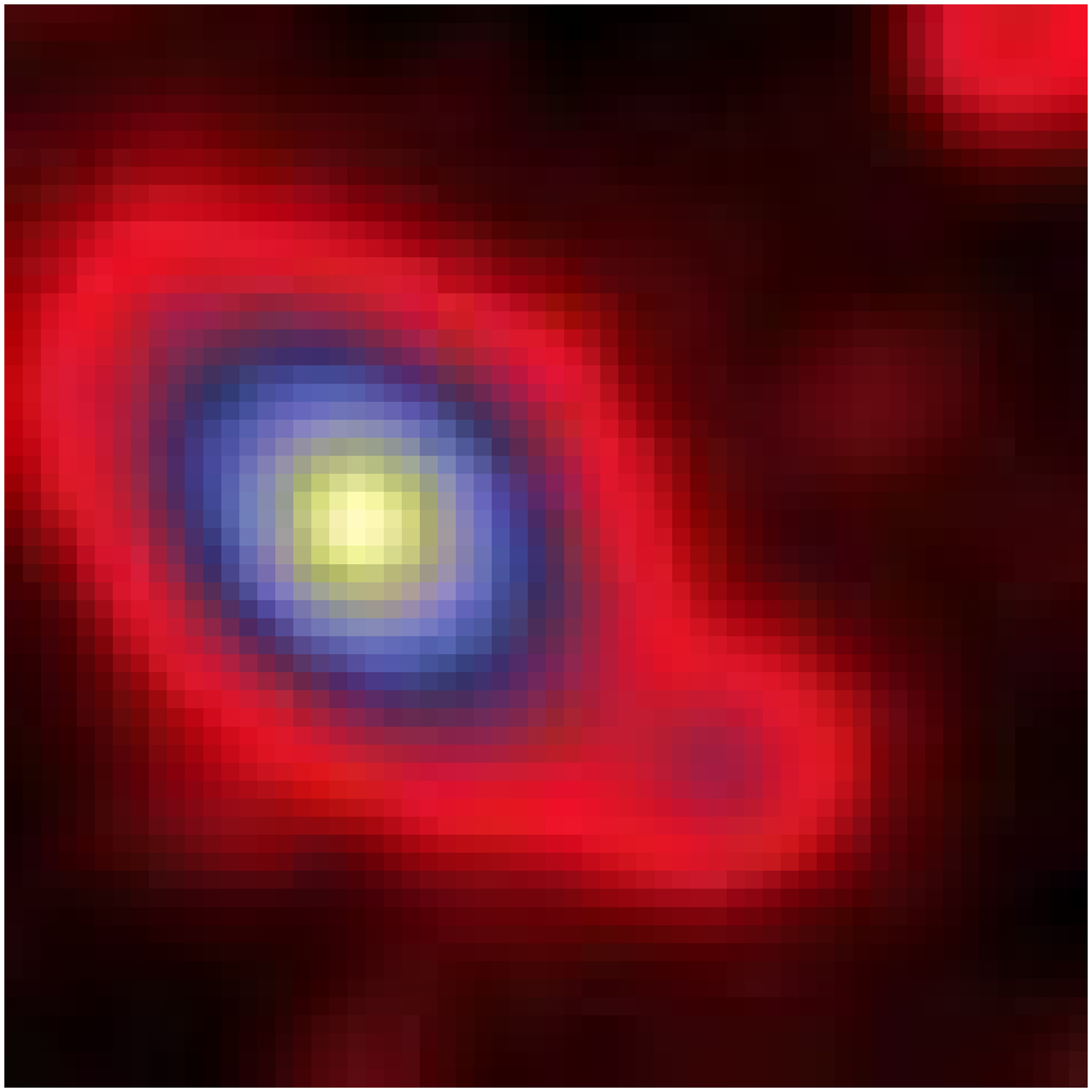}  \FGb{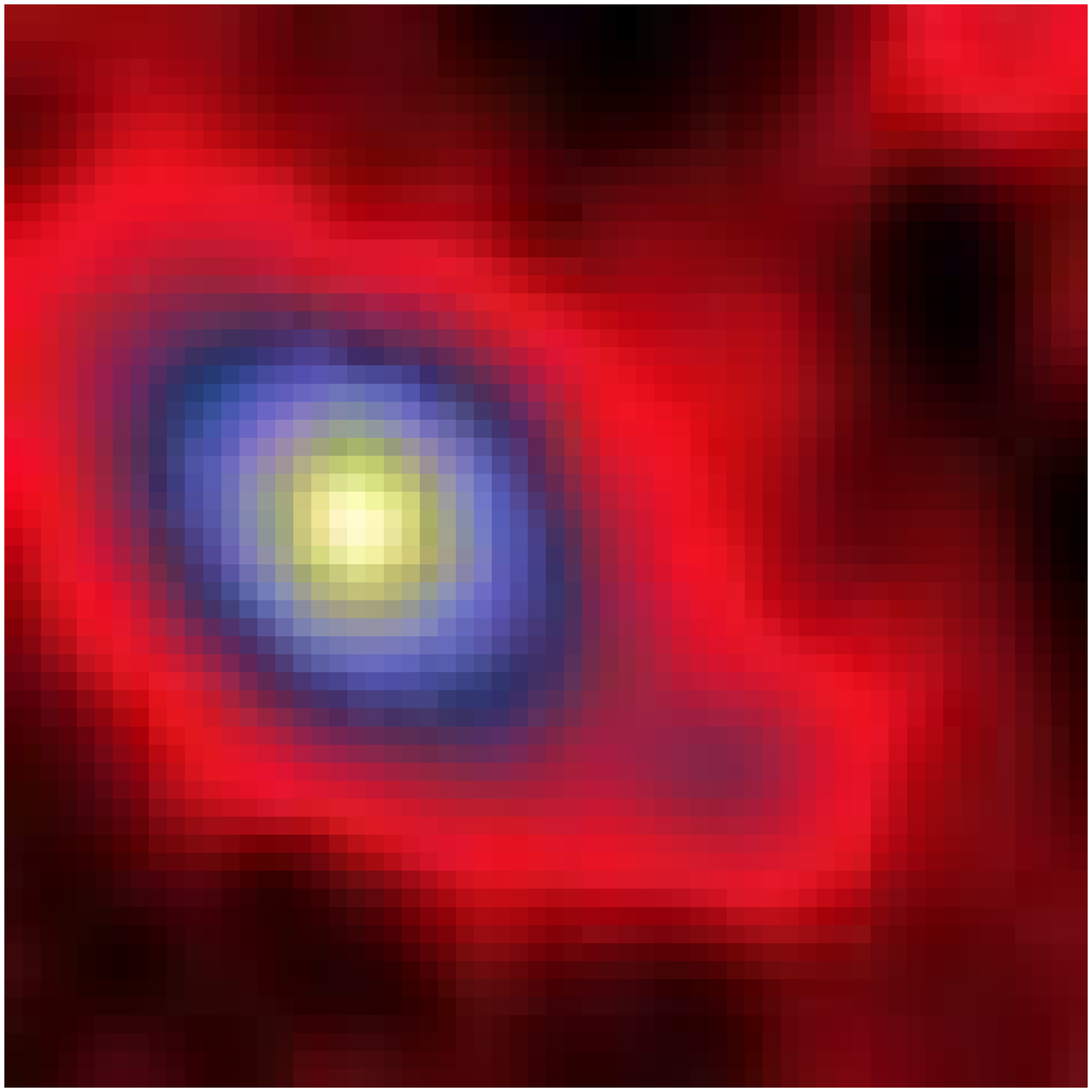}  \FGb{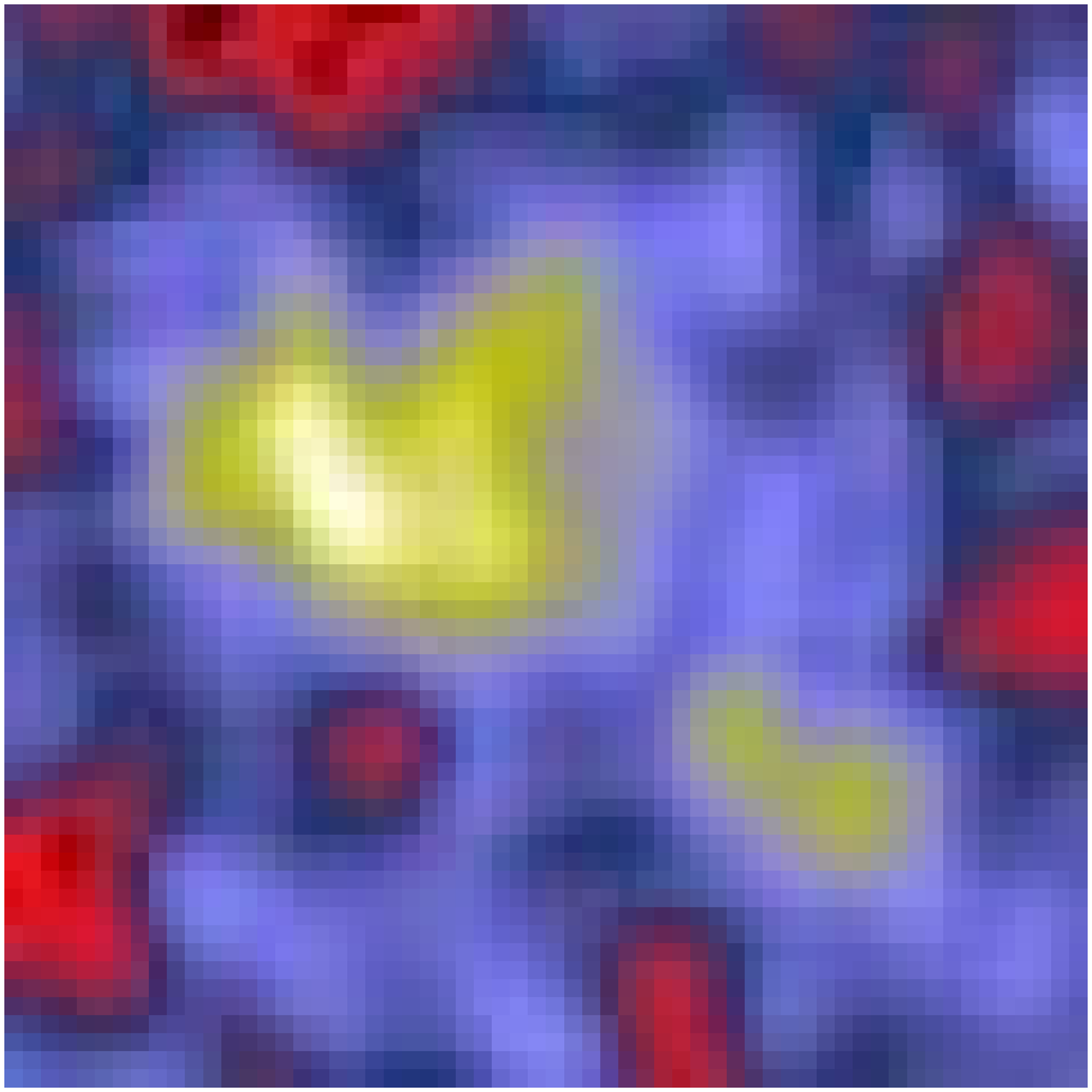}  \FGb{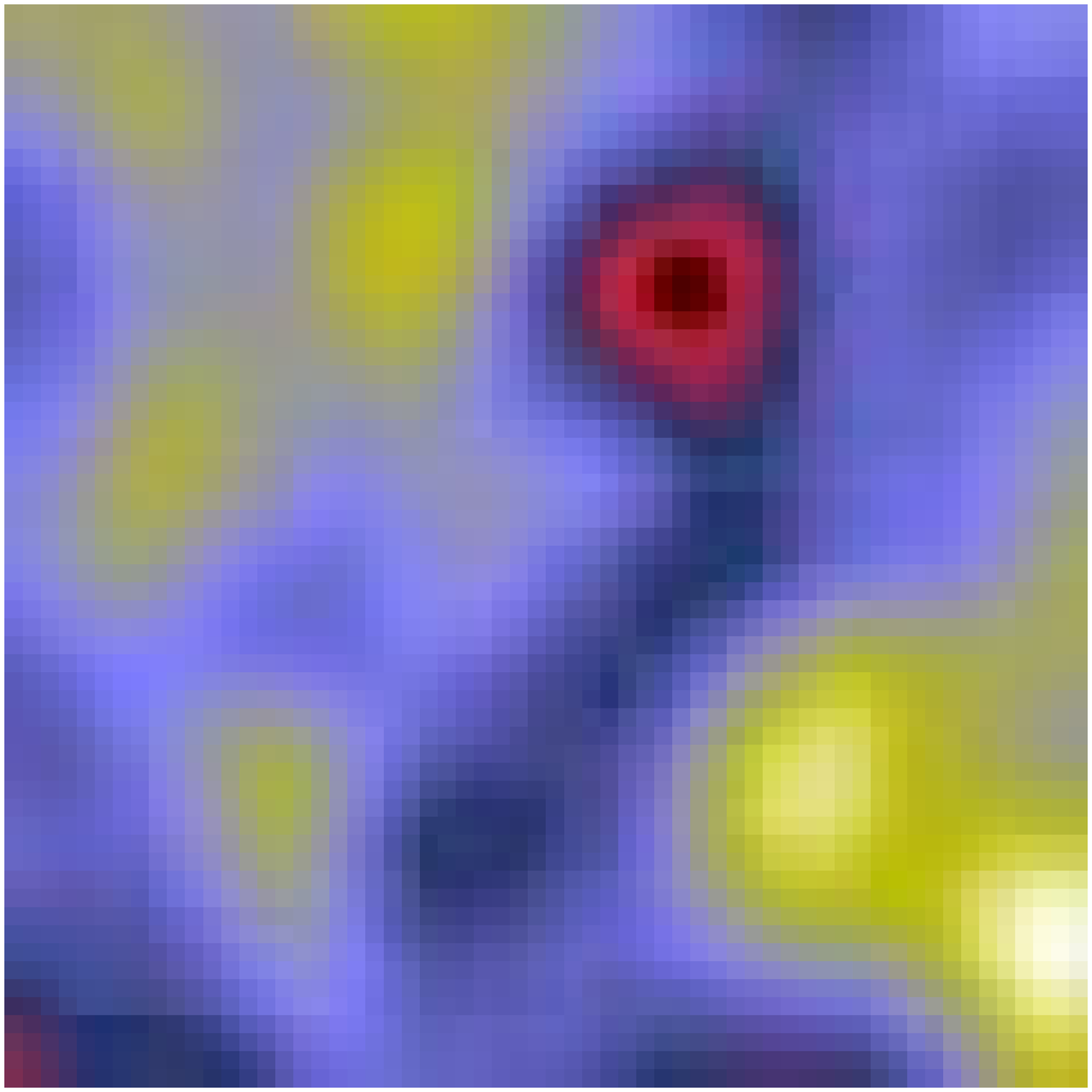} \\  {\#63   NCIA000991}  \\
  \FGb{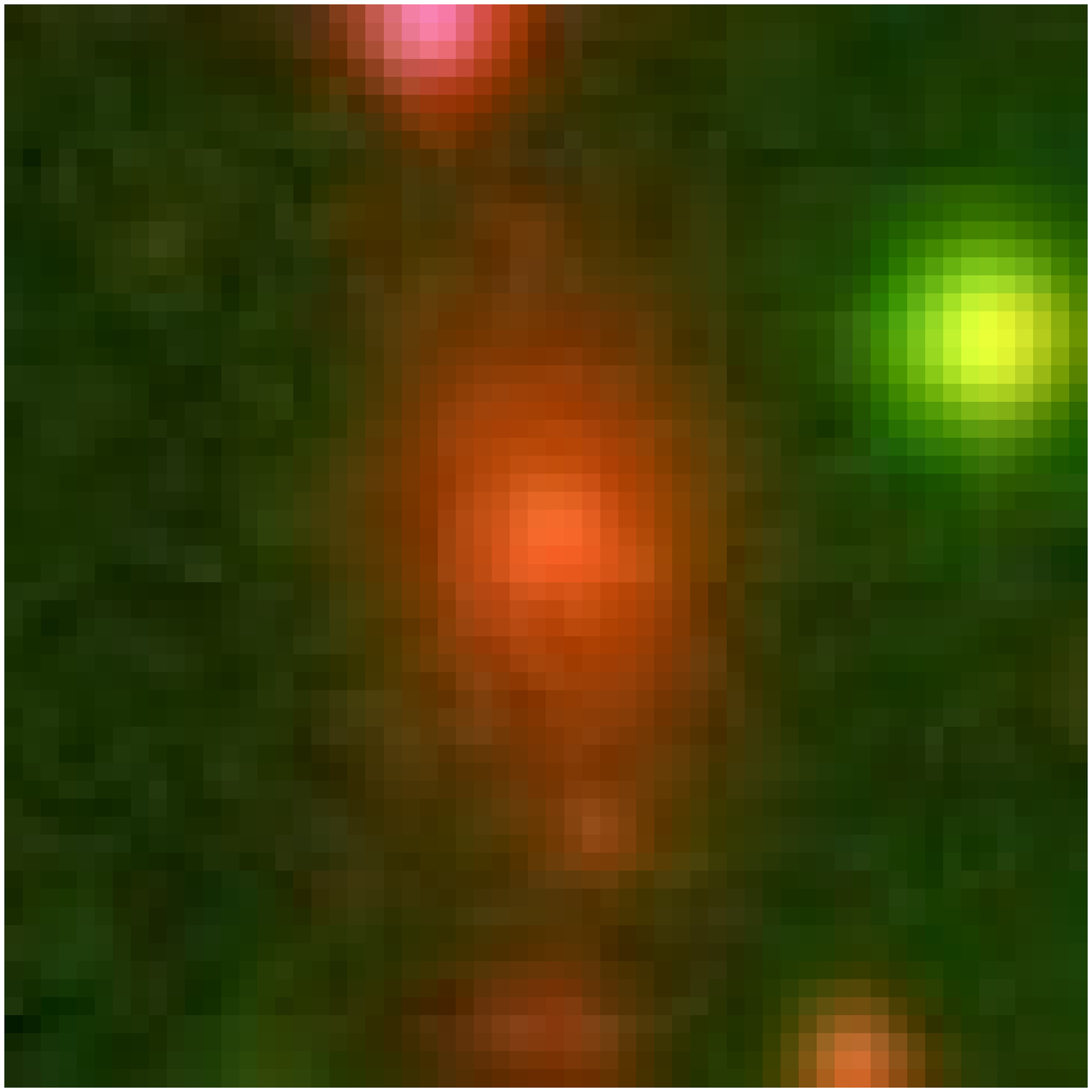}  \FGb{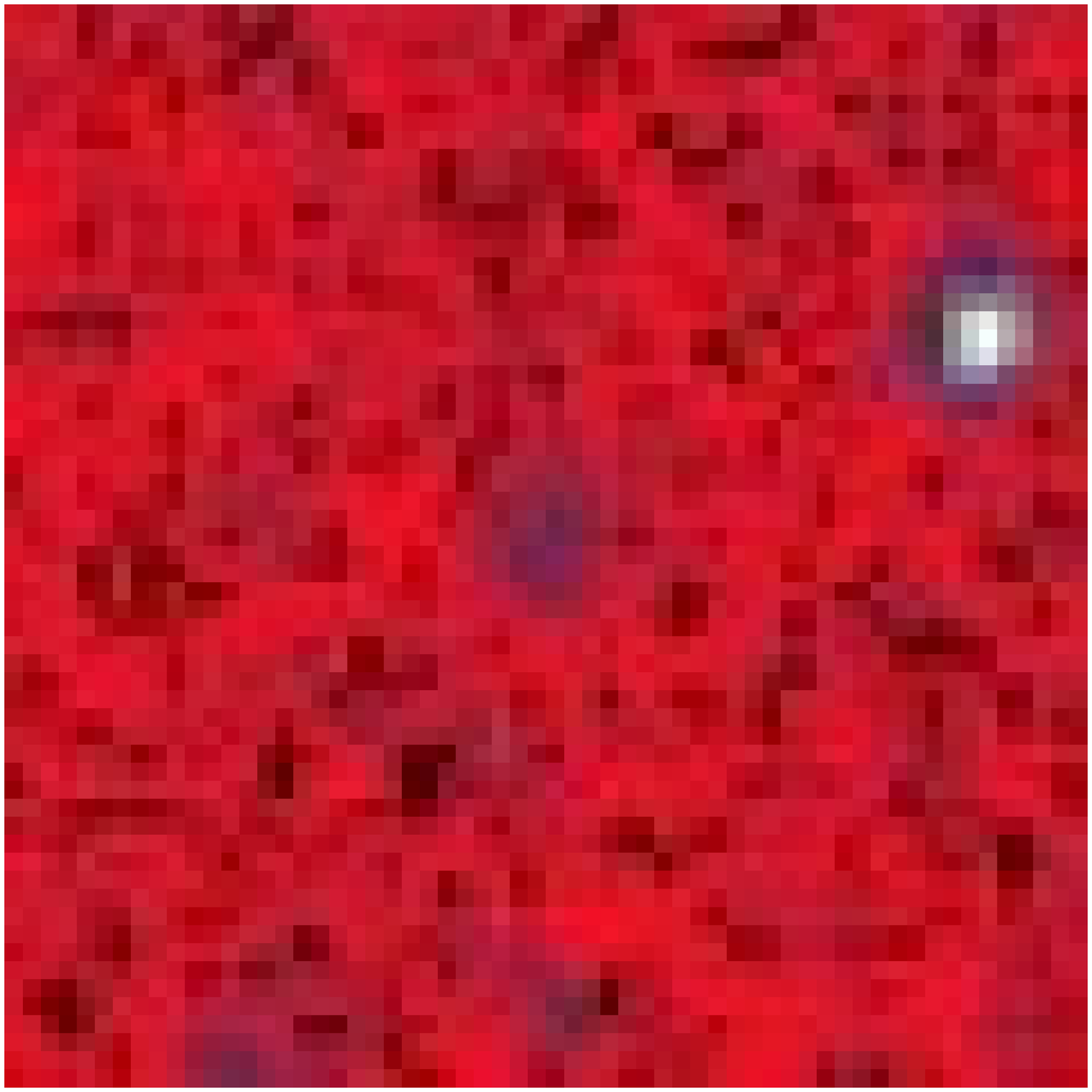}  \FGb{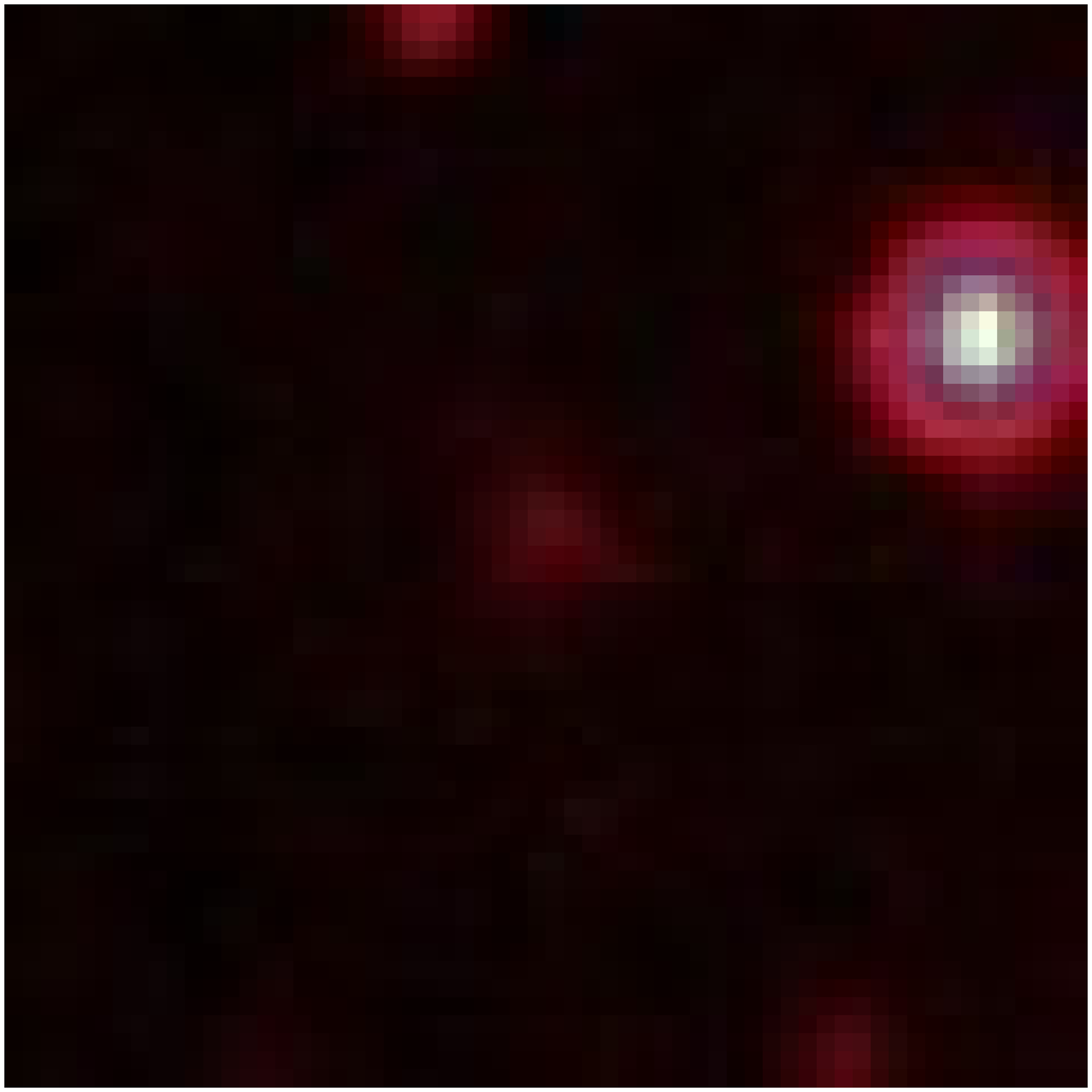}  \FGb{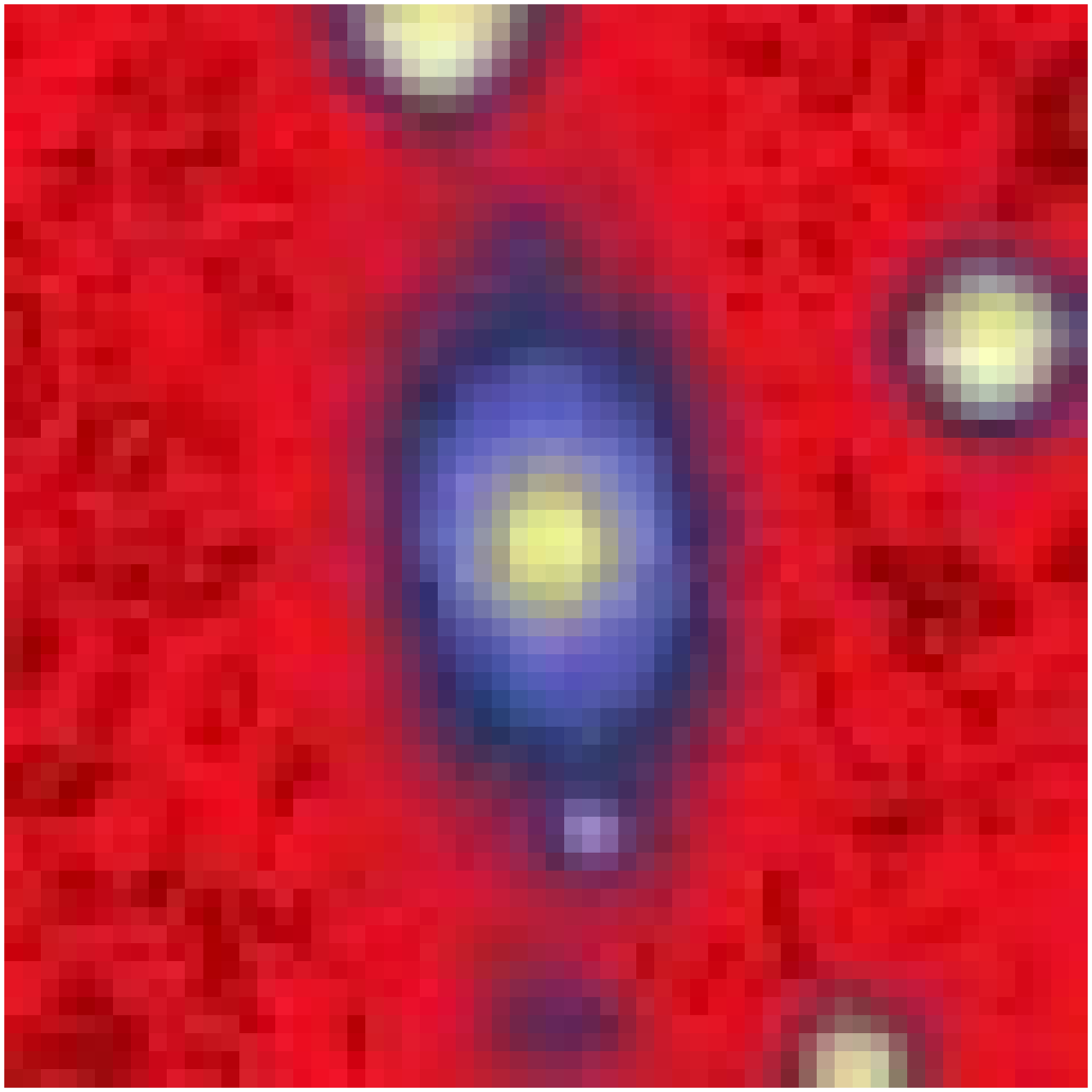}  \FGb{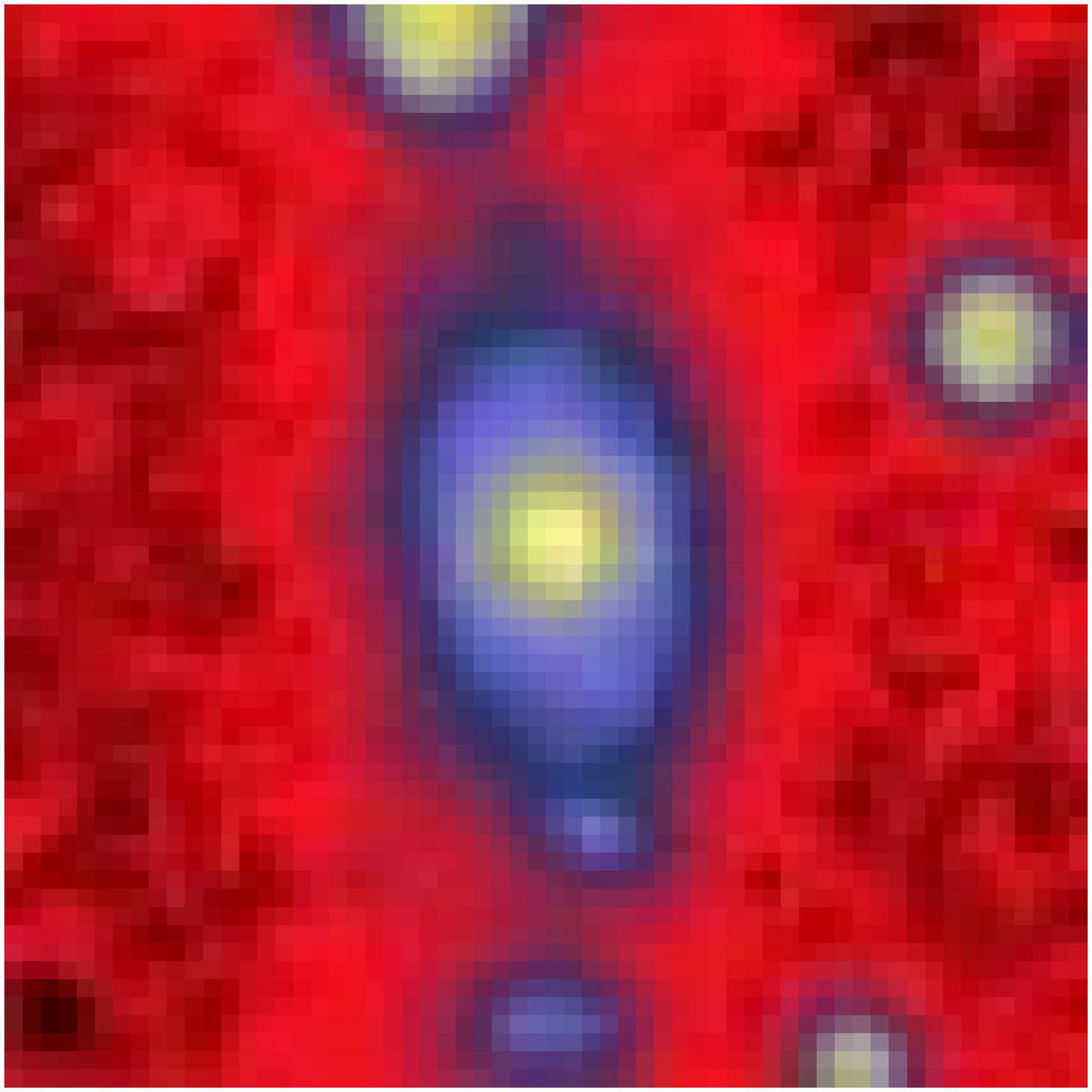}  \FGb{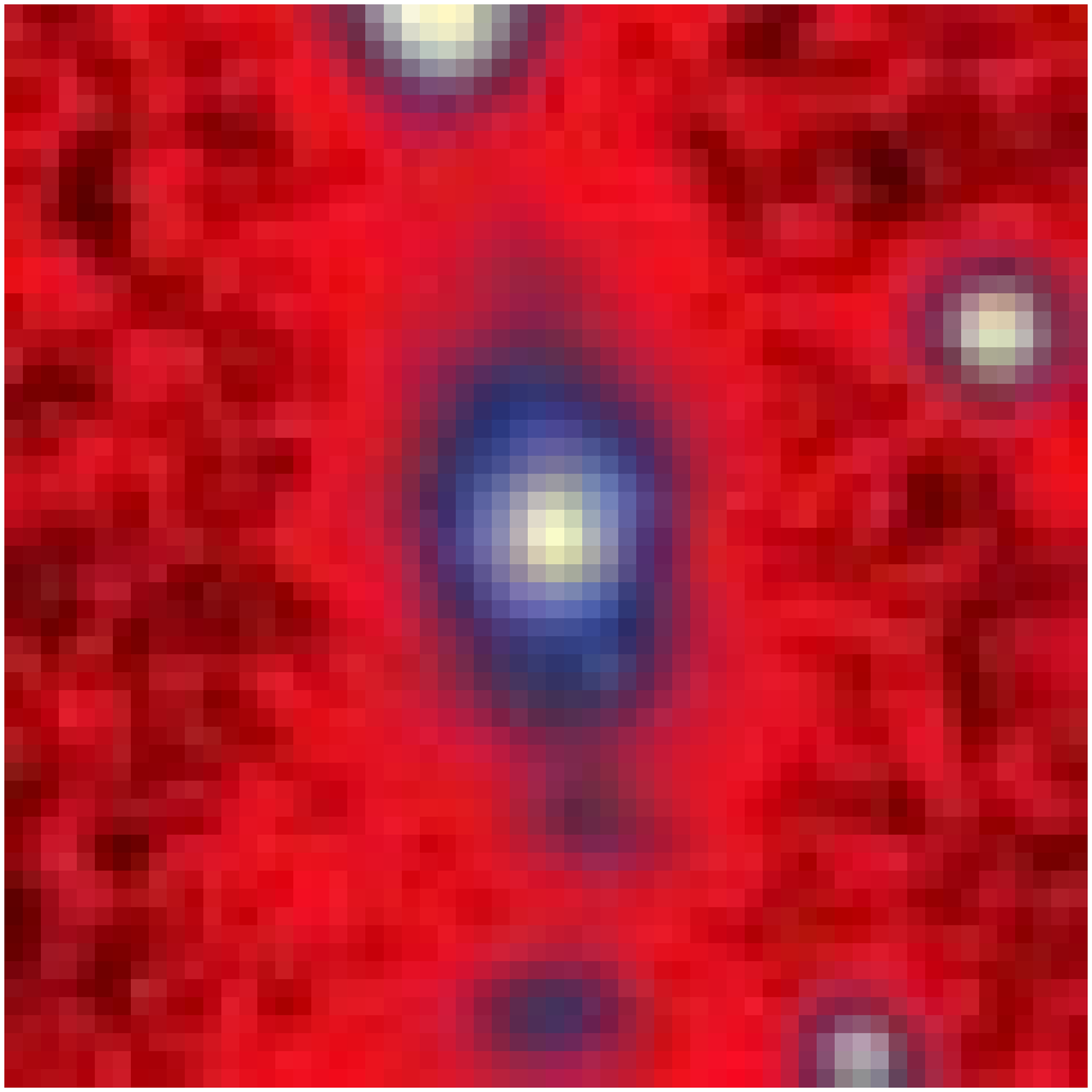}  \FGb{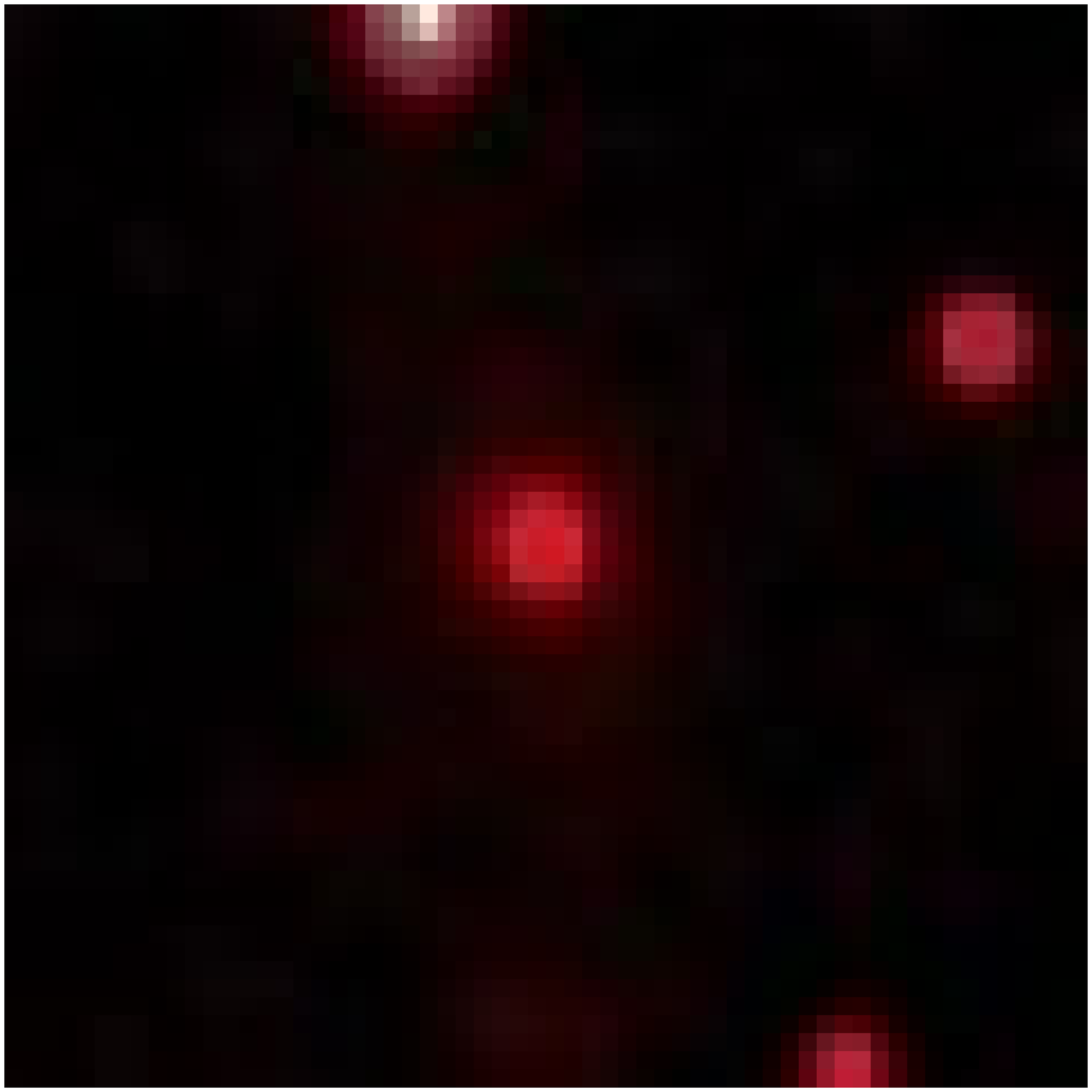}  \FGb{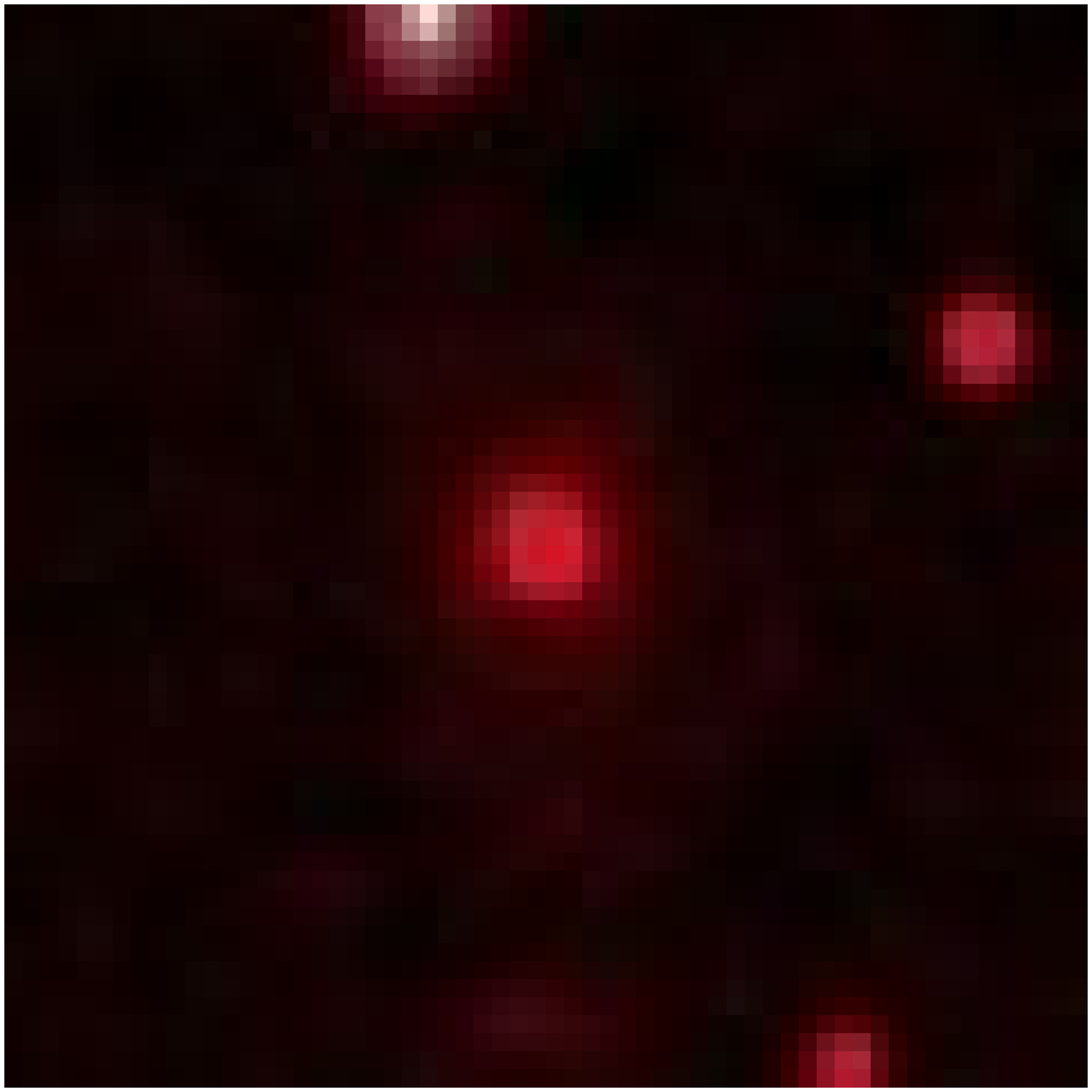}  \FGb{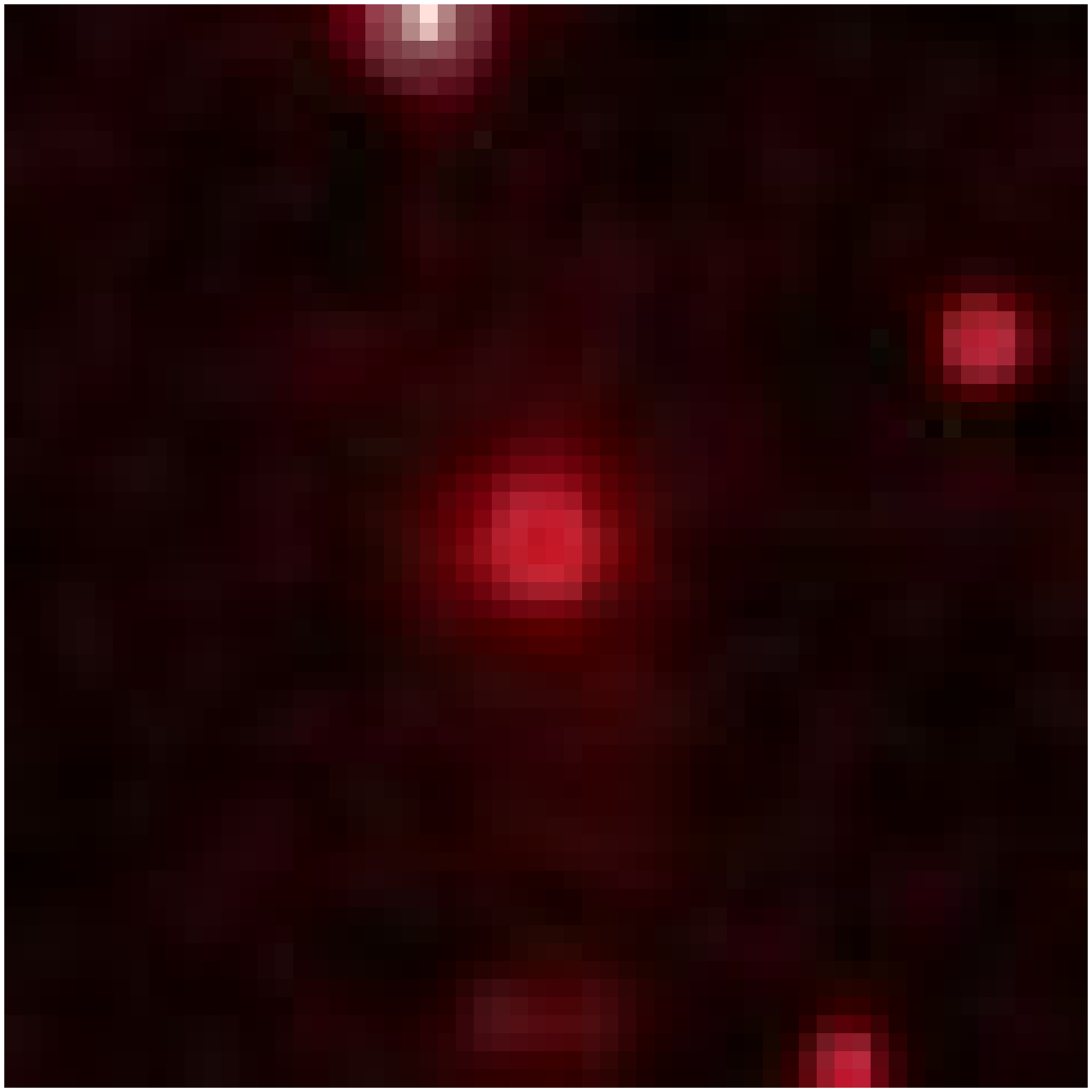}  \FGb{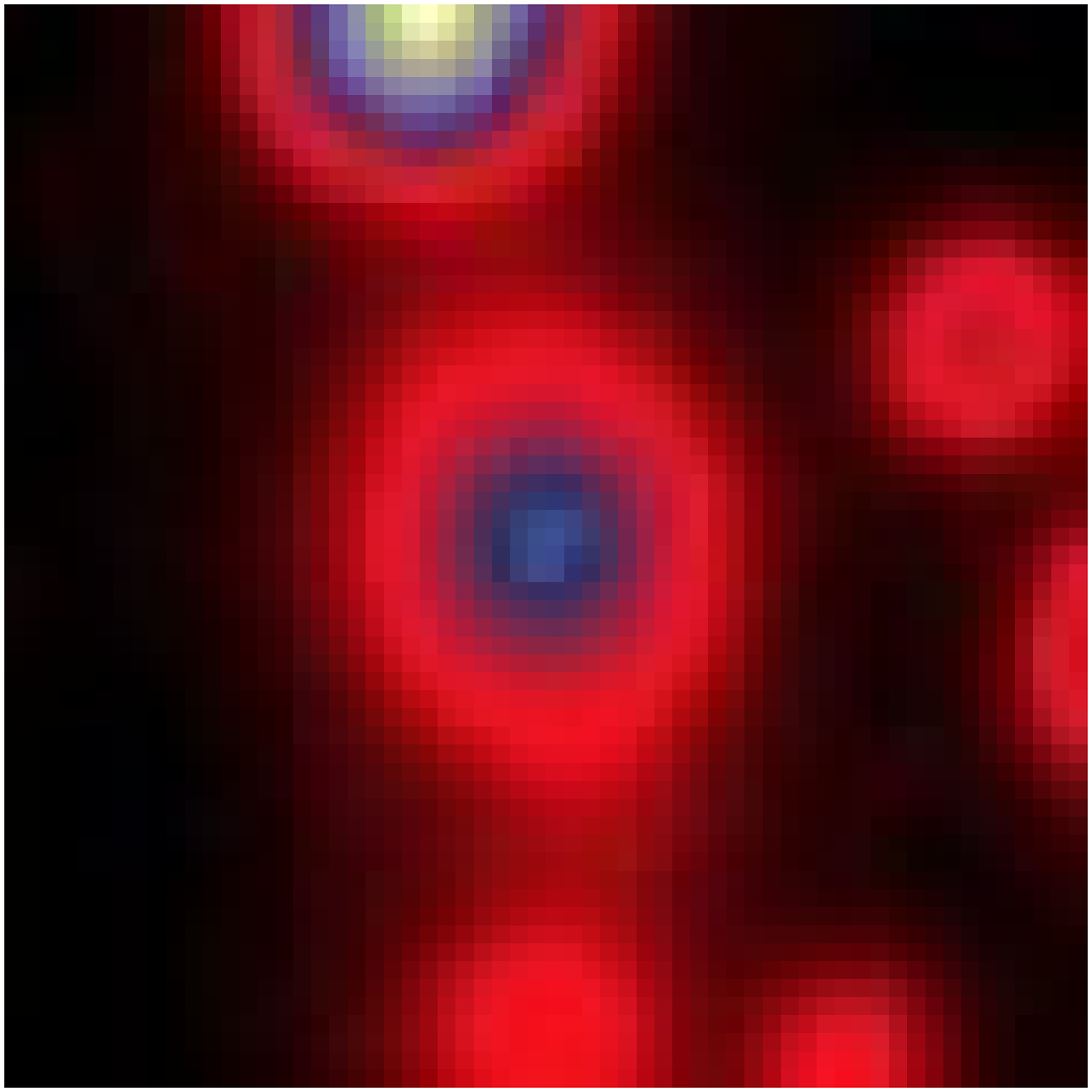}  \FGb{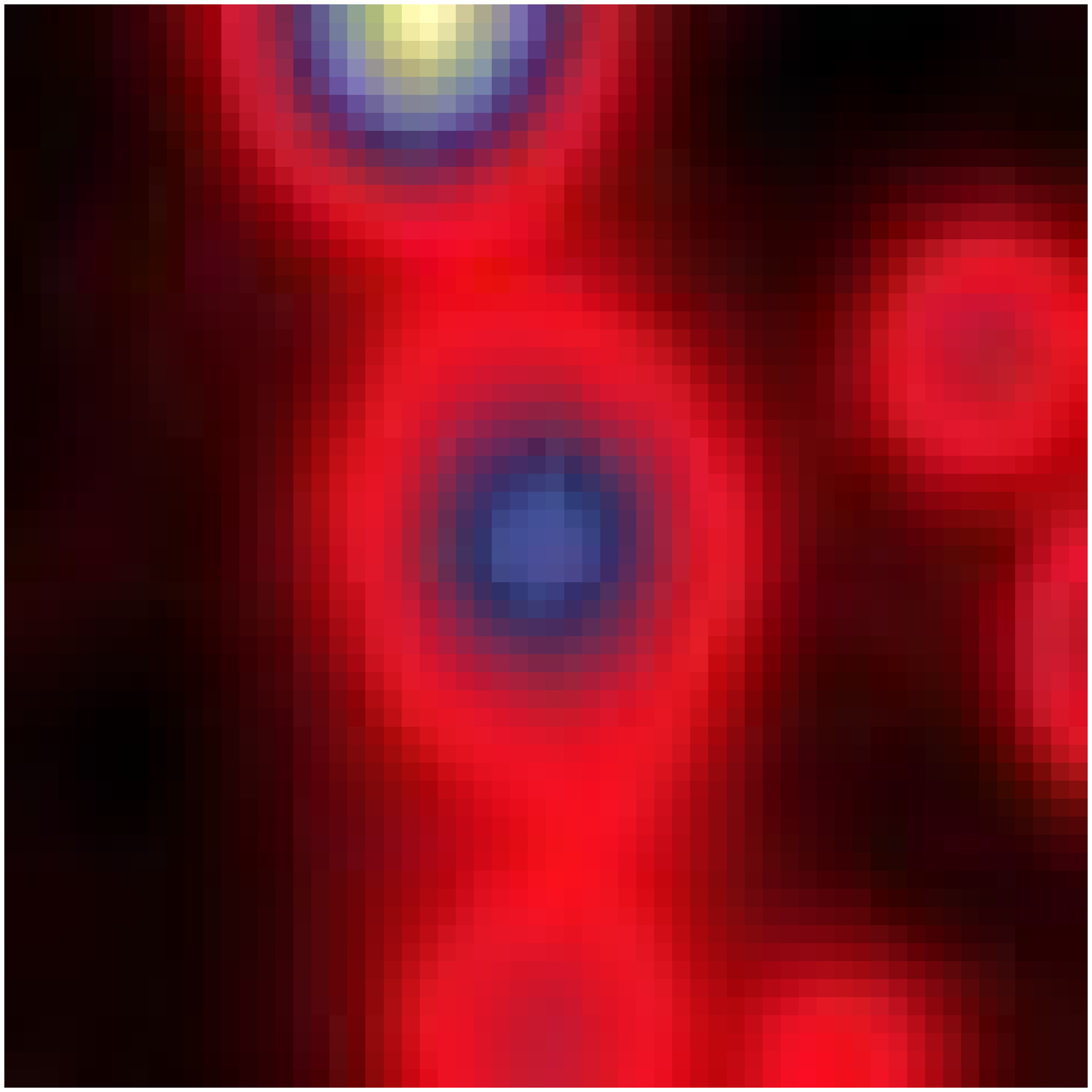}  \FGb{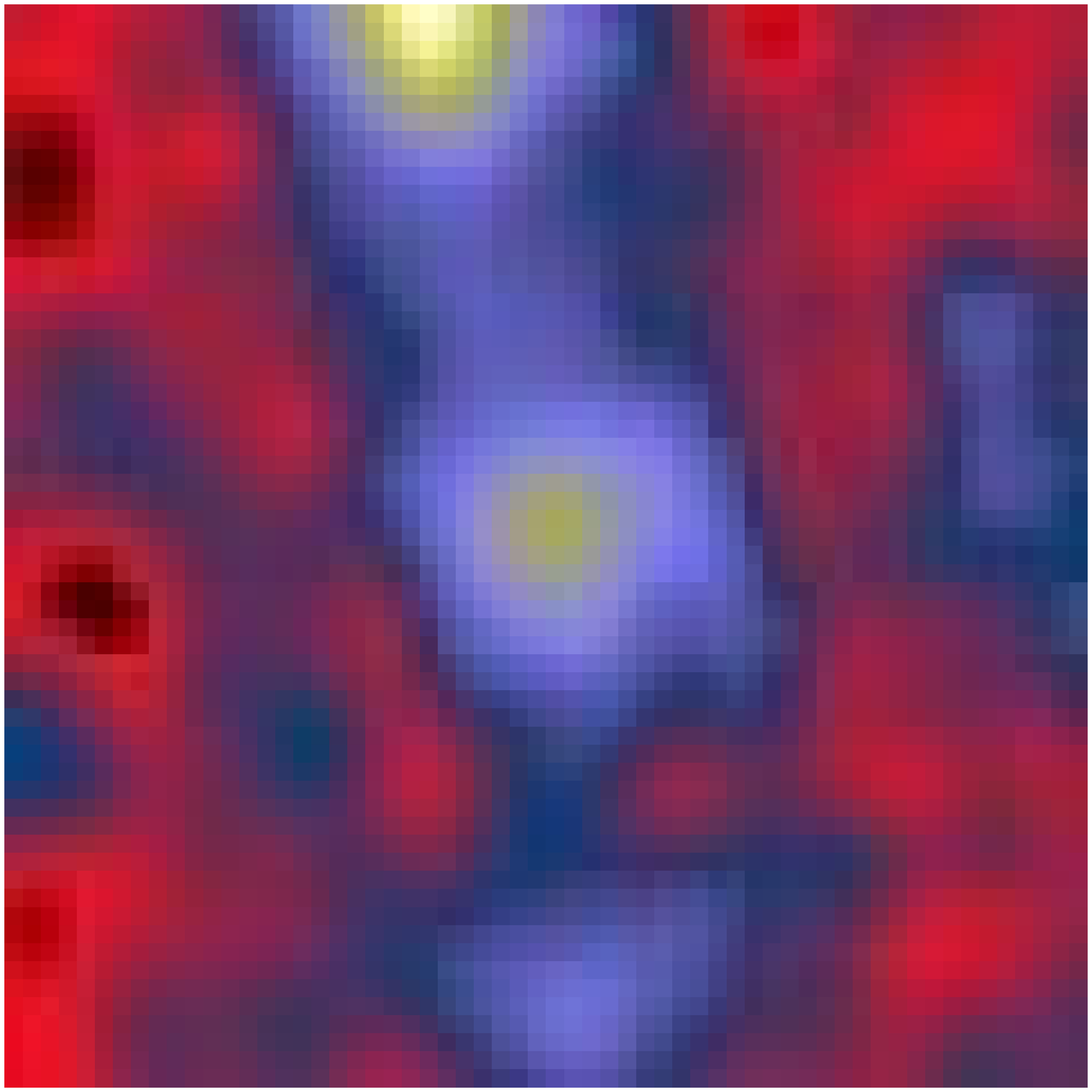}  \FGb{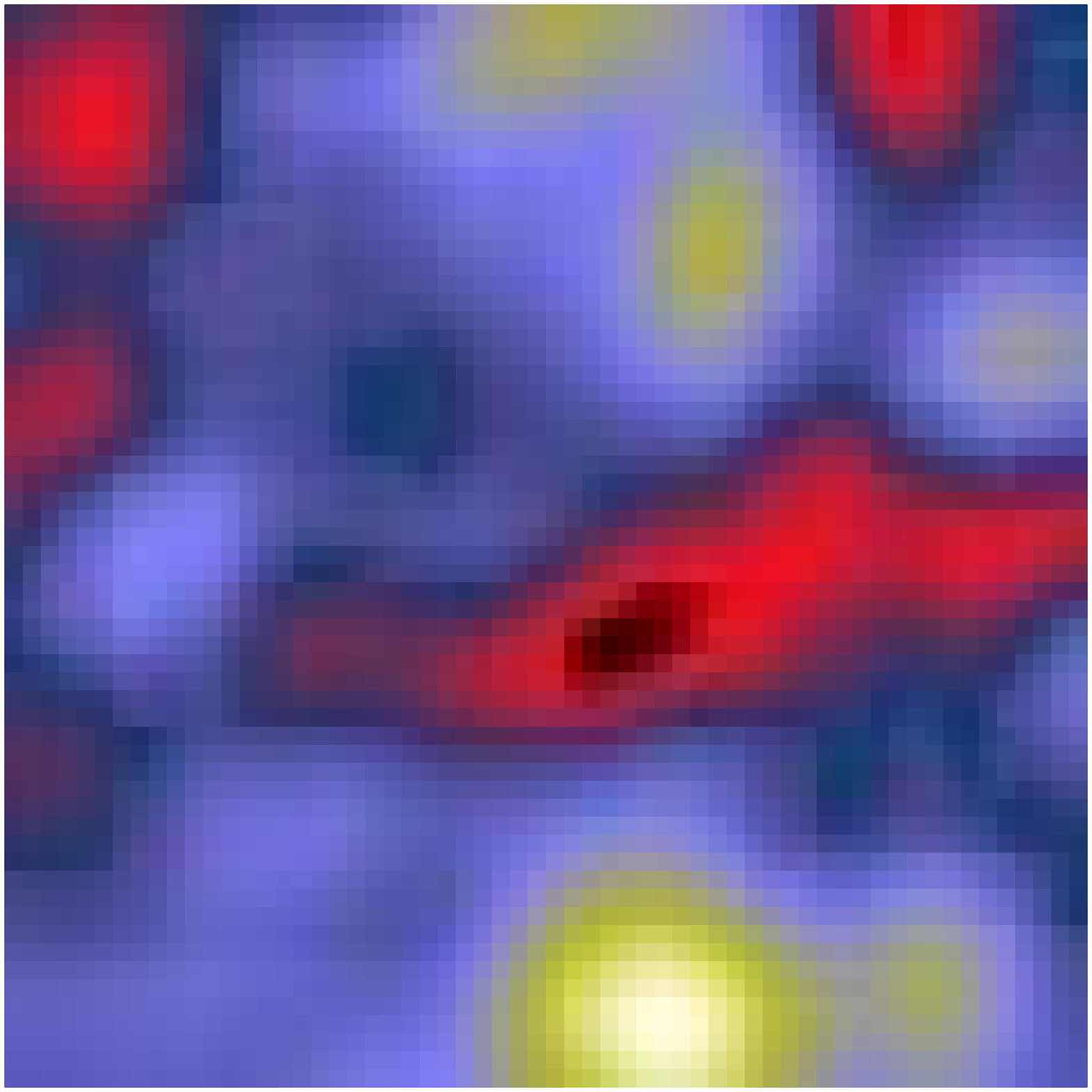} \\  {\#64   NCIA000260}  \\
  \FGb{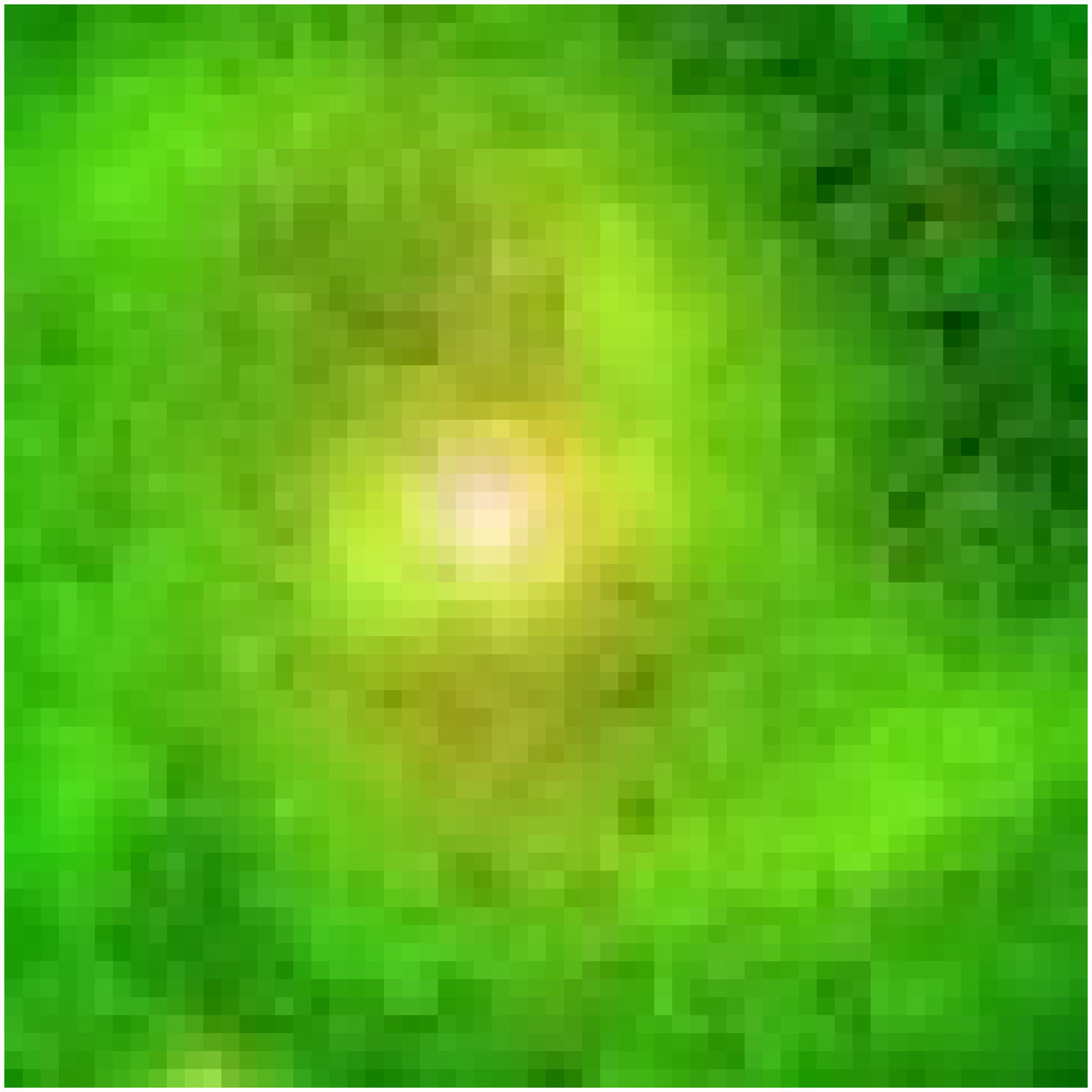}  \FGb{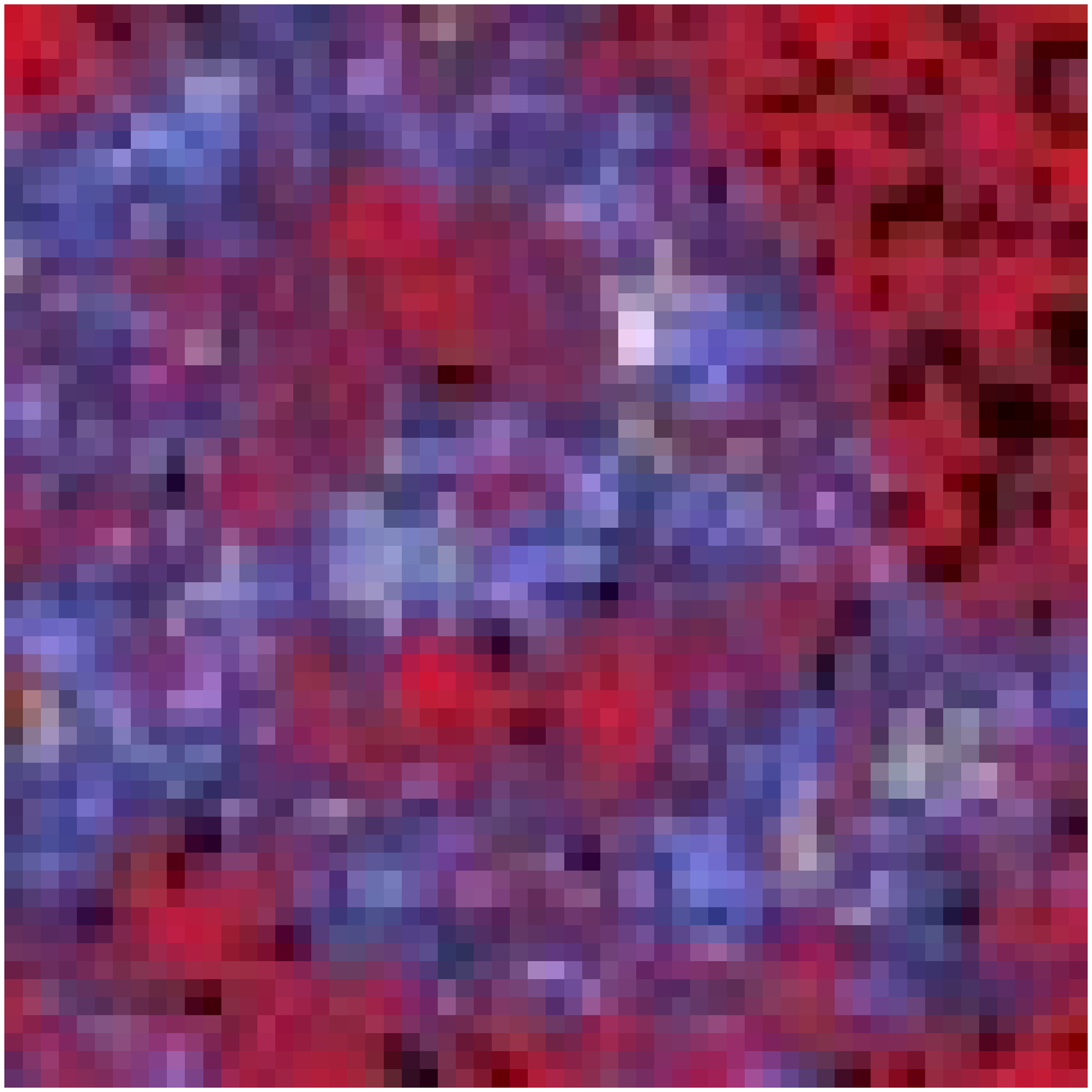}  \FGb{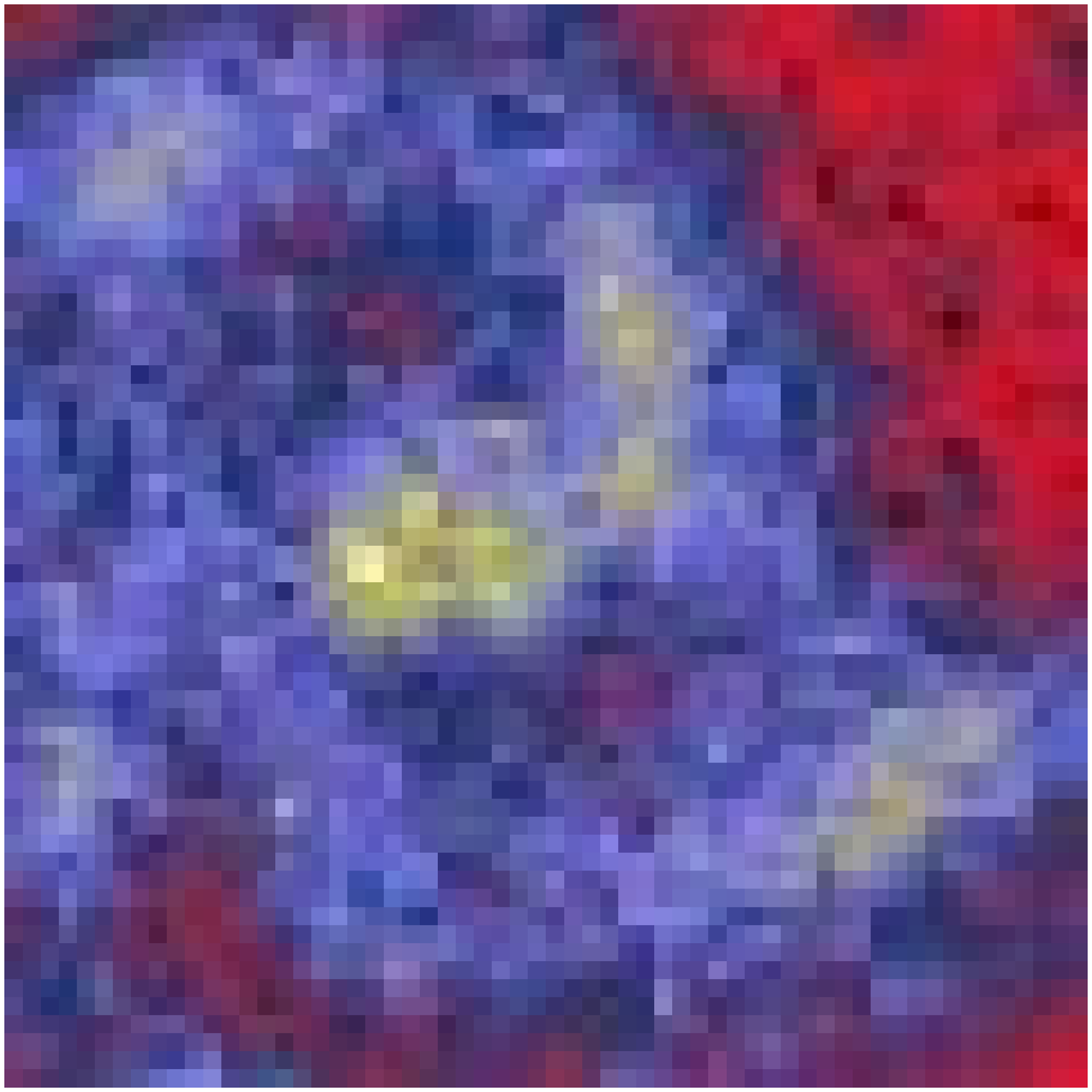}  \FGb{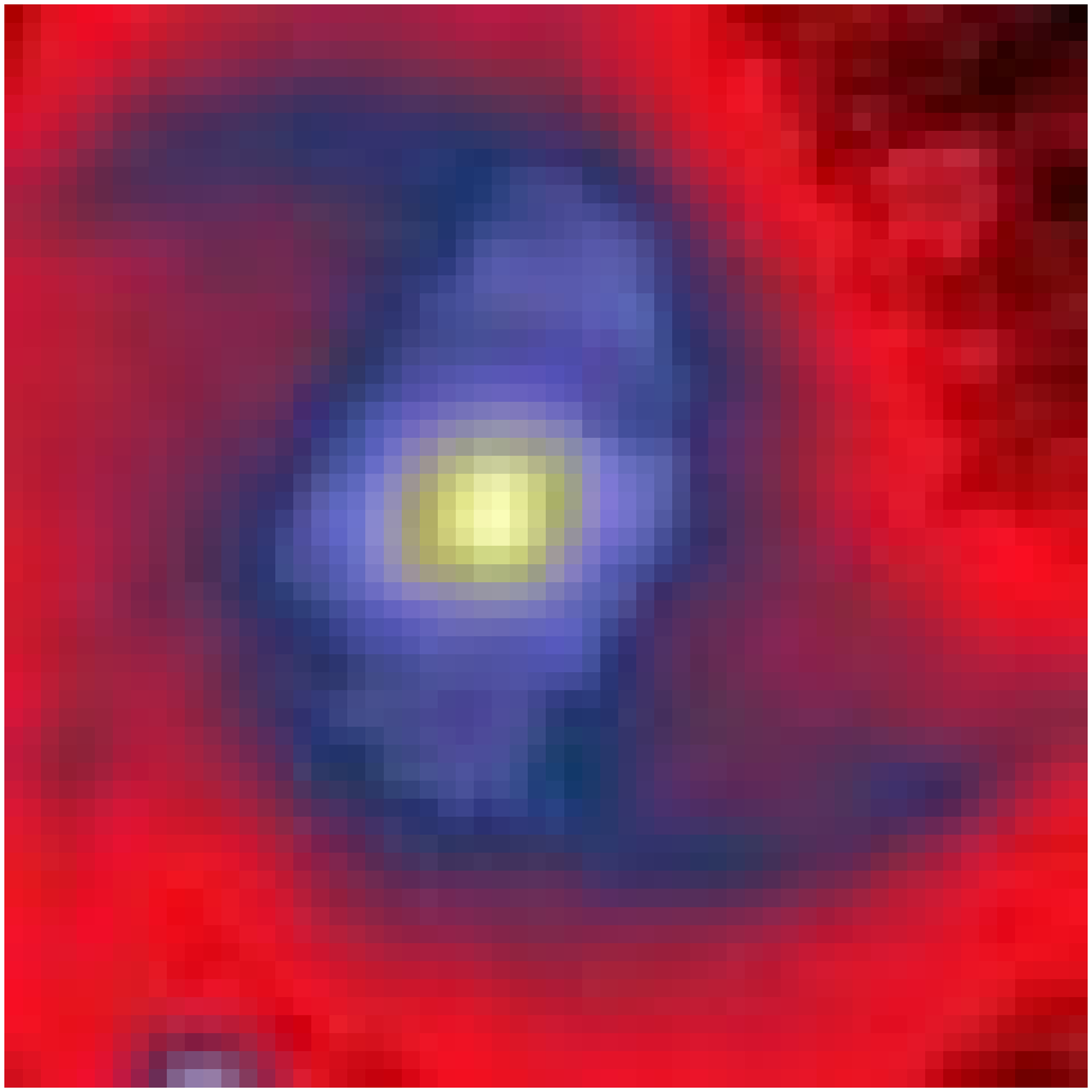}  \FGb{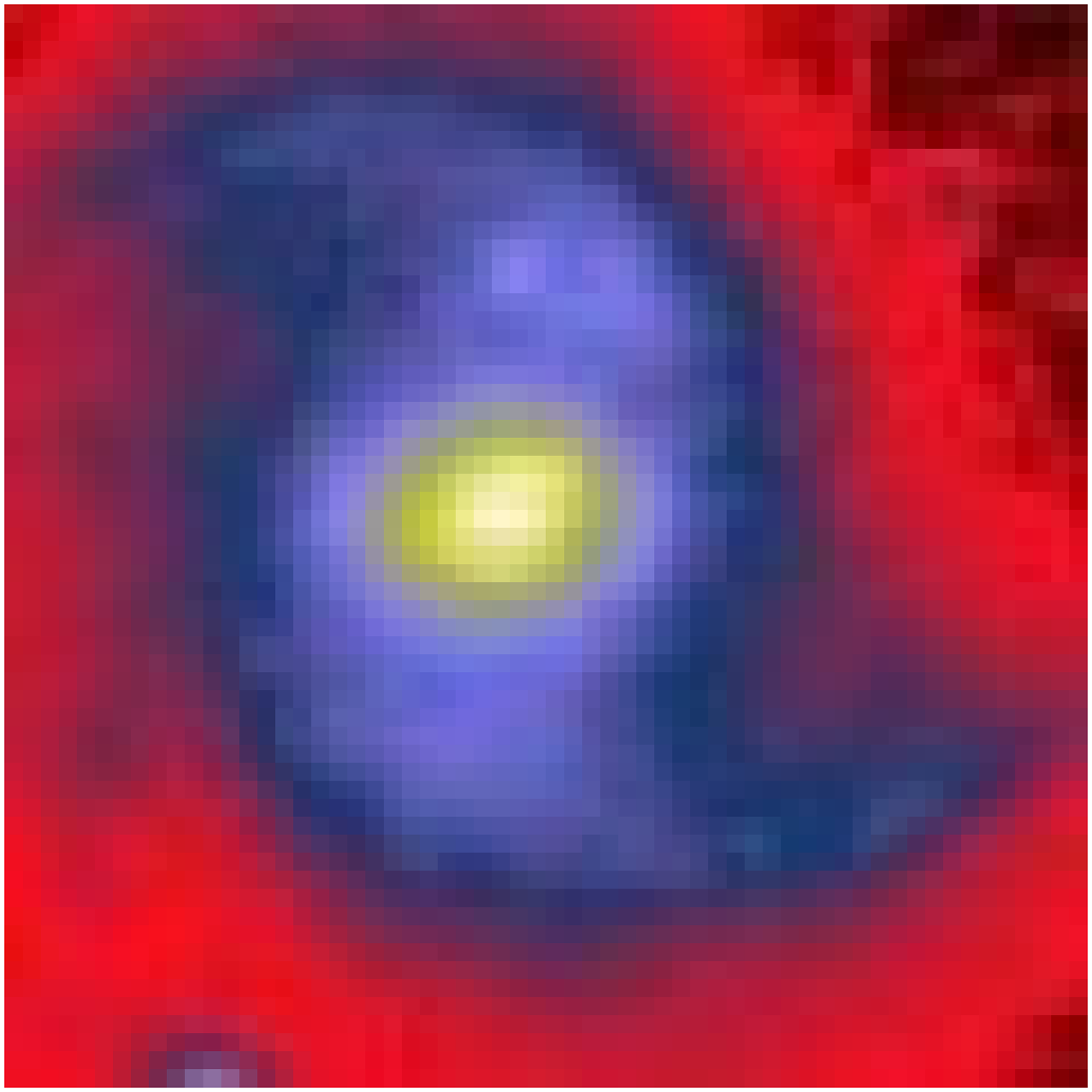}  \FGb{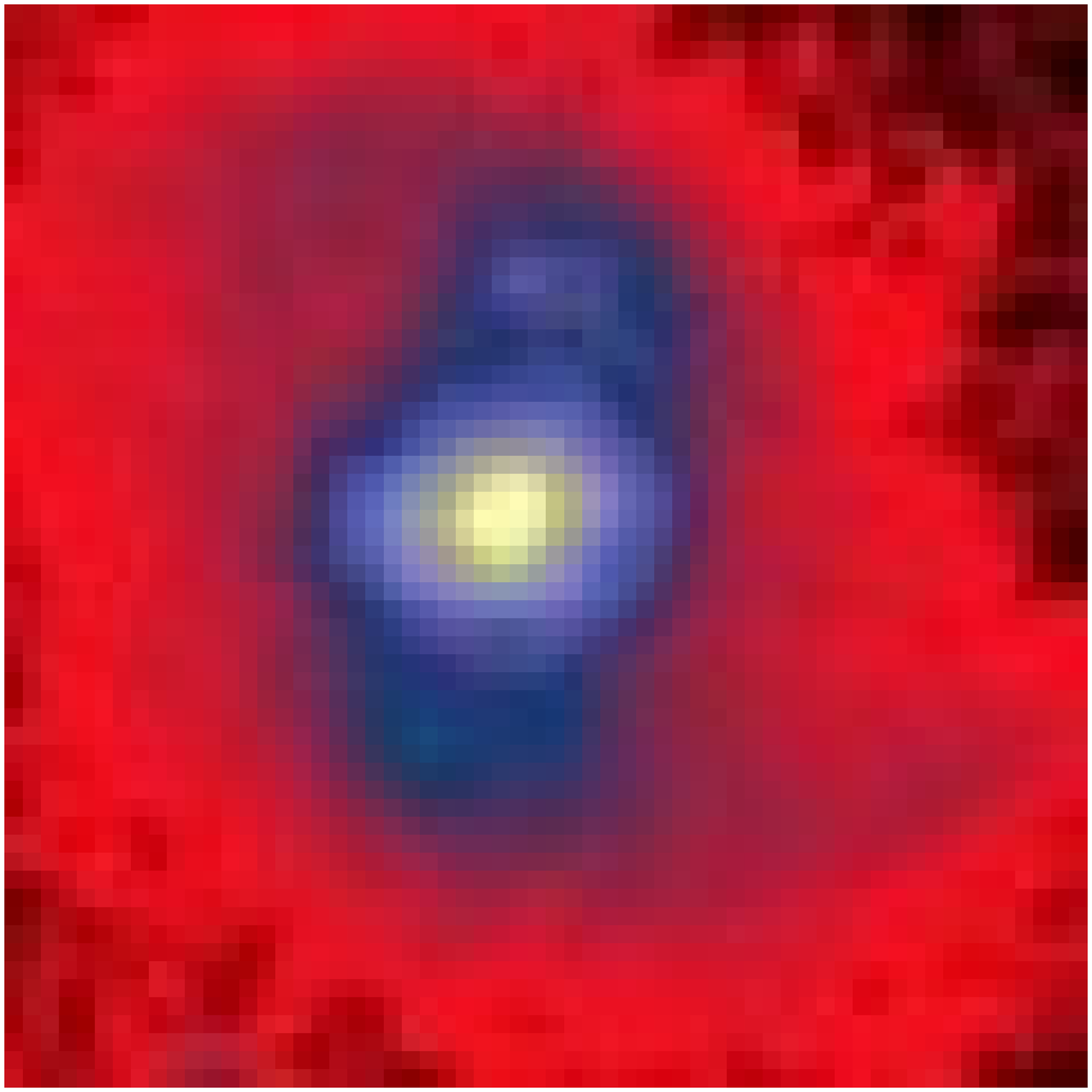}  \FGb{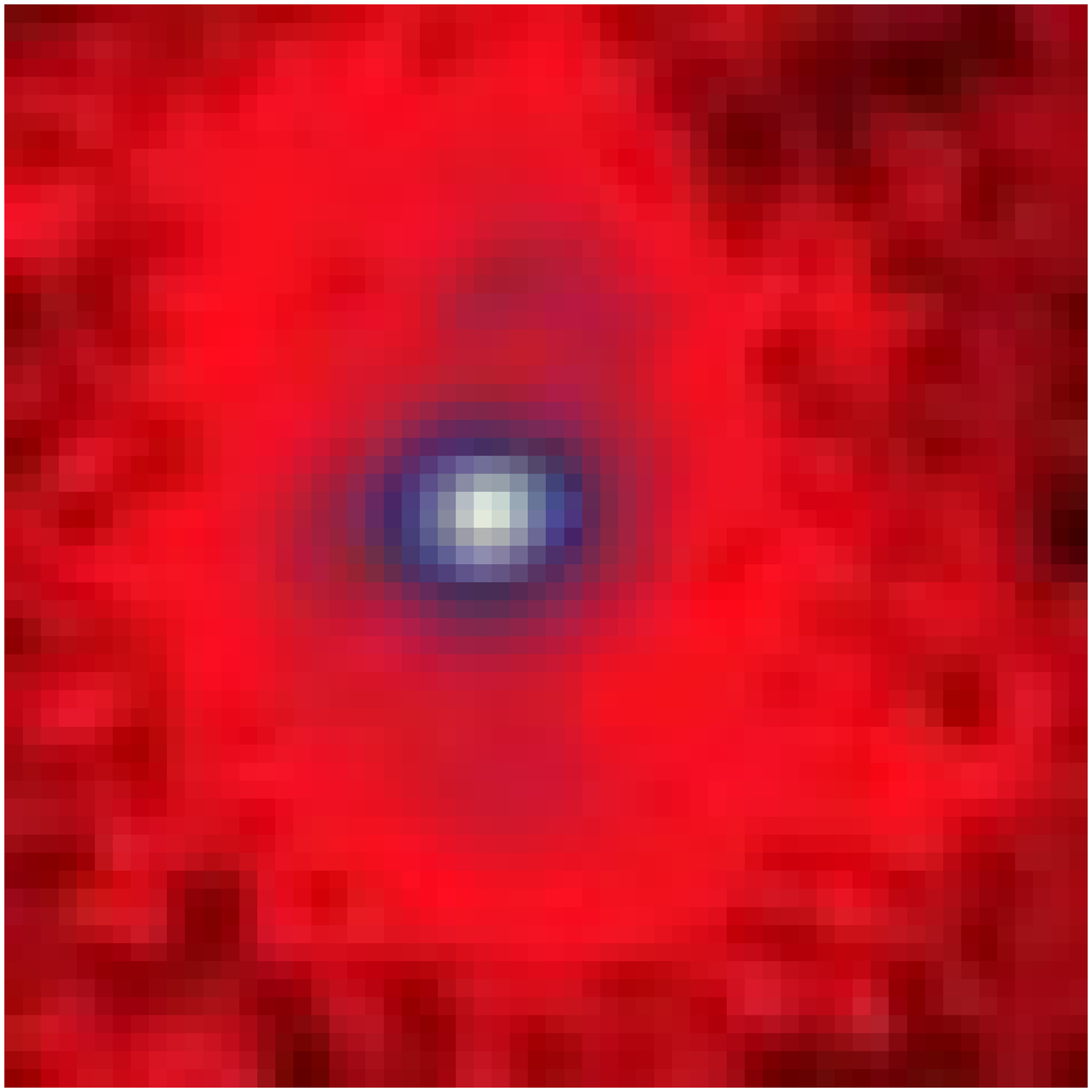}  \FGb{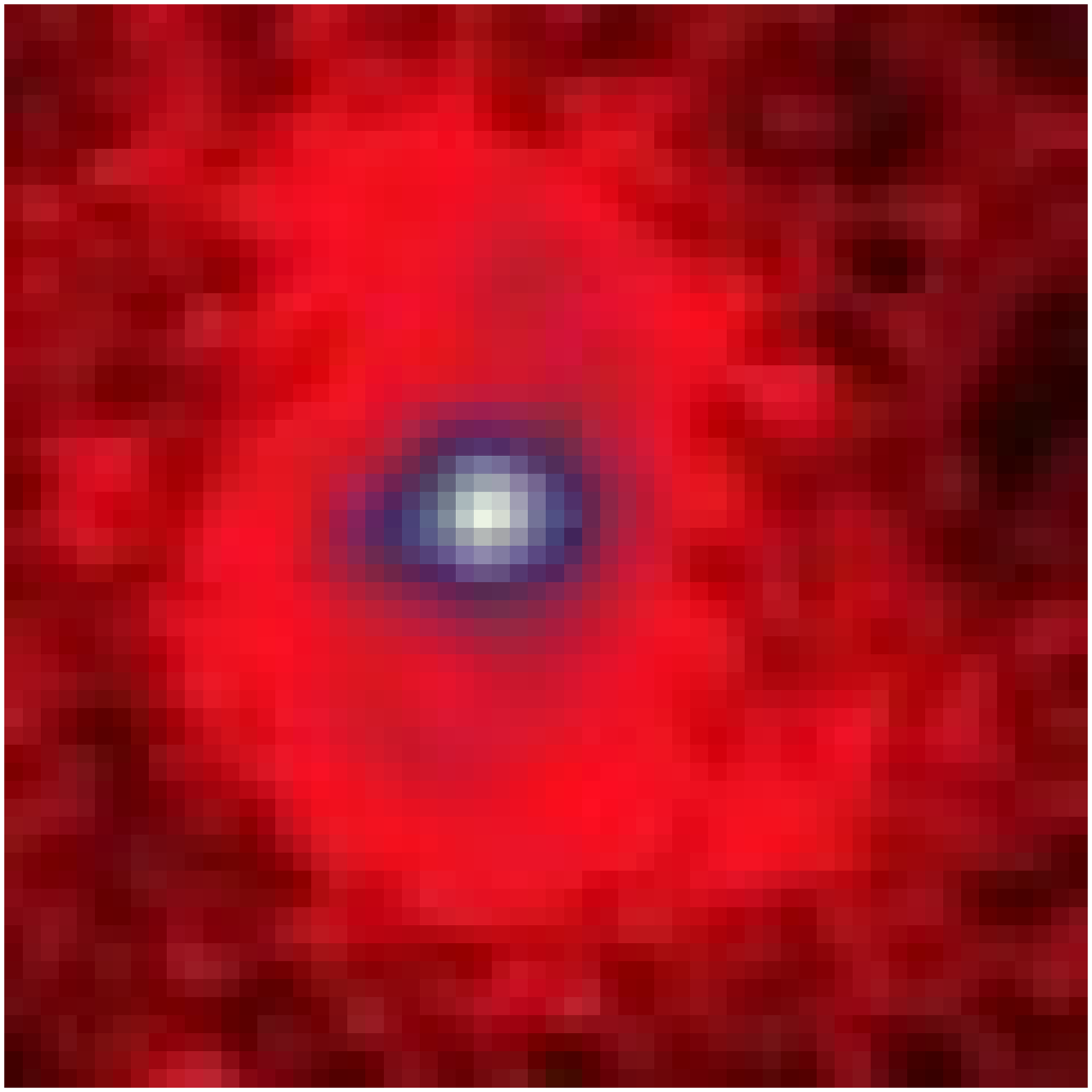}  \FGb{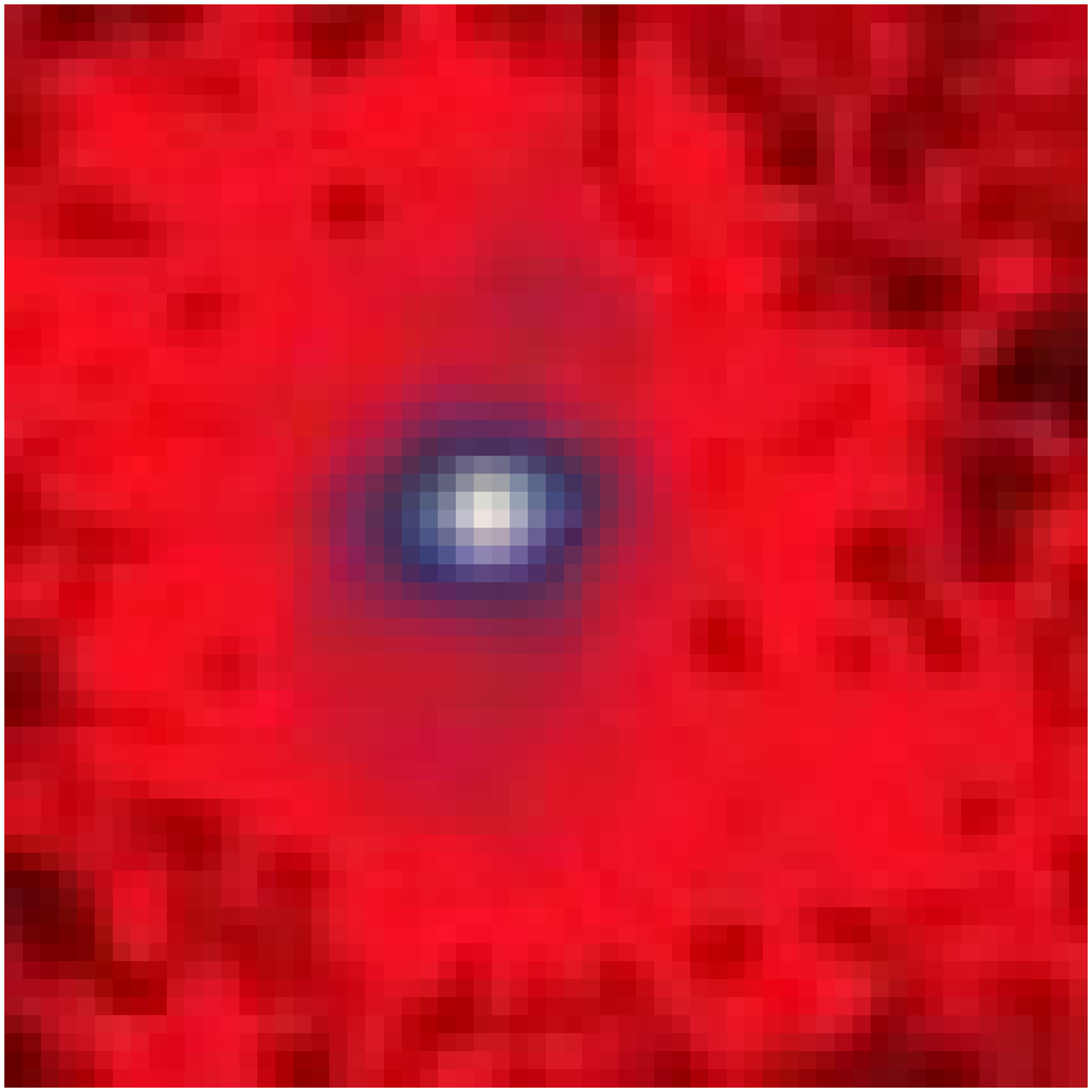}  \FGb{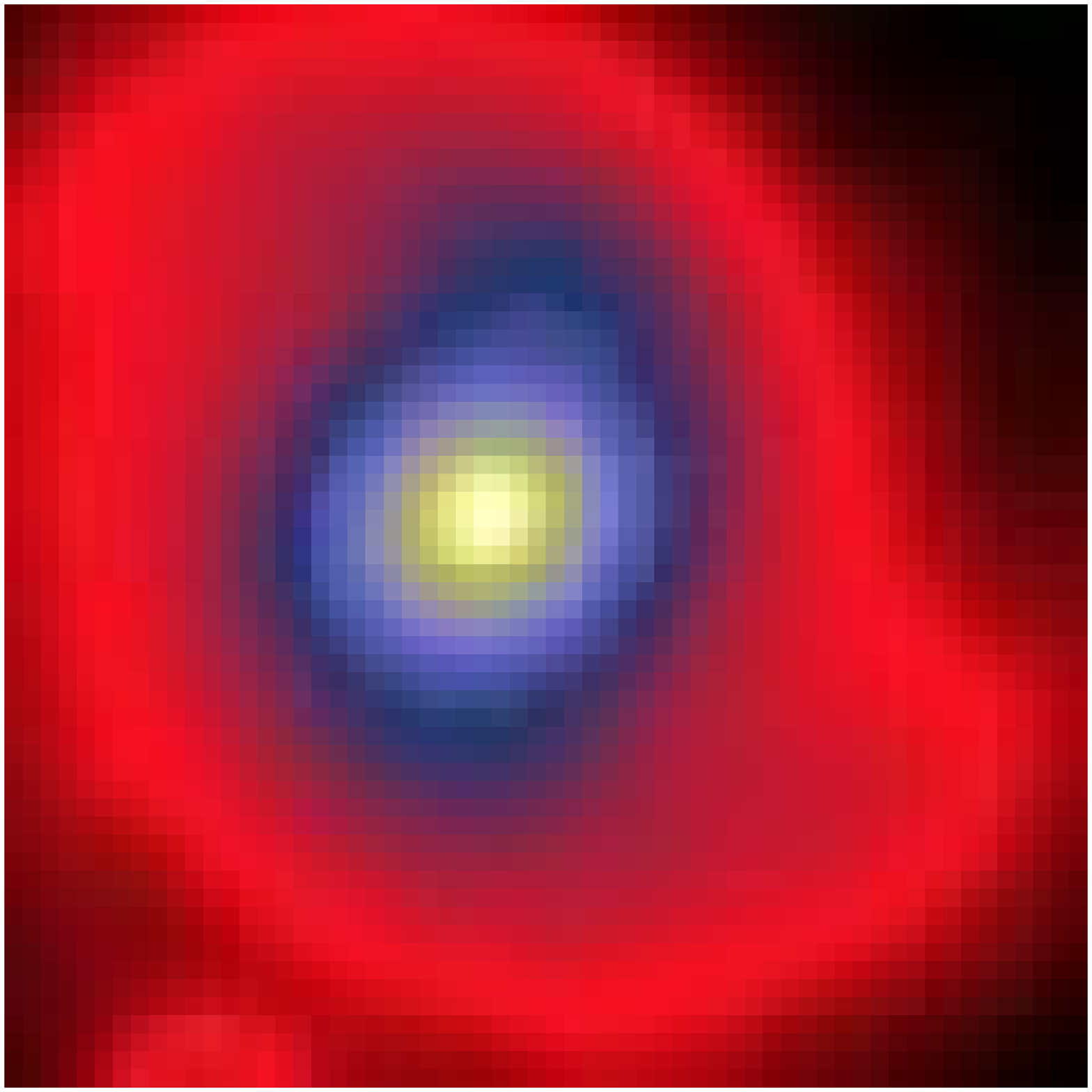}  \FGb{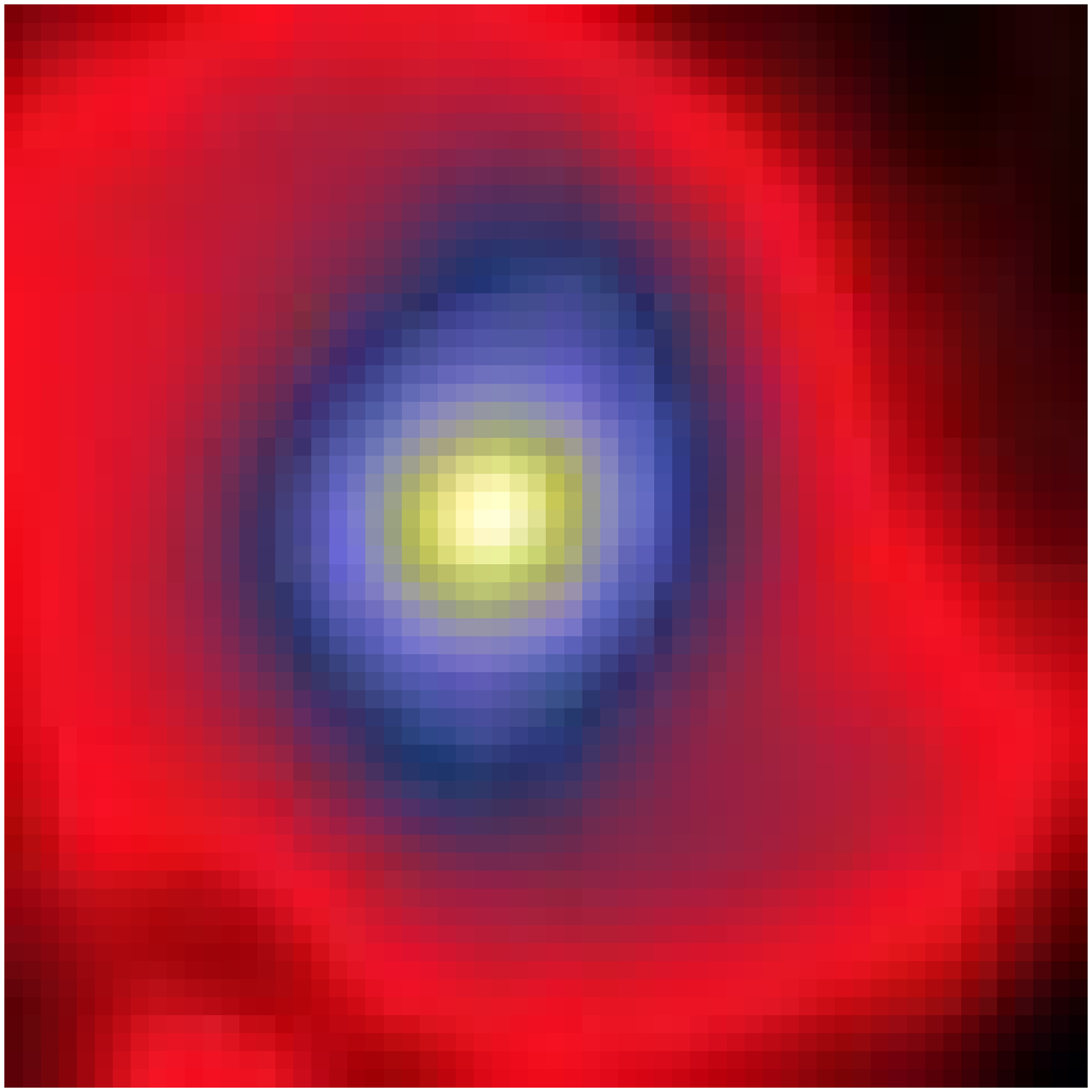}  \FGb{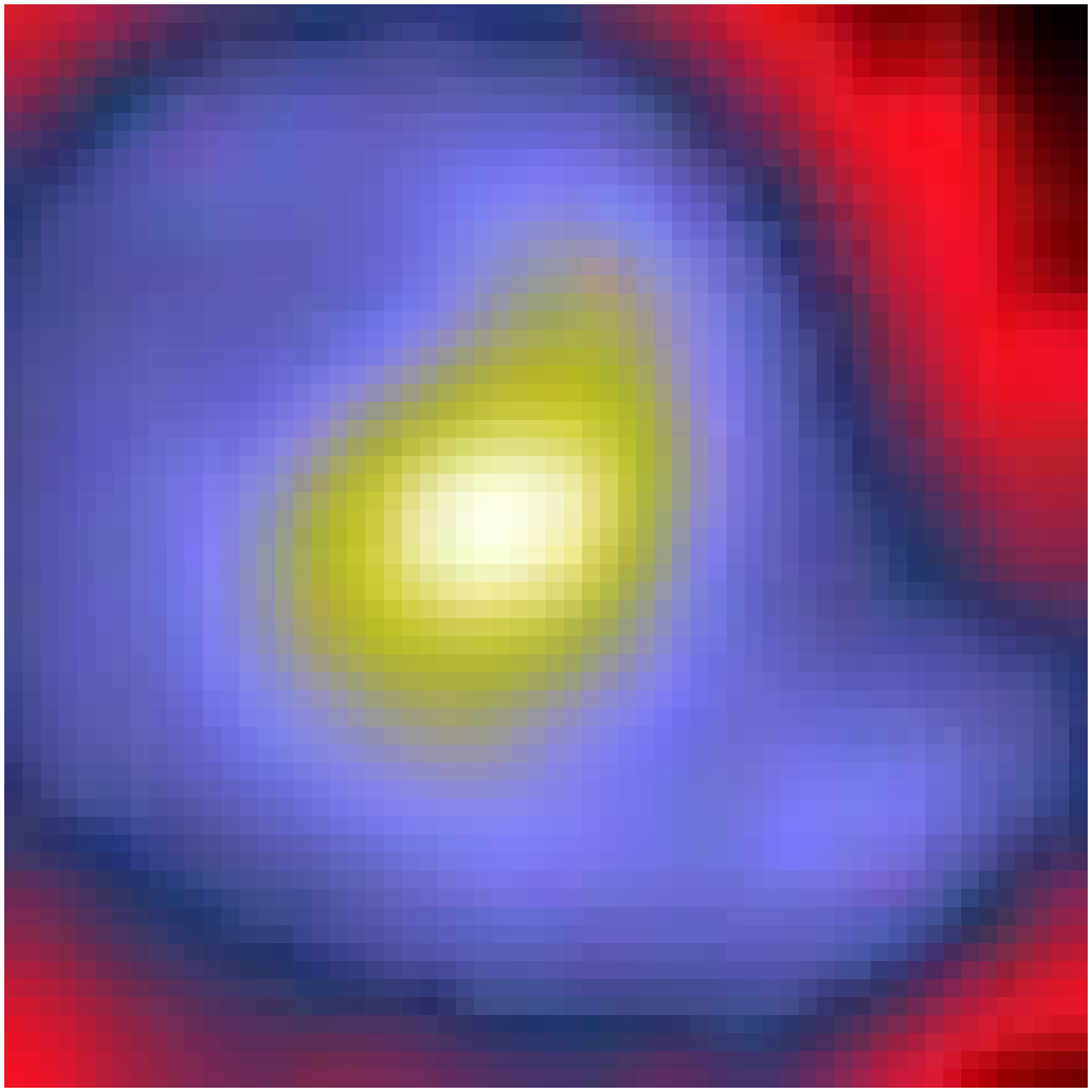}  \FGb{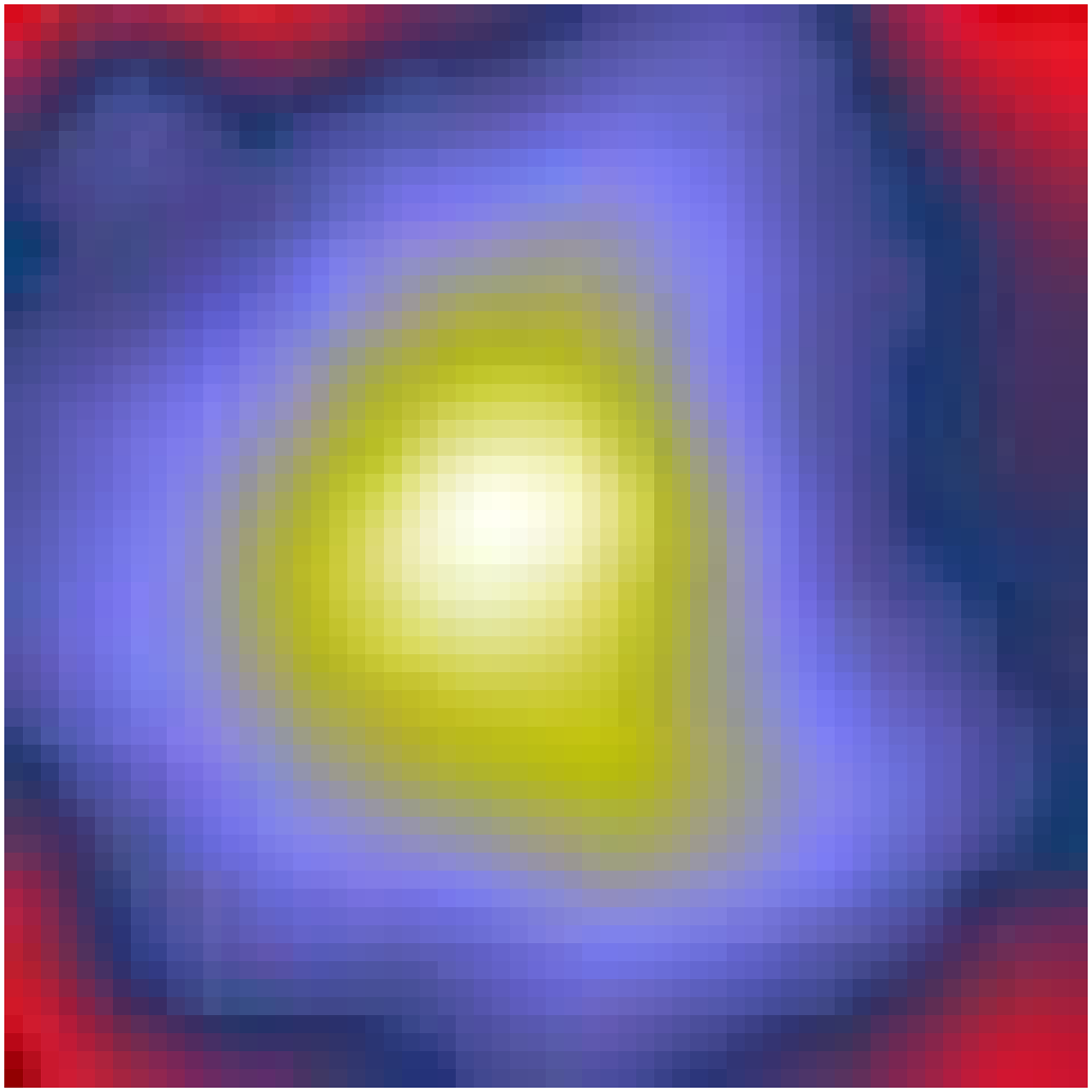} \\  {\#65   NCIA001051}  \\
{ \\ \textbf{Figure \ref{fig:allgal}.(cont)} ~~Visible Galaxies within 40\min\ of NGC~891 } \\
\end{figure}

\clearpage
\begin{figure}[ht]
  \HDb{rgb}  \HDb{FUV}  \HDb{NUV}   \HDb{B}     \HDb{R}    \HDb{IR}   \HDb{J}    \HDb{H}     \HDb{K}     \HDb{W1}    \HDb{W2}    \HDb{W3}    \HDb{W4}  \\
  \FGb{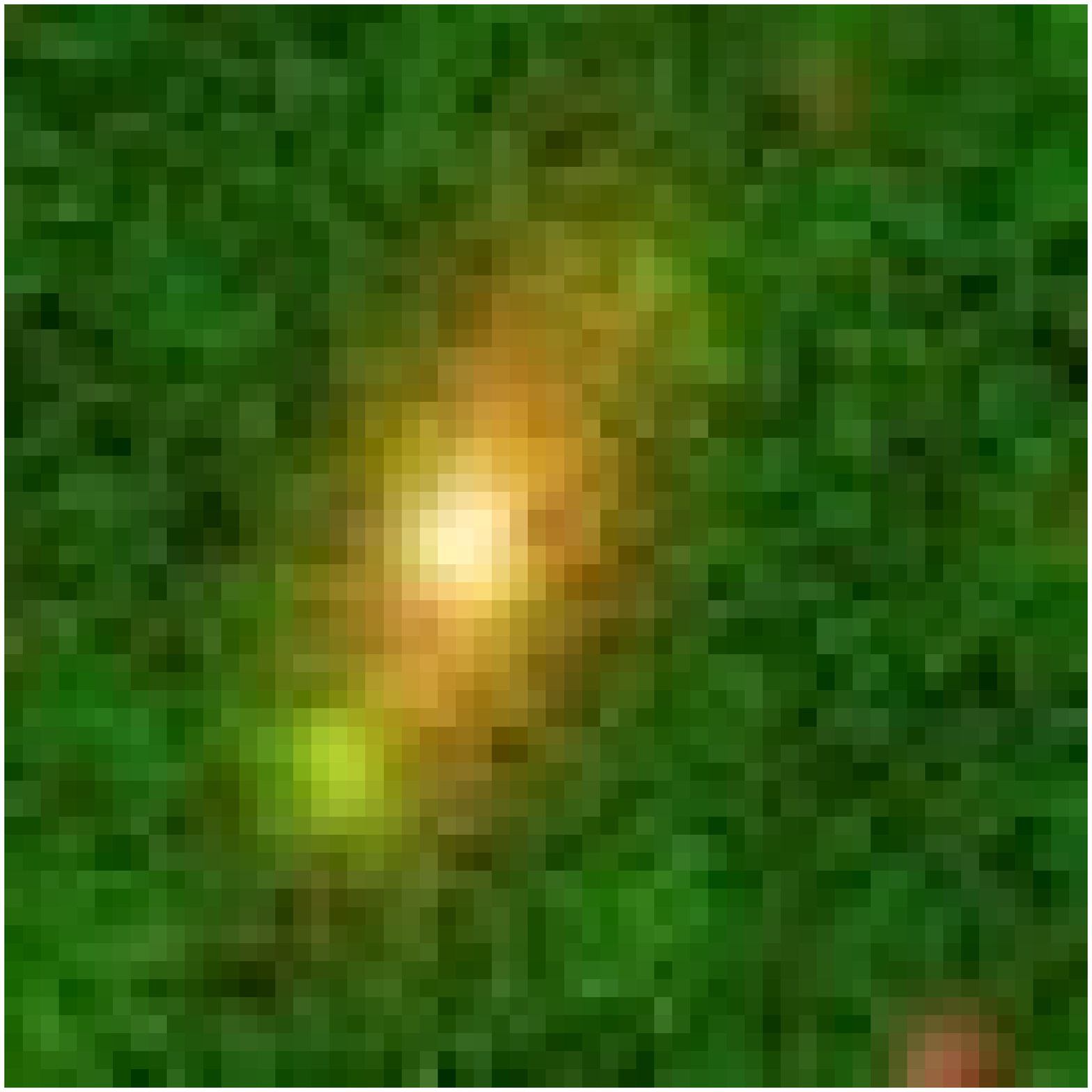}  \FGb{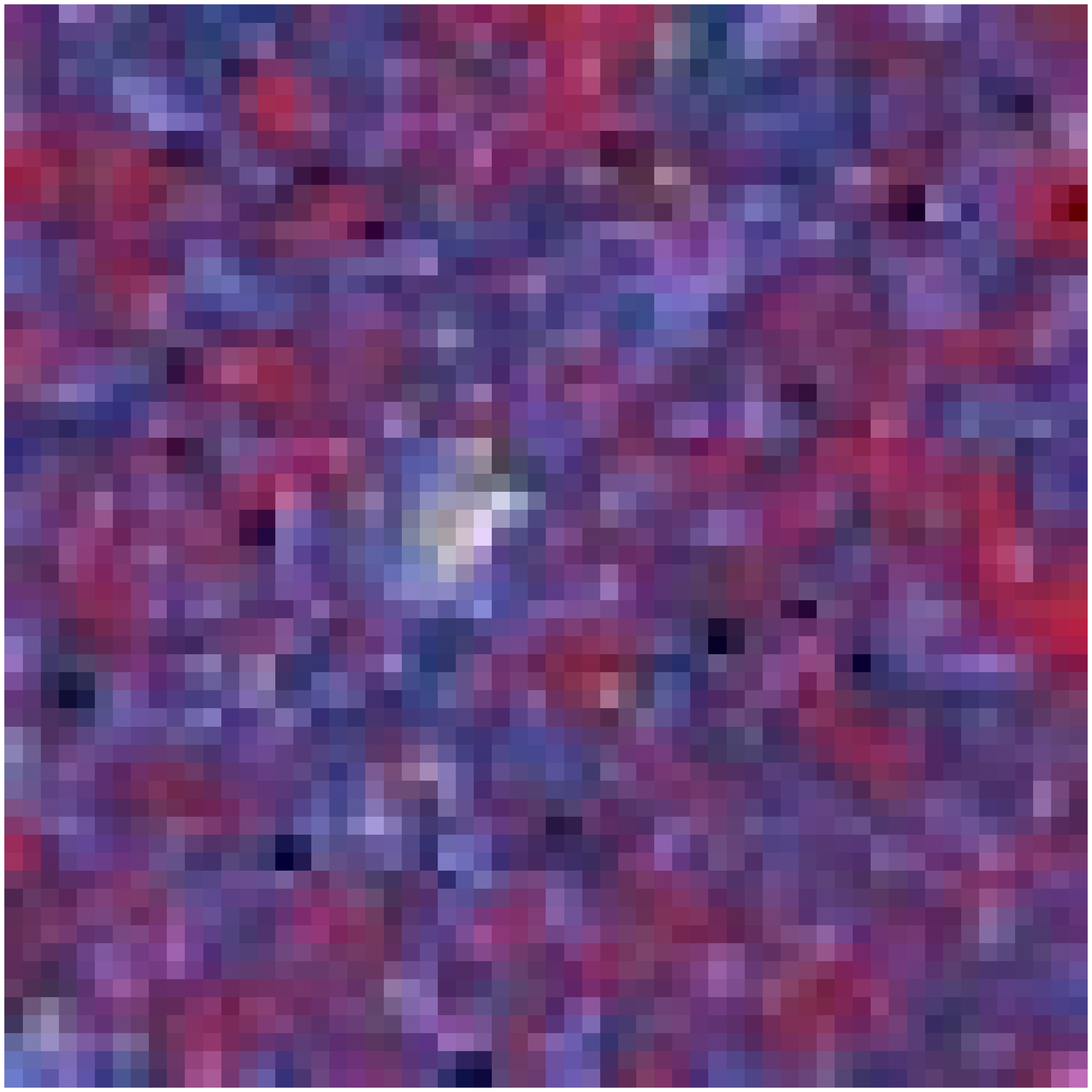}  \FGb{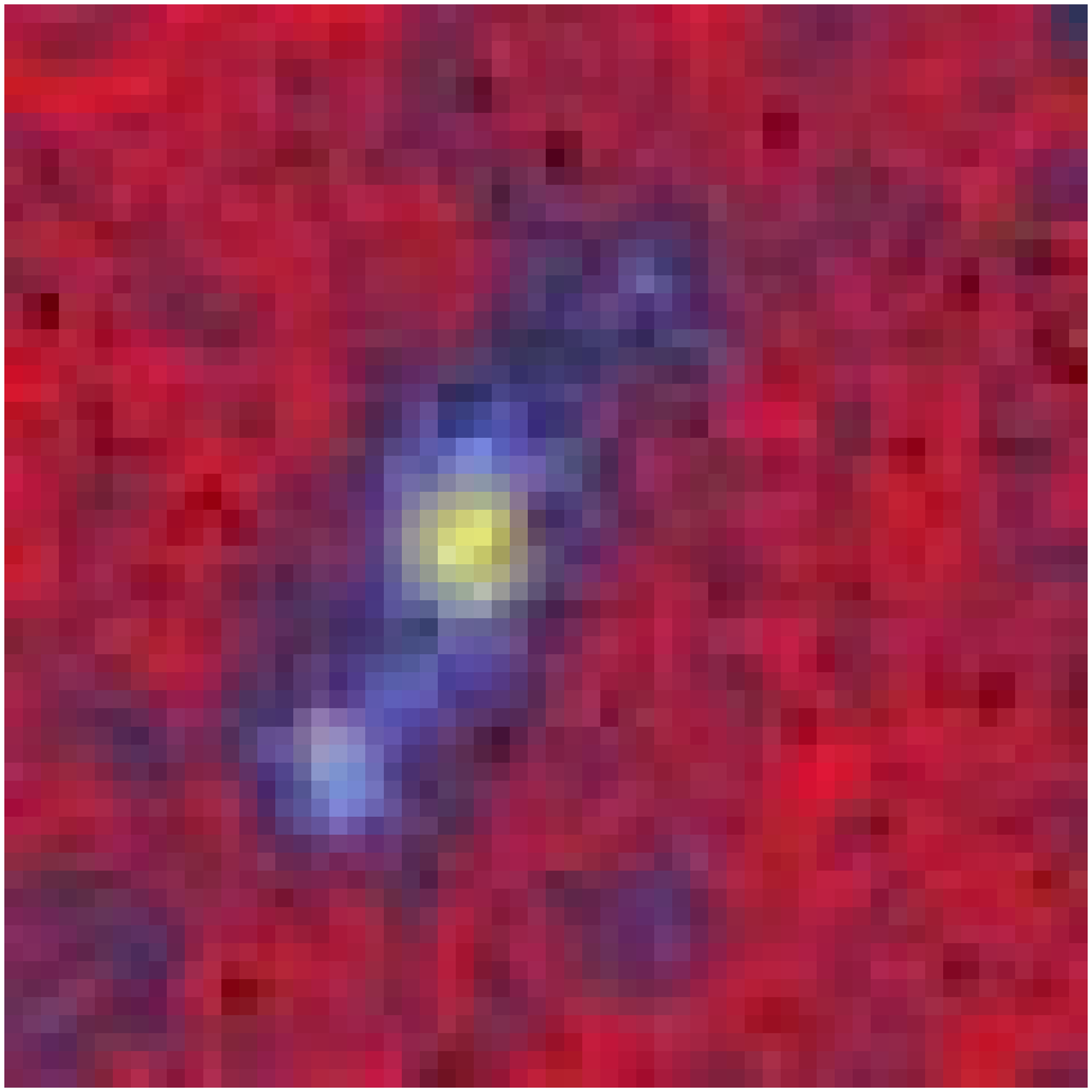}  \FGb{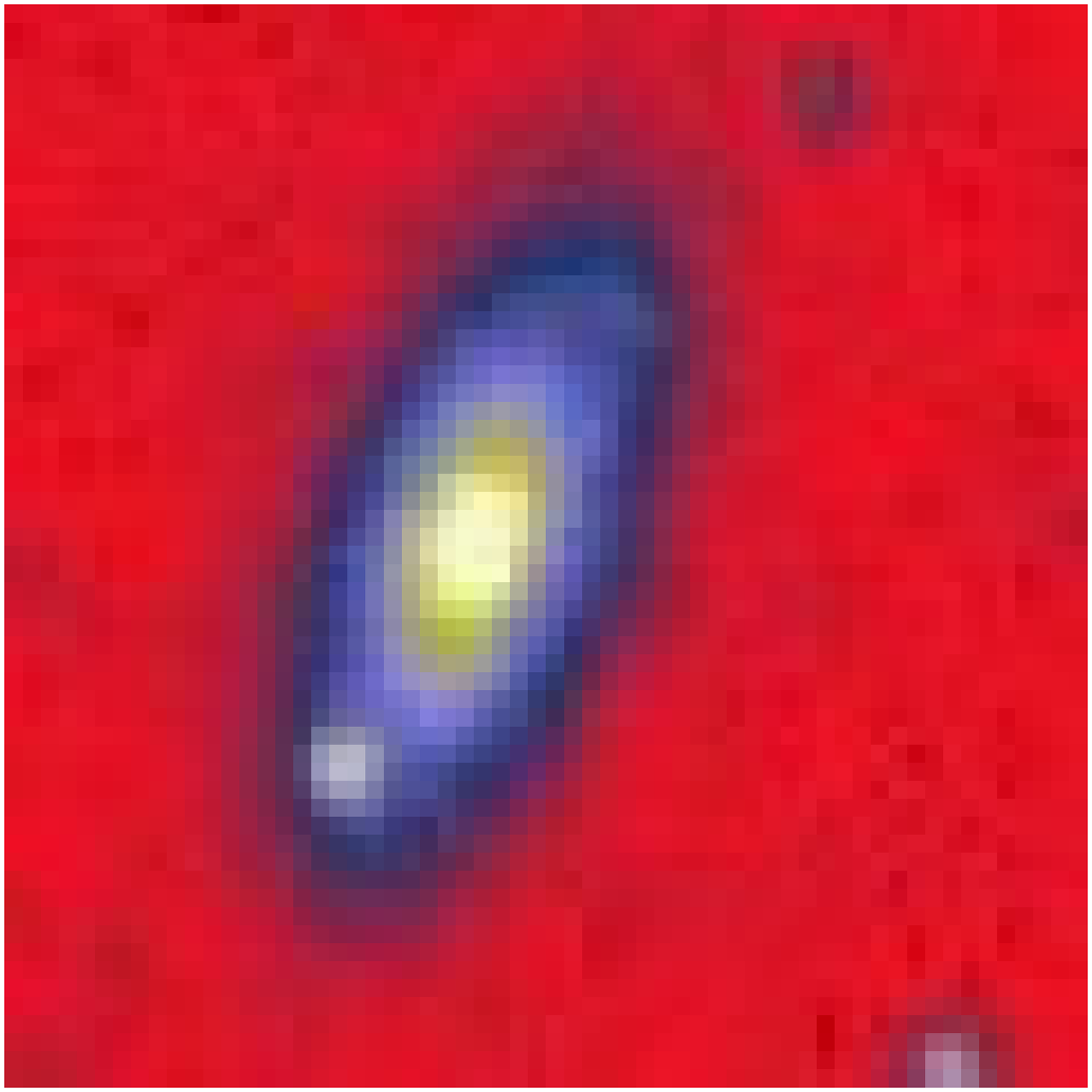}  \FGb{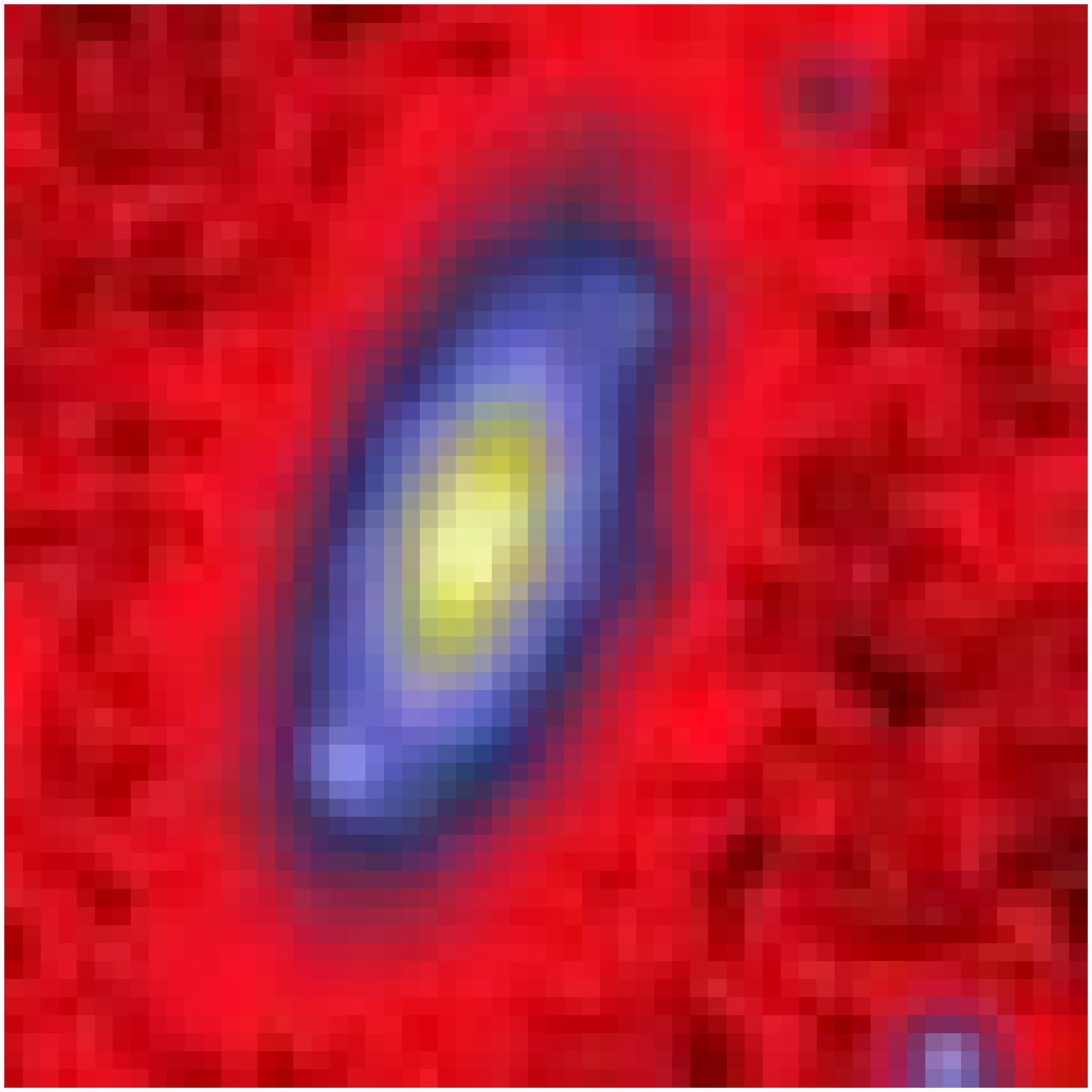}  \FGb{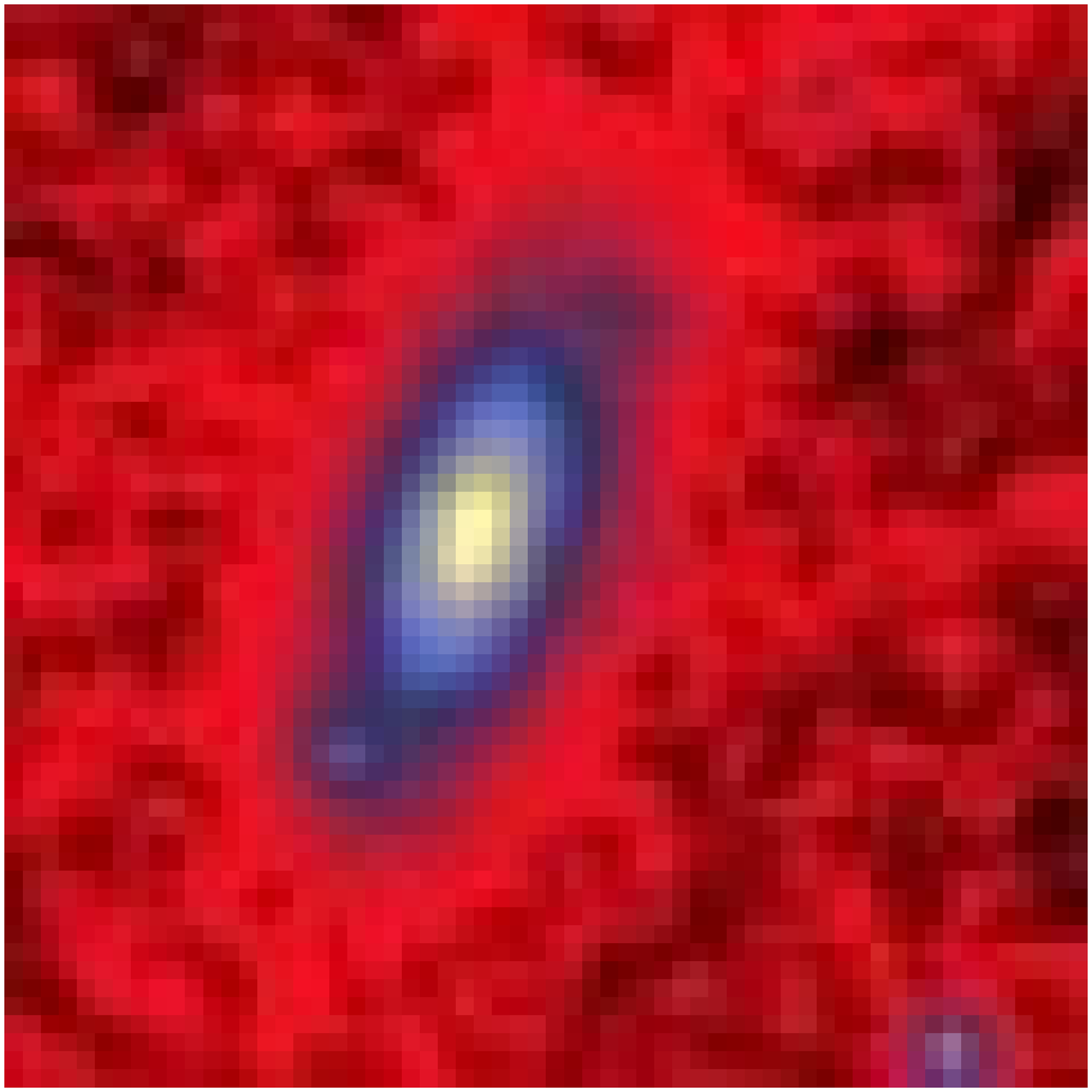}  \FGb{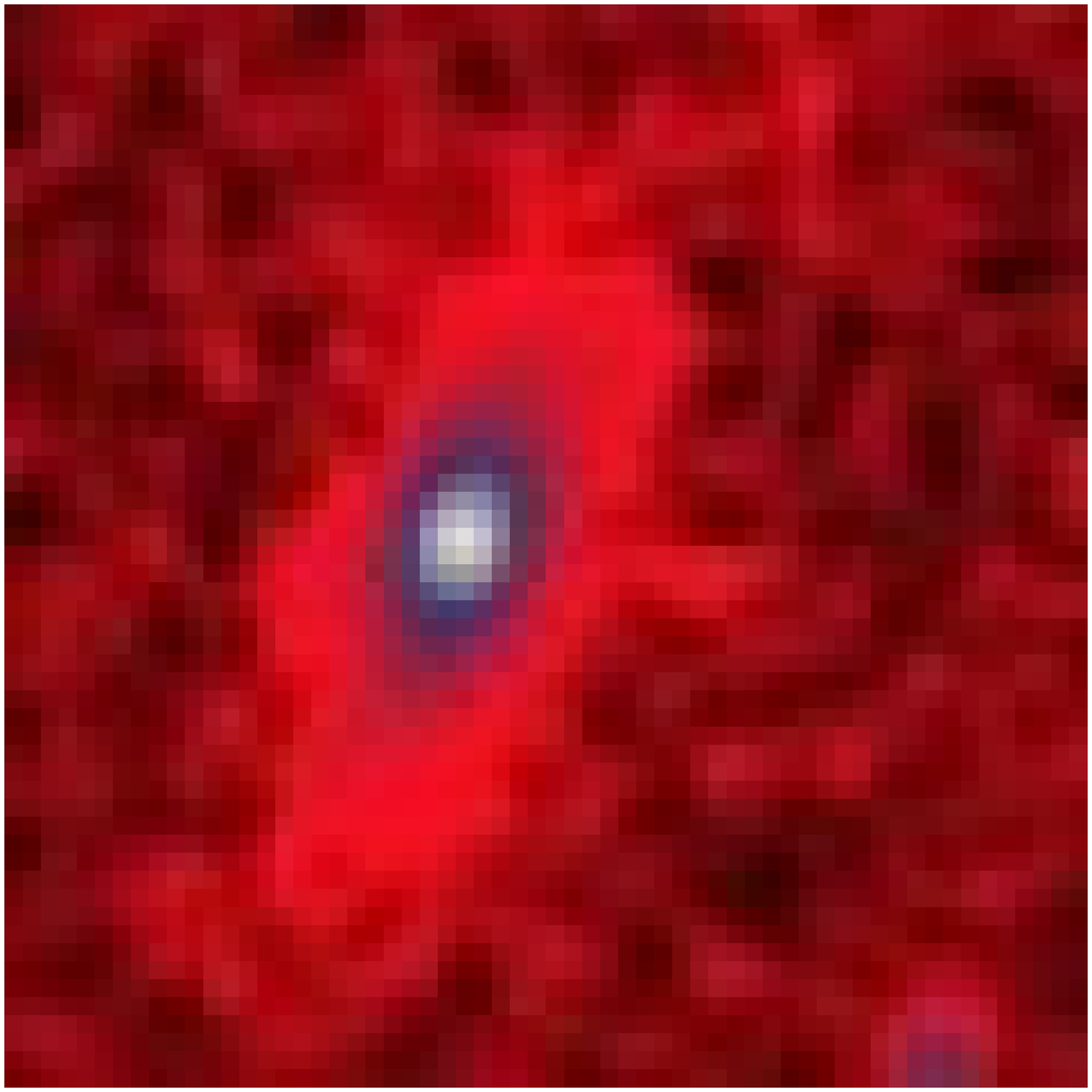}  \FGb{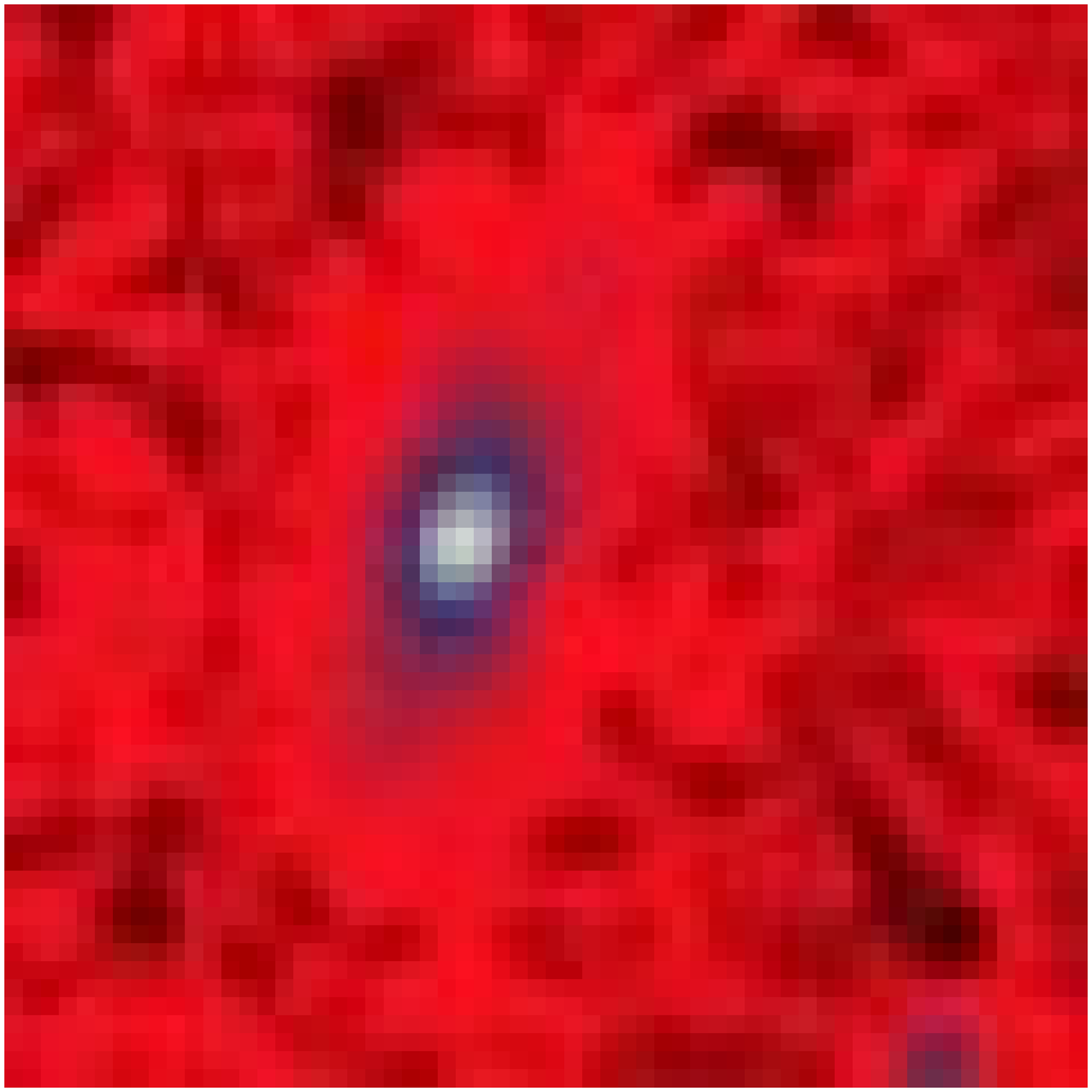}  \FGb{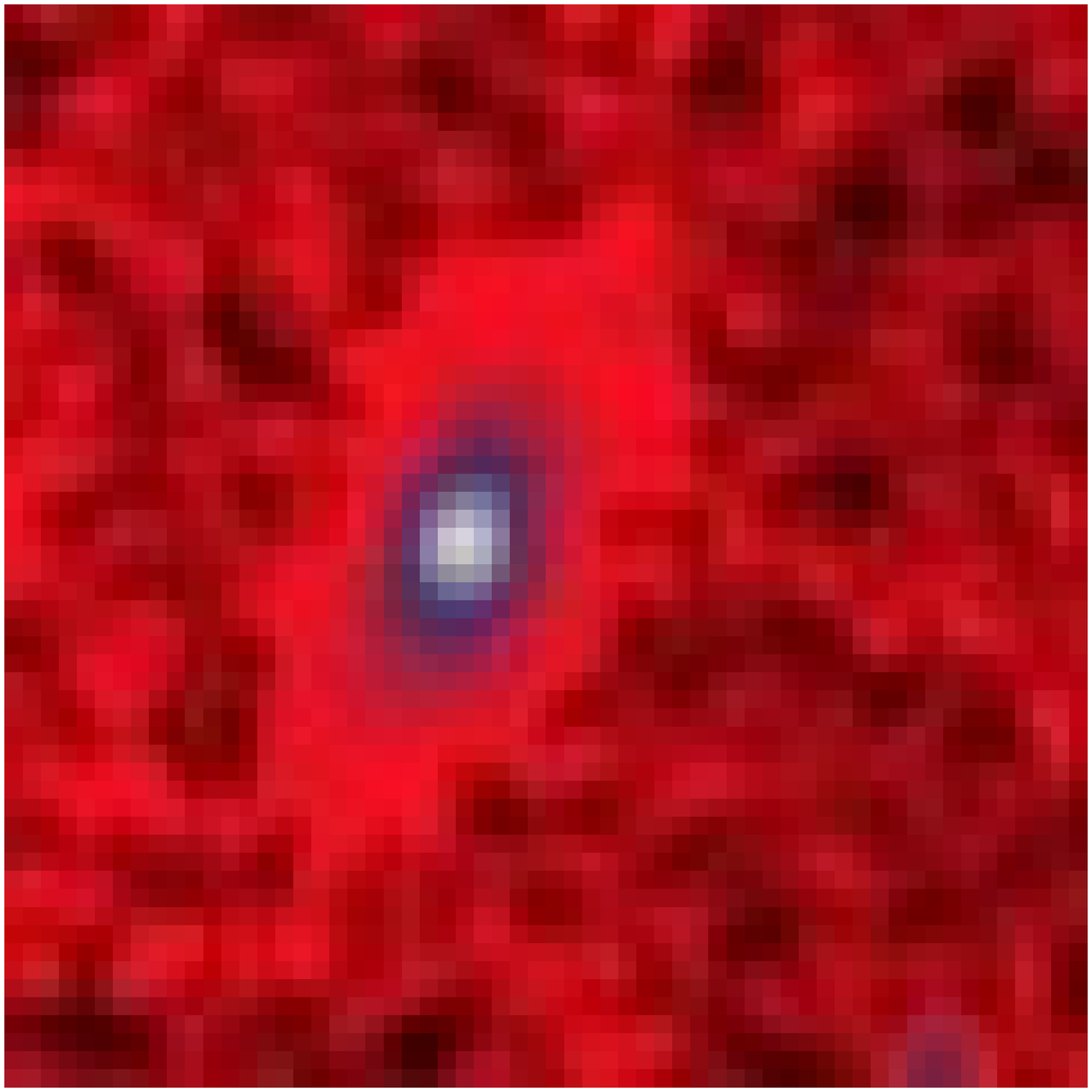}  \FGb{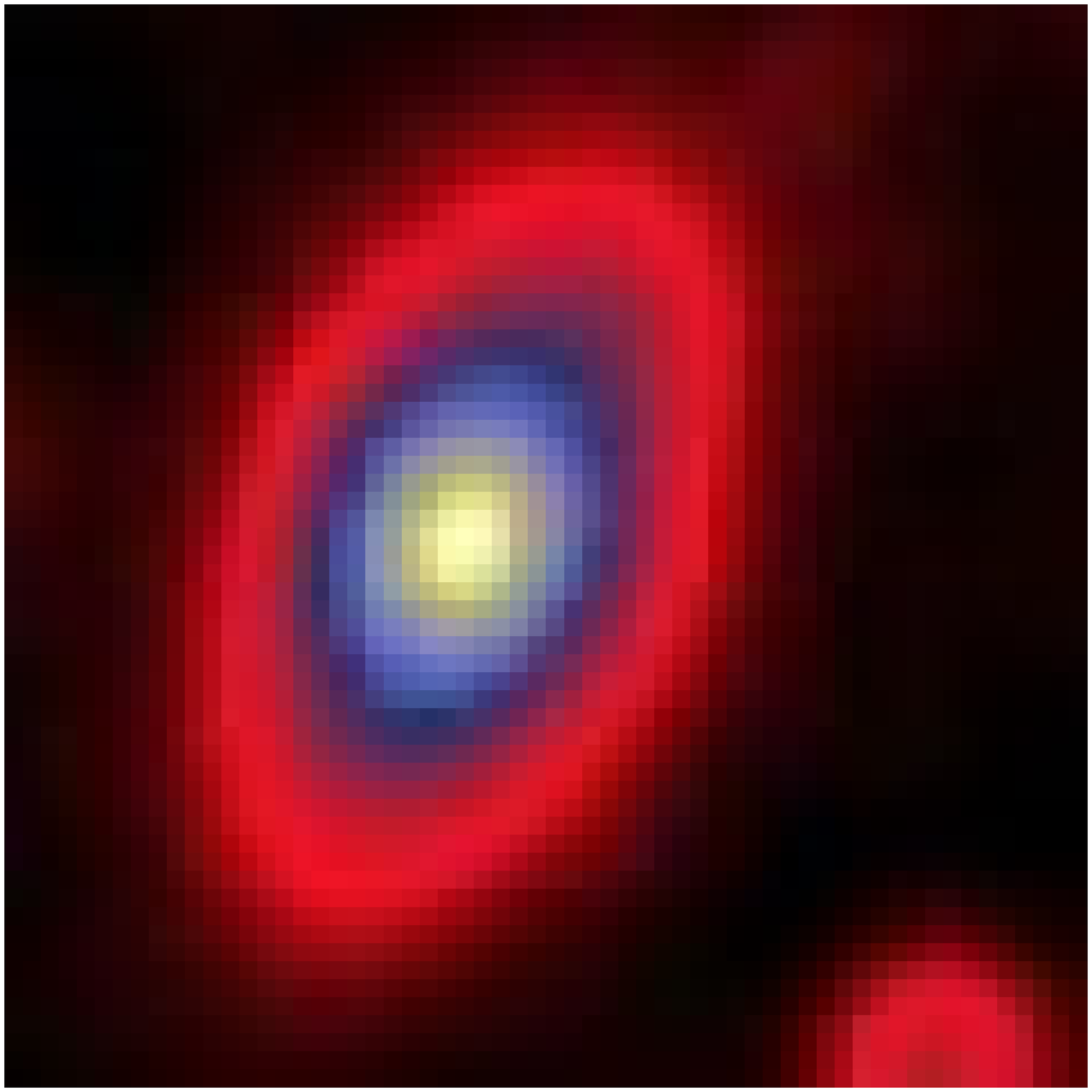}  \FGb{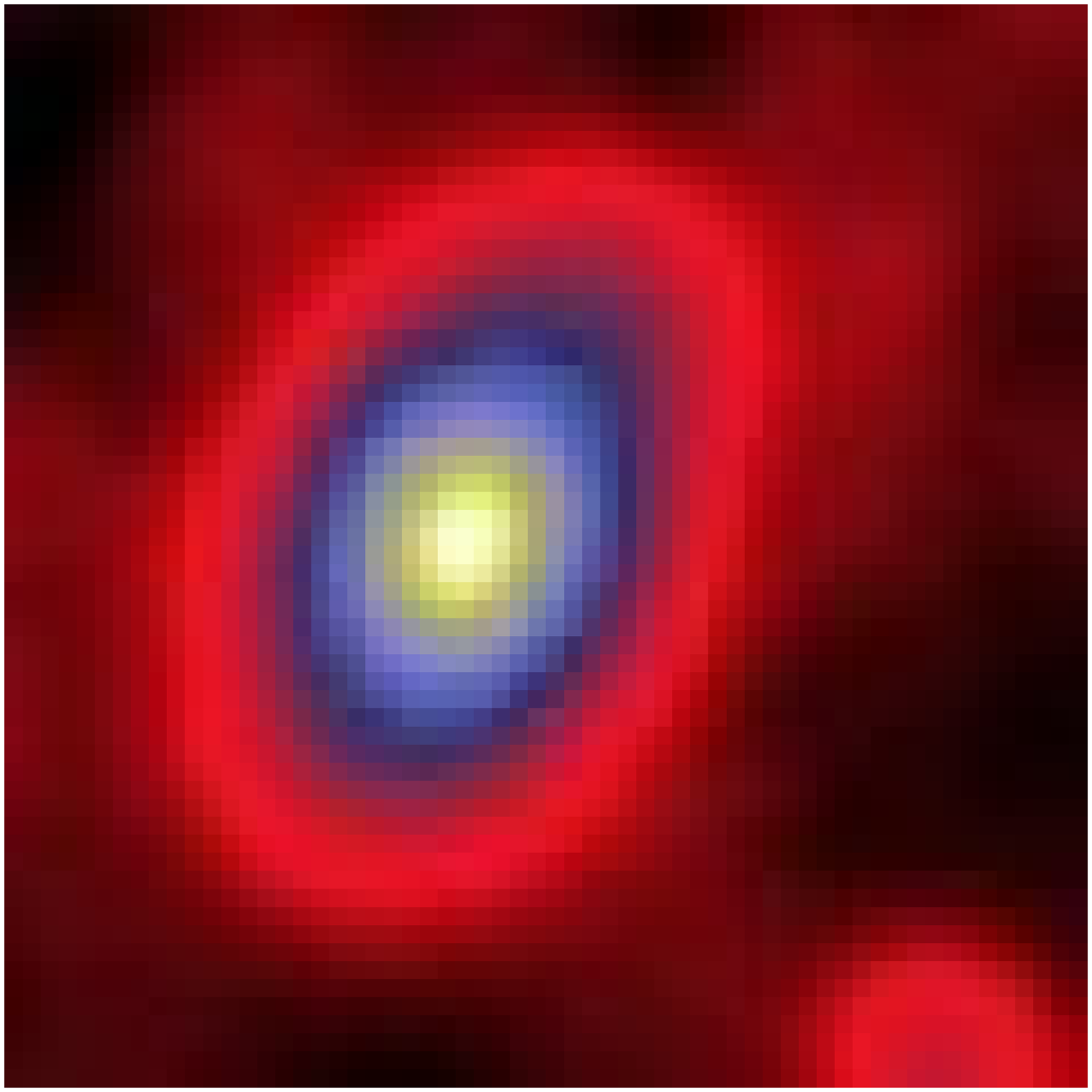}  \FGb{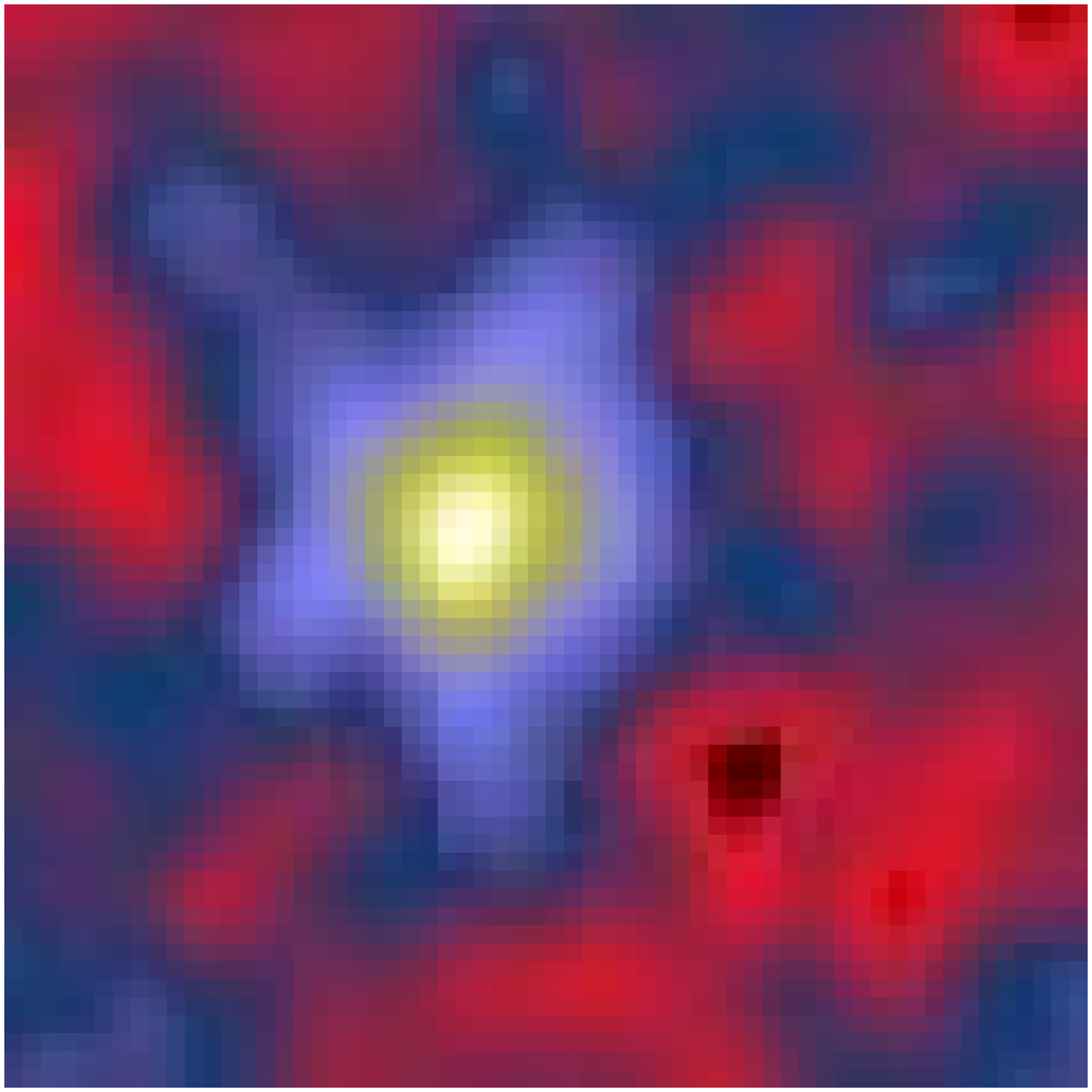}  \FGb{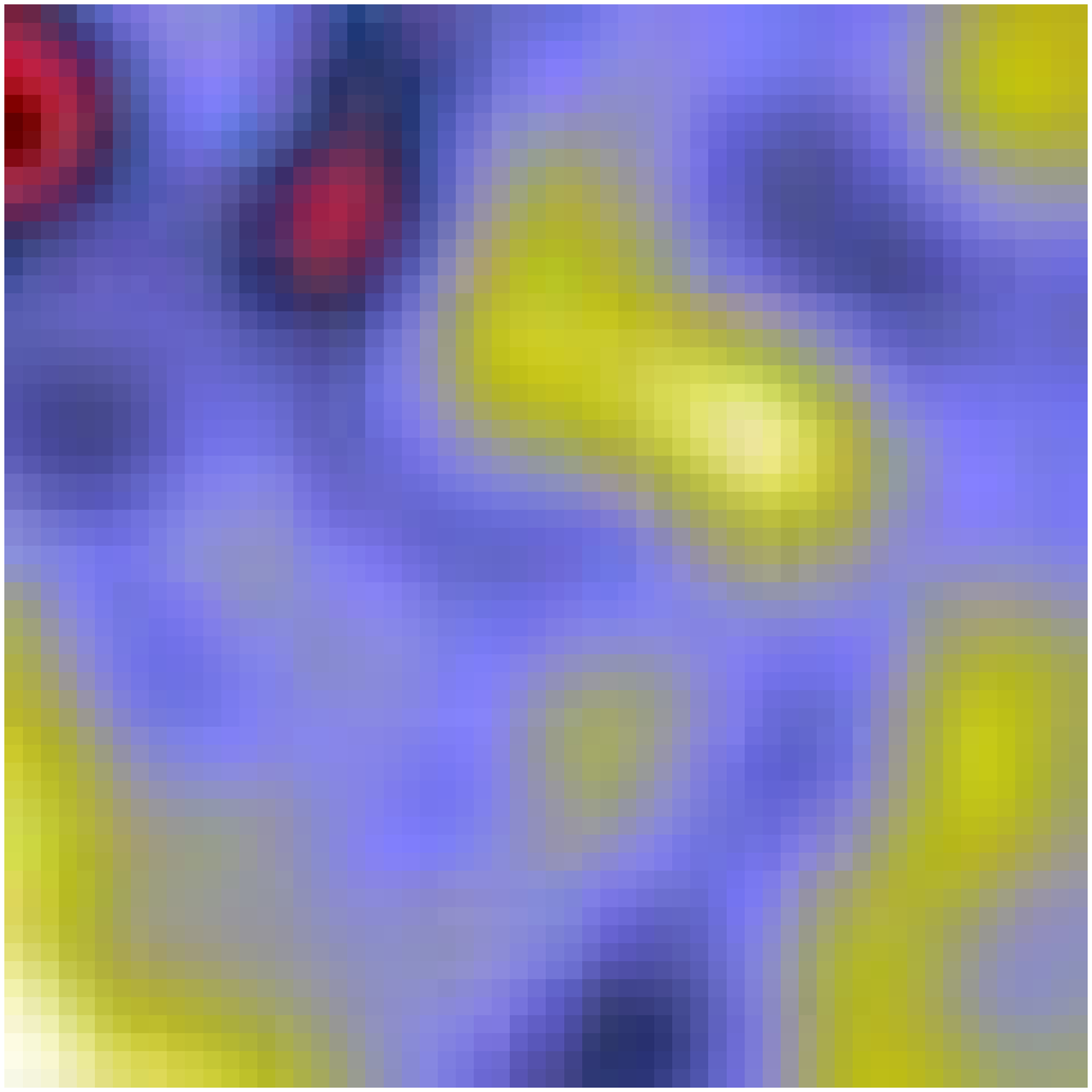} \\  {\#66   NCIA000248}  \\
  \FGb{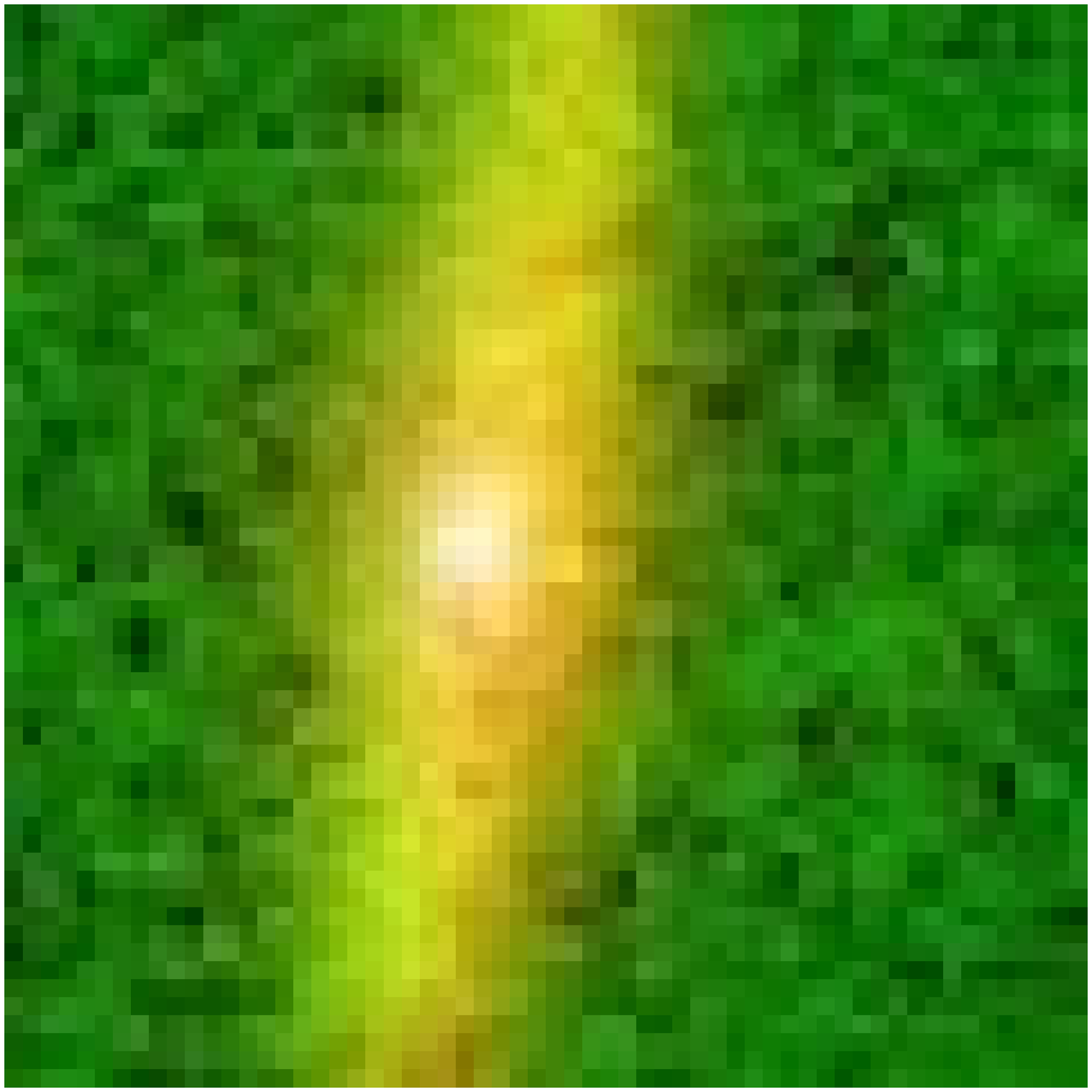}  \FGb{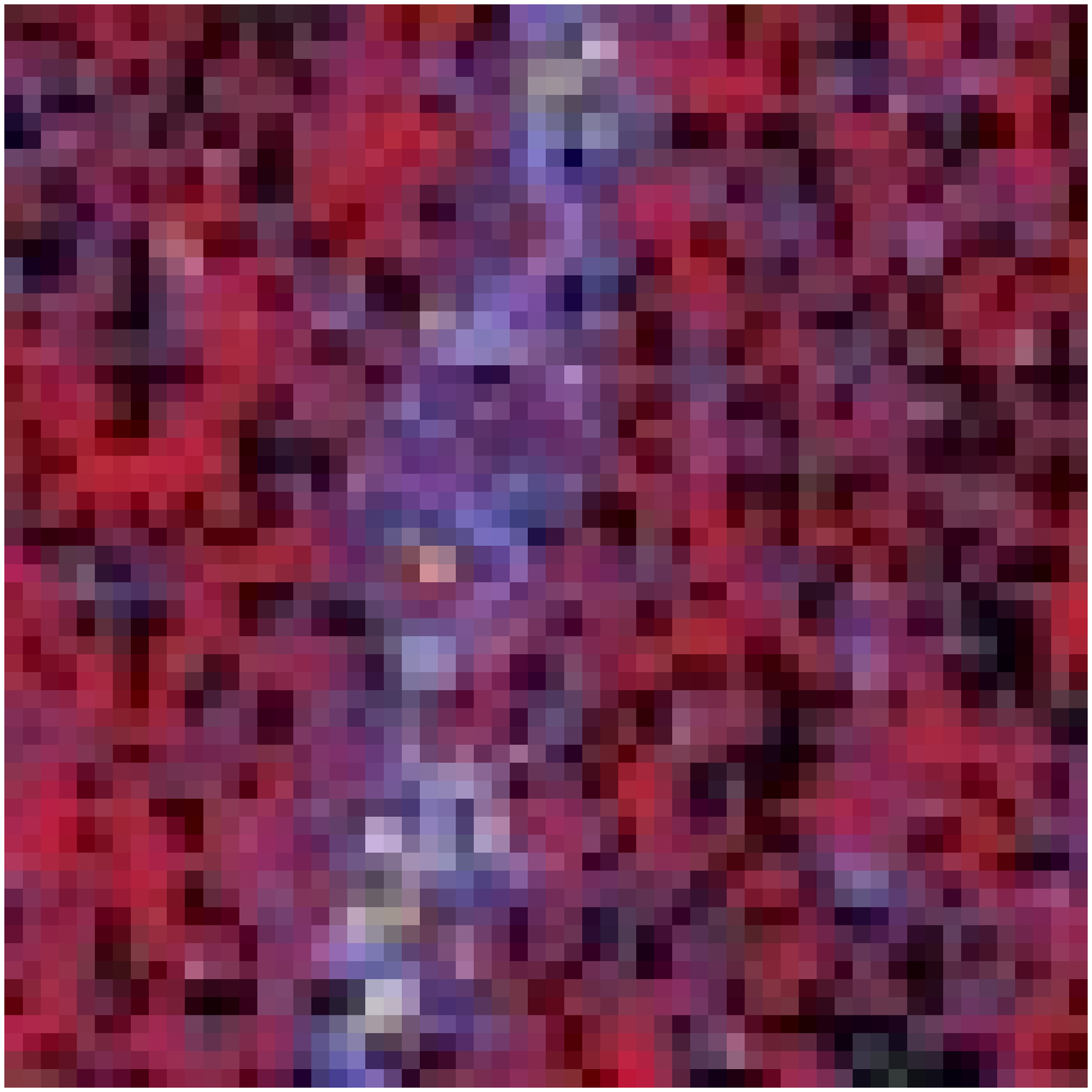}  \FGb{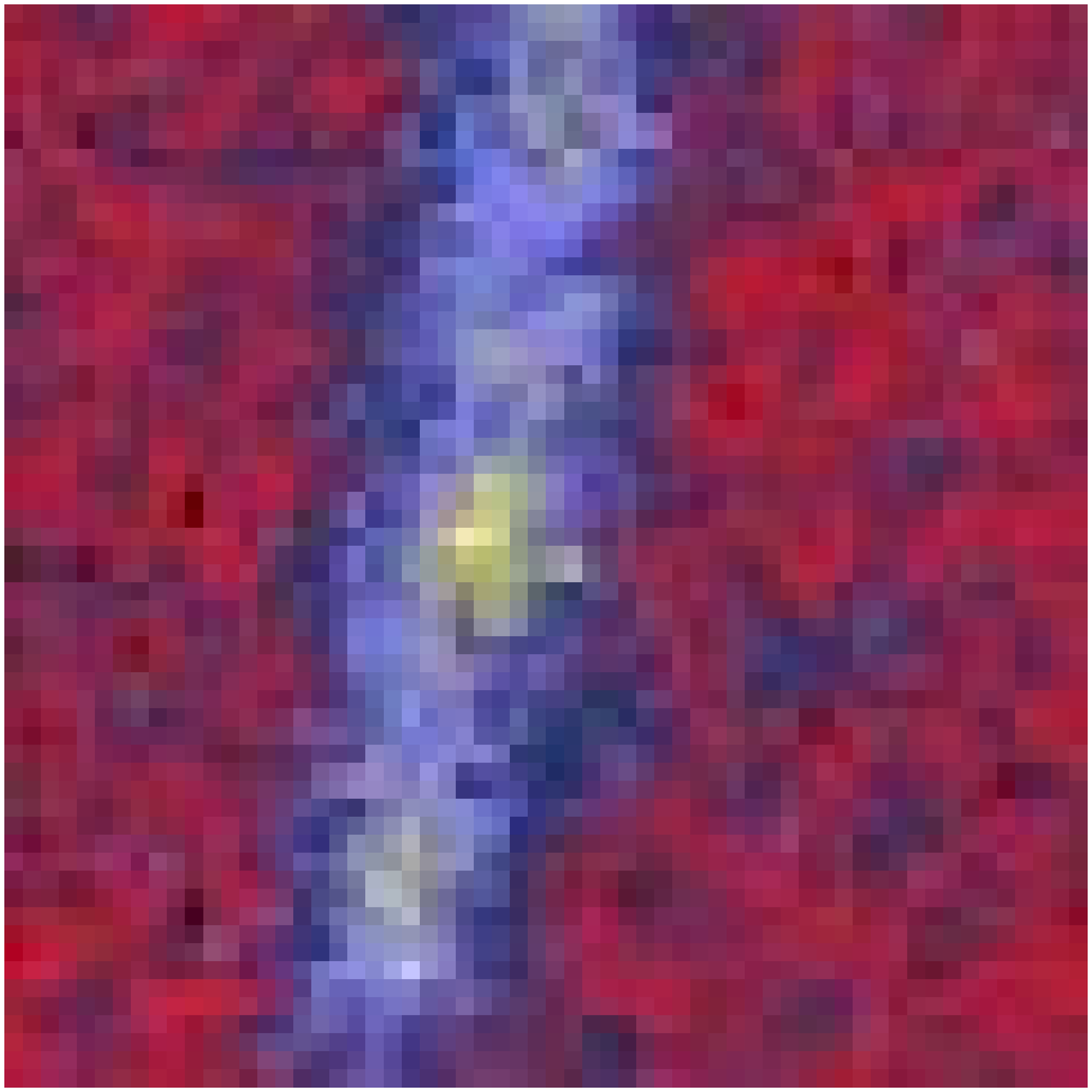}  \FGb{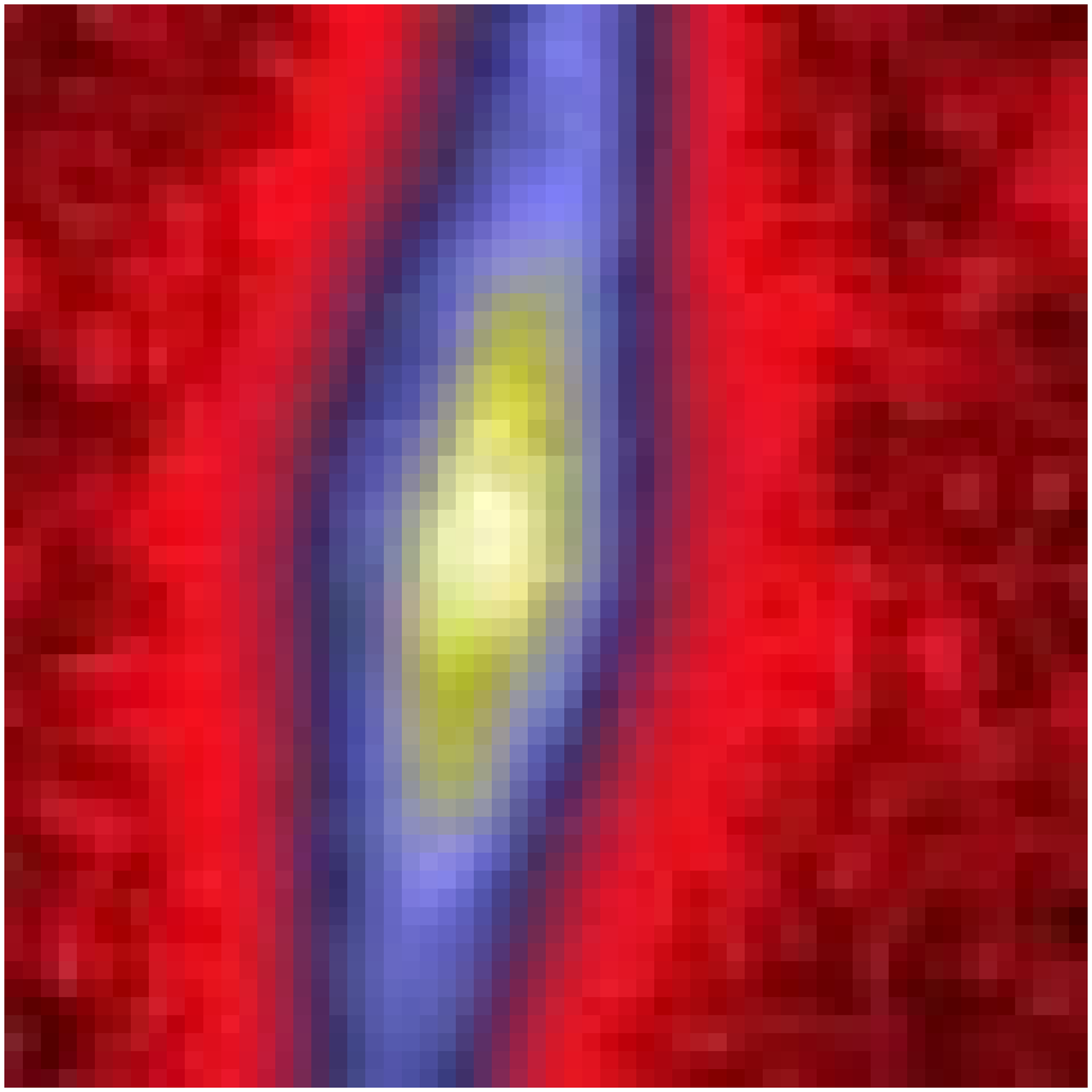}  \FGb{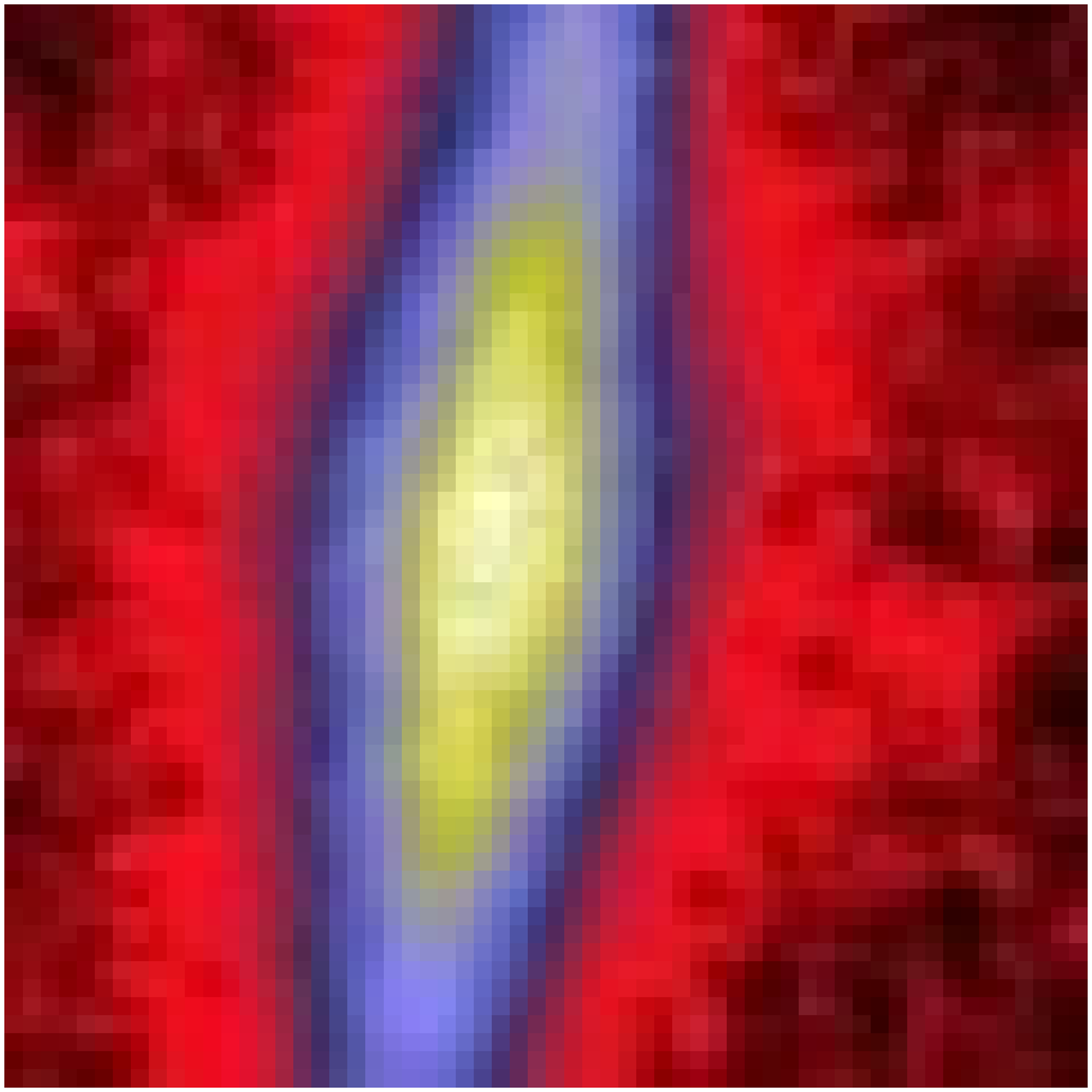}  \FGb{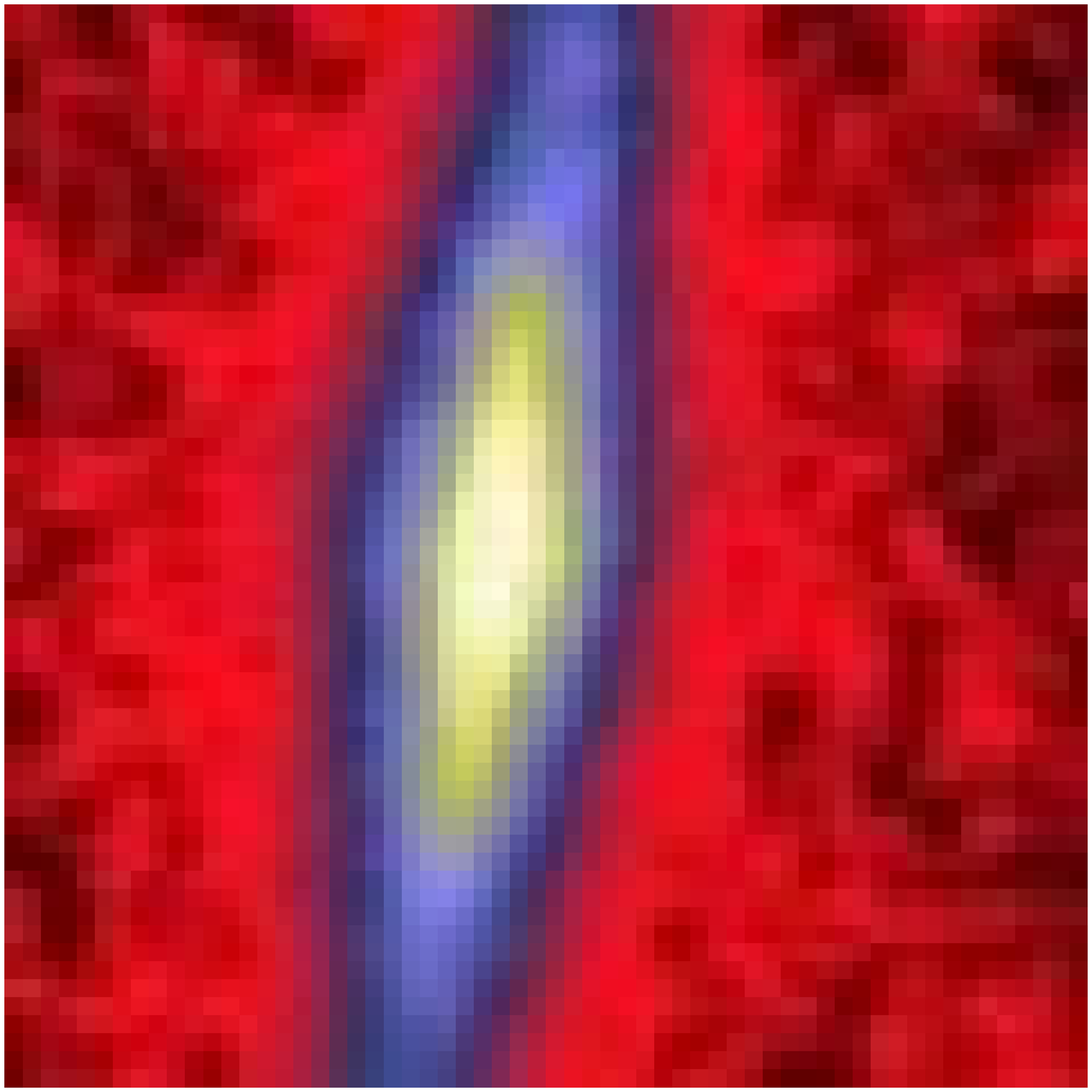}  \FGb{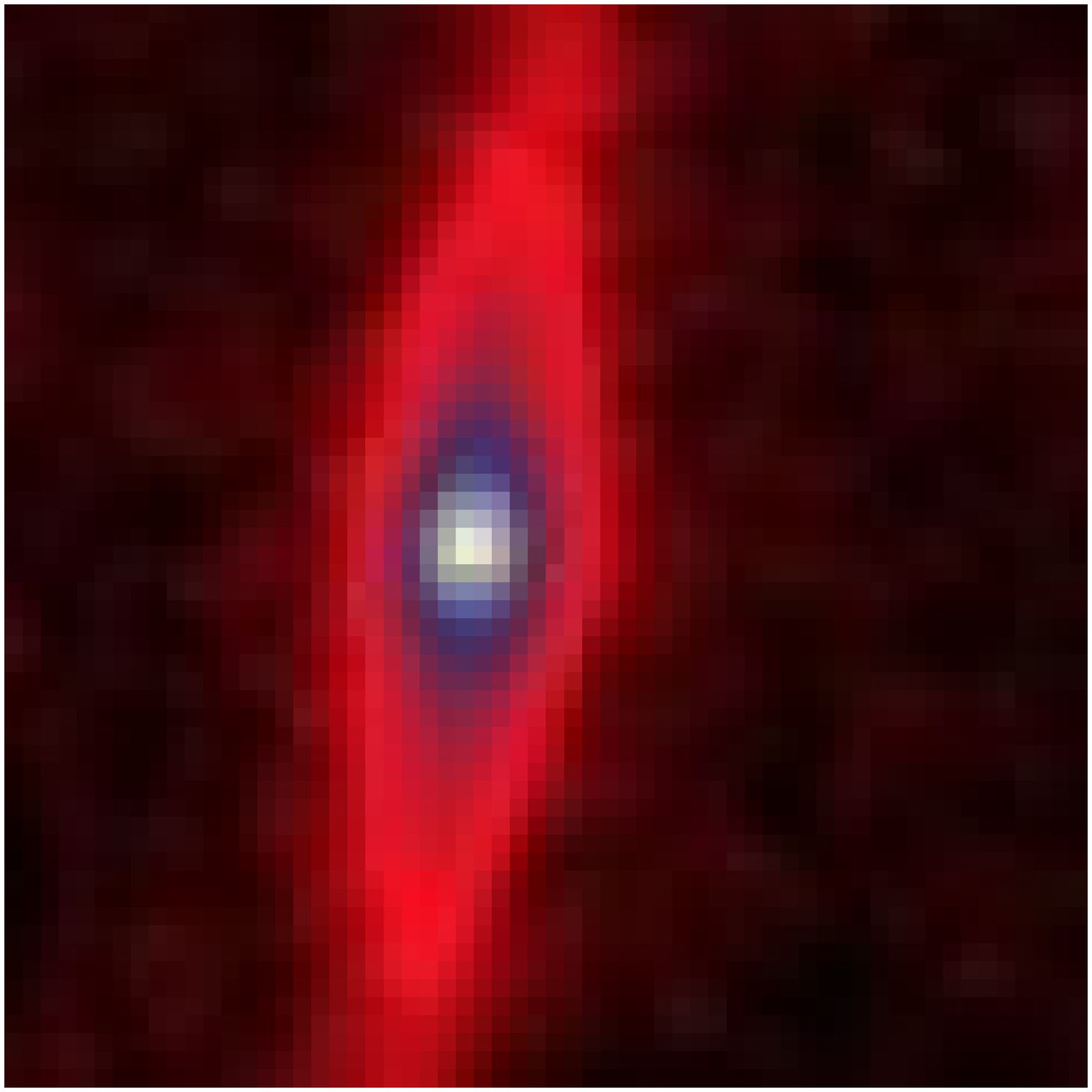}  \FGb{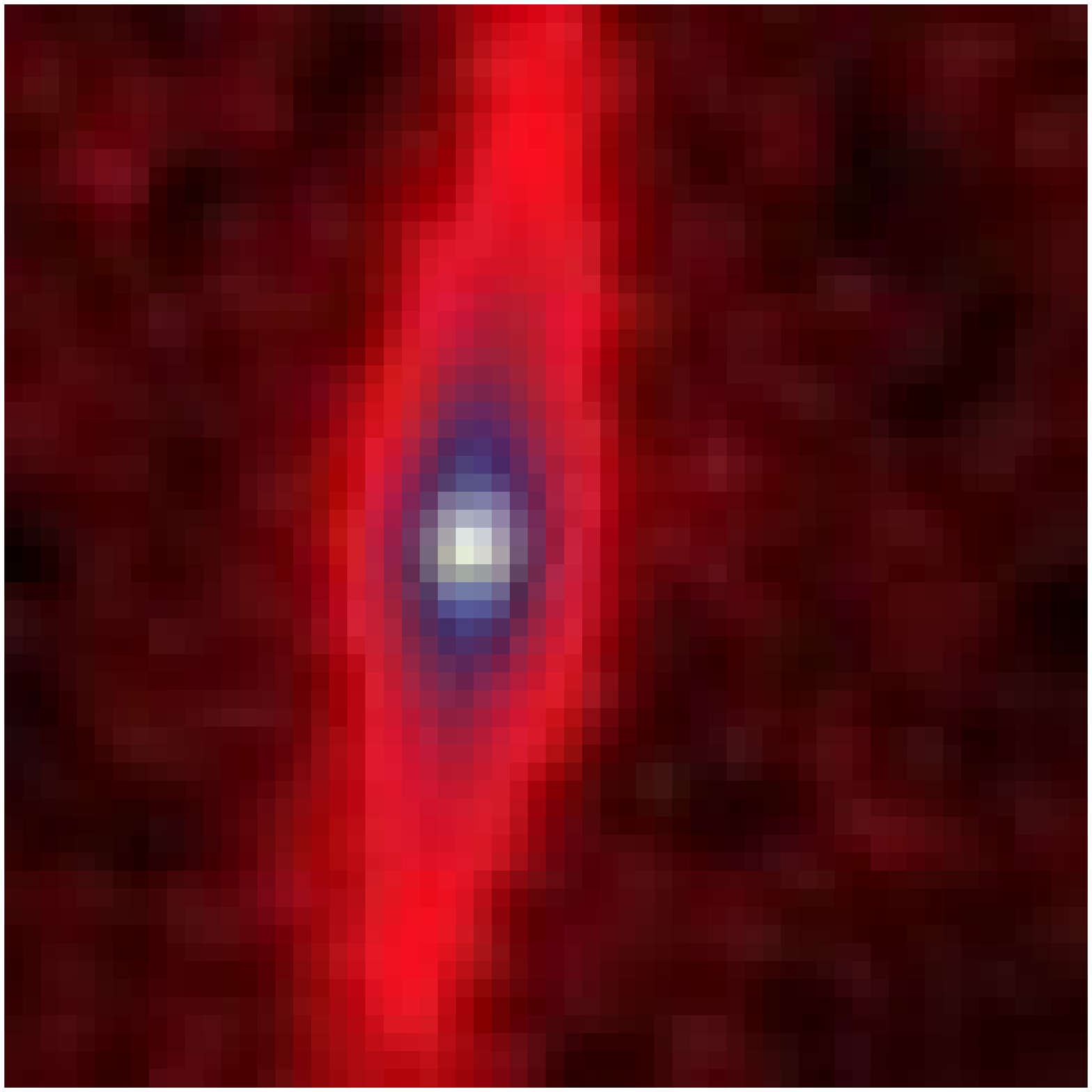}  \FGb{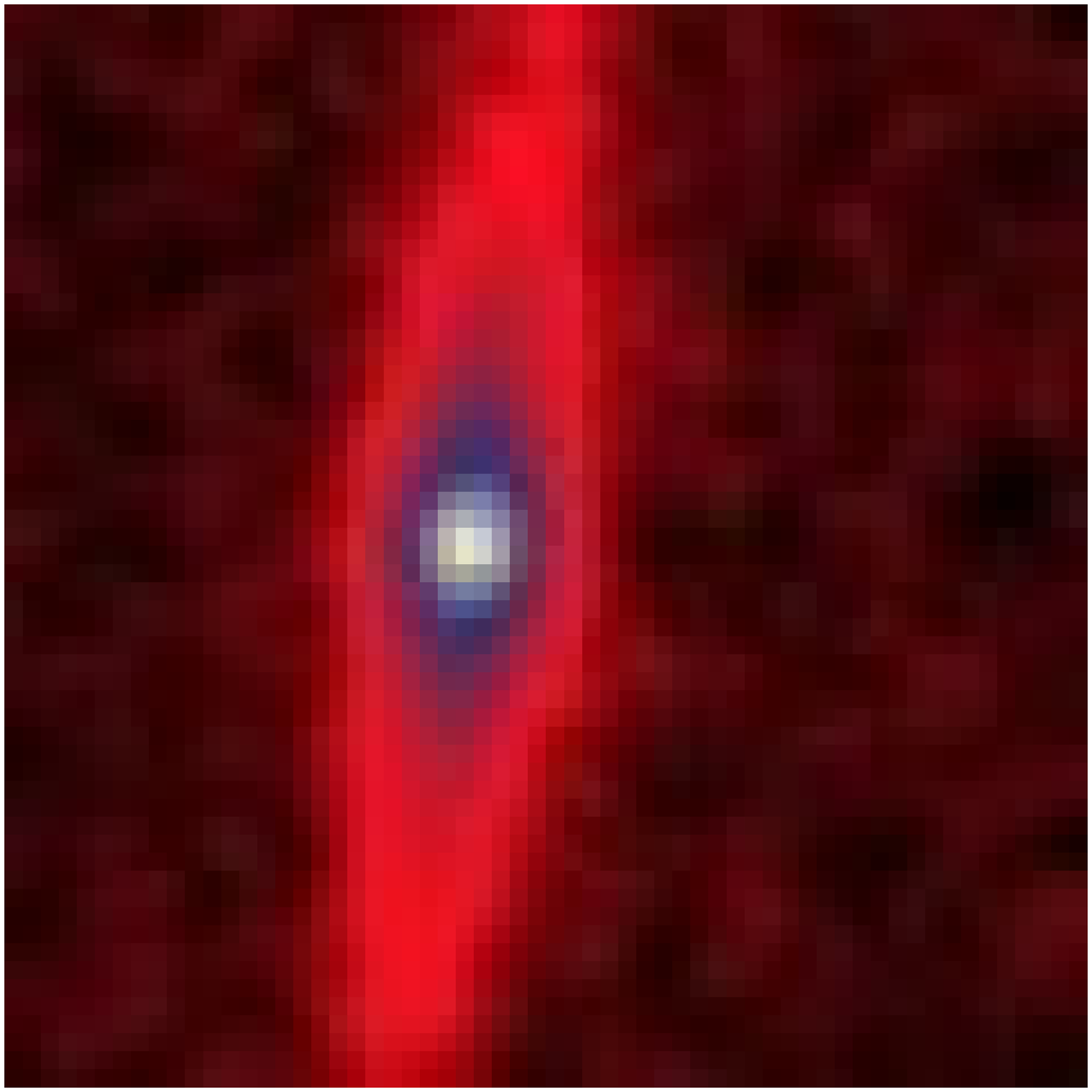}  \FGb{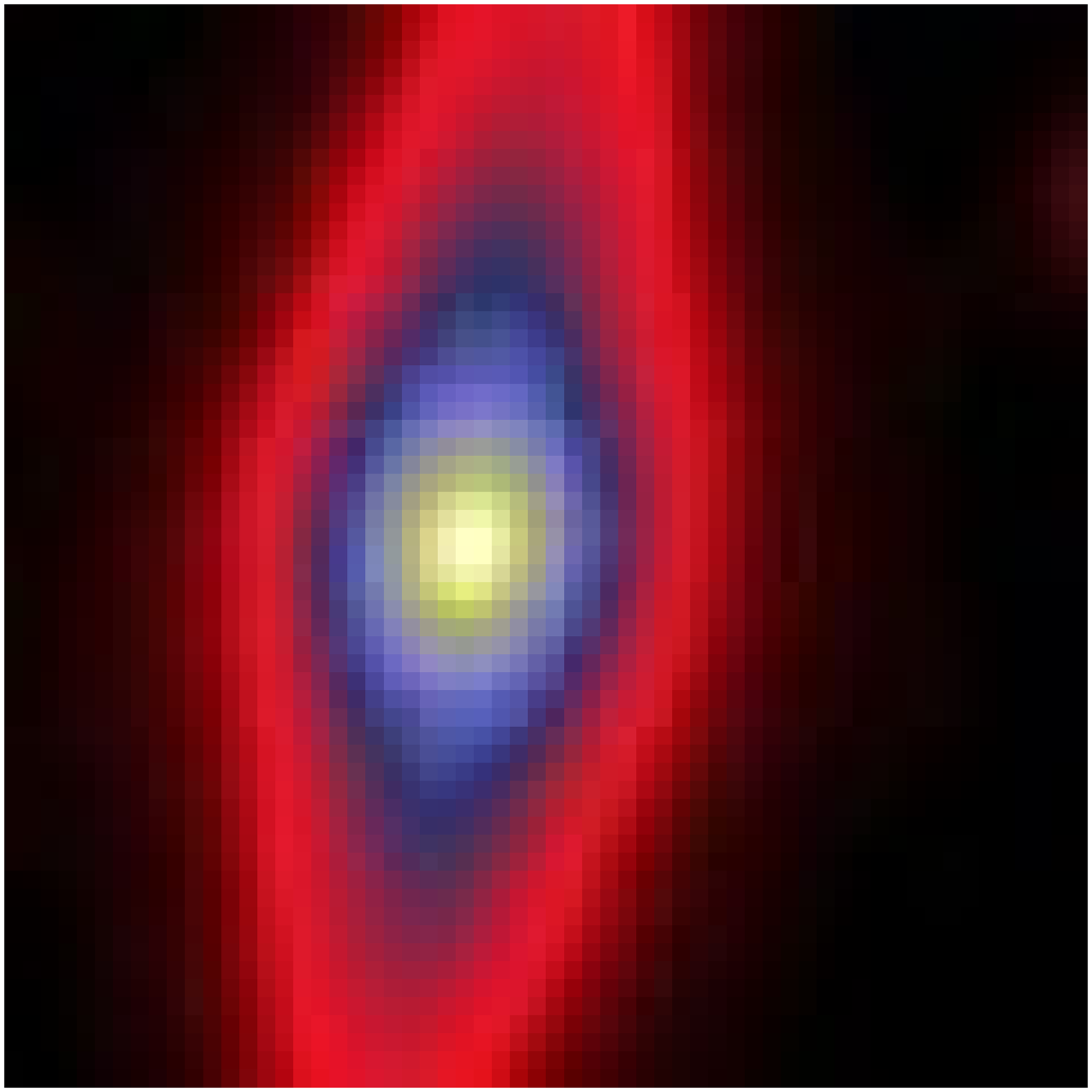}  \FGb{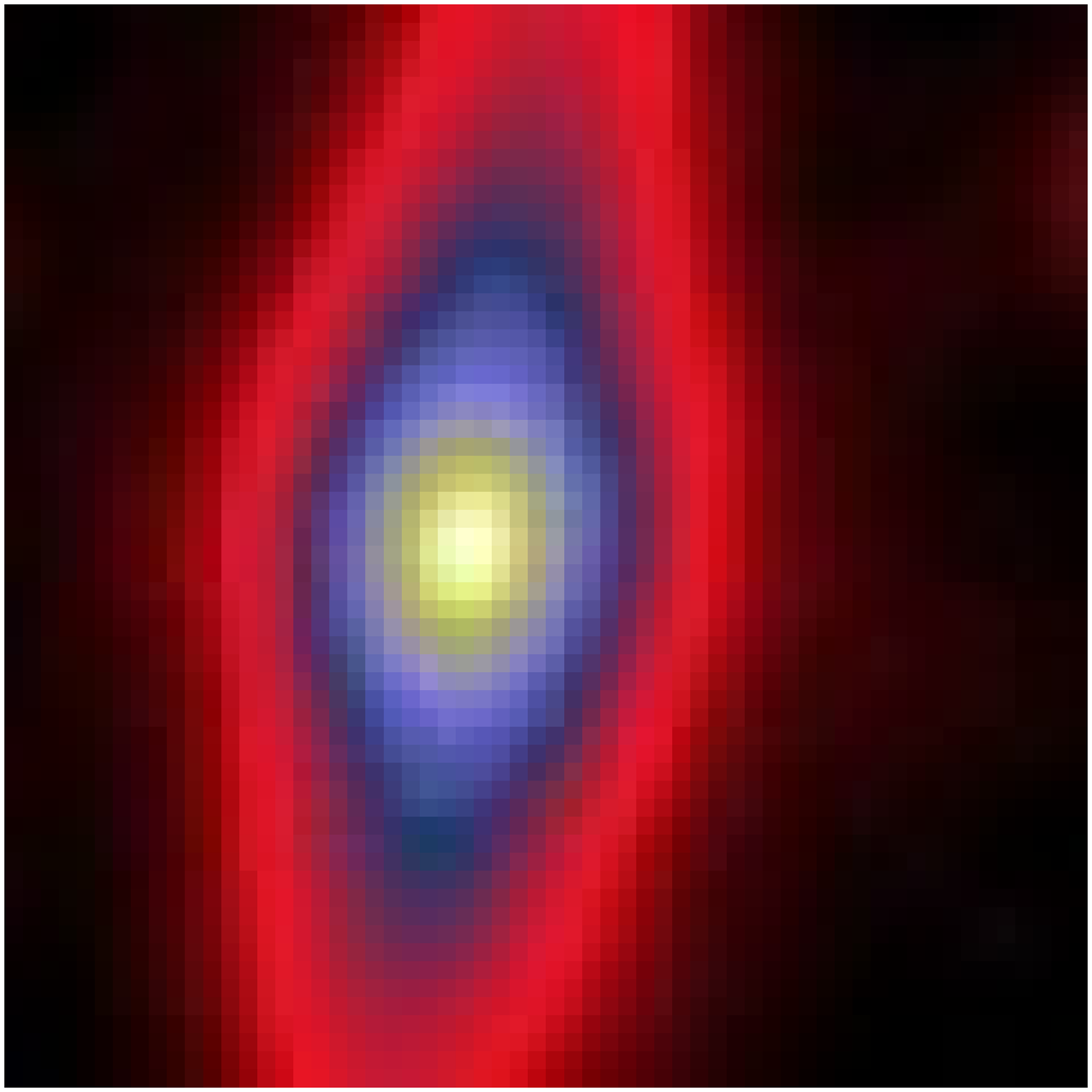}  \FGb{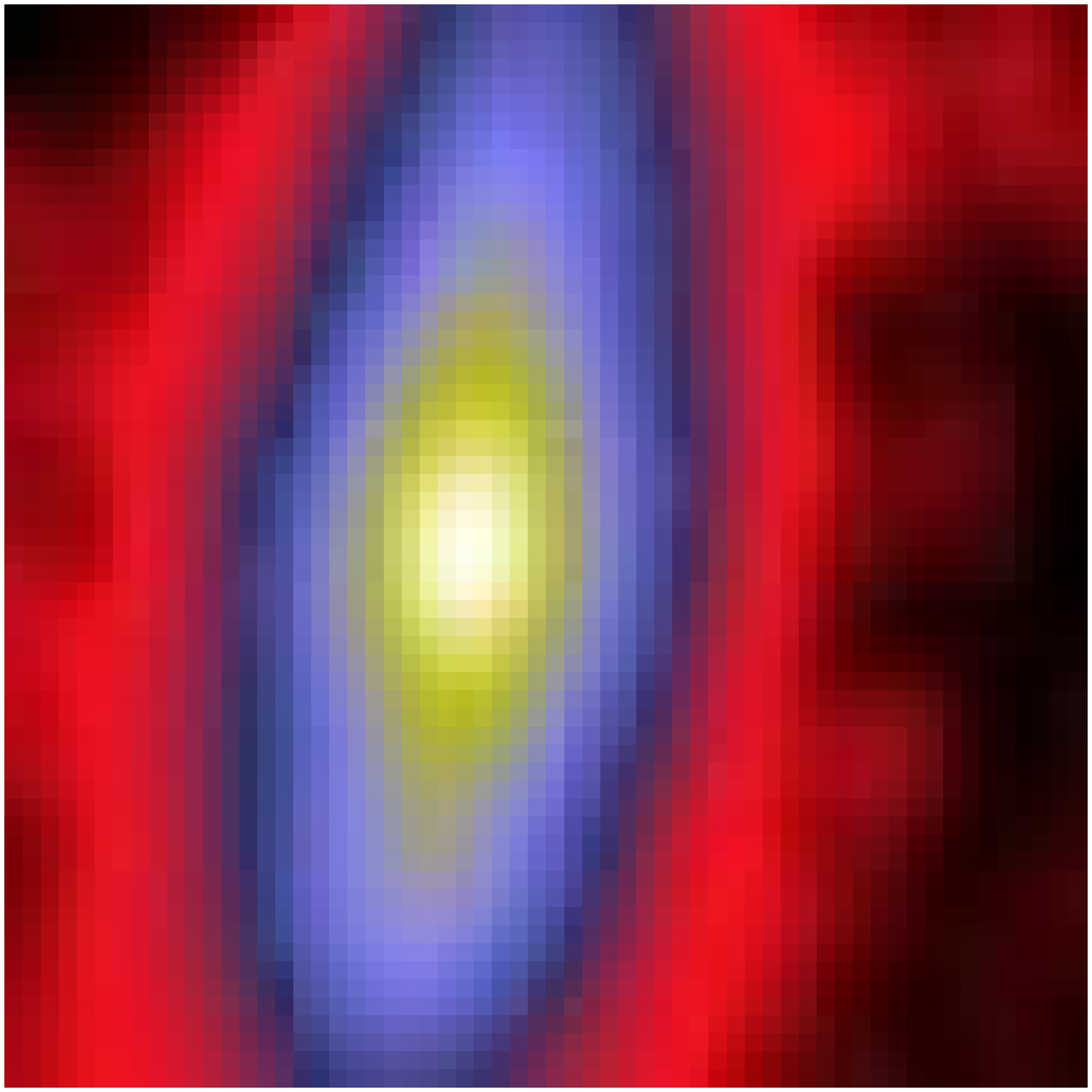}  \FGb{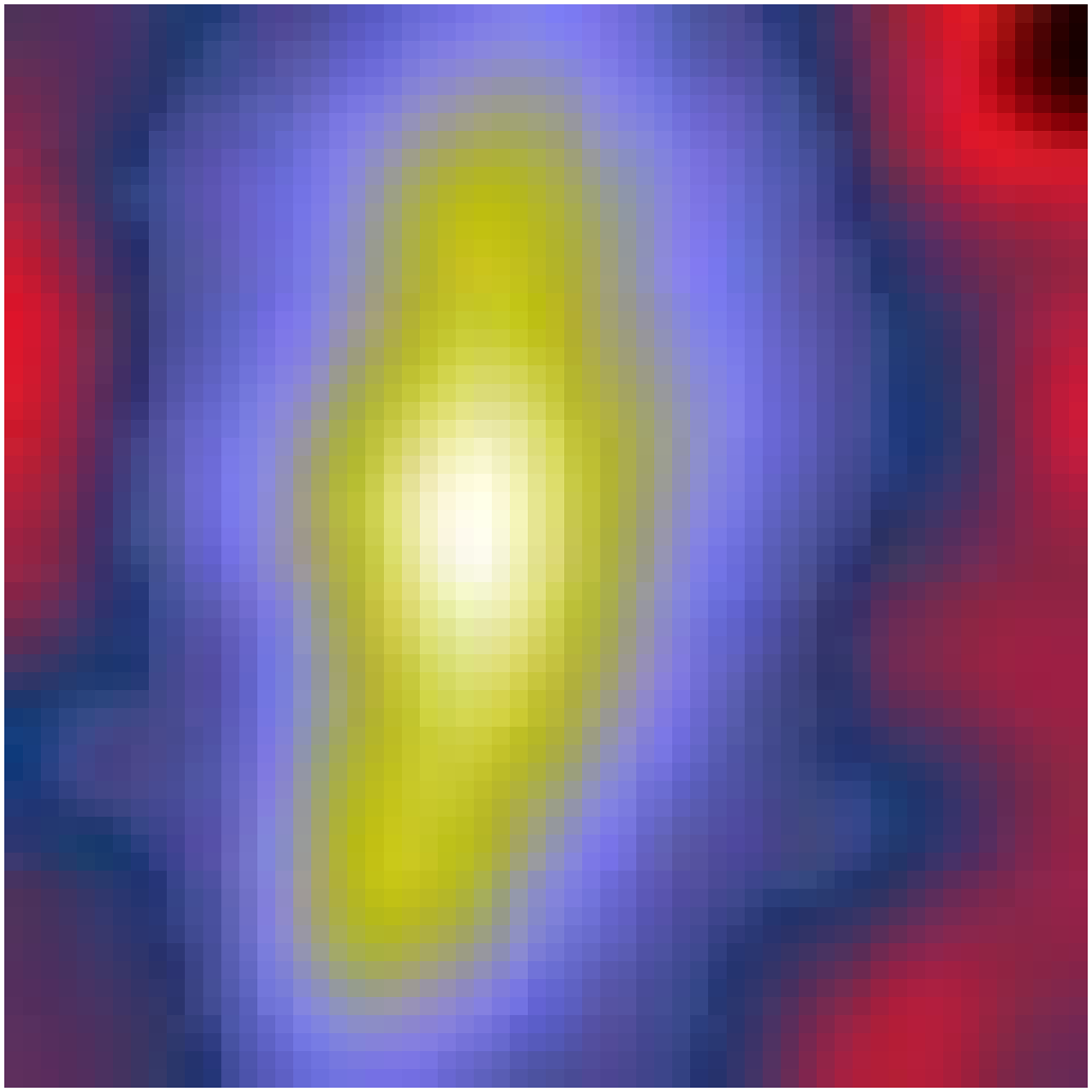} \\  {\#67   NCIA001220}  \\
  \FGb{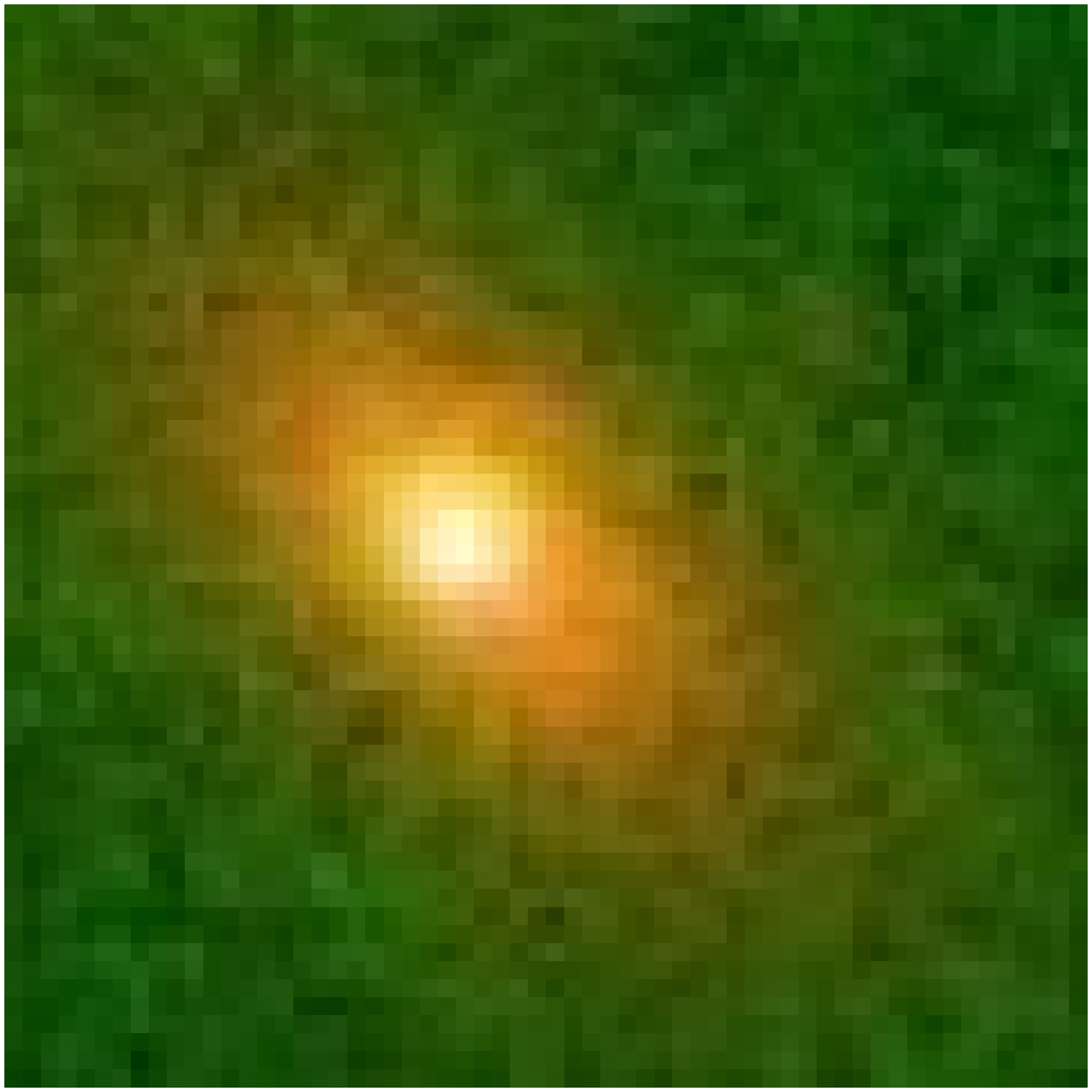}  \FGb{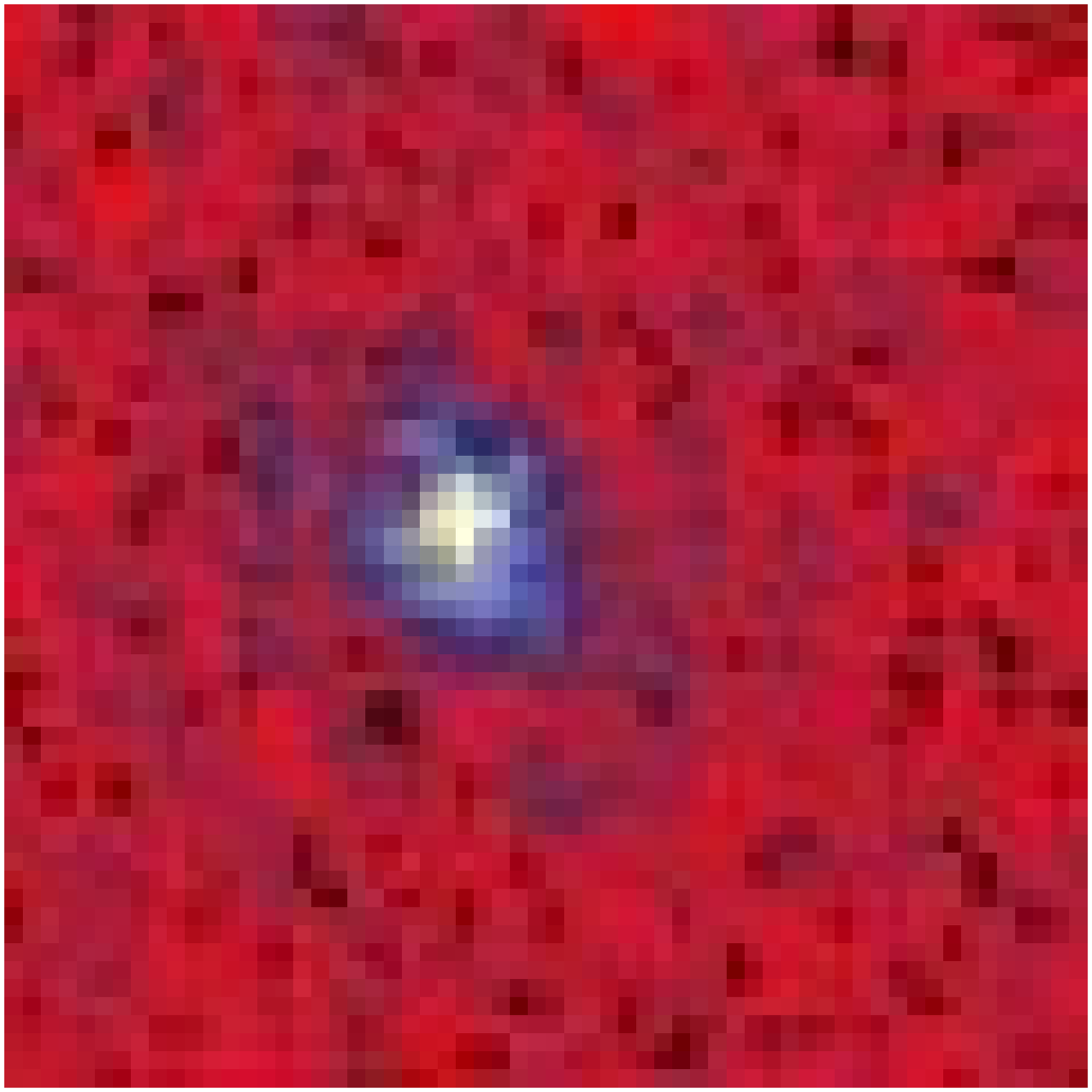}  \FGb{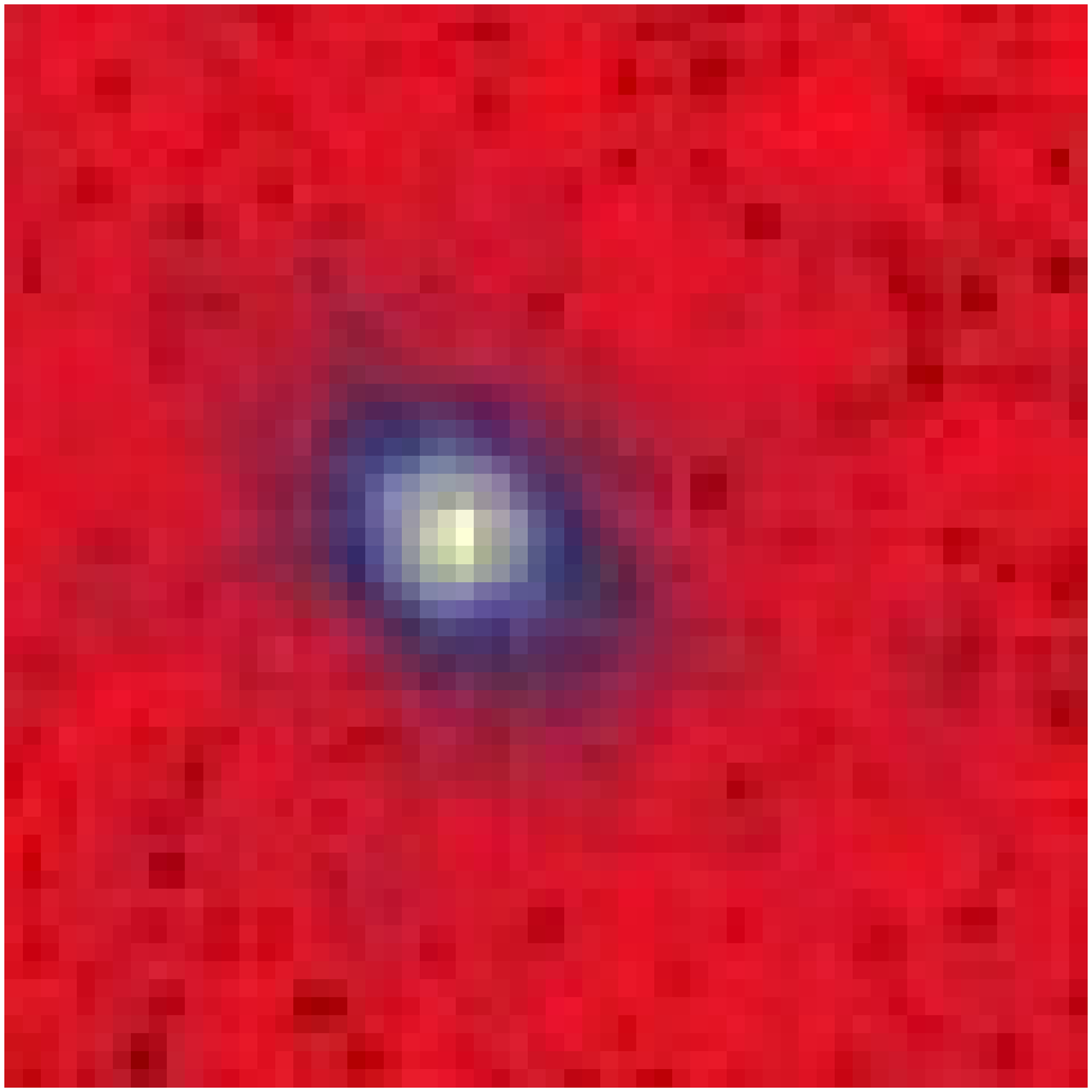}  \FGb{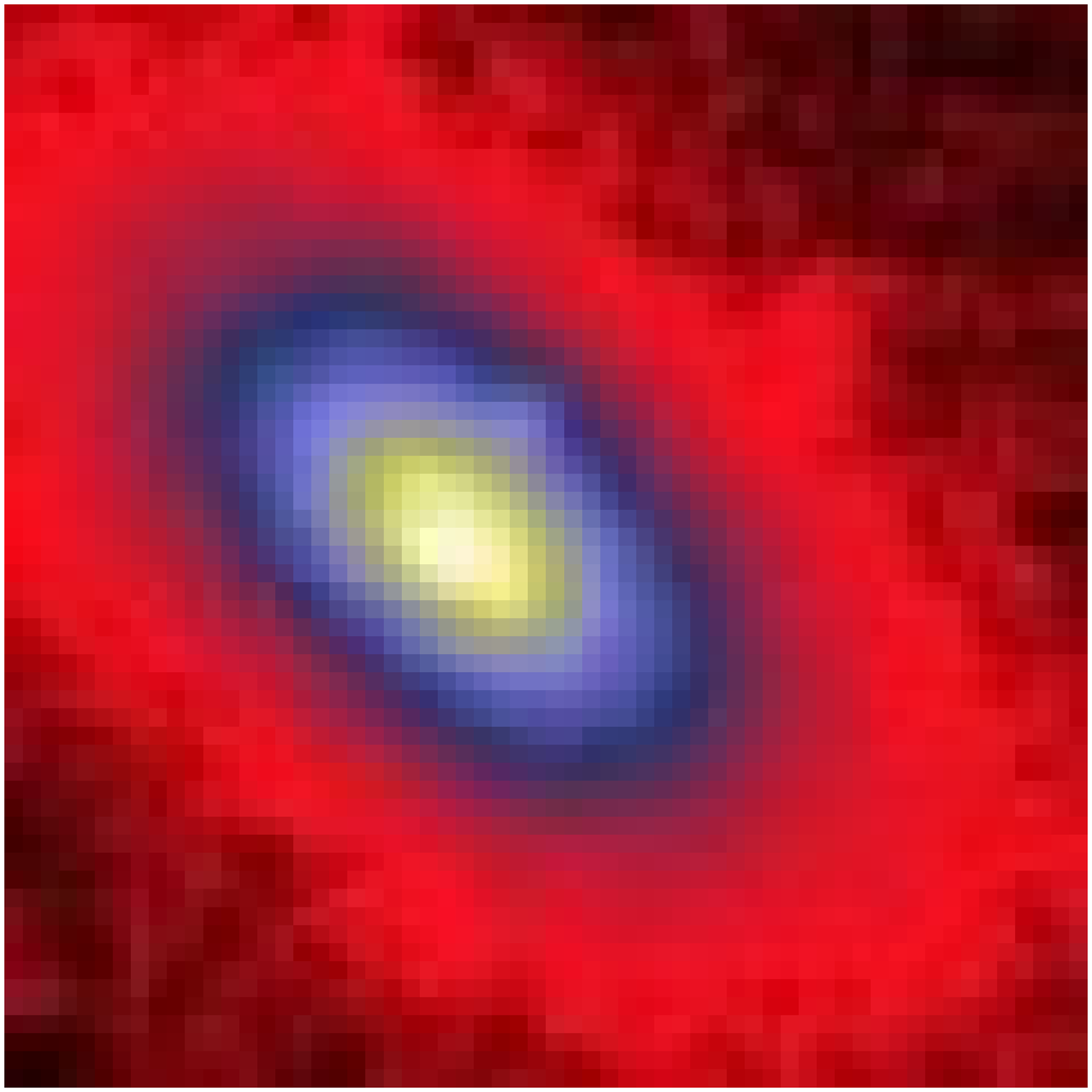}  \FGb{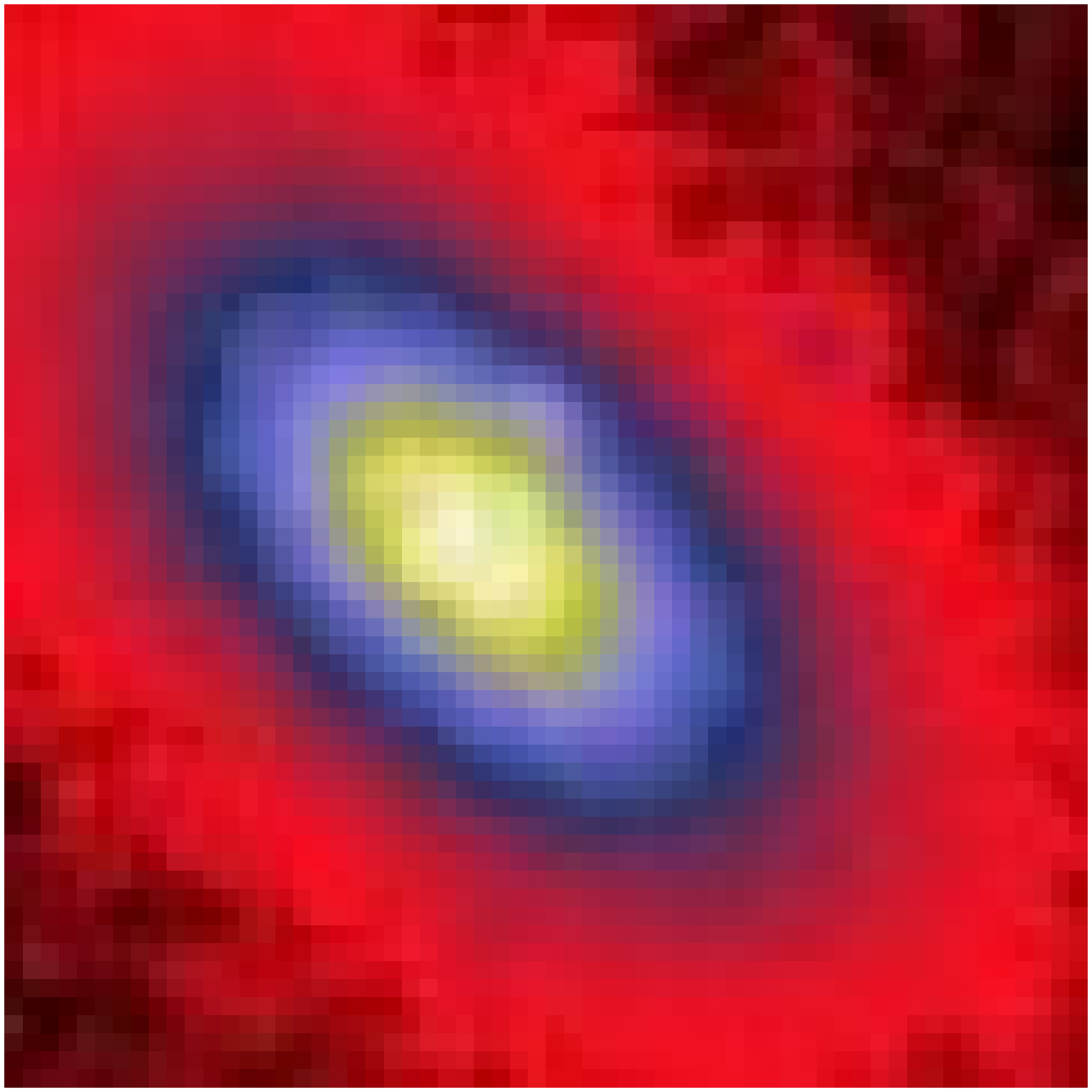}  \FGb{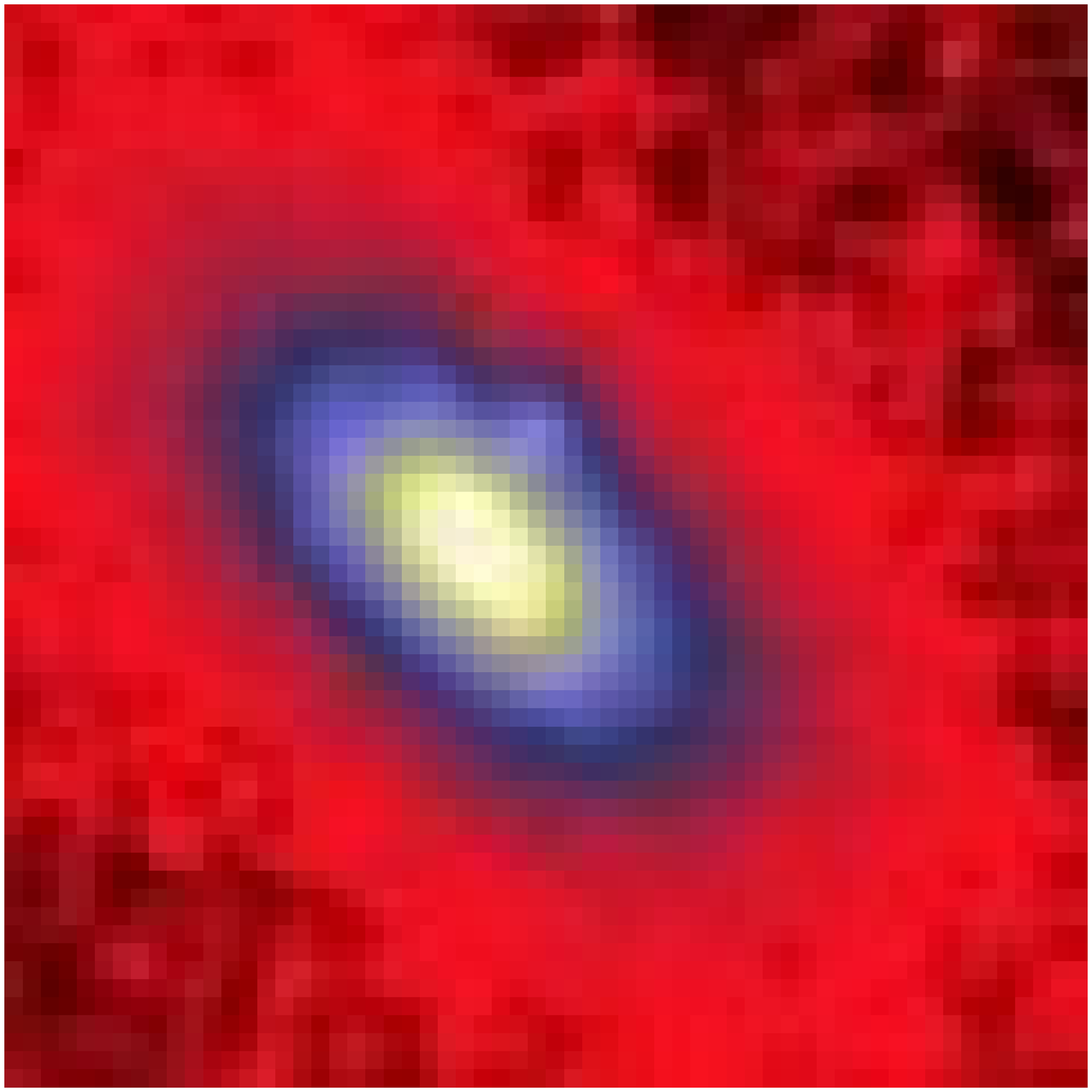}  \FGb{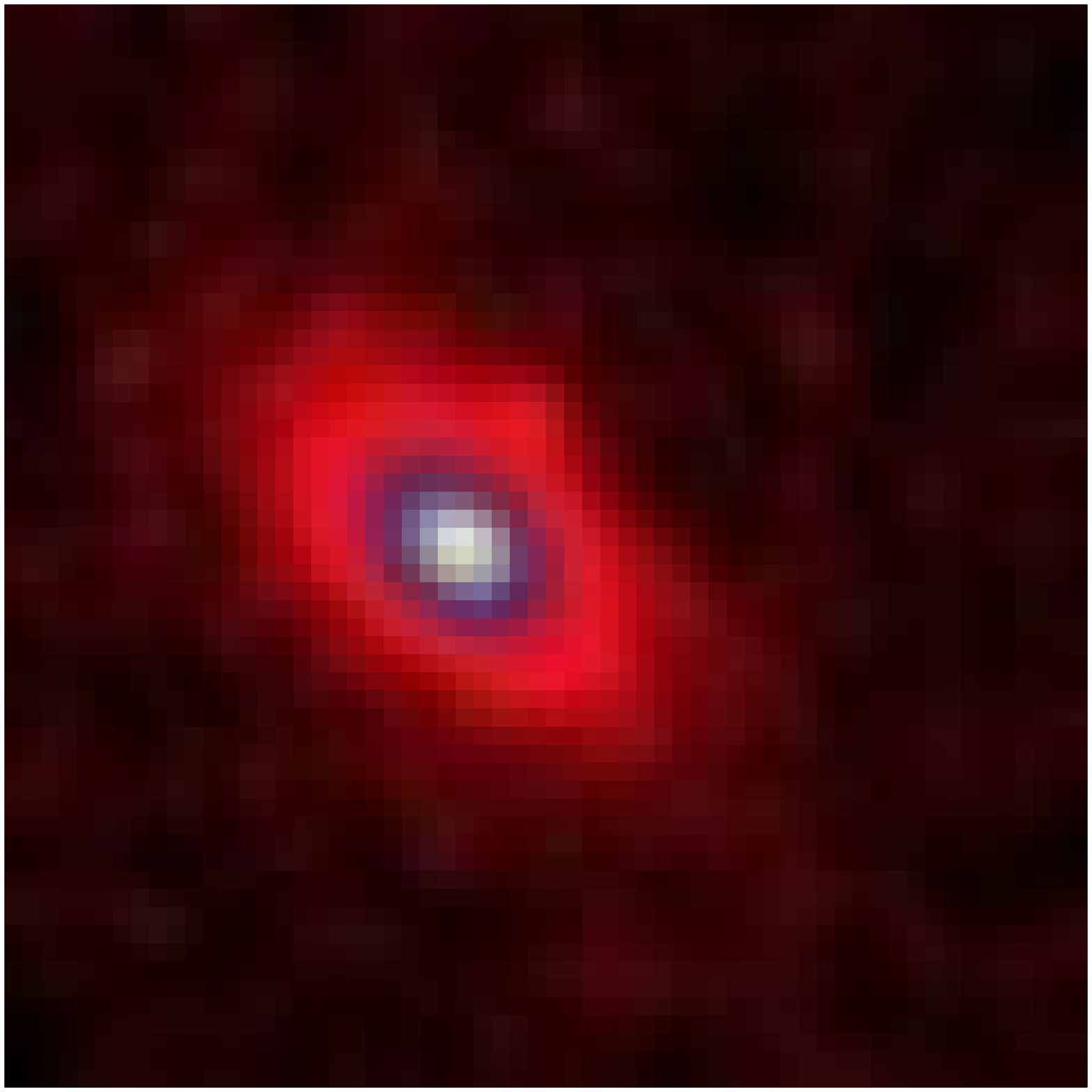}  \FGb{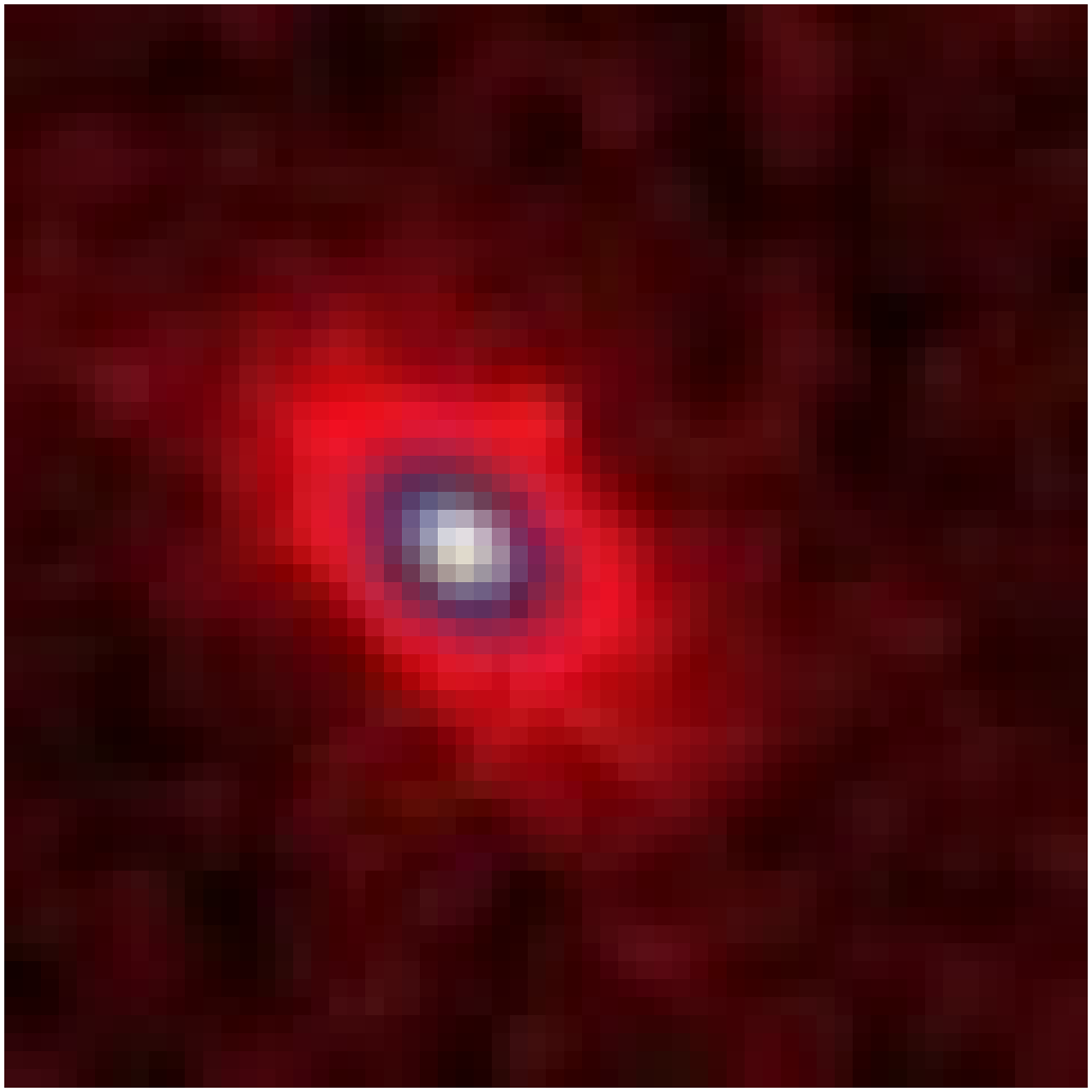}  \FGb{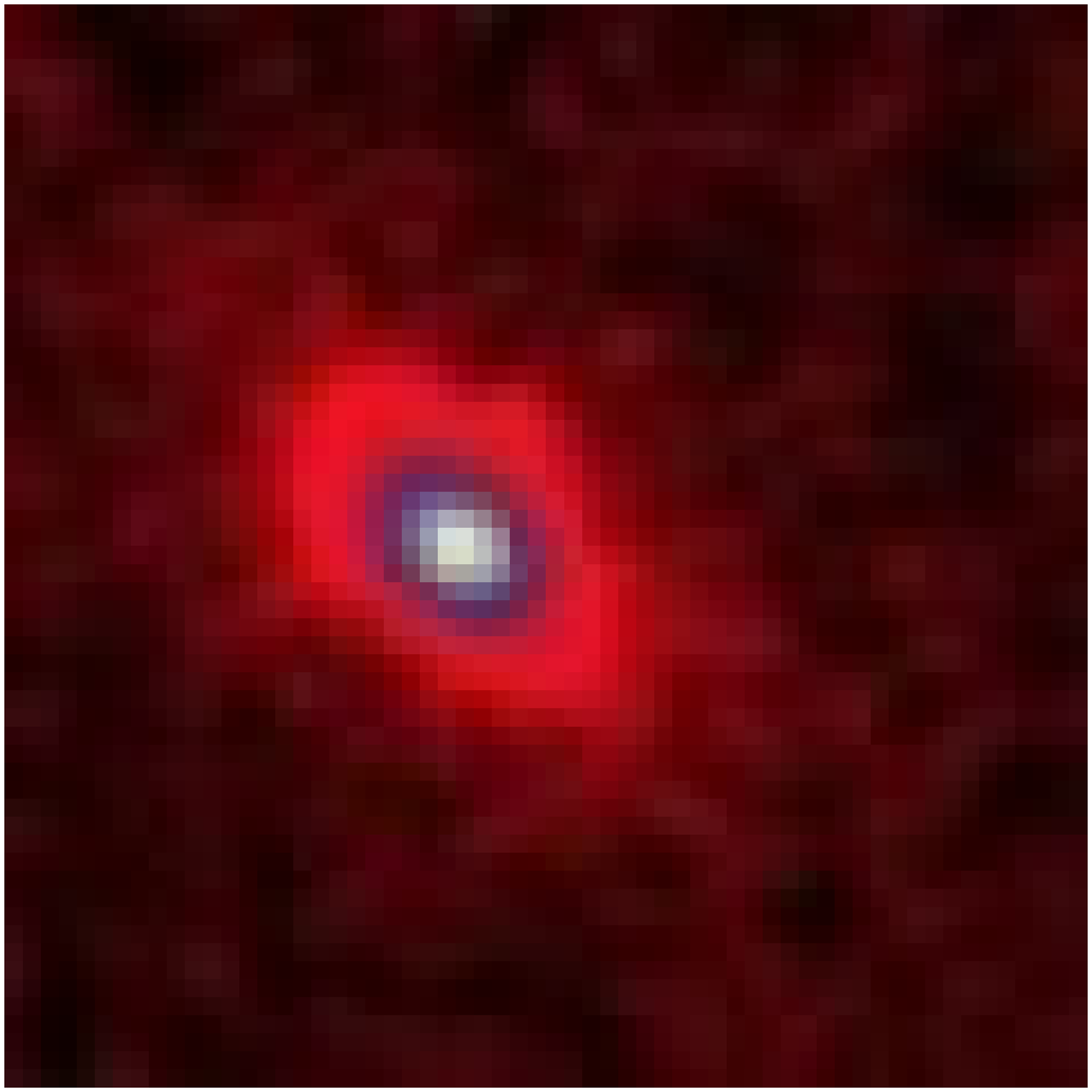}  \FGb{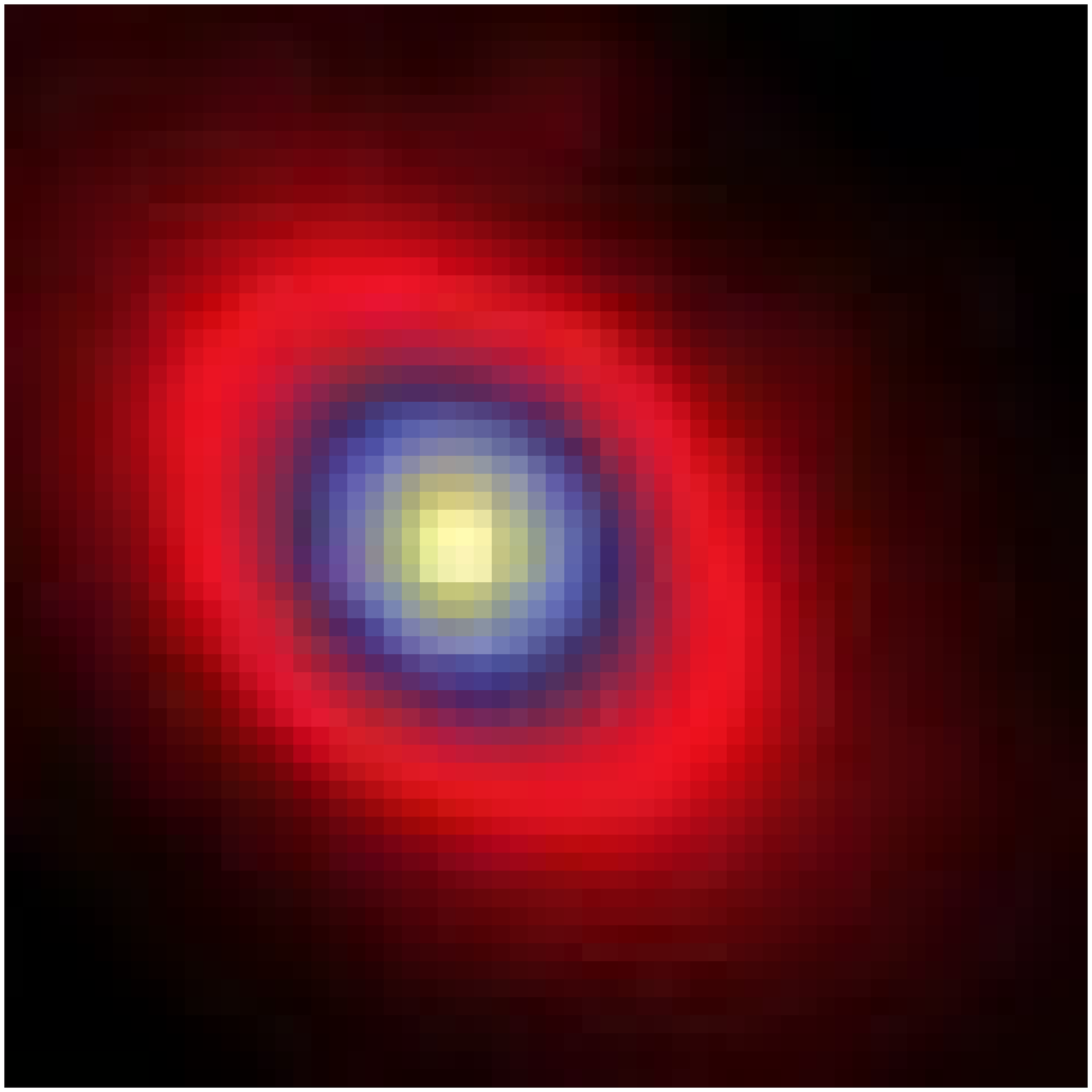}  \FGb{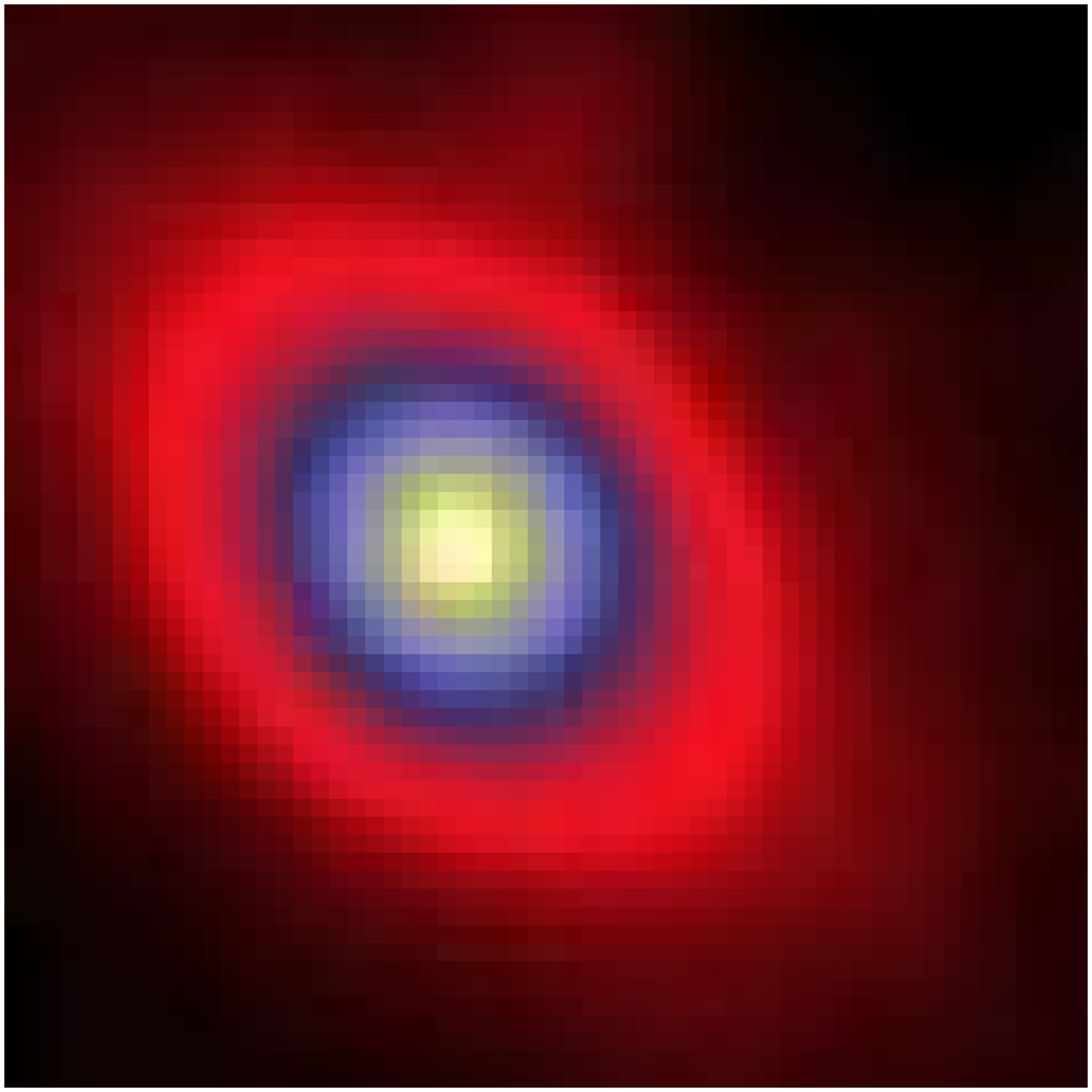}  \FGb{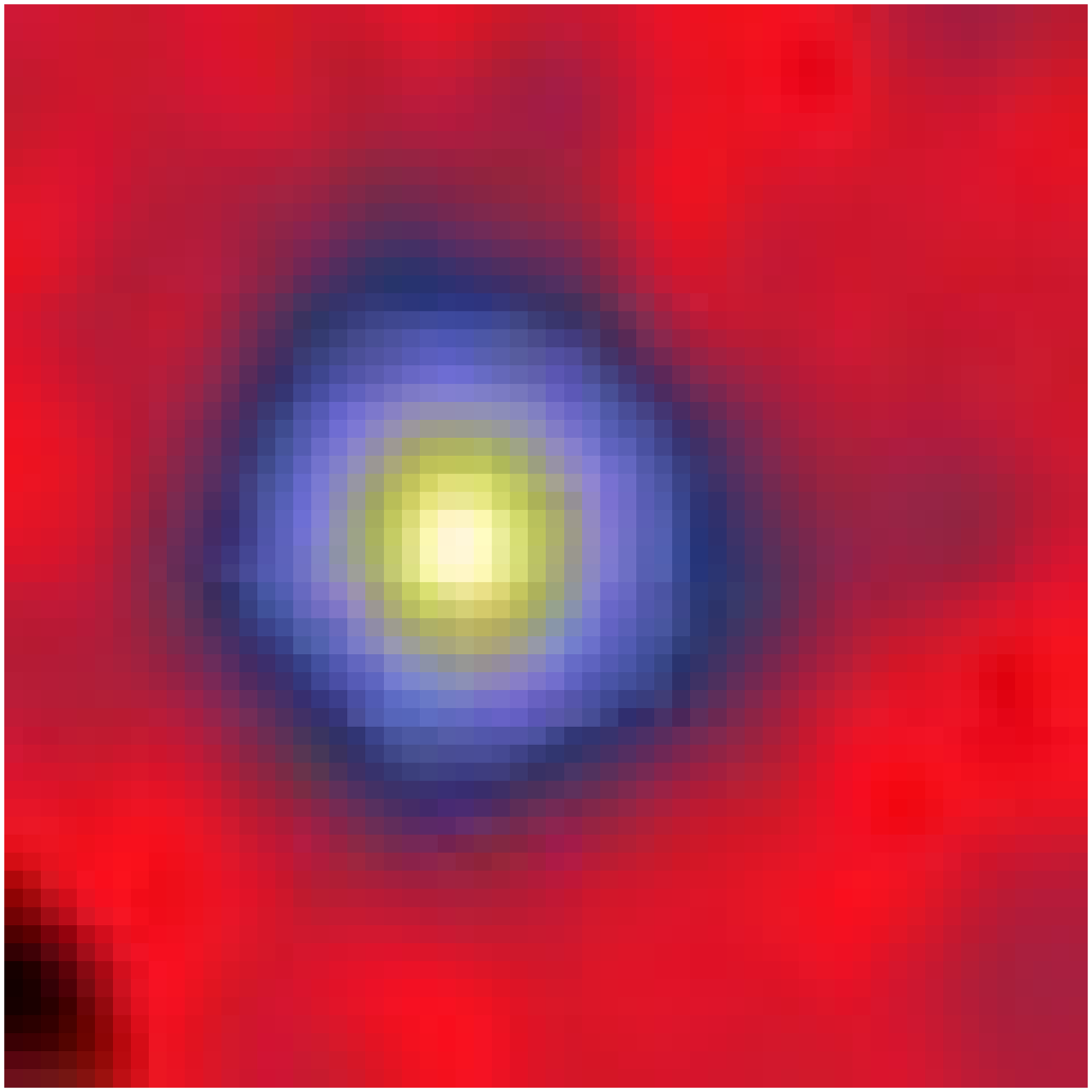}  \FGb{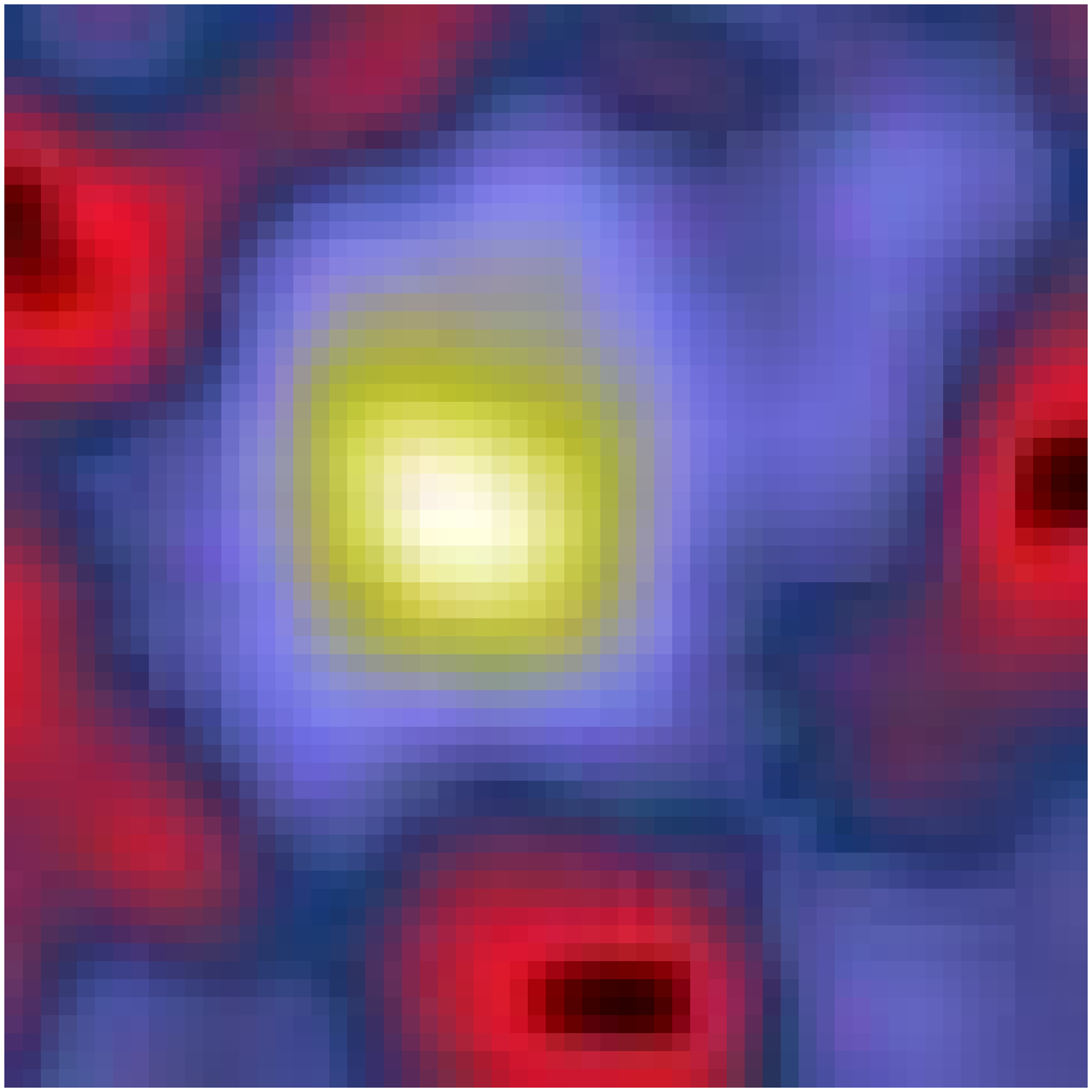} \\  {\#68   NCIA000407}  \\
  \FGb{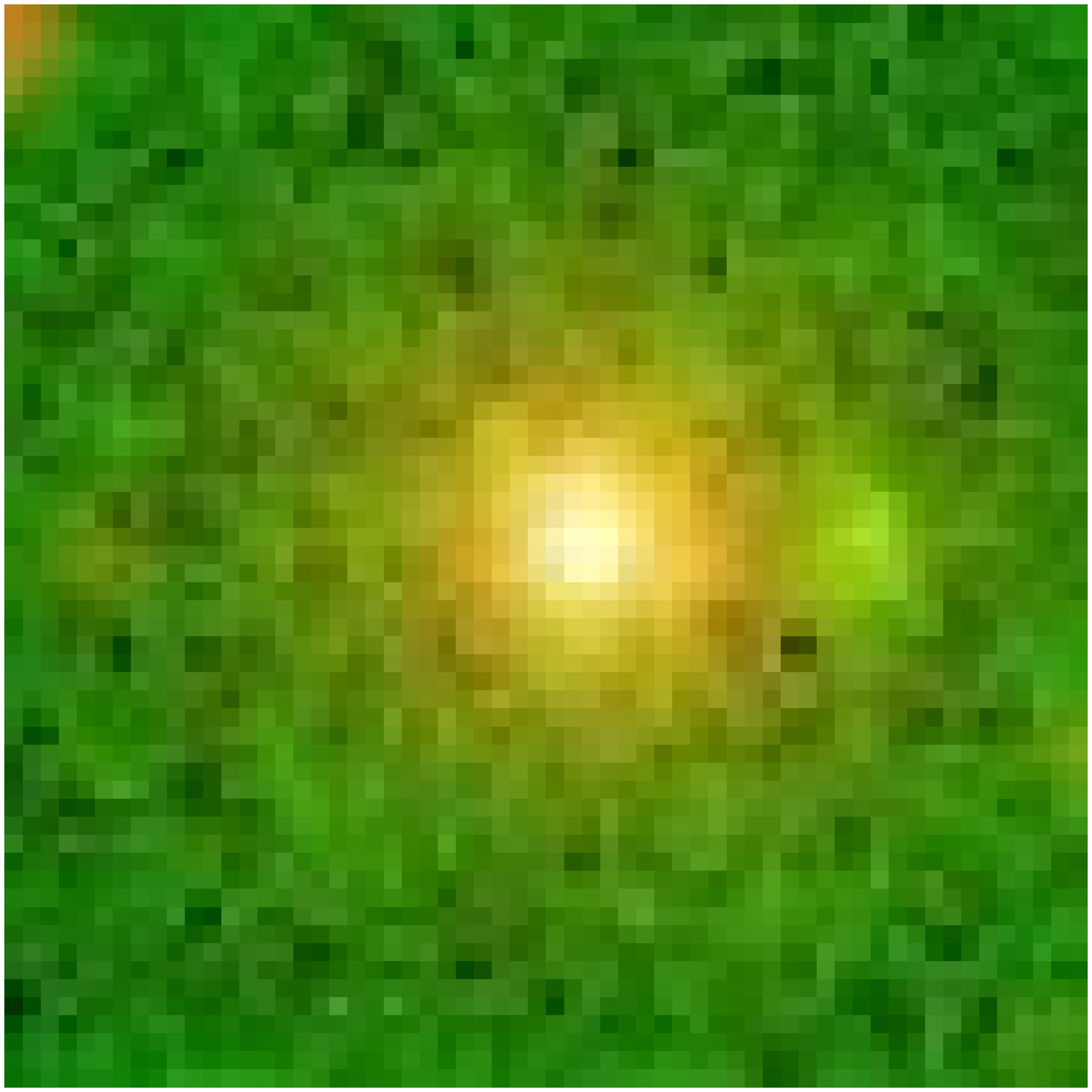}  \FGb{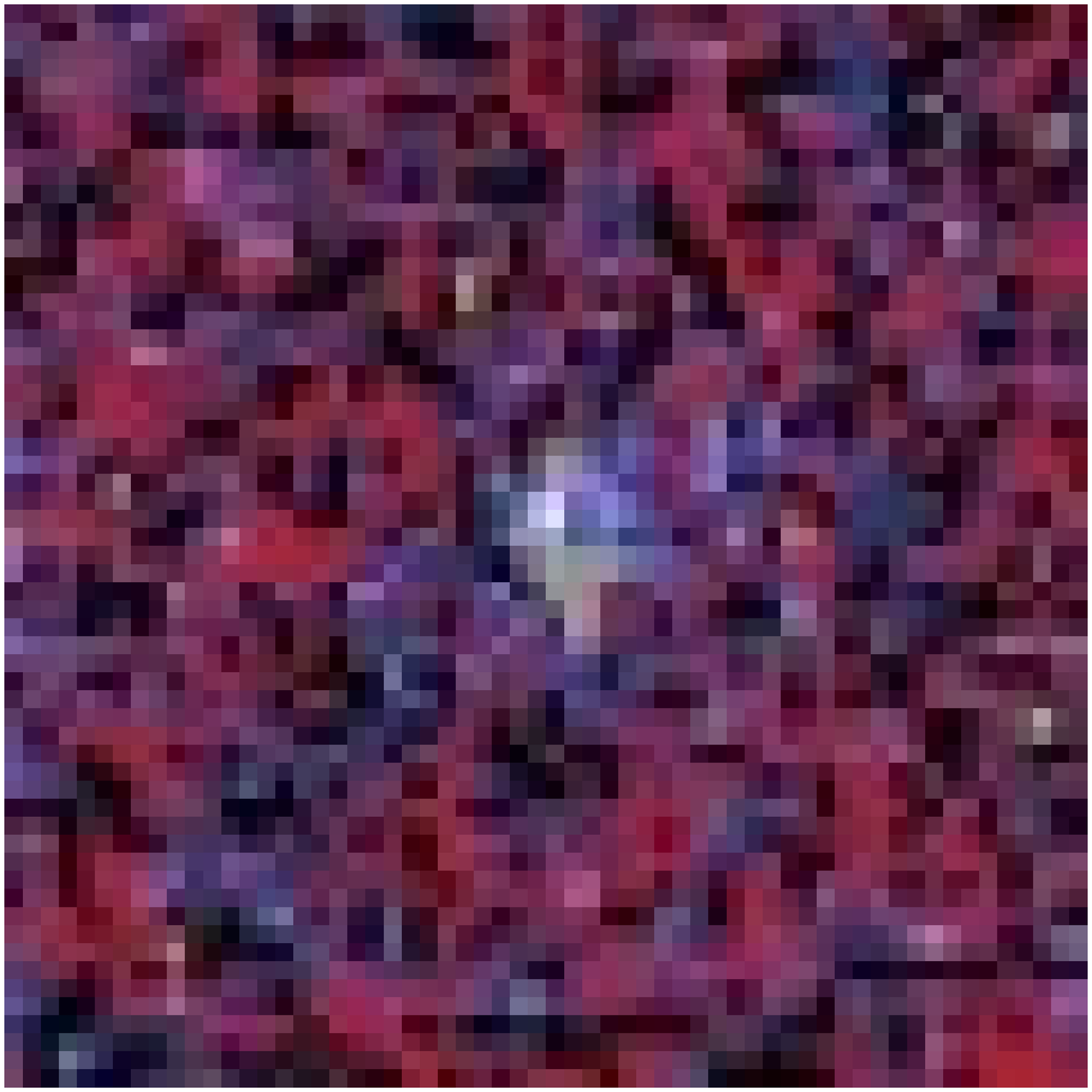}  \FGb{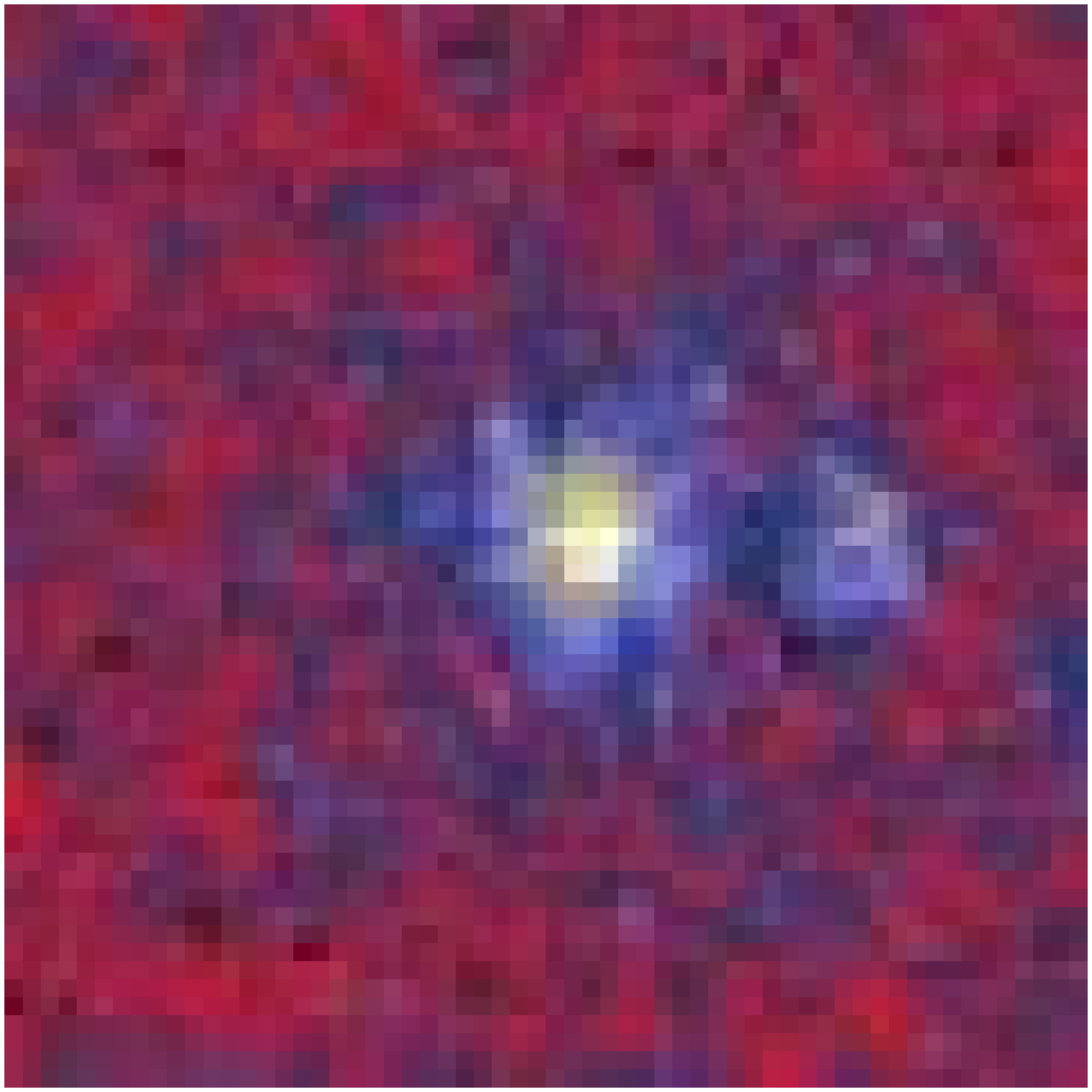}  \FGb{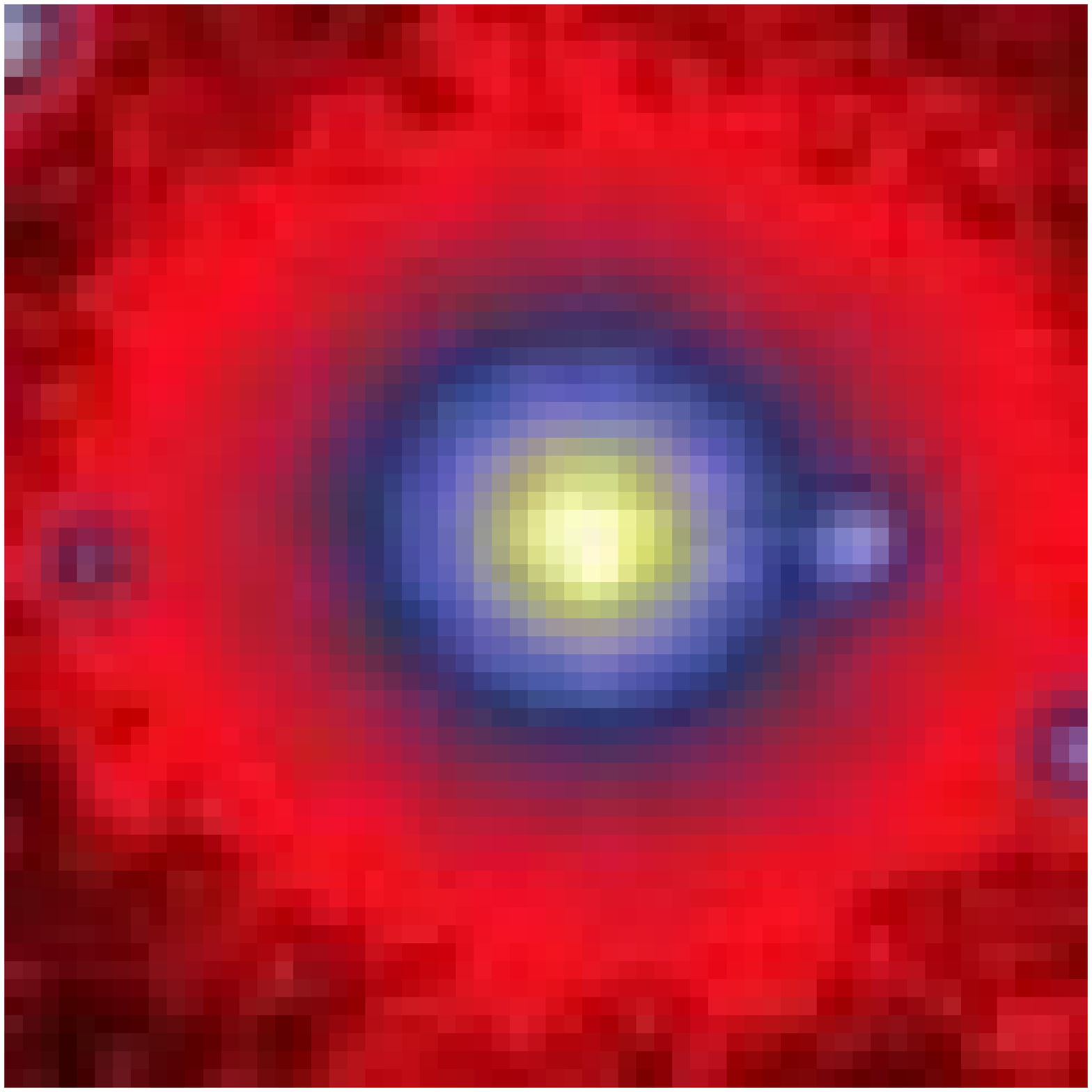}  \FGb{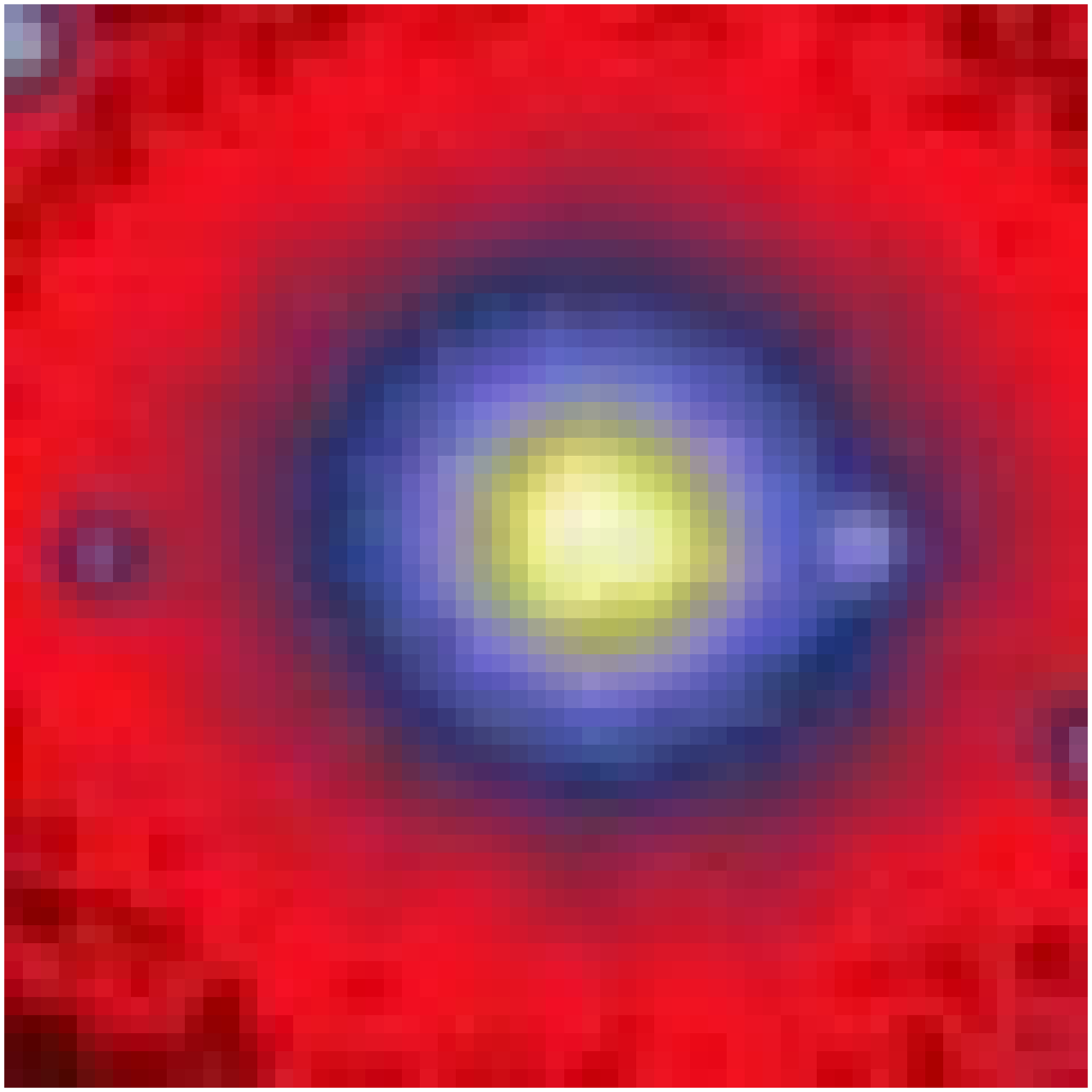}  \FGb{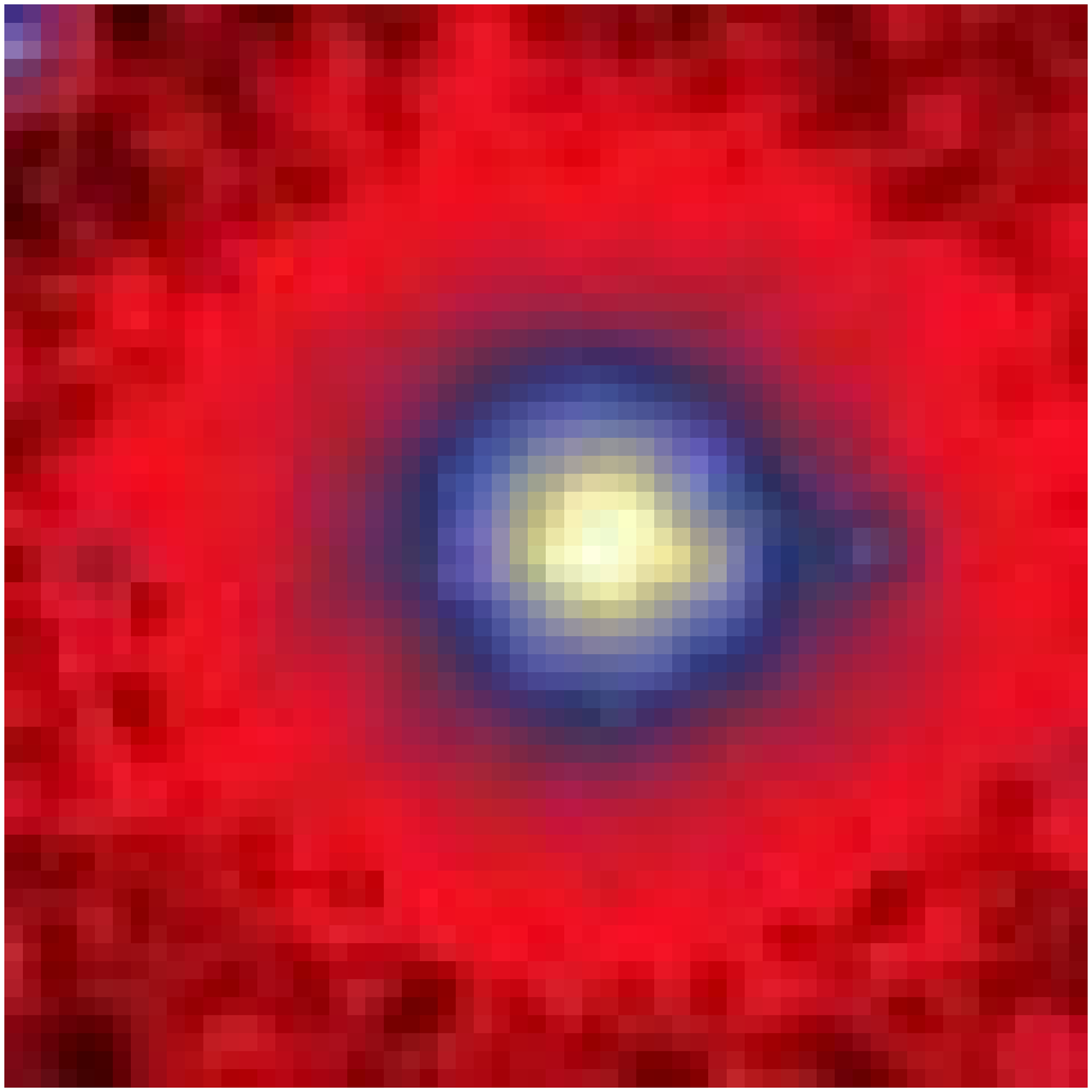}  \FGb{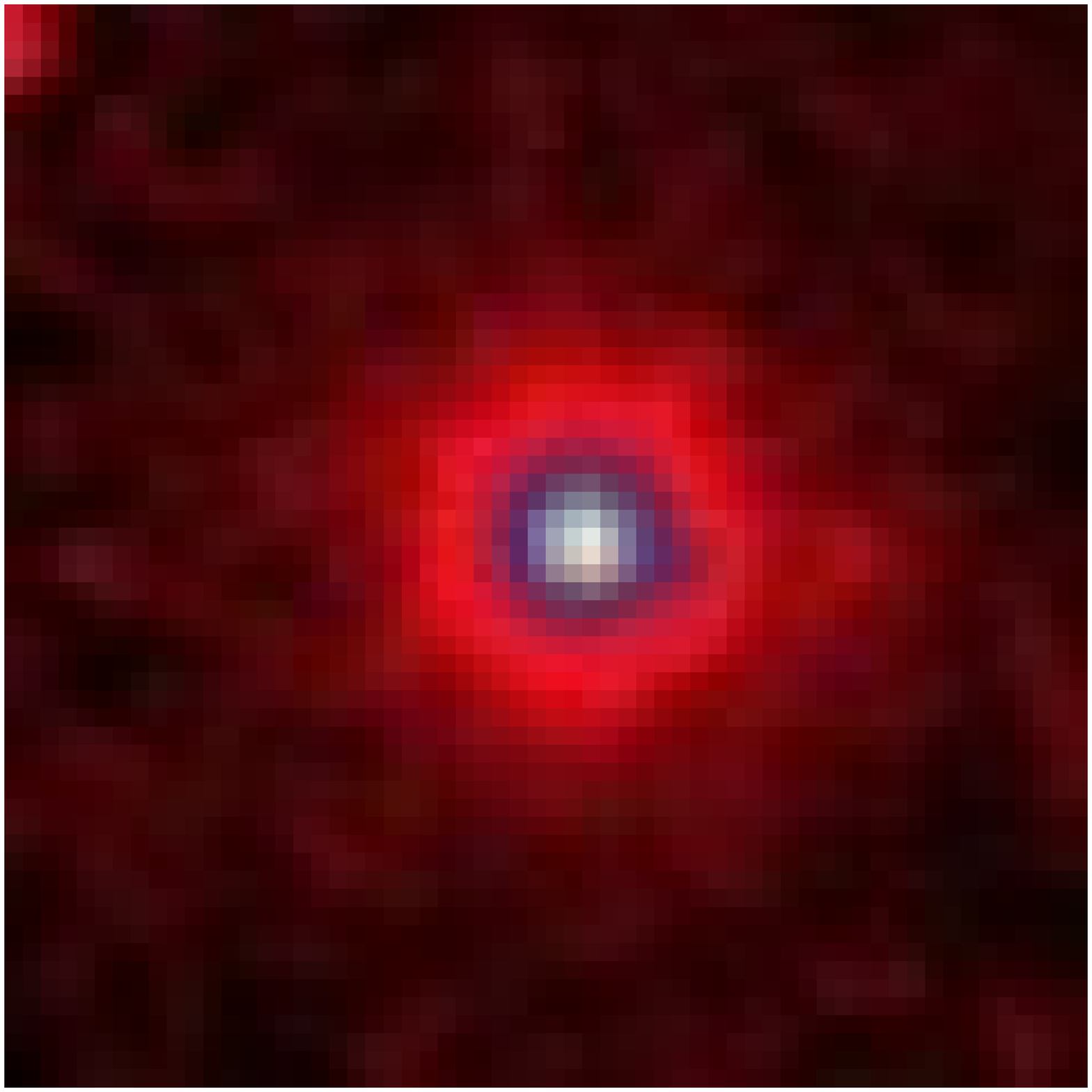}  \FGb{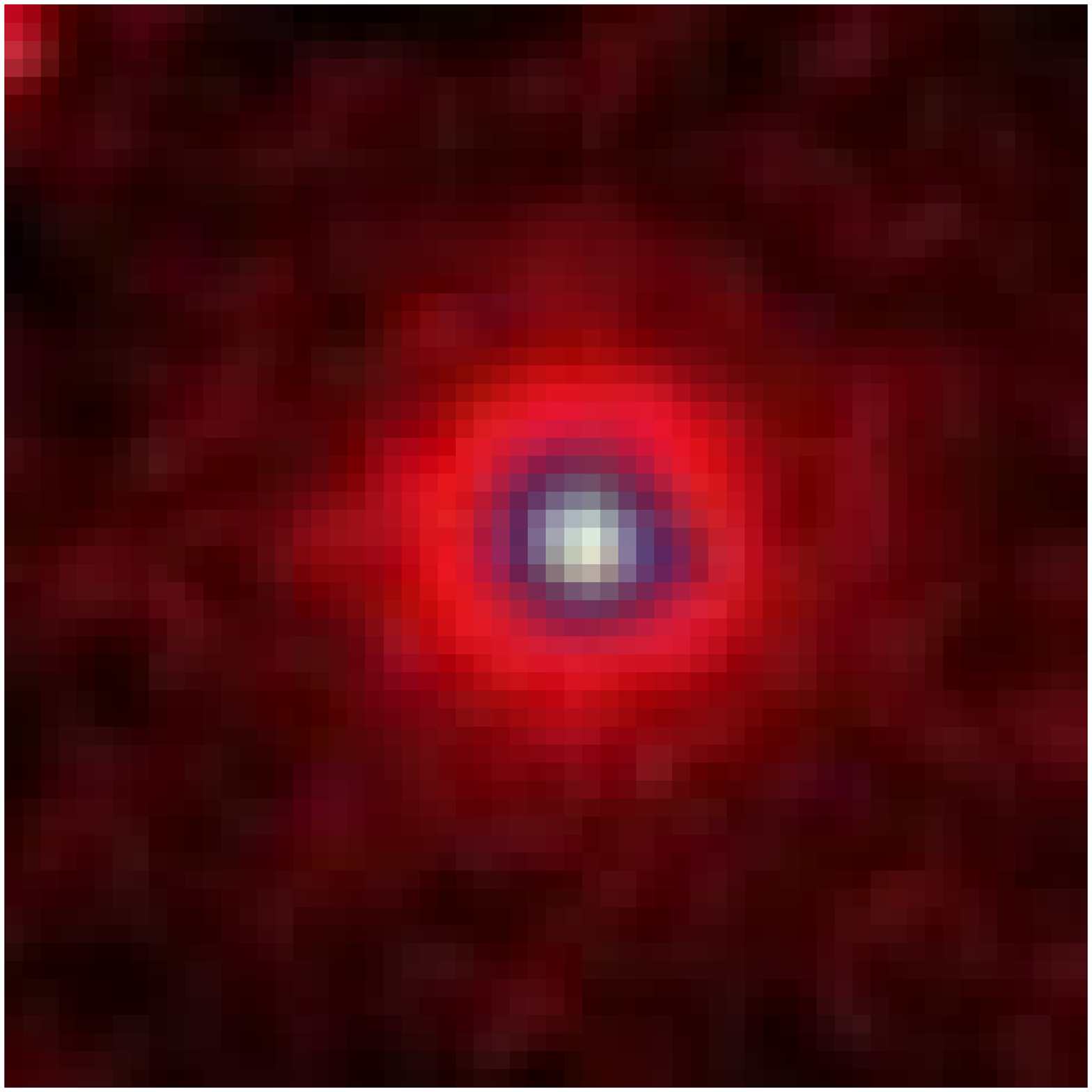}  \FGb{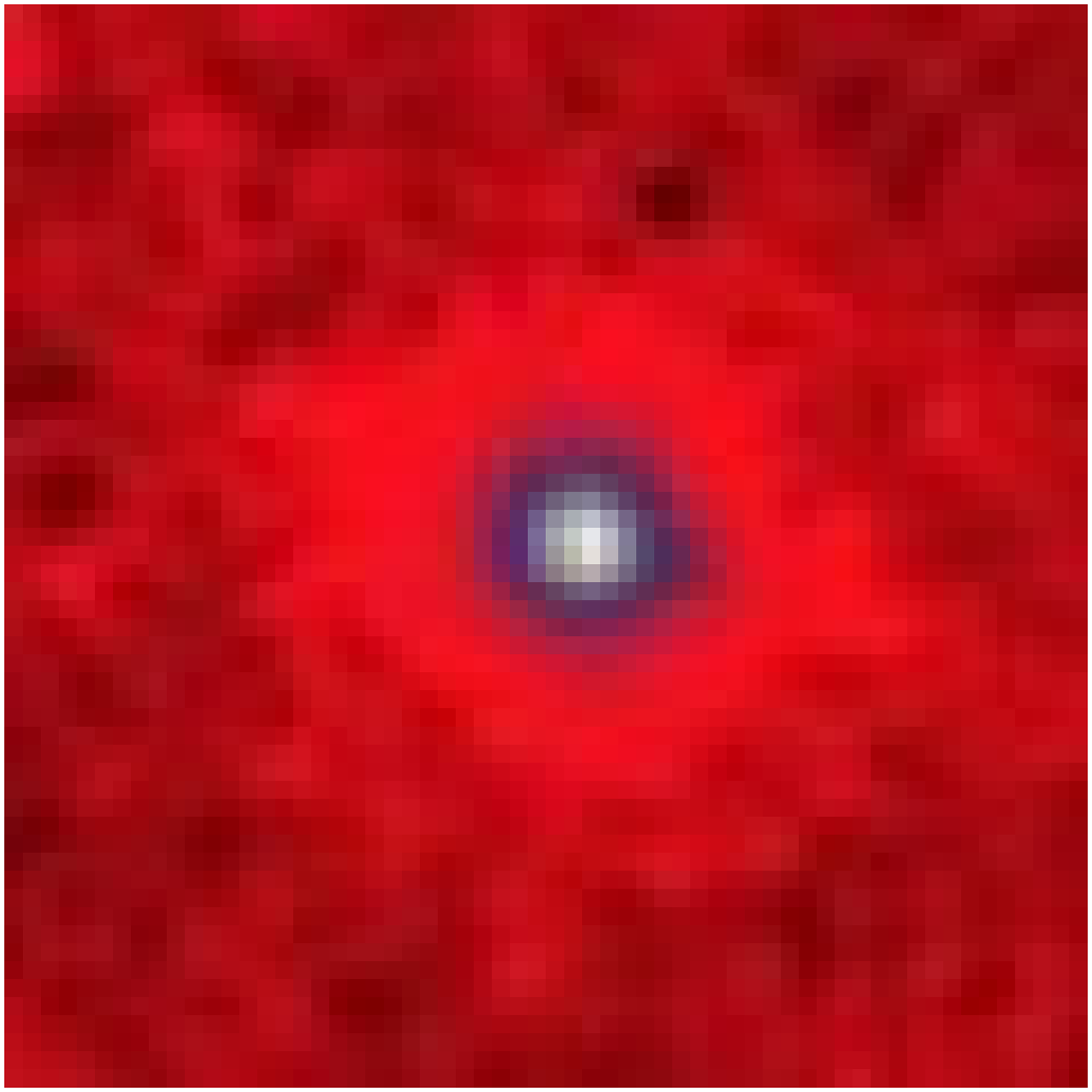}  \FGb{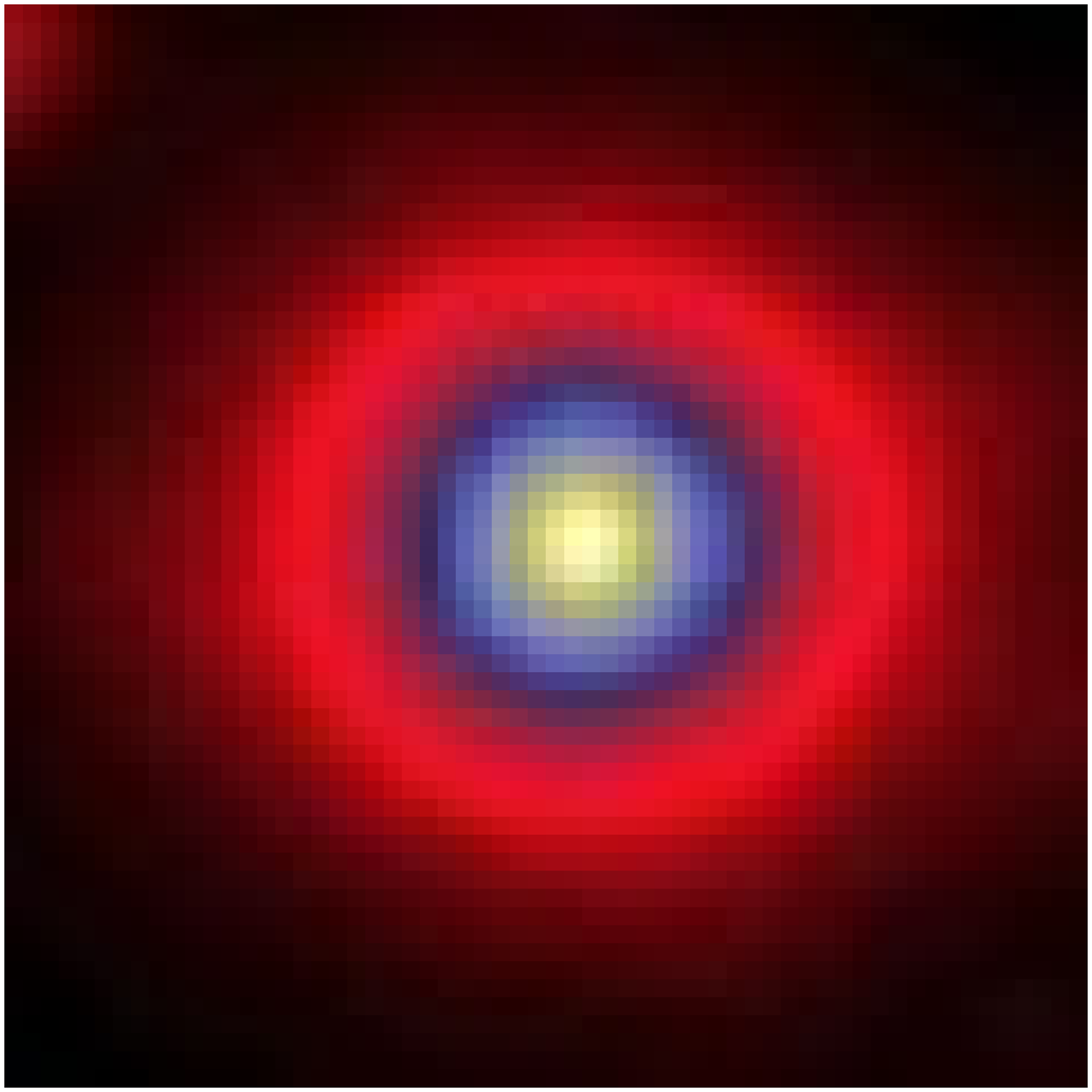}  \FGb{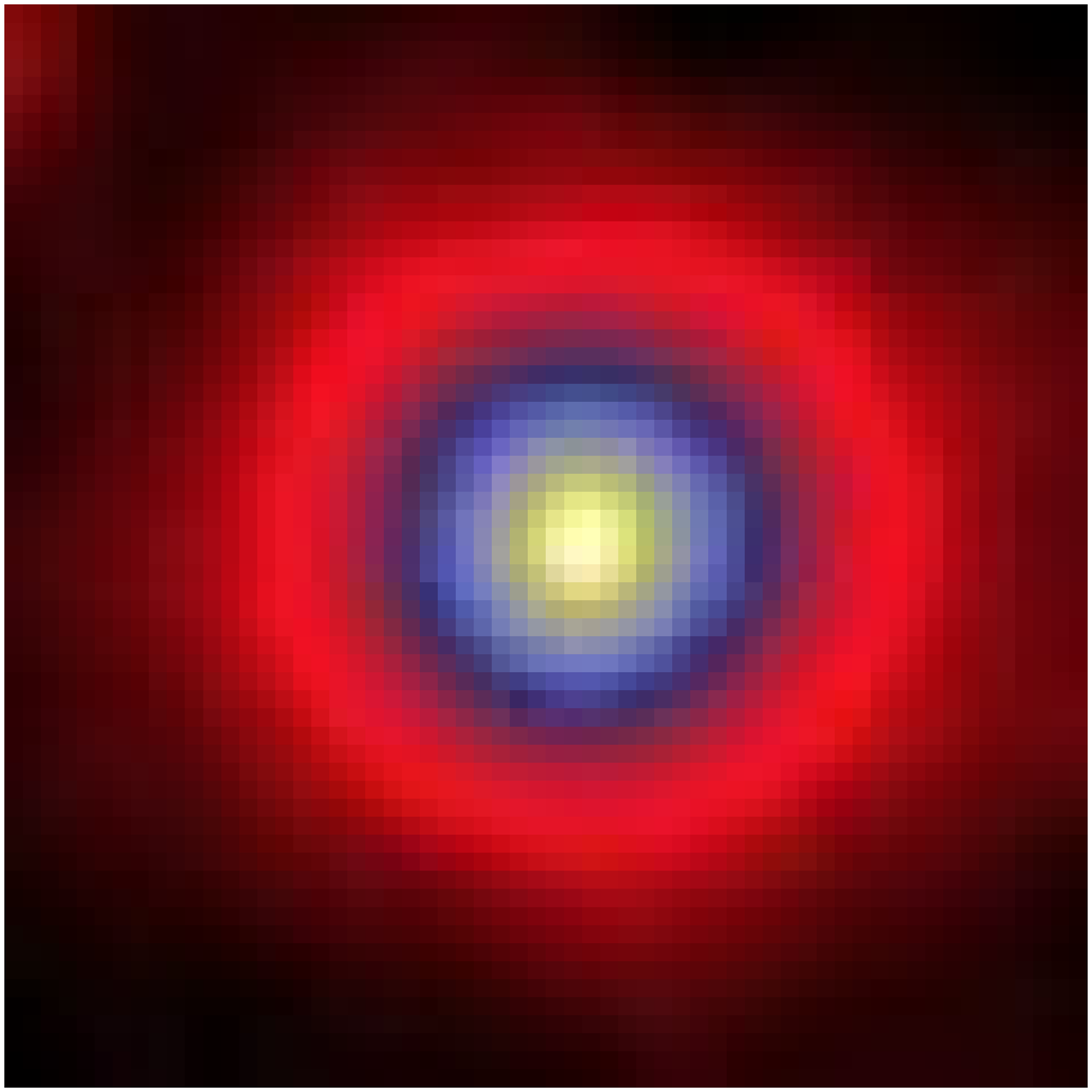}  \FGb{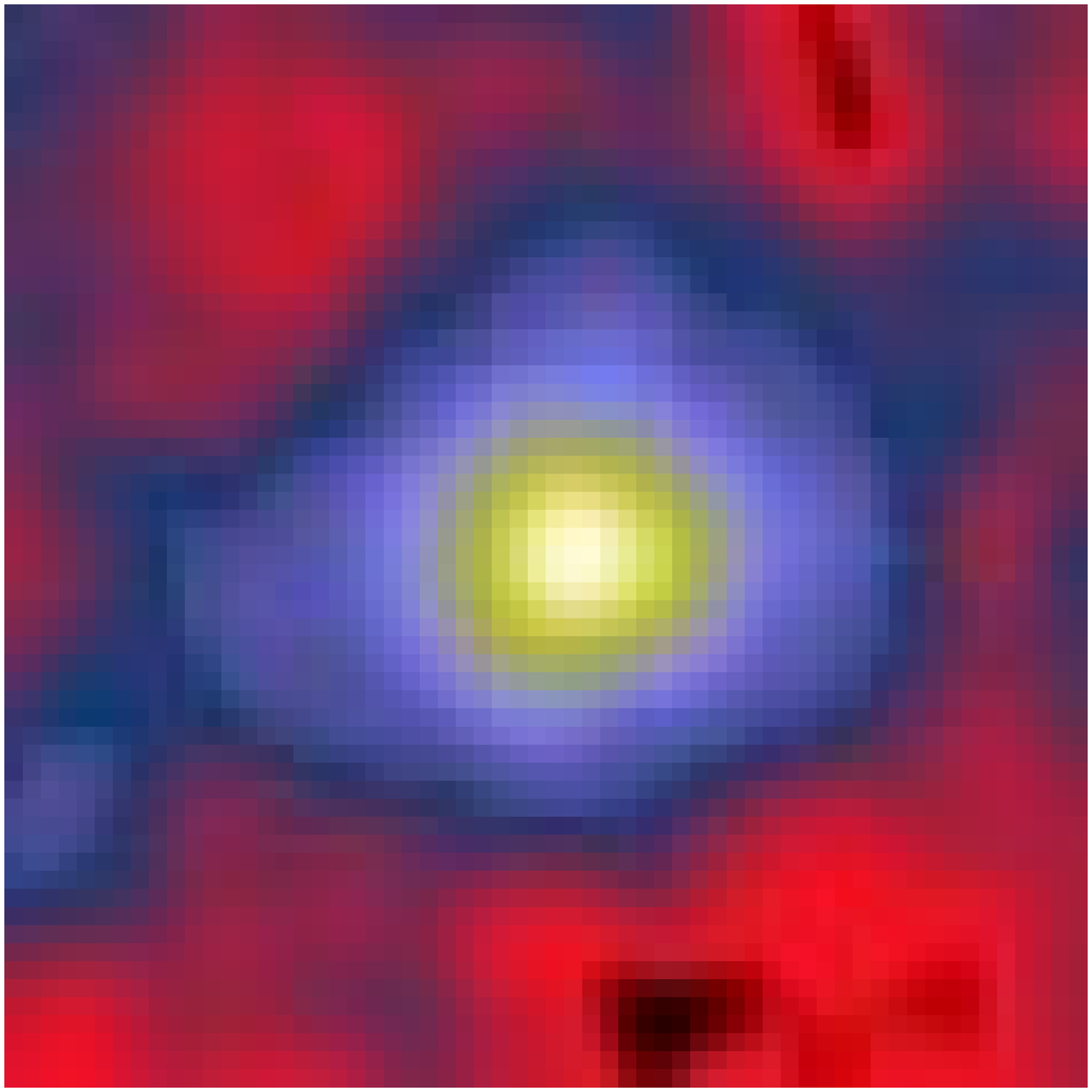}  \FGb{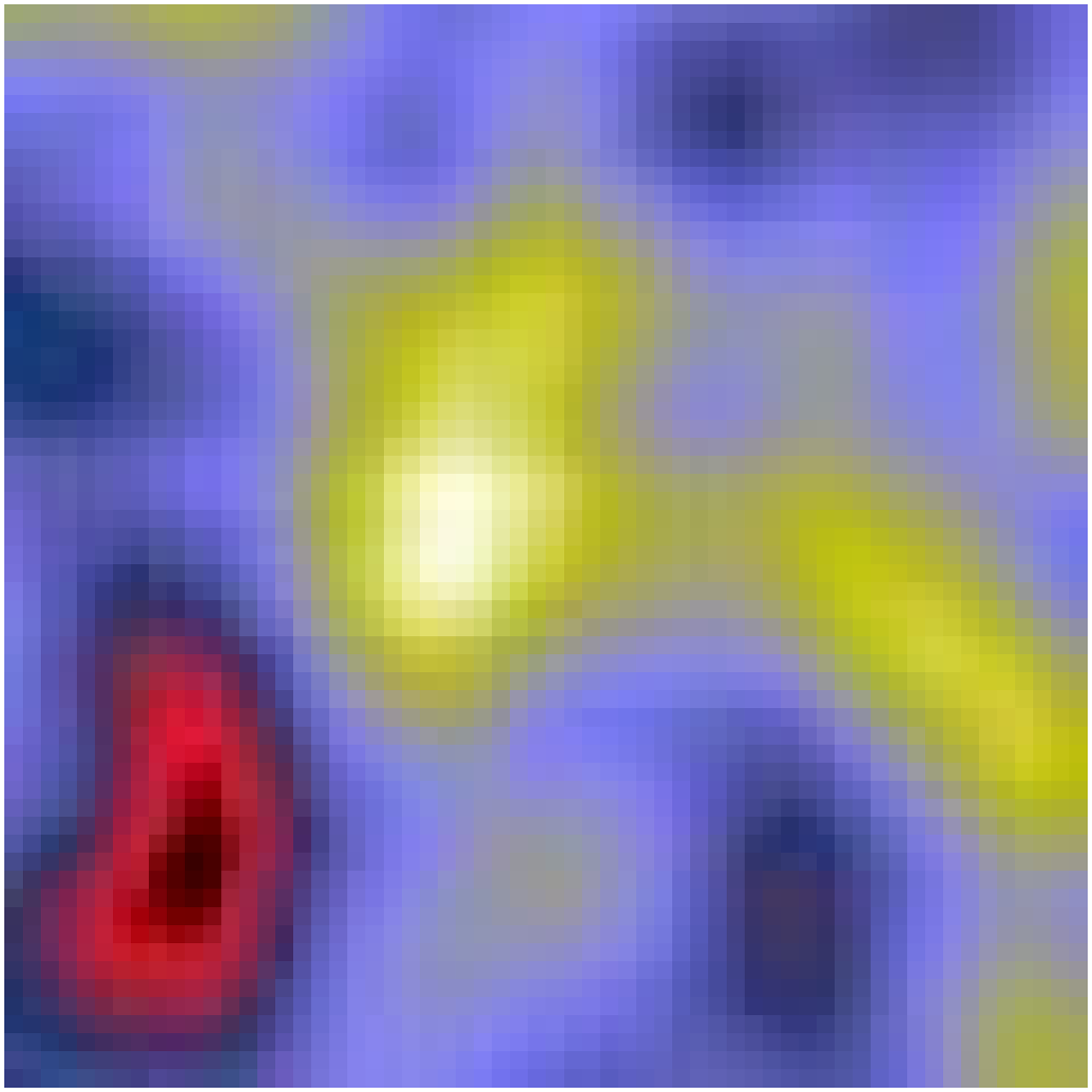} \\  {\#69   NCIA001111}  \\
  \FGb{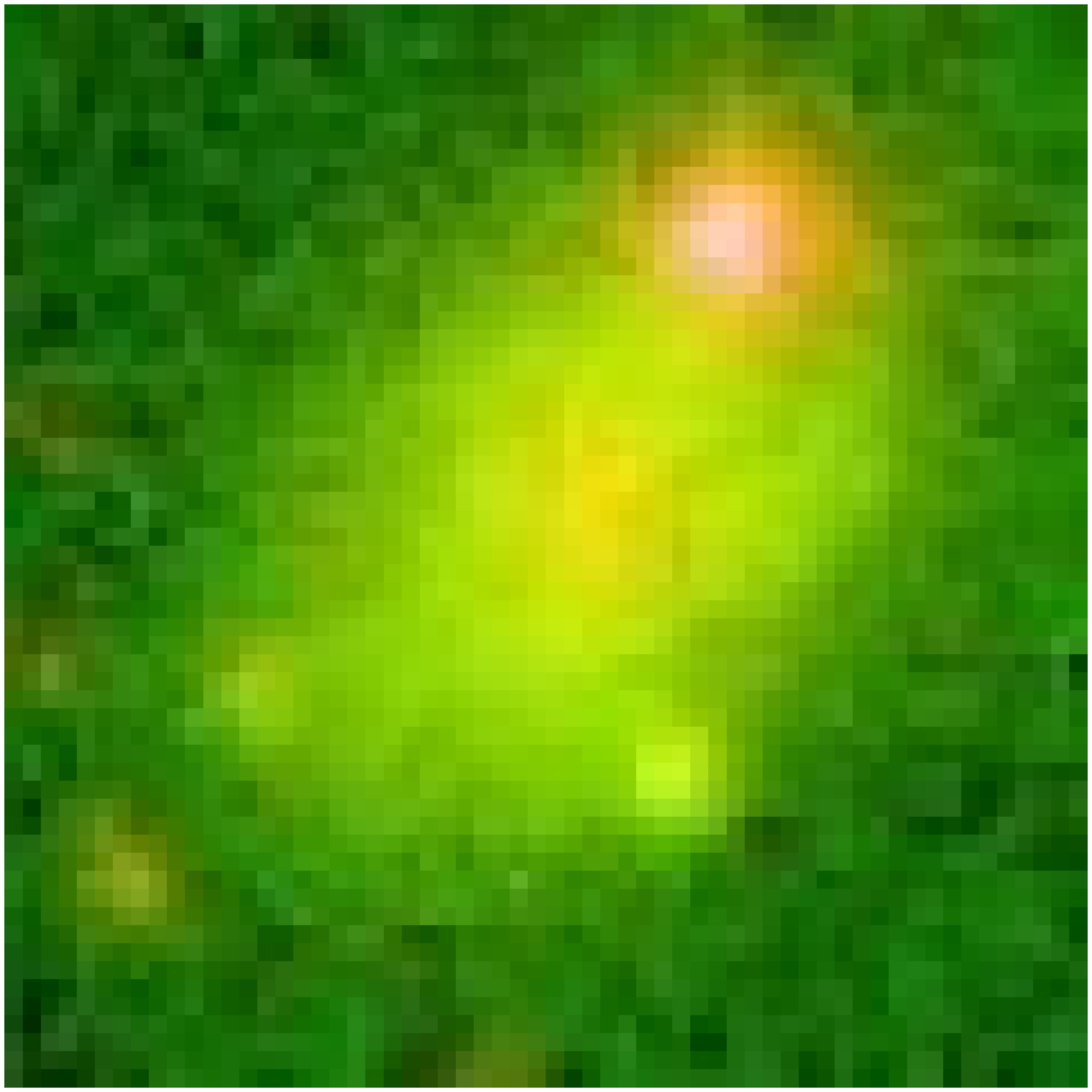}  \FGb{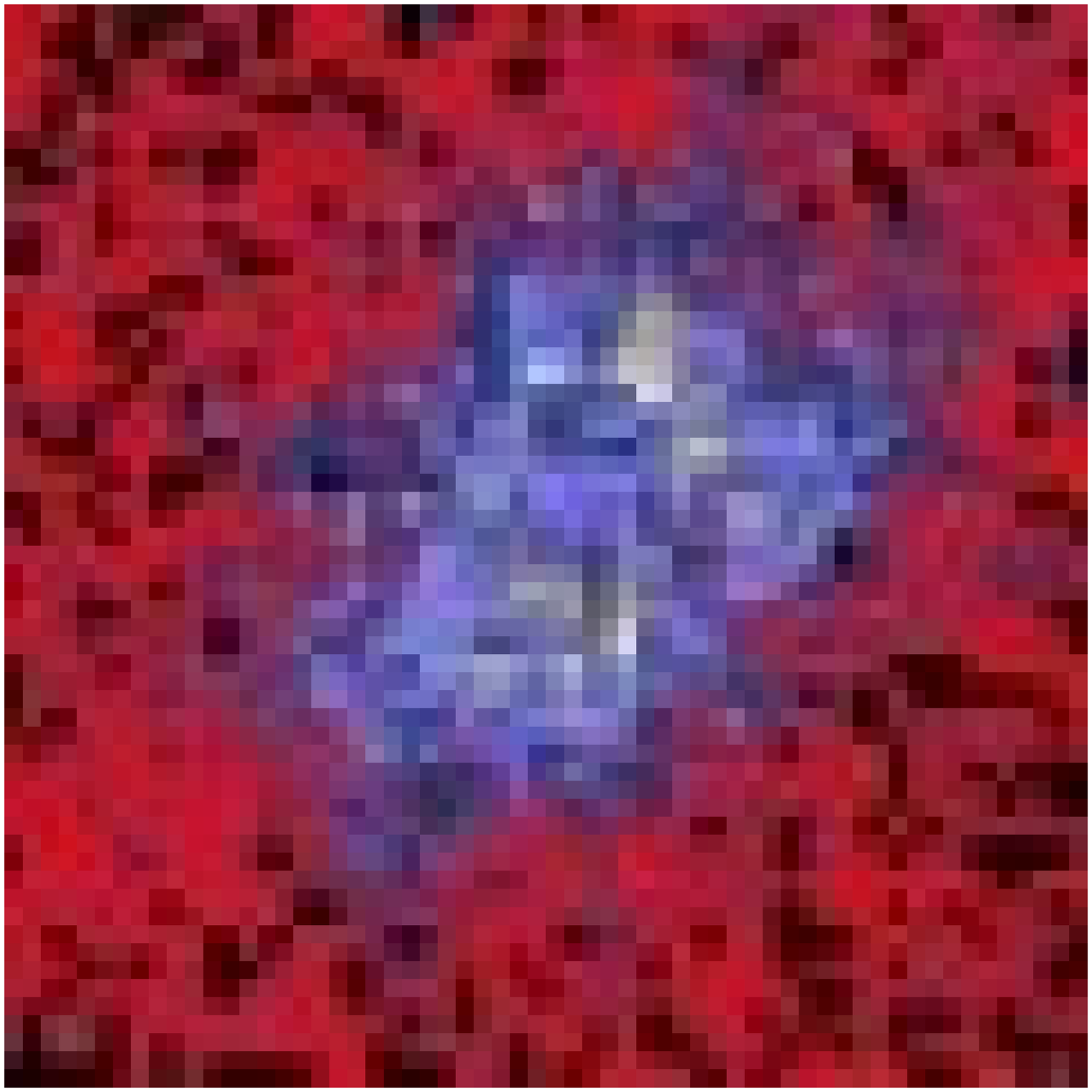}  \FGb{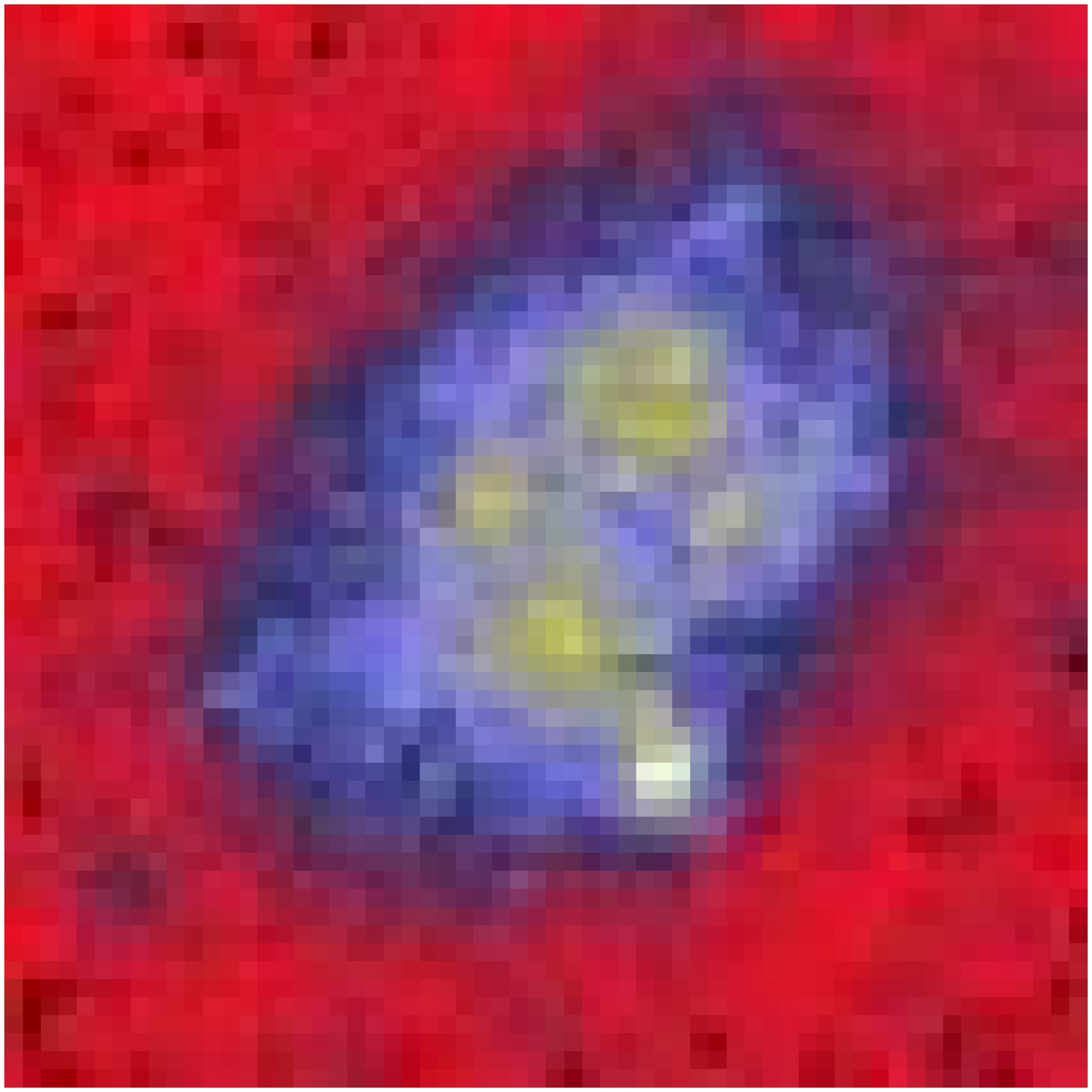}  \FGb{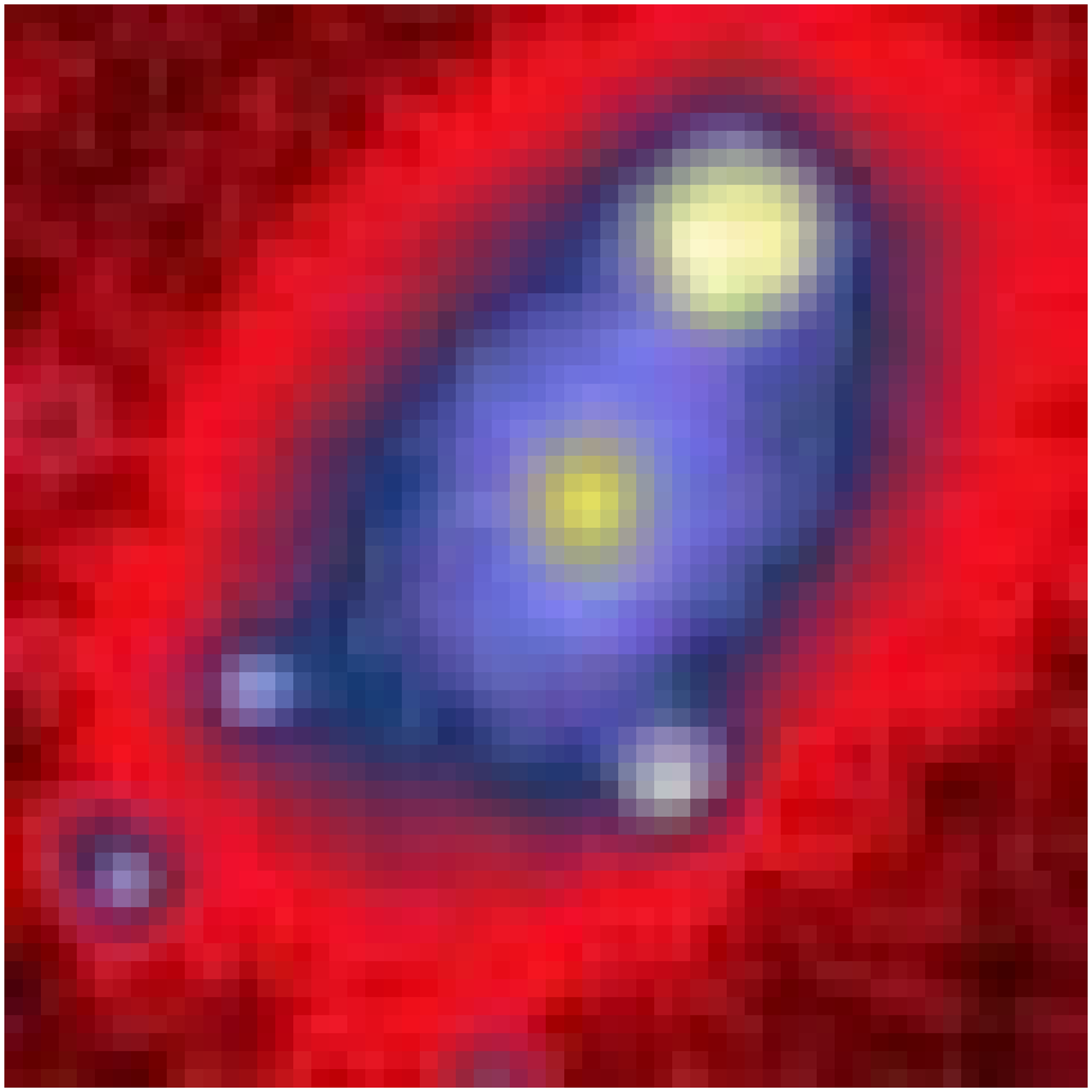}  \FGb{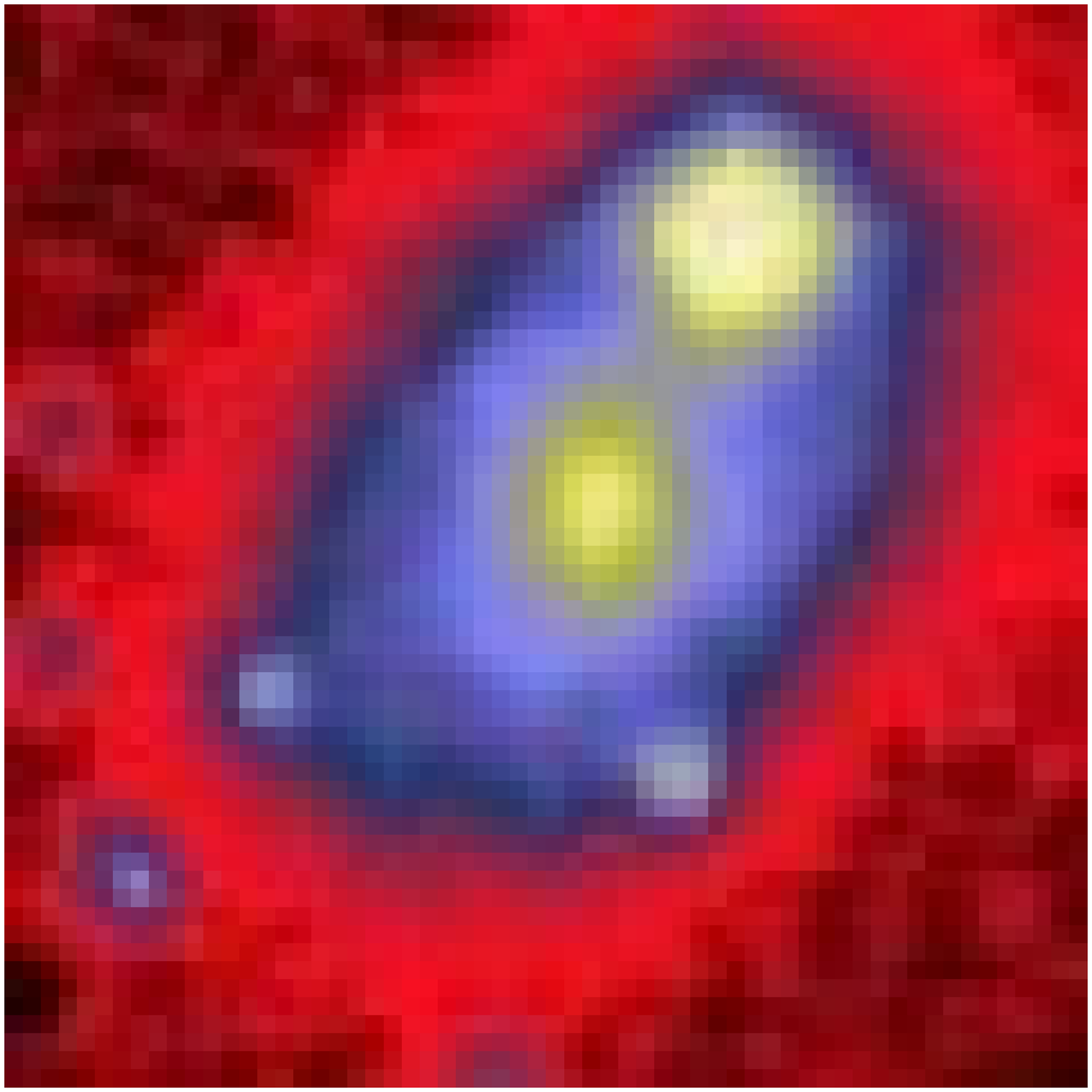}  \FGb{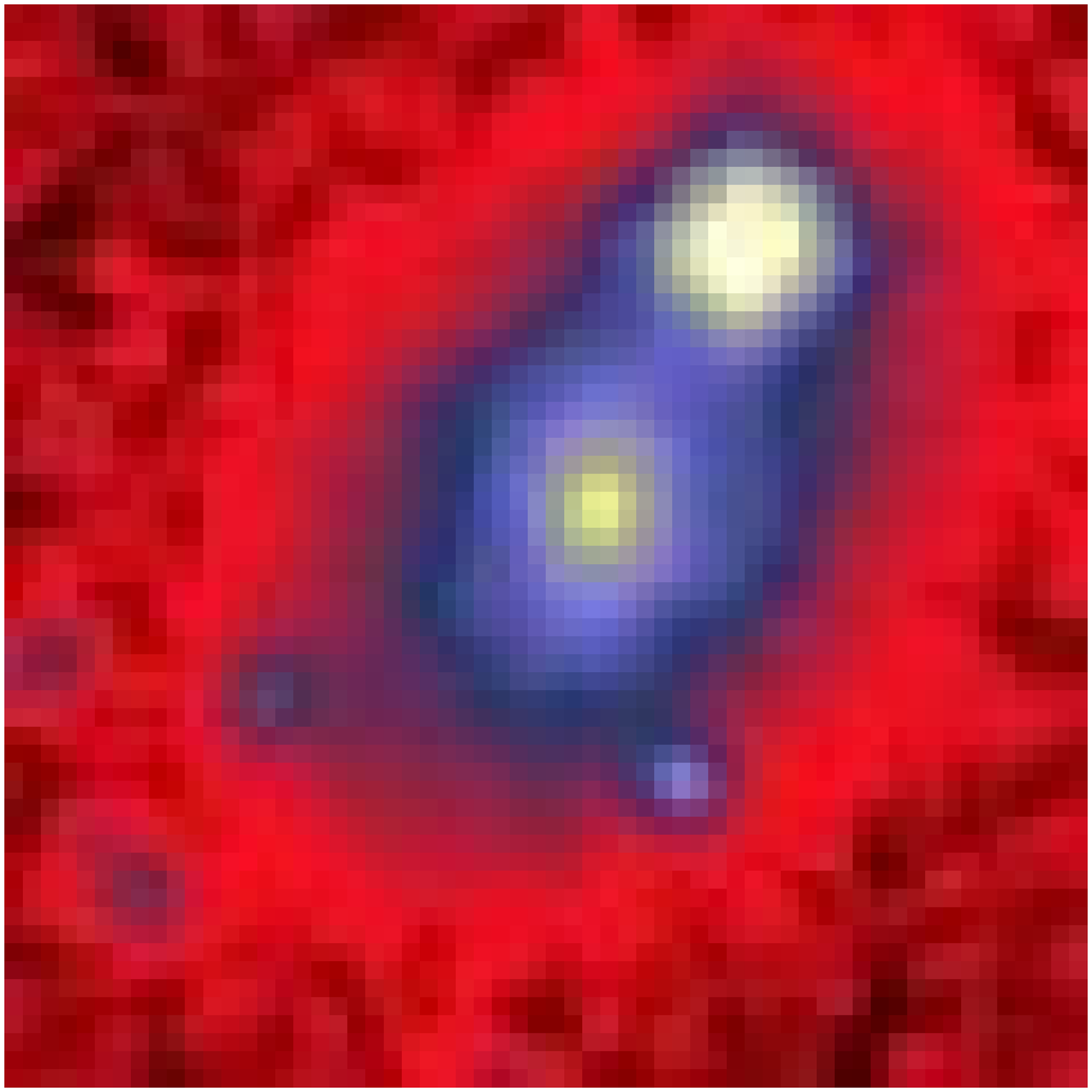}  \FGb{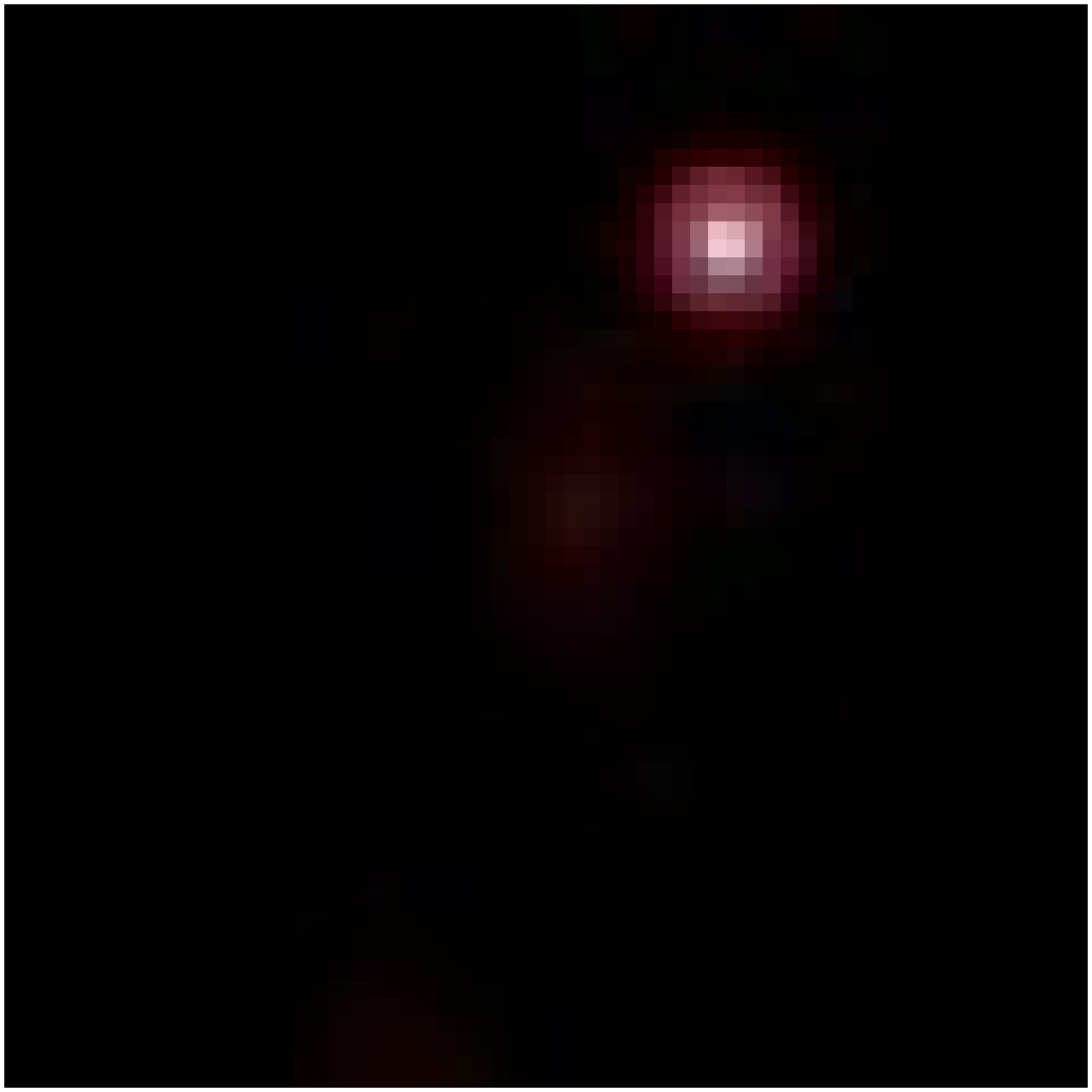}  \FGb{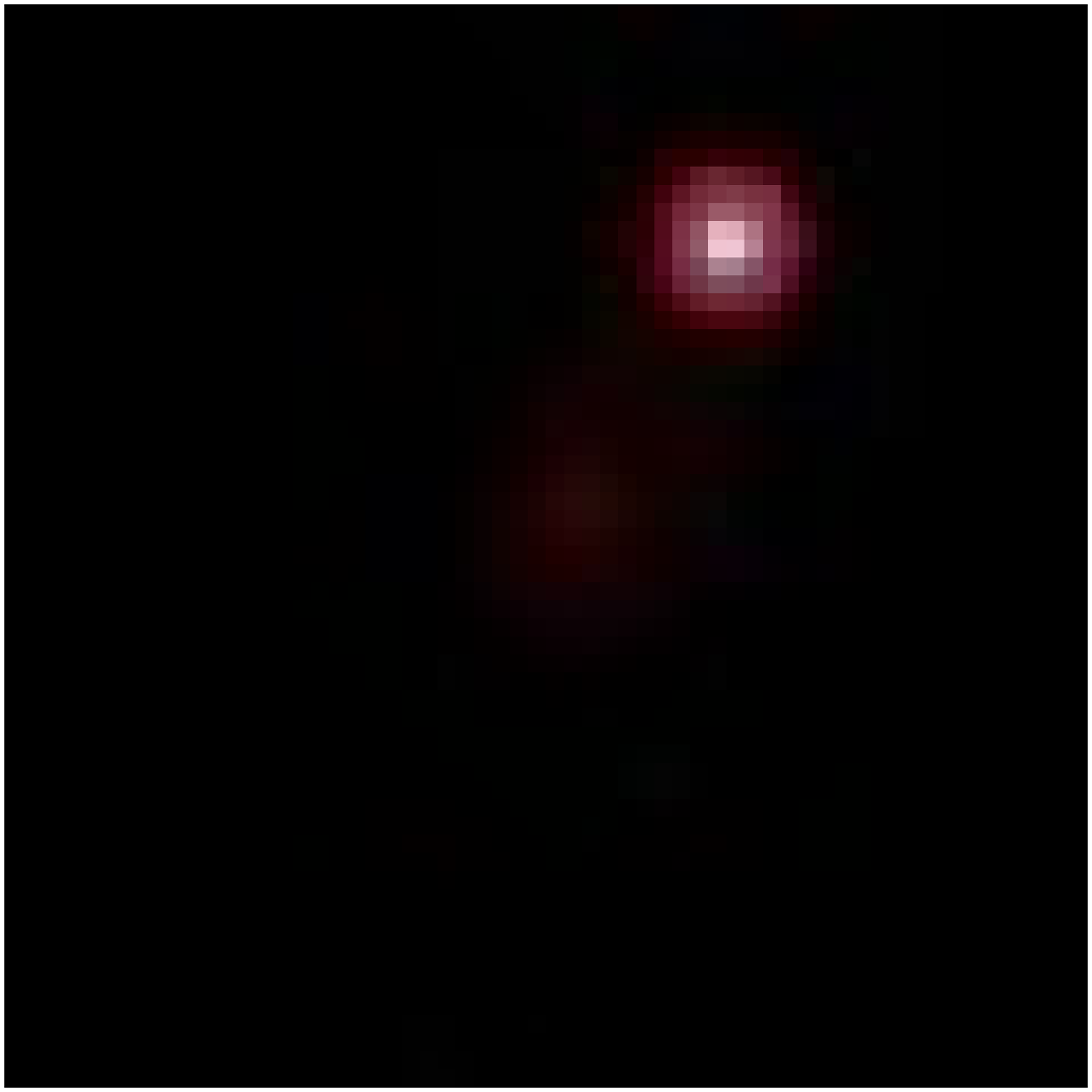}  \FGb{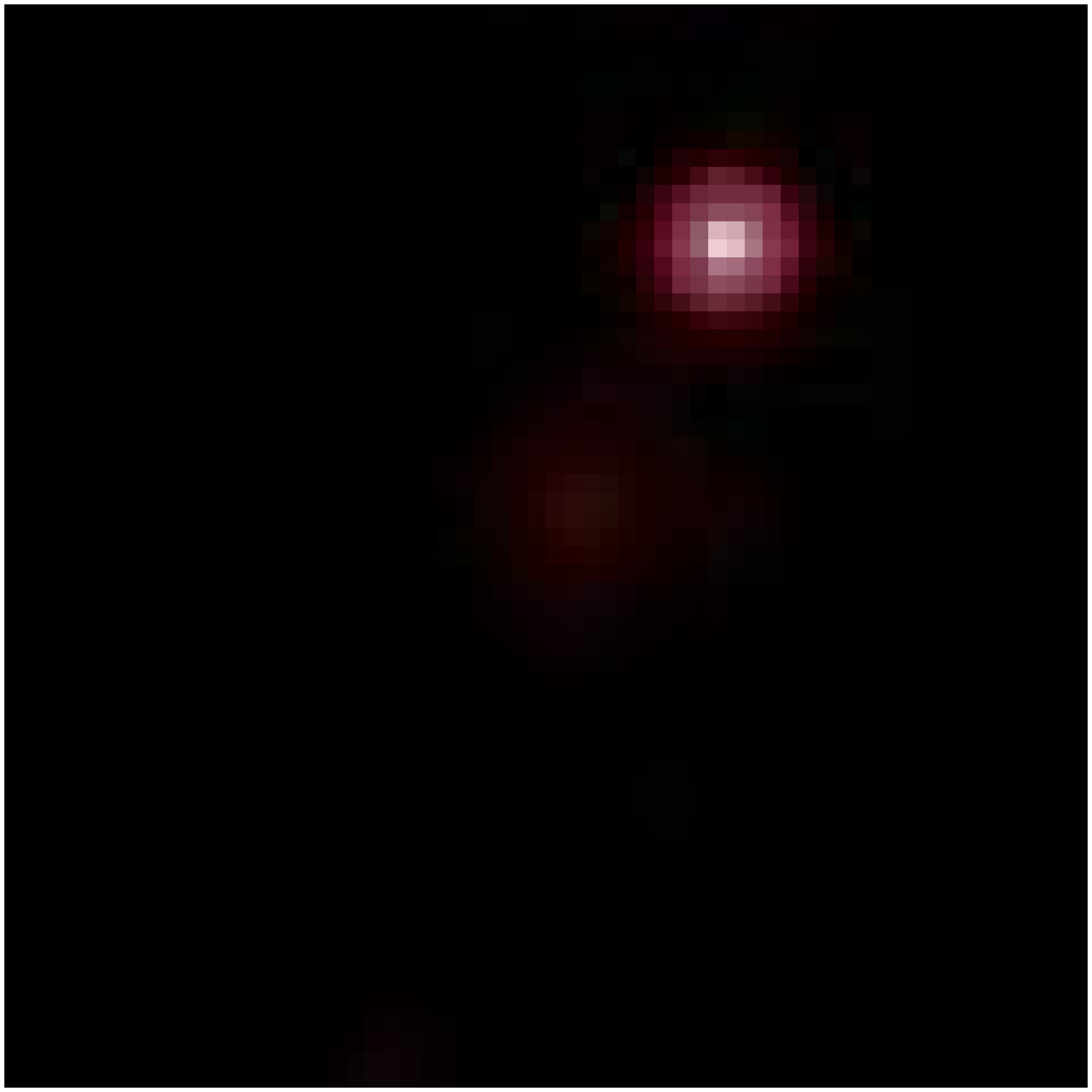}  \FGb{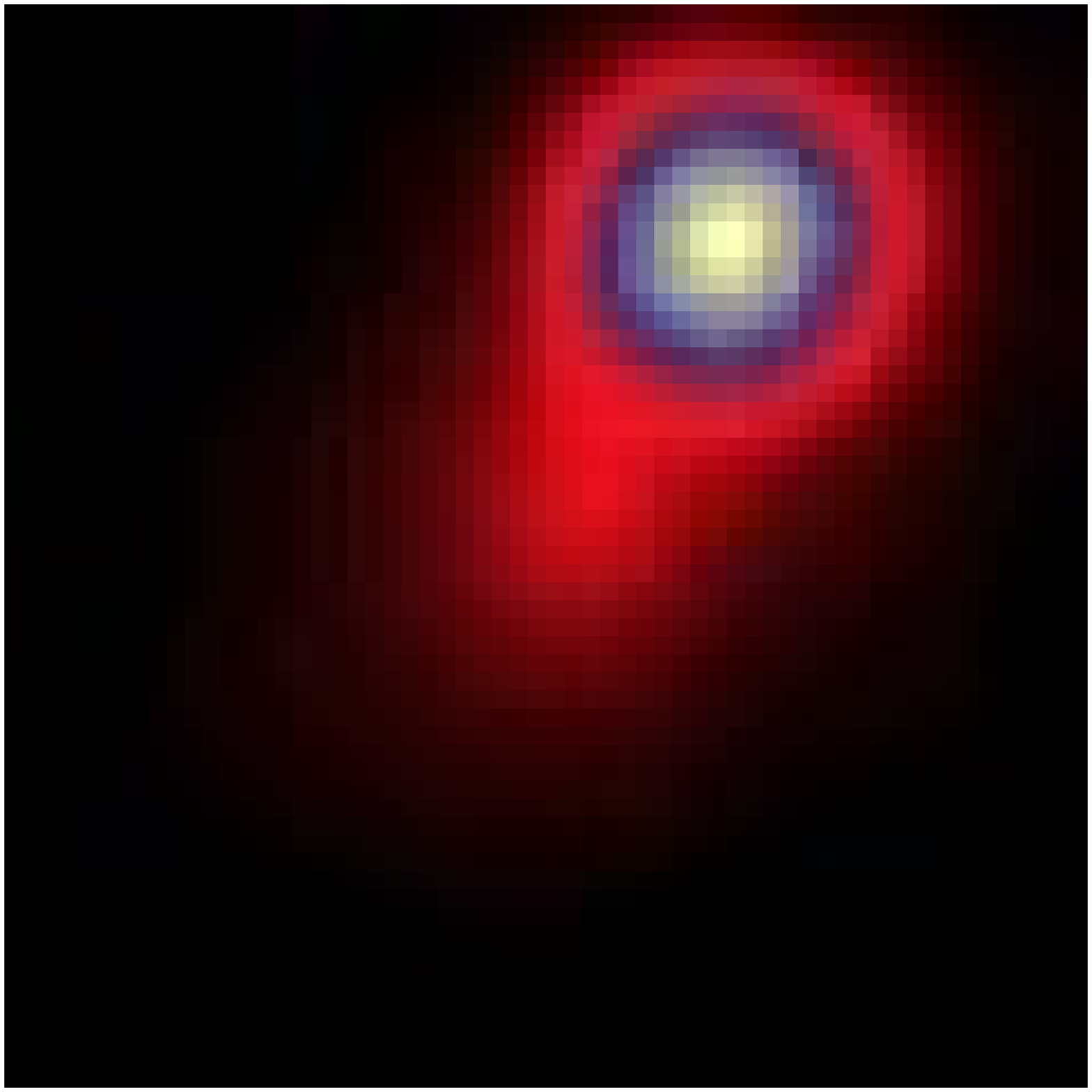}  \FGb{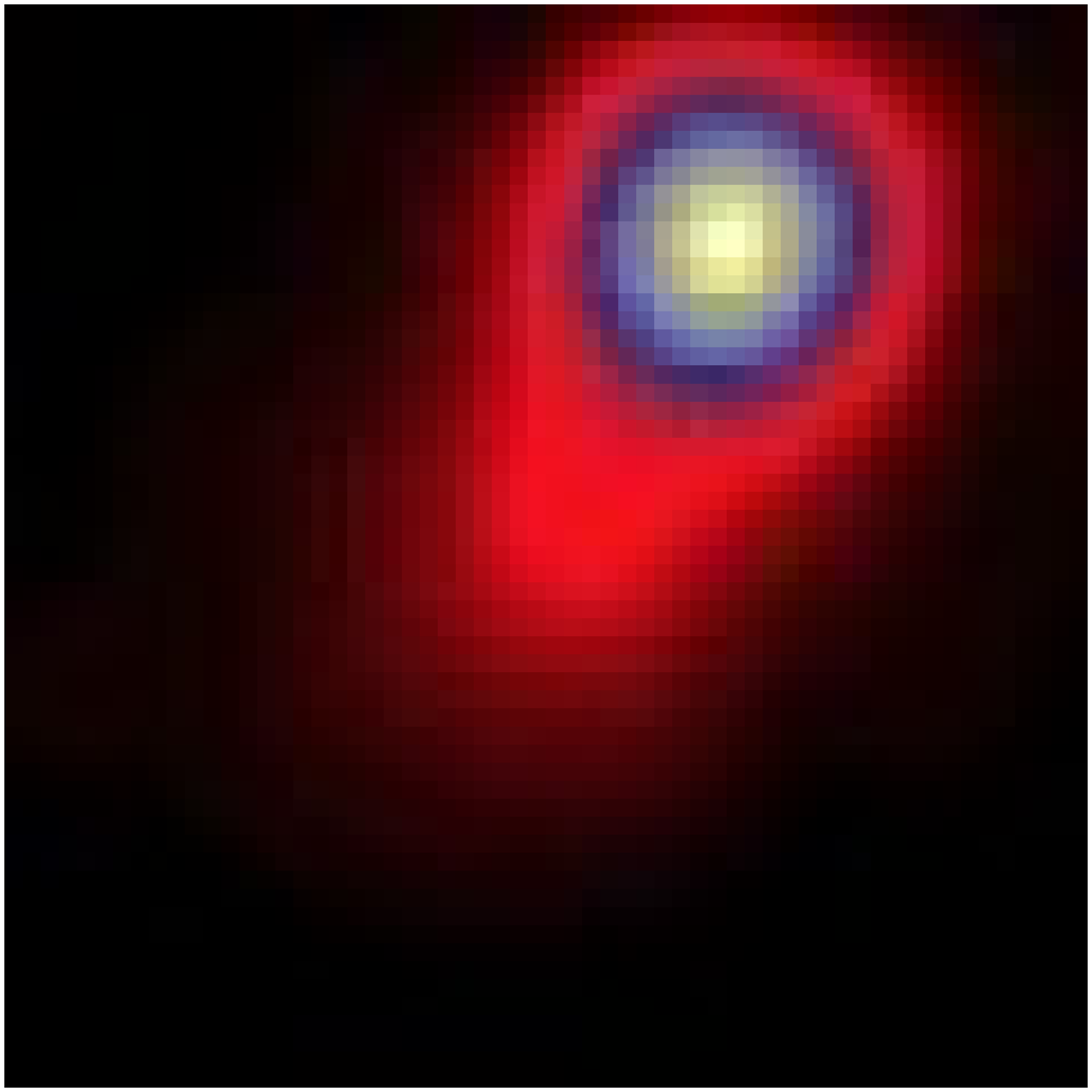}  \FGb{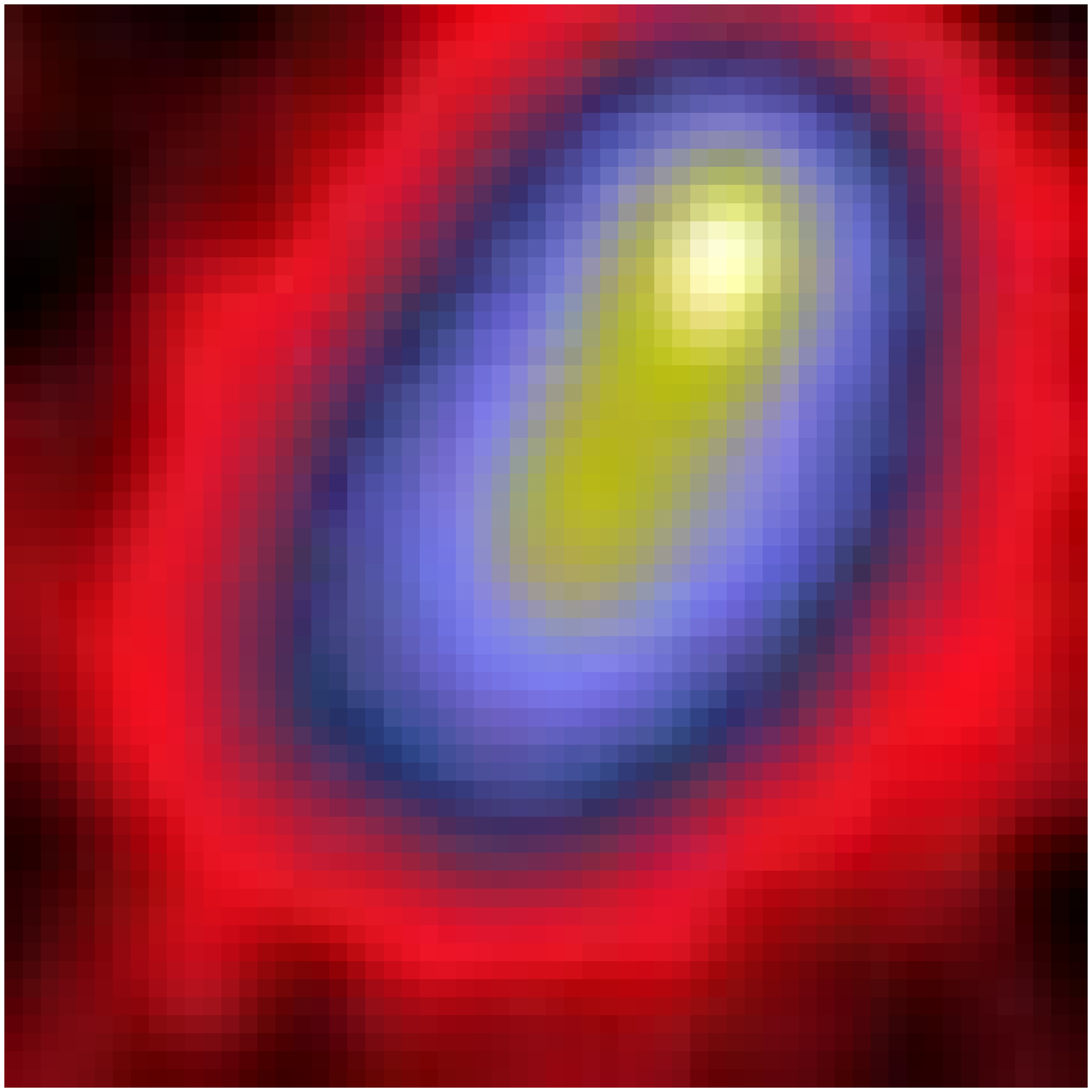}  \FGb{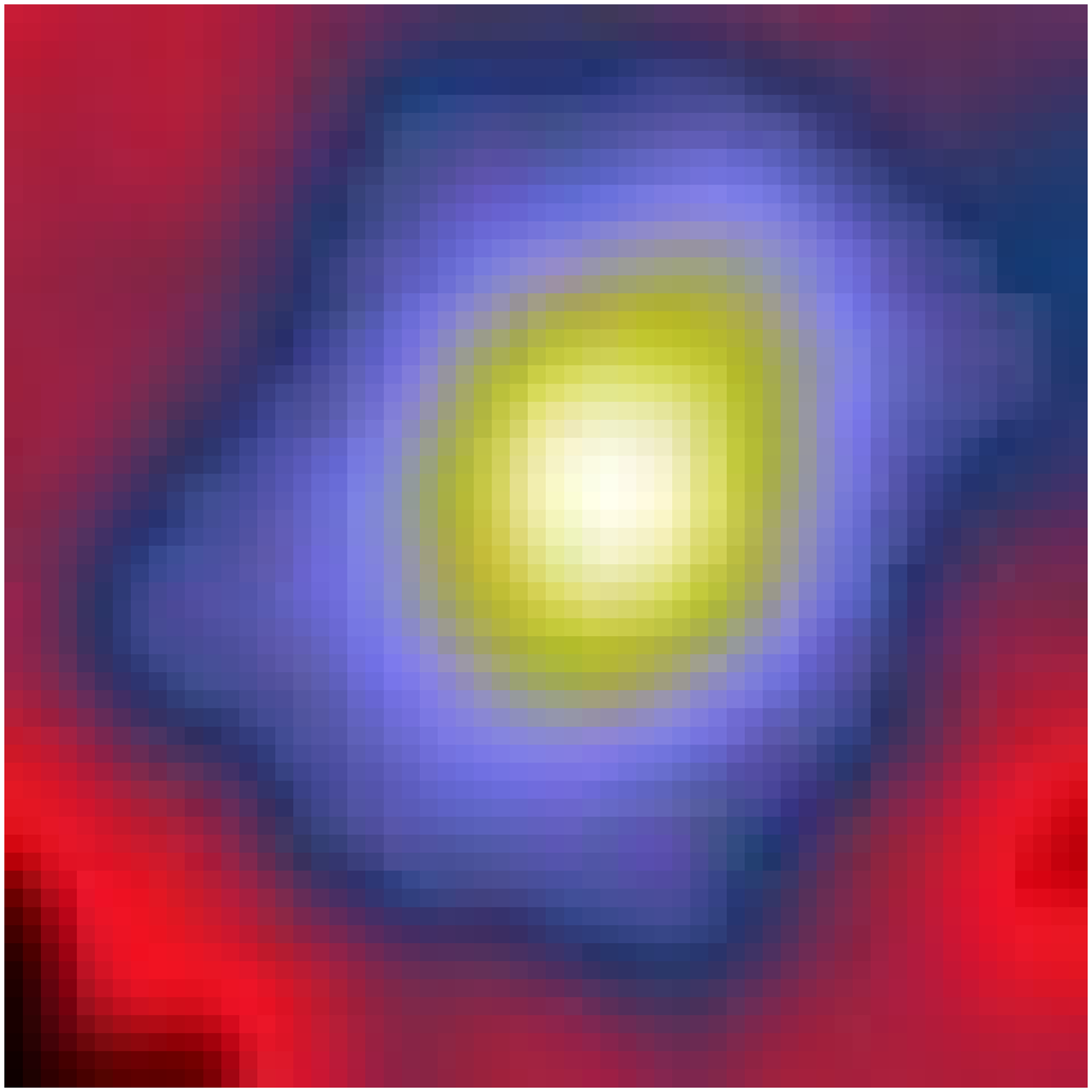} \\  {\#70   NCIA001162}  \\
  \FGb{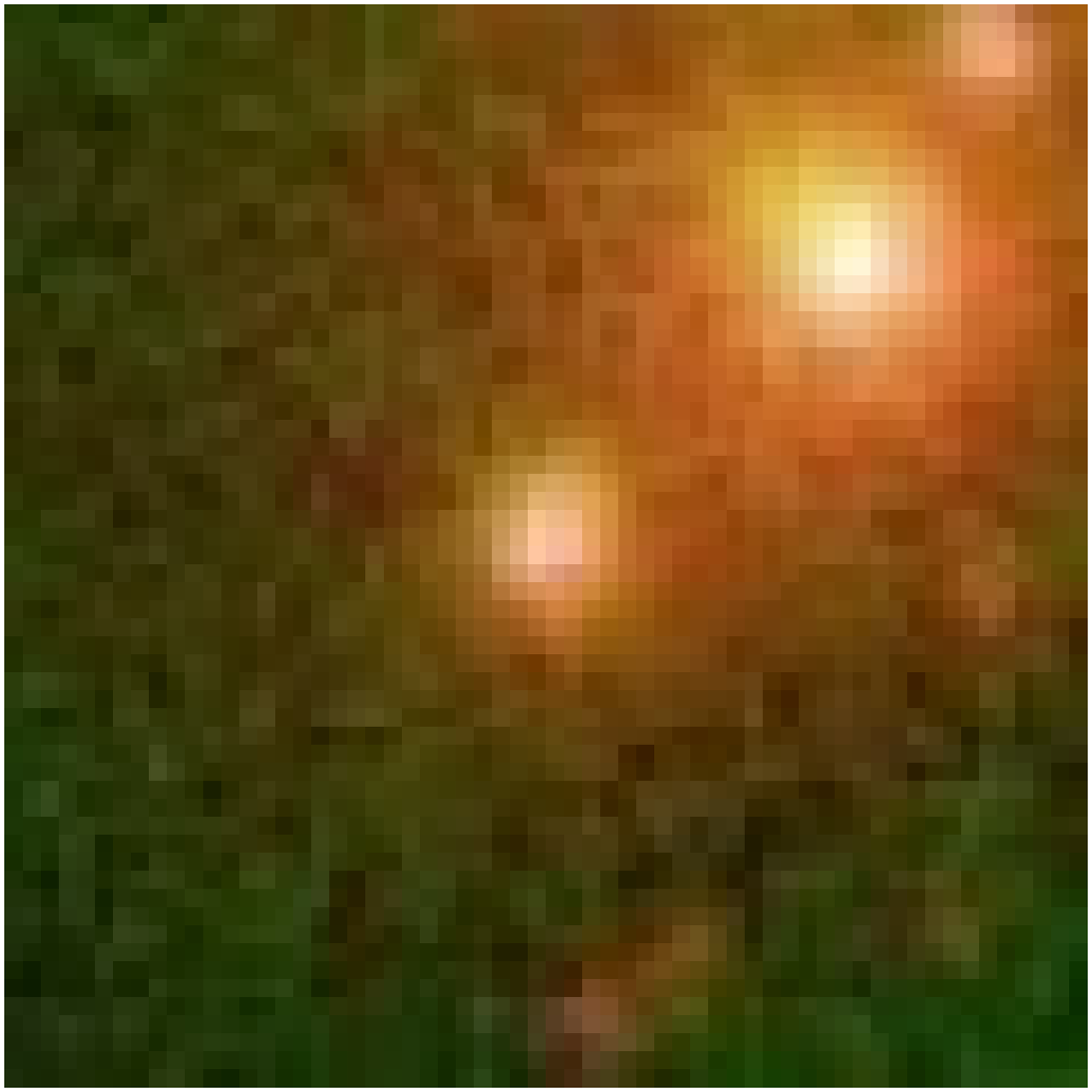}  \FGb{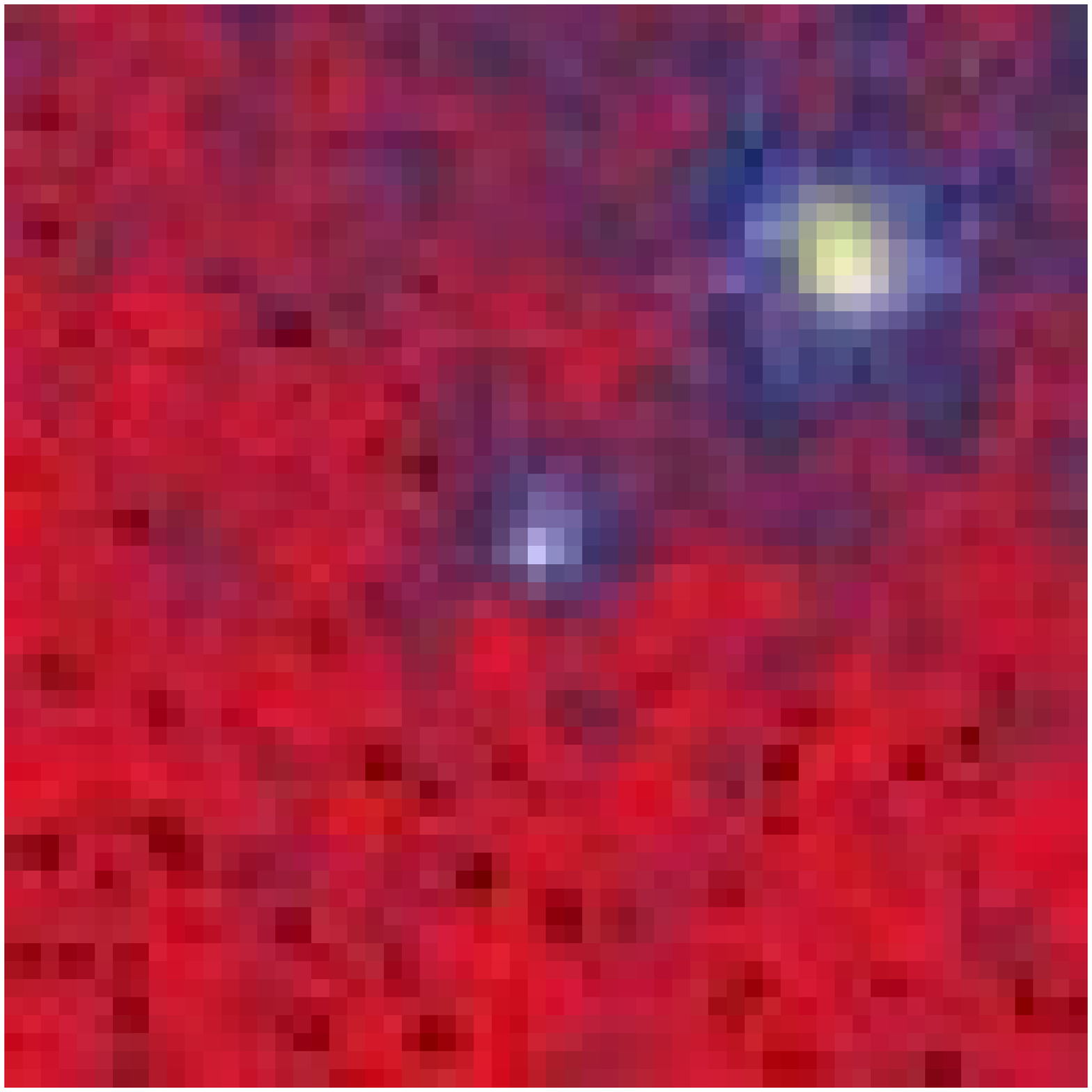}  \FGb{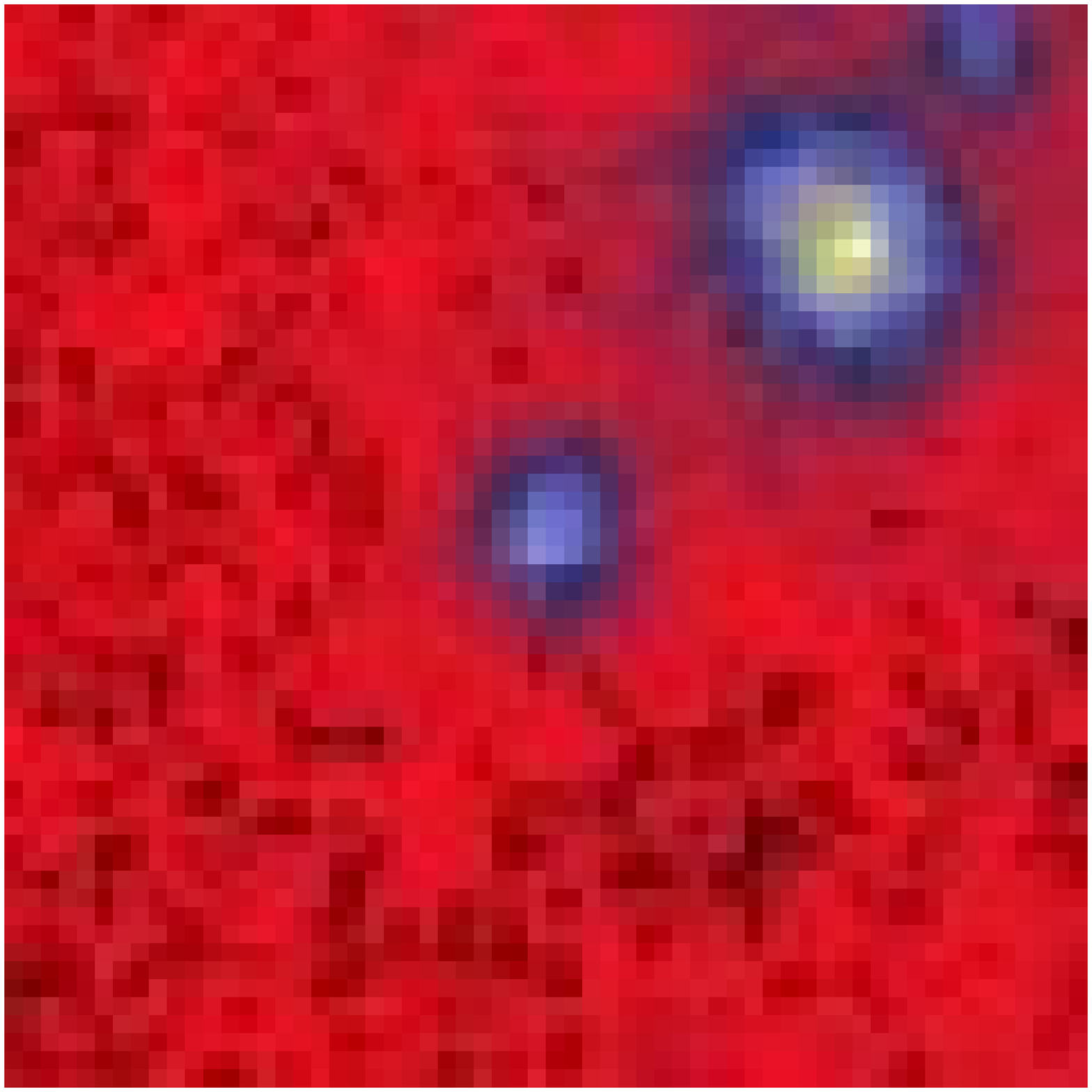}  \FGb{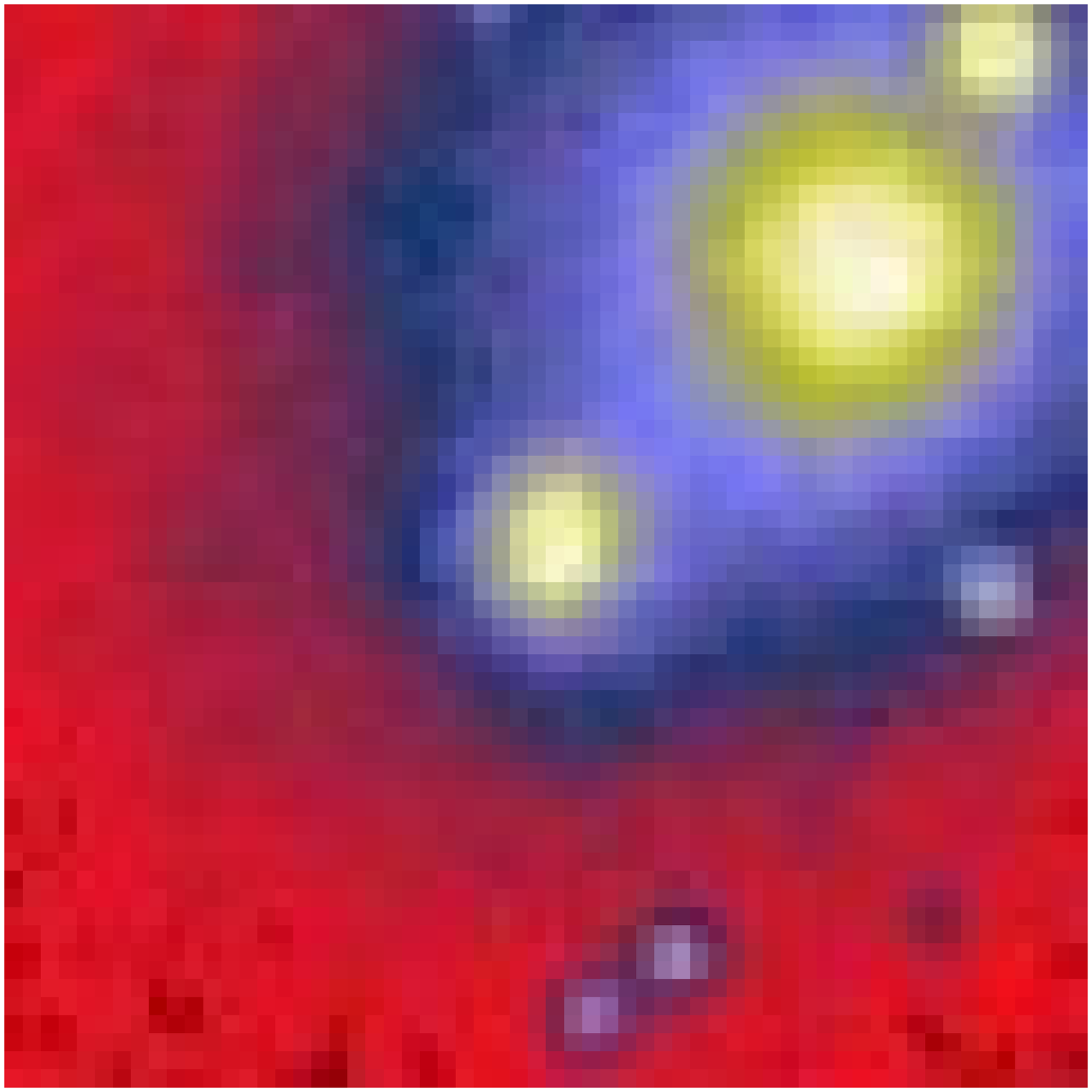}  \FGb{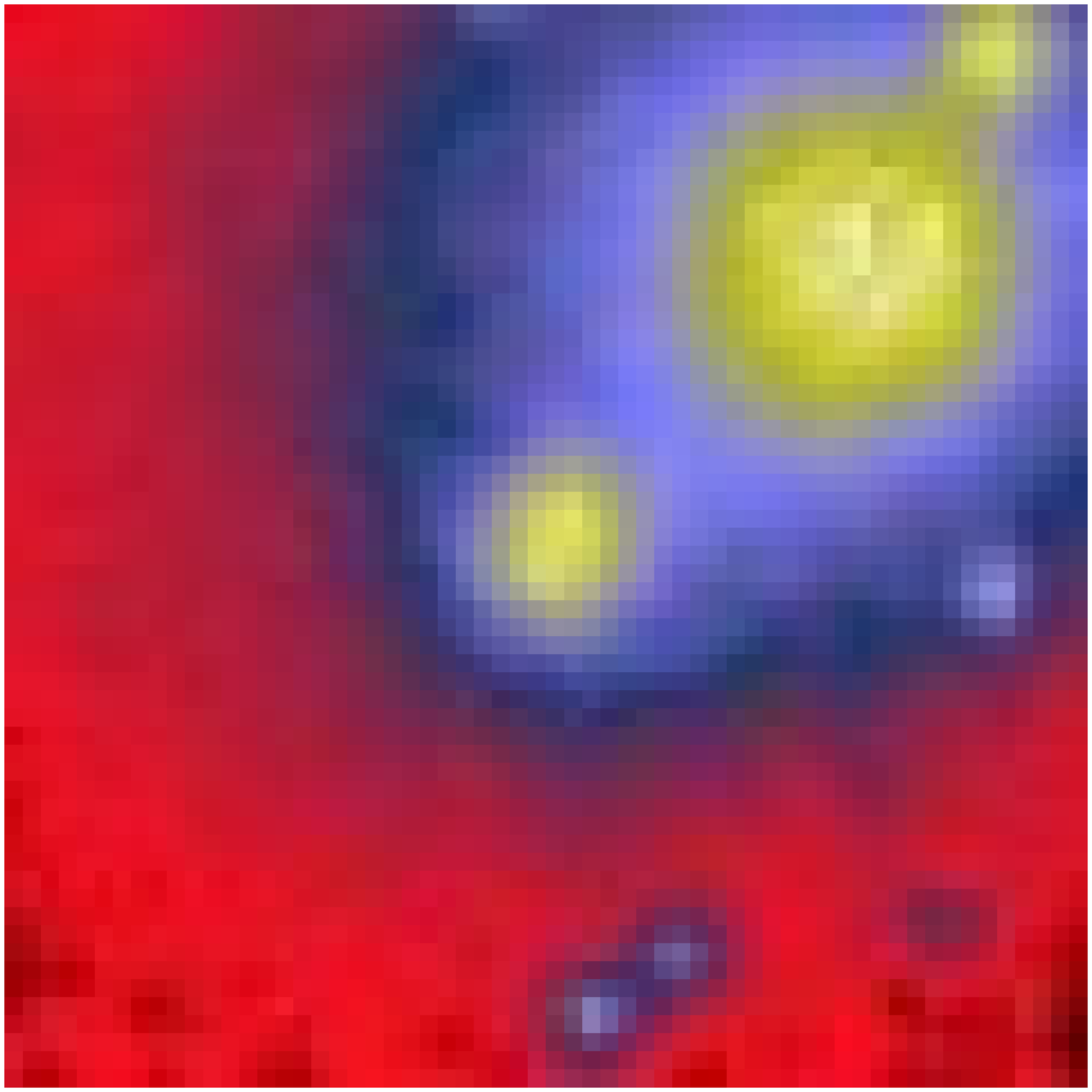}  \FGb{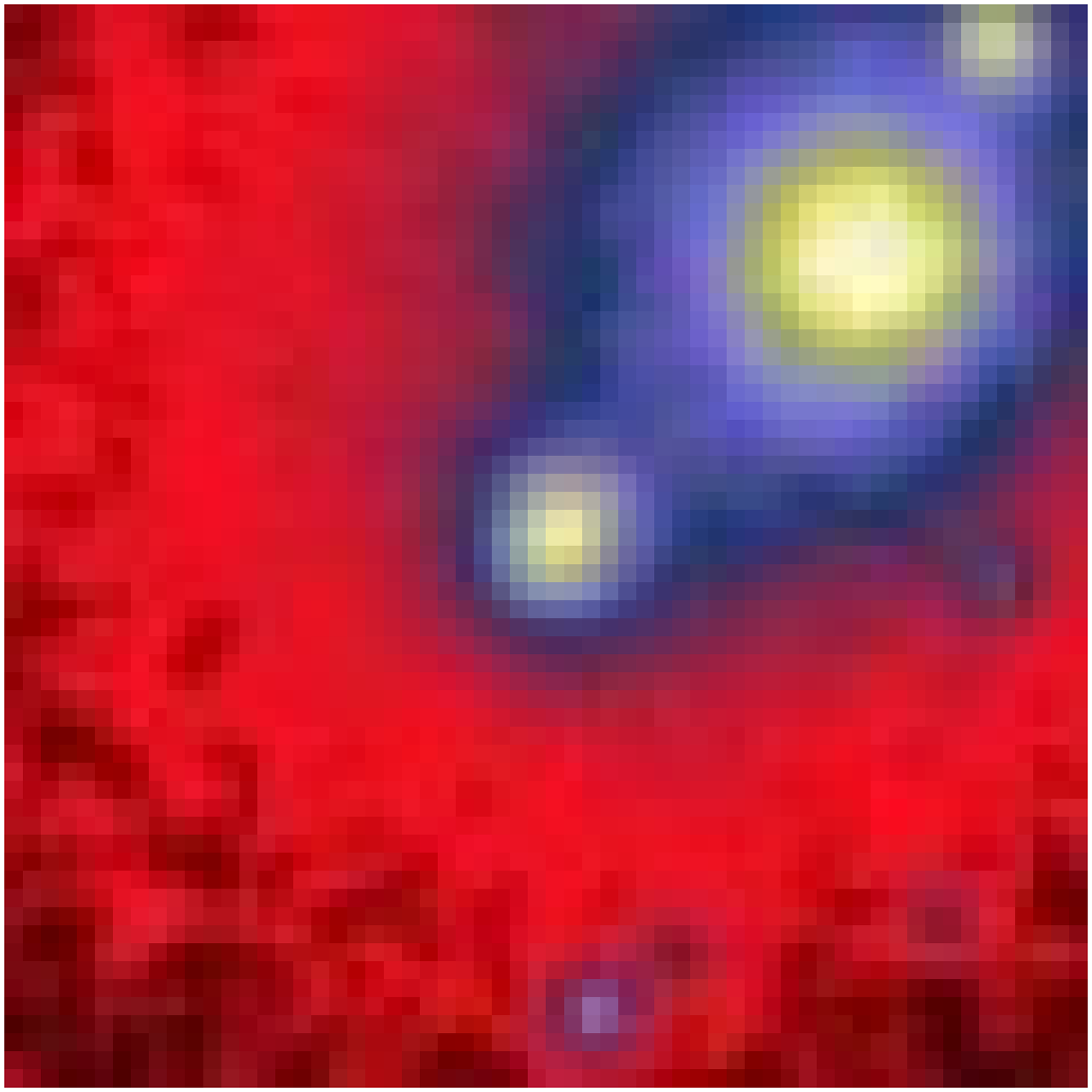}  \FGb{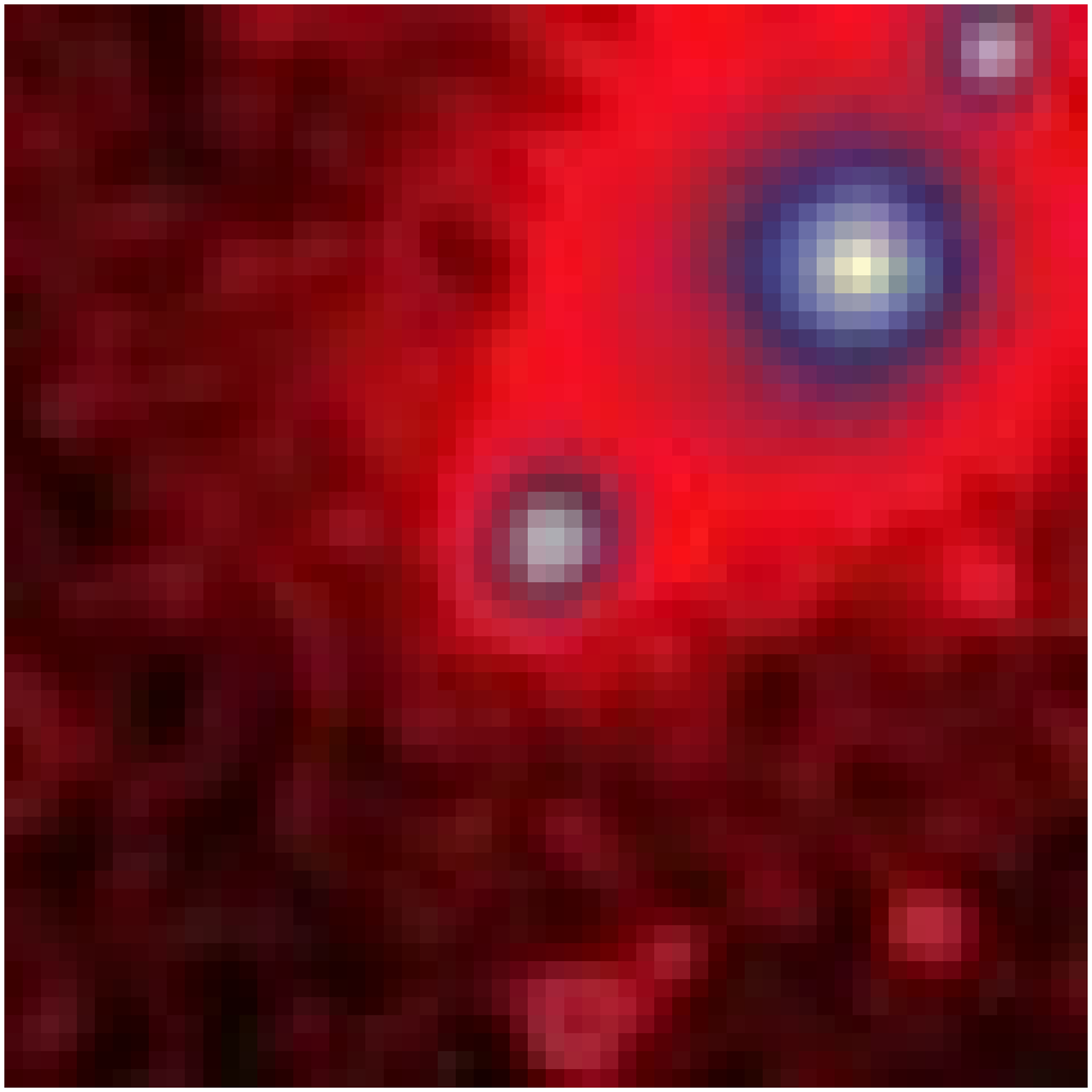}  \FGb{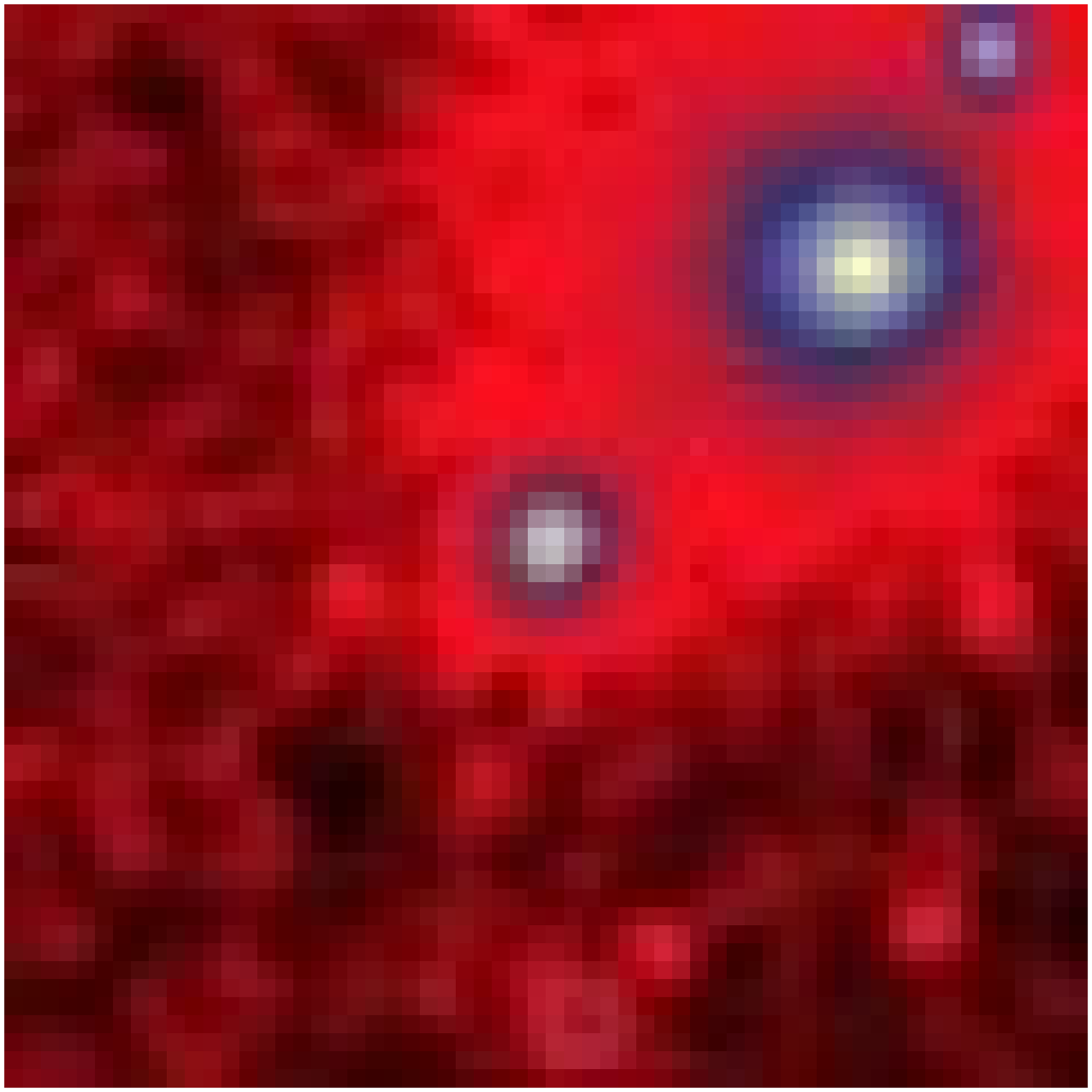}  \FGb{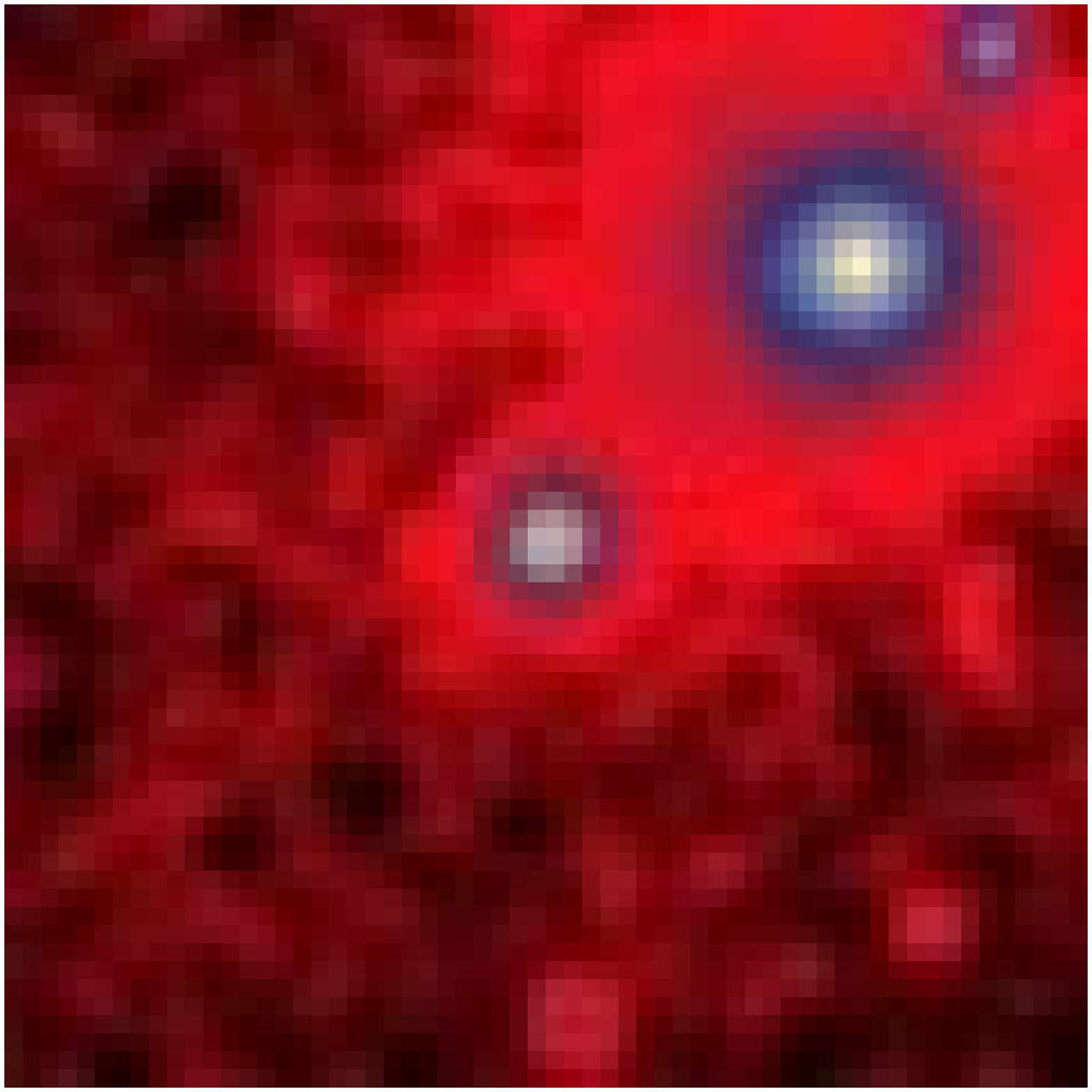}  \FGb{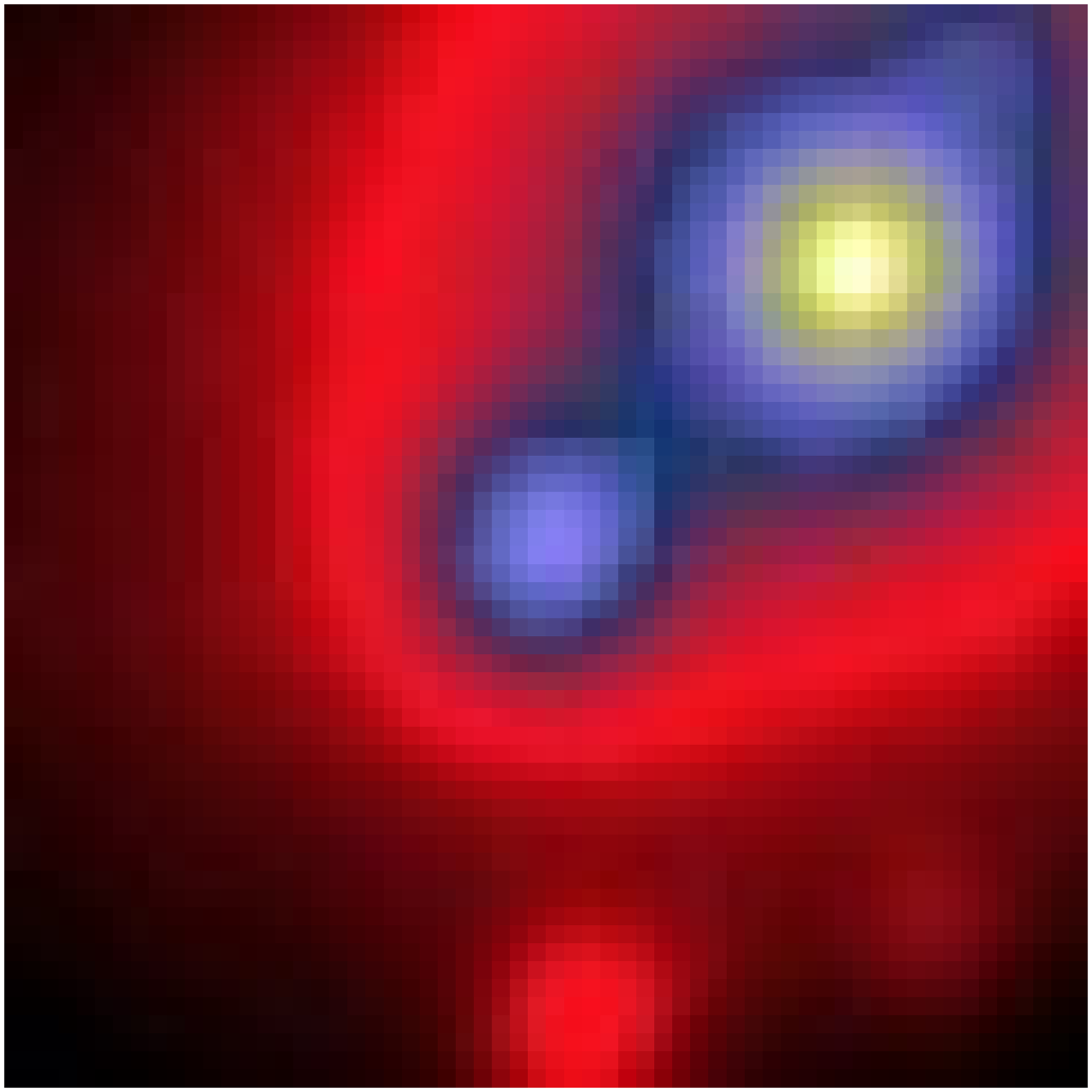}  \FGb{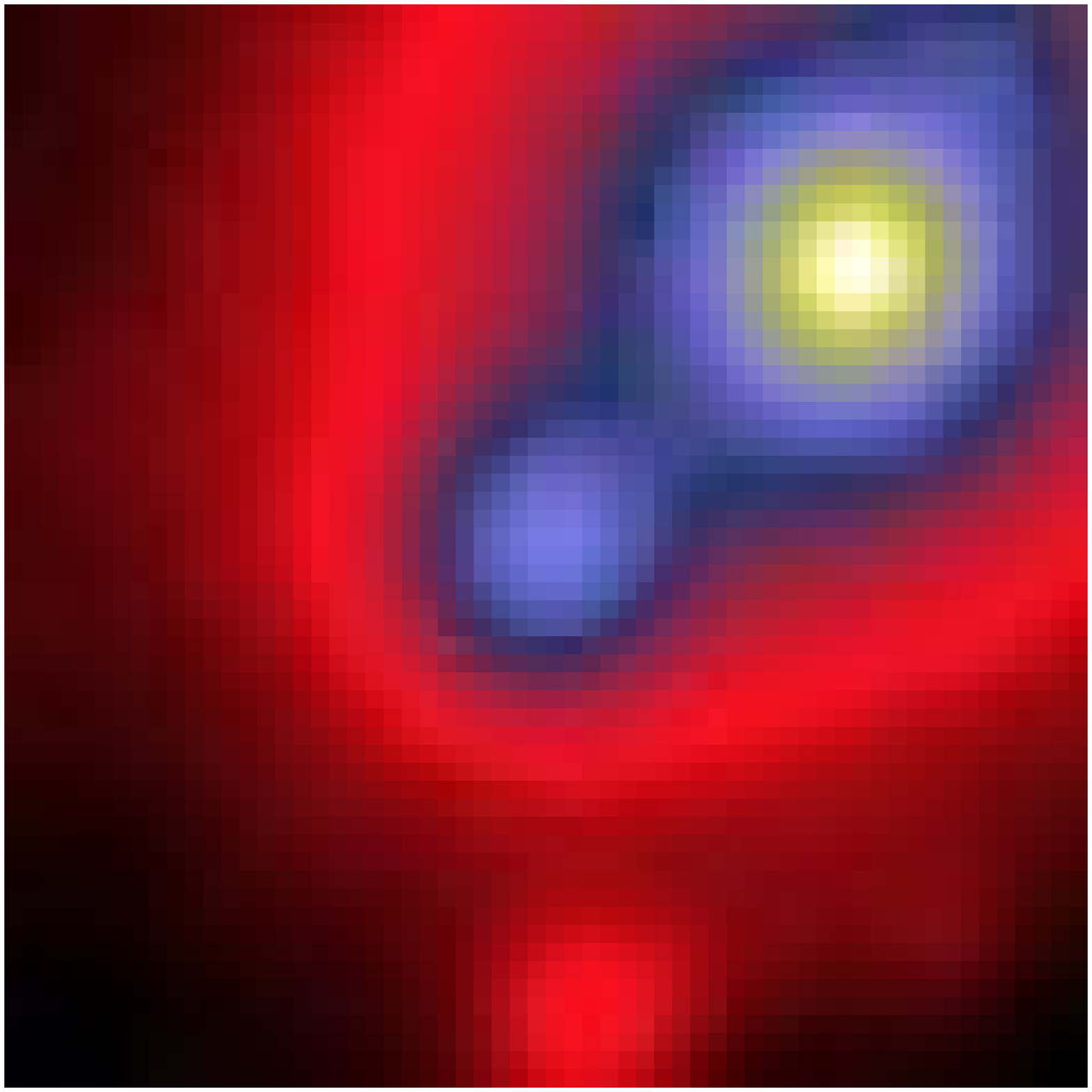}  \FGb{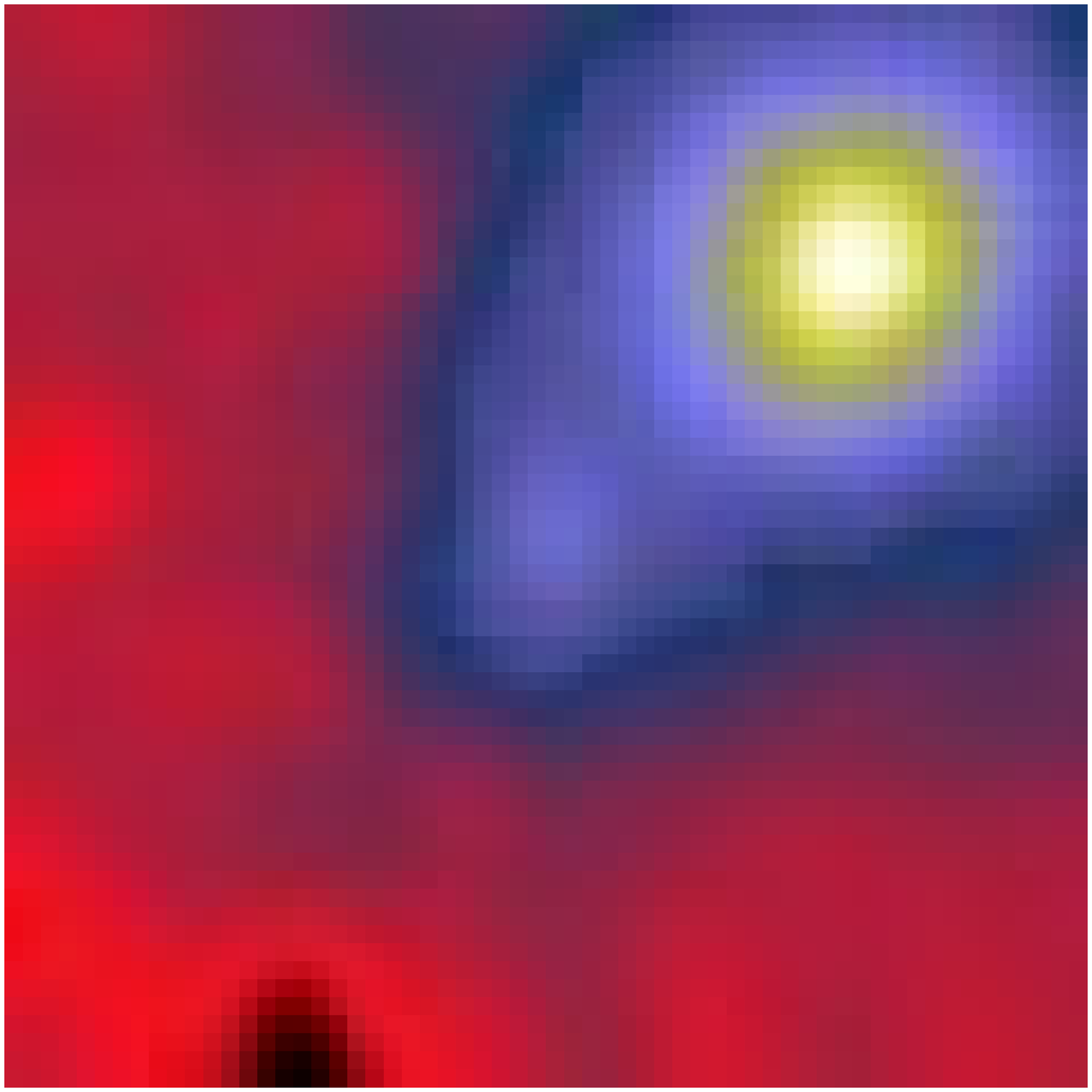}  \FGb{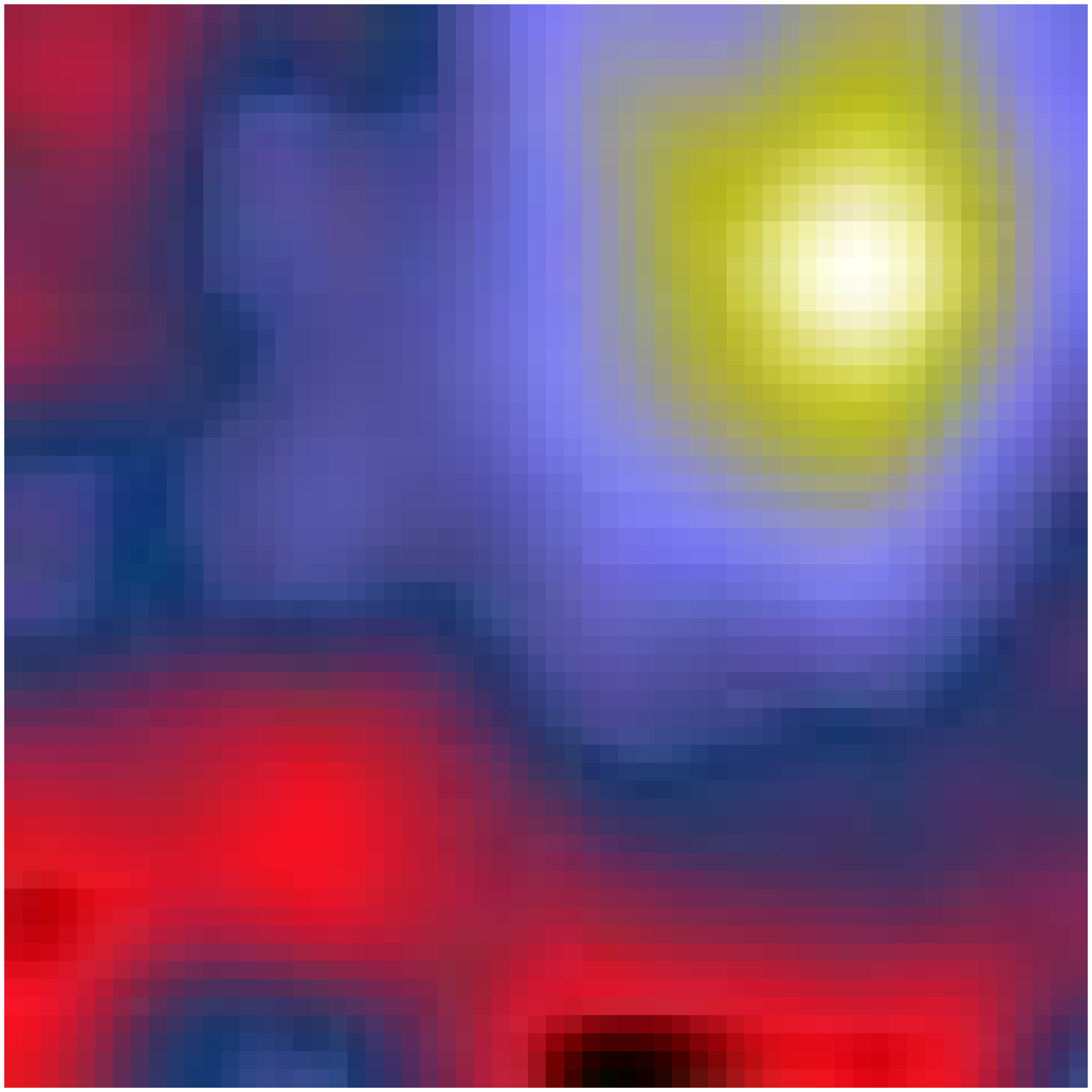} \\  {\#71   NCIA000110}  \\
{ \\ \textbf{Figure \ref{fig:allgal}.(cont)} ~~Visible Galaxies within 40\min\ of NGC~891 } \\
\end{figure}

\clearpage
\begin{figure*}[t]
   \fbox{\plotone{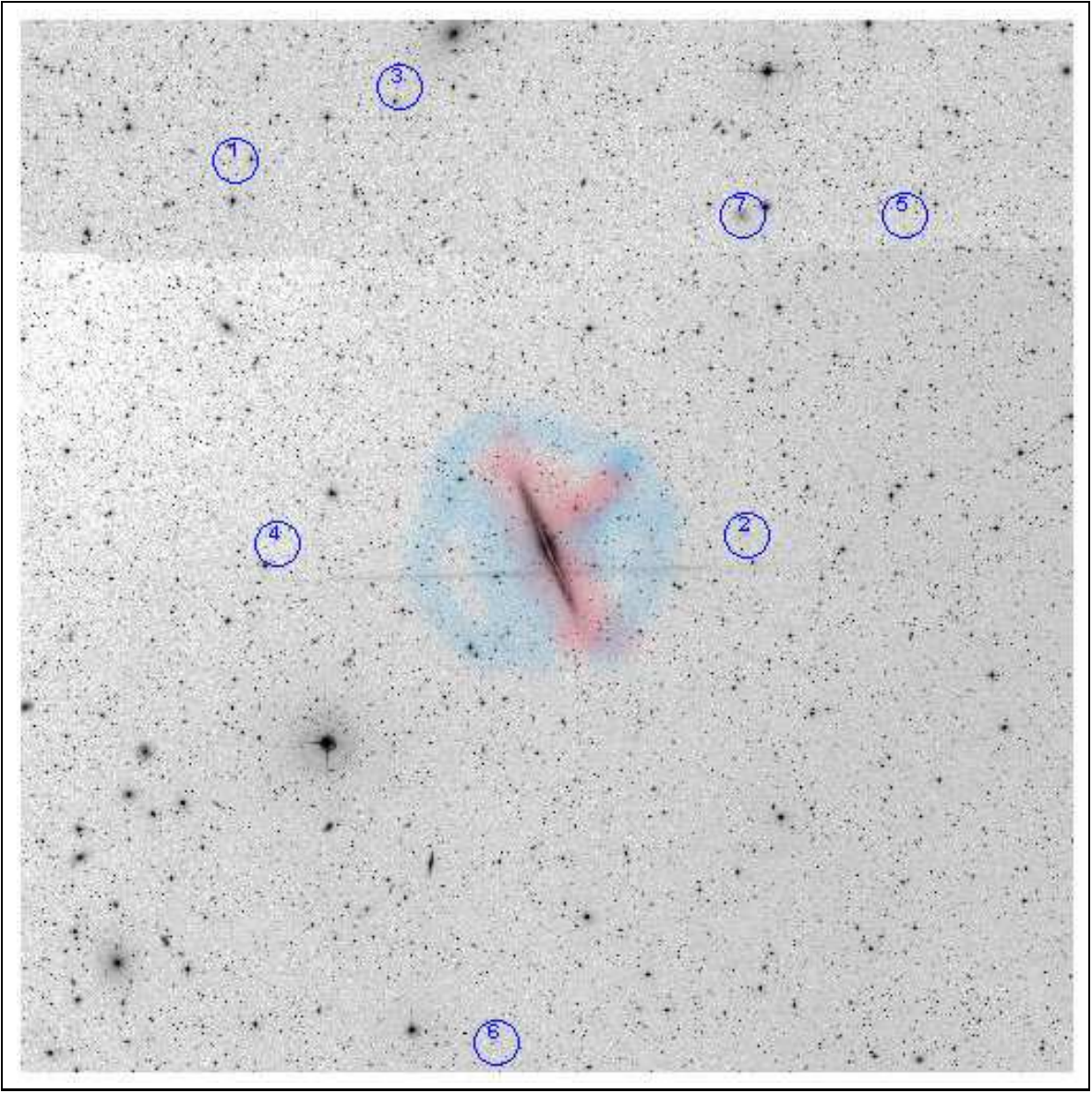}}
   \caption{DSS2 R image overlaid with the positions of the 7 dwarf galaxies identified in this paper using the numbering in Table \ref{tbl:allgal}.  The scale of this figure is 80\arcmin\ on a side which is about 228 \kpc\ at the distance of NGC 891.  The blue highlighted region is the approximate extent of the streams found in \cite{mou10::1} and the red highlighted region is the approximate extent of the disturbance to the \HI\ surrounding the disk found by the \cite{oos07}.   }
   \label{fig:NGC891_with_dSphs}
\end{figure*}

\begin{figure*}[t]
   \fbox{\plotone{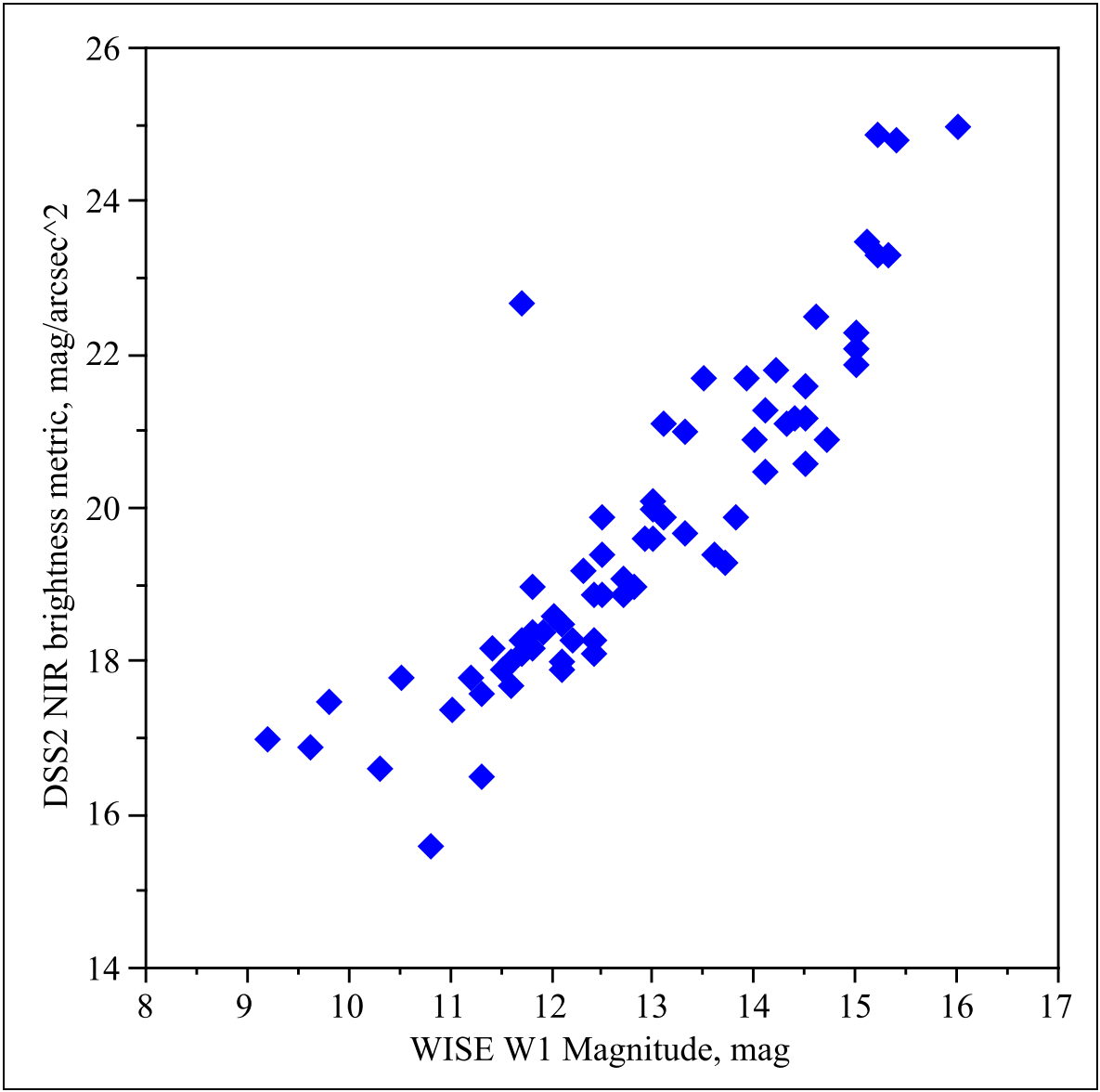}}
   \caption{The brightness parameter NIR $\mu$ is based on the GSC 2.3.2 NIR magnitude and size is a smooth function of the WISE W1 apparent magnitude.  This plot shows that despite the known systematic errors in the GSC magnitudes, the cataloged information is sufficient to identify low mass galaxies. }
   \label{fig:NIRmu}
\end{figure*}

\begin{figure*}[t]
   \fbox{\plotone{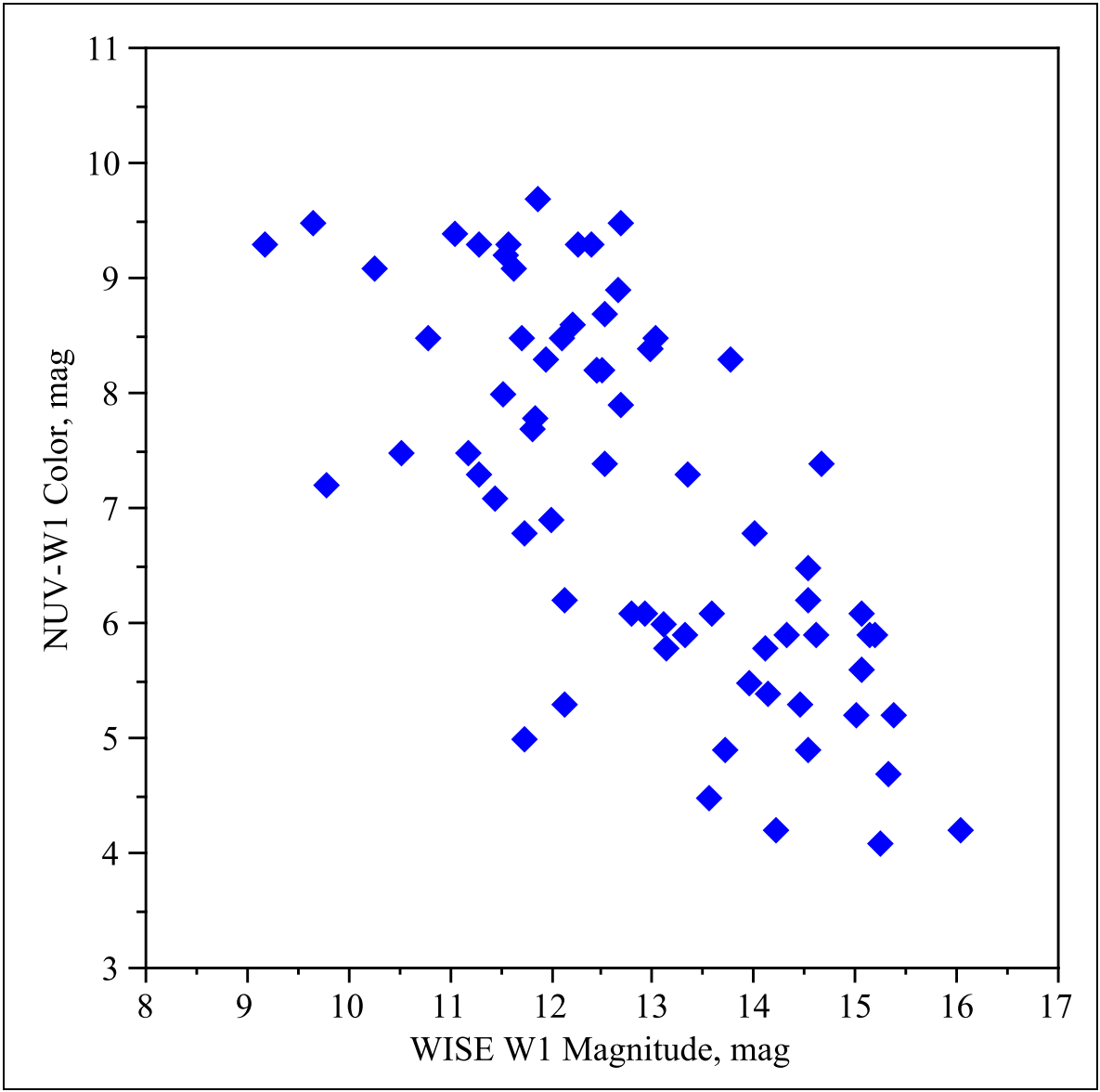}}
   \caption{Galex NUV- Wise W1 plotted vs W1 apparent magnitude shows that there is a large UV excess for low mass galaxies in comparison to higher mass galaxies.  The large scatter in this relation is due to the real effect of differences in SFR, not measurement error.   }
   \label{fig:NUV-W1}
\end{figure*}

\end{document}